# The *Gaia*-ESO Survey: Calibrating the lithium–age relation with open clusters and associations ⋆ ⋆⋆

## I. Cluster age range and initial membership selections

M.L. Gutiérrez Albarrán[1], D. Montes[1], M. Gómez Garrido[1,2], H.M. Tabernero[1,3], J.I. González Hernández[4,5], E. Marfil[1], A. Frasca[6], A. C. Lanzafame[7,6], A. Klutsch[8,6], E. Franciosini[9], S. Randich[9], R. Smiljanic[10], A. J. Korn[11], G. Gilmore[12], E. J. Alfaro[23], M. Baratella[24], A. Bayo[20,21], T. Bensby[13], R. Bonito[14], G. Carraro[15], E. Delgado Mena[3], S. Feltzing[13], A. Gonneau[12], U. Heiter[11], A. Hourihane[12], F. Jiménez Esteban[16], P. Jofre[22], T. Masseron[4], L. Monaco[17], L. Morbidelli[9], L. Prisinzano[14], V. Roccatagliata[18], S. Sousa[3], M. Van der Swaelmen[9], C. C. Worley[12] and S. Zaggia[19]

*(Affiliations can be found after the references)*

**ABSTRACT**

*Context.* Previous studies of open clusters have shown that lithium depletion is not only strongly age dependent but also shows a complex pattern with other parameters that is not yet understood. For pre- and main-sequence late-type stars, these parameters include metallicity, mixing mechanisms, convection structure, rotation, and magnetic activity.
*Aims.* We perform a thorough membership analysis for a large number of stars observed within the *Gaia*-ESO survey (GES) in the field of 20 open clusters, ranging in age from young clusters and associations, to intermediate-age and old open clusters.
*Methods.* Based on the parameters derived from the GES spectroscopic observations, we obtained lists of candidate members for each of the clusters in the sample by deriving radial velocity distributions and studying the position of the kinematic selections in the $EW$(Li)-versus-$T_{\mathrm{eff}}$ plane to obtain lithium members. We used gravity indicators to discard field contaminants and studied [Fe/H] metallicity to further confirm the membership of the candidates. We also made use of studies using recent data from the *Gaia* DR1 and DR2 releases to assess our member selections.
*Results.* We identified likely member candidates for the sample of 20 clusters observed in GES (iDR4) with UVES and GIRAFFE, and conducted a comparative study that allowed us to characterize the properties of these members as well as identify field contaminant stars, both lithium-rich giants and non-giant outliers.
*Conclusions.* This work is the first step towards the calibration of the lithium–age relation and its dependence on other GES parameters. During this project we aim to use this relation to infer the ages of GES field stars, and identify their potential membership to young associations and stellar kinematic groups of different ages.

## 1. Introduction

Lithium is a very fragile element that is easily destroyed in stellar interiors, burning at temperatures above ∼ 2.5 x $10^6$ K, corresponding to the temperature at the base of the convective zone of a solar-mass star on the zero-age main sequence (ZAMS; Siess et al. 2000). For this reason, lithium is slowly being depleted and its surface abundance decreases over time in solar-type and lower mass stars (Jeffries et al. 2014; Bouvier et al. 2016; Lyubimkov 2016). According to standard stellar models, low-mass stars show lithium depletion increasing with decreasing mass, while stars more massive than the Sun undergo little or no depletion, and very low-mass stars show no depletion at all, given that their central temperature never reaches the Li burning point (Jones et al. 1999). An additional contribution of surface lithium abundance can also be detected for some stars, such as Li-rich giants. Given the low stellar temperatures necessary to destroy lithium in stellar interiors, these Li-rich stars would require extra non-standard mixing mechanisms to account for the additional lithium detected on their surfaces (see Sect. 4).

Because it only survives in the outer layers of a star[1], lithium is a very sensitive tracer of stellar evolution and non-standard mixing mechanisms in stellar interiors (see, e.g., Sestito & Randich 2005; Castro et al. 2016), and is particularly relevant for studies of the evolution of low-mass stars and for the determination of the age of stellar clusters. Cluster ages determined in this way are less subject to systematic uncertainties than ages derived from other methods (e.g., Hobbs & Pilachowski 1986; Oliveira et al. 2003; Soderblom et al. 2014).

As most stars do not form individually, but inside clusters and associations, the study of clusters of different ages (from a few Myr to several Gyr) and chemical compositions is essential to understand star formation and evolution. In addition to this, open clusters are very useful tracers when studying the formation and evolution of the Galaxy, especially the spatial distribution of elemental abundances in the Galactic thin disc and their evolution with time (e.g., Friel 1995; Smiljanic et al. 2014; Magrini et al. 2015; Netopil et al. 2016; Casali et al. 2019).

---
⋆ Based on observations collected with ESO telescopes at the La Silla Paranal Observatory in Chile, for the *Gaia*-ESO Large Public Spectroscopic Survey (188.B-3002, 193.B-0936).
⋆⋆ All tables in Appendix C available in electronic form at the CDS via ...

[1] And in fully convective stars the surface abundance of Li is rapidly depleted when the core reaches the Li-burning temperature.

While standard models of stellar evolution including convection as the only mixing mechanism (e.g., Soderblom et al. 1990) predict that stellar Li abundances should only be a function of effective temperature and age, observations of solar and late-type stars in open clusters of different ages show that lithium depletion depends also on a series of other factors, such as metallicity, rotation, mixing mechanisms, convection structure, mass loss and magnetic activity (e.g., Deliyannis et al. 1990; Soderblom et al. 1993; Ventura et al. 1998; Jones et al. 1999; Randich et al. 2002; Charbonnel & Talon 2005; Pallavicini et al. 2005; Bouvier 2008). This indicates the presence of additional non-standard mixing processes, such as rotational mixing, diffusion, mass loss or gravitational waves, in addition to convection (e.g., Duncan 1981; Soderblom et al. 1995; Pallavicini et al. 1997). Even though a large amount of theoretical and observational work has been dedicated to the understanding of Li and its evolution (e.g., Sestito & Randich 2005), current available data have shown a complex pattern of Li depletion in pre- and main-sequence stars that is not yet understood. The most precise way to calibrate these effects is to conduct a comprehensive study of stellar groups with similar ages, such as open clusters, associations, and kinematic groups.

In the present work we use lithium, among other criteria, to constrain the cluster membership of a series of open clusters and associations using data from the *Gaia*-ESO Survey (GES). The membership analysis and calibration of the ages of open clusters and associations is of great importance to study the lithium–age relation (Soderblom 1983, 2010), which will allow us to use lithium as an effective age indicator for the field stars from GES whose age is still unknown. Thus, the ultimate aim of this project, which will continue in a separate forthcoming paper (see Sect. 7), is to use the spectroscopic observations obtained by GES for a large number of stars in a wide sample of open clusters and associations in order to apply this analysis of cluster membership to calibrate this lithium–age relation and establish its dependence on other parameters that can also be derived from the GES observations. With this ultimate aim, we focus in this work on presenting an analysis of membership for a data sample of 20 GES open clusters.

The *Gaia*-ESO Survey (GES – Gilmore et al. 2012; Randich et al. 2013)[2] is a large public spectroscopic survey that provides a homogeneous overview of the distribution of kinematics, dynamical structure and chemical compositions in the Galaxy (Bergemann et al. 2014; Smiljanic et al. 2014). The survey uses the multi-object spectrograph FLAMES on the Very Large Telescope (ESO, Chile) to obtain high-quality, uniformly calibrated spectroscopy of about $10^5$ stars, plus a sample of about 100 open clusters (OCs) and star-forming regions (SFRs) of all ages, metallicities and stellar masses [3]. GES is unique among other surveys for its depth, its UVES observations, and its comprehensive data for open clusters. Combined with precision astrometry provided by *Gaia*, delivering accurate parallaxes and proper motions, GES provides a rich dataset yielding 3D spatial distributions, 3D kinematics, chemical abundances, and improved fundamental parameters for all target objects (e.g., Beccari et al. 2018; Cantat-Gaudin et al. 2018; Randich et al. 2018; Roc-

---

[2] https://www.gaia-eso.eu/
[3] In the end GES observed 65 clusters, as well as analysing ESO archive data for about 20 additional open clusters

cataliata et al. 2018; Soubiran et al. 2018; Cánovas et al. 2019; Bossini et al. 2019).

This paper is organised as follows. In Sect. 2 we discuss the GES target selection and describe the spectral measurements we initially took using the available GES data. Section 3 describes our criteria on radial velocities (*RV*), atmospheric lithium content, surface gravity, and metallicity to identify likely cluster members. In Sect. 4 we discuss the selection of giant and non-giant (NG) outlier contaminants also obtained during the membership process. In Sect. 5 we present our lists of candidate members for all clusters studied here (individual cluster notes with more detailed information on the membership process for each of the pre-selected clusters can be also found in Appendix A). In Sect. 6 we present some further discussion of our results. Finally, we summarise our results and discuss our future work as part of this project in Sect. 7.

## 2. Data

GES observations are performed with the optical spectrograph FLAMES at the VLT (Pasquini et al. 2002), providing both high-resolution spectra with UVES (R=47, 000) of mainly single FGK stars (e.g., Smiljanic et al. 2014; Frasca et al. 2015; Lanzafame et al. 2015), and medium resolution spectra with GIRAFFE (R=5, 500–6, 500) of late-type (F to M) stars in the PMS (pre-main sequence) or MS phase. The GIRAFFE/HR15N setup is particularly useful when it comes to the study of young stars considering that it covers both H$\alpha$ and Li (6707.84 Å) spectral regions. However, fundamental parameters such as $T_{\text{eff}}$, $\log g$ and [Fe/H] are less well determined in this wavelength range than in other settings (e.g., Lanzafame et al. 2015). The WG10 and WG11 GES working groups (WGs) are focused on the spectroscopic analysis of the GIRAFFE and UVES FGK stars, respectively (e.g., Gilmore et al. 2012; Sacco et al. 2015), while WG12 is dedicated to the analysis of stars in the fields of young clusters using both UVES and GIRAFFE data.

The analysis is performed in cycles, after the reduction of new spectra observed by GES. Recommended parameters are defined by improving upon each new analysis by means of updated input and methods using a calibration strategy described in Pancino et al. (2017). At the end of each cycle, and after additional internal checks are made following the data homogenisation, an internal data release (iDR) is produced and made available to the GES consortium (e.g., Lanzafame et al. 2015). The last internal data releases (iDR5, iDR6) include all the data derived from the observations collected until the completion of the survey, in January 2018.

For all the following analysis presented in this paper we used the data provided by the fourth internal data release of GES (iDR4). iDR4 is a full internal release within the GES consortium available since February 2016, containing recommended parameters, derived products (such as individual element abundances or chromospheric activity), and radial and rotational velocities for 38 clusters, both open and globular. We also note that we decided to wait until iDR6 had been fully released instead of upgrading our analysis with the iDR5 data (see Sect. 7). We consider that there is not an appreciable difference between iDR4 and iDR5 (as opposed to iDR4 vs. iDR6) in regards to for example the number of clusters present in the release: there are 47 clusters in iDR5 versus 38 in iDR4 - only 9 additional clusters, compared to the total of 80 in iDR6, 42 more than in iDR4. We also note that, seeing as we focus in this study on FGK stars, we discarded all stars with $T_{\text{eff}} > 7500$ K from the iDR4 sample.

The output parameters resulting from the spectroscopic analysis of GES WGs are divided into raw, fundamental and derived parameters: Raw parameters such as H$\alpha$ emission and Li equivalent widths ($EW$s) are directly measured on the input spectra. Their values are used in the case of groups such as WG12 to optimise the evaluation of the fundamental parameters ($T_{\rm eff}$, log $g$, [Fe/H], projected rotational velocity ($v \sin i$), veiling ($r$), and the gravity-sensitive spectral index $\gamma$ (Damiani et al. 2014)). Lastly, derived parameters (such as elemental abundances and chromospheric activity indices), are those that require prior knowledge of the fundamental parameters. Smiljanic et al. (2014) derived parameters for UVES spectra of FGK stars, while Lanzafame et al. (2015) did the same specifically for PMS stellar spectra.

As part of the analysis of iDR4 data conducted by the GES UCM node (Lanzafame et al. 2015), during the course of this study we analysed the UVES spectra and manually measured the $EW$s of Li I $\lambda$6707.76 and adjacent Fe I $\lambda$6707.43 lines ($EW$(Fe)). The initial $EW$s of these lines were measured with the automatic tool TAME (Tool for Automatic Measurement of Equivalent Widths – Kang & Lee 2012; Tabernero et al. 2019). This tool allowed us to discard all spectra with $EW$(Li)<5 mÅ. We then performed an individual analysis of each of the remaining spectra by measuring the $EW$(Li) and $EW$(Fe) manually with the IRAF task splot (e.g., Smiljanic et al. 2014; Lanzafame et al. 2015), using the TAME values for comparison purposes. With enough resolution (among other factors such as the lack of broadening because of rotation), the Li line and the nearby blends are distinguishable, and $EW$(Li) and $EW$(Fe) can be measured individually, deblending and adopting a Gaussian fitting to the line profile. However, in the case of lower resolution spectra only $EW$(Li I + Fe I) can be measured. $EW$s were corrected as $EW$(Li) = $EW$(Li I + Fe I) − $EW$(Fe) in those cases where the Li and Fe lines could not be resolved. $EW$(Fe) was estimated using the *ewfind* driver within MOOG code (Sneden 1973) and adopting the recommended stellar parameters[4]. We also made use of the lithium measurements derived by the OACT (Osservatorio Astrofisico di Catania) node, adding to our cluster calibration analysis a number of GIRAFFE stars with no recommended $EW$(Li) values in iDR4. This has been especially important regarding the intermediate-age and old clusters considered in this study, for which only a few UVES Li values were listed in the iDR4 sample.

Our present sample from iDR4 includes 12493 UVES and GIRAFFE spectra of 20 open clusters of ages ranging from 1 Myr to 5 Gyr. Given the nature of our particular study, we discarded the 12 old globular clusters out of the 38 clusters in iDR4, as lithium cannot be used as a youth indicator in those cases. Of the remaining 26 we also discarded six young and intermediate open clusters during our analysis (NGC 2264, NGC 2451, NGC 3532, NGC 3293, NGC 6530, and Trumpler 14) as a result of the data suffering from the contamination of nebular lines, which could affect the RV distributions and therefore our membership analysis (Klutsch et al. 2020).[5] The remaining 20 clusters that constitute our sample include two SFRs (1–3 Myr) and five young clusters (10–38 Myr) with no nebulosity issues, along with three intermediate clusters (251–500 Myr), and ten old clusters (0.8–5 Gyr). In Sect. 3 we discuss the membership criteria followed to present the lists of initial candidate members of all the clusters included in the sample.

A number of membership studies have already been conducted and the authors identified potential members from the GES data for most of the 20 clusters selected in the present paper (Table 1). These studies have been of great use to evaluate the goodness of our membership analysis by comparing our final candidates with previous membership lists. Table 1 also lists the age estimates and mean metallicities from the literature for all clusters. We divided the sample clusters into groups according to age: young (1–50 Myr), intermediate (50–700 Myr), and old clusters (> 700 Myr). As shown in the table, we differentiate between publications that include membership studies, and the few that only mention them and/or study them without taking membership analysis primarily into account. In the individual notes of Appendix A, where we present our results of cluster membership, we reference these studies in more detail for each of the clusters. These previous GES studies also provided their mean properties. In particular we made use of their mean ages (Table 1), $RV$s (Table 2), and metallicities (Tables 1 and 3)

### 3. Selection criteria and membership analysis

To obtain final lists of candidate members for the 20 clusters in our sample, we conducted a homogeneous and coherent analysis of their membership according to the following criteria:

- $RV$ analysis (Sect. 3.1): We selected the $RV$ candidates by fitting the radial velocity distributions derived from GES for each cluster using a two-sigma clipping method.
- Li content (Sect. 3.2): Any $RV$ candidate is considered a potential lithium member according to its locus in the $EW$(Li)-versus-$T_{\rm eff}$ diagrams.
- Gravity indicators (Sect. 3.3): We use the Kiel (log $g$-versus-$T_{\rm eff}$) diagram to identify outliers, such as lithium-rich giant stars and other field contaminants, which we disregard hereafter during our analysis. In the case of young clusters, we mainly use the gravity indicator $\gamma$ to effectively discard giant contaminants.
- Metallicity (Sect. 3.4): An analysis of the metallicity distributions for each cluster provides confirmation of the membership of the candidate stars.
- *Gaia* studies (Sects. 3.5 and 5): Finally, we made use of additional studies conducted from *Gaia* DR1 and DR2 data (Cantat-Gaudin et al. 2018; Randich et al. 2018; Soubiran et al. 2018; Bossini et al. 2019; Cánovas et al. 2019) to further confirm our candidate selections.

Regarding the order of criteria, for the young clusters, due to their appreciable field contamination, we discarded all giant contaminants using the $\gamma$ index before performing the $RV$ analysis and obtaining lithium members. In addition, for the intermediate-age and old clusters, we relied more on the study of their metallicity distribution to ascertain final members from the initial $RV$ candidates, as a result of the increasing difficulty in using lithium as a relevant criterion in this age range.

We also note that for all clusters we identified and discarded a series of SB1 and SB2/3/4 binary stars, which can add significant contamination to our analysis. SB1s were excluded from our kinematic analysis as they can strongly affect the observed RV distributions, but we included them in the rest of our membership analysis as lithium measurements are not affected. On the other hand, SB2/3/4s were fully discarded from our data sample for all clusters. These binary stars were identified using the iDR4 data release metadata, as well as existing studies (Merle et al. 2017, 2020)

---

[4] More details about how the recommended $EW$s were determined, as well as the associated errors, can be found in, e.g. Smiljanic et al. (2014), Lanzafame et al. (2015), and Tabernero et al. (2019)

[5] The reason for excluding those clusters with high differential nebulosity from this study is the fact that the survey is fiber-fed, and thus subtraction of the nebular sky background is not a straightforward procedure (Bonito et al. 2013, 2019)

**Table 1.** Age estimates, mean metallicity, distance to the Sun, and GES membership studies from the literature for the 20 clusters in our sample.

| Cluster | Age (Myr) | [Fe/H] (dex) | Distance (kpc) | References Ages | References [Fe/H] | References Distance | GES membership studies |
|---|---|---|---|---|---|---|---|
| $\rho$ Oph | 1–3 | $-0.08 \pm 0.02$ | $0.13 \pm 0.01$ | 1, 2 | 2 | 2 | 1, 2[a], 3 |
| Cha I | 2 | $-0.07 \pm 0.04$ | $0.16 \pm 0.02$ | 4, 5, 6 | 2, 5 | 4, 5, 7, 8 | 2, 4, 5, 7, 8 |
| $\gamma$ Vel | 10–20 | $-0.06 \pm 0.02$ | 0.35–0.40 | 2, 5, 7, 9, 11, 12, 26 | 2, 11 | 5, 7, 9, 12, 26 | 2, 5, 7, 9, 10, 11, 12, 13, 26 |
| NGC 2547 | 35–45 | $-0.03 \pm 0.06$ | $0.36 \pm 0.02$ | 2, 5, 15, 16, 17, 26, 39 | 2 | 2, 5, 26 | 2, 5, 14, 15, 17, 22 |
| IC 2391 | $36 \pm 2$[b] | $-0.03 \pm 0.02$ | $0.16 \pm 0.01$ | 2, 15, 18, 19, 21, 23 | 2, 20, 21, 23 | 2, 19, 20, 21, 39 | 2, 14, 15, 22 |
| IC 2602 | $35 \pm 1$[b] | $-0.02 \pm 0.02$ | $0.15 \pm 0.01$ | 2, 15, 18, 21, 23 | 2, 21, 23 | 2, 21, 39 | 2, 14, 15, 22 |
| IC 4665 | $38 \pm 3$[b] | $0.00 \pm 0.02$ | $0.36 \pm 0.01$ | 2, 15, 18, 21, 24, 25 | 2, 39 | 2, 24, 25 | 2, 14, 15, 22 |
| NGC 2516 | $251 \pm 3$[b] | $-0.06 \pm 0.05$ | 0.41 | 16, 18, 27, 28, 29 | 27, 28 | 29, 39 | 14, 15, 17, 27, 28 |
| NGC 6705 | $300 \pm 50$ | $+0.16 \pm 0.04$ | 1.88 | 27, 28 | 27, 28, 48 | 30, 39 | 14, 27, 28, 31, 32[a], 33 |
| NGC 4815 | $570 \pm 70$ | $+0.11 \pm 0.01$ | 2.40–2.90 | 27, 28, 34, 48 | 27, 28, 34, 48 | 34, 39 | 14, 27, 28, 31, 32, 33, 34 |
| NGC 6633 | $773 \pm 10$[b] | $-0.01 \pm 0.11$ | 0.39 | 16, 18, 27, 28, 48 | 16, 27, 28, 35 48 | 30, 35, 39 | 14, 15, 27, 28 |
| Trumpler 23 | $800 \pm 100$ | $+0.21 \pm 0.04$ | 2.20 | 27, 28, 36, 48 | 27, 28, 36, 48 | 36, 39 | 27, 28, 36 |
| Berkeley 81 | $860 \pm 100$ | $+0.22 \pm 0.07$ | 3.00 | 27, 28, 32, 37, 48 | 27, 28, 37, 48 | 37, 39 | 14, 27, 28, 32 |
| NGC 6005 | $973 \pm 4$[b] | $+0.19 \pm 0.02$ | 2.70 | 18, 27, 28, 48 | 27, 28, 48 | 30, 39 | 14, 27, 28 |
| NGC 6802 | $1000 \pm 100$ | $+0.10 \pm 0.02$ | 1.80 | 27, 28, 38, 48 | 27, 28, 38, 48 | 39 | 14, 27, 28, 38 |
| Pismis 18 | $1200 \pm 400$ | $+0.22 \pm 0.04$ | 2.20 | 27, 28, 40, 48 | 27, 28, 48 | 39, 40 | 27, 28, 41 |
| Trumpler 20 | $1500 \pm 150$ | $+0.10 \pm 0.05$ | 3.00 | 27, 28, 42 | 27, 28 | 39, 42 | 14, 27, 28, 31, 32, 33, 42, 43 |
| Berkeley 44 | $1600 \pm 300$ | $+0.27 \pm 0.06$ | 1.80–3.10 | 27, 28, 44, 48 | 27, 28, 48 | 39, 44 | 14, 27, 28 |
| M67 | 4000–4500 | $-0.01 \pm 0.04$ | 0.90 | 16, 45, 46 | 16, 48, 49, 50 | 30, 39, 47 | … |
| NGC 2243 | $4000 \pm 120$ | $-0.38 \pm 0.04$ | 4.50 | 28, 48, 49, 51, 52 | 28, 48, 49, 51 | 31, 39, 51 | 14, 28 |

**Notes.** [a] GES studies that reference the clusters and/or study them without taking membership analysis primarily into account. [b] Updated cluster ages using *Gaia* data, as listed by Bossini et al. (2019).

**References.** For cluster ages, metallicities, distances, and membership studies. For the reference values shown here we chose the latest or most robust estimates for each cluster, while larger ranges taking into account more than one literature value are additionally cited in the individual notes of Appendix A: (1) Rigliaco et al. (2016); (2) Spina et al. (2017); (3) Cánovas et al. (2019); (4) Spina et al. (2014 a); (5) Sacco et al. (2015); (6) López Martí et al. (2013); (7) Frasca et al. (2015); (8) Roccatagliata et al. (2018); (9) Jeffries et al. (2014); (10) Damiani et al. (2014); (11) Spina et al. (2014 b); (12) Franciosini et al. (2018); (13) Prisinzano et al. (2016); (14) Cantat-Gaudin et al. (2018); (15) Randich et al. (2018); (16) Sestito & Randich (2005); (17) Jackson et al. (2016); (18) Bossini et al. (2019); (19) Platais et al. (2007); (20) De Silva et al. (2013); (21) Smiljanic et al. (2011); (22) Bravi et al. (2018); (23) Randich et al. (2001); (24) Martin & Montes (1997); (25) Jeffries et al. (2009 b); (26) Beccari et al. (2018); (27) Jacobson et al. (2016); (28) Magrini et al. (2017); (29) Jeffries et al. (2001); (30) Kharchenko et al. (2005); (31) Magrini et al. (2014); (32) Magrini et al. (2015); (33) Tautvaišienė et al. (2015); (34) Friel et al. (2014); (35) Jeffries et al. (2002); (36) Overbeek et al. (2017); (37) Donati et al. (2014 b); (38) Tang et al. (2017); (39) Dias et al. (2002); (40) Piatti et al. (1998); (41) Hatzidimitriou et al. (2019); (42) Donati et al. (2014 a); (43) Smiljanic et al. (2016); (44) Hayes & Friel (2014); (45) Pallavicini et al. (2005); (46) Richer et al. (1998); (47) Friel et al. (2010); (48) Magrini et al. (2018); (49) Heiter et al. (2014); (50) Overbeek et al. (2016); (51) Jacobson et al. (2011); (52) Friel & Janes (1993).

## 3.1. Kinematic selection

Despite the fact that the spectroscopic targets in the field of the clusters we are studying were photometrically selected to be likely members, the GES sample also suffers from significant field star contamination. Many of these outliers can be separated from the cluster stars on the basis of their *RV*s (e.g., Cantat-Gaudin et al. 2014; Friel et al. 2014). Thus, the analysis of the distributions of radial velocity is decisive for estimating the cluster membership on the basis of a first selection of their potential kinematic candidates.

We obtained *RV* candidates for each cluster by studying each of the velocity distributions of the *RV* measurements that were derived from both the UVES and GIRAFFE spectra. First, we discarded initial field outliers at the tails of each of the *RV* distributions using the RStudio[6] *boxplot* command[7]. We then fitted a Gaussian curve to the resulting distribution by applying an iterative two-sigma clipping procedure on the median (e.g., Donati et al. 2014 a; Friel et al. 2014). In the same way as the analysis carried out by studies such as Friel et al. (2014), we minimise the influence of the field star contaminants that could affect the estimate of the average value by relying on the median of the distribution, a more robust measure of the cluster velocity than the mean, which is more significantly affected by the presence of outliers in the distribution. After convergence is reached[8], this method results in final average velocities and dispersions for each cluster. We consider as *RV* members all stars with *RV*s lying within $2\sigma$ from the average cluster velocity provided by the fit. In the specific case of those clusters that display two peaks in their *RV* distributions ($\gamma$ Vel and NGC 2547), we note that we relied primarily on the member selections presented in the literature (Damiani et al. 2014; Jeffries et al. 2014; Spina et al. 2014 b; Frasca et al. 2015; Sacco et al. 2015; Prisinzano et al. 2016; Cantat-Gaudin et al. 2018; Randich et al. 2018), as explained in Sect. 5.

For the remaining 18 clusters analysed here, Table 2 presents the mean velocity, dispersion, and *RV* membership intervals rendered by each fit, as well as the number of resulting *RV* members. After applying all membership criteria we also obtained mean *RV*s and dispersions for the velocity distributions of the final candidates for all clusters. For each cluster studied, our mean *RV* value is in agreement with that reported in the literature, as shown in Table 1. While every study made use of their own methods and criteria, a comparison with these previous literature values can be useful for further assessing the goodness of our results. We also note that, depending on a series of factors, from the number of stars in the sample to the quality of the data, some distributions exhibit more dispersion than others, and thus, larger values of $\sigma$, even after discarding field outliers with the two-sigma-clipping procedure. However, the velocity intervals, defined by $\sigma$ to ascertain whether or not the stars of the sample are *RV* members, do include in all cases the reference *RV* values from the literature for each cluster. Along with this, we subsequently identified and discarded any contaminants among our kinematic selections by applying additional membership criteria.

---

[6] RStudio is an integrated development environment (IDE) for R, a programming language for statistical computing and graphics.
[7] This tool shows the interquartile range (IQR) in a box-and-whisker plot, indicating the spread of the values in the distribution and the most probable outliers. The demarcation line for outliers is 1.5xIQR – any value lying more than 1.5 times the length of the box from either end is considered to be a clear outlier of the distribution.

[8] The $2\sigma$ clipping algorithm proceeds as follows: We fitted the distribution with a Gaussian curve to calculate its median ($m$) and standard deviation ($\sigma$). All points smaller or larger than $m\pm2\sigma$ are then disregarded. This is repeated in an iterative manner until convergence is reached and the obtained $\sigma$ remains within a certain tolerance level of the previous one. In each iteration, the range of input data decreases and so outliers can be effectively removed from the distribution.

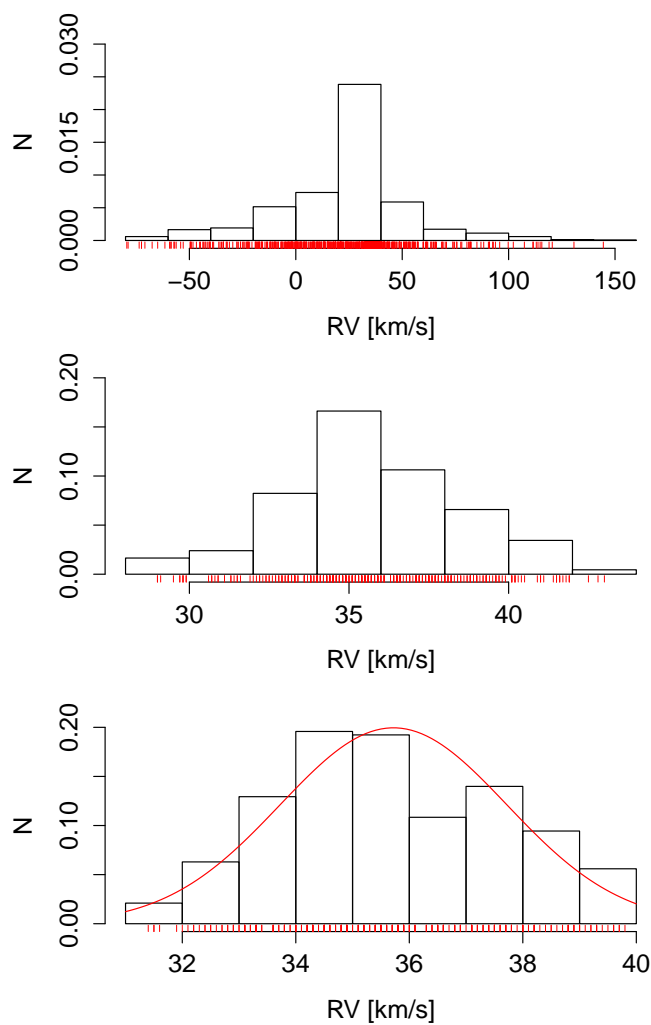

**Fig. 1.** Distribution of radial velocities and *RV* selection for stars in the field of the cluster NGC 6705. The top panel shows the initial *RV* distribution for all the GES sources. We discard a few contaminants at the tails using the RStudio *boxplot* command (middle panel), and we show the Gaussian fit of the peak of the distribution using the $2\sigma$ clipping procedure around the median (bottom panel).

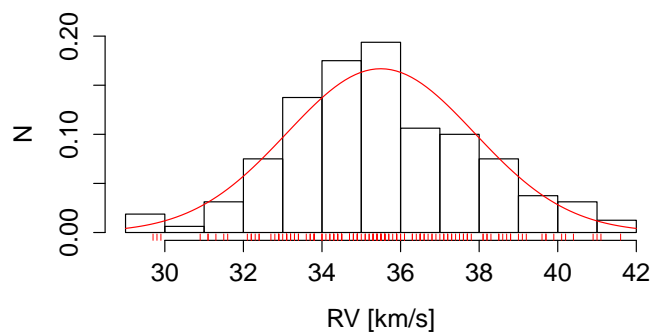

**Fig. 2.** Gaussian fit of the *RV* distribution of the members of the cluster NGC 6705 resulting from our membership analysis.

As an example, Fig. 1 displays the *RV* histogram at different stages of the analysis leading to the selection of sources belonging to the one of the clusters in our sample, intermediate-age cluster NGC 6705. The top panel shows the initial *RV* distribution for all GES targets in this cluster field. The distribution is broad, an indicator of contamination by field outliers, and presents an increasing dispersion with distance far away from the cluster centre, where the contaminants dominate. The red markings provide an additional visualisation of how the data points are spread out. The middle panel exemplifies how we discarded a series of field stars with the RStudio *boxplot* tool, as a result of which the distribution becomes less broad at smaller distances, with the cluster members starting to predominate over the field contaminants (e.g., Friel et al. 2014). Finally, the bottom panel shows the fitted *RV* distribution following the two-sigma clipping procedure around the median. The solid line indicates the Gaussian fit of the peak of the distribution, which identifies the signature of the cluster with respect to the field contaminants, and gives the central mean velocity and dispersion $\sigma$. Figure 2 displays the *RV* distribution and Gaussian fit of the final candidate selection for NGC 6705, after applying all criteria resulting from our membership analysis. We report all the mean *RV* values and their associated dispersions in Table 2, and we also present the kinematic distributions for all clusters in our sample in Appendix B.

Regarding the young clusters, we also recall that we discarded all giant contaminants using gravity indicators (Sect 3.3) before applying any other criteria, due to the large number of outliers in these fields. Thus, we only took the NG stars in the sample into account to study the velocity distribution and obtain *RV* members[9]. This initial filter minimised the presence of field outliers and appreciably reduced the dispersions of the velocity distributions, which resulted in improved values of $\sigma$ obtained from the Gaussian fits.

### 3.2. Lithium members

As mentioned in the introduction, lithium is a powerful membership indicator and of great use in determining the age of clusters. Given that Li starts to be depleted during the PMS phase and that young FGK stars seem to always show a strong lithium feature (e.g., Soderblom 2010), the presence of lithium in stellar spectra is a relevant indicator of youth in late-type

---

[9] In this study we discarded evolved stars in the field of young clusters without taking into account small effects, such as different initial accretion patterns which potentially lead to gravity spreads in this age range and thus to the possibility of biasing the sample to only objects with a particular accretion history.

**Table 2.** Fit parameters and *RV* members for the sample clusters.

| Cluster[a] | $RV^b$ (km s$^{-1}$) | 2σ clipping ⟨$RV$⟩ (km s$^{-1}$) | σ (km s$^{-1}$) | 2σ membership intervals | No. *RV* members | Final fit of member list ⟨$RV$⟩ (km s$^{-1}$) | σ (km s$^{-1}$) |
|---|---|---|---|---|---|---|---|
| ρ Oph | −7.0 ± 0.2 | −6.0 | 2.0 | [−10.0, −2.0] | 48 | −6.3 | 1.7 |
| Cha I | 14.6 ± 1.2 | 16.0 | 1.5 | [12.7, 19.7] | 100 | 15.7 | 1.2 |
| IC 2391 | 15.3 ± 0.2 | 15.7 | 3.0 | [9.7, 21.7] | 51 | 14.9 | 0.6 |
| IC 2602 | 17.4 ± 0.2 | 15.8 | 13.7 | [−11.6, 43.2] | 325 | 17.8 | 0.7 |
| IC 4665 | −14.4 ± 0.8 | −13.5 | 14.2 | [−41.9, 14.9] | 237 | −13.7 | 0.6 |
| NGC 2516 | 23.8 ± 0.2 | 23.9 | 0.6 | [21.9, 25.9] | 430 | 24.0 | 0.8 |
| NGC 6705 | 36.0 ± 0.2 | 35.7 | 2.0 | [31.7, 39.7] | 305 | 35.5 | 2.0 |
| NGC 4815 | −29.8 ± 0.3 | −27.1 | 5.6 | [−38.3, −15.9] | 119 | −27.5 | 5.7 |
| NGC 6633 | −28.6 ± 0.1 | −21.8 | 12.8 | [−47.4, 3.8] | 685 | −25.9 | 7.4 |
| Trumpler 23 | −61.4 ± 0.5 | −61.3 | 1.9 | [−65, −57.4] | 57 | −61.3 | 1.4 |
| Berkeley 81 | 50.0 ± 0.7 | 48.0 | 2.4 | [43.2, 52.8] | 74 | 48.0 | 1.0 |
| NGC 6005 | −25.6 ± 0.5 | −25.2 | 4.0 | [−33.2, −17.2] | 190 | −25.1 | 2.3 |
| NGC 6802 | 11.8 ± 0.4 | 13.2 | 1.7 | [9.8, 16.6] | 93 | 12.5 | 1.2 |
| Pismis 18 | −28.5 ± 0.6 | −30.1 | 2.8 | [−35.7, −24.5] | 47 | −27.8 | 0.6 |
| Trumpler 20 | −39.8 ± 0.2 | −39.6 | 1.7 | [−43.0, −36.2] | 515 | −40.0 | 1.3 |
| Berkeley 44 | −7.6 ± 0.4 | −8.6 | 0.7 | [−10.1, −7.3] | 34 | −8.5 | 0.7 |
| M67 | 34.1 ± 0.1 | 34.6 | 0.9 | [32.8, 36.3] | 18 | 34.6 | 0.9 |
| NGC 2243 | 59.6 ± 0.5 | 59.7 | 0.8 | [58.1, 61.3] | 400 | 59.9 | 0.7 |

**Notes.** [a] Regarding the cluster γ Vel, we directly used the selections obtained by Jeffries et al. (2014), Damiani et al. (2014), Spina et al. (2014 b), Frasca et al. (2015), and Prisinzano et al. (2016). Similarly, we use those done by Sacco et al. (2015), Cantat-Gaudin et al. (2018), and Randich et al. (2018) for the cluster NGC 2547. [b] References for the mean cluster *RV*s: we adopted those from Soubiran et al. (2018) for all clusters except for ρ Oph (Rigliaco et al. 2016), Cha I (López Martí et al. 2013) and M67 (Gaia Collaboration et al. 2018).

stars[10]. However, a few G/K giants may also have large Li content, and contamination by Li-rich field giants therefore remains possible (e.g., Smith et al. 1995). In the case of clusters older than 50 Myr, we subsequently discarded these giants with the aid of gravity indicators (see Sects. 3.3 and 4).

In this study we consider $EW$(Li) as one of the principal criteria in our analysis to select probable cluster members. We obtain the Li members of each cluster by studying the position of the *RV* candidates in $EW$(Li)-versus-$T_{\rm eff}$ figures with a series of Li envelopes as a guide. We use the upper lithium envelope of IC 2602 (35 Myr) (Montes et al. 2001), the upper (Neuhaeuser et al. 1997) and lower (Soderblom et al. 1993) envelopes of the Pleiades (78–125 Myr), and the upper envelope of the Hyades (750 Myr) (Soderblom et al. 1993). These envelopes delimit the region populated by member stars in well-known open clusters covering a large range of ages. Given that various studies have already obtained age estimates for the clusters we are studying, we can distinguish the bona-fide cluster members from the Li-rich contaminants and other field stars by studying their position in the $EW$(Li)-versus-$T_{\rm eff}$ diagram with respect to the Li envelopes.

As an example, Fig. 3 shows the $EW$(Li)-versus-$T_{\rm eff}$ diagram for the 35 Myr young cluster IC 2602. As described above, Li members were selected by studying the position of the *RV* selection with respect to the IC 2602 envelope. We disregard the stars lying above the IC 2602 envelope, which are younger than 35 Myr, and those at the bottom of the figure, which are older than the cluster members. In this specific case, we can also compare the position of the Li candidates and final selection for IC 2602 with the corresponding Li envelope for the same cluster. We later applied our gravity criteria to distinguish the giant and NG outliers with Li, as shown in the figure for completeness (see Sects. 3.3 and 4). We present the $EW$(Li)-versus-$T_{\rm eff}$ diagrams for all clusters in our sample in Appendix B.

---

[10] For the Li membership analysis in this study we have generally not taken into account small Li variations and anomalies caused by a series of effects, such as planet engulfment or the influence of parameters such as chromospheric activity or rotation. These effects can also cause gravity spreads as well as variations on the metallicity in some cases. We plan on studying these effects and the dependence of Li on these stellar parameters in greater depth in our future work.

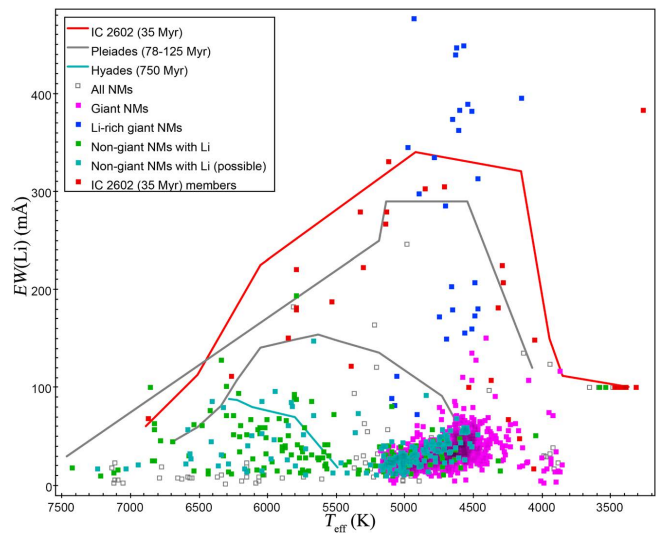

**Fig. 3.** $EW$(Li)-versus-$T_{\rm eff}$ diagram showing the final candidate selection (red squares) for IC 2602, a 35-Myr-old cluster. The upper envelope of $EW$(Li) for the cluster IC 2602 is shown in red; the upper and lower envelopes of the Pleiades cluster are shown in grey; and the turquoise line represents the upper envelope of the Hyades cluster. For completeness we show here all field contaminants of interest, colour-coded as follows: *RV* non-members (open grey squares), Li-rich giants (blue), giant outliers which are not Li-rich, and finally (fuchsia), NG non-members (green) and possible candidates (turquoise). For more details, we refer the readers to Sects. 3.3 and 4).

We note that, in the case of the SFRs (age ≤ 5 Myr), $EW$(Li) values can be underestimated in stars with a strong mass accretion rate due to the veiling factor (Frasca et al. 2015). Following the criterion applied by Sacco et al. (2017), for the two SFRs in our sample we considered as members all accretors with H$\alpha$10% > 270–300 km s$^{-1}$ (White & Basri 2003)[11], regardless of whether they are Li members or not. On the other hand, for intermediate-age and especially old clusters it becomes harder to ascertain the membership of the cluster candidates by relying on their lithium content. In these cases, the criteria on surface gravity and metallicity take on even greater importance when carrying out our membership analysis.

### 3.3. Gravity indicators: Kiel diagram and $\gamma$ index

We made use of the $T_{\rm eff}$ and log $g$ GES spectroscopic parameters to study the ($T_{\rm eff}$, log $g$) plane (also known as the Kiel diagrams) for each of the 20 pre-selected clusters. Among our candidates we identified the giant stars in the field of the clusters thanks to their locus on the Kiel diagrams. This is especially helpful to exclude evolved field contaminants for which we were not able to establish a secure membership based on lithium. We consider all stars with log $g$ < 3.5 to be likely giants, while the Li-rich giants are giant stars with A(Li) > 1.5 (see Sect. 4) for more details). For all clusters we made use of the PARSEC isochrones (Bressan et al. 2012), with Z=0.019 (except for the very low metallicity cluster NGC 2243, where we used isochrones with Z=0.006), and ages ranging from 1 Myr to 5 Gyr. As an example, in Fig. 4 we present the Kiel diagram for the cluster NGC 6705, while the Kiel diagrams for all clusters in our sample are in Appendix B. As the recommended log $g$ values are often missing for the young stars in the field of clusters younger than 50 Myr when observed with the GIRAFFE setups, we made use of the $\gamma$ index defined by Damiani et al. (2014) and provided by the Consortium. This index is another efficient gravity indicator for the GIRAFFE targets observed with HR15N, allowing a clear separation between low-gravity giants ($\gamma \geq 1$), and higher gravity stars for spectral types later than G in the MS and PMS ($\gamma \leq 1$), as shown by Spina et al. (2017). By plotting the $\gamma$ index of the Li candidate members as a function of the stellar effective temperature, we have an alternative method to identify giant background stars that we excluded before applying the other membership criteria. As in previous works (e.g., Damiani et al. 2014; Delgado Mena et al. 2016; Spina et al. 2017), we classify as Li-rich background giants all stars with effective temperatures lower than 5200 K, $A(Li) > 1.5$, and $\gamma > 1.01$. In Fig. 5 we show as an example the $\gamma$-versus-$T_{\rm eff}$ diagram for the young cluster Cha I, where the dashed lines (at $T_{\rm eff}$=5200 and $\gamma$=1.01) delimit the locus of the giant background stars. This region is clearly sep-arated from the main sequence and pre-main sequence member stars. The other $\gamma$-versus-$T_{\rm eff}$ diagrams of the young clusters in our sample are displayed in Appendix B.

---
[11] A tracer of accretion and youth indicator in young PMS stars, H$\alpha$10% refers to the width of the H$\alpha$ emission line at 10% peak intensity. As already mentioned in Sect. 2 when discussing the cluster sample, H$\alpha$ measurements are reliable only for clusters with no dominant nebular contribution to the emission.

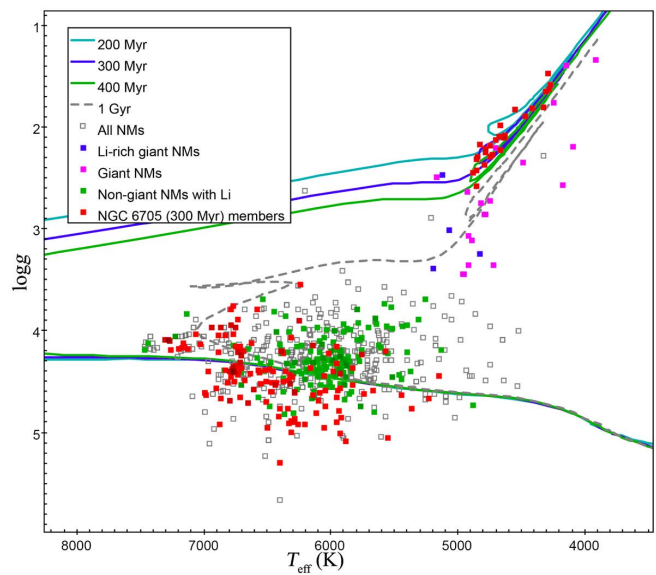

**Fig. 4.** Kiel diagram of the GES sources (open squares) in the field of the cluster NGC 6705 (300 Myr). We indicate the candidate members with red squares, while the other coloured squares denote additional field contaminants of interest: Li-rich giants (blue), (non-Li-rich) giant outliers (fuchsia), and NG non-members with Li (green). We overplot the PARSEC isochrones for a metallicity of Z=0.019 at 200 Myr (turquoise curve), 300 Myr (blue curve), 400 Myr (green curve), and 1 Gyr (gray dashed curve).

### 3.4. Metallicity

We have also taken the metallicity of the clusters into account to identify additional non-members. We use the spectroscopic index [Fe/H] derived from the GES analysis as a proxy of the metallicity. Using the [Fe/H] histograms. we search for stars with metallicities too far away from the mean cluster value. Given the homogeneity of cluster member stars, these stars are likely outliers. As an example we show the metallicity distribution for cluster NGC 6705 plotted in Fig. 6. As we mention in Sect. 3.2, the lithium criterion is less relevant when analysing older clusters, as it becomes harder to ascertain the membership of $RV$ candidates based on their position in the $EW$(Li)-versus-$T_{\rm eff}$ diagram. Therefore, for these clusters we relied more heavily on their metallicity distributions to discard outliers.

Similarly to the selection of $RV$ members (see Sect. 3.1), for the metallicity analysis we fitted the initial [Fe/H] distribution for each cluster (including all stars in the field before applying other membership criteria), to obtain probable metallicity candidates. For the young clusters we only took the NG stars into account to study the [Fe/H] distribution and obtain metallicity members, because of the high field contamination. We fitted each of the distributions by applying a $2\sigma$ clipping procedure on the median and adopting a $2\sigma$ limit about the cluster mean [Fe/H] yielded by the Gaussian fit to identify the most likely metallicity members. Thus, stars that are members on the basis of all the former criteria but show [Fe/H] values visibly far from the cluster average in the distribution were classified as non-members and disregarded afterwards. Figure 7 shows an example of the [Fe/H] distribution analysis for cluster NGC 6705, comparing the initial fit following the $2\sigma$ clipping procedure, from which we obtain metallicity membership limits, with the final distribution of the metallicities of the final candidates for the cluster.

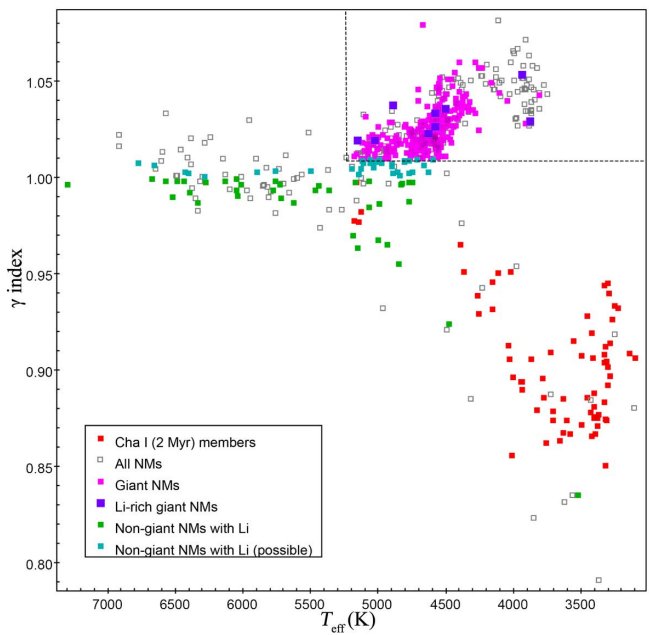

**Fig. 5.** Gravity-sensitive spectral index $\gamma$ as a function of $T_{\rm eff}$ for the sources (open squares) in the field of the SFR Cha I. The candidate members of Cha I are marked in red squares, while the other coloured squares denote the field contaminants of interest: Li-rich giants (blue), (non-Li-rich) giant outliers (fuchsia), NG non-members with Li (green), and potential NG outliers (turquoise – i.e. those in the $1.01 > \gamma > 1.0$ range, see Sect. 4). As indicated by the dashed lines, we classified any stars with $T_{\rm eff} < 5200$ K and $\gamma > 1.01$. as giants.

Unlike the UVES metallicities, the [Fe/H] values derived from the GIRAFFE spectra are widely dispersed and subject to larger uncertainties that contribute to broadening the distributions, which is a result of the lower resolution spectroscopy of the setups selected during the Survey (e.g., Spina et al. 2014 b). Because the metallicity and gravity criteria are less reliable for GIRAFFE targets, we also accepted as candidates for the intermediate-age and old clusters a number of GIRAFFE Li members with [Fe/H] values outside the $2\sigma$ limit from the cluster mean provided by our fit (this was done case by case, as described in more detail in the individual notes of Appendix A, although we also consider a maximum threshold for all instances). For this reason, we also used existing membership studies from the literature to ascertain the membership of possible GIRAFFE candidates. For more details regarding specific clusters, we refer the reader to the individual notes in Appendix A and the tables in Appendix C.

We show the results of the analysis of the metallicity distributions for all clusters in Table 3, including the mean [Fe/H], dispersion and membership intervals rendered by each fit. As in the case of $RV$ distributions, we then fitted the metallicity distributions of our final selections of candidate members for each cluster and compared the central mean [Fe/H] and its dispersion with those present in the literature (also shown in Table 1). We find that our estimates mostly agree with the literature, with the exception of a few clusters ($\rho$ Oph, Cha I, Pismis 18 and M67, see Table 3). A possible explanation for this could be related to the lower accuracy on the [Fe/H] values derived from GIRAFFE spectra (see the individual notes in Appendix A for more details on this matter). However, it is worth noting that the literature values of [Fe/H] are obtained with different methods and from different datasets. Thus, we only conduct qualitative comparisons of these measures with those obtained from the homogeneously measured iDR4 sample, which does not consist of a means of firmly assessing the membership of our final candidates.

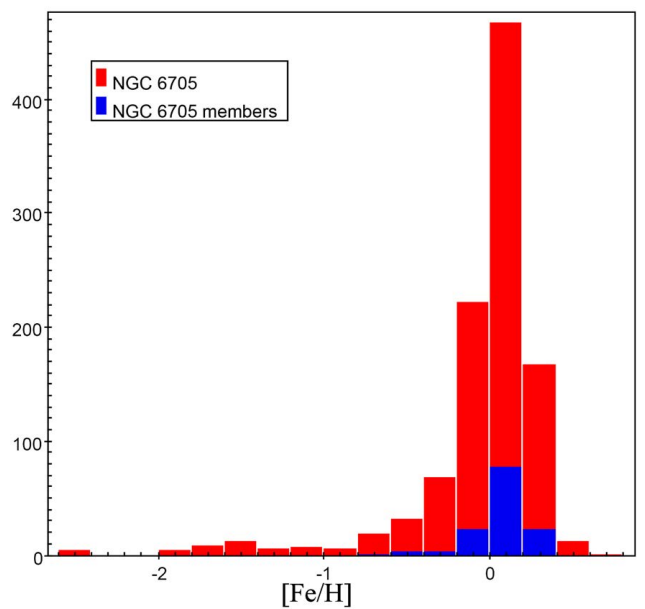

**Fig. 6.** Histogram of [Fe/H] values for all the GES stars (red) in the field of the cluster NGC 6705, as well as the candidate members (blue) resulting from our analysis. The histogram shows an increasing dispersion towards the tails, confirming that this initial distribution is dominated by field stars.

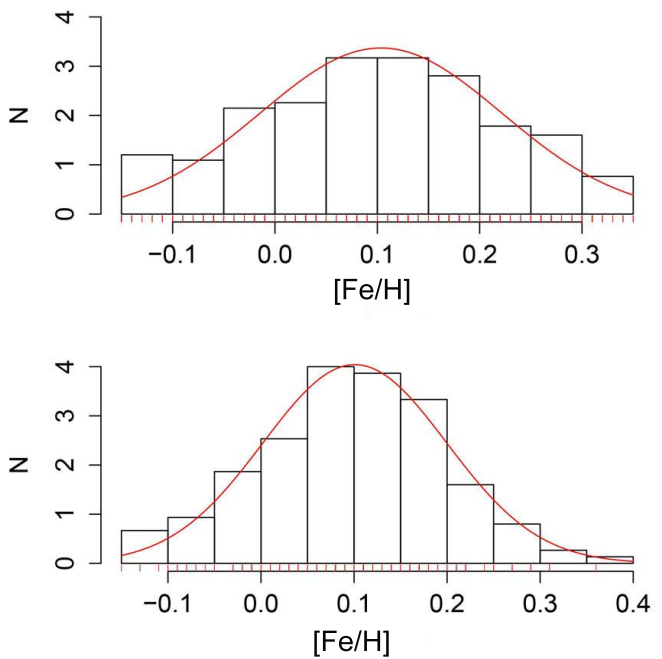

**Fig. 7.** Distributions of [Fe/H] and Gaussian fits for the intermediate-age cluster NGC 6705. We display the histogram both for sources resulting from the $2\sigma$ clipping procedure on all the GES sources in this field (top panel), and for likely cluster members after applying all of our membership criteria (bottom panel).

**Table 3.** Fit parameters and metallicity membership for the sample clusters.

| Cluster[a] | [Fe/H][b] | 2σ clipping | | 2σ membership | Final fit of member list | |
|---|---|---|---|---|---|---|
| | (dex) | ⟨[Fe/H]⟩ (dex) | σ (dex) | intervals | ⟨[Fe/H]⟩ (dex) | σ (dex) |
| ρ Oph | −0.08 ± 0.02 | −0.16 | 0.10 | [−0.36, +0.04] | −0.19[c] | 0.09 |
| Cha I | −0.07 ± 0.04 | −0.09 | 0.13 | [−0.35, +0.17] | −0.18[c] | 0.08 |
| IC 2391 | −0.03 ± 0.02 | −0.09 | 0.09 | [−0.27, +0.09] | −0.11 | 0.09 |
| IC 2602 | −0.02 ± 0.02 | −0.06 | 0.11 | [−0.28, +0.16] | −0.09 | 0.11 |
| IC 4665 | 0.00 ± 0.02 | −0.08 | 0.12 | [−0.32, +0.16] | −0.04 | 0.10 |
| NGC 2516 | −0.06 ± 0.05 | −0.01 | 0.09 | [−0.19, +0.17] | −0.02 | 0.09 |
| NGC 6705 | +0.16 ± 0.04 | +0.10 | 0.12 | [−0.14, +0.34] | +0.10 | 0.10 |
| NGC 4815 | +0.11 ± 0.01 | +0.05 | 0.17 | [−0.29, +0.39] | +0.01 | 0.13 |
| NGC 6633 | −0.01 ± 0.11 | −0.02 | 0.17 | [−0.36, +0.32] | −0.01 | 0.17 |
| Trumpler 23 | +0.21 ± 0.04 | +0.13 | 0.10 | [−0.07, +0.33] | +0.14 | 0.04 |
| Berkeley 81 | +0.22 ± 0.07 | +0.16 | 0.13 | [−0.10, +0.42] | +0.20 | 0.06 |
| NGC 6005 | +0.19 ± 0.02 | +0.08 | 0.14 | [−0.20, +0.36] | +0.13 | 0.06 |
| NGC 6802 | +0.10 ± 0.02 | −0.01 | 0.11 | [−0.23, +0.21] | +0.00 | 0.11 |
| Pismis 18 | +0.22 ± 0.04 | +0.05 | 0.13 | [−0.21, +0.31] | +0.05[c] | 0.09 |
| Trumpler 20 | +0.10 ± 0.05 | +0.07 | 0.18 | [−0.29, +0.43] | +0.10 | 0.09 |
| Berkeley 44 | +0.27 ± 0.06 | +0.13 | 0.13 | [−0.17, +0.43] | +0.17 | 0.07 |
| M67 | −0.01 ± 0.04 | −0.03 | 0.06 | [−0.09, +0.03] | −0.02 | 0.03 |
| NGC 2243 | −0.38 ± 0.04 | −0.60 | 0.12 | [−0.84, −0.36] | −0.49 | 0.14 |

**Notes.** [a] Regarding the cluster γ Vel, we directly used the selections obtained by Jeffries et al. (2014), Damiani et al. (2014), Spina et al. (2014 b), Frasca et al. (2015), and Prisinzano et al. (2016). Similarly, we use those done by Sacco et al. (2015), Cantat-Gaudin et al. (2018), and Randich et al. (2018) for the cluster NGC 2547. [b] See references on cluster [Fe/H] metallicities in Table 1 [c] In the cases of these clusters, the final mean [Fe/H] values obtained deviate appreciably from the literature values, potentially as a result of the lower resolution of the GIRAFFE spectra.

### 3.5. Gaia studies

To assess the relevance of our selections and to aid in the confirmation of our final candidates, we made use of the recent membership studies that were conducted from the *Gaia*-DR1 (Randich et al. 2018) and *Gaia*-DR2 (Cantat-Gaudin et al. 2018; Cánovas et al. 2019) data.

Randich et al. (2018) combined the parallaxes and proper motions in the *Gaia*-DR1 TGAS catalogue and the spectroscopic information from the iDR4 GES data for eight open clusters to calibrate stellar evolution and ages. Six of them are included in our data sample (namely, NGC 2547, IC 2391, IC 2602, IC 4665, NGC 2516 and NGC 6633). As well as considering an astrometric membership selection, these latter authors derived the cluster membership probabilities for the GES targets and used several spectroscopic tracers similar to ours: GES stars are required to have values of $T_{\rm eff}$, *RV*s (and v sin *i* when possible), and log *g* or γ index, as well as *EW*(Li) measurements. Candidates were selected based on criteria such as lithium content and *RV* membership probabilities, and contaminants were also discarded based on gravity, low metallicity or slow rotation. Cantat-Gaudin et al. (2018) used the astrometry data provided by the *Gaia*-DR2 release and applied an unsupervised membership assignment code (UPMASK) to list members and derived mean properties for 1229 clusters. Fourteen of the preselected 20 clusters are included in Cantat-Gaudin et al. (2018) (namely, NGC 2547, IC 2391, IC 2602, IC 4665, NGC 2516, NGC 6705, NGC 4815, NGC 6633, Berkeley 81, NGC 6005, NGC 6802, Trumpler 20, Berkeley 44, and NGC 2243). In contrast to Randich et al. (2018), this study makes use of the *Gaia* data alone and does not consider any spectroscopic criteria during the membership analysis. Cánovas et al. (2019) also used *Gaia*-DR2 astrometric measurements to study the young cluster ρ Oph, applying density-based clustering algorithms to select candidate members and identify potential new members.

In the end, we took advantage of these three studies to indirectly consider *Gaia* astrometry as a criterion to confirm the candidates from our spectroscopic analysis as members of each of the 17 clusters in common. For each of the clusters considered in these studies, see the cluster individual notes in Appendix A for more details regarding the comparison between the candidates listed in the *Gaia* studies and our final member selections. In the tables of Appendix C we also include for reference these *Gaia* membership selections alongside the columns listing the results of our membership analysis criteria.

Finally, when available, we adopted the ages revised by Bossini et al. (2019) for six of our clusters, which are mainly the young ones, and the *RV*s from Soubiran et al. (2018) for 17 of them, as reported in Table 1 and Table 2, respectively. We also observed that, judging by the updated *Gaia* ages, as well as the apparent Li envelopes of our candidates, it is possible that some of the former age estimates for the pre-selected intermediate-age and old clusters could be overestimates - NGC 6005, for example, had a former age estimate of 1.2 Gyr, while Bossini et al. (2019) gives a lower age of 973 ± 5 Myr, which is more in accordance with the Li envelope of our candidate selection.

### 4. Identification of giant and NG contaminants

The study of Li-rich giants is of great interest because even though they can be found ubiquitously in a number of different environments – from open clusters, to globular clusters, the Galactic Bulge, and even dwarf galaxies (e.g., Smiljanic et al. 2016, and references therein) –, most of them are still not well understood. The existence and properties of these stars contradict expectations from standard stellar evolution models (those which only include convection as a mixing mechanism), and would need additional mechanisms that explained a supplemental contribution of surface Li abundance (e.g., Casey et al. 2016;

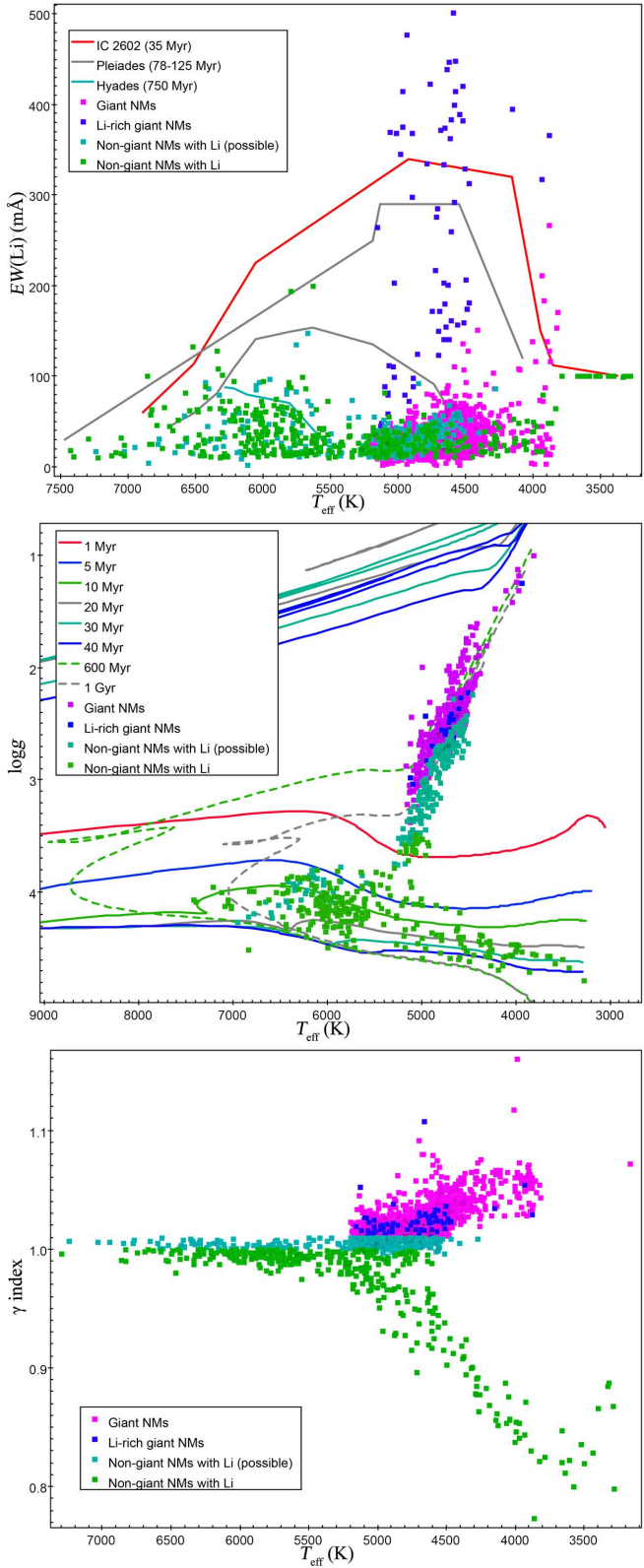

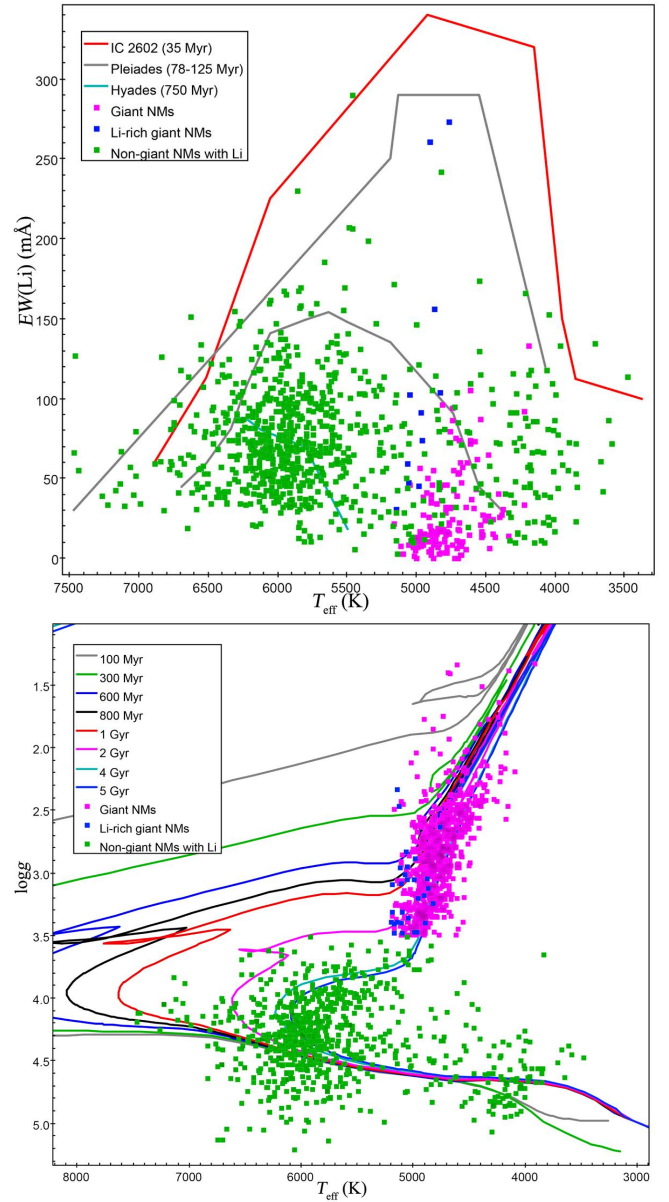

**Fig. 8.** Panels from top to bottom: $EW$(Li), $\log g$, and $\gamma$ as a function of $T_{\rm eff}$ for the Li-rich giant (blue squares), (non-Li-rich) giant (fuchsia squares), and NG (green squares) outliers in the field of the young clusters, in all cases without taking into account the rest of field stars. Turquoise squares indicate potential NGs in the $1.01 > \gamma > 1.0$ range.

**Fig. 9.** $EW$(Li)-versus-$T_{\rm eff}$ and Kiel diagrams for the Li-rich giant (blue squares), (non-Li-rich) giant (fuchsia squares), and NG (green squares) outliers in the field of the intermediate-age and old clusters in the sample.

Smiljanic et al. 2018). Li-rich giants comprise approximately 1–2% of FGK giants (e.g., Lyubimkov 2016; Smiljanic et al. 2016; Gao et al. 2019, and references therein), and are defined as those that have A(Li)≥1.5 (Iben 1967; Cameron, & Fowler 1971; Lagarde et al. 2012; Delgado Mena et al. 2016; Gao et al. 2019). According to standard evolutionary models, this limit refers to the post-dredge up Li abundance of a low-mass star (Iben 1967; Lagarde et al. 2012).

As discussed in Sect. 3.3, gravity indicators help identify giant contaminants in the field of the clusters by plotting the sample stars in the Kiel diagram and the ($\gamma$, $T_{\rm eff}$) plane. Given their interest (e.g., Smiljanic et al. 2018), we also aim to study these giant outliers, specifically potential Li-rich giants with $A(Li) > 1.5$. We consider as likely giants any source with $\log g < 3.5$ (Spina et al. 2014 a, 2017) and/or with $\gamma > 1.01$ (Damiani et al. 2014; Casey et al. 2016; Sacco et al. 2015; Spina et al. 2017). We also consider Li-rich giants to have $T_{\rm eff} < 5200$ K (Casey et al. 2016; Spina et al. 2017) and, in the case of stars in the field of young clusters, a lack of Hα emission, given that this is a youth indicator for PMS stars (Casey et al. 2016). We note that the classification of Li-rich giant stars in this study (see Table 5) is only preliminary. We find a large number of potential Li-rich giants in the field of some clusters (e.g. IC 2602) and, while these stars fulfil the adopted criteria ($T_{eff}$ < 5200 K and $A(Li)$ > 1.5), given the rare nature of these objects, further confirmation would be required to accept them as Li-rich giants.

In addition to Li-rich giants, we also included giant contaminants which are not Li rich ($A(Li)$ < 1.5), as well as a series of NG contaminants, outliers which have not yet been studied in detail in previous GES works. These stars are classified as non-members during the membership analysis for each cluster, and their Li content makes them interesting targets. We consider as NG outliers with Li all non-member stars with $\log g$ > 3.5 and a Li limit of $EW(Li)$ > 10 mÅ (in order to exclude stars with very low values of Li). In the case of young clusters, we consider those non-member stars with $\gamma$ < 1.01 to be definite NG contaminants, but decided to mark those stars in the 1.01 > $\gamma$ > 1.0 range, as well as a small number of stars with $\gamma$ < 1.0, as potential NG outliers only, as we find some $\log g$ < 3.5 measurements in the iDR4 sample for these young clusters in this $\gamma$ range, which would indicate that these stars are giants. However, given the lack of data, $\log g$ is not the most reliable gravity indicator for young clusters, and thus we consider all these stars as potential NGs following our $\gamma$ index criterion for giant contaminants; see Table 4 for a summary of the criteria considered for all giant and NG contaminants. The outliers found for each of the pre-selected clusters are indicated in Tables 5 and 6, as well as in the tables of Appendix C.

In the diagrams of $EW(Li)$, $\log g$, and $\gamma$ as a function of $T_{eff}$, for both the young clusters (Fig. 8) and the intermediate-age and old ones (Fig. 9), we display the locus of outlier contaminants, both giants and NGs, identified during our analysis in the field of all 20 clusters in our sample. In Sect. 5 we also specify, when discussing the individual notes on each cluster of the sample, the number of giant and NG contaminants found in the field of each cluster. Finally, we note that all non-members found in this study are marked as 'Li-rich G' (Li-rich giant), 'G' ((Non-Li-rich) Giant), and 'NG' (non-giant) in the Tables of Appendix C.

**Table 4.** Criteria for giant and NG outliers.

| Outlier type | Criteria |
|---|---|
| Li-rich Gs | $\log g$ < 3.5, $\gamma$ > 1.01, $A(Li)$ > 1.5, $T_{eff}$ < 5200 K |
| Gs | $\log g$ < 3.5, $\gamma$ > 1.01, $A(Li)$ < 1.5 |
| NGs | $\log g$ > 3.5, $\gamma$ < 1.01, $EW(Li)$ > 10 mÅ |
| Possible NGs | 1.01 > $\gamma$ > 1.0, $EW(Li)$ > 10 mÅ |

## 5. Results: Cluster member selections

In this section we present our results from the membership analysis of each cluster, as summarised in Table 5. For each cluster, we report *i*) the number of stars from the iDR4 sample observed with both UVES and GIRAFFE; *ii*) those with measured values of $EW(Li)$; *iii*) the number of stars selected as candidate members; and *iv*) the number of potential outlier contaminants. Readers are directed to the individual notes of Appendix A, where we offer a more detailed discussion of the membership analysis for each cluster, as well as commenting on features of interest regarding individual stars in the selection, and comparing our candidate lists with former membership studies (also listed in Table 1).

The full tables resulting from our membership analysis are provided in Appendix C. In the tables of Appendix C we show the results of our analysis and the final selections of candidate members for each of the 20 open clusters analysed, as well as the lists of giant and NG contaminants of interest also studied in this work. We indicate here for reference the columns for each table: The star ID, the GES object name from coordinates (CNAME), the RV of each star with its error, $T_{eff}$ with error, $\log g$ with error, $\gamma$ index with error (for the young clusters), [Fe/H] metallicity with error, corrected values of $EW(Li)$ with error, and flags for $EW(Li)$ errors (see footnotes in the tables for more details). These are followed by the membership analysis columns, listing all candidates following all our criteria (RV, $EW(Li)$, $\log g$ and/or $\gamma$ index, [Fe/H]), as well as additional columns, whenever possible, listing candidates according to different studies from the literature. Each table ends with the list of final candidate members and non-members, as well as a final column listing the giant and NG contaminants with Li.

We also show our final selections in the following figures: Figure 10 shows the $EW(Li)$-vs-$T_{eff}$ diagrams subdivided into young, intermediate-age, and old clusters; Fig. 11 shows the $\gamma$-vs-$T_{eff}$ diagram for the young clusters in our sample; and Fig. 12 shows the Kiel diagram for all clusters in the sample, for these same age ranges. Additionally, Appendix B shows all the individual figures for the pre-selected clusters, including both candidate member as well as giant and NG contaminants of interest.

## 6. Discussion

In Table 6 we show some further results of our membership analysis for the 20 clusters studied. As in Table 5, we show the number of stars in the field of each cluster from the initial iDR4 sample, and the number of candidate stars for all clusters, as well as the giant and NG outlier contaminants obtained as a parallel result during the membership analysis of each cluster. With these results we derived percentages of the candidate members and contaminants, which we used to assess the number of members and outliers found for different age ranges and clusters. Regarding the candidate members, these percentages are considered firstly with respect to all stars in the field of each cluster, and also with respect to all stars that present Li in the initial sample (as mentioned in Sect. 2, we also consider values from the OACT node for GIRAFFE stars in clusters older than 50 Myr). We consider, for each cluster, only percentages of any outlier class with respect to stars with $EW(Li)$ > 10 mÅ, as explained in Sect. 4. However, in the case of the candidate members, percentages with respect to all stars with only $EW(Li)$ > 10 mÅ are not considered given that there are, in the case of some of the older clusters, a series of attested members with Li values below this limit. We also note that the percentages of the giant contaminants only correspond to those stars with a $EW(Li)$ value, given that for the giant outliers our selection is based more on $A(Li)$ than on $EW(Li)$ (see Sect. 4).

These percentages allow us to to rank the clusters and age ranges in terms of the percentage of candidate members and contaminants identified, with respect to all GES stars in the field of each cluster considered in this study. First considering all clusters, M67 is the cluster which has the highest percentage of candidate members (90%), followed by NGC 2516 (65%); then NGC 4815, Trumpler 20 and 23, and NGC 6802 (49–53%),

**Table 5.** Main results for the 20 open clusters analysed, indicating, for each cluster, the number of stars from the sample detected in UVES and GIRAFFE; the number of stars with *EW*(Li) values; the number of stars selected as candidate members (including *RV* and Li members, as well as the final and possible members); and the number of Li-rich giant and NG field contaminants. Here, 'Li-rich G' refers to the Li-rich giants, 'G' to the (non-Li-rich) giants, and 'NG' to the non-giants.

| Cluster | UVES | | GIRAFFE | | Membership | | | | Outliers | | |
|---|---|---|---|---|---|---|---|---|---|---|---|
| | All stars | With Li | All stars | With Li | *RV* mem | Li mem | Final | Possible | NG[a] | G | Li-rich G |
| $\rho$ Oph | 23 | 23 | 288 | 213 | 48 | 45 | 45 | … | 48(75) | 90 | 2 |
| Cha I | 40 | 39 | 673 | 473 | 100 | 84 | 85 | … | 44(76) | 247 | 9 |
| $\gamma$ Vel | 60 | 60 | 1213 | 855 | … | … | 210 | … | 13(14) | 506 | 14 |
| NGC 2547 | 25 | 25 | 431 | 278 | … | … | 107 | … | … | 122 | 3 |
| IC 2391 | 24 | 23 | 397 | 360 | 51 | 63 | 27 | … | 18(67) | 253 | 10 |
| IC 2602 | 115 | 115 | 1716 | 1465 | 325 | 52 | 32 | … | 138(303) | 1212 | 28 |
| IC 4665 | 32 | 32 | 534 | 534 | 237 | 91 | 37 | 3 | 168(244) | 133 | 2 |
| NGC 2516 | 50 | 33 | 714 | 429 | 430 | 298 | 298 | … | 59 | 70 | 0 |
| NGC 6705 | 31 | 31 | 726 | 309 | 305 | 163 | 163 | … | 166 | 19 | 6 |
| NGC 4815 | 12 | 12 | 190 | 46 | 120 | 29 | 29 | … | 23 | 5 | 0 |
| NGC 6633 | 53 | 42 | 1594 | 354 | 685 | 132 | 101 | 17 | 186 | 590 | 13 |
| Trumpler 23 | 16 | 15 | 132 | 17 | 59 | 17 | 17 | … | 11 | 4 | 1 |
| Berkeley 81 | 14 | 14 | 257 | 77 | 74 | 27 | 28 | … | 60 | 0 | 0 |
| NGC 6005 | 19 | 19 | 538 | 89 | 190 | 41 | 38 | … | 62 | 17 | 1 |
| NGC 6802 | 13 | 13 | 178 | 29 | 93 | 24 | 22 | … | 14 | 2 | 3 |
| Pismis 18 | 10 | 10 | 114 | 30 | 47 | 15 | 15 | … | 23 | 1 | 0 |
| Trumpler 20 | 42 | 41 | 1352 | 214 | 515 | 127 | 124 | … | 122 | 15 | … |
| Berkeley 44 | 7 | 7 | 86 | 43 | 34 | 23 | 22 | … | 28 | 4 | 0 |
| M67 | 22 | 20 | 3 | 0 | 18 | 19 | 18 | 1 | 1 | 0 | 0 |
| NGC 2243 | 27 | 26 | 722 | 108 | 400 | 36 | 36 | … | 90 | 7 | 7 |

**Notes.** [a] Between parentheses we indicate the total of NG outliers, taking the potential NGs in the $1.01 > \gamma > 1.0$ range into account.

NGC 6705 (48%), and Berkeley 44 (44%). Next we have Pismis 18 (38%), NGC 6005, and NGC 2547 (35%) (the young cluster with the highest percentage), Berkeley 81 (31%), and NGC 2243 and NGC 6633 (26–27%). So far, we see that, apart from NGC 2547, the clusters that provide the highest candidate percentages are either in the old or intermediate range. As to the remaining young clusters, $\gamma$ Velorum (23%) and the SFRs are the ones with the highest percentages regarding the number of cluster candidates ($\rho$ Oph with 19% and Cha I with 16%), followed by IC IC2391, IC 2602 and IC 4665, which are the clusters with the lowest percentages of cluster candidates in our list (2–9%).

Regarding age ranges separately, the lowest percentages for young clusters are found for IC 2391, IC 2602 and IC 4665, followed by the SFRs ($\rho$ Oph and Cha I), $\gamma$ Velorum, and NGC 2547, which presents the highest percentage. For intermediate clusters, NGC 6705 has the lowest member percentages, followed closely by NGC 4815, while NGC 2516 is the cluster for which we obtain the most members. In the case of the clusters in the old range, we note that M67 is the cluster with the highest percentage (90%), followed by Trumpler 23, NGC 6802, and Trumpler 20 and Berkeley 44 (44–53%); while the clusters with the lowest candidate percentages are Berkeley 81, NGC 6005, and Pismis 18 (31–38%), and NGC 6633 and NGC 2243 (26–27%).

Similarly, considering the outlier contaminants, we firstly note that we only obtained Li-rich contaminants in the field of 13 of the 20 clusters in our sample, and all percentages are low (1–3%) (as could be expected, considering that these stars only comprise 1–2% of FGK giants, as discussed in Sect. 4). Additionally, given that the selection criteria for giant outliers relies on A(Li) and not *EW*(Li), in the case of some clusters the Li-rich giants do not present Li values (or very few), and thus the percentages are listed as 0%. We find the highest percentages of Li-rich giants in the field of clusters Trumpler 23 and IC 2391 (3%), followed by 2% for Cha I, $\gamma$ Vel, IC 2602, and NGC 6633. We find the lowest percentages in the field of $\rho$ Oph, NGC 2547, and NGC 6005 (1%). We find Li-rich giants in all young clusters, with their presence becoming scarcer in the field of intermediate and old clusters. As for those giant outliers that are not Li-rich, which we listed for 18 of the 20 clusters in the sample (however, we again note that in some cases the percentages are negligible for lack of *EW*(Li) values), the highest percentages of these are found in the field of the young clusters: from 68–73% for IC 2391 and IC 2602, to 45–50% for Cha I and $\gamma$ Vel, and 30–39% for the rest. We then find percentages in the 14–17% range for NGC 2516, NGC 6633, Trumpler 23, and NGC 6005; and the lowest percentages can be found in the field of NGC 6705, NGC 6802, and Pismis 18 (1–5%). Finally, regarding the NG contaminants (which we obtain for all 20 clusters of the sample), we find the highest percentages in the field of NGC 2243, Pismis 18, Berkeley 81, NGC 6005, and (IC 4665) (63–74%), followed by Berkeley 44, Trumpler 20 and NGC 6633 (50–56%), and the intermediate clusters NGC 6705 and NGC 4815 (43–49%). We then find percentages in the 20-40% range in the field of clusters such as NGC 6802, ($\rho$ Oph), Trumpler 23, and (IC 2602). We find the lowest percentages (1–18%) for M67, NGC 2516, and young clusters such as Cha I or $\gamma$ Vel. In the case of these NG outliers, we note that in Table 6 we list two percentages, the second one indicating that we take the stars in the $1.01 > \gamma > 1.0$ range into account as well (see Sect. 4).

If we consider individual age ranges for the contaminant stars, we see that for young clusters the highest percentages for Li-rich and non-Li-rich giants are found in the fields of IC 2391 and IC 2602, followed by $\gamma$ Vel and Cha I. On the other hand,

**Table 6.** Results for the 20 open clusters in the sample, indicating, for each cluster: all stars (both UVES and GIRAFFE) from the GES sample; the number of stars detected with measured $EW(Li)$ values and those with $EW(Li) > 10$; and the number of stars selected as candidate members, and giant and NG field contaminants. Regarding member stars, we provide their percentages with respect to all GES stars, and to those with a $EW(Li)$ measurement in the field of each cluster. Percentages in parenthesis also take into account the number of possible candidates for each cluster (see Table 5 and the individual notes in Appendix A.). In the case of the giant and NG contaminants, we consider percentages with respect to all stars with $EW(Li) > 10$ mÅ only, given the criteria specified in Sect. 4 to select outliers.

| Cluster | iDR4 stars | | | Members | | Giant outliers | | Li-rich giants | | NG outliers | |
|---|---|---|---|---|---|---|---|---|---|---|---|
| | All | With Li | $Li > 10$ | # | %(All) | %(with Li) | #(All) | %($Li > 10$)[a] | #(All) | %($Li > 10$)[a] | #(All) | %($Li > 10$) |
| $\rho$ Oph | 311 | 236 | 228 | 45 | 14 | 19 | 90 | 39 | 2 | 1 | 48(75)[b] | 21(33) |
| Cha I | 713 | 512 | 490 | 85 | 12 | 16 | 247 | 50 | 9 | 2 | 44(76) | 9(16) |
| $\gamma$ Vel | 1273 | 915 | 877 | 210 | 16 | 23 | 506 | 45 | 14 | 2 | 13(14) | 1(2) |
| NGC 2547 | 456 | 306 | 297 | 107 | 23 | 35 | 122 | 37 | 3 | 1 | ... | 0 |
| IC 2391 | 421 | 383 | 368 | 27 | 6 | 7 | 253 | 68 | 10 | 3 | 18(67) | 5(18) |
| IC 2602 | 1831 | 1580 | 1545 | 32 | 2 | 2 | 1212 | 73 | 28 | 2 | 138(303) | 9(20) |
| IC 4665 | 566 | 446 | 406 | 37(40) | 7(7) | 8(9) | 133 | 30 | 2 | 0 | 168(244) | 41(60) |
| NGC 2516 | 764 | 462 | 378 | 298 | 40 | 65 | 70 | 17 | 0 | 0 | 59 | 16 |
| NGC 6705 | 757 | 340 | 339 | 163 | 22 | 48 | 19 | 1 | 6 | 0 | 166 | 49 |
| NGC 4815 | 202 | 58 | 53 | 29 | 14 | 50 | 5 | 0 | 0 | 0 | 23 | 43 |
| NGC 6633 | 1647 | 396 | 377 | 101(118) | 6(7) | 26(30) | 590 | 15 | 13 | 2 | 186 | 49 |
| Trumpler 23 | 148 | 32 | 29 | 17 | 11 | 53 | 4 | 14 | 1 | 3 | 11 | 38 |
| Berkeley 81 | 271 | 91 | 88 | 28 | 10 | 31 | 0 | 0 | 0 | 0 | 60 | 68 |
| NGC 6005 | 557 | 108 | 98 | 38 | 7 | 35 | 17 | 17 | 1 | 1 | 62 | 63 |
| NGC 6802 | 191 | 42 | 41 | 22 | 12 | 52 | 2 | 5 | 3 | 0 | 14 | 34 |
| Pismis 18 | 124 | 40 | 35 | 15 | 12 | 38 | 1 | 3 | 0 | 0 | 23 | 66 |
| Trumpler 20 | 1394 | 255 | 223 | 124 | 9 | 49 | 15 | 0 | 0 | 0 | 122 | 55 |
| Berkeley 44 | 93 | 50 | 50 | 22 | 24 | 44 | 4 | 0 | 0 | 0 | 28 | 56 |
| M67 | 25 | 20 | 9 | 18(19) | 72(76) | 90(95) | 0 | 0 | 0 | 0 | 1 | 11 |
| NGC 2243 | 749 | 134 | 121 | 36 | 5 | 27 | 7 | 0 | 7 | 0 | 90 | 74 |

**Notes.** [a] We note that, given that the selection criteria for Li-rich and non-Li-rich giants makes use of A(Li) instead of $EW(Li)$, the percentages of the giant contaminants only correspond to those stars dislaying a lithium line, and not to the totality of the giant contaminant selection. [b] Between parentheses we indicate the total of NG outliers, taking the potential NGs in the $1.01 > \gamma > 1.0$ range into account.

the clusters with the lowest Li-rich giant percentages - such as IC 4665 and $\rho$ Oph - seem to also present the highest percentages for NG contaminants. Regarding intermediate clusters, we find the highest percentage of non-Li-rich giants for NGC 2516, and we only find Li-rich giants in the field of NGC 6705, albeit with no $EW(Li)$ values. We find similar percentages of NG outliers in the field of NGC 6705 and NGC 4815, with the lowest value for NGC 2516. We note that NGC 2516 is also the intermediate cluster with the highest member percentage, while NGC 4815 and NGC 6705 have the lowest member percentages (and the highest values for outlier contaminants). Finally, we only find Li-rich giants for five of the ten old clusters, with the clusters NGC 6633 and Trumpler 23 presenting the highest percentages. These two latter clusters, alongside NGC 6005, also present the highest percentages for non-Li-rich giant outliers. Regarding NG contaminants, we find the highest percentage in the field of cluster NGC 2243, closely followed by Berkeley 81, Pismis 18, and NGC 6005. We find the lowest values for M67, followed by NGC 6802, and Trumpler 23. We note that these clusters present both the highest member percentages and the lowest NG outlier values. This potential correlation of higher member percentages paired with lower contaminant percentages (especially the percentages for NG outliers) is something that we observe both in the intermediate and old age ranges, and to a lesser extent also for the young clusters.

## 7. Summary and future work

We used the GES-derived data provided in iDR4 to conduct an analysis of the membership and Li abundance of a series of 20 young, intermediate-age, and old clusters and associations ranging from 1 Myr to 5 Gyr.

We summarise our analysis and the main results of this work as follows:

– During the membership selection process we used the iDR4 survey-derived radial velocities, stellar atmospheric parameters, lithium $EW$s, and metallicities to obtain lists of candidate members for the selected clusters.
– We started using radial velocities to derive a series of initial $RV$ candidates for each cluster. The position of these $RV$ candidates in the $EW(Li)$-versus-$T_{eff}$ diagram then provided a series of lithium members. These members were analysed using gravity criteria (Kiel diagrams and the gravity index $\gamma$) that allow us to discard giant field contaminants. Finally, we used the [Fe/H] metallicity values to further confirm the membership of our candidates for each cluster. We also make use of recent studies based on *Gaia* DR1 and DR2 data to assess our candidate selections.
– As a result of this membership study, we identified potential members for each cluster, as shown in Table 5 of this paper. The results of our membership analysis are discussed in more detail in the individual notes of Appendix A. Additionally, Appendix C presents the associated tables in which we also specify which membership criteria are fulfilled by each of the stars studied in all selected clusters.
– Furthermore, as an additional result of our membership analysis we obtained a series of Li-rich giant and NG outliers with Li, which we aim to study in more detail. As in the case of the candidate members, the number of outliers for each cluster is also given in Table 5, and the list for each cluster is presented in Appendix C.
– We find that our lists of cluster candidates agree with those of previous GES studies, when available. However, we also consider a series of candidate stars to be members despite not appearing in former lists, given that they fulfil all our membership criteria. We also note that we studied the membership of old cluster M67, which has not been previously studied using GES data.

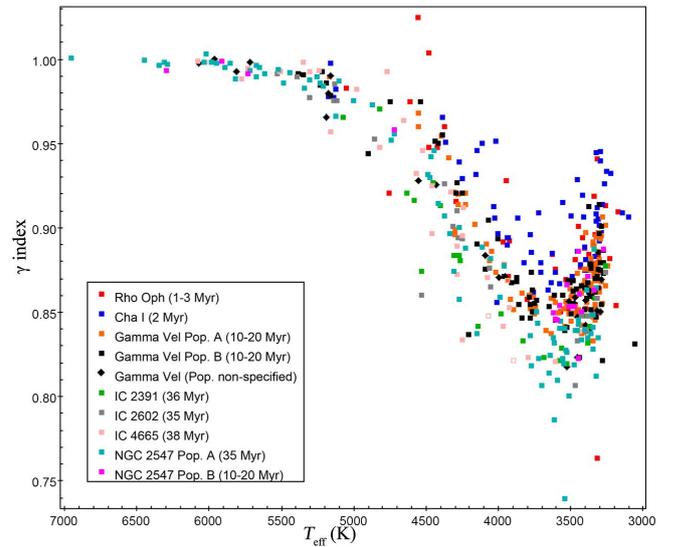

**Fig. 11.** Gravity index $\gamma$ as a function of $T_{\rm eff}$ for the young members in the young clusters of our sample.

Regarding our future work, we aim to calibrate the lithium–age relation and obtain its dependence on other stellar parameters derived from the GES spectroscopic observations, such as the level of chromospheric activity (H$\alpha$), accretion indicators, rotation ($v \sin i$), metallicity ([Fe/H]), and other parameters available from the literature (e.g. photometric rotational period).

- The age of each cluster will be revised using all this information, the lithium depletion boundary when possible, and other methods. For each cluster observed within GES, we also aim to include as part of our membership analysis all the $EW$(Li) measurements provided by other authors, as well as other well-known open clusters studied in the literature but not observed by GES. This will help to extend the age coverage.
- We also aim to update the analysis and cluster member selections in the present study with the upgraded measurements and parameters of the last internal GES data release (iDR6), as well as astrometry from *Gaia*. This will also allow us to add new clusters to our calibration and therefore contribute to better constraining the lithium–age relation.
- We plan to use these upgraded cluster candidate selections in a separate forthcoming paper in continuation of the work presented here in order to derive the lithium–age relation and its dependence on other stellar parameters. This will allow us to infer the ages of GES field stars whose age is still unknown, and to study the potential membership of these field stars to young associations and stellar kinematic groups of different ages.
- Finally, another aspect of our future work involves studying in more detail some of the unknown non-member contaminants in the field of the clusters presented in this paper (Li-rich stars, and giant and NG outliers with Li), possibly including new young field stars or Li-rich giants.

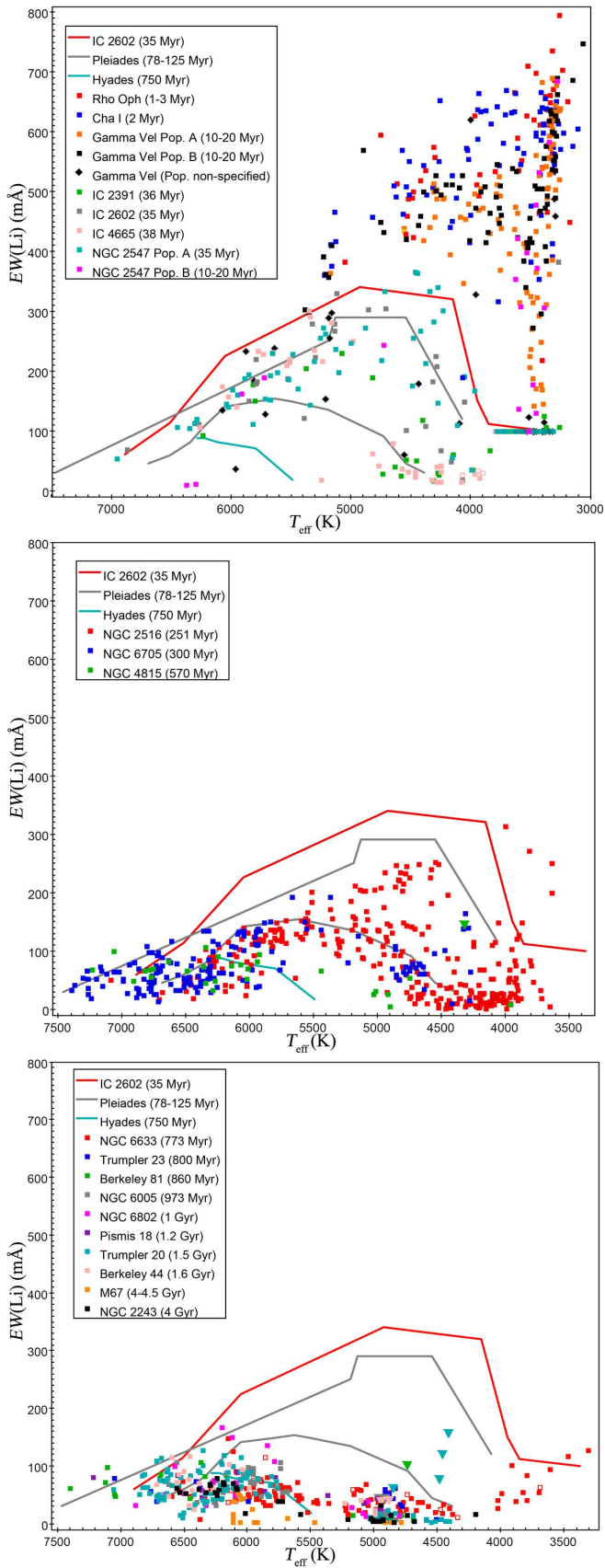

**Fig. 10.** $EW$(Li)-versus-$T_{\rm eff}$ diagrams for the candidate members of the young clusters (1–50 Myr; top panel), as well as the intermediate-age (50–700 Myr; middle panel) and old clusters (> 700 Myr; bottom panel). Open squares indicate possible members only, while inverted triangles refer to Li-rich members.

*Acknowledgements.* Financial support was provided by the Universidad Complutense de Madrid and by the Spanish Ministry of Economy and Competitiveness (MINECO) from project AYA2016-79425-C3-1-P. We acknowledge the support from INAF and Ministero dell' Istruzione, dell' Universitá' e della

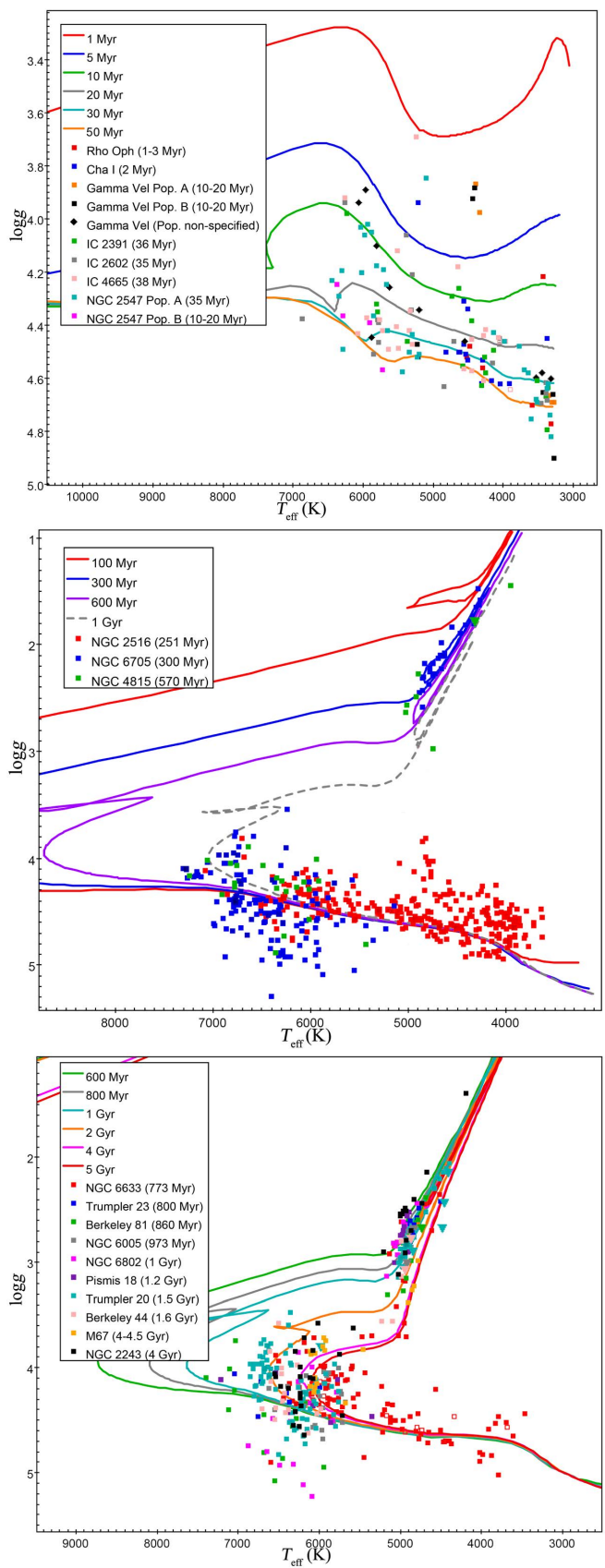

**Fig. 12.** Kiel diagrams for the candidate members of the young clusters (1–50 Myr; top panel), as well as the intermediate-age (50–700 Myr; middle panel) and old clusters (> 700 Myr; bottom panel). We overplot the PARSEC isochrones in a similar age range, with a metallicity of Z=0.019. Open squares indicate possible members only, while inverted triangles refer to Li-rich members.

Ricerca (MIUR) in the form of the grant "Premiale VLT 2012". T.B. was funded by the project grant "The New Milky Way" from the Knut and Alice Wallenberg Foundation. J.I.G.H. acknowledges financial support from the Spanish Ministry of Science, Innovation and Universities (MICIU) under the 2003 Ramón y Cajal program RYC-2013-14875, and also from the Spanish Ministry project MICIU AYA2017-86389-P. E.M. acknowledges financial support from the Spanish Ministerio de Ciencia e Innovación through fellowship FPU15/01476. A.G. acknowledges support from the European Union FP7 programme from the UK space agency. U.H. acknowledges support from the Swedish National Space Agency (SNSA/Rymdstyrelsen). F.J.E. acknowledges financial support from the Spanish MINECO/FEDER through the grant AyA2017-84089. S.G.S acknowledges the support of Fundação para a Ciência e Tecnologia (FCT) through national funds and research grant (project ref. UID/FIS/04434/2013, and PTDC/FIS-AST/7073/2014). S.G.S also acknowledges the support from FCT through Investigador FCT contract of reference IF/00028/2014 and POPH/FSE (EC) by FEDER funding through the program "Programa Operacional de Factores de Competitividad"- COMPETE MT also acknowledges support from the FCT - Fundação para a Ciência e a Tecnologia through national funds (PTDC/FIS-AST/28953/2017) and by FEDER - Fundo Europeu de Desenvolvimento Regional through COMPETE2020 - Programa Operacional Competitividade e Internacionalização (POCI-01-0145-FEDER-028953). Based on data products from observations made with ESO Telescopes at the La Silla Paranal Observatory under programme focusID 188.B-3002. These data products have been processed by the Cambridge Astronomy Survey Unit (CASU) at the Institute of Astronomy, University of Cambridge, and by the FLAMES/UVES reduction team at INAF–Osservatorio Astrofisico di Arcetri. These data have been obtained from the GES Data Archive, prepared and hosted by the Wide Field Astronomy Unit, Institute for Astronomy, University of Edinburgh, which is funded by the UK Science and Technology Facilities Council. This work was partly supported by the European Union FP7 programme through ERC grant number 320360 and by the Leverhulme Trust through grant RPG-2012-541. The results presented here benefit from discussions held during GES workshops and conferences supported by the ESF (European Science Foundation) through the GREAT Research Network Programme. This work was also supported by Fundação para a Ciência e Tecnologia (FCT) through the research grants UID/FIS/04434/2019, UIDB/04434/2020 and UIDP/04434/2020. This work has made use of data from the European Space Agency (ESA) mission *Gaia* (https://www.cosmos.esa.int/gaia), processed by the *Gaia* Data Processing and Analysis Consortium (DPAC, https://www.cosmos.esa.int/web/gaia/dpac/consortium). Funding for the DPAC has been provided by national institutions, in particular the institutions participating in the *Gaia* Multilateral Agreement. This publication makes use of the VizieR database (Ochsenbein et al. 2000) and the SIMBAD database (Wenger et al. 2000), both operated at CDS, Centre de Données astronomiques de Strasbourg, France. This research also made use of the WEBDA database, operated at the Department of Theoretical Physics and Astrophysics of the Masaryk University, and the interactive graphical viewer and editor for tabular data TOPCAT (Taylor 2005). For the analysis of the distributions of *RV* and metallicity we used RStudio Team (2015). Integrated Development for R. RStudio, Inc., Boston, MA (http://www.rstudio.com/). Finally, we would like to thank the anonymous referee for helpful comments and suggestions.

[1] Departamento de Física de la Tierra y Astrofísica and IPARCOS-UCM (Instituto de Física de Partículas y del Cosmos de la UCM), Facultad de Ciencias Físicas, Universidad Complutense de Madrid, E-28040, Madrid, Spain e-mail: mlgutierrez@ucm.es
[2] Observatorio Astronómico Nacional (OAN-IGN), Apdo 112, 28803 Alcalá de Henares, Madrid, Spain
[3] Instituto de Astrofísica e Ciências do Espaço, Universidade do Porto, CAUP, Rua das Estrelas, 4150-762 Porto, Portugal
[4] Instituto de Astrofísica de Canarias (IAC), E-38205 La Laguna, Tenerife, Spain
[5] Universidad de la Laguna, Dept. Astrofísica, E-38206 La Laguna, Tenerife, Spain
[6] INAF - Osservatorio Astrofisico di Catania, via S. Sofia, 78, 95123 Catania, Italy
[7] Università di Catania, Dipartimento di Fisica e Astronomia, Sezione Astrofisica, Via S. Sofia 78, I-95123 Catania, Italy
[8] Institut für Astronomie und Astrophysik, Eberhard Karls Universität, Sand 1, D-72076 Tübingen, Germany
[9] INAF - Osservatorio Astrofisico di Arcetri, Largo E. Fermi 5, 50125, Firenze, Italy
[10] Nicolaus Copernicus Astronomical Center, Polish Academy of Sciences, ul. Bartycka 18, 00-716, Warsaw, Poland
[11] Observational Astrophysics, Division of Astronomy and Space Physics, Department of Physics and Astronomy, Uppsala University, Box 516, SE-751 20 Uppsala, Sweden
[12] Institute of Astronomy, University of Cambridge, Madingley Road, Cambridge CB3 0HA, United Kingdom
[13] Lund Observatory, Department of Astronomy and Theoretical Physics, Box 43, SE-221 00 Lund, Sweden
[14] INAF - Osservatorio Astronomico di Palermo, Piazza del Parlamento, 1, 90134 Palermo, Italy
[15] Dipartimento di Fisica e Astronomia, Università di Padova, Vicolo dell'Osservatorio 3, 35122 Padova, Italy
[16] Spanish Virtual Observatory, Centro de Astrobiología (INTA-CSIC), P.O. Box 78, 28691 Villanueva de la Cañada, Madrid, Spain
[17] Departamento de Ciencias Físicas, Universidad Andrés Bello, Fernández Concha 700, Las Condes, Santiago, Chile
[18] Dipartimento di Fisica "E. Fermi", Università di Pisa, Largo Bruno Pontecorvo 3, 56127 Pisa, Italy
[19] INAF - Padova Observatory, Vicolo dell'Osservatorio 5, 35122 Padova, Italy
[20] Instituto de Física y Astronomía, Universidad de Valparaíso, Chile
[21] Núcleo Milenio Formación Planetaria - NPF, Universidad de Valparaíso
[22] Núcleo de Astronomía, Facultad de Ingeniería y Ciencias, Universidad Diego Portales (UDP), Santiago de Chile
[23] Instituto de Astrofísica de Andalucía (CSIC), Glorieta de la Astronomía s/n, Granada 18008, Spain
[24] Dipartimento di Fisica e Astronomia Galileo Galilei, Vicolo Osservatorio 3, I-35122 Padova, Italy

## Appendix A: Individual cluster notes.

*Appendix A.1: SFRs (age ≤ 5 Myr) and young open clusters (age ≤ 50 Myr)*

### Appendix A.1.1: $\rho$ Ophiuchi

Of the final members in $\rho$ Oph, 29 belong to the SFR L1688 (Rigliaco et al. 2016), and 17 of them are also strong accretors with H$\alpha$10% > 270–300 km s$^{-1}$. Of these 17 accreting stars, only one (J16273311-2441152) does not pass our gravity criteria. This star, with $\gamma$=1.026 and A(Li)=3.18, could be listed as a potential Li-rich giant, but due to the fact that it is also a strong accretor (with H$\alpha$10% = 456 km s$^{-1}$) we counted it as a likely member of $\rho$ Oph, also in accordance with Rigliaco et al. (2016). Cánovas et al. (2019) listed 38 candidates that are common to our selection (for the remaining two stars in their study the iDR4 sample does not include Li measurements). We note that one of the stars in our final selection (J16270456-2442140) presents a [Fe/H] value which deviates appreciably from the rest of candidates, while another four candidates are $RV$ non-members with $RV$s deviating 2.5$\sigma$ (or 5 km s$^{-1}$) from the mean of the cluster and in one case up to 4$\sigma$ (or 8 km s$^{-1}$). We included these stars given that they fulfil the rest of the membership criteria (especially regarding gravity indicators and lithium), and they are also listed as candidates by studies such as Cánovas et al. (2019).

### Appendix A.1.2: Chamaeleon I

As a result of the analysis of Cha I we find 35 accreting stars, of which only one (J11092578-7623207) is not a Li member. This is probably due to possible veiling suppressing the absorption Li line, and so we classified this strong accretor as an additional likely member. We also note that three of the stars in our final selection are $RV$ non-members, with $RV$s deviating 2.7–4$\sigma$ (or 4–6 km s$^{-1}$) from the mean of the cluster and, in one case, up to 6.7$\sigma$ (or 10 km s$^{-1}$). These stars fulfil the rest of the membership criteria (especially regarding gravity indicators and lithium), and they are also listed as candidates by Sacco et al. (2017) and Robrade & Schmitt (2007). Regarding previous selections, we find all the UVES members observed by Spina et al. (2014 a) in our selection, except for J10555973-7724399 and J11092378-7623207, for which several parameters are not released in the iDR4 catalog. On the other hand, all our candidate stars are included in the member list of Sacco et al. (2017), except for one (J11110238-7613327), which is listed in Robrade & Schmitt (2007). Regarding field contaminants, one Li-rich giant (J11000515-7623259) is listed in Casey et al. (2016).

### Appendix A.1.3: Vela OB2 association: $\gamma$ Velorum and NGC 2547

The cluster membership selections for both of these clusters consist of 210 stars in $\gamma$ Vel (104 in $\gamma$ Vel A and 83 in $\gamma$ Vel B, as well as 23 additional candidate stars that are not associated to a specific population), and 107 stars in NGC 2547 (88 in NGC 2547 A and 19 in NGC 2547 B). As mentioned in Sect. 3, for this study we used the membership selections obtained by a series of former studies in the literature: Regarding $\gamma$ Vel, we first used Jeffries et al. (2014), the study specifying the members of Pop. A and B, as well as a series of other GES studies (Damiani et al. 2014; Spina et al. 2014 b; Frasca et al. 2015; Prisinzano et al. 2016). For NGC 2547 we used Sacco et al. (2015), together with the *Gaia* studies conducted by Cantat-Gaudin et al. (2018) and Randich et al. (2018). All three of these latter papers list the membership probabilities of each candidate to Pop. A/B. For $\gamma$ Vel, we find 23 additional candidates that were not listed in the previous GES studies. We added them to the candidates not associated with one of the two $\gamma$ Vel populations. We note that Frasca et al. (2015) used the data from Jeffries et al. (2014), and in this case we consider five stars listed as members which had no listed Pop. A/B in Jeffries et al. (2014).

Recent studies Cantat-Gaudin et al. (2018) and Randich et al. (2018) furthermore offer updated membership probabilities for NGC 2547 (as well as a series of other clusters in our sample), using data from *Gaia*. Randich et al. (2018) listed a series of new members with respect to Sacco et al. (2015). Although their membership probabilities for each population are generally consistent with each other, six stars were associated with different populations in Sacco et al. (2015) and Randich et al. (2018). In this case, we adopted the membership from Randich et al. (2018), as this is the most recent study which used both the GES iDR4 data and *Gaia* DR2 data. For this cluster we also compared the member stars in Sacco et al. (2015), Cantat-Gaudin et al. (2018) and Randich et al. (2018) with the candidate members in Bravi et al. (2018). For all our candidates, Bravi et al. (2018) find high probabilities, namely ranging from 60 to 100 %, of being *RV* members of the cluster.

Finally, as to field contaminants, we also find two of the $\gamma$ Vel Li-rich giants in our list (J08095783-4701385 and J08102116-4740125), as well as one of the Li-rich giants from NGC 2547 (J08110403-4852137), in Casey et al. (2016). Another one, (J08110403-4658057) in $\gamma$ Vel, is listed in Smiljanic et al. (2018). Regarding the NG non-members, we included only the contaminants marked as such by the aforementioned studies. The reason for this is that without an additional membership analysis (as we have done for the rest of the clusters) we cannot confirm many NG stars not listed by these studies as either members or contaminants.

Appendix A.1.4: IC 2391

We discarded but 34 Li members of IC 2391 for not having any measured values of chromospheric activity, taking into account both the accretion and chromospheric $EW$(H$\alpha$). From these 34 stars, we discarded seven as non-members for not fulfilling our *RV* and metallicity criteria. Some of them have very large *RV*s deviating up to $150\sigma$ (or 450 km s$^{-1}$) from the mean of the cluster. Four of these seven stars are also listed as non-members by (Randich et al. 2018). We also note that we consider three candidates which are *RV* non-members, with *RV*s deviating from the mean by up to $4.3\sigma$ (or 13 km s$^{-1}$), as they are good spectroscopic candidates, fulfilling the rest of the membership criteria (especially regarding gravity indicators and lithium).

We first compared our selection with a series of non-GES studies, finding ten of our candidates in one or more previous lists (Patten & Simon 1996; Barrado y Navascués et al. 2001, 2004; Platais et al. 2007; Marsden et al. 2009; Spezzi et al. 2009; Messina et al. 2011; De Silva et al. 2013; Elliott et al. 2015). Regarding GES studies, Bravi et al. (2018) derived *RV* membership probabilities and lists of candidate members for this cluster (alongside IC 2602, IC 4665 and NGC 2547) using iDR4 data, which were selected from their $EW$(Li), gravity, and metallicity values reported in the GES iDR4 data. Comparing our final selection of 27 members with the list of 53 candidate stars of Bravi et al. (2018), we find 30 kinematic candidates and 17 final members in common (all of the common members stars have high *RV* membership probabilities of at least 0.95, except for two stars in the 0.7–0.9 range). We note that, for many stars Bravi et al. (2018) used values of $EW$(Li) and/or $T_{\mathrm{eff}}$ which were derived from one of the WG12 nodes and do not appear in the iDR4 sample. As a result, we excluded these stars from our membership analysis and only consider those stars in Bravi et al. (2018) with $EW$(Li) values in the iDR4 sample. Our mean *RV* and $\sigma$ for IC 2391 also agree with the estimates in Bravi et al. (2018).

Finally, Cantat-Gaudin et al. (2018) and Randich et al. (2018) also list updated membership probabilities for IC 2391. We have 19 common stars listed as members with Randich et al. (2018), as well as 13 stars with Cantat-Gaudin et al. (2018). As we mention in Sect. 3.5, in this study we relied more heavily on Randich et al. (2018), as they used the same iDR4 sample and their membership criteria are similar to ours. We relied less heavily on Cantat-Gaudin et al. (2018) from a comparative standpoint, given that they primarily base their membership analysis on *Gaia* astrometry and do not use spectroscopic criteria. We note that we discarded a small number of possible candidate stars which were listed as non-members by both Randich et al. (2018) and Bravi et al. (2018), and we also did not consider a small number of stars listed by Randich et al. (2018) and Cantat-Gaudin et al. (2018) for not having a measured H$\alpha$ in the iDR4 sample (the latter also applies to the clusters IC 2602 and IC 4665 listed below). We also mention here a series of recent studies that used *Gaia*-DR2 data to study the spacial-kinematic distribution and cluster membership of IC 2391 (Postnikova et al. 2019 a,b; Vereshchagin et al. 2019).

Appendix A.1.5: IC 2602

We discarded all but 42 of the 52 Li candidates in IC 2602 for not having measured values of chromospheric activity. Of these 42 Li members we finally discarded ten stars, all of which are also listed as non-members by Randich et al. (2018) and Bravi et al. (2018). Comparing our selection with former studies, we find four of our candidates in one or more of the member lists of these non-GES studies (Randich et al. 1997; Stauffer et al. 1997; Marsden et al. 2009; Randich et al. 2001; Smiljanic et al. 2011). Regarding GES studies, we find 55 kinematic candidates and 27 final members in common with the list of 101 candidates in Bravi et al. (2018). As in the case of IC 2391, many of the stars in Bravi et al. (2018) have no $EW$(Li) values in the iDR4 sample, and therefore we excluded them in our own membership analysis. The mean *RV* and $\sigma$ derived in Bravi et al. (2018) for IC 2602 are also in agreement with the ones obtained in this paper. As for *Gaia* studies, we have 28 common stars listed as members with Randich et al. (2018), as well as 19 stars with Cantat-Gaudin et al. (2018).

Appendix A.1.6: IC 4665

We discarded all but 51 Li members in IC 4665 for not having measured values of $EW$(H$\alpha$) in the GES sample, and finally discarded 11 of these 51 stars as non-members for not fulfilling our *RV* and metallicity criteria. Five of these stars are also listed as non-members by both Randich et al. (2018) and Bravi et al. (2018). Finally, three of the final Li members are also listed as non-members by Randich et al. (2018), and therefore we decided to consider them as possible candidates only. Comparing our selection with former studies, we find six of our member stars in Jeffries et al. (2009 b). Regarding GES studies, we find 30 kinematic candidates and 19 final members in common with the list of 122 candidates in Bravi et al. (2018). Our mean *RV* and $\sigma$ are also consistent with the previous estimates in this work. As for *Gaia*, we find 21 common stars listed as members with both Randich et al. (2018) and Cantat-Gaudin et al. (2018).

*Appendix A.2: Intermediate-age clusters (age= 50–700 Myr)*

Appendix A.2.1: NGC 2516

In our selection of candidate members of NGC 2516 we listed 30 Li members with *RV*s deviating up to $13\sigma$ (or 8 km s$^{-1}$) from the mean of the cluster. All of these stars fulfil our main criteria and are also included as members by Randich et al. (2018). Regarding all the intermediate clusters of our sample, as already mentioned in Sects. 3.2 and 3.4, we also note that, given the lower resolution of the GIRAFFE spectra, we accepted as candidates a number of Li members with [Fe/H] values outside the limit of $2\sigma$ from the cluster mean provided by our fit.

Comparing our UVES selection with existing GES studies, we find all of our final UVES candidates to be the same as those in the list of NGC 2516 stars classified as members by (Jacobson et al. 2016), and all but two of our candidates are also listed as members by (Magrini et al. 2017). Regarding the GIRAFFE candidates, we find 211 of our 273 candidates listed as members in the membership list of Jeffries et al. (2001), as well as 25 in Terndrup et al. (2002), 24 in Irwin et al. (2007), 19 in Jackson & Jeffries (2010), two in Heiter et al. (2014), 23 in Jackson et al. (2016), 49 in Sampedro et al. (2017), and 45 candidates in Bailey et al. (2018). Another candidate of interest is J07572938-6050104, which seems to deviate slightly more from the rest in the $EW$(Li)-versus-$T_{\rm eff}$ diagram, but is also listed as a member of NGC 2516 in existing studies (Jeffries et al. 2001; Sampedro et al. 2017). Regarding the studies using data from *Gaia*, we have 280 and 230 candidates in common with Randich et al. (2018) and Cantat-Gaudin et al. (2018), respectively. We also note that, in order to help confirm the membership of the stars in the field of this cluster, we further made use of the additional members of Jeffries et al. (1998).

We also find among our candidates a series of K type stars (evolutionary stage unknown) with high values of Li and in the 3500–4000 K temperature range, which are listed as members by a series of former studies (Jeffries et al. 2001; Irwin et al. 2007; Jackson & Jeffries 2010; Jackson et al. 2016), and which seem to be lower mass, lower luminosity PMS stars, chromospherically active and rapidly rotating, which have not yet depleted their original Li content (Pallavicini et al. 1997). These stars can be helpful to study the LDB for this cluster, as well as the age–LDB luminosity relationship, a reliable and sensitive age calibration method for clusters in the 20–200 Myr range which requires few assumptions and is not model-dependent (Barrado y Navascués et al. 1999; Soderblom 2010; Soderblom et al. 2014).

Appendix A.2.2: NGC 6705

Comparing our selection of candidate members for NGC 6705 with existing studies, we note that all of our 27 final UVES candidates are listed as NGC 6705 member stars by GES studies Tautvaišienė et al. (2015) and Jacobson et al. (2016). We also find 21 and 15 of our UVES candidate stars in the membership GES studies of Magrini et al. (2014) and Magrini et al. (2017), respectively. On the other hand, we find 119 of our GIRAFFE candidates listed as members by Cantat-Gaudin et al. (2014), as well as 55 common candidates in the membership study of Sampedro et al. (2017), and 77 common members in the list of Cantat-Gaudin et al. (2018). We also note that both Cantat-Gaudin et al. (2018) and Sampedro et al. (2017) list a number of our member stars as non-members. We however decided to classify them as candidate members because they are included as members in other studies (Messina et al. 2010; Cantat-Gaudin et al. 2014), and fulfil our membership criteria[12].

Appendix A.2.3: NGC 4815

Of 29 Li candidates in NGC 4815 we discarded one star (J12583456-6453419) for not fulfilling our gravity indicator criterion, as well as for having a [Fe/H] value far from the mean of the cluster. In addition to the remaining 28 Li members, which fulfil the rest of the criteria, we also accepted the UVES non-Li member J12572442-6455173, given that it is listed as a candidate of NGC 4815 by a series of previous studies (Magrini et al. 2015; Tautvaišienė et al. 2015; Jacobson et al. 2016). This star, with a Li value significantly far from the other candidates, seems to be a Li-rich member. Comparing our selection with other GES studies, we note that five of our eight UVES candidates are included in the lists of NGC 4815 stars classified as members in several studies (Friel et al. 2014; Magrini et al. 2015; Tautvaišienė et al. 2015; Jacobson et al. 2016). Three of our eight UVES candidates are also listed in the membership study of Magrini et al. (2017). Although not listed in these studies, we consider three additional UVES stars, which fulfilled all of our membership criteria, as candidates of the cluster. We also find six of our GIRAFFE candidates listed in Cantat-Gaudin et al. (2018). Finally, this study includes two of the stars in our candidate list as non-members, but we consider them because they fulfil our membership criteria.

*Appendix A.3: Old clusters (age > 700 Myr)*

Appendix A.3.1: NGC 6633

Our *RV* analysis of NGC 6633 reveals a large contaminant population with positive *RV*s in the middle of the distribution, which could not be discarded with the aid of the $2\sigma$ clipping procedure (as we did for most *RV* contaminants in the tails for the rest of the clusters in the sample). The presence of this contaminant population effectively affected the mean *RV* rendered by the Gaussian fit and also gave a very high dispersion even after the final convergence of the clipping procedure. When comparing the final fit with literature values, we saw that taking all the contaminant positive *RV*s into consideration caused the mean *RV* to deviate considerably from the reference estimate for this cluster ($-28.6 \pm 0.1$ km s$^{-1}$). For this reason, we decided to filter this contaminant population with positive *RV*s before re-analysing the *RV* distribution for the cluster and obtaining a mean and dispersion which were more probable, and consistent with the literature values as well. A series of stars from this outlier population could also be discarded when studying the Li, gravity and/or metallicity criteria, but in other cases it was the *RV* criterion which helped discard them, given that they seemed to fulfil all other criteria.

After the *RV* analysis, of the 131 Li members obtained we further discarded 12 stars. Three of them we discarded for not fulfilling the gravity and/or metallicity criteria, and the remaining nine were marginal *RV* members which presented positive *RV*s far from the mean of the cluster rendered by the fit. We considered these stars to be probable contaminants from the aforementioned outlier population, given that they significantly affected the *RV* mean rendered by the final fit of the candidate selection. Five of these stars are additionally listed as non-members by Randich et al. (2018). Finally, we consider 17 Li members to be possible members only as they are listed as non-members by Randich et al. (2018) but also fulfil our membership criteria. Similarly to the previously discussed intermediate clusters, given the lower resolution of the GIRAFFE spectra, we note that for the clusters in the old age-range we similarly accepted as probable candidates most Li members with [Fe/H] values outside the $\sigma$-limit from the cluster mean provided by our fit (see Sects. 3.2 and 3.4). As an additional feature of interest, we also note that the lower metallicity of this cluster, ranging from −0.10 to −0.01 dex (Jeffries et al. 2002; Jacobson et al. 2016), and of −0.01 ± 0.11 in Magrini et al. (2018), can explain why in our analysis the *EW*(Li) envelope for NGC 6633 is above the Hyades (and thus, the fact that Li is being depleted at a slower pace in the case of NGC 6633), despite the very similar age of these two clusters (Umezu & Saio 2000).

---

[12] As mentioned above, we relied less heavily on Cantat-Gaudin et al. (2018) from a comparative viewpoint because the study bases the membership analysis primarily on *Gaia* astrometry (parallaxes and proper motions as well as velocity) and does not use the spectroscopic information or lithium as main criteria, as we do in this study.

Comparing our selection with existing studies, we find 11 of our 13 UVES members in the member list of Jacobson et al. (2016), and also eight of these UVES candidates in the list of stars classified as members by Magrini et al. (2017). We also list the remaining two UVES stars in our selection, which are not included in any of these studies (Jeffries 1997; Jacobson et al. 2016; Magrini et al. 2017), given that they fulfil all of our membership criteria. Regarding our GIRAFFE candidates, we find one star (J18275896+0629050) in Jeffries (1997), as well as another (J18275187+0624499) in Peña et al. (2017). Regarding the studies using data from *Gaia*, we have 35 and 17 candidates in common with Randich et al. (2018) and Cantat-Gaudin et al. (2018), respectively. We also note that, in order to help ascertain the potential membership of the stars in the field of this cluster, we also used the additional members in Jeffries (1997), Heiter et al. (2014), Magrini et al. (2017), and Sampedro et al. (2017).

Focusing on a couple of particular cases for this cluster, we firstly note that three of our selection candidates present a higher Li than the rest, but seem to be consistent with the *EW*(Li) envelope in Jeffries (1997). One of them (J18274267+0639082) is listed as a non-member in the list of Randich et al. (2018), but we consider it a member given that it fulfils our membership criteria and is also listed as a member by other studies (Jacobson et al. 2016; Magrini et al. 2017). Secondly, we note that among our candidates we find a series of K stars with high values of Li, in the 3300–4000 K temperature range, some of which are listed as members by Randich et al. (2018), and in agreement with other studies such as (Jeffries 1997). As with the case of NGC 2516, these seem to be lower mass, lower luminosity PMS stars which have not yet depleted their original Li content (Pallavicini et al. 1997). These stars can be helpful to study the LDB for this cluster, as well as the age–LDB luminosity relationship (Barrado y Navascués et al. 1999; Soderblom 2010; Soderblom et al. 2014). In this case, the study of the LDB of the Hyades, of very similar age, can be helpful as well (Martín et al. 2018). Finally, regarding non-members, one of the Li-rich giants in our list (J18265248+0627259) is listed in Smiljanic et al. (2018).

Appendix A.3.2: Trumpler 23

Regarding the old clusters in our sample, as already mentioned in Sects. 3.2 and 3.4, given the lower resolution of the GIRAFFE spectra we accepted as candidates a number of Li members with [Fe/H] values outside the limit of $2\sigma$ from the cluster mean provided by our fit. Comparing our final selection for Trumpler 23 with other existing GES studies, we find all 11 of our final UVES candidates to be classified as members of Trumpler 23 by several studies (Jacobson et al. 2016; Magrini et al. 2017; Overbeek et al. 2017). In spite of being listed as a non-member by Overbeek et al. (2017), we consider the UVES star J16004025-5329439 as a member, as opposed to a Li-rich giant non-member, given that it fulfils all our membership criteria, as in Magrini et al. (2017). Regarding our six GIRAFFE candidates, all are listed as members by Overbeek et al. (2017), and some also by Sampedro et al. (2017).

Appendix A.3.3: Berkeley 81

As well as the 27 Li members in Berkeley 81, we also consider one UVES Li-rich giant star (J19014498-0027496) as an additional candidate instead of a giant contaminant, not only because it fulfils all of our membership criteria, but also because other studies (Jacobson et al. 2016; Magrini et al. 2017) considered it to be a member of this cluster. Comparing our selection with existing studies, we find that our selection of UVES candidates coincides with those stars classified as final members by both Jacobson et al. (2016) and Magrini et al. (2017). As for our GIRAFFE candidates, we find seven of these candidate stars in the list of stars classified as *RV* members of Magrini et al. (2015), and we also have six stars in common with *Gaia* study Cantat-Gaudin et al. (2018).

Appendix A.3.4: NGC 6005

Of the 41 Li candidates in NGC 6005 we discarded three stars for not fulfilling our gravity and/or metallicity criteria. Comparing our selection with other existing GES studies, we find all but two of our 14 UVES candidates in the list of NGC 6005 stars classified as members by Jacobson et al. (2016). The remaining two UVES stars in our candidate list (J15553294-5725298 and J15555069-5726255) are not included in this study but we included them as potential members, as they fulfil all of our membership criteria. Regarding the GIRAFFE candidates, we find 11 common stars in Cantat-Gaudin et al. (2018).

Appendix A.3.5: NGC 6802

Of the 24 Li candidates in NGC 6802 we discarded two UVES stars for not fulfilling our gravity criterion, and for having [Fe/H] values far from the mean of the cluster. We also note that we included two Li members with *RV*s deviating up to $2.4\sigma$ (or 4 km s$^{-1}$) from the mean of the cluster. These stars fulfil our main criteria and are also included as members by Tang et al. (2017). Comparing our selection with other GES studies, we find that our final UVES candidates are the same as those in the list of NGC 6802 stars classified as members by both Jacobson et al. (2016) and Magrini et al. (2017). As for our ten GIRAFFE candidates, we find all of them in the list of stars classified as members of NGC 6802 by Tang et al. (2017), and we also have six common stars with Cantat-Gaudin et al. (2018). Regarding non-members, of the three Li-rich giants found, one of them (J19303773+2016196) is marginal, with a Li abundance very close to our criterion of $A(Li) > 1.5$, and we consider it a a Li-rich giant for this reason. Finally, one of the Li-rich giants in our list (J19304281+2016107) is also listed in Casey et al. (2016).

### Appendix A.3.6: Pismis 18

During the analysis for Pismis 18 we note that we included three Li members with *RV*s deviating up to $4.3\sigma$ (or 12 km s$^{-1}$) from the mean of the cluster. These stars fulfil our main criteria and are also included as members by a series of studies (Sampedro et al. 2017; Hatzidimitriou et al. 2019). Comparing our selection with other existing GES studies, we find all but one of our final UVES candidates in the list of Pismis 18 stars classified as members by Jacobson et al. (2016). The remaining star, J13365001-6205376, is included as a member in Hatzidimitriou et al. (2019). We also find eight of our candidates listed as high confidence members in a recent study by Hatzidimitriou et al. (2019). As for GIRAFFE candidates, we find three of our candidate stars in Sampedro et al. (2017).

### Appendix A.3.7: Trumpler 20

Of the 125 Li candidates in Trumpler 20 we discarded one star for not fulfilling our gravity criterion and having [Fe/H] values far from the mean of the cluster. We note that as part of our final selection we included eight Li members with *RV*s deviating up to $4\sigma$ (or 7 km s$^{-1}$) from the mean of the cluster. These stars fulfil our main criteria and are also included as members by a series of studies (Sampedro et al. 2017; Cantat-Gaudin et al. 2018). Additionally, three of our final UVES candidates are Li-rich member stars, in agreement with existing GES studies such as Smiljanic et al. (2016), who discussed two of these three Li-rich giant members (J12400449-6036566 and J12395566-6035233) in their analysis. Comparing the members in our selection with other studies, we firstly note that the 41 UVES candidates in our list are the same as those in the list of stars classified as members of Trumpler 20 by both Jacobson et al. (2016) and Smiljanic et al. (2016). We also find some of our GIRAFFE candidates in GES studies (Donati et al. 2014 a; Tautvaišienė et al. 2015; Merle et al. 2017), and in the non-GES study Sampedro et al. (2017), as well as 45 common stars with *Gaia* study Cantat-Gaudin et al. (2018). On the other hand, we also note that a couple of stars in our selection, which fulfil all our criteria, are nevertheless listed as non-members by Sampedro et al. (2017) and Cantat-Gaudin et al. (2018). Finally, we also find two stars (J12392452-6035361 and J12393024-6037097) which we listed as potential Li-rich members and fulfil the rest of our criteria. Given the attested Li-rich members in this clusters, we have listed them as a Li-rich candidates rather than as Li-rich giant contaminants.

### Appendix A.3.8: Berkeley 44

Of the 23 Li candidates in Berkeley 44 we discarded one star for not fulfilling our gravity indicator criterion, and for having [Fe/H] values far from the mean metallicity of the cluster. We note that as part of our final selection we included five Li members with *RV*s deviating up to 2 km s$^{-1}$ from the 2- interval (and up to 4 km s$^{-1}$ from the mean of the cluster). These stars fulfil our main criteria and are also included as members by a series of studies (Sampedro et al. 2017; Cantat-Gaudin et al. 2018). Comparing our selection with other studies, we note that all UVES stars classified as members by Jacobson et al. (2016) are among our final candidates. We also find some of our GIRAFFE candidates in Sampedro et al. (2017), and 14 common candidates with Cantat-Gaudin et al. (2018).

### Appendix A.3.9: M67

Of the 19 Li members in M67, we consider all but one as candidate members of the cluster, in agreement with earlier studies which include membership lists for this cluster (Pace et al. 2012; Pasquini et al. 2012; Carlberg 2014; Geller et al. 2015; Brucalassi et al. 2017). The remaining Li member (J08505891+1148192) we listed as a possible member, given that, while it fulfils all of our membership criteria, it is listed as a non-member by (Geller et al. 2015). We note that we accepted as a final candidate a star deviating by up to $7.8\sigma$ (or 7 km s$^{-1}$) from the mean of the cluster, also in agreement with non-GES membership studies (Pace et al. 2012; Pasquini et al. 2012; Carlberg 2014; Geller et al. 2015). We also used lists of M67 candidates found in a series of non-GES studies (Hobbs & Pilachowski 1986; Balachandran 1995; Pallavicini et al. 1997; Pasquini et al. 1997; Jones et al. 1999; Randich et al. 2002) to further help us analyse the stars in our GES sample.

### Appendix A.3.10: NGC 2243

For NGC 2243, a cluster with particularly low metallicity, we firstly note that for the individual figures of this cluster, shown in Appendix B, we considered PARSEC isochrones with Z=0.006 instead of the usual near-solar metallicity of Z=0.019 used for the rest of the clusters in our sample. Regarding the GIRAFFE stars with Li in the field of NGC 2243 found in the OACT node (used in addition to the iDR4 sample), we discarded as non-members about 35 *RV* members with appreciable values of Li higher than the upper envelope of the Hyades cluster by up to 40 mÅ. We suggest that these contaminant field stars might be part of a younger cluster close to NGC 2243. Comparing our selection of candidates with previous membership studies, we find most of our UVES member stars in the lists provided by GES study Magrini et al. (2017), as well as in a couple of non-GES studies (Jacobson et al. 2011; Heiter et al. 2014). We also find one of our GIRAFFE candidates (J06292133-3118094) in Sampedro et al. (2017), and 25 common stars with Cantat-Gaudin et al. (2018). Given that NGC 2243 and M67 are close age-wise, we have additionally made use of the Li envelope created by our candidate selection for M67 (as well as former M67 attested members) to help confirm the potential membership of the GIRAFFE stars for NGC 2243. We also note that for this cluster we observe a larger dispersion among the attested UVES and GIRAFFE candidates, with some members higher in the $EW$(Li)-versus-$T_{\rm eff}$ diagram than the rest, also in comparison with our selection for M67.

# Appendix B: Individual cluster figures.

Appendix B.0.1: $\rho$ Ophiuchi

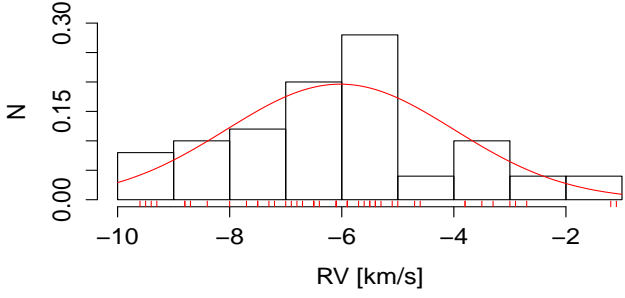

**Fig. B.1.** Gaussian fit of the $RV$ distribution for $\rho$ Oph after the $2\sigma$ clipping procedure

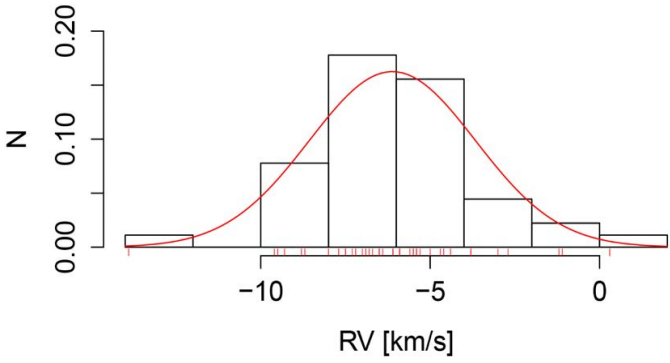

**Fig. B.2.** Gaussian fit of the $RV$ distribution for the final selection of $\rho$ Oph

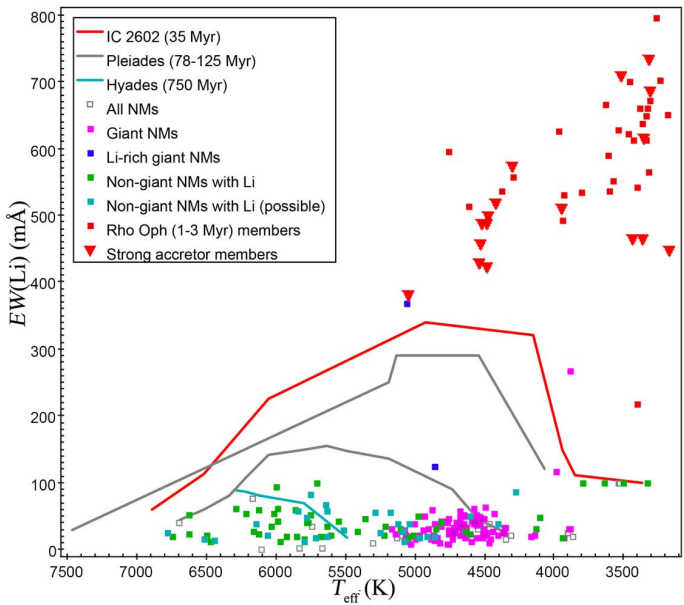

**Fig. B.3.** $EW$(Li)-versus-$T_\mathrm{eff}$ figure for the final candidates of $\rho$ Oph. All Li non-members are shown in open squares, of which we select giant and NG field contaminants of interest.

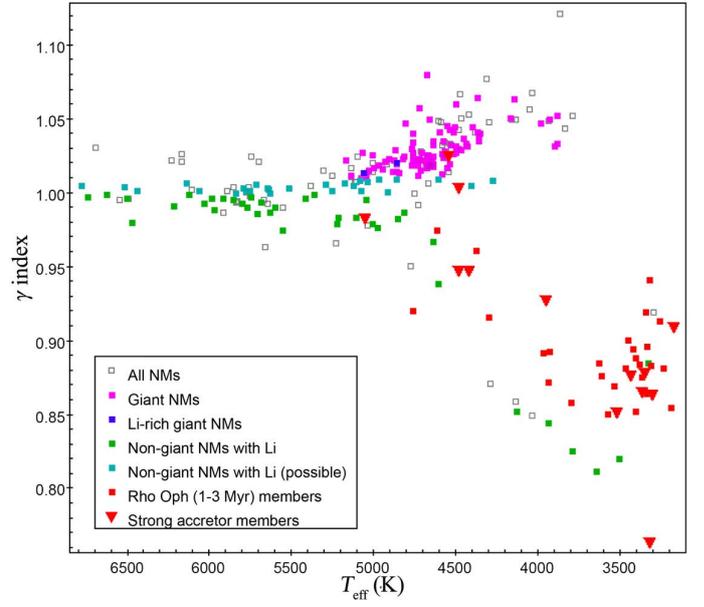

**Fig. B.4.** Gravity index $\gamma$ as a function of $T_\mathrm{eff}$ for $\rho$ Oph.

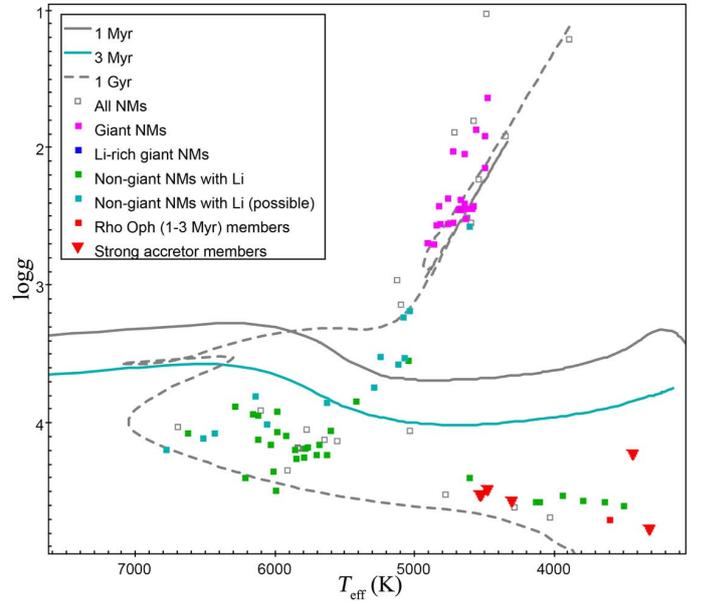

**Fig. B.5.** Kiel diagram for $\rho$ Oph. We use the PARSEC isochrones with a metallicity of Z=0.019.

Appendix B.0.2: Chamaeleon I

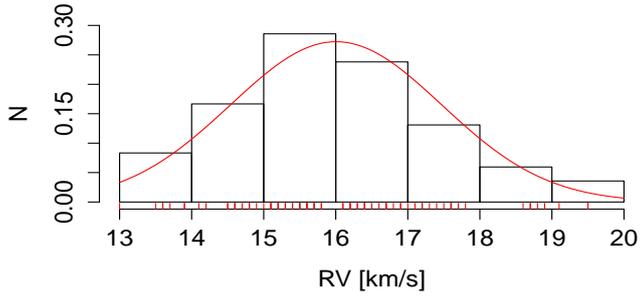

**Fig. B.6.** Gaussian fit of the *RV* distribution for Cha I after the $2\sigma$ clipping procedure

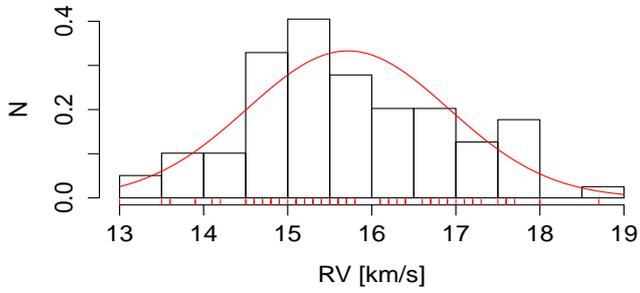

**Fig. B.7.** Gaussian fit of the *RV* distribution for the final selection of Cha I

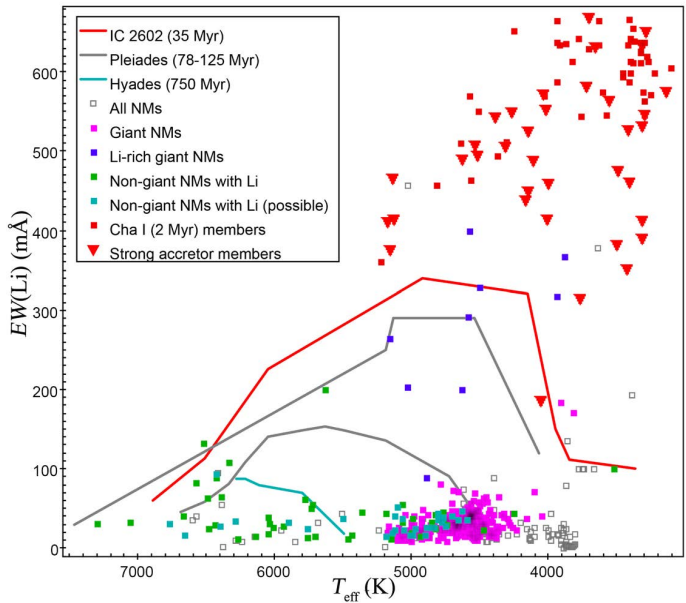

**Fig. B.8.** *EW*(Li)-versus-$T_{\rm eff}$ figure for Cha I

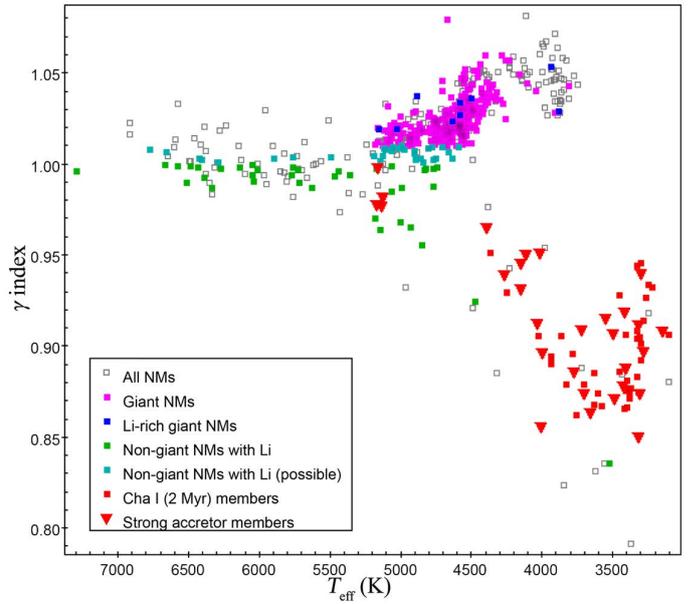

**Fig. B.9.** Gravity index $\gamma$ as a function of $T_{\rm eff}$ for Cha I

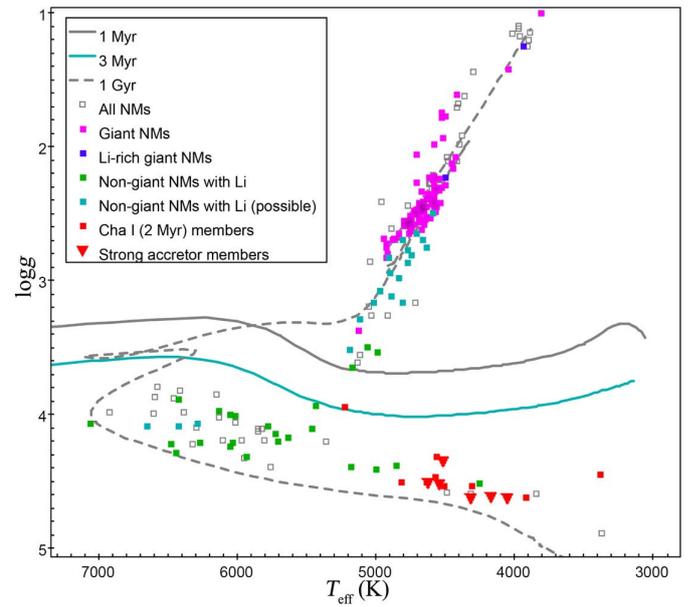

**Fig. B.10.** Kiel diagram for Cha I

Appendix B.0.3: γ Velorum

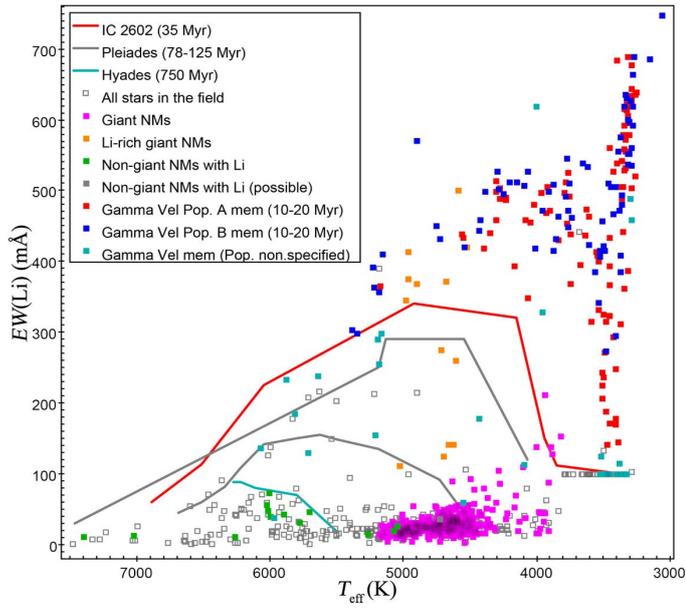

**Fig. B.11.** $EW$(Li)-versus-$T_{\rm eff}$ figure for γ Velorum

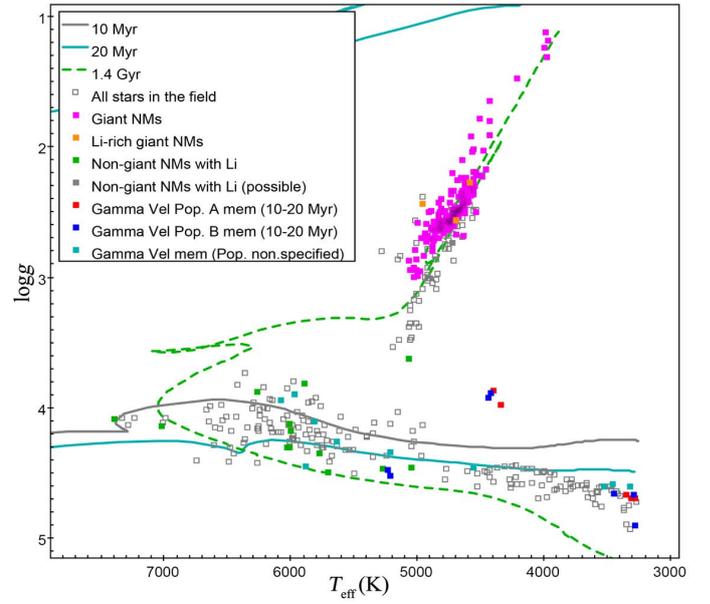

**Fig. B.13.** Kiel diagram for γ Velorum

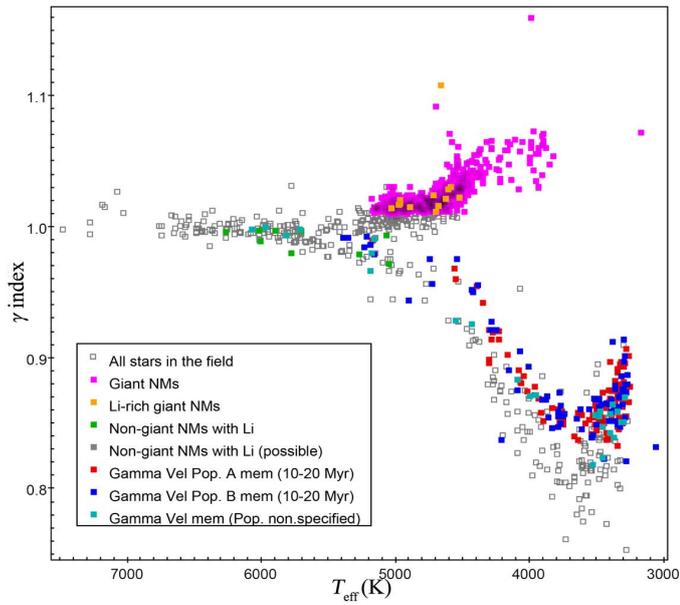

**Fig. B.12.** Gravity index γ as a function of $T_{\rm eff}$ for γ Velorum

## Appendix B.0.4: NGC 2547

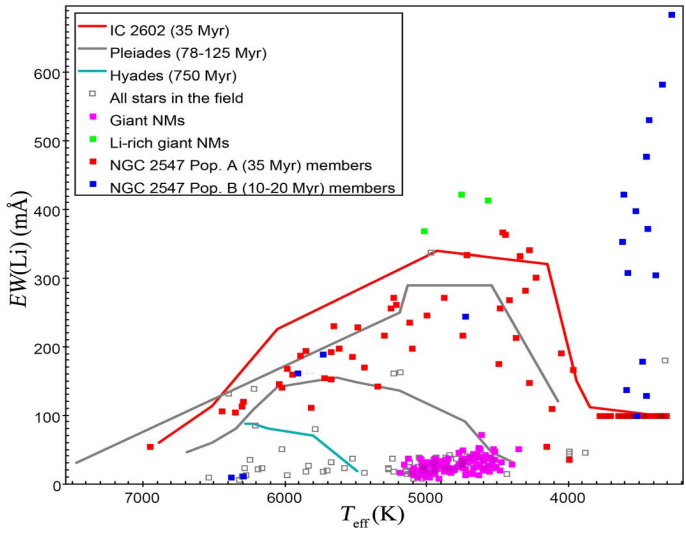

**Fig. B.14.** $EW$(Li)-versus-$T_{\rm eff}$ figure for NGC 2547

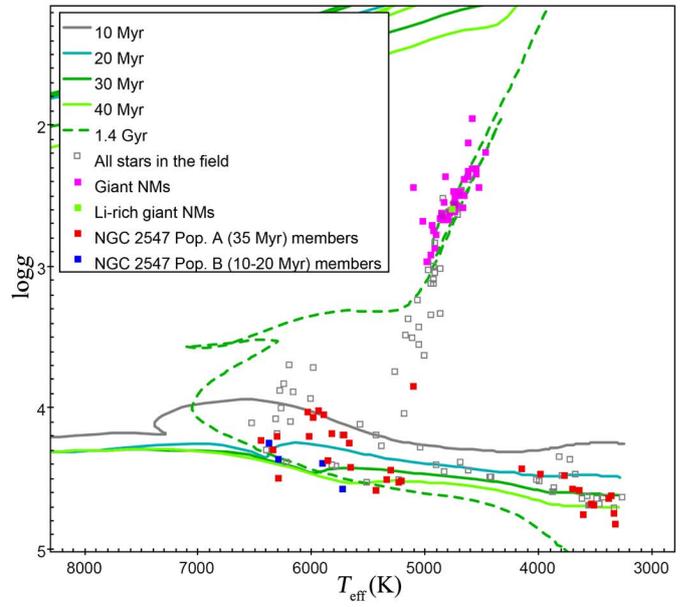

**Fig. B.16.** Kiel diagram for NGC 2547

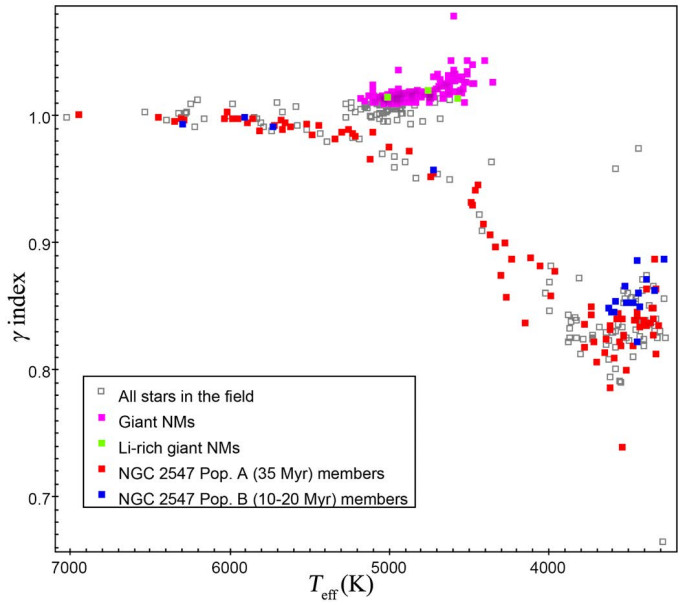

**Fig. B.15.** Gravity index $\gamma$ as a function of $T_{\rm eff}$ for NGC 2547

Appendix B.0.5: IC 2391

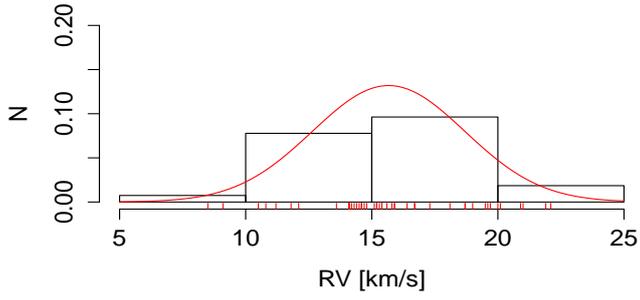

**Fig. B.17.** Gaussian fit of the $RV$ distribution for IC 2391 after the $2\sigma$ clipping procedure

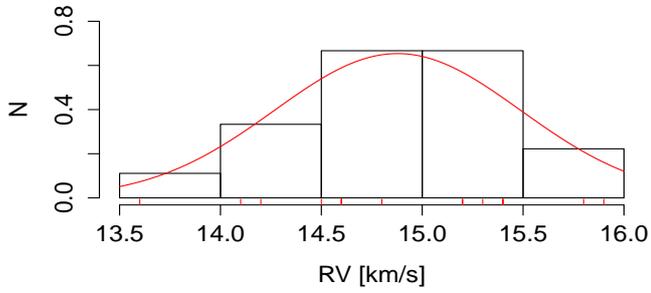

**Fig. B.18.** Gaussian fit of the $RV$ distribution for the final selection of IC 2391

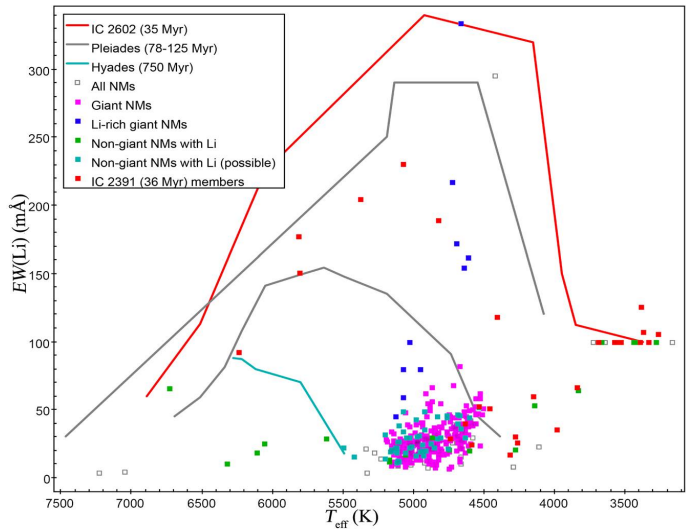

**Fig. B.19.** $EW$(Li)-versus-$T_{\text{eff}}$ figure for IC 2391

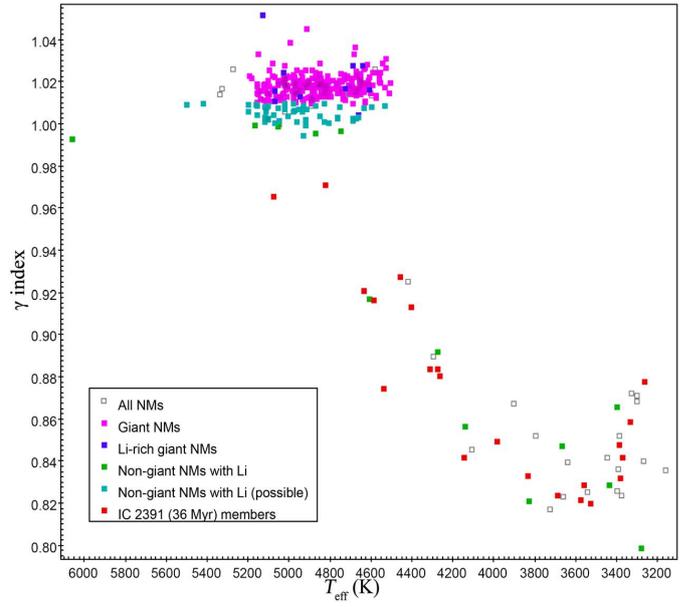

**Fig. B.20.** Gravity index $\gamma$ as a function of $T_{\text{eff}}$ for IC 2391

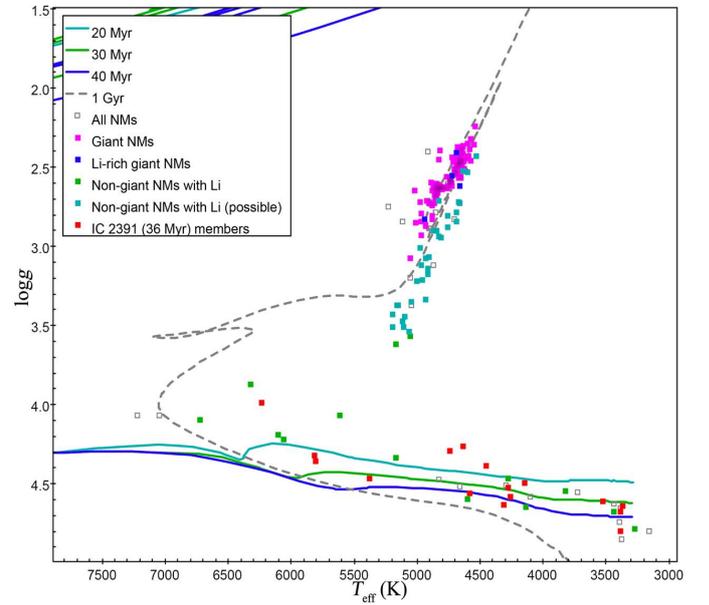

**Fig. B.21.** HR diagram for IC 2391

## Appendix B.0.6: IC 2602

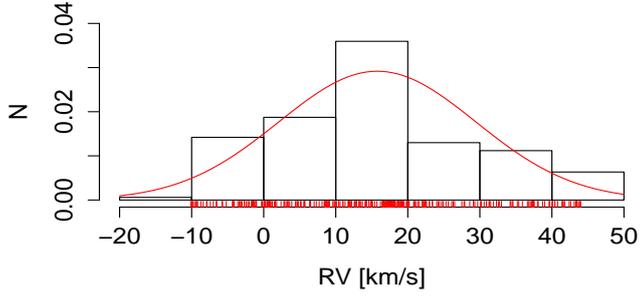

**Fig. B.22.** Gaussian fit of the $RV$ distribution for IC 2602 after the $2\sigma$ clipping procedure

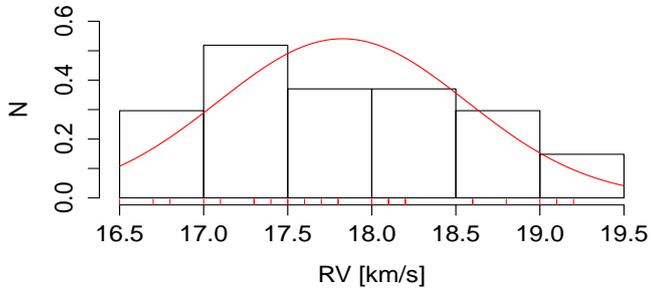

**Fig. B.23.** Gaussian fit of the $RV$ distribution for the final selection of IC 2602

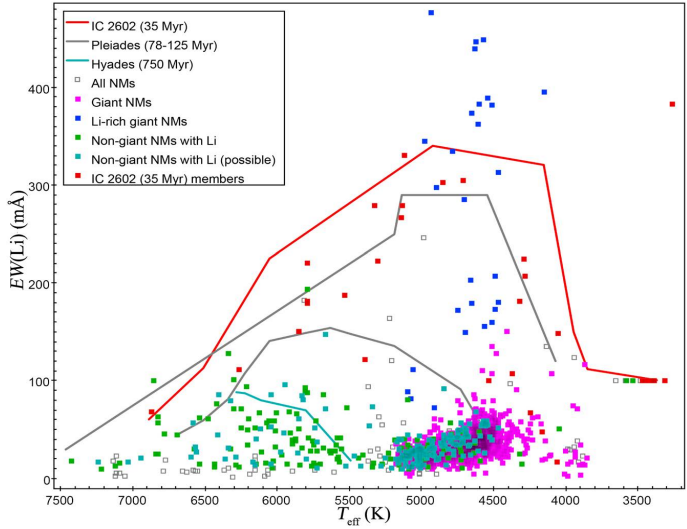

**Fig. B.24.** $EW$(Li)-versus-$T_{\rm eff}$ figure for IC 2602

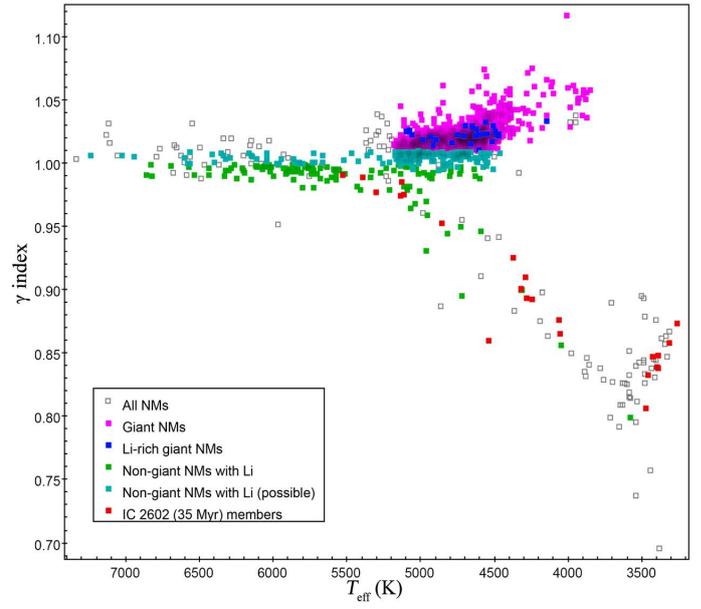

**Fig. B.25.** Gravity index $\gamma$ as a function of $T_{\rm eff}$ for IC 2602

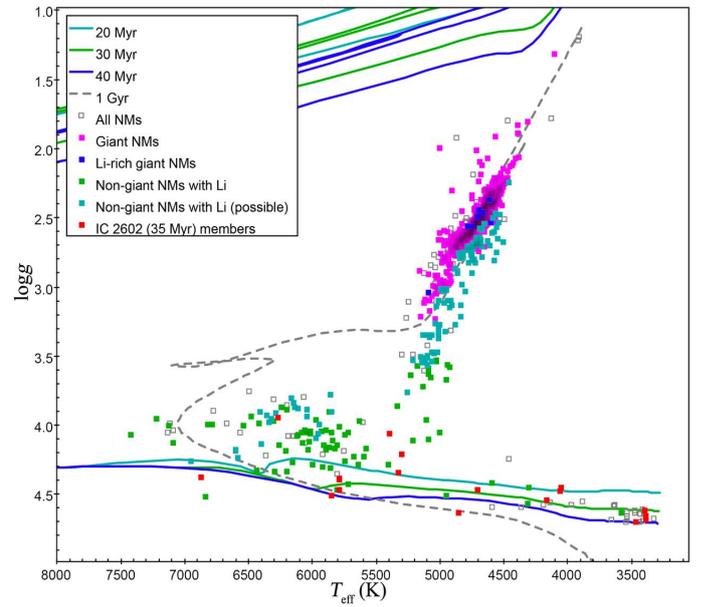

**Fig. B.26.** Kiel diagram for IC 2602

Appendix B.0.7: IC 4665

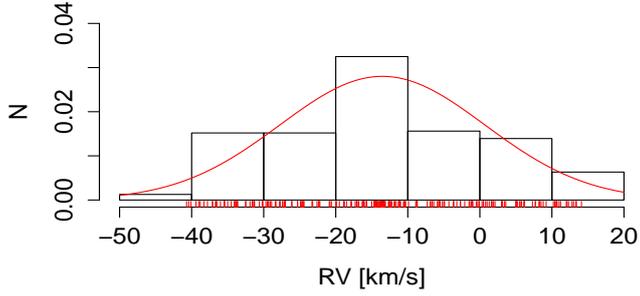

**Fig. B.27.** Gaussian fit of the *RV* distribution for IC 4665 after the $2\sigma$ clipping procedure

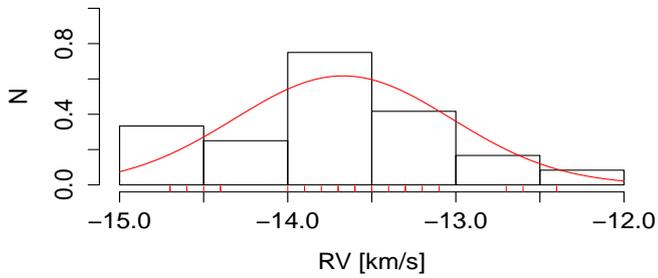

**Fig. B.28.** Gaussian fit of the *RV* distribution for the final selection of IC 4665

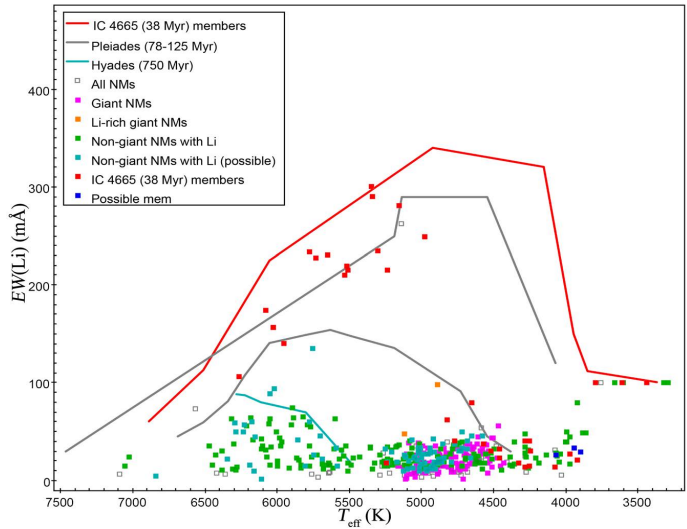

**Fig. B.29.** *EW*(Li)-versus-$T_{\rm eff}$ figure for IC 4665

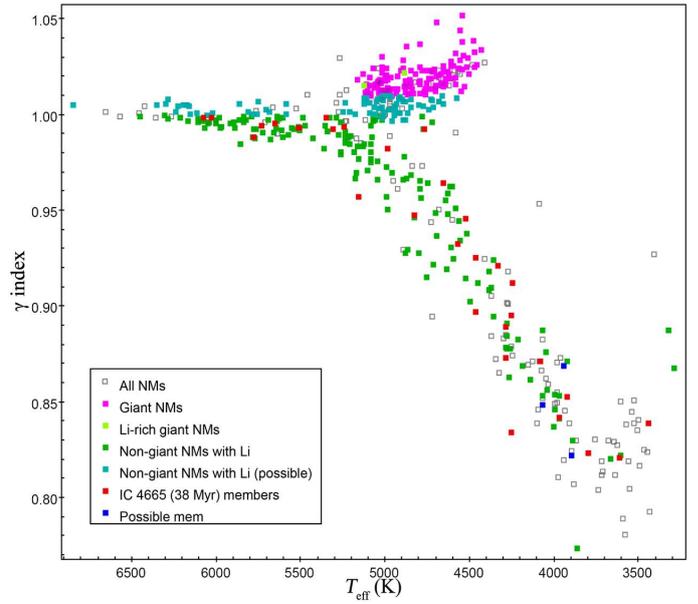

**Fig. B.30.** Gravity index $\gamma$ as a function of $T_{\rm eff}$ for IC 4665

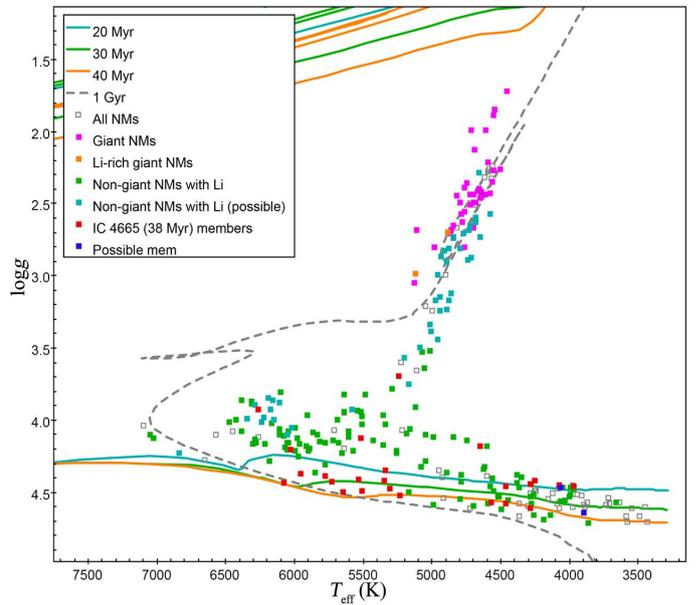

**Fig. B.31.** Kiel diagram for IC 4665

## Appendix B.0.8: NGC 2516

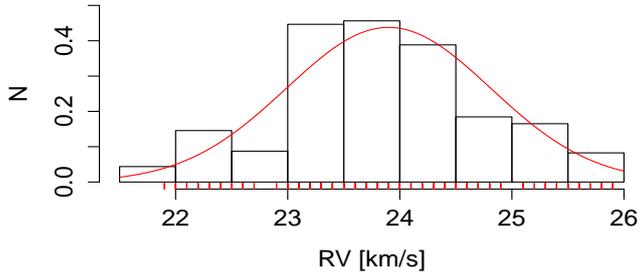

**Fig. B.32.** Gaussian fit of the *RV* distribution for NGC 2516

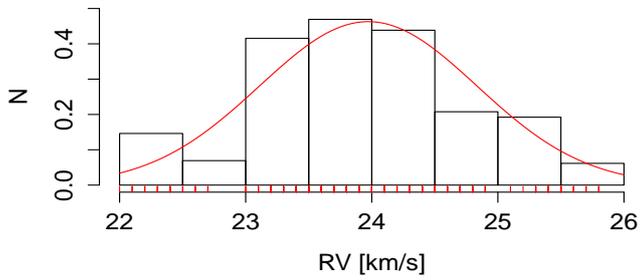

**Fig. B.33.** Gaussian fit of the *RV* distribution for the final selection of NGC 2516

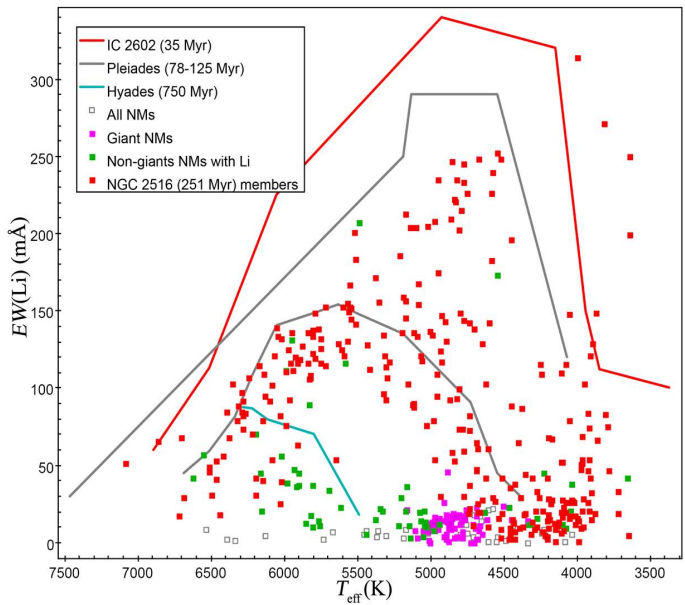

**Fig. B.34.** *EW*(Li)-versus-$T_{\rm eff}$ figure for NGC 2516

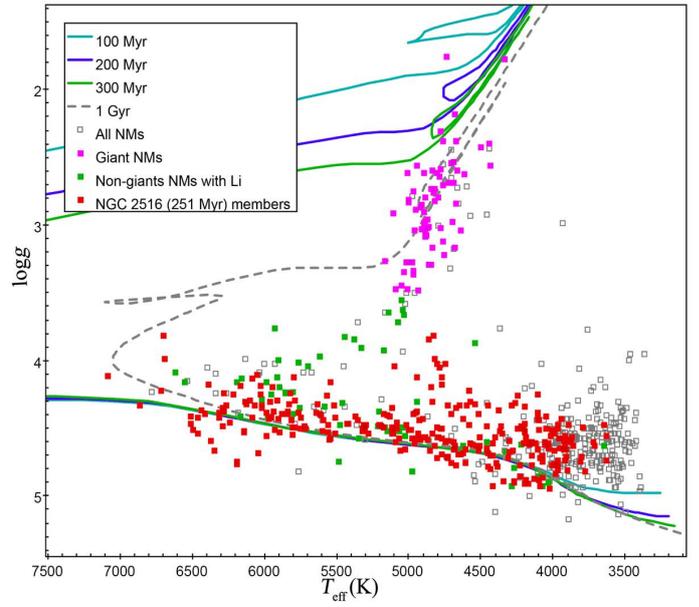

**Fig. B.35.** Kiel diagram for NGC 2516

Appendix B.0.9: NGC 6705

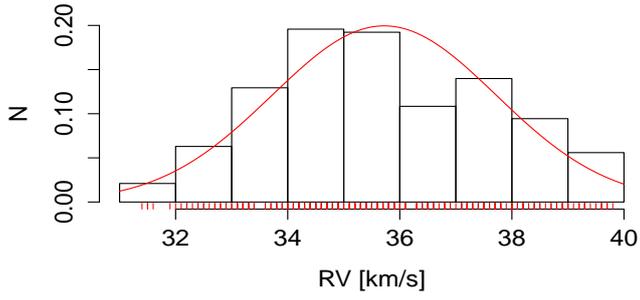

**Fig. B.36.** Gaussian fit of the *RV* distribution for NGC 6705

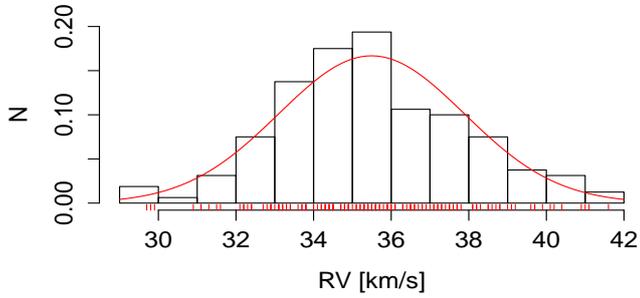

**Fig. B.37.** Gaussian fit of the *RV* distribution for the final selection of NGC 6705

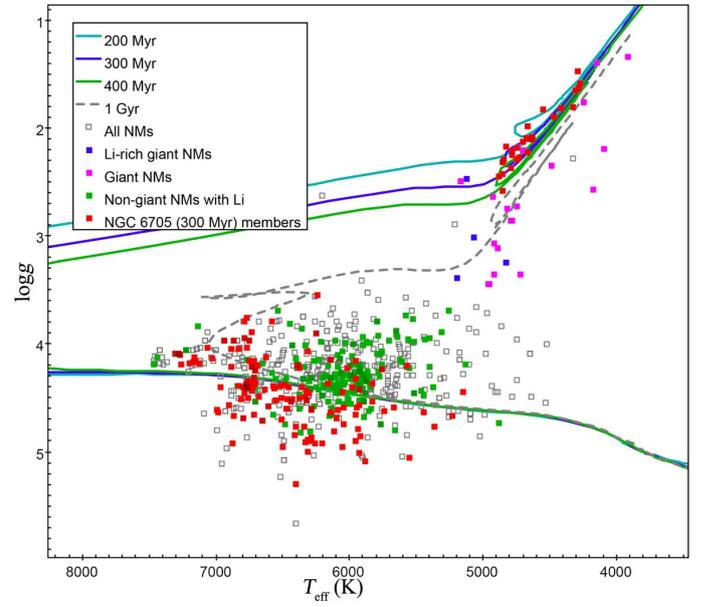

**Fig. B.39.** Kiel diagram for NGC 6705

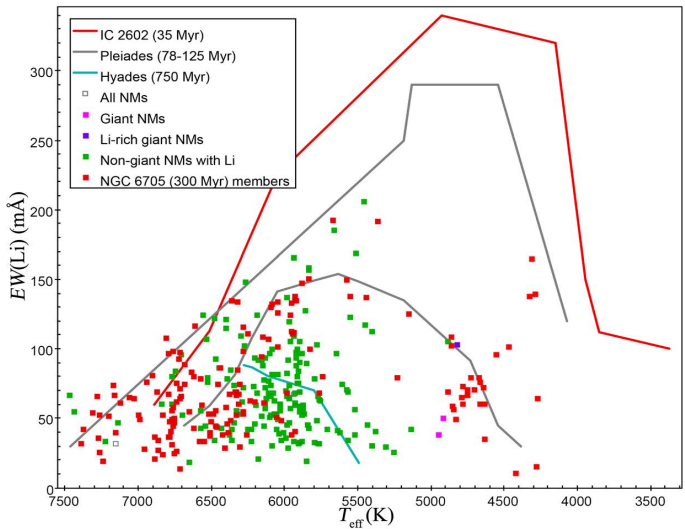

**Fig. B.38.** *EW*(Li)-versus-$T_{\rm eff}$ figure for NGC 6705

Appendix B.0.10: NGC 4815

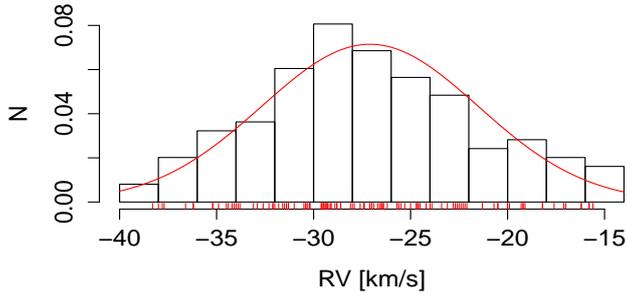

**Fig. B.40.** Gaussian fit of the *RV* distribution for NGC 4815

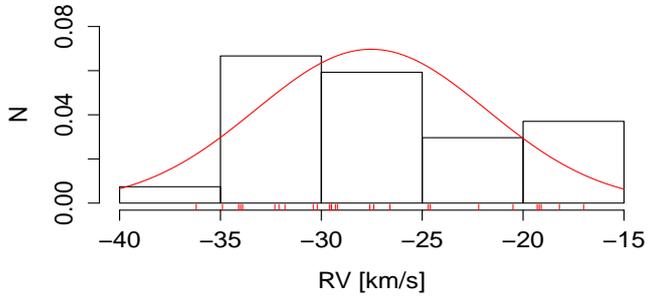

**Fig. B.41.** Gaussian fit of the *RV* distribution for the final selection of NGC 4815

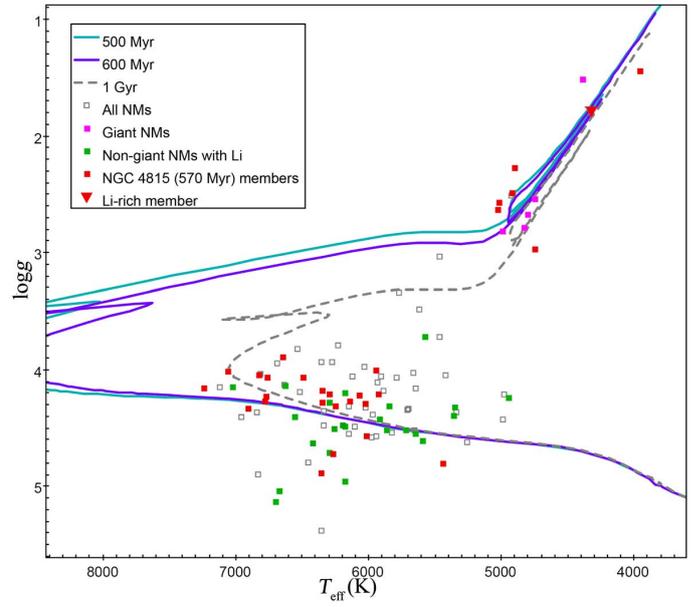

**Fig. B.43.** Kiel diagram for NGC 4815

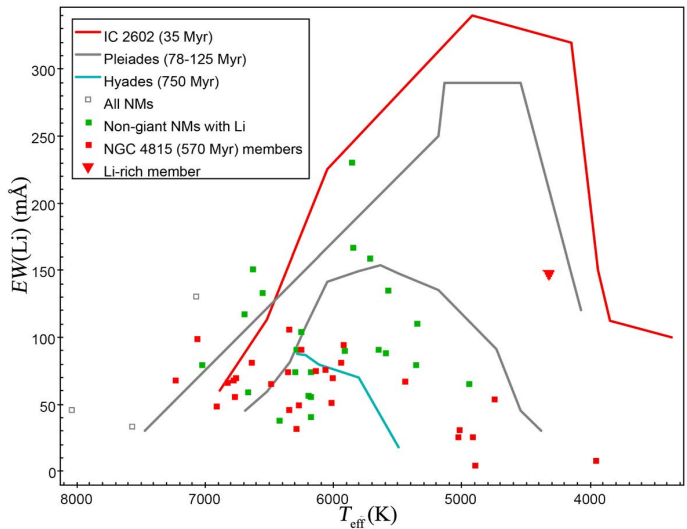

**Fig. B.42.** *EW*(Li)-versus-$T_{\rm eff}$ figure for NGC 4815

Appendix B.0.11: NGC 6633

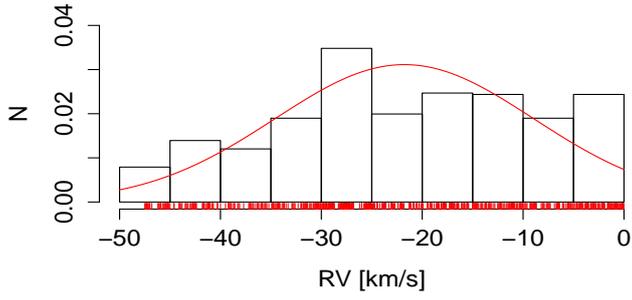

**Fig. B.44.** Gaussian fit of the *RV* distribution for NGC 6633

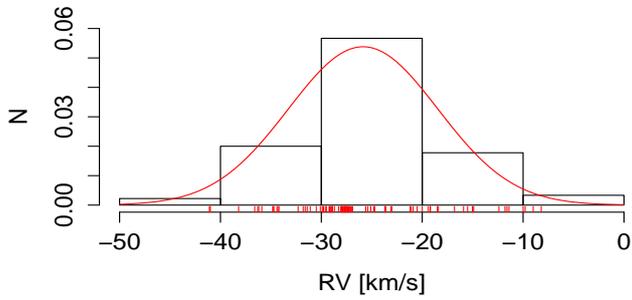

**Fig. B.45.** Gaussian fit of the *RV* distribution for the final selection of NGC 6633

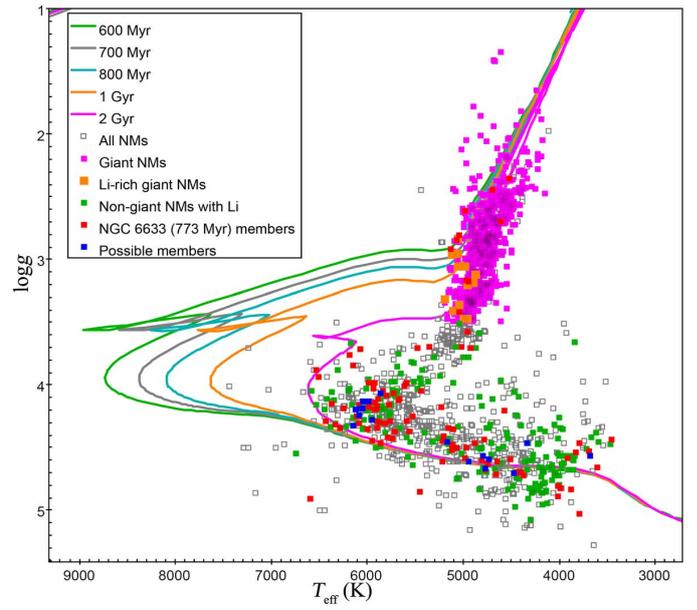

**Fig. B.47.** Kiel diagram for NGC 6633

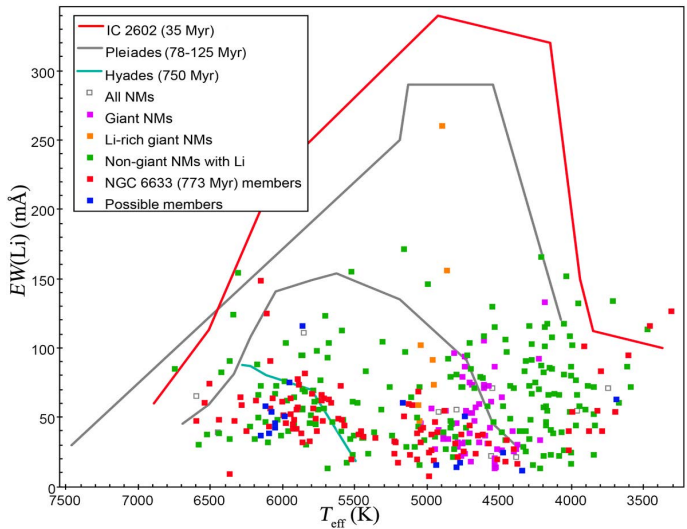

**Fig. B.46.** $EW(\text{Li})$-versus-$T_{\text{eff}}$ figure for NGC 6633

Appendix B.0.12: Trumpler 23

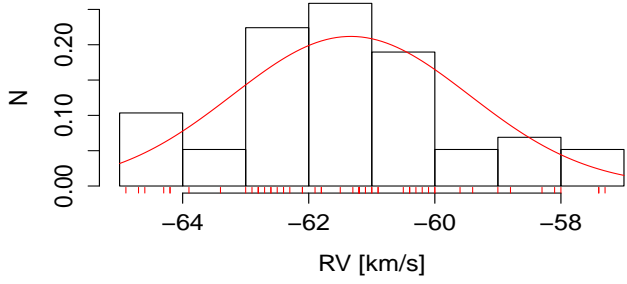

**Fig. B.48.** Gaussian fit of the $RV$ distribution for Trumpler 23

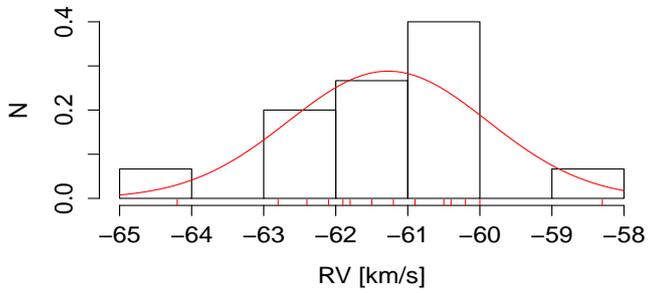

**Fig. B.49.** Gaussian fit of the $RV$ distribution for the final selection of Trumpler 23

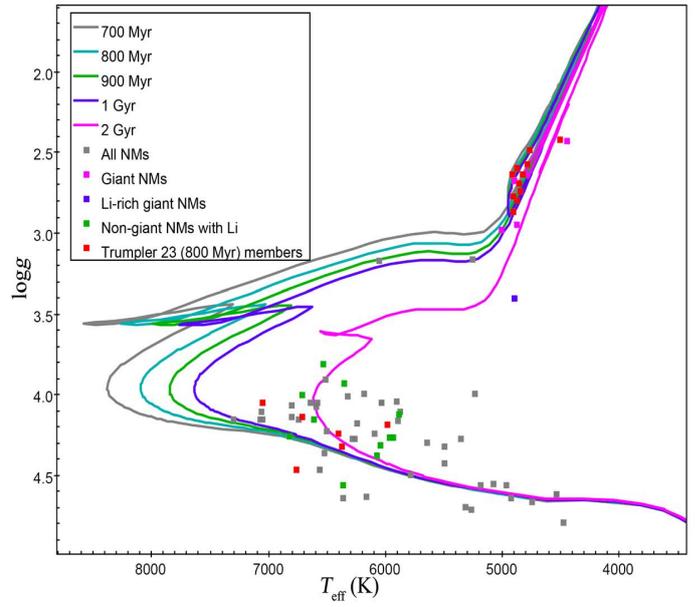

**Fig. B.51.** Kiel diagram for Trumpler 23

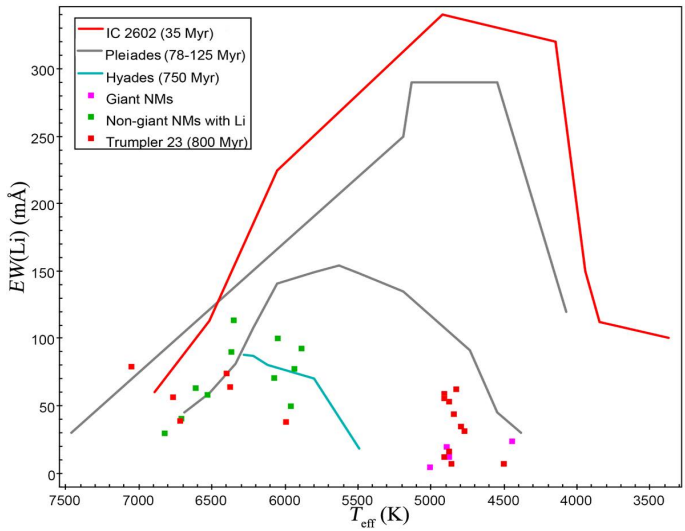

**Fig. B.50.** $EW$(Li)-versus-$T_{\rm eff}$ figure for Trumpler 23

## Appendix B.0.13: Berkeley 81

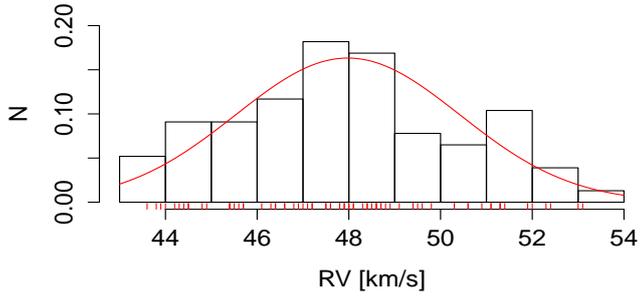

**Fig. B.52.** Gaussian fit of the *RV* distribution for Berkeley 81

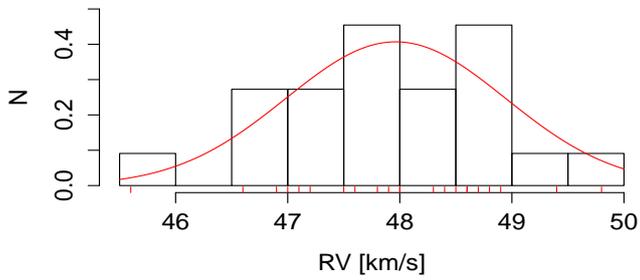

**Fig. B.53.** Gaussian fit of the *RV* distribution for the final selection of Berkeley 81

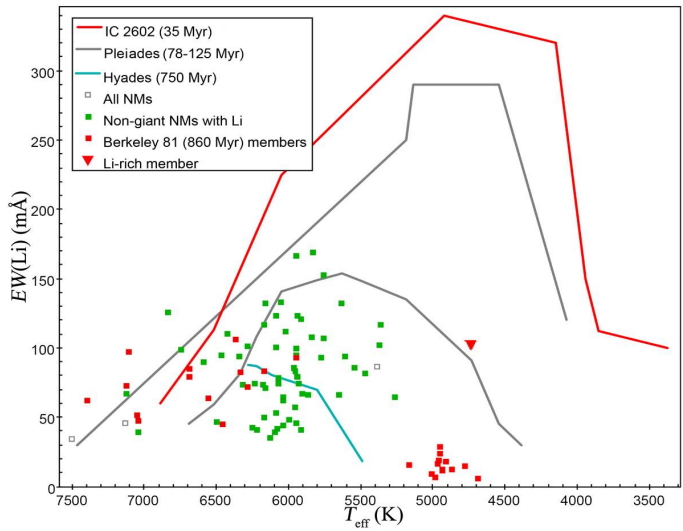

**Fig. B.54.** *EW*(Li)-versus-$T_{\rm eff}$ figure for Berkeley 81

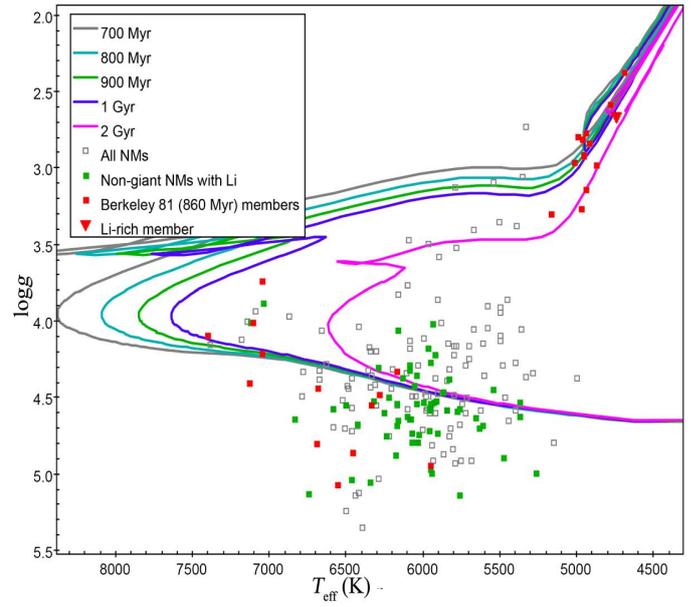

**Fig. B.55.** Kiel diagram for Berkeley 81

Appendix B.0.14: NGC 6005

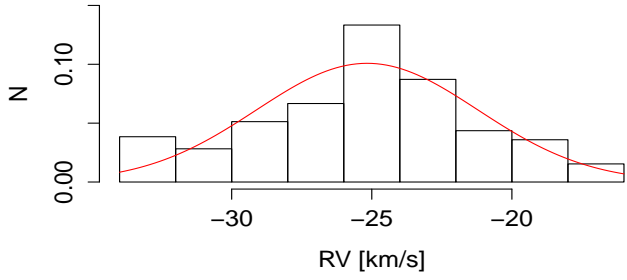

**Fig. B.56.** Gaussian fit of the *RV* distribution for NGC 6005

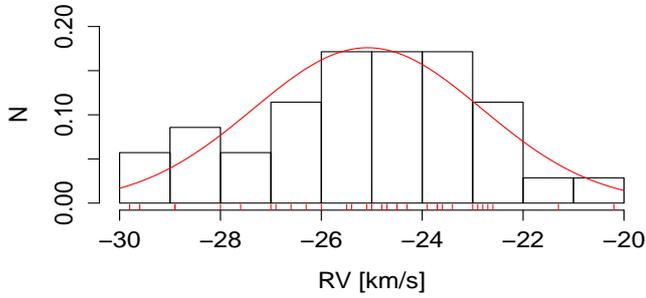

**Fig. B.57.** Gaussian fit of the *RV* distribution for the final selection of NGC 6005

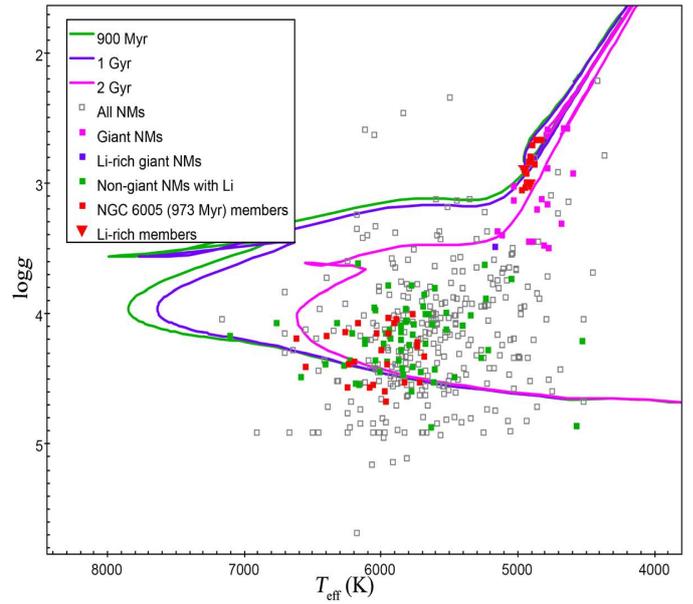

**Fig. B.59.** Kiel diagram for NGC 6005

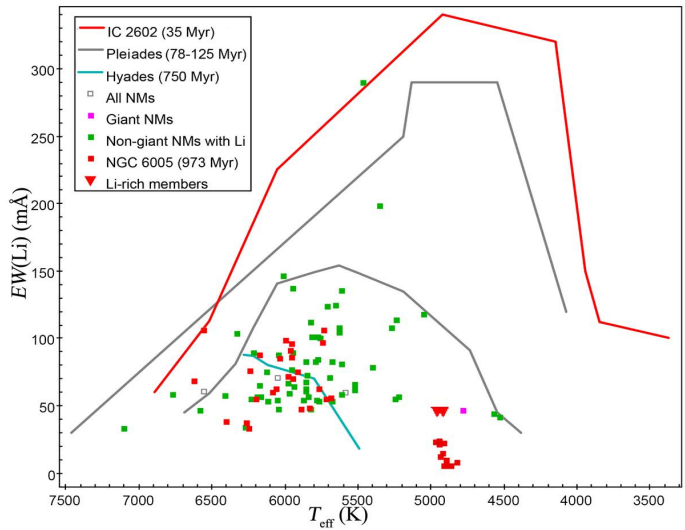

**Fig. B.58.** *EW*(Li)-versus-$T_{\text{eff}}$ figure for NGC 6005

Appendix B.0.15: NGC 6802

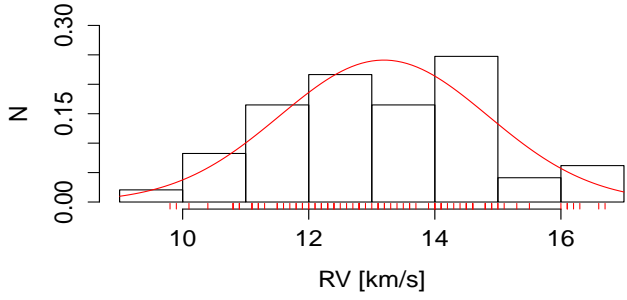

**Fig. B.60.** Gaussian fit of the *RV* distribution for NGC 6802

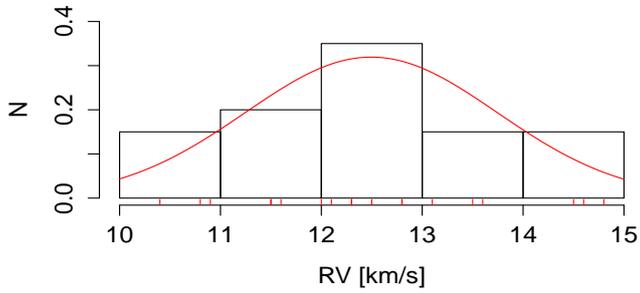

**Fig. B.61.** Gaussian fit of the *RV* distribution for the final selection of NGC 6802

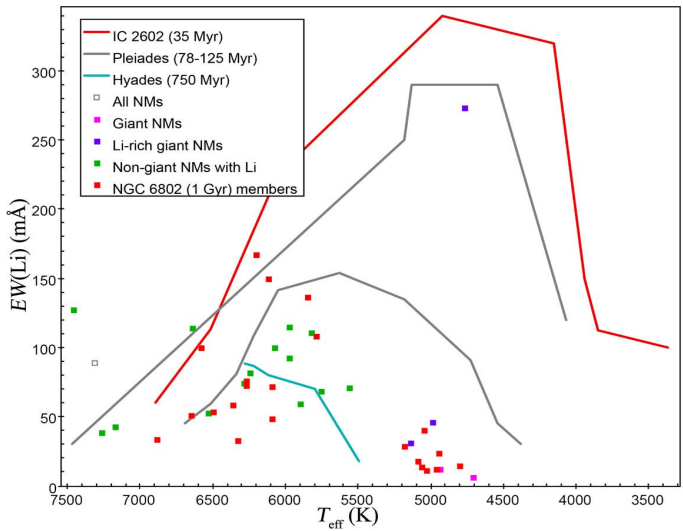

**Fig. B.62.** $EW$(Li)-versus-$T_\mathrm{eff}$ figure for NGC 6802

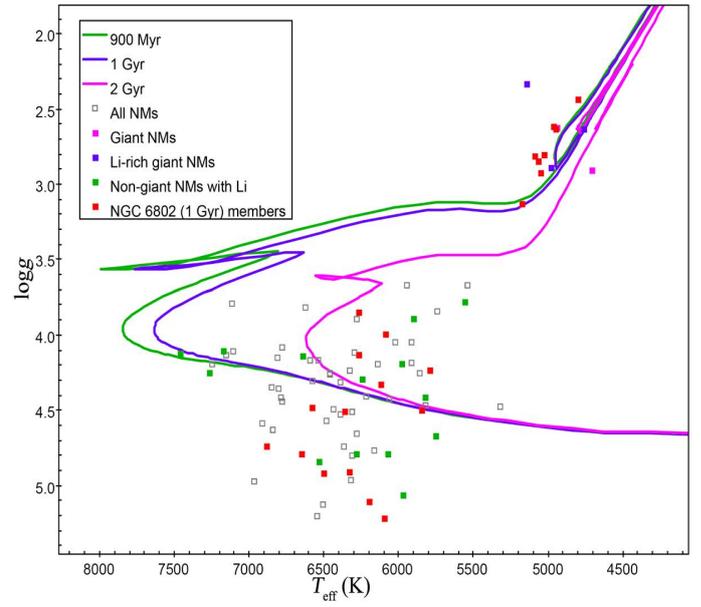

**Fig. B.63.** Kiel diagram for NGC 6802

Appendix B.0.16: Pismis 18

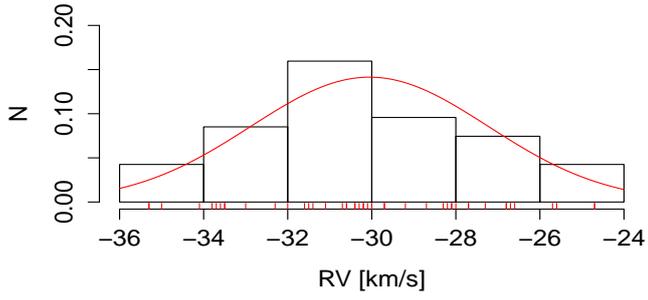

**Fig. B.64.** Gaussian fit of the *RV* distribution for Pismis 18

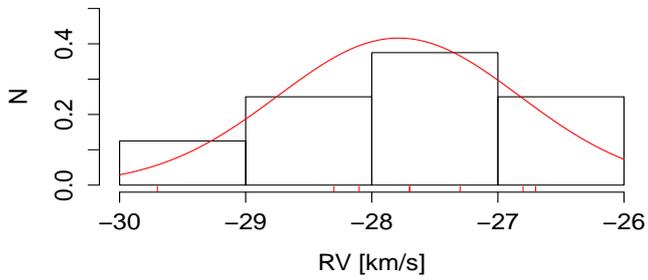

**Fig. B.65.** Gaussian fit of the *RV* distribution for the final selection of Pismis 18

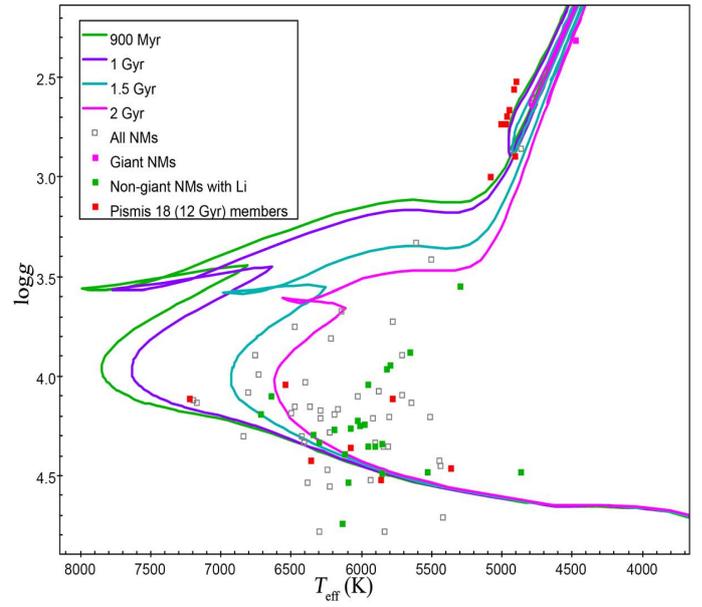

**Fig. B.67.** Kiel diagram for Pismis 18

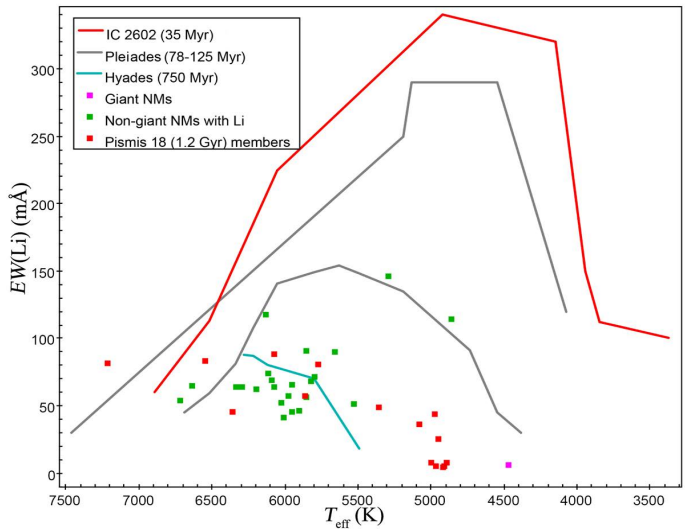

**Fig. B.66.** *EW*(Li)-versus-$T_{\rm eff}$ figure for Pismis 18

Appendix B.0.17: Trumpler 20

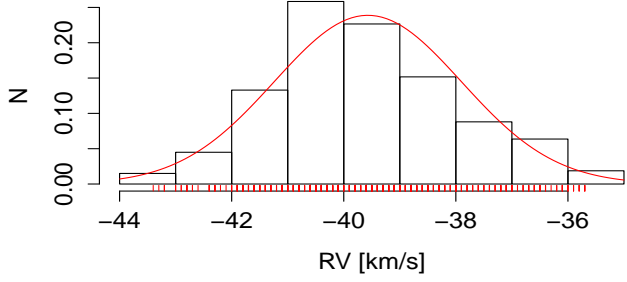

**Fig. B.68.** Gaussian fit of the *RV* distribution for Trumpler 20

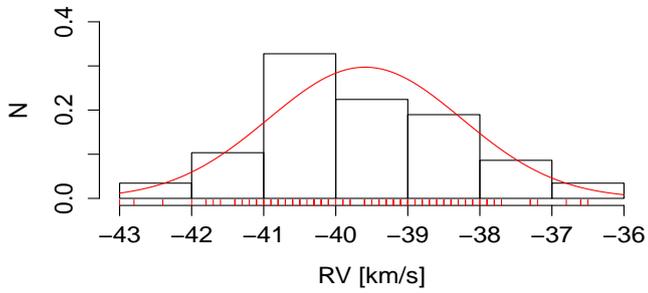

**Fig. B.69.** Gaussian fit of the *RV* distribution for the final selection of Trumpler 20

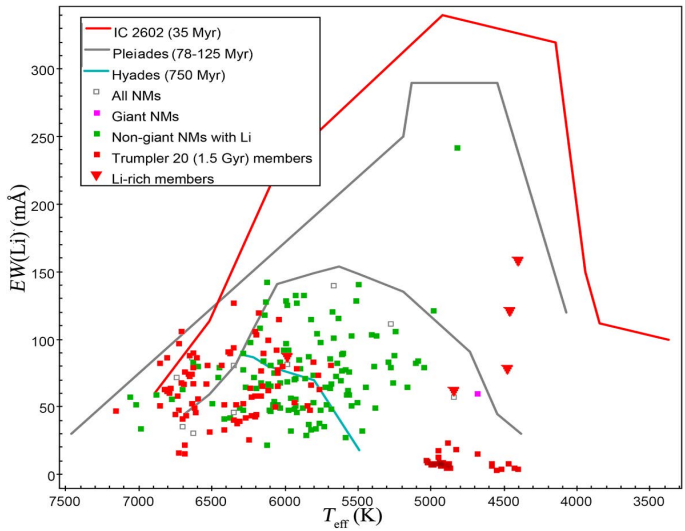

**Fig. B.70.** *EW*(Li)-versus-$T_{\rm eff}$ figure for Trumpler 20

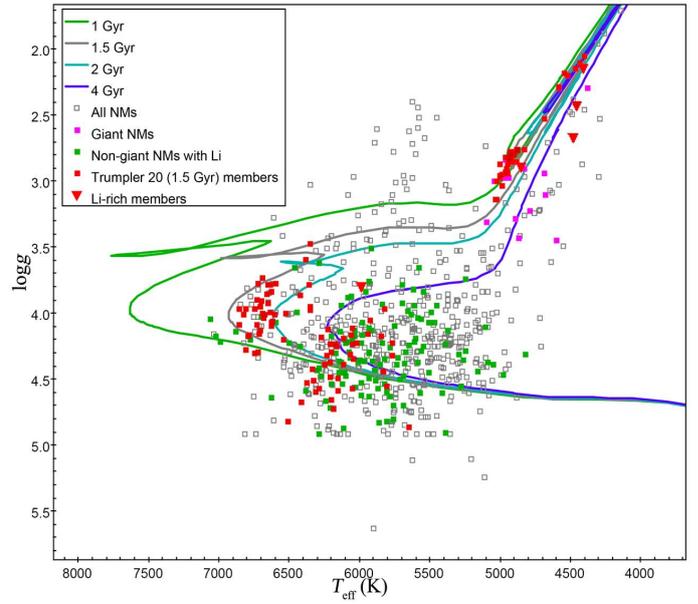

**Fig. B.71.** Kiel diagram for Trumpler 20

Appendix B.0.18: Berkeley 44

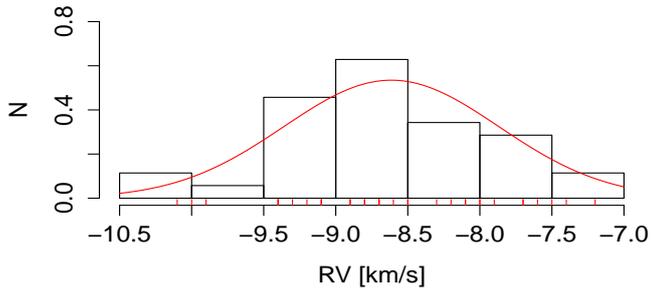

**Fig. B.72.** Gaussian fit of the *RV* distribution for Berkeley 44

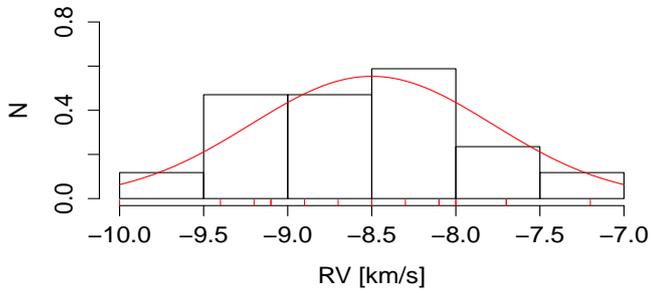

**Fig. B.73.** Gaussian fit of the *RV* distribution for the final selection of Berkeley 44

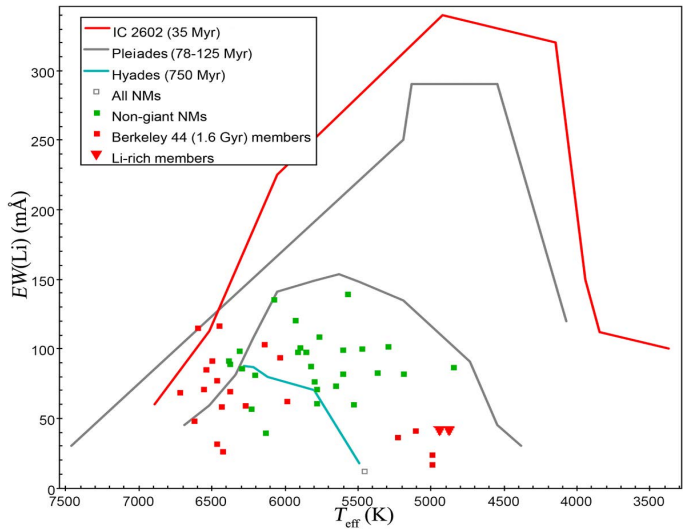

**Fig. B.74.** *EW*(Li)-versus-$T_{\rm eff}$ figure for Berkeley 44

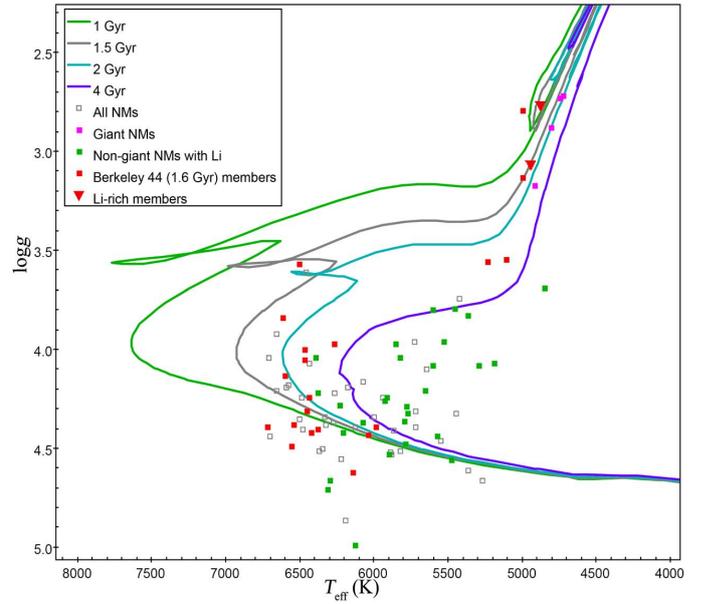

**Fig. B.75.** Kiel diagram for Berkeley 44

## Appendix B.0.19: M67

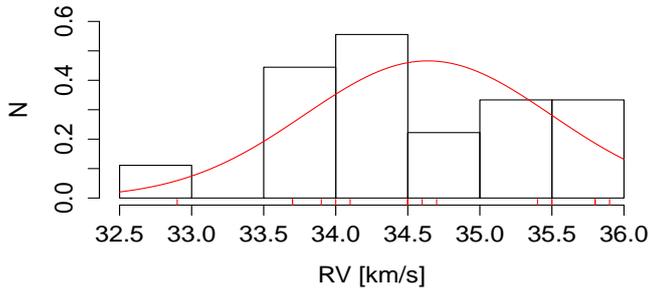

**Fig. B.76.** Gaussian fit of the *RV* distribution for M67

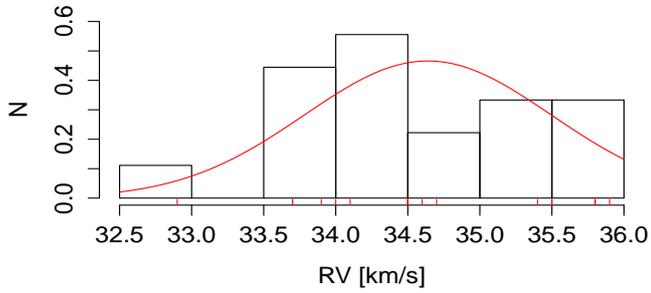

**Fig. B.77.** Gaussian fit of the *RV* distribution for the final selection of M67

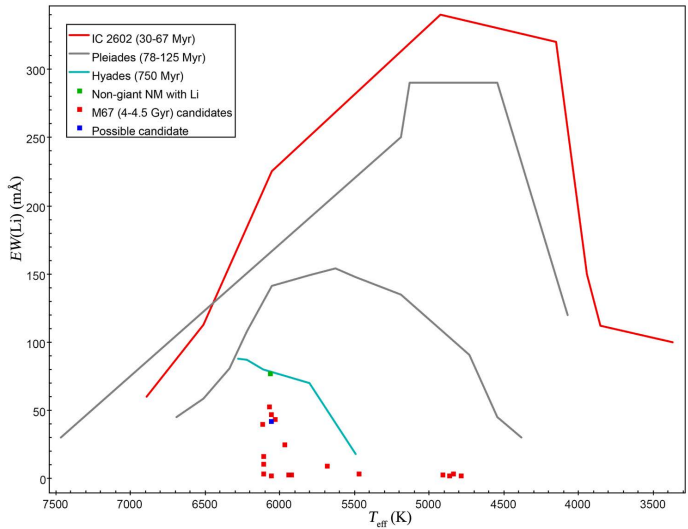

**Fig. B.78.** $EW(\mathrm{Li})$-versus-$T_{\mathrm{eff}}$ figure for M67

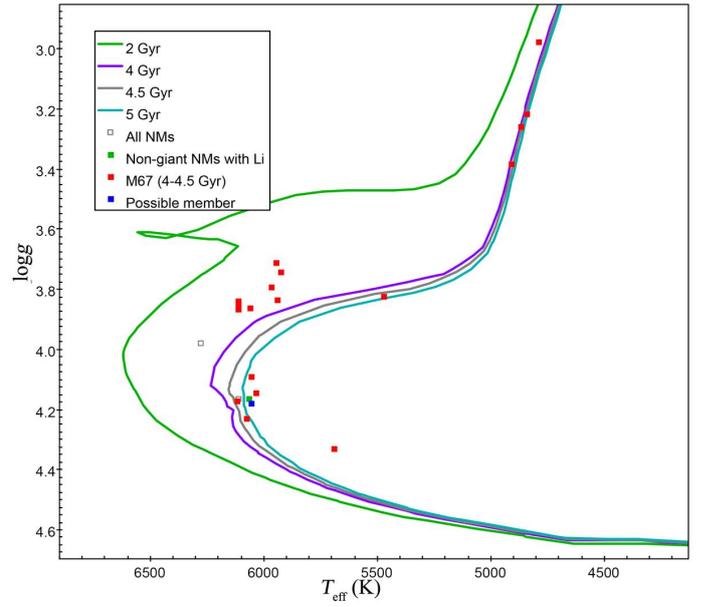

**Fig. B.79.** Kiel diagram for M67

Appendix B.0.20: NGC 2243

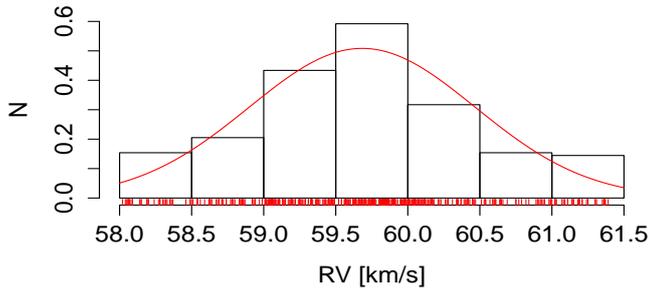

**Fig. B.80.** Gaussian fit of the *RV* distribution for NGC 2243

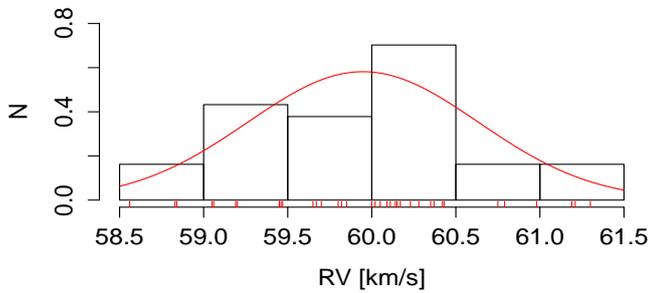

**Fig. B.81.** Gaussian fit of the *RV* distribution for the final selection of NGC 2243

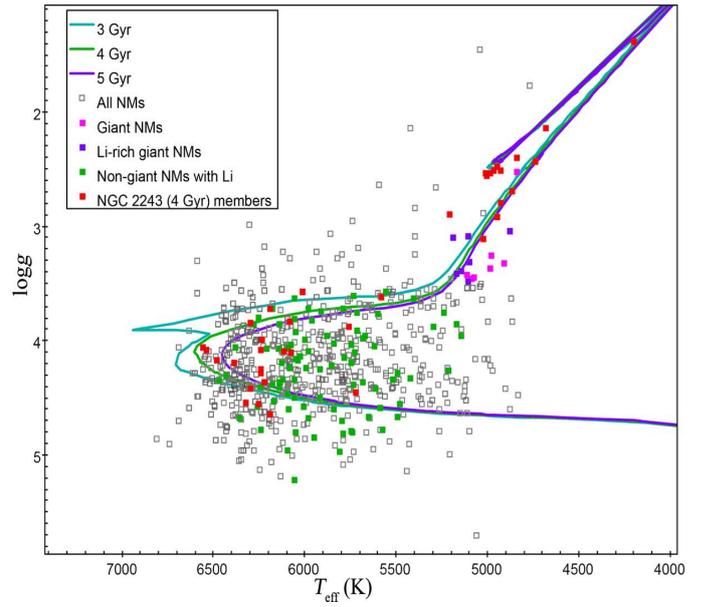

**Fig. B.83.** Kiel diagram for NGC 2243. Due to the metal-poor metallicity of this cluster we have considered PARSEC isochrones with Z=0.006 instead of the usual near-solar metallicity of Z=0.019

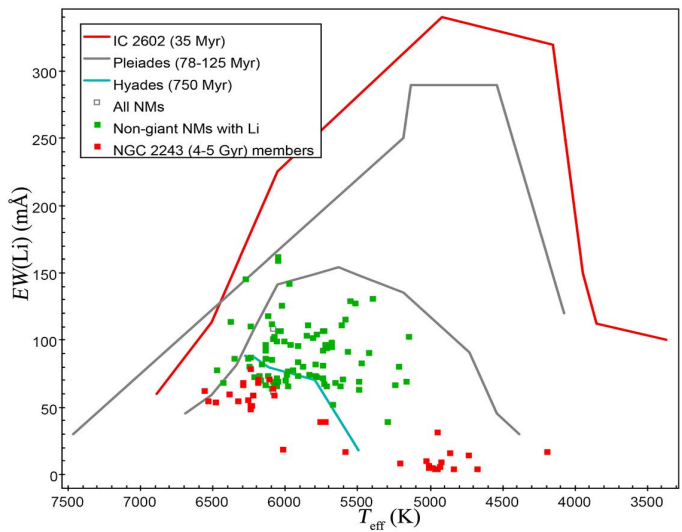

**Fig. B.82.** $EW$(Li)-versus-$T_{\rm eff}$ figure for NGC 2243

# Appendix C: Tables with the full lists of cluster stars with membership analysis

**Table C.1.** Rho Ophiuchi

| ID | CNAME | RV (km s$^{-1}$) | $T_{\text{eff}}$ (K) | $logg$ (dex) | $\gamma^a$ | [Fe/H] (dex) | $EW(\text{Li})^b$ (mÅ) | $EW(\text{Li})$ error flag$^c$ | $\gamma$ | $logg$ | RV | Li | [Fe/H] | Cánovas 2019$^d$ | Final$^e$ | Non-mem with Li$^f$ |
|---|---|---|---|---|---|---|---|---|---|---|---|---|---|---|---|---|
| 51483 | 16271005-2445343 | -36.2 ± 0.3 | 4035 ± 28 | 4.68 ± 0.14 | 0.850 ± 0.006 | -0.21 ± 0.05 | ... | ... | Y | Y | N | ... | Y | ... | n | ... |
| 55421 | 16271055-2457445 | -18.3 ± 0.2 | 4588 ± 76 | 2.44 ± 0.15 | 1.013 ± 0.003 | 0.03 ± 0.14 | <50 | 3 | Y | N | ... | ... | ... | ... | n | G |
| 55422 | 16271064-2456539 | -34.8 ± 0.3 | 5517 ± 68 | ... | 1.003 ± 0.005 | ... | 29 ± 7 | 1 | Y | ... | N | N | ... | ... | n | NG? |
| 55423 | 16271097-2455213 | -26.7 ± 0.2 | 5064 ± 342 | ... | 1.014 ± 0.004 | -0.20 ± 0.32 | 369 ± 8 | 1 | N | ... | ... | ... | ... | ... | n | Li-rich G |
| 51484 | 16271172-2400140 | 97.8 ± 0.3 | 5008 ± 245 | ... | 1.026 ± 0.012 | -0.36 ± 0.31 | <39 | 3 | N | ... | ... | ... | ... | ... | n | G |
| 55424 | 16271292-2459170 | -23.8 ± 0.2 | 5078 ± 71 | 3.52 ± 0.15 | 1.001 ± 0.003 | -0.16 ± 0.13 | <12 | 3 | Y | Y | N | N | Y | ... | n | NG? |
| 55425 | 16271482-2453337 | -49.2 ± 0.3 | 5107 ± 68 | ... | 0.983 ± 0.007 | -0.03 ± 0.06 | <32 | 3 | Y | ... | N | N | Y | ... | n | NG |
| 55426 | 16271487-2459400 | -76.8 ± 0.2 | 4917 ± 227 | ... | 1.001 ± 0.003 | -0.39 ± 0.10 | <20 | 3 | Y | ... | N | N | N | ... | n | NG? |
| 3317 | 16271513-2451388 | -6.1 ± 0.6 | 3602 ± 140 | 4.70 ± 0.12 | ... | -0.22 ± 0.11 | 537 ± 24 | 2 | ... | Y | Y | Y | Y | Y | Y | ... |
| 55427 | 16271555-2453497 | -38.3 ± 0.2 | 4577 ± 40 | 2.42 ± 0.18 | 1.024 ± 0.002 | 0.01 ± 0.03 | <45 | 3 | N | N | ... | ... | ... | ... | n | G |
| 55428 | 16271558-2452204 | -87.7 ± 0.3 | 5747 ± 141 | ... | 1.025 ± 0.011 | -0.29 ± 0.13 | <34 | 3 | N | ... | ... | ... | ... | ... | n | ... |
| 55429 | 16271631-2449065 | 17.9 ± 0.3 | 4405 ± 107 | ... | 1.005 ± 0.012 | -0.35 ± 0.12 | <38 | 3 | Y | ... | N | N | Y | ... | n | NG? |
| 55430 | 16271675-2454358 | 14.6 ± 0.2 | 6221 ± 68 | 4.39 ± 0.15 | 0.991 ± 0.004 | 0.18 ± 0.04 | 54 ± 7 | 1 | Y | Y | N | N | N | ... | n | NG |
| 51485 | 16271692-2406480 | -72.5 ± 0.2 | 4362 ± 175 | ... | 1.039 ± 0.006 | -0.13 ± 0.02 | <34 | 3 | N | ... | ... | ... | ... | ... | n | G |
| 55431 | 16271707-2447111 | 55.5 ± 0.2 | 4136 ± 362 | ... | 1.050 ± 0.007 | -0.35 ± 0.20 | ... | ... | N | ... | ... | ... | ... | ... | n | ... |
| 51486 | 16271716-2400281 | -83.8 ± 0.3 | 4713 ± 167 | ... | 1.022 ± 0.008 | -0.08 ± 0.09 | <29 | 3 | N | ... | ... | ... | ... | ... | n | G |
| 51487 | 16271781-2358454 | 11.8 ± 0.2 | 3794 ± 89 | ... | 1.053 ± 0.005 | ... | ... | ... | N | ... | ... | ... | ... | ... | n | ... |
| 55432 | 16271792-2458585 | -52.9 ± 0.2 | 5365 ± 93 | ... | 0.999 ± 0.004 | -0.04 ± 0.03 | 46 ± 7 | 1 | Y | ... | N | N | Y | ... | n | NG |
| 51488 | 16271832-2451141 | -34.8 ± 0.5 | 4758 ± 236 | ... | 1.000 ± 0.022 | -0.23 ± 0.19 | ... | ... | Y | ... | N | ... | Y | ... | n | ... |
| 55433 | 16271836-2454537$^g$ | -7.5 ± 0.2 | 3435 ± 73 | 4.22 ± 0.20 | 0.877 ± 0.002 | -0.31 ± 0.18 | 465 ± 16 | 1 | Y | Y | Y | Y | Y | Y | Y | ... |
| 55434 | 16271873-2457291 | 8.4 ± 0.2 | 5671 ± 79 | ... | 0.995 ± 0.002 | -0.44 ± 0.13 | <3 | 3 | Y | ... | N | N | N | ... | n | ... |
| 55435 | 16272033-2501019 | -27.0 ± 0.3 | 5796 ± 96 | ... | 1.003 ± 0.005 | 0.11 ± 0.05 | <13 | 3 | Y | ... | N | N | N | ... | n | NG? |
| 55436 | 16272219-2457085 | -27.6 ± 0.2 | 5636 ± 56 | 3.85 ± 0.11 | 1.000 ± 0.004 | 0.10 ± 0.04 | <19 | 3 | Y | Y | N | N | N | ... | n | NG? |
| 51489 | 16272270-2401594 | -63.7 ± 0.3 | 4426 ± 186 | ... | 1.053 ± 0.012 | -0.17 ± 0.05 | ... | ... | N | ... | ... | ... | ... | ... | n | ... |
| 3327 | 16291704-2424080 | -26.1 ± 0.6 | 4629 ± 9 | 2.51 ± 0.05 | ... | 0.09 ± 0.01 | <20 | 3 | ... | N | ... | ... | ... | ... | n | ... |
| 51551 | 16291858-2458507 | -55.1 ± 0.3 | 4369 ± 223 | ... | 1.064 ± 0.013 | -0.36 ± 0.19 | <25 | 3 | N | ... | ... | ... | ... | ... | n | G |
| 51552 | 16292139-2443106 | 163.2 ± 0.3 | 5011 ± 183 | ... | 1.015 ± 0.011 | ... | ... | ... | N | ... | ... | ... | ... | ... | n | ... |
| 3328 | 16292145-2457366 | 2.2 ± 0.6 | 5988 ± 24 | 3.91 ± 0.08 | ... | -0.22 ± 0.05 | 43 ± 9 | 2 | ... | Y | N | N | Y | ... | n | NG |
| 51553 | 16292156-2445094 | -46.1 ± 0.2 | 4433 ± 162 | ... | 1.032 ± 0.004 | 0.04 ± 0.08 | <53 | 3 | N | ... | ... | ... | ... | ... | n | G |
| 3329 | 16292199-2427194 | -22.1 ± 0.6 | 6002 ± 79 | 4.49 ± 0.09 | ... | -0.02 ± 0.03 | 94 ± 9 | 2 | ... | Y | N | N | Y | ... | n | NG |
| 51554 | 16292304-2420273 | 23.3 ± 0.3 | 5256 ± 187 | ... | 1.012 ± 0.007 | -0.29 ± 0.17 | <26 | 3 | N | ... | ... | ... | ... | ... | n | ... |
| 51555 | 16292339-2449462 | 9.8 ± 0.2 | 4637 ± 58 | 2.44 ± 0.14 | 1.017 ± 0.003 | -0.05 ± 0.09 | <27 | 3 | N | N | ... | ... | ... | ... | n | G |
| 51556 | 16292364-2443548 | -96.0 ± 0.3 | 4772 ± 86 | ... | 1.019 ± 0.007 | -0.04 ± 0.03 | <43 | 3 | N | ... | ... | ... | ... | ... | n | G |
| 51557 | 16292448-2422123 | -160.2 ± 0.2 | 4824 ± 14 | 2.42 ± 0.06 | 1.023 ± 0.006 | -0.16 ± 0.02 | <10 | 3 | N | N | ... | ... | ... | ... | n | G |
| 51558 | 16292703-2423183 | 167.3 ± 134.7 | ... | ... | ... | ... | ... | ... | ... | ... | ... | ... | ... | ... | n | ... |
| 51559 | 16292740-2448521 | -16.3 ± 0.2 | 4166 ± 318 | ... | 1.050 ± 0.005 | -0.19 ± 0.07 | <20 | 3 | N | ... | ... | ... | ... | ... | n | G |
| 51560 | 16292812-2419570 | 20.9 ± 0.3 | 4763 ± 186 | ... | 1.034 ± 0.007 | -0.07 ± 0.16 | <47 | 3 | N | ... | ... | ... | ... | ... | n | G |
| 51561 | 16293077-2503172 | -3.5 ± 0.7 | 4138 ± 94 | ... | 0.859 ± 0.018 | 0.00 ± 0.16 | ... | ... | Y | ... | Y | ... | Y | ... | n | ... |
| 3330 | 16293171-2500120 | -29.7 ± 0.6 | 3899 ± 71 | 1.21 ± 0.12 | ... | -0.07 ± 0.13 | <26 | 3 | ... | N | ... | ... | ... | ... | n | ... |
| 51418 | 16253492-2356280 | 67.1 ± 0.3 | 4144 ± 303 | ... | 1.063 ± 0.010 | -0.28 ± 0.07 | <21 | 3 | N | ... | ... | ... | ... | ... | n | G |
| 51419 | 16253553-2434011 | 34.3 ± 0.2 | 3603 ± 111 | ... | ... | ... | ... | ... | ... | ... | ... | ... | ... | ... | n | ... |
| 51420 | 16253610-2404033 | -32.4 ± 0.3 | 3983 ± 113 | ... | 1.047 ± 0.013 | ... | 117 ± 8 | 1 | N | ... | ... | ... | ... | ... | n | G |
| 51421 | 16253616-2500283 | -77.3 ± 0.2 | 4297 ± 265 | ... | 1.048 ± 0.004 | -0.36 ± 0.21 | ... | ... | N | ... | ... | ... | ... | ... | n | ... |
| 51422 | 16253780-2413436 | -16.5 ± 0.6 | 3542 ± 107 | ... | ... | ... | <100 | 3 | ... | ... | ... | ... | ... | ... | n | ... |
| 51423 | 16253789-2443065 | 12.3 ± 0.3 | 4673 ± 30 | 2.37 ± 0.09 | 1.019 ± 0.007 | -0.14 ± 0.15 | <38 | 3 | N | N | ... | ... | ... | ... | n | G |
| 51424 | 16253881-2436149 | 44.7 ± 0.5 | 5987 ± 199 | 4.06 ± 0.14 | 0.997 ± 0.010 | -0.13 ± 0.14 | 61 ± 11 | 1 | Y | Y | N | N | Y | ... | n | NG |
| 51425 | 16253958-2426349 | -3.8 ± 1.5 | ... | ... | ... | ... | ... | ... | ... | ... | ... | ... | ... | ... | n | ... |
| 51426 | 16254043-2356134 | -32.3 ± 0.3 | 4708 ± 186 | ... | 1.028 ± 0.009 | -0.38 ± 0.01 | <41 | 3 | N | ... | ... | ... | ... | ... | n | G |
| 51427 | 16254112-2453512 | -157.3 ± 0.3 | 4731 ± 114 | 2.02 ± 0.03 | 1.019 ± 0.007 | -0.51 ± 0.04 | <21 | 3 | N | N | ... | ... | ... | ... | n | G |
| 51428 | 16254136-2445446 | -31.6 ± 0.3 | 5836 ± 38 | 4.18 ± 0.14 | 0.995 ± 0.007 | -0.04 ± 0.02 | ... | ... | Y | Y | N | ... | Y | ... | n | ... |
| 51429 | 16254148-2448598 | -55.8 ± 0.2 | 5170 ± 112 | ... | 1.022 ± 0.005 | ... | <30 | 3 | N | ... | ... | ... | ... | ... | n | G |
| 51430 | 16254338-2449121 | 22.1 ± 0.2 | 6128 ± 123 | 4.12 ± 0.02 | 0.999 ± 0.003 | 0.06 ± 0.06 | <23 | 3 | Y | Y | N | N | N | ... | n | NG |
| 51431 | 16254432-2441062 | -32.3 ± 0.3 | 4487 ± 159 | ... | 1.043 ± 0.012 | 0.01 ± 0.04 | ... | ... | N | ... | ... | ... | ... | ... | n | ... |
| 51432 | 16254767-2437394 | -5.4 ± 0.3 | 3470 ± 19 | ... | 0.881 ± 0.009 | -0.26 ± 0.13 | 621 ± 10 | 1 | Y | ... | Y | Y | Y | Y | Y | ... |
| 51433 | 16254825-2442209 | -40.2 ± 0.2 | 4637 ± 63 | ... | 1.025 ± 0.006 | -0.08 ± 0.09 | <33 | 3 | N | ... | ... | ... | ... | ... | n | G |
| 51434 | 16254960-2452244 | -64.1 ± 0.2 | 4655 ± 84 | ... | 1.022 ± 0.008 | ... | <26 | 3 | N | ... | ... | ... | ... | ... | n | G |
| 51435 | 16255032-2400083 | -595.1 ± 377.0 | ... | ... | ... | ... | ... | ... | ... | ... | ... | ... | ... | ... | n | ... |
| 51436 | 16255056-2447358 | 99.5 ± 0.2 | 4761 ± 331 | ... | 1.025 ± 0.003 | -0.53 ± 0.34 | <8 | 3 | N | ... | ... | ... | ... | ... | n | G |
| 51437 | 16255104-2448284 | 37.3 ± 0.3 | 5975 ± 137 | ... | 0.989 ± 0.007 | -0.22 ± 0.03 | <13 | 3 | Y | ... | N | N | Y | ... | n | NG |

**Table C.1.** continued.

| ID | CNAME | RV (km s$^{-1}$) | $T_{\rm eff}$ (K) | logg (dex) | $\gamma^a$ | [Fe/H] (dex) | EW(Li)$^b$ (mÅ) | EW(Li) error flag$^c$ | Membership $\gamma$ | logg | RV | Li | [Fe/H] | Cánovas 2019$^d$ | Final$^e$ | Non-mem with Li$^f$ |
|---|---|---|---|---|---|---|---|---|---|---|---|---|---|---|---|---|
| 51438 | 16255241-2456260 | -13.6 ± 0.2 | 4470 ± 141 | … | 1.029 ± 0.005 | -0.38 ± 0.16 | <11 | 3 | N | … | … | … | … | … | n | G |
| 51439 | 16255277-2358185 | 3.2 ± 0.2 | 4859 ± 174 | … | 1.009 ± 0.004 | -0.11 ± 0.19 | <19 | 3 | Y | … | N | N | Y | … | n | NG? |
| 51440 | 16255447-2458396 | -21.2 ± 0.2 | 4528 ± 134 | … | 1.032 ± 0.004 | 0.06 ± 0.11 | <46 | 3 | N | … | … | … | … | … | n | G |
| 51441 | 16255754-2442082 | -26.0 ± 0.3 | 4906 ± 6 | 2.69 ± 0.10 | 1.023 ± 0.007 | -0.04 ± 0.06 | <32 | 3 | N | N | … | … | … | … | n | G |
| 51372 | 16245715-2444473 | -101.2 ± 0.2 | 4679 ± 163 | … | 1.017 ± 0.005 | -0.01 ± 0.01 | <33 | 3 | N | … | … | … | … | … | n | G |
| 51373 | 16245763-2455015 | 4.1 ± 0.3 | 4533 ± 153 | … | 1.029 ± 0.007 | -0.16 ± 0.15 | <36 | 3 | N | … | … | … | … | … | n | … |
| 51374 | 16245974-2456008$^g$ | -5.4 ± 0.3 | 3365 ± 15 | … | 0.866 ± 0.009 | -0.24 ± 0.14 | 466 ± 8 | 1 | Y | … | Y | Y | Y | … | Y | … |
| 51375 | 16250148-2428588 | -48.9 ± 0.3 | 3526 ± 112 | … | … | … | <100 | 3 | … | … | … | … | … | … | n | … |
| 51376 | 16250245-2403091 | -65.3 ± 0.4 | 4641 ± 53 | … | 0.967 ± 0.017 | 0.09 ± 0.25 | <27 | 3 | Y | … | N | N | N | … | n | NG |
| 51377 | 16250278-2458473 | -147.8 ± 0.3 | 4714 ± 142 | 1.88 ± 0.20 | 1.028 ± 0.007 | -0.55 ± 0.10 | … | … | N | N | … | … | … | … | n | … |
| 51378 | 16250320-2500570 | 40.4 ± 0.2 | 4513 ± 160 | … | 1.026 ± 0.005 | -0.14 ± 0.20 | <24 | 3 | N | … | … | … | … | … | n | G |
| 51379 | 16250382-2359196 | -75.3 ± 0.5 | 4815 ± 124 | 2.55 ± 0.08 | 1.025 ± 0.016 | -0.11 ± 0.14 | <42 | 3 | N | … | … | … | … | … | n | G |
| 51380 | 16250626-2453469 | -11.1 ± 0.3 | 4585 ± 35 | 1.80 ± 0.06 | 1.033 ± 0.010 | -0.37 ± 0.07 | … | … | N | N | … | … | … | … | n | … |
| 51381 | 16250674-2440291 | 13.7 ± 0.7 | 6031 ± 41 | 4.15 ± 0.16 | 0.993 ± 0.013 | -0.11 ± 0.08 | <34 | 3 | Y | Y | N | N | Y | … | n | NG |
| 51382 | 16250737-2457014 | -32.9 ± 0.2 | 4762 ± 268 | … | 1.022 ± 0.007 | -0.16 ± 0.22 | <29 | 3 | N | … | … | … | … | … | n | G |
| 51383 | 16250739-2458302 | 81.4 ± 0.3 | 4514 ± 126 | … | 1.041 ± 0.009 | -0.32 ± 0.27 | <18 | 3 | N | … | … | … | … | … | n | G |
| 51384 | 16250769-2358022 | 35.0 ± 0.3 | 5103 ± 10 | … | 1.025 ± 0.008 | -0.10 ± 0.12 | … | … | N | … | … | … | … | … | n | … |
| 51385 | 16250887-2452163 | -136.3 ± 0.3 | 4641 ± 226 | … | 1.035 ± 0.008 | -0.43 ± 0.19 | <17 | 3 | N | … | … | … | … | … | n | G |
| 51386 | 16251174-2457360 | -40.7 ± 0.3 | 4674 ± 293 | … | 1.026 ± 0.008 | -0.29 ± 0.17 | <22 | 3 | N | … | … | … | … | … | n | G |
| 51387 | 16251188-2437081 | -23.0 ± 0.2 | 4055 ± 147 | … | 1.056 ± 0.009 | -0.09 ± 0.10 | … | … | N | … | … | … | … | … | n | … |
| 51388 | 16251207-2448591 | -14.5 ± 0.3 | 4353 ± 189 | … | 1.040 ± 0.010 | -0.45 ± 0.12 | 32 ± 5 | 1 | N | … | … | … | … | … | n | G |
| 51389 | 16251208-2358303 | -51.7 ± 0.3 | 4557 ± 170 | … | 1.017 ± 0.009 | -0.01 ± 0.07 | <51 | 3 | N | … | … | … | … | … | n | G |
| 3308 | 16251282-2449168 | 10.4 ± 0.6 | 6017 ± 109 | 4.35 ± 0.05 | … | 0.29 ± 0.08 | 54 ± 6 | 2 | … | Y | N | N | N | … | n | NG |
| 51390 | 16251430-2455450 | -131.9 ± 0.3 | 4812 ± 217 | … | 1.047 ± 0.009 | … | 34 ± 6 | 1 | N | … | … | … | … | … | n | G |
| 3309 | 16251469-2456069$^g$ | -1.1 ± 0.6 | 4307 ± 107 | 4.56 ± 0.16 | … | -0.08 ± 0.14 | 575 ± 27 | 2 | … | Y | N | Y | Y | Y | Y | … |
| 51391 | 16251661-2404405 | -83.2 ± 0.3 | 4575 ± 267 | … | 1.035 ± 0.012 | -0.05 ± 0.15 | <62 | 3 | N | … | … | … | … | … | n | G |
| 51392 | 16251710-2443247 | 28.3 ± 0.2 | 4163 ± 318 | … | 1.050 ± 0.007 | -0.26 ± 0.13 | … | … | N | … | … | … | … | … | n | … |
| 51393 | 16251732-2441369 | 12.8 ± 0.3 | 5083 ± 5 | 3.23 ± 0.10 | 1.009 ± 0.007 | -0.03 ± 0.05 | <28 | 3 | Y | N | N | N | Y | … | n | NG? |
| 51442 | 16255761-2441014 | -132.4 ± 0.3 | 4725 ± 281 | … | 1.058 ± 0.013 | -0.36 ± 0.09 | <33 | 3 | N | … | … | … | … | … | n | G |
| 51443 | 16255766-2358163 | 63.7 ± 0.2 | 5639 ± 130 | … | 1.003 ± 0.004 | -0.37 ± 0.17 | 59 ± 9 | 1 | Y | … | N | N | N | … | n | NG? |
| 51444 | 16255893-2452483 | -3.8 ± 0.6 | 3313 ± 25 | … | 0.883 ± 0.013 | -0.27 ± 0.14 | 671 ± 13 | 1 | Y | … | Y | Y | Y | … | Y | … |
| 51445 | 16255920-2501142 | -63.6 ± 0.2 | 4598 ± 85 | … | 1.041 ± 0.003 | … | <12 | 3 | N | … | … | … | … | … | n | G |
| 51446 | 16255965-2421223 | -5.5 ± 0.3 | 3318 ± 64 | … | 0.941 ± 0.005 | -0.26 ± 0.14 | 565 ± 22 | 1 | Y | … | Y | Y | Y | … | Y | … |
| 51447 | 16255992-2450060 | -144.3 ± 0.2 | 4659 ± 220 | … | 1.050 ± 0.006 | -0.26 ± 0.14 | <24 | 3 | N | … | … | … | … | … | n | G |
| 51448 | 16260207-2402032 | -24.7 ± 0.2 | 3580 ± 131 | … | … | … | … | … | … | … | … | … | … | … | n | … |
| 51449 | 16260289-2408474 | -25.1 ± 0.5 | 6170 ± 35 | … | 1.021 ± 0.010 | -0.28 ± 0.30 | 76 ± 6 | 1 | N | … | … | … | … | … | n | … |
| 51450 | 16260371-2408290 | -52.8 ± 1.1 | … | … | … | … | … | … | … | … | … | … | … | … | n | … |
| 51451 | 16260544-2355408 | -7.2 ± 0.2 | 3338 ± 29 | … | 0.896 ± 0.008 | -0.28 ± 0.14 | 649 ± 20 | 1 | Y | … | Y | Y | Y | … | Y | … |
| 51452 | 16260558-2452339 | 13.1 ± 0.3 | 4567 ± 72 | 1.87 ± 0.14 | 1.028 ± 0.008 | -0.35 ± 0.15 | <42 | 3 | N | N | … | … | … | … | n | G |
| 51453 | 16260894-2401280 | -119.5 ± 0.2 | 4724 ± 264 | … | 1.023 ± 0.004 | -0.32 ± 0.10 | <25 | 3 | N | … | … | … | … | … | n | G |
| 3310 | 16261089-2452154 | 80.3 ± 0.6 | 4491 ± 51 | 1.02 ± 0.05 | … | -1.31 ± 0.11 | <22 | 3 | … | N | … | … | … | … | n | … |
| 51454 | 16261154-2459460 | -28.9 ± 0.2 | 5845 ± 104 | … | 1.000 ± 0.003 | … | 58 ± 7 | 1 | Y | … | N | N | … | … | n | NG? |
| 51455 | 16261348-2400410 | -76.3 ± 0.3 | 4931 ± 159 | … | 1.022 ± 0.009 | -0.44 ± 0.20 | <50 | 3 | N | … | … | … | … | … | n | G |
| 51456 | 16261383-2448469 | -99.0 ± 0.4 | … | … | … | … | … | … | … | … | … | … | … | … | n | … |
| 51457 | 16261671-2400069 | -31.7 ± 0.2 | 4761 ± 39 | 2.55 ± 0.12 | 1.027 ± 0.004 | -0.07 ± 0.04 | <23 | 3 | N | N | … | … | … | … | n | G |
| 3311 | 16261706-2420216$^g$ | -5.4 ± 0.6 | 4482 ± 56 | 4.48 ± 0.22 | … | -0.06 ± 0.02 | 501 ± 15 | 2 | … | Y | Y | Y | Y | Y | Y | … |
| 51458 | 16261836-2453479 | 43.9 ± 0.2 | 4759 ± 369 | … | 1.019 ± 0.005 | -0.67 ± 0.28 | … | … | N | … | … | … | … | … | n | … |
| 51459 | 16261877-2407190 | -9.6 ± 0.2 | 3536 ± 27 | … | 0.870 ± 0.004 | -0.24 ± 0.13 | 627 ± 36 | 1 | Y | … | Y | Y | Y | Y | Y | … |
| 51460 | 16261909-2403485 | -6.9 ± 0.3 | 4609 ± 218 | … | 1.049 ± 0.011 | -0.34 ± 0.16 | … | … | N | … | … | … | … | … | n | … |
| 55373 | 16261996-2449486 | -90.7 ± 0.2 | 5646 ± 109 | … | 1.003 ± 0.003 | … | 68 ± 4 | 1 | Y | … | N | N | … | … | n | NG? |
| 51461 | 16262061-2501367 | 2.9 ± 0.3 | 4129 ± 229 | 4.57 ± 0.05 | 0.852 ± 0.011 | 0.05 ± 0.23 | <31 | 3 | Y | Y | N | N | N | … | n | NG |
| 55374 | 16262122-2451042 | -59.5 ± 0.2 | 4885 ± 215 | … | 1.014 ± 0.002 | -0.35 ± 0.04 | <15 | 3 | N | … | … | … | … | … | n | G |
| 55388 | 16263576-2450032 | 33.2 ± 0.2 | 4548 ± 85 | 2.23 ± 0.18 | 1.014 ± 0.003 | -0.17 ± 0.21 | <31 | 3 | N | N | … | … | … | … | n | … |
| 55389 | 16263602-2453479 | 48.2 ± 0.3 | 5049 ± 245 | 3.54 ± 0.12 | 0.996 ± 0.005 | -0.23 ± 0.25 | <18 | 3 | Y | Y | N | N | Y | … | n | NG |
| 55390 | 16263652-2455309 | 24.6 ± 0.3 | 5920 ± 85 | 4.34 ± 0.19 | 0.987 ± 0.006 | -0.12 ± 0.10 | … | … | Y | Y | … | … | Y | … | n | … |
| 55391 | 16263686-2456591 | -56.9 ± 0.3 | 5123 ± 86 | 3.57 ± 0.18 | 1.005 ± 0.007 | -0.20 ± 0.04 | <39 | 3 | Y | Y | N | N | Y | … | n | NG? |
| 55392 | 16263781-2458228 | -28.9 ± 0.2 | 5898 ± 136 | … | 1.002 ± 0.004 | -0.06 ± 0.03 | … | … | Y | … | N | … | Y | … | n | … |
| 51471 | 16263867-2355497 | -97.4 ± 0.3 | 4717 ± 134 | … | 1.018 ± 0.012 | -0.15 ± 0.08 | … | … | N | … | … | … | … | … | n | … |
| 55393 | 16263941-2447498 | -47.5 ± 0.3 | 5778 ± 191 | 4.04 ± 0.17 | 1.000 ± 0.008 | 0.01 ± 0.08 | … | … | Y | Y | N | … | Y | … | n | … |

| ID | CNAME | RV (km s$^{-1}$) | $T_{\text{eff}}$ (K) | $\log g$ (dex) | $\gamma^a$ | [Fe/H] (dex) | $EW(\text{Li})^b$ (mÅ) | $EW(\text{Li})$ error flag$^c$ | $\gamma$ | $\log g$ | RV | Li | [Fe/H] | Cánovas 2019$^d$ | Final$^e$ | Non-mem with Li$^f$ |
|---|---|---|---|---|---|---|---|---|---|---|---|---|---|---|---|---|
| 55394 | 16264286-2448282 | -186.8 ± 0.3 | 4734 ± 157 | … | 1.022 ± 0.008 | -0.54 ± 0.05 | <15 | 3 | N | … | … | … | … | … | n | G |
| 51472 | 16264294-2443062 | 5.9 ± 2.5 | 6495 ± 348 | … | 0.996 ± 0.018 | -0.24 ± 0.30 | … | … | Y | … | N | … | Y | … | n | … |
| 51473 | 16264310-2411095 | -8.0 ± 0.2 | 3938 ± 120 | … | 0.872 ± 0.005 | -0.13 ± 0.02 | 493 ± 17 | 1 | Y | … | Y | Y | Y | Y | Y | … |
| 55395 | 16264348-2452248 | 118.4 ± 0.2 | 3883 ± 95 | … | 1.034 ± 0.003 | -0.07 ± 0.05 | <31 | 3 | N | … | … | … | … | … | n | G |
| 51474 | 16264429-2443141 | -7.3 ± 0.3 | 3455 ± 101 | … | 0.901 ± 0.004 | -0.27 ± 0.11 | 700 ± 22 | 1 | Y | … | Y | Y | Y | Y | Y | … |
| 55396 | 16264441-2447138 | -6.5 ± 0.2 | 3367 ± 77 | … | 0.876 ± 0.004 | -0.26 ± 0.14 | 637 ± 20 | 1 | Y | … | Y | Y | Y | Y | Y | … |
| 55397 | 16264484-2456278 | 52.6 ± 0.2 | 4438 ± 124 | … | 1.033 ± 0.004 | -0.49 ± 0.12 | <27 | 3 | N | … | … | … | … | … | n | G |
| 55398 | 16264595-2456503 | -28.8 ± 0.2 | 5760 ± 58 | … | 0.998 ± 0.004 | -0.15 ± 0.03 | … | … | Y | … | N | … | Y | … | n | … |
| 55399 | 16264602-2443535 | -29.8 ± 0.3 | 4290 ± 291 | 4.61 ± 0.08 | 0.871 ± 0.011 | 0.03 ± 0.17 | … | … | Y | Y | N | … | Y | … | n | … |
| 55400 | 16264646-2459576 | 44.3 ± 0.2 | 4968 ± 213 | … | 1.009 ± 0.005 | -0.16 ± 0.22 | <17 | 3 | Y | … | N | N | Y | … | n | NG? |
| 55401 | 16264654-2453286 | 30.7 ± 1.2 | 6501 ± 170 | … | 0.996 ± 0.008 | -0.45 ± 0.16 | <22 | 3 | Y | … | N | N | N | … | n | NG |
| 55402 | 16264692-2451252 | 2.4 ± 0.3 | 4541 ± 45 | … | 1.033 ± 0.008 | -0.23 ± 0.29 | <50 | 3 | N | … | … | … | … | … | n | … |
| 55403 | 16264705-2444298 | -6.5 ± 0.3 | 3425 ± 5 | … | 0.895 ± 0.010 | -0.26 ± 0.15 | 612 ± 32 | 1 | Y | … | Y | Y | Y | Y | Y | … |
| 51475 | 16264720-2502158 | 46.3 ± 0.2 | 4729 ± 57 | 2.55 ± 0.10 | 1.012 ± 0.003 | -0.10 ± 0.08 | <23 | 3 | N | N | … | … | … | … | n | G |
| 55404 | 16264801-2457245 | 42.7 ± 0.4 | 3936 ± 129 | 4.53 ± 0.18 | 0.844 ± 0.005 | -0.18 ± 0.03 | <17 | 3 | Y | Y | N | N | Y | … | n | NG |
| 51476 | 16264864-2356341$^g$ | -8.7 ± 0.2 | 3953 ± 153 | … | 0.928 ± 0.004 | -0.17 ± 0.11 | 512 ± 17 | 1 | Y | … | Y | Y | Y | Y | Y | … |
| 51477 | 16265048-2413522 | -5.9 ± 0.7 | 3381 ± 99 | … | 0.884 ± 0.016 | -0.27 ± 0.14 | 660 ± 21 | 1 | Y | … | Y | Y | Y | Y | Y | … |
| 3318 | 16272291-2453412 | 37.3 ± 0.6 | 5853 ± 54 | 4.25 ± 0.02 | … | 0.24 ± 0.05 | <19 | 3 | … | Y | N | N | N | … | n | NG |
| 55437 | 16272297-2448071 | -4.6 ± 0.2 | 3407 ± 47 | … | 0.888 ± 0.003 | -0.23 ± 0.16 | 541 ± 5 | 1 | Y | … | Y | Y | Y | Y | Y | … |
| 55438 | 16272343-2454366 | 43.9 ± 0.2 | 5839 ± 8 | 4.17 ± 0.10 | 0.994 ± 0.002 | -0.14 ± 0.11 | <2 | 3 | Y | Y | N | N | Y | … | n | … |
| 51490 | 16272497-2355538 | 17.8 ± 0.3 | 4557 ± 117 | … | 1.045 ± 0.005 | … | <17 | 3 | N | … | … | … | … | … | n | G |
| 55440 | 16272535-2459087 | 14.6 ± 0.2 | 5748 ± 163 | … | 0.997 ± 0.003 | 0.00 ± 0.11 | 51 ± 6 | 1 | Y | … | N | Y | Y | … | n | NG |
| 55441 | 16272579-2458522 | -28.0 ± 0.2 | 5037 ± 103 | 4.05 ± 0.03 | 0.978 ± 0.003 | -0.16 ± 0.04 | <8 | 3 | Y | Y | N | N | Y | … | n | … |
| 55442 | 16272873-2454318 | -3.3 ± 0.2 | 3330 ± 16 | … | 0.885 ± 0.004 | -0.28 ± 0.14 | 100 ± 19 | 1 | Y | … | Y | N | Y | … | n | NG |
| 51491 | 16272877-2409379 | -40.2 ± 0.4 | 6628 ± 140 | … | 0.998 ± 0.010 | 0.51 ± 0.11 | <23 | 3 | Y | … | N | N | N | … | n | NG |
| 55443 | 16272998-2453142 | -2.9 ± 0.2 | 6111 ± 92 | 3.90 ± 0.12 | 1.003 ± 0.003 | -0.16 ± 0.05 | <1 | 3 | Y | Y | N | N | Y | … | n | … |
| 55444 | 16273095-2500187 | -68.7 ± 0.2 | 5109 ± 161 | … | 1.007 ± 0.004 | … | … | … | Y | … | N | … | … | … | n | … |
| 51492 | 16273134-2359082 | -25.1 ± 0.2 | 3790 ± 29 | 4.56 ± 0.09 | 0.825 ± 0.003 | -0.21 ± 0.10 | <100 | 3 | Y | Y | N | N | Y | … | n | NG |
| 51493 | 16273284-2403063 | -34.3 ± 0.3 | 5221 ± 56 | … | 0.979 ± 0.010 | … | <22 | 3 | Y | … | N | … | … | … | n | NG |
| 55445 | 16273311-2441152$^g$ | -4.4 ± 0.5 | 4543 ± 123 | … | 1.026 ± 0.014 | -0.10 ± 0.11 | 430 ± 14 | 1 | N | … | Y | Y | Y | Y | Y | … |
| 55446 | 16273325-2453150 | 28.2 ± 0.2 | 3868 ± 132 | … | 1.122 ± 0.006 | -0.15 ± 0.05 | <20 | 3 | N | … | … | … | … | … | n | … |
| 55447 | 16273466-2455237 | -25.9 ± 0.2 | 5143 ± 106 | … | 1.011 ± 0.004 | … | … | … | N | … | … | … | … | … | n | … |
| 55448 | 16273526-2438334 | -2.7 ± 0.4 | 3343 ± 74 | … | 0.919 ± 0.006 | -0.25 ± 0.16 | 612 ± 18 | 1 | Y | … | Y | Y | Y | Y | Y | … |
| 55449 | 16273667-2457162 | -49.8 ± 0.2 | 4607 ± 230 | 4.40 ± 0.11 | 0.938 ± 0.005 | -0.08 ± 0.08 | <24 | 3 | Y | Y | N | N | Y | … | n | NG |
| 51494 | 16273714-2359330 | 1.9 ± 0.2 | 5253 ± 103 | 3.51 ± 0.08 | 1.001 ± 0.003 | -0.12 ± 0.13 | <24 | 3 | Y | Y | N | N | Y | … | n | NG? |
| 51495 | 16273797-2357238$^g$ | -7.7 ± 0.4 | 3524 ± 179 | … | 0.853 ± 0.007 | -0.15 ± 0.11 | 710 ± 26 | 1 | Y | … | Y | Y | Y | Y | Y | … |
| 3319 | 16273832-2357324$^g$ | -5.9 ± 0.6 | 4524 ± 187 | 4.52 ± 0.15 | … | -0.08 ± 0.13 | 489 ± 19 | 2 | … | Y | Y | Y | Y | Y | Y | … |
| 55450 | 16273832-2451283 | -18.6 ± 1.4 | … | … | … | … | … | … | … | … | … | … | … | … | n | … |
| 51496 | 16273833-2404013 | -6.1 ± 0.2 | 4376 ± 295 | … | 0.961 ± 0.006 | -0.10 ± 0.01 | 536 ± 13 | 1 | Y | … | Y | Y | Y | Y | Y | … |
| 51497 | 16273851-2359451 | -89.4 ± 0.4 | 4279 ± 123 | … | 1.009 ± 0.020 | 0.36 ± 0.13 | <87 | 3 | Y | … | N | N | N | … | n | NG? |
| 51509 | 16281475-2423225 | -149.6 ± 6.9 | … | … | … | … | … | … | … | … | … | … | … | … | n | … |
| 51510 | 16281673-2405142$^g$ | -9.5 ± 0.3 | 4426 ± 359 | … | 0.948 ± 0.008 | -0.01 ± 0.13 | 519 ± 8 | 1 | Y | … | Y | Y | Y | Y | Y | … |
| 51511 | 16281922-2457340 | -4.7 ± 0.2 | 3633 ± 101 | … | 0.885 ± 0.006 | -0.23 ± 0.14 | 666 ± 23 | 1 | Y | … | Y | Y | Y | Y | Y | … |
| 51512 | 16282151-2421549 | -8.8 ± 0.3 | 3612 ± 13 | … | 0.876 ± 0.015 | -0.24 ± 0.13 | 590 ± 19 | 1 | Y | … | Y | Y | Y | Y | Y | … |
| 51513 | 16282207-2428421 | -33.4 ± 0.4 | 5137 ± 99 | … | 1.017 ± 0.011 | … | … | … | N | … | … | … | … | … | n | … |
| 51514 | 16282333-2422405$^g$ | -9.3 ± 0.4 | 4487 ± 137 | … | 0.948 ± 0.011 | -0.03 ± 0.05 | 489 ± 24 | 1 | Y | … | Y | Y | Y | Y | Y | … |
| 51515 | 16282399-2454503 | 35.8 ± 0.3 | 6063 ± 75 | 4.01 ± 0.03 | 1.001 ± 0.003 | -0.07 ± 0.03 | <21 | 3 | Y | Y | N | N | Y | … | n | NG? |
| 51516 | 16282430-2409316 | -6.9 ± 0.3 | 3969 ± 254 | … | 0.892 ± 0.014 | -0.19 ± 0.13 | 627 ± 26 | 1 | Y | … | Y | Y | Y | Y | Y | … |
| 51517 | 16282480-2435434 | -5.2 ± 0.7 | … | … | … | … | … | … | … | … | … | … | … | Y | n | … |
| 51518 | 16282509-2428198 | -15.5 ± 0.2 | 3883 ± 143 | … | 1.052 ± 0.005 | -0.02 ± 0.13 | 267 ± 15 | 1 | N | … | … | … | … | … | n | G |
| 51519 | 16282831-2452214 | 12.8 ± 0.2 | 3896 ± 156 | … | 1.032 ± 0.008 | -0.16 ± 0.05 | <31 | 3 | N | … | … | … | … | … | n | G |
| 51520 | 16282887-2454334 | 65.7 ± 0.2 | 5058 ± 118 | … | 1.012 ± 0.005 | -0.16 ± 0.26 | … | … | N | … | … | … | … | … | n | … |
| 51521 | 16283101-2402335 | -3.5 ± 0.3 | 4316 ± 190 | … | 1.077 ± 0.010 | … | <22 | 3 | N | … | … | … | … | … | n | … |
| 51522 | 16283288-2447550 | -38.9 ± 0.5 | 6520 ± 116 | 4.11 ± 0.24 | 1.004 ± 0.006 | -0.07 ± 0.09 | <15 | 3 | Y | Y | N | N | Y | … | n | NG? |
| 51523 | 16283526-2448419 | -38.2 ± 0.3 | 4675 ± 62 | 2.45 ± 0.15 | 1.025 ± 0.003 | -0.07 ± 0.14 | <33 | 3 | N | N | … | … | … | … | n | G |
| 51524 | 16283601-2457146 | -96.6 ± 0.2 | 4872 ± 297 | … | 1.029 ± 0.003 | -0.49 ± 0.06 | <21 | 3 | N | … | … | … | … | … | n | G |
| 51525 | 16283602-2447354 | 8.1 ± 0.8 | 6236 ± 279 | … | 1.022 ± 0.016 | -0.16 ± 0.21 | … | … | N | … | … | … | … | … | n | … |
| 51526 | 16283632-2432225 | -12.1 ± 0.3 | 4542 ± 106 | … | 1.019 ± 0.011 | … | <42 | 3 | N | … | … | … | … | … | n | G |
| 51527 | 16283787-2355459 | 102.6 ± 0.5 | … | … | … | … | … | … | … | … | … | … | … | … | n | … |

**Table C.1.** continued.

| ID | CNAME | RV (km s$^{-1}$) | $T_{\text{eff}}$ (K) | logg (dex) | $\gamma^a$ | [Fe/H] (dex) | EW(Li)$^b$ (mÅ) | EW(Li) error flag$^c$ | Membership $\gamma$ | logg | RV | Li | [Fe/H] | Cánovas 2019$^d$ | Final$^e$ | Non-mem with Li$^f$ |
|---|---|---|---|---|---|---|---|---|---|---|---|---|---|---|---|---|
| 51528 | 16283834-2426371 | -27.1 ± 0.2 | 4637 ± 8 | 2.52 ± 0.10 | 1.019 ± 0.006 | 0.08 ± 0.04 | <50 | 3 | N | N | … | … | … | … | n | G |
| 51529 | 16283844-2425073 | -44.5 ± 0.3 | 6551 ± 87 | … | 0.995 ± 0.006 | 0.35 ± 0.07 | … | … | Y | … | N | … | N | … | n | … |
| 51530 | 16284153-2433239 | -93.8 ± 0.3 | 4860 ± 177 | … | 1.020 ± 0.008 | -0.10 ± 0.14 | 125 ± 6 | 1 | N | … | … | … | … | … | n | Li-rich G |
| 51531 | 16284304-2422522$^g$ | -5.3 ± 0.4 | 3320 ± 15 | 4.77 ± 0.15 | 0.764 ± 0.013 | -0.27 ± 0.15 | 735 ± 34 | 1 | Y | Y | Y | Y | Y | Y | Y | … |
| 51532 | 16284311-2450190 | -34.7 ± 0.3 | 5291 ± 147 | 3.74 ± 0.11 | 1.005 ± 0.008 | 0.04 ± 0.02 | 55 ± 4 | 1 | Y | Y | N | N | N | … | n | NG? |
| 51536 | 16284666-2446367 | -42.0 ± 1.5 | 6170 ± 293 | … | 1.027 ± 0.017 | -0.02 ± 0.21 | … | … | N | … | … | … | … | … | n | … |
| 51537 | 16284704-2500399 | 52.4 ± 0.2 | 4455 ± 150 | … | 1.036 ± 0.004 | -0.25 ± 0.13 | <16 | 3 | N | … | … | … | … | … | n | G |
| 3321 | 16284724-2404046 | -34.9 ± 0.6 | 5099 ± 212 | 3.13 ± 0.22 | … | 0.11 ± 0.04 | <21 | 3 | … | N | … | … | … | … | n | … |
| 51538 | 16284836-2404410 | -38.8 ± 0.3 | 4681 ± 72 | 2.44 ± 0.16 | 1.080 ± 0.010 | 0.27 ± 0.27 | 58 ± 19 | 1 | N | N | … | … | … | … | n | G |
| 51539 | 16284843-2424505 | -36.2 ± 0.3 | 4782 ± 179 | 4.51 ± 0.04 | 0.950 ± 0.007 | -0.17 ± 0.07 | … | … | Y | Y | N | … | Y | … | n | … |
| 51540 | 16285113-2458479 | -40.7 ± 0.2 | 3654 ± 129 | … | … | … | … | … | … | … | … | … | … | … | n | … |
| 51541 | 16285199-2447399 | 57.9 ± 0.2 | 4542 ± 14 | … | 1.043 ± 0.006 | -0.06 ± 0.03 | <38 | 3 | N | … | … | … | … | … | n | G |
| 51542 | 16285671-2445533 | 33.1 ± 0.7 | 5699 ± 69 | … | 1.021 ± 0.015 | -0.38 ± 0.10 | … | … | N | … | … | … | … | … | n | … |
| 51543 | 16285679-2438101 | 4.5 ± 1.0 | … | … | … | … | … | … | … | … | … | … | … | … | n | … |
| 3322 | 16285779-2428046 | -74.3 ± 0.6 | 5784 ± 11 | 4.18 ± 0.11 | … | 0.18 ± 0.08 | 41 ± 13 | 2 | … | Y | N | N | N | … | n | NG |
| 51544 | 16285837-2443549 | -27.6 ± 0.7 | 6700 ± 186 | 4.03 ± 0.37 | 1.031 ± 0.014 | 0.63 ± 0.16 | <41 | 3 | N | Y | N | N | N | … | n | … |
| 3323 | 16285929-2449089 | 30.9 ± 0.6 | 5795 ± 15 | 4.25 ± 0.04 | … | 0.09 ± 0.02 | <16 | 3 | … | Y | N | N | N | … | n | NG |
| 51545 | 16290213-2501481 | -17.8 ± 0.2 | 4387 ± 127 | … | 1.041 ± 0.006 | -0.40 ± 0.03 | … | … | N | … | … | … | … | … | n | … |
| 51546 | 16290288-2427494 | -1.2 ± 0.8 | 3188 ± 185 | … | 0.855 ± 0.012 | -0.24 ± 0.16 | 652 ± 18 | 1 | Y | … | N | Y | Y | … | Y | … |
| 51547 | 16290392-2451414 | -3.8 ± 0.4 | 4759 ± 112 | … | 0.921 ± 0.011 | … | 594 ± 14 | 1 | Y | … | Y | Y | … | Y | Y | … |
| 51548 | 16290654-2426503 | -39.2 ± 0.2 | 4563 ± 105 | … | 1.024 ± 0.004 | -0.08 ± 0.08 | <36 | 3 | N | … | … | … | … | … | n | G |
| 3324 | 16290795-2432162 | -18.7 ± 0.6 | 6291 ± 127 | 3.88 ± 0.12 | … | -0.16 ± 0.14 | <62 | 3 | … | Y | N | N | Y | … | n | NG |
| 51549 | 16290865-2501154 | -52.1 ± 0.2 | 4576 ± 93 | … | 1.025 ± 0.003 | 0.00 ± 0.08 | <39 | 3 | N | … | … | … | … | … | n | G |
| 3325 | 16290921-2426425 | 4.7 ± 0.6 | 4352 ± 65 | 1.91 ± 0.08 | … | -0.24 ± 0.05 | <16 | 3 | … | N | … | … | … | … | n | … |
| 3326 | 16291158-2501595 | -87.1 ± 0.6 | 4598 ± 14 | 2.55 ± 0.09 | … | 0.21 ± 0.02 | <40 | 3 | … | N | … | … | … | … | n | … |
| 51550 | 16291234-2445347 | -130.1 ± 0.4 | 4732 ± 183 | … | 0.992 ± 0.018 | -0.10 ± 0.14 | … | … | Y | … | N | … | Y | … | n | … |
| 51362 | 16244251-2440177 | -41.7 ± 0.5 | 4667 ± 383 | … | 1.007 ± 0.017 | -0.14 ± 0.29 | … | … | Y | … | N | … | Y | … | n | … |
| 51363 | 16244390-2447544 | -59.3 ± 0.3 | 4655 ± 89 | … | 1.018 ± 0.009 | … | <52 | 3 | N | … | … | … | … | … | n | G |
| 51366 | 16244941-2459388 | -3.0 ± 0.3 | 3404 ± 32 | … | 0.853 ± 0.015 | -0.24 ± 0.15 | 219 ± 52 | 1 | Y | … | Y | Y | Y | … | Y | … |
| 51367 | 16245058-2438359 | 1.8 ± 1.2 | 4036 ± 145 | … | 1.068 ± 0.020 | -0.37 ± 0.22 | … | … | N | … | … | … | … | … | n | … |
| 51368 | 16245096-2453363 | -3.8 ± 0.3 | 4478 ± 156 | … | 1.067 ± 0.011 | … | … | … | N | … | … | … | … | … | n | … |
| 51369 | 16245271-2357480 | -110.2 ± 0.3 | 4572 ± 407 | … | 1.034 ± 0.012 | -0.21 ± 0.33 | … | … | N | … | … | … | … | … | n | … |
| 51370 | 16245426-2429403 | -67.0 ± 0.4 | 4977 ± 296 | … | 0.976 ± 0.017 | -0.02 ± 0.09 | <51 | 3 | Y | … | N | N | Y | … | n | NG |
| 51371 | 16245462-2444065 | 32.3 ± 0.3 | 4875 ± 184 | … | 1.022 ± 0.005 | -0.04 ± 0.05 | <60 | 3 | N | … | … | … | … | … | n | G |
| 51394 | 16251812-2449275 | -56.8 ± 0.2 | 4686 ± 8 | 2.45 ± 0.07 | 1.022 ± 0.007 | -0.03 ± 0.04 | <36 | 3 | N | N | … | … | … | … | n | G |
| 51395 | 16252215-2430139 | -25.2 ± 0.2 | 4401 ± 177 | … | 1.045 ± 0.003 | -0.08 ± 0.03 | <21 | 3 | N | … | … | … | … | … | n | G |
| 51396 | 16252243-2402057 | -6.8 ± 0.2 | 4616 ± 276 | … | 0.975 ± 0.003 | 0.01 ± 0.07 | 514 ± 8 | 1 | Y | … | Y | Y | Y | … | Y | … |
| 51397 | 16252248-2440038 | -13.3 ± 0.3 | 5755 ± 70 | … | 1.002 ± 0.010 | 0.08 ± 0.09 | 83 ± 5 | 1 | Y | … | N | N | N | … | n | NG? |
| 51398 | 16252293-2457196 | 0.9 ± 0.2 | 4606 ± 73 | 2.57 ± 0.09 | 1.009 ± 0.003 | 0.05 ± 0.13 | <51 | 3 | Y | N | N | N | N | … | n | NG? |
| 51399 | 16252314-2442092 | 24.6 ± 0.3 | 4504 ± 41 | … | 1.060 ± 0.008 | -0.38 ± 0.10 | <47 | 3 | N | … | … | … | … | … | n | G |
| 51400 | 16252348-2422541 | -32.1 ± 0.4 | 5554 ± 143 | … | 0.975 ± 0.015 | 0.09 ± 0.10 | <43 | 3 | Y | … | N | N | N | … | n | NG |
| 51401 | 16252429-2415401 | -13.9 ± 2.2 | 3263 ± 5 | … | 0.913 ± 0.015 | -0.27 ± 0.15 | 796 ± 22 | 1 | Y | … | N | Y | Y | Y | Y | … |
| 51402 | 16252506-2446119 | 40.4 ± 0.3 | 3837 ± 103 | … | 1.043 ± 0.009 | … | … | … | N | … | … | … | … | … | n | … |
| 51403 | 16252546-2407276 | -66.3 ± 0.3 | 4766 ± 34 | 2.36 ± 0.08 | 1.040 ± 0.012 | -0.05 ± 0.14 | <41 | 3 | N | N | … | … | … | … | n | G |
| 51404 | 16252612-2401058 | -47.9 ± 0.2 | 3930 ± 64 | … | 1.049 ± 0.009 | … | <19 | 3 | N | … | … | … | … | … | n | G |
| 51405 | 16252622-2423566 | 40.7 ± 0.8 | 5762 ± 291 | … | 1.004 ± 0.019 | -0.10 ± 0.20 | … | … | Y | … | N | … | Y | … | n | … |
| 51406 | 16252652-2356519 | -31.5 ± 0.2 | 4845 ± 147 | 2.56 ± 0.13 | 1.014 ± 0.006 | -0.13 ± 0.17 | <18 | 3 | N | N | … | … | … | … | n | G |
| 51407 | 16252700-2444477 | -47.4 ± 0.3 | 5718 ± 335 | … | 1.005 ± 0.009 | 0.32 ± 0.21 | <17 | 3 | Y | … | N | N | N | … | n | NG? |
| 51408 | 16252885-2422599 | 28.2 ± 0.3 | 3503 ± 28 | 4.60 ± 0.18 | 0.820 ± 0.011 | -0.24 ± 0.16 | <100 | 3 | Y | Y | N | N | Y | … | n | NG |
| 51409 | 16252989-2439157 | 76.9 ± 2.3 | … | … | … | … | … | … | … | … | … | … | … | … | n | … |
| 51410 | 16252993-2357436 | 10.4 ± 0.3 | 4534 ± 113 | … | 1.032 ± 0.014 | -0.21 ± 0.23 | <49 | 3 | N | … | … | … | … | … | n | G |
| 51411 | 16253084-2457295 | -7.2 ± 0.3 | 4863 ± 131 | 2.70 ± 0.12 | 1.014 ± 0.007 | -0.01 ± 0.06 | <39 | 3 | N | N | … | … | … | … | n | G |
| 51412 | 16253156-2402229 | 30.3 ± 0.6 | 5922 ± 161 | 4.09 ± 0.27 | 0.997 ± 0.008 | -0.25 ± 0.15 | <20 | 3 | Y | Y | N | N | Y | … | n | NG |
| 51413 | 16253198-2417524 | -42.6 ± 0.9 | 5230 ± 269 | … | 0.966 ± 0.019 | -0.53 ± 0.23 | … | … | Y | … | N | … | N | … | n | … |
| 51414 | 16253223-2419590 | -0.4 ± 0.5 | 5010 ± 236 | … | 1.021 ± 0.021 | -0.12 ± 0.18 | … | … | N | … | … | … | … | … | n | … |
| 51415 | 16253292-2454478 | -5.1 ± 0.2 | 5605 ± 178 | 4.05 ± 0.12 | 0.990 ± 0.005 | 0.15 ± 0.01 | <22 | 3 | Y | Y | Y | N | N | … | n | NG |
| 51416 | 16253379-2401189 | 35.1 ± 0.3 | 5142 ± 94 | … | 1.012 ± 0.010 | … | <23 | 3 | N | … | … | … | … | … | n | G |
| 51417 | 16253406-2436319 | 3.2 ± 0.3 | 4946 ± 184 | … | 1.016 ± 0.007 | -0.10 ± 0.17 | <29 | 3 | N | … | … | … | … | … | n | G |
| 3312 | 16262138-2459123 | -46.3 ± 0.6 | 5132 ± 25 | 2.96 ± 0.04 | … | -0.07 ± 0.03 | <17 | 3 | … | N | … | … | … | … | n | … |

| ID | CNAME | RV (km s$^{-1}$) | $T_\text{eff}$ (K) | logg (dex) | $\gamma^a$ | [Fe/H] (dex) | EW(Li)$^b$ (mÅ) | EW(Li) error flag$^c$ | Membership | | | | | | Final$^e$ | Non-mem with Li$^f$ |
| | | | | | | | | | $\gamma$ | logg | RV | Li | [Fe/H] | Cánovas 2019$^d$ | | |
|---|---|---|---|---|---|---|---|---|---|---|---|---|---|---|---|---|
| 55375 | 16262353-2453426 | 1.9 ± 0.3 | 5707 ± 139 | 4.23 ± 0.01 | 0.986 ± 0.005 | -0.14 ± 0.10 | 100 ± 5 | 1 | Y | Y | N | N | Y | … | n | NG |
| 51462 | 16262407-2416134$^g$ | -5.6 ± 0.3 | 4483 ± 123 | … | 1.004 ± 0.011 | … | 424 ± 13 | 1 | Y | … | Y | Y | … | Y | Y | … |
| 51463 | 16262549-2501120 | 6.7 ± 0.2 | 5072 ± 118 | … | 1.028 ± 0.004 | … | <22 | 3 | N | … | … | … | … | … | n | G |
| 51464 | 16262657-2357334 | -10.3 ± 0.3 | 4862 ± 474 | … | 1.019 ± 0.007 | -0.74 ± 0.16 | <13 | 3 | N | … | … | … | … | … | n | G |
| 55377 | 16262700-2447476 | -20.0 ± 0.2 | 5647 ± 277 | 4.12 ± 0.28 | 0.993 ± 0.006 | 0.23 ± 0.08 | … | … | Y | Y | N | … | N | … | n | … |
| 55378 | 16262744-2449592 | 11.5 ± 0.2 | 4713 ± 131 | … | 1.016 ± 0.003 | -0.08 ± 0.20 | <26 | 3 | N | … | … | … | … | … | n | G |
| 3313 | 16262750-2358416 | 13.0 ± 0.4 | 6130 ± 76 | 3.94 ± 0.04 | … | -0.45 ± 0.05 | 60 ± 8 | 2 | … | Y | N | N | N | … | n | NG |
| 51465 | 16262898-2355215 | -87.3 ± 0.2 | 4769 ± 395 | … | 1.030 ± 0.005 | -0.56 ± 0.23 | <8 | 3 | N | … | … | … | … | … | n | G |
| 55379 | 16262906-2448428 | 52.6 ± 0.2 | … | … | … | … | … | … | … | … | … | … | … | … | n | … |
| 55380 | 16262933-2453355 | -25.3 ± 0.2 | 5801 ± 79 | … | 0.992 ± 0.002 | … | <12 | 3 | Y | … | N | N | … | … | n | NG |
| 55381 | 16262948-2451195 | 57.1 ± 0.2 | 5075 ± 118 | … | 1.010 ± 0.002 | -0.14 ± 0.22 | <29 | 3 | N | … | … | … | … | … | n | G |
| 51466 | 16262953-2359380 | -147.8 ± 0.3 | 4592 ± 309 | … | 1.048 ± 0.016 | -0.11 ± 0.08 | … | … | N | … | … | … | … | … | n | … |
| 55382 | 16262996-2450391 | 144.9 ± 0.3 | 5388 ± 298 | … | 1.005 ± 0.005 | -0.58 ± 0.26 | … | … | Y | … | N | … | N | … | n | … |
| 55383 | 16263056-2454326 | -27.7 ± 0.3 | 6144 ± 14 | 3.80 ± 0.09 | 1.007 ± 0.003 | -0.27 ± 0.06 | 38 ± 8 | 1 | Y | Y | N | N | Y | … | n | NG? |
| 55384 | 16263149-2456578 | -79.6 ± 0.2 | 5000 ± 179 | … | 1.017 ± 0.005 | -0.28 ± 0.08 | <17 | 3 | N | … | … | … | … | … | n | G |
| 51467 | 16263297-2400168$^g$ | -6.7 ± 0.3 | 3352 ± 62 | … | 0.879 ± 0.010 | -0.23 ± 0.16 | 616 ± 14 | 1 | Y | … | Y | Y | Y | Y | Y | … |
| 55385 | 16263331-2455375 | -18.0 ± 0.2 | 5219 ± 193 | … | 0.983 ± 0.005 | 0.16 ± 0.11 | <34 | 3 | Y | … | N | N | N | … | n | NG |
| 55386 | 16263366-2447530 | -8.4 ± 0.3 | 4817 ± 23 | … | 0.987 ± 0.008 | 0.01 ± 0.11 | <40 | 3 | Y | … | Y | N | Y | … | n | NG |
| 51468 | 16263403-2500011 | 5.7 ± 0.2 | 3642 ± 83 | 4.57 ± 0.20 | 0.812 ± 0.004 | -0.21 ± 0.14 | <100 | 3 | Y | Y | N | N | Y | … | n | NG |
| 51469 | 16263416-2423282 | -2.2 ± 1.2 | … | … | … | … | … | … | … | … | … | … | … | … | n | … |
| 55387 | 16263561-2445209 | -36.4 ± 0.3 | 5635 ± 178 | 4.23 ± 0.07 | 0.987 ± 0.009 | -0.06 ± 0.14 | <34 | 3 | Y | Y | N | N | Y | … | n | NG |
| 51470 | 16263561-2500174 | -26.6 ± 0.3 | 5040 ± 25 | 3.18 ± 0.06 | 1.007 ± 0.007 | -0.02 ± 0.05 | <40 | 3 | Y | N | N | N | Y | … | n | NG? |
| 55405 | 16265128-2448406 | 0.1 ± 0.6 | 5667 ± 30 | … | 0.963 ± 0.013 | -0.43 ± 0.11 | … | … | Y | … | N | … | N | … | n | … |
| 51478 | 16265166-2403538 | -106.1 ± 0.3 | 3545 ± 114 | … | … | … | … | … | … | … | … | … | … | … | n | … |
| 55406 | 16265277-2449248 | -134.6 ± 0.3 | 4363 ± 112 | … | 1.035 ± 0.013 | -0.33 ± 0.13 | <24 | 3 | N | … | … | … | … | … | n | G |
| 55407 | 16265402-2456262 | -11.3 ± 0.3 | 6441 ± 24 | 4.07 ± 0.08 | 1.001 ± 0.003 | -0.25 ± 0.17 | <15 | 3 | Y | Y | N | N | Y | … | n | NG? |
| 55408 | 16265461-2458095 | -9.0 ± 0.2 | 5033 ± 155 | … | 1.015 ± 0.004 | -0.15 ± 0.20 | <9 | 3 | N | … | … | … | … | … | n | G |
| 55409 | 16265744-2449495 | -37.7 ± 0.4 | 4467 ± 209 | … | 1.051 ± 0.011 | … | <39 | 3 | N | … | … | … | … | … | n | … |
| 3314 | 16265752-2446060 | 41.5 ± 0.6 | 4106 ± 142 | 4.57 ± 0.11 | … | -0.17 ± 0.10 | <49 | 3 | … | Y | N | N | Y | … | n | NG |
| 55410 | 16265789-2457518 | -52.7 ± 0.3 | 5309 ± 183 | … | 1.015 ± 0.006 | -0.90 ± 0.05 | <10 | 3 | N | … | … | … | … | … | n | … |
| 55411 | 16265847-2452197 | -42.4 ± 0.3 | 4853 ± 33 | … | 0.982 ± 0.008 | 0.03 ± 0.07 | <31 | 3 | Y | … | N | N | Y | … | n | NG |
| 55412 | 16265850-2445368 | -5.0 ± 0.2 | 5055 ± 101 | … | 0.983 ± 0.002 | … | 382 ± 2 | 1 | Y | … | Y | Y | … | … | Y | … |
| 3315 | 16270007-2501426 | -23.6 ± 0.4 | 6165 ± 162 | 3.93 ± 0.16 | … | -0.43 ± 0.14 | <27 | 3 | … | Y | N | N | N | … | n | NG |
| 55413 | 16270114-2458568 | -47.7 ± 0.2 | 5684 ± 149 | 4.16 ± 0.11 | 0.993 ± 0.004 | 0.13 ± 0.04 | <19 | 3 | Y | Y | N | N | Y | … | n | NG |
| 55414 | 16270167-2456342 | 8.1 ± 0.3 | 5858 ± 3 | 4.19 ± 0.19 | 0.995 ± 0.005 | -0.01 ± 0.11 | 60 ± 5 | 1 | Y | Y | N | N | Y | … | n | NG |
| 51479 | 16270276-2502437 | -90.2 ± 0.2 | 4372 ± 260 | … | 1.041 ± 0.004 | -0.09 ± 0.13 | <26 | 3 | N | … | … | … | … | … | n | G |
| 51480 | 16270405-2409318 | -8.8 ± 0.3 | 3802 ± 45 | … | 0.858 ± 0.008 | -0.19 ± 0.13 | 535 ± 20 | 1 | Y | … | Y | Y | Y | Y | Y | … |
| 3316 | 16270429-2359485 | -12.0 ± 0.4 | 6631 ± 276 | 4.07 ± 0.17 | … | -0.14 ± 0.18 | <52 | 3 | … | Y | N | N | Y | … | n | NG |
| 51481 | 16270431-2403280 | 1.9 ± 0.2 | 4583 ± 76 | … | 1.027 ± 0.006 | … | <42 | 3 | N | … | … | … | … | … | n | G |
| 55415 | 16270442-2500229 | 11.0 ± 0.3 | 5778 ± 27 | … | 1.001 ± 0.005 | -0.30 ± 0.11 | 48 ± 4 | 1 | Y | … | N | N | Y | … | n | NG? |
| 55416 | 16270451-2442596 | -6.1 ± 0.3 | 4299 ± 365 | … | 0.916 ± 0.006 | -0.07 ± 0.08 | 557 ± 7 | 1 | Y | … | Y | Y | Y | Y | Y | … |
| 55417 | 16270456-2442140 | -6.5 ± 0.3 | 3927 ± 34 | … | 0.893 ± 0.012 | 0.27 ± 0.10 | 530 ± 14 | 1 | Y | … | Y | Y | N | Y | Y | … |
| 55418 | 16270570-2452272 | -5.9 ± 0.5 | 6746 ± 104 | … | 0.997 ± 0.005 | -0.17 ± 0.10 | <20 | 3 | Y | … | Y | N | Y | … | n | NG |
| 51482 | 16270616-2453492 | 39.8 ± 0.3 | 4503 ± 58 | 1.91 ± 0.01 | 1.032 ± 0.009 | -0.26 ± 0.16 | <31 | 3 | N | N | … | … | … | … | n | G |
| 55419 | 16270659-2441488$^g$ | -5.5 ± 0.3 | 3174 ± 39 | … | 0.910 ± 0.013 | … | 449 ± 14 | 1 | Y | … | Y | Y | … | Y | Y | … |
| 55420 | 16270748-2452544 | -22.8 ± 0.3 | 6782 ± 83 | 4.19 ± 0.16 | 1.005 ± 0.004 | -0.10 ± 0.08 | <26 | 3 | Y | Y | N | N | Y | … | n | NG? |
| 3320 | 16273901-2358187$^g$ | -5.9 ± 0.6 | 4532 ± 184 | 4.52 ± 0.16 | … | -0.08 ± 0.13 | 459 ± 15 | 2 | … | Y | Y | Y | Y | Y | Y | … |
| 55451 | 16274181-2457385 | -72.1 ± 0.3 | 5769 ± 125 | 4.17 ± 0.07 | 0.990 ± 0.005 | -0.06 ± 0.18 | 40 ± 3 | 1 | Y | Y | N | N | Y | … | n | NG |
| 51498 | 16274187-2404272 | -7.5 ± 0.2 | 3572 ± 48 | … | 0.851 ± 0.005 | -0.24 ± 0.13 | 552 ± 27 | 1 | Y | … | Y | Y | Y | Y | Y | … |
| 55452 | 16274308-2453583 | 30.3 ± 0.2 | 4482 ± 123 | 1.64 ± 0.12 | 1.029 ± 0.003 | -0.39 ± 0.11 | <25 | 3 | N | N | … | … | … | … | n | G |
| 55453 | 16274359-2451259 | -160.8 ± 0.2 | 4508 ± 349 | … | 1.044 ± 0.004 | -0.23 ± 0.06 | <21 | 3 | N | … | … | … | … | … | n | G |
| 51499 | 16274387-2355470 | -61.5 ± 0.2 | 4641 ± 113 | 2.04 ± 0.14 | 1.032 ± 0.007 | -0.24 ± 0.09 | <33 | 3 | N | N | … | … | … | … | n | G |
| 55454 | 16274882-2453381 | 12.2 ± 0.3 | 5853 ± 156 | … | 1.004 ± 0.009 | 0.26 ± 0.14 | … | … | Y | … | N | … | N | … | n | … |
| 55455 | 16274924-2452449 | 106.7 ± 0.3 | … | … | … | … | … | … | … | … | … | … | … | … | n | … |
| 55456 | 16275008-2457375 | -38.8 ± 0.3 | 5118 ± 80 | … | 1.008 | … | … | … | Y | … | N | … | … | … | n | … |
| 51500 | 16275170-2358413 | -28.3 ± 0.3 | 5010 ± 138 | … | 0.979 ± 0.008 | -0.02 ± 0.02 | <21 | 3 | Y | … | N | N | Y | … | n | NG |
| 55457 | 16275206-2455517 | -9.4 ± 0.3 | 5557 ± 39 | 4.13 ± 0.14 | 0.990 ± 0.008 | 0.13 ± 0.10 | … | … | Y | Y | Y | … | N | … | n | … |
| 55458 | 16275433-2451073 | 9.0 ± 0.2 | 4971 ± 183 | … | 1.019 ± 0.004 | -0.11 ± 0.22 | <21 | 3 | N | … | … | … | … | … | n | G |
| 55459 | 16275440-2452008 | -13.4 ± 0.2 | 4497 ± 174 | 2.15 ± 0.15 | 1.028 ± 0.003 | 0.14 ± 0.06 | <63 | 3 | N | N | … | … | … | … | n | G |

**Table C.1.** continued.

| ID | CNAME | RV (km s$^{-1}$) | $T_{\text{eff}}$ (K) | logg (dex) | $\gamma^a$ | [Fe/H] (dex) | $EW$(Li)$^b$ (mÅ) | $EW$(Li) error flag$^c$ | Membership $\gamma$ | logg | RV | Li | [Fe/H] | Cánovas 2019$^d$ | Final$^e$ | Non-mem with Li$^f$ |
|---|---|---|---|---|---|---|---|---|---|---|---|---|---|---|---|---|
| 51501 | 16275547-2400325 | -23.3 ± 0.3 | 4923 ± 144 | … | 1.019 ± 0.010 | -0.11 ± 0.16 | … | … | N | … | … | … | … | … | n | … |
| 55460 | 16275669-2455284 | -49.7 ± 0.3 | 5420 ± 34 | 3.84 ± 0.15 | 0.996 ± 0.008 | -0.29 ± 0.13 | <27 | 3 | Y | Y | N | N | Y | … | n | NG |
| 55461 | 16275996-2448193$^g$ | 0.3 ± 0.3 | 3307 ± 81 | … | 0.864 ± 0.004 | -0.27 ± 0.18 | 687 ± 70 | 1 | Y | … | N | Y | Y | Y | Y | … |
| 55462 | 16280011-2453427 | -5.7 ± 0.2 | 3295 ± 37 | … | 0.919 ± 0.004 | -0.25 ± 0.14 | … | … | Y | … | Y | Y | Y | Y | n | … |
| 51502 | 16280080-2400517 | -6.4 ± 0.4 | 3235 ± 120 | … | 0.882 ± 0.008 | … | 703 ± 9 | 1 | Y | … | Y | Y | … | Y | Y | … |
| 51503 | 16280118-2402103 | -124.2 ± 0.3 | 5044 ± 53 | … | 1.012 ± 0.009 | -0.28 ± 0.12 | <20 | 3 | N | … | … | … | … | … | n | G |
| 51504 | 16280802-2400033 | -108.6 ± 0.5 | … | … | … | … | … | … | … | … | … | … | … | … | n | … |
| 51505 | 16280948-2401556 | 50.5 ± 0.4 | 5178 ± 96 | … | 1.006 ± 0.015 | … | <56 | 3 | Y | … | N | N | … | … | n | NG? |
| 51506 | 16281099-2406177 | -7.0 ± 0.4 | 3333 ± 53 | … | 0.864 ± 0.013 | -0.27 ± 0.14 | 661 ± 14 | 1 | Y | … | Y | Y | Y | … | Y | … |
| 51507 | 16281108-2429197 | 23.4 ± 2.3 | … | … | … | … | … | … | … | … | … | … | … | … | n | … |
| 51508 | 16281181-2401503 | 38.8 ± 0.3 | 6470 ± 107 | … | 0.980 ± 0.008 | 0.48 ± 0.08 | <13 | 3 | Y | … | N | N | N | … | n | NG |
| 51533 | 16284345-2452189 | 40.1 ± 0.2 | 4674 ± 82 | … | 1.018 ± 0.002 | … | <47 | 3 | N | … | … | … | … | … | n | G |
| 51534 | 16284365-2406406 | 46.0 ± 0.2 | 3937 ± 63 | … | 1.049 ± 0.009 | … | … | … | N | … | … | … | … | … | n | … |
| 51535 | 16284530-2453080 | -25.1 ± 0.2 | 4648 ± 8 | 2.41 ± 0.13 | 1.021 ± 0.003 | -0.11 ± 0.07 | <38 | 3 | N | N | … | … | … | … | n | G |

**Notes.** $^{(a)}$ Empirical gravity indicator defined by Damiani et al. (2014). $^{(b)}$ The values of $EW$(Li) for this cluster are corrected (subtracted adjacent Fe (6707.43 Å) line). $^{(c)}$ Flags for the errors of the corrected $EW$(Li) values, as follows: 1=$EW$(Li) corrected by blends contribution using models; 2=$EW$(Li) measured separately (Li line resolved - UVES only); and 3=Upper limit (no error for $EW$(Li) is given). $^{(d)}$ Cánovas et al. (2019) $^{(e)}$ The letters "Y" and "N" indicate if the star is a cluster member or not. $^{(f)}$ 'Li-rich G', 'G' and 'NG' indicate "Li-rich giant", "giant" and "non-giant" Li field contaminants, respectively. $^{(g)}$ Strong accretor members.

**Table C.2.** Chamaeleon I

| ID | CNAME | RV (km s$^{-1}$) | $T_{\rm eff}$ (K) | $logg$ (dex) | $\gamma^a$ | [Fe/H] (dex) | EW(Li)$^b$ (mÅ) | EW(Li) error flag$^c$ | $\gamma$ | $logg$ | RV | Li | [Fe/H] | Final$^d$ | Non-mem with Li$^e$ |
|---|---|---|---|---|---|---|---|---|---|---|---|---|---|---|---|
| 6942 | 11133030-7807023 | 9.8 ± 0.2 | 3310 ± 85 | ... | ... | ... | ... | ... | ... | ... | ... | ... | ... | n | ... |
| 6943 | 11133199-7550375 | 2.1 ± 0.2 | 4473 ± 215 | ... | 1.020 ± 0.003 | 0.16 ± 0.02 | <63 | 3 | N | ... | ... | ... | ... | n | G |
| 6467 | 10580551-7728239 | 46.9 ± 0.2 | 5109 ± 69 | ... | 1.008 ± 0.002 | -0.09 ± 0.09 | 51 ± 9 | 1 | Y | ... | N | N | Y | n | NG? |
| 6944 | 11133356-7635374 | 19.9 ± 0.3 | 3314 ± 24 | ... | 0.905 ± 0.009 | -0.19 ± 0.14 | 574 ± 54 | 1 | Y | ... | Y | Y | Y | Y | ... |
| 6945 | 11133423-7559461 | 13.0 ± 0.2 | 4542 ± 135 | ... | 1.025 ± 0.005 | -0.09 ± 0.05 | <38 | 3 | N | ... | ... | ... | ... | n | ... |
| 6946 | 11133759-7740154 | 14.0 ± 0.2 | 4756 ± 78 | 2.57 ± 0.05 | 1.024 ± 0.005 | 0.01 ± 0.06 | <42 | 3 | N | N | ... | ... | ... | n | G |
| 6947 | 11133884-7720445 | 3.6 ± 0.2 | 4518 ± 141 | ... | 1.033 ± 0.002 | 0.00 ± 0.06 | <47 | 3 | N | ... | ... | ... | ... | n | G |
| 6948 | 11134244-7728284 | 7.8 ± 0.3 | 4587 ± 1 | 2.33 ± 0.10 | 1.021 ± 0.015 | -0.10 ± 0.05 | <40 | 3 | N | N | ... | ... | ... | n | G |
| 6949 | 11134256-7604228 | 17.7 ± 0.2 | 4626 ± 102 | ... | 1.018 ± 0.005 | ... | <57 | 3 | N | ... | ... | ... | ... | n | ... |
| 6950 | 11134338-7540243 | 144.5 ± 0.2 | 4880 ± 257 | ... | 1.012 ± 0.002 | -0.30 ± 0.23 | 43 ± 4 | 1 | N | ... | ... | ... | ... | n | G |
| 6951 | 11134951-7728422 | 67.0 ± 1.1 | ... | ... | ... | ... | ... | ... | ... | ... | ... | ... | ... | n | ... |
| 6952 | 11134979-7706416 | 54.2 ± 0.2 | 4556 ± 115 | ... | 1.047 ± 0.002 | ... | <18 | 3 | N | ... | ... | ... | ... | n | G |
| 648 | 11135757-7818460 | ... | ... | ... | ... | ... | ... | ... | ... | ... | ... | ... | ... | n | ... |
| 6953 | 11140162-7608594 | 31.6 ± 0.2 | 4563 ± 70 | 2.42 ± 0.09 | 1.011 ± 0.003 | -0.06 ± 0.05 | <40 | 3 | N | N | ... | ... | ... | n | G |
| 6954 | 11140220-7811445 | 30.7 ± 0.2 | 4595 ± 24 | 2.49 ± 0.02 | 1.010 ± 0.002 | -0.03 ± 0.10 | <36 | 3 | Y | N | N | N | Y | n | NG? |
| 6955 | 11140335-7621573 | 6.7 ± 0.2 | 6050 ± 36 | 4.00 ± 0.02 | 0.999 ± 0.003 | -0.08 ± 0.07 | 39 ± 1 | 1 | Y | Y | N | N | Y | n | NG |
| 649 | 11140585-7729058 | -2.1 ± 0.6 | 3915 ± 79 | 1.24 ± 0.20 | ... | 0.06 ± 0.16 | <27 | 3 | ... | N | N | ... | Y | n | ... |
| 6956 | 11140914-7718576 | 41.6 ± 0.2 | 4734 ± 134 | 2.51 ± 0.20 | 1.020 ± 0.002 | -0.07 ± 0.14 | ... | ... | N | N | ... | ... | ... | n | ... |
| 650 | 11140941-7714492 | -24.0 ± 0.6 | 4363 ± 57 | 1.62 ± 0.11 | ... | -0.31 ± 0.11 | <18 | 3 | ... | N | N | ... | Y | n | ... |
| 6957 | 11141044-7549438 | 9.7 ± 0.2 | 4695 ± 15 | 2.53 ± 0.09 | 1.021 ± 0.001 | 0.03 ± 0.09 | <33 | 3 | N | N | ... | ... | ... | n | ... |
| 6958 | 11141073-7531068 | -27.2 ± 0.2 | 5041 ± 70 | ... | 1.019 ± 0.001 | -0.05 ± 0.07 | <18 | 3 | N | ... | ... | ... | ... | n | G |
| 6959 | 11141565-7627364 | 17.2 ± 0.3 | 3372 ± 110 | ... | 0.877 ± 0.006 | -0.27 ± 0.14 | 640 ± 10 | 1 | Y | ... | Y | Y | Y | Y | ... |
| 651 | 11141568-7738326 | 16.9 ± 0.4 | 7057 ± 221 | 4.07 ± 0.21 | ... | -0.13 ± 0.17 | <33 | 3 | ... | Y | Y | N | Y | n | NG |
| 6960 | 11142120-7733403 | -3.7 ± 0.2 | 6656 ± 36 | 4.09 ± 0.07 | 1.007 ± 0.002 | 0.05 ± 0.03 | <16 | 3 | Y | Y | N | N | Y | n | NG? |
| 6500 | 10594559-7808218 | -11.5 ± 0.3 | 5133 ± 3 | 3.61 ± 0.04 | 1.000 ± 0.006 | -0.06 ± 0.02 | ... | ... | Y | Y | N | ... | Y | n | ... |
| 6961 | 11142454-7733062$^f$ | 19.5 ± 1.3 | 3315 ± 68 | ... | 0.874 ± 0.017 | -0.23 ± 0.16 | 532 ± 17 | 1 | Y | ... | Y | Y | Y | Y | ... |
| 6501 | 10594801-7644496 | 22.9 ± 0.2 | 4539 ± 138 | 2.42 ± 0.20 | 1.016 ± 0.005 | 0.14 ± 0.05 | <70 | 3 | N | N | ... | ... | ... | n | G |
| 7012 | 11214759-7633205 | -41.5 ± 0.2 | 4627 ± 116 | ... | 1.026 ± 0.003 | -0.14 ± 0.14 | <31 | 3 | N | ... | ... | ... | ... | n | G |
| 7013 | 11215273-7623406 | 52.0 ± 0.2 | 3915 ± 81 | ... | 1.035 ± 0.004 | ... | <5 | 3 | N | ... | ... | ... | ... | n | ... |
| 6502 | 10595255-7622512 | 12.3 ± 0.3 | 3299 ± 86 | ... | ... | ... | ... | ... | ... | ... | ... | ... | ... | n | ... |
| 7014 | 11215361-7635140 | 41.2 ± 0.2 | 4552 ± 204 | ... | 1.037 ± 0.006 | -0.22 ± 0.19 | ... | ... | N | ... | ... | ... | ... | n | ... |
| 6503 | 10595574-7537587 | 300.7 ± 0.2 | 3391 ± 115 | ... | ... | ... | 194 ± 22 | 1 | ... | ... | ... | ... | ... | n | ... |
| 6504 | 10595657-7726008 | -12.8 ± 0.3 | 5064 ± 82 | ... | 1.008 ± 0.007 | ... | <37 | 3 | Y | ... | N | N | ... | n | NG? |
| 6525 | 11004830-7735206 | 20.3 ± 0.7 | ... | ... | ... | ... | ... | ... | ... | ... | ... | ... | ... | n | ... |
| 6526 | 11005105-7531281 | 58.9 ± 0.2 | 4141 ± 341 | ... | 1.046 ± 0.002 | -0.29 ± 0.18 | <14 | 3 | N | ... | ... | ... | ... | n | ... |
| 7020 | 11230503-7544307 | -11.8 ± 0.2 | 5114 ± 36 | ... | 1.014 ± 0.001 | -0.04 ± 0.03 | ... | ... | N | ... | ... | ... | ... | n | ... |
| 6527 | 11005455-7823327 | -68.5 ± 0.2 | 6385 ± 31 | 3.98 ± 0.07 | 1.007 ± 0.002 | 0.32 ± 0.02 | ... | ... | Y | Y | N | ... | N | n | ... |
| 7021 | 11231005-7615239 | 27.6 ± 0.2 | 4659 ± 43 | 2.57 ± 0.10 | 1.012 ± 0.002 | 0.10 ± 0.02 | <44 | 3 | N | N | ... | ... | ... | n | G |
| 6528 | 11005526-7655476 | 55.5 ± 0.3 | 4545 ± 74 | 2.25 ± 0.17 | 1.017 ± 0.009 | -0.21 ± 0.05 | ... | ... | N | N | ... | ... | ... | n | ... |
| 6529 | 11005728-7750445 | 9.9 ± 0.2 | 4945 ± 212 | ... | 1.011 ± 0.004 | -0.19 ± 0.24 | <16 | 3 | N | ... | ... | ... | ... | n | G |
| 6530 | 11005794-7545151 | 12.5 ± 0.2 | 4929 ± 98 | 2.75 ± 0.12 | 1.016 ± 0.002 | -0.05 ± 0.09 | <24 | 3 | N | N | ... | ... | ... | n | G |
| 6531 | 11005841-7750099 | 33.3 ± 0.2 | 5950 ± 124 | 4.32 ± 0.04 | 0.990 ± 0.004 | 0.13 ± 0.05 | ... | ... | Y | Y | N | ... | Y | n | ... |
| 6583 | 11032619-7720037 | -26.2 ± 0.3 | 4686 ± 110 | ... | 1.025 ± 0.009 | -0.08 ± 0.11 | <32 | 3 | N | ... | ... | ... | ... | n | G |
| 6584 | 11032892-7740518 | -20.3 ± 0.2 | 4289 ± 185 | ... | 1.060 ± 0.004 | -0.16 ± 0.05 | <32 | 3 | N | ... | ... | ... | ... | n | G |
| 6585 | 11033587-7743146 | 10.5 ± 0.4 | 4998 ± 214 | ... | 1.014 ± 0.018 | -0.09 ± 0.15 | ... | ... | N | ... | ... | ... | ... | n | ... |
| 671 | 11033599-7628242 | 13.7 ± 0.6 | 4398 ± 163 | 1.98 ± 0.41 | ... | -0.23 ± 0.05 | 49 ± 9 | 2 | ... | N | Y | ... | Y | n | ... |
| 6586 | 11034013-7713102 | 87.1 ± 0.9 | ... | ... | ... | ... | ... | ... | ... | ... | ... | ... | ... | n | ... |
| 6587 | 11034035-7633264 | 45.6 ± 0.2 | 4633 ± 94 | 2.44 ± 0.19 | 1.030 ± 0.004 | 0.07 ± 0.05 | <43 | 3 | N | N | ... | ... | ... | n | G |
| 6588 | 11034263-7633122 | 32.4 ± 0.2 | 5001 ± 180 | 4.40 ± 0.19 | 0.968 ± 0.005 | -0.02 ± 0.03 | <19 | 3 | Y | Y | N | N | Y | n | NG |
| 7046 | 11274126-7552017 | 18.0 ± 0.2 | 4653 ± 105 | 2.49 ± 0.19 | 1.020 ± 0.003 | 0.02 ± 0.09 | <33 | 3 | N | N | ... | ... | ... | n | G |
| 6607 | 11043186-7714508 | 46.3 ± 0.6 | ... | ... | ... | ... | ... | ... | ... | ... | ... | ... | ... | n | ... |
| 6608 | 11043290-7718034 | -14.0 ± 0.3 | 5731 ± 62 | ... | 0.997 ± 0.009 | ... | <51 | 3 | Y | ... | N | N | ... | n | NG |
| 6609 | 11043347-7818110 | -18.8 ± 0.2 | 4691 ± 68 | 2.42 ± 0.15 | 1.029 ± 0.005 | -0.05 ± 0.03 | <29 | 3 | N | N | ... | ... | ... | n | G |
| 6610 | 11043352-7812254 | -11.1 ± 0.3 | 6394 ± 126 | ... | 0.993 ± 0.008 | 0.15 ± 0.09 | 65 ± 4 | 1 | Y | ... | N | N | Y | n | NG |
| 6611 | 11043354-7532339 | -27.7 ± 0.2 | 4520 ± 160 | ... | 1.018 ± 0.002 | 0.08 ± 0.05 | ... | ... | N | ... | ... | ... | ... | n | ... |
| 6389 | 10522837-7727256 | 64.2 ± 0.3 | 5024 ± 248 | ... | 1.021 ± 0.006 | -0.53 ± 0.23 | <19 | 3 | N | ... | ... | ... | ... | n | G |
| 6612 | 11043417-7536599 | 56.5 ± 0.2 | 4582 ± 74 | 2.38 ± 0.16 | 1.021 ± 0.003 | -0.02 ± 0.04 | <35 | 3 | N | N | ... | ... | ... | n | G |
| 6613 | 11043473-7634285 | 14.4 ± 0.2 | 3912 ± 253 | ... | 1.072 ± 0.002 | -0.40 ± 0.10 | ... | ... | N | ... | ... | ... | ... | n | ... |
| 6614 | 11043504-7509043 | 28.0 ± 0.2 | 3919 ± 60 | ... | 1.039 ± 0.004 | ... | ... | ... | N | ... | ... | ... | ... | n | ... |

**Table C.2.** continued.

| ID | CNAME | RV (km s$^{-1}$) | $T_{\text{eff}}$ (K) | $\log g$ (dex) | $\gamma^a$ | [Fe/H] (dex) | $EW(\text{Li})^b$ (mÅ) | $EW(\text{Li})$ error flag$^c$ | $\gamma$ | $\log g$ | RV | Li | [Fe/H] | Final$^d$ | Non-mem with Li$^e$ |
|---|---|---|---|---|---|---|---|---|---|---|---|---|---|---|---|
| 6615 | 11043528-7627399 | 11.3 ± 0.2 | 4743 ± 60 | 2.62 ± 0.11 | 1.014 ± 0.002 | 0.03 ± 0.07 | <38 | 3 | N | N | … | … | … | n | G |
| 6616 | 11043610-7509532 | 14.4 ± 0.2 | 4631 ± 108 | … | 1.025 ± 0.003 | … | <40 | 3 | N | … | … | … | … | n | G |
| 6617 | 11043663-7702139 | -2.5 ± 0.2 | 3873 ± 81 | … | … | … | <8 | 3 | … | … | … | … | … | n | … |
| 6390 | 10523794-7711535 | 45.7 ± 0.3 | 4719 ± 226 | … | … | … | … | … | N | … | … | … | … | n | … |
| 6391 | 10523799-7718477 | -24.8 ± 0.2 | 5026 ± 155 | … | 1.013 ± 0.004 | -0.20 ± 0.21 | <10 | 3 | N | … | … | … | … | n | G |
| 6618 | 11044101-7640133 | 0.9 ± 0.2 | 4688 ± 24 | 2.53 ± 0.06 | 1.010 ± 0.004 | -0.08 ± 0.04 | <29 | 3 | N | N | … | … | … | n | G |
| 6691 | 11070546-7748375 | -17.0 ± 0.3 | 5376 ± 21 | … | 1.004 ± 0.006 | -0.16 ± 0.10 | … | … | Y | … | N | … | Y | n | … |
| 6692 | 11070584-7753387 | 21.6 ± 0.2 | 4942 ± 111 | 2.68 ± 0.11 | 1.015 ± 0.002 | -0.13 ± 0.13 | <22 | 3 | N | N | … | … | … | n | G |
| 6693 | 11070718-7559353 | -26.5 ± 0.2 | 4566 ± 93 | 2.44 ± 0.17 | 1.017 ± 0.004 | 0.10 ± 0.03 | <57 | 3 | N | N | … | … | … | n | G |
| 6694 | 11070869-7650351 | 59.5 ± 0.3 | 4336 ± 247 | … | 1.035 ± 0.014 | -0.18 ± 0.04 | <26 | 3 | N | … | … | … | … | n | G |
| 6447 | 10565940-7729312 | 6.3 ± 0.2 | 4706 ± 171 | 2.05 ± 0.04 | 1.025 ± 0.005 | -0.27 ± 0.07 | <19 | 3 | N | N | … | … | … | n | G |
| 6695 | 11070874-7621003 | -36.5 ± 0.2 | 4238 ± 270 | … | 1.057 ± 0.002 | -0.18 ± 0.05 | <23 | 3 | N | … | … | … | … | n | G |
| 6448 | 10565981-7617043 | 37.3 ± 0.2 | 4794 ± 136 | 2.58 ± 0.15 | 1.013 ± 0.003 | -0.06 ± 0.16 | <24 | 3 | N | N | … | … | … | n | G |
| 6449 | 10570065-7635360 | 7.9 ± 0.2 | 3902 ± 81 | … | 1.048 ± 0.001 | … | <17 | 3 | N | … | … | … | … | n | … |
| 6696 | 11070919-7723049 | 445.2 ± 13.0 | … | … | … | … | … | … | … | … | … | … | … | n | … |
| 6450 | 10570085-7644110 | -20.8 ± 0.2 | 6415 ± 33 | 3.82 ± 0.07 | 1.011 ± 0.002 | 0.06 ± 0.02 | 96 ± 4 | 1 | N | Y | … | … | … | n | … |
| 6697 | 11070974-7646224 | -9.8 ± 0.3 | 3871 ± 81 | … | … | … | … | … | … | … | … | … | … | n | … |
| 6698 | 11071148-7746394$^f$ | 16.9 ± 0.3 | 3708 ± 58 | … | 0.874 ± 0.006 | -0.17 ± 0.10 | 670 ± 28 | 1 | Y | … | Y | Y | Y | Y | … |
| 6699 | 11071206-7632232$^f$ | 16.9 ± 0.2 | 4039 ± 215 | … | 0.913 ± 0.003 | -0.16 ± 0.03 | 573 ± 17 | 1 | Y | … | Y | Y | Y | Y | … |
| 6451 | 10570418-7735190 | 71.6 ± 0.2 | 3832 ± 234 | … | 1.057 ± 0.002 | -0.30 ± 0.10 | <3 | 3 | N | … | … | … | … | n | … |
| 6452 | 10570623-7653321 | -13.9 ± 0.2 | 4777 ± 47 | 2.86 ± 0.06 | 1.009 ± 0.004 | 0.00 ± 0.02 | <32 | 3 | Y | N | N | N | Y | n | NG? |
| 6700 | 11071330-7743498 | 15.1 ± 0.4 | 3411 ± 28 | … | 0.876 ± 0.017 | … | 667 ± 19 | 1 | Y | … | Y | Y | … | Y | … |
| 6453 | 10570802-7703158 | 32.5 ± 0.4 | … | … | … | … | … | … | … | … | … | … | … | n | … |
| 6701 | 11071445-7750266 | 8.4 ± 0.3 | 5114 ± 18 | 3.55 ± 0.13 | 0.997 ± 0.010 | -0.10 ± 0.19 | … | … | Y | Y | N | … | Y | n | … |
| 6702 | 11071530-7754060 | 4.7 ± 0.2 | 6384 ± 31 | … | 0.995 ± 0.002 | 0.35 ± 0.02 | <2 | 3 | Y | … | N | N | N | n | … |
| 6703 | 11071590-7608063 | -15.5 ± 0.2 | 3316 ± 103 | … | … | … | … | … | … | … | … | … | … | n | … |
| 6704 | 11071622-7723068 | 511.9 ± 29.6 | … | … | … | … | … | … | … | … | … | … | … | n | … |
| 6705 | 11071656-7726215 | 44.2 ± 0.2 | 4003 ± 184 | … | 1.066 ± 0.002 | -0.15 ± 0.09 | <15 | 3 | N | … | … | … | … | n | … |
| 6706 | 11071915-7603048$^f$ | 15.6 ± 0.3 | 3658 ± 40 | … | 0.864 ± 0.008 | -0.19 ± 0.12 | 633 ± 39 | 1 | Y | … | Y | Y | Y | Y | … |
| 6707 | 11072022-7738111$^f$ | 15.5 ± 0.3 | 3423 ± 89 | … | 0.919 ± 0.010 | -0.26 ± 0.14 | 528 ± 23 | 1 | Y | … | Y | Y | Y | Y | … |
| 6755 | 11083044-7552317 | -57.3 ± 0.2 | 4523 ± 153 | 1.93 ± 0.19 | 1.036 ± 0.003 | -0.12 ± 0.15 | <22 | 3 | N | N | … | … | … | n | G |
| 6756 | 11083055-7630341 | -6.0 ± 0.7 | 6406 ± 130 | … | 1.002 ± 0.011 | 0.11 ± 0.20 | <28 | 3 | Y | … | N | N | Y | n | NG? |
| 6757 | 11083105-7740324 | 15.6 ± 0.5 | 6039 ± 6 | 4.21 ± 0.19 | 0.991 ± 0.013 | -0.02 ± 0.12 | <32 | 3 | Y | Y | Y | N | Y | n | NG |
| 6758 | 11083248-7716537 | -20.3 ± 0.3 | 4502 ± 105 | … | 1.029 ± 0.009 | -0.09 ± 0.11 | <43 | 3 | N | … | … | … | … | n | … |
| 6759 | 11083651-7759535 | 84.0 ± 0.2 | 4751 ± 49 | 2.55 ± 0.06 | 1.018 ± 0.006 | -0.09 ± 0.05 | <36 | 3 | N | N | … | … | … | n | G |
| 6760 | 11083905-7716042$^f$ | 14.1 ± 0.2 | 4272 ± 310 | … | 0.939 ± 0.004 | -0.05 ± 0.08 | 551 ± 5 | 1 | Y | … | Y | Y | Y | Y | … |
| 6761 | 11083952-7734166 | 14.6 ± 0.4 | … | … | … | … | … | … | … | … | … | … | … | n | … |
| 635 | 11084041-7756310 | -5.1 ± 0.6 | 6429 ± 171 | 3.88 ± 0.12 | … | -0.01 ± 0.12 | 89 ± 17 | 2 | … | Y | N | N | Y | n | NG |
| 6762 | 11084069-7636078 | 13.6 ± 0.2 | 3872 ± 187 | … | 0.906 ± 0.004 | -0.15 ± 0.14 | 635 ± 7 | 1 | Y | … | Y | Y | Y | Y | … |
| 6763 | 11084266-7510426 | 113.6 ± 0.2 | 3810 ± 134 | … | 1.036 ± 0.005 | -0.06 ± 0.15 | <18 | 3 | N | … | … | … | … | n | … |
| 6764 | 11084296-7743500 | -164.1 ± 59.8 | … | … | … | … | … | … | … | … | … | … | … | n | … |
| 6765 | 11084302-7627468 | 5.9 ± 0.3 | 4516 ± 73 | … | 1.031 ± 0.013 | -0.15 ± 0.09 | … | … | N | … | … | … | … | n | … |
| 6766 | 11084406-7646108 | -35.9 ± 0.3 | 4383 ± 244 | … | 1.034 ± 0.014 | -0.15 ± 0.04 | <14 | 3 | N | … | … | … | … | n | G |
| 6767 | 11084442-7736499 | 104.0 ± 0.3 | 4940 ± 177 | … | 1.013 ± 0.008 | -0.22 ± 0.35 | <25 | 3 | N | … | … | … | … | n | G |
| 6828 | 11100369-7633291$^f$ | 16.4 ± 0.6 | 4014 ± 219 | … | 0.856 ± 0.013 | -0.07 ± 0.07 | 415 ± 10 | 1 | Y | … | Y | Y | Y | Y | … |
| 6829 | 11100469-7635452$^f$ | 16.6 ± 0.2 | 4157 ± 322 | … | 0.932 ± 0.003 | -0.10 ± 0.10 | 527 ± 11 | 1 | Y | … | Y | Y | Y | Y | … |
| 6830 | 11100523-7756292 | 15.8 ± 0.3 | 4536 ± 143 | … | 1.011 ± 0.010 | -0.16 ± 0.25 | <27 | 3 | N | … | … | … | … | n | G |
| 642 | 11100704-7629377$^f$ | 14.7 ± 0.6 | 4519 ± 353 | 4.34 ± 0.46 | … | -0.28 ± 0.23 | 496 ± 13 | 2 | … | Y | Y | Y | Y | Y | … |
| 6831 | 11100864-7746177 | 22.6 ± 0.3 | 4746 ± 54 | … | 0.998 ± 0.008 | -0.05 ± 0.15 | <30 | 3 | Y | … | N | N | Y | n | NG |
| 6832 | 11101141-7635292$^f$ | 17.5 ± 0.3 | 4394 ± 146 | … | 0.966 ± 0.007 | -0.06 ± 0.05 | 544 ± 6 | 1 | Y | … | Y | Y | Y | Y | … |
| 6833 | 11101153-7733522 | 16.4 ± 0.4 | 3305 ± 15 | … | 0.902 ± 0.009 | -0.26 ± 0.14 | 563 ± 17 | 1 | Y | … | Y | Y | Y | Y | … |
| 6834 | 11101596-7757556 | -32.7 ± 0.2 | 4571 ± 52 | … | 1.044 ± 0.003 | -0.07 ± 0.06 | … | … | N | … | … | … | … | n | … |
| 6835 | 11102633-7728394 | 19.8 ± 2.8 | 6773 ± 289 | … | 1.008 ± 0.016 | 0.11 ± 0.25 | <31 | 3 | Y | … | N | N | Y | n | NG? |
| 6836 | 11102819-7626164 | 26.3 ± 0.2 | 4501 ± 136 | … | 1.036 ± 0.002 | … | <37 | 3 | N | … | … | … | … | n | G |
| 6837 | 11102852-7716596 | 17.6 ± 0.4 | 3271 ± 80 | … | 0.927 ± 0.009 | -0.26 ± 0.14 | 613 ± 7 | 1 | Y | … | Y | Y | Y | Y | … |
| 6838 | 11103110-7742001 | -24.7 ± 0.3 | 4852 ± 215 | … | 1.020 ± 0.011 | -0.10 ± 0.20 | <26 | 3 | N | … | … | … | … | n | G |
| 6839 | 11103115-7820052 | -42.5 ± 0.2 | 3918 ± 84 | … | 1.046 ± 0.005 | -0.17 ± 0.17 | … | … | N | … | … | … | … | n | … |
| 6840 | 11103416-7750203 | 1.5 ± 0.2 | 4585 ± 77 | 2.46 ± 0.16 | 1.018 ± 0.003 | 0.09 ± 0.04 | <49 | 3 | N | N | … | … | … | n | G |
| 6841 | 11103481-7722053 | 15.7 ± 1.3 | … | … | … | … | … | … | … | … | … | … | … | n | … |

**Table C.2.** continued.

| ID | CNAME | RV (km s$^{-1}$) | $T_{\text{eff}}$ (K) | logg (dex) | $\gamma^a$ | [Fe/H] (dex) | EW(Li)$^b$ (mÅ) | EW(Li) error flag$^c$ | Membership $\gamma$ | logg | RV | Li | [Fe/H] | Final$^d$ | Non-mem with Li$^e$ |
|---|---|---|---|---|---|---|---|---|---|---|---|---|---|---|---|
| 6842 | 11103801-7732399$^f$ | 15.5 ± 0.2 | 5177 ± 91 | ... | 0.978 ± 0.002 | ... | 412 ± 14 | 1 | Y | ... | Y | Y | ... | Y | ... |
| 6843 | 11103996-7647079 | 44.7 ± 0.4 | 4972 ± 229 | ... | 0.933 ± 0.022 | 0.06 ± 0.15 | ... | ... | Y | ... | N | ... | Y | n | ... |
| 6844 | 11104343-7641132 | 328.7 ± 528.3 | ... | ... | ... | ... | ... | ... | ... | ... | ... | ... | ... | n | ... |
| 6845 | 11104959-7717517 | -700.0 ± 0.2 | 3726 ± 73 | ... | 0.888 ± 0.005 | -0.15 ± 0.11 | ... | ... | Y | ... | N | ... | Y | n | ... |
| 6846 | 11105038-7631441 | -3.3 ± 0.2 | 4573 ± 86 | ... | 1.047 ± 0.002 | ... | <45 | 3 | N | ... | ... | ... | ... | n | G |
| 6847 | 11105076-7718032 | 15.2 ± 0.3 | 3332 ± 13 | ... | 0.883 ± 0.013 | -0.26 ± 0.14 | 626 ± 12 | 1 | Y | ... | Y | Y | Y | Y | ... |
| 6848 | 11105215-7709286 | 1.8 ± 0.2 | 4661 ± 21 | 2.41 ± 0.01 | 1.023 ± 0.003 | -0.02 ± 0.06 | <33 | 3 | N | N | ... | ... | ... | n | G |
| 6849 | 11105333-7634319 | -700.0 ± 0.2 | 3254 ± 48 | ... | 0.919 ± 0.005 | -0.17 ± 0.17 | ... | ... | Y | ... | N | ... | Y | n | ... |
| 6850 | 11105597-7645325 | 19.3 ± 0.5 | ... | ... | ... | ... | ... | ... | ... | ... | ... | ... | ... | n | ... |
| 6386 | 10515793-7711491 | 90.3 ± 0.4 | 4953 ± 171 | ... | 1.025 ± 0.012 | -0.10 ± 0.04 | ... | ... | N | ... | ... | ... | ... | n | ... |
| 6387 | 10520512-7726476 | 249.1 ± 0.2 | 4204 ± 268 | ... | 1.059 ± 0.004 | -0.23 ± 0.06 | ... | ... | N | ... | ... | ... | ... | n | ... |
| 6392 | 10531969-7704399 | 7.1 ± 0.3 | 4590 ± 106 | ... | 1.035 ± 0.014 | ... | <25 | 3 | N | ... | ... | ... | ... | n | ... |
| 6393 | 10532373-7721256 | 5.4 ± 0.2 | 3970 ± 21 | 1.17 ± 0.17 | 1.027 ± 0.009 | 0.10 ± 0.24 | ... | ... | N | N | ... | ... | ... | n | ... |
| 6394 | 10532392-7734035 | -12.5 ± 0.2 | 3560 ± 115 | ... | ... | ... | ... | ... | ... | ... | ... | ... | ... | n | ... |
| 6395 | 10532815-7710268 | 26.8 ± 0.4 | 5685 ± 70 | ... | 1.012 ± 0.014 | 0.25 ± 0.09 | <33 | 3 | N | ... | ... | ... | ... | n | ... |
| 6510 | 11001129-7756318 | -25.2 ± 0.2 | 4696 ± 17 | 2.56 ± 0.01 | 1.011 ± 0.006 | 0.00 ± 0.01 | <35 | 3 | N | N | ... | ... | ... | n | G |
| 6413 | 10550051-7624268 | 16.3 ± 0.2 | 4444 ± 143 | ... | 1.051 ± 0.002 | ... | <9 | 3 | N | ... | ... | ... | ... | n | G |
| 6414 | 10550245-7736205 | -11.7 ± 0.3 | 4458 ± 168 | ... | 1.027 ± 0.009 | -0.13 ± 0.05 | <39 | 3 | N | ... | ... | ... | ... | n | G |
| 6415 | 10550312-7620073 | -0.9 ± 0.2 | 4634 ± 53 | 2.39 ± 0.07 | 1.015 ± 0.002 | -0.08 ± 0.03 | <26 | 3 | N | N | ... | ... | ... | n | G |
| 6416 | 10550964-7730540 | 16.2 ± 0.6 | ... | ... | ... | ... | ... | ... | ... | ... | ... | ... | ... | n | ... |
| 6417 | 10551035-7621480 | -61.6 ± 0.2 | 4951 ± 213 | ... | 1.016 ± 0.002 | -0.33 ± 0.03 | ... | ... | N | ... | ... | ... | ... | n | ... |
| 6520 | 11002857-7629164 | 0.7 ± 0.3 | 6429 ± 45 | 4.08 ± 0.09 | 1.003 ± 0.003 | -0.02 ± 0.03 | 93 ± 2 | 1 | Y | Y | N | N | Y | n | NG? |
| 6418 | 10551478-7659216 | -10.5 ± 0.2 | 4962 ± 174 | ... | 1.014 ± 0.007 | -0.06 ± 0.06 | ... | ... | N | ... | ... | ... | ... | n | ... |
| 6419 | 10551638-7659406 | 5.2 ± 0.2 | 4436 ± 172 | ... | 1.055 ± 0.005 | -0.17 ± 0.13 | <15 | 3 | N | ... | ... | ... | ... | n | G |
| 6420 | 10551988-7733186 | 46.1 ± 0.2 | 4435 ± 199 | ... | 1.035 ± 0.002 | -0.15 ± 0.17 | <22 | 3 | N | ... | ... | ... | ... | n | G |
| 6421 | 10552521-7717275 | 1.8 ± 0.2 | 5012 ± 173 | ... | 1.010 ± 0.006 | -0.18 ± 0.24 | ... | ... | N | ... | ... | ... | ... | n | ... |
| 6422 | 10553321-7700312 | 51.1 ± 0.2 | 4728 ± 57 | 2.50 ± 0.16 | 1.020 ± 0.004 | -0.11 ± 0.09 | <24 | 3 | N | N | ... | ... | ... | n | G |
| 6423 | 10553487-7720502 | 30.7 ± 0.2 | 4093 ± 301 | ... | 1.051 ± 0.005 | -0.25 ± 0.02 | ... | ... | N | ... | ... | ... | ... | n | ... |
| 6424 | 10553587-7720013 | 21.5 ± 0.3 | 4580 ± 182 | ... | 1.018 ± 0.009 | -0.03 ± 0.15 | <42 | 3 | N | ... | ... | ... | ... | n | G |
| 6425 | 10553718-7732243 | -41.2 ± 0.2 | 5072 ± 75 | ... | 1.021 ± 0.002 | ... | <12 | 3 | N | ... | ... | ... | ... | n | G |
| 6383 | 10501991-7713482 | 36.6 ± 0.3 | 5733 ± 6 | ... | 0.997 ± 0.007 | -0.03 ± 0.04 | ... | ... | Y | ... | N | ... | Y | n | ... |
| 6426 | 10554074-7736258 | 12.5 ± 0.3 | 5778 ± 35 | 4.08 ± 0.10 | 0.994 ± 0.005 | -0.26 ± 0.20 | 62 ± 3 | 1 | Y | Y | N | N | Y | n | NG |
| 6427 | 10554584-7701193 | 27.5 ± 0.2 | 4371 ± 275 | ... | 1.037 ± 0.004 | -0.10 ± 0.07 | <27 | 3 | N | ... | ... | ... | ... | n | G |
| 6522 | 11003689-7726089 | 6.0 ± 0.3 | 6491 ± 58 | ... | 0.999 ± 0.003 | -0.28 ± 0.05 | 64 ± 2 | 1 | Y | ... | N | N | Y | n | NG |
| 6428 | 10554589-7713053 | 33.2 ± 0.8 | ... | ... | ... | ... | ... | ... | ... | ... | ... | ... | ... | n | ... |
| 6521 | 11003216-7811239 | -2.6 ± 0.3 | 4565 ± 33 | 2.30 ± 0.10 | 1.024 ± 0.008 | -0.06 ± 0.07 | <41 | 3 | N | N | ... | ... | ... | n | G |
| 6429 | 10554599-7716229 | -33.5 ± 0.3 | 4976 ± 163 | ... | 1.010 ± 0.012 | -0.17 ± 0.15 | <20 | 3 | N | ... | ... | ... | ... | n | G |
| 6430 | 10554668-7713597 | -6.5 ± 0.5 | 6520 ± 172 | ... | 0.990 ± 0.012 | 0.46 ± 0.13 | 133 ± 8 | 1 | Y | ... | N | N | N | n | NG |
| 667 | 11010007-7738516 | 108.5 ± 0.6 | 3937 ± 90 | 1.24 ± 0.13 | 1.054 ± 0.003 | -0.27 ± 0.12 | 318 ± 1 | 2 | N | N | ... | ... | ... | n | Li-rich G |
| 6431 | 10554967-7634237 | 26.5 ± 0.2 | 3893 ± 79 | ... | 1.035 ± 0.002 | ... | <23 | 3 | N | ... | ... | ... | ... | n | ... |
| 6432 | 10555220-7736062 | 117.5 ± 0.2 | 4453 ± 290 | ... | 1.050 ± 0.005 | -0.68 ± 0.21 | <8 | 3 | N | ... | ... | ... | ... | n | ... |
| 6433 | 10555930-7718207 | 27.9 ± 0.2 | 4529 ± 109 | ... | 1.036 ± 0.006 | -0.06 ± 0.02 | <25 | 3 | N | ... | ... | ... | ... | n | G |
| 631 | 11075588-7727257 | 18.0 ± 0.6 | 4640 ± 210 | 4.50 ± 0.15 | ... | -0.07 ± 0.13 | 510 ± 20 | 2 | ... | Y | Y | Y | Y | Y | ... |
| 661 | 10555973-7724399 | ... | ... | ... | ... | ... | ... | ... | ... | ... | ... | ... | ... | n | ... |
| 6730 | 11075699-7741558 | 25.5 ± 0.2 | 4047 ± 113 | 1.42 ± 0.12 | 1.040 ± 0.007 | 0.22 ± 0.34 | <41 | 3 | N | N | ... | ... | ... | n | G |
| 6434 | 10560172-7635289 | -18.3 ± 0.2 | 4870 ± 240 | ... | 1.022 ± 0.003 | -0.17 ± 0.21 | <15 | 3 | N | ... | ... | ... | ... | n | G |
| 6731 | 11075792-7738449 | -700.0 ± 0.2 | 5624 ± 140 | ... | 1.000 ± 0.007 | ... | ... | ... | Y | ... | N | ... | ... | n | ... |
| 6732 | 11075809-7742413$^f$ | 20.5 ± 0.7 | 3434 ± 155 | ... | 0.878 ± 0.012 | -0.19 ± 0.14 | 353 ± 32 | 1 | Y | ... | N | Y | Y | Y | ... |
| 6532 | 11010697-7734544 | 12.9 ± 0.3 | 4460 ± 110 | 2.12 ± 0.11 | 1.024 ± 0.014 | 0.06 ± 0.01 | <65 | 3 | N | N | ... | ... | ... | n | G |
| 6733 | 11075845-7653442 | 60.4 ± 0.3 | 5767 ± 31 | ... | 0.999 ± 0.008 | -0.09 ± 0.02 | <14 | 3 | Y | ... | N | N | Y | n | NG |
| 6734 | 11075850-7518295 | 102.9 ± 0.2 | 3815 ± 84 | ... | 1.045 ± 0.003 | 0.01 ± 0.11 | <4 | 3 | N | ... | ... | ... | ... | n | ... |
| 6533 | 11010798-7658326 | 4.4 ± 0.2 | 4624 ± 59 | 2.35 ± 0.11 | 1.020 ± 0.004 | -0.09 ± 0.04 | <28 | 3 | N | N | ... | ... | ... | n | G |
| 6454 | 10572460-7621289 | 51.4 ± 0.2 | 4621 ± 3 | 2.27 ± 0.16 | 1.021 ± 0.004 | -0.26 ± 0.12 | ... | ... | N | N | ... | ... | ... | n | ... |
| 6740 | 11080394-7508458 | 17.8 ± 0.2 | 5017 ± 74 | 3.16 ± 0.18 | 1.009 ± 0.002 | -0.08 ± 0.12 | <17 | 3 | Y | N | Y | N | Y | n | NG? |
| 6534 | 11010913-7733092 | 28.9 ± 1.7 | ... | ... | ... | ... | ... | ... | ... | ... | ... | ... | ... | n | ... |
| 6735 | 11075919-7611099 | 8.2 ± 0.2 | 4332 ± 215 | ... | 1.038 ± 0.003 | 0.01 ± 0.05 | <45 | 3 | N | ... | ... | ... | ... | n | G |
| 6455 | 10572504-7704387 | 26.9 ± 0.2 | 4636 ± 125 | 2.52 ± 0.19 | 1.013 ± 0.003 | 0.02 ± 0.10 | <29 | 3 | N | N | ... | ... | ... | n | G |
| 6736 | 11075993-7715317 | 19.1 ± 2.9 | ... | ... | ... | ... | ... | ... | ... | ... | ... | ... | ... | n | ... |
| 633 | 11080412-7513273 | -13.6 ± 0.6 | 4419 ± 57 | 2.10 ± 0.10 | ... | -0.02 ± 0.11 | 35 ± 9 | 2 | ... | N | N | ... | Y | n | ... |

**Table C.2.** continued.

| ID | CNAME | RV (km s$^{-1}$) | $T_{\rm eff}$ (K) | logg (dex) | $\gamma$[a] | [Fe/H] (dex) | EW(Li)[b] (mÅ) | EW(Li) error flag[c] | $\gamma$ | logg | RV | Li | [Fe/H] | Final[d] | Non-mem with Li[e] |
|---|---|---|---|---|---|---|---|---|---|---|---|---|---|---|---|
| 6737 | 11080002-7717304 | 16.5 ± 0.6 | … | … | … | … | … | … | … | … | … | … | … | n | … |
| 632 | 11080148-7742288 | 11.9 ± 0.6 | … | … | … | … | … | … | … | … | … | … | … | n | … |
| 6535 | 11011356-7752378 | 37.0 ± 0.3 | 5139 ± 145 | … | 1.013 ± 0.010 | … | … | … | N | … | … | … | … | n | … |
| 6738 | 11080151-7552330 | 30.5 ± 0.2 | 4563 ± 109 | … | 1.030 ± 0.006 | … | <39 | 3 | N | … | … | … | … | n | G |
| 6536 | 11011426-7809320 | 15.3 ± 0.2 | 6052 ± 149 | 4.23 ± 0.05 | 0.994 ± 0.003 | 0.19 ± 0.09 | <18 | 3 | Y | Y | Y | N | N | n | NG |
| 6768 | 11084830-7753001 | 73.1 ± 0.3 | 4463 ± 199 | … | 1.018 ± 0.011 | -0.30 ± 0.23 | <32 | 3 | N | … | … | … | … | n | G |
| 6769 | 11084887-7715174 | 50.9 ± 0.3 | 4772 ± 43 | 2.77 ± 0.17 | 1.003 ± 0.015 | -0.07 ± 0.23 | <39 | 3 | Y | N | N | N | Y | n | NG? |
| 6544 | 11013026-7536490 | 81.6 ± 0.2 | 3837 ± 88 | … | … | … | <2 | 3 | … | … | … | … | … | n | … |
| 6739 | 11080297-7738425 | 18.9 ± 0.5 | 3983 ± 212 | … | 0.954 ± 0.010 | -0.21 ± 0.09 | … | … | Y | … | Y | … | Y | n | … |
| 6456 | 10573004-7620097 | 40.6 ± 0.2 | 4606 ± 47 | 2.36 ± 0.13 | 1.022 ± 0.001 | -0.07 ± 0.01 | … | … | N | N | … | … | … | n | … |
| 6770 | 11085090-7625135[f] | 14.7 ± 0.3 | 3147 ± 258 | … | 0.909 ± 0.012 | -0.24 ± 0.15 | 577 ± 14 | 1 | Y | … | Y | Y | Y | Y | … |
| 6545 | 11013177-7742095 | 21.8 ± 0.2 | 4505 ± 48 | 1.76 ± 0.08 | 1.028 ± 0.006 | -0.38 ± 0.07 | <15 | 3 | N | N | … | … | … | n | G |
| 6771 | 11085109-7628214 | 17.6 ± 0.2 | 5367 ± 45 | 4.20 ± 0.12 | 0.984 ± 0.003 | -0.01 ± 0.14 | … | … | Y | Y | Y | … | Y | n | … |
| 6457 | 10573075-7743370 | -19.3 ± 0.2 | 4001 ± 201 | … | 1.051 ± 0.003 | -0.21 ± 0.09 | <22 | 3 | N | … | … | … | … | n | … |
| 636 | 11085231-7743329 | … | … | … | … | … | … | … | … | … | … | … | … | n | … |
| 6546 | 11013297-7817390 | -21.2 ± 0.2 | 4670 ± 229 | … | 1.030 ± 0.005 | -0.17 ± 0.24 | <14 | 3 | N | … | … | … | … | n | G |
| 6772 | 11085242-7519027[f] | 14.9 ± 0.2 | 3559 ± 162 | … | 0.916 ± 0.004 | -0.23 ± 0.14 | 565 ± 5 | 1 | Y | … | Y | Y | Y | Y | … |
| 637 | 11085326-7519374 | 53.0 ± 0.6 | … | … | … | … | … | … | … | … | … | … | … | n | … |
| 6773 | 11085367-7521359[f] | 15.1 ± 0.2 | 3774 ± 49 | … | 0.886 ± 0.004 | -0.19 ± 0.13 | 316 ± 4 | 1 | Y | … | Y | Y | Y | Y | … |
| 6547 | 11013437-7817558 | 33.3 ± 0.2 | 4869 ± 118 | 2.68 ± 0.16 | 1.018 ± 0.004 | -0.08 ± 0.09 | <22 | 3 | N | N | … | … | … | n | G |
| 7016 | 11223400-7630315 | -3.6 ± 0.2 | 4692 ± 69 | 2.42 ± 0.13 | 1.028 ± 0.002 | -0.02 ± 0.03 | <36 | 3 | N | N | … | … | … | n | G |
| 6774 | 11085373-7649301 | 0.8 ± 0.3 | 5467 ± 278 | 4.10 ± 0.15 | 0.994 ± 0.007 | 0.06 ± 0.09 | <12 | 3 | Y | Y | N | N | Y | n | NG |
| 6460 | 10574219-7659356 | 16.8 ± 0.2 | 3455 ± 174 | … | 0.886 ± 0.005 | -0.24 ± 0.14 | 598 ± 25 | 1 | Y | … | Y | Y | Y | Y | … |
| 6549 | 11013493-7538118 | -0.6 ± 0.2 | 4628 ± 110 | 2.39 ± 0.13 | 1.022 ± 0.002 | 0.01 ± 0.01 | <37 | 3 | N | N | … | … | … | n | G |
| 6775 | 11085422-7732115 | 13.9 ± 0.4 | 3255 ± 17 | … | 0.934 ± 0.018 | … | 572 ± 13 | 1 | Y | … | Y | Y | … | Y | … |
| 6776 | 11085433-7651285 | 54.9 ± 0.2 | 4809 ± 152 | … | 1.027 ± 0.005 | -0.27 ± 0.22 | <11 | 3 | N | … | … | … | … | n | G |
| 6550 | 11015081-7543551 | -45.4 ± 0.2 | 4620 ± 70 | 2.23 ± 0.17 | 1.033 ± 0.001 | -0.11 ± 0.08 | <21 | 3 | N | N | … | … | … | n | G |
| 6461 | 10574416-7638073 | -5.3 ± 0.2 | 4419 ± 171 | … | 1.039 ± 0.001 | -0.18 ± 0.04 | <20 | 3 | N | … | … | … | … | n | G |
| 6777 | 11085464-7702129[f] | 15.6 ± 0.3 | 4005 ± 237 | … | 0.897 ± 0.006 | -0.10 ± 0.04 | 461 ± 27 | 1 | Y | … | Y | Y | Y | Y | … |
| 6551 | 11015483-7744445 | -10.0 ± 0.4 | 6441 ± 193 | 4.29 ± 0.40 | 0.998 ± 0.012 | 0.20 ± 0.08 | <29 | 3 | Y | Y | N | N | N | n | NG |
| 7017 | 11223654-7626375 | -23.1 ± 0.2 | 4158 ± 11 | … | 1.049 ± 0.004 | -0.18 ± 0.01 | <25 | 3 | N | … | … | … | … | n | … |
| 6778 | 11085500-7647433 | -22.2 ± 0.3 | 5605 ± 147 | … | 1.000 ± 0.008 | -0.49 ± 0.20 | … | … | Y | … | N | … | N | n | … |
| 6552 | 11020137-7748096 | 34.3 ± 0.5 | 4497 ± 223 | … | 1.003 ± 0.020 | -0.79 ± 0.29 | … | … | Y | … | N | … | N | n | … |
| 6462 | 10574434-7624334 | 48.3 ± 0.2 | 4677 ± 107 | … | 1.027 ± 0.003 | -0.46 ± 0.08 | <13 | 3 | N | … | … | … | … | n | G |
| 669 | 11020524-7525093 | -26.3 ± 0.6 | 4382 ± 168 | 1.91 ± 0.41 | … | -0.26 ± 0.08 | <23 | 3 | … | N | N | … | Y | n | … |
| 6553 | 11021315-7812133 | 58.0 ± 0.2 | 4505 ± 85 | … | 1.044 ± 0.005 | -0.17 ± 0.03 | <28 | 3 | N | … | … | … | … | n | G |
| 6779 | 11085527-7704502 | -26.8 ± 0.2 | 3930 ± 107 | … | 1.059 ± 0.003 | -0.09 ± 0.14 | … | … | N | … | … | … | … | n | … |
| 6554 | 11021336-7531193 | 11.2 ± 0.2 | 4133 ± 312 | … | 1.057 ± 0.003 | -0.22 ± 0.03 | <22 | 3 | N | … | … | … | … | n | … |
| 6780 | 11085632-7731518 | -9.3 ± 0.2 | 5062 ± 88 | … | 1.010 ± 0.002 | -0.08 ± 0.12 | <23 | 3 | Y | … | N | N | Y | n | NG? |
| 6555 | 11021608-7720280 | 49.3 ± 0.2 | 4284 ± 306 | … | 1.034 ± 0.009 | -0.21 ± 0.05 | <30 | 3 | N | … | … | … | … | n | G |
| 6556 | 11021927-7536576 | 16.2 ± 0.5 | … | … | … | … | … | … | … | … | … | … | … | n | … |
| 663 | 10574797-7617429 | 5.9 ± 0.6 | 5046 ± 26 | 2.85 ± 0.03 | … | 0.17 ± 0.05 | <14 | 3 | … | N | N | … | N | n | … |
| 6781 | 11085822-7733560 | 52.6 ± 0.3 | 5182 ± 40 | … | 1.011 ± 0.010 | -0.07 ± 0.05 | <20 | 3 | N | … | … | … | … | n | G |
| 6557 | 11022361-7753229 | -42.8 ± 1.8 | … | … | … | … | … | … | … | … | … | … | … | n | … |
| 6796 | 11091380-7628396 | 18.7 ± 0.4 | 3384 ± 167 | 4.45 ± 0.18 | 0.875 ± 0.007 | -0.24 ± 0.21 | 656 ± 37 | 1 | Y | Y | Y | Y | Y | Y | … |
| 6558 | 11022401-7713549 | 2.0 ± 0.3 | 4674 ± 152 | 2.33 ± 0.08 | 1.080 ± 0.014 | 0.07 ± 0.23 | <36 | 3 | N | N | … | … | … | n | G |
| 670 | 11022491-7733357[f] | 15.7 ± 0.6 | 4543 ± 174 | 4.51 ± 0.16 | … | -0.07 ± 0.13 | 509 ± 18 | 2 | … | Y | Y | Y | Y | Y | … |
| 6797 | 11091577-7509107 | 84.0 ± 0.2 | 4544 ± 116 | … | 1.052 ± 0.004 | … | <10 | 3 | N | … | … | … | … | n | G |
| 6559 | 11023265-7729129 | 16.2 ± 0.3 | 3638 ± 18 | … | 0.868 ± 0.008 | -0.19 ± 0.13 | 634 ± 8 | 1 | Y | … | Y | Y | Y | Y | … |
| 7022 | 11233269-7546306 | 14.7 ± 0.2 | 4660 ± 12 | 2.42 ± 0.17 | 1.027 ± 0.001 | -0.06 ± 0.09 | <28 | 3 | N | N | … | … | … | n | G |
| 7023 | 11233317-7616492 | -36.2 ± 0.2 | 4695 ± 57 | 2.51 ± 0.15 | 1.021 ± 0.002 | -0.03 ± 0.05 | <31 | 3 | N | N | … | … | … | n | … |
| 6589 | 11034805-7823570 | 25.6 ± 0.2 | 4924 ± 110 | 3.25 ± 0.12 | 0.998 ± 0.004 | -0.03 ± 0.09 | <37 | 3 | Y | N | N | N | Y | n | … |
| 6798 | 11091703-7639542 | 16.8 ± 0.3 | 6287 ± 142 | … | 1.019 ± 0.006 | -0.04 ± 0.02 | <11 | 3 | N | … | … | … | … | n | … |
| 6463 | 10575208-7653198 | -14.4 ± 0.2 | 3752 ± 130 | … | … | … | <100 | 3 | … | … | … | … | … | n | … |
| 639 | 11091769-7627578 | 15.7 ± 0.6 | 4515 ± 170 | 4.53 ± 0.16 | … | -0.05 ± 0.13 | 551 ± 17 | 2 | … | Y | Y | Y | Y | Y | … |
| 7024 | 11233619-7626297 | 61.3 ± 0.2 | 3873 ± 81 | … | 1.035 ± 0.002 | … | <2 | 3 | N | … | … | … | … | n | … |
| 6799 | 11091812-7630292[f] | 17.1 ± 0.4 | 4024 ± 144 | … | 0.952 ± 0.017 | -0.23 ± 0.08 | 554 ± 43 | 1 | Y | … | Y | Y | Y | Y | … |
| 623 | 11034945-7700101 | -40.4 ± 0.6 | 3895 ± 72 | 1.13 ± 0.12 | … | 0.03 ± 0.11 | <30 | 3 | … | N | N | … | Y | n | … |
| 6800 | 11092068-7713591 | -5.1 ± 0.2 | 4800 ± 67 | 2.55 ± 0.07 | 1.028 ± 0.004 | -0.06 ± 0.03 | <25 | 3 | N | N | … | … | … | n | G |

**Table C.2.** continued.

| ID | CNAME | RV (km s$^{-1}$) | $T_{\text{eff}}$ (K) | $\log g$ (dex) | $\gamma^a$ | [Fe/H] (dex) | $EW$(Li)$^b$ (mÅ) | $EW$(Li) error flag$^c$ | $\gamma$ | $\log g$ | Membership RV | Li | [Fe/H] | Final$^d$ | Non-mem with Li$^e$ |
|---|---|---|---|---|---|---|---|---|---|---|---|---|---|---|---|
| 6384 | 10513384-7709421 | 233.9 ± 235.9 | ... | ... | ... | ... | ... | ... | ... | ... | ... | ... | ... | n | ... |
| 6590 | 11035098-7543535 | -37.3 ± 0.2 | 4759 ± 47 | 2.62 ± 0.18 | 1.025 ± 0.002 | -0.03 ± 0.01 | <35 | 3 | N | N | ... | ... | ... | n | G |
| 6469 | 10581517-7734538 | 72.2 ± 0.8 | ... | ... | ... | ... | ... | ... | ... | ... | ... | ... | ... | n | ... |
| 6801 | 11092293-7631133 | -621.3 ± 28.1 | ... | ... | ... | ... | ... | ... | ... | ... | ... | ... | ... | n | ... |
| 7028 | 11235979-7624004 | 21.2 ± 0.2 | 6134 ± 103 | 4.02 ± 0.12 | 1.002 ± 0.002 | 0.10 ± 0.08 | ... | ... | Y | Y | N | ... | Y | n | ... |
| 6591 | 11035138-7631403 | -36.4 ± 0.3 | 4971 ± 108 | ... | 1.022 ± 0.008 | -0.16 ± 0.21 | <23 | 3 | N | ... | ... | ... | ... | n | G |
| 640 | 11092378-7623207$^f$ | ... | 4059 ± 159 | 4.62 ± 0.15 | ... | -0.18 ± 0.12 | 187 ± 13 | 1 | ... | Y | N | ... | Y | Y | ... |
| 6592 | 11035144-7540335 | 12.5 ± 0.7 | ... | ... | ... | ... | ... | ... | ... | ... | ... | ... | ... | n | ... |
| 6470 | 10581724-7817139 | -32.7 ± 0.3 | 5121 ± 72 | 3.28 ± 0.11 | 1.009 ± 0.009 | -0.03 ± 0.09 | <22 | 3 | Y | N | N | N | Y | n | NG? |
| 6471 | 10581736-7629397 | 38.6 ± 0.2 | 4802 ± 130 | 2.57 ± 0.17 | 1.020 ± 0.004 | -0.04 ± 0.13 | <23 | 3 | N | N | ... | ... | ... | n | G |
| 6802 | 11092482-7642073 | -42.8 ± 0.3 | 4708 ± 58 | 2.27 ± 0.11 | 1.046 ± 0.010 | -0.09 ± 0.05 | <22 | 3 | N | N | ... | ... | ... | n | G |
| 7029 | 11240017-7543069 | -5.1 ± 0.2 | 4604 ± 70 | 2.40 ± 0.18 | 1.030 ± 0.003 | 0.02 ± 0.11 | <35 | 3 | N | N | ... | ... | ... | n | G |
| 6803 | 11092582-7726243 | 8.7 ± 0.6 | 4396 ± 169 | ... | 1.050 ± 0.020 | -0.01 ± 0.16 | ... | ... | N | ... | ... | ... | ... | n | ... |
| 6804 | 11092723-7509125 | -46.9 ± 0.2 | 5091 ± 88 | ... | 1.018 ± 0.002 | ... | <18 | 3 | N | ... | ... | ... | ... | n | G |
| 7030 | 11240583-7554484 | 404.9 ± 0.2 | 3475 ± 87 | ... | ... | ... | ... | ... | ... | ... | ... | ... | ... | n | ... |
| 6851 | 11105707-7715510 | -20.9 ± 0.3 | 5097 ± 138 | ... | 1.032 ± 0.010 | -0.12 ± 0.14 | <26 | 3 | N | ... | ... | ... | ... | n | G |
| 7031 | 11241037-7623105 | -18.6 ± 0.2 | 3997 ± 115 | ... | 1.065 ± 0.002 | -0.12 ± 0.10 | ... | ... | N | ... | ... | ... | ... | n | ... |
| 6852 | 11105801-7656557 | 39.3 ± 0.7 | ... | ... | ... | ... | ... | ... | ... | ... | ... | ... | ... | n | ... |
| 6472 | 10581962-7647562 | 31.2 ± 0.2 | 4299 ± 221 | ... | 1.036 ± 0.002 | -0.07 ± 0.04 | <37 | 3 | N | ... | ... | ... | ... | n | G |
| 7032 | 11242981-7554237 | 13.0 ± 0.3 | 3301 ± 12 | ... | 0.946 ± 0.007 | -0.24 ± 0.16 | 563 ± 10 | 1 | Y | ... | Y | Y | Y | Y | ... |
| 6853 | 11105890-7751001 | -30.2 ± 0.2 | 4492 ± 80 | ... | 1.044 ± 0.006 | -0.20 ± 0.15 | <31 | 3 | N | ... | ... | ... | ... | n | G |
| 6473 | 10582075-7643457 | 23.2 ± 0.2 | 4660 ± 92 | 2.46 ± 0.17 | 1.018 ± 0.006 | -0.01 ± 0.16 | <29 | 3 | N | N | ... | ... | ... | n | G |
| 7033 | 11243521-7550357 | -49.7 ± 0.2 | 4629 ± 111 | ... | 1.023 ± 0.002 | 0.02 ± 0.08 | <41 | 3 | N | ... | ... | ... | ... | n | G |
| 6593 | 11035682-7721329 | 5.7 ± 0.3 | 3399 ± 118 | ... | 0.867 ± 0.011 | -0.28 ± 0.14 | 599 ± 10 | 1 | Y | ... | N | Y | Y | Y | ... |
| 6854 | 11110193-7816084 | 41.0 ± 0.2 | 4163 ± 271 | ... | 1.049 ± 0.005 | -0.17 ± 0.09 | <38 | 3 | N | ... | ... | ... | ... | n | G |
| 7034 | 11244111-7534508 | -31.0 ± 0.2 | 5010 ± 88 | ... | 1.019 ± 0.002 | -0.13 ± 0.11 | <12 | 3 | N | ... | ... | ... | ... | n | G |
| 6855 | 11110238-7613327 | 22.6 ± 0.2 | 5032 ± 56 | 3.26 ± 0.09 | 1.005 ± 0.007 | -0.04 ± 0.07 | 458 ± 6 | 1 | Y | N | N | Y | Y | Y | ... |
| 6474 | 10582134-7641231 | 6.8 ± 0.2 | 4940 ± 86 | ... | 1.014 ± 0.005 | -0.08 ± 0.13 | <20 | 3 | N | ... | ... | ... | ... | n | G |
| 7035 | 11251664-7535425 | 50.0 ± 0.2 | 4130 ± 360 | ... | 1.053 ± 0.003 | -0.26 ± 0.12 | <11 | 3 | N | ... | ... | ... | ... | n | G |
| 6594 | 11035785-7536251 | 40.5 ± 0.2 | 4634 ± 98 | ... | 1.023 ± 0.002 | ... | 200 ± 2 | 1 | N | ... | ... | ... | ... | n | Li-rich G |
| 6856 | 11110383-7706261 | 41.5 ± 0.2 | 4755 ± 90 | 2.54 ± 0.16 | 1.014 ± 0.006 | -0.14 ± 0.14 | <27 | 3 | N | N | ... | ... | ... | n | G |
| 6475 | 10582244-7744094 | 19.2 ± 0.2 | 4576 ± 95 | 2.48 ± 0.20 | 1.017 ± 0.005 | 0.06 ± 0.04 | <41 | 3 | N | N | ... | ... | ... | n | G |
| 657 | 11252677-7553273 | -3.6 ± 0.2 | 6482 ± 247 | 4.22 ± 0.36 | ... | -0.02 ± 0.10 | <25 | 3 | ... | Y | N | ... | Y | n | NG |
| 6857 | 11110537-7609583 | -53.8 ± 0.5 | ... | ... | ... | ... | ... | ... | ... | ... | ... | ... | ... | n | ... |
| 6476 | 10582380-7640460 | 103.4 ± 0.2 | 4383 ± 224 | ... | 1.046 ± 0.004 | -0.36 ± 0.25 | <12 | 3 | N | ... | ... | ... | ... | n | ... |
| 7036 | 11254094-7553279 | 36.6 ± 0.2 | 4518 ± 176 | ... | 1.021 ± 0.004 | 0.12 ± 0.05 | <58 | 3 | N | ... | ... | ... | ... | n | G |
| 6598 | 11040993-7716402 | 34.4 ± 0.3 | 5170 ± 14 | 3.65 ± 0.20 | 0.998 ± 0.008 | -0.03 ± 0.08 | <15 | 3 | Y | Y | N | N | Y | n | NG |
| 6858 | 11110549-7638285 | -8.2 ± 0.2 | 4662 ± 81 | 2.45 ± 0.10 | 1.016 ± 0.006 | -0.03 ± 0.03 | <38 | 3 | N | N | ... | ... | ... | n | G |
| 7037 | 11255544-7534492 | 65.4 ± 0.4 | 3514 ± 97 | ... | ... | ... | ... | ... | ... | ... | ... | ... | ... | n | ... |
| 6859 | 11110714-7722056 | 5.9 ± 0.2 | 4708 ± 7 | 2.47 ± 0.08 | 1.025 ± 0.005 | -0.04 ± 0.01 | <27 | 3 | N | N | ... | ... | ... | n | G |
| 7038 | 11255706-7547367 | -6.9 ± 0.2 | 3976 ± 3 | 1.11 ± 0.13 | 1.068 ± 0.001 | 0.06 ± 0.26 | ... | ... | N | N | ... | ... | ... | n | ... |
| 6860 | 11111017-7605258 | 118.3 ± 0.3 | ... | ... | ... | ... | ... | ... | ... | ... | ... | ... | ... | n | ... |
| 7039 | 11261403-7539566 | 8.1 ± 0.2 | 5080 ± 51 | ... | 1.012 ± 0.003 | -0.02 ± 0.06 | <24 | 3 | N | ... | ... | ... | ... | n | G |
| 6861 | 11111062-7657205 | 5.1 ± 0.2 | 4603 ± 81 | 2.31 ± 0.14 | 1.023 ± 0.005 | -0.09 ± 0.03 | ... | ... | N | N | ... | ... | ... | n | ... |
| 7040 | 11263011-7543191 | 3.9 ± 0.2 | 4445 ± 172 | 2.16 ± 0.19 | 1.027 ± 0.003 | 0.17 ± 0.03 | <59 | 3 | N | N | ... | ... | ... | n | G |
| 6862 | 11111069-7721118 | -7.2 ± 0.2 | 4768 ± 104 | 2.57 ± 0.04 | 1.026 ± 0.007 | -0.04 ± 0.10 | <31 | 3 | N | N | ... | ... | ... | n | G |
| 7041 | 11264490-7542323 | 22.0 ± 0.2 | 4760 ± 56 | 2.60 ± 0.12 | 1.016 ± 0.002 | -0.02 ± 0.10 | <23 | 3 | N | N | ... | ... | ... | n | G |
| 643 | 11111333-7731178 | 23.3 ± 0.6 | 4809 ± 42 | 3.16 ± 0.12 | 1.009 ± 0.005 | 0.16 ± 0.02 | <36 | 3 | Y | N | N | N | Y | n | NG? |
| 6863 | 11111917-7628162 | 39.2 ± 0.2 | 4327 ± 216 | ... | 1.040 ± 0.002 | -0.16 ± 0.06 | <39 | 3 | N | ... | ... | ... | ... | n | G |
| 6599 | 11041245-7743144 | 674.1 ± 63.2 | ... | ... | ... | ... | ... | ... | ... | ... | ... | ... | ... | n | ... |
| 6864 | 11111994-7721348 | 67.8 ± 0.4 | 4847 ± 252 | ... | 1.023 ± 0.014 | -0.14 ± 0.16 | ... | ... | N | ... | ... | ... | ... | n | ... |
| 6488 | 10590573-7703380 | -10.2 ± 0.2 | 4492 ± 149 | ... | 1.031 ± 0.003 | -0.03 ± 0.11 | <34 | 3 | N | ... | ... | ... | ... | n | G |
| 6865 | 11112005-7632241 | 21.2 ± 0.2 | 4109 ± 346 | ... | 1.044 ± 0.002 | -0.23 ± 0.02 | 71 ± 8 | 1 | N | ... | ... | ... | ... | n | G |
| 7047 | 11274909-7547065 | -55.6 ± 0.2 | 4021 ± 70 | 1.15 ± 0.15 | 1.060 ± 0.002 | -0.04 ± 0.16 | <26 | 3 | N | N | ... | ... | ... | n | G |
| 6866 | 11112247-7813148 | 14.1 ± 0.2 | 5105 ± 125 | ... | 1.014 ± 0.003 | ... | <11 | 3 | N | ... | ... | ... | ... | n | G |
| 6489 | 10590699-7701404$^f$ | 17.7 ± 0.2 | 5164 ± 84 | ... | 0.998 ± 0.001 | ... | 377 ± 1 | 1 | Y | ... | Y | Y | ... | Y | ... |
| 6867 | 11112260-7705538 | 14.6 ± 0.3 | 3331 ± 7 | ... | 0.908 ± 0.008 | -0.26 ± 0.14 | 638 ± 13 | 1 | Y | ... | Y | Y | Y | Y | ... |
| 6868 | 11112305-7701048 | 7.5 ± 0.4 | 5271 ± 228 | ... | 0.984 ± 0.020 | 0.27 ± 0.13 | ... | ... | Y | ... | N | ... | N | n | ... |
| 644 | 11112801-7749213 | -27.2 ± 0.6 | 3902 ± 63 | 1.20 ± 0.11 | ... | -0.05 ± 0.13 | ... | ... | ... | N | N | ... | Y | n | ... |

**Table C.2.** continued.

| ID | CNAME | RV (km s$^{-1}$) | $T_{\text{eff}}$ (K) | logg (dex) | $\gamma^a$ | [Fe/H] (dex) | EW(Li)$^b$ (mÅ) | EW(Li) error flag$^c$ | $\gamma$ | logg | RV | Li | [Fe/H] | Final$^d$ | Non-mem with Li$^e$ |
|---|---|---|---|---|---|---|---|---|---|---|---|---|---|---|---|
| 6869 | 11112906-7609292 | 65.9 ± 1.2 | 3295 ± 86 | ... | ... | ... | ... | ... | ... | ... | ... | ... | ... | n | ... |
| 6490 | 10591333-7534238 | 16.6 ± 0.2 | 4426 ± 165 | 2.08 ± 0.20 | 1.028 ± 0.003 | 0.07 ± 0.10 | <49 | 3 | N | N | ... | ... | ... | n | G |
| 6870 | 11112997-7718253 | 49.0 ± 0.2 | 4750 ± 30 | 2.80 ± 0.14 | 1.003 ± 0.003 | -0.10 ± 0.07 | <27 | 3 | Y | N | N | N | Y | n | NG? |
| 6491 | 10591513-7741087 | 12.6 ± 0.2 | 4893 ± 84 | 3.11 ± 0.25 | 1.003 ± 0.002 | -0.01 ± 0.05 | <26 | 3 | Y | N | N | N | Y | n | NG? |
| 6871 | 11113045-7712269 | -16.6 ± 0.3 | 4907 ± 216 | ... | 1.027 ± 0.010 | -0.16 ± 0.19 | ... | ... | N | ... | ... | ... | ... | n | ... |
| 6600 | 11041689-7717385 | 21.1 ± 0.2 | 4581 ± 35 | 1.97 ± 0.06 | 1.021 ± 0.004 | -0.30 ± 0.02 | <35 | 3 | N | N | ... | ... | ... | n | G |
| 6872 | 11113168-7712003 | 61.5 ± 0.3 | 4487 ± 278 | ... | 1.030 ± 0.010 | -0.27 ± 0.26 | <22 | 3 | N | ... | ... | ... | ... | n | G |
| 6896 | 11120829-7540038 | 15.3 ± 0.2 | 4675 ± 108 | 2.62 ± 0.17 | 1.010 ± 0.001 | -0.08 ± 0.11 | <24 | 3 | N | N | ... | ... | ... | n | G |
| 6492 | 10591813-7702316 | 16.7 ± 0.2 | 4486 ± 203 | ... | 1.024 ± 0.006 | -0.02 ± 0.08 | ... | ... | N | ... | ... | ... | ... | n | ... |
| 6897 | 11120984-7634366$^f$ | 17.3 ± 0.2 | 3300 ± 39 | ... | 0.940 ± 0.006 | -0.22 ± 0.13 | 547 ± 29 | 1 | Y | ... | Y | Y | Y | Y | ... |
| 6898 | 11121096-7658073 | 9.7 ± 0.6 | 6924 ± 109 | 3.98 ± 0.19 | 1.023 ± 0.006 | 0.21 ± 0.10 | ... | ... | N | Y | ... | ... | ... | n | ... |
| 6493 | 10591980-7704527 | 39.7 ± 0.2 | 4889 ± 210 | ... | 1.006 ± 0.006 | -0.26 ± 0.13 | ... | ... | Y | ... | N | ... | Y | n | ... |
| 6899 | 11121150-7826439 | 3.3 ± 0.2 | 5858 ± 33 | 4.12 ± 0.13 | 0.996 ± 0.001 | 0.05 ± 0.07 | ... | ... | Y | Y | N | ... | Y | n | ... |
| 6494 | 10592015-7757388 | 19.5 ± 0.2 | 4666 ± 52 | 2.69 ± 0.01 | 1.007 ± 0.007 | 0.10 ± 0.01 | <36 | 3 | Y | N | Y | N | Y | n | NG? |
| 6900 | 11121167-7644595 | 101.6 ± 0.4 | 4116 ± 144 | ... | 1.082 ± 0.020 | -0.23 ± 0.19 | ... | ... | N | ... | ... | ... | ... | n | ... |
| 6495 | 10592098-7745025 | -24.5 ± 0.2 | 4663 ± 34 | 2.44 ± 0.11 | 1.023 ± 0.002 | -0.01 ± 0.02 | <47 | 3 | N | N | ... | ... | ... | n | G |
| 6901 | 11121243-7602251 | 15.8 ± 0.2 | 4381 ± 241 | ... | 1.039 ± 0.003 | -0.17 ± 0.15 | <24 | 3 | N | ... | ... | ... | ... | n | G |
| 6619 | 11044241-7719041 | -13.2 ± 0.3 | 4916 ± 141 | ... | 1.029 ± 0.008 | -0.12 ± 0.18 | <49 | 3 | N | ... | ... | ... | ... | n | G |
| 6902 | 11121400-7546194 | -29.1 ± 0.2 | 4239 ± 253 | ... | 1.053 ± 0.006 | -0.04 ± 0.09 | ... | ... | N | ... | ... | ... | ... | n | ... |
| 6620 | 11044309-7508171 | 12.6 ± 0.2 | 3670 ± 99 | ... | ... | ... | <100 | 3 | ... | ... | ... | N | ... | n | ... |
| 6903 | 11121617-7733007 | 18.8 ± 0.3 | 6138 ± 36 | 3.97 ± 0.12 | 0.998 ± 0.003 | -0.30 ± 0.15 | <14 | 3 | Y | Y | Y | N | Y | n | NG |
| 6904 | 11121702-7630272 | 27.8 ± 0.2 | 4607 ± 132 | ... | 1.018 ± 0.006 | 0.05 ± 0.07 | <40 | 3 | N | ... | ... | ... | ... | n | G |
| 6905 | 11121923-7656280 | -53.0 ± 0.8 | ... | ... | ... | ... | ... | ... | ... | ... | ... | ... | ... | n | ... |
| 6621 | 11044346-7530503 | 28.3 ± 0.2 | 4546 ± 159 | ... | 1.027 ± 0.003 | 0.08 ± 0.10 | <48 | 3 | N | ... | ... | ... | ... | n | G |
| 6906 | 11122100-7655556 | 134.3 ± 0.7 | ... | ... | ... | ... | ... | ... | ... | ... | ... | ... | ... | n | ... |
| 6496 | 10592441-7742221 | -18.9 ± 0.2 | 4661 ± 24 | 2.46 ± 0.03 | 1.014 ± 0.003 | -0.03 ± 0.01 | <30 | 3 | N | N | ... | ... | ... | n | G |
| 6907 | 11122220-7824415 | 44.7 ± 0.2 | 4411 ± 162 | ... | 1.052 ± 0.004 | -0.46 ± 0.05 | ... | ... | N | ... | ... | ... | ... | n | ... |
| 624 | 11044460-7706240 | 29.8 ± 0.6 | 4304 ± 14 | 1.43 ± 0.04 | ... | -0.27 ± 0.04 | <17 | 3 | ... | N | N | ... | Y | n | ... |
| 6908 | 11122270-7610122 | -15.4 ± 0.3 | 4560 ± 223 | ... | 1.020 ± 0.011 | 0.05 ± 0.06 | <49 | 3 | N | ... | ... | ... | ... | n | G |
| 6909 | 11122441-7637064$^f$ | 15.0 ± 0.2 | 5130 ± 116 | ... | 0.983 ± 0.002 | ... | 416 ± 10 | 1 | Y | ... | Y | Y | ... | Y | ... |
| 6505 | 10595716-7812270 | -0.2 ± 0.3 | 6323 ± 74 | 4.22 ± 0.16 | 0.998 ± 0.005 | 0.20 ± 0.05 | ... | ... | Y | Y | N | ... | N | n | ... |
| 6910 | 11122501-7559376 | -26.0 ± 0.2 | 4360 ± 252 | ... | 1.029 ± 0.002 | 0.07 ± 0.01 | ... | ... | N | ... | ... | ... | ... | n | ... |
| 6622 | 11044791-7711320 | 0.3 ± 0.3 | 5128 ± 64 | 3.37 ± 0.19 | 1.010 ± 0.012 | -0.04 ± 0.12 | <22 | 3 | N | N | ... | ... | ... | n | G |
| 6911 | 11122675-7735183 | 11.4 ± 0.3 | 5632 ± 123 | 4.17 ± 0.07 | 0.987 ± 0.006 | 0.07 ± 0.06 | 200 ± 8 | 1 | Y | Y | N | N | Y | n | NG |
| 7051 | 11282127-7540372 | -42.2 ± 0.2 | 4548 ± 166 | ... | 1.034 ± 0.002 | -0.06 ± 0.04 | <29 | 3 | N | ... | ... | ... | ... | n | G |
| 6912 | 11122775-7625293 | -2.2 ± 0.3 | 3565 ± 73 | ... | 0.836 ± 0.006 | -0.23 ± 0.13 | ... | ... | Y | ... | N | ... | Y | n | ... |
| 6913 | 11123294-7727006 | 12.8 ± 0.2 | 4580 ± 76 | ... | 1.027 ± 0.006 | ... | 400 ± 6 | 1 | N | ... | ... | ... | ... | n | Li-rich G |
| 625 | 11045100-7625240 | 15.6 ± 0.6 | 4575 ± 355 | 4.47 ± 0.09 | ... | -0.08 ± 0.14 | 569 ± 13 | 2 | ... | Y | Y | Y | Y | Y | ... |
| 6914 | 11123453-7614501 | 64.7 ± 0.3 | 5068 ± 21 | 3.49 ± 0.11 | 0.999 ± 0.008 | -0.01 ± 0.06 | <32 | 3 | Y | N | N | N | Y | n | NG |
| 6915 | 11123643-7826339 | 6.5 ± 0.2 | 5123 ± 49 | ... | 1.019 ± 0.002 | -0.05 ± 0.08 | <18 | 3 | N | ... | ... | ... | ... | n | G |
| 6916 | 11123996-7650403 | -40.9 ± 0.2 | 4602 ± 98 | ... | 1.044 ± 0.005 | ... | ... | ... | N | ... | ... | ... | ... | n | ... |
| 6623 | 11045285-7625514 | 14.6 ± 0.2 | 4031 ± 266 | ... | 0.906 ± 0.003 | -0.05 ± 0.10 | 571 ± 24 | 1 | Y | ... | Y | Y | Y | Y | ... |
| 6917 | 11124087-7546567 | 35.8 ± 0.2 | 4719 ± 115 | 2.50 ± 0.15 | 1.020 ± 0.003 | 0.00 ± 0.09 | <34 | 3 | N | N | ... | ... | ... | n | G |
| 6624 | 11045366-7819351 | 5.5 ± 0.2 | 3986 ± 26 | ... | 1.031 ± 0.002 | 0.11 ± 0.25 | <44 | 3 | N | ... | ... | ... | ... | n | ... |
| 6918 | 11124210-7658400 | 15.8 ± 0.2 | 4371 ± 228 | ... | 0.951 ± 0.002 | -0.05 ± 0.05 | 495 ± 12 | 1 | Y | ... | Y | Y | Y | Y | ... |
| 6506 | 11000515-7623259 | -15.9 ± 0.2 | 4505 ± 135 | 2.22 ± 0.18 | 1.036 ± 0.003 | 0.06 ± 0.10 | 329 ± 9 | 1 | N | N | ... | ... | ... | n | Li-rich G |
| 646 | 11124268-7722230 | 14.9 ± 0.6 | 5226 ± 26 | 3.94 ± 0.09 | ... | -0.07 ± 0.01 | 360 ± 20 | 2 | ... | Y | Y | Y | Y | Y | ... |
| 652 | 11142964-7707063 | -16.6 ± 0.6 | 5938 ± 33 | 4.31 ± 0.08 | ... | 0.04 ± 0.02 | 27 ± 5 | 2 | ... | Y | N | ... | Y | n | NG |
| 6962 | 11143150-7615578 | 60.2 ± 0.2 | 4463 ± 195 | ... | 1.051 ± 0.008 | -0.24 ± 0.13 | <31 | 3 | N | ... | ... | ... | ... | n | G |
| 653 | 11143515-7539288 | -7.1 ± 0.6 | 4424 ± 40 | 1.70 ± 0.02 | ... | -0.32 ± 0.04 | <13 | 3 | ... | N | N | ... | Y | n | ... |
| 6625 | 11045701-7715569 | 283.7 ± 12.2 | ... | ... | ... | ... | ... | ... | ... | ... | ... | ... | ... | n | ... |
| 6963 | 11143740-7608349 | 16.2 ± 0.3 | 5152 ± 87 | ... | 0.964 ± 0.006 | 0.04 ± 0.03 | <11 | 3 | Y | ... | Y | N | Y | n | NG |
| 6507 | 11000766-7701110 | -5.5 ± 0.2 | 4573 ± 86 | ... | 1.030 ± 0.007 | ... | <45 | 3 | N | ... | ... | ... | ... | n | G |
| 6964 | 11144093-7828094 | 33.7 ± 0.2 | 4529 ± 54 | 1.74 ± 0.30 | 1.026 ± 0.003 | -0.40 ± 0.05 | <28 | 3 | N | N | ... | ... | ... | n | G |
| 6648 | 11053588-7638034 | -5.4 ± 0.2 | 5086 ± 60 | ... | 1.010 ± 0.002 | -0.13 ± 0.12 | <23 | 3 | N | ... | ... | ... | ... | n | G |
| 6508 | 11000911-7729383 | 5.2 ± 0.6 | ... | ... | ... | ... | ... | ... | ... | ... | ... | ... | ... | n | ... |
| 6965 | 11144757-7806332 | 32.2 ± 0.2 | 4854 ± 299 | 4.38 ± 0.07 | 0.955 ± 0.004 | 0.07 ± 0.07 | <32 | 3 | Y | Y | N | N | Y | n | NG |
| 6649 | 11053871-7607036 | 0.9 ± 0.2 | 4678 ± 39 | 2.47 ± 0.07 | 1.017 ± 0.003 | -0.02 ± 0.10 | <28 | 3 | N | N | ... | ... | ... | n | G |
| 6509 | 11000947-7622406 | 30.7 ± 0.2 | 4555 ± 147 | ... | 1.015 ± 0.004 | 0.06 ± 0.13 | <42 | 3 | N | ... | ... | ... | ... | n | G |

| ID | CNAME | RV (km s$^{-1}$) | $T_{eff}$ (K) | logg (dex) | $\gamma^a$ | [Fe/H] (dex) | $EW$(Li)$^b$ (mÅ) | $EW$(Li) error flag$^c$ | $\gamma$ | logg | RV | Li | [Fe/H] | Final$^d$ | Non-mem with Li$^e$ |
|---|---|---|---|---|---|---|---|---|---|---|---|---|---|---|---|
| 6966 | 11144785-7829132 | 4.5 ± 0.2 | 4626 ± 102 | … | 1.018 ± 0.003 | … | <75 | 3 | N | … | … | … | … | n | … |
| 6650 | 11054296-7745213 | 48.5 ± 0.3 | 5901 ± 145 | … | 1.003 ± 0.008 | 0.08 ± 0.11 | <32 | 3 | Y | … | N | N | Y | n | NG? |
| 6967 | 11144829-7534570 | 25.7 ± 0.2 | 4538 ± 144 | … | 1.018 ± 0.002 | 0.11 ± 0.07 | <41 | 3 | N | … | … | … | … | n | G |
| 6651 | 11054300-7726517 | 16.7 ± 0.3 | 3227 ± 93 | … | 0.932 ± 0.008 | -0.23 ± 0.16 | 599 ± 34 | 1 | Y | … | Y | Y | Y | Y | … |
| 6968 | 11145031-7733390 | 14.8 ± 0.2 | 3710 ± 37 | … | 0.879 ± 0.003 | -0.19 ± 0.08 | 629 ± 6 | 1 | Y | … | Y | Y | Y | Y | … |
| 6969 | 11145091-7616451 | -3.3 ± 0.2 | 4209 ± 286 | … | 1.049 ± 0.005 | -0.10 ± 0.02 | <32 | 3 | N | … | … | … | … | n | … |
| 6652 | 11054308-7754348 | 24.2 ± 0.6 | 7297 ± 243 | … | 0.997 ± 0.012 | … | <31 | 3 | Y | … | N | N | … | n | NG |
| 6970 | 11150161-7737158 | 28.9 ± 0.2 | 6015 ± 42 | 4.01 ± 0.06 | 0.997 ± 0.003 | -0.42 ± 0.24 | 27 ± 2 | 1 | Y | Y | N | N | N | n | NG |
| 6653 | 11054690-7522435 | -2.7 ± 0.2 | 4517 ± 85 | … | 1.035 ± 0.004 | -0.16 ± 0.14 | <14 | 3 | N | … | … | … | … | n | G |
| 6971 | 11150691-7809226 | 16.3 ± 0.3 | 3525 ± 58 | … | 0.836 ± 0.007 | -0.22 ± 0.15 | <100 | 3 | Y | … | Y | N | Y | n | NG |
| 6654 | 11054702-7743278 | 16.5 ± 0.6 | 6289 ± 249 | 4.07 ± 0.53 | 1.001 ± 0.014 | -0.19 ± 0.20 | <34 | 3 | Y | Y | Y | N | Y | n | NG? |
| 6972 | 11150814-7714027 | 18.6 ± 0.2 | 3853 ± 113 | … | 1.046 ± 0.003 | -0.07 ± 0.15 | <4 | 3 | N | … | … | … | … | n | … |
| 6655 | 11054774-7707059 | -42.2 ± 0.2 | 3892 ± 115 | … | 1.064 ± 0.003 | -0.13 ± 0.16 | <20 | 3 | N | … | … | … | … | n | … |
| 6973 | 11151121-7546329 | 41.5 ± 0.2 | 4527 ± 97 | … | 1.046 ± 0.003 | … | <24 | 3 | N | … | … | … | … | n | … |
| 6656 | 11054972-7640462 | 46.3 ± 0.2 | 5083 ± 208 | … | 1.013 ± 0.002 | -0.25 ± 0.25 | <15 | 3 | N | … | … | … | … | n | G |
| 6657 | 11055261-7618255 | 15.1 ± 0.2 | 4257 ± 537 | … | 0.929 ± 0.005 | 0.00 ± 0.15 | 653 ± 6 | 1 | Y | … | Y | Y | Y | Y | … |
| 6974 | 11151520-7739406 | 58.2 ± 0.2 | 4031 ± 241 | … | 1.044 ± 0.005 | -0.32 ± 0.10 | … | … | N | … | … | … | … | n | … |
| 6658 | 11055528-7551530 | 16.8 ± 0.2 | 4420 ± 172 | … | 1.031 ± 0.002 | 0.07 ± 0.12 | <34 | 3 | N | … | … | … | … | n | G |
| 6975 | 11151671-7543246 | -17.3 ± 0.2 | 4429 ± 165 | … | 1.039 ± 0.001 | -0.18 ± 0.09 | <22 | 3 | N | … | … | … | … | n | G |
| 6659 | 11055683-7656438 | -16.5 ± 0.2 | 5167 ± 43 | … | 1.013 ± 0.006 | -0.13 ± 0.10 | <27 | 3 | N | … | … | … | … | n | G |
| 6976 | 11152815-7806272 | 23.2 ± 0.2 | 4474 ± 114 | … | 1.031 ± 0.004 | -0.15 ± 0.13 | <22 | 3 | N | … | … | … | … | n | G |
| 627 | 11055780-7607489 | … | … | … | … | … | … | … | … | … | … | … | … | n | … |
| 6977 | 11153096-7729089 | 23.8 ± 0.2 | 4263 ± 278 | … | 1.057 ± 0.005 | -0.11 ± 0.05 | <27 | 3 | N | … | … | … | … | n | G |
| 6660 | 11060011-7507252 | 14.7 ± 0.6 | … | … | … | … | … | … | … | … | … | … | … | n | … |
| 6978 | 11153304-7528574 | 193.0 ± 0.2 | … | … | … | … | … | … | … | … | … | … | … | n | … |
| 6661 | 11060466-7710063 | -15.7 ± 0.4 | … | … | … | … | … | … | … | … | … | … | … | n | … |
| 6979 | 11154638-7816279 | -47.7 ± 0.2 | 3871 ± 202 | … | 1.048 ± 0.002 | -0.27 ± 0.10 | … | … | N | … | … | … | … | n | … |
| 628 | 11060511-7511454 | 79.2 ± 0.6 | 4963 ± 41 | 2.40 ± 0.06 | … | -0.47 ± 0.01 | <13 | 3 | … | N | N | … | N | n | … |
| 6980 | 11154883-7530482 | -16.0 ± 0.2 | 4680 ± 9 | 2.44 ± 0.08 | 1.022 ± 0.001 | -0.05 ± 0.02 | <29 | 3 | N | N | … | … | … | n | G |
| 6662 | 11060581-7641289 | 15.5 ± 0.2 | 4702 ± 9 | 2.50 ± 0.07 | 1.019 ± 0.008 | 0.01 ± 0.09 | <43 | 3 | N | N | … | … | … | n | G |
| 6981 | 11160186-7533520 | 143.0 ± 0.2 | 4380 ± 226 | … | 1.048 ± 0.002 | -0.29 ± 0.13 | <12 | 3 | N | … | … | … | … | n | … |
| 6663 | 11060737-7755027 | -35.6 ± 0.7 | 4901 ± 291 | … | 1.035 ± 0.022 | -0.70 ± 0.30 | … | … | N | … | … | … | … | n | … |
| 6982 | 11060542-7827391 | 28.8 ± 0.2 | 3892 ± 79 | … | 1.027 ± 0.001 | … | <1 | 3 | N | … | … | … | … | n | … |
| 6664 | 11060755-7555091 | -11.6 ± 0.2 | 5077 ± 10 | … | 1.016 ± 0.001 | -0.01 ± 0.03 | <23 | 3 | N | … | … | … | … | n | G |
| 6983 | 11161131-7813050 | 66.9 ± 0.2 | 4507 ± 80 | … | 1.027 ± 0.003 | -0.20 ± 0.06 | <21 | 3 | N | … | … | … | … | n | G |
| 6665 | 11060772-7631400 | 23.8 ± 0.2 | 4484 ± 252 | … | 1.028 ± 0.006 | -0.10 ± 0.18 | <28 | 3 | N | … | … | … | … | n | G |
| 6666 | 11060908-7640424 | 52.5 ± 0.2 | 3845 ± 40 | … | 1.041 ± 0.006 | 0.01 ± 0.11 | <15 | 3 | N | … | … | … | … | n | … |
| 6667 | 11060917-7715111 | -1.8 ± 0.7 | 5961 ± 50 | … | 1.030 ± 0.016 | -0.39 ± 0.25 | <23 | 3 | N | … | … | … | … | n | … |
| 7006 | 11210377-7612209 | -21.3 ± 0.2 | 5048 ± 117 | … | 1.023 ± 0.001 | -0.16 ± 0.14 | <18 | 3 | N | … | … | … | … | n | G |
| 6668 | 11061724-7614592 | 77.5 ± 0.2 | 4563 ± 118 | … | 1.040 ± 0.005 | … | … | … | N | … | … | … | … | n | … |
| 7007 | 11210632-7627233 | 6.1 ± 0.2 | 4577 ± 115 | … | 1.036 ± 0.002 | -0.09 ± 0.03 | 31 ± 3 | 1 | N | … | … | … | … | n | G |
| 6669 | 11062555-7633418 | 14.8 ± 2.4 | … | … | … | … | … | … | … | … | … | … | … | n | … |
| 6708 | 11072040-7729403 | 15.2 ± 0.2 | 3286 ± 12 | … | 0.914 ± 0.005 | -0.25 ± 0.14 | 619 ± 17 | 1 | Y | … | Y | Y | Y | Y | … |
| 6709 | 11072314-7649014 | 39.2 ± 0.2 | 4509 ± 150 | … | 1.041 ± 0.007 | -0.16 ± 0.21 | <41 | 3 | N | … | … | … | … | n | G |
| 6710 | 11072446-7742265 | 34.8 ± 0.3 | 5500 ± 78 | … | 1.004 ± 0.008 | … | <37 | 3 | Y | … | N | N | … | n | NG? |
| 6711 | 11072586-7701531 | 44.2 ± 0.2 | 4481 ± 68 | … | 1.047 ± 0.005 | -0.38 ± 0.04 | … | … | N | … | … | … | … | n | … |
| 6712 | 11072825-7652119$^f$ | 15.3 ± 0.2 | 3495 ± 84 | … | 0.872 ± 0.004 | -0.25 ± 0.13 | 477 ± 5 | 1 | Y | … | Y | Y | Y | Y | … |
| 6714 | 11073063-7624549 | 66.0 ± 0.2 | 4684 ± 232 | … | 1.021 ± 0.005 | -0.25 ± 0.18 | <22 | 3 | N | … | … | … | … | n | G |
| 6715 | 11073279-7645173 | 46.9 ± 0.7 | 6339 ± 180 | … | 0.983 ± 0.020 | 0.86 ± 0.14 | … | … | Y | … | N | … | N | n | … |
| 6716 | 11073302-7728277 | 19.1 ± 0.5 | 3376 ± 24 | 4.89 ± 0.16 | 0.792 ± 0.016 | -0.22 ± 0.13 | … | … | Y | Y | Y | … | Y | n | … |
| 6717 | 11073475-7509203 | -3.4 ± 0.2 | 4976 ± 67 | 3.07 ± 0.08 | 1.008 ± 0.002 | -0.01 ± 0.06 | <25 | 3 | Y | N | N | N | Y | n | NG? |
| 6718 | 11073519-7734493 | 14.8 ± 0.3 | 3402 ± 16 | … | 0.882 ± 0.013 | -0.26 ± 0.13 | 642 ± 23 | 1 | Y | … | Y | Y | Y | Y | … |
| 7015 | 11220810-7618365 | -7.6 ± 0.2 | 4321 ± 209 | 4.59 ± 0.11 | 0.886 ± 0.002 | -0.05 ± 0.10 | … | … | Y | Y | N | … | Y | n | … |
| 659 | 10505592-7707285 | … | … | … | … | … | … | … | … | … | … | … | … | n | … |
| 6385 | 10514312-7714273 | 47.3 ± 1.8 | … | … | … | … | … | … | … | … | … | … | … | n | … |
| 6388 | 10522648-7726595 | -1.5 ± 0.2 | 3871 ± 81 | … | 1.035 ± 0.004 | … | 79 ± 3 | 1 | N | … | … | … | … | n | … |
| 6396 | 10533569-7714420 | 11.4 ± 0.4 | 5700 ± 69 | … | 0.997 ± 0.012 | 0.09 ± 0.12 | … | … | Y | … | N | … | Y | n | … |
| 6397 | 10533750-7735272 | 165.3 ± 0.2 | 5197 ± 257 | … | 1.006 ± 0.003 | -0.54 ± 0.23 | <3 | 3 | Y | … | N | N | N | n | … |
| 6398 | 10533776-7735128 | 132.1 ± 0.2 | 3643 ± 122 | … | … | … | 379 ± 18 | 1 | … | … | … | … | … | n | … |

**Table C.2.** continued.

| ID | CNAME | RV (km s$^{-1}$) | $T_{\text{eff}}$ (K) | logg (dex) | $\gamma^a$ | [Fe/H] (dex) | EW(Li)$^b$ (mÅ) | EW(Li) error flag$^c$ | $\gamma$ | logg | RV | Li | [Fe/H] | Final$^d$ | Non-mem with Li$^e$ |
|---|---|---|---|---|---|---|---|---|---|---|---|---|---|---|---|
| 6399 | 10534571-7717036 | 29.2 ± 0.5 | 6243 ± 224 | ... | 1.021 ± 0.013 | -0.01 ± 0.16 | <9 | 3 | N | ... | ... | ... | ... | n | ... |
| 6497 | 10592866-7657022 | 180.5 ± 0.2 | 4890 ± 42 | ... | 1.038 ± 0.004 | -0.63 ± 0.21 | 88 ± 5 | 1 | N | ... | ... | ... | ... | n | Li-rich G |
| 6400 | 10534929-7718110 | 18.7 ± 0.2 | 5663 ± 28 | ... | 0.993 ± 0.005 | 0.07 ± 0.02 | ... | ... | Y | ... | Y | ... | Y | n | ... |
| 6401 | 10535365-7713222 | 4.8 ± 0.3 | 4710 ± 279 | ... | 1.040 ± 0.014 | -0.10 ± 0.16 | <47 | 3 | N | ... | ... | ... | ... | n | G |
| 6402 | 10535752-7718202 | -17.9 ± 0.5 | 6390 ± 93 | ... | 1.021 ± 0.006 | 0.13 ± 0.07 | 55 ± 5 | 1 | N | ... | ... | ... | ... | n | ... |
| 6403 | 10542423-7723500 | -41.0 ± 0.2 | 4583 ± 91 | ... | 1.034 ± 0.005 | ... | 292 ± 8 | 1 | N | ... | ... | ... | ... | n | Li-rich G |
| 6404 | 10543141-7710130 | -2.9 ± 1.6 | 6358 ± 390 | ... | 0.990 ± 0.020 | -0.28 ± 0.33 | ... | ... | Y | ... | N | ... | Y | n | ... |
| 6405 | 10543209-7724165 | -6.4 ± 0.3 | 4558 ± 127 | ... | 1.037 ± 0.012 | -0.31 ± 0.13 | <32 | 3 | N | ... | ... | ... | ... | n | G |
| 6498 | 10593357-7726412 | 6.5 ± 0.2 | 4516 ± 156 | ... | 1.033 ± 0.004 | 0.15 ± 0.02 | <47 | 3 | N | ... | ... | ... | ... | n | G |
| 6406 | 10543791-7723065 | 7.0 ± 0.2 | 4146 ± 292 | ... | 1.046 ± 0.006 | -0.30 ± 0.09 | <13 | 3 | N | ... | ... | ... | ... | n | ... |
| 6407 | 10543976-7720513 | -33.5 ± 0.3 | 6498 ± 58 | ... | 1.002 ± 0.003 | 0.05 ± 0.04 | ... | ... | Y | ... | N | ... | Y | n | ... |
| 6499 | 10593676-7654230 | -41.1 ± 0.2 | 4860 ± 84 | 2.70 ± 0.09 | 1.028 ± 0.005 | -0.04 ± 0.01 | ... | ... | N | N | ... | ... | ... | n | ... |
| 6408 | 10544368-7715284 | 21.5 ± 0.3 | 4548 ± 63 | 2.23 ± 0.19 | 1.031 ± 0.012 | -0.10 ± 0.01 | <60 | 3 | N | N | ... | ... | ... | n | G |
| 6409 | 10544570-7630540 | 9.1 ± 0.3 | 3851 ± 64 | 4.59 ± 0.11 | 0.824 ± 0.007 | -0.37 ± 0.33 | ... | ... | Y | Y | N | ... | N | n | ... |
| 666 | 10593816-7822421 | 11.2 ± 0.6 | 4486 ± 30 | 2.10 ± 0.12 | ... | -0.10 ± 0.04 | <18 | 3 | ... | N | N | ... | Y | n | ... |
| 6410 | 10544738-7654406 | -44.6 ± 0.2 | 4519 ± 212 | ... | 1.031 ± 0.003 | -0.19 ± 0.20 | <19 | 3 | N | ... | ... | ... | ... | n | G |
| 6411 | 10544923-7737216 | 5.8 ± 0.2 | 4349 ± 169 | ... | 1.044 ± 0.007 | -0.08 ± 0.04 | <43 | 3 | N | ... | ... | ... | ... | n | G |
| 6412 | 10545678-7709007 | -30.2 ± 4.2 | ... | ... | ... | ... | ... | ... | ... | ... | ... | ... | ... | n | ... |
| 6435 | 10560917-7727236 | 44.4 ± 0.2 | 3910 ± 149 | ... | 1.046 ± 0.003 | 0.01 ± 0.11 | <10 | 3 | N | ... | ... | ... | ... | n | ... |
| 6436 | 10561277-7723286 | -27.4 ± 0.2 | 4932 ± 69 | ... | 1.021 ± 0.006 | -0.02 ± 0.08 | <37 | 3 | N | ... | ... | ... | ... | n | G |
| 6437 | 10561638-7630530 | 12.8 ± 0.4 | ... | ... | ... | ... | ... | ... | ... | ... | ... | ... | ... | n | ... |
| 6438 | 10561938-7735230 | -1.4 ± 0.3 | 5193 ± 105 | 3.52 ± 0.12 | 1.005 ± 0.005 | -0.19 ± 0.19 | <15 | 3 | Y | Y | N | N | Y | n | NG? |
| 6439 | 10562463-7739117 | -2.8 ± 0.2 | 4884 ± 268 | ... | 1.016 ± 0.003 | -0.12 ± 0.20 | <13 | 3 | N | ... | ... | ... | ... | n | G |
| 6440 | 10562715-7632348 | 70.4 ± 0.2 | 4793 ± 54 | 2.59 ± 0.13 | 1.013 ± 0.003 | -0.20 ± 0.09 | <19 | 3 | N | N | ... | ... | ... | n | G |
| 6441 | 10562843-7717312 | -11.8 ± 0.3 | 5070 ± 76 | ... | 1.014 ± 0.007 | -0.23 ± 0.26 | <17 | 3 | N | ... | ... | ... | ... | n | G |
| 6442 | 10563044-7711393$^f$ | 17.6 ± 0.3 | 4154 ± 246 | ... | 0.946 ± 0.005 | -0.12 ± 0.06 | 450 ± 3 | 1 | Y | ... | Y | Y | Y | Y | ... |
| 6511 | 11001181-7536224 | 54.4 ± 0.2 | 3870 ± 91 | ... | 1.040 ± 0.003 | -0.09 ± 0.17 | <4 | 3 | N | ... | ... | ... | ... | n | ... |
| 6443 | 10563146-7618334 | 17.2 ± 0.3 | 3319 ± 25 | ... | 0.875 ± 0.010 | -0.24 ± 0.16 | 635 ± 26 | 1 | Y | ... | Y | Y | Y | Y | ... |
| 6444 | 10563672-7652520 | 39.0 ± 0.2 | 4886 ± 106 | 2.70 ± 0.11 | 1.014 ± 0.005 | -0.06 ± 0.09 | <21 | 3 | N | N | ... | ... | ... | n | G |
| 6512 | 11001294-7757351 | -40.2 ± 0.4 | 5242 ± 151 | ... | 1.011 ± 0.011 | -0.76 ± 0.03 | <23 | 3 | N | ... | ... | ... | ... | n | ... |
| 6445 | 10563968-7655496 | 45.3 ± 0.2 | 4107 ± 164 | ... | 1.050 ± 0.003 | -0.10 ± 0.15 | <27 | 3 | N | ... | ... | ... | ... | n | ... |
| 662 | 10564115-7744292 | -0.7 ± 0.6 | 4407 ± 43 | 1.67 ± 0.07 | ... | -0.35 ± 0.05 | 89 ± 6 | 2 | ... | N | N | ... | Y | n | ... |
| 6513 | 11001408-7644155 | 78.5 ± 0.3 | ... | ... | ... | ... | ... | ... | ... | ... | ... | ... | ... | n | ... |
| 6446 | 10564871-7619133 | -35.6 ± 0.7 | 5367 ± 274 | ... | 0.994 ± 0.019 | -0.41 ± 0.21 | <45 | 3 | Y | ... | N | N | N | n | NG |
| 6515 | 11001558-7814102 | -49.8 ± 0.2 | 4923 ± 172 | 2.79 ± 0.15 | 1.011 ± 0.003 | -0.16 ± 0.10 | <19 | 3 | N | N | ... | ... | ... | n | G |
| 6516 | 11001679-7649504 | 38.9 ± 0.2 | 3984 ± 27 | ... | 1.031 ± 0.006 | -0.13 ± 0.16 | <11 | 3 | N | ... | ... | ... | ... | n | ... |
| 6517 | 11001719-7645222 | -2.6 ± 0.2 | 4742 ± 165 | ... | 1.024 ± 0.006 | -0.30 ± 0.11 | <9 | 3 | N | ... | ... | ... | ... | n | G |
| 6518 | 11001949-7627212 | 5.1 ± 0.2 | 4830 ± 43 | ... | 0.997 ± 0.003 | 0.03 ± 0.02 | <24 | 3 | Y | ... | N | N | Y | n | NG |
| 6673 | 11064003-7624103 | 2.8 ± 0.2 | 4600 ± 120 | ... | 1.019 ± 0.004 | 0.05 ± 0.09 | <47 | 3 | N | ... | ... | ... | ... | n | G |
| 6674 | 11064110-7652598 | 22.2 ± 0.3 | 4336 ± 179 | ... | 1.035 ± 0.008 | -0.34 ± 0.04 | ... | ... | N | ... | ... | ... | ... | n | ... |
| 6519 | 11001984-7736575 | -4.1 ± 0.2 | 5737 ± 41 | ... | 1.005 ± 0.002 | -0.28 ± 0.04 | ... | ... | Y | ... | N | ... | Y | n | ... |
| 6675 | 11064180-7635489$^f$ | 15.7 ± 0.3 | 3321 ± 2 | ... | 0.912 ± 0.013 | -0.25 ± 0.14 | 414 ± 11 | 1 | Y | ... | Y | Y | Y | Y | ... |
| 6676 | 11064235-7632450 | 15.0 ± 0.2 | 4259 ± 254 | ... | 1.025 ± 0.008 | 0.05 ± 0.09 | <43 | 3 | N | ... | ... | ... | ... | n | G |
| 6677 | 11064369-7758081 | -33.7 ± 0.2 | 5071 ± 75 | ... | 1.019 ± 0.003 | ... | ... | ... | N | ... | ... | ... | ... | n | ... |
| 6678 | 11064383-7728228 | -6.7 ± 0.3 | 5434 ± 247 | ... | 0.974 ± 0.011 | 0.05 ± 0.07 | ... | ... | Y | ... | N | ... | Y | n | ... |
| 6679 | 11064435-7748178 | 99.6 ± 0.6 | 4702 ± 224 | ... | 1.016 ± 0.019 | -0.51 ± 0.22 | <72 | 3 | N | ... | ... | ... | ... | n | G |
| 629 | 11064510-7727023$^f$ | 14.9 ± 0.6 | 4316 ± 104 | 4.62 ± 0.13 | ... | 0.00 ± 0.12 | 507 ± 54 | 2 | ... | Y | Y | Y | Y | Y | ... |
| 6680 | 11064543-7718317 | 73.0 ± 0.2 | 4481 ± 155 | ... | 0.925 ± 0.003 | 0.07 ± 0.19 | <23 | 3 | Y | ... | N | N | Y | n | NG |
| 6458 | 10573758-7733233 | -15.6 ± 0.2 | 4600 ± 101 | ... | 1.030 ± 0.007 | ... | <52 | 3 | N | ... | ... | ... | ... | n | G |
| 6681 | 11064586-7625318 | -52.2 ± 0.2 | 5056 ± 107 | ... | 1.018 ± 0.002 | -0.24 ± 0.18 | <11 | 3 | N | ... | ... | ... | ... | n | G |
| 6682 | 11064634-7750338 | 106.1 ± 0.3 | 4531 ± 2 | 1.78 ± 0.18 | 1.041 ± 0.008 | -0.42 ± 0.03 | <45 | 3 | N | N | ... | ... | ... | n | G |
| 6683 | 11064863-7614346 | 46.1 ± 0.2 | 4557 ± 115 | ... | 1.049 ± 0.006 | ... | 14 ± 3 | 1 | N | ... | ... | ... | ... | n | G |
| 6684 | 11065340-7749484 | 35.2 ± 0.3 | 5069 ± 204 | ... | 0.985 ± 0.009 | 0.13 ± 0.07 | <55 | 3 | Y | ... | N | N | Y | n | NG |
| 6523 | 11004022-7619280 | 16.1 ± 0.4 | 3459 ± 30 | ... | 0.929 ± 0.017 | ... | 594 ± 12 | 1 | Y | ... | Y | Y | ... | Y | ... |
| 6685 | 11065733-7742106 | 13.5 ± 0.3 | 3401 ± 54 | ... | 0.876 ± 0.007 | -0.25 ± 0.13 | 614 ± 19 | 1 | Y | ... | Y | Y | Y | Y | ... |
| 630 | 11065856-7713326 | -7.6 ± 0.6 | 3815 ± 89 | 1.00 ± 0.20 | 1.043 ± 0.003 | 0.02 ± 0.11 | 171 ± 11 | 2 | N | N | ... | ... | ... | n | G |
| 6459 | 10574007-7726259 | 27.0 ± 0.3 | 4585 ± 98 | ... | 1.021 ± 0.008 | -0.04 ± 0.16 | <43 | 3 | N | ... | ... | ... | ... | n | G |
| 6686 | 11065906-7718535 | -640.0 ± 0.3 | 3109 ± 38 | ... | 0.881 ± 0.008 | ... | ... | ... | Y | ... | N | ... | ... | n | ... |
| 6687 | 11070021-7526067 | -19.5 ± 0.2 | 4523 ± 137 | ... | 1.020 ± 0.004 | 0.05 ± 0.12 | <49 | 3 | N | ... | ... | ... | ... | n | G |

**Table C.2.** continued.

| ID | CNAME | RV (km s$^{-1}$) | $T_{\text{eff}}$ (K) | $\log g$ (dex) | $\gamma^a$ | [Fe/H] (dex) | EW(Li)$^b$ (mÅ) | EW(Li) error flag$^c$ | $\gamma$ | $\log g$ | RV | Li | [Fe/H] | Final$^d$ | Non-mem with Li$^e$ |
|---|---|---|---|---|---|---|---|---|---|---|---|---|---|---|---|
| 6688 | 11070305-7622179 | 18.8 ± 0.2 | 4609 ± 142 | ... | 1.012 ± 0.005 | 0.06 ± 0.11 | ... | ... | N | ... | ... | ... | ... | n | ... |
| 6689 | 11070350-7631443 | 52.3 ± 0.2 | 3936 ± 63 | ... | 1.046 ± 0.007 | ... | <19 | 3 | N | ... | ... | ... | ... | n | ... |
| 6690 | 11070380-7635440 | -1.7 ± 0.2 | 6300 ± 37 | 4.10 ± 0.08 | 1.001 ± 0.002 | -0.02 ± 0.03 | ... | ... | Y | Y | N | ... | Y | n | ... |
| 6719 | 11073832-7747168 | 14.8 ± 0.3 | 3304 ± 31 | ... | 0.892 ± 0.013 | -0.26 ± 0.14 | 637 ± 16 | 1 | Y | ... | Y | Y | Y | Y | ... |
| 6524 | 11004342-7808403 | 84.3 ± 0.6 | ... | ... | ... | ... | ... | ... | ... | ... | ... | ... | ... | n | ... |
| 6720 | 11074028-7554221 | 51.4 ± 0.2 | 4579 ± 124 | ... | 1.037 ± 0.003 | ... | <15 | 3 | N | ... | ... | ... | ... | n | G |
| 6464 | 10575316-7634459 | 1.2 ± 0.2 | 5029 ± 159 | ... | 1.019 ± 0.002 | -0.08 ± 0.16 | 203 ± 3 | 1 | N | ... | ... | ... | ... | n | Li-rich G |
| 7008 | 11212359-7617551 | 7.9 ± 0.2 | 4574 ± 86 | ... | 1.029 ± 0.005 | ... | <55 | 3 | N | ... | ... | ... | ... | n | ... |
| 6465 | 10575376-7724495 | 16.4 ± 0.3 | 3421 ± 45 | ... | 0.866 ± 0.013 | -0.27 ± 0.14 | 635 ± 20 | 1 | Y | ... | Y | Y | Y | Y | ... |
| 6721 | 11074245-7733593 | 14.9 ± 0.8 | ... | ... | ... | ... | ... | ... | ... | ... | ... | ... | ... | n | ... |
| 6722 | 11074352-7635002 | 23.3 ± 0.2 | 4906 ± 132 | 2.83 ± 0.02 | 1.007 ± 0.005 | -0.09 ± 0.17 | <17 | 3 | Y | N | N | N | Y | n | NG? |
| 6723 | 11074366-7739411$^f$ | 15.7 ± 0.2 | 4120 ± 166 | ... | 0.951 ± 0.005 | -0.09 ± 0.10 | 490 ± 16 | 1 | Y | ... | Y | Y | Y | Y | ... |
| 656 | 11213017-7616098 | -29.5 ± 0.6 | 4893 ± 7 | 2.61 ± 0.02 | ... | 0.02 ± 0.04 | <18 | 3 | ... | N | N | ... | Y | n | ... |
| 7009 | 11213079-7633351 | 15.5 ± 0.2 | 3632 ± 7 | ... | 0.885 ± 0.006 | -0.21 ± 0.12 | 665 ± 22 | 1 | Y | ... | Y | Y | Y | Y | ... |
| 6724 | 11074726-7516029 | -11.3 ± 0.2 | 4613 ± 113 | 2.53 ± 0.11 | 1.011 ± 0.002 | 0.09 ± 0.03 | <49 | 3 | N | N | ... | ... | ... | n | G |
| 6725 | 11074763-7711156 | -16.1 ± 0.2 | 5003 ± 106 | ... | 1.027 ± 0.004 | -0.04 ± 0.07 | <29 | 3 | N | ... | ... | ... | ... | n | G |
| 6537 | 11011846-7527366 | 33.2 ± 0.2 | 4563 ± 129 | 2.24 ± 0.20 | 1.020 ± 0.003 | -0.13 ± 0.12 | <27 | 3 | N | N | ... | ... | ... | n | G |
| 6726 | 11074962-7755197 | 68.8 ± 0.3 | 4596 ± 28 | 2.21 ± 0.16 | 1.033 ± 0.008 | -0.21 ± 0.02 | <28 | 3 | N | N | ... | ... | ... | n | G |
| 6538 | 11011875-7627025 | 13.9 ± 0.2 | 3943 ± 232 | ... | 0.894 ± 0.004 | -0.13 ± 0.12 | 588 ± 3 | 1 | Y | ... | Y | Y | Y | Y | ... |
| 6727 | 11075088-7506200 | 39.8 ± 0.2 | 4573 ± 82 | ... | 1.013 ± 0.003 | ... | <57 | 3 | N | ... | ... | ... | ... | n | G |
| 7010 | 11213578-7614179 | 10.6 ± 0.2 | 4422 ± 117 | 1.61 ± 0.14 | 1.039 ± 0.003 | -0.37 ± 0.03 | <14 | 3 | N | N | ... | ... | ... | n | G |
| 6539 | 11011884-7809462 | -1.7 ± 0.2 | 4595 ± 18 | 2.43 ± 0.11 | 1.020 ± 0.006 | 0.04 ± 0.02 | <44 | 3 | N | N | ... | ... | ... | n | G |
| 6728 | 11075225-7736569 | 16.3 ± 0.8 | ... | ... | ... | ... | ... | ... | ... | ... | ... | ... | ... | n | ... |
| 6729 | 11075241-7659279 | -47.9 ± 0.5 | 5827 ± 150 | ... | 1.006 ± 0.012 | -0.36 ± 0.05 | ... | ... | Y | ... | N | ... | N | n | ... |
| 6540 | 11012040-7627398 | 46.7 ± 0.2 | 4608 ± 42 | 2.26 ± 0.15 | 1.025 ± 0.006 | -0.13 ± 0.09 | ... | ... | N | N | ... | ... | ... | n | ... |
| 6541 | 11012580-7729201 | 228.6 ± 0.6 | ... | ... | ... | ... | ... | ... | ... | ... | ... | ... | ... | n | ... |
| 6542 | 11012712-7655307 | 0.9 ± 0.7 | ... | ... | ... | ... | ... | ... | ... | ... | ... | ... | ... | n | ... |
| 7011 | 11214199-7632457 | -31.7 ± 0.2 | 3837 ± 88 | ... | ... | ... | ... | ... | ... | ... | ... | ... | ... | n | ... |
| 6466 | 10580155-7702388 | 27.0 ± 0.2 | 4775 ± 40 | ... | 0.998 ± 0.003 | 0.04 ± 0.03 | <36 | 3 | Y | ... | N | N | Y | n | NG |
| 6543 | 11012813-7725552 | -2.3 ± 0.3 | 5852 ± 97 | 4.10 ± 0.15 | 0.996 ± 0.004 | -0.63 ± 0.12 | <9 | 3 | Y | Y | N | N | N | n | ... |
| 6741 | 11081069-7637425 | -10.1 ± 0.2 | 4755 ± 23 | 2.51 ± 0.04 | 1.023 ± 0.005 | -0.04 ± 0.01 | <35 | 3 | N | N | ... | ... | ... | n | G |
| 6477 | 10582648-7734491 | -5.3 ± 0.4 | 6462 ± 93 | 3.88 ± 0.19 | 1.015 ± 0.007 | 0.43 ± 0.07 | ... | ... | N | Y | ... | ... | ... | n | ... |
| 6382 | 10500113-7719550 | 0.4 ± 0.2 | 4837 ± 25 | 2.67 ± 0.13 | 1.018 ± 0.002 | -0.07 ± 0.10 | <29 | 3 | N | N | ... | ... | ... | n | G |
| 6742 | 11081092-7713489 | -24.8 ± 0.2 | 5146 ± 98 | ... | 1.002 ± 0.004 | -0.05 ± 0.04 | <19 | 3 | Y | ... | N | N | Y | n | NG? |
| 668 | 11012887-7539520 | ... | ... | ... | ... | ... | ... | ... | ... | ... | ... | ... | ... | n | ... |
| 6478 | 10583143-7732360 | -3.8 ± 0.2 | 6275 ± 54 | 4.21 ± 0.12 | 0.998 ± 0.003 | 0.16 ± 0.04 | <12 | 3 | Y | Y | N | N | Y | n | NG |
| 6560 | 11023663-7724195 | 23.4 ± 0.3 | 5974 ± 163 | 4.19 ± 0.06 | 0.994 ± 0.006 | -0.04 ± 0.11 | ... | ... | Y | Y | N | ... | Y | n | ... |
| 6561 | 11023752-7817531 | 7.7 ± 0.2 | 4401 ± 220 | ... | 1.032 ± 0.006 | -0.17 ± 0.12 | <14 | 3 | N | ... | ... | ... | ... | n | G |
| 6562 | 11023785-7717507 | -3.9 ± 0.3 | 5798 ± 81 | ... | 0.993 ± 0.005 | ... | ... | ... | Y | ... | N | ... | ... | n | ... |
| 6744 | 11081509-7733531$^f$ | 14.5 ± 0.2 | 5143 ± 86 | ... | 0.978 ± 0.004 | ... | 467 ± 3 | 1 | Y | ... | Y | Y | ... | Y | ... |
| 6563 | 11024256-7717108 | 14.0 ± 1.1 | 5515 ± 238 | ... | 1.024 ± 0.015 | -0.34 ± 0.18 | <44 | 3 | N | ... | ... | ... | ... | n | ... |
| 6564 | 11025147-7826440 | 7.2 ± 0.3 | ... | ... | ... | ... | ... | ... | ... | ... | ... | ... | ... | n | ... |
| 6479 | 10583795-7616481 | 106.8 ± 0.2 | 3926 ± 172 | ... | 1.052 ± 0.002 | -0.29 ± 0.10 | <16 | 3 | N | ... | ... | ... | ... | n | ... |
| 6565 | 11025225-7533401 | -20.9 ± 0.2 | 3538 ± 107 | ... | ... | ... | ... | ... | ... | ... | ... | ... | ... | n | ... |
| 6745 | 11081648-7744371 | 15.8 ± 0.2 | 3415 ± 38 | ... | 0.907 ± 0.005 | -0.27 ± 0.14 | 587 ± 12 | 1 | Y | ... | Y | Y | Y | Y | ... |
| 6480 | 10583860-7737111 | -27.6 ± 0.2 | 5003 ± 184 | ... | 1.021 ± 0.003 | -0.19 ± 0.20 | <23 | 3 | N | ... | ... | ... | ... | n | G |
| 6566 | 11025363-7743055 | -40.9 ± 0.8 | 6578 ± 249 | 3.79 ± 0.46 | 1.034 ± 0.016 | 0.31 ± 0.19 | <40 | 3 | N | Y | ... | ... | ... | n | ... |
| 6746 | 11081703-7744118 | 15.4 ± 0.3 | 3328 ± 125 | ... | 0.945 ± 0.014 | ... ± ... | 612 ± 27 | 1 | Y | ... | Y | Y | ... | Y | ... |
| 6567 | 11025413-7632597 | 8.7 ± 0.2 | 4488 ± 199 | ... | 1.022 ± 0.007 | 0.09 ± 0.12 | <57 | 3 | N | ... | ... | ... | ... | n | G |
| 6747 | 11081916-7656525 | 19.7 ± 0.3 | 4930 ± 130 | 2.72 ± 0.08 | 1.013 ± 0.007 | -0.08 ± 0.05 | <12 | 3 | N | N | ... | ... | ... | n | G |
| 6568 | 11025463-7824427 | -5.6 ± 0.2 | 4489 ± 145 | ... | 1.052 ± 0.008 | ... | <29 | 3 | N | ... | ... | ... | ... | n | G |
| 6748 | 11082047-7705249 | 1.8 ± 0.3 | 4577 ± 71 | 2.32 ± 0.08 | 1.016 ± 0.009 | -0.02 ± 0.02 | <44 | 3 | N | N | ... | ... | ... | n | G |
| 6569 | 11025466-7701032 | 17.6 ± 0.2 | 4361 ± 157 | ... | 1.043 ± 0.008 | 0.16 ± 0.03 | <66 | 3 | N | ... | ... | ... | ... | n | G |
| 6481 | 10584371-7732453 | -3.3 ± 1.0 | ... | ... | ... | ... | ... | ... | ... | ... | ... | ... | ... | n | ... |
| 6749 | 11082131-7737040 | -4.1 ± 0.3 | 4745 ± 180 | ... | 1.017 ± 0.010 | 0.00 ± 0.05 | <38 | 3 | N | ... | ... | ... | ... | n | G |
| 6570 | 11025504-7721508$^f$ | 16.2 ± 0.5 | 3319 ± 125 | ... | 0.851 ± 0.013 | ... | 391 ± 20 | 1 | Y | ... | Y | Y | ... | Y | ... |
| 6482 | 10584598-7623480 | 2.8 ± 0.2 | 5038 ± 127 | ... | 1.017 ± 0.004 | -0.09 ± 0.11 | <23 | 3 | N | ... | ... | ... | ... | n | G |
| 6571 | 11025538-7713502 | 84.0 ± 0.2 | 3936 ± 65 | ... | 1.052 ± 0.006 | ... | ... | ... | N | ... | ... | ... | ... | n | ... |
| 6483 | 10584820-7754466 | 47.1 ± 0.2 | 5006 ± 154 | ... | 1.016 ± 0.003 | -0.18 ± 0.18 | <9 | 3 | N | ... | ... | ... | ... | n | G |

**Table C.2.** continued.

| ID | CNAME | RV (km s$^{-1}$) | $T_{\text{eff}}$ (K) | logg (dex) | $\gamma^a$ | [Fe/H] (dex) | EW(Li)$^b$ (mÅ) | EW(Li) error flag$^c$ | $\gamma$ | logg | RV | Li | [Fe/H] | Final$^d$ | Non-mem with Li$^e$ |
|---|---|---|---|---|---|---|---|---|---|---|---|---|---|---|---|
| 6750 | 11082237-7730277 | 14.3 ± 0.8 | … | … | … | … | … | … | … | … | … | … | … | n | … |
| 6572 | 11025797-7633127 | 22.2 ± 0.2 | 4724 ± 119 | 2.57 ± 0.13 | 1.012 ± 0.005 | 0.01 ± 0.09 | <32 | 3 | N | N | … | … | … | n | G |
| 6573 | 11030208-7713097 | -1.7 ± 0.4 | 5167 ± 25 | … | 1.020 ± 0.013 | -0.26 ± 0.27 | <22 | 3 | N | … | … | … | … | n | G |
| 6751 | 11082270-7604369 | 1.8 ± 0.2 | 4557 ± 81 | … | 1.046 ± 0.002 | … | <39 | 3 | N | … | … | … | … | n | G |
| 6574 | 11030227-7817121 | -3.0 ± 0.2 | 5811 ± 175 | 4.18 ± 0.09 | 0.990 ± 0.005 | 0.26 ± 0.11 | … | … | Y | Y | N | … | N | n | … |
| 6484 | 10585008-7818065 | 36.9 ± 0.3 | 4741 ± 253 | … | 1.022 ± 0.010 | -0.42 ± 0.13 | <20 | 3 | N | … | … | … | … | n | G |
| 6752 | 11082341-7556342 | 4.4 ± 0.2 | 3534 ± 107 | … | … | … | … | … | … | … | … | … | … | n | … |
| 6575 | 11030712-7524472 | -15.3 ± 0.2 | 4575 ± 42 | 2.33 ± 0.15 | 1.029 ± 0.001 | -0.03 ± 0.01 | … | … | N | N | … | … | … | n | … |
| 6485 | 10585043-7754292 | 34.9 ± 0.2 | 4520 ± 198 | … | 1.037 ± 0.006 | -0.48 ± 0.13 | <10 | 3 | N | … | … | … | … | n | G |
| 6753 | 11082410-7741473 | 14.5 ± 0.4 | 3106 ± 107 | … | 0.907 ± 0.019 | … | 605 ± 20 | 1 | Y | … | Y | Y | … | Y | … |
| 6576 | 11031157-7528197 | 6.2 ± 0.2 | 5056 ± 91 | 3.18 ± 0.12 | 1.008 ± 0.002 | -0.04 ± 0.08 | … | … | Y | N | N | … | Y | n | … |
| 664 | 10585418-7743115 | 11.8 ± 0.6 | 4723 ± 54 | 3.16 ± 0.17 | … | 0.12 ± 0.04 | <22 | 3 | … | N | N | … | Y | n | … |
| 634 | 11082577-7648315 | 16.7 ± 0.6 | 4491 ± 30 | 2.07 ± 0.11 | … | -0.07 ± 0.05 | <19 | 3 | … | N | Y | … | Y | n | … |
| 6577 | 11031160-7721042 | 28.0 ± 0.2 | 5569 ± 59 | … | 1.002 ± 0.001 | 0.14 ± 0.05 | … | … | Y | … | N | … | Y | n | … |
| 6486 | 10585795-7734183 | -16.6 ± 0.3 | 6649 ± 112 | … | 1.005 ± 0.006 | 0.13 ± 0.09 | … | … | Y | … | N | … | Y | n | … |
| 6754 | 11082854-7753272 | -12.4 ± 0.2 | 4765 ± 121 | 2.65 ± 0.10 | 1.011 ± 0.005 | 0.02 ± 0.04 | <23 | 3 | N | N | … | … | … | n | G |
| 6578 | 11031416-7811298 | -0.9 ± 0.2 | 4936 ± 237 | … | 0.966 ± 0.003 | 0.04 ± 0.03 | <14 | 3 | Y | … | N | N | Y | n | NG |
| 6579 | 11031843-7725341 | 40.4 ± 0.3 | 4504 ± 162 | 2.28 ± 0.12 | 1.011 ± 0.012 | 0.10 ± 0.08 | <41 | 3 | N | N | … | … | … | n | G |
| 6580 | 11031996-7811348 | 53.7 ± 0.2 | 4866 ± 116 | … | 1.019 ± 0.002 | -0.16 ± 0.10 | <30 | 3 | N | … | … | … | … | n | G |
| 6782 | 11090151-7646372 | -10.9 ± 0.4 | 5186 ± 170 | 4.39 ± 0.15 | 0.970 ± 0.010 | 0.08 ± 0.22 | <43 | 3 | Y | Y | N | N | Y | n | NG |
| 665 | 10590108-7722407$^f$ | 16.7 ± 0.6 | 4171 ± 102 | 4.61 ± 0.13 | … | -0.18 ± 0.11 | 440 ± 17 | 2 | … | Y | Y | Y | Y | Y | … |
| 6581 | 11032037-7744028 | -13.8 ± 0.5 | 6338 ± 180 | … | 0.987 ± 0.011 | 0.02 ± 0.13 | 108 ± 6 | 1 | Y | … | N | N | Y | n | NG |
| 6783 | 11090154-7509594 | -17.3 ± 0.2 | 4679 ± 89 | 2.47 ± 0.04 | 1.019 ± 0.002 | -0.02 ± 0.03 | <33 | 3 | N | N | … | … | … | n | G |
| 6582 | 11032074-7737550 | 26.7 ± 1.9 | … | … | … | … | … | … | … | … | … | … | … | n | … |
| 6784 | 11090332-7700495 | 95.6 ± 0.2 | 5026 ± 128 | … | 1.010 ± 0.004 | -0.21 ± 0.28 | <15 | 3 | N | … | … | … | … | n | G |
| 6487 | 10590331-7533397 | 8.8 ± 0.2 | 4428 ± 171 | … | 1.042 ± 0.002 | -0.17 ± 0.13 | <17 | 3 | N | … | … | … | … | n | G |
| 6785 | 11090367-7707456 | 37.9 ± 0.2 | 4651 ± 60 | 2.54 ± 0.17 | 1.018 ± 0.007 | 0.05 ± 0.07 | <33 | 3 | N | N | … | … | … | n | G |
| 6786 | 11090369-7701533 | -36.2 ± 0.5 | 5114 ± 260 | … | 1.033 ± 0.020 | -0.26 ± 0.19 | … | … | N | … | … | … | … | n | … |
| 7018 | 11224454-7632094 | 0.2 ± 0.2 | 4574 ± 86 | … | 1.045 ± 0.003 | … | <21 | 3 | N | … | … | … | … | n | … |
| 6787 | 11090512-7709580 | 17.5 ± 1.8 | … | … | … | … | … | … | … | … | … | … | … | n | … |
| 6595 | 11035900-7728362 | 33.3 ± 1.1 | … | … | … | … | … | … | … | … | … | … | … | n | … |
| 6596 | 11035903-7743349 | 6.0 ± 1.6 | … | … | … | … | … | … | … | … | … | … | … | n | … |
| 6788 | 11090651-7711246 | 9.7 ± 0.3 | 5725 ± 174 | 4.14 ± 0.08 | 0.990 ± 0.010 | 0.07 ± 0.13 | 56 ± 6 | 1 | Y | Y | N | N | Y | n | NG |
| 6789 | 11090689-7509177 | 4.6 ± 0.2 | 4577 ± 94 | 2.42 ± 0.17 | 1.016 ± 0.002 | 0.05 ± 0.10 | <38 | 3 | N | N | … | … | … | n | G |
| 6791 | 11090915-7553477 | 14.6 ± 0.3 | 3609 ± 33 | … | 0.874 ± 0.010 | -0.21 ± 0.14 | 574 ± 18 | 1 | Y | … | Y | Y | Y | Y | … |
| 7019 | 11225032-7622478 | -4.8 ± 0.2 | 4565 ± 130 | … | 1.031 ± 0.003 | -0.01 ± 0.02 | <43 | 3 | N | … | … | … | … | n | … |
| 6792 | 11090976-7609132 | 43.3 ± 0.3 | 4554 ± 151 | … | 1.014 ± 0.015 | -0.47 ± 0.15 | <35 | 3 | N | … | … | … | … | n | G |
| 6793 | 11091038-7505496 | -3.4 ± 0.2 | 4569 ± 82 | … | 1.031 ± 0.004 | … | <37 | 3 | N | … | … | … | … | n | G |
| 638 | 11091172-7729124 | 15.3 ± 0.6 | 3919 ± 142 | 4.62 ± 0.13 | … | -0.16 ± 0.11 | 635 ± 25 | 2 | … | Y | Y | Y | Y | Y | … |
| 6794 | 11091297-7729115 | 13.6 ± 0.2 | 3830 ± 175 | … | 0.879 ± 0.004 | -0.18 ± 0.12 | 613 ± 12 | 1 | Y | … | Y | Y | Y | Y | … |
| 6597 | 11040959-7623031 | -2.7 ± 0.2 | 4899 ± 240 | … | 1.029 ± 0.003 | -0.32 ± 0.18 | <16 | 3 | N | … | … | … | … | n | G |
| 6601 | 11042061-7536138 | -20.9 ± 0.2 | 4580 ± 11 | 2.37 ± 0.12 | 1.021 ± 0.003 | -0.04 ± 0.03 | <41 | 3 | N | N | … | … | … | n | G |
| 6602 | 11042076-7709551 | -15.8 ± 0.2 | 4900 ± 51 | … | 1.021 ± 0.003 | -0.02 ± 0.05 | … | … | N | … | … | … | … | n | … |
| 6795 | 11091379-7637531 | 234.6 ± 0.2 | 5155 ± 259 | … | 1.013 ± 0.003 | -0.49 ± 0.24 | … | … | N | … | … | … | … | n | … |
| 6805 | 11092756-7736513 | -39.6 ± 0.5 | 6493 ± 189 | … | 1.002 ± 0.009 | -0.93 ± 0.25 | … | … | Y | … | N | … | N | n | … |
| 6806 | 11092855-7633281 | 604.5 ± 278.3 | … | … | … | … | … | … | … | … | … | … | … | n | … |
| 6807 | 11093697-7615232 | 110.5 ± 0.2 | 4319 ± 190 | … | 1.051 ± 0.004 | … | … | … | N | … | … | … | … | n | … |
| 6603 | 11042375-7724239 | 4.2 ± 1.7 | 6609 ± 241 | 3.99 ± 0.48 | 1.009 ± 0.014 | 0.08 ± 0.19 | <37 | 3 | Y | Y | N | N | Y | n | … |
| 6808 | 11093859-7652263 | 28.5 ± 0.6 | … | … | … | … | … | … | … | … | … | … | … | n | … |
| 6809 | 11094006-7628392 | 15.7 ± 0.2 | 3939 ± 219 | … | 0.890 ± 0.005 | -0.15 ± 0.14 | 637 ± 12 | 1 | Y | … | Y | Y | Y | Y | … |
| 6810 | 11094525-7740332 | 14.0 ± 0.7 | … | … | … | … | … | … | … | … | … | … | … | n | … |
| 6812 | 11094621-7634463$^f$ | 15.2 ± 0.4 | 3410 ± 119 | … | 0.889 ± 0.016 | -0.23 ± 0.13 | 462 ± … | 1 | Y | … | Y | Y | Y | Y | … |
| 6813 | 11094646-7628575 | -5.0 ± 0.2 | 7144 ± 23 | … | … | … | … | … | … | … | … | … | … | n | … |
| 6605 | 11042548-7720430 | 266.8 ± 0.3 | 5047 ± 123 | … | 1.017 ± 0.008 | -0.25 ± 0.25 | <43 | 3 | N | … | … | … | … | n | G |
| 6814 | 11094715-7749565 | 26.4 ± 0.3 | 5824 ± 28 | 4.10 ± 0.14 | 0.996 ± 0.007 | -0.40 ± 0.12 | … | … | Y | Y | N | … | N | n | … |
| 6815 | 11094775-7557577 | -50.3 ± 0.2 | 4235 ± 274 | … | 1.054 ± 0.002 | -0.16 ± 0.03 | <19 | 3 | N | … | … | … | … | n | … |
| 6606 | 11042663-7711568 | 22.3 ± 0.3 | 5764 ± 21 | 4.39 ± 0.06 | 0.982 ± 0.006 | 0.10 ± 0.06 | … | … | Y | Y | N | … | Y | n | … |
| 6817 | 11095016-7601498 | 24.0 ± 0.2 | 3886 ± 170 | … | 1.052 ± 0.004 | -0.18 ± 0.05 | <5 | 3 | N | … | … | … | … | n | … |
| 641 | 11095119-7658568 | 35.9 ± 0.6 | 5707 ± 53 | 4.20 ± 0.09 | … | 0.43 ± 0.06 | <15 | 3 | … | Y | N | … | N | n | NG |

| ID | CNAME | RV (km s$^{-1}$) | $T_{\text{eff}}$ (K) | $\log g$ (dex) | $\gamma^a$ | [Fe/H] (dex) | EW(Li)$^b$ (mÅ) | EW(Li) error flag$^c$ | $\gamma$ | $\log g$ | RV | Li | [Fe/H] | Final$^d$ | Non-mem with Li$^e$ |
|---|---|---|---|---|---|---|---|---|---|---|---|---|---|---|---|
| 7025 | 11234743-7548064 | 12.2 ± 0.2 | 4524 ± 165 | ... | 1.025 ± 0.003 | 0.05 ± 0.11 | <36 | 3 | N | ... | ... | ... | ... | n | G |
| 6818 | 11095146-7645452 | 31.4 ± 0.4 | 4985 ± 167 | ... | 1.010 ± 0.015 | 0.03 ± 0.03 | <41 | 3 | N | ... | ... | ... | ... | n | G |
| 6819 | 11095205-7657587 | 26.5 ± 0.4 | 4772 ± 127 | ... | 0.988 ± 0.017 | 0.11 ± 0.03 | <46 | 3 | Y | ... | N | N | Y | n | NG |
| 6820 | 11095258-7756312 | 40.1 ± 0.2 | 4985 ± 186 | ... | 1.017 ± 0.004 | -0.20 ± 0.17 | <30 | 3 | N | ... | ... | ... | ... | n | G |
| 6821 | 11095307-7614573 | 23.0 ± 0.2 | 4643 ± 121 | ... | 1.037 ± 0.003 | ... | <16 | 3 | N | ... | ... | ... | ... | n | G |
| 7026 | 11234998-7626199 | 45.0 ± 0.2 | 4533 ± 105 | ... | 1.052 ± 0.005 | ... | <26 | 3 | N | ... | ... | ... | ... | n | ... |
| 6626 | 11045890-7656550 | -13.4 ± 0.2 | 4614 ± 115 | ... | 1.027 ± 0.004 | -0.14 ± 0.17 | <15 | 3 | N | ... | ... | ... | ... | n | G |
| 6822 | 11095340-7634255 | -700.0 ± 0.2 | 4384 ± 162 | ... | 0.977 ± 0.009 | -0.19 ± 0.23 | ... | ... | Y | ... | N | ... | Y | n | ... |
| 6627 | 11050290-7811362 | 18.6 ± 0.2 | 4635 ± 9 | 2.75 ± 0.11 | 1.003 ± 0.007 | 0.13 ± 0.03 | <48 | 3 | Y | N | Y | N | Y | n | NG? |
| 6823 | 11095407-7629253$^f$ | 15.4 ± 0.6 | 3726 ± 20 | ... | 0.909 ± 0.019 | ... | 582 ± 36 | 1 | Y | ... | Y | Y | ... | Y | ... |
| 6628 | 11050373-7722359 | -27.0 ± 0.6 | 5910 ± 309 | ... | 1.010 ± 0.018 | -0.33 ± 0.24 | ... | ... | N | ... | ... | ... | ... | n | ... |
| 6824 | 11095760-7616205 | 174.4 ± 0.2 | 3779 ± 149 | ... | 1.048 ± 0.005 | 0.00 ± 0.12 | <100 | 3 | N | ... | ... | ... | ... | n | ... |
| 6629 | 11050433-7517093 | 130.3 ± 0.2 | 4523 ± 109 | ... | 1.041 ± 0.004 | ... | <13 | 3 | N | ... | ... | ... | ... | n | G |
| 6825 | 11095840-7610290 | -9.1 ± 0.2 | 5008 ± 65 | ... | 1.015 ± 0.003 | -0.07 ± 0.12 | <33 | 3 | N | ... | ... | ... | ... | n | G |
| 6630 | 11050573-7823472 | 6.0 ± 0.2 | 4497 ± 123 | 4.58 ± 0.04 | 0.921 ± 0.002 | 0.00 ± 0.07 | ... | ... | Y | Y | N | ... | Y | n | ... |
| 6826 | 11095873-7737088 | -700.0 ± 0.2 | 4233 ± 123 | ... | 0.943 ± 0.003 | ... | ... | ... | Y | ... | N | ... | ... | n | ... |
| 6631 | 11050752-7812063 | 16.5 ± 0.9 | ... | ... | ... | ... | ... | ... | ... | ... | ... | ... | ... | n | ... |
| 6827 | 11100010-7634578 | -700.0 ± 0.2 | 5162 ± 84 | ... | 0.988 ± 0.002 | ... | ... | ... | Y | ... | N | ... | ... | n | ... |
| 6632 | 11050773-7635492 | 90.4 ± 0.3 | 4415 ± 162 | ... | 1.030 ± 0.008 | -0.51 ± 0.18 | ... | ... | N | ... | ... | ... | ... | n | ... |
| 7027 | 11235458-7535480 | -19.2 ± 0.2 | 4841 ± 71 | 2.64 ± 0.07 | 1.018 ± 0.001 | 0.00 ± 0.03 | <33 | 3 | N | N | ... | ... | ... | n | G |
| 6873 | 11113194-7728100 | 13.4 ± 0.3 | 4500 ± 151 | ... | 1.030 ± 0.014 | -0.04 ± 0.03 | <54 | 3 | N | ... | ... | ... | ... | n | G |
| 6633 | 11050937-7706578 | -48.5 ± 0.2 | 4551 ± 181 | ... | 1.039 ± 0.003 | -0.17 ± 0.19 | <20 | 3 | N | ... | ... | ... | ... | n | G |
| 6874 | 11113327-7607399 | -21.0 ± 0.3 | 6150 ± 94 | 3.85 ± 0.12 | 1.010 ± 0.006 | 0.01 ± 0.09 | <15 | 3 | N | Y | ... | ... | ... | n | ... |
| 6634 | 11051259-7733489 | 50.3 ± 1.6 | ... | ... | ... | ... | ... | ... | ... | ... | ... | ... | ... | n | ... |
| 6875 | 11113337-7608402 | 6.0 ± 0.2 | 4546 ± 39 | 2.32 ± 0.10 | 1.028 ± 0.006 | 0.09 ± 0.06 | <43 | 3 | N | N | ... | ... | ... | n | G |
| 6635 | 11051467-7711290 | 17.4 ± 0.3 | 3436 ± 203 | ... | 0.885 ± 0.010 | -0.27 ± 0.18 | ... | ... | Y | ... | Y | ... | Y | n | ... |
| 6876 | 11113474-7636211 | 15.3 ± 0.2 | 3937 ± 249 | ... | 0.894 ± 0.003 | -0.13 ± 0.12 | 664 ± 14 | 1 | Y | ... | Y | Y | Y | Y | ... |
| 6636 | 11051521-7752545 | 11.0 ± 0.6 | ... | ... | ... | ... | ... | ... | ... | ... | ... | ... | ... | n | ... |
| 7042 | 11265011-7553125 | 64.3 ± 0.2 | 4904 ± 146 | 2.94 ± 0.12 | 1.006 ± 0.001 | -0.14 ± 0.18 | <16 | 3 | Y | N | N | N | Y | n | NG? |
| 6877 | 11113965-7620152$^f$ | 17.0 ± 0.2 | 3503 ± 155 | ... | 0.908 ± 0.003 | -0.23 ± 0.15 | 383 ± 19 | 1 | Y | ... | Y | Y | Y | Y | ... |
| 6637 | 11051798-7706565 | 24.7 ± 0.2 | 4102 ± 356 | ... | 1.043 ± 0.003 | -0.27 ± 0.04 | <15 | 3 | N | ... | ... | ... | ... | n | ... |
| 6878 | 11114071-7746466 | -41.3 ± 0.2 | 4771 ± 159 | ... | 1.024 ± 0.003 | -0.13 ± 0.17 | <13 | 3 | N | ... | ... | ... | ... | n | G |
| 6638 | 11051849-7607205 | -9.6 ± 0.2 | 4904 ± 111 | ... | 1.019 ± 0.002 | -0.05 ± 0.09 | <26 | 3 | N | ... | ... | ... | ... | n | G |
| 6879 | 11114370-7645523 | 14.2 ± 0.2 | 4411 ± 223 | ... | 1.037 ± 0.007 | -0.14 ± 0.07 | <33 | 3 | N | ... | ... | ... | ... | n | G |
| 645 | 11114632-7620092$^f$ | 17.6 ± 0.6 | 4630 ± 162 | 4.50 ± 0.18 | ... | -0.06 ± 0.14 | 492 ± 8 | 2 | ... | Y | Y | Y | Y | Y | ... |
| 6639 | 11051929-7640122 | 2.1 ± 0.3 | 3625 ± 72 | ... | 0.832 ± 0.013 | -0.21 ± 0.14 | ... | ... | Y | ... | N | ... | Y | n | ... |
| 7043 | 11265370-7552519 | 30.0 ± 0.2 | 3873 ± 81 | ... | 1.029 ± 0.003 | ... | <25 | 3 | N | ... | ... | ... | ... | n | ... |
| 6880 | 11114762-7537364 | 9.9 ± 0.2 | 4492 ± 184 | ... | 1.028 ± 0.004 | -0.08 ± 0.14 | <34 | 3 | N | ... | ... | ... | ... | n | ... |
| 6640 | 11052121-7604372 | 48.4 ± 0.2 | 4556 ± 117 | ... | 1.044 ± 0.003 | ... | <34 | 3 | N | ... | ... | ... | ... | n | G |
| 7044 | 11265386-7544468 | 77.3 ± 0.2 | 3863 ± 167 | ... | 1.059 ± 0.004 | -0.24 ± 0.01 | 135 ± 3 | 1 | N | ... | ... | ... | ... | n | ... |
| 6881 | 11115000-7620360 | -21.6 ± 0.2 | 4457 ± 134 | ... | 1.048 ± 0.001 | -0.09 ± 0.10 | <28 | 3 | N | ... | ... | ... | ... | n | G |
| 6641 | 11052272-7709290 | 14.5 ± 1.1 | ... | ... | ... | ... | ... | ... | ... | ... | ... | ... | ... | n | ... |
| 6882 | 11115140-7827154 | 13.7 ± 0.2 | 4550 ± 131 | ... | 1.016 ± 0.004 | 0.12 ± 0.07 | <40 | 3 | N | ... | ... | ... | ... | n | G |
| 6642 | 11052355-7607438 | 51.9 ± 0.2 | 3542 ± 107 | ... | ... | ... | ... | ... | ... | ... | ... | ... | ... | n | ... |
| 6643 | 11052456-7558428 | -4.6 ± 0.2 | 4579 ± 113 | 2.29 ± 0.17 | 1.023 ± 0.006 | -0.09 ± 0.02 | <18 | 3 | N | N | ... | ... | ... | n | G |
| 6883 | 11115302-7610568 | 16.4 ± 0.3 | 6673 ± 74 | ... | 1.000 ± 0.005 | 0.51 ± 0.06 | 40 ± 3 | 1 | Y | ... | Y | N | N | n | NG |
| 6644 | 11052472-7626209 | 15.6 ± 0.3 | 3581 ± 79 | ... | 0.867 ± 0.007 | -0.25 ± 0.13 | 545 ± 8 | 1 | Y | ... | Y | Y | Y | Y | ... |
| 6884 | 11115367-7712411 | 37.4 ± 0.3 | 4705 ± 149 | ... | 1.015 ± 0.012 | -0.15 ± 0.20 | <57 | 3 | N | ... | ... | ... | ... | n | G |
| 6645 | 11052851-7639489 | 1.0 ± 0.2 | 4950 ± 104 | ... | 1.015 ± 0.002 | -0.04 ± 0.09 | ... | ... | N | ... | ... | ... | ... | n | ... |
| 6885 | 11115400-7619311 | 16.3 ± 0.2 | 3785 ± 53 | ... | 0.896 ± 0.002 | -0.18 ± 0.13 | 642 ± 22 | 1 | Y | ... | Y | Y | Y | Y | ... |
| 6646 | 11052879-7604365 | 9.8 ± 0.2 | 4600 ± 62 | 2.43 ± 0.16 | 1.023 ± 0.003 | 0.05 ± 0.03 | <31 | 3 | N | N | ... | ... | ... | n | G |
| 6886 | 11115590-7608299 | 14.8 ± 0.3 | 6576 ± 57 | ... | 0.999 ± 0.003 | -0.17 ± 0.05 | 82 ± 5 | 1 | Y | ... | Y | N | Y | n | NG |
| 6647 | 11053041-7713365 | 59.0 ± 0.9 | ... | ... | ... | ... | ... | ... | ... | ... | ... | ... | ... | n | ... |
| 6887 | 11115711-7630121 | 127.5 ± 0.2 | 3916 ± 61 | ... | 1.029 ± 0.003 | 0.01 ± 0.11 | 184 ± 3 | 1 | N | ... | ... | ... | ... | n | G |
| 626 | 11053303-7700120 | 10.1 ± 0.6 | 4253 ± 146 | 4.51 ± 0.11 | ... | -0.10 ± 0.14 | <44 | 3 | ... | Y | N | N | Y | n | NG |
| 6888 | 11115799-7746379 | 42.8 ± 0.2 | 4644 ± 121 | ... | 1.018 ± 0.003 | -0.13 ± 0.13 | <36 | 3 | N | ... | ... | ... | ... | n | G |
| 6670 | 11063531-7754021 | -21.1 ± 0.3 | 5126 ± 2 | ... | 1.005 ± 0.010 | -0.19 ± 0.10 | <40 | 3 | Y | ... | N | N | Y | n | NG? |
| 6889 | 11115972-7714534 | -22.9 ± 0.3 | 4402 ± 177 | ... | 1.060 ± 0.010 | -0.10 ± 0.03 | <43 | 3 | N | ... | ... | ... | ... | n | G |
| 6671 | 11063841-7612032 | -24.5 ± 0.2 | 3535 ± 107 | ... | ... | ... | ... | ... | ... | ... | ... | ... | ... | n | ... |

**Table C.2.** continued.

| ID | CNAME | RV (km s$^{-1}$) | $T_{\text{eff}}$ (K) | $\log g$ (dex) | $\gamma^a$ | [Fe/H] (dex) | $EW(\text{Li})^b$ (mÅ) | $EW(\text{Li})$ error flag$^c$ | $\gamma$ | $\log g$ | RV | Li | [Fe/H] | Final$^d$ | Non-mem with Li$^e$ |
|---|---|---|---|---|---|---|---|---|---|---|---|---|---|---|---|
| 6890 | 11120038-7614106 | 38.3 ± 0.2 | 4690 ± 157 | ... | 1.016 ± 0.002 | -0.05 ± 0.14 | <29 | 3 | N | ... | ... | ... | ... | n | G |
| 6672 | 11063873-7628521 | 23.3 ± 0.2 | 4830 ± 78 | ... | 0.997 ± 0.008 | -0.04 ± 0.14 | <42 | 3 | Y | ... | N | N | Y | n | NG |
| 7045 | 11270345-7552047 | 23.4 ± 0.2 | 4607 ± 75 | ... | 1.022 ± 0.004 | 0.08 ± 0.05 | <38 | 3 | N | ... | ... | ... | ... | n | G |
| 6891 | 11120074-7613444 | 19.5 ± 0.2 | 3873 ± 81 | ... | 1.034 ± 0.003 | ... | <3 | 3 | N | ... | ... | ... | ... | n | ... |
| 6892 | 11120327-7637034 | 18.0 ± 0.9 | 3304 ± 103 | ... | ... | ... | 545 ± 44 | 1 | ... | ... | Y | Y | ... | Y | ... |
| 6893 | 11120384-7650542 | 35.6 ± 0.4 | ... | ... | ... | ... | ... | ... | ... | ... | ... | ... | ... | n | ... |
| 6894 | 11120518-7712408 | -7.7 ± 0.2 | 4474 ± 110 | ... | 1.035 ± 0.002 | -0.10 ± 0.02 | <32 | 3 | N | ... | ... | ... | ... | n | G |
| 6895 | 11120721-7821392 | 6.3 ± 0.2 | 4580 ± 15 | 2.21 ± 0.09 | 1.022 ± 0.002 | -0.14 ± 0.13 | <19 | 3 | N | N | ... | ... | ... | n | G |
| 647 | 11124299-7637049 | 15.1 ± 0.6 | 4570 ± 32 | 4.31 ± 0.34 | ... | -0.11 ± 0.06 | 463 ± 11 | 2 | ... | Y | Y | Y | Y | Y | ... |
| 6919 | 11124380-7726482 | 33.5 ± 0.2 | 3751 ± 106 | ... | 1.044 ± 0.009 | ... | <100 | 3 | N | ... | ... | ... | ... | n | ... |
| 6920 | 11124722-7604391 | 33.2 ± 0.2 | 4547 ± 122 | ... | 1.048 ± 0.002 | ... | <13 | 3 | N | ... | ... | ... | ... | n | G |
| 7048 | 11280460-7544500 | -36.4 ± 0.2 | 4673 ± 167 | ... | 1.022 ± 0.003 | -0.11 ± 0.12 | <26 | 3 | N | ... | ... | ... | ... | n | G |
| 6921 | 11125066-7711437 | 35.2 ± 0.2 | 6105 ± 97 | 4.19 ± 0.11 | 0.995 ± 0.002 | 0.03 ± 0.03 | ... | ... | Y | Y | N | ... | Y | n | ... |
| 6922 | 11125879-7613505 | 61.4 ± 0.3 | 3882 ± 89 | ... | 1.029 ± 0.017 | ... | 367 ± 29 | 1 | N | ... | ... | ... | ... | n | Li-rich G |
| 6923 | 11125939-7616533 | -23.4 ± 0.2 | 3814 ± 82 | ... | ... | ... | <6 | 3 | ... | ... | ... | ... | ... | n | ... |
| 6924 | 11130285-7736078 | -16.8 ± 0.3 | 4995 ± 184 | 3.53 ± 0.12 | 0.987 ± 0.014 | 0.00 ± 0.03 | <38 | 3 | Y | Y | N | N | Y | n | NG |
| 6925 | 11130309-7621003 | 126.4 ± 0.2 | 4378 ± 229 | ... | 1.042 ± 0.001 | -0.32 ± 0.18 | <21 | 3 | N | ... | ... | ... | ... | n | G |
| 6926 | 11130450-7534369$^f$ | 16.1 ± 0.4 | 3290 ± 5 | ... | 0.897 ± 0.011 | -0.27 ± 0.14 | 652 ± 51 | 1 | Y | ... | Y | Y | Y | Y | ... |
| 6927 | 11130526-7617396 | 71.7 ± 0.2 | 5158 ± 91 | ... | 1.020 ± 0.003 | ... | 264 ± 6 | 1 | N | ... | ... | ... | ... | n | Li-rich G |
| 6928 | 11130884-7815229 | 75.8 ± 0.2 | 3749 ± 106 | ... | ... | ... | <100 | 3 | ... | ... | ... | ... | ... | n | ... |
| 6929 | 11131106-7613480 | 13.3 ± 0.3 | 4657 ± 99 | ... | 1.029 ± 0.008 | ... | <72 | 3 | N | ... | ... | ... | ... | n | G |
| 6930 | 11131526-7721210 | 98.1 ± 0.3 | ... | ... | ... | ... | ... | ... | ... | ... | ... | ... | ... | n | ... |
| 6931 | 11131652-7531491 | -20.7 ± 0.2 | 4711 ± 8 | 2.65 ± 0.01 | 1.010 ± 0.002 | -0.01 ± 0.03 | <41 | 3 | Y | N | N | N | Y | n | NG? |
| 6932 | 11131728-7559520 | -0.9 ± 0.2 | 4590 ± 100 | 2.24 ± 0.16 | 1.029 ± 0.002 | -0.08 ± 0.02 | <27 | 3 | N | N | ... | ... | ... | n | G |
| 6933 | 11132246-7625568 | 39.0 ± 0.4 | ... | ... | ... | ... | ... | ... | ... | ... | ... | ... | ... | n | ... |
| 7049 | 11281215-7542322 | 7.9 ± 0.2 | 4909 ± 247 | ... | 1.014 ± 0.002 | -0.16 ± 0.27 | <26 | 3 | N | ... | ... | ... | ... | n | G |
| 6934 | 11132420-7808469 | -25.0 ± 0.2 | 4930 ± 15 | 2.83 ± 0.12 | 1.017 ± 0.003 | -0.02 ± 0.08 | <32 | 3 | N | N | ... | ... | ... | n | G |
| 6935 | 11132427-7734468 | -7.6 ± 0.3 | 6923 ± 70 | ... | 1.017 ± 0.003 | -0.52 ± 0.08 | ... | ... | N | ... | ... | ... | ... | n | ... |
| 6936 | 11132446-7629227 | 15.5 ± 0.2 | 3385 ± 119 | ... | 0.872 ± 0.006 | -0.26 ± 0.14 | 612 ± 12 | 1 | Y | ... | Y | Y | Y | Y | ... |
| 7050 | 11281394-7542525 | 69.4 ± 0.2 | 4944 ± 168 | ... | 1.015 ± 0.002 | -0.19 ± 0.17 | <20 | 3 | N | ... | ... | ... | ... | n | G |
| 6937 | 11132627-7619233 | -8.2 ± 0.2 | 6018 ± 192 | 4.06 ± 0.03 | 1.001 ± 0.002 | 0.01 ± 0.12 | ... | ... | Y | Y | N | ... | Y | n | ... |
| 6938 | 11132737-7634165 | 14.2 ± 0.2 | 3765 ± 79 | ... | 0.862 ± 0.002 | -0.19 ± 0.13 | 544 ± 5 | 1 | Y | ... | Y | Y | Y | Y | ... |
| 6939 | 11132747-7620175 | 37.2 ± 0.2 | 5439 ± 44 | 3.94 ± 0.20 | 0.996 ± 0.002 | 0.03 ± 0.03 | <17 | 3 | Y | Y | N | N | Y | n | NG |
| 6940 | 11132915-7722497 | 21.9 ± 0.2 | 4858 ± 33 | ... | 1.003 ± 0.004 | 0.05 ± 0.06 | <27 | 3 | Y | ... | N | N | Y | n | NG? |
| 6941 | 11132971-7629012 | 16.7 ± 0.2 | 3328 ± 10 | ... | 0.905 ± 0.007 | -0.22 ± 0.15 | 625 ± 8 | 1 | Y | ... | Y | Y | Y | Y | ... |
| 6984 | 11161338-7545591 | 52.0 ± 0.2 | 3745 ± 106 | ... | ... | ... | <100 | 3 | ... | ... | ... | ... | ... | n | ... |
| 6985 | 11162170-7716024 | -16.1 ± 0.2 | 4544 ± 113 | ... | 1.049 ± 0.002 | ... | ... | ... | N | ... | ... | ... | ... | n | ... |
| 6986 | 11162439-7712299 | 46.3 ± 0.2 | 4810 ± 9 | 2.69 ± 0.04 | 1.009 ± 0.004 | -0.13 ± 0.12 | <44 | 3 | Y | N | N | N | Y | n | NG? |
| 6987 | 11164057-7533327 | -21.4 ± 0.2 | 3976 ± 42 | 1.09 ± 0.17 | 1.045 ± 0.002 | 0.01 ± 0.24 | <18 | 3 | N | N | ... | ... | ... | n | ... |
| 6988 | 11164757-7820038 | -19.9 ± 0.2 | 4764 ± 23 | 2.55 ± 0.06 | 1.022 ± 0.002 | -0.04 ± 0.03 | <36 | 3 | N | N | ... | ... | ... | n | G |
| 654 | 11170509-7538518 | 8.9 ± 0.6 | 4773 ± 46 | 2.44 ± 0.03 | ... | 0.07 ± 0.06 | <16 | 3 | ... | N | N | ... | Y | n | ... |
| 6989 | 11171177-7532156 | 14.4 ± 0.2 | 4659 ± 9 | 2.46 ± 0.09 | 1.018 ± 0.003 | -0.07 ± 0.07 | <25 | 3 | N | N | ... | ... | ... | n | G |
| 6990 | 11171216-7809366 | -2.3 ± 0.2 | 4400 ± 200 | ... | 1.045 ± 0.002 | -0.13 ± 0.05 | <19 | 3 | N | ... | ... | ... | ... | n | G |
| 6991 | 11171448-7540272 | 11.5 ± 0.2 | 4441 ± 158 | ... | 1.031 ± 0.002 | -0.19 ± 0.16 | 44 ± 2 | 1 | N | ... | ... | ... | ... | n | G |
| 7052 | 11285199-7545491 | -5.5 ± 0.2 | 4994 ± 186 | ... | 1.010 ± 0.002 | -0.09 ± 0.14 | <16 | 3 | Y | ... | N | N | Y | n | NG? |
| 6992 | 11174601-7534220 | -46.2 ± 0.2 | 4531 ± 121 | 2.30 ± 0.15 | 1.023 ± 0.002 | 0.08 ± 0.01 | <48 | 3 | N | N | ... | ... | ... | n | G |
| 6993 | 11175040-7535004 | 11.9 ± 0.2 | 6599 ± 18 | 3.87 ± 0.04 | 1.014 ± 0.001 | 0.11 ± 0.01 | ... | ... | N | Y | ... | ... | ... | n | ... |
| 658 | 11291261-7546263 | 17.6 ± 0.6 | 4823 ± 98 | 4.50 ± 0.15 | ... | -0.07 ± 0.15 | 458 ± 19 | 2 | ... | Y | Y | Y | Y | Y | ... |
| 6994 | 11175053-7536579 | 24.5 ± 0.2 | 4839 ± 90 | 2.98 ± 0.15 | 1.001 ± 0.002 | 0.01 ± 0.09 | <26 | 3 | Y | N | N | N | Y | n | NG? |
| 6995 | 11175236-7542593 | -18.3 ± 0.2 | 4751 ± 10 | 2.58 ± 0.11 | 1.020 ± 0.002 | -0.06 ± 0.04 | <27 | 3 | N | N | ... | ... | ... | n | G |
| 6996 | 11180482-7540475 | 2.8 ± 0.2 | 5763 ± 97 | ... | 1.004 ± 0.001 | -0.35 ± 0.10 | 24 ± 3 | 1 | Y | ... | N | N | N | n | NG? |
| 655 | 11182024-7621576 | 14.8 ± 0.6 | 4307 ± 113 | 4.53 ± 0.11 | ... | -0.13 ± 0.11 | 511 ± 22 | 2 | ... | Y | Y | Y | Y | Y | ... |
| 6997 | 11182702-7625206 | 11.9 ± 0.2 | 4793 ± 105 | ... | 1.022 ± 0.003 | -0.13 ± 0.14 | ... | ... | N | ... | ... | ... | ... | n | ... |
| 6998 | 11184547-7617020 | -12.3 ± 0.2 | 3755 ± 105 | ... | ... | ... | ... | ... | ... | ... | ... | ... | ... | n | ... |
| 6999 | 11192619-7614098 | 71.7 ± 0.2 | 4552 ± 124 | ... | 1.043 ± 0.003 | ... | ... | ... | N | ... | ... | ... | ... | n | ... |
| 7000 | 11194733-7614224 | -37.4 ± 0.3 | 4485 ± 136 | ... | 1.030 ± 0.003 | -0.24 ± 0.21 | <23 | 3 | N | ... | ... | ... | ... | n | G |
| 7001 | 11195060-7628492 | 444.8 ± 0.2 | 2950 ± 100 | ... | ... | ... | ... | ... | ... | ... | ... | ... | ... | n | ... |
| 7002 | 11195899-7613389 | 23.6 ± 0.2 | 4596 ± 106 | ... | 1.033 ± 0.004 | ... | <24 | 3 | N | ... | ... | ... | ... | n | G |
| 7003 | 11200309-7615231 | -22.3 ± 0.2 | 4732 ± 16 | 2.47 ± 0.09 | 1.025 ± 0.002 | -0.05 ± 0.04 | <32 | 3 | N | N | ... | ... | ... | n | G |

| ID | CNAME | RV (km s$^{-1}$) | $T_{\text{eff}}$ (K) | $logg$ (dex) | $\gamma^a$ | [Fe/H] (dex) | $EW(Li)^b$ (mÅ) | $EW(Li)$ error flag$^c$ | Membership $\gamma$ | Membership $logg$ | Membership RV | Membership Li | Membership [Fe/H] | Final$^d$ | Non-mem with Li$^e$ |
|---|---|---|---|---|---|---|---|---|---|---|---|---|---|---|---|
| 7004 | 11201287-7624396 | 38.7 ± 0.2 | 4787 ± 109 | 2.55 ± 0.16 | 1.017 ± 0.002 | -0.08 ± 0.16 | 81 ± 4 | 1 | N | N | ... | ... | ... | n | G |
| 7005 | 11202496-7619262 | 53.1 ± 0.2 | 4873 ± 131 | ... | 1.021 ± 0.001 | -0.20 ± 0.17 | <13 | 3 | N | ... | ... | ... | ... | n | G |

**Notes.** $^{(a)}$ Empirical gravity indicator defined by Damiani et al. (2014). $^{(b)}$ The values of $EW(Li)$ for this cluster are corrected (subtracted adjacent Fe (6707.43 Å) line). $^{(c)}$ Flags for the errors of the corrected $EW(Li)$ values, as follows: 1=$EW(Li)$ corrected by blends contribution using models; 2=$EW(Li)$ measured separately (Li line resolved - UVES only); and 3=Upper limit (no error for $EW(Li)$ is given). $^{(d)}$ The letters "Y" and "N" indicate if the star is a cluster member or not. $^{(e)}$ 'Li-rich G', 'G' and 'NG' indicate "Li-rich giant", ""giant" and "non-giant" Li field contaminants, respectively. $^{(f)}$ Strong accretor members.

**Table C.3.** Gamma Velorum

| ID | CNAME | RV (km s$^{-1}$) | $T_{\text{eff}}$ (K) | $\log g$ (dex) | $\gamma^a$ | [Fe/H] (dex) | $EW(\text{Li})^b$ (mÅ) | $EW(\text{Li})$ error flag$^c$ | Jeffries 2014$^d$ | Damiani 2014$^d$ | Spina 2014$^d$ | Frasca 2015$^d$ | Prisinzano 2016$^d$ | Final members$^e$ | Non-mem with Li$^f$ |
|---|---|---|---|---|---|---|---|---|---|---|---|---|---|---|---|
| 8924 | 08073015-4732578 | 128.2 ± 0.2 | 4538 ± 190 | … | 1.061 ± 0.003 | -0.17 ± 0.13 | … | … | … | … | … | … | … | … | G |
| 9546 | 08101934-4652487 | 157.7 ± 0.2 | 4602 ± 94 | … | 1.046 ± 0.006 | … | <32 | 3 | … | … | … | … | … | … | G |
| 8925 | 08073047-4654400 | 114.3 ± 0.3 | 4540 ± 113 | … | 1.047 ± 0.004 | … | <36 | 3 | … | … | … | … | … | … | G |
| 8926 | 08073093-4659051 | 48.4 ± 0.2 | 4420 ± 207 | … | 1.055 ± 0.002 | -0.07 ± 0.03 | <27 | 3 | … | … | … | … | … | … | G |
| 8927 | 08073103-4720559 | 84.8 ± 0.2 | 4716 ± 92 | … | 1.012 ± 0.004 | -0.20 ± 0.13 | … | … | … | … | … | … | … | … | G |
| 8928 | 08073145-4704464 | 57.0 ± 0.2 | 5007 ± 144 | … | 1.014 ± 0.002 | -0.18 ± 0.17 | <22 | 3 | … | … | … | … | … | … | G |
| 8929 | 08073161-4708016 | 129.9 ± 0.2 | 4798 ± 173 | … | 1.014 ± 0.002 | -0.15 ± 0.13 | <23 | 3 | … | … | … | … | … | … | G |
| 9547 | 08102067-4740253 | 56.8 ± 0.2 | 5127 ± 124 | … | 0.999 ± 0.003 | -0.13 ± 0.11 | … | … | … | … | … | … | … | … | … |
| 8930 | 08073178-4724258 | -6.4 ± 0.2 | 4138 ± 195 | 4.46 ± 0.09 | 0.875 ± 0.003 | -0.08 ± 0.13 | … | … | … | … | … | … | … | … | … |
| 9548 | 08102076-4723336 | 75.3 ± 0.2 | 4608 ± 91 | … | 1.016 ± 0.002 | … | <55 | 3 | … | … | … | … | … | … | G |
| 3445 | 08073209-4746433 | 12.8 ± 0.6 | 6003 ± 22 | 4.17 ± 0.03 | … | 0.06 ± 0.02 | <40 | 3 | … | … | N | … | … | n | NG |
| 8931 | 08073216-4742456 | 92.5 ± 0.2 | 4587 ± 79 | … | 1.037 ± 0.003 | -0.15 ± 0.11 | <32 | 3 | … | … | … | … | … | … | G |
| 3446 | 08073237-4722119 | -16.7 ± 0.4 | 7286 ± 298 | 4.13 ± 0.16 | … | 0.00 ± 0.14 | <7 | 3 | … | … | N | … | … | n | … |
| 9549 | 08102116-4740125 | 71.0 ± 0.2 | 4591 ± 66 | 2.27 ± 0.18 | 1.030 ± 0.002 | -0.12 ± 0.10 | 501 ± 3 | 1 | … | … | … | … | … | … | Li-rich G |
| 8932 | 08073244-4746270 | 61.7 ± 0.2 | 5119 ± 152 | … | 1.017 ± 0.003 | -0.15 ± 0.09 | <18 | 3 | … | … | … | … | … | … | G |
| 9550 | 08102135-4701025 | 54.6 ± 0.2 | 3898 ± 174 | … | 1.063 ± 0.003 | -0.20 ± 0.05 | 138 ± 2 | 1 | … | … | … | … | … | … | G |
| 8933 | 08073251-4734419 | 21.6 ± 0.2 | 3786 ± 85 | … | 0.847 ± 0.005 | -0.17 ± 0.14 | … | … | B | Y | … | Y | Y | Y | … |
| 9551 | 08102135-4701381 | 37.1 ± 0.2 | 4832 ± 188 | 2.63 ± 0.20 | 1.016 ± 0.003 | -0.08 ± 0.12 | <22 | 3 | … | … | … | … | … | … | G |
| 8934 | 08073257-4727143 | 64.5 ± 0.2 | 5009 ± 129 | … | 1.005 ± 0.003 | -0.12 ± 0.15 | <21 | 3 | … | … | … | … | … | … | … |
| 8935 | 08073288-4722268 | 75.4 ± 0.2 | 5505 ± 172 | … | 1.006 ± 0.002 | -0.10 ± 0.14 | <13 | 3 | … | … | … | … | … | … | … |
| 8936 | 08073310-4729141 | 13.6 ± 0.2 | 4437 ± 339 | … | 0.926 ± 0.003 | -0.03 ± 0.05 | 179 ± 3 | 1 | … | Y | … | … | … | Y | … |
| 3447 | 08073315-4744513 | 30.3 ± 0.6 | 5162 ± 9 | 2.86 ± 0.03 | … | 0.02 ± 0.02 | <15 | 3 | … | … | N | … | … | n | … |
| 9552 | 08102201-4747145 | 52.6 ± 0.2 | 4757 ± 77 | 2.78 ± 0.15 | 1.003 ± 0.005 | -0.08 ± 0.13 | <38 | 3 | … | … | … | … | … | … | … |
| 8937 | 08073363-4703355 | 17.2 ± 0.2 | 3965 ± 187 | … | 0.878 ± 0.003 | -0.08 ± 0.07 | 497 ± 5 | 1 | A | Y | … | Y | Y | Y | … |
| 9008 | 08075141-4658394 | 53.4 ± 0.2 | 4569 ± 82 | … | 1.026 ± 0.003 | -0.22 ± 0.15 | <15 | 3 | … | … | … | … | … | … | G |
| 9553 | 08102218-4723295 | 120.6 ± 0.2 | 4644 ± 142 | … | 1.024 ± 0.003 | -0.20 ± 0.19 | <23 | 3 | … | … | … | … | … | … | G |
| 9554 | 08102227-4727157 | 17.3 ± 0.2 | 4236 ± 197 | … | 0.914 ± 0.003 | -0.03 ± 0.09 | 488 ± 2 | 1 | A | Y | … | Y | Y | Y | … |
| 3454 | 08075167-4706085 | 44.7 ± 0.6 | 4638 ± 56 | 2.68 ± 0.10 | 1.012 ± 0.002 | 0.17 ± 0.01 | 39 ± 11 | 1 | … | … | N | … | … | n | G |
| 9555 | 08102235-4715235 | 692.0 ± 11.9 | … | … | … | … | … | … | … | … | … | … | … | … | … |
| 9009 | 08075168-4722517 | 110.7 ± 0.2 | 4974 ± 131 | … | 1.024 ± 0.004 | -0.16 ± 0.21 | <25 | 3 | … | … | … | … | … | … | G |
| 9010 | 08075210-4740176 | 21.9 ± 0.3 | 3440 ± 30 | … | 0.814 ± 0.017 | … | … | … | … | … | … | … | … | … | … |
| 9556 | 08102242-4741441 | 89.2 ± 0.2 | 4429 ± 167 | … | 1.043 ± 0.003 | -0.25 ± 0.06 | … | … | … | … | … | … | … | … | G |
| 9011 | 08075279-4745290 | 4.1 ± 0.2 | 4777 ± 130 | 4.46 ± 0.11 | 0.956 ± 0.004 | -0.03 ± 0.02 | <21 | 3 | … | … | … | … | … | … | … |
| 9557 | 08102278-4733384 | 19.3 ± 0.7 | 3498 ± 111 | … | 0.895 ± 0.012 | -0.25 ± 0.15 | … | … | … | … | … | … | … | … | … |
| 9012 | 08075290-4656036 | 48.9 ± 0.2 | 6641 ± 40 | … | 0.999 ± 0.002 | -0.10 ± 0.03 | <5 | 3 | … | … | … | … | … | … | … |
| 9013 | 08075314-4726322 | 19.2 ± 0.2 | 3509 ± 36 | … | 0.868 ± 0.007 | -0.23 ± 0.15 | 457 ± 18 | 1 | B | Y | … | Y | Y | Y | … |
| 9611 | 08103927-4655305 | 165.2 ± 0.2 | 4531 ± 134 | … | 1.034 ± 0.007 | -0.22 ± 0.09 | <23 | 3 | … | … | … | … | … | … | G |
| 9612 | 08103927-4716476 | 17.6 ± 0.8 | 3326 ± 18 | … | 0.869 ± 0.009 | -0.27 ± 0.14 | 623 ± 11 | 1 | A | Y | … | Y | Y | Y | … |
| 9014 | 08075334-4657022 | 49.4 ± 0.2 | 5131 ± 43 | … | 1.014 ± 0.003 | -0.05 ± 0.10 | … | … | … | … | … | … | … | … | G |
| 9015 | 08075349-4715517 | 83.4 ± 0.2 | 4942 ± 76 | 3.00 ± 0.01 | 1.005 ± 0.004 | -0.16 ± 0.12 | <22 | 3 | … | … | … | … | … | … | … |
| 9613 | 08103948-4718465 | 504.8 ± 0.2 | 3150 ± 155 | … | … | … | 687 ± 124 | 1 | B | Y | … | Y | Y | Y | … |
| 9016 | 08075432-4716081 | 23.9 ± 0.2 | 5833 ± 157 | 4.23 ± 0.16 | 0.995 ± 0.003 | 0.09 ± 0.03 | 178 ± 2 | 1 | … | … | … | … | … | … | … |
| 9017 | 08075481-4739165 | 21.2 ± 0.2 | 4735 ± 27 | 2.36 ± 0.10 | 1.012 ± 0.002 | -0.25 ± 0.02 | <14 | 3 | … | … | … | … | … | … | G |
| 9614 | 08104004-4722162 | 18.7 ± 0.5 | 3474 ± 96 | … | 0.859 ± 0.010 | -0.21 ± 0.15 | <100 | 3 | … | Y | … | … | Y | Y | … |
| 9615 | 08104042-4707541 | 60.5 ± 0.2 | 3936 ± 88 | … | 1.061 ± 0.002 | -0.09 ± 0.16 | … | … | … | … | … | … | … | … | G |
| 9018 | 08075510-4742586 | 96.1 ± 0.2 | 4451 ± 157 | … | 1.050 ± 0.004 | -0.25 ± 0.14 | <16 | 3 | … | … | … | … | … | … | G |
| 9019 | 08075546-4707460 | 20.0 ± 0.2 | 4545 ± 3 | … | 0.975 ± 0.002 | 0.09 ± 0.08 | 421 ± 2 | 1 | B | Y | … | Y | Y | Y | … |
| 9616 | 08104067-4655470 | 21.8 ± 0.2 | 6004 ± 206 | 4.18 ± 0.08 | 0.998 ± 0.001 | -0.02 ± 0.07 | 139 ± 4 | 1 | … | … | … | … | … | … | … |
| 9617 | 08104074-4659310 | 21.4 ± 0.5 | 3736 ± 66 | … | 0.762 ± 0.006 | -0.16 ± 0.10 | … | … | … | … | … | … | Y | … | … |
| 9020 | 08075548-4714560 | 115.0 ± 0.2 | 5020 ± 103 | … | 0.945 ± 0.004 | … | <24 | 3 | … | … | … | … | … | … | G |
| 9618 | 08104075-4734202 | 17.8 ± 0.3 | 3452 ± 96 | … | 0.850 ± 0.007 | -0.24 ± 0.16 | 323 ± 10 | 1 | A | … | … | Y | Y | Y | … |
| 9021 | 08075561-4655059 | 74.5 ± 0.2 | 4761 ± 86 | 2.57 ± 0.09 | 1.020 ± 0.002 | -0.08 ± 0.01 | <22 | 3 | … | … | … | … | … | … | G |
| 9022 | 08075571-4746355 | 51.6 ± 0.2 | 4627 ± 63 | 2.43 ± 0.16 | 1.022 ± 0.005 | 0.02 ± 0.09 | <37 | 3 | … | … | … | … | … | … | G |
| 9619 | 08104076-4655315 | 31.2 ± 0.2 | 5041 ± 141 | … | 1.021 ± 0.002 | -0.06 ± 0.13 | <18 | 3 | … | … | … | … | … | … | G |
| 9023 | 08075596-4732550 | 45.9 ± 0.2 | 4597 ± 100 | … | 1.034 ± 0.002 | -0.16 ± 0.14 | <19 | 3 | … | … | … | … | … | … | G |
| 9620 | 08104090-4713015 | 94.8 ± 0.2 | 4785 ± 79 | … | 1.016 ± 0.004 | -0.25 ± 0.20 | <18 | 3 | … | … | … | … | … | … | G |
| 9024 | 08075625-4723375 | 68.9 ± 0.2 | 4464 ± 194 | … | 1.040 ± 0.004 | -0.19 ± 0.03 | <23 | 3 | … | … | … | … | … | … | G |
| 9025 | 08075637-4742283 | 109.8 ± 0.2 | 4474 ± 121 | … | 1.029 ± 0.002 | -0.21 ± 0.17 | … | … | … | … | … | … | … | … | G |
| 9621 | 08104135-4719047 | 71.7 ± 0.2 | 4675 ± 60 | 2.53 ± 0.09 | 1.011 ± 0.003 | -0.09 ± 0.04 | <29 | 3 | … | … | … | … | … | … | G |

| ID | CNAME | RV (km s$^{-1}$) | $T_{\rm eff}$ (K) | logg (dex) | $\gamma^a$ | [Fe/H] (dex) | EW(Li)$^b$ (mÅ) | EW(Li) error flag$^c$ | Jeffries 2014$^d$ | Damiani 2014$^d$ | Literature members Spina 2014$^d$ | Frasca 2015$^d$ | Prisinzano 2016$^d$ | Final members$^e$ | Non-mem with Li$^f$ |
|---|---|---|---|---|---|---|---|---|---|---|---|---|---|---|---|
| 9622 | 08104172-4711481 | 63.9 ± 0.2 | 4715 ± 1 | 2.55 ± 0.08 | 1.014 ± 0.004 | -0.08 ± 0.05 | <45 | 3 | … | … | … | … | … | … | G |
| 9026 | 08075685-4744289 | 35.5 ± 0.2 | 4489 ± 183 | … | 1.020 ± 0.003 | -0.08 ± 0.10 | <29 | 3 | … | … | … | … | … | … | G |
| 9027 | 08075705-4708002 | 22.0 ± 0.6 | 3285 ± 11 | 4.90 ± 0.13 | 0.821 ± 0.013 | -0.26 ± 0.14 | 593 ± 18 | 1 | B | Y | … | … | Y | Y | … |
| 9623 | 08104261-4653382 | 12.7 ± 0.2 | 5729 ± 57 | … | 0.998 ± 0.002 | -0.02 ± 0.03 | 19 ± 2 | 1 | … | … | … | … | … | … | … |
| 9624 | 08104268-4714349 | 167.0 ± 0.2 | 4605 ± 139 | … | 1.027 ± 0.005 | -0.30 ± 0.10 | <30 | 3 | … | … | … | … | … | … | G |
| 9097 | 08082021-4720259 | 17.3 ± 0.3 | 3327 ± 19 | … | 0.875 ± 0.009 | -0.27 ± 0.14 | … | … | A | Y | … | Y | Y | Y | … |
| 9625 | 08104301-4722363 | 79.0 ± 0.2 | 4645 ± 74 | 2.68 ± 0.04 | 1.005 ± 0.005 | -0.04 ± 0.01 | <38 | 3 | … | … | … | … | … | … | … |
| 9098 | 08082138-4701422 | 110.2 ± 0.2 | 4375 ± 233 | … | 1.056 ± 0.002 | -0.20 ± 0.07 | 66 ± 8 | 1 | … | … | … | … | … | … | G |
| 9099 | 08082236-4710596 | 21.0 ± 0.3 | 4392 ± 138 | … | 0.956 ± 0.005 | -0.06 ± 0.06 | 500 ± 4 | 1 | B | Y | … | … | Y | Y | … |
| 9685 | 08105880-4718529 | 17.7 ± 0.3 | 3487 ± 7 | … | 0.854 ± 0.007 | -0.22 ± 0.15 | 314 ± 15 | 1 | A | Y | … | Y | Y | Y | … |
| 9100 | 08082244-4726160 | 93.7 ± 0.2 | 4971 ± 119 | … | 1.014 ± 0.003 | -0.18 ± 0.13 | <18 | 3 | … | … | … | … | … | … | G |
| 3501 | 08105888-4700140 | 58.3 ± 0.4 | … | … | … | … | … | … | … | … | N | … | … | n | … |
| 9101 | 08082282-4732033 | -0.7 ± 0.2 | 4167 ± 340 | 4.47 ± 0.05 | 0.857 ± 0.004 | -0.17 ± 0.01 | … | … | … | … | … | … | … | … | … |
| 9686 | 08105895-4725009 | 36.0 ± 0.2 | 4782 ± 36 | 2.74 ± 0.06 | 1.008 ± 0.004 | 0.02 ± 0.06 | <44 | 3 | … | … | … | … | … | … | … |
| 9102 | 08082311-4658267 | 51.0 ± 0.3 | 3614 ± 126 | 4.69 ± 0.05 | 0.785 ± 0.008 | -0.21 ± 0.15 | … | … | … | … | … | … | … | … | … |
| 9103 | 08082394-4709166 | 22.6 ± 0.2 | 5717 ± 93 | 4.24 ± 0.17 | 0.992 ± 0.002 | 0.08 ± 0.04 | 208 ± 2 | 1 | … | … | … | … | … | … | … |
| 9687 | 08105937-4731234 | 66.9 ± 0.2 | 4998 ± 139 | … | 1.015 ± 0.003 | -0.20 ± 0.15 | <17 | 3 | … | … | … | … | … | … | G |
| 9104 | 08082419-4656033 | 41.9 ± 0.2 | 5153 ± 190 | … | 1.014 ± 0.004 | -0.26 ± 0.20 | <18 | 3 | … | … | … | … | … | … | G |
| 9105 | 08082433-4723280 | 15.4 ± 0.2 | 5237 ± 230 | … | 0.998 ± 0.004 | -0.12 ± 0.20 | <23 | 3 | … | … | … | … | … | … | … |
| 9688 | 08105959-4709096 | 12.3 ± 0.3 | 3461 ± 86 | 4.58 ± 0.20 | 0.823 ± 0.008 | -0.22 ± 0.15 | <100 | 3 | … | … | … | … | Y | Y | … |
| 9106 | 08082450-4714200 | 20.2 ± 0.2 | 4877 ± 9 | 2.57 ± 0.13 | 1.013 ± 0.002 | -0.06 ± 0.01 | <20 | 3 | … | … | … | … | … | … | … |
| 9689 | 08105959-4740596 | 23.4 ± 0.2 | 6488 ± 25 | … | 0.996 ± 0.001 | 0.12 ± 0.02 | … | … | … | … | … | … | … | … | … |
| 9107 | 08082455-4701342 | 113.3 ± 0.2 | 4594 ± 159 | … | 1.028 ± 0.005 | -0.34 ± 0.16 | <25 | 3 | … | … | … | … | … | … | G |
| 9108 | 08082556-4721377 | 48.8 ± 0.2 | 4802 ± 8 | 2.57 ± 0.01 | 1.012 ± 0.002 | -0.07 ± 0.11 | … | … | … | … | … | … | … | … | … |
| 9690 | 08105996-4717226 | 41.5 ± 0.2 | 4799 ± 165 | 2.59 ± 0.20 | 1.017 ± 0.004 | -0.10 ± 0.12 | … | … | … | … | … | … | … | … | … |
| 9110 | 08082583-4736406 | 49.4 ± 0.2 | 4679 ± 51 | 2.49 ± 0.13 | 1.019 ± 0.003 | -0.06 ± 0.02 | 18 ± 2 | 1 | … | … | … | … | … | … | … |
| 9111 | 08082619-4658483 | 78.4 ± 0.2 | 4719 ± 169 | … | 1.022 ± 0.004 | -0.24 ± 0.16 | … | … | … | … | … | … | … | … | G |
| 9691 | 08110042-4732281 | 83.3 ± 0.2 | 4664 ± 32 | 2.49 ± 0.07 | 1.012 ± 0.004 | -0.09 ± 0.10 | <31 | 3 | … | … | … | … | … | … | G |
| 9112 | 08082639-4717149 | 14.8 ± 0.3 | 6017 ± 154 | … | 0.990 ± 0.003 | -0.12 ± 0.02 | 126 ± 11 | 1 | … | … | … | … | … | … | … |
| 9113 | 08082641-4728511 | 60.6 ± 0.2 | 4722 ± 131 | 2.48 ± 0.18 | 1.016 ± 0.003 | -0.14 ± 0.12 | <23 | 3 | … | … | … | … | … | … | G |
| 9692 | 08110132-4735379 | 32.6 ± 0.2 | 6768 ± 26 | … | 1.001 ± 0.002 | 0.33 ± 0.02 | <2 | 3 | … | … | … | … | … | … | … |
| 9114 | 08082652-4730435 | -1.2 ± 0.2 | 5772 ± 67 | 3.98 ± 0.01 | 0.997 ± 0.002 | 0.09 ± 0.01 | 18 ± 6 | 1 | … | … | … | … | … | … | … |
| 9115 | 08082737-4716557 | 0.9 ± 0.2 | 4076 ± 205 | … | 0.882 ± 0.005 | 0.10 ± 0.16 | 80 ± 3 | 1 | … | … | … | … | … | … | … |
| 9116 | 08082803-4656008 | 37.1 ± 0.3 | 6188 ± 37 | 4.00 ± 0.14 | 0.999 ± 0.003 | -0.22 ± 0.13 | <6 | 3 | … | … | … | … | … | … | … |
| 9693 | 08110241-4712555 | 28.7 ± 0.2 | 4726 ± 98 | 2.52 ± 0.18 | 1.022 ± 0.003 | -0.02 ± 0.11 | 49 ± 2 | 1 | … | … | … | … | … | … | … |
| 9179 | 08084784-4729598 | 24.4 ± 0.2 | 6455 ± 31 | 4.35 ± 0.07 | 0.997 ± 0.002 | 0.18 ± 0.02 | 49 ± 2 | 1 | … | … | … | … | … | … | … |
| 9694 | 08110242-4728257 | 82.6 ± 0.2 | 3825 ± 9 | … | 1.054 ± 0.001 | … | 153 ± 1 | 1 | … | … | … | … | … | … | G |
| 9180 | 08084787-4652335 | 68.4 ± 0.2 | 5015 ± 27 | 2.94 ± 0.20 | 1.014 ± 0.005 | -0.18 ± 0.13 | <35 | 3 | … | … | … | … | … | … | G |
| 9761 | 08111703-4721309 | 17.5 ± 0.4 | 3656 ± 27 | 4.68 ± 0.01 | 0.780 ± 0.011 | -0.21 ± 0.12 | <100 | 3 | … | … | … | … | … | … | … |
| 9181 | 08084854-4728191 | 83.9 ± 2.0 | … | … | … | … | … | … | … | … | … | … | Y | … | … |
| 9762 | 08111743-4743014 | 62.9 ± 0.3 | 5003 ± 138 | … | 1.008 ± 0.004 | -0.17 ± 0.11 | <16 | 3 | … | … | … | … | … | … | G |
| 9182 | 08084873-4656489 | 44.9 ± 0.2 | 4999 ± 69 | … | 1.018 ± 0.002 | -0.05 ± 0.06 | <24 | 3 | … | … | … | … | … | … | G |
| 9763 | 08111762-4735589 | 8.3 ± 0.2 | 4921 ± 121 | 2.67 ± 0.10 | 1.015 ± 0.003 | -0.10 ± 0.10 | <13 | 3 | … | … | … | … | … | … | G |
| 9183 | 08084881-4653424 | 17.3 ± 0.2 | 3773 ± 95 | … | 0.851 ± 0.004 | -0.17 ± 0.14 | 454 ± 17 | 1 | A | Y | … | Y | Y | Y | … |
| 9764 | 08111767-4724341 | 21.7 ± 0.2 | 5011 ± 206 | … | 1.015 ± 0.004 | -0.23 ± 0.12 | 43 ± 3 | 1 | … | … | … | … | … | … | G |
| 9184 | 08084914-4704100 | 132.9 ± 0.2 | 4483 ± 163 | … | 1.026 ± 0.005 | -0.15 ± 0.10 | <38 | 3 | … | … | … | … | … | … | G |
| 9765 | 08111770-4703214 | 73.5 ± 0.2 | 4606 ± 91 | … | 1.022 ± 0.004 | … | <50 | 3 | … | … | … | … | … | … | G |
| 9185 | 08084951-4729065 | 6.6 ± 0.2 | 6030 ± 116 | 4.05 ± 0.02 | 0.999 ± 0.003 | 0.08 ± 0.04 | 59 ± 2 | 1 | … | … | … | … | … | … | … |
| 9766 | 08111784-4723095 | 17.0 ± 0.3 | 3675 ± 37 | … | 0.840 ± 0.007 | -0.19 ± 0.13 | 396 ± 5 | 1 | A | Y | … | Y | Y | Y | … |
| 9767 | 08111801-4701176 | 25.7 ± 0.2 | 4178 ± 195 | 4.58 ± 0.04 | 0.866 ± 0.005 | 0.02 ± 0.20 | … | … | … | … | … | … | … | … | … |
| 9768 | 08111816-4737083 | 28.6 ± 0.2 | 3954 ± 5 | … | 1.035 ± 0.002 | -0.11 ± 0.22 | <33 | 3 | … | … | … | … | … | … | G |
| 9186 | 08084996-4736435 | 102.3 ± 0.2 | 4776 ± 102 | … | 1.021 ± 0.003 | -0.29 ± 0.28 | <13 | 3 | … | … | … | … | … | … | G |
| 3505 | 08111853-4704187 | 41.6 ± 0.6 | 4988 ± 43 | 3.38 ± 0.08 | 1.000 ± 0.001 | -0.09 ± 0.02 | 16 ± 2 | 1 | … | … | N | … | … | n | … |
| 9187 | 08085000-4658374 | 1.9 ± 0.2 | 6310 ± 53 | 4.00 ± 0.12 | 1.003 ± 0.003 | -0.29 ± 0.04 | … | … | … | … | … | … | … | … | … |
| 9769 | 08111853-4731109 | 123.2 ± 0.2 | 4226 ± 230 | … | 1.053 ± 0.002 | -0.25 ± 0.08 | … | … | … | … | … | … | … | … | G |
| 9188 | 08085078-4701176 | 16.9 ± 0.3 | 5269 ± 22 | … | 1.013 ± 0.005 | -0.16 ± 0.12 | <16 | 3 | … | … | … | … | … | … | … |
| 9770 | 08111877-4710195 | 55.2 ± 0.2 | 4778 ± 10 | 2.60 ± 0.12 | 1.021 ± 0.004 | -0.08 ± 0.01 | <22 | 3 | … | … | … | … | … | … | G |
| 9189 | 08085117-4716075 | 18.1 ± 0.2 | 4307 ± 420 | … | 0.896 ± 0.002 | -0.01 ± 0.10 | 500 ± 5 | 1 | A | Y | … | Y | Y | Y | … |
| 9771 | 08111902-4733209 | 19.0 ± 0.5 | … | … | … | … | … | … | … | … | … | … | … | … | … |

Gutiérrez Albarrán et al.: Calibrating the lithium–age relation I.

**Table C.3.** continued.

| ID | CNAME | RV (km s$^{-1}$) | $T_{eff}$ (K) | log g (dex) | $\gamma^a$ | [Fe/H] (dex) | EW(Li)$^b$ (mÅ) | EW(Li) error flag$^c$ | Jeffries 2014$^d$ | Damiani 2014$^d$ | Spina 2014$^d$ | Frasca 2015$^d$ | Prisinzano 2016$^d$ | Final members$^e$ | Non-mem with Li$^f$ |
|---|---|---|---|---|---|---|---|---|---|---|---|---|---|---|---|
| 9190 | 08085128-4701428 | 28.4 ± 0.2 | 4819 ± 179 | 2.64 ± 0.17 | 1.016 ± 0.003 | -0.04 ± 0.07 | <26 | 3 | ... | ... | ... | ... | ... | ... | G |
| 9772 | 08111911-4724120 | 13.3 ± 0.3 | 4098 ± 220 | ... | 0.827 ± 0.008 | -0.30 ± 0.20 | <40 | 3 | ... | ... | ... | ... | ... | ... | ... |
| 9191 | 08085160-4720529 | 53.1 ± 0.2 | 4887 ± 65 | 2.76 ± 0.17 | 1.016 ± 0.004 | -0.13 ± 0.07 | <16 | 3 | ... | ... | ... | ... | ... | ... | G |
| 9773 | 08111956-4721068 | 14.4 ± 0.3 | 7286 ± 37 | ... | ... | ... | ... | ... | ... | ... | ... | ... | ... | ... | ... |
| 9774 | 08112006-4709209 | 94.5 ± 0.2 | 4628 ± 118 | ... | 1.029 ± 0.003 | -0.31 ± 0.04 | <19 | 3 | ... | ... | ... | ... | ... | ... | G |
| 9192 | 08085166-4717374 | 58.7 ± 0.2 | 5099 ± 128 | ... | 1.006 ± 0.003 | -0.18 ± 0.14 | ... | ... | ... | ... | ... | ... | ... | ... | ... |
| 9775 | 08112037-4652100 | 124.7 ± 0.2 | 4732 ± 70 | 2.59 ± 0.18 | 1.013 ± 0.005 | -0.15 ± 0.12 | 4 ± 2 | 1 | ... | ... | ... | ... | ... | ... | G |
| 9776 | 08112045-4732486 | 8.2 ± 0.3 | 4273 ± 351 | ... | 0.903 ± 0.006 | 0.06 ± 0.16 | ... | ... | ... | ... | ... | ... | ... | ... | ... |
| 9777 | 08112057-4708363 | 28.0 ± 0.3 | 6671 ± 32 | 4.01 ± 0.06 | 1.009 ± 0.002 | -0.06 ± 0.03 | ... | ... | ... | ... | ... | ... | ... | ... | ... |
| 9193 | 08085214-4722579 | 17.3 ± 0.2 | 4304 ± 421 | ... | 0.899 ± 0.002 | -0.02 ± 0.09 | 458 ± 5 | 1 | A | Y | ... | Y | Y | Y | ... |
| 9194 | 08085219-4711004 | 121.0 ± 0.2 | 4970 ± 36 | ... | 1.020 ± 0.003 | -0.10 ± 0.09 | <22 | 3 | ... | ... | ... | ... | ... | ... | G |
| 9195 | 08085231-4713596 | 17.5 ± 0.4 | 3344 ± 43 | ... | 0.889 ± 0.014 | -0.28 ± 0.14 | 616 ± 15 | 1 | A | ... | ... | ... | Y | Y | ... |
| 9778 | 08112077-4719208 | 89.0 ± 0.2 | 4657 ± 49 | 2.38 ± 0.14 | 1.017 ± 0.002 | -0.17 ± 0.11 | <22 | 3 | ... | ... | ... | ... | ... | ... | G |
| 9196 | 08085239-4727170 | -4.5 ± 0.2 | 5049 ± 117 | 4.29 ± 0.16 | 0.966 ± 0.004 | -0.09 ± 0.03 | <27 | 3 | ... | ... | ... | ... | ... | ... | ... |
| 9779 | 08112086-4737042 | 76.3 ± 0.2 | 3939 ± 18 | ... | 1.031 ± 0.003 | -0.11 ± 0.22 | 45 ± 11 | 1 | ... | ... | ... | ... | ... | ... | G |
| 9197 | 08085245-4655441 | 43.1 ± 0.2 | 4563 ± 114 | 2.49 ± 0.10 | 1.009 ± 0.002 | 0.09 ± 0.08 | <47 | 3 | ... | ... | ... | ... | ... | ... | ... |
| 9847 | 08113695-4720032 | 42.0 ± 0.2 | 5094 ± 174 | ... | 1.021 ± 0.003 | -0.21 ± 0.13 | <23 | 3 | ... | ... | ... | ... | ... | ... | G |
| 9848 | 08113716-4738170 | 3.7 ± 0.2 | 5165 ± 51 | ... | 1.011 ± 0.005 | -0.05 ± 0.07 | <27 | 3 | ... | ... | ... | ... | ... | ... | G |
| 9256 | 08090867-4658110 | 23.7 ± 0.2 | 6350 ± 42 | ... | 0.996 ± 0.002 | 0.02 ± 0.03 | 47 ± 4 | 1 | ... | ... | ... | ... | ... | ... | ... |
| 9849 | 08113740-4745109 | 29.7 ± 1.1 | ... | ... | ... | ... | ... | ... | ... | ... | ... | ... | ... | ... | ... |
| 9850 | 08113770-4727236 | 97.8 ± 0.2 | 4786 ± 109 | ... | 1.010 ± 0.002 | -0.24 ± 0.24 | <23 | 3 | ... | ... | ... | ... | ... | ... | G |
| 9257 | 08090875-4707441 | 19.0 ± 0.3 | 3327 ± 3 | ... | 0.873 ± 0.008 | -0.27 ± 0.15 | ... | ... | A | Y | ... | Y | Y | ... | ... |
| 9851 | 08113781-4726376 | 13.9 ± 0.8 | 3402 ± 179 | ... | 0.897 ± 0.009 | -0.23 ± 0.17 | 684 ± 11 | 1 | A | Y | ... | Y | Y | Y | ... |
| 9852 | 08113815-4713236 | 116.2 ± 0.2 | 3914 ± 120 | ... | 1.070 ± 0.003 | -0.10 ± 0.16 | 88 ± 8 | 1 | ... | ... | ... | ... | ... | ... | G |
| 9258 | 08090909-4744540 | 23.7 ± 0.3 | 4094 ± 270 | 4.47 ± 0.14 | 0.859 ± 0.007 | -0.19 ± 0.02 | <12 | 3 | ... | ... | ... | ... | ... | ... | ... |
| 9853 | 08113846-4711536 | 19.0 ± 0.5 | 3390 ± 109 | ... | 0.875 ± 0.010 | -0.28 ± 0.14 | ... | ... | ... | ... | ... | ... | Y | ... | ... |
| 9854 | 08113846-4732118 | 12.8 ± 0.3 | 3708 ± 64 | 4.59 ± 0.14 | 0.802 ± 0.010 | -0.21 ± 0.14 | ... | ... | ... | ... | ... | ... | ... | ... | ... |
| 9259 | 08090915-4745105 | 20.2 ± 0.3 | 3505 ± 21 | ... | 0.849 ± 0.010 | -0.27 ± 0.13 | 421 ± 27 | 1 | B | Y | ... | Y | Y | Y | ... |
| 9855 | 08113963-4702371 | 78.4 ± 0.2 | 4575 ± 109 | 2.04 ± 0.14 | 1.024 ± 0.003 | -0.38 ± 0.15 | 16 ± 4 | 1 | ... | ... | ... | ... | ... | ... | G |
| 9260 | 08090943-4732585 | 9.7 ± 0.2 | 4750 ± 9 | 2.59 ± 0.05 | 1.016 ± 0.002 | 0.03 ± 0.05 | <41 | 3 | ... | ... | ... | ... | ... | ... | G |
| 9856 | 08113966-4652220 | 32.9 ± 0.2 | 5136 ± 32 | ... | 1.014 ± 0.002 | -0.04 ± 0.07 | <26 | 3 | ... | ... | ... | ... | ... | ... | G |
| 9261 | 08090950-4742334 | 72.2 ± 0.2 | 4491 ± 166 | ... | 1.031 ± 0.004 | -0.09 ± 0.01 | <47 | 3 | ... | ... | ... | ... | ... | ... | G |
| 9857 | 08113973-4655494 | -16.9 ± 0.2 | 5034 ± 140 | ... | 1.020 ± 0.002 | -0.15 ± 0.12 | <14 | 3 | ... | ... | ... | ... | ... | ... | G |
| 9262 | 08090966-4724525 | 31.3 ± 0.2 | 3911 ± 213 | ... | 0.887 ± 0.003 | -0.17 ± 0.14 | ... | ... | ... | ... | Y | ... | Y | ... | ... |
| 9858 | 08114027-4721397 | 72.3 ± 0.2 | 4989 ± 168 | ... | 1.009 ± 0.004 | -0.18 ± 0.13 | <13 | 3 | ... | ... | ... | ... | ... | ... | G |
| 9263 | 08090978-4726305 | 16.5 ± 0.3 | 3459 ± 29 | ... | 0.850 ± 0.009 | -0.24 ± 0.15 | 394 ± 9 | 1 | A | ... | ... | Y | Y | Y | ... |
| 9264 | 08090983-4727549 | 13.1 ± 0.2 | 4432 ± 100 | 4.53 ± 0.03 | 0.920 ± 0.003 | 0.04 ± 0.14 | <23 | 3 | ... | ... | ... | ... | ... | ... | ... |
| 9265 | 08091002-4726342 | 16.0 ± 0.8 | 3278 ± 40 | ... | 0.880 ± 0.010 | -0.28 ± 0.18 | 664 ± 15 | 1 | B | Y | ... | Y | Y | Y | ... |
| 9859 | 08114056-4722239 | 55.5 ± 0.2 | 4728 ± 15 | 2.53 ± 0.09 | 1.019 ± 0.004 | -0.08 ± 0.05 | <30 | 3 | ... | ... | ... | ... | ... | ... | G |
| 9860 | 08114058-4655574 | 27.6 ± 0.2 | 6166 ± 122 | 3.93 ± 0.02 | 1.004 ± 0.002 | -0.17 ± 0.02 | 59 ± 1 | 1 | ... | ... | ... | ... | ... | ... | ... |
| 9861 | 08114059-4653451 | 58.2 ± 0.2 | 4867 ± 113 | ... | 0.994 ± 0.004 | -0.09 ± 0.03 | 26 ± ... | 3 | ... | ... | ... | ... | ... | ... | ... |
| 9266 | 08091036-4720250 | 16.8 ± 0.2 | 3765 ± 66 | ... | 0.859 ± 0.003 | -0.19 ± 0.13 | 525 ± 6 | 1 | A | Y | ... | Y | Y | Y | ... |
| 9862 | 08114064-4708453 | 87.4 ± 0.2 | 4613 ± 46 | ... | 1.013 ± 0.005 | -0.01 ± 0.14 | <33 | 3 | ... | ... | ... | ... | ... | ... | G |
| 9863 | 08114072-4659305 | 1.8 ± 0.2 | 3792 ± 60 | 4.58 ± 0.15 | 0.814 ± 0.004 | -0.17 ± 0.14 | <100 | 3 | ... | ... | ... | ... | ... | ... | ... |
| 9267 | 08091101-4734316 | 16.8 ± 0.4 | 3320 ± 34 | ... | 0.871 ± 0.013 | -0.25 ± 0.14 | 620 ± 16 | 1 | A | Y | ... | ... | Y | Y | ... |
| 3510 | 08114116-4720345 | 33.3 ± 0.6 | 5007 ± 2 | 2.73 ± 0.02 | ... | 0.12 ± 0.05 | <17 | 3 | ... | ... | N | ... | ... | n | ... |
| 9268 | 08091107-4654060 | 126.6 ± 0.2 | 4624 ± 107 | ... | 1.033 ± 0.004 | -0.19 ± 0.15 | <28 | 3 | ... | ... | ... | ... | ... | ... | G |
| 9939 | 08120116-4702107 | -5.2 ± 0.2 | 4863 ± 166 | ... | 1.021 ± 0.002 | -0.05 ± 0.09 | <22 | 3 | ... | ... | ... | ... | ... | ... | G |
| 9940 | 08120127-4654045 | 0.4 ± 0.2 | 6246 ± 28 | 4.01 ± 0.06 | 1.002 ± 0.002 | -0.04 ± 0.02 | ... | ... | ... | ... | ... | ... | ... | ... | ... |
| 9941 | 08120128-4653030 | 62.6 ± 0.2 | 4758 ± 14 | 2.55 ± 0.09 | 1.019 ± 0.002 | -0.09 ± 0.04 | <21 | 3 | ... | ... | ... | ... | ... | ... | G |
| 9324 | 08092751-4747219 | 92.9 ± 0.2 | 4626 ± 38 | 2.50 ± 0.13 | 1.014 ± 0.004 | -0.05 ± 0.03 | 11 ± 3 | 1 | ... | ... | ... | ... | ... | ... | G |
| 9942 | 08120253-4717273 | 52.8 ± 0.2 | 5057 ± 182 | ... | 1.012 ± 0.003 | -0.08 ± 0.09 | <25 | 3 | ... | ... | ... | ... | ... | ... | G |
| 9943 | 08120284-4722391 | 11.8 ± 0.3 | 3457 ± 3 | ... | 0.847 ± 0.010 | -0.23 ± 0.14 | <100 | 3 | ... | ... | ... | ... | Y | Y | ... |
| 9325 | 08092842-4748036 | 134.1 ± 0.3 | 5078 ± 49 | ... | 1.023 ± 0.006 | -0.14 ± 0.11 | <22 | 3 | ... | ... | ... | ... | ... | ... | G |
| 9944 | 08120311-4651530 | 38.8 ± 0.2 | 5917 ± 75 | 4.06 ± 0.03 | 0.997 ± 0.003 | -0.23 ± 0.06 | 53 ± 5 | 1 | ... | ... | ... | ... | ... | ... | ... |
| 9326 | 08092860-4720178 | 15.4 ± 0.2 | 4349 ± 50 | 3.97 ± 0.02 | 0.942 ± 0.003 | -0.14 ± 0.19 | 440 ± 3 | 1 | A | ... | ... | Y | Y | Y | ... |
| 9945 | 08120315-4711018 | 25.3 ± 0.2 | 5104 ± 77 | ... | 1.022 ± 0.006 | -0.03 ± 0.05 | ... | ... | ... | ... | ... | ... | ... | ... | G |
| 9946 | 08120336-4704255 | 83.6 ± 0.2 | 3371 ± 144 | ... | ... | ... | ... | ... | ... | ... | ... | ... | ... | ... | ... |
| 9327 | 08092938-4720012 | -700.0 ± 0.3 | ... | ... | ... | ... | ... | ... | ... | ... | ... | ... | ... | ... | ... |

| ID | CNAME | RV (km s$^{-1}$) | $T_{\rm eff}$ (K) | log g (dex) | $\gamma^a$ | [Fe/H] (dex) | EW(Li)$^b$ (mÅ) | EW(Li) error flag$^c$ | Jeffries 2014$^d$ | Damiani 2014$^d$ | Literature members Spina 2014$^d$ | Frasca 2015$^d$ | Prisinzano 2016$^d$ | Final members$^e$ | Non-mem with Li$^f$ |
|---|---|---|---|---|---|---|---|---|---|---|---|---|---|---|---|
| 9947 | 08120343-4739341 | 82.1 ± 0.2 | 5249 ± 118 | … | 1.005 ± 0.004 | -0.37 ± 0.05 | <18 | 3 | … | … | … | … | … | … | … |
| 9328 | 08092962-4706252 | 60.2 ± 0.2 | 4813 ± 163 | … | 1.014 ± 0.005 | -0.18 ± 0.13 | 34 ± 5 | 1 | … | … | … | … | … | … | G |
| 9948 | 08120367-4714136 | 48.9 ± 0.2 | 5185 ± 117 | … | 1.021 ± 0.005 | -0.06 ± 0.03 | <14 | 3 | … | … | … | … | … | … | G |
| 9949 | 08120368-4730364 | 0.6 ± 0.2 | 4597 ± 74 | 2.42 ± 0.13 | 1.019 ± 0.005 | 0.03 ± 0.10 | <31 | 3 | … | … | … | … | … | … | G |
| 9329 | 08093012-4657559 | 19.1 ± 0.2 | 5162 ± 93 | … | 0.979 ± 0.002 | … | 411 ± 8 | 1 | B | Y | … | Y | Y | Y | … |
| 9330 | 08093020-4737525 | -1.5 ± 0.2 | 5156 ± 13 | 4.02 ± 0.09 | 0.980 ± 0.002 | 0.06 ± 0.06 | <23 | 3 | … | … | … | … | … | … | … |
| 9331 | 08093028-4734086 | 17.1 ± 0.3 | 3371 ± 85 | … | 0.861 ± 0.009 | -0.27 ± 0.14 | 312 ± 15 | 1 | A | Y | … | Y | Y | Y | … |
| 9332 | 08093040-4653454 | 54.5 ± 0.2 | 5085 ± 81 | … | 1.017 ± 0.004 | -0.06 ± 0.10 | 33 ± 5 | 1 | … | … | … | … | … | … | … |
| 9950 | 08120388-4702062 | 12.7 ± 0.2 | 5794 ± 21 | 4.17 ± 0.14 | 0.993 ± 0.002 | 0.01 ± 0.05 | 32 ± 4 | 1 | … | … | … | … | … | … | G |
| 9951 | 08120420-4726104 | 55.8 ± 0.2 | 4158 ± 274 | … | 1.048 ± 0.002 | -0.22 ± 0.03 | … | … | … | … | … | … | … | … | G |
| 9952 | 08120474-4706108 | 73.8 ± 0.2 | 4663 ± 144 | 2.46 ± 0.19 | 1.016 ± 0.005 | -0.09 ± 0.05 | <19 | 3 | … | … | … | … | … | … | G |
| 9333 | 08093135-4723124 | 17.3 ± 0.2 | 4066 ± 218 | … | 0.881 ± 0.004 | -0.07 ± 0.09 | 348 ± 4 | 1 | A | Y | … | Y | Y | Y | … |
| 9335 | 08093154-4737066 | 20.7 ± 0.2 | 3621 ± 34 | … | 0.861 ± 0.005 | -0.21 ± 0.12 | 535 ± 18 | 1 | B | … | … | Y | Y | Y | … |
| 9336 | 08093186-4728542 | 190.8 ± 0.2 | 4511 ± 21 | 1.78 ± 0.18 | 1.033 ± 0.004 | -0.45 ± 0.07 | 22 ± 4 | 1 | … | … | … | … | … | … | G |
| 9953 | 08120601-4737113 | 22.4 ± 0.4 | 3301 ± 14 | … | 0.851 ± 0.017 | … | 489 ± 23 | 1 | … | … | … | … | Y | Y | … |
| 9337 | 08093188-4742203 | 16.2 ± 0.2 | 4835 ± 70 | 2.53 ± 0.06 | 1.011 ± 0.003 | -0.19 ± 0.03 | <28 | 3 | … | … | … | … | … | … | G |
| 9954 | 08120688-4723433 | 93.7 ± 0.2 | 3884 ± 136 | … | 1.060 ± 0.003 | -0.20 ± 0.06 | 128 ± 10 | 1 | … | … | … | … | … | … | G |
| 9338 | 08093221-4701104 | 65.8 ± 0.2 | 5279 ± 117 | … | 1.015 ± 0.004 | -0.30 ± 0.11 | <23 | 3 | … | … | … | … | … | … | … |
| 9955 | 08120699-4704512 | 81.5 ± 0.2 | 3975 ± 27 | 1.18 ± 0.16 | 1.073 ± 0.001 | -0.04 ± 0.10 | <45 | 3 | … | … | … | … | … | … | G |
| 9339 | 08093240-4724395 | -3.9 ± 0.2 | 5041 ± 126 | … | 0.981 ± 0.003 | -0.01 ± 0.05 | <16 | 3 | … | … | … | … | … | … | … |
| 9397 | 08094692-4731389 | 17.0 ± 0.2 | 3591 ± 33 | … | 0.850 ± 0.005 | -0.21 ± 0.14 | 316 ± 23 | 1 | A | Y | … | Y | Y | Y | … |
| 9398 | 08094702-4744298 | 30.1 ± 0.2 | 5166 ± 106 | … | 0.991 ± 0.001 | 0.02 ± 0.07 | 299 ± 14 | 1 | … | … | … | … | Y | Y | … |
| 3483 | 08094711-4740038 | 27.6 ± 0.6 | 4761 ± 39 | 2.63 ± 0.09 | … | 0.12 ± 0.01 | <17 | 3 | … | … | … | … | … | … | … |
| 9399 | 08094766-4708371 | 14.8 ± 0.2 | 3767 ± 73 | … | 0.863 ± 0.003 | -0.19 ± 0.13 | 472 ± 17 | 1 | B | Y | … | Y | Y | Y | … |
| 9400 | 08094780-4717006 | 19.4 ± 0.4 | 3364 ± 32 | … | 0.879 ± 0.015 | -0.27 ± 0.14 | 494 ± 43 | 1 | B | Y | … | … | Y | Y | … |
| 9401 | 08094808-4722208 | 14.3 ± 0.2 | 6008 ± 37 | 4.25 ± 0.12 | 0.990 ± 0.003 | 0.13 ± 0.03 | … | … | … | … | … | … | … | … | … |
| 9402 | 08094811-4740323 | 19.2 ± 0.2 | 3561 ± 43 | … | 0.861 ± 0.006 | -0.25 ± 0.14 | 385 ± 5 | 1 | B | Y | … | Y | Y | Y | … |
| 9403 | 08094852-4719418 | 16.6 ± 0.2 | 3459 ± 116 | … | 0.874 ± 0.006 | -0.23 ± 0.15 | 434 ± 17 | 1 | A | Y | … | Y | Y | Y | … |
| 9404 | 08094856-4733102 | 122.0 ± 0.2 | 4541 ± 135 | … | 1.028 ± 0.003 | -0.18 ± 0.13 | … | … | … | … | … | … | … | … | G |
| 9406 | 08094884-4703530 | 65.9 ± 0.2 | 4001 ± 3 | 1.23 ± 0.13 | 1.053 ± 0.002 | -0.06 ± 0.19 | 16 ± 11 | 1 | … | … | … | … | … | … | … |
| 9407 | 08094939-4725072 | 19.3 ± 0.2 | 5970 ± 91 | 3.89 ± 0.04 | 1.000 ± 0.003 | -0.25 ± 0.15 | 37 ± 6 | 1 | … | … | … | Y | … | Y | … |
| 9408 | 08094951-4712079 | 19.9 ± 0.2 | 4072 ± 181 | … | 0.905 ± 0.003 | -0.05 ± 0.12 | 527 ± 4 | 1 | B | Y | … | Y | Y | Y | … |
| 9409 | 08094975-4706195 | 65.9 ± 0.2 | 5105 ± 167 | … | 1.012 ± 0.004 | -0.17 ± 0.14 | … | … | … | … | … | … | … | … | G |
| 9410 | 08094981-4720129 | 17.3 ± 0.2 | 4559 ± 131 | … | 0.968 ± 0.002 | -0.04 ± 0.09 | 439 ± 2 | 1 | A | Y | … | Y | Y | Y | … |
| 3484 | 08095042-4657080 | 40.7 ± 0.4 | … | … | … | … | … | … | … | … | … | … | … | … | … |
| 9411 | 08095048-4723123 | 20.2 ± 0.3 | 3488 ± 5 | … | 0.869 ± 0.005 | -0.25 ± 0.13 | 415 ± 22 | 1 | B | … | … | Y | Y | Y | … |
| 9412 | 08095062-4728064 | 35.1 ± 0.2 | … | … | … | … | … | … | … | … | … | Y | … | … | … |
| 8814 | 08070001-4740443 | 65.0 ± 0.2 | 4945 ± 173 | … | 1.019 ± 0.003 | -0.25 ± 0.18 | … | … | … | … | … | … | … | … | G |
| 8815 | 08070008-4720258 | 70.6 ± 0.2 | 4653 ± 65 | … | 1.016 ± 0.004 | -0.31 ± 0.02 | … | … | … | … | … | … | … | … | … |
| 9413 | 08095071-4716433 | 18.0 ± 0.3 | 3478 ± 101 | … | 0.839 ± 0.007 | -0.24 ± 0.15 | … | … | … | … | Y | … | Y | … | … |
| 8816 | 08070034-4654063 | 52.7 ± 0.2 | 5011 ± 202 | … | 1.023 ± 0.004 | -0.22 ± 0.17 | <33 | 3 | … | … | … | … | … | … | … |
| 9475 | 08100311-4713178 | 46.4 ± 0.2 | 4642 ± 81 | 2.45 ± 0.15 | 1.025 ± 0.004 | 0.07 ± 0.04 | <45 | 3 | … | … | … | … | … | … | G |
| 8817 | 08070036-4745250 | 17.6 ± 0.8 | 3351 ± 59 | 4.66 ± 0.17 | 0.833 ± 0.012 | -0.27 ± 0.14 | 442 ± 37 | 1 | A | Y | … | Y | Y | Y | … |
| 9476 | 08100325-4656071 | 39.4 ± 0.3 | 5202 ± 277 | … | 1.011 ± 0.009 | -0.32 ± 0.18 | <19 | 3 | … | … | … | … | … | … | … |
| 8818 | 08070053-4732093 | 34.5 ± 0.2 | 5683 ± 175 | 4.37 ± 0.17 | 0.989 ± 0.002 | 0.07 ± 0.03 | … | … | … | … | … | … | … | … | … |
| 9477 | 08100341-4706041 | 80.1 ± 0.2 | 4514 ± 159 | … | 1.026 ± 0.003 | -0.13 ± 0.10 | 27 ± 1 | 1 | … | … | … | … | … | … | G |
| 8819 | 08070135-4738460 | 56.9 ± 0.3 | 7483 ± 28 | … | 0.998 ± 0.001 | … | <1 | 3 | … | … | … | … | … | … | … |
| 9478 | 08100355-4656113 | 29.7 ± 0.2 | 4640 ± 100 | 2.43 ± 0.16 | 1.019 ± 0.003 | 0.00 ± 0.11 | <36 | 3 | … | … | … | … | … | … | G |
| 8820 | 08070164-4703343 | 33.0 ± 0.3 | 4226 ± 252 | 4.52 ± 0.01 | 0.874 ± 0.006 | -0.20 ± 0.03 | <25 | 3 | … | … | … | … | … | … | … |
| 9479 | 08100370-4710433 | 139.5 ± 0.2 | 4284 ± 259 | … | 1.065 ± 0.002 | -0.25 ± 0.12 | 86 ± 8 | 1 | … | … | … | … | … | … | G |
| 8821 | 08070165-4655570 | 1.2 ± 0.2 | 6220 ± 38 | 3.99 ± 0.08 | 1.003 ± 0.002 | 0.05 ± 0.03 | 51 ± 2 | 1 | … | … | … | … | … | … | … |
| 8822 | 08070184-4652549 | 51.2 ± 0.2 | 4216 ± 141 | 4.58 ± 0.05 | 0.879 ± 0.004 | 0.01 ± 0.20 | <15 | 3 | … | … | … | … | … | … | … |
| 9480 | 08100377-4716535 | 3.3 ± 0.8 | 5870 ± 111 | … | 1.013 ± 0.015 | … | … | … | … | … | … | … | … | … | … |
| 8823 | 08070215-4716156 | 27.5 ± 0.3 | 3924 ± 154 | 4.63 ± 0.06 | 0.817 ± 0.009 | -0.16 ± 0.14 | … | … | … | … | … | … | … | … | … |
| 8824 | 08070240-4704300 | 4.4 ± 0.2 | 5862 ± 79 | 3.95 ± 0.10 | 1.001 ± 0.002 | -0.10 ± 0.06 | 37 ± 9 | 1 | … | … | … | … | … | … | … |
| 9481 | 08100408-4722341 | 95.8 ± 0.2 | 4656 ± 156 | … | 1.017 ± 0.003 | -0.17 ± 0.18 | <18 | 3 | … | … | … | … | … | … | G |
| 8825 | 08070243-4737589 | 88.6 ± 0.2 | 4631 ± 97 | 2.48 ± 0.14 | 1.012 ± 0.004 | -0.02 ± 0.04 | … | … | … | … | … | … | … | … | G |
| 9482 | 08100421-4716546 | 17.9 ± 0.2 | 3538 ± 107 | … | … | … | … | … | … | … | … | … | … | … | … |
| 8826 | 08070309-4705198 | 37.9 ± 0.2 | 4260 ± 274 | 4.45 ± 0.15 | 0.891 ± 0.005 | -0.05 ± 0.12 | <22 | 3 | … | … | … | … | … | … | … |

Gutiérrez Albarrán et al.: Calibrating the lithium–age relation I.

**Table C.3.** continued.

| ID | CNAME | RV (km s$^{-1}$) | $T_{\text{eff}}$ (K) | logg (dex) | $\gamma^a$ | [Fe/H] (dex) | EW(Li)$^b$ (mÅ) | EW(Li) error flag$^c$ | Jeffries 2014$^d$ | Damiani 2014$^d$ | Literature members Spina 2014$^d$ | Frasca 2015$^d$ | Prisinzano 2016$^d$ | Final members$^e$ | Non-mem with Li$^f$ |
|---|---|---|---|---|---|---|---|---|---|---|---|---|---|---|---|
| 9483 | 08100438-4745317 | 77.8 ± 0.2 | 4701 ± 20 | 2.56 ± 0.05 | 1.012 ± 0.005 | -0.06 ± 0.01 | 124 ± 3 | 1 | … | … | … | … | … | … | Li-rich G |
| 8827 | 08070328-4722546 | 77.1 ± 0.2 | 4544 ± 93 | 2.48 ± 0.02 | 1.007 ± 0.004 | -0.03 ± 0.02 | <43 | 3 | … | … | … | … | … | … | … |
| 8828 | 08070352-4738124 | 44.5 ± 0.2 | 4763 ± 11 | 2.57 ± 0.10 | 1.020 ± 0.003 | -0.05 ± 0.06 | <30 | 3 | … | … | … | … | … | … | G |
| 8829 | 08070429-4747175 | 17.9 ± 0.2 | 5260 ± 105 | … | 1.010 ± 0.002 | 0.01 ± 0.09 | <12 | 3 | … | … | … | … | … | … | … |
| 9484 | 08100460-4724497 | 32.8 ± 0.2 | 4996 ± 326 | … | 1.007 ± 0.003 | -0.27 ± 0.18 | <16 | 3 | … | … | … | … | … | … | … |
| 8830 | 08070431-4653184 | 136.3 ± 0.2 | 4145 ± 295 | … | 1.063 ± 0.003 | -0.32 ± 0.12 | <19 | 3 | … | … | … | … | … | … | G |
| 9485 | 08100491-4737599 | 75.6 ± 0.2 | 5007 ± 35 | 2.86 ± 0.13 | 1.011 ± 0.003 | -0.34 ± 0.07 | <8 | 3 | … | … | … | … | … | … | G |
| 8831 | 08070483-4741597 | 55.6 ± 0.2 | 5054 ± 202 | … | 1.030 ± 0.003 | -0.08 ± 0.06 | … | … | … | … | … | … | … | … | G |
| 8832 | 08070490-4740418 | 47.0 ± 0.2 | 4533 ± 153 | … | 1.021 ± 0.004 | 0.04 ± 0.12 | <43 | 3 | … | … | … | … | … | … | G |
| 8833 | 08070500-4740463 | 87.3 ± 0.2 | 4392 ± 217 | … | 1.038 ± 0.003 | -0.30 ± 0.15 | <16 | 3 | … | … | … | … | … | … | G |
| 3438 | 08070521-4734401 | 65.9 ± 0.6 | 5576 ± 35 | 4.32 ± 0.07 | 0.990 ± 0.001 | 0.26 ± 0.02 | <14 | 3 | … | … | … | … | … | … | … |
| 8834 | 08070522-4735313 | -22.1 ± 0.2 | 5021 ± 87 | … | 1.027 ± 0.002 | -0.09 ± 0.10 | <16 | 3 | … | … | … | … | … | … | G |
| 9486 | 08100586-4737170 | 51.2 ± 0.2 | 4598 ± 30 | 2.28 ± 0.06 | 1.033 ± 0.003 | 0.02 ± 0.06 | 28 ± 2 | 1 | … | … | … | … | … | … | … |
| 8835 | 08070576-4734593 | 45.8 ± 0.2 | 4379 ± 249 | … | 1.041 ± 0.001 | 0.11 ± 0.09 | <50 | 3 | … | … | … | … | … | … | G |
| 9487 | 08100589-4715074 | 38.7 ± 0.2 | 4902 ± 117 | … | 1.014 ± 0.003 | -0.10 ± 0.09 | <26 | 3 | … | … | … | … | … | … | G |
| 9488 | 08100599-4706129 | 63.9 ± 0.2 | 5281 ± 148 | … | 0.996 ± 0.002 | -0.35 ± 0.07 | 53 ± 2 | 1 | … | … | … | … | … | … | … |
| 8836 | 08070630-4658247 | 67.4 ± 0.2 | 4420 ± 222 | … | 1.041 ± 0.002 | -0.01 ± 0.03 | <42 | 3 | … | … | … | … | … | … | G |
| 9489 | 08100600-4713273 | 88.8 ± 0.2 | 4585 ± 63 | 2.44 ± 0.03 | 1.008 ± 0.004 | -0.12 ± 0.01 | <22 | 3 | … | … | … | … | … | … | … |
| 9046 | 08080239-4726092 | 14.9 ± 0.2 | 5359 ± 45 | 4.38 ± 0.08 | 0.978 ± 0.002 | -0.05 ± 0.06 | <14 | 3 | … | … | … | … | … | … | … |
| 9062 | 08080881-4732336 | 17.2 ± 0.2 | 3751 ± 81 | … | 0.854 ± 0.006 | -0.19 ± 0.13 | 439 ± 7 | 1 | A | Y | … | Y | Y | Y | … |
| 9074 | 08081292-4735206 | 80.4 ± 0.2 | 4626 ± 9 | 2.28 ± 0.16 | 1.031 ± 0.004 | -0.17 ± 0.02 | 37 ± 3 | 1 | … | … | … | … | … | … | … |
| 3459 | 08080882-4736515 | 29.7 ± 0.6 | 4616 ± 6 | 2.37 ± 0.12 | 1.018 ± 0.001 | 0.03 ± 0.06 | 32 ± 6 | 1 | … | … | N | … | … | n | G |
| 9075 | 08081302-4722228 | 148.7 ± 0.2 | 4484 ± 88 | … | 1.037 ± 0.004 | -0.32 ± 0.14 | 10 ± 2 | 1 | … | … | … | … | … | … | G |
| 9063 | 08080927-4717121 | 37.4 ± 0.2 | 4875 ± 157 | … | 1.024 ± 0.003 | -0.15 ± 0.17 | 31 ± 2 | 1 | … | … | … | … | … | … | G |
| 9064 | 08080943-4732250 | 17.8 ± 0.2 | 3504 ± 43 | … | 0.860 ± 0.005 | -0.21 ± 0.14 | 237 ± 28 | 1 | A | … | … | Y | Y | Y | … |
| 9076 | 08081368-4713411 | 14.0 ± 0.2 | 4993 ± 157 | … | 1.008 ± 0.003 | -0.10 ± 0.19 | <17 | 3 | … | … | … | … | … | … | G |
| 9077 | 08081434-4741166 | 57.0 ± 0.2 | 4738 ± 56 | … | 1.014 ± 0.004 | -0.15 ± 0.12 | <17 | 3 | … | … | … | … | … | … | G |
| 9065 | 08080973-4745072 | 93.5 ± 0.2 | 4167 ± 310 | … | 1.065 ± 0.003 | -0.28 ± 0.15 | … | … | … | … | … | … | … | … | G |
| 3461 | 08081445-4701498 | 25.6 ± 0.6 | 5806 ± 5 | 3.92 ± 0.12 | 0.996 ± 0.002 | -0.10 ± 0.05 | 79 ± 1 | 2 | … | … | … | … | … | … | … |
| 9066 | 08081010-4658053 | 28.7 ± 0.2 | 4718 ± 71 | 2.87 ± 0.18 | 1.000 ± 0.005 | 0.04 ± 0.05 | <35 | 3 | … | … | … | … | … | … | … |
| 9117 | 08082819-4706243 | 34.9 ± 0.2 | 5579 ± 57 | … | 0.995 ± 0.003 | -0.01 ± 0.14 | 48 ± 7 | 1 | … | … | … | … | … | … | … |
| 9067 | 08081027-4705319 | 134.8 ± 0.2 | 4512 ± 125 | … | 1.022 ± 0.004 | -0.13 ± 0.06 | … | … | … | … | … | … | … | … | G |
| 9118 | 08082846-4716020 | 18.2 ± 0.2 | 4422 ± 146 | 3.88 ± 0.07 | 0.951 ± 0.004 | -0.07 ± 0.25 | 459 ± 4 | 1 | B | Y | … | Y | Y | Y | … |
| 9068 | 08081029-4718178 | 38.1 ± 0.2 | 4627 ± 36 | 2.34 ± 0.07 | 1.017 ± 0.004 | -0.09 ± 0.07 | <44 | 3 | … | … | … | … | … | … | … |
| 9119 | 08082863-4715394 | 18.9 ± 0.3 | 3370 ± 18 | … | 0.879 ± 0.013 | -0.22 ± 0.15 | 577 ± 21 | 1 | B | Y | … | … | Y | Y | … |
| 9069 | 08081156-4709452 | 68.1 ± 0.2 | 4602 ± 149 | … | 1.016 ± 0.004 | 0.00 ± 0.07 | <39 | 3 | … | … | … | … | … | … | G |
| 9120 | 08082888-4715140 | 21.4 ± 0.4 | 3298 ± 13 | … | 0.875 ± 0.018 | … | 489 ± 18 | 1 | B | … | … | … | Y | Y | … |
| 9121 | 08082904-4730262 | 57.6 ± 0.2 | 5706 ± 3 | … | 0.997 ± 0.002 | -0.26 ± 0.12 | <2 | 3 | … | … | … | … | … | … | … |
| 9122 | 08082926-4702305 | 35.1 ± 0.2 | … | … | … | … | … | … | … | … | … | … | Y | … | … |
| 9070 | 08081215-4703592 | 37.5 ± 0.2 | 4877 ± 148 | … | 1.015 ± 0.003 | -0.05 ± 0.13 | <25 | 3 | … | … | … | … | … | … | G |
| 9123 | 08082948-4656598 | 73.4 ± 0.3 | 4594 ± 85 | 2.30 ± 0.15 | 1.016 ± 0.008 | -0.16 ± 0.11 | <55 | 3 | … | … | … | … | … | … | G |
| 9071 | 08081217-4739008 | 27.9 ± 1.5 | 3587 ± 29 | … | 0.833 ± 0.019 | … | … | … | … | … | … | … | … | … | … |
| 3460 | 08081245-4705593 | … | … | … | … | … | … | … | … | … | … | … | … | … | … |
| 9124 | 08082970-4709329 | 18.4 ± 0.4 | 3347 ± 9 | … | 0.848 ± 0.016 | -0.23 ± 0.13 | 593 ± 65 | 1 | A | … | … | … | Y | Y | … |
| 8769 | 08064390-4731532 | 16.8 ± 0.3 | 3413 ± 43 | … | 0.866 ± 0.012 | -0.23 ± 0.16 | 177 ± 7 | 1 | A | Y | … | Y | Y | Y | … |
| 9072 | 08081264-4732022 | 76.2 ± 0.2 | 4955 ± 59 | 2.83 ± 0.19 | 1.022 ± 0.004 | -0.10 ± 0.10 | <22 | 3 | … | … | … | … | … | … | G |
| 9125 | 08083030-4733305 | -502.1 ± 3.8 | 3290 ± 86 | … | … | … | … | … | … | Y | … | … | Y | … | … |
| 8770 | 08064496-4715591 | 44.7 ± 0.3 | 6373 ± 96 | 4.27 ± 0.21 | 0.997 ± 0.005 | -0.13 ± 0.08 | … | … | … | … | … | … | … | … | … |
| 9073 | 08081271-4714074 | 624.9 ± 13.3 | 3564 ± 98 | … | … | … | … | … | … | … | … | Y | … | Y | … |
| 9126 | 08083108-4728494 | 30.5 ± 0.3 | 3796 ± 68 | … | 0.833 ± 0.008 | -0.17 ± 0.14 | … | … | … | … | … | … | … | … | … |
| 8771 | 08064498-4706095 | 36.7 ± 0.2 | 4739 ± 145 | … | 1.024 ± 0.002 | -0.05 ± 0.10 | … | … | … | … | … | … | … | … | G |
| 9127 | 08083192-4704002 | 41.3 ± 0.2 | 4903 ± 99 | 3.00 ± 0.07 | 1.006 ± 0.003 | -0.07 ± 0.08 | <29 | 3 | … | … | … | … | … | … | … |
| 8772 | 08064509-4719414 | 100.6 ± 0.2 | 4657 ± 139 | … | 1.026 ± 0.003 | -0.20 ± 0.12 | … | … | … | … | … | … | … | … | … |
| 3463 | 08083232-4722553 | 5.1 ± 0.6 | 4655 ± 34 | 2.35 ± 0.10 | 1.021 ± 0.001 | 0.15 ± 0.05 | 34 ± 9 | 1 | … | … | N | … | … | n | G |
| 8773 | 08064531-4702090 | 70.2 ± 0.2 | 4939 ± 165 | … | 1.013 ± 0.003 | -0.40 ± 0.24 | <4 | 3 | … | … | … | … | … | … | G |
| 9128 | 08083248-4655559 | 57.0 ± 0.2 | 5158 ± 91 | … | 1.005 ± 0.003 | … | … | … | … | … | … | … | … | … | … |
| 9354 | 08093639-4703223 | 31.5 ± 0.2 | 4944 ± 166 | … | 1.020 ± 0.003 | -0.10 ± 0.18 | <24 | 3 | … | … | … | … | … | … | G |
| 8774 | 08064620-4734401 | 20.2 ± 0.8 | 3352 ± 26 | … | 0.822 ± 0.020 | … | … | … | … | … | … | … | Y | … | … |
| 9371 | 08094018-4730406 | 0.0 ± 0.6 | … | … | … | … | … | … | … | … | … | … | … | … | … |

**Table C.3.** continued.

| ID | CNAME | RV (km s$^{-1}$) | $T_{\text{eff}}$ (K) | logg (dex) | $\gamma^a$ | [Fe/H] (dex) | EW(Li)$^b$ (mÅ) | EW(Li) error flag$^c$ | Jeffries 2014$^d$ | Damiani 2014$^d$ | Literature members Spina 2014$^d$ | Frasca 2015$^d$ | Prisinzano 2016$^d$ | Final members$^e$ | Non-mem with Li$^f$ |
|---|---|---|---|---|---|---|---|---|---|---|---|---|---|---|---|
| 8775 | 08064752-4747060 | 24.5 ± 0.2 | 5061 ± 23 | … | 0.975 ± 0.004 | 0.08 ± 0.06 | <38 | 3 | … | … | … | … | … | … | … |
| 3435 | 08064772-4659492 | 11.7 ± 0.6 | 5786 ± 4 | 4.20 ± 0.05 | 0.996 ± 0.002 | 0.04 ± 0.01 | 10 ± 1 | 1 | … | … | N | … | … | n | … |
| 9129 | 08083324-4734505 | 32.4 ± 0.2 | 5987 ± 168 | 4.24 ± 0.08 | 0.996 ± 0.002 | 0.13 ± 0.05 | … | … | … | … | … | … | … | … | … |
| 9372 | 08094028-4654111 | 73.2 ± 0.2 | 4057 ± 234 | … | 1.065 ± 0.003 | -0.18 ± 0.06 | … | … | … | … | … | … | … | … | G |
| 8776 | 08064823-4735289 | 12.3 ± 0.9 | 3533 ± 7 | 4.60 ± 0.18 | 0.818 ± 0.016 | -0.26 ± 0.13 | <100 | 3 | … | … | … | … | Y | Y | … |
| 9130 | 08083328-4716048 | 19.7 ± 0.6 | 3294 ± 12 | … | 0.901 ± 0.018 | … | 535 ± 66 | 1 | B | … | … | … | Y | Y | … |
| 8777 | 08064867-4733258 | 29.9 ± 0.3 | 3981 ± 183 | 4.68 ± 0.15 | 0.810 ± 0.008 | -0.15 ± 0.12 | … | … | … | … | … | … | … | … | … |
| 8778 | 08064972-4716144 | 100.5 ± 0.2 | 4780 ± 119 | … | 1.023 ± 0.006 | -0.37 ± 0.12 | … | … | … | … | … | … | … | … | G |
| 9373 | 08094046-4728324 | 30.1 ± 0.6 | 3306 ± 78 | … | 0.914 ± 0.006 | -0.29 ± 0.18 | 553 ± 59 | 1 | B | Y | … | Y | Y | Y | … |
| 9131 | 08083354-4711111 | 77.5 ± 0.2 | 4524 ± 154 | … | 1.022 ± 0.003 | 0.06 ± 0.01 | 421 ± 3 | 1 | … | … | … | … | … | … | Li-rich G |
| 8779 | 08065007-4732221 | 21.0 ± 0.4 | … | … | … | … | … | … | B | … | … | … | … | Y | … |
| 9132 | 08083354-4726084 | 134.9 ± 0.2 | 5049 ± 79 | … | 1.019 ± 0.003 | -0.13 ± 0.19 | <15 | 3 | … | … | … | … | … | … | G |
| 9381 | 08094345-4714196 | 115.1 ± 0.2 | 4838 ± 77 | 2.56 ± 0.20 | 1.014 ± 0.005 | -0.24 ± 0.24 | <20 | 3 | … | … | … | … | … | … | G |
| 8780 | 08065067-4707056 | 110.2 ± 0.2 | 3454 ± 151 | … | … | … | … | … | … | … | … | … | … | … | … |
| 9374 | 08094065-4721466 | 17.3 ± 0.3 | 3583 ± 73 | … | 0.873 ± 0.006 | -0.21 ± 0.14 | … | … | … | … | … | … | … | … | … |
| 9375 | 08094068-4723539 | 37.4 ± 86.8 | … | … | … | … | … | … | … | … | … | … | … | … | … |
| 9382 | 08094380-4741233 | -20.9 ± 0.3 | 3548 ± 33 | … | 0.820 ± 0.007 | -0.23 ± 0.15 | <100 | 3 | … | … | … | … | … | … | … |
| 8781 | 08065080-4740222 | 1.8 ± 0.2 | 4902 ± 182 | … | 1.014 ± 0.003 | -0.08 ± 0.14 | <21 | 3 | … | … | … | … | … | … | G |
| 9133 | 08083414-4703445 | 56.0 ± 0.2 | 4226 ± 343 | 4.44 ± 0.12 | 0.876 ± 0.004 | -0.09 ± 0.13 | <22 | 3 | … | … | … | … | … | … | … |
| 3482 | 08094390-4703094 | 25.6 ± 0.4 | 7029 ± 202 | 4.14 ± 0.16 | … | -0.16 ± 0.15 | <12 | 3 | … | … | N | … | … | n | NG |
| 9376 | 08094097-4726411 | 19.0 ± 0.3 | 3411 ± 34 | … | 0.852 ± 0.010 | -0.25 ± 0.14 | 296 ± 7 | 1 | B | Y | … | Y | Y | Y | … |
| 8782 | 08065119-4716538 | -25.0 ± 0.2 | 6216 ± 38 | 3.84 ± 0.08 | 1.006 ± 0.002 | -0.16 ± 0.03 | 49 ± 9 | 1 | … | … | … | … | … | … | … |
| 9147 | 08083846-4737039 | 0.9 ± 0.3 | 3779 ± 57 | 4.60 ± 0.10 | 0.805 ± 0.007 | -0.19 ± 0.13 | … | … | … | … | … | … | … | … | … |
| 9383 | 08094434-4741370 | 152.2 ± 0.2 | 4448 ± 203 | … | 1.038 ± 0.003 | -0.25 ± 0.16 | <17 | 3 | … | … | … | … | … | … | G |
| 8783 | 08065206-4728486 | 22.3 ± 0.6 | … | … | … | … | … | … | … | … | … | … | … | … | … |
| 9148 | 08083847-4711230 | 19.5 ± 1.5 | 3063 ± 108 | … | 0.832 ± 0.018 | … | 747 ± 30 | 1 | B | … | … | … | Y | Y | … |
| 9149 | 08083854-4728208 | -655.2 ± 5.2 | … | … | … | … | … | … | … | … | … | … | … | … | … |
| 8784 | 08065218-4701410 | 31.1 ± 0.2 | 5885 ± 59 | 4.20 ± 0.14 | 0.992 ± 0.002 | -0.13 ± 0.18 | 27 ± 2 | 1 | … | … | … | … | … | … | … |
| 9377 | 08094171-4726420 | 16.6 ± 0.6 | 3317 ± 18 | … | 0.897 ± 0.018 | … | 690 ± 65 | 1 | A | … | … | … | Y | Y | … |
| 9150 | 08083873-4734584 | 47.6 ± 0.2 | 5003 ± 140 | … | 1.021 ± 0.003 | -0.10 ± 0.12 | 7 ± 2 | 1 | … | … | … | … | … | … | … |
| 8785 | 08065228-4712370 | 33.1 ± 0.2 | 3483 ± 121 | 4.60 ± 0.11 | 0.807 ± 0.008 | -0.15 ± 0.11 | … | … | … | … | … | … | Y | … | … |
| 9378 | 08094182-4703145 | 34.9 ± 0.2 | 4541 ± 63 | … | 1.027 ± 0.003 | -0.13 ± 0.10 | <25 | 3 | … | … | … | … | … | … | G |
| 9415 | 08095080-4740450 | 20.0 ± 0.2 | 3409 ± 14 | … | 0.865 ± 0.011 | -0.25 ± 0.15 | 506 ± 22 | 1 | B | … | … | Y | Y | Y | … |
| 9379 | 08094199-4703317 | 17.0 ± 0.2 | 3935 ± 175 | … | 0.869 ± 0.003 | -0.01 ± 0.15 | 469 ± 4 | 1 | A | Y | … | Y | Y | Y | … |
| 8786 | 08065248-4713522 | 100.4 ± 0.2 | 4373 ± 234 | … | 1.039 ± 0.002 | -0.29 ± 0.23 | … | … | … | … | … | … | … | … | … |
| 9151 | 08083966-4717093 | 123.5 ± 0.2 | 4682 ± 35 | … | 1.020 ± 0.003 | -0.31 ± 0.11 | … | … | … | … | … | … | … | … | G |
| 8787 | 08065330-4745598 | 28.7 ± 0.4 | 3838 ± 167 | 4.63 ± 0.05 | 0.794 ± 0.015 | -0.16 ± 0.13 | <47 | 3 | … | … | … | … | … | … | … |
| 9380 | 08094250-4706583 | 160.5 ± 0.2 | 4549 ± 124 | … | 1.054 ± 0.003 | … | … | … | … | … | … | … | … | … | G |
| 9416 | 08095155-4657029 | 127.4 ± 0.2 | 4531 ± 102 | … | 1.037 ± 0.006 | -0.27 ± 0.18 | … | … | … | … | … | … | … | … | G |
| 8788 | 08065358-4723040 | 18.7 ± 0.2 | 5453 ± 10 | … | 1.000 ± 0.001 | -0.08 ± 0.11 | 37 ± 4 | 1 | … | … | … | … | … | … | … |
| 3466 | 08083990-4741513 | 16.4 ± 0.6 | 5620 ± 31 | 4.18 ± 0.10 | 0.982 ± 0.002 | -0.36 ± 0.01 | <9 | 3 | … | … | N | … | … | n | … |
| 9417 | 08095190-4656394 | 54.4 ± 0.3 | 3598 ± 40 | … | 0.835 ± 0.008 | -0.21 ± 0.14 | … | … | … | … | … | … | … | … | … |
| 8789 | 08065376-4658294 | 26.3 ± 0.2 | 4710 ± 92 | 2.56 ± 0.08 | 1.011 ± 0.003 | 0.00 ± 0.13 | <32 | 3 | … | … | … | … | … | … | G |
| 9152 | 08084003-4729476 | 18.0 ± 0.4 | 3347 ± 34 | … | 0.892 ± 0.013 | -0.26 ± 0.14 | 530 ± 19 | 1 | A | … | … | … | Y | Y | … |
| 9418 | 08095203-4715417 | 64.6 ± 0.3 | 3933 ± 133 | … | 0.853 ± 0.006 | -0.30 ± 0.24 | <27 | 3 | … | … | … | … | … | … | … |
| 9638 | 08104633-4723563 | 82.7 ± 0.3 | 5505 ± 157 | 4.50 ± 0.06 | 0.978 ± 0.007 | -0.11 ± 0.09 | <14 | 3 | … | … | … | … | … | … | … |
| 9639 | 08104649-4742216 | 19.7 ± 0.2 | 3961 ± 151 | … | 0.871 ± 0.004 | -0.06 ± 0.10 | 472 ± 12 | 1 | B | Y | … | Y | Y | Y | … |
| 8790 | 08065411-4719194 | 65.9 ± 0.2 | 4701 ± 22 | 2.50 ± 0.12 | 1.022 ± 0.003 | -0.06 ± 0.10 | <24 | 3 | … | … | … | … | … | … | … |
| 9153 | 08084029-4727265 | 21.9 ± 0.2 | 5793 ± 101 | 4.28 ± 0.09 | 0.990 ± 0.001 | 0.13 ± 0.03 | 63 ± 1 | 1 | … | … | … | … | … | … | … |
| 9640 | 08104712-4706030 | 102.5 ± 0.2 | 4705 ± 133 | … | 1.092 ± 0.002 | … | 44 ± 7 | 1 | … | … | … | … | … | … | G |
| 9654 | 08105161-4721076 | 66.8 ± 0.2 | 4959 ± 141 | … | 1.012 ± 0.003 | -0.19 ± 0.12 | <14 | 3 | … | … | … | … | … | … | … |
| 8791 | 08065420-4735127 | 144.3 ± 0.2 | 4722 ± 26 | 2.33 ± 0.19 | 1.021 ± 0.004 | -0.38 ± 0.13 | … | … | … | … | … | … | … | … | G |
| 9419 | 08095241-4735486 | 41.4 ± 0.2 | 5032 ± 99 | … | 1.018 ± 0.004 | -0.07 ± 0.07 | <22 | 3 | … | … | … | … | … | … | G |
| 3498 | 08105180-4746394 | 7.8 ± 0.6 | 4974 ± 34 | 2.70 ± 0.09 | 1.015 ± 0.002 | -0.11 ± 0.04 | <18 | 3 | … | … | N | … | … | n | G |
| 8837 | 08070652-4719207 | 88.2 ± 0.2 | 4620 ± 160 | … | 1.025 ± 0.007 | -0.24 ± 0.23 | <26 | 3 | … | … | … | … | … | … | G |
| 9420 | 08095247-4745501 | 60.1 ± 0.3 | 4269 ± 288 | 4.58 ± 0.10 | 0.869 ± 0.010 | -0.13 ± 0.02 | … | … | … | … | … | … | … | … | … |
| 9154 | 08084136-4700373 | 46.9 ± 0.3 | 3554 ± 48 | … | 0.816 ± 0.011 | -0.22 ± 0.15 | … | … | … | … | … | … | … | … | … |
| 8838 | 08070678-4710137 | 84.8 ± 0.2 | 4707 ± 106 | … | 0.996 ± 0.008 | … | … | … | … | … | … | … | … | … | … |
| 9421 | 08095250-4742526 | 294.3 ± 0.2 | 5132 ± 175 | … | 1.016 ± 0.004 | -0.83 ± 0.18 | … | … | … | … | … | … | … | … | G |

Gutiérrez Albarrán et al.: Calibrating the lithium–age relation I.

**Table C.3.** continued.

| ID | CNAME | $RV$ (km s$^{-1}$) | $T_\text{eff}$ (K) | $\log g$ (dex) | $\gamma^a$ | [Fe/H] (dex) | $EW(\text{Li})^b$ (mÅ) | $EW(\text{Li})$ error flag$^c$ | Jeffries 2014$^d$ | Damiani 2014$^d$ | Literature members Spina 2014$^d$ | Frasca 2015$^d$ | Prisinzano 2016$^d$ | Final members$^e$ | Non-mem with Li$^f$ |
|---|---|---|---|---|---|---|---|---|---|---|---|---|---|---|---|
| 9655 | 08105267-4704157 | 19.5 ± 0.5 | 3335 ± 19 | ... | 0.862 ± 0.016 | -0.26 ± 0.14 | 637 ± 25 | 1 | B | Y | ... | ... | Y | Y | ... |
| 3439 | 08070717-4721463 | 36.8 ± 0.6 | 5283 ± 24 | 2.79 ± 0.07 | ... | 0.08 ± 0.04 | 49 ± 10 | 2 | ... | ... | N | ... | ... | n | ... |
| 9422 | 08095265-4717121 | 18.1 ± 0.2 | 4400 ± 71 | 3.87 ± 0.09 | 0.955 ± 0.003 | -0.10 ± 0.22 | 418 ± 7 | 1 | A | ... | ... | Y | Y | Y | ... |
| 9155 | 08084142-4730035 | 105.3 ± 0.2 | 4524 ± 23 | ... | 1.029 ± 0.003 | -0.30 ± 0.18 | ... | ... | ... | ... | ... | ... | ... | ... | G |
| 9656 | 08105268-4706451 | 72.5 ± 1.0 | 3485 ± 33 | ... | 0.807 ± 0.017 | ... | ... | ... | ... | ... | ... | ... | ... | ... | ... |
| 8839 | 08070778-4743451 | 11.7 ± 0.5 | 3439 ± 33 | ... | 0.811 ± 0.019 | ... | ... | ... | ... | ... | ... | ... | ... | ... | ... |
| 9423 | 08095284-4659518 | 32.4 ± 0.2 | 4980 ± 144 | ... | 1.008 ± 0.005 | -0.07 ± 0.13 | ... | ... | ... | ... | ... | ... | ... | ... | ... |
| 9156 | 08084197-4738293 | 56.7 ± 0.2 | 4793 ± 49 | 2.66 ± 0.06 | 1.011 ± 0.004 | -0.02 ± 0.07 | <31 | 3 | ... | ... | ... | ... | ... | ... | G |
| 8840 | 08070787-4745273 | 57.1 ± 0.2 | 4627 ± 88 | ... | 1.028 ± 0.003 | -0.19 ± 0.13 | <21 | 3 | ... | ... | ... | ... | ... | ... | G |
| 9157 | 08084223-4726299 | 17.1 ± 0.2 | 4994 ± 24 | ... | 1.001 ± 0.003 | 0.07 ± 0.05 | 27 ± 5 | 1 | ... | ... | ... | ... | ... | ... | ... |
| 8841 | 08070788-4702206 | -0.1 ± 0.2 | ... | ... | ... | ... | ... | ... | ... | ... | ... | ... | Y | ... | ... |
| 9424 | 08095301-4725267 | 104.4 ± 0.2 | 5029 ± 124 | ... | 1.011 ± 0.003 | -0.18 ± 0.21 | <16 | 3 | ... | ... | ... | ... | ... | ... | G |
| 9657 | 08105273-4735100 | -48.4 ± 0.2 | 5555 ± 10 | ... | 1.000 ± 0.002 | -0.52 ± 0.04 | 34 ± 3 | 1 | ... | ... | ... | ... | ... | ... | ... |
| 8842 | 08070803-4714331 | 85.6 ± 0.8 | ... | ... | ... | ... | ... | ... | ... | ... | ... | ... | ... | ... | ... |
| 9425 | 08095320-4746264 | 60.2 ± 0.2 | 4663 ± 77 | ... | 1.014 ± 0.005 | ... | <47 | 3 | ... | ... | ... | ... | ... | ... | G |
| 9158 | 08084242-4716492 | 86.0 ± 0.2 | 4501 ± 143 | ... | 1.040 ± 0.003 | -0.17 ± 0.02 | <25 | 3 | ... | ... | ... | ... | ... | ... | G |
| 8843 | 08070844-4712290 | 22.1 ± 0.2 | 5992 ± 164 | 4.13 ± 0.18 | 1.002 ± 0.002 | 0.27 ± 0.12 | <14 | 3 | ... | ... | ... | ... | ... | ... | ... |
| 9426 | 08095348-4734441 | 8.6 ± 0.2 | 6154 ± 116 | 4.18 ± 0.15 | 0.995 ± 0.002 | 0.03 ± 0.07 | 44 ± 1 | 1 | ... | ... | ... | ... | ... | ... | ... |
| 9159 | 08084259-4711096 | 35.1 ± 0.2 | 5510 ± 125 | 4.32 ± 0.11 | 0.977 ± 0.003 | 0.04 ± 0.04 | 203 ± 12 | 1 | ... | ... | ... | ... | ... | ... | ... |
| 8844 | 08070846-4725464 | 103.9 ± 0.2 | 4638 ± 98 | ... | 1.017 ± 0.004 | -0.30 ± 0.09 | 32 ± 3 | 1 | ... | ... | ... | ... | ... | ... | G |
| 9427 | 08095355-4707554 | 62.8 ± 0.2 | 4901 ± 190 | ... | 1.017 ± 0.003 | -0.10 ± 0.11 | <28 | 3 | ... | ... | ... | ... | ... | ... | G |
| 8845 | 08070855-4719165 | 116.5 ± 0.4 | 3601 ± 73 | 4.73 ± 0.13 | 0.776 ± 0.016 | -0.26 ± 0.15 | <100 | 3 | ... | ... | ... | ... | ... | ... | ... |
| 9658 | 08105364-4743233 | 13.2 ± 0.2 | 5182 ± 194 | ... | 0.980 ± 0.003 | 0.03 ± 0.08 | 255 ± 3 | 1 | ... | Y | ... | ... | ... | Y | ... |
| 9160 | 08084304-4654513 | 30.7 ± 0.2 | 5984 ± 136 | 4.21 ± 0.05 | 0.995 ± 0.002 | 0.25 ± 0.09 | <18 | 3 | ... | ... | ... | ... | ... | ... | ... |
| 8846 | 08070929-4742009 | 11.0 ± 0.2 | 5408 ± 55 | 4.22 ± 0.21 | 0.981 ± 0.003 | 0.09 ± 0.01 | 94 ± 2 | 1 | ... | ... | ... | ... | ... | ... | ... |
| 9659 | 08105365-4725088 | 12.2 ± 0.2 | 3469 ± 137 | ... | 0.864 ± 0.004 | -0.23 ± 0.14 | 510 ± 18 | 1 | B | Y | ... | Y | Y | Y | ... |
| 9428 | 08095370-4716085 | 16.3 ± 0.2 | 3640 ± 13 | ... | 0.858 ± 0.003 | -0.19 ± 0.13 | ... | ... | A | Y | ... | Y | Y | ... | ... |
| 8847 | 08070943-4702194 | 138.8 ± 0.2 | 4539 ± 40 | ... | 1.030 ± 0.004 | -0.32 ± 0.14 | ... | ... | ... | ... | ... | ... | ... | ... | G |
| 9429 | 08095396-4729589 | 79.3 ± 0.2 | 4685 ± 43 | 2.49 ± 0.13 | 1.021 ± 0.003 | -0.06 ± 0.01 | <23 | 3 | ... | ... | ... | ... | ... | ... | G |
| 9660 | 08105380-4722425 | 93.9 ± 0.2 | 4543 ± 113 | ... | 1.052 ± 0.002 | ... | 32 ± 12 | 1 | ... | ... | ... | ... | ... | ... | G |
| 9161 | 08084346-4654174 | 115.2 ± 0.2 | 3695 ± 95 | ... | ... | ... | ... | ... | ... | ... | ... | ... | ... | ... | ... |
| 8848 | 08070988-4707448 | 115.5 ± 0.2 | 4712 ± 14 | 2.49 ± 0.01 | 1.014 ± 0.003 | -0.07 ± 0.05 | <37 | 3 | ... | ... | ... | ... | ... | ... | G |
| 3485 | 08095427-4721419 | 20.9 ± 0.6 | 5884 ± 147 | 4.45 ± 0.10 | ... | 0.10 ± 0.03 | 234 ± 11 | 2 | ... | ... | Y | Y | ... | Y | ... |
| 9661 | 08105418-4730347 | 46.8 ± 0.2 | 4888 ± 18 | 2.79 ± 0.18 | 1.019 ± 0.002 | -0.07 ± 0.09 | ... | ... | ... | ... | ... | ... | ... | ... | ... |
| 9198 | 08085256-4746446 | 70.9 ± 0.2 | 4977 ± 177 | 4.36 ± 0.02 | 0.964 ± 0.003 | -0.18 ± 0.01 | <16 | 3 | ... | ... | ... | ... | ... | ... | ... |
| 8849 | 08070996-4746574 | -8.9 ± 0.2 | 5430 ± 104 | ... | 1.004 ± 0.003 | -0.10 ± 0.03 | <21 | 3 | ... | ... | ... | ... | ... | ... | ... |
| 3486 | 08095429-4710344 | 38.5 ± 0.6 | 5005 ± 32 | 2.58 ± 0.12 | 1.013 ± 0.001 | -0.27 ± 0.02 | 16 ± 4 | 1 | ... | ... | N | ... | ... | n | G |
| 9662 | 08105511-4731510 | 42.7 ± 0.2 | 4837 ± 62 | 2.62 ± 0.03 | 1.014 ± 0.002 | -0.01 ± 0.07 | <29 | 3 | ... | ... | ... | ... | ... | ... | G |
| 9199 | 08085268-4710381 | 20.1 ± 0.3 | 7031 ± 40 | ... | 1.011 ± 0.002 | ... | <6 | 3 | ... | ... | ... | ... | ... | ... | ... |
| 8850 | 08071027-4656352 | 71.4 ± 0.2 | 4959 ± 63 | ... | 1.017 ± 0.004 | -0.13 ± 0.13 | <17 | 3 | ... | ... | ... | ... | ... | ... | G |
| 3487 | 08095783-4701385 | 26.0 ± 0.6 | 4965 ± 6 | 2.43 ± 0.03 | 1.018 ± 0.002 | -0.15 ± 0.07 | 375 ± 6 | 2 | ... | ... | N | ... | ... | n | Li-rich G |
| 8851 | 08071048-4730387 | 69.1 ± 0.2 | 3169 ± 5 | ... | 1.072 ± 0.001 | ... | ... | ... | ... | ... | ... | ... | ... | ... | G |
| 8852 | 08071075-4713342 | 137.9 ± 0.2 | 4374 ± 289 | ... | 1.060 ± 0.003 | -0.21 ± 0.07 | 36 ± 2 | 1 | ... | ... | ... | ... | ... | ... | G |
| 9663 | 08105525-4733439 | 125.1 ± 0.2 | 4532 ± 121 | ... | 1.030 ± 0.006 | -0.17 ± 0.10 | ... | ... | ... | ... | ... | ... | ... | ... | G |
| 9200 | 08085276-4745146 | 68.6 ± 0.2 | 4979 ± 18 | 2.95 ± 0.12 | 1.016 ± 0.003 | -0.03 ± 0.01 | <24 | 3 | ... | ... | ... | ... | ... | ... | G |
| 9444 | 08095786-4720085 | 17.5 ± 0.2 | 4086 ± 172 | ... | 0.891 ± 0.002 | -0.07 ± 0.09 | 459 ± 4 | 1 | A | Y | ... | Y | Y | Y | ... |
| 8853 | 08071087-4721535 | 97.2 ± 0.2 | 4642 ± 92 | ... | 1.020 ± 0.003 | -0.36 ± 0.20 | <20 | 3 | ... | ... | ... | ... | ... | ... | G |
| 3467 | 08085306-4704067 | 22.8 ± 0.6 | 6014 ± 1 | 4.30 ± 0.04 | 0.989 ± 0.001 | 0.08 ± 0.01 | 73 ± 1 | 2 | ... | ... | N | ... | ... | n | NG |
| 8854 | 08071090-4739487 | 46.0 ± 0.3 | 5996 ± 199 | 4.23 ± 0.14 | 0.998 ± 0.007 | -0.13 ± 0.08 | <19 | 3 | ... | ... | ... | ... | ... | ... | ... |
| 9445 | 08095807-4737443 | 22.0 ± 0.2 | 3612 ± 36 | ... | 0.846 ± 0.005 | -0.21 ± 0.12 | 424 ± 5 | 1 | B | Y | ... | Y | Y | Y | ... |
| 9201 | 08085320-4712385 | 26.8 ± 0.4 | 3545 ± 28 | ... | 0.814 ± 0.015 | ... | <100 | 3 | ... | ... | ... | ... | ... | ... | ... |
| 8855 | 08071112-4715003 | 165.4 ± 0.2 | 3289 ± 86 | ... | ... | ... | 104 ± 18 | 1 | ... | ... | ... | ... | ... | ... | ... |
| 9446 | 08095822-4703093 | 106.1 ± 0.3 | 4638 ± 318 | ... | 1.027 ± 0.008 | -0.19 ± 0.24 | ... | ... | ... | ... | ... | ... | ... | ... | ... |
| 9202 | 08085391-4715075 | 17.2 ± 0.2 | 3918 ± 212 | ... | 0.858 ± 0.004 | -0.13 ± 0.12 | 505 ± 15 | 1 | A | Y | ... | Y | Y | Y | ... |
| 8856 | 08071142-4657325 | -10.2 ± 0.2 | 4727 ± 14 | 2.66 ± 0.10 | 1.007 ± 0.003 | -0.06 ± 0.04 | <28 | 3 | ... | ... | ... | ... | ... | ... | ... |
| 9447 | 08095823-4658198 | 9.0 ± 0.2 | 4722 ± 271 | ... | 1.021 ± 0.004 | -0.27 ± 0.10 | <14 | 3 | ... | ... | ... | ... | ... | ... | G |
| 9203 | 08085400-4717236 | 17.4 ± 0.2 | 4019 ± 117 | ... | 0.874 ± 0.003 | -0.06 ± 0.09 | 480 ± 10 | 1 | A | Y | ... | Y | Y | Y | ... |
| 9664 | 08105557-4724264 | 25.3 ± 0.2 | 4829 ± 35 | ... | 0.963 ± 0.003 | 0.09 ± 0.02 | <25 | 3 | ... | ... | ... | ... | ... | ... | ... |
| 8857 | 08071155-4719512 | 17.6 ± 0.2 | 4229 ± 161 | ... | 0.921 ± 0.003 | -0.05 ± 0.12 | 490 ± 16 | 1 | A | Y | ... | Y | Y | Y | ... |

| ID | CNAME | RV (km s$^{-1}$) | $T_{\text{eff}}$ (K) | logg (dex) | $\gamma^a$ | [Fe/H] (dex) | EW(Li)$^b$ (mÅ) | EW(Li) error flag$^c$ | Jeffries 2014$^d$ | Damiani 2014$^d$ | Literature members Spina 2014$^d$ | Frasca 2015$^d$ | Prisinzano 2016$^d$ | Final members$^e$ | Non-mem with Li$^f$ |
|---|---|---|---|---|---|---|---|---|---|---|---|---|---|---|---|
| 9448 | 08095842-4715483 | 17.0 ± 0.2 | 3840 ± 114 | ... | 0.862 ± 0.003 | -0.13 ± 0.12 | 431 ± 5 | 1 | A | Y | ... | Y | Y | Y | ... |
| 9204 | 08085418-4740417 | 90.6 ± 0.2 | 4583 ± 93 | 1.92 ± 0.15 | 1.029 ± 0.003 | -0.42 ± 0.14 | <13 | 3 | ... | ... | ... | ... | ... | ... | G |
| 8858 | 08071165-4716106 | 56.1 ± 0.2 | 4650 ± 152 | 2.39 ± 0.18 | 1.018 ± 0.004 | -0.07 ± 0.08 | <22 | 3 | ... | ... | ... | ... | ... | ... | G |
| 9665 | 08105564-4706527 | 19.0 ± 0.3 | 3399 ± 104 | 4.67 ± 0.06 | 0.838 ± 0.008 | -0.25 ± 0.16 | <100 | 3 | ... | ... | ... | ... | ... | ... | ... |
| 9449 | 08095903-4715230 | 17.6 ± 0.2 | 3753 ± 67 | ... | 0.855 ± 0.004 | -0.19 ± 0.13 | 369 ± 5 | 1 | A | Y | ... | Y | Y | Y | ... |
| 8859 | 08071205-4742055 | 80.5 ± 0.2 | 4684 ± 139 | ... | 1.016 ± 0.003 | -0.20 ± 0.11 | 372 ± 6 | 1 | ... | ... | ... | ... | ... | ... | Li-rich G |
| 9666 | 08105574-4704580 | 68.5 ± 0.2 | 4909 ± 194 | ... | 1.014 ± 0.003 | -0.15 ± 0.19 | <15 | 3 | ... | ... | ... | ... | ... | ... | G |
| 9205 | 08085439-4725176 | 99.9 ± 0.2 | 4806 ± 111 | ... | 1.022 ± 0.004 | -0.19 ± 0.09 | 39 ± 6 | 1 | ... | ... | ... | ... | ... | ... | G |
| 9450 | 08095922-4716215 | 17.3 ± 0.2 | 4050 ± 222 | ... | 0.887 ± 0.003 | -0.05 ± 0.12 | 455 ± 12 | 1 | A | Y | ... | Y | Y | Y | ... |
| 8881 | 08071762-4702538 | -4.3 ± 0.3 | 3349 ± 11 | 4.89 ± 0.18 | 0.803 ± 0.009 | -0.25 ± 0.15 | ... | ... | ... | ... | ... | ... | ... | ... | ... |
| 9206 | 08085439-4734352 | 97.9 ± 0.2 | 4664 ± 58 | ... | 1.027 ± 0.004 | -0.37 ± 0.22 | <20 | 3 | ... | ... | ... | ... | ... | ... | G |
| 9667 | 08105577-4718066 | 15.9 ± 0.2 | 3306 ± 19 | ... | 0.881 ± 0.010 | -0.20 ± 0.14 | 634 ± 27 | 1 | A | Y | ... | Y | Y | Y | ... |
| 8882 | 08071788-4724040 | 42.6 ± 0.2 | 4914 ± 117 | ... | 1.012 ± 0.004 | -0.17 ± 0.14 | <20 | 3 | ... | ... | ... | ... | ... | ... | G |
| 9451 | 08095939-4657470 | 37.7 ± 0.2 | 4593 ± 154 | ... | 1.024 ± 0.002 | 0.07 ± 0.04 | <36 | 3 | ... | ... | ... | ... | ... | ... | G |
| 8883 | 08071870-4653568 | 26.4 ± 0.2 | 4532 ± 148 | ... | 0.959 ± 0.003 | -0.02 ± 0.05 | 107 ± 4 | 1 | ... | ... | ... | ... | ... | ... | ... |
| 3502 | 08110285-4724405 | 17.8 ± 0.6 | 5233 ± 81 | 4.47 ± 0.15 | 0.985 ± 0.002 | -0.03 ± 0.12 | 392 ± 3 | 2 | B | Y | Y | Y | Y | Y | ... |
| 8884 | 08071906-4747018 | 20.8 ± 0.9 | 3409 ± 140 | ... | 0.863 ± 0.014 | -0.27 ± 0.14 | <100 | 3 | ... | ... | ... | ... | ... | ... | ... |
| 9452 | 08095956-4745567 | -20.5 ± 0.3 | 3492 ± 58 | ... | 0.855 ± 0.015 | -0.22 ± 0.15 | <100 | 3 | ... | ... | ... | ... | ... | ... | ... |
| 3468 | 08085455-4700053 | 51.8 ± 0.6 | 4961 ± 28 | 2.38 ± 0.05 | ... | -0.49 ± 0.02 | <11 | 3 | ... | ... | N | ... | ... | n | ... |
| 9695 | 08110319-4652394 | 28.8 ± 0.3 | 4096 ± 238 | 4.58 ± 0.02 | 0.844 ± 0.007 | 0.07 ± 0.26 | ... | ... | ... | ... | ... | ... | ... | ... | ... |
| 8885 | 08071923-4701540 | 61.8 ± 0.7 | 3472 ± 113 | ... | 0.849 ± 0.016 | -0.22 ± 0.15 | ... | ... | ... | ... | ... | ... | ... | ... | ... |
| 9453 | 08095964-4714441 | 52.1 ± 0.2 | 4991 ± 62 | 3.18 ± 0.05 | 1.005 ± 0.003 | -0.08 ± 0.10 | <29 | 3 | ... | ... | ... | ... | ... | ... | ... |
| 3443 | 08071937-4710143 | 50.3 ± 0.6 | 4581 ± 29 | 2.37 ± 0.11 | 1.024 ± 0.002 | 0.02 ± 0.05 | 32 ± 8 | 1 | ... | ... | N | ... | ... | n | ... |
| 3488 | 08095967-4726048 | 11.6 ± 0.4 | 5214 ± 117 | 4.34 ± 0.27 | ... | 0.00 ± 0.13 | 154 ± 29 | 2 | ... | ... | Y | Y | Y | Y | ... |
| 9207 | 08085468-4745256 | -3.2 ± 0.3 | 6549 ± 43 | 3.97 ± 0.09 | 1.008 ± 0.002 | -0.09 ± 0.04 | <19 | 3 | ... | ... | ... | ... | ... | ... | ... |
| 9454 | 08095986-4654056 | 19.2 ± 0.3 | 3862 ± 167 | ... | 0.855 ± 0.008 | -0.14 ± 0.12 | 433 ± 18 | 1 | B | ... | ... | Y | Y | Y | ... |
| 8886 | 08071951-4654400 | 25.1 ± 0.2 | 7191 ± 37 | ... | 1.017 ± 0.002 | ... | ... | ... | ... | ... | ... | ... | ... | ... | ... |
| 9208 | 08085492-4702448 | 207.4 ± 0.2 | 4537 ± 123 | ... | 1.051 ± 0.003 | ... | ... | ... | ... | ... | ... | ... | ... | ... | ... |
| 9696 | 08110328-4716357 | 13.1 ± 0.3 | 4094 ± 523 | ... | 0.876 ± 0.003 | -0.19 ± 0.14 | 507 ± 24 | 1 | B | Y | ... | Y | Y | Y | ... |
| 9455 | 08100002-4713309 | 22.2 ± 0.2 | 4812 ± 68 | 2.47 ± 0.06 | 1.013 ± 0.001 | -0.12 ± 0.02 | ... | ... | ... | ... | ... | ... | ... | ... | ... |
| 9209 | 08085509-4708362 | 112.5 ± 0.2 | 4233 ± 214 | ... | 1.043 ± 0.003 | -0.15 ± 0.12 | 37 ± 7 | 1 | ... | ... | ... | ... | ... | ... | G |
| 8887 | 08071976-4728499 | 120.6 ± 0.2 | 4396 ± 202 | ... | 1.048 ± 0.002 | -0.26 ± 0.18 | ... | ... | ... | ... | ... | ... | ... | ... | G |
| 9697 | 08110335-4742214 | 125.1 ± 0.2 | 4356 ± 191 | ... | 1.038 ± 0.003 | -0.26 ± 0.15 | <28 | 3 | ... | ... | ... | ... | ... | ... | G |
| 9456 | 08100014-4723395 | 28.4 ± 159.4 | ... | ... | ... | ... | ... | ... | ... | ... | ... | ... | ... | ... | ... |
| 8888 | 08072080-4713598 | 58.0 ± 0.2 | 4818 ± 155 | ... | 1.018 ± 0.004 | -0.19 ± 0.14 | ... | ... | ... | ... | ... | ... | ... | ... | G |
| 9698 | 08110370-4712592 | 30.1 ± 0.3 | 6387 ± 48 | ... | 0.985 ± 0.003 | 0.46 ± 0.03 | ... | ... | ... | ... | ... | ... | ... | ... | ... |
| 9457 | 08100014-4731317 | 73.5 ± 0.2 | 5019 ± 55 | ... | 1.013 ± 0.005 | -0.35 ± 0.11 | <12 | 3 | ... | ... | ... | ... | ... | ... | G |
| 9210 | 08085585-4703140 | -277.6 ± 507.8 | ... | ... | ... | ... | ... | ... | ... | ... | ... | ... | ... | ... | ... |
| 8889 | 08072086-4746423 | 51.6 ± 0.2 | 4782 ± 57 | ... | 1.021 ± 0.003 | -0.12 ± 0.19 | <25 | 3 | ... | ... | ... | ... | ... | ... | G |
| 9490 | 08100658-4708524 | 59.8 ± 0.2 | 5061 ± 75 | 3.25 ± 0.20 | 1.009 ± 0.005 | -0.04 ± 0.02 | <29 | 3 | ... | ... | ... | ... | ... | ... | ... |
| 9699 | 08110375-4700324 | 32.1 ± 0.2 | 5402 ± 54 | 4.22 ± 0.08 | 0.981 ± 0.003 | 0.05 ± 0.01 | <32 | 3 | ... | ... | ... | ... | ... | ... | ... |
| 9211 | 08085592-4654421 | 20.3 ± 0.8 | 3523 ± 2 | ... | 0.869 ± 0.013 | -0.22 ± 0.15 | ... | ... | ... | ... | ... | ... | Y | ... | ... |
| 8890 | 08072119-4740003 | 15.7 ± 0.3 | ... | ... | ... | ... | ... | ... | ... | ... | ... | ... | ... | ... | ... |
| 9700 | 08110392-4729259 | 118.0 ± 0.2 | 4711 ± 14 | 2.59 ± 0.03 | 1.009 ± 0.003 | -0.13 ± 0.12 | <22 | 3 | ... | ... | ... | ... | ... | ... | G |
| 3469 | 08085599-4659333 | 28.4 ± 0.6 | 5136 ± 30 | 2.83 ± 0.08 | ... | -0.01 ± 0.03 | <13 | 3 | ... | ... | N | ... | ... | n | ... |
| 8891 | 08072123-4655177 | 124.7 ± 0.2 | 4408 ± 188 | ... | 1.050 ± 0.004 | -0.25 ± 0.16 | ... | ... | ... | ... | ... | ... | ... | ... | G |
| 9491 | 08100726-4729418 | 48.5 ± 0.7 | 3531 ± 30 | ... | 0.839 ± 0.017 | ... | ... | ... | ... | ... | ... | ... | ... | ... | ... |
| 9701 | 08110403-4658057 | 80.0 ± 0.2 | 4891 ± 149 | ... | 1.012 ± 0.004 | -0.14 ± 0.15 | 35 ± 5 | 1 | ... | ... | ... | ... | ... | ... | Li-rich G |
| 9212 | 08085604-4743319 | 21.8 ± 0.5 | 3276 ± 13 | ... | 0.887 ± 0.021 | ... | 690 ± 43 | 1 | B | ... | ... | ... | Y | Y | ... |
| 8892 | 08072198-4711230 | 21.8 ± 0.5 | 3377 ± 117 | ... | 0.863 ± 0.008 | -0.27 ± 0.18 | 408 ± 39 | 1 | B | Y | ... | Y | ... | Y | ... |
| 9492 | 08100729-4744407 | 16.8 ± 0.2 | 3543 ± 37 | ... | 0.842 ± 0.006 | -0.22 ± 0.14 | 408 ± 20 | 1 | A | Y | ... | Y | Y | Y | ... |
| 9226 | 08085938-4732131 | 47.7 ± 0.2 | 4579 ± 135 | ... | 1.065 ± 0.002 | ... | <21 | 3 | ... | ... | ... | ... | ... | ... | G |
| 8893 | 08072219-4711001 | 99.1 ± 0.2 | 4520 ± 32 | ... | 1.028 ± 0.004 | -0.22 ± 0.04 | 39 ± 8 | 1 | ... | ... | ... | ... | ... | ... | G |
| 9227 | 08085946-4709571 | 98.4 ± 0.2 | 3991 ± 5 | ... | 1.160 ± 0.002 | -0.03 ± 0.16 | ... | ... | ... | ... | ... | ... | ... | ... | G |
| 9702 | 08110441-4733029 | 69.3 ± 0.2 | 4684 ± 87 | 2.61 ± 0.03 | 1.008 ± 0.004 | -0.13 ± 0.04 | <23 | 3 | ... | ... | ... | ... | ... | ... | ... |
| 8894 | 08072225-4706599 | 43.1 ± 0.2 | 4816 ± 6 | 2.62 ± 0.03 | 1.015 ± 0.004 | -0.01 ± 0.08 | <36 | 3 | ... | ... | ... | ... | ... | ... | G |
| 9228 | 08085966-4718298 | 82.7 ± 0.2 | 4927 ± 105 | 2.93 ± 0.09 | 1.010 ± 0.004 | -0.15 ± 0.09 | <36 | 3 | ... | ... | ... | ... | ... | ... | ... |
| 9703 | 08110453-4734475 | 17.2 ± 0.4 | 3333 ± 23 | 4.74 ± 0.11 | 0.822 ± 0.006 | -0.29 ± 0.13 | ... | ... | ... | ... | ... | ... | Y | ... | ... |
| 8895 | 08072237-4719444 | -8.1 ± 0.2 | 6437 ± 25 | 4.06 ± 0.05 | 1.005 ± 0.001 | 0.12 ± 0.02 | <2 | 3 | ... | ... | ... | ... | ... | ... | ... |

**Table C.3.** continued.

| ID | CNAME | RV (km s$^{-1}$) | $T_{\rm eff}$ (K) | logg (dex) | $\gamma^a$ | [Fe/H] (dex) | EW(Li)$^b$ (mÅ) | EW(Li) error flag$^c$ | Jeffries 2014$^d$ | Damiani 2014$^d$ | Spina 2014$^d$ | Frasca 2015$^d$ | Prisinzano 2016$^d$ | Final members$^e$ | Non-mem with Li$^f$ |
|---|---|---|---|---|---|---|---|---|---|---|---|---|---|---|---|
| 9493 | 08100768-4725336 | 23.4 ± 0.2 | 5808 ± 46 | … | 1.002 ± 0.002 | -0.30 ± 0.12 | <15 | 3 | … | … | … | … | … | … | … |
| 9704 | 08110464-4710596 | 135.2 ± 1.4 | … | … | … | … | … | … | … | … | … | … | … | … | … |
| 8896 | 08072245-4700074 | 373.9 ± 0.2 | … | … | … | … | … | … | … | … | … | … | … | … | … |
| 9705 | 08110476-4654557 | 44.8 ± 0.2 | 5138 ± 102 | … | 1.014 ± 0.003 | … | 18 ± 4 | 1 | … | … | … | … | … | … | G |
| 8897 | 08072292-4655555 | 30.1 ± 0.2 | 4342 ± 460 | … | … | 0.00 ± 0.12 | … | … | … | Y | … | … | Y | … | … |
| 8898 | 08072340-4741482 | 82.8 ± 0.2 | 4659 ± 54 | … | 1.036 ± 0.004 | -0.19 ± 0.06 | 30 ± 7 | 1 | … | … | … | … | … | … | G |
| 9494 | 08100797-4656025 | 54.7 ± 0.2 | 4557 ± 142 | 2.32 ± 0.16 | 1.023 ± 0.004 | 0.12 ± 0.04 | <47 | 3 | … | … | … | … | … | … | G |
| 9230 | 08090034-4730260 | 15.7 ± 0.2 | 5000 ± 164 | … | 1.008 ± 0.003 | -0.06 ± 0.14 | <21 | 3 | … | … | … | … | … | … | … |
| 9706 | 08110504-4652299 | 53.2 ± 0.2 | 4887 ± 61 | 2.63 ± 0.10 | 1.020 ± 0.003 | -0.09 ± 0.07 | <24 | 3 | … | … | … | … | … | … | G |
| 8899 | 08072364-4731091 | 29.9 ± 0.2 | 6595 ± 23 | … | 0.994 ± 0.001 | 0.34 ± 0.02 | <2 | 3 | … | … | … | … | … | … | … |
| 9495 | 08100798-4745268 | 8.0 ± 0.2 | 4825 ± 117 | 2.61 ± 0.11 | 1.017 ± 0.004 | 0.01 ± 0.10 | <24 | 3 | … | … | … | … | … | … | G |
| 9231 | 08090035-4725410 | 94.5 ± 0.2 | 3900 ± 161 | … | 1.066 ± 0.003 | -0.20 ± 0.09 | … | … | … | … | … | … | … | … | G |
| 9707 | 08110519-4728308 | 56.9 ± 0.2 | 4571 ± 99 | 2.54 ± 0.08 | 1.008 ± 0.003 | 0.08 ± 0.06 | <37 | 3 | … | … | … | … | … | … | … |
| 9496 | 08100801-4722080 | 13.6 ± 0.2 | 6159 ± 98 | 4.15 ± 0.11 | 0.996 ± 0.002 | -0.02 ± 0.03 | … | … | … | … | … | … | … | … | … |
| 9232 | 08090039-4709300 | 122.4 ± 0.2 | 4776 ± 129 | … | 1.009 ± 0.003 | -0.23 ± 0.22 | … | … | … | … | … | … | … | … | … |
| 9497 | 08100804-4703141 | 176.8 ± 0.2 | 3939 ± 63 | … | 1.056 ± 0.003 | … | 211 ± 3 | 1 | … | … | … | … | … | … | G |
| 8900 | 08072426-4732493 | 25.7 ± 0.2 | 5273 ± 24 | … | 0.973 ± 0.003 | 0.02 ± 0.02 | <17 | 3 | … | … | … | … | … | … | … |
| 9708 | 08110549-4729304 | -15.0 ± 0.2 | 6631 ± 32 | … | … | -0.14 ± 0.03 | … | … | … | … | … | … | … | … | … |
| 8901 | 08072429-4718571 | 24.6 ± 0.4 | 3564 ± 56 | 4.61 ± 0.15 | 0.805 ± 0.014 | -0.22 ± 0.14 | <100 | 3 | … | … | … | … | … | … | … |
| 9233 | 08090060-4725023 | 198.2 ± 0.2 | 4562 ± 34 | … | 1.023 ± 0.007 | -0.27 ± 0.18 | <29 | 3 | … | … | … | … | … | … | G |
| 9709 | 08110585-4654105 | 27.5 ± 0.2 | 5611 ± 185 | 4.06 ± 0.10 | 0.990 ± 0.003 | 0.22 ± 0.08 | <20 | 3 | … | … | … | … | … | … | … |
| 9498 | 08100813-4717566 | 32.6 ± 0.2 | 4759 ± 18 | 2.58 ± 0.05 | 1.016 ± 0.003 | 0.04 ± 0.06 | <29 | 3 | … | … | … | … | … | … | G |
| 9234 | 08090115-4707376 | 134.7 ± 0.2 | 3616 ± 131 | … | … | … | … | … | … | … | … | … | … | … | … |
| 9499 | 08100833-4712458 | 0.9 ± 0.3 | 6574 ± 50 | … | 1.000 ± 0.003 | -0.03 ± 0.04 | … | … | … | … | … | … | … | … | … |
| 8938 | 08073402-4740513 | 21.3 ± 0.3 | 3362 ± 20 | … | 0.852 ± 0.012 | -0.27 ± 0.14 | … | … | B | Y | … | … | Y | Y | … |
| 9235 | 08090157-4717069 | 21.2 ± 0.2 | 3655 ± 73 | … | 0.854 ± 0.006 | -0.19 ± 0.13 | 539 ± 18 | 1 | B | Y | … | Y | Y | Y | … |
| 9500 | 08100835-4740002 | 14.6 ± 0.2 | 5024 ± 82 | 2.79 ± 0.02 | 1.011 ± 0.003 | -0.18 ± 0.08 | … | … | … | … | … | … | … | … | G |
| 8939 | 08073409-4700411 | 78.2 ± 0.2 | 4520 ± 171 | … | 1.017 ± 0.004 | 0.08 ± 0.10 | <55 | 3 | … | … | … | … | … | … | G |
| 9236 | 08090159-4709020 | 22.1 ± 0.5 | 3952 ± 282 | 4.49 ± 0.11 | 0.820 ± 0.015 | -0.19 ± 0.13 | <46 | 3 | … | … | … | … | … | … | … |
| 9501 | 08100859-4709118 | 16.6 ± 0.2 | 3793 ± 72 | … | 0.861 ± 0.003 | -0.17 ± 0.14 | 497 ± 18 | 1 | A | Y | … | Y | Y | Y | … |
| 8940 | 08073416-4720436 | 24.2 ± 0.6 | 3328 ± 14 | 4.92 ± 0.19 | 0.800 ± 0.011 | -0.29 ± 0.14 | … | … | … | … | … | … | Y | … | … |
| 9237 | 08090208-4734554 | 130.6 ± 0.2 | 4869 ± 143 | … | 1.021 ± 0.004 | -0.17 ± 0.21 | <21 | 3 | … | … | … | … | … | … | … |
| 9728 | 08111018-4741363 | 109.7 ± 0.2 | 4541 ± 113 | … | 1.045 ± 0.004 | … | … | … | … | … | … | … | … | … | G |
| 8941 | 08073417-4705351 | 20.0 ± 0.3 | 3452 ± 43 | … | 0.862 ± 0.008 | -0.23 ± 0.15 | … | … | … | … | … | Y | … | … | … |
| 9502 | 08100885-4655317 | 68.6 ± 0.3 | 3635 ± 38 | … | 0.846 ± 0.009 | -0.21 ± 0.14 | … | … | … | … | … | … | … | … | … |
| 8942 | 08073420-4725072 | 19.8 ± 0.2 | 3777 ± 76 | … | 0.874 ± 0.003 | -0.19 ± 0.13 | 547 ± 16 | 1 | B | … | … | … | Y | Y | … |
| 9238 | 08090253-4728341 | 66.5 ± 0.2 | 4975 ± 135 | … | 1.014 ± 0.003 | -0.22 ± 0.13 | <22 | 3 | … | … | … | … | … | … | G |
| 9729 | 08111020-4721112 | 8.5 ± 0.2 | 4581 ± 150 | … | 1.023 ± 0.003 | 0.03 ± 0.07 | <34 | 3 | … | … | … | … | … | … | G |
| 9503 | 08100905-4738100 | -10.2 ± 0.2 | 4674 ± 19 | 2.38 ± 0.08 | 1.025 ± 0.002 | -0.07 ± 0.03 | 12 ± 2 | 1 | … | … | … | … | … | … | G |
| 8943 | 08073442-4654016 | 18.1 ± 0.3 | 3328 ± 33 | … | 0.895 ± 0.010 | -0.26 ± 0.15 | 614 ± 9 | 1 | A | … | … | Y | Y | Y | … |
| 9514 | 08101337-4727547 | 46.8 ± 0.2 | 4331 ± 274 | … | 1.041 ± 0.001 | 0.05 ± 0.14 | <49 | 3 | … | … | … | … | … | … | G |
| 9239 | 08090351-4733270 | 49.3 ± 0.2 | 5515 ± 148 | … | 0.998 ± 0.003 | 0.08 ± 0.01 | 38 ± 10 | 1 | … | … | … | … | … | … | … |
| 9730 | 08111105-4703509 | 93.6 ± 0.2 | 4863 ± 25 | 2.67 ± 0.06 | 1.014 ± 0.005 | -0.12 ± 0.02 | <26 | 3 | … | … | … | … | … | … | G |
| 9515 | 08101369-4716536 | 19.2 ± 0.3 | 3316 ± 28 | … | 0.886 ± 0.014 | -0.25 ± 0.14 | 600 ± 12 | 1 | B | … | … | … | Y | Y | … |
| 3448 | 08073447-4716569 | 15.8 ± 0.6 | 6030 ± 27 | 4.30 ± 0.04 | … | 0.15 ± 0.03 | 56 ± 4 | 2 | … | … | N | … | … | n | NG |
| 9240 | 08090362-4713418 | 16.8 ± 0.5 | 3309 ± 6 | … | 0.862 ± 0.012 | -0.27 ± 0.14 | 579 ± 18 | 1 | A | … | … | … | Y | Y | … |
| 9731 | 08111106-4744307 | 23.3 ± 0.2 | 6367 ± 41 | 4.24 ± 0.09 | 0.998 ± 0.002 | -0.10 ± 0.03 | 37 ± 1 | 1 | … | … | … | … | … | … | … |
| 9241 | 08090379-4742157 | 17.3 ± 0.2 | 3782 ± 93 | … | 0.852 ± 0.004 | -0.17 ± 0.14 | 512 ± 16 | 1 | A | Y | … | Y | Y | Y | … |
| 9516 | 08101401-4721563 | 22.0 ± 1.6 | 3281 ± 33 | … | 0.754 ± 0.014 | -0.24 ± 0.15 | … | … | … | … | … | … | … | … | … |
| 9732 | 08111107-4735329 | 19.9 ± 0.4 | 3732 ± 90 | 4.59 ± 0.17 | 0.805 ± 0.016 | -0.19 ± 0.13 | <100 | 3 | … | … | … | … | … | … | … |
| 8944 | 08073467-4724355 | 95.0 ± 0.2 | 3991 ± 198 | … | 1.064 ± 0.003 | -0.26 ± 0.06 | <16 | 3 | … | … | … | … | … | … | G |
| 9517 | 08101443-4655059 | 115.8 ± 0.2 | 4309 ± 257 | … | 1.046 ± 0.004 | -0.25 ± 0.12 | 60 ± 8 | 1 | … | … | … | … | … | … | G |
| 8945 | 08073470-4726315 | 143.8 ± 0.2 | 4163 ± 317 | … | 1.048 ± 0.004 | -0.30 ± 0.17 | 69 ± 8 | 1 | … | … | … | … | … | … | G |
| 9733 | 08111130-4722129 | 35.2 ± 0.2 | 4606 ± 155 | … | 1.029 ± 0.002 | 0.00 ± 0.09 | 260 ± 5 | 1 | … | … | … | … | … | … | Li-rich G |
| 8946 | 08073487-4658279 | 0.3 ± 0.2 | 5150 ± 94 | … | 0.992 ± 0.003 | … | 54 ± 4 | 1 | … | … | … | … | … | … | … |
| 9518 | 08101482-4708279 | 16.9 ± 0.2 | 5175 ± 88 | … | 0.990 ± 0.002 | … | 365 ± 3 | 1 | A | … | … | Y | Y | Y | … |
| 8947 | 08073523-4731052 | 24.5 ± 0.2 | 5541 ± 37 | 4.32 ± 0.15 | 0.985 ± 0.003 | 0.10 ± 0.03 | 149 ± 2 | 1 | … | … | … | … | … | … | … |
| 9269 | 08091216-4656127 | 57.0 ± 0.2 | 4418 ± 227 | … | 1.027 ± 0.005 | 0.03 ± 0.10 | <42 | 3 | … | … | … | … | … | … | G |
| 9734 | 08111135-4655454 | 102.3 ± 0.2 | 4498 ± 83 | … | 1.044 ± 0.003 | -0.24 ± 0.12 | 37 ± 5 | 1 | … | … | … | … | … | … | G |

| ID | CNAME | $RV$ (km s$^{-1}$) | $T_{\rm eff}$ (K) | $\log g$ (dex) | $\gamma^a$ | [Fe/H] (dex) | $EW({\rm Li})^b$ (mÅ) | $EW({\rm Li})$ error flag$^c$ | Jeffries 2014$^d$ | Damiani 2014$^d$ | Literature members Spina 2014$^d$ | Frasca 2015$^d$ | Prisinzano 2016$^d$ | Final members$^e$ | Non-mem with Li$^f$ |
|---|---|---|---|---|---|---|---|---|---|---|---|---|---|---|---|
| 8948 | 08073548-4707070 | 61.3 ± 0.2 | 5050 ± 213 | … | 1.010 ± 0.005 | -0.30 ± 0.17 | … | … | … | … | … | … | … | … | G |
| 9519 | 08101508-4656074 | 71.1 ± 0.2 | 4598 ± 164 | … | 1.012 ± 0.006 | -0.12 ± 0.09 | <37 | 3 | … | … | … | … | … | … | G |
| 9735 | 08111144-4727377 | 17.7 ± 0.2 | 5352 ± 47 | … | 0.991 ± 0.001 | 0.07 ± 0.04 | 298 ± 4 | 1 | B | Y | … | Y | Y | Y | … |
| 8949 | 08073567-4718855 | 86.5 ± 0.2 | 4611 ± 140 | … | 1.022 ± 0.004 | 0.00 ± 0.08 | … | … | … | … | … | … | … | … | G |
| 9520 | 08101525-4727582 | 76.9 ± 0.2 | 5060 ± 73 | 3.21 ± 0.17 | 1.008 ± 0.002 | -0.16 ± 0.09 | <15 | 3 | … | … | … | … | … | … | … |
| 9736 | 08111185-4729447 | 19.7 ± 0.3 | 3293 ± 60 | … | 0.870 ± 0.007 | -0.29 ± 0.18 | 459 ± 11 | 1 | … | Y | … | Y | Y | Y | … |
| 9270 | 08091267-4707338 | 90.3 ± 0.2 | 4828 ± 138 | 2.50 ± 0.19 | 1.018 ± 0.003 | -0.23 ± 0.15 | <24 | 3 | … | … | … | … | … | … | G |
| 9521 | 08101527-4737159 | 37.1 ± 0.2 | 6036 ± 126 | 3.85 ± 0.15 | 1.008 ± 0.002 | -0.10 ± 0.02 | … | … | … | … | … | … | … | … | … |
| 9737 | 08111195-4719311 | 157.2 ± 0.2 | 4703 ± 230 | … | 1.033 ± 0.006 | -0.38 ± 0.13 | <23 | 3 | … | … | … | … | … | … | G |
| 8950 | 08073666-4722278 | 50.0 ± 0.2 | 4929 ± 134 | … | 1.016 ± 0.003 | -0.16 ± 0.18 | 29 ± 4 | 1 | … | … | … | … | … | … | … |
| 9738 | 08111198-4700100 | 73.6 ± 0.2 | 4871 ± 85 | … | 1.016 ± 0.004 | -0.12 ± 0.05 | … | … | … | … | … | … | … | … | G |
| 8951 | 08073674-4707347 | 34.6 ± 0.2 | 4691 ± 19 | 2.25 ± 0.04 | 1.015 ± 0.002 | -0.27 ± 0.04 | <20 | 3 | … | … | … | … | … | … | … |
| 9739 | 08111208-4721439 | 18.4 ± 0.2 | 3840 ± 169 | … | 0.847 ± 0.005 | -0.18 ± 0.12 | 465 ± 22 | 1 | B | Y | … | Y | Y | Y | … |
| 8952 | 08073685-4730159 | 64.9 ± 0.2 | 4627 ± 192 | … | 1.013 ± 0.005 | -0.13 ± 0.08 | <16 | 3 | … | … | … | … | … | … | G |
| 9271 | 08091335-4721216 | 42.5 ± 0.8 | … | … | … | … | … | … | … | … | … | … | Y | … | … |
| 9522 | 08101610-4658333 | 55.3 ± 0.3 | 5346 ± 42 | 4.11 ± 0.15 | 0.986 ± 0.005 | 0.03 ± 0.01 | <13 | 3 | … | … | … | … | … | … | … |
| 9523 | 08101652-4711333 | 30.6 ± 0.2 | 4378 ± 234 | 4.45 ± 0.17 | 0.912 ± 0.006 | 0.03 ± 0.15 | <24 | 3 | … | … | … | … | … | … | … |
| 8953 | 08073697-4711044 | 88.5 ± 0.2 | 4888 ± 47 | 2.76 ± 0.10 | 1.012 ± 0.003 | -0.15 ± 0.07 | <17 | 3 | … | … | … | … | … | … | … |
| 3503 | 08111229-4704491 | -1.8 ± 0.6 | 4686 ± 25 | 2.45 ± 0.11 | … | -0.32 ± 0.01 | <10 | 3 | … | … | N | … | … | n | … |
| 9272 | 08091375-4727388 | 30.1 ± 0.2 | 4818 ± 89 | 2.58 ± 0.16 | 1.023 ± 0.002 | -0.06 ± 0.09 | 11 ± 2 | 1 | … | … | … | … | … | … | G |
| 8954 | 08073724-4654115 | -1.5 ± 0.3 | 6646 ± 40 | 3.98 ± 0.08 | 1.008 ± 0.002 | -0.16 ± 0.04 | <13 | 3 | … | … | … | … | … | … | … |
| 9740 | 08111258-4708072 | 20.0 ± 0.3 | 3305 ± 7 | … | 0.897 ± 0.010 | -0.25 ± 0.16 | 598 ± 12 | 1 | B | Y | … | Y | Y | Y | … |
| 8955 | 08073729-4726446 | 133.0 ± 0.2 | 3635 ± 135 | … | … | … | … | … | … | … | … | … | … | … | … |
| 9741 | 08111267-4725139 | 19.3 ± 0.2 | 4543 ± 106 | 2.19 ± 0.13 | 1.032 ± 0.002 | -0.02 ± 0.04 | <27 | 3 | … | … | … | … | … | … | G |
| 8970 | 08074293-4737322 | 66.4 ± 0.2 | 4482 ± 172 | … | 1.041 ± 0.003 | -0.20 ± 0.05 | <32 | 3 | … | … | … | … | … | … | G |
| 9780 | 08112114-4654523 | 99.0 ± 0.2 | 4562 ± 58 | 2.24 ± 0.19 | 1.026 ± 0.006 | -0.18 ± 0.03 | <20 | 3 | … | … | … | … | … | … | G |
| 9524 | 08101699-4703590 | 25.1 ± 0.2 | 3324 ± 1 | 4.60 ± 0.16 | 0.851 ± 0.012 | -0.27 ± 0.15 | <100 | 3 | … | … | … | … | Y | Y | … |
| 8971 | 08074304-4711071 | 20.2 ± 0.2 | 3755 ± 93 | … | 0.855 ± 0.004 | -0.19 ± 0.13 | 510 ± 22 | 1 | B | Y | … | Y | Y | Y | … |
| 9781 | 08112115-4715064 | 58.0 ± 0.5 | 3486 ± 117 | … | 0.842 ± 0.013 | -0.25 ± 0.13 | <100 | 3 | … | … | … | … | … | … | … |
| 3474 | 08091397-4722030 | 11.3 ± 0.6 | 4632 ± 8 | 2.29 ± 0.08 | 1.027 ± 0.001 | -0.09 ± 0.05 | 90 ± 5 | 2 | … | … | N | … | … | n | G |
| 9525 | 08101706-4744165 | 23.0 ± 0.2 | 4834 ± 33 | 2.52 ± 0.08 | 1.013 ± 0.003 | -0.09 ± 0.05 | <25 | 3 | … | … | … | … | … | … | G |
| 8972 | 08074342-4737517 | 65.7 ± 0.2 | 4649 ± 89 | 2.44 ± 0.14 | 1.022 ± 0.003 | 0.02 ± 0.04 | <52 | 3 | … | … | … | … | … | … | … |
| 3506 | 08112119-4654274 | 60.7 ± 0.6 | 6017 ± 15 | 4.12 ± 0.01 | 0.997 ± 0.002 | 0.17 ± 0.01 | 48 ± 2 | 2 | … | … | N | … | … | n | NG |
| 9526 | 08101733-4658193 | 41.6 ± 0.2 | 5010 ± 221 | … | 1.010 ± 0.003 | -0.17 ± 0.13 | … | … | … | … | … | … | … | … | … |
| 8973 | 08074349-4706116 | 59.6 ± 0.2 | 5518 ± 41 | … | 0.997 ± 0.003 | -0.14 ± 0.08 | <22 | 3 | … | … | … | … | … | … | … |
| 9274 | 08091415-4747417 | -9.2 ± 0.2 | 5042 ± 159 | 4.17 ± 0.07 | 0.976 ± 0.003 | -0.22 ± 0.07 | <13 | 3 | … | … | … | … | … | … | … |
| 9527 | 08101733-4658524 | 70.7 ± 0.3 | 3624 ± 69 | … | 0.837 ± 0.007 | -0.21 ± 0.14 | <100 | 3 | … | … | … | … | … | … | … |
| 8974 | 08074361-4722095 | 17.5 ± 0.2 | 4282 ± 186 | … | 0.920 ± 0.003 | -0.10 ± 0.09 | 488 ± 7 | 1 | A | Y | … | Y | Y | Y | … |
| 9275 | 08091447-4742219 | 67.2 ± 0.2 | 5039 ± 110 | … | 1.025 ± 0.003 | -0.20 ± 0.18 | <20 | 3 | … | … | … | … | … | … | … |
| 9558 | 08102293-4702132 | 44.0 ± 0.2 | 5075 ± 234 | … | 1.012 ± 0.003 | -0.25 ± 0.22 | <18 | 3 | … | … | … | … | … | … | G |
| 9782 | 08112142-4746299 | 19.4 ± 0.5 | 3281 ± 12 | … | 0.873 ± 0.020 | … | 619 ± 18 | 1 | B | … | … | … | Y | Y | G |
| 8975 | 08074449-4659298 | 77.8 ± 0.2 | 4396 ± 217 | … | 1.055 ± 0.004 | -0.20 ± 0.01 | … | … | … | … | … | … | … | … | … |
| 9276 | 08091486-4735390 | 16.8 ± 0.2 | 4769 ± 97 | 2.49 ± 0.15 | 1.009 ± 0.004 | -0.06 ± 0.06 | <25 | 3 | … | … | … | … | … | … | … |
| 8976 | 08074489-4734134 | 43.9 ± 0.3 | 3769 ± 56 | … | 0.875 ± 0.011 | -0.19 ± 0.13 | <100 | 3 | … | … | … | … | … | … | … |
| 9277 | 08091506-4704454 | 41.5 ± 0.2 | 5119 ± 169 | … | 1.015 ± 0.003 | -0.23 ± 0.18 | <4 | 3 | … | … | … | … | … | … | G |
| 3507 | 08112188-4711281 | 24.3 ± 0.4 | 7341 ± 341 | 4.07 ± 0.21 | … | -0.12 ± 0.17 | <8 | 3 | … | … | … | … | … | … | … |
| 8977 | 08074501-4726527 | 37.1 ± 0.2 | 4927 ± 130 | … | 1.023 ± 0.002 | -0.15 ± 0.13 | 54 ± 4 | 1 | … | … | … | … | … | … | G |
| 9278 | 08091534-4714263 | 508.6 ± 0.5 | 3319 ± 139 | … | … | … | … | … | … | … | … | … | Y | … | … |
| 9559 | 08102379-4735091 | -1.5 ± 0.3 | 6365 ± 131 | … | 0.997 ± 0.002 | 0.04 ± 0.15 | … | … | … | … | … | … | … | … | … |
| 9783 | 08112193-4653343 | 59.8 ± 0.2 | 4763 ± 30 | 2.64 ± 0.11 | 1.014 ± 0.003 | -0.09 ± 0.03 | 54 ± 4 | 1 | … | … | … | … | … | … | G |
| 9279 | 08091543-4726105 | 22.5 ± 0.4 | 3338 ± 63 | … | 0.870 ± 0.008 | -0.27 ± 0.14 | … | … | B | Y | … | Y | Y | Y | … |
| 9560 | 08102429-4715177 | 96.5 ± 0.9 | … | … | … | … | … | … | … | … | … | … | … | … | … |
| 9784 | 08112210-4659299 | -14.4 ± 0.2 | 4896 ± 140 | … | 0.971 ± 0.002 | 0.09 ± 0.05 | <27 | 3 | … | … | … | … | … | … | … |
| 9295 | 08092052-4730344 | 22.3 ± 0.2 | 6082 ± 136 | 4.19 ± 0.06 | 0.995 ± 0.002 | -0.09 ± 0.08 | … | … | … | … | … | … | … | … | … |
| 8978 | 08074542-4720203 | 67.9 ± 0.2 | 4640 ± 110 | … | 1.019 ± 0.004 | -0.19 ± 0.15 | <37 | 3 | … | … | … | … | … | … | G |
| 9561 | 08102438-4735264 | 54.6 ± 0.2 | 5052 ± 92 | … | 1.009 ± 0.004 | -0.06 ± 0.04 | <28 | 3 | … | … | … | … | … | … | … |
| 9296 | 08092105-4708586 | 29.6 ± 0.2 | 4820 ± 38 | … | 0.992 ± 0.004 | 0.07 ± 0.05 | <29 | 3 | … | … | … | … | … | … | … |
| 9562 | 08102451-4736423 | 17.2 ± 0.2 | 3454 ± 38 | … | 0.852 ± 0.006 | -0.26 ± 0.14 | 171 ± 22 | 1 | A | Y | … | Y | Y | Y | … |
| 9785 | 08112238-4652168 | 13.6 ± 0.2 | 6149 ± 134 | 3.94 ± 0.03 | 1.005 ± 0.002 | -0.08 ± 0.01 | 67 ± 2 | 1 | … | … | … | … | … | … | … |

**Table C.3.** continued.

| ID | CNAME | $RV$ (km s$^{-1}$) | $T_{\rm eff}$ (K) | $\log g$ (dex) | $\gamma^a$ | [Fe/H] (dex) | $EW({\rm Li})^b$ (mÅ) | $EW({\rm Li})$ error flag$^c$ | Jeffries 2014$^d$ | Damiani 2014$^d$ | Spina 2014$^d$ | Frasca 2015$^d$ | Prisinzano 2016$^d$ | Final members$^e$ | Non-mem with Li$^f$ |
|---|---|---|---|---|---|---|---|---|---|---|---|---|---|---|---|
| 9297 | 08092124-4733324 | 80.2 ± 0.2 | 4653 ± 90 | ... | 1.023 ± 0.003 | -0.11 ± 0.07 | <39 | 3 | ... | ... | ... | ... | ... | ... | G |
| 9563 | 08102484-4726483 | 17.3 ± 0.3 | 3338 ± 41 | ... | 0.846 ± 0.009 | -0.26 ± 0.14 | ... | ... | A | Y | ... | ... | Y | Y | ... |
| 9298 | 08092126-4656557 | -2.7 ± 0.3 | 3411 ± 33 | ... | 0.837 ± 0.012 | -0.23 ± 0.15 | ... | ... | ... | ... | ... | ... | ... | ... | ... |
| 9786 | 08112255-4730288 | 62.0 ± 0.2 | 4719 ± 56 | 2.51 ± 0.08 | 1.019 ± 0.002 | 0.02 ± 0.02 | <34 | 3 | ... | ... | ... | ... | ... | ... | G |
| 8980 | 08074648-4711496 | 19.4 ± 0.3 | 3390 ± 56 | ... | 0.868 ± 0.008 | -0.25 ± 0.14 | 556 ± 10 | 1 | B | Y | ... | Y | Y | Y | ... |
| 9564 | 08102502-4739351 | 37.3 ± 0.2 | 4730 ± 117 | 2.48 ± 0.20 | 1.022 ± 0.003 | -0.04 ± 0.17 | <30 | 3 | ... | ... | ... | ... | ... | ... | G |
| 8981 | 08074648-4727066 | 34.4 ± 0.2 | 5586 ± 82 | ... | 0.989 ± 0.002 | 0.00 ± 0.03 | <8 | 3 | ... | ... | ... | ... | ... | ... | ... |
| 9787 | 08112262-4747495 | 31.1 ± 0.2 | 5069 ± 166 | 3.38 ± 0.13 | 1.002 ± 0.004 | -0.19 ± 0.11 | <14 | 3 | ... | ... | ... | ... | ... | ... | ... |
| 8982 | 08074649-4743324 | 24.1 ± 0.3 | 3632 ± 88 | ... | 0.821 ± 0.013 | -0.21 ± 0.14 | <100 | 3 | ... | ... | ... | ... | ... | ... | ... |
| 9299 | 08092139-4657095 | -8.1 ± 0.2 | 3965 ± 60 | 4.54 ± 0.14 | 0.857 ± 0.003 | -0.15 ± 0.01 | ... | ... | ... | ... | ... | ... | ... | ... | ... |
| 9565 | 08102545-4722406 | 132.0 ± 0.2 | 3623 ± 125 | ... | ... | ... | ... | ... | ... | ... | ... | ... | ... | ... | ... |
| 3453 | 08074671-4658173 | 61.1 ± 0.6 | 4733 ± 46 | 2.45 ± 0.08 | ... | 0.28 ± 0.04 | <22 | 3 | ... | ... | N | ... | ... | n | ... |
| 9300 | 08092142-4708066 | 24.3 ± 0.3 | 3349 ± 15 | ... | 0.859 ± 0.008 | -0.24 ± 0.15 | <100 | 3 | ... | ... | ... | ... | Y | Y | ... |
| 9566 | 08102583-4736247 | 17.0 ± 0.3 | 3334 ± 15 | ... | 0.852 ± 0.011 | -0.27 ± 0.14 | 573 ± 36 | 1 | A | Y | ... | ... | Y | Y | ... |
| 9789 | 08112283-4709048 | 51.7 ± 0.2 | 4527 ± 180 | ... | 1.022 ± 0.003 | 0.08 ± 0.07 | <54 | 3 | ... | ... | ... | ... | ... | ... | G |
| 8983 | 08074740-4711102 | 23.3 ± 0.2 | 5639 ± 35 | 4.29 ± 0.20 | 0.990 ± 0.002 | 0.04 ± 0.01 | 216 ± 2 | 1 | ... | ... | ... | ... | ... | ... | ... |
| 9301 | 08092151-4704091 | 64.2 ± 0.2 | 5130 ± 129 | ... | 1.015 ± 0.002 | -0.11 ± 0.14 | ... | ... | ... | ... | ... | ... | ... | ... | G |
| 9567 | 08102583-4739583 | 16.5 ± 0.2 | 6205 ± 130 | 4.09 ± 0.13 | 0.999 ± 0.003 | -0.05 ± 0.09 | ... | ... | ... | ... | ... | ... | ... | ... | ... |
| 8984 | 08074754-4707321 | 58.7 ± 0.2 | 4606 ± 66 | 2.38 ± 0.08 | 1.012 ± 0.003 | -0.09 ± 0.02 | <27 | 3 | ... | ... | ... | ... | ... | ... | G |
| 9302 | 08092183-4735370 | 20.1 ± 0.3 | 3396 ± 43 | ... | 0.859 ± 0.013 | -0.27 ± 0.16 | ... | ... | B | Y | ... | Y | Y | Y | ... |
| 9568 | 08102587-4701453 | 14.1 ± 0.2 | 5021 ± 99 | ... | 1.021 ± 0.003 | -0.06 ± 0.13 | <20 | 3 | ... | ... | ... | ... | ... | ... | G |
| 8985 | 08074792-4724147 | 27.9 ± 0.4 | 3427 ± 30 | ... | 0.774 ± 0.017 | ... | ... | ... | ... | ... | ... | ... | ... | ... | ... |
| 9303 | 08092191-4705413 | 90.1 ± 0.2 | 4621 ± 129 | ... | 1.026 ± 0.003 | -0.43 ± 0.02 | <21 | 3 | ... | ... | ... | ... | ... | ... | G |
| 9569 | 08102595-4700132 | 96.0 ± 0.2 | 4183 ± 287 | ... | 1.058 ± 0.002 | -0.23 ± 0.06 | ... | ... | ... | ... | ... | ... | ... | ... | G |
| 8986 | 08074812-4725319 | 92.4 ± 0.2 | 4663 ± 70 | ... | 1.034 ± 0.004 | -0.31 ± 0.05 | ... | ... | ... | ... | ... | ... | ... | ... | ... |
| 3508 | 08112306-4733019 | 63.7 ± 0.6 | 4664 ± 26 | 2.47 ± 0.07 | 1.019 ± 0.001 | 0.03 ± 0.02 | 32 ± 7 | 1 | ... | ... | N | ... | ... | n | G |
| 9570 | 08102624-4657222 | 62.0 ± 0.2 | 6489 ± 25 | 4.41 ± 0.04 | 0.996 ± 0.002 | 0.36 ± 0.02 | ... | ... | ... | ... | ... | ... | ... | ... | ... |
| 8987 | 08074815-4738056 | 30.9 ± 0.2 | 5117 ± 215 | ... | 1.016 ± 0.003 | -0.22 ± 0.08 | <24 | 3 | ... | ... | ... | ... | ... | ... | ... |
| 9304 | 08092234-4717333 | 14.6 ± 0.3 | 3294 ± 3 | 4.66 ± 0.18 | 0.855 ± 0.012 | -0.27 ± 0.14 | 560 ± 38 | 1 | B | Y | ... | Y | Y | Y | ... |
| 3509 | 08112313-4737105 | 10.9 ± 0.6 | 5048 ± 17 | 4.46 ± 0.05 | 0.972 ± 0.003 | 0.04 ± 0.04 | 27 ± 3 | 2 | ... | ... | N | ... | ... | n | NG |
| 8988 | 08074827-4733433 | 44.8 ± 0.2 | 5244 ± 232 | 4.36 ± 0.10 | 0.976 ± 0.004 | -0.31 ± 0.12 | ... | ... | ... | ... | ... | ... | ... | ... | ... |
| 9571 | 08102633-4701114 | 18.5 ± 0.3 | 3424 ± 6 | ... | 0.854 ± 0.009 | -0.25 ± 0.15 | ... | ... | ... | ... | ... | ... | Y | ... | ... |
| 9790 | 08112320-4652335 | 19.9 ± 0.2 | 5190 ± 120 | ... | 0.986 ± 0.002 | ... | 357 ± 2 | 1 | B | ... | ... | Y | Y | Y | ... |
| 9028 | 08075724-4707364 | -3.3 ± 0.2 | 5855 ± 98 | 4.17 ± 0.11 | 0.997 ± 0.002 | 0.23 ± 0.09 | 22 ± 1 | 1 | ... | ... | ... | ... | ... | ... | ... |
| 9572 | 08102633-4743433 | 35.8 ± 0.2 | 5214 ± 311 | ... | 1.011 ± 0.004 | -0.35 ± 0.16 | <15 | 3 | ... | ... | ... | ... | ... | ... | ... |
| 9791 | 08112383-4739074 | 104.5 ± 0.2 | 4716 ± 71 | ... | 1.024 ± 0.004 | -0.14 ± 0.14 | 276 ± 12 | 1 | ... | ... | ... | ... | ... | ... | Li-rich G |
| 9305 | 08092317-4710376 | 35.0 ± 0.2 | 5157 ± 55 | ... | 1.015 ± 0.003 | -0.10 ± 0.12 | <11 | 3 | ... | ... | ... | ... | ... | ... | G |
| 9029 | 08075757-4743462 | 17.1 ± 0.3 | 4309 ± 405 | ... | 0.921 ± 0.004 | -0.06 ± 0.08 | 513 ± 4 | 1 | A | Y | ... | Y | Y | Y | ... |
| 9792 | 08112427-4658462 | 123.3 ± 0.3 | ... | ... | ... | ... | ... | ... | ... | ... | ... | ... | ... | ... | ... |
| 9306 | 08092325-4700479 | 45.3 ± 0.2 | 5009 ± 190 | 3.35 ± 0.20 | 1.002 ± 0.005 | -0.12 ± 0.20 | <35 | 3 | ... | ... | ... | ... | ... | ... | ... |
| 9030 | 08075778-4653151 | 72.5 ± 0.3 | 5147 ± 111 | ... | 1.008 ± 0.005 | -0.12 ± 0.09 | <36 | 3 | ... | ... | ... | ... | ... | ... | ... |
| 9573 | 08102707-4656221 | 24.7 ± 0.2 | 4651 ± 52 | 2.44 ± 0.18 | 1.023 ± 0.003 | -0.05 ± 0.10 | <24 | 3 | ... | ... | ... | ... | ... | ... | G |
| 9307 | 08092375-4735049 | 20.8 ± 0.3 | 3348 ± 74 | ... | 0.849 ± 0.012 | -0.27 ± 0.14 | 619 ± 13 | 1 | B | Y | ... | Y | Y | Y | ... |
| 9813 | 08112894-4747050 | 34.4 ± 0.2 | 4880 ± 49 | 2.69 ± 0.08 | 1.017 ± 0.004 | -0.02 ± 0.07 | <20 | 3 | ... | ... | ... | ... | ... | ... | G |
| 9588 | 08103274-4722479 | 73.7 ± 0.2 | 4627 ± 124 | ... | 1.022 ± 0.003 | -0.10 ± 0.04 | 141 ± 2 | 1 | ... | ... | ... | ... | ... | ... | Li-rich G |
| 9308 | 08092384-4701077 | -11.4 ± 0.2 | 5011 ± 118 | ... | 1.030 ± 0.003 | -0.15 ± 0.13 | <21 | 3 | ... | ... | ... | ... | ... | ... | G |
| 9032 | 08075818-4659544 | 351.9 ± 0.2 | ... | ... | ... | ... | ... | ... | ... | ... | ... | ... | ... | ... | ... |
| 9589 | 08103297-4715175 | 66.7 ± 0.2 | 4856 ± 101 | ... | 1.018 ± 0.003 | -0.16 ± 0.08 | <26 | 3 | ... | ... | ... | ... | ... | ... | ... |
| 9309 | 08092398-4744090 | 16.7 ± 0.3 | 3486 ± 9 | ... | 0.850 ± 0.008 | -0.24 ± 0.15 | 188 ± 26 | 1 | A | Y | ... | Y | Y | Y | ... |
| 9814 | 08112895-4659189 | 36.9 ± 0.2 | 4585 ± 163 | ... | 1.017 ± 0.001 | 0.05 ± 0.08 | <40 | 3 | ... | ... | ... | ... | ... | ... | G |
| 9815 | 08112895-4711070 | 86.6 ± 0.3 | 4915 ± 335 | 3.00 ± 0.15 | 1.001 ± 0.010 | -0.27 ± 0.22 | ... | ... | ... | ... | ... | ... | ... | ... | G |
| 9033 | 08075919-4730264 | 47.7 ± 0.2 | 4601 ± 247 | ... | 1.071 ± 0.003 | -0.04 ± 0.07 | 32 ± 1 | 1 | ... | ... | ... | ... | ... | ... | ... |
| 9590 | 08103328-4656196 | 66.6 ± 0.2 | 4811 ± 6 | 2.61 ± 0.10 | 1.019 ± 0.003 | -0.12 ± 0.05 | <19 | 3 | ... | ... | ... | ... | ... | ... | G |
| 9816 | 08112900-4720271 | 119.8 ± 0.2 | 4474 ± 123 | ... | 1.037 ± 0.003 | -0.19 ± 0.06 | ... | ... | ... | ... | ... | ... | ... | ... | G |
| 9034 | 08075927-4702559 | 120.0 ± 0.2 | 4332 ± 230 | ... | 1.051 ± 0.004 | -0.20 ± 0.05 | ... | ... | ... | ... | ... | ... | ... | ... | G |
| 9591 | 08103379-4656471 | 52.2 ± 0.2 | 5933 ± 101 | ... | 1.001 ± 0.002 | -0.07 ± 0.09 | 68 ± 2 | 1 | ... | ... | ... | ... | ... | ... | ... |
| 9340 | 08093270-4740357 | -4.6 ± 0.2 | 3803 ± 67 | ... | 0.827 ± 0.005 | -0.17 ± 0.14 | ... | ... | ... | ... | ... | ... | ... | ... | ... |
| 9817 | 08112966-4727457 | 57.0 ± 0.2 | 4866 ± 91 | 2.64 ± 0.08 | 1.016 ± 0.004 | -0.06 ± 0.06 | 39 ± 12 | 1 | ... | ... | ... | ... | ... | ... | G |
| 9035 | 08075941-4731478 | 48.8 ± 0.2 | 4617 ± 158 | ... | 1.023 ± 0.003 | -0.17 ± 0.20 | <23 | 3 | ... | ... | ... | ... | ... | ... | G |

| ID | CNAME | RV (km s$^{-1}$) | $T_{\text{eff}}$ (K) | $\log g$ (dex) | $\gamma^a$ | [Fe/H] (dex) | $EW(\text{Li})^b$ (mÅ) | $EW(\text{Li})$ error flag$^c$ | Jeffries 2014$^d$ | Damiani 2014$^d$ | Spina 2014$^d$ | Frasca 2015$^d$ | Prisinzano 2016$^d$ | Final members$^e$ | Non-mem with Li$^f$ |
|---|---|---|---|---|---|---|---|---|---|---|---|---|---|---|---|
| 9592 | 08103402-4728363 | 83.0 ± 0.2 | 4782 ± 133 | 2.49 ± 0.19 | 1.013 ± 0.003 | -0.25 ± 0.17 | <21 | 3 | … | … | … | … | … | … | G |
| 9341 | 08093286-4726540 | 17.3 ± 0.3 | 3384 ± 102 | … | 0.864 ± 0.008 | -0.26 ± 0.14 | 115 ± 29 | 1 | … | Y | … | … | Y | Y | … |
| 9818 | 08112974-4744035 | 17.7 ± 0.4 | 3272 ± 31 | 4.72 ± 0.13 | 0.852 ± 0.017 | -0.26 ± 0.14 | … | … | … | B | … | … | … | … | … |
| 9593 | 08103418-4657332 | 19.9 ± 0.2 | 3875 ± 156 | … | 0.871 ± 0.002 | -0.13 ± 0.13 | 415 ± 4 | 1 | B | Y | … | Y | Y | Y | … |
| 9036 | 08075952-4737209 | 16.8 ± 0.3 | 3332 ± 16 | … | 0.853 ± 0.013 | -0.26 ± 0.14 | 533 ± 14 | 1 | A | Y | … | … | Y | Y | … |
| 9342 | 08093288-4729264 | 95.2 ± 0.2 | 5150 ± 90 | … | 1.007 ± 0.003 | -0.46 ± 0.26 | <15 | 3 | … | … | … | … | … | … | … |
| 9819 | 08113025-4710494 | 71.4 ± 0.2 | 5114 ± 137 | … | 1.010 ± 0.004 | -0.13 ± 0.19 | … | … | … | … | … | … | … | … | … |
| 9594 | 08103422-4740535 | 80.2 ± 0.2 | 4431 ± 185 | 1.91 ± 0.13 | 1.033 ± 0.004 | -0.04 ± 0.04 | <45 | 3 | … | … | … | … | … | … | G |
| 3478 | 08093304-4737066 | -27.9 ± 0.6 | 5640 ± 119 | 4.26 ± 0.18 | … | -0.01 ± 0.10 | 239 ± 6 | 2 | … | … | Y | … | … | Y | … |
| 9037 | 08075956-4657304 | 81.8 ± 0.2 | 4640 ± 110 | 2.45 ± 0.15 | 1.017 ± 0.003 | -0.04 ± 0.01 | <40 | 3 | … | … | … | … | … | … | G |
| 9343 | 08093321-4707215 | 2.2 ± 0.2 | 5698 ± 101 | … | 1.011 ± 0.002 | -0.18 ± 0.02 | … | … | … | … | … | … | … | … | … |
| 9038 | 08075969-4657162 | 55.6 ± 0.2 | 4812 ± 212 | … | 1.026 ± 0.005 | -0.22 ± 0.17 | <8 | 3 | … | … | … | … | … | … | G |
| 9595 | 08103439-4745297 | 17.1 ± 0.3 | 3322 ± 2 | … | 0.880 ± 0.013 | -0.27 ± 0.14 | 387 ± 21 | 1 | A | … | … | … | Y | Y | … |
| 9344 | 08093321-4722596 | 17.8 ± 0.3 | 3368 ± 25 | … | 0.867 ± 0.014 | -0.22 ± 0.15 | 542 ± 42 | 1 | A | Y | … | … | Y | Y | … |
| 9820 | 08113046-4715554 | 51.6 ± 0.2 | 4825 ± 26 | 2.58 ± 0.01 | 1.016 ± 0.003 | -0.07 ± 0.12 | 25 ± 7 | 1 | … | … | … | … | … | … | G |
| 9821 | 08113075-4733594 | 27.7 ± 0.2 | 5008 ± 90 | … | 1.013 ± 0.003 | -0.11 ± 0.15 | <13 | 3 | … | … | … | … | … | … | … |
| 9596 | 08103450-4708372 | 43.4 ± 0.2 | 4989 ± 213 | … | 1.018 ± 0.002 | -0.12 ± 0.17 | … | … | … | … | … | … | … | … | G |
| 9039 | 08080021-4702491 | 33.5 ± 0.2 | 5176 ± 17 | … | 0.998 ± 0.003 | -0.06 ± 0.14 | 23 ± 2 | 1 | … | … | … | … | … | … | G |
| 9345 | 08093331-4737542 | 122.3 ± 0.2 | 4516 ± 123 | … | 1.031 ± 0.003 | -0.24 ± 0.14 | 21 ± 5 | 1 | … | … | … | … | … | … | G |
| 9597 | 08103465-4701472 | 17.0 ± 0.3 | 3343 ± 11 | … | 0.874 ± 0.013 | -0.25 ± 0.14 | 559 ± 11 | 1 | A | Y | … | … | Y | Y | … |
| 9040 | 08080039-4731527 | 10.0 ± 0.2 | 5488 ± 106 | 4.17 ± 0.04 | 0.987 ± 0.003 | 0.12 ± 0.04 | <16 | 3 | … | … | … | … | … | … | … |
| 9346 | 08093332-4718502 | 18.9 ± 0.2 | 3788 ± 74 | … | 0.859 ± 0.003 | -0.17 ± 0.14 | 492 ± 5 | 1 | B | Y | … | Y | Y | Y | … |
| 3455 | 08080053-4702145 | 6.6 ± 0.6 | 5709 ± 39 | 4.49 ± 0.04 | 0.997 ± 0.002 | -0.01 ± 0.02 | 46 ± 5 | 2 | … | … | N | … | … | n | NG |
| 9347 | 08093364-4722285 | 18.1 ± 0.3 | 3516 ± 61 | … | 0.856 ± 0.007 | -0.20 ± 0.14 | 124 ± 27 | 1 | … | … | … | … | Y | Y | … |
| 9822 | 08113090-4718026 | 38.3 ± 0.2 | 5184 ± 97 | … | 0.989 ± 0.002 | … | 390 ± 3 | 1 | … | … | … | … | … | … | G |
| 9041 | 08080158-4730190 | 86.9 ± 0.2 | 4505 ± 141 | … | 1.029 ± 0.004 | -0.20 ± 0.06 | <25 | 3 | … | … | … | … | … | … | … |
| 9598 | 08103511-4729559 | -13.4 ± 0.2 | 5770 ± 82 | … | 1.002 ± 0.002 | … | <9 | 3 | … | … | … | … | … | … | … |
| 9348 | 08093365-4746184 | 83.8 ± 0.2 | 5902 ± 134 | 4.09 ± 0.01 | 0.995 ± 0.001 | -0.63 ± 0.27 | <14 | 3 | … | … | … | … | … | … | … |
| 9823 | 08113092-4742285 | 122.3 ± 0.2 | 4654 ± 30 | 2.23 ± 0.08 | 1.012 ± 0.004 | -0.34 ± 0.06 | … | … | … | … | … | … | … | … | … |
| 9042 | 08080181-4731175 | 53.0 ± 0.2 | 3990 ± 43 | 1.12 ± 0.16 | 1.060 ± 0.001 | -0.09 ± 0.18 | 46 ± 10 | 1 | … | … | … | … | … | … | G |
| 9349 | 08093367-4652265 | 66.1 ± 0.2 | 4755 ± 39 | 2.63 ± 0.10 | 1.011 ± 0.004 | -0.13 ± 0.08 | <28 | 3 | … | … | … | … | … | … | … |
| 9824 | 08113104-4732326 | 3.7 ± 0.7 | … | … | … | … | … | … | … | … | … | … | … | … | … |
| 9043 | 08080182-4715246 | 43.9 ± 0.2 | 4939 ± 157 | … | 1.015 ± 0.004 | -0.10 ± 0.10 | <21 | 3 | … | … | … | … | … | … | G |
| 9825 | 08113128-4652297 | 2.8 ± 0.2 | 5734 ± 90 | 4.13 ± 0.09 | 0.994 ± 0.002 | 0.17 ± 0.02 | <19 | 3 | … | … | … | … | … | … | … |
| 9044 | 08080228-4709227 | 52.6 ± 0.2 | 4677 ± 73 | 2.46 ± 0.13 | 1.018 ± 0.003 | -0.06 ± 0.05 | <28 | 3 | … | … | … | … | … | … | … |
| 9045 | 08080233-4740262 | 71.2 ± 0.2 | 4877 ± 54 | 2.72 ± 0.12 | 1.013 ± 0.003 | -0.15 ± 0.07 | … | … | … | … | … | … | … | … | G |
| 9350 | 08093450-4740527 | 21.4 ± 0.2 | 4155 ± 435 | … | 0.890 ± 0.002 | -0.02 ± 0.10 | 512 ± 3 | 1 | B | Y | … | … | Y | Y | … |
| 9351 | 08093499-4743294 | 49.2 ± 0.3 | 6600 ± 36 | 4.15 ± 0.07 | 1.004 ± 0.003 | 0.02 ± 0.03 | … | … | … | … | … | … | … | … | … |
| 9599 | 08103612-4726395 | 104.5 ± 0.2 | 4502 ± 112 | … | 1.029 ± 0.002 | -0.15 ± 0.08 | <26 | 3 | … | … | … | … | … | … | G |
| 9826 | 08113205-4747213 | 63.3 ± 0.2 | 4562 ± 81 | 2.28 ± 0.17 | 1.026 ± 0.005 | -0.06 ± 0.04 | <28 | 3 | … | … | … | … | … | … | G |
| 9827 | 08113214-4735340 | 17.1 ± 0.3 | 3350 ± 52 | … | 0.890 ± 0.016 | -0.28 ± 0.14 | 579 ± 14 | 1 | A | … | … | … | Y | Y | … |
| 9352 | 08093506-4725141 | 17.4 ± 0.3 | 3287 ± 11 | … | 0.907 ± 0.016 | … | 627 ± 24 | 1 | A | … | … | … | Y | Y | … |
| 3493 | 08103648-4659502 | -1.8 ± 0.6 | 4894 ± 33 | 2.61 ± 0.08 | 1.019 ± 0.001 | -0.01 ± 0.03 | 27 ± 7 | 1 | … | … | N | … | … | n | G |
| 9828 | 08113220-4745567 | 30.1 ± 0.2 | 4002 ± 92 | … | 0.871 ± 0.009 | … | 620 ± 40 | 1 | … | Y | … | … | Y | Y | … |
| 9626 | 08104358-4653127 | 28.9 ± 3.9 | 3285 ± 86 | … | … | … | … | … | … | Y | … | … | Y | … | … |
| 9864 | 08114123-4703033 | 17.4 ± 0.3 | 3391 ± 19 | … | 0.854 ± 0.011 | -0.24 ± 0.15 | … | … | … | … | … | … | Y | … | … |
| 9353 | 08093564-4735144 | 44.8 ± 0.2 | 4732 ± 93 | 2.53 ± 0.17 | 1.018 ± 0.003 | -0.03 ± 0.15 | <28 | 3 | … | … | … | … | … | … | G |
| 9627 | 08104379-4704598 | 88.7 ± 0.2 | 4442 ± 168 | … | 1.043 ± 0.003 | -0.17 ± 0.01 | <26 | 3 | … | … | … | … | … | … | G |
| 9628 | 08104382-4747548 | 44.4 ± 0.2 | 4783 ± 2 | 2.78 ± 0.05 | 1.007 ± 0.004 | -0.07 ± 0.01 | <32 | 3 | … | … | … | … | … | … | … |
| 9865 | 08114135-4746051 | -11.2 ± 0.2 | 5980 ± 142 | 3.80 ± 0.07 | 1.005 ± 0.002 | -0.47 ± 0.07 | … | … | … | … | … | … | … | … | … |
| 9629 | 08104419-4651569 | 59.1 ± 0.3 | 3526 ± 62 | … | 0.847 ± 0.009 | -0.23 ± 0.15 | … | … | … | … | … | … | … | … | … |
| 9866 | 08114144-4704333 | 82.8 ± 0.2 | 4883 ± 165 | … | 1.016 ± 0.003 | -0.14 ± 0.11 | … | … | … | … | … | … | … | … | G |
| 9630 | 08104429-4726310 | 64.3 ± 0.2 | 4741 ± 142 | … | 1.016 ± 0.003 | -0.22 ± 0.15 | … | … | … | … | … | … | … | … | … |
| 9867 | 08114151-4659324 | 90.0 ± 0.2 | 4779 ± 216 | … | 1.016 ± 0.004 | -0.15 ± 0.10 | … | … | … | … | … | … | … | … | G |
| 9631 | 08104454-4727056 | 18.4 ± 0.3 | 3322 ± 7 | … | 0.873 ± 0.006 | -0.28 ± 0.14 | 595 ± 8 | 1 | A | Y | … | Y | Y | Y | … |
| 9868 | 08114165-4653241 | 61.6 ± 0.2 | 5048 ± 219 | … | 1.010 ± 0.004 | -0.50 ± 0.18 | … | … | … | … | … | … | … | … | G |
| 9632 | 08104492-4714159 | 68.2 ± 0.2 | 4617 ± 102 | … | 1.033 ± 0.003 | … | <31 | 3 | … | … | … | … | … | … | G |
| 3497 | 08104495-4723015 | -1.2 ± 0.6 | 4831 ± 17 | 2.57 ± 0.06 | 1.022 ± 0.001 | -0.03 ± 0.02 | 23 ± 5 | 1 | … | … | … | … | … | … | G |
| 9869 | 08114234-4710564 | 15.2 ± 0.2 | 5137 ± 94 | … | 1.001 ± 0.003 | … | 18 ± 5 | 1 | … | … | … | … | … | … | … |

**Table C.3.** continued.

| ID | CNAME | RV (km s$^{-1}$) | $T_{\rm eff}$ (K) | $\log g$ (dex) | $\gamma^a$ | [Fe/H] (dex) | EW(Li)$^b$ (mÅ) | EW(Li) error flag$^c$ | Jeffries 2014$^d$ | Damiani 2014$^d$ | Spina 2014$^d$ | Frasca 2015$^d$ | Prisinzano 2016$^d$ | Final members$^e$ | Non-mem with Li$^f$ |
|---|---|---|---|---|---|---|---|---|---|---|---|---|---|---|---|
| 9633 | 08104510-4741432 | 26.5 ± 0.3 | 3998 ± 210 | 4.49 ± 0.11 | 0.843 ± 0.009 | -0.21 ± 0.06 | <19 | 3 | … | … | … | … | … | … | … |
| 9870 | 08114254-4709323 | 73.6 ± 0.2 | 4071 ± 332 | … | 1.058 ± 0.003 | -0.28 ± 0.07 | … | … | … | … | … | … | … | … | G |
| 9871 | 08114284-4729504 | 23.3 ± 0.2 | 5719 ± 140 | … | 0.998 ± 0.001 | -0.01 ± 0.13 | 129 ± 6 | 1 | … | … | … | … | Y | Y | … |
| 9872 | 08114296-4717582 | -0.1 ± 0.2 | 5751 ± 25 | … | 1.004 ± 0.001 | -0.31 ± 0.06 | <16 | 3 | … | … | … | … | … | … | … |
| 9873 | 08114332-4730000 | 20.8 ± 0.2 | 4556 ± 290 | 4.46 ± 0.16 | 0.928 ± 0.004 | -0.02 ± 0.08 | 60 ± 3 | 1 | … | … | … | … | Y | Y | … |
| 9634 | 08104517-4659531 | -3.2 ± 0.3 | 7328 ± 30 | … | … | … | … | … | … | … | … | … | … | … | … |
| 9874 | 08114355-4740489 | 111.3 ± 0.2 | 4509 ± 108 | … | 1.022 ± 0.005 | -0.26 ± 0.17 | … | … | … | … | … | … | … | … | G |
| 9875 | 08114356-4718172 | 55.3 ± 0.2 | 5172 ± 57 | … | 1.004 ± 0.003 | -0.13 ± 0.06 | <16 | 3 | … | … | … | … | … | … | … |
| 9876 | 08114373-4705023 | 57.8 ± 0.2 | 4648 ± 93 | 2.55 ± 0.07 | 1.010 ± 0.005 | -0.04 ± 0.02 | <30 | 3 | … | … | … | … | … | … | … |
| 9877 | 08114426-4723223 | 33.6 ± 0.2 | 4005 ± 87 | … | 1.049 ± 0.002 | -0.13 ± 0.15 | 138 ± 1 | 1 | … | … | … | … | … | … | G |
| 9635 | 08104600-4736516 | 23.6 ± 0.2 | 4965 ± 212 | 4.13 ± 0.17 | 0.967 ± 0.004 | -0.02 ± 0.05 | 100 ± 6 | 1 | … | … | … | … | … | … | … |
| 9636 | 08104602-4721380 | 39.6 ± 0.2 | 5621 ± 98 | … | 0.991 ± 0.003 | -0.05 ± 0.06 | … | … | … | … | … | … | … | … | … |
| 9878 | 08114439-4713150 | 121.6 ± 0.2 | 5008 ± 190 | … | 1.012 ± 0.004 | -0.17 ± 0.26 | <26 | 3 | … | … | … | … | … | … | G |
| 9879 | 08114441-4742525 | 6.6 ± 0.2 | 5554 ± 191 | 4.22 ± 0.06 | 0.985 ± 0.003 | 0.20 ± 0.07 | <21 | 3 | … | … | … | … | … | … | … |
| 9637 | 08104620-4705396 | 62.1 ± 0.2 | 4908 ± 94 | … | 1.019 ± 0.004 | -0.14 ± 0.07 | <29 | 3 | … | … | … | … | … | … | G |
| 9880 | 08114451-4735287 | 2.9 ± 0.2 | 5482 ± 175 | … | 1.001 ± 0.003 | -0.12 ± 0.02 | 51 ± 5 | 1 | … | … | … | … | … | … | … |
| 9881 | 08114456-4657516 | 19.9 ± 0.2 | 4904 ± 172 | … | 0.944 ± 0.003 | … | 570 ± 3 | 1 | B | Y | … | Y | Y | Y | … |
| 9882 | 08114462-4659463 | 92.2 ± 0.3 | 4967 ± 128 | … | 1.018 ± 0.003 | -0.20 ± 0.16 | <19 | 3 | … | … | … | … | … | … | G |
| 9883 | 08114469-4658096 | 0.8 ± 0.4 | 3374 ± 20 | 4.78 ± 0.04 | 0.815 ± 0.013 | -0.28 ± 0.14 | <100 | 3 | … | … | … | … | … | … | … |
| 9904 | 08115098-4651540 | 36.9 ± 0.2 | 4728 ± 82 | 2.48 ± 0.14 | 1.028 ± 0.003 | 0.01 ± 0.06 | 20 ± 7 | 1 | … | … | … | … | … | … | G |
| 9905 | 08115105-4707386 | 41.0 ± 0.2 | 5753 ± 172 | 4.20 ± 0.20 | 0.996 ± 0.002 | 0.19 ± 0.01 | … | … | … | … | … | … | … | … | … |
| 9906 | 08115144-4736575 | 16.8 ± 0.2 | 4913 ± 46 | 2.49 ± 0.09 | 1.019 ± 0.003 | -0.13 ± 0.02 | <11 | 3 | … | … | … | … | … | … | G |
| 9907 | 08115175-4744118 | 87.4 ± 0.3 | 4591 ± 88 | 2.32 ± 0.13 | 1.024 ± 0.008 | -0.05 ± 0.06 | <57 | 3 | … | … | … | … | … | … | G |
| 9908 | 08115262-4744063 | 54.7 ± 0.2 | 5759 ± 145 | 4.30 ± 0.03 | 0.988 ± 0.003 | 0.11 ± 0.10 | 25 ± 3 | 1 | … | … | … | … | … | … | … |
| 9909 | 08115272-4721471 | 32.0 ± 0.2 | 4956 ± 259 | … | 1.012 ± 0.003 | -0.22 ± 0.21 | <15 | 3 | … | … | … | … | … | … | G |
| 9910 | 08115296-4713123 | 29.2 ± 0.2 | 5069 ± 132 | … | 1.007 ± 0.005 | -0.08 ± 0.13 | <23 | 3 | … | … | … | … | … | … | … |
| 9911 | 08115302-4737498 | 154.9 ± 0.2 | 4220 ± 231 | … | 1.037 ± 0.007 | -0.10 ± 0.08 | … | … | … | … | … | … | … | … | G |
| 9912 | 08115363-4734096 | 14.4 ± 0.2 | 4896 ± 19 | 2.49 ± 0.17 | 1.016 ± 0.003 | -0.10 ± 0.04 | <19 | 3 | … | … | … | … | … | … | G |
| 9913 | 08115418-4701002 | 17.0 ± 0.3 | 3403 ± 33 | … | 0.876 ± 0.011 | -0.24 ± 0.15 | 540 ± 19 | 1 | A | … | … | … | Y | Y | … |
| 9914 | 08115447-4700164 | 29.0 ± 0.2 | 5703 ± 85 | … | 0.994 ± 0.002 | -0.17 ± 0.11 | <12 | 3 | … | … | … | … | … | … | … |
| 3513 | 08115451-4655430 | -6.5 ± 0.6 | 6517 ± 23 | 4.07 ± 0.07 | … | -0.05 ± 0.02 | <5 | 3 | … | … | N | … | … | n | … |
| 9915 | 08115473-4737129 | 28.9 ± 0.3 | 3584 ± 86 | 4.64 ± 0.11 | 0.795 ± 0.007 | -0.22 ± 0.14 | … | … | … | … | … | … | … | … | … |
| 9916 | 08115484-4702217 | 22.6 ± 0.2 | 4787 ± 256 | 4.55 ± 0.01 | 0.944 ± 0.003 | -0.15 ± 0.09 | <29 | 3 | … | … | … | … | … | … | … |
| 9917 | 08115542-4744534 | 5.2 ± 0.2 | 4678 ± 35 | 2.49 ± 0.13 | 1.021 ± 0.003 | -0.03 ± 0.04 | <27 | 3 | … | … | … | … | … | … | G |
| 9918 | 08115543-4654318 | 34.1 ± 0.2 | 6749 ± 22 | 4.40 ± 0.04 | 1.001 ± 0.001 | 0.44 ± 0.02 | … | … | … | … | … | … | … | … | … |
| 9919 | 08115543-4720421 | 48.6 ± 0.2 | 4991 ± 230 | … | 1.017 ± 0.004 | -0.23 ± 0.17 | <23 | 3 | … | … | … | … | … | … | G |
| 9920 | 08115553-4742435 | 112.6 ± 0.2 | 4464 ± 194 | … | 1.032 ± 0.003 | -0.31 ± 0.25 | … | … | … | … | … | … | … | … | G |
| 9956 | 08120705-4737413 | 59.0 ± 0.2 | 4798 ± 24 | … | 1.004 ± 0.005 | -0.09 ± 0.05 | <18 | 3 | … | … | … | … | … | … | … |
| 9957 | 08120811-4731428 | 50.0 ± 0.2 | 5136 ± 125 | … | 1.014 ± 0.004 | -0.07 ± 0.07 | … | … | … | … | … | … | … | … | G |
| 9958 | 08120840-4717374 | 58.2 ± 0.2 | 5954 ± 127 | 4.25 ± 0.16 | 0.996 ± 0.002 | -0.03 ± 0.05 | <11 | 3 | … | … | … | … | … | … | … |
| 9959 | 08120914-4732186 | 26.4 ± 0.2 | 4728 ± 208 | 2.55 ± 0.19 | 1.014 ± 0.004 | -0.03 ± 0.04 | <35 | 3 | … | … | … | … | … | … | G |
| 9960 | 08120971-4724550 | 46.6 ± 0.2 | 4558 ± 119 | 2.37 ± 0.16 | 1.017 ± 0.002 | 0.07 ± 0.07 | <51 | 3 | … | … | … | … | … | … | G |
| 8991 | 08074889-4729195 | 86.9 ± 0.2 | 4575 ± 186 | … | 1.019 ± 0.003 | 0.01 ± 0.06 | <45 | 3 | … | … | … | … | … | … | G |
| 8992 | 08074900-4719027 | 91.8 ± 0.2 | 4636 ± 100 | … | 1.029 ± 0.004 | -0.34 ± 0.07 | … | … | … | … | … | … | … | … | G |
| 8993 | 08074909-4744364 | 16.6 ± 0.2 | 3494 ± 14 | … | 0.860 ± 0.008 | -0.23 ± 0.14 | 271 ± 6 | 1 | A | Y | … | Y | Y | Y | … |
| 8994 | 08074910-4656248 | 92.9 ± 0.2 | 4919 ± 224 | … | 1.010 ± 0.003 | -0.30 ± 0.23 | <14 | 3 | … | … | … | … | … | … | … |
| 8995 | 08074928-4729547 | 72.7 ± 0.2 | 5067 ± 146 | … | 1.012 ± 0.003 | -0.25 ± 0.17 | <13 | 3 | … | … | … | … | … | … | G |
| 8996 | 08074935-4727119 | 3.9 ± 0.5 | … | … | … | … | … | … | … | … | … | … | … | … | … |
| 8997 | 08074943-4731559 | 5.8 ± 0.3 | 3473 ± 16 | … | 0.828 ± 0.016 | -0.22 ± 0.15 | … | … | … | … | Y | … | … | … | … |
| 8998 | 08074989-4736441 | 9.1 ± 0.2 | 6166 ± 61 | 4.11 ± 0.18 | 0.996 ± 0.003 | 0.02 ± 0.05 | <23 | 3 | … | … | … | … | … | … | … |
| 9312 | 08092546-4725396 | 15.4 ± 0.2 | 4267 ± 165 | 4.61 ± 0.06 | 0.883 ± 0.003 | 0.09 ± 0.23 | <23 | 3 | … | … | … | … | … | … | … |
| 8999 | 08074992-4710580 | 8.0 ± 0.3 | 5836 ± 96 | 3.94 ± 0.10 | 1.001 ± 0.003 | -0.12 ± 0.08 | 59 ± 6 | 1 | … | … | … | … | … | … | … |
| 9000 | 08075005-4735314 | -0.9 ± 0.4 | 3547 ± 35 | 4.60 ± 0.19 | 0.814 ± 0.016 | -0.22 ± 0.15 | … | … | … | … | … | … | … | … | … |
| 9313 | 08092576-4730559 | 18.6 ± 0.3 | 3430 ± 92 | … | 0.837 ± 0.010 | -0.24 ± 0.16 | 528 ± 32 | 1 | A | Y | … | Y | Y | Y | … |
| 9001 | 08075027-4701514 | 5.5 ± 0.2 | 4981 ± 133 | … | 1.025 ± 0.004 | -0.13 ± 0.14 | <28 | 3 | … | … | … | … | … | … | G |
| 9314 | 08092603-4707582 | 119.2 ± 0.2 | 5063 ± 212 | … | 1.000 ± 0.004 | -0.44 ± 0.21 | … | … | … | … | … | … | … | … | … |
| 9002 | 08075036-4725207 | 28.5 ± 0.3 | 3871 ± 126 | 4.60 ± 0.10 | 0.815 ± 0.008 | -0.16 ± 0.13 | … | … | … | … | … | … | … | … | … |
| 3477 | 08092627-4731001 | 18.1 ± 0.6 | 5220 ± 93 | 4.52 ± 0.12 | 0.993 ± 0.001 | -0.05 ± 0.11 | 363 ± 6 | 2 | B | Y | … | Y | Y | Y | … |
| 9003 | 08075041-4657297 | 35.0 ± 0.2 | 4984 ± 179 | … | 1.016 ± 0.005 | -0.13 ± 0.16 | <18 | 3 | … | … | … | … | … | … | G |

| ID | CNAME | $RV$ (km s$^{-1}$) | $T_{\text{eff}}$ (K) | $\log g$ (dex) | $\gamma^a$ | [Fe/H] (dex) | $EW(\text{Li})^b$ (mÅ) | $EW(\text{Li})$ error flag$^c$ | Jeffries 2014$^d$ | Damiani 2014$^d$ | Literature members Spina 2014$^d$ | Frasca 2015$^d$ | Prisinzano 2016$^d$ | Final members$^e$ | Non-mem with Li$^f$ |
|---|---|---|---|---|---|---|---|---|---|---|---|---|---|---|---|
| 9315 | 08092630-4703410 | 84.9 ± 0.3 | 4754 ± 48 | … | 1.020 ± 0.008 | -0.18 ± 0.08 | … | … | … | … | … | … | … | … | G |
| 9004 | 08075041-4708098 | 38.6 ± 0.5 | 5416 ± 100 | … | 0.997 ± 0.002 | … | … | … | … | … | … | … | … | … | … |
| 9316 | 08092634-4737267 | 16.8 ± 0.3 | 3483 ± 126 | … | 0.826 ± 0.006 | -0.24 ± 0.15 | … | … | … | … | … | … | Y | … | … |
| 9005 | 08075058-4732530 | -2.7 ± 0.2 | 4891 ± 190 | … | 1.018 ± 0.003 | -0.05 ± 0.08 | <18 | 3 | … | … | … | … | … | … | G |
| 9317 | 08092665-4720398 | 23.9 ± 0.2 | 4861 ± 86 | 2.83 ± 0.05 | 1.007 ± 0.005 | -0.05 ± 0.05 | <35 | 3 | … | … | … | … | … | … | … |
| 9006 | 08075093-4710226 | 94.5 ± 0.2 | 4676 ± 49 | … | 1.028 ± 0.003 | -0.28 ± 0.04 | 21 ± 4 | 1 | … | … | … | … | … | … | G |
| 9318 | 08092667-4716547 | 55.7 ± 0.2 | 4774 ± 191 | … | 1.020 ± 0.003 | -0.11 ± 0.05 | <25 | 3 | … | … | … | … | … | … | G |
| 9007 | 08075108-4744027 | 2.6 ± 0.2 | 4902 ± 153 | … | 1.019 ± 0.005 | -0.05 ± 0.09 | 215 ± 4 | 1 | … | … | … | … | … | … | … |
| 9601 | 08103682-4728489 | 15.8 ± 0.2 | 3504 ± 38 | … | 0.845 ± 0.004 | -0.25 ± 0.14 | 325 ± 15 | 1 | A | Y | … | Y | Y | Y | … |
| 9319 | 08092677-4734519 | 89.3 ± 0.2 | 4390 ± 217 | … | 1.040 ± 0.002 | -0.22 ± 0.11 | … | … | … | … | … | … | … | … | G |
| 9047 | 08080314-4741495 | 21.8 ± 0.2 | 4253 ± 225 | … | 0.921 ± 0.003 | -0.03 ± 0.17 | 495 ± 24 | 1 | B | Y | … | Y | Y | Y | … |
| 9602 | 08103686-4730257 | 46.4 ± 0.2 | 4650 ± 67 | 2.35 ± 0.06 | 1.019 ± 0.002 | -0.05 ± 0.04 | <35 | 3 | … | … | … | … | … | … | G |
| 9048 | 08080384-4707160 | -7.2 ± 0.2 | 5263 ± 136 | … | 1.002 ± 0.003 | -0.21 ± 0.14 | <3 | 3 | … | … | … | … | … | … | … |
| 9049 | 08080388-4720433 | 17.2 ± 0.2 | 3756 ± 114 | … | 0.849 ± 0.006 | -0.17 ± 0.14 | 493 ± 28 | 1 | A | Y | … | Y | Y | Y | … |
| 9603 | 08103732-4722165 | 28.4 ± 0.4 | 3611 ± 25 | … | 0.816 ± 0.016 | … | <100 | 3 | … | … | … | … | … | … | … |
| 9320 | 08092707-4724277 | 17.2 ± 0.3 | 3356 ± 53 | … | 0.877 ± 0.011 | -0.28 ± 0.14 | 506 ± 30 | 1 | A | Y | … | Y | Y | Y | … |
| 3456 | 08080431-4716272 | 23.5 ± 0.6 | 5272 ± 16 | 4.47 ± 0.03 | 0.980 ± 0.002 | -0.03 ± 0.03 | <15 | 3 | … | … | N | … | … | n | NG |
| 9604 | 08103733-4728518 | 64.6 ± 0.2 | 6385 ± 26 | … | 0.998 ± 0.002 | 0.06 ± 0.02 | <15 | 3 | … | … | … | … | … | … | … |
| 9321 | 08092708-4718330 | 43.9 ± 0.2 | 5162 ± 324 | … | 1.015 ± 0.003 | -0.26 ± 0.21 | <12 | 3 | … | … | … | … | … | … | G |
| 9050 | 08080519-4725405 | 52.9 ± 0.2 | 4754 ± 9 | 2.72 ± 0.11 | 1.007 ± 0.003 | -0.03 ± 0.08 | <53 | 3 | … | … | … | … | … | … | … |
| 3457 | 08080526-4722060 | … | … | … | … | … | … | … | … | … | N | … | … | n | … |
| 9605 | 08103769-4725532 | 24.3 ± 0.3 | 3431 ± 62 | 4.64 ± 0.12 | 0.817 ± 0.007 | -0.25 ± 0.15 | … | … | … | … | … | … | … | … | … |
| 9322 | 08092747-4741193 | 91.1 ± 0.2 | 4800 ± 51 | … | 1.000 ± 0.004 | -0.13 ± 0.09 | <40 | 3 | … | … | … | … | … | … | … |
| 9051 | 08080555-4706544 | 95.2 ± 0.2 | 4463 ± 108 | … | 1.032 ± 0.003 | -0.26 ± 0.11 | <22 | 3 | … | … | … | … | … | … | … |
| 3494 | 08103782-4702584 | 8.6 ± 0.6 | 4747 ± 10 | 2.42 ± 0.05 | 1.018 ± 0.001 | 0.03 ± 0.02 | 28 ± 7 | 2 | … | … | N | … | … | n | … |
| 9323 | 08092749-4723072 | 18.6 ± 0.6 | 3375 ± 80 | … | 0.860 ± 0.009 | -0.26 ± 0.15 | 385 ± 17 | 1 | B | Y | … | Y | Y | Y | … |
| 9052 | 08080579-4747307 | 82.8 ± 0.2 | 5029 ± 38 | … | 1.014 ± 0.003 | -0.20 ± 0.15 | … | … | … | … | … | … | … | … | G |
| 9606 | 08103784-4707024 | 74.9 ± 0.2 | 5077 ± 179 | … | 1.004 ± 0.004 | -0.10 ± 0.14 | <19 | 3 | … | … | … | … | … | … | … |
| 9355 | 08093642-4717442 | 16.7 ± 0.2 | 3877 ± 137 | … | 0.859 ± 0.003 | -0.16 ± 0.13 | … | … | … | Y | … | … | Y | … | … |
| 9053 | 08080586-4706536 | 18.7 ± 0.2 | 5754 ± 59 | 4.29 ± 0.01 | 0.988 ± 0.004 | -0.07 ± 0.13 | … | … | … | … | … | … | … | … | … |
| 9356 | 08093675-4714276 | 88.1 ± 0.2 | 4700 ± 13 | … | 1.025 ± 0.004 | -0.32 ± 0.05 | 48 ± 7 | 1 | … | … | … | … | … | … | G |
| 9357 | 08093681-4717040 | 16.9 ± 0.2 | 4284 ± 434 | … | 0.914 ± 0.003 | -0.02 ± 0.12 | 503 ± 5 | 1 | A | Y | … | Y | Y | Y | … |
| 9924 | 08115645-4652487 | 125.0 ± 0.2 | 4579 ± 87 | … | 1.024 ± 0.005 | -0.01 ± 0.08 | … | … | … | … | … | … | … | … | … |
| 9054 | 08080644-4725586 | 17.2 ± 0.3 | 3325 ± 7 | … | 0.878 ± 0.006 | -0.22 ± 0.15 | 617 ± 47 | 1 | A | Y | … | … | Y | Y | … |
| 3495 | 08103799-4723323 | -0.8 ± 0.6 | 6995 ± 255 | 4.07 ± 0.23 | … | -0.29 ± 0.14 | 25 ± 6 | 2 | … | … | … | … | … | … | … |
| 9925 | 08115654-4652093 | 8.0 ± 0.2 | 4977 ± 70 | … | 1.021 ± 0.003 | -0.02 ± 0.04 | <24 | 3 | … | … | … | … | … | … | … |
| 9358 | 08093745-4724469 | 42.1 ± 0.2 | 5152 ± 96 | 3.47 ± 0.19 | 1.005 ± 0.004 | -0.12 ± 0.01 | <23 | 3 | … | … | … | … | … | … | … |
| 9055 | 08080673-4703323 | 50.7 ± 0.2 | 4018 ± 113 | 4.59 ± 0.06 | 0.844 ± 0.005 | -0.08 ± 0.10 | … | … | … | … | … | … | … | … | … |
| 9926 | 08115704-4659426 | 28.3 ± 0.2 | 4715 ± 83 | 2.51 ± 0.14 | 1.019 ± 0.003 | 0.00 ± 0.12 | 11 ± 10 | 1 | … | … | … | … | … | … | G |
| 9359 | 08093748-4659186 | 118.1 ± 0.2 | 3905 ± 146 | … | 1.071 ± 0.004 | -0.11 ± 0.12 | … | … | … | … | … | … | … | … | G |
| 9607 | 08103865-4658456 | 125.2 ± 0.2 | 3311 ± 85 | … | … | … | … | … | … | … | … | … | … | … | … |
| 9927 | 08115737-4705283 | 17.3 ± 0.2 | 4277 ± 256 | 4.55 ± 0.01 | 0.880 ± 0.005 | -0.11 ± 0.04 | <27 | 3 | … | … | … | … | … | … | … |
| 9360 | 08093782-4704373 | 63.7 ± 0.2 | 4853 ± 117 | … | 1.022 ± 0.004 | -0.15 ± 0.10 | <35 | 3 | … | … | … | … | … | … | G |
| 3458 | 08080690-4715075 | 50.8 ± 0.4 | … | … | … | … | … | … | … | … | N | … | … | n | … |
| 9608 | 08103868-4737297 | 5.6 ± 0.2 | 5060 ± 30 | 2.94 ± 0.18 | 1.015 ± 0.002 | -0.19 ± 0.12 | <20 | 3 | … | … | … | … | … | … | G |
| 9056 | 08080708-4741589 | 21.6 ± 0.4 | 3446 ± 25 | 4.65 ± 0.12 | 0.823 ± 0.013 | -0.22 ± 0.15 | 505 ± 44 | 1 | B | Y | … | … | Y | Y | … |
| 9609 | 08103890-4707124 | 125.3 ± 0.2 | 4379 ± 231 | … | 1.058 ± 0.003 | -0.18 ± 0.05 | <27 | 3 | … | … | … | … | … | … | G |
| 9610 | 08103912-4744052 | 97.1 ± 0.2 | 4651 ± 55 | 2.33 ± 0.20 | 1.021 ± 0.003 | -0.20 ± 0.16 | 36 ± 10 | 1 | … | … | … | … | … | … | G |
| 9361 | 08093832-4702312 | 67.0 ± 0.2 | 4526 ± 113 | … | 1.030 ± 0.004 | -0.18 ± 0.03 | 32 ± 13 | 1 | … | … | … | … | … | … | G |
| 9928 | 08115754-4741195 | 24.7 ± 0.3 | 5698 ± 128 | 4.41 ± 0.10 | 0.981 ± 0.003 | 0.17 ± 0.07 | … | … | … | … | … | … | … | … | … |
| 9057 | 08080712-4733112 | 33.3 ± 4.0 | … | … | … | … | … | … | … | … | … | … | … | … | … |
| 9641 | 08104714-4742076 | 48.5 ± 0.3 | 3784 ± 69 | 4.55 ± 0.20 | 0.819 ± 0.007 | -0.17 ± 0.14 | <100 | 3 | … | … | … | … | … | … | … |
| 9362 | 08093839-4737325 | 11.9 ± 0.2 | 5069 ± 62 | 2.86 ± 0.08 | 1.014 ± 0.002 | -0.08 ± 0.13 | <30 | 3 | … | … | … | … | … | … | G |
| 9929 | 08115762-4734123 | 27.0 ± 0.3 | 3615 ± 49 | … | 0.822 ± 0.014 | -0.21 ± 0.14 | <100 | 3 | … | … | … | … | … | … | … |
| 9058 | 08080774-4659130 | 19.8 ± 0.2 | 3881 ± 151 | … | 0.868 ± 0.004 | -0.16 ± 0.11 | 508 ± 6 | 1 | B | … | … | Y | Y | Y | … |
| 9363 | 08093840-4710161 | 82.5 ± 0.2 | 4708 ± 153 | … | 1.026 ± 0.003 | -0.39 ± 0.13 | … | … | … | … | … | … | … | … | G |
| 9930 | 08115781-4702205 | 74.0 ± 0.2 | 4849 ± 161 | 2.97 ± 0.06 | 1.004 ± 0.003 | -0.04 ± 0.03 | 23 ± 8 | 1 | … | … | … | … | … | … | … |
| 9059 | 08080787-4726533 | 21.4 ± 0.3 | 3322 ± 36 | … | 0.871 ± 0.012 | -0.26 ± 0.14 | 538 ± 14 | 1 | B | Y | … | … | Y | Y | … |
| 9642 | 08104745-4703503 | 18.0 ± 0.3 | 3406 ± 128 | … | 0.869 ± 0.005 | -0.27 ± 0.14 | 169 ± 6 | 1 | A | Y | … | Y | Y | Y | … |

**Table C.3.** continued.

| ID | CNAME | RV (km s$^{-1}$) | $T_{\rm eff}$ (K) | $\log g$ (dex) | $\gamma^a$ | [Fe/H] (dex) | $EW({\rm Li})^b$ (mÅ) | $EW({\rm Li})$ error flag$^c$ | Jeffries 2014$^d$ | Damiani 2014$^d$ | Spina 2014$^d$ | Frasca 2015$^d$ | Prisinzano 2016$^d$ | Final members$^e$ | Non-mem with Li$^f$ |
|---|---|---|---|---|---|---|---|---|---|---|---|---|---|---|---|
| 9364 | 08093868-4737070 | 21.0 ± 0.3 | 3532 ± 5 | … | 0.841 ± 0.007 | -0.25 ± 0.14 | 415 ± 7 | 1 | B | … | … | Y | Y | Y | … |
| 9060 | 08080817-4742304 | 58.8 ± 0.2 | 4949 ± 150 | … | 1.010 ± 0.004 | -0.15 ± 0.05 | … | … | … | … | … | … | … | … | G |
| 9643 | 08104751-4706259 | 124.0 ± 0.2 | 4739 ± 204 | … | 1.016 ± 0.005 | -0.23 ± 0.15 | 44 ± 9 | 1 | … | … | … | … | … | … | G |
| 9365 | 08093874-4742045 | 59.9 ± 0.2 | 4899 ± 217 | … | 1.015 ± 0.003 | -0.11 ± 0.09 | 369 ± 5 | 1 | … | … | … | … | … | … | Li-rich G |
| 9931 | 08115872-4735093 | 49.1 ± 0.2 | 4461 ± 171 | 2.17 ± 0.18 | 1.024 ± 0.003 | 0.16 ± 0.03 | 17 ± 5 | 1 | … | … | … | … | … | … | G |
| 9366 | 08093894-4717051 | 120.9 ± 0.2 | 4524 ± 105 | … | 1.027 ± 0.005 | -0.13 ± 0.06 | … | … | … | … | … | … | … | … | G |
| 9061 | 08080856-4708003 | 12.0 ± 0.2 | 4697 ± 18 | 2.67 ± 0.04 | 1.010 ± 0.007 | 0.06 ± 0.07 | <49 | 3 | … | … | … | … | … | … | G |
| 9644 | 08104790-4707148 | 9.6 ± 0.2 | 5062 ± 109 | … | 1.018 ± 0.002 | -0.05 ± 0.11 | <15 | 3 | … | … | … | … | … | … | G |
| 9933 | 08115921-4703251 | 17.7 ± 0.3 | 5273 ± 19 | … | 0.996 ± 0.005 | -0.24 ± 0.12 | <19 | 3 | … | … | … | … | … | … | … |
| 9078 | 08081447-4720181 | 10.2 ± 0.2 | 4944 ± 36 | 2.66 ± 0.08 | 1.011 ± 0.002 | -0.10 ± 0.12 | <29 | 3 | … | … | … | … | … | … | G |
| 9934 | 08115927-4729172 | 29.9 ± 0.2 | 4713 ± 215 | … | 1.029 ± 0.004 | -0.03 ± 0.08 | <42 | 3 | … | … | … | … | … | … | G |
| 9367 | 08093920-4721387 | 17.4 ± 0.2 | 5390 ± 7 | … | 0.992 ± 0.001 | 0.09 ± 0.01 | 303 ± 2 | 1 | B | … | Y | … | Y | Y | Y | … |
| 9935 | 08115956-4657100 | 17.3 ± 0.3 | 3470 ± 11 | … | 0.855 ± 0.009 | -0.23 ± 0.15 | 141 ± 19 | 1 | A | … | … | Y | Y | Y | … |
| 9079 | 08081474-4745221 | 79.7 ± 0.2 | 4533 ± 92 | … | 1.032 ± 0.004 | -0.31 ± 0.02 | <36 | 3 | … | … | … | … | … | … | G |
| 9645 | 08104829-4746049 | 17.6 ± 0.3 | 3640 ± 107 | … | 0.838 ± 0.007 | -0.21 ± 0.14 | 437 ± 6 | 1 | A | Y | … | Y | Y | Y | … |
| 9368 | 08093936-4739060 | 14.6 ± 0.2 | 3512 ± 38 | … | 0.864 ± 0.005 | -0.24 ± 0.14 | 225 ± 27 | 1 | A | … | … | Y | Y | Y | … |
| 9080 | 08081496-4708181 | 9.7 ± 0.2 | 5440 ± 174 | … | 1.004 ± 0.003 | -0.26 ± 0.04 | … | … | … | … | … | … | … | … | … |
| 9936 | 08120000-4710418 | 136.4 ± 0.2 | 4110 ± 153 | … | 1.066 ± 0.004 | -0.18 ± 0.06 | 109 ± 9 | 1 | … | … | … | … | … | … | G |
| 9081 | 08081498-4715380 | 19.3 ± 0.2 | 3914 ± 144 | … | 0.868 ± 0.004 | -0.02 ± 0.13 | 499 ± 12 | 1 | B | Y | … | Y | Y | Y | … |
| 9937 | 08120059-4730425 | 108.4 ± 0.2 | 4475 ± 92 | … | 1.035 ± 0.003 | -0.31 ± 0.19 | <19 | 3 | … | … | … | … | … | … | G |
| 9082 | 08081544-4726285 | 14.7 ± 0.2 | 4406 ± 101 | 4.56 ± 0.13 | 0.915 ± 0.003 | 0.01 ± 0.12 | <26 | 3 | … | … | … | … | … | … | … |
| 9369 | 08093963-4731103 | 20.9 ± 0.4 | 3313 ± 26 | … | 0.876 ± 0.014 | -0.27 ± 0.14 | … | … | B | … | … | … | Y | Y | … |
| 9370 | 08093983-4737212 | 73.8 ± 0.2 | 4169 ± 177 | … | 1.058 ± 0.002 | -0.16 ± 0.11 | 52 ± 2 | 1 | … | … | … | … | … | … | G |
| 9938 | 08120109-4737113 | 77.0 ± 0.3 | 5207 ± 199 | … | 1.006 ± 0.008 | -0.40 ± 0.18 | … | … | B | … | … | … | … | … | … |
| 9646 | 08104962-4713314 | 16.5 ± 0.3 | 3482 ± 92 | … | 0.856 ± 0.007 | -0.25 ± 0.15 | <100 | 3 | … | Y | … | … | Y | Y | … |
| 9083 | 08081611-4735232 | 70.4 ± 0.2 | 4375 ± 28 | … | 1.046 ± 0.003 | -0.16 ± 0.03 | <25 | 3 | … | … | … | … | … | … | G |
| 9384 | 08094453-4655232 | 90.5 ± 0.3 | 5059 ± 75 | … | 1.011 ± 0.006 | -0.15 ± 0.12 | <10 | 3 | … | … | … | … | … | … | G |
| 9084 | 08081615-4746089 | 52.3 ± 0.2 | 6021 ± 62 | 4.13 ± 0.11 | 0.994 ± 0.002 | -0.32 ± 0.26 | 43 ± 5 | 1 | … | … | … | … | … | … | … |
| 9085 | 08081645-4702264 | 34.9 ± 0.2 | 6477 ± 32 | 4.18 ± 0.07 | 1.001 ± 0.002 | -0.07 ± 0.03 | <11 | 3 | … | … | … | … | … | … | … |
| 9385 | 08094478-4720441 | 18.3 ± 0.2 | 4297 ± 315 | … | 0.921 ± 0.003 | -0.06 ± 0.14 | 527 ± 8 | 1 | B | Y | … | Y | Y | Y | … |
| 9647 | 08104993-4707477 | 23.4 ± 1.5 | 3382 ± 5 | … | 0.913 ± 0.010 | -0.23 ± 0.14 | 473 ± 15 | 1 | B | … | … | Y | Y | Y | … |
| 9086 | 08081668-4724444 | 17.2 ± 0.4 | 3321 ± 3 | … | 0.886 ± 0.017 | -0.26 ± 0.14 | 606 ± 22 | 1 | A | … | … | … | Y | Y | … |
| 9386 | 08094490-4747136 | 37.4 ± 0.2 | 5048 ± 101 | 3.46 ± 0.04 | 0.998 ± 0.003 | -0.02 ± 0.09 | <26 | 3 | … | … | … | … | … | … | … |
| 9087 | 08081723-4740092 | 20.8 ± 0.2 | 4439 ± 78 | 3.92 ± 0.01 | 0.952 ± 0.003 | -0.13 ± 0.21 | 444 ± 4 | 1 | B | Y | … | Y | Y | Y | … |
| 9648 | 08105006-4741592 | 39.6 ± 2.6 | … | … | … | … | … | … | … | … | … | … | … | … | … |
| 9088 | 08081726-4721353 | 79.1 ± 0.2 | 5027 ± 42 | 3.00 ± 0.19 | 1.012 ± 0.003 | -0.17 ± 0.11 | <12 | 3 | … | … | … | … | … | … | G |
| 9387 | 08094519-4719061 | 15.6 ± 0.5 | 3365 ± 62 | … | 0.864 ± 0.016 | -0.29 ± 0.13 | … | … | A | Y | … | … | Y | … | … |
| 9649 | 08105105-4725512 | -245.5 ± 329.3 | … | … | … | … | … | … | … | … | … | … | … | … | … |
| 9388 | 08094519-4719499 | 8.8 ± 0.2 | 6886 ± 34 | … | … | 0.19 ± 0.03 | … | … | … | … | … | … | … | … | … |
| 9650 | 08105119-4714368 | -430.9 ± 1.5 | … | … | … | … | … | … | … | … | … | … | … | … | … |
| 9090 | 08081780-4722457 | 18.7 ± 0.4 | 3294 ± 12 | … | 0.896 ± 0.018 | … | 645 ± 20 | 1 | A | … | … | … | Y | Y | … |
| 9389 | 08094536-4721101 | 16.9 ± 0.3 | 3303 ± 5 | … | 0.893 ± 0.011 | -0.23 ± 0.15 | 626 ± 11 | 1 | A | Y | … | Y | Y | Y | … |
| 9651 | 08105128-4725499 | 134.9 ± 0.2 | 3529 ± 123 | … | … | … | … | … | … | … | … | … | … | … | … |
| 9091 | 08081781-4745064 | 60.4 ± 0.3 | 3918 ± 81 | … | … | … | … | … | … | … | … | … | … | … | … |
| 9652 | 08105146-4725482 | 416.5 ± 3.0 | … | … | … | … | … | … | … | … | … | … | … | … | … |
| 9092 | 08081788-4744531 | 169.0 ± 19.2 | … | … | … | … | … | … | … | … | … | … | … | … | … |
| 9390 | 08094568-4656238 | 112.9 ± 0.2 | 4471 ± 128 | … | 1.041 ± 0.004 | -0.24 ± 0.12 | 57 ± 11 | 1 | … | … | … | … | … | … | G |
| 9653 | 08105161-4704579 | 17.8 ± 0.3 | 3286 ± 9 | 4.69 ± 0.18 | 0.855 ± 0.015 | -0.26 ± 0.14 | 678 ± 18 | 1 | A | Y | … | … | Y | Y | … |
| 9093 | 08081831-4654064 | 57.3 ± 0.3 | 3674 ± 115 | 4.62 ± 0.11 | 0.808 ± 0.010 | -0.22 ± 0.12 | … | … | … | … | … | … | … | … | … |
| 9391 | 08094598-4713431 | 50.2 ± 0.2 | 4974 ± 160 | … | 1.009 ± 0.004 | -0.13 ± 0.19 | 24 ± 2 | 1 | … | … | … | … | … | … | … |
| 9668 | 08105597-4714184 | 92.8 ± 0.2 | 4934 ± 2 | 2.72 ± 0.03 | 1.012 ± 0.005 | -0.15 ± 0.10 | <20 | 3 | … | … | … | … | … | … | G |
| 9094 | 08081920-4716505 | 92.7 ± 0.2 | 4731 ± 43 | … | 1.021 ± 0.004 | -0.32 ± 0.05 | 24 ± 3 | 1 | … | … | … | … | … | … | G |
| 9392 | 08094609-4657202 | 38.0 ± 0.2 | 5010 ± 83 | … | 1.016 ± 0.003 | -0.08 ± 0.14 | <11 | 3 | … | … | … | … | … | … | G |
| 9669 | 08105600-4740069 | 21.2 ± 0.3 | 3426 ± 149 | … | 0.866 ± 0.011 | -0.25 ± 0.14 | 468 ± 10 | 1 | B | Y | … | Y | Y | Y | … |
| 9095 | 08081956-4722133 | 40.8 ± 209.5 | … | … | … | … | … | … | … | … | … | … | … | … | … |
| 9393 | 08094617-4735385 | 41.1 ± 0.2 | 4917 ± 140 | … | 1.010 ± 0.003 | -0.22 ± 0.09 | <23 | 3 | … | … | … | … | … | … | … |
| 3434 | 08063616-4748206 | 15.2 ± 0.4 | … | … | … | … | … | … | … | … | N | … | … | n | … |
| 9670 | 08105628-4708018 | 224.3 ± 0.2 | 4233 ± 318 | … | 1.051 ± 0.003 | -0.38 ± 0.24 | … | … | … | … | … | … | … | … | G |
| 9096 | 08081991-4714244 | 153.6 ± 0.2 | 5113 ± 162 | … | 1.012 ± 0.003 | -0.19 ± 0.30 | <11 | 3 | … | … | … | … | … | … | G |

**Table C.3.** continued.

| ID | CNAME | RV (km s$^{-1}$) | $T_{\text{eff}}$ (K) | log g (dex) | $\gamma^a$ | [Fe/H] (dex) | EW(Li)$^b$ (mÅ) | EW(Li) error flag$^c$ | Jeffries 2014$^d$ | Damiani 2014$^d$ | Literature members Spina 2014$^d$ | Frasca 2015$^d$ | Prisinzano 2016$^d$ | Final members$^e$ | Non-mem with Li$^f$ |
|---|---|---|---|---|---|---|---|---|---|---|---|---|---|---|---|
| 9394 | 08094617-4745295 | 23.5 ± 0.3 | 3486 ± 41 | ... | 0.863 ± 0.007 | -0.26 ± 0.14 | 273 ± 16 | 1 | B | ... | ... | ... | Y | Y | ... |
| 8758 | 08063889-4742312 | 10.4 ± 0.2 | 4843 ± 94 | 2.61 ± 0.10 | 1.017 ± 0.002 | -0.01 ± 0.14 | <27 | 3 | ... | ... | ... | ... | ... | ... | G |
| 9671 | 08105647-4656329 | 29.7 ± 0.2 | 6489 ± 29 | 4.30 ± 0.06 | 0.999 ± 0.002 | 0.26 ± 0.02 | <4 | 3 | ... | ... | ... | ... | ... | ... | ... |
| 9134 | 08083417-4658067 | 31.2 ± 0.2 | 5173 ± 120 | ... | 1.007 ± 0.003 | -0.19 ± 0.14 | <10 | 3 | ... | ... | ... | ... | ... | ... | ... |
| 9672 | 08105656-4657560 | 26.4 ± 0.2 | 4722 ± 242 | ... | 1.013 ± 0.004 | -0.10 ± 0.18 | <23 | 3 | ... | ... | ... | ... | ... | ... | G |
| 8759 | 08063901-4734403 | 11.3 ± 0.2 | 4433 ± 169 | 1.65 ± 0.08 | 1.033 ± 0.001 | -0.31 ± 0.06 | ... | ... | ... | ... | ... | ... | ... | ... | G |
| 9135 | 08083418-4717201 | 44.6 ± 0.2 | 4558 ± 119 | 2.34 ± 0.19 | 1.024 ± 0.002 | 0.06 ± 0.09 | <48 | 3 | ... | ... | ... | ... | ... | ... | G |
| 9673 | 08105684-4653427 | 19.7 ± 0.3 | 3315 ± 11 | ... | 0.865 ± 0.015 | -0.26 ± 0.14 | 553 ± 13 | 1 | B | Y | ... | ... | Y | Y | ... |
| 8760 | 08063957-4735493 | 8.5 ± 0.3 | 3825 ± 44 | 4.55 ± 0.20 | 0.832 ± 0.005 | -0.17 ± 0.14 | ... | ... | ... | ... | ... | ... | ... | ... | ... |
| 9136 | 08083445-4702536 | 175.4 ± 0.2 | 3635 ± 135 | ... | ... | ... | ... | ... | ... | ... | ... | ... | ... | ... | ... |
| 9395 | 08094655-4711042 | 14.9 ± 0.2 | 3410 ± 132 | ... | 0.876 ± 0.006 | -0.28 ± 0.14 | 248 ± 6 | 1 | A | Y | ... | Y | Y | Y | ... |
| 9674 | 08105686-4739150 | 17.4 ± 0.3 | 3347 ± 73 | ... | 0.871 ± 0.013 | -0.26 ± 0.14 | 363 ± 13 | 1 | A | Y | ... | ... | Y | Y | ... |
| 9675 | 08105703-4655092 | 22.3 ± 0.2 | 5196 ± 31 | 3.53 ± 0.07 | 0.997 ± 0.005 | -0.13 ± 0.16 | <10 | 3 | ... | ... | ... | ... | ... | ... | ... |
| 8761 | 08064077-4736441 | 60.4 ± 0.2 | 4970 ± 116 | ... | 1.020 ± 0.002 | -0.11 ± 0.09 | 414 ± 3 | 1 | ... | ... | ... | ... | ... | ... | G |
| 9137 | 08083455-4715203 | 3.1 ± 0.3 | 7278 ± 40 | ... | 0.994 ± 0.002 | ... | ... | ... | ... | ... | ... | ... | ... | ... | ... |
| 8762 | 08064093-4746195 | 66.7 ± 0.2 | 5402 ± 229 | 4.21 ± 0.13 | 0.982 ± 0.002 | 0.09 ± 0.09 | <28 | 3 | ... | ... | ... | ... | ... | ... | ... |
| 8763 | 08064098-4731259 | 25.4 ± 0.2 | 4644 ± 198 | 2.60 ± 0.07 | 1.006 ± 0.003 | 0.01 ± 0.12 | <31 | 3 | ... | ... | ... | ... | ... | ... | G |
| 9138 | 08083489-4705594 | 94.0 ± 0.2 | 4491 ± 100 | ... | 1.039 ± 0.004 | -0.21 ± 0.08 | 29 ± 6 | 1 | ... | ... | ... | ... | ... | ... | G |
| 9676 | 08105738-4713021 | 30.4 ± 0.3 | 3545 ± 16 | 4.75 ± 0.03 | 0.782 ± 0.015 | -0.23 ± 0.15 | ... | ... | ... | ... | ... | ... | ... | ... | ... |
| 8764 | 08064166-4723463 | 39.3 ± 0.2 | 4715 ± 113 | ... | 1.023 ± 0.004 | -0.08 ± 0.13 | 40 ± 5 | 1 | ... | ... | ... | ... | ... | ... | G |
| 8765 | 08064167-4721374 | 21.2 ± 0.2 | 5744 ± 85 | 4.30 ± 0.09 | 0.989 ± 0.003 | 0.09 ± 0.05 | <29 | 3 | ... | ... | ... | ... | ... | ... | ... |
| 9677 | 08105740-4725130 | 19.0 ± 6.1 | ... | ... | ... | ... | ... | ... | ... | ... | ... | ... | ... | ... | ... |
| 9139 | 08083502-4740586 | 113.9 ± 0.2 | 4284 ± 150 | ... | 1.060 ± 0.004 | -0.19 ± 0.09 | ... | ... | ... | ... | ... | ... | ... | ... | G |
| 9396 | 08094687-4708478 | 83.3 ± 0.2 | 4742 ± 34 | 2.58 ± 0.03 | 1.016 ± 0.003 | -0.04 ± 0.02 | <36 | 3 | ... | ... | ... | ... | ... | ... | G |
| 8766 | 08064265-4703190 | 80.2 ± 0.2 | 4503 ± 82 | ... | 1.037 ± 0.002 | -0.17 ± 0.02 | ... | ... | ... | ... | ... | ... | ... | ... | G |
| 8767 | 08064307-4705345 | 23.3 ± 0.2 | 5742 ± 92 | ... | 1.003 ± 0.002 | -0.01 ± 0.12 | ... | ... | ... | ... | ... | ... | ... | ... | ... |
| 9678 | 08105748-4743056 | 162.6 ± 0.2 | 4553 ± 122 | ... | 1.043 ± 0.007 | ... | <14 | 3 | ... | ... | ... | ... | ... | ... | G |
| 9140 | 08083505-4745068 | 94.7 ± 0.2 | 4696 ± 17 | ... | 1.026 ± 0.003 | -0.29 ± 0.07 | 37 ± 9 | 1 | ... | ... | ... | ... | ... | ... | G |
| 8768 | 08064322-4746397 | 3.2 ± 0.2 | 4930 ± 142 | ... | 1.014 ± 0.002 | -0.08 ± 0.15 | <16 | 3 | ... | ... | ... | ... | ... | ... | ... |
| 9679 | 08105759-4656102 | 12.8 ± 0.3 | 3377 ± 52 | ... | 0.839 ± 0.015 | -0.23 ± 0.13 | <100 | 3 | ... | ... | ... | ... | Y | Y | ... |
| 8792 | 08065421-4730211 | 18.7 ± 0.2 | 5058 ± 97 | ... | 1.008 ± 0.002 | -0.15 ± 0.15 | <20 | 3 | ... | ... | ... | ... | ... | ... | ... |
| 9680 | 08105785-4732051 | 115.2 ± 0.2 | 4691 ± 117 | 2.51 ± 0.14 | 1.012 ± 0.003 | -0.08 ± 0.15 | <36 | 3 | ... | ... | ... | ... | ... | ... | G |
| 9430 | 08095457-4734423 | 9.9 ± 0.2 | 3433 ± 125 | ... | 0.843 ± 0.004 | -0.23 ± 0.16 | ... | ... | A | ... | ... | Y | ... | ... | ... |
| 8793 | 08065422-4708484 | 23.1 ± 0.2 | 4831 ± 62 | 2.42 ± 0.11 | 1.014 ± 0.005 | -0.09 ± 0.10 | ... | ... | ... | ... | ... | ... | ... | ... | G |
| 8794 | 08065427-4703376 | -13.9 ± 0.2 | 4606 ± 96 | 2.42 ± 0.18 | 1.022 ± 0.002 | -0.01 ± 0.03 | <38 | 3 | ... | ... | ... | ... | ... | ... | G |
| 9681 | 08105815-4738466 | 20.0 ± 0.3 | 3630 ± 112 | 4.66 ± 0.08 | 0.793 ± 0.011 | -0.24 ± 0.15 | ... | ... | ... | ... | ... | ... | ... | ... | ... |
| 9141 | 08083622-4653341 | 426.6 ± 3.2 | ... | ... | ... | ... | ... | ... | ... | ... | ... | ... | ... | ... | ... |
| 8795 | 08065465-4655016 | -12.8 ± 0.2 | 5036 ± 111 | ... | 1.030 ± 0.004 | ... | <30 | 3 | ... | ... | ... | ... | ... | ... | G |
| 9432 | 08095470-4734543 | 19.0 ± 0.2 | 4780 ± 59 | ... | 1.000 ± 0.004 | 0.03 ± 0.09 | <23 | 3 | ... | ... | ... | ... | ... | ... | ... |
| 9682 | 08105821-4739576 | 35.9 ± 0.2 | 4822 ± 71 | 2.63 ± 0.09 | 1.016 ± 0.002 | -0.04 ± 0.07 | <24 | 3 | ... | ... | ... | ... | ... | ... | ... |
| 9142 | 08083656-4731392 | 18.7 ± 0.2 | 4733 ± 160 | 2.34 ± 0.11 | 1.024 ± 0.004 | -0.10 ± 0.07 | <23 | 3 | ... | ... | ... | ... | ... | ... | G |
| 8796 | 08065469-4657241 | 19.4 ± 0.2 | 4752 ± 276 | ... | 0.975 ± 0.002 | 0.02 ± 0.07 | 450 ± 7 | 1 | B | ... | ... | Y | Y | Y | ... |
| 9683 | 08105833-4705225 | 86.1 ± 0.2 | 4536 ± 95 | ... | 1.037 ± 0.003 | -0.19 ± 0.05 | 34 ± 3 | 1 | ... | ... | ... | ... | ... | ... | G |
| 8797 | 08065507-4735336 | 20.0 ± 0.3 | 6467 ± 97 | ... | 0.999 ± 0.002 | -0.06 ± 0.04 | 71 ± 10 | 1 | ... | ... | ... | ... | ... | ... | ... |
| 9433 | 08095479-4707280 | 50.6 ± 0.2 | 5027 ± 102 | ... | 1.009 ± 0.003 | -0.06 ± 0.09 | <30 | 3 | ... | ... | ... | ... | ... | ... | ... |
| 9684 | 08105843-4724442 | 79.0 ± 0.2 | 4924 ± 198 | 2.98 ± 0.18 | 1.004 ± 0.002 | -0.41 ± 0.17 | ... | ... | ... | ... | ... | ... | ... | ... | ... |
| 3464 | 08083707-4727371 | ... | 7402 ± 320 | 4.08 ± 0.22 | ... | -0.15 ± 0.18 | <11 | 3 | ... | ... | N | ... | ... | n | NG |
| 8798 | 08065515-4746488 | 87.1 ± 0.2 | 4433 ± 167 | 1.91 ± 0.13 | 1.027 ± 0.002 | -0.09 ± 0.05 | <41 | 3 | ... | ... | ... | ... | ... | ... | G |
| 9710 | 08110601-4726209 | 16.9 ± 0.2 | 3900 ± 192 | ... | 0.852 ± 0.003 | -0.13 ± 0.13 | 472 ± 5 | 1 | A | Y | ... | Y | Y | Y | ... |
| 8799 | 08065527-4724416 | 131.7 ± 0.3 | 4987 ± 155 | ... | 1.007 ± 0.006 | -0.21 ± 0.31 | ... | ... | ... | ... | ... | ... | ... | ... | ... |
| 9434 | 08095519-4739497 | 76.1 ± 0.2 | 4356 ± 214 | ... | 1.033 ± 0.005 | -0.08 ± 0.09 | <30 | 3 | ... | ... | ... | ... | ... | ... | G |
| 9711 | 08110659-4746173 | 25.3 ± 0.2 | 5221 ± 105 | 4.12 ± 0.04 | 0.977 ± 0.004 | 0.10 ± 0.06 | 212 ± 6 | 1 | ... | ... | ... | ... | ... | ... | ... |
| 3436 | 08065592-4704528 | 101.0 ± 0.6 | 4543 ± 8 | 2.50 ± 0.09 | 1.014 ± 0.002 | 0.09 ± 0.05 | 45 ± 12 | 1 | ... | ... | N | ... | ... | n | G |
| 3465 | 08083759-4736010 | 9.7 ± 0.6 | 5898 ± 22 | 3.81 ± 0.01 | 0.997 ± 0.001 | -0.34 ± 0.01 | 42 ± 4 | 2 | ... | ... | N | ... | ... | n | NG |
| 8800 | 08065618-4721015 | 17.4 ± 0.5 | 3263 ± 14 | ... | 0.902 ± 0.018 | ... | 637 ± 94 | 1 | A | ... | ... | ... | Y | Y | ... |
| 9712 | 08110660-4724259 | 46.4 ± 0.2 | 5078 ± 188 | ... | 1.015 ± 0.003 | -0.18 ± 0.07 | 27 ± 2 | 1 | ... | ... | ... | ... | ... | ... | G |
| 9144 | 08083772-4703454 | 86.7 ± 0.2 | 4839 ± 78 | 2.69 ± 0.05 | 1.016 ± 0.003 | -0.13 ± 0.09 | ... | ... | ... | ... | ... | ... | ... | ... | G |
| 9435 | 08095552-4711225 | 17.7 ± 0.3 | 3363 ± 40 | ... | 0.875 ± 0.010 | -0.25 ± 0.14 | 489 ± 18 | 1 | A | Y | ... | ... | Y | Y | ... |
| 8801 | 08065664-4703000 | 13.4 ± 0.2 | 5192 ± 79 | ... | 0.966 ± 0.004 | ... | 290 ± 8 | 1 | ... | Y | ... | ... | ... | Y | ... |

**Table C.3.** continued.

| ID | CNAME | RV (km s$^{-1}$) | $T_{\rm eff}$ (K) | $\log g$ (dex) | $\gamma^a$ | [Fe/H] (dex) | $EW({\rm Li})^b$ (mÅ) | $EW({\rm Li})$ error flag$^c$ | Jeffries 2014$^d$ | Damiani 2014$^d$ | Literature members Spina 2014$^d$ | Frasca 2015$^d$ | Prisinzano 2016$^d$ | Final members$^e$ | Non-mem with Li$^f$ |
|---|---|---|---|---|---|---|---|---|---|---|---|---|---|---|---|
| 9145 | 08083827-4745000 | 17.3 ± 0.3 | 3453 ± 56 | … | 0.864 ± 0.012 | -0.27 ± 0.13 | 561 ± 12 | 1 | A | Y | … | Y | Y | Y | … |
| 9436 | 08095576-4724194 | 27.9 ± 0.2 | 5871 ± 154 | 4.33 ± 0.01 | 0.989 ± 0.002 | 0.07 ± 0.05 | … | … | … | … | … | … | … | … | … |
| 8802 | 08065672-4712133 | 16.3 ± 0.6 | 3263 ± 14 | … | 0.866 ± 0.017 | … | 521 ± 15 | 1 | A | … | … | … | Y | Y | … |
| 8803 | 08065682-4715493 | 33.2 ± 0.2 | 4670 ± 93 | 2.48 ± 0.15 | 1.020 ± 0.004 | 0.03 ± 0.09 | <41 | 3 | … | … | … | … | … | … | G |
| 9713 | 08110686-4724254 | 31.6 ± 55.1 | … | … | … | … | … | … | … | … | … | … | … | … | … |
| 9437 | 08095581-4715426 | 16.5 ± 0.3 | 3545 ± 39 | … | 0.846 ± 0.007 | -0.22 ± 0.14 | … | … | … | … | … | … | Y | … | … |
| 9714 | 08110698-4657247 | 110.1 ± 0.2 | 4531 ± 14 | … | 1.040 ± 0.005 | -0.25 ± 0.09 | <15 | 3 | … | … | … | … | … | … | G |
| 9146 | 08083838-4728187 | 20.1 ± 0.2 | 4725 ± 360 | … | 0.957 ± 0.004 | 0.00 ± 0.08 | 433 ± 4 | 1 | B | Y | … | Y | Y | Y | … |
| 8804 | 08065702-4712414 | 33.4 ± 0.2 | 5165 ± 173 | … | 1.015 ± 0.005 | -0.15 ± 0.06 | … | … | … | … | … | … | … | … | G |
| 9438 | 08095600-4702028 | 27.5 ± 0.2 | 5265 ± 207 | … | 0.999 ± 0.003 | -0.33 ± 0.15 | <15 | 3 | … | … | … | … | … | … | … |
| 9439 | 08095611-4717335 | 17.5 ± 0.4 | 3294 ± 13 | … | 0.855 ± 0.017 | … | 504 ± 15 | 1 | A | … | … | … | Y | Y | … |
| 8805 | 08065712-4744354 | 111.6 ± 0.2 | 3850 ± 178 | … | 1.048 ± 0.003 | -0.18 ± 0.17 | … | … | … | … | … | … | … | … | G |
| 9440 | 08095623-4704353 | 17.6 ± 0.2 | 3555 ± 67 | … | 0.853 ± 0.005 | -0.21 ± 0.14 | 413 ± 16 | 1 | A | Y | … | Y | Y | Y | … |
| 8806 | 08065722-4731514 | 59.6 ± 0.2 | 6204 ± 28 | 4.27 ± 0.06 | 0.995 ± 0.002 | 0.42 ± 0.02 | 50 ± 7 | 1 | … | … | … | … | … | … | … |
| 9715 | 08110763-4701114 | 73.9 ± 0.2 | 4549 ± 80 | … | 1.029 ± 0.003 | -0.20 ± 0.07 | <21 | 3 | … | … | … | … | … | … | G |
| 9441 | 08095637-4713351 | 17.5 ± 0.6 | 3252 ± 19 | … | 0.878 ± 0.018 | … | 639 ± 98 | 1 | A | … | … | … | Y | Y | … |
| 8807 | 08065724-4653085 | 105.2 ± 0.2 | 4627 ± 79 | … | 1.033 ± 0.006 | -0.20 ± 0.15 | <10 | 3 | … | … | … | … | … | … | G |
| 8808 | 08065781-4744443 | 34.9 ± 0.5 | 5191 ± 44 | … | 0.945 ± 0.017 | -0.27 ± 0.04 | <50 | 3 | … | … | … | … | … | … | … |
| 9162 | 08084354-4658456 | 38.2 ± 0.4 | 3452 ± 89 | … | 0.854 ± 0.015 | -0.27 ± 0.14 | … | … | … | … | … | … | … | … | … |
| 8809 | 08065823-4721241 | 47.7 ± 0.2 | 4918 ± 176 | … | 1.010 ± 0.004 | -0.21 ± 0.08 | <28 | 3 | … | … | … | … | … | … | G |
| 9716 | 08110799-4734217 | 33.3 ± 0.2 | 4276 ± 233 | 4.55 ± 0.04 | 0.885 ± 0.015 | -0.02 ± 0.11 | … | … | … | … | … | … | Y | … | … |
| 9163 | 08084356-4744423 | 38.0 ± 0.2 | 4954 ± 134 | … | 1.016 ± 0.002 | -0.13 ± 0.15 | <14 | 3 | … | … | … | … | … | … | G |
| 8810 | 08065939-4731072 | 95.1 ± 0.2 | 5136 ± 87 | … | 1.009 ± 0.003 | -0.25 ± 0.21 | <13 | 3 | … | … | … | … | … | … | G |
| 9717 | 08110823-4741364 | 32.1 ± 0.2 | 5129 ± 154 | … | 1.004 ± 0.003 | -0.31 ± 0.08 | <15 | 3 | … | … | … | … | … | … | G |
| 9164 | 08084378-4744247 | -4.1 ± 0.2 | 7285 ± 42 | … | 1.003 ± 0.002 | … | … | … | … | … | … | … | … | … | … |
| 8811 | 08065948-4729191 | 11.6 ± 1.1 | 3349 ± 22 | … | 0.858 ± 0.017 | … | … | … | … | … | … | … | Y | … | … |
| 9718 | 08110838-4737320 | 46.1 ± 0.2 | 5065 ± 55 | … | 1.015 ± 0.004 | -0.09 ± 0.07 | <24 | 3 | … | … | … | … | … | … | G |
| 9165 | 08084383-4703324 | 55.8 ± 0.2 | 5024 ± 87 | … | 1.017 ± 0.003 | -0.10 ± 0.13 | <20 | 3 | … | … | … | … | … | … | G |
| 8812 | 08065968-4712149 | 39.4 ± 0.2 | 6260 ± 42 | 4.14 ± 0.09 | 0.999 ± 0.002 | -0.17 ± 0.03 | 41 ± 6 | 1 | … | … | … | … | … | … | … |
| 9719 | 08110843-4703243 | 104.9 ± 0.2 | 4474 ± 121 | … | 1.048 ± 0.005 | -0.22 ± 0.10 | 78 ± 3 | 1 | … | … | … | … | … | … | G |
| 8813 | 08065984-4658511 | 32.1 ± 0.3 | 3296 ± 7 | … | 0.902 ± 0.011 | -0.25 ± 0.14 | … | … | … | … | … | … | Y | … | … |
| 9443 | 08095750-4715287 | 50.9 ± 0.2 | 4853 ± 97 | 2.86 ± 0.14 | 1.010 ± 0.004 | -0.07 ± 0.06 | 42 ± 10 | 1 | … | … | … | … | … | … | … |
| 9720 | 08110849-4658146 | 29.0 ± 0.2 | 7169 ± 38 | … | 1.015 ± 0.002 | … | … | … | … | … | … | … | … | … | … |
| 8860 | 08071233-4659264 | 13.5 ± 0.2 | 4098 ± 189 | … | 0.884 ± 0.004 | -0.08 ± 0.07 | 114 ± 11 | 1 | … | … | … | … | Y | Y | … |
| 9458 | 08100015-4700080 | 17.1 ± 0.6 | 3292 ± 4 | … | 0.869 ± 0.015 | -0.26 ± 0.14 | … | … | A | … | … | … | Y | Y | … |
| 8861 | 08071235-4659397 | 86.5 ± 0.2 | 4546 ± 116 | … | 1.051 ± 0.003 | … | … | … | … | … | … | … | … | … | G |
| 9166 | 08084464-4707197 | -2.6 ± 0.2 | 4534 ± 202 | 4.50 ± 0.09 | 0.927 ± 0.004 | -0.06 ± 0.04 | <5 | 3 | … | … | … | … | … | … | … |
| 9459 | 08100043-4702239 | 20.7 ± 0.2 | 4556 ± 33 | 2.01 ± 0.03 | 1.029 ± 0.002 | -0.23 ± 0.04 | <17 | 3 | … | … | … | … | … | … | G |
| 9721 | 08110894-4706522 | 16.7 ± 0.4 | 3309 ± 35 | 4.69 ± 0.12 | 0.850 ± 0.017 | -0.25 ± 0.14 | 597 ± 56 | 1 | A | Y | … | … | Y | Y | … |
| 9167 | 08084469-4706591 | 17.1 ± 0.2 | 3627 ± 9 | … | 0.858 ± 0.005 | -0.19 ± 0.12 | 474 ± 14 | 1 | A | Y | … | Y | Y | Y | … |
| 8862 | 08071240-4705115 | 135.3 ± 0.2 | 4732 ± 109 | 2.20 ± 0.11 | 1.022 ± 0.004 | -0.35 ± 0.01 | <16 | 3 | … | … | … | … | … | … | G |
| 9460 | 08100047-4724148 | 27.4 ± 0.3 | 3607 ± 53 | … | 0.818 ± 0.008 | -0.21 ± 0.14 | <100 | 3 | … | … | … | … | … | … | … |
| 9722 | 08110913-4727000 | 74.4 ± 0.2 | 4652 ± 75 | … | 1.016 ± 0.003 | … | <44 | 3 | … | … | … | … | … | … | G |
| 9168 | 08084481-4731317 | 164.9 ± 0.2 | 4575 ± 102 | … | 1.042 ± 0.005 | -0.60 ± 0.20 | <11 | 3 | … | … | … | … | … | … | … |
| 8863 | 08071254-4722415 | -8.0 ± 0.2 | 5010 ± 121 | … | 1.017 ± 0.004 | -0.08 ± 0.04 | <21 | 3 | … | … | … | … | … | … | G |
| 9461 | 08100053-4717581 | 16.9 ± 0.2 | 3612 ± 23 | … | 0.856 ± 0.005 | -0.21 ± 0.12 | 464 ± 5 | 1 | A | Y | … | Y | Y | Y | … |
| 9169 | 08084502-4715479 | 91.5 ± 0.2 | 4577 ± 80 | … | 1.040 ± 0.004 | -0.23 ± 0.12 | … | … | … | … | … | … | … | … | G |
| 8864 | 08071271-4708276 | 36.5 ± 0.2 | 4760 ± 84 | 2.57 ± 0.11 | 1.016 ± 0.001 | 0.01 ± 0.12 | <32 | 3 | … | … | … | … | … | … | G |
| 9462 | 08100058-4657467 | 81.2 ± 0.2 | 5034 ± 24 | 2.93 ± 0.18 | 1.015 ± 0.005 | -0.18 ± 0.16 | <22 | 3 | … | … | … | … | … | … | G |
| 9723 | 08110934-4718501 | 100.2 ± 0.2 | 4507 ± 161 | … | 1.029 ± 0.005 | -0.15 ± 0.08 | <46 | 3 | … | … | … | … | … | … | G |
| 9170 | 08084532-4701292 | 88.7 ± 0.2 | 4981 ± 133 | … | 1.017 ± 0.003 | -0.09 ± 0.13 | 346 ± 2 | 1 | … | … | … | … | … | … | Li-rich G |
| 8865 | 08071311-4721123 | 30.5 ± 0.4 | 3556 ± 34 | … | 0.861 ± 0.019 | … | … | … | … | … | … | … | … | … | … |
| 9724 | 08110963-4738540 | 23.0 ± 0.2 | 4811 ± 99 | 2.40 ± 0.04 | 1.014 ± 0.003 | -0.20 ± 0.03 | <17 | 3 | … | … | … | … | … | … | G |
| 9171 | 08084583-4709032 | 14.6 ± 0.2 | 5818 ± 135 | 4.10 ± 0.09 | 0.993 ± 0.002 | 0.09 ± 0.06 | 185 ± 2 | 1 | … | Y | … | … | … | Y | … |
| 3440 | 08071363-4725156 | 14.2 ± 0.6 | 5778 ± 3 | 4.35 ± 0.04 | 0.980 ± 0.002 | 0.02 ± 0.03 | 31 ± 1 | 2 | … | … | N | … | … | n | NG |
| 9463 | 08100066-4702503 | 16.3 ± 0.2 | 5670 ± 2 | 4.26 ± 0.11 | 0.988 ± 0.003 | 0.05 ± 0.11 | <14 | 3 | … | … | … | … | … | … | … |
| 9725 | 08110976-4657363 | 3.0 ± 0.4 | 3473 ± 90 | 4.63 ± 0.18 | 0.815 ± 0.015 | -0.22 ± 0.13 | <100 | 3 | … | … | … | … | … | … | … |
| 9464 | 08100066-4744550 | 17.1 ± 0.3 | 3534 ± 56 | … | 0.856 ± 0.007 | -0.18 ± 0.17 | 332 ± 20 | 1 | A | Y | … | Y | Y | Y | … |
| 9465 | 08100070-4706262 | 107.9 ± 0.2 | 4596 ± 158 | … | 1.026 ± 0.003 | -0.30 ± 0.08 | <31 | 3 | … | … | … | … | … | … | G |

| ID | CNAME | RV (km s$^{-1}$) | $T_{\rm eff}$ (K) | logg (dex) | $\gamma^a$ | [Fe/H] (dex) | EW(Li)$^b$ (mÅ) | EW(Li) error flag$^c$ | Jeffries 2014$^d$ | Damiani 2014$^d$ | Literature members Spina 2014$^d$ | Frasca 2015$^d$ | Prisinzano 2016$^d$ | Final members$^e$ | Non-mem with Li$^f$ |
|---|---|---|---|---|---|---|---|---|---|---|---|---|---|---|---|
| 9726 | 08110978-4711085 | 84.0 ± 0.2 | 5108 ± 4 | … | 1.015 ± 0.003 | -0.12 ± 0.10 | <27 | 3 | … | … | … | … | … | … | G |
| 8866 | 08071377-4735166 | 51.5 ± 0.2 | 4882 ± 64 | … | 1.015 ± 0.003 | -0.11 ± 0.14 | <27 | 3 | … | … | … | … | … | … | G |
| 9172 | 08084651-4734447 | 106.0 ± 0.2 | 4784 ± 1 | 2.68 ± 0.07 | 1.010 ± 0.003 | -0.17 ± 0.15 | <23 | 3 | … | … | … | … | … | … | … |
| 9466 | 08100079-4744038 | 17.9 ± 0.3 | 3369 ± 118 | 4.76 ± 0.01 | 0.836 ± 0.008 | -0.28 ± 0.13 | … | … | … | … | … | … | Y | … | … |
| 3441 | 08071383-4736156 | 49.0 ± 0.6 | 4479 ± 28 | 2.03 ± 0.12 | 1.030 ± 0.001 | -0.09 ± 0.06 | 32 ± 11 | 1 | … | … | N | … | … | n | G |
| 9742 | 08111270-4726419 | 34.3 ± 0.2 | 4753 ± 166 | 2.57 ± 0.19 | 1.018 ± 0.002 | -0.04 ± 0.07 | 52 ± 2 | 1 | … | … | … | … | … | … | G |
| 8867 | 08071405-4656309 | 52.0 ± 0.2 | 4599 ± 171 | 2.28 ± 0.20 | 1.029 ± 0.004 | -0.01 ± 0.08 | <37 | 3 | … | … | … | … | … | … | G |
| 9743 | 08111277-4729236 | 77.2 ± 0.2 | 4531 ± 125 | … | 1.029 ± 0.003 | -0.28 ± 0.07 | 44 ± 10 | 1 | … | … | … | … | … | … | G |
| 9467 | 08100116-4710106 | 56.0 ± 0.2 | 4720 ± 226 | … | 1.021 ± 0.004 | -0.21 ± 0.13 | … | … | … | … | … | … | … | … | G |
| 8868 | 08071427-4706532 | 28.9 ± 0.2 | 4083 ± 211 | … | 1.028 ± 0.002 | 0.00 ± 0.15 | <38 | 3 | … | … | … | … | … | … | G |
| 9744 | 08111295-4733374 | 26.3 ± 0.3 | 6744 ± 40 | … | 0.993 ± 0.002 | 0.04 ± 0.03 | … | … | … | … | … | … | … | … | … |
| 9174 | 08084700-4739180 | 21.1 ± 0.3 | 3542 ± 37 | … | 0.840 ± 0.007 | -0.22 ± 0.14 | 341 ± 15 | 1 | B | Y | … | Y | Y | Y | … |
| 3442 | 08071501-4658153 | 25.2 ± 0.4 | 6300 ± 113 | 3.96 ± 0.15 | … | 0.00 ± 0.12 | <7 | 3 | … | … | N | … | … | n | … |
| 9745 | 08111297-4701320 | 77.7 ± 0.2 | 4772 ± 145 | … | 1.021 ± 0.003 | -0.16 ± 0.03 | 31 ± 11 | 1 | … | … | … | … | … | … | G |
| 9175 | 08084729-4708067 | 19.3 ± 0.3 | 3296 ± 3 | … | 0.880 ± 0.014 | -0.26 ± 0.14 | 628 ± 15 | 1 | B | … | … | … | Y | Y | … |
| 9468 | 08100149-4653398 | 124.1 ± 0.3 | … | … | … | … | … | … | … | … | … | … | … | … | … |
| 8869 | 08071520-4725599 | 108.7 ± 0.2 | 4525 ± 80 | … | 1.031 ± 0.003 | -0.24 ± 0.15 | … | … | … | … | … | … | … | … | … |
| 9176 | 08084729-4711458 | 50.6 ± 0.2 | 5214 ± 26 | … | 1.011 ± 0.004 | -0.16 ± 0.10 | <28 | 3 | … | … | … | … | … | … | … |
| 9469 | 08100153-4714148 | 555.8 ± 706.1 | … | … | … | … | … | … | … | … | … | … | … | … | … |
| 8870 | 08071539-4656122 | 84.8 ± 0.2 | 5079 ± 138 | … | 1.018 ± 0.003 | -0.19 ± 0.24 | … | … | … | … | … | … | … | … | G |
| 9747 | 08111345-4709172 | 43.3 ± 0.2 | 4611 ± 303 | 4.48 ± 0.07 | 0.928 ± 0.004 | -0.11 ± 0.09 | … | … | … | … | … | … | … | … | … |
| 9177 | 08084740-4719302 | 18.8 ± 0.3 | 3400 ± 119 | … | 0.857 ± 0.007 | -0.27 ± 0.14 | <100 | 3 | … | Y | … | … | Y | Y | … |
| 8871 | 08071540-4711440 | 21.0 ± 0.3 | 4210 ± 149 | … | 0.837 ± 0.004 | … | 513 ± 25 | 1 | B | Y | … | … | Y | Y | … |
| 8872 | 08071557-4720087 | 46.7 ± 0.2 | 4932 ± 106 | … | 1.016 ± 0.002 | -0.18 ± 0.12 | <17 | 3 | … | … | … | … | … | … | G |
| 9748 | 08111362-4653542 | -7.1 ± 0.2 | 5936 ± 209 | … | 1.003 ± 0.002 | -0.29 ± 0.03 | 47 ± 3 | 1 | … | … | … | … | … | … | … |
| 9178 | 08084743-4742349 | 18.8 ± 0.3 | 3366 ± 87 | … | 0.876 ± 0.006 | -0.25 ± 0.14 | 480 ± 27 | 1 | B | Y | … | Y | Y | Y | … |
| 8873 | 08071599-4657574 | -5.4 ± 0.2 | 5255 ± 137 | … | 0.997 ± 0.004 | -0.02 ± 0.05 | <23 | 3 | … | … | … | … | … | … | … |
| 9470 | 08100188-4739493 | 25.2 ± 0.5 | 3595 ± 96 | … | 0.829 ± 0.014 | -0.22 ± 0.14 | … | … | … | … | … | … | … | … | … |
| 9749 | 08111383-4732503 | 45.4 ± 0.2 | 5014 ± 7 | 3.13 ± 0.05 | 1.006 ± 0.003 | -0.04 ± 0.06 | <28 | 3 | … | … | … | … | … | … | … |
| 9471 | 08100201-4742041 | 20.9 ± 0.2 | 3767 ± 77 | … | 0.850 ± 0.014 | -0.17 ± 0.14 | 448 ± 5 | 1 | B | Y | … | Y | Y | Y | … |
| 8874 | 08071603-4700080 | 140.6 ± 0.2 | 4508 ± 49 | … | 1.031 ± 0.004 | -0.31 ± 0.12 | … | … | … | … | … | … | … | … | G |
| 8875 | 08071617-4715126 | 56.6 ± 0.2 | 4674 ± 60 | 2.48 ± 0.15 | 1.018 ± 0.003 | -0.09 ± 0.06 | <39 | 3 | … | … | … | … | … | … | G |
| 9750 | 08111383-4740113 | 4.2 ± 0.2 | 4987 ± 60 | … | 1.015 ± 0.002 | -0.02 ± 0.05 | <34 | 3 | … | … | … | … | … | … | G |
| 9213 | 08085623-4708133 | 23.6 ± 0.2 | 5583 ± 118 | … | 0.985 ± 0.004 | 0.08 ± 0.05 | <17 | 3 | … | … | … | … | … | … | … |
| 8876 | 08071630-4703346 | 28.8 ± 0.2 | 6270 ± 41 | 4.41 ± 0.09 | 0.991 ± 0.003 | 0.35 ± 0.03 | 42 ± 4 | 1 | … | … | … | … | … | … | … |
| 9751 | 08111405-4654493 | 71.0 ± 0.3 | 4708 ± 162 | … | 1.025 ± 0.004 | -0.24 ± 0.07 | … | … | … | … | … | … | … | … | … |
| 9472 | 08100229-4745123 | 16.6 ± 0.2 | 3535 ± 28 | … | 0.857 ± 0.006 | -0.22 ± 0.14 | 401 ± 5 | 1 | A | Y | … | Y | Y | Y | G |
| 9752 | 08111406-4729045 | 63.2 ± 0.2 | 4773 ± 55 | 2.54 ± 0.14 | 1.019 ± 0.002 | -0.13 ± 0.07 | <28 | 3 | … | … | … | … | … | … | G |
| 9214 | 08085655-4653342 | 25.7 ± 0.2 | 5189 ± 104 | … | 1.005 ± 0.004 | -0.28 ± 0.17 | <8 | 3 | … | … | … | … | … | … | … |
| 8877 | 08071669-4656141 | 82.7 ± 0.2 | 4548 ± 116 | … | 1.047 ± 0.004 | … | … | … | … | … | … | … | … | … | … |
| 8878 | 08071673-4722458 | 125.7 ± 0.2 | 4660 ± 64 | … | 1.022 ± 0.004 | -0.26 ± 0.03 | <30 | 3 | … | … | … | … | … | … | G |
| 9473 | 08100280-4736372 | 17.3 ± 0.3 | 3363 ± 68 | … | 0.882 ± 0.009 | -0.23 ± 0.15 | 505 ± 37 | 1 | A | Y | … | Y | Y | Y | … |
| 9215 | 08085661-4730350 | 423.1 ± 3.2 | … | … | … | … | … | … | … | … | … | … | Y | … | … |
| 9474 | 08100293-4709301 | 57.6 ± 0.3 | 5073 ± 26 | … | 1.030 ± 0.005 | 0.01 ± 0.04 | … | … | … | … | … | … | … | … | … |
| 8879 | 08071682-4655235 | 37.6 ± 0.3 | 3906 ± 75 | 4.61 ± 0.07 | 0.829 ± 0.007 | -0.31 ± 0.23 | … | … | … | … | … | … | … | … | … |
| 9753 | 08111443-4719345 | 83.5 ± 0.2 | 5016 ± 137 | … | 0.991 ± 0.004 | -0.35 ± 0.13 | <5 | 3 | … | … | … | … | … | … | … |
| 8880 | 08071727-4748277 | 45.7 ± 0.2 | 4561 ± 99 | 2.19 ± 0.18 | 1.036 ± 0.002 | -0.05 ± 0.04 | <29 | 3 | … | … | … | … | … | … | G |
| 9754 | 08111467-4659062 | 62.5 ± 0.2 | 4928 ± 55 | 3.02 ± 0.03 | 1.007 ± 0.002 | -0.07 ± 0.05 | <22 | 3 | … | … | … | … | … | … | … |
| 9216 | 08085678-4719079 | 48.9 ± 0.2 | 4797 ± 37 | 2.56 ± 0.03 | 1.021 ± 0.002 | -0.06 ± 0.04 | <27 | 3 | … | … | … | … | … | … | G |
| 8902 | 08072511-4735356 | 40.4 ± 0.2 | 5027 ± 43 | 3.16 ± 0.04 | 1.005 ± 0.003 | -0.02 ± 0.08 | <36 | 3 | … | … | … | … | … | … | … |
| 9755 | 08111480-4705207 | 16.3 ± 0.3 | 3396 ± 31 | … | 0.856 ± 0.010 | -0.24 ± 0.15 | 145 ± 8 | 1 | A | … | … | Y | Y | Y | … |
| 9217 | 08085702-4720151 | 114.0 ± 0.2 | 4664 ± 154 | … | 1.108 ± 0.005 | … | 141 ± 14 | 1 | … | … | … | … | … | … | Li-rich G |
| 9504 | 08100938-4701588 | 5.5 ± 0.2 | 6565 ± 114 | 4.15 ± 0.06 | 1.002 ± 0.001 | -0.44 ± 0.10 | <3 | 3 | … | … | … | … | … | … | … |
| 8903 | 08072538-4728178 | 20.8 ± 0.5 | 3330 ± 50 | … | 0.871 ± 0.012 | -0.27 ± 0.14 | … | … | B | … | … | Y | Y | Y | … |
| 9756 | 08111569-4729087 | 64.4 ± 0.2 | 4761 ± 124 | … | 1.019 ± 0.002 | -0.20 ± 0.14 | <16 | 3 | … | … | … | … | … | … | G |
| 9218 | 08085711-4746437 | 17.9 ± 0.3 | 3581 ± 81 | 4.64 ± 0.12 | 0.795 ± 0.014 | -0.22 ± 0.15 | … | … | … | … | … | … | … | … | … |
| 8904 | 08072539-4702334 | 82.5 ± 0.2 | 4929 ± 155 | … | 1.016 ± 0.003 | -0.16 ± 0.14 | … | … | … | … | … | … | … | … | G |
| 8905 | 08072565-4656260 | 30.2 ± 0.2 | 3976 ± 22 | 1.31 ± 0.19 | 1.043 ± 0.003 | 0.02 ± 0.18 | <23 | 3 | … | … | … | … | … | … | G |
| 9757 | 08111608-4711036 | -12.1 ± 0.2 | 4961 ± 134 | … | 1.013 ± 0.002 | -0.06 ± 0.10 | 30 ± 3 | 1 | … | … | … | … | … | … | G |

**Table C.3.** continued.

| ID | CNAME | $RV$ (km s$^{-1}$) | $T_{\rm eff}$ (K) | $logg$ (dex) | $\gamma^a$ | [Fe/H] (dex) | $EW({\rm Li})^b$ (mÅ) | $EW({\rm Li})$ error flag$^c$ | Jeffries 2014$^d$ | Damiani 2014$^d$ | Literature members Spina 2014$^d$ | Frasca 2015$^d$ | Prisinzano 2016$^d$ | Final members$^e$ | Non-mem with Li$^f$ |
|---|---|---|---|---|---|---|---|---|---|---|---|---|---|---|---|
| 8906 | 08072582-4654462 | 110.3 ± 0.2 | 4673 ± 48 | … | 1.032 ± 0.003 | -0.30 ± 0.09 | 27 ± 3 | 1 | … | … | … | … | … | … | G |
| 3504 | 08111619-4713188 | … | … | … | … | … | … | … | … | … | … | … | … | … | … |
| 8907 | 08072586-4702212 | 70.6 ± 0.2 | 4787 ± 162 | … | 1.022 ± 0.002 | -0.25 ± 0.24 | <16 | 3 | … | … | … | … | … | … | G |
| 9506 | 08100979-4723379 | 65.9 ± 0.2 | 4706 ± 78 | 2.46 ± 0.19 | 1.026 ± 0.002 | -0.10 ± 0.04 | <26 | 3 | … | … | … | … | … | … | G |
| 9758 | 08111643-4708121 | 5.1 ± 0.4 | 3607 ± 124 | … | 0.813 ± 0.011 | -0.23 ± 0.15 | <100 | 3 | … | … | … | … | … | … | … |
| 9219 | 08085810-4732507 | 92.5 ± 0.2 | 4722 ± 52 | 2.51 ± 0.14 | 1.016 ± 0.003 | -0.14 ± 0.09 | 13 ± 3 | 1 | … | … | … | … | … | … | G |
| 8908 | 08072621-4721354 | 10.1 ± 0.5 | 3419 ± 5 | 4.65 ± 0.11 | 0.824 ± 0.012 | -0.25 ± 0.15 | … | … | … | … | … | … | Y | … | … |
| 9759 | 08111666-4659318 | 96.8 ± 0.2 | 4643 ± 44 | 2.40 ± 0.07 | 1.019 ± 0.003 | -0.06 ± 0.03 | 14 ± 8 | 1 | … | … | … | … | … | … | G |
| 8909 | 08072635-4703565 | 81.7 ± 0.2 | 4668 ± 62 | … | 1.022 ± 0.002 | -0.22 ± 0.09 | <18 | 3 | … | … | … | … | … | … | G |
| 9220 | 08085831-4726491 | -1.6 ± 0.2 | 5942 ± 55 | 4.14 ± 0.09 | 0.993 ± 0.002 | 0.08 ± 0.01 | 26 ± 1 | 1 | … | … | … | … | … | … | … |
| 8910 | 08072639-4655181 | 83.3 ± 0.2 | 4619 ± 22 | 2.24 ± 0.20 | 1.048 ± 0.005 | -0.15 ± 0.02 | 39 ± 3 | 1 | … | … | … | … | … | … | G |
| 9221 | 08085856-4704343 | 57.9 ± 0.2 | 5036 ± 151 | 4.26 ± 0.02 | 0.971 ± 0.005 | -0.02 ± 0.03 | <29 | 3 | … | … | … | … | … | … | … |
| 9507 | 08101040-4730470 | 17.6 ± 0.2 | 4555 ± 133 | … | 0.961 ± 0.002 | -0.04 ± 0.10 | 434 ± 5 | 1 | A | Y | … | Y | Y | Y | … |
| 8911 | 08072677-4655080 | 18.1 ± 0.4 | 3402 ± 30 | … | 0.867 ± 0.009 | -0.24 ± 0.15 | … | … | … | … | … | … | Y | … | … |
| 9760 | 08111700-4716338 | 12.4 ± 0.3 | 4311 ± 402 | 4.45 ± 0.08 | 0.880 ± 0.006 | -0.16 ± 0.04 | <20 | 3 | … | … | … | … | … | … | … |
| 8912 | 08072680-4709333 | 46.1 ± 0.2 | 5421 ± 18 | … | 1.011 ± 0.004 | -0.27 ± 0.05 | … | … | … | … | … | … | … | … | … |
| 9793 | 08112431-4711176 | 11.1 ± 0.2 | 4718 ± 94 | … | 0.969 ± 0.002 | 0.16 ± 0.11 | <35 | 3 | … | … | … | … | … | … | … |
| 8913 | 08072682-4712412 | 83.4 ± 0.2 | 4751 ± 113 | … | 1.018 ± 0.003 | -0.24 ± 0.15 | 26 ± 2 | 1 | … | … | … | … | … | … | G |
| 9794 | 08112485-4652181 | 76.6 ± 0.2 | 4726 ± 10 | 2.53 ± 0.10 | 1.023 ± 0.003 | -0.06 ± 0.10 | <27 | 3 | … | … | … | … | … | … | G |
| 9222 | 08085883-4654063 | -16.7 ± 0.2 | 4918 ± 152 | … | 1.024 ± 0.003 | -0.18 ± 0.13 | <13 | 3 | … | … | … | … | … | … | G |
| 9508 | 08101138-4739211 | 120.9 ± 0.2 | 4320 ± 244 | … | 1.048 ± 0.003 | -0.22 ± 0.08 | <21 | 3 | … | … | … | … | … | … | G |
| 8914 | 08072690-4718542 | -3.8 ± 0.2 | 5077 ± 80 | … | 1.014 ± 0.001 | … | <14 | 3 | … | … | … | … | … | … | G |
| 9795 | 08112487-4701583 | 62.6 ± 0.2 | 4927 ± 146 | … | 1.014 ± 0.004 | -0.20 ± 0.11 | <14 | 3 | … | … | … | … | … | … | G |
| 9223 | 08085898-4709130 | 36.5 ± 0.2 | 5034 ± 152 | … | 1.014 ± 0.002 | -0.09 ± 0.17 | 111 ± 5 | 1 | … | … | … | … | … | … | Li-rich G |
| 9509 | 08101152-4655508 | 158.5 ± 0.3 | … | … | … | … | … | … | … | … | … | … | … | … | … |
| 8915 | 08072699-4715577 | 78.4 ± 0.2 | 3682 ± 87 | … | … | … | 442 ± 7 | 1 | … | … | … | … | … | … | … |
| 9796 | 08112514-4720330 | 26.8 ± 0.2 | 7078 ± 35 | … | 1.027 ± 0.002 | … | … | … | … | … | … | … | … | … | … |
| 9224 | 08085909-4744124 | 11.7 ± 0.5 | 6403 ± 53 | … | … | 0.04 ± 0.04 | … | … | … | … | … | … | … | … | … |
| 3489 | 08101160-4737555 | 11.1 ± 0.6 | 4431 ± 29 | 1.80 ± 0.05 | 1.028 ± 0.001 | -0.31 ± 0.04 | 17 ± 3 | 1 | … | … | N | … | … | n | G |
| 9797 | 08112532-4720463 | 67.1 ± 0.2 | 5350 ± 138 | 4.19 ± 0.09 | 0.982 ± 0.002 | -0.03 ± 0.02 | <23 | 3 | … | … | … | … | … | … | … |
| 8916 | 08072714-4726452 | 28.5 ± 0.2 | 4651 ± 67 | 2.09 ± 0.13 | 1.030 ± 0.003 | -0.29 ± 0.05 | 31 ± 7 | 1 | … | … | … | … | … | … | G |
| 9225 | 08085928-4658453 | 111.9 ± 0.3 | 4896 ± 7 | 2.62 ± 0.13 | 1.030 ± 0.007 | -0.11 ± 0.02 | <31 | 3 | … | … | … | … | … | … | G |
| 8917 | 08072777-4657433 | 51.4 ± 0.2 | 4475 ± 215 | 4.48 ± 0.13 | 0.922 ± 0.003 | -0.03 ± 0.11 | <15 | 3 | … | … | … | … | … | … | … |
| 9798 | 08112543-4735057 | 94.0 ± 0.2 | 4722 ± 96 | 2.60 ± 0.17 | 1.010 ± 0.004 | -0.21 ± 0.14 | … | … | … | … | … | … | … | … | … |
| 9242 | 08090402-4702002 | 26.4 ± 0.5 | 3576 ± 74 | … | 0.816 ± 0.012 | -0.22 ± 0.15 | … | … | … | … | … | … | … | … | … |
| 9510 | 08101196-4744433 | 17.5 ± 0.7 | … | … | … | … | … | … | A | … | … | … | Y | Y | … |
| 8918 | 08072787-4743328 | 116.9 ± 0.2 | 4505 ± 87 | … | 1.023 ± 0.003 | -0.29 ± 0.23 | <20 | 3 | … | … | … | … | … | … | G |
| 9799 | 08112544-4738220 | 17.4 ± 0.2 | 6446 ± 90 | … | 0.998 ± 0.002 | -0.15 ± 0.09 | <14 | 3 | … | … | … | … | … | … | … |
| 9243 | 08090421-4728203 | 17.8 ± 0.5 | 3330 ± 26 | … | 0.847 ± 0.014 | -0.27 ± 0.14 | 631 ± 20 | 1 | B | … | … | … | Y | Y | … |
| 3490 | 08101201-4720464 | 57.9 ± 0.6 | 6270 ± 67 | 3.87 ± 0.03 | 0.997 ± 0.001 | -0.47 ± 0.04 | 12 ± 5 | 1 | … | … | N | … | … | n | NG |
| 9800 | 08112548-4701386 | 47.5 ± 0.2 | 4741 ± 65 | … | 1.022 ± 0.003 | -0.13 ± 0.09 | <31 | 3 | … | … | … | … | … | … | G |
| 8919 | 08072824-4746146 | -0.2 ± 0.2 | 5867 ± 90 | 4.10 ± 0.06 | 0.994 ± 0.003 | 0.15 ± 0.05 | <11 | 3 | … | … | … | … | … | … | … |
| 3491 | 08101211-4718398 | 64.2 ± 0.6 | 4679 ± 48 | 2.25 ± 0.26 | … | -0.04 ± 0.08 | <38 | 3 | … | … | N | … | … | n | … |
| 8920 | 08072830-4734093 | 49.0 ± 0.2 | 4221 ± 189 | 1.48 ± 0.08 | 1.034 ± 0.002 | 0.05 ± 0.11 | <45 | 3 | … | … | … | … | … | … | G |
| 9511 | 08101211-4731437 | 18.5 ± 0.2 | 6098 ± 117 | 4.18 ± 0.11 | 0.995 ± 0.001 | -0.05 ± 0.13 | 38 ± 3 | 1 | … | … | … | … | … | … | … |
| 8921 | 08072905-4743285 | 60.4 ± 0.2 | 6246 ± 35 | 4.05 ± 0.08 | 1.001 ± 0.002 | -0.03 ± 0.03 | <2 | 3 | … | … | … | … | … | … | … |
| 8922 | 08072980-4748293 | 65.1 ± 0.2 | 4514 ± 150 | 2.21 ± 0.19 | 1.026 ± 0.004 | 0.03 ± 0.07 | <55 | 3 | … | … | … | … | … | … | G |
| 9244 | 08090493-4717425 | 134.0 ± 0.2 | 4513 ± 49 | … | 1.027 ± 0.004 | -0.30 ± 0.11 | <31 | 3 | … | … | … | … | … | … | G |
| 8923 | 08072982-4713172 | 11.6 ± 1.7 | … | … | … | … | … | … | … | … | … | … | … | … | … |
| 9801 | 08112609-4717017 | 115.2 ± 0.2 | 3936 ± 54 | … | 1.029 ± 0.003 | -0.13 ± 0.16 | 62 ± 17 | 1 | … | … | … | … | … | … | … |
| 9245 | 08090495-4717512 | 45.4 ± 0.2 | 4564 ± 81 | … | 1.039 ± 0.002 | -0.19 ± 0.03 | <25 | 3 | … | … | … | … | … | … | G |
| 9512 | 08101236-4731088 | 43.7 ± 0.2 | 5384 ± 98 | … | 1.000 ± 0.004 | -0.25 ± 0.11 | <23 | 3 | … | … | … | … | … | … | … |
| 9802 | 08112672-4725528 | 107.3 ± 0.2 | 5008 ± 87 | … | 1.020 ± 0.002 | -0.15 ± 0.20 | … | … | … | … | … | … | … | … | G |
| 9246 | 08090509-4721490 | 29.0 ± 0.3 | 6540 ± 25 | 4.09 ± 0.05 | 1.005 ± 0.001 | 0.03 ± 0.02 | … | … | … | … | … | … | … | … | … |
| 9513 | 08101291-4743261 | 1.3 ± 0.2 | 5031 ± 113 | … | 1.005 ± 0.003 | -0.08 ± 0.10 | <24 | 3 | … | … | … | … | … | … | … |
| 8956 | 08073795-4709273 | 20.2 ± 0.3 | 3483 ± 1 | … | 0.838 ± 0.008 | -0.27 ± 0.13 | … | … | … | … | … | … | Y | … | … |
| 9803 | 08112702-4657438 | 66.5 ± 0.2 | 4715 ± 6 | 2.54 ± 0.07 | 1.017 ± 0.002 | -0.05 ± 0.09 | <32 | 3 | … | … | … | … | … | … | G |
| 3471 | 08090542-4740261 | 2.3 ± 0.6 | 5022 ± 4 | 2.56 ± 0.09 | 1.012 ± 0.001 | -0.10 ± 0.03 | 17 ± 3 | 1 | … | … | N | … | … | n | … |
| 3492 | 08101307-4722227 | 28.8 ± 0.6 | 4855 ± 109 | 3.08 ± 0.23 | … | 0.19 ± 0.07 | <22 | 3 | … | … | N | … | … | n | … |

| ID | CNAME | RV (km s$^{-1}$) | $T_{\text{eff}}$ (K) | log g (dex) | $\gamma^a$ | [Fe/H] (dex) | EW(Li)$^b$ (mÅ) | EW(Li) error flag$^c$ | Jeffries 2014$^d$ | Damiani 2014$^d$ | Literature members Spina 2014$^d$ | Frasca 2015$^d$ | Prisinzano 2016$^d$ | Final members$^e$ | Non-mem with Li$^f$ |
|---|---|---|---|---|---|---|---|---|---|---|---|---|---|---|---|
| 8957 | 08073887-4737172 | 166.3 ± 0.2 | 4253 ± 54 | … | 1.044 ± 0.007 | -0.53 ± 0.08 | … | … | … | … | … | … | … | … | G |
| 9804 | 08112722-4729152 | 31.3 ± 0.2 | 6368 ± 18 | 3.73 ± 0.04 | 1.013 ± 0.001 | 0.00 ± 0.01 | 91 ± 6 | 1 | … | … | … | … | … | … | … |
| 9805 | 08112733-4739305 | 49.9 ± 0.2 | 5953 ± 95 | … | 1.001 ± 0.003 | -0.07 ± 0.04 | <17 | 3 | … | … | … | … | … | … | … |
| 9247 | 08090662-4746249 | 24.3 ± 0.3 | 6608 ± 50 | 4.28 ± 0.10 | 1.001 ± 0.003 | -0.01 ± 0.04 | 54 ± 8 | 1 | … | … | … | … | … | … | … |
| 9529 | 08101746-4746136 | 17.2 ± 0.5 | 3338 ± 24 | … | 0.855 ± 0.021 | … | 556 ± 23 | 1 | A | … | … | … | Y | Y | … |
| 8958 | 08073910-4718269 | 102.6 ± 0.2 | 4377 ± 232 | … | 1.042 ± 0.003 | -0.29 ± 0.15 | … | … | … | … | … | … | … | … | G |
| 9806 | 08112751-4659054 | 50.2 ± 0.2 | 4589 ± 148 | … | 1.035 ± 0.002 | -0.18 ± 0.15 | … | … | … | … | … | … | … | … | G |
| 9248 | 08090679-4652105 | 74.3 ± 0.2 | 4638 ± 25 | 2.38 ± 0.08 | 1.017 ± 0.004 | -0.14 ± 0.02 | <29 | 3 | … | … | … | … | … | … | G |
| 9530 | 08101751-4730320 | 51.0 ± 0.2 | 4601 ± 137 | 2.27 ± 0.15 | 1.024 ± 0.001 | -0.05 ± 0.05 | <35 | 3 | … | … | … | … | … | … | G |
| 8959 | 08073975-4654042 | -2.1 ± 0.2 | 5095 ± 85 | … | 0.973 ± 0.005 | … | 53 ± 8 | 1 | … | … | … | … | … | … | … |
| 9807 | 08112790-4717291 | 20.4 ± 0.2 | 4964 ± 149 | 3.09 ± 0.17 | 1.006 ± 0.004 | -0.11 ± 0.18 | <24 | 3 | … | … | … | … | … | … | … |
| 9531 | 08101766-4705359 | 96.3 ± 0.2 | 5014 ± 142 | … | 1.021 ± 0.004 | -0.19 ± 0.24 | <15 | 3 | … | … | … | … | … | … | … |
| 8960 | 08073978-4714129 | 99.0 ± 0.2 | 4675 ± 66 | 2.48 ± 0.17 | 1.019 ± 0.003 | -0.10 ± 0.07 | <32 | 3 | … | … | … | … | … | … | G |
| 9808 | 08112821-4747057 | 81.3 ± 0.2 | 5004 ± 32 | 2.98 ± 0.13 | 1.010 ± 0.005 | -0.19 ± 0.08 | <16 | 3 | … | … | … | … | … | … | G |
| 9532 | 08101768-4741364 | 24.1 ± 0.3 | 3669 ± 49 | 4.57 ± 0.18 | 0.810 ± 0.010 | -0.21 ± 0.12 | … | … | … | … | … | … | … | … | … |
| 8961 | 08073983-4715458 | 48.9 ± 0.2 | 4172 ± 285 | … | 0.893 ± 0.004 | -0.03 ± 0.14 | … | … | … | … | … | … | … | … | … |
| 9533 | 08101782-4741494 | 81.2 ± 0.2 | 4921 ± 69 | … | 1.016 ± 0.003 | -0.23 ± 0.14 | <18 | 3 | … | … | … | … | … | … | … |
| 9809 | 08112836-4708259 | 90.5 ± 0.2 | 4548 ± 154 | 2.29 ± 0.18 | 1.012 ± 0.004 | -0.13 ± 0.06 | <2 | 3 | … | … | … | … | … | … | G |
| 9249 | 08090725-4732490 | 50.1 ± 0.2 | 4972 ± 174 | … | 1.015 ± 0.003 | -0.18 ± 0.13 | <25 | 3 | … | … | … | … | … | … | G |
| 9534 | 08101786-4703188 | 81.7 ± 0.2 | 4519 ± 120 | … | 1.018 ± 0.005 | -0.20 ± 0.07 | … | … | … | … | … | … | … | … | … |
| 8962 | 08074010-4720518 | 4.2 ± 0.2 | 3959 ± 202 | … | 0.872 ± 0.004 | -0.09 ± 0.05 | 329 ± 4 | 1 | … | … | … | … | Y | Y | … |
| 9810 | 08112863-4708491 | 39.5 ± 0.2 | 5156 ± 5 | … | 1.014 ± 0.003 | -0.01 ± 0.05 | <23 | 3 | … | … | … | … | … | … | … |
| 3472 | 08090738-4738136 | … | … | … | … | … | … | … | … | … | … | … | … | … | … |
| 9535 | 08101791-4703555 | 16.6 ± 0.3 | 3429 ± 59 | … | 0.862 ± 0.007 | -0.27 ± 0.13 | 293 ± 8 | 1 | A | Y | … | Y | Y | Y | … |
| 3450 | 08074016-4721020 | 21.6 ± 0.4 | 7017 ± 114 | 4.11 ± 0.13 | … | -0.02 ± 0.12 | <6 | 3 | … | … | N | … | … | n | … |
| 9811 | 08112875-4709475 | 66.7 ± 0.2 | 4036 ± 18 | … | 0.876 ± 0.005 | 0.08 ± 0.04 | <28 | 3 | … | … | … | … | … | … | … |
| 9250 | 08090754-4740499 | -0.1 ± 0.2 | 4856 ± 127 | 3.01 ± 0.18 | 1.007 ± 0.002 | -0.04 ± 0.08 | <32 | 3 | … | … | … | … | … | … | … |
| 3451 | 08074019-4730403 | 1.9 ± 0.6 | 5072 ± 28 | 3.62 ± 0.07 | 0.993 ± 0.002 | -0.08 ± 0.02 | 19 ± 3 | 2 | … | … | N | … | … | n | NG |
| 9812 | 08112885-4735348 | 5.5 ± 0.2 | 4927 ± 117 | 2.78 ± 0.16 | 1.018 ± 0.003 | -0.04 ± 0.07 | <21 | 3 | … | … | … | … | … | … | … |
| 9251 | 08090758-4718422 | 18.6 ± 0.2 | 4290 ± 171 | … | 0.927 ± 0.003 | -0.06 ± 0.13 | 503 ± 5 | 1 | B | Y | … | Y | Y | Y | … |
| 9252 | 08090779-4701137 | 18.1 ± 0.2 | 5005 ± 143 | … | 1.019 ± 0.002 | -0.15 ± 0.17 | <15 | 3 | … | … | … | … | … | … | … |
| 8963 | 08074035-4717452 | 7.6 ± 0.2 | 5304 ± 149 | … | 0.995 ± 0.004 | 0.00 ± 0.13 | … | … | … | … | … | … | … | … | … |
| 9536 | 08101826-4704539 | -3.7 ± 0.3 | 3667 ± 34 | 4.59 ± 0.17 | 0.806 ± 0.010 | -0.21 ± 0.12 | <100 | 3 | … | … | … | … | … | … | … |
| 9253 | 08090804-4728514 | 15.7 ± 0.8 | … | … | … | … | … | … | … | … | … | … | Y | … | … |
| 9254 | 08090819-4721081 | 104.8 ± 0.2 | 4549 ± 30 | … | 1.031 ± 0.003 | -0.28 ± 0.07 | … | … | … | … | … | … | … | … | G |
| 9537 | 08101827-4710138 | 67.3 ± 0.4 | 4805 ± 82 | … | 1.026 ± 0.007 | -0.49 ± 0.07 | … | … | … | … | … | … | … | … | G |
| 9829 | 08113254-4711333 | 5.9 ± 0.3 | 5736 ± 2 | … | 0.997 ± 0.002 | -0.47 ± 0.22 | <4 | 3 | … | … | … | … | … | … | G |
| 9255 | 08090831-4740056 | 107.0 ± 0.2 | 4292 ± 208 | … | 1.060 ± 0.004 | -0.24 ± 0.06 | 104 ± 3 | 1 | … | … | … | … | … | … | G |
| 9538 | 08101835-4747323 | -13.3 ± 0.3 | 6427 ± 47 | 3.82 ± 0.10 | 1.010 ± 0.003 | -0.17 ± 0.04 | … | … | … | … | … | … | … | … | … |
| 8964 | 08074117-4716085 | 19.8 ± 0.3 | 5301 ± 77 | … | 1.003 ± 0.005 | -0.19 ± 0.13 | 22 ± 3 | 1 | … | … | … | … | … | … | … |
| 3473 | 08090850-4701407 | 20.3 ± 0.6 | 6650 ± 201 | 4.10 ± 0.13 | … | -0.06 ± 0.14 | <6 | 3 | … | Y | … | … | … | … | … |
| 9280 | 08091606-4738476 | 19.6 ± 0.3 | 3401 ± 35 | … | 0.843 ± 0.009 | -0.28 ± 0.14 | <100 | 3 | … | … | … | … | Y | Y | … |
| 8965 | 08074176-4720313 | 81.4 ± 0.2 | 4662 ± 65 | … | 1.017 ± 0.003 | -0.31 ± 0.10 | 51 ± 3 | 1 | … | … | … | … | … | … | G |
| 9830 | 08113261-4658306 | 61.9 ± 0.2 | 4636 ± 141 | … | 1.019 ± 0.003 | -0.19 ± 0.11 | … | … | … | … | … | … | … | … | G |
| 3475 | 08091638-4713374 | -12.4 ± 0.4 | … | … | … | … | … | … | … | … | … | … | … | … | … |
| 8966 | 08074177-4747392 | 29.7 ± 1.8 | 3290 ± 86 | … | … | … | … | … | … | Y | … | … | Y | … | … |
| 9831 | 08113270-4701132 | 56.8 ± 0.2 | 4632 ± 97 | 2.33 ± 0.08 | 1.020 ± 0.003 | -0.04 ± 0.02 | <30 | 3 | … | … | … | … | … | … | G |
| 8967 | 08074238-4703287 | 51.5 ± 0.2 | 5076 ± 119 | … | 1.020 ± 0.006 | -0.27 ± 0.21 | … | … | … | … | … | … | … | … | G |
| 9539 | 08101859-4719367 | 71.0 ± 0.2 | 4641 ± 119 | … | 1.026 ± 0.003 | -0.16 ± 0.09 | 33 ± 5 | 1 | … | … | … | … | … | … | G |
| 8968 | 08074265-4712071 | 71.6 ± 0.2 | 5250 ± 6 | … | 1.004 ± 0.002 | -0.10 ± 0.08 | <20 | 3 | … | … | … | … | … | … | … |
| 9540 | 08101877-4714065 | 19.1 ± 0.2 | 3786 ± 116 | … | 0.855 ± 0.004 | -0.18 ± 0.12 | 491 ± 4 | 1 | B | … | … | Y | Y | Y | … |
| 9832 | 08113284-4721004 | 15.8 ± 0.2 | 4932 ± 16 | 2.59 ± 0.19 | 1.012 ± 0.002 | -0.15 ± 0.02 | <15 | 3 | … | … | … | … | … | … | G |
| 8969 | 08074266-4715386 | 11.9 ± 0.2 | 4012 ± 175 | … | 0.869 ± 0.004 | -0.02 ± 0.14 | … | … | … | … | … | … | … | … | … |
| 9541 | 08101887-4740417 | 68.1 ± 1.5 | … | … | … | … | … | … | … | … | … | … | … | … | … |
| 9833 | 08113286-4738058 | 105.2 ± 0.2 | 4395 ± 216 | … | 1.038 ± 0.005 | -0.17 ± 0.11 | 80 ± 9 | 1 | … | … | … | … | … | … | G |
| 9281 | 08091746-4741422 | 57.6 ± 0.2 | 4058 ± 257 | 4.49 ± 0.09 | 0.848 ± 0.004 | -0.05 ± 0.14 | … | … | … | … | … | … | … | … | … |
| 3452 | 08074278-4659566 | … | 7241 ± 301 | 4.07 ± 0.23 | … | -0.19 ± 0.17 | <18 | 3 | … | … | … | … | … | … | … |
| 9542 | 08101890-4713018 | 47.0 ± 0.3 | 4775 ± 117 | 2.53 ± 0.11 | 1.014 ± 0.006 | -0.08 ± 0.11 | <27 | 3 | … | … | … | … | … | … | G |
| 9834 | 08113304-4723186 | 58.9 ± 0.2 | 6254 ± 37 | 4.17 ± 0.08 | 0.998 ± 0.002 | -0.07 ± 0.03 | … | … | … | … | … | … | … | … | … |

**Table C.3.** continued.

| ID | CNAME | RV (km s$^{-1}$) | $T_{eff}$ (K) | $logg$ (dex) | $\gamma^a$ | [Fe/H] (dex) | $EW(Li)^b$ (mÅ) | $EW(Li)$ error flag$^c$ | Jeffries 2014$^d$ | Damiani 2014$^d$ | Literature members Spina 2014$^d$ | Frasca 2015$^d$ | Prisinzano 2016$^d$ | Final members$^e$ | Non-mem with Li$^f$ |
|---|---|---|---|---|---|---|---|---|---|---|---|---|---|---|---|
| 9282 | 08091748-4715558 | 18.8 ± 0.2 | 3526 ± 56 | … | 0.869 ± 0.004 | -0.22 ± 0.15 | 406 ± 21 | 1 | B | Y | … | Y | Y | Y | … |
| 8989 | 08074837-4705591 | 13.7 ± 0.2 | 5063 ± 83 | 3.35 ± 0.19 | 1.006 ± 0.003 | -0.04 ± 0.11 | <22 | 3 | … | … | … | … | … | … | … |
| 9543 | 08101902-4740401 | 68.5 ± 0.2 | 3910 ± 96 | … | 1.062 ± 0.003 | -0.22 ± 0.15 | <19 | 3 | … | … | … | … | … | … | G |
| 9835 | 08113359-4718257 | 44.5 ± 0.2 | 5778 ± 109 | … | 1.032 ± 0.001 | -0.02 ± 0.02 | … | … | … | … | … | … | … | … | … |
| 8990 | 08074838-4702588 | 99.7 ± 0.2 | 4521 ± 142 | … | 1.028 ± 0.003 | -0.17 ± 0.11 | 35 ± 7 | 1 | … | … | … | … | … | … | G |
| 9836 | 08113408-4742034 | 47.6 ± 0.3 | 3609 ± 117 | … | 0.819 ± 0.012 | -0.23 ± 0.14 | <100 | 3 | … | … | … | … | … | … | … |
| 9544 | 08101904-4653324 | 72.6 ± 0.2 | 4471 ± 187 | … | 1.049 ± 0.003 | -0.22 ± 0.09 | <17 | 3 | … | … | … | … | … | … | G |
| 9283 | 08091826-4735302 | 59.7 ± 0.2 | 3905 ± 11 | … | 1.039 ± 0.001 | 0.00 ± 0.12 | … | … | … | … | … | … | … | … | G |
| 9837 | 08113498-4732325 | 32.3 ± 0.2 | 4981 ± 156 | … | 1.009 ± 0.004 | -0.25 ± 0.13 | … | … | … | … | … | … | … | … | … |
| 9545 | 08101923-4734111 | 16.4 ± 0.2 | 4532 ± 56 | 4.60 ± 0.07 | 0.929 ± 0.003 | -0.10 ± 0.05 | <11 | 3 | … | … | … | … | … | … | … |
| 9284 | 08091845-4653440 | 25.5 ± 0.2 | 5955 ± 120 | 4.03 ± 0.15 | 0.998 ± 0.002 | -0.25 ± 0.10 | <10 | 3 | … | … | … | … | … | … | … |
| 9838 | 08113502-4727271 | -3.8 ± 0.2 | 3554 ± 76 | … | 0.827 ± 0.004 | -0.21 ± 0.14 | … | … | … | … | … | … | … | … | … |
| 9285 | 08091846-4718420 | 16.7 ± 0.2 | 3511 ± 41 | … | 0.854 ± 0.005 | -0.22 ± 0.14 | 206 ± 21 | 1 | A | Y | … | Y | Y | Y | … |
| 3476 | 08091875-4708534 | 13.7 ± 0.6 | 6078 ± 132 | 3.94 ± 0.16 | 0.998 ± 0.002 | -0.04 ± 0.13 | 136 ± 8 | 2 | … | … | … | Y | … | Y | … |
| 9839 | 08113513-4719263 | 87.4 ± 0.2 | 4433 ± 158 | … | 1.051 ± 0.003 | -0.25 ± 0.08 | <29 | 3 | … | … | … | … | … | … | G |
| 9286 | 08091903-4733168 | 161.0 ± 0.2 | 4910 ± 200 | … | 1.013 ± 0.004 | -0.29 ± 0.27 | <31 | 3 | … | … | … | … | … | … | G |
| 9574 | 08102764-4716419 | 53.7 ± 2.6 | … | … | … | … | … | … | … | … | … | … | … | … | … |
| 9287 | 08091914-4713569 | 18.8 ± 0.4 | 3307 ± 12 | … | 0.909 ± 0.014 | … | … | … | A | … | … | … | Y | … | … |
| 9840 | 08113585-4722573 | 42.7 ± 0.2 | 4703 ± 142 | 2.53 ± 0.17 | 1.014 ± 0.004 | 0.00 ± 0.10 | <41 | 3 | … | … | … | … | … | … | G |
| 9575 | 08102766-4716387 | 53.6 ± 0.2 | 4138 ± 21 | … | 1.036 ± 0.003 | -0.01 ± 0.03 | <33 | 3 | … | … | … | … | … | … | G |
| 9841 | 08113597-4742110 | 110.5 ± 0.2 | 5031 ± 106 | … | 1.007 ± 0.003 | -0.22 ± 0.28 | … | … | … | … | … | … | … | … | … |
| 9576 | 08102778-4717245 | 17.5 ± 0.5 | 3288 ± 10 | … | 0.874 ± 0.013 | -0.26 ± 0.14 | … | … | A | … | … | … | Y | Y | … |
| 9289 | 08091939-4705458 | 32.4 ± 1.3 | 4205 ± 63 | … | 0.841 ± 0.009 | … | <41 | 3 | … | … | … | … | … | … | … |
| 9842 | 08113602-4656299 | 53.4 ± 0.2 | 4566 ± 52 | 2.30 ± 0.13 | 1.024 ± 0.003 | -0.04 ± 0.03 | <31 | 3 | … | … | … | … | … | … | G |
| 9290 | 08091973-4718118 | 52.3 ± 0.2 | 4919 ± 184 | … | 1.017 ± 0.003 | -0.09 ± 0.09 | <22 | 3 | … | … | … | … | … | … | … |
| 9843 | 08113606-4705323 | 11.8 ± 0.2 | 6351 ± 34 | 4.15 ± 0.07 | 1.001 ± 0.002 | 0.13 ± 0.02 | <23 | 3 | … | … | … | … | … | … | … |
| 9577 | 08102899-4654434 | 129.5 ± 0.2 | 4497 ± 67 | … | 1.036 ± 0.003 | -0.28 ± 0.08 | <16 | 3 | … | … | … | … | … | … | G |
| 9291 | 08091979-4720483 | 18.8 ± 0.2 | 3743 ± 31 | … | 0.857 ± 0.004 | -0.19 ± 0.12 | 462 ± 6 | 1 | B | … | … | Y | Y | Y | … |
| 9844 | 08113606-4708265 | 60.3 ± 0.2 | 4491 ± 168 | 2.23 ± 0.10 | 1.011 ± 0.003 | 0.07 ± 0.09 | <40 | 3 | … | … | … | … | … | … | G |
| 9578 | 08102926-4737566 | 21.7 ± 0.3 | 4884 ± 128 | 3.04 ± 0.13 | 1.001 ± 0.007 | -0.01 ± 0.10 | <26 | 3 | … | … | … | … | … | … | … |
| 9292 | 08092015-4717309 | 79.6 ± 0.2 | 4011 ± 113 | 4.49 ± 0.08 | 0.858 ± 0.005 | -0.11 ± 0.07 | <36 | 3 | … | … | … | … | … | … | … |
| 9845 | 08113611-4740089 | 43.3 ± 0.2 | 5170 ± 316 | … | 1.012 ± 0.003 | -0.26 ± 0.15 | 12 ± 2 | 1 | … | … | … | … | … | … | G |
| 9579 | 08102950-4737279 | 62.1 ± 0.2 | 4538 ± 117 | 2.22 ± 0.20 | 1.024 ± 0.003 | -0.18 ± 0.09 | … | … | … | … | … | … | … | … | G |
| 9293 | 08092021-4742080 | 61.6 ± 0.2 | 4572 ± 95 | … | 1.028 ± 0.004 | -0.18 ± 0.04 | <26 | 3 | … | … | … | … | … | … | G |
| 9846 | 08113618-4712420 | 37.9 ± 0.2 | 4977 ± 199 | … | 1.010 ± 0.003 | -0.07 ± 0.14 | <20 | 3 | … | … | … | … | … | … | G |
| 9884 | 08114510-4718487 | 46.5 ± 0.2 | 6267 ± 42 | 4.16 ± 0.09 | 0.999 ± 0.002 | -0.01 ± 0.03 | <3 | 3 | … | … | … | … | … | … | … |
| 9580 | 08103014-4726139 | 16.7 ± 0.2 | 4170 ± 281 | … | 0.902 ± 0.005 | -0.06 ± 0.05 | 394 ± 5 | 1 | A | Y | … | Y | Y | Y | … |
| 9294 | 08092039-4713359 | 4.3 ± 0.2 | 4871 ± 147 | 2.70 ± 0.14 | 1.016 ± 0.003 | -0.02 ± 0.05 | <19 | 3 | … | … | … | … | … | … | … |
| 9885 | 08114538-4726422 | 96.1 ± 0.2 | 4736 ± 62 | … | 1.018 ± 0.005 | -0.16 ± 0.13 | 20 ± 3 | 1 | … | … | … | … | … | … | G |
| 9581 | 08103074-4726219 | 14.5 ± 0.2 | 4011 ± 191 | … | 0.894 ± 0.002 | -0.03 ± 0.15 | 419 ± 9 | 1 | B | Y | … | Y | Y | Y | … |
| 9310 | 08092456-4746025 | 9.0 ± 0.3 | 4048 ± 114 | 4.69 ± 0.17 | 0.838 ± 0.008 | -0.22 ± 0.13 | … | … | … | … | … | … | … | … | … |
| 9886 | 08114595-4702353 | 18.9 ± 0.2 | 4428 ± 271 | 4.47 ± 0.14 | 0.913 ± 0.004 | -0.02 ± 0.12 | … | … | … | … | … | … | … | … | … |
| 9582 | 08103086-4739041 | 91.8 ± 0.2 | 3363 ± 96 | … | … | … | … | … | … | … | … | … | … | … | … |
| 9887 | 08114630-4726476 | 32.0 ± 0.3 | 6654 ± 56 | … | 0.995 ± 0.003 | 0.24 ± 0.04 | … | … | … | … | … | … | … | … | … |
| 9311 | 08092546-4700250 | 87.4 ± 0.2 | 4885 ± 114 | … | 1.018 ± 0.006 | -0.18 ± 0.14 | … | … | … | … | … | … | … | … | G |
| 9888 | 08114649-4739162 | 95.9 ± 0.2 | 4792 ± 195 | … | 1.012 ± 0.004 | -0.25 ± 0.26 | … | … | … | … | … | … | … | … | G |
| 9889 | 08114705-4721543 | 32.1 ± 0.3 | 6576 ± 73 | … | 0.993 ± 0.004 | 0.17 ± 0.06 | … | … | … | … | … | … | … | … | … |
| 9583 | 08103101-4704056 | 66.6 ± 0.2 | 4736 ± 69 | … | 1.021 ± 0.003 | -0.16 ± 0.06 | <25 | 3 | … | … | … | … | … | … | G |
| 9890 | 08114774-4741283 | 21.5 ± 0.4 | 3580 ± 60 | … | 0.817 ± 0.017 | -0.23 ± 0.14 | <100 | 3 | … | … | … | … | … | … | … |
| 9891 | 08114788-4704551 | 59.8 ± 0.2 | 4453 ± 310 | 4.49 ± 0.10 | 0.910 ± 0.004 | -0.03 ± 0.10 | <14 | 3 | … | … | … | … | … | … | … |
| 9892 | 08114831-4736476 | 79.6 ± 0.3 | 3917 ± 107 | 4.51 ± 0.20 | 0.847 ± 0.005 | -0.24 ± 0.13 | <31 | 3 | … | … | … | … | … | … | … |
| 9584 | 08103171-4704161 | 43.9 ± 0.2 | 4905 ± 104 | 2.70 ± 0.08 | 1.017 ± 0.002 | -0.03 ± 0.01 | <28 | 3 | … | … | … | … | … | … | G |
| 9585 | 08103203-4655016 | 20.0 ± 0.2 | 6424 ± 24 | … | 0.996 ± 0.001 | -0.23 ± 0.02 | 56 ± 2 | 1 | … | … | … | … | … | … | … |
| 9586 | 08103224-4708296 | 122.4 ± 0.2 | 3956 ± 249 | … | 1.056 ± 0.003 | -0.26 ± 0.06 | … | … | … | … | … | … | … | … | G |
| 9893 | 08114909-4705100 | 102.2 ± 0.2 | 4679 ± 106 | … | 1.023 ± 0.004 | -0.17 ± 0.08 | … | … | … | … | … | … | … | … | G |
| 9587 | 08103242-4658174 | 39.3 ± 0.2 | 4923 ± 182 | … | 1.018 ± 0.004 | -0.08 ± 0.13 | <31 | 3 | … | … | … | … | … | … | G |
| 9894 | 08114946-4723173 | 30.3 ± 0.2 | 5108 ± 170 | … | 1.002 ± 0.004 | -0.15 ± 0.16 | <17 | 3 | … | … | … | … | … | … | … |
| 9895 | 08114953-4718483 | 73.9 ± 0.2 | 4749 ± 178 | … | 1.015 ± 0.006 | -0.22 ± 0.13 | … | … | … | … | … | … | … | … | G |
| 9600 | 08103670-4745404 | 79.6 ± 0.2 | 4730 ± 43 | 2.57 ± 0.08 | 1.012 ± 0.004 | -0.13 ± 0.03 | <31 | 3 | … | … | … | … | … | … | G |

| ID | CNAME | RV (km s$^{-1}$) | $T_{\text{eff}}$ (K) | $\log g$ (dex) | $\gamma^a$ | [Fe/H] (dex) | EW(Li)$^b$ (mÅ) | EW(Li) error flag$^c$ | Jeffries 2014$^d$ | Damiani 2014$^d$ | Spina 2014$^d$ | Frasca 2015$^d$ | Prisinzano 2016$^d$ | Final members$^e$ | Non-mem with Li$^f$ |
|---|---|---|---|---|---|---|---|---|---|---|---|---|---|---|---|
| 9896 | 08114962-4742259 | 119.4 ± 0.2 | 4542 ± 113 | … | 1.046 ± 0.005 | … | … | … | … | … | … | … | … | … | G |
| 9897 | 08114967-4719524 | 99.7 ± 0.2 | 5056 ± 93 | 3.48 ± 0.13 | 1.000 ± 0.004 | -0.16 ± 0.19 | … | … | … | … | … | … | … | … | … |
| 9898 | 08114969-4727534 | 40.5 ± 0.2 | 4956 ± 101 | … | 1.019 ± 0.003 | -0.11 ± 0.11 | <25 | 3 | … | … | … | … | … | … | G |
| 3511 | 08114979-4700130 | 58.1 ± 0.6 | 4726 ± 11 | 2.74 ± 0.05 | 1.007 ± 0.001 | 0.18 ± 0.01 | 35 ± 9 | 1 | … | … | N | … | … | n | NG? |
| 9899 | 08115019-4707262 | 43.3 ± 0.3 | 3798 ± 61 | 4.62 ± 0.10 | 0.806 ± 0.009 | -0.18 ± 0.14 | <100 | 3 | … | … | … | … | … | … | … |
| 9900 | 08115050-4659395 | 125.4 ± 0.2 | 4804 ± 23 | 2.60 ± 0.17 | 1.025 ± 0.006 | -0.14 ± 0.02 | 26 ± 8 | 1 | … | … | … | … | … | … | G |
| 9901 | 08115075-4706238 | 102.7 ± 0.2 | 4500 ± 113 | … | 1.035 ± 0.003 | -0.12 ± 0.03 | 94 ± 9 | 1 | … | … | … | … | … | … | G |
| 9902 | 08115083-4725277 | 3.1 ± 0.2 | 6061 ± 20 | 4.08 ± 0.15 | 0.995 ± 0.002 | -0.07 ± 0.05 | 71 ± 2 | 1 | … | … | … | … | … | … | … |
| 9903 | 08115086-4714077 | 94.9 ± 0.2 | 4512 ± 129 | … | 1.035 ± 0.003 | -0.19 ± 0.05 | … | … | … | … | … | … | … | … | G |
| 9921 | 08115562-4701388 | 100.5 ± 0.2 | 4707 ± 23 | 2.50 ± 0.13 | 1.020 ± 0.002 | -0.09 ± 0.06 | <29 | 3 | … | … | … | … | … | … | G |
| 9922 | 08115579-4731508 | 16.9 ± 0.3 | 3456 ± 43 | … | 0.851 ± 0.009 | -0.25 ± 0.15 | 366 ± 27 | 1 | A | … | … | Y | Y | Y | … |
| 9923 | 08115586-4657472 | 56.4 ± 0.2 | 4323 ± 208 | … | 1.065 ± 0.001 | -0.15 ± 0.08 | <25 | 3 | … | … | … | … | … | … | G |

**Notes.** $^{(a)}$ Empirical gravity indicator defined by Damiani et al. (2014). $^{(b)}$ The values of EW(Li) for this cluster are corrected (subtracted adjacent Fe (6707.43 Å) line). $^{(c)}$ Flags for the errors of the corrected EW(Li) values, as follows: 1=EW(Li) corrected by blends contribution using models; 2=EW(Li) measured separately (Li line resolved - UVES only); and 3=Upper limit (no error for EW(Li) is given). $^d$ For this cluster we have made use of the membership selections obtained by Jeffries et al. (2014) (where 'A' and 'B' refer to Pop. A and Pop. B of the cluster, respectively) Damiani et al. (2014), Spina et al. (2014 b), Frasca et al. (2015), and Prisinzano et al. (2016). $^{(e)}$ The letters "Y" and "N" indicate if the star is a cluster member or not $^{(f)}$ 'Li-rich G', 'G' and 'NG' indicate "Li-rich giant", "giant" and "non-giant" Li field contaminants, respectively.

**Table C.4.** NGC 2547

| ID | CNAME | RV (km s$^{-1}$) | $T_{\text{eff}}$ (K) | $\log g$ (dex) | $\gamma^a$ | [Fe/H] (dex) | EW(Li)$^b$ (mÅ) | EW(Li) error flag$^c$ | Pop. A/B (Sacco)$^d$ | Pop. (Randich) | Pop. (Cantat-Gaudin) | Pop. (adopted final)$^e$ | Non-mem with Li$^f$ |
|---|---|---|---|---|---|---|---|---|---|---|---|---|---|
| 44255 | 08073009-4911229 | 505.0 ± 2.4 | 3607 ± 25 | … | 0.831 ± 0.017 | … | … | … | … | … | … | … | … |
| 44256 | 08073022-4901579 | 19.7 ± 0.3 | 3369 ± 220 | … | … | -0.24 ± 0.14 | … | … | … | … | … | … | … |
| 44505 | 08101944-4907444 | -34.0 ± 0.2 | 5093 ± 118 | … | 1.012 ± 0.002 | -0.29 ± 0.15 | <13 | 3 | … | … | … | … | G |
| 44257 | 08073081-4901156 | 40.5 ± 0.3 | 5281 ± 158 | … | 1.000 ± 0.005 | -0.36 ± 0.25 | … | … | … | … | … | … | … |
| 44506 | 08101987-4856000 | 17.3 ± 0.2 | 4726 ± 397 | … | 0.958 ± 0.003 | -0.01 ± 0.01 | 244 ± 5 | 1 | B | … | B | B | … |
| 44258 | 08073081-4905422 | 149.5 ± 0.2 | 4849 ± 18 | 2.67 ± 0.03 | 1.013 ± 0.003 | 0.00 ± 0.04 | <38 | 3 | … | … | … | … | G |
| 44507 | 08102002-4849510 | 509.1 ± 0.3 | 3326 ± 45 | … | … | -0.27 ± 0.15 | <181 | 3 | … | … | … | … | … |
| 44508 | 08102002-4850165 | 11.3 ± 1.0 | 3379 ± 99 | … | 0.837 ± 0.014 | -0.28 ± 0.14 | … | … | … | … | … | … | … |
| 44509 | 08102029-4902254 | 13.7 ± 0.3 | 3538 ± 21 | … | 0.863 ± 0.011 | -0.22 ± 0.14 | … | … | … | … | … | … | … |
| 44510 | 08102103-4910448 | 12.6 ± 0.3 | 3520 ± 17 | … | 0.826 ± 0.010 | … | … | … | … | … | … | … | … |
| 44511 | 08102106-4911465 | 27.9 ± 0.2 | 5834 ± 78 | … | 0.998 ± 0.002 | -0.03 ± 0.19 | 26 ± 4 | 1 | … | … | … | … | … |
| 44512 | 08102150-4901339 | 12.3 ± 0.2 | 3730 ± 29 | … | 0.826 ± 0.006 | -0.21 ± 0.12 | … | … | … | … | … | … | … |
| 44513 | 08102172-4845417 | 44.1 ± 0.2 | 4574 ± 318 | … | 1.015 ± 0.005 | -0.07 ± 0.22 | 415 ± 5 | 1 | … | … | … | … | Li-rich G |
| 44514 | 08102187-4900068 | 12.7 ± 5.2 | 3351 ± 26 | … | 0.866 ± 0.020 | … | … | … | … | … | … | … | … |
| 44515 | 08102202-4918051 | 38.2 ± 0.2 | 4928 ± 100 | 3.12 ± 0.02 | 1.003 ± 0.003 | -0.07 ± 0.09 | <30 | 3 | … | … | … | … | … |
| 44516 | 08102210-4911230 | 13.3 ± 0.3 | 3618 ± 33 | … | 0.832 ± 0.007 | -0.23 ± 0.14 | <100 | 3 | A | … | A | A | … |
| 44277 | 08075151-4919093 | 469.3 ± 2.1 | 3453 ± 52 | … | 0.841 ± 0.015 | -0.25 ± 0.15 | … | … | … | … | … | … | … |
| 44552 | 08103930-4929046 | 13.0 ± 0.2 | 3965 ± 124 | … | 0.878 ± 0.004 | -0.04 ± 0.13 | 167 ± 16 | 1 | A | A | A | A | … |
| 44278 | 08075394-4858450 | 64.1 ± 0.3 | 4983 ± 68 | … | 1.014 ± 0.006 | -0.19 ± 0.09 | … | … | … | … | … | … | G |
| 44553 | 08103983-4904377 | -0.7 ± 0.2 | 6210 ± 59 | 3.69 ± 0.15 | 1.013 ± 0.001 | -0.21 ± 0.12 | 85 ± 2 | 1 | … | … | … | … | … |
| 44554 | 08104061-4907056 | 477.5 ± 1.6 | 3376 ± 16 | … | 0.851 ± 0.014 | -0.27 ± 0.14 | <100 | 3 | … | … | … | … | … |
| 44555 | 08104108-4907525 | 12.5 ± 0.3 | 3358 ± 67 | … | 0.849 ± 0.012 | -0.27 ± 0.14 | <100 | 3 | A | A | … | A | … |
| 44556 | 08104252-4924552 | 478.6 ± 4.7 | … | … | … | … | … | … | … | … | … | … | … |
| 44279 | 08075719-4910559 | 63.6 ± 0.2 | 5095 ± 23 | … | 1.012 ± 0.004 | -0.17 ± 0.13 | <23 | 3 | … | … | … | … | G |
| 44297 | 08082095-4902029 | 20.1 ± 0.2 | 3612 ± 54 | … | 0.846 ± 0.006 | -0.21 ± 0.14 | 422 ± 24 | 1 | B | B | … | B | … |
| 44557 | 08104335-4901152 | 13.1 ± 0.2 | 4493 ± 259 | … | 0.932 ± 0.003 | -0.01 ± 0.07 | 175 ± 5 | 1 | A | A | … | A | … |
| 44558 | 08104343-4930158 | 68.4 ± 0.2 | 4981 ± 39 | 3.02 ± 0.03 | 1.007 ± 0.005 | -0.11 ± 0.14 | <18 | 3 | … | … | … | … | … |
| 44586 | 08105917-4904277 | 12.7 ± 1.0 | … | … | … | … | … | … | … | … | … | … | … |
| 44587 | 08105940-4851272 | 12.7 ± 0.5 | 3482 ± 31 | … | 0.858 ± 0.017 | … | … | … | … | … | … | … | … |
| 44588 | 08105964-4932338 | 44.5 ± 0.2 | 4931 ± 13 | 2.74 ± 0.04 | 1.014 ± 0.005 | -0.11 ± 0.06 | <30 | 3 | … | … | … | … | G |
| 44589 | 08110007-4904413 | 5.7 ± 0.3 | 3622 ± 7 | 4.64 ± 0.09 | 0.795 ± 0.010 | -0.22 ± 0.14 | <100 | 3 | … | … | … | … | … |
| 44590 | 08110009-4906442 | 12.1 ± 0.2 | 5723 ± 6 | 4.19 ± 0.19 | 0.993 ± 0.002 | 0.01 ± 0.07 | 154 ± 2 | 1 | A | A | … | A | … |
| 44591 | 08110044-4919466 | 69.0 ± 0.2 | 5276 ± 98 | 3.74 ± 0.08 | 1.000 ± 0.003 | -0.01 ± 0.02 | <22 | 3 | … | … | … | … | … |
| 44592 | 08110059-4920466 | 11.9 ± 0.7 | … | … | … | … | … | … | … | … | … | … | … |
| 2971 | 08110139-4900089 | 13.0 ± 0.6 | 5445 ± 73 | 4.58 ± 0.06 | 0.993 ± 0.002 | -0.05 ± 0.02 | 170 ± 3 | 2 | A | A | … | A | … |
| 44298 | 08082690-4859131 | 13.0 ± 0.2 | 4481 ± 278 | … | 0.930 ± 0.003 | -0.04 ± 0.05 | 256 ± 3 | 1 | A | A | … | A | … |
| 44593 | 08110155-4856150 | 91.3 ± 0.4 | 3999 ± 387 | … | 0.870 ± 0.013 | -0.18 ± 0.12 | <45 | 3 | … | … | … | … | … |
| 44299 | 08082691-4857134 | 24.3 ± 0.2 | 4827 ± 105 | 2.36 ± 0.06 | 1.020 ± 0.004 | -0.18 ± 0.15 | <23 | 3 | … | … | … | … | G |
| 2972 | 08110201-4916095 | 54.8 ± 0.2 | 5583 ± 234 | 4.11 ± 0.09 | 0.993 ± 0.002 | 0.13 ± 0.12 | <23 | 3 | … | … | … | … | … |
| 44594 | 08110234-4916249 | 13.2 ± 0.2 | 5681 ± 17 | … | 0.997 ± 0.002 | 0.00 ± 0.02 | 193 ± 4 | 1 | A | A | … | A | … |
| 44595 | 08110259-4928323 | 99.0 ± 0.2 | 4845 ± 123 | … | 1.020 ± 0.003 | -0.19 ± 0.16 | <24 | 3 | … | … | … | … | G |
| 44328 | 08084960-4909584 | -14.2 ± 0.2 | 4987 ± 126 | … | 1.019 ± 0.003 | -0.22 ± 0.14 | <17 | 3 | … | … | … | … | G |
| 2983 | 08084975-4924561 | 6.6 ± 0.6 | 6284 ± 53 | 3.87 ± 0.06 | 1.002 ± 0.002 | -0.25 ± 0.01 | 24 ± 6 | 1 | … | … | … | … | … |
| 44621 | 08111831-4848127 | 62.6 ± 0.2 | 4810 ± 41 | 2.64 ± 0.07 | 1.017 ± 0.006 | -0.11 ± 0.04 | <20 | 3 | … | … | … | … | G |
| 44329 | 08085036-4916162 | 10.6 ± 0.2 | 6521 ± 13 | … | … | -0.06 ± 0.01 | … | … | … | … | … | … | … |
| 44622 | 08111885-4849246 | 33.8 ± 1.2 | 3459 ± 117 | … | 0.855 ± 0.015 | -0.26 ± 0.14 | … | … | … | … | … | … | … |
| 44330 | 08085162-4938230 | 35.4 ± 0.2 | 5031 ± 149 | … | 1.013 ± 0.004 | -0.10 ± 0.13 | <24 | 3 | … | … | … | … | G |
| 44331 | 08085197-4930499 | -9.1 ± 0.2 | 3872 ± 120 | 4.56 ± 0.16 | 0.826 ± 0.004 | -0.16 ± 0.13 | … | … | … | … | … | … | … |
| 44623 | 08112069-4849225 | 11.9 ± 0.7 | … | … | … | … | … | … | … | … | … | … | … |
| 44624 | 08112074-4913100 | 27.6 ± 0.2 | 6405 ± 23 | 4.30 ± 0.05 | 0.997 ± 0.001 | 0.28 ± 0.02 | 132 ± 2 | 1 | … | … | … | … | … |
| 44649 | 08113647-4909057 | 33.6 ± 0.2 | 4958 ± 178 | 3.34 ± 0.30 | 0.998 ± 0.005 | -0.06 ± 0.11 | <28 | 3 | … | … | … | … | … |
| 44357 | 08090851-4855307 | 20.6 ± 3.6 | … | … | … | … | … | … | … | … | … | … | … |
| 44358 | 08090863-4938126 | 103.7 ± 0.2 | 4831 ± 79 | … | 1.021 ± 0.003 | -0.21 ± 0.16 | <16 | 3 | … | … | … | … | G |
| 44359 | 08090869-4935130 | 95.3 ± 1.9 | 3277 ± 34 | 4.63 ± 0.20 | 0.857 ± 0.015 | -0.18 ± 0.10 | … | … | … | … | … | … | … |
| 44360 | 08090911-4922405 | 12.2 ± 0.3 | 3368 ± 50 | 4.62 ± 0.20 | 0.838 ± 0.010 | -0.25 ± 0.14 | <100 | 3 | A | A | … | A | … |
| 44650 | 08113861-4914234 | 459.6 ± 2.0 | … | … | … | … | … | … | … | … | … | … | … |
| 44361 | 08090914-4858463 | 124.4 ± 0.2 | 4911 ± 80 | … | 1.019 ± 0.003 | -0.17 ± 0.20 | … | … | … | … | … | … | G |
| 44651 | 08113879-4847231 | 151.9 ± 0.2 | 5159 ± 91 | … | 1.012 ± 0.004 | … | <28 | 3 | … | … | … | … | G |
| 44652 | 08113903-4912281 | 125.7 ± 0.2 | 4566 ± 73 | … | 1.030 ± 0.004 | -0.29 ± 0.14 | <27 | 3 | … | … | … | … | G |

**Table C.4.** continued.

| ID | CNAME | RV (km s$^{-1}$) | $T_{\rm eff}$ (K) | $logg$ (dex) | $\gamma^a$ | [Fe/H] (dex) | $EW$(Li)$^b$ (mÅ) | $EW$(Li) error flag$^c$ | Pop. A/B (Sacco)$^d$ | Pop. (Randich) | Pop. (Cantat-Gaudin) | Pop. (adopted final)$^e$ | Non-mem with Li$^f$ |
|---|---|---|---|---|---|---|---|---|---|---|---|---|---|
| 44653 | 08113996-4903256 | 472.8 ± 2.8 | ... | ... | ... | ... | ... | ... | ... | ... | ... | ... | ... |
| 44362 | 08090970-4936209 | 47.7 ± 0.2 | 4744 ± 36 | 2.54 ± 0.11 | 1.016 ± 0.003 | -0.11 ± 0.08 | <38 | 3 | ... | ... | ... | ... | ... |
| 44654 | 08114045-4920115 | 13.3 ± 0.3 | 3885 ± 106 | ... | 0.844 ± 0.006 | -0.08 ± 0.10 | ... | ... | ... | ... | ... | ... | ... |
| 44363 | 08091025-4902250 | 16.3 ± 0.2 | 4309 ± 590 | ... | 0.875 ± 0.004 | -0.06 ± 0.14 | 283 ± 4 | 1 | B | A | ... | A | ... |
| 44364 | 08091082-4929167 | 11.9 ± 0.2 | 3648 ± 41 | ... | 0.824 ± 0.006 | -0.19 ± 0.16 | <100 | 3 | A | A | ... | A | ... |
| 44365 | 08091111-4914183 | 11.4 ± 0.5 | 3352 ± 80 | ... | 0.841 ± 0.011 | -0.27 ± 0.14 | <100 | 3 | A | A | ... | A | ... |
| 44366 | 08091164-4857019 | 74.7 ± 0.2 | 4919 ± 100 | 3.04 ± 0.02 | 1.003 ± 0.003 | -0.16 ± 0.12 | <22 | 3 | ... | ... | ... | ... | ... |
| 44367 | 08091190-4914427 | 10.2 ± 3.2 | ... | ... | ... | ... | ... | ... | ... | ... | ... | ... | ... |
| 44673 | 08120146-4936078 | 12.6 ± 0.6 | 3457 ± 29 | ... | 0.857 ± 0.016 | ... | ... | ... | ... | ... | ... | ... | ... |
| 44401 | 08092759-4852241 | 140.9 ± 0.2 | 4536 ± 120 | ... | 1.027 ± 0.003 | -0.08 ± 0.06 | <35 | 3 | ... | ... | ... | ... | G |
| 44402 | 08092916-4901240 | 9.1 ± 0.2 | 5051 ± 119 | 4.27 ± 0.03 | 0.970 ± 0.003 | -0.04 ± 0.01 | <19 | 3 | ... | ... | ... | ... | ... |
| 44403 | 08093010-4855077 | 13.4 ± 0.6 | 3388 ± 3 | 4.64 ± 0.16 | 0.835 ± 0.013 | -0.27 ± 0.14 | <100 | 3 | A | A | ... | A | ... |
| 44674 | 08120383-4912045 | 15.3 ± 2.3 | ... | ... | ... | ... | ... | ... | ... | ... | ... | ... | ... |
| 44675 | 08120386-4903579 | 109.7 ± 0.2 | 4520 ± 116 | ... | 1.044 ± 0.007 | -0.22 ± 0.09 | <48 | 3 | ... | ... | ... | ... | G |
| 44676 | 08120387-4918367 | 49.8 ± 0.2 | 4944 ± 106 | ... | 1.019 ± 0.003 | -0.13 ± 0.13 | <20 | 3 | ... | ... | ... | ... | G |
| 44404 | 08093053-4920443 | 14.2 ± 0.2 | 7025 ± 13 | ... | 0.999 ± 0.001 | -0.08 ± 0.01 | ... | ... | ... | ... | ... | ... | ... |
| 44405 | 08093097-4855153 | 42.3 ± 0.2 | 4946 ± 35 | 2.71 ± 0.20 | 1.037 ± 0.003 | -0.02 ± 0.09 | <37 | 3 | ... | ... | ... | ... | G |
| 44677 | 08120453-4935333 | 72.5 ± 0.2 | 4699 ± 7 | ... | 1.033 ± 0.005 | -0.33 ± 0.01 | 38 ± 8 | 1 | ... | ... | ... | ... | G |
| 44406 | 08093105-4853166 | 99.3 ± 0.2 | 4519 ± 39 | ... | 1.036 ± 0.004 | -0.21 ± 0.10 | <35 | 3 | ... | ... | ... | ... | G |
| 44678 | 08120487-4914070 | 99.1 ± 0.2 | 4959 ± 150 | ... | 1.011 ± 0.003 | -0.67 ± 0.20 | <18 | 3 | ... | ... | ... | ... | G |
| 44679 | 08120501-4900055 | 51.4 ± 0.2 | 5064 ± 230 | ... | 1.014 ± 0.004 | -0.21 ± 0.17 | <12 | 3 | ... | ... | ... | ... | G |
| 44407 | 08093246-4911128 | 11.3 ± 0.7 | ... | ... | ... | ... | ... | ... | ... | ... | ... | ... | ... |
| 44437 | 08094807-4936237 | 29.2 ± 0.2 | 5190 ± 37 | 4.04 ± 0.02 | 0.982 ± 0.005 | 0.03 ± 0.01 | 164 ± 3 | 1 | ... | ... | ... | ... | ... |
| 44438 | 08095021-4927232 | 12.5 ± 0.3 | 3654 ± 136 | 4.58 ± 0.16 | 0.815 ± 0.010 | -0.19 ± 0.13 | <100 | 3 | A | A | ... | A | ... |
| 44439 | 08095046-4846112 | 474.3 ± 1.6 | 3305 ± 16 | ... | 0.831 ± 0.019 | ... | ... | ... | ... | ... | ... | ... | ... |
| 44440 | 08095054-4851029 | 98.6 ± 0.2 | 4921 ± 13 | 2.86 ± 0.18 | 1.014 ± 0.004 | -0.14 ± 0.08 | 7 ± 2 | 1 | ... | ... | ... | ... | G |
| 44466 | 08100373-4924574 | 13.1 ± 0.2 | 5395 ± 37 | 4.27 ± 0.06 | 0.980 ± 0.002 | -0.01 ± 0.07 | ... | ... | ... | ... | ... | ... | ... |
| 3001 | 08100380-4901071 | 13.9 ± 0.4 | 5914 ± 113 | 4.39 ± 0.12 | 0.999 ± 0.002 | -0.01 ± 0.13 | 162 ± 7 | 2 | A | B | ... | B | ... |
| 44467 | 08100442-4850544 | 27.7 ± 0.2 | 4717 ± 89 | 2.63 ± 0.12 | 1.012 ± 0.003 | 0.03 ± 0.08 | <41 | 3 | ... | ... | ... | ... | ... |
| 44468 | 08100542-4907585 | 11.8 ± 1.1 | 3329 ± 27 | 4.82 ± 0.07 | 0.813 ± 0.013 | -0.27 ± 0.14 | <100 | 3 | A | A | ... | A | ... |
| 44469 | 08100558-4914581 | 35.6 ± 0.2 | 5011 ± 168 | ... | 1.006 ± 0.003 | -0.17 ± 0.16 | ... | ... | ... | ... | ... | ... | ... |
| 44470 | 08100561-4926254 | 15.0 ± 1.8 | ... | ... | ... | ... | ... | ... | ... | ... | ... | ... | ... |
| 44471 | 08100575-4927298 | 34.7 ± 0.3 | 3883 ± 107 | 4.59 ± 0.10 | 0.823 ± 0.007 | -0.21 ± 0.08 | <46 | 3 | ... | ... | ... | ... | ... |
| 44472 | 08100616-4925589 | 12.7 ± 0.5 | ... | ... | ... | ... | ... | ... | ... | ... | ... | ... | ... |
| 44282 | 08080263-4903144 | 61.7 ± 0.2 | 4601 ± 135 | ... | 1.033 ± 0.004 | -0.15 ± 0.12 | <34 | 3 | ... | ... | ... | ... | G |
| 44283 | 08080299-4925420 | 8.1 ± 0.2 | 5019 ± 120 | ... | 1.012 ± 0.003 | -0.14 ± 0.17 | <21 | 3 | ... | ... | ... | ... | G |
| 44292 | 08081347-4927509 | 72.8 ± 0.2 | 4651 ± 75 | ... | 1.026 ± 0.004 | -0.38 ± 0.17 | <25 | 3 | ... | ... | ... | ... | G |
| 44288 | 08080969-4904154 | 21.7 ± 0.2 | 5217 ± 65 | ... | 1.003 ± 0.004 | -0.25 ± 0.16 | ... | ... | ... | ... | ... | ... | ... |
| 44289 | 08081095-4917234 | 118.1 ± 0.2 | 4731 ± 29 | ... | 1.032 ± 0.003 | -0.20 ± 0.12 | <12 | 3 | ... | ... | ... | ... | G |
| 44300 | 08082881-4924354 | -5.2 ± 0.2 | 4909 ± 209 | 4.40 ± 0.16 | 0.964 ± 0.004 | -0.04 ± 0.03 | <27 | 3 | ... | ... | ... | ... | ... |
| 44290 | 08081164-4909164 | 13.3 ± 1.0 | 3333 ± 21 | ... | 0.841 ± 0.018 | ... | ... | ... | ... | ... | ... | ... | ... |
| 44291 | 08081196-4930268 | 19.0 ± 0.4 | 3342 ± 21 | ... | 0.863 ± 0.017 | ... | 583 ± 43 | 1 | B | B | ... | B | ... |
| 44301 | 08082965-4855452 | -15.2 ± 302.6 | ... | ... | ... | ... | ... | ... | ... | ... | ... | ... | ... |
| 44414 | 08093593-4927191 | 12.5 ± 0.2 | 3992 ± 149 | ... | 0.882 ± 0.003 | -0.06 ± 0.09 | ... | ... | ... | ... | ... | ... | ... |
| 44302 | 08083266-4926106 | 69.2 ± 0.2 | 4932 ± 121 | ... | 1.017 ± 0.004 | -0.18 ± 0.15 | <23 | 3 | ... | ... | ... | ... | G |
| 44303 | 08083306-4904133 | 125.7 ± 0.2 | 4569 ± 109 | ... | 1.020 ± 0.003 | -0.17 ± 0.16 | <25 | 3 | ... | ... | ... | ... | G |
| 44421 | 08094020-4846264 | -18.4 ± 0.3 | 5706 ± 13 | ... | 1.011 ± 0.005 | -0.33 ± 0.08 | <19 | 3 | ... | ... | ... | ... | ... |
| 44422 | 08094039-4916447 | 62.1 ± 0.2 | 5105 ± 219 | ... | 1.015 ± 0.002 | -0.26 ± 0.15 | <20 | 3 | ... | ... | ... | ... | G |
| 44304 | 08083331-4935136 | 75.5 ± 0.2 | 4837 ± 185 | ... | 1.018 ± 0.003 | -0.22 ± 0.19 | ... | ... | ... | ... | ... | ... | G |
| 44427 | 08094318-4914040 | 95.3 ± 0.2 | 4589 ± 172 | ... | 1.033 ± 0.003 | -0.35 ± 0.10 | 42 ± 6 | 1 | ... | ... | ... | ... | G |
| 44423 | 08094048-4915064 | 17.4 ± 0.2 | 5275 ± 179 | ... | 1.011 ± 0.003 | -0.10 ± 0.19 | <23 | 3 | ... | ... | ... | ... | ... |
| 44424 | 08094053-4930204 | 14.1 ± 0.3 | 3438 ± 25 | 4.63 ± 0.19 | 0.827 ± 0.014 | -0.25 ± 0.15 | ... | ... | ... | ... | ... | ... | ... |
| 44305 | 08083389-4907176 | 10.1 ± 0.2 | 4970 ± 180 | ... | 0.960 ± 0.004 | ... | 337 ± 12 | 1 | ... | ... | ... | ... | ... |
| 2994 | 08094083-4923503 | 11.3 ± 0.6 | 5658 ± 143 | 4.42 ± 0.13 | 0.995 ± 0.002 | 0.04 ± 0.14 | 230 ± 10 | 2 | A | A | ... | A | ... |
| 44425 | 08094157-4926399 | 330.3 ± 0.2 | 4948 ± 157 | ... | 1.022 ± 0.004 | -0.76 ± 0.20 | <26s | 3 | ... | ... | ... | ... | G |
| 44441 | 08095071-4910383 | -7.1 ± 0.2 | 6021 ± 219 | 3.93 ± 0.11 | 1.001 ± 0.001 | -0.34 ± 0.07 | 52 ± 5 | 1 | ... | ... | ... | ... | ... |
| 44315 | 08083910-4904025 | 19.1 ± 0.3 | 3583 ± 125 | ... | 0.854 ± 0.007 | -0.22 ± 0.14 | 308 ± 17 | 1 | B | B | ... | B | ... |
| 44442 | 08095108-4928057 | 7.5 ± 0.2 | 5808 ± 83 | ... | 0.995 ± 0.003 | ... | ... | ... | ... | ... | ... | ... | ... |
| 44443 | 08095145-4932300 | 43.6 ± 0.2 | 4938 ± 101 | 3.00 ± 0.08 | 1.007 ± 0.004 | -0.10 ± 0.09 | <21 | 3 | ... | ... | ... | ... | ... |

Article number, page 85 of 264

**Table C.4.** continued.

| ID | CNAME | RV (km s$^{-1}$) | $T_{\rm eff}$ (K) | $\log g$ (dex) | $\gamma^a$ | [Fe/H] (dex) | EW(Li)$^b$ (mÅ) | EW(Li) error flag$^c$ | Pop. A/B (Sacco)$^d$ | Pop. (Randich) | Pop. (Cantat-Gaudin) | Pop. (adopted final)$^e$ | Non-mem with Li$^f$ |
|---|---|---|---|---|---|---|---|---|---|---|---|---|---|
| 44316 | 08083983-4855229 | 18.5 ± 0.8 | … | … | … | … | … | … | … | … | … | … | … |
| 44426 | 08094267-4847084 | -30.3 ± 2.7 | … | … | … | … | … | … | … | … | … | … | … |
| 44444 | 08095212-4857590 | 13.5 ± 0.4 | 3312 ± 15 | … | 0.835 ± 0.016 | … | <100 | 3 | A | A | … | A | … |
| 44317 | 08084064-4853377 | 14.1 ± 0.8 | … | … | … | … | … | … | … | … | … | … | … |
| 44318 | 08084089-4858085 | -0.9 ± 0.2 | 6275 ± 32 | 4.00 ± 0.09 | 1.003 ± 0.002 | -0.05 ± 0.02 | <12 | 3 | … | … | … | … | … |
| 44576 | 08105210-4921138 | 12.6 ± 0.3 | 3588 ± 10 | … | 0.831 ± 0.006 | … | … | … | … | … | … | … | … |
| 44319 | 08084139-4847153 | 12.8 ± 0.2 | 4740 ± 28 | 2.57 ± 0.04 | 1.019 ± 0.004 | 0.03 ± 0.02 | <35 | 3 | … | … | … | … | … |
| 2981 | 08084225-4915153 | 48.2 ± 0.6 | 5069 ± 8 | 3.23 ± 0.01 | 1.009 ± 0.002 | -0.12 ± 0.03 | <15 | 3 | … | … | … | … | … |
| 2969 | 08105280-4936425 | 34.7 ± 0.6 | 4869 ± 13 | 3.01 ± 0.07 | … | 0.07 ± 0.03 | <20 | 3 | … | … | … | … | … |
| 44445 | 08095342-4926105 | 38.9 ± 0.2 | 5036 ± 107 | … | 1.010 ± 0.004 | -0.06 ± 0.09 | … | … | … | … | … | … | G |
| 44577 | 08105292-4849203 | 40.4 ± 0.2 | 4940 ± 142 | … | 1.010 ± 0.004 | -0.09 ± 0.11 | <31 | 3 | … | … | … | … | … |
| 44578 | 08105300-4852322 | 23.5 ± 0.7 | … | … | … | … | … | … | … | … | … | … | … |
| 44320 | 08084294-4903502 | 19.1 ± 1.5 | … | … | … | … | … | … | … | … | … | … | … |
| 44446 | 08095355-4911523 | 59.7 ± 0.2 | 4747 ± 52 | 2.56 ± 0.17 | 1.019 ± 0.002 | -0.13 ± 0.08 | 27 ± 9 | 1 | … | … | … | … | G |
| 44321 | 08084313-4854304 | 25.5 ± 0.2 | 5154 ± 9 | 3.37 ± 0.20 | 1.009 ± 0.003 | -0.08 ± 0.12 | <20 | 3 | … | … | … | … | … |
| 44579 | 08105523-4926365 | 27.6 ± 0.4 | 3321 ± 19 | … | 0.825 ± 0.017 | … | … | … | … | … | … | … | … |
| 44332 | 08085272-4848140 | 44.8 ± 0.2 | 5160 ± 81 | … | 1.008 ± 0.004 | -0.18 ± 0.13 | … | … | … | … | … | … | … |
| 44580 | 08105529-4901015 | 83.5 ± 0.2 | 4707 ± 24 | 2.48 ± 0.08 | 1.022 ± 0.004 | -0.10 ± 0.01 | <31 | 3 | … | … | … | … | G |
| 44457 | 08095786-4849182 | 14.0 ± 0.2 | 5257 ± 1 | … | 0.990 ± 0.003 | 0.05 ± 0.01 | 256 ± 3 | 1 | A | A | … | A | … |
| 44581 | 08105543-4931088 | 44.5 ± 0.2 | 4971 ± 132 | … | 1.015 ± 0.004 | -0.14 ± 0.19 | <28 | 3 | … | … | … | … | G |
| 44582 | 08105561-4852342 | -7.7 ± 0.2 | 5021 ± 142 | … | 1.004 ± 0.003 | -0.20 ± 0.23 | <16 | 3 | … | … | … | … | … |
| 44458 | 08095873-4934465 | 12.2 ± 1.6 | … | … | … | … | … | … | … | … | … | … | … |
| 44333 | 08085430-4924494 | 402.5 ± 1.5 | … | … | … | … | … | … | … | … | … | … | … |
| 44334 | 08085446-4929553 | 23.5 ± 0.2 | 4025 ± 98 | 4.51 ± 0.07 | 0.862 ± 0.003 | -0.04 ± 0.15 | … | … | … | … | … | … | … |
| 44596 | 08110268-4911084 | 13.1 ± 2.4 | … | … | … | … | … | … | … | … | … | … | … |
| 3000 | 08095940-4859111 | 29.7 ± 0.6 | 5239 ± 4 | 4.50 ± 0.02 | 0.983 ± 0.002 | 0.06 ± 0.07 | 162 ± 5 | 2 | … | … | … | … | … |
| 44335 | 08085448-4859096 | 7.7 ± 2.6 | 3344 ± 69 | … | 0.887 ± 0.013 | -0.27 ± 0.14 | <100 | 3 | … | A | … | A | … |
| 44336 | 08085463-4927033 | 14.2 ± 0.7 | … | … | … | … | … | … | … | … | … | … | … |
| 44597 | 08110321-4930298 | 25.7 ± 0.2 | 5122 ± 54 | 3.50 ± 0.03 | 1.000 ± 0.005 | -0.08 ± 0.16 | <32 | 3 | … | … | … | … | … |
| 44337 | 08085572-4857167 | 37.0 ± 0.2 | 4873 ± 24 | … | 0.994 ± 0.004 | 0.05 ± 0.03 | <38 | 3 | … | … | … | … | … |
| 44473 | 08100708-4912597 | 12.8 ± 0.5 | 3399 ± 26 | … | 0.839 ± 0.016 | … | … | … | … | … | … | … | … |
| 44598 | 08110403-4852137 | 54.1 ± 0.2 | 4762 ± 31 | 2.59 ± 0.18 | 1.021 ± 0.005 | -0.12 ± 0.07 | 423 ± 8 | 1 | … | … | … | … | Li-rich G |
| 44474 | 08100749-4910447 | 12.0 ± 0.2 | 4233 ± 402 | … | 0.888 ± 0.004 | -0.05 ± 0.10 | 302 ± 3 | 1 | A | A | … | A | … |
| 44599 | 08110435-4853491 | 62.0 ± 0.2 | 5017 ± 285 | … | 1.015 ± 0.003 | -0.15 ± 0.21 | 369 ± 5 | 1 | … | … | … | … | Li-rich G |
| 44475 | 08100755-4856334 | 31.4 ± 0.2 | 4535 ± 154 | 2.44 ± 0.18 | 1.011 ± 0.002 | 0.11 ± 0.06 | <50 | 3 | … | … | … | … | G |
| 44476 | 08100759-4900035 | 140.0 ± 0.2 | 4607 ± 1 | … | 1.028 ± 0.003 | -0.25 ± 0.13 | <20 | 3 | … | … | … | … | G |
| 44344 | 08085993-4849091 | 124.8 ± 0.2 | 4637 ± 96 | … | 1.020 ± 0.005 | -0.34 ± 0.15 | … | … | … | … | … | … | G |
| 2957 | 08100777-4855099 | 1.7 ± 0.6 | 6247 ± 86 | 3.82 ± 0.16 | 1.010 ± 0.002 | -0.03 ± 0.11 | <36 | 3 | … | … | … | … | … |
| 44345 | 08090022-4920332 | 12.6 ± 0.2 | 5127 ± 172 | … | 0.966 ± 0.004 | … | 237 ± 8 | 1 | A | A | … | A | … |
| 44600 | 08110496-4859465 | 12.4 ± 0.3 | 3543 ± 10 | … | 0.841 ± 0.011 | -0.23 ± 0.14 | <100 | 3 | A | A | … | A | … |
| 44252 | 08072373-4902233 | 48.1 ± 0.2 | 4703 ± 37 | 2.49 ± 0.08 | 1.021 ± 0.004 | 0.03 ± 0.03 | <39 | 3 | … | … | … | … | G |
| 44601 | 08110541-4920544 | 12.3 ± 0.7 | … | … | … | … | … | … | … | … | … | … | … |
| 44346 | 08090041-4912594 | 38.7 ± 1.2 | … | … | … | … | … | … | … | … | … | … | … |
| 2959 | 08100810-4912157 | 13.9 ± 0.6 | 5347 ± 22 | 4.50 ± 0.09 | 0.983 ± 0.002 | -0.17 ± 0.03 | 143 ± 6 | 2 | A | … | … | A | … |
| 44602 | 08110558-4939029 | 89.0 ± 0.2 | 4755 ± 49 | … | 1.021 ± 0.004 | -0.17 ± 0.08 | <34 | 3 | … | … | … | … | G |
| 44253 | 08072494-4901368 | 79.2 ± 0.2 | 5738 ± 200 | 4.19 ± 0.14 | 0.989 ± 0.004 | 0.21 ± 0.04 | <19 | 3 | … | … | … | … | … |
| 44347 | 08090090-4847206 | 80.4 ± 0.2 | 4939 ± 157 | … | 1.012 ± 0.003 | -0.37 ± 0.15 | … | … | … | … | … | … | G |
| 44477 | 08100826-4919565 | 30.7 ± 0.2 | 6309 ± 23 | 4.18 ± 0.05 | 0.999 ± 0.001 | 0.05 ± 0.02 | <11 | 3 | … | … | … | … | … |
| 44259 | 08073400-4902576 | 0.8 ± 0.2 | 5108 ± 105 | … | 1.020 ± 0.003 | … | … | … | … | … | … | … | G |
| 44609 | 08110989-4940283 | 55.5 ± 0.2 | 5068 ± 192 | … | 1.016 ± 0.004 | -0.21 ± 0.21 | <10 | 3 | … | … | … | … | G |
| 44610 | 08110996-4857056 | 149.0 ± 0.2 | 4617 ± 94 | … | 1.044 ± 0.004 | … | <73 | 3 | … | … | … | … | G |
| 44478 | 08100859-4849348 | 111.5 ± 0.2 | 4543 ± 84 | … | 1.022 ± 0.003 | -0.20 ± 0.08 | <30 | 3 | … | … | … | … | G |
| 44348 | 08090208-4849580 | 57.3 ± 0.2 | 4674 ± 43 | … | 1.024 ± 0.004 | -0.29 ± 0.07 | 24 ± 2 | 1 | … | … | … | … | G |
| 2986 | 08090250-4858172 | 23.5 ± 0.6 | 5732 ± 5 | 4.57 ± 0.11 | 0.992 ± 0.002 | 0.05 ± 0.05 | 189 ± 4 | 2 | B | … | … | B | … |
| 44611 | 08111019-4925258 | 77.5 ± 0.2 | 4911 ± 44 | 2.77 ± 0.08 | 1.011 ± 0.005 | -0.01 ± 0.08 | <33 | 3 | … | … | … | … | G |
| 44479 | 08100893-4915413 | 13.5 ± 0.2 | 4419 ± 372 | … | 0.915 ± 0.003 | -0.01 ± 0.08 | 268 ± 6 | 1 | A | … | … | A | … |
| 44260 | 08073423-4858044 | 21.5 ± 0.5 | … | … | … | … | … | … | … | … | … | … | … |
| 44349 | 08090287-4906022 | 14.3 ± 0.2 | 6451 ± 18 | 4.23 ± 0.04 | 1.000 ± 0.001 | 0.07 ± 0.01 | 107 ± 5 | 1 | A | A | … | A | … |
| 44612 | 08111100-4930549 | 42.6 ± 0.2 | 5182 ± 50 | 3.48 ± 0.18 | 1.006 ± 0.004 | -0.15 ± 0.11 | … | … | … | … | … | … | … |

| ID | CNAME | RV (km s$^{-1}$) | $T_{\text{eff}}$ (K) | $\log g$ (dex) | $\gamma^a$ | [Fe/H] (dex) | EW(Li)$^b$ (mÅ) | EW(Li) error flag$^c$ | Pop. A/B (Sacco)$^d$ | Pop. (Randich) | Pop. (Cantat-Gaudin) | Pop. (adopted final)$^e$ | Non-mem with Li$^f$ |
|---|---|---|---|---|---|---|---|---|---|---|---|---|---|
| 44490 | 08101374-4917112 | 12.2 ± 0.3 | 3535 ± 82 | ... | 0.828 ± 0.008 | -0.20 ± 0.14 | <100 | 3 | A | A | ... | A | ... |
| 44261 | 08073463-4904107 | 60.1 ± 0.3 | 4366 ± 268 | 4.59 ± 0.13 | 0.965 ± 0.005 | -0.16 ± 0.10 | 382 ± 12 | ... | ... | ... | ... | ... | ... |
| 44350 | 08090392-4845560 | 27.9 ± 0.2 | 5594 ± 3 | ... | 1.008 ± 0.004 | -0.21 ± 0.04 | ... | ... | ... | ... | ... | ... | ... |
| 44491 | 08101431-4928268 | 12.5 ± 0.6 | 3594 ± 20 | ... | 0.810 ± 0.014 | ... | <100 | 3 | A | A | A | A | ... |
| 44613 | 08111121-4928358 | 6.2 ± 0.2 | 5063 ± 111 | 3.42 ± 0.11 | 1.002 ± 0.005 | -0.03 ± 0.07 | <25 | 3 | ... | ... | ... | ... | ... |
| 44368 | 08091194-4926542 | 14.1 ± 0.3 | 3557 ± 164 | ... | 0.823 ± 0.010 | -0.12 ± 0.15 | <100 | 3 | A | A | ... | A | ... |
| 44369 | 08091206-4910152 | 11.8 ± 0.6 | ... | ... | ... | ... | ... | ... | ... | ... | ... | ... | ... |
| 44492 | 08101474-4912320 | 108.1 ± 0.2 | 5042 ± 51 | ... | 1.003 ± 0.002 | -0.30 ± 0.23 | <8 | 3 | ... | ... | ... | ... | ... |
| 44614 | 08111134-4904442 | 14.1 ± 0.2 | 6955 ± 13 | ... | 1.001 ± 0.001 | 0.04 ± 0.01 | 54 ± 1 | 1 | A | A | ... | A | ... |
| 44370 | 08091210-4927299 | 411.8 ± 1.7 | ... | ... | ... | ... | ... | ... | ... | ... | ... | ... | ... |
| 44493 | 08101502-4851129 | 12.7 ± 0.2 | 3642 ± 44 | ... | 0.830 ± 0.005 | -0.20 ± 0.16 | ... | ... | ... | ... | ... | ... | ... |
| 44371 | 08091223-4911255 | 64.0 ± 0.2 | ... | ... | ... | ... | ... | ... | ... | ... | ... | ... | ... |
| 44372 | 08091240-4937351 | 21.5 ± 0.2 | 4630 ± 6 | 4.44 ± 0.03 | 0.951 ± 0.003 | 0.03 ± 0.11 | ... | ... | ... | ... | ... | ... | ... |
| 44262 | 08073643-4922200 | 67.9 ± 0.2 | 4988 ± 63 | ... | 1.016 ± 0.004 | -0.25 ± 0.19 | ... | ... | ... | ... | ... | ... | G |
| 2987 | 08091307-4925401 | 13.9 ± 0.6 | 5234 ± 77 | 4.52 ± 0.12 | 0.986 ± 0.003 | -0.08 ± 0.11 | 273 ± 11 | 2 | A | A | ... | A | ... |
| 44494 | 08101542-4911095 | 62.6 ± 0.2 | 5852 ± 72 | ... | 0.999 ± 0.002 | 0.11 ± 0.02 | <17 | 3 | ... | ... | ... | ... | ... |
| 44373 | 08091332-4857599 | 11.4 ± 0.2 | 5106 ± 84 | 3.85 ± 0.11 | 0.987 ± 0.003 | 0.00 ± 0.03 | 198 ± 5 | 1 | A | A | ... | A | ... |
| 44495 | 08101546-4905487 | 13.1 ± 0.2 | 5822 ± 240 | 4.18 ± 0.03 | 0.989 ± 0.001 | 0.05 ± 0.13 | 111 ± 3 | 1 | A | A | ... | A | ... |
| 44615 | 08111213-4901226 | 15.6 ± 2.2 | ... | ... | ... | ... | ... | ... | ... | ... | ... | ... | ... |
| 44374 | 08091341-4929232 | 10.8 ± 0.7 | ... | ... | ... | ... | ... | ... | ... | ... | ... | ... | ... |
| 44375 | 08091348-4913304 | 54.3 ± 0.2 | 4756 ± 10 | 2.56 ± 0.07 | 1.015 ± 0.003 | -0.08 ± 0.08 | <29 | 3 | ... | ... | ... | ... | G |
| 44496 | 08101679-4913171 | 93.5 ± 0.2 | 4879 ± 73 | 2.66 ± 0.10 | 1.014 ± 0.003 | -0.14 ± 0.08 | <25 | 3 | ... | ... | ... | ... | G |
| 44616 | 08111239-4850242 | 81.4 ± 0.2 | 5030 ± 77 | ... | 1.014 ± 0.003 | -0.35 ± 0.16 | <25 | 3 | ... | ... | ... | ... | G |
| 44497 | 08101691-4856292 | 24.9 ± 0.2 | 5097 ± 104 | ... | 1.016 ± 0.003 | ... | <8 | 3 | ... | ... | ... | ... | ... |
| 44376 | 08091385-4900132 | 58.1 ± 0.2 | 4862 ± 91 | 2.62 ± 0.09 | 1.015 ± 0.004 | -0.08 ± 0.10 | <31 | 3 | ... | ... | ... | ... | G |
| 44498 | 08101697-4854194 | 12.1 ± 0.2 | 5678 ± 9 | 4.24 ± 0.15 | 0.990 ± 0.003 | 0.04 ± 0.03 | 152 ± 4 | 1 | A | A | ... | A | ... |
| 44377 | 08091385-4933181 | 17.3 ± 0.3 | 3414 ± 22 | ... | 0.872 ± 0.012 | ... | ... | ... | ... | ... | ... | ... | ... |
| 44499 | 08101697-4925294 | 61.9 ± 0.2 | 4980 ± 96 | 2.96 ± 0.05 | 1.008 ± 0.003 | -0.14 ± 0.09 | <26 | 3 | ... | ... | ... | ... | ... |
| 44378 | 08091401-4904029 | 0.3 ± 0.2 | 6847 ± 14 | ... | ... | 0.42 ± 0.01 | ... | ... | ... | ... | ... | ... | ... |
| 44625 | 08112121-4913280 | 53.7 ± 0.2 | 6226 ± 38 | 4.38 ± 0.09 | 0.992 ± 0.003 | 0.25 ± 0.02 | 139 ± 3 | 1 | ... | ... | ... | ... | ... |
| 44517 | 08102311-4902454 | 14.2 ± 0.3 | 3736 ± 9 | ... | 0.844 ± 0.008 | ... | <100 | 3 | A | A | ... | A | ... |
| 44626 | 08112186-4907446 | 41.7 ± 0.2 | 4787 ± 90 | 2.62 ± 0.12 | 1.020 ± 0.003 | -0.09 ± 0.03 | <26 | 3 | ... | ... | ... | ... | G |
| 44518 | 08102343-4855483 | 79.0 ± 0.2 | 5017 ± 62 | ... | 1.018 ± 0.004 | -0.22 ± 0.14 | <29 | 3 | ... | ... | ... | ... | G |
| 2976 | 08074522-4912215 | -2.5 ± 0.6 | 5794 ± 34 | 4.41 ± 0.01 | ... | -0.06 ± 0.04 | 80 ± 6 | 2 | ... | ... | ... | ... | ... |
| 44627 | 08112213-4921470 | 14.2 ± 0.2 | 4056 ± 187 | ... | 0.882 ± 0.004 | -0.08 ± 0.11 | 192 ± 7 | 1 | A | A | ... | A | ... |
| 44270 | 08074563-4924438 | 51.0 ± 0.2 | 5107 ± 157 | ... | 1.025 ± 0.003 | -0.16 ± 0.13 | <16 | 3 | ... | ... | ... | ... | G |
| 44271 | 08074609-4915345 | 13.1 ± 0.2 | 3995 ± 140 | 4.46 ± 0.16 | 0.859 ± 0.004 | 0.02 ± 0.18 | 35 ± 3 | 1 | ... | A | ... | A | ... |
| 44628 | 08112240-4854481 | 48.4 ± 0.2 | 5013 ± 183 | ... | 1.005 ± 0.005 | -0.09 ± 0.14 | <28 | 3 | ... | ... | ... | ... | ... |
| 44519 | 08102486-4901495 | 13.4 ± 0.2 | 5950 ± 42 | 4.02 ± 0.10 | 0.998 ± 0.002 | 0.00 ± 0.02 | 160 ± 4 | 1 | A | A | ... | A | ... |
| 44386 | 08092129-4900412 | 6.6 ± 0.2 | 4975 ± 237 | ... | 0.968 ± 0.003 | 0.03 ± 0.04 | <21 | 3 | ... | ... | ... | ... | ... |
| 44520 | 08102491-4851482 | 12.2 ± 0.2 | ... | ... | ... | ... | ... | ... | ... | ... | ... | ... | ... |
| 44387 | 08092131-4905431 | 12.6 ± 0.3 | 3830 ± 107 | ... | 0.840 ± 0.006 | -0.16 ± 0.13 | ... | ... | ... | ... | ... | ... | ... |
| 44629 | 08112261-4850565 | 12.6 ± 1.8 | ... | ... | ... | ... | ... | ... | ... | ... | ... | ... | ... |
| 44388 | 08092131-4905540 | 10.7 ± 0.2 | 3570 ± 67 | ... | 0.836 ± 0.005 | -0.24 ± 0.14 | ... | ... | ... | ... | ... | ... | ... |
| 44521 | 08102529-4923024 | 29.1 ± 0.2 | 3499 ± 83 | 4.64 ± 0.14 | 0.824 ± 0.008 | -0.26 ± 0.15 | <100 | 3 | ... | ... | ... | ... | ... |
| 44272 | 08074750-4911094 | 17.4 ± 0.8 | ... | ... | ... | ... | ... | ... | ... | ... | ... | ... | ... |
| 44630 | 08112291-4933131 | 83.9 ± 0.2 | 4863 ± 28 | 2.64 ± 0.10 | 1.017 ± 0.004 | -0.15 ± 0.07 | <36 | 3 | ... | ... | ... | ... | G |
| 44631 | 08112300-4902284 | 73.2 ± 0.2 | 4839 ± 116 | 2.54 ± 0.12 | 1.011 ± 0.003 | -0.25 ± 0.20 | <22 | 3 | ... | ... | ... | ... | G |
| 44389 | 08092201-4931420 | 12.8 ± 0.2 | 4903 ± 171 | ... | 1.017 ± 0.004 | -0.04 ± 0.08 | <25 | 3 | ... | ... | ... | ... | G |
| 44522 | 08102629-4937253 | 21.9 ± 0.2 | 4838 ± 330 | 4.45 ± 0.11 | 0.951 ± 0.004 | -0.02 ± 0.04 | 42 ± 5 | 1 | ... | ... | ... | ... | ... |
| 2988 | 08092276-4916309 | 12.8 ± 0.6 | 5307 ± 78 | 4.43 ± 0.15 | 0.988 ± 0.002 | -0.08 ± 0.01 | 217 ± 1 | 2 | A | A | ... | A | ... |
| 2989 | 08092293-4907575 | 13.2 ± 0.6 | 6294 ± 101 | 4.49 ± 0.18 | 0.997 ± 0.001 | 0.00 ± 0.03 | 120 ± 1 | 2 | A | A | ... | A | ... |
| 44523 | 08102704-4908487 | 12.6 ± 0.2 | 5523 ± 39 | ... | 0.994 ± 0.001 | -0.04 ± 0.02 | 186 ± 3 | 1 | A | A | ... | A | ... |
| 44635 | 08112887-4923364 | 12.1 ± 0.3 | 3705 ± 94 | 4.57 ± 0.01 | 0.807 ± 0.007 | -0.21 ± 0.12 | <100 | 3 | A | A | ... | A | ... |
| 44390 | 08092351-4923416 | 121.7 ± 0.2 | 4485 ± 102 | ... | 1.040 ± 0.002 | -0.26 ± 0.17 | <21 | 3 | ... | ... | ... | ... | G |
| 44636 | 08112894-4921190 | 20.0 ± 0.4 | 3388 ± 2 | ... | 0.871 ± 0.013 | -0.27 ± 0.14 | 306 ± 11 | 1 | B | B | ... | B | ... |
| 44280 | 08075918-4913102 | 100.9 ± 0.2 | 5087 ± 129 | ... | 1.002 ± 0.003 | -0.57 ± 0.24 | ... | ... | ... | ... | ... | ... | ... |
| 44531 | 08103301-4915066 | 2.5 ± 0.2 | 4876 ± 154 | ... | 1.019 ± 0.002 | -0.06 ± 0.09 | <23 | 3 | ... | ... | ... | ... | G |
| 44391 | 08092422-4921455 | 29.1 ± 0.2 | 5041 ± 52 | ... | 0.997 ± 0.003 | 0.01 ± 0.07 | <35 | 3 | ... | ... | ... | ... | ... |

**Table C.4.** continued.

| ID | CNAME | RV (km s$^{-1}$) | $T_{\text{eff}}$ (K) | $\log g$ (dex) | $\gamma^a$ | [Fe/H] (dex) | EW(Li)$^b$ (mÅ) | EW(Li) error flag$^c$ | Pop. A/B (Sacco)$^d$ | Pop. (Randich) | Pop. (Cantat-Gaudin) | Pop. (adopted final)$^e$ | Non-mem with Li$^f$ |
|---|---|---|---|---|---|---|---|---|---|---|---|---|---|
| 44392 | 08092438-4906282 | 11.9 ± 0.2 | 3725 ± 47 | ... | 0.828 ± 0.006 | -0.20 ± 0.12 | ... | ... | ... | ... | ... | ... | ... |
| 44637 | 08112917-4907229 | 13.1 ± 0.3 | 3471 ± 5 | ... | 0.840 ± 0.014 | -0.22 ± 0.15 | <100 | 3 | A | A | ... | A | ... |
| 44281 | 08075933-4910436 | 42.4 ± 0.2 | 4750 ± 70 | 2.56 ± 0.14 | 1.016 ± 0.005 | -0.05 ± 0.14 | <34 | 3 | ... | ... | ... | ... | G |
| 44532 | 08103397-4922186 | 12.9 ± 0.3 | 3533 ± 23 | ... | 0.827 ± 0.010 | -0.22 ± 0.15 | <100 | 3 | A | A | A | A | ... |
| 44408 | 08093277-4855312 | 26.6 ± 0.3 | 3626 ± 21 | ... | 0.809 ± 0.015 | ... | <100 | 3 | ... | ... | ... | ... | ... |
| 44533 | 08103432-4900496 | -9.5 ± 0.2 | 6540 ± 21 | 4.11 ± 0.05 | 1.004 ± 0.001 | -0.24 ± 0.02 | <9 | 3 | ... | ... | ... | ... | ... |
| 44638 | 08113045-4915475 | 99.9 ± 0.2 | 4551 ± 86 | 2.35 ± 0.17 | 1.019 ± 0.003 | -0.05 ± 0.02 | <38 | 3 | ... | ... | ... | ... | G |
| 44639 | 08113082-4933416 | 23.6 ± 0.2 | 4950 ± 71 | ... | 1.018 ± 0.004 | -0.05 ± 0.08 | <37 | 3 | ... | ... | ... | ... | G |
| 44534 | 08103470-4908400 | 13.6 ± 0.2 | 5990 ± 76 | 4.06 ± 0.03 | 0.999 ± 0.001 | 0.06 ± 0.04 | 169 ± 2 | 1 | A | A | ... | A | ... |
| 44535 | 08103536-4936171 | 35.9 ± 0.2 | 5026 ± 164 | ... | 1.008 ± 0.003 | -0.10 ± 0.19 | <19 | 3 | ... | ... | ... | ... | ... |
| 44536 | 08103549-4905001 | 4.9 ± 0.2 | 4598 ± 183 | ... | ... | 0.03 ± 0.05 | ... | ... | ... | ... | ... | ... | ... |
| 44409 | 08093393-4938453 | 472.7 ± 5.1 | ... | ... | ... | ... | ... | ... | ... | ... | ... | ... | ... |
| 44537 | 08103571-4927495 | 12.5 ± 0.3 | 4154 ± 546 | 4.43 ± 0.20 | 0.838 ± 0.008 | -0.12 ± 0.11 | <55 | 3 | ... | A | A | A | ... |
| 44410 | 08093443-4906027 | 20.4 ± 0.5 | 3279 ± 13 | ... | 0.887 ± 0.020 | ... | 685 ± 23 | 1 | B | B | ... | B | ... |
| 44538 | 08103572-4909008 | 13.6 ± 0.2 | 4375 ± 413 | ... | 0.907 ± 0.003 | -0.02 ± 0.11 | 214 ± 6 | 1 | A | A | ... | A | ... |
| 44640 | 08113168-4900035 | -8.7 ± 0.4 | 3583 ± 79 | 4.62 ± 0.15 | 0.801 ± 0.011 | -0.24 ± 0.14 | <100 | 3 | ... | ... | ... | ... | ... |
| 44539 | 08103575-4937078 | 73.7 ± 0.2 | 5083 ± 28 | ... | 1.010 ± 0.004 | -0.18 ± 0.19 | 16 ± 3 | 1 | ... | ... | ... | ... | ... |
| 44641 | 08113190-4930135 | 20.0 ± 0.3 | 3480 ± 8 | ... | 0.853 ± 0.010 | -0.24 ± 0.15 | 178 ± 27 | 1 | B | B | ... | B | ... |
| 44411 | 08093458-4936318 | 83.9 ± 0.2 | 4604 ± 104 | ... | 1.036 ± 0.004 | ... | <35 | 3 | ... | ... | ... | ... | G |
| 44540 | 08103582-4920054 | 12.8 ± 0.3 | 3404 ± 22 | ... | 0.838 ± 0.012 | -0.27 ± 0.14 | ... | ... | ... | ... | ... | ... | ... |
| 44642 | 08113198-4917493 | 162.2 ± 0.2 | 4873 ± 82 | ... | 1.019 ± 0.003 | -0.19 ± 0.16 | ... | ... | ... | ... | ... | ... | G |
| 44541 | 08103637-4847086 | 148.6 ± 0.2 | 4910 ± 34 | ... | 1.019 ± 0.005 | -0.14 ± 0.07 | ... | ... | ... | ... | ... | ... | G |
| 44412 | 08093547-4913033 | 11.2 ± 1.4 | ... | ... | ... | ... | ... | ... | ... | ... | ... | ... | ... |
| 44413 | 08093552-4854210 | 144.9 ± 0.2 | 4643 ± 87 | ... | 1.024 ± 0.004 | ... | <42 | 3 | ... | ... | ... | ... | G |
| 2967 | 08104410-4913153 | 63.0 ± 0.6 | 5109 ± 24 | 2.44 ± 0.09 | 1.011 ± 0.002 | -0.37 ± 0.02 | <15 | 3 | ... | ... | ... | ... | G |
| 44655 | 08114140-4859550 | 24.2 ± 0.3 | 3585 ± 82 | ... | 0.819 ± 0.007 | -0.23 ± 0.14 | ... | ... | ... | ... | ... | ... | ... |
| 44559 | 08104437-4939001 | 20.5 ± 2.5 | ... | ... | ... | ... | ... | ... | ... | ... | ... | ... | ... |
| 44560 | 08104494-4929069 | 80.4 ± 0.2 | 5062 ± 45 | 3.55 ± 0.02 | 0.999 ± 0.005 | -0.06 ± 0.06 | <38 | 3 | ... | ... | ... | ... | ... |
| 44656 | 08114166-4911327 | 47.9 ± 0.2 | 5011 ± 306 | ... | 1.016 ± 0.005 | -0.26 ± 0.28 | <25 | 3 | ... | ... | ... | ... | G |
| 44561 | 08104546-4901068 | 11.8 ± 0.2 | 6311 ± 21 | 4.19 ± 0.05 | 0.998 ± 0.001 | -0.04 ± 0.02 | 114 ± 3 | 1 | A | A | ... | A | ... |
| 44562 | 08104554-4905204 | 10.2 ± 0.2 | 4879 ± 276 | ... | 0.973 ± 0.002 | 0.00 ± 0.07 | 273 ± 2 | 1 | ... | A | ... | A | ... |
| 44563 | 08104554-4931459 | 14.4 ± 1.2 | ... | ... | ... | ... | ... | ... | ... | ... | ... | ... | ... |
| 44564 | 08104570-4906171 | 11.1 ± 0.2 | 3624 ± 27 | ... | 0.835 ± 0.006 | -0.24 ± 0.13 | <100 | 3 | ... | A | ... | A | ... |
| 44657 | 08114433-4854471 | 17.1 ± 0.2 | 5150 ± 14 | ... | 1.010 ± 0.004 | -0.05 ± 0.11 | <19 | 3 | ... | ... | ... | ... | ... |
| 44565 | 08104618-4855013 | 12.7 ± 0.4 | 3340 ± 22 | 4.71 ± 0.10 | 0.831 ± 0.016 | -0.26 ± 0.14 | ... | ... | ... | ... | ... | ... | ... |
| 44658 | 08114495-4904558 | 36.5 ± 0.2 | 5063 ± 136 | ... | 1.015 ± 0.005 | -0.07 ± 0.12 | <21 | 3 | ... | ... | ... | ... | G |
| 44662 | 08115117-4850296 | 46.2 ± 0.2 | 4683 ± 69 | ... | 0.998 ± 0.004 | 0.00 ± 0.06 | <34 | 3 | ... | ... | ... | ... | ... |
| 44663 | 08115134-4911410 | 16.3 ± 0.2 | 5491 ± 145 | ... | 0.986 ± 0.002 | 0.01 ± 0.04 | 229 ± 16 | 1 | B | A | ... | A | ... |
| 44664 | 08115256-4851583 | 28.4 ± 0.2 | 5052 ± 241 | ... | 1.017 ± 0.004 | -0.19 ± 0.17 | <37 | 3 | ... | ... | ... | ... | G |
| 44665 | 08115345-4924253 | 403.2 ± 1.3 | ... | ... | ... | ... | ... | ... | ... | ... | ... | ... | ... |
| 44666 | 08115398-4919461 | 55.8 ± 0.2 | 4699 ± 123 | 4.38 ± 0.10 | 0.954 ± 0.004 | 0.01 ± 0.08 | <40 | 3 | ... | ... | ... | ... | ... |
| 2977 | 08074892-4917421 | 16.6 ± 0.6 | 4875 ± 58 | 3.32 ± 0.13 | 1.003 ± 0.003 | 0.26 ± 0.06 | 33 ± 6 | 1 | ... | ... | ... | ... | ... |
| 44394 | 08092574-4903156 | 13.3 ± 1.1 | 3407 ± 34 | ... | 0.820 ± 0.021 | ... | ... | ... | ... | ... | ... | ... | ... |
| 44274 | 08075006-4925445 | 69.0 ± 0.2 | 4636 ± 98 | ... | 1.027 ± 0.004 | ... | 52 ± 3 | 1 | ... | ... | ... | ... | G |
| 44395 | 08092586-4932443 | 20.2 ± 0.3 | 3625 ± 65 | ... | 0.850 ± 0.013 | -0.21 ± 0.14 | 354 ± 18 | 1 | B | B | ... | B | ... |
| 44396 | 08092592-4909585 | 12.1 ± 0.5 | 3347 ± 23 | 4.74 ± 0.02 | 0.828 ± 0.014 | -0.27 ± 0.14 | <100 | 3 | A | A | ... | A | ... |
| 44275 | 08075034-4924122 | 404.9 ± 1.7 | ... | ... | ... | ... | ... | ... | ... | ... | ... | ... | ... |
| 44276 | 08075124-4926449 | 77.2 ± 0.2 | 4479 ± 138 | 2.19 ± 0.19 | 1.026 ± 0.004 | 0.01 ± 0.12 | <33 | 3 | ... | ... | ... | ... | G |
| 44397 | 08092678-4932554 | 11.8 ± 0.3 | 3793 ± 70 | ... | 0.825 ± 0.006 | -0.17 ± 0.14 | ... | ... | ... | ... | ... | ... | ... |
| 44543 | 08103713-4916587 | 12.0 ± 0.2 | 4002 ± 216 | 4.52 ± 0.19 | 0.847 ± 0.004 | -0.03 ± 0.12 | 47 ± 5 | 1 | ... | ... | ... | ... | ... |
| 44398 | 08092683-4900187 | 53.7 ± 0.2 | 4878 ± 99 | ... | 1.011 ± 0.004 | -0.19 ± 0.18 | <19 | 3 | ... | ... | ... | ... | G |
| 44284 | 08080444-4927511 | 69.1 ± 0.2 | 4626 ± 41 | 2.36 ± 0.10 | 1.019 ± 0.004 | -0.11 ± 0.03 | 50 ± 5 | 1 | ... | ... | ... | ... | G |
| 44544 | 08103735-4928104 | 12.2 ± 0.3 | 3460 ± 62 | ... | 0.840 ± 0.011 | -0.22 ± 0.13 | <100 | 3 | A | A | ... | A | ... |
| 44399 | 08092732-4916061 | 100.7 ± 0.2 | 4728 ± 26 | ... | 1.020 ± 0.003 | -0.23 ± 0.12 | 22 ± 5 | 1 | ... | ... | ... | ... | G |
| 44545 | 08103764-4904185 | -9.4 ± 0.3 | 3644 ± 9 | ... | 0.820 ± 0.007 | ... | <100 | 3 | ... | ... | ... | ... | ... |
| 44400 | 08092734-4915305 | 105.0 ± 0.2 | 4950 ± 112 | ... | 1.017 ± 0.003 | -0.23 ± 0.22 | 14 ± 2 | 1 | ... | ... | ... | ... | G |
| 44546 | 08103784-4846175 | 1.4 ± 0.2 | 5439 ± 144 | 4.18 ± 0.05 | 0.987 ± 0.003 | 0.03 ± 0.02 | <17 | 3 | ... | ... | ... | ... | ... |
| 44285 | 08080591-4906006 | 16.3 ± 0.2 | 5006 ± 137 | ... | 1.012 ± 0.003 | -0.24 ± 0.12 | <32 | 3 | ... | ... | ... | ... | G |
| 44547 | 08103793-4914225 | 9.6 ± 0.2 | 5676 ± 189 | ... | 1.003 ± 0.001 | -0.26 ± 0.17 | 32 ± 7 | 1 | ... | ... | ... | ... | ... |

| ID | CNAME | RV (km s$^{-1}$) | $T_{\text{eff}}$ (K) | log g (dex) | $\gamma^a$ | [Fe/H] (dex) | EW(Li)$^b$ (mÅ) | EW(Li) error flag$^c$ | Pop. A/B (Sacco)$^d$ | Pop. (Randich) | Pop. (Cantat-Gaudin) | Pop. (adopted final)$^e$ | Non-mem with Li$^f$ |
|---|---|---|---|---|---|---|---|---|---|---|---|---|---|
| 44415 | 08093716-4905595 | 29.5 ± 1.7 | … | … | … | … | … | … | … | … | … | … | … |
| 44286 | 08080644-4920336 | 12.8 ± 0.5 | 3416 ± 27 | … | 0.837 ± 0.016 | … | … | … | … | … | … | … | … |
| 44548 | 08103812-4912285 | 13.4 ± 0.2 | 5006 ± 234 | … | 0.976 ± 0.002 | 0.00 ± 0.07 | 247 ± 6 | 1 | A | … | … | A | … |
| 44667 | 08115645-4908049 | 80.1 ± 0.2 | 4956 ± 31 | 2.92 ± 0.12 | 1.013 ± 0.005 | -0.11 ± 0.02 | <20 | 3 | … | … | … | … | G |
| 44416 | 08093744-4902028 | 12.2 ± 0.6 | 3551 ± 38 | 4.68 ± 0.04 | 0.791 ± 0.016 | -0.22 ± 0.14 | … | … | … | … | … | … | … |
| 2978 | 08080667-4911212 | 63.0 ± 0.6 | 6189 ± 94 | 4.09 ± 0.05 | … | 0.11 ± 0.08 | <21 | 3 | … | … | … | … | … |
| 44549 | 08103823-4845447 | 394.0 ± 2.2 | … | … | … | … | … | … | … | … | … | … | … |
| 44550 | 08103836-4851570 | 93.3 ± 0.2 | 4477 ± 178 | … | 1.026 ± 0.005 | -0.26 ± 0.16 | … | … | … | … | … | … | G |
| 44287 | 08080675-4910331 | 468.0 ± 5.1 | 3292 ± 13 | … | 0.665 ± 0.018 | … | … | … | … | … | … | … | … |
| 44551 | 08103840-4906404 | 53.6 ± 0.2 | 4692 ± 35 | 2.46 ± 0.16 | 1.024 ± 0.003 | -0.10 ± 0.07 | <39 | 3 | … | … | … | … | G |
| 44668 | 08115737-4911322 | 13.1 ± 0.2 | 5621 ± 36 | … | 0.992 ± 0.002 | 0.00 ± 0.04 | 197 ± 7 | 1 | A | A | A | A | … |
| 44417 | 08093818-4846001 | 95.8 ± 0.2 | 5008 ± 81 | … | 1.011 ± 0.004 | -0.29 ± 0.17 | <20 | 3 | … | … | … | … | G |
| 44669 | 08115745-4858503 | 18.0 ± 0.3 | 3509 ± 65 | … | 0.861 ± 0.010 | -0.26 ± 0.15 | … | … | … | … | … | … | … |
| 44566 | 08104728-4938406 | 50.0 ± 0.2 | 4622 ± 24 | 2.32 ± 0.08 | 1.024 ± 0.004 | -0.07 ± 0.01 | <39 | 3 | … | … | … | … | G |
| 44670 | 08115815-4933283 | 24.3 ± 2.1 | … | … | … | … | … | … | … | … | … | … | … |
| 2979 | 08080817-4913590 | 10.3 ± 0.6 | 6167 ± 32 | 3.88 ± 0.02 | 0.998 ± 0.002 | 0.16 ± 0.02 | 24 ± 4 | 1 | … | A | … | … | … |
| 44567 | 08104768-4859070 | 34.3 ± 0.2 | 5243 ± 11 | … | 1.006 ± 0.004 | -0.10 ± 0.10 | <37 | 3 | … | … | … | … | … |
| 44418 | 08093897-4932407 | 70.2 ± 0.2 | 4671 ± 40 | 2.58 ± 0.11 | 1.014 ± 0.003 | 0.08 ± 0.02 | <44 | 3 | … | … | … | … | G |
| 44568 | 08104793-4917078 | 12.9 ± 0.3 | 3862 ± 144 | … | 0.838 ± 0.005 | -0.15 ± 0.11 | … | … | … | … | … | … | … |
| 44293 | 08081455-4909326 | 12.4 ± 0.3 | 3479 ± 122 | … | 0.819 ± 0.010 | -0.25 ± 0.16 | <100 | 3 | A | A | … | A | … |
| 44569 | 08104805-4916295 | 476.0 ± 4.5 | … | … | … | … | … | … | … | … | … | … | … |
| 44419 | 08093925-4928494 | 14.9 ± 0.3 | 3688 ± 238 | 4.46 ± 0.19 | 0.834 ± 0.010 | -0.14 ± 0.11 | … | … | … | … | … | … | … |
| 44671 | 08115959-4937231 | -32.3 ± 0.4 | 4730 ± 33 | 2.56 ± 0.10 | 1.019 ± 0.005 | -0.06 ± 0.07 | … | … | … | … | … | … | … |
| 44570 | 08104831-4912292 | 13.3 ± 0.4 | 3276 ± 10 | … | 0.826 ± 0.015 | … | … | … | … | … | … | … | … |
| 44420 | 08093944-4854110 | 21.6 ± 1.7 | … | … | … | … | … | … | … | … | … | … | … |
| 44571 | 08104835-4915249 | 13.3 ± 0.4 | 3447 ± 42 | … | 0.844 ± 0.010 | -0.22 ± 0.13 | … | … | … | … | … | … | … |
| 44672 | 08120078-4939353 | 17.0 ± 2.1 | … | … | … | … | … | … | … | … | … | … | … |
| 44572 | 08104858-4906056 | 68.9 ± 0.2 | 4559 ± 86 | 2.30 ± 0.13 | 1.020 ± 0.003 | -0.04 ± 0.01 | 18 ± 15 | 1 | … | … | … | … | G |
| 44294 | 08081575-4902254 | 26.8 ± 0.2 | 4933 ± 163 | 3.09 ± 0.17 | 1.006 ± 0.004 | -0.12 ± 0.13 | <39 | 3 | … | … | … | … | … |
| 44573 | 08104978-4908198 | 391.1 ± 2.8 | … | … | … | … | … | … | … | … | … | … | … |
| 44428 | 08094477-4915078 | 75.2 ± 0.2 | 4665 ± 65 | 2.50 ± 0.12 | 1.018 ± 0.003 | 0.01 ± 0.02 | <34 | 3 | … | … | … | … | G |
| 44574 | 08104984-4911258 | 11.1 ± 0.2 | 6352 ± 14 | 4.29 ± 0.03 | 0.996 ± 0.001 | -0.16 ± 0.01 | 104 ± 2 | 1 | A | A | … | A | … |
| 44429 | 08094482-4902195 | 60.3 ± 0.2 | 4980 ± 76 | … | 1.017 ± 0.003 | -0.19 ± 0.09 | <9 | 3 | … | … | … | … | G |
| 44575 | 08104995-4906215 | 7.3 ± 0.3 | 4422 ± 249 | 4.48 ± 0.07 | 0.910 ± 0.005 | -0.07 ± 0.08 | … | … | … | … | … | … | … |
| 2995 | 08094507-4856307 | 14.9 ± 0.6 | 6025 ± 96 | 4.20 ± 0.14 | 1.003 ± 0.002 | -0.02 ± 0.12 | 140 ± 4 | 2 | A | A | … | A | … |
| 44430 | 08094510-4846520 | 13.5 ± 0.3 | 3783 ± 467 | 4.48 ± 0.13 | 0.836 ± 0.010 | -0.23 ± 0.13 | <100 | 3 | A | A | … | A | … |
| 44295 | 08081758-4914493 | 16.8 ± 0.3 | 3432 ± 16 | … | 0.851 ± 0.010 | -0.25 ± 0.16 | 532 ± 35 | 1 | B | B | … | B | … |
| 44431 | 08094556-4917365 | 12.5 ± 0.3 | 4275 ± 582 | … | 0.858 ± 0.005 | -0.01 ± 0.15 | 341 ± 20 | 1 | A | … | … | A | … |
| 44583 | 08105610-4918399 | 41.1 ± 2.1 | 3442 ± 279 | … | 0.975 ± 0.008 | -0.18 ± 0.10 | … | … | … | … | … | … | … |
| 44296 | 08081986-4915361 | -4.3 ± 0.2 | 5018 ± 127 | 3.62 ± 0.05 | 0.996 ± 0.004 | 0.01 ± 0.06 | <26 | 3 | … | … | … | … | … |
| 44432 | 08094634-4848247 | 70.6 ± 0.2 | 4638 ± 97 | … | 1.032 ± 0.003 | -0.29 ± 0.05 | <14 | 3 | … | … | … | … | G |
| 44584 | 08105658-4912074 | 91.8 ± 0.2 | 4747 ± 21 | 2.55 ± 0.11 | 1.015 ± 0.003 | -0.12 ± 0.08 | 49 ± 6 | 1 | … | … | … | … | G |
| 44433 | 08094645-4849506 | 98.0 ± 0.2 | 4757 ± 95 | 2.47 ± 0.19 | 1.015 ± 0.003 | -0.28 ± 0.14 | 27 ± 11 | 1 | … | … | … | … | G |
| 44434 | 08094673-4902546 | 18.4 ± 1.4 | 3532 ± 29 | … | 0.854 ± 0.017 | … | … | … | … | … | … | … | … |
| 44585 | 08105720-4923478 | 11.1 ± 0.4 | 3352 ± 40 | … | 0.849 ± 0.014 | -0.25 ± 0.14 | <100 | 3 | … | A | A | A | … |
| 44435 | 08094675-4931067 | 45.6 ± 5.2 | 3535 ± 240 | … | … | … | … | … | … | … | … | … | … |
| 44306 | 08083501-4916124 | 22.0 ± 0.2 | 4845 ± 111 | 2.51 ± 0.19 | 1.009 ± 0.003 | -0.15 ± 0.09 | … | … | … | … | … | … | … |
| 44436 | 08094684-4907198 | 12.1 ± 0.2 | 5894 ± 65 | 4.05 ± 0.02 | 0.995 ± 0.001 | 0.05 ± 0.02 | 187 ± 1 | 1 | A | … | … | A | … |
| 44447 | 08095429-4908418 | 13.1 ± 0.3 | 3453 ± 102 | … | 0.846 ± 0.013 | -0.24 ± 0.14 | <100 | 3 | A | … | … | A | … |
| 44307 | 08083553-4858570 | 85.0 ± 0.2 | 4588 ± 83 | 1.95 ± 0.18 | 1.028 ± 0.005 | -0.44 ± 0.03 | <20 | 3 | … | … | … | … | G |
| 44448 | 08095446-4851279 | 74.7 ± 0.2 | 4730 ± 17 | 2.56 ± 0.05 | 1.016 ± 0.004 | -0.07 ± 0.02 | 33 ± 3 | 1 | … | … | … | … | G |
| 44308 | 08083560-4909433 | 15.6 ± 0.8 | 3330 ± 51 | … | 0.865 ± 0.012 | -0.27 ± 0.14 | <100 | 3 | B | A | … | A | … |
| 44309 | 08083609-4852238 | -7.7 ± 0.3 | 3837 ± 138 | … | 0.841 ± 0.005 | -0.17 ± 0.10 | … | … | … | … | … | … | … |
| 44449 | 08095476-4933509 | 28.6 ± 0.2 | 5796 ± 83 | … | 0.992 ± 0.004 | … | … | … | … | … | … | … | … |
| 44450 | 08095507-4858140 | 13.0 ± 0.2 | 4117 ± 148 | … | 0.889 ± 0.003 | -0.03 ± 0.13 | 109 ± 10 | 1 | A | A | … | A | … |
| 44310 | 08083727-4906373 | 1.3 ± 0.2 | 5119 ± 87 | … | 0.999 ± 0.004 | … | <16 | 3 | … | … | … | … | … |
| 44451 | 08095536-4919182 | 56.1 ± 0.2 | 4793 ± 39 | … | 0.997 ± 0.002 | 0.04 ± 0.02 | <38 | 3 | … | … | … | … | … |
| 44603 | 08110683-4934269 | 30.0 ± 0.2 | 5239 ± 43 | … | 1.011 ± 0.003 | -0.06 ± 0.09 | <19 | 3 | … | … | … | … | … |
| 44311 | 08083830-4932320 | 12.0 ± 0.3 | 3625 ± 6 | … | 0.824 ± 0.008 | -0.21 ± 0.12 | … | … | … | … | … | … | … |

**Table C.4.** continued.

| ID | CNAME | RV (km s$^{-1}$) | $T_{eff}$ (K) | logg (dex) | $\gamma^a$ | [Fe/H] (dex) | EW(Li)$^b$ (mÅ) | EW(Li) error flag$^c$ | Pop. A/B (Sacco)$^d$ | Pop. (Randich) | Pop. (Cantat-Gaudin) | Pop. (adopted final)$^e$ | Non-mem with Li$^f$ |
|---|---|---|---|---|---|---|---|---|---|---|---|---|---|
| 44452 | 08095581-4922407 | 35.1 ± 0.2 | 5044 ± 98 | … | … | … | … | … | … | … | … | … | … |
| 44604 | 08110722-4853061 | 12.8 ± 0.3 | 3578 ± 35 | … | 0.841 ± 0.006 | -0.23 ± 0.15 | <100 | 3 | A | A | A | A | … |
| 44312 | 08083841-4934401 | 18.7 ± 0.2 | 4923 ± 103 | 2.83 ± 0.01 | 1.009 ± 0.004 | -0.06 ± 0.11 | <23 | 3 | … | … | … | … | … |
| 44605 | 08110759-4921423 | 19.9 ± 0.3 | 3598 ± 26 | … | 0.846 ± 0.006 | -0.21 ± 0.12 | 137 ± 21 | 1 | B | B | … | B | … |
| 44313 | 08083843-4931303 | 12.3 ± 0.3 | 3431 ± 61 | … | 0.834 ± 0.014 | -0.27 ± 0.14 | <100 | 3 | A | A | … | A | … |
| 44314 | 08083843-4939320 | 15.3 ± 2.2 | … | … | … | … | … | … | … | … | … | … | … |
| 44606 | 08110781-4920585 | 108.5 ± 0.2 | 4684 ± 87 | … | 1.029 ± 0.003 | -0.21 ± 0.15 | <15 | 3 | … | … | … | … | G |
| 44453 | 08095640-4932446 | 12.5 ± 0.6 | 3461 ± 34 | … | 0.833 ± 0.020 | … | … | … | … | … | … | … | … |
| 44454 | 08095644-4922109 | 12.3 ± 0.2 | 4279 ± 328 | … | 0.901 ± 0.004 | -0.01 ± 0.12 | 148 ± 3 | 1 | A | … | … | A | … |
| 44455 | 08095669-4938474 | 73.4 ± 0.2 | 4744 ± 49 | 2.55 ± 0.18 | 1.017 ± 0.003 | -0.15 ± 0.11 | <37 | 3 | … | … | … | … | G |
| 44456 | 08095677-4934370 | 66.4 ± 0.8 | … | … | … | … | … | … | … | … | … | … | … |
| 44607 | 08110842-4853314 | 133.8 ± 0.2 | 4628 ± 29 | 2.12 ± 0.40 | 1.018 ± 0.004 | -0.38 ± 0.05 | <29 | 3 | … | … | … | … | G |
| 44322 | 08084386-4922097 | 34.0 ± 3.7 | … | … | … | … | … | … | … | … | … | … | … |
| 44323 | 08084406-4932166 | 74.5 ± 0.2 | 4805 ± 34 | 2.67 ± 0.08 | 1.016 ± 0.003 | -0.12 ± 0.01 | <22 | 3 | … | … | … | … | G |
| 44324 | 08084459-4934202 | 12.1 ± 0.3 | 4724 ± 350 | … | 0.956 ± 0.003 | -0.02 ± 0.07 | 335 ± 14 | 1 | A | A | … | A | … |
| 44608 | 08110933-4911533 | 13.6 ± 0.8 | … | … | … | … | … | … | … | … | … | … | … |
| 44459 | 08100063-4853392 | 13.0 ± 0.3 | 3449 ± 43 | … | 0.843 ± 0.010 | -0.23 ± 0.13 | <100 | 3 | A | A | … | A | … |
| 44325 | 08084618-4916240 | 16.2 ± 0.3 | 3519 ± 11 | 4.69 ± 0.05 | 0.801 ± 0.008 | -0.22 ± 0.15 | <100 | 3 | B | A | … | A | … |
| 44460 | 08100083-4909099 | 23.5 ± 3.9 | … | … | … | … | … | … | … | … | … | … | … |
| 44326 | 08084695-4907245 | 51.5 ± 0.2 | 4749 ± 45 | 2.61 ± 0.11 | 1.011 ± 0.004 | -0.16 ± 0.12 | <31 | 3 | … | … | … | … | G |
| 44461 | 08100138-4847439 | 83.6 ± 0.2 | 4593 ± 61 | 2.30 ± 0.14 | 1.019 ± 0.003 | -0.17 ± 0.02 | … | … | … | … | … | … | G |
| 44462 | 08100180-4915454 | 13.3 ± 0.3 | 3777 ± 184 | … | 0.818 ± 0.009 | -0.19 ± 0.16 | <100 | 3 | A | A | … | A | … |
| 44327 | 08084760-4855204 | 491.7 ± 3.3 | 3445 ± 130 | … | … | -0.30 ± 0.14 | … | … | … | … | … | … | … |
| 44463 | 08100204-4939564 | 12.6 ± 0.3 | 4451 ± 235 | … | 0.946 ± 0.007 | 0.00 ± 0.11 | 364 ± 8 | 1 | A | A | … | A | … |
| 44338 | 08085636-4851415 | 41.1 ± 0.2 | 4359 ± 237 | … | 1.028 ± 0.004 | 0.13 ± 0.13 | <51 | 3 | … | … | … | … | G |
| 44617 | 08111420-4848595 | 56.0 ± 0.2 | 5036 ± 174 | … | 1.015 ± 0.005 | -0.12 ± 0.20 | <13 | 3 | … | … | … | … | G |
| 44464 | 08100282-4932511 | 43.5 ± 0.2 | 4961 ± 111 | 3.12 ± 0.13 | 1.006 ± 0.004 | -0.10 ± 0.12 | <20 | 3 | … | … | … | … | … |
| 44618 | 08111420-4924449 | 38.7 ± 0.2 | 5187 ± 291 | … | 1.015 ± 0.004 | -0.31 ± 0.21 | <16 | 3 | … | … | … | … | G |
| 44339 | 08085667-4859420 | 101.7 ± 0.2 | 4744 ± 39 | 2.51 ± 0.14 | 1.014 ± 0.003 | -0.17 ± 0.17 | <23 | 3 | … | … | … | … | G |
| 44340 | 08085677-4926054 | 39.7 ± 0.3 | 3734 ± 162 | 4.36 ± 0.10 | 0.825 ± 0.010 | -0.27 ± 0.20 | … | … | … | … | … | … | … |
| 44465 | 08100297-4927293 | 42.5 ± 0.2 | 5145 ± 104 | … | 1.007 ± 0.003 | -0.18 ± 0.16 | <15 | 3 | … | … | … | … | … |
| 44480 | 08100928-4920539 | -16.9 ± 0.2 | 5989 ± 195 | 3.71 ± 0.16 | 1.010 ± 0.001 | -0.41 ± 0.09 | <14 | 3 | … | … | … | … | … |
| 44619 | 08111579-4908139 | 11.1 ± 0.3 | 3528 ± 36 | … | 0.866 ± 0.012 | -0.21 ± 0.12 | <100 | 3 | … | A | … | A | … |
| 44341 | 08085758-4911261 | 12.1 ± 0.3 | 3476 ± 9 | 4.67 ± 0.07 | 0.811 ± 0.010 | -0.22 ± 0.15 | … | … | … | … | … | … | … |
| 44481 | 08100938-4900565 | 13.0 ± 0.2 | 5143 ± 48 | 3.37 ± 0.26 | … | -0.03 ± 0.01 | 222 ± 4 | … | … | A | … | A | … |
| 44342 | 08085779-4901171 | 12.5 ± 0.3 | 3725 ± 100 | … | 0.823 ± 0.008 | -0.19 ± 0.13 | <100 | 3 | A | A | … | A | … |
| 44482 | 08100955-4856343 | 71.2 ± 0.2 | 4624 ± 87 | 2.38 ± 0.08 | 1.014 ± 0.003 | -0.08 ± 0.02 | <49 | 3 | … | … | … | … | … |
| 2984 | 08085790-4854189 | 21.7 ± 0.6 | 6377 ± 109 | 4.25 ± 0.10 | … | 0.28 ± 0.05 | <10 | 3 | … | … | B | B | … |
| 44483 | 08100995-4936336 | 12.8 ± 0.2 | 4584 ± 76 | … | 1.019 ± 0.005 | … | … | … | … | … | … | … | … |
| 44484 | 08101006-4853042 | 33.3 ± 0.2 | 3877 ± 63 | 4.56 ± 0.12 | 0.833 ± 0.003 | -0.18 ± 0.03 | … | … | … | … | … | … | … |
| 44620 | 08111669-4922290 | 12.4 ± 0.7 | 3393 ± 26 | … | 0.864 ± 0.016 | … | <100 | 3 | A | … | A | A | … |
| 44343 | 08085856-4911171 | 15.3 ± 0.4 | 3409 ± 27 | … | 0.825 ± 0.016 | … | … | … | … | … | … | … | … |
| 44485 | 08101041-4858052 | 12.2 ± 0.5 | … | … | … | … | … | … | … | … | … | … | … |
| 2985 | 08085860-4908550 | 42.8 ± 0.6 | 5032 ± 41 | 2.68 ± 0.07 | 1.013 ± 0.002 | -0.25 ± 0.07 | 20 ± 3 | 1 | … | … | … | … | G |
| 44486 | 08101083-4858477 | 78.5 ± 0.2 | 4989 ± 120 | 2.96 ± 0.19 | 1.010 ± 0.003 | -0.18 ± 0.13 | <15 | 3 | … | … | … | … | G |
| 44487 | 08101165-4922274 | 23.2 ± 0.2 | 6298 ± 14 | 4.36 ± 0.03 | 0.994 ± 0.001 | 0.15 ± 0.01 | 10 ± 1 | 1 | … | B | … | B | … |
| 44254 | 08072806-4901038 | 184.9 ± 0.2 | 4516 ± 123 | … | 1.033 ± 0.004 | -0.26 ± 0.15 | <26 | 3 | … | … | … | … | G |
| 44632 | 08112560-4927206 | 19.6 ± 0.3 | 3454 ± 21 | … | 0.887 ± 0.012 | -0.24 ± 0.16 | 478 ± 13 | 1 | B | B | … | B | … |
| 44351 | 08090470-4924396 | 13.6 ± 0.3 | 3550 ± 14 | … | 0.820 ± 0.007 | -0.23 ± 0.13 | <100 | 3 | A | A | … | A | … |
| 44352 | 08090471-4933274 | 16.6 ± 0.2 | 3519 ± 7 | … | 0.854 ± 0.007 | -0.22 ± 0.14 | <100 | 3 | B | B | … | B | … |
| 44488 | 08101213-4904315 | 391.2 ± 2.1 | … | … | … | … | … | … | … | … | … | … | … |
| 44633 | 08112590-4901101 | 75.9 ± 0.2 | 5040 ± 113 | … | 1.014 ± 0.005 | -0.11 ± 0.17 | <19 | 3 | … | … | … | … | G |
| 44489 | 08101223-4913056 | 12.9 ± 0.2 | 6042 ± 63 | 4.03 ± 0.04 | 0.999 ± 0.001 | 0.05 ± 0.06 | 146 ± 3 | 1 | A | A | … | A | … |
| 44263 | 08073784-4907170 | 14.8 ± 0.2 | 4957 ± 213 | … | 1.008 ± 0.005 | -0.22 ± 0.24 | <35 | 3 | … | … | … | … | … |
| 44353 | 08090654-4846474 | 74.8 ± 0.2 | 4656 ± 3 | 2.38 ± 0.08 | 1.025 ± 0.006 | -0.08 ± 0.01 | <41 | 3 | … | … | … | … | G |
| 44264 | 08073903-4915112 | 87.4 ± 0.2 | 4777 ± 2 | 2.61 ± 0.13 | 1.017 ± 0.005 | -0.12 ± 0.06 | 47 ± 10 | 1 | … | … | … | … | G |
| 44354 | 08090679-4915347 | 459.0 ± 3.1 | 3346 ± 26 | … | 0.838 ± 0.020 | … | … | … | … | … | … | … | … |
| 44355 | 08090720-4919144 | 12.3 ± 0.2 | 4748 ± 317 | … | 0.952 ± 0.004 | 0.05 ± 0.09 | 218 ± 6 | 1 | A | A | … | A | … |
| 44634 | 08112830-4906341 | 116.4 ± 0.2 | 4552 ± 6 | … | 1.032 ± 0.005 | -0.32 ± 0.14 | <39 | 3 | … | … | … | … | G |

| ID | CNAME | RV (km s$^{-1}$) | $T_{\rm eff}$ (K) | $logg$ (dex) | $\gamma^a$ | [Fe/H] (dex) | $EW$(Li)$^b$ (mÅ) | $EW$(Li) error flag$^c$ | Pop. A/B (Sacco)$^d$ | Pop. (Randich) | Pop. (Cantat-Gaudin) | Pop. (adopted final)$^e$ | Non-mem with Li$^f$ |
|---|---|---|---|---|---|---|---|---|---|---|---|---|---|
| 44356 | 08090722-4847280 | 0.8 ± 0.2 | 4439 ± 148 | 4.48 ± 0.12 | 0.923 ± 0.003 | 0.01 ± 0.11 | <15 | 3 | … | … | … | … | … |
| 44265 | 08073999-4903023 | 20.5 ± 0.3 | 3531 ± 7 | … | 0.867 ± 0.009 | -0.23 ± 0.14 | 399 ± 9 | 1 | … | B | … | B | … |
| 2960 | 08101799-4923503 | 14.4 ± 0.6 | 5214 ± 72 | 4.51 ± 0.12 | 0.984 ± 0.002 | -0.07 ± 0.11 | 262 ± 1 | 2 | A | A | … | A | … |
| 44500 | 08101811-4858252 | 20.5 ± 0.3 | 3445 ± 14 | … | 0.861 ± 0.008 | … | 373 ± 21 | 1 | B | B | … | B | … |
| 44643 | 08113236-4850197 | 91.3 ± 0.2 | 4602 ± 94 | … | 1.079 ± 0.006 | … | 24 ± 6 | 1 | … | … | … | … | G |
| 44644 | 08113238-4848010 | 12.3 ± 0.3 | 3572 ± 64 | … | 0.845 ± 0.010 | -0.23 ± 0.14 | <100 | 3 | A | A | A | A | … |
| 44266 | 08074077-4927238 | 402.7 ± 1.5 | 3541 ± 29 | … | 0.835 ± 0.017 | … | … | … | … | … | … | … | … |
| 44267 | 08074101-4919554 | 125.1 ± 0.2 | 4405 ± 204 | … | 1.045 ± 0.004 | -0.20 ± 0.18 | <29 | 3 | … | … | … | … | G |
| 44645 | 08113260-4924367 | 112.7 ± 0.2 | 4774 ± 19 | 2.60 ± 0.15 | 1.018 ± 0.003 | -0.13 ± 0.09 | <28 | 3 | … | … | … | … | G |
| 44501 | 08101836-4906461 | 25.6 ± 0.2 | 6322 ± 12 | 4.07 ± 0.03 | 1.003 ± 0.001 | 0.17 ± 0.01 | 7 ± 1 | 1 | … | … | … | … | … |
| 44268 | 08074194-4908306 | -8.1 ± 0.5 | 3563 ± 32 | 4.69 ± 0.06 | 0.792 ± 0.014 | -0.25 ± 0.15 | 100 ± 12 | 1 | … | … | … | … | … |
| 44502 | 08101841-4926312 | 14.7 ± 1.8 | … | … | … | … | … | … | … | … | … | … | … |
| 44379 | 08091645-4907239 | 12.6 ± 0.3 | 3528 ± 13 | … | 0.857 ± 0.006 | -0.29 ± 0.14 | … | … | … | … | … | … | … |
| 44269 | 08074232-4907549 | 96.1 ± 0.3 | 4959 ± 137 | … | 1.010 ± 0.007 | -0.43 ± 0.29 | <28 | 3 | … | … | … | … | G |
| 44503 | 08101842-4902145 | 2.5 ± 0.3 | 3818 ± 85 | 4.67 ± 0.11 | 0.873 ± 0.004 | -0.19 ± 0.12 | 293 ± 5 | … | … | … | … | … | … |
| 44646 | 08113271-4940061 | 28.3 ± 0.3 | 5133 ± 86 | … | 1.017 ± 0.005 | -0.25 ± 0.19 | <37 | 3 | … | … | … | … | G |
| 44380 | 08091651-4846157 | 52.8 ± 0.2 | 4953 ± 186 | … | 1.015 ± 0.003 | -0.13 ± 0.20 | 32 ± 3 | 1 | … | … | … | … | G |
| 44647 | 08113276-4925588 | 150.6 ± 0.2 | 5007 ± 122 | … | 1.015 ± 0.003 | -0.27 ± 0.26 | <9 | 3 | … | … | … | … | G |
| 44381 | 08091657-4909309 | 11.9 ± 0.2 | 4341 ± 460 | … | 0.897 ± 0.003 | -0.06 ± 0.07 | 333 ± 10 | 1 | A | A | … | A | … |
| 44382 | 08091713-4925277 | 12.9 ± 0.9 | 3617 ± 186 | 4.75 ± 0.15 | 0.786 ± 0.011 | -0.20 ± 0.12 | <100 | 3 | A | A | … | A | … |
| 44383 | 08091770-4908344 | -655.7 ± 4.5 | … | … | … | … | … | … | … | … | … | … | … |
| 44504 | 08101902-4859112 | 44.7 ± 0.2 | 5042 ± 69 | … | 1.012 ± 0.003 | -0.08 ± 0.10 | <29 | 3 | … | … | … | … | G |
| 44273 | 08074854-4900573 | 92.5 ± 0.2 | 4688 ± 109 | 2.46 ± 0.19 | 1.014 ± 0.003 | -0.16 ± 0.16 | <24 | 3 | … | … | … | … | … |
| 44384 | 08091822-4916256 | 83.5 ± 0.2 | 4861 ± 97 | 2.62 ± 0.13 | 1.018 ± 0.003 | -0.13 ± 0.07 | <23 | 3 | … | … | … | … | G |
| 2962 | 08102725-4909509 | … | … | … | … | … | … | … | … | … | … | … | … |
| 44524 | 08102741-4920196 | 12.2 ± 0.3 | 3405 ± 21 | … | 0.840 ± 0.013 | … | <100 | 3 | A | A | A | A | … |
| 44648 | 08113562-4912217 | 21.6 ± 0.4 | … | … | … | … | … | … | … | … | … | … | … |
| 2963 | 08102854-4856518 | 12.4 ± 0.6 | 5861 ± 95 | 4.37 ± 0.02 | 0.998 ± 0.001 | 0.05 ± 0.02 | 194 ± 1 | 2 | A | A | … | A | … |
| 44525 | 08103001-4903218 | 13.7 ± 0.5 | 3391 ± 25 | … | 0.875 ± 0.016 | … | … | … | … | … | … | … | … |
| 44385 | 08092021-4849575 | 118.6 ± 0.2 | 4521 ± 56 | … | 1.028 ± 0.003 | -0.21 ± 0.08 | <23 | 3 | … | … | … | … | G |
| 44393 | 08092475-4911138 | 18.5 ± 1.3 | 3588 ± 32 | … | 0.959 ± 0.020 | … | … | … | … | … | … | … | … |
| 44526 | 08103088-4910443 | 13.5 ± 0.4 | 3805 ± 229 | 4.47 ± 0.12 | 0.813 ± 0.007 | -0.18 ± 0.10 | … | … | … | … | … | … | … |
| 44527 | 08103100-4851071 | -5.6 ± 0.3 | 3831 ± 186 | 4.35 ± 0.10 | 0.825 ± 0.008 | -0.10 ± 0.15 | … | … | … | … | … | … | … |
| 44528 | 08103139-4921590 | 232.2 ± 0.2 | … | … | … | … | … | … | … | … | … | … | … |
| F | 08114859-4928089 | 15.5 ± 4.2 | 3453 ± 28 | … | 0.823 ± 0.017 | … | <129 | 3 | B | … | B | B | … |
| 44529 | 08103186-4921296 | 15.3 ± 0.3 | 4467 ± 295 | … | 0.942 ± 0.002 | -0.02 ± 0.12 | 367 ± 16 | 1 | B | A | … | A | … |
| 44660 | 08114867-4919306 | 12.7 ± 0.3 | 3739 ± 23 | … | 0.851 ± 0.006 | -0.19 ± 0.12 | <100 | 3 | A | A | … | A | … |
| 44661 | 08114894-4928258 | 77.2 ± 0.2 | 4966 ± 148 | … | 1.013 ± 0.004 | -0.21 ± 0.17 | 30 ± 6 | 1 | … | … | … | … | G |
| 44530 | 08103259-4909436 | 9.9 ± 0.2 | 3546 ± 48 | 4.68 ± 0.11 | 0.740 ± 0.011 | -0.22 ± 0.15 | <100 | 3 | … | A | … | A | … |
| 2965 | 08103672-4851270 | 7.4 ± 0.6 | 5523 ± 27 | 4.52 ± 0.04 | 0.990 ± 0.003 | -0.02 ± 0.02 | <38 | 3 | … | … | … | … | … |
| 44542 | 08103676-4858389 | 11.3 ± 0.4 | 3421 ± 30 | … | 0.863 ± 0.018 | … | … | … | … | … | … | … | … |

**Notes.** $^{(a)}$ Empirical gravity indicator defined by Damiani et al. (2014). $^{(b)}$ The values of $EW$(Li) for this cluster are corrected (subtracted adjacent Fe (6707.43 Å) line). $^{(c)}$ Flags for the errors of the corrected $EW$(Li) values, as follows: 1=$EW$(Li) corrected by blends contribution using models; 2=$EW$(Li) measured separately (Li line resolved - UVES only); and 3=Upper limit (no error for $EW$(Li) is given). $^{d}$ For this cluster we have made use of the membership selection obtained by Sacco et al. (2015), Cantat-Gaudin et al. (2018) and Randich et al. (2018). $^{(e)}$ The letters "Y" and "N" indicate if the star is a cluster member or not. $^{(f)}$ "Li-rich G" and "G" indicate "Li-rich giant" and "giant" Li field contaminants, respectively.

**Table C.5.** IC 2391

| ID | CNAME | RV (km s$^{-1}$) | $T_{\text{eff}}$ (K) | $logg$ (dex) | $\gamma^a$ | [Fe/H] (dex) | EW(Li)$^b$ (mÅ) | EW(Li) error flag$^c$ | $\gamma$ | $logg$ | Membership RV | Li | H$\alpha$ | [Fe/H] | Gaia studies Randich$^d$ | Cantat-Gaudin$^d$ | Final$^e$ | NMs with Li$^f$ |
|---|---|---|---|---|---|---|---|---|---|---|---|---|---|---|---|---|---|---|
| 37058 | 08385985-5217393 | 32.2 ± 0.2 | 4740 ± 77 | 2.60 ± 0.09 | 1.013 ± 0.003 | 0.03 ± 0.04 | <45 | 3 | N | N | … | … | … | … | … | … | n | G |
| 37059 | 08390075-5231025 | 140.2 ± 0.2 | 4975 ± 36 | 2.93 ± 0.18 | 1.015 ± 0.004 | -0.08 ± 0.08 | <26 | 3 | N | N | … | … | … | … | … | … | n | G |
| 37060 | 08390452-5312153 | 47.7 ± 0.2 | 5104 ± 133 | … | 1.021 ± 0.004 | -0.22 ± 0.20 | <19 | 3 | N | … | … | … | … | … | … | … | n | G |
| 37061 | 08390566-5232465 | 15.5 ± 0.2 | 5070 ± 124 | … | 1.011 ± 0.004 | -0.14 ± 0.21 | <13 | 3 | N | … | … | … | … | … | … | … | n | G |
| 37062 | 08390629-5225062 | 79.4 ± 0.2 | 4614 ± 135 | … | 1.021 ± 0.002 | -0.28 ± 0.07 | <24 | 3 | N | … | … | … | … | … | … | … | n | G |
| 37063 | 08390640-5334370 | 35.5 ± 0.2 | 4967 ± 143 | … | 1.025 ± 0.003 | -0.08 ± 0.14 | <18 | 3 | N | … | … | … | … | … | … | … | n | G |
| 37064 | 08390665-5343226 | 26.4 ± 0.2 | 5058 ± 119 | … | 1.008 ± 0.002 | -0.06 ± 0.13 | <19 | 3 | Y | … | N | N | N | Y | … | … | n | NG? |
| 37065 | 08390670-5310121 | 29.1 ± 0.3 | 4144 ± 157 | 4.64 ± 0.13 | 0.857 ± 0.012 | -0.16 ± 0.02 | <53 | 3 | Y | Y | N | Y | N | Y | N | … | n | NG |
| 37066 | 08390676-5352167 | 15.4 ± 0.2 | 4583 ± 61 | 2.32 ± 0.15 | 1.026 ± 0.004 | -0.03 ± 0.02 | <29 | 3 | N | … | … | … | … | … | … | … | n | … |
| 37067 | 08390692-5317258 | 18.7 ± 0.2 | 5026 ± 160 | … | 1.021 ± 0.003 | -0.11 ± 0.22 | <16 | 3 | N | … | … | … | … | … | … | … | n | G |
| 37068 | 08390725-5223138 | 9.2 ± 0.2 | 4738 ± 103 | 2.59 ± 0.12 | 1.016 ± 0.004 | 0.04 ± 0.07 | <28 | 3 | N | N | … | … | … | … | … | … | n | G |
| 37069 | 08390757-5348280 | 50.7 ± 0.2 | 4608 ± 132 | … | 1.023 ± 0.003 | 0.05 ± 0.07 | <49 | 3 | N | … | … | … | … | … | … | … | n | G |
| 37070 | 08390797-5257481 | 41.2 ± 0.2 | 4811 ± 68 | 2.94 ± 0.18 | 1.001 ± 0.003 | -0.07 ± 0.06 | <30 | 3 | Y | N | N | Y | N | Y | … | … | n | NG? |
| 37071 | 08390859-5319164 | 69.0 ± 0.2 | 4816 ± 171 | … | 1.013 ± 0.004 | -0.15 ± 0.11 | 28 ± 3 | 1 | N | … | … | … | … | … | … | … | n | G |
| 37072 | 08390890-5218280 | 37.4 ± 0.2 | 4643 ± 87 | … | 1.028 ± 0.004 | … | 155 ± 2 | 1 | N | … | … | … | … | … | … | … | n | Li-rich G |
| 37073 | 08390890-5341042 | 9.5 ± 0.2 | 5021 ± 103 | … | 1.028 ± 0.004 | -0.07 ± 0.13 | <31 | 3 | N | … | … | … | … | … | … | … | n | G |
| 37074 | 08391120-5311494 | 44.8 ± 0.2 | 4954 ± 211 | … | 1.020 ± 0.003 | -0.15 ± 0.22 | <21 | 3 | N | … | … | … | … | … | … | … | n | G |
| 37075 | 08391133-5328374 | 33.4 ± 0.2 | 4661 ± 58 | 2.56 ± 0.13 | 1.017 ± 0.004 | 0.09 ± 0.01 | 21 ± 3 | 1 | N | N | … | … | … | … | … | … | n | G |
| 37076 | 08391137-5216592 | 68.2 ± 0.2 | 5046 ± 132 | … | 1.021 ± 0.003 | -0.09 ± 0.11 | <18 | 3 | N | … | … | … | … | … | … | … | n | G |
| 37077 | 08391241-5232102 | 28.9 ± 0.2 | 4601 ± 144 | 2.52 ± 0.09 | 1.008 ± 0.003 | 0.03 ± 0.12 | <44 | 3 | Y | N | N | Y | N | Y | … | … | n | NG? |
| 37112 | 08393395-5347234 | -4.7 ± 0.2 | 4526 ± 145 | … | 1.027 ± 0.001 | 0.11 ± 0.06 | 62 ± 6 | 1 | N | … | … | … | … | … | … | … | n | G |
| 37113 | 08393421-5253230 | 100.0 ± 0.2 | 4880 ± 123 | … | 1.013 ± 0.004 | -0.14 ± 0.16 | 29 ± 8 | 1 | N | … | … | … | … | … | … | … | n | G |
| 2494 | 08393496-5318577 | 25.5 ± 0.6 | … | … | … | … | … | … | … | … | … | … | … | … | … | … | n | … |
| 2495 | 08393574-5343310 | 50.8 ± 0.6 | 5615 ± 129 | 4.06 ± 0.25 | … | -0.01 ± 0.03 | <29 | 3 | … | Y | N | N | … | Y | … | … | n | NG |
| 37114 | 08393635-5227205 | 38.7 ± 0.2 | 5024 ± 119 | … | 1.019 ± 0.003 | -0.05 ± 0.07 | <21 | 3 | N | … | … | … | … | … | … | … | n | G |
| 37115 | 08393802-5242491 | 51.0 ± 0.2 | 5026 ± 133 | … | 1.010 ± 0.004 | -0.13 ± 0.20 | <22 | 3 | Y | … | N | N | … | Y | … | … | n | NG? |
| 37116 | 08393802-5336544 | 33.1 ± 0.2 | 5008 ± 85 | … | 1.018 ± 0.003 | -0.06 ± 0.10 | <24 | 3 | N | … | … | … | … | … | … | … | n | G |
| 37137 | 08395519-5353230 | 55.6 ± 0.2 | 4770 ± 4 | 2.56 ± 0.09 | 1.020 ± 0.003 | -0.08 ± 0.04 | <27 | 3 | N | N | … | … | … | … | … | … | n | G |
| 37138 | 08395641-5244121 | 16.4 ± 0.2 | 5057 ± 144 | … | 1.019 ± 0.003 | -0.07 ± 0.14 | <12 | 3 | N | … | … | … | … | … | … | … | n | G |
| 37139 | 08395644-5346333 | 53.8 ± 0.2 | 4743 ± 25 | 2.56 ± 0.07 | 1.020 ± 0.002 | -0.05 ± 0.09 | <28 | 3 | N | N | … | … | … | … | … | … | n | G |
| 37140 | 08395830-5223448 | 7.6 ± 0.2 | 4642 ± 83 | 2.41 ± 0.14 | 1.020 ± 0.002 | -0.02 ± 0.09 | <39 | 3 | N | N | … | … | … | … | … | … | n | G |
| 37156 | 08401405-5318248 | 49.0 ± 0.2 | 5031 ± 178 | … | 1.012 ± 0.003 | -0.15 ± 0.24 | <14 | 3 | N | … | … | … | … | … | … | … | n | G |
| 37157 | 08401421-5226240 | 52.9 ± 0.2 | 4712 ± 11 | 2.53 ± 0.14 | 1.019 ± 0.002 | -0.05 ± 0.02 | <32 | 3 | N | N | … | … | … | … | … | … | n | G |
| 37158 | 08401533-5325310 | 14.4 ± 0.2 | 4639 ± 21 | 2.36 ± 0.04 | 1.026 ± 0.004 | -0.04 ± 0.09 | <27 | 3 | N | N | … | … | … | … | … | … | n | G |
| 37159 | 08401609-5325476 | 15.2 ± 0.6 | 3269 ± 12 | … | 0.840 ± 0.017 | … | … | … | Y | … | Y | … | Y | … | Y | Y | n | … |
| 37180 | 08403950-5300371 | 96.9 ± 0.2 | 4556 ± 75 | 2.35 ± 0.16 | 1.025 ± 0.006 | 0.01 ± 0.06 | <58 | 3 | N | N | … | … | … | … | … | … | n | G |
| 37181 | 08403980-5245588 | 40.0 ± 0.2 | 4597 ± 85 | 2.34 ± 0.10 | 1.013 ± 0.002 | -0.09 ± 0.07 | <42 | 3 | N | N | … | … | … | … | … | … | n | G |
| 37182 | 08404084-5313317 | 17.8 ± 2.7 | … | … | … | … | … | … | … | … | … | … | … | … | Y | Y | n | … |
| 37183 | 08404085-5258402 | 28.3 ± 0.3 | 4686 ± 71 | … | 1.033 ± 0.007 | -0.09 ± 0.10 | <31 | 3 | N | … | … | … | … | … | … | … | n | G |
| 37184 | 08404412-5316241 | 41.2 ± 0.2 | 4848 ± 63 | 2.65 ± 0.08 | 1.015 ± 0.003 | -0.06 ± 0.05 | <25 | 3 | N | N | … | … | … | … | … | … | n | G |
| 37185 | 08404413-5233010 | 86.7 ± 0.2 | 4831 ± 128 | 2.71 ± 0.08 | 1.010 ± 0.004 | -0.11 ± 0.06 | <29 | 3 | Y | N | N | N | … | Y | … | … | n | NG? |
| 37186 | 08404513-5248521 | -25.1 ± 0.2 | … | … | … | … | … | … | … | … | … | … | … | … | … | Y | n | … |
| 37187 | 08404643-5348082 | 76.0 ± 0.2 | 5106 ± 154 | … | 1.002 ± 0.003 | -0.42 ± 0.18 | 11 ± 5 | 1 | Y | … | N | N | … | N | … | … | n | NG? |
| 37188 | 08404734-5302553 | 77.9 ± 0.2 | 4684 ± 11 | 2.48 ± 0.02 | 1.014 ± 0.006 | -0.06 ± 0.03 | 31 ± 10 | 1 | N | N | … | … | … | … | … | … | n | G |
| 37189 | 08404796-5335083 | 52.1 ± 0.2 | 4672 ± 32 | 2.43 ± 0.07 | 1.022 ± 0.004 | -0.02 ± 0.01 | 82 ± 14 | 1 | N | N | … | … | … | … | … | … | n | G |
| 37190 | 08404810-5304466 | 21.5 ± 0.2 | 4981 ± 126 | … | 1.019 ± 0.003 | -0.02 ± 0.06 | <28 | 3 | N | … | … | … | … | … | … | … | n | G |
| 37191 | 08404934-5250223 | 49.4 ± 0.2 | 4583 ± 71 | 2.38 ± 0.16 | 1.025 ± 0.005 | 0.03 ± 0.02 | <38 | 3 | N | N | … | … | … | … | … | … | n | G |
| 37279 | 08421944-5229314 | 19.0 ± 0.2 | 4853 ± 60 | 2.67 ± 0.19 | 1.024 ± 0.005 | -0.05 ± 0.10 | <37 | 3 | N | N | … | … | … | … | … | … | n | G |
| 37280 | 08422062-5257515 | 47.2 ± 0.2 | 4900 ± 103 | … | 1.013 ± 0.004 | -0.08 ± 0.06 | 54 ± 10 | 1 | N | … | … | … | … | … | … | … | n | G |
| 37281 | 08422456-5259570 | 10.5 ± 0.2 | 4963 ± 80 | 3.21 ± 0.08 | 1.001 ± 0.002 | -0.01 ± 0.06 | <31 | 3 | Y | N | Y | Y | N | Y | … | … | n | NG? |
| 37282 | 08422781-5322568 | 9.1 ± 0.2 | 4265 ± 215 | 4.58 ± 0.01 | 0.881 ± 0.004 | 0.04 ± 0.18 | <26 | 3 | Y | Y | Y | Y | Y | Y | … | … | Y | … |
| 2514 | 08422990-5330347 | 58.3 ± 0.6 | 4665 ± 8 | 2.44 ± 0.05 | … | 0.15 ± 0.04 | <15 | 3 | … | N | … | … | … | … | … | … | n | … |
| 37283 | 08422992-5239568 | 39.3 ± 0.2 | 4966 ± 206 | … | 1.015 ± 0.003 | -0.13 ± 0.21 | 30 ± 2 | 1 | N | … | … | … | … | … | … | … | n | G |
| 37284 | 08423068-5257345 | 14.6 ± 0.3 | 4147 ± 372 | 4.49 ± 0.02 | 0.842 ± 0.007 | -0.06 ± 0.12 | <60 | 3 | Y | Y | Y | Y | Y | Y | Y | Y | Y | … |
| 37285 | 08423199-5301043 | 20.9 ± 0.2 | 4313 ± 190 | 4.63 ± 0.12 | 0.884 ± 0.004 | -0.12 ± 0.07 | <17 | 3 | Y | Y | Y | Y | Y | Y | … | … | Y | … |
| 37286 | 08423411-5312542 | 46.6 ± 0.3 | 4845 ± 107 | 2.64 ± 0.16 | 1.022 ± 0.006 | -0.08 ± 0.02 | <12 | 3 | N | N | … | … | … | … | … | … | n | G |
| 37287 | 08423530-5250392 | 20.2 ± 0.3 | 4881 ± 249 | … | 1.016 ± 0.006 | -0.06 ± 0.12 | 13 ± 3 | 1 | N | … | … | … | … | … | … | … | n | G |
| 37288 | 08423995-5239416 | 4.7 ± 0.2 | 5201 ± 33 | 3.51 ± 0.13 | 1.006 ± 0.003 | -0.07 ± 0.10 | <14 | 3 | Y | Y | N | N | … | Y | … | … | n | NG? |
| 37289 | 08424071-5311202 | 3.1 ± 0.2 | 5141 ± 94 | … | 1.016 ± 0.005 | … | <33 | 3 | N | … | … | … | … | … | … | … | n | G |

**Table C.5.** continued.

| ID | CNAME | RV (km s$^{-1}$) | $T_{\rm eff}$ (K) | logg (dex) | $\gamma^a$ | [Fe/H] (dex) | EW(Li)$^b$ (mÅ) | EW(Li) error flag$^c$ | $\gamma$ | logg | \multicolumn{4}{c}{Membership} | [Fe/H] | \multicolumn{2}{c}{Gaia studies} | Final$^e$ | NMs with Li$^f$ |
|---|---|---|---|---|---|---|---|---|---|---|---|---|---|---|---|---|---|---|---|
| | | | | | | | | | | | RV | Li | H$\alpha$ | | Randich$^d$ | Cantat-Gaudin$^d$ | | |
| 37290 | 08424099-5330359 | 14.7 ± 0.2 | 3664 ± 47 | ... | 0.823 ± 0.003 | -0.19 ± 0.13 | ... | ... | Y | ... | Y | ... | Y | Y | ... | Y | n | ... |
| 37291 | 08424120-5319510 | 52.7 ± 0.2 | 4870 ± 138 | ... | 1.021 ± 0.003 | -0.10 ± 0.09 | <22 | 3 | N | ... | ... | ... | ... | ... | ... | ... | n | G |
| 37292 | 08424366-5326384 | 44.9 ± 0.2 | 4559 ± 198 | ... | 1.020 ± 0.004 | 0.04 ± 0.08 | <40 | 3 | N | ... | ... | ... | ... | ... | ... | ... | n | G |
| 37293 | 08424385-5234272 | 19.7 ± 0.2 | 4943 ± 205 | ... | 1.016 ± 0.003 | -0.07 ± 0.13 | <18 | 3 | N | ... | ... | ... | ... | ... | ... | ... | n | G |
| 37294 | 08424419-5331431 | 33.7 ± 0.2 | 4970 ± 43 | 2.84 ± 0.04 | 1.011 ± 0.003 | -0.01 ± 0.05 | <25 | 3 | N | N | ... | ... | ... | ... | ... | ... | n | G |
| 37295 | 08424440-5312319 | 61.9 ± 0.3 | 4750 ± 45 | ... | 0.997 ± 0.006 | -0.10 ± 0.04 | <21 | 3 | Y | ... | N | N | ... | Y | ... | ... | n | NG |
| 37296 | 08424803-5257010 | 75.2 ± 0.3 | 4822 ± 126 | ... | 1.018 ± 0.007 | -0.22 ± 0.19 | <14 | 3 | N | ... | ... | ... | ... | ... | ... | ... | n | G |
| 37297 | 08424824-5301577 | 44.2 ± 0.2 | 4507 ± 206 | ... | 1.020 ± 0.007 | 0.07 ± 0.09 | <51 | 3 | N | ... | ... | ... | ... | ... | ... | ... | n | G |
| 37298 | 08424903-5252157 | 15.6 ± 0.4 | 3397 ± 147 | 4.74 ± 0.01 | 0.836 ± 0.009 | -0.26 ± 0.13 | ... | ... | Y | Y | Y | ... | Y | Y | Y | Y | n | ... |
| 2515 | 08424965-5246269 | 46.9 ± 0.6 | 6108 ± 85 | 4.19 ± 0.12 | ... | 0.19 ± 0.11 | <19 | 3 | ... | Y | N | N | ... | N | N | ... | n | NG |
| 37299 | 08424966-5251439 | 12.3 ± 0.2 | 4657 ± 62 | 2.56 ± 0.14 | 1.019 ± 0.006 | 0.12 ± 0.03 | <37 | 3 | N | N | ... | ... | ... | ... | ... | ... | n | G |
| 2497 | 08394304-5257510 | 22.1 ± 0.6 | 6732 ± 213 | 4.09 ± 0.15 | ... | -0.05 ± 0.14 | 66 ± 18 | 2 | ... | Y | N | Y | N | Y | N | Y | n | NG |
| 37128 | 08394843-5313583 | 14.6 ± 0.5 | 3335 ± 18 | ... | 0.859 ± 0.015 | ... | <100 | 3 | Y | ... | Y | Y | Y | ... | Y | Y | Y | ... |
| 37130 | 08395111-5310539 | 63.5 ± 0.2 | 4613 ± 148 | ... | 1.019 ± 0.004 | -0.12 ± 0.11 | <38 | 3 | N | ... | ... | ... | ... | ... | ... | ... | n | G |
| 36971 | 08351399-5308560 | 94.9 ± 0.2 | 4896 ± 52 | 2.88 ± 0.16 | 1.009 ± 0.005 | -0.21 ± 0.20 | 8 ± 3 | 1 | Y | N | N | N | ... | Y | ... | ... | n | ... |
| 37129 | 08395023-5335503 | 57.9 ± 0.2 | 4631 ± 124 | ... | 1.027 ± 0.004 | -0.14 ± 0.13 | 38 ± 4 | 1 | N | ... | ... | ... | ... | ... | ... | ... | n | G |
| 36972 | 08351459-5311068 | 27.5 ± 0.2 | 5080 ± 16 | ... | 1.017 ± 0.004 | 0.01 ± 0.03 | <27 | 3 | N | ... | ... | ... | ... | ... | ... | ... | n | G |
| 36973 | 08351787-5301421 | 33.1 ± 0.2 | 5500 ± 50 | ... | 1.009 ± 0.004 | 0.03 ± 0.15 | <22 | 3 | Y | ... | N | N | N | Y | ... | ... | n | NG? |
| 37131 | 08395152-5315159 | 27.0 ± 0.2 | 4726 ± 112 | 2.55 ± 0.15 | 1.017 ± 0.003 | 0.01 ± 0.08 | 217 ± 10 | 1 | N | N | ... | ... | ... | ... | ... | ... | n | Li-rich G |
| 36974 | 08351898-5315189 | 71.7 ± 0.2 | 4638 ± 171 | 2.52 ± 0.11 | 1.008 ± 0.006 | -0.12 ± 0.13 | <29 | 3 | Y | N | ... | ... | ... | Y | ... | ... | n | NG? |
| 36975 | 08352291-5307301 | 112.8 ± 0.2 | 4888 ± 142 | 2.82 ± 0.09 | 1.010 ± 0.004 | -0.11 ± 0.06 | <21 | 3 | N | N | ... | ... | ... | ... | ... | ... | n | G |
| 37321 | 08432495-5248169 | 9.7 ± 0.2 | 4931 ± 97 | 2.71 ± 0.09 | 1.020 ± 0.004 | -0.02 ± 0.05 | <29 | 3 | N | N | ... | ... | ... | ... | ... | ... | n | G |
| 36976 | 08353587-5315069 | 30.0 ± 0.2 | 5074 ± 219 | ... | 1.011 ± 0.003 | -0.21 ± 0.16 | 80 ± 4 | 1 | N | ... | ... | ... | ... | ... | ... | ... | n | Li-rich G |
| 37141 | 08395838-5218197 | 35.2 ± 0.2 | 4957 ± 172 | ... | 1.015 ± 0.002 | -0.08 ± 0.13 | <22 | 3 | N | ... | ... | ... | ... | ... | ... | ... | n | G |
| 37322 | 08432584-5302201 | 34.4 ± 0.2 | 5084 ± 80 | ... | 1.023 ± 0.005 | -0.07 ± 0.14 | 14 ± 4 | 1 | N | ... | ... | ... | ... | ... | ... | ... | n | G |
| 36977 | 08354078-5315506 | 6.4 ± 0.2 | 4532 ± 154 | ... | 1.018 ± 0.004 | 0.09 ± 0.05 | 58 ± 6 | 1 | N | ... | ... | ... | ... | ... | ... | ... | n | G |
| 37323 | 08432587-5256248 | 34.8 ± 0.2 | 5105 ± 75 | ... | 1.009 ± 0.005 | -0.02 ± 0.04 | <27 | 3 | Y | ... | N | Y | N | Y | ... | ... | n | NG? |
| 36978 | 08354546-5301099 | -6.4 ± 0.2 | 5100 ± 5 | ... | 1.025 ± 0.005 | -0.15 ± 0.12 | <15 | 3 | N | ... | ... | ... | ... | ... | ... | ... | n | G |
| 37142 | 08395870-5253047 | 4.1 ± 0.2 | 5103 ± 17 | ... | 1.014 ± 0.001 | -0.01 ± 0.04 | <20 | 3 | N | ... | ... | ... | ... | ... | ... | ... | n | G |
| 2519 | 08443450-5255325 | 19.6 ± 0.6 | 5175 ± 35 | 4.33 ± 0.05 | ... | 0.21 ± 0.05 | <13 | 3 | ... | Y | Y | N | Y | N | ... | ... | n | NG |
| 2481 | 08354551-5315242 | -5.4 ± 0.6 | 4667 ± 213 | 4.51 ± 0.18 | ... | -0.07 ± 0.15 | <11 | 3 | ... | Y | N | N | ... | Y | ... | Y | n | ... |
| 37345 | 08443655-5246405 | 124.7 ± 0.2 | 4879 ± 170 | ... | 1.019 ± 0.004 | -0.13 ± 0.18 | <35 | 3 | N | ... | ... | ... | ... | ... | ... | ... | n | G |
| 36979 | 08354969-5315311 | -0.8 ± 0.2 | 5028 ± 96 | ... | 1.024 ± 0.003 | -0.06 ± 0.07 | <22 | 3 | N | ... | ... | ... | ... | ... | ... | ... | n | G |
| 37346 | 08444075-5243070 | 42.5 ± 0.2 | 5159 ± 91 | ... | 1.019 ± 0.006 | ... | <20 | 3 | N | ... | ... | ... | ... | ... | ... | ... | n | G |
| 36980 | 08355028-5300325 | 20.1 ± 0.2 | 4891 ± 99 | 2.59 ± 0.01 | 1.012 ± 0.003 | -0.11 ± 0.10 | <15 | 3 | N | N | ... | ... | ... | ... | ... | ... | n | G |
| 37347 | 08444164-5242033 | 22.4 ± 0.2 | 4975 ± 173 | 3.11 ± 0.14 | 1.006 ± 0.005 | -0.04 ± 0.08 | 30 ± 8 | 1 | Y | N | N | Y | N | Y | ... | ... | n | NG? |
| 36981 | 08355195-5318258 | 83.5 ± 0.7 | ... | ... | ... | ... | ... | ... | ... | ... | ... | ... | ... | N | ... | ... | n | ... |
| 37143 | 08395933-5231595 | 21.1 ± 0.2 | 5025 ± 162 | ... | 1.020 ± 0.002 | -0.08 ± 0.16 | <24 | 3 | N | ... | ... | ... | ... | ... | ... | ... | n | G |
| 37348 | 08444334-5248432 | 59.7 ± 0.2 | 4655 ± 35 | 2.46 ± 0.13 | 1.021 ± 0.003 | -0.06 ± 0.01 | <36 | 3 | N | N | ... | ... | ... | ... | ... | ... | n | G |
| 36982 | 08360510-5315216 | 41.6 ± 0.2 | 4594 ± 146 | ... | 1.025 ± 0.004 | -0.19 ± 0.15 | <25 | 3 | N | ... | ... | ... | ... | ... | ... | ... | n | G |
| 37349 | 08444805-5247228 | 32.6 ± 0.2 | 4917 ± 241 | ... | 1.020 ± 0.004 | -0.12 ± 0.22 | 62 ± 3 | 1 | N | ... | ... | ... | ... | ... | ... | ... | n | G |
| 36983 | 08360676-5314162 | 11.2 ± 0.2 | 4914 ± 162 | 3.06 ± 0.20 | 1.007 ± 0.003 | -0.07 ± 0.12 | 20 ± 5 | 1 | Y | N | Y | Y | N | Y | ... | ... | n | NG? |
| 37350 | 08445267-5248145 | 34.9 ± 0.2 | 4909 ± 225 | ... | 1.019 ± 0.003 | -0.10 ± 0.19 | 45 ± 2 | 1 | N | ... | ... | ... | ... | ... | ... | ... | n | G |
| 36984 | 08360961-5317270 | 41.3 ± 0.2 | 5116 ± 164 | ... | 1.006 ± 0.004 | -0.22 ± 0.11 | 11 ± 3 | 1 | Y | ... | N | N | N | Y | ... | ... | n | NG? |
| 37362 | 08452722-5250419 | 10.9 ± 0.2 | 4973 ± 143 | ... | 1.019 ± 0.002 | -0.03 ± 0.06 | <17 | 3 | N | ... | ... | ... | ... | ... | ... | ... | n | G |
| 37351 | 08450338-5254307 | 9.6 ± 0.2 | 5110 ± 71 | ... | 1.015 ± 0.003 | -0.07 ± 0.14 | 10 ± 2 | 1 | N | ... | ... | ... | ... | ... | ... | ... | n | G |
| 36985 | 08361153-5312360 | 53.0 ± 0.2 | 4279 ± 257 | 4.46 ± 0.11 | 0.892 ± 0.005 | -0.15 ± 0.04 | 21 ± 3 | 1 | Y | Y | N | Y | N | Y | ... | ... | n | NG |
| 37352 | 08450414-5249582 | 73.7 ± 0.2 | 4855 ± 114 | 2.78 ± 0.18 | 1.013 ± 0.003 | -0.09 ± 0.09 | <23 | 3 | N | N | ... | ... | ... | ... | ... | ... | n | ... |
| 36986 | 08361220-5300045 | 2.1 ± 0.2 | 4995 ± 166 | ... | 1.039 ± 0.002 | -0.16 ± 0.15 | <26 | 3 | N | ... | ... | ... | ... | ... | ... | ... | n | G |
| 37353 | 08450525-5250236 | 44.9 ± 0.2 | 4918 ± 86 | ... | 1.045 ± 0.004 | -0.06 ± 0.09 | <27 | 3 | N | ... | ... | ... | ... | ... | ... | ... | n | G |
| 36987 | 08361799-5312103 | 98.4 ± 0.2 | 5275 ± 72 | ... | 1.026 ± 0.003 | -0.07 ± 0.09 | <19 | 3 | N | ... | ... | ... | ... | ... | ... | ... | n | ... |
| 37354 | 08451201-5253060 | 40.2 ± 0.2 | 5012 ± 184 | ... | 1.017 ± 0.002 | -0.11 ± 0.21 | <20 | 3 | N | ... | ... | ... | ... | ... | ... | ... | n | G |
| 36988 | 08362283-5307387 | 14.1 ± 0.2 | 5010 ± 68 | 3.21 ± 0.01 | 1.004 ± 0.004 | -0.03 ± 0.08 | <23 | 3 | Y | N | Y | N | N | Y | ... | ... | n | NG? |
| 37355 | 08451348-5303020 | 8.8 ± 0.3 | 4982 ± 163 | ... | 1.021 ± 0.007 | -0.08 ± 0.16 | 17 ± 5 | 1 | N | ... | ... | ... | ... | ... | ... | ... | n | G |
| 36989 | 08362680-5314386 | 83.5 ± 0.2 | 4674 ± 45 | ... | 1.025 ± 0.003 | -0.31 ± 0.04 | 12 ± 6 | 1 | N | ... | ... | ... | ... | ... | ... | ... | n | G |
| 37356 | 08451623-5244251 | 27.5 ± 0.3 | 5060 ± 12 | 3.07 ± 0.16 | 1.012 ± 0.010 | -0.04 ± 0.07 | 48 ± 7 | 1 | N | N | ... | ... | ... | ... | ... | ... | n | G |
| 37144 | 08400085-5319044 | -11.1 ± 0.2 | 5049 ± 76 | 3.37 ± 0.07 | 1.000 ± 0.002 | -0.03 ± 0.06 | ... | ... | Y | N | N | ... | ... | Y | ... | ... | n | ... |
| 36990 | 08363139-5320463 | 31.5 ± 0.2 | 4950 ± 58 | 2.82 ± 0.06 | 1.013 ± 0.002 | 0.02 ± 0.07 | 80 ± 2 | 1 | N | N | ... | ... | ... | ... | ... | ... | n | Li-rich G |
| 37357 | 08451786-5253011 | 27.6 ± 0.2 | 4751 ± 28 | 2.56 ± 0.19 | 1.024 ± 0.003 | -0.08 ± 0.12 | 7 ± 4 | 1 | N | N | ... | ... | ... | ... | ... | ... | n | G |

**Table C.5.** continued.

| ID | CNAME | RV (km s$^{-1}$) | $T_{\text{eff}}$ (K) | $\log g$ (dex) | $\gamma^a$ | [Fe/H] (dex) | EW(Li)$^b$ (mÅ) | EW(Li) error flag$^c$ | $\gamma$ | $\log g$ | Membership RV | Li | H$\alpha$ | [Fe/H] | Gaia studies Randich$^d$ | Cantat-Gaudin$^d$ | Final$^e$ | NMs with Li$^f$ |
|---|---|---|---|---|---|---|---|---|---|---|---|---|---|---|---|---|---|---|
| 36991 | 08363862-5309557 | 40.2 ± 0.2 | 4608 ± 307 | 4.59 ± 0.08 | 0.917 ± 0.003 | -0.15 ± 0.05 | 20 ± 4 | 1 | Y | Y | N | Y | N | Y | … | … | n | NG |
| 37358 | 08451883-5259257 | 14.6 ± 0.4 | 3387 ± 152 | 4.79 ± 0.07 | 0.832 ± 0.012 | -0.26 ± 0.13 | 125 ± 26 | 1 | Y | Y | Y | Y | Y | Y | Y | Y | Y | … |
| 36992 | 08363926-5305042 | 81.4 ± 0.2 | 4730 ± 24 | 2.61 ± 0.16 | 1.012 ± 0.005 | -0.04 ± 0.03 | <36 | 3 | N | N | … | … | … | … | … | … | n | G |
| 37359 | 08452277-5246539 | -1.9 ± 0.2 | 4845 ± 87 | 2.66 ± 0.12 | 1.019 ± 0.005 | -0.03 ± 0.04 | 7 ± 3 | 1 | N | N | … | … | … | … | … | … | n | G |
| 37015 | 08380733-5300459 | 58.1 ± 0.2 | 5123 ± 4 | 3.47 ± 0.04 | 1.001 ± 0.004 | -0.05 ± 0.02 | <22 | 3 | Y | N | N | N | N | Y | … | … | n | NG? |
| 37360 | 08452470-5301226 | 14.1 ± 0.3 | 3446 ± 119 | 4.62 ± 0.16 | 0.842 ± 0.010 | -0.23 ± 0.15 | … | … | Y | Y | Y | … | Y | Y | Y | Y | n | … |
| 37016 | 08381065-5221484 | 41.6 ± 0.2 | 5139 ± 94 | … | 1.016 ± 0.004 | … | <27 | 3 | N | … | … | … | … | … | … | … | n | G |
| 37361 | 08452692-5252020 | 15.8 ± 0.2 | 4827 ± 352 | … | 0.971 ± 0.002 | -0.03 ± 0.03 | 189 ± 2 | 1 | Y | … | Y | Y | Y | Y | Y | Y | Y | … |
| 37017 | 08381085-5221124 | 2.8 ± 0.2 | 5158 ± 3 | … | 1.015 ± 0.001 | -0.17 ± 0.13 | <9 | 3 | N | … | … | … | … | … | … | … | n | G |
| 37018 | 08381110-5301192 | -10.6 ± 0.2 | 4954 ± 185 | … | 1.007 ± 0.003 | -0.11 ± 0.14 | <19 | 3 | Y | … | N | N | … | Y | … | … | n | NG? |
| 37145 | 08400203-5248404 | -1.3 ± 0.2 | 5127 ± 98 | … | 1.011 ± 0.002 | … | 26 ± 2 | 1 | N | … | … | … | … | … | … | … | n | G |
| 37019 | 08381473-5303362 | 30.8 ± 0.2 | 4960 ± 158 | … | 1.021 ± 0.003 | -0.06 ± 0.12 | <20 | 3 | N | … | … | … | … | … | … | … | n | G |
| 37020 | 08381700-5320496 | 17.9 ± 0.2 | 4740 ± 42 | 2.59 ± 0.05 | 1.012 ± 0.001 | 0.05 ± 0.05 | <41 | 3 | N | N | … | … | … | … | … | … | n | G |
| 2501 | 08400571-5305599 | -5.6 ± 0.6 | 4876 ± 50 | 3.12 ± 0.12 | 1.013 ± 0.002 | 0.14 ± 0.02 | 33 ± 7 | 1 | N | N | … | … | … | … | … | … | n | … |
| 37021 | 08381822-5225023 | 20.7 ± 0.2 | 4857 ± 199 | 2.68 ± 0.19 | 1.013 ± 0.003 | -0.07 ± 0.13 | <21 | 3 | N | N | … | … | … | … | … | … | n | G |
| 37148 | 08400576-5354437 | 88.5 ± 0.2 | 5022 ± 118 | … | 1.006 ± 0.002 | -0.24 ± 0.21 | <9 | 3 | Y | … | N | N | … | Y | … | … | n | … |
| 37022 | 08381824-5329435 | 66.9 ± 0.2 | 4982 ± 141 | … | 1.006 ± 0.004 | -0.40 ± 0.17 | 16 ± 3 | 1 | Y | … | N | … | N | … | … | … | n | NG? |
| 37023 | 08381913-5322283 | 11.8 ± 0.2 | 4916 ± 13 | 2.71 ± 0.05 | 1.017 ± 0.003 | -0.03 ± 0.09 | 31 ± … | 3 | N | N | … | … | … | … | … | … | n | G |
| 37149 | 08400623-5324475 | 21.9 ± 0.2 | 3728 ± 116 | 4.55 ± 0.20 | 0.817 ± 0.004 | -0.17 ± 0.14 | <100 | 3 | Y | Y | N | Y | Y | Y | N | … | n | … |
| 37024 | 08381987-5310515 | 10.5 ± 0.2 | 4688 ± 164 | 2.54 ± 0.12 | 1.010 ± 0.005 | -0.04 ± 0.13 | <26 | 3 | N | N | … | … | … | … | … | … | n | G |
| 37025 | 08382349-5316545 | 30.8 ± 0.3 | 3644 ± 81 | … | 0.840 ± 0.009 | -0.21 ± 0.14 | <100 | 3 | Y | … | N | Y | Y | Y | N | … | n | … |
| 2502 | 08400624-5338069 | 15.4 ± 0.6 | 5808 ± 90 | 4.35 ± 0.15 | … | -0.05 ± 0.14 | 150 ± 22 | 2 | … | Y | Y | Y | Y | Y | Y | … | Y | … |
| 37026 | 08382370-5317454 | 64.2 ± 0.2 | 4958 ± 115 | … | 1.016 ± 0.003 | -0.11 ± 0.12 | <18 | 3 | N | … | … | … | … | … | … | … | n | G |
| 2484 | 08382373-5257322 | 20.0 ± 0.6 | 4742 ± 74 | 4.29 ± 0.38 | … | -0.02 ± 0.06 | <29 | 3 | … | Y | Y | Y | Y | Y | … | … | Y | … |
| 37027 | 08382523-5305362 | 13.6 ± 0.7 | … | … | … | … | … | … | … | … | … | … | … | … | … | … | n | … |
| 37028 | 08382536-5333363 | 17.1 ± 0.3 | 5073 ± 98 | … | 1.016 ± 0.006 | -0.06 ± 0.12 | 59 ± 5 | 1 | N | … | … | … | … | … | … | … | n | Li-rich G |
| 37029 | 08382671-5310072 | 17.3 ± 0.3 | 3439 ± 76 | 4.67 ± 0.11 | 0.829 ± 0.015 | -0.26 ± 0.13 | <100 | 3 | Y | Y | Y | Y | N | Y | Y | … | n | NG |
| 37030 | 08382711-5325105 | 13.4 ± 0.8 | … | … | … | … | … | … | … | … | … | … | … | … | … | Y | n | … |
| 37031 | 08382726-5221190 | 33.4 ± 0.2 | 5141 ± 94 | … | 1.012 ± 0.004 | … | <17 | 3 | N | … | … | … | … | … | … | … | n | G |
| 37032 | 08382788-5252313 | 0.6 ± 0.2 | 5128 ± 18 | … | 1.019 ± 0.002 | -0.06 ± 0.05 | … | … | N | … | … | … | … | … | … | … | n | G |
| 37033 | 08382965-5314035 | 105.8 ± 0.3 | 5424 ± 29 | … | 1.010 ± 0.008 | -0.34 ± 0.11 | 16 ± … | 3 | Y | … | N | N | … | N | … | … | n | NG? |
| 37150 | 08400812-5324165 | 44.8 ± 0.2 | 4886 ± 9 | 2.74 ± 0.02 | 1.013 ± 0.002 | 0.01 ± 0.04 | 28 ± … | 3 | N | N | … | … | … | … | … | … | n | G |
| 37034 | 08383076-5228568 | 44.3 ± 0.3 | 5025 ± 164 | … | 1.007 ± 0.004 | -0.09 ± 0.18 | 30 ± 3 | 1 | Y | … | N | Y | N | Y | … | … | n | NG? |
| 37035 | 08383462-5331497 | 106.2 ± 0.2 | 4812 ± 103 | … | 1.023 ± 0.005 | -0.22 ± 0.17 | 55 ± 3 | 1 | N | … | … | … | … | … | … | … | n | G |
| 37036 | 08383609-5257532 | 38.8 ± 0.2 | 4937 ± 90 | … | 1.012 ± 0.004 | -0.07 ± 0.09 | 41 ± 4 | 1 | N | … | … | … | … | … | … | … | n | G |
| 37037 | 08383620-5322169 | 11.8 ± 0.2 | 5120 ± 15 | … | 1.008 ± 0.003 | -0.07 ± 0.11 | <22 | 3 | Y | … | Y | Y | N | Y | … | … | n | NG? |
| 37078 | 08391484-5322599 | 46.9 ± 0.2 | 5157 ± 3 | 3.37 ± 0.18 | 1.009 ± 0.003 | -0.05 ± 0.11 | <20 | 3 | Y | N | N | N | N | Y | … | … | n | NG? |
| 2489 | 08391618-5316486 | -536.1 ± 0.4 | … | … | … | … | … | … | … | … | … | … | … | … | … | … | n | … |
| 37160 | 08401782-5336393 | 61.2 ± 0.2 | 4676 ± 120 | 2.72 ± 0.06 | 1.003 ± 0.005 | -0.10 ± 0.04 | <46 | 3 | Y | N | N | Y | N | Y | … | … | n | NG? |
| 37161 | 08401829-5330288 | 14.6 ± 0.2 | 4458 ± 225 | 4.38 ± 0.18 | 0.927 ± 0.003 | 0.01 ± 0.12 | 51 ± 9 | 1 | Y | Y | Y | Y | Y | Y | … | Y | Y | … |
| 37079 | 08391713-5355032 | 59.3 ± 0.2 | 4531 ± 106 | … | 1.031 ± 0.005 | -0.17 ± 0.01 | <23 | 3 | N | … | … | … | … | … | … | … | n | G |
| 2490 | 08391723-5218236 | 11.9 ± 0.6 | 5121 ± 21 | 2.84 ± 0.02 | … | -0.03 ± 0.03 | <10 | 3 | … | N | … | … | … | … | … | … | n | … |
| 37080 | 08391725-5258482 | 36.6 ± 0.2 | 5117 ± 95 | … | 1.011 ± 0.003 | -0.06 ± 0.10 | <25 | 3 | N | … | … | … | … | … | … | … | n | G |
| 2504 | 08401890-5345161 | 19.2 ± 0.6 | 5234 ± 56 | 2.75 ± 0.10 | … | -0.02 ± 0.01 | <14 | 3 | … | N | … | … | … | … | … | … | n | … |
| 37162 | 08401905-5219417 | 56.2 ± 0.2 | 5029 ± 95 | … | 1.015 ± 0.004 | -0.05 ± 0.07 | <26 | 3 | N | … | … | … | … | … | … | … | n | G |
| 37081 | 08391826-5330035 | 86.4 ± 0.2 | 4862 ± 151 | … | 1.019 ± 0.002 | -0.21 ± 0.20 | 9 ± 4 | 1 | N | … | … | … | … | … | … | … | n | G |
| 37082 | 08392002-5307562 | 32.4 ± 0.2 | 4920 ± 99 | 3.17 ± 0.14 | 0.999 ± 0.003 | 0.03 ± 0.04 | 36 ± … | 3 | Y | N | N | Y | N | Y | … | … | n | NG? |
| 37083 | 08392017-5323466 | 43.9 ± 0.2 | 5096 ± 193 | … | 1.018 ± 0.004 | -0.24 ± 0.24 | <24 | 3 | N | … | … | … | … | … | … | … | n | G |
| 37084 | 08392028-5249544 | 65.0 ± 0.2 | 4952 ± 122 | … | 1.016 ± 0.003 | -0.07 ± 0.06 | <13 | 3 | N | … | … | … | … | … | … | … | n | G |
| 37092 | 08392440-5223424 | 35.5 ± 0.2 | 4719 ± 82 | 2.46 ± 0.15 | 1.022 ± 0.003 | -0.08 ± 0.07 | 68 ± 7 | 1 | N | … | … | … | … | … | … | … | n | G |
| 37163 | 08402079-5324355 | 44.1 ± 0.2 | 5025 ± 86 | … | 1.022 ± 0.005 | -0.05 ± 0.05 | <26 | 3 | N | … | … | … | … | … | … | … | n | G |
| 37093 | 08392496-5230364 | 94.7 ± 0.2 | 4746 ± 57 | 2.59 ± 0.17 | 1.017 ± 0.003 | -0.11 ± 0.07 | <19 | 3 | N | N | … | … | … | … | … | … | n | G |
| 37094 | 08392578-5246292 | 57.8 ± 0.2 | 4764 ± 58 | 2.61 ± 0.01 | 1.015 ± 0.002 | -0.05 ± 0.09 | <34 | 3 | N | N | … | … | … | … | … | … | n | G |
| 37169 | 08402890-5225398 | 28.7 ± 0.2 | 4560 ± 189 | … | 1.023 ± 0.004 | 0.05 ± 0.07 | <46 | 3 | N | … | … | … | … | … | … | … | n | G |
| 37170 | 08402939-5310496 | 50.5 ± 0.2 | 4664 ± 77 | 2.48 ± 0.16 | 1.016 ± 0.003 | -0.09 ± 0.08 | <28 | 3 | N | N | … | … | … | … | … | … | n | G |
| 37095 | 08392714-5221433 | 55.9 ± 0.2 | 4660 ± 53 | 2.39 ± 0.12 | 1.021 ± 0.003 | -0.08 ± 0.06 | <27 | 3 | N | N | … | … | … | … | … | … | n | G |
| 37171 | 08402993-5329079 | 35.8 ± 0.2 | 4762 ± 31 | 2.60 ± 0.14 | 1.020 ± 0.005 | -0.08 ± 0.01 | 32 ± 4 | 1 | N | N | … | … | … | … | … | … | n | … |
| 37096 | 08392848-5304563 | 49.0 ± 0.2 | 5120 ± 139 | … | 1.003 ± 0.002 | -0.20 ± 0.17 | <19 | 3 | Y | … | N | N | … | Y | … | … | n | NG? |
| 37097 | 08392896-5250239 | 52.2 ± 0.2 | 5108 ± 75 | 3.51 ± 0.17 | 1.003 ± 0.003 | -0.10 ± 0.14 | <26 | 3 | Y | Y | N | N | N | Y | … | … | n | NG? |

**Table C.5.** continued.

| ID | CNAME | RV (km s$^{-1}$) | $T_{\text{eff}}$ (K) | $logg$ (dex) | $\gamma^a$ | [Fe/H] (dex) | EW(Li)$^b$ (mÅ) | EW(Li) error flag$^c$ | $\gamma$ | $logg$ | Membership RV | Li | H$\alpha$ | [Fe/H] | Gaia studies Randich$^d$ | Cantat-Gaudin$^d$ | Final$^e$ | NMs with Li$^f$ |
|---|---|---|---|---|---|---|---|---|---|---|---|---|---|---|---|---|---|---|
| 37172 | 08403120-5311443 | 47.7 ± 0.2 | 4807 ± 34 | 2.63 ± 0.10 | 1.020 ± 0.004 | -0.09 ± 0.03 | 21 ± 9 | 1 | N | N | … | … | … | … | … | … | n | G |
| 37098 | 08392908-5311059 | 34.2 ± 0.2 | 4936 ± 181 | 3.07 ± 0.15 | 1.006 ± 0.003 | -0.08 ± 0.09 | <20 | 3 | Y | N | N | N | N | Y | … | … | n | NG? |
| 37117 | 08393932-5311091 | 18.0 ± 0.2 | 5019 ± 15 | 2.85 ± 0.02 | 1.016 ± 0.002 | 0.00 ± 0.02 | <29 | 3 | N | N | … | … | … | … | … | … | n | G |
| 37192 | 08405129-5255423 | -5.4 ± 0.2 | 4931 ± 169 | … | 1.010 ± 0.004 | -0.03 ± 0.06 | <23 | 3 | N | … | … | … | … | … | … | … | n | G |
| 37193 | 08405170-5341409 | 116.0 ± 0.2 | 4914 ± 8 | 2.72 ± 0.06 | 1.018 ± 0.005 | -0.08 ± 0.02 | <29 | 3 | N | N | … | … | … | … | … | … | n | G |
| 37194 | 08405232-5246469 | 5.7 ± 0.2 | 4978 ± 143 | … | 1.013 ± 0.004 | -0.11 ± 0.18 | 40 ± 4 | 1 | N | … | … | … | … | … | … | … | n | G |
| 37118 | 08394052-5336412 | -1.5 ± 0.2 | 4629 ± 135 | 2.50 ± 0.20 | 1.018 ± 0.005 | 0.05 ± 0.06 | <45 | 3 | N | N | … | … | … | … | … | … | n | G |
| 37195 | 08405259-5333576 | 41.2 ± 0.2 | 4609 ± 91 | … | 1.016 ± 0.004 | … | 161 ± 3 | 1 | N | … | … | … | … | … | … | … | n | Li-rich G |
| 37196 | 08405272-5346247 | 26.5 ± 0.2 | 4831 ± 243 | … | 1.020 ± 0.005 | -0.12 ± 0.20 | 30 ± 8 | 1 | N | … | … | … | … | … | … | … | n | G |
| 37119 | 08394056-5306074 | 16.0 ± 1.9 | … | … | … | … | … | … | … | … | … | … | … | … | Y | Y | n | … |
| 37197 | 08405408-5328089 | -10.6 ± 0.2 | 4654 ± 3 | 2.39 ± 0.10 | 1.024 ± 0.003 | -0.09 ± 0.04 | <34 | 3 | N | N | … | … | … | … | … | … | n | G |
| 37198 | 08405414-5307492 | 34.7 ± 0.2 | 4889 ± 108 | 2.73 ± 0.08 | 1.012 ± 0.003 | -0.05 ± 0.07 | <27 | 3 | N | N | … | … | … | … | … | … | n | G |
| 37199 | 08405519-5234499 | 44.9 ± 19.2 | … | … | … | … | … | … | … | … | … | … | … | … | Y | Y | n | … |
| 37200 | 08405588-5247408 | 32.4 ± 0.2 | 4544 ± 25 | 2.23 ± 0.08 | 1.029 ± 0.003 | -0.04 ± 0.08 | <47 | 3 | N | N | … | … | … | … | … | … | n | G |
| 37120 | 08394141-5304035 | 19.7 ± 0.4 | 3399 ± 24 | … | 0.826 ± 0.015 | … | … | … | Y | … | Y | … | Y | … | … | … | n | … |
| 37201 | 08405643-5308309 | 55.0 ± 0.3 | 4666 ± 149 | 2.62 ± 0.02 | 1.004 ± 0.008 | -0.13 ± 0.20 | 334 ± 6 | 1 | N | … | … | … | … | … | … | … | n | Li-rich G |
| 37202 | 08405697-5244127 | 47.8 ± 0.2 | 4968 ± 193 | … | 1.014 ± 0.003 | -0.19 ± 0.21 | <28 | 3 | N | … | … | … | … | … | … | … | n | G |
| 37121 | 08394166-5342159 | 25.5 ± 0.2 | 4807 ± 7 | 2.62 ± 0.08 | 1.017 ± 0.002 | -0.03 ± 0.12 | <28 | 3 | N | N | … | … | … | … | … | … | n | G |
| 37203 | 08405752-5238207 | 14.2 ± 0.2 | 3690 ± 12 | … | 0.824 ± 0.004 | -0.19 ± 0.12 | <100 | 3 | Y | … | Y | Y | Y | Y | Y | Y | Y | … |
| 37204 | 08410002-5244215 | 43.5 ± 0.2 | 5030 ± 121 | … | 1.020 ± 0.002 | -0.05 ± 0.09 | <18 | 3 | N | … | … | … | … | … | … | … | n | G |
| 37122 | 08394203-5357189 | 6.4 ± 0.8 | 3905 ± 54 | … | 0.868 ± 0.018 | -0.07 ± 0.22 | … | … | Y | … | N | … | Y | N | … | … | n | … |
| 37205 | 08410030-5241328 | 12.1 ± 0.2 | 5010 ± 183 | … | 1.013 ± 0.003 | -0.08 ± 0.16 | <11 | 3 | N | … | … | … | … | … | … | … | n | G |
| 37123 | 08394221-5342412 | -1.1 ± 0.2 | 4922 ± 140 | 3.13 ± 0.01 | 1.002 ± 0.002 | 0.00 ± 0.04 | <49 | 3 | Y | N | N | N | N | Y | … | … | n | NG? |
| 37206 | 08410228-5250141 | 52.8 ± 0.2 | 4109 ± 237 | 4.58 ± 0.05 | 0.846 ± 0.004 | -0.03 ± 0.17 | <23 | 3 | Y | Y | N | Y | Y | Y | N | … | n | … |
| 37124 | 08394268-5252358 | -15.8 ± 0.2 | 5130 ± 106 | … | 1.052 ± 0.006 | … | <45 | 3 | N | … | … | … | … | … | … | … | n | Li-rich G |
| 37207 | 08410275-5249304 | 64.9 ± 0.2 | 4638 ± 112 | 2.46 ± 0.17 | 1.015 ± 0.005 | -0.07 ± 0.03 | <18 | 3 | N | N | … | … | … | … | … | … | n | G |
| 37208 | 08410435-5351301 | 47.8 ± 0.2 | 4760 ± 36 | 2.63 ± 0.14 | 1.015 ± 0.003 | -0.10 ± 0.05 | <17 | 3 | N | N | … | … | … | … | … | … | n | G |
| 37209 | 08410525-5307502 | 66.7 ± 0.2 | 5158 ± 91 | … | 1.004 ± 0.005 | … | <21 | 3 | Y | … | N | N | N | … | … | … | n | NG? |
| 37210 | 08410532-5248111 | 34.7 ± 0.2 | 4717 ± 124 | … | 1.023 ± 0.004 | -0.02 ± 0.10 | 37 ± 4 | 1 | N | … | … | … | … | … | … | … | n | G |
| 37211 | 08410558-5235207 | 19.0 ± 0.2 | 5162 ± 6 | 3.37 ± 0.16 | 1.009 ± 0.004 | -0.02 ± 0.05 | 19 ± 5 | 1 | Y | N | Y | Y | N | Y | … | … | n | NG? |
| 37212 | 08410648-5308128 | 150.9 ± 0.2 | 4888 ± 145 | … | 1.017 ± 0.003 | -0.20 ± 0.19 | <10 | 3 | N | … | … | … | … | … | … | … | n | G |
| 37213 | 08410701-5251041 | 46.8 ± 0.2 | 4607 ± 128 | 2.35 ± 0.17 | 1.015 ± 0.003 | -0.10 ± 0.11 | <44 | 3 | N | N | … | … | … | … | … | … | n | G |
| 37214 | 08410736-5258276 | 48.6 ± 0.2 | 4680 ± 18 | 2.37 ± 0.13 | 1.036 ± 0.009 | -0.04 ± 0.11 | <44 | 3 | N | N | … | … | … | … | … | … | n | G |
| 37237 | 08412773-5252326 | 68.2 ± 0.3 | 5084 ± 118 | … | 1.029 ± 0.010 | -0.08 ± 0.08 | <20 | 3 | N | … | … | … | … | … | … | … | n | G |
| 37238 | 08412915-5316223 | 15.6 ± 1.6 | … | … | … | … | … | … | … | … | … | … | … | … | … | Y | n | … |
| 37239 | 08412932-5237376 | 19.5 ± 0.2 | 5057 ± 125 | 3.56 ± 0.11 | 0.999 ± 0.002 | -0.07 ± 0.14 | <17 | 3 | Y | Y | Y | N | N | Y | … | … | n | NG |
| 37240 | 08412979-5324555 | 50.2 ± 0.2 | 5043 ± 136 | … | 1.013 ± 0.003 | -0.05 ± 0.08 | 39 ± 6 | 1 | N | … | … | … | … | … | … | … | n | G |
| 37241 | 08412998-5240379 | 15.2 ± 0.2 | 3830 ± 102 | 4.54 ± 0.17 | 0.821 ± 0.006 | -0.15 ± 0.13 | 64 ± 24 | 1 | Y | Y | Y | Y | N | Y | Y | … | n | NG |
| 37242 | 08413027-5323563 | 88.8 ± 0.2 | 4653 ± 58 | 2.41 ± 0.08 | 1.021 ± 0.005 | -0.02 ± 0.03 | <46 | 3 | N | N | … | … | … | … | … | … | n | G |
| 37243 | 08413234-5321557 | 16.4 ± 0.2 | 5200 ± 119 | 3.43 ± 0.10 | 1.009 ± 0.006 | 0.01 ± 0.02 | <31 | 3 | Y | N | Y | Y | N | Y | … | … | n | NG? |
| 37244 | 08413335-5339049 | 22.6 ± 0.2 | 4707 ± 46 | 2.52 ± 0.03 | 1.019 ± 0.005 | -0.01 ± 0.03 | 44 ± 5 | 1 | N | N | … | … | … | … | … | … | n | G |
| 37245 | 08413388-5247453 | 24.2 ± 0.2 | 4949 ± 176 | … | 1.018 ± 0.002 | -0.07 ± 0.12 | <22 | 3 | N | … | … | … | … | … | … | … | n | G |
| 37246 | 08413413-5332464 | 114.6 ± 0.2 | 4803 ± 208 | … | 1.015 ± 0.003 | -0.19 ± 0.26 | <20 | 3 | N | … | … | … | … | … | … | … | n | G |
| 37247 | 08413585-5300063 | 12.7 ± 0.2 | 4668 ± 79 | 2.45 ± 0.11 | 1.024 ± 0.004 | 0.08 ± 0.02 | <40 | 3 | N | N | … | … | … | … | … | … | n | G |
| 37248 | 08413595-5309268 | 471.5 ± 1.3 | 3165 ± 163 | 4.79 ± 0.01 | 0.836 ± 0.012 | -0.24 ± 0.14 | <100 | 3 | Y | N | N | Y | Y | Y | … | Y | n | … |
| 37249 | 08413967-5259340 | 15.9 ± 0.2 | 4406 ± 482 | … | 0.913 ± 0.004 | -0.02 ± 0.08 | 118 ± 4 | 1 | Y | … | Y | Y | Y | Y | Y | … | Y | … |
| 37250 | 08413988-5314076 | 30.8 ± 0.2 | 5157 ± 3 | … | 1.011 ± 0.004 | -0.04 ± 0.10 | <34 | 3 | N | … | … | … | … | … | … | … | n | G |
| 37251 | 08414179-5331128 | 25.2 ± 0.2 | 5050 ± 9 | 3.34 ± 0.11 | 1.001 ± 0.002 | 0.06 ± 0.01 | <29 | 3 | Y | N | N | Y | N | Y | … | … | n | NG? |
| 37252 | 08414208-5241281 | 51.5 ± 0.2 | 5153 ± 100 | … | 1.016 ± 0.004 | -0.16 ± 0.17 | <24 | 3 | N | … | … | … | … | … | … | … | n | G |
| 37253 | 08414278-5321005 | -11.4 ± 0.2 | 5054 ± 98 | … | 1.021 ± 0.004 | -0.10 ± 0.12 | <22 | 3 | N | … | … | … | … | … | … | … | n | G |
| 37254 | 08414395-5314068 | 13.6 ± 0.3 | 3391 ± 138 | 4.67 ± 0.09 | 0.848 ± 0.010 | -0.25 ± 0.13 | <100 | 3 | Y | Y | Y | Y | Y | Y | Y | … | Y | … |
| 37255 | 08414532-5315424 | 49.3 ± 0.2 | 4772 ± 224 | … | 1.018 ± 0.003 | -0.20 ± 0.17 | <25 | 3 | N | … | … | … | … | … | … | … | n | G |
| 37256 | 08414828-5331558 | 16.6 ± 0.2 | 5075 ± 114 | … | 1.011 ± 0.002 | -0.05 ± 0.09 | 24 ± 10 | 1 | N | … | … | … | … | … | … | … | n | G |
| 37257 | 08415154-5320597 | 14.6 ± 0.2 | 3578 ± 60 | … | 0.822 ± 0.004 | -0.22 ± 0.14 | <100 | 3 | Y | … | Y | Y | Y | Y | Y | Y | Y | … |
| 37258 | 08415196-5325195 | -6.4 ± 0.2 | 4855 ± 53 | 2.68 ± 0.07 | 1.019 ± 0.003 | 0.02 ± 0.01 | <32 | 3 | N | N | … | … | … | … | … | … | n | G |
| 37259 | 08415197-5318197 | 33.4 ± 0.2 | 5027 ± 182 | … | 1.025 ± 0.003 | -0.06 ± 0.10 | 100 ± 3 | 1 | N | … | … | … | … | … | … | … | n | Li-rich G |
| 37300 | 08425161-5324277 | 0.2 ± 0.2 | 4766 ± 32 | 2.80 ± 0.19 | 1.003 ± 0.003 | -0.03 ± 0.05 | <44 | 3 | Y | N | N | Y | N | Y | … | … | n | NG? |
| 37301 | 08425286-5247441 | 44.7 ± 0.2 | 4297 ± 238 | 4.50 ± 0.02 | 0.890 ± 0.005 | -0.18 ± 0.01 | <8 | 3 | Y | Y | N | N | Y | Y | … | … | n | … |
| 37302 | 08425386-5251207 | 45.4 ± 0.2 | 4799 ± 42 | 2.63 ± 0.07 | 1.013 ± 0.005 | -0.03 ± 0.08 | <37 | 3 | N | N | … | … | … | … | … | … | n | G |

**Table C.5.** continued.

| ID | CNAME | RV (km s$^{-1}$) | $T_{\text{eff}}$ (K) | $\log g$ (dex) | $\gamma^a$ | [Fe/H] (dex) | EW(Li)$^b$ (mÅ) | EW(Li) error flag$^c$ | $\gamma$ | $\log g$ | RV | Li | H$\alpha$ | [Fe/H] | Randich$^d$ | Cantat-Gaudin$^d$ | Final$^e$ | NMs with Li$^f$ |
|---|---|---|---|---|---|---|---|---|---|---|---|---|---|---|---|---|---|---|
| 37303 | 08425529-5252233 | 8.1 ± 0.2 | 4565 ± 114 | … | 1.024 ± 0.003 | 0.01 ± 0.08 | <40 | 3 | N | … | … | … | … | … | … | … | n | G |
| 37304 | 08425572-5320349 | 0.6 ± 0.2 | 5054 ± 89 | … | 1.019 ± 0.002 | -0.04 ± 0.09 | <29 | 3 | N | … | … | … | … | … | … | … | n | G |
| 37305 | 08425631-5259002 | 48.4 ± 0.2 | 5106 ± 53 | 3.44 ± 0.09 | 1.003 ± 0.005 | -0.06 ± 0.10 | <38 | 3 | Y | N | N | N | N | Y | … | … | n | NG? |
| 37306 | 08425825-5302233 | 46.7 ± 0.3 | 4677 ± 50 | 2.48 ± 0.14 | 1.016 ± 0.007 | -0.11 ± 0.08 | <37 | 3 | N | N | … | … | … | … | … | … | n | G |
| 37307 | 08430093-5239058 | 81.7 ± 0.2 | 4691 ± 31 | 2.84 ± 0.16 | 1.002 ± 0.005 | 0.03 ± 0.01 | <47 | 3 | Y | N | N | Y | N | Y | … | … | n | NG? |
| 37308 | 08430354-5235549 | -3.1 ± 0.2 | 5057 ± 102 | … | 1.012 ± 0.003 | -0.10 ± 0.14 | 8 ± 3 | 1 | N | … | … | … | … | … | … | … | n | G |
| 37309 | 08430372-5324148 | 0.1 ± 0.2 | 4594 ± 72 | 2.43 ± 0.12 | 1.019 ± 0.003 | 0.07 ± 0.01 | <41 | 3 | N | N | … | … | … | … | … | … | n | G |
| 37310 | 08430821-5244002 | -32.6 ± 0.2 | 4847 ± 191 | … | 1.018 ± 0.002 | -0.08 ± 0.14 | 22 | 3 | N | … | … | … | … | … | … | … | n | G |
| 37311 | 08431030-5324577 | 30.4 ± 0.2 | 4859 ± 147 | 2.63 ± 0.19 | 1.024 ± 0.002 | -0.06 ± 0.09 | <24 | 3 | N | N | … | … | … | … | … | … | n | G |
| 37312 | 08431049-5245204 | 25.9 ± 0.2 | 5097 ± 88 | … | 1.014 ± 0.004 | -0.05 ± 0.11 | 29 ± 3 | 1 | N | … | … | … | … | … | … | … | n | G |
| 37313 | 08431058-5250134 | 14.8 ± 0.2 | 3564 ± 57 | … | 0.829 ± 0.004 | -0.23 ± 0.14 | <100 | 3 | Y | … | Y | Y | Y | Y | Y | Y | Y | … |
| 37314 | 08431303-5329082 | 22.6 ± 0.3 | 3400 ± 130 | … | 0.866 ± 0.009 | -0.27 ± 0.13 | <100 | 3 | Y | … | N | Y | Y | N | N | … | n | NG |
| 37315 | 08431362-5324105 | 14.3 ± 0.2 | 4616 ± 7 | 2.44 ± 0.06 | 1.020 ± 0.004 | 0.02 ± 0.08 | 38 ± 3 | 1 | N | N | … | … | … | … | … | … | n | G |
| 37316 | 08431537-5251231 | 25.2 ± 0.2 | 5107 ± 14 | … | 1.015 ± 0.003 | 0.01 ± 0.02 | <26 | 3 | N | … | … | … | … | … | … | … | n | G |
| 37317 | 08431548-5238493 | 15.2 ± 0.4 | 3380 ± 134 | 4.84 ± 0.13 | 0.824 ± 0.011 | -0.26 ± 0.13 | … | … | Y | Y | Y | … | Y | Y | Y | Y | n | … |
| 37318 | 08432115-5304048 | 47.9 ± 0.2 | 4766 ± 10 | 2.87 ± 0.19 | 1.003 ± 0.005 | 0.02 ± 0.09 | <45 | 3 | Y | N | N | Y | N | Y | … | … | n | NG? |
| 37319 | 08432355-5301079 | -6.0 ± 0.2 | 4765 ± 44 | 2.63 ± 0.03 | 1.017 ± 0.005 | -0.01 ± 0.04 | <33 | 3 | N | N | … | … | … | … | … | … | n | G |
| 37320 | 08432464-5303553 | 51.2 ± 0.3 | 4676 ± 53 | 2.49 ± 0.13 | 1.023 ± 0.010 | -0.02 ± 0.04 | 14 ± 7 | 1 | N | N | … | … | … | … | … | … | n | G |
| 37100 | 08393005-5310243 | 12.4 ± 0.2 | 5035 ± 85 | … | 1.014 ± 0.002 | -0.02 ± 0.05 | <26 | 3 | N | … | … | … | … | … | … | … | n | G |
| 37101 | 08393041-5319416 | 56.5 ± 0.2 | 4927 ± 104 | … | 1.018 ± 0.004 | -0.10 ± 0.10 | <16 | 3 | N | … | … | … | … | … | … | … | n | G |
| 37102 | 08393041-5350097 | 51.2 ± 0.2 | 4851 ± 92 | 2.89 ± 0.11 | 1.004 ± 0.002 | -0.08 ± 0.02 | <22 | 3 | Y | N | N | N | N | Y | … | … | n | NG? |
| 37263 | 08415760-5251531 | 18.1 ± 0.5 | 3798 ± 28 | … | 0.852 ± 0.018 | … | … | … | Y | … | Y | … | N | … | Y | … | n | … |
| 37103 | 08393049-5348377 | 48.6 ± 0.2 | 4732 ± 78 | 2.59 ± 0.19 | 1.021 ± 0.004 | -0.07 ± 0.01 | <29 | 3 | N | N | … | … | … | … | … | … | n | G |
| 37264 | 08415898-5312367 | 15.0 ± 0.3 | … | … | … | … | … | … | … | … | … | … | … | … | … | … | n | … |
| 37104 | 08393083-5237500 | 56.5 ± 0.2 | 4665 ± 47 | 2.46 ± 0.04 | 1.015 ± 0.003 | -0.01 ± 0.05 | 23 ± 3 | 1 | N | N | … | … | … | … | … | … | n | G |
| 37265 | 08415997-5315298 | 73.8 ± 0.2 | 4650 ± 90 | 2.50 ± 0.16 | 1.014 ± 0.003 | -0.03 ± 0.08 | <29 | 3 | N | N | … | … | … | … | … | … | n | G |
| 37266 | 08420160-5321234 | 14.1 ± 0.2 | 5043 ± 126 | … | 1.013 ± 0.003 | -0.03 ± 0.06 | 23 ± 4 | 1 | N | … | … | … | … | … | … | … | n | G |
| 37267 | 08420212-5252167 | 35.3 ± 0.3 | 5150 ± 12 | … | 1.033 ± 0.007 | -0.08 ± 0.13 | <11 | 3 | N | … | … | … | … | … | … | … | n | G |
| 37105 | 08393196-5320524 | 2.9 ± 0.2 | 4278 ± 250 | 4.52 ± 0.01 | 0.884 ± 0.004 | -0.12 ± 0.08 | 31 ± 3 | 1 | Y | Y | N | Y | Y | Y | … | … | Y | … |
| 37268 | 08420262-5239511 | 49.4 ± 0.2 | 4515 ± 176 | … | 1.015 ± 0.004 | 0.12 ± 0.05 | <47 | 3 | N | … | … | … | … | … | … | … | n | G |
| 37106 | 08393202-5301440 | 34.4 ± 0.2 | 4830 ± 35 | 2.63 ± 0.04 | 1.016 ± 0.001 | 0.01 ± 0.08 | <24 | 3 | N | N | … | … | … | … | … | … | n | G |
| 37269 | 08420373-5259491 | -11.1 ± 0.2 | 5153 ± 106 | … | 1.014 ± 0.003 | … | <11 | 3 | N | … | … | … | … | … | … | … | n | G |
| 37107 | 08393234-5258030 | 80.3 ± 0.2 | 4662 ± 78 | 2.72 ± 0.09 | 1.003 ± 0.004 | -0.04 ± 0.02 | <36 | 3 | Y | N | N | Y | N | Y | … | … | n | NG? |
| 37270 | 08420463-5232265 | 466.3 ± 3.1 | … | … | … | … | … | … | … | … | … | … | … | … | … | … | n | … |
| 37271 | 08420492-5253539 | 14.3 ± 0.3 | 3546 ± 31 | … | 0.825 ± 0.015 | -0.24 ± 0.14 | … | … | Y | … | Y | … | Y | Y | … | Y | n | … |
| 2508 | 08420561-5239291 | 16.2 ± 0.6 | 5061 ± 24 | 3.19 ± 0.03 | … | -0.02 ± 0.01 | <12 | 3 | … | N | … | … | … | … | … | … | n | … |
| 37108 | 08393322-5323324 | 54.9 ± 0.2 | 4967 ± 202 | … | 1.014 ± 0.004 | -0.13 ± 0.18 | <27 | 3 | N | … | … | … | … | … | … | … | n | G |
| 37272 | 08420803-5313061 | 72.9 ± 0.2 | 4692 ± 9 | 2.41 ± 0.08 | 1.028 ± 0.004 | -0.04 ± 0.09 | 172 ± 4 | 1 | N | N | … | … | … | … | … | … | n | Li-rich G |
| 37109 | 08393339-5228168 | 48.0 ± 0.2 | 4832 ± 75 | 2.61 ± 0.11 | 1.019 ± 0.003 | -0.07 ± 0.07 | <33 | 3 | N | N | … | … | … | … | … | … | n | G |
| 37273 | 08420879-5244496 | 29.9 ± 0.2 | 3667 ± 31 | … | 0.847 ± 0.003 | -0.19 ± 0.13 | <100 | 3 | Y | … | N | Y | Y | Y | N | Y | n | NG |
| 37110 | 08393355-5242415 | 50.2 ± 0.2 | 4981 ± 175 | … | 1.012 ± 0.003 | -0.11 ± 0.17 | <16 | 3 | N | … | … | … | … | … | … | … | n | G |
| 37111 | 08393362-5249381 | 46.7 ± 0.2 | 4641 ± 87 | … | 1.019 ± 0.004 | … | <49 | 3 | N | … | … | … | … | … | … | … | n | G |
| 2510 | 08421230-5306038 | 10.8 ± 0.4 | 6239 ± 163 | 3.98 ± 0.16 | … | -0.09 ± 0.15 | 92 ± 19 | 2 | … | Y | Y | Y | Y | Y | … | … | Y | … |
| 37274 | 08421272-5317203 | 8.9 ± 0.2 | 5020 ± 53 | … | 1.022 ± 0.005 | -0.03 ± 0.06 | 11 ± 6 | 1 | N | … | … | … | … | … | … | … | n | G |
| 37275 | 08421357-5300004 | 39.3 ± 0.2 | 4631 ± 59 | 2.51 ± 0.05 | 1.011 ± 0.006 | 0.01 ± 0.10 | <53 | 3 | N | N | … | … | … | … | … | … | n | G |
| 37276 | 08421638-5246234 | 41.2 ± 0.2 | 4707 ± 18 | 2.51 ± 0.04 | 1.015 ± 0.004 | -0.02 ± 0.01 | <38 | 3 | N | N | … | … | … | … | … | … | n | G |
| 37277 | 08421665-5247042 | 28.5 ± 0.2 | 4659 ± 27 | 2.46 ± 0.10 | 1.020 ± 0.002 | -0.01 ± 0.03 | <33 | 3 | N | N | … | … | … | … | … | … | n | G |
| 2511 | 08421690-5240336 | 26.2 ± 0.6 | 4920 ± 16 | 2.39 ± 0.06 | … | -0.26 ± 0.01 | <11 | 3 | … | N | … | … | … | … | … | … | n | … |
| 37125 | 08394454-5321441 | 58.5 ± 0.2 | 4655 ± 59 | 2.42 ± 0.06 | 1.018 ± 0.004 | -0.01 ± 0.03 | <36 | 3 | N | N | … | … | … | … | … | … | n | G |
| 37278 | 08421776-5238242 | 23.3 ± 0.2 | 4813 ± 118 | 2.62 ± 0.13 | 1.016 ± 0.003 | -0.02 ± 0.07 | <38 | 3 | N | N | … | … | … | … | … | … | n | G |
| 37126 | 08394459-5337538 | 29.2 ± 0.3 | 4588 ± 329 | 4.55 ± 0.02 | 0.917 ± 0.011 | -0.14 ± 0.01 | <25 | 3 | Y | Y | N | Y | Y | Y | … | … | Y | … |
| 37324 | 08433277-5247541 | 33.2 ± 0.2 | 4745 ± 70 | 2.59 ± 0.09 | 1.015 ± 0.002 | 0.03 ± 0.04 | <47 | 3 | N | N | … | … | … | … | … | … | n | G |
| 36993 | 08364253-5312255 | 52.9 ± 0.2 | 4578 ± 107 | 2.46 ± 0.16 | 1.012 ± 0.003 | 0.09 ± 0.06 | <50 | 3 | N | N | … | … | … | … | … | … | n | G |
| 37325 | 08433712-5247063 | -129.3 ± 4.0 | … | … | … | … | … | … | … | … | … | … | … | … | … | … | n | … |
| 36994 | 08364271-5307586 | -11.3 ± 1.4 | … | … | … | … | … | … | … | … | … | … | … | … | … | … | n | … |
| 37127 | 08394524-5311441 | 26.0 ± 0.2 | 4980 ± 124 | … | 1.024 ± 0.005 | -0.07 ± 0.13 | <39 | 3 | N | … | … | … | … | … | … | … | n | G |
| 37326 | 08433841-5250551 | 14.1 ± 0.5 | 3391 ± 145 | 4.65 ± 0.12 | 0.852 ± 0.014 | -0.25 ± 0.13 | … | … | Y | Y | Y | … | Y | Y | Y | Y | n | … |
| 36995 | 08364355-5320436 | 21.0 ± 0.2 | 4872 ± 28 | … | 0.996 ± 0.004 | 0.05 ± 0.08 | <30 | 3 | Y | … | Y | Y | N | Y | … | … | n | NG |
| 37327 | 08433979-5258309 | 7.4 ± 0.2 | 4538 ± 191 | … | 0.874 ± 0.003 | … | 53 ± 3 | 1 | Y | … | N | Y | Y | … | … | … | Y | … |

| ID | CNAME | RV (km s$^{-1}$) | $T_{\rm eff}$ (K) | logg (dex) | $\gamma^a$ | [Fe/H] (dex) | EW(Li)$^b$ (mÅ) | EW(Li) error flag$^c$ | \multicolumn{6}{c|}{Membership} | \multicolumn{2}{c|}{Gaia studies} | Final$^e$ | NMs with Li$^f$ |
| | | | | | | | | | $\gamma$ | logg | RV | Li | H$\alpha$ | [Fe/H] | Randich$^d$ | Cantat-Gaudin$^d$ | | |
|---|---|---|---|---|---|---|---|---|---|---|---|---|---|---|---|---|---|---|
| 36996 | 08364479-5306491 | 57.4 ± 0.2 | 4875 ± 40 | 2.90 ± 0.06 | 1.009 ± 0.005 | -0.05 ± 0.06 | <25 | 3 | Y | N | N | N | N | Y | … | … | n | NG? |
| 37328 | 08433989-5301596 | 26.5 ± 0.2 | 4757 ± 53 | … | 1.021 ± 0.005 | -0.02 ± 0.04 | <31 | 3 | N | … | … | … | … | … | … | … | n | G |
| 36997 | 08364573-5311329 | 15.4 ± 0.6 | 3263 ± 13 | … | 0.878 ± 0.017 | … | 106 ± 17 | 1 | Y | … | Y | Y | Y | … | Y | Y | Y | … |
| 37329 | 08434131-5238566 | 64.1 ± 0.3 | 4940 ± 2 | 2.87 ± 0.20 | 1.022 ± 0.006 | -0.09 ± 0.02 | <24 | 3 | N | N | … | … | … | … | … | … | n | G |
| 36999 | 08365118-5304510 | 12.5 ± 0.2 | 4558 ± 122 | … | 1.020 ± 0.003 | 0.11 ± 0.02 | 53 ± 8 | 1 | N | … | … | … | … | … | … | … | n | G |
| 37330 | 08434376-5301221 | 82.9 ± 0.2 | 4847 ± 86 | 2.69 ± 0.04 | 1.019 ± 0.006 | -0.08 ± 0.01 | <31 | 3 | N | N | … | … | … | … | … | … | n | G |
| 2482 | 08365498-5308342 | 15.2 ± 0.6 | 5379 ± 49 | 4.46 ± 0.05 | … | -0.08 ± 0.01 | 205 ± 4 | 2 | … | Y | Y | Y | Y | Y | Y | Y | Y | … |
| 37331 | 08434636-5257031 | 49.0 ± 0.2 | 4578 ± 89 | 2.32 ± 0.10 | 1.014 ± 0.003 | -0.09 ± 0.03 | <40 | 3 | N | N | … | … | … | … | … | … | n | G |
| 37000 | 08370869-5309423 | 61.2 ± 0.2 | 4968 ± 142 | … | 1.011 ± 0.004 | -0.22 ± 0.13 | 27 ± 7 | 1 | N | … | … | … | … | … | … | … | n | G |
| 37332 | 08434694-5259351 | 29.2 ± 0.2 | 4815 ± 119 | 2.62 ± 0.15 | 1.018 ± 0.005 | -0.03 ± 0.09 | <17 | 3 | N | N | … | … | … | … | … | … | n | G |
| 37333 | 08434941-5259502 | 59.8 ± 0.2 | 4939 ± 93 | 2.86 ± 0.20 | 1.016 ± 0.004 | -0.08 ± 0.09 | <37 | 3 | N | N | … | … | … | … | … | … | n | G |
| 37001 | 08371709-5305201 | 56.9 ± 0.2 | 4855 ± 91 | … | 1.017 ± 0.004 | -0.08 ± 0.07 | <20 | 3 | N | … | … | … | … | … | … | … | n | G |
| 37334 | 08435119-5256089 | 64.5 ± 0.2 | 4873 ± 135 | … | 1.022 ± 0.005 | -0.17 ± 0.14 | 67 ± 8 | 1 | N | … | … | … | … | … | … | … | n | G |
| 37002 | 08372022-5311224 | 5.7 ± 0.2 | 4935 ± 170 | 3.33 ± 0.16 | 0.995 ± 0.004 | 0.01 ± 0.03 | <25 | 3 | Y | N | N | N | N | Y | … | … | n | NG? |
| 37335 | 08435167-5242455 | 4.6 ± 0.2 | 4672 ± 73 | 2.56 ± 0.20 | 1.014 ± 0.002 | -0.10 ± 0.11 | <22 | 3 | N | N | … | … | … | … | … | … | n | G |
| 37003 | 08372149-5308312 | -12.8 ± 0.2 | 4977 ± 153 | … | 1.018 ± 0.004 | -0.10 ± 0.08 | <25 | 3 | N | … | … | … | … | … | … | … | n | G |
| 2516 | 08435595-5241541 | 54.8 ± 0.6 | 4714 ± 18 | 2.82 ± 0.10 | 1.014 ± 0.002 | 0.04 ± 0.01 | <39 | 3 | N | N | … | … | … | … | … | … | n | … |
| 37004 | 08372460-5310028 | 26.4 ± 0.2 | 4832 ± 86 | 2.45 ± 0.06 | 1.015 ± 0.003 | -0.13 ± 0.01 | <25 | 3 | N | N | … | … | … | … | … | … | n | G |
| 37336 | 08440210-5256239 | 40.2 ± 0.2 | 5156 ± 82 | … | 1.018 ± 0.003 | -0.05 ± 0.08 | 23 ± 4 | 1 | N | … | … | … | … | … | … | … | n | G |
| 37005 | 08372611-5307006 | 32.4 ± 2.7 | … | … | … | … | … | … | N | … | … | … | … | … | … | … | n | … |
| 37132 | 08395196-5329155 | 5.8 ± 0.2 | 4658 ± 33 | 2.46 ± 0.12 | 1.017 ± 0.003 | -0.04 ± 0.08 | <26 | 3 | N | N | … | … | … | … | … | … | n | G |
| 2517 | 08440521-5253171 | 14.1 ± 0.6 | 5819 ± 36 | 4.32 ± 0.09 | … | 0.02 ± 0.06 | 177 ± 9 | 2 | … | Y | Y | Y | Y | Y | Y | … | Y | … |
| 37133 | 08395199-5305526 | 22.5 ± 0.2 | 4622 ± 63 | 2.52 ± 0.20 | 1.028 ± 0.006 | 0.13 ± 0.03 | <44 | 3 | N | N | … | … | … | … | … | … | n | G |
| 37006 | 08372675-5305321 | 83.6 ± 0.2 | 4787 ± 13 | 2.62 ± 0.13 | 1.017 ± 0.004 | -0.17 ± 0.06 | <41 | 3 | N | N | … | … | … | … | … | … | n | G |
| 37337 | 08441732-5302586 | 6.6 ± 0.2 | 4917 ± 224 | … | 1.018 ± 0.002 | -0.12 ± 0.20 | <19 | 3 | N | … | … | … | … | … | … | … | n | G |
| 37007 | 08372955-5304084 | 91.2 ± 0.2 | 4698 ± 124 | … | 1.018 ± 0.005 | -0.18 ± 0.09 | <35 | 3 | N | … | … | … | … | … | … | … | n | G |
| 37338 | 08441984-5259424 | 17.1 ± 0.2 | 4966 ± 187 | … | 1.019 ± 0.003 | -0.09 ± 0.17 | <17 | 3 | N | … | … | … | … | … | … | … | n | G |
| 37008 | 08373280-5303334 | 16.7 ± 0.4 | 3306 ± 16 | … | 0.868 ± 0.018 | … | … | … | Y | … | Y | … | Y | … | Y | … | … | … |
| 37339 | 08442238-5248588 | 3.1 ± 0.2 | 4887 ± 299 | … | 1.016 ± 0.003 | -0.11 ± 0.22 | 50 ± 8 | 1 | N | … | … | … | … | … | … | … | n | G |
| 37009 | 08373689-5305481 | 51.6 ± 0.2 | 4823 ± 31 | 2.62 ± 0.02 | 1.017 ± 0.004 | -0.04 ± 0.08 | <34 | 3 | N | N | … | … | … | … | … | … | n | G |
| 37340 | 08442273-5247262 | 2.1 ± 0.2 | 5073 ± 75 | … | 1.022 ± 0.004 | -0.06 ± 0.08 | <22 | 3 | N | … | … | … | … | … | … | … | n | G |
| 37010 | 08375033-5306249 | 62.6 ± 0.2 | 4695 ± 72 | 2.48 ± 0.06 | 1.015 ± 0.003 | 0.00 ± 0.04 | 20 ± 4 | 1 | N | N | … | … | … | … | … | … | n | G |
| 37341 | 08442554-5246494 | 17.6 ± 0.2 | 4975 ± 72 | 2.79 ± 0.08 | 1.018 ± 0.003 | -0.02 ± 0.04 | <25 | 3 | N | N | … | … | … | … | … | … | n | G |
| 37011 | 08375621-5311032 | 20.3 ± 0.2 | 4827 ± 139 | 2.39 ± 0.06 | 1.015 ± 0.005 | -0.10 ± 0.08 | <17 | 3 | N | N | … | … | … | … | … | … | n | G |
| 37342 | 08442762-5253431 | 22.1 ± 0.2 | 5093 ± 184 | … | 1.011 ± 0.003 | -0.22 ± 0.18 | 12 ± 2 | 1 | N | … | … | … | … | … | … | … | n | G |
| 37012 | 08375639-5310052 | 19.2 ± 0.2 | 4983 ± 137 | 2.72 ± 0.15 | 1.016 ± 0.003 | -0.35 ± 0.10 | <14 | 3 | N | N | … | … | … | … | … | … | n | G |
| 37013 | 08375991-5258388 | 8.7 ± 0.2 | 5007 ± 188 | … | 1.022 ± 0.004 | -0.10 ± 0.19 | <15 | 3 | N | … | … | … | … | … | … | … | n | G |
| 37343 | 08442842-5301478 | 73.4 ± 0.2 | 4841 ± 135 | 2.64 ± 0.13 | 1.013 ± 0.006 | -0.11 ± 0.06 | <9 | 3 | N | N | … | … | … | … | … | … | n | G |
| 37014 | 08380503-5224195 | 47.1 ± 0.2 | 4872 ± 57 | 2.63 ± 0.10 | 1.025 ± 0.004 | -0.12 ± 0.08 | 22 ± 6 | 1 | N | N | … | … | … | … | … | … | n | G |
| 37344 | 08443396-5246277 | 35.9 ± 0.2 | 4870 ± 40 | 2.72 ± 0.03 | 1.012 ± 0.003 | 0.00 ± 0.05 | 56 ± 2 | 1 | N | N | … | … | … | … | … | … | n | G |
| 37038 | 08383696-5252424 | 9.1 ± 4.3 | … | … | … | … | … | … | … | … | … | … | … | … | Y | … | n | … |
| 2518 | 08443402-5244022 | 16.7 ± 0.6 | 4637 ± 420 | 4.26 ± 0.56 | 0.921 ± 0.005 | -0.06 ± 0.05 | 40 ± 1 | 1 | Y | Y | Y | Y | Y | Y | … | … | Y | … |
| 37039 | 08383903-5316000 | 59.4 ± 0.3 | 5076 ± 90 | … | 1.004 ± 0.006 | -0.16 ± 0.16 | <48 | 3 | Y | … | N | N | N | Y | … | … | n | NG? |
| 37134 | 08395367-5318036 | 14.5 ± 0.2 | 3837 ± 97 | … | 0.833 ± 0.003 | -0.13 ± 0.13 | 67 ± 25 | 1 | Y | … | Y | Y | Y | Y | Y | Y | Y | … |
| 37040 | 08384081-5331283 | 7.5 ± 0.2 | 5186 ± 90 | … | 1.022 ± 0.004 | -0.08 ± 0.10 | 26 ± 3 | 1 | N | … | … | … | … | … | … | … | n | G |
| 37041 | 08384138-5255039 | 18.7 ± 0.3 | 3985 ± 275 | … | 0.850 ± 0.013 | -0.16 ± 0.12 | <36 | 3 | Y | … | Y | Y | Y | Y | … | … | Y | … |
| 37135 | 08395378-5339075 | 30.1 ± 0.2 | 5141 ± 94 | … | 1.012 ± 0.004 | … | 12 ± 4 | 1 | N | … | … | … | … | … | … | … | n | G |
| 37042 | 08384269-5305160 | 15.9 ± 0.2 | 6060 ± 30 | 4.21 ± 0.06 | 0.993 ± 0.002 | 0.16 ± 0.06 | 25 ± 4 | 1 | Y | Y | Y | N | N | N | Y | … | n | NG |
| 37043 | 08384401-5322511 | 28.3 ± 1.3 | … | … | … | … | … | … | … | … | … | … | … | … | N | … | n | … |
| 37044 | 08384465-5315117 | -16.5 ± 0.2 | 5330 ± 103 | … | 1.017 ± 0.003 | -0.39 ± 0.01 | <4 | 3 | N | … | … | … | … | … | … | … | n | … |
| 37045 | 08384653-5315458 | 96.1 ± 0.2 | 4655 ± 77 | 2.52 ± 0.19 | 1.022 ± 0.003 | 0.01 ± 0.02 | 21 ± 4 | 1 | N | N | … | … | … | … | … | … | n | G |
| 37046 | 08384657-5319114 | 43.7 ± 0.2 | 5066 ± 107 | … | 1.013 ± 0.003 | -0.06 ± 0.09 | 27 ± 4 | 1 | N | … | … | … | … | … | … | … | n | G |
| 37047 | 08384955-5215080 | 39.2 ± 0.2 | 5117 ± 130 | … | 1.017 ± 0.005 | … | <25 | 3 | N | … | … | … | … | … | … | … | n | G |
| 37136 | 08395474-5336040 | 7.8 ± 0.2 | 5106 ± 75 | … | 1.015 ± 0.003 | -0.06 ± 0.13 | <17 | 3 | N | … | … | … | … | … | … | … | n | G |
| 37048 | 08384961-5344092 | 25.5 ± 0.2 | 4891 ± 182 | … | 1.025 ± 0.004 | -0.10 ± 0.16 | 17 ± 3 | 1 | N | … | … | … | … | … | … | … | n | G |
| 37049 | 08384981-5255460 | 60.0 ± 0.2 | 4534 ± 151 | 2.43 ± 0.10 | 1.009 ± 0.006 | 0.09 ± 0.05 | <51 | 3 | Y | N | N | Y | N | Y | … | … | n | NG? |
| 2486 | 08385073-5225382 | 33.1 ± 0.6 | 7047 ± 183 | 4.06 ± 0.19 | … | -0.08 ± 0.13 | <4 | 3 | … | Y | N | N | N | Y | N | … | n | … |
| 37050 | 08385117-5232285 | 52.7 ± 0.2 | 4981 ± 88 | 3.00 ± 0.13 | 1.010 ± 0.003 | -0.07 ± 0.06 | 42 ± 11 | 1 | Y | N | N | N | N | Y | … | … | n | NG? |
| 37051 | 08385149-5321086 | 25.6 ± 0.2 | 4845 ± 32 | 2.60 ± 0.07 | 1.014 ± 0.004 | -0.07 ± 0.20 | 36 ± 3 | 1 | N | N | … | … | … | … | … | … | n | G |

**Table C.5.** continued.

| ID | CNAME | RV (km s$^{-1}$) | $T_{\text{eff}}$ (K) | $logg$ (dex) | $\gamma^a$ | [Fe/H] (dex) | $EW(\text{Li})^b$ (mÅ) | $EW(\text{Li})$ error flag$^c$ | $\gamma$ | $logg$ | Membership RV | Li | H$\alpha$ | [Fe/H] | Gaia studies Randich$^d$ | Cantat-Gaudin$^d$ | Final$^e$ | NMs with Li$^f$ |
|---|---|---|---|---|---|---|---|---|---|---|---|---|---|---|---|---|---|---|
| 37052 | 08385249-5215400 | -8.3 ± 0.2 | 4969 ± 123 | … | 1.015 ± 0.003 | -0.08 ± 0.12 | 17 ± 6 | 1 | N | … | … | … | … | … | … | … | n | G |
| 37053 | 08385343-5320250 | 19.7 ± 0.2 | 5058 ± 140 | … | 1.019 ± 0.003 | -0.09 ± 0.17 | 16 ± 3 | 1 | N | … | … | … | … | … | … | … | n | G |
| 2487 | 08385545-5222391 | 1.6 ± 0.6 | 4702 ± 19 | 2.53 ± 0.06 | 1.022 ± 0.001 | 0.04 ± 0.04 | 32 ± 6 | 1 | N | N | … | … | … | … | … | … | n | G |
| 2488 | 08385566-5257516 | 44.7 ± 0.6 | … | … | … | … | … | … | … | … | … | … | … | … | N | Y | n | … |
| 2499 | 08400303-5351064 | -15.2 ± 0.6 | 7228 ± 291 | 4.06 ± 0.19 | … | -0.14 ± 0.15 | <4 | 3 | … | Y | N | N | … | Y | Y | … | n | … |
| 37054 | 08385666-5315138 | 41.7 ± 0.2 | 4887 ± 131 | 2.65 ± 0.19 | 1.022 ± 0.006 | -0.10 ± 0.11 | <23 | 3 | N | N | … | … | … | … | … | … | n | G |
| 37055 | 08385666-5330275 | 8.6 ± 0.2 | 4986 ± 125 | … | 1.016 ± 0.004 | -0.04 ± 0.07 | <33 | 3 | N | … | … | … | … | … | … | … | n | G |
| 2500 | 08400345-5306273 | 6.7 ± 0.6 | 5025 ± 9 | 2.64 ± 0.05 | 1.019 ± 0.001 | -0.27 ± 0.02 | 21 ± 6 | 1 | N | N | … | … | … | … | … | … | n | G |
| 37056 | 08385882-5222237 | -34.9 ± 0.2 | 4919 ± 86 | … | 1.026 ± 0.003 | -0.01 ± 0.07 | <29 | 3 | N | … | … | … | … | … | … | … | n | G |
| 37146 | 08400357-5330531 | 18.7 ± 0.2 | 5108 ± 176 | … | 1.008 ± 0.002 | -0.29 ± 0.22 | <9 | 3 | Y | … | Y | N | N | N | … | … | n | … |
| 37057 | 08385965-5252425 | 4.6 ± 0.2 | 4906 ± 179 | … | 1.020 ± 0.004 | -0.07 ± 0.10 | <12 | 3 | N | … | … | … | … | … | … | … | n | G |
| 2492 | 08392060-5234475 | 0.2 ± 0.6 | 6322 ± 135 | 3.87 ± 0.12 | … | -0.03 ± 0.12 | <10 | 3 | … | Y | N | N | … | Y | N | … | n | NG |
| 37085 | 08392067-5220224 | 51.2 ± 0.2 | 4737 ± 68 | 2.57 ± 0.18 | 1.017 ± 0.003 | -0.13 ± 0.07 | <29 | 3 | N | N | … | … | … | … | … | … | n | G |
| 37086 | 08392095-5329375 | 52.3 ± 0.2 | 4817 ± 181 | … | 1.013 ± 0.005 | -0.16 ± 0.11 | <27 | 3 | N | … | … | … | … | … | … | … | n | G |
| 37087 | 08392149-5234136 | 51.1 ± 0.2 | 4693 ± 127 | … | 1.021 ± 0.003 | -0.15 ± 0.10 | <29 | 3 | N | … | … | … | … | … | … | … | n | G |
| 37147 | 08400511-5234116 | 76.6 ± 0.2 | 4754 ± 145 | … | 1.008 ± 0.003 | -0.20 ± 0.14 | <22 | 3 | Y | … | N | Y | N | Y | … | … | n | NG? |
| 37088 | 08392258-5355056 | 15.4 ± 0.2 | 5075 ± 170 | … | 0.966 ± 0.002 | … | 230 ± 3 | 1 | Y | … | Y | Y | Y | … | Y | Y | Y | … |
| 37089 | 08392367-5235095 | 14.8 ± 0.2 | 5338 ± 132 | … | 1.014 ± 0.003 | -0.16 ± 0.09 | <22 | 3 | N | … | … | … | … | … | … | … | n | … |
| 37090 | 08392367-5253563 | 55.4 ± 0.3 | 5029 ± 100 | … | 1.021 ± 0.006 | -0.09 ± 0.09 | <37 | 3 | N | … | … | … | … | … | … | … | n | G |
| 37091 | 08392411-5229291 | 70.0 ± 0.2 | 4714 ± 9 | 2.51 ± 0.10 | 1.017 ± 0.003 | -0.10 ± 0.07 | 37 ± 3 | 1 | N | N | … | … | … | … | … | … | n | G |
| 37151 | 08401000-5235114 | 17.6 ± 1.8 | … | … | … | … | … | … | … | … | … | … | … | … | Y | Y | n | … |
| 37099 | 08392921-5228419 | 44.3 ± 0.2 | 4637 ± 84 | 2.37 ± 0.14 | 1.018 ± 0.003 | -0.09 ± 0.09 | <35 | 3 | N | N | … | … | … | … | … | … | n | G |
| 37152 | 08401022-5345021 | 54.9 ± 0.2 | 4612 ± 150 | … | 1.018 ± 0.003 | -0.12 ± 0.11 | <32 | 3 | N | … | … | … | … | … | … | … | n | G |
| 37153 | 08401060-5251512 | 45.9 ± 0.2 | 4909 ± 127 | … | 1.012 ± 0.003 | -0.11 ± 0.12 | <23 | 3 | N | … | … | … | … | … | … | … | n | G |
| 37154 | 08401064-5336239 | 19.1 ± 0.2 | 4894 ± 154 | … | 1.011 ± 0.003 | -0.09 ± 0.14 | <22 | 3 | N | … | … | … | … | … | … | … | n | G |
| 37155 | 08401102-5242524 | 15.0 ± 0.2 | 4727 ± 169 | … | 1.016 ± 0.003 | -0.07 ± 0.13 | <26 | 3 | N | … | … | … | … | … | … | … | n | … |
| 37164 | 08402180-5340155 | 28.7 ± 0.2 | 4972 ± 112 | … | 1.019 ± 0.002 | -0.03 ± 0.05 | <26 | 3 | N | … | … | … | … | … | … | … | n | G |
| 37165 | 08402189-5342053 | 22.4 ± 0.2 | 4935 ± 187 | … | 1.019 ± 0.004 | -0.07 ± 0.11 | <18 | 3 | N | … | … | … | … | … | … | … | n | … |
| 37166 | 08402295-5336456 | 43.3 ± 0.2 | 4866 ± 84 | 2.67 ± 0.12 | 1.018 ± 0.002 | -0.08 ± 0.02 | <21 | 3 | N | N | … | … | … | … | … | … | n | G |
| 37167 | 08402313-5345000 | -3.1 ± 0.2 | 4893 ± 191 | … | 1.019 ± 0.004 | -0.09 ± 0.15 | <28 | 3 | N | … | … | … | … | … | … | … | n | G |
| 37168 | 08402728-5243577 | 40.3 ± 0.2 | 4930 ± 256 | … | 1.009 ± 0.004 | -0.20 ± 0.21 | <16 | 3 | Y | … | N | N | N | Y | … | … | n | NG? |
| 37173 | 08403453-5303058 | 2.3 ± 0.2 | 5013 ± 67 | … | 1.023 ± 0.004 | -0.02 ± 0.04 | 17 ± 3 | 1 | N | … | … | … | … | … | … | … | n | G |
| 37174 | 08403479-5245265 | 17.5 ± 0.2 | 4906 ± 232 | … | 1.021 ± 0.004 | -0.10 ± 0.19 | <21 | 3 | N | … | … | … | … | … | … | … | n | G |
| 37175 | 08403578-5314571 | 76.6 ± 0.2 | 4672 ± 54 | 2.53 ± 0.12 | 1.015 ± 0.004 | 0.00 ± 0.03 | <38 | 3 | N | N | … | … | … | … | … | … | n | G |
| 37176 | 08403748-5307427 | 467.5 ± 0.7 | 3280 ± 255 | 4.78 ± 0.03 | 0.799 ± 0.008 | -0.27 ± 0.13 | <100 | 3 | Y | Y | N | Y | Y | N | … | Y | n | NG |
| 37177 | 08403826-5317417 | 47.8 ± 0.2 | 4632 ± 94 | 2.45 ± 0.05 | 1.010 ± 0.003 | -0.06 ± 0.05 | <33 | 3 | N | N | … | … | … | … | … | … | n | G |
| 37178 | 08403841-5306296 | -14.3 ± 0.2 | 5014 ± 168 | … | 1.019 ± 0.003 | -0.14 ± 0.16 | 18 ± 3 | 1 | N | … | … | … | … | … | … | … | n | G |
| 37179 | 08403847-5329247 | 15.3 ± 0.2 | 4845 ± 78 | 2.64 ± 0.13 | 1.021 ± 0.003 | -0.02 ± 0.07 | <29 | 3 | N | N | … | … | … | … | … | … | n | G |
| 37215 | 08410766-5237039 | 42.5 ± 0.2 | 4905 ± 197 | … | 1.020 ± 0.004 | -0.11 ± 0.13 | <29 | 3 | N | … | … | … | … | … | … | … | n | G |
| 37216 | 08410820-5238056 | 15.3 ± 0.3 | 3531 ± 5 | 4.61 ± 0.20 | 0.820 ± 0.008 | -0.25 ± 0.13 | <100 | 3 | Y | Y | Y | Y | Y | Y | Y | Y | Y | … |
| 37217 | 08410844-5243529 | 47.9 ± 0.2 | 4944 ± 91 | … | 1.025 ± 0.004 | -0.08 ± 0.07 | <32 | 3 | N | … | … | … | … | … | … | … | n | G |
| 37218 | 08410847-5331579 | 30.1 ± 0.2 | 4786 ± 251 | … | 1.021 ± 0.003 | -0.09 ± 0.14 | <17 | 3 | N | … | … | … | … | … | … | … | n | G |
| 37219 | 08410934-5302130 | 15.1 ± 0.6 | 3302 ± 13 | … | 0.871 ± 0.016 | … | … | … | Y | … | Y | … | Y | … | Y | … | n | … |
| 37220 | 08410989-5327366 | 12.1 ± 0.2 | 4831 ± 128 | 2.93 ± 0.16 | 1.001 ± 0.003 | 0.00 ± 0.06 | 45 ± 3 | 1 | Y | N | Y | Y | N | Y | … | … | n | NG? |
| 37221 | 08411053-5239166 | 13.1 ± 0.2 | 5122 ± 27 | … | 1.019 ± 0.003 | -0.01 ± 0.04 | <22 | 3 | N | … | … | … | … | … | … | … | n | G |
| 37222 | 08411411-5316430 | 35.2 ± 0.3 | 5024 ± 190 | … | 1.019 ± 0.006 | -0.16 ± 0.27 | <25 | 3 | N | … | … | … | … | … | … | … | n | G |
| 37223 | 08411437-5312587 | 28.8 ± 0.3 | 5072 ± 120 | 3.54 ± 0.13 | 1.001 ± 0.006 | -0.01 ± 0.05 | <34 | 3 | Y | Y | N | Y | N | Y | … | … | n | NG? |
| 37224 | 08411530-5241534 | 52.2 ± 0.2 | 5169 ± 63 | 3.61 ± 0.04 | 0.999 ± 0.003 | -0.10 ± 0.14 | <12 | 3 | Y | Y | N | N | … | Y | … | … | n | NG |
| 37225 | 08411544-5300144 | 160.1 ± 0.2 | 5113 ± 41 | … | 1.006 ± 0.003 | -0.19 ± 0.26 | <12 | 3 | Y | … | N | N | … | Y | … | … | n | NG? |
| 2505 | 08411617-5304260 | … | … | … | … | … | … | … | … | … | … | … | … | … | … | … | n | … |
| 37226 | 08412078-5331166 | 21.4 ± 0.2 | 5012 ± 120 | … | 1.018 ± 0.003 | -0.06 ± 0.13 | <19 | 3 | N | … | … | … | … | … | … | … | n | G |
| 37227 | 08412223-5304468 | 15.2 ± 0.3 | 3372 ± 25 | 4.63 ± 0.15 | 0.842 ± 0.007 | -0.26 ± 0.13 | 107 ± 27 | 1 | Y | Y | Y | Y | Y | Y | Y | … | Y | … |
| 37228 | 08412305-5301544 | 67.2 ± 0.3 | 4851 ± 12 | 2.60 ± 0.17 | 1.029 ± 0.006 | -0.12 ± 0.07 | <26 | 3 | N | N | … | … | … | … | … | … | n | G |
| 37229 | 08412314-5259579 | 4.6 ± 0.2 | 4692 ± 28 | 2.51 ± 0.08 | 1.017 ± 0.003 | 0.04 ± 0.10 | <39 | 3 | N | N | … | … | … | … | … | … | n | G |
| 37230 | 08412444-5336332 | 73.3 ± 0.2 | 4690 ± 35 | 2.78 ± 0.15 | 1.003 ± 0.003 | 0.06 ± 0.01 | <40 | 3 | Y | N | N | Y | N | Y | … | … | n | NG? |
| 37231 | 08412526-5333573 | 8.5 ± 0.2 | 4421 ± 348 | … | 0.925 ± 0.003 | 0.00 ± 0.08 | 295 ± 2 | 1 | Y | … | N | N | Y | Y | N | … | n | … |
| 2506 | 08412588-5322415 | 20.1 ± 0.4 | 4835 ± 119 | 4.47 ± 0.13 | … | -0.06 ± 0.15 | … | … | … | Y | Y | … | Y | Y | … | Y | n | … |
| 37232 | 08412598-5326349 | 14.4 ± 0.3 | 3332 ± 82 | … | 0.872 ± 0.011 | -0.28 ± 0.14 | … | … | Y | … | Y | … | Y | N | Y | Y | n | … |
| 37233 | 08412642-5320332 | 30.0 ± 0.2 | 4825 ± 182 | … | 1.017 ± 0.004 | -0.09 ± 0.14 | <20 | 3 | N | … | … | … | … | … | … | … | n | G |

**Table C.5.** continued.

| ID | CNAME | $RV$ (km s$^{-1}$) | $T_{\text{eff}}$ (K) | $logg$ (dex) | $\gamma^a$ | [Fe/H] (dex) | $EW$(Li)$^b$ (mÅ) | $EW$(Li) error flag$^c$ | Membership $\gamma$ | Membership $logg$ | Gaia studies RV | Gaia studies Li | Gaia studies H$\alpha$ | Gaia studies [Fe/H] | Gaia studies Randich$^d$ | Gaia studies Cantat-Gaudin$^d$ | Final$^e$ | NMs with Li$^f$ |
|---|---|---|---|---|---|---|---|---|---|---|---|---|---|---|---|---|---|---|
| 37234 | 08412662-5304262 | 13.1 ± 0.2 | 4731 ± 80 | 2.44 ± 0.20 | 1.023 ± 0.005 | -0.09 ± 0.21 | <8 | 3 | N | N | … | … | … | … | … | … | n | G |
| 37235 | 08412722-5239356 | 66.7 ± 0.2 | 4854 ± 106 | 2.62 ± 0.11 | 1.018 ± 0.003 | -0.06 ± 0.07 | <22 | 3 | N | N | … | … | … | … | … | … | n | G |
| 37236 | 08412746-5238258 | 24.7 ± 0.2 | 4976 ± 129 | … | 1.022 ± 0.005 | -0.07 ± 0.14 | <17 | 3 | N | … | … | … | … | … | … | … | n | G |
| 37260 | 08415455-5301214 | 116.9 ± 0.2 | 4681 ± 27 | 2.44 ± 0.05 | 1.017 ± 0.004 | -0.06 ± 0.03 | <28 | 3 | N | N | … | … | … | … | … | … | n | G |
| 37261 | 08415463-5251080 | 17.4 ± 0.3 | 5196 ± 52 | … | 1.023 ± 0.007 | 0.00 ± 0.02 | 28 ± 4 | 1 | N | … | … | … | … | … | … | … | n | G |
| 37262 | 08415507-5237485 | 7.7 ± 0.2 | 4887 ± 8 | 2.80 ± 0.05 | 1.010 ± 0.003 | 0.02 ± 0.02 | 53 ± 2 | 1 | N | N | … | … | … | … | … | … | n | G |

**Notes.** $^{(a)}$ Empirical gravity indicator defined by Damiani et al. (2014). $^{(b)}$ The values of $EW$(Li) for this cluster are corrected (subtracted adjacent Fe (6707.43 Å) line). $^{(c)}$ Flags for the errors of the corrected $EW$(Li) values, as follows: 1=$EW$(Li) corrected by blends contribution using models; 2=$EW$(Li) measured separately (Li line resolved - UVES only); and 3=Upper limit (no error for $EW$(Li) is given). $^{(d)}$ Randich et al. (2018), Cantat-Gaudin et al. (2018) $^{(e)}$ The letters "Y" and "N" indicate if the star is a cluster member or not. $^{(f)}$ 'Li-rich G', 'G' and 'NG' indicate "Li-rich giant", "giant" and "non-giant" Li field contaminants, respectively.

**Table C.6.** IC 2602

| ID | CNAME | RV (km s$^{-1}$) | $T_{\text{eff}}$ (K) | logg (dex) | $\gamma^a$ | [Fe/H] (dex) | EW(Li)$^b$ (mÅ) | EW(Li) error flag$^c$ | Membership $\gamma$ | logg | RV | Li | H$\alpha$ | [Fe/H] | Gaia studies Randich$^d$ | Cantat-Gaudin$^d$ | Final$^e$ | NMs with Li$^f$ |
|---|---|---|---|---|---|---|---|---|---|---|---|---|---|---|---|---|---|---|
| 37609 | 10345408-6408452 | 18.6 ± 0.2 | 4058 ± 185 | 4.45 ± 0.14 | 0.865 ± 0.002 | -0.05 ± 0.13 | 148 ± 12 | 1 | Y | Y | Y | Y | Y | Y | Y | Y | Y | … |
| 37610 | 10345416-6407241 | 48.7 ± 0.3 | 3660 ± 110 | 4.68 ± 0.04 | 0.792 ± 0.006 | -0.24 ± 0.15 | <100 | 3 | Y | Y | N | Y | Y | Y | N | … | n | … |
| 37611 | 10345587-6407385 | 20.1 ± 0.2 | 3472 ± 7 | 4.69 ± 0.04 | 0.807 ± 0.003 | -0.24 ± 0.14 | <100 | 3 | Y | Y | Y | Y | Y | Y | Y | Y | Y | … |
| 37612 | 10350061-6418105 | 3.7 ± 0.2 | 4890 ± 95 | 2.72 ± 0.10 | 1.016 ± 0.004 | -0.01 ± 0.07 | <38 | 3 | N | N | … | … | … | … | … | … | n | G |
| 37628 | 10353536-6420129 | 22.0 ± 0.2 | 4496 ± 198 | … | 1.027 ± 0.002 | 0.10 ± 0.09 | <51 | 3 | N | … | … | … | … | … | … | … | n | G |
| 37629 | 10353774-6339074 | 64.6 ± 0.2 | 4716 ± 92 | … | 1.017 ± 0.003 | -0.19 ± 0.18 | <23 | 3 | N | … | … | … | … | … | … | … | n | G |
| 37639 | 10355387-6403385 | 2.6 ± 0.2 | 4768 ± 178 | 2.55 ± 0.18 | 1.021 ± 0.005 | 0.01 ± 0.08 | <35 | 3 | N | N | … | … | … | … | … | … | n | G |
| 37644 | 10360982-6337353 | -12.8 ± 0.2 | 4651 ± 16 | 2.37 ± 0.07 | 1.028 ± 0.003 | -0.03 ± 0.04 | <32 | 3 | N | N | … | … | … | … | … | … | n | G |
| 2541 | 10361111-6409576 | -16.2 ± 0.6 | 4982 ± 47 | 2.78 ± 0.11 | 1.020 ± 0.003 | -0.05 ± 0.02 | 24 ± 5 | 2 | N | N | … | … | … | … | … | … | n | G |
| 37363 | 10272866-6442045 | 12.2 ± 0.2 | 4792 ± 32 | 2.59 ± 0.02 | 1.014 ± 0.004 | 0.01 ± 0.09 | <26 | 3 | N | N | … | … | … | … | … | … | n | G |
| 37364 | 10273151-6440514 | 0.2 ± 0.2 | 4669 ± 146 | 2.41 ± 0.19 | 1.021 ± 0.003 | -0.09 ± 0.07 | <29 | 3 | N | N | … | … | … | … | … | … | n | G |
| 37645 | 10361230-6416354 | 11.0 ± 0.2 | 4630 ± 102 | … | 1.020 ± 0.004 | … | <57 | 3 | N | … | … | … | … | … | … | … | n | G |
| 37365 | 10273316-6441553 | 31.3 ± 0.3 | 4471 ± 150 | 4.24 ± 0.19 | 0.942 ± 0.009 | 0.09 ± 0.20 | … | … | Y | Y | Y | … | Y | … | … | … | n | … |
| 37366 | 10273559-6445310 | 109.5 ± 0.3 | 4647 ± 21 | 2.29 ± 0.13 | 1.029 ± 0.008 | -0.15 ± 0.02 | … | … | N | N | … | … | … | … | … | … | n | G |
| 37367 | 10274286-6440250 | 24.0 ± 0.2 | 4692 ± 23 | 2.46 ± 0.11 | 1.023 ± 0.003 | -0.06 ± 0.11 | <50 | 3 | N | N | … | … | … | … | … | … | n | G |
| 37660 | 10364786-6330489 | 10.6 ± 0.2 | 4954 ± 66 | 3.02 ± 0.10 | 1.009 ± 0.003 | 0.01 ± 0.03 | <40 | 3 | Y | Y | Y | N | N | Y | … | … | n | NG? |
| 37661 | 10365057-6335352 | 17.5 ± 0.2 | 4617 ± 62 | 2.51 ± 0.06 | 1.010 ± 0.002 | 0.02 ± 0.07 | <49 | 3 | N | N | … | … | … | … | … | … | n | G |
| 2521 | 10284372-6435392 | 19.0 ± 0.6 | 6876 ± 36 | 4.38 ± 0.03 | … | 0.01 ± 0.06 | 68 ± 4 | 2 | … | Y | Y | Y | Y | Y | Y | … | Y | … |
| 37402 | 10284440-6450234 | -3.2 ± 0.2 | 4613 ± 55 | 2.36 ± 0.17 | 1.019 ± 0.003 | -0.12 ± 0.10 | <30 | 3 | N | N | … | … | … | … | … | … | n | G |
| 37403 | 10284622-6434417 | 39.0 ± 0.3 | 5849 ± 159 | 4.11 ± 0.05 | 0.994 ± 0.005 | -0.55 ± 0.15 | 15 ± 3 | 1 | Y | Y | Y | N | N | N | … | … | n | NG |
| 37662 | 10365685-6337119 | -8.8 ± 0.2 | 4681 ± 74 | … | 1.026 ± 0.002 | -0.09 ± 0.07 | <24 | 3 | N | … | … | … | … | … | … | … | n | G |
| 37404 | 10284793-6445441 | -40.5 ± 0.2 | 5106 ± 83 | … | 1.015 ± 0.003 | -0.27 ± 0.15 | <14 | 3 | N | … | … | … | … | … | … | … | n | G |
| 37405 | 10284803-6452372 | 37.2 ± 0.2 | 4443 ± 179 | 2.11 ± 0.16 | 1.022 ± 0.004 | 0.17 ± 0.03 | <65 | 3 | N | N | … | … | … | … | … | … | n | G |
| 37663 | 10365837-6327389 | -11.6 ± 0.2 | 4734 ± 15 | 2.54 ± 0.03 | 1.017 ± 0.004 | 0.00 ± 0.02 | <25 | 3 | N | N | … | … | … | … | … | … | n | G |
| 37406 | 10285048-6453215 | -38.8 ± 0.2 | 4737 ± 80 | … | 1.021 ± 0.005 | -0.11 ± 0.12 | <29 | 3 | N | … | … | … | … | … | … | … | n | G |
| 2523 | 10301225-6435258 | 80.5 ± 0.6 | 4701 ± 31 | 2.23 ± 0.06 | 1.017 ± 0.003 | -0.09 ± 0.04 | 28 ± 6 | 1 | N | N | … | … | … | … | … | … | n | G |
| 37722 | 10392646-6351507 | -18.0 ± 0.2 | 5057 ± 135 | … | 1.018 ± 0.002 | -0.07 ± 0.10 | <21 | 3 | N | … | … | … | … | … | … | … | n | G |
| 37483 | 10301350-6444201 | 32.3 ± 0.2 | 4958 ± 168 | 4.50 ± 0.01 | 0.959 ± 0.004 | -0.06 ± 0.01 | <12 | 3 | Y | Y | Y | N | N | Y | … | … | n | NG |
| 53541 | 10392655-6406589 | -42.9 ± 0.2 | 4860 ± 105 | 2.67 ± 0.13 | 1.013 ± 0.003 | -0.10 ± 0.11 | <25 | 3 | N | N | … | … | … | … | … | … | n | G |
| 37484 | 10301366-6330283 | -21.6 ± 0.2 | 4879 ± 18 | 2.67 ± 0.11 | 1.024 ± 0.004 | -0.03 ± 0.06 | … | … | N | N | … | … | … | … | … | … | n | G |
| 37485 | 10301501-6443407 | 17.4 ± 0.2 | 4675 ± 8 | 2.58 ± 0.02 | 1.009 ± 0.003 | -0.04 ± 0.07 | <34 | 3 | Y | N | Y | N | N | Y | … | … | n | NG? |
| 37486 | 10301541-6447306 | 117.4 ± 0.2 | 4927 ± 264 | … | 1.012 ± 0.004 | -0.21 ± 0.27 | <21 | 3 | N | … | … | … | … | … | … | … | n | G |
| 37723 | 10392774-6426023 | -1.6 ± 0.2 | 4610 ± 75 | 2.37 ± 0.09 | 1.021 ± 0.004 | -0.01 ± 0.06 | <43 | 3 | N | N | … | … | … | … | … | … | n | G |
| 37487 | 10301598-6434506 | -8.2 ± 0.2 | 4898 ± 181 | … | 1.024 ± 0.004 | -0.09 ± 0.15 | <26 | 3 | N | … | … | … | … | … | … | … | n | G |
| 37488 | 10301641-6325292 | 265.2 ± 0.2 | … | … | … | … | … | … | … | … | … | … | … | … | … | … | n | … |
| 37489 | 10301709-6448566 | 118.5 ± 0.2 | 4864 ± 296 | … | 1.011 ± 0.005 | -0.25 ± 0.31 | <30 | 3 | N | … | … | … | … | … | … | … | n | G |
| 37490 | 10301861-6439147 | -14.5 ± 0.2 | 4622 ± 58 | 2.37 ± 0.09 | 1.028 ± 0.004 | -0.05 ± 0.01 | <43 | 3 | N | N | … | … | … | … | … | … | n | G |
| 53542 | 10393131-6410435 | 14.2 ± 0.2 | 6025 ± 40 | 4.12 ± 0.14 | 0.994 ± 0.002 | -0.17 ± 0.03 | 60 ± 4 | 1 | Y | Y | Y | N | N | Y | … | … | n | NG |
| 37491 | 10302044-6442103 | 5.5 ± 0.2 | 4526 ± 155 | … | 1.028 ± 0.004 | 0.12 ± 0.04 | <54 | 3 | N | … | … | … | … | … | … | … | n | G |
| 2524 | 10302405-6318366 | -10.7 ± 0.6 | 4842 ± 31 | 2.62 ± 0.02 | 1.019 ± 0.003 | 0.03 ± 0.02 | <15 | 3 | N | N | … | … | … | … | … | … | n | G |
| 37724 | 10393250-6428016 | -5.1 ± 0.2 | 4718 ± 84 | 2.51 ± 0.13 | 1.023 ± 0.003 | -0.01 ± 0.01 | <31 | 3 | N | N | … | … | … | … | … | … | n | G |
| 53616 | 10405794-6424527 | -6.1 ± 0.2 | 4618 ± 25 | 2.41 ± 0.06 | 1.017 ± 0.003 | -0.04 ± 0.04 | <38 | 3 | N | N | … | … | … | … | … | … | n | G |
| 37492 | 10302545-6441288 | 12.6 ± 0.3 | 4958 ± 99 | 3.00 ± 0.16 | 1.010 ± 0.006 | -0.03 ± 0.08 | <33 | 3 | N | N | … | … | … | … | … | … | n | G |
| 37493 | 10302616-6435162 | -20.3 ± 0.2 | 5005 ± 1 | 2.97 ± 0.18 | 1.021 ± 0.005 | -0.01 ± 0.06 | <33 | 3 | N | N | … | … | … | … | … | … | n | G |
| 37725 | 10393405-6427100 | -16.2 ± 0.2 | 4707 ± 91 | 2.44 ± 0.16 | 1.028 ± 0.002 | -0.06 ± 0.02 | … | … | N | N | … | … | … | … | … | … | n | G |
| 53617 | 10405798-6416196 | -2.5 ± 0.2 | 5049 ± 52 | … | 1.013 ± 0.003 | 0.00 ± 0.01 | <23 | 3 | N | … | … | … | … | … | … | … | n | G |
| 2526 | 10302710-6450088 | -0.2 ± 0.6 | 4792 ± 18 | 2.59 ± 0.06 | … | -0.02 ± 0.01 | <18 | 3 | … | N | … | … | … | … | … | … | n | … |
| 37494 | 10302772-6435266 | -25.9 ± 0.2 | 4970 ± 12 | 2.78 ± 0.08 | 1.019 ± 0.006 | -0.05 ± 0.08 | <39 | 3 | N | N | … | … | … | … | … | … | n | G |
| 37726 | 10393455-6431008 | -0.8 ± 0.2 | 4538 ± 101 | 2.32 ± 0.19 | 1.022 ± 0.003 | -0.04 ± 0.10 | <34 | 3 | N | N | … | … | … | … | … | … | n | G |
| 53618 | 10405816-6411525 | 79.6 ± 0.2 | 4397 ± 206 | … | 1.037 ± 0.004 | -0.25 ± 0.08 | 69 ± 3 | 1 | N | … | … | … | … | … | … | … | n | G |
| 37495 | 10302795-6447174 | -6.7 ± 0.2 | 4891 ± 264 | … | 1.012 ± 0.003 | -0.11 ± 0.17 | <18 | 3 | N | … | … | … | … | … | … | … | n | G |
| 53619 | 10405860-6410494 | -27.0 ± 0.2 | 4678 ± 37 | … | 1.029 ± 0.001 | -0.10 ± 0.09 | … | … | N | … | … | … | … | … | … | … | n | G |
| 37496 | 10303004-6441037 | 16.5 ± 0.2 | 4485 ± 197 | … | 1.024 ± 0.004 | 0.14 ± 0.04 | <54 | 3 | N | … | … | … | … | … | … | … | n | G |
| 37497 | 10303397-6443068 | 48.4 ± 0.2 | 4643 ± 95 | … | 0.996 ± 0.005 | 0.00 ± 0.03 | 26 ± 3 | 1 | Y | … | N | N | N | Y | … | … | n | NG |
| 37727 | 10393502-6413486 | -6.5 ± 0.2 | 3979 ± 46 | 4.57 ± 0.09 | 0.851 ± 0.004 | -0.10 ± 0.09 | <29 | 3 | Y | Y | Y | Y | Y | Y | N | … | n | … |
| 38036 | 10442108-6441410 | 11.3 ± 0.2 | 4803 ± 2 | 2.60 ± 0.05 | 1.016 ± 0.004 | -0.03 ± 0.10 | <20 | 3 | N | N | … | … | … | … | … | … | n | G |
| 37797 | 10405924-6531348 | 14.6 ± 0.3 | 4944 ± 191 | 3.09 ± 0.17 | 1.007 ± 0.007 | -0.03 ± 0.05 | <31 | 3 | Y | N | Y | N | N | Y | … | … | n | NG? |
| 37498 | 10303498-6438353 | -5.6 ± 0.2 | 4664 ± 5 | 2.53 ± 0.04 | 1.015 ± 0.006 | 0.05 ± 0.06 | … | … | N | N | … | … | … | … | … | … | n | G |
| 53912 | 10453686-6429109 | 20.2 ± 0.3 | 6005 ± 115 | 4.28 ± 0.13 | 0.990 ± 0.005 | 0.07 ± 0.04 | 68 ± 4 | 1 | Y | Y | Y | N | N | Y | … | … | n | NG |

| ID | CNAME | RV (km s$^{-1}$) | $T_{\text{eff}}$ (K) | $\log g$ (dex) | $\gamma^a$ | [Fe/H] (dex) | EW(Li)$^b$ (mÅ) | EW(Li) error flag$^c$ | $\gamma$ | $\log g$ | Membership RV | Li | H$\alpha$ | [Fe/H] | Gaia studies Randich$^d$ | Cantat-Gaudin$^d$ | Final$^e$ | NMs with Li$^f$ |
|---|---|---|---|---|---|---|---|---|---|---|---|---|---|---|---|---|---|---|
| 53543 | 10393553-6413383 | -10.7 ± 0.3 | 7238 ± 64 | ... | ... | ... | ... | ... | ... | ... | ... | ... | ... | ... | ... | ... | n | ... |
| 53620 | 10405928-6405019 | 8.5 ± 0.2 | 6339 ± 65 | ... | 0.991 ± 0.004 | 0.03 ± 0.05 | 129 ± 7 | 1 | Y | ... | Y | Y | N | Y | ... | ... | n | NG |
| 37499 | 10303522-6446477 | -10.9 ± 0.2 | 4597 ± 105 | 2.44 ± 0.08 | 1.011 ± 0.007 | -0.03 ± 0.01 | <31 | 3 | N | N | ... | ... | ... | ... | ... | ... | n | G |
| 53806 | 10442114-6423514 | -8.7 ± 0.4 | 6168 ± 87 | 3.80 ± 0.07 | 1.008 ± 0.005 | -0.17 ± 0.11 | <17 | 3 | Y | Y | Y | N | N | Y | ... | ... | n | NG? |
| 53621 | 10405933-6412226 | -37.2 ± 0.2 | 4275 ± 219 | ... | 1.073 ± 0.002 | -0.17 ± 0.05 | <32 | 3 | N | ... | ... | ... | ... | ... | ... | ... | n | G |
| 38100 | 10453700-6445476 | -12.6 ± 0.2 | 4572 ± 171 | ... | 1.075 ± 0.003 | -0.10 ± 0.10 | 24 ± ... | 3 | N | ... | ... | ... | ... | ... | ... | ... | n | G |
| 37500 | 10303560-6442074 | 21.5 ± 0.3 | 5425 ± 157 | ... | 1.002 ± 0.005 | ... | 13 ± 3 | 1 | Y | ... | Y | N | N | ... | ... | ... | n | NG? |
| 53807 | 10442124-6430068 | 12.2 ± 1.4 | 3490 ± 178 | ... | 0.894 ± 0.015 | -0.22 ± 0.13 | <100 | 3 | Y | ... | Y | Y | Y | Y | ... | Y | n | ... |
| 37798 | 10405941-6430326 | 14.8 ± 0.2 | 4919 ± 114 | 2.74 ± 0.08 | 1.014 ± 0.002 | -0.02 ± 0.04 | <32 | 3 | N | N | ... | ... | ... | ... | ... | ... | n | G |
| 53561 | 10400095-6419586 | -0.7 ± 0.2 | 4576 ± 181 | ... | 1.024 ± 0.001 | -0.02 ± 0.02 | 449 ± 2 | 1 | N | ... | ... | ... | ... | ... | ... | ... | n | Li-rich G |
| 38101 | 10453718-6421542 | 25.6 ± 0.2 | 5140 ± 94 | ... | 1.013 ± 0.002 | ... | 24 ± 6 | 1 | N | ... | ... | ... | ... | ... | ... | ... | n | G |
| 37799 | 10410000-6543218 | 9.0 ± 0.3 | 5020 ± 115 | ... | 1.010 ± 0.008 | -0.10 ± 0.19 | 16 ± 4 | 1 | N | ... | ... | ... | ... | ... | ... | ... | n | G |
| 37501 | 10303590-6444138 | 8.7 ± 0.2 | 5009 ± 166 | ... | 1.007 ± 0.005 | -0.14 ± 0.19 | <31 | 3 | Y | ... | Y | N | N | Y | ... | ... | n | NG? |
| 53562 | 10400112-6407386 | 9.9 ± 0.2 | 3908 ± 8 | ... | 1.038 ± 0.001 | -0.02 ± 0.13 | <14 | 3 | N | ... | ... | ... | ... | ... | ... | ... | n | G |
| 38102 | 10453719-6416088 | 15.4 ± 0.3 | 3542 ± 31 | ... | 0.841 ± 0.004 | -0.24 ± 0.13 | ... | ... | Y | ... | Y | ... | N | Y | Y | ... | n | ... |
| 53622 | 10410051-6410386 | 116.0 ± 0.2 | 4856 ± 258 | ... | 1.062 ± 0.001 | -0.16 ± 0.20 | <20 | 3 | N | ... | ... | ... | ... | ... | ... | ... | n | G |
| 37502 | 10303701-6444432 | -37.2 ± 0.3 | 5089 ± 68 | ... | 1.008 ± 0.007 | -0.15 ± 0.13 | <12 | 3 | Y | ... | N | N | N | Y | ... | ... | n | NG? |
| 53808 | 10442164-6418310 | -20.9 ± 0.3 | 5436 ± 180 | ... | 0.982 ± 0.006 | ... | <22 | 3 | Y | ... | N | N | N | ... | ... | ... | n | NG |
| 53623 | 10410067-6414582 | -1.0 ± 0.2 | 6155 ± 67 | 3.84 ± 0.13 | 1.007 ± 0.002 | -0.05 ± 0.01 | <14 | 3 | Y | Y | Y | N | N | Y | ... | ... | n | NG? |
| 37591 | 10335200-6411257 | 17.9 ± 0.4 | 3372 ± 62 | ... | 0.862 ± 0.009 | -0.26 ± 0.13 | ... | ... | Y | ... | Y | ... | N | Y | Y | Y | n | ... |
| 53624 | 10410076-6406494 | -2.7 ± 0.6 | 7033 ± 88 | ... | 1.006 ± 0.004 | -0.18 ± 0.09 | <17 | 3 | Y | ... | Y | N | N | Y | ... | ... | n | NG? |
| 37592 | 10335370-6409135 | 43.4 ± 0.2 | 4578 ± 181 | ... | 1.021 ± 0.002 | 0.06 ± 0.06 | <41 | 3 | N | ... | ... | ... | ... | ... | ... | ... | n | G |
| 53563 | 10400215-6413493 | -1.4 ± 0.2 | 5093 ± 53 | 4.10 ± 0.09 | 0.980 ± 0.002 | -0.05 ± 0.03 | 10 ± 6 | 1 | Y | Y | Y | N | Y | Y | ... | ... | n | NG |
| 53625 | 10410082-6405211 | -5.0 ± 0.2 | 4936 ± 38 | 2.68 ± 0.02 | 1.026 ± 0.005 | 0.01 ± 0.03 | <27 | 3 | N | N | ... | ... | ... | ... | ... | ... | n | G |
| 37593 | 10335575-6414176 | 12.0 ± 0.2 | 5027 ± 7 | 2.91 ± 0.04 | 1.011 ± 0.003 | -0.02 ± 0.04 | <33 | 3 | N | N | ... | ... | ... | ... | ... | ... | n | G |
| 37747 | 10400215-6426338 | 17.7 ± 0.2 | 4790 ± 150 | ... | 1.019 ± 0.002 | -0.07 ± 0.15 | ... | ... | N | ... | ... | ... | ... | ... | ... | ... | n | G |
| 37594 | 10335905-6407008 | 29.7 ± 0.2 | 4575 ± 151 | ... | 1.025 ± 0.002 | -0.03 ± 0.08 | <39 | 3 | N | ... | ... | ... | ... | ... | ... | ... | n | G |
| 38037 | 10442223-6416597 | -3.8 ± 0.2 | 4664 ± 40 | 2.45 ± 0.12 | 1.021 ± 0.002 | -0.03 ± 0.05 | <25 | 3 | N | N | ... | ... | ... | ... | ... | ... | n | G |
| 37800 | 10410147-6535443 | -14.8 ± 0.2 | 4596 ± 68 | 2.60 ± 0.06 | 1.005 ± 0.004 | 0.05 ± 0.02 | <44 | 3 | Y | N | N | N | N | Y | ... | ... | n | NG? |
| 37595 | 10340013-6402265 | 21.5 ± 0.2 | 4661 ± 84 | 2.51 ± 0.16 | 1.018 ± 0.002 | 0.02 ± 0.08 | <42 | 3 | N | N | ... | ... | ... | ... | ... | ... | n | G |
| 37748 | 10400281-6356013 | 43.5 ± 0.2 | 4539 ± 92 | 2.40 ± 0.14 | 1.015 ± 0.002 | 0.13 ± 0.04 | <51 | 3 | N | N | ... | ... | ... | ... | ... | ... | n | G |
| 38038 | 10442243-6402382 | 33.4 ± 0.2 | 4558 ± 117 | ... | 1.056 ± 0.004 | ... | ... | ... | N | ... | ... | ... | ... | ... | ... | ... | n | G |
| 53626 | 10410166-6413209 | 69.3 ± 0.2 | 4545 ± 113 | ... | 1.036 ± 0.003 | ... | <28 | 3 | N | ... | ... | ... | ... | ... | ... | ... | n | G |
| 37801 | 10410166-6436311 | 17.0 ± 0.2 | 5175 ± 111 | ... | 1.016 ± 0.003 | -0.16 ± 0.14 | ... | ... | N | ... | ... | ... | ... | ... | ... | ... | n | G |
| 53913 | 10453771-6426478 | 16.6 ± 0.2 | 6016 ± 243 | 4.05 ± 0.12 | 0.998 ± 0.003 | -0.33 ± 0.19 | 33 ± 3 | 1 | Y | Y | Y | N | N | N | ... | ... | n | NG |
| 37596 | 10340396-6410030 | 12.3 ± 0.2 | 4690 ± 53 | 2.48 ± 0.07 | 1.016 ± 0.003 | 0.01 ± 0.07 | <43 | 3 | N | N | ... | ... | ... | ... | ... | ... | n | G |
| 53564 | 10400322-6415029 | 33.4 ± 0.2 | 4681 ± 103 | 2.19 ± 0.12 | 1.024 ± 0.005 | -0.27 ± 0.05 | <42 | 3 | N | N | ... | ... | ... | ... | ... | ... | n | G |
| 2568 | 10442256-6415301 | 17.5 ± 0.6 | 5854 ± 33 | 4.51 ± 0.09 | ... | 0.03 ± 0.06 | 151 ± 7 | 2 | ... | Y | Y | Y | Y | Y | Y | Y | Y | ... |
| 53627 | 10410225-6413434 | 5.8 ± 0.2 | 4581 ± 76 | ... | 1.027 ± 0.004 | ... | <63 | 3 | N | ... | ... | ... | ... | ... | ... | ... | n | G |
| 37597 | 10340443-6410035 | 44.3 ± 0.4 | 5807 ± 231 | 4.46 ± 0.20 | 0.982 ± 0.013 | 0.09 ± 0.05 | 89 ± 8 | 1 | Y | Y | N | N | N | Y | ... | ... | n | NG |
| 53628 | 10410292-6409211 | -32.7 ± 0.2 | 5040 ± 87 | ... | 1.013 ± 0.006 | -0.10 ± 0.16 | <13 | 3 | N | ... | ... | ... | ... | ... | ... | ... | n | G |
| 53629 | 10410296-6414107 | 57.6 ± 0.2 | 4604 ± 94 | ... | 1.045 ± 0.003 | ... | <33 | 3 | N | ... | ... | ... | ... | ... | ... | ... | n | G |
| 38103 | 10453832-6328585 | 7.7 ± 0.2 | 4619 ± 59 | 2.46 ± 0.12 | 1.015 ± 0.002 | 0.03 ± 0.05 | <46 | 3 | N | N | ... | ... | ... | ... | ... | ... | n | G |
| 53630 | 10410300-6420203 | 17.8 ± 0.2 | 4675 ± 140 | 2.71 ± 0.02 | 1.006 ± 0.005 | 0.10 ± 0.03 | <43 | 3 | Y | N | Y | N | N | Y | ... | ... | n | NG? |
| 37802 | 10410348-6450130 | -7.9 ± 0.2 | 4696 ± 56 | 2.50 ± 0.17 | 1.019 ± 0.004 | -0.10 ± 0.09 | <43 | 3 | N | N | ... | ... | ... | ... | ... | ... | n | G |
| 37749 | 10400486-6404399 | 55.5 ± 0.2 | 4653 ± 66 | 2.43 ± 0.09 | 1.018 ± 0.002 | -0.02 ± 0.02 | <37 | 3 | N | N | ... | ... | ... | ... | ... | ... | n | G |
| 37846 | 10414099-6458367 | 11.8 ± 0.2 | 4467 ± 201 | ... | 1.041 ± 0.002 | -0.01 ± 0.04 | <38 | 3 | N | ... | ... | ... | ... | ... | ... | ... | n | G |
| 53914 | 10453837-6436166 | -30.1 ± 0.2 | 4981 ± 129 | ... | 1.020 ± 0.001 | -0.11 ± 0.08 | <17 | 3 | N | ... | ... | ... | ... | ... | ... | ... | n | G |
| 53809 | 10442301-6426062 | 19.5 ± 0.3 | 5683 ± 45 | 4.22 ± 0.09 | 0.989 ± 0.005 | 0.00 ± 0.27 | 86 ± 5 | 1 | Y | Y | Y | N | N | Y | ... | ... | n | NG |
| 37847 | 10414161-6532408 | 73.7 ± 0.3 | 4924 ± 243 | ... | 1.015 ± 0.007 | -0.28 ± 0.24 | <12 | 3 | N | ... | ... | ... | ... | ... | ... | ... | n | G |
| 53915 | 10453858-6433393 | 0.5 ± 0.2 | 5185 ± 51 | 3.57 ± 0.07 | 0.995 ± 0.004 | -0.05 ± 0.08 | <40 | 3 | Y | Y | Y | N | N | Y | ... | ... | n | NG |
| 37848 | 10414274-6445336 | -2.0 ± 0.2 | 4729 ± 10 | 2.55 ± 0.12 | 1.020 ± 0.002 | -0.06 ± 0.03 | <32 | 3 | N | N | ... | ... | ... | ... | ... | ... | n | G |
| 53674 | 10414306-6407118 | -5.6 ± 0.2 | 4972 ± 71 | 2.96 ± 0.19 | 1.017 ± 0.003 | 0.01 ± 0.04 | <30 | 3 | N | N | ... | ... | ... | ... | ... | ... | n | G |
| 38104 | 10453903-6413029 | 17.0 ± 0.6 | 3388 ± 170 | ... | 0.696 ± 0.005 | -0.27 ± 0.13 | ... | ... | Y | ... | Y | ... | N | Y | Y | ... | n | ... |
| 53818 | 10443001-6422577 | 41.8 ± 0.2 | 5746 ± 102 | ... | 0.994 ± 0.004 | -0.19 ± 0.16 | <9 | 3 | Y | ... | Y | N | N | Y | ... | ... | n | ... |
| 37750 | 10400568-6434489 | -6.7 ± 0.2 | 4917 ± 155 | ... | 1.024 ± 0.002 | -0.06 ± 0.12 | <25 | 3 | N | ... | ... | ... | ... | ... | ... | ... | n | G |
| 37849 | 10414398-6536442 | 15.2 ± 0.2 | 4637 ± 102 | ... | 1.024 ± 0.004 | ... | <62 | 3 | N | ... | ... | ... | ... | ... | ... | ... | n | G |
| 37598 | 10341159-6409269 | -0.5 ± 0.2 | 4635 ± 32 | 2.41 ± 0.08 | 1.020 ± 0.002 | -0.03 ± 0.02 | <28 | 3 | N | N | ... | ... | ... | ... | ... | ... | n | G |
| 53675 | 10414468-6421293 | 43.8 ± 0.2 | 3991 ± 179 | ... | 1.045 ± 0.002 | -0.14 ± 0.11 | <16 | 3 | N | ... | ... | ... | ... | ... | ... | ... | n | G |

**Table C.6.** continued.

| ID | CNAME | RV (km s$^{-1}$) | $T_{\rm eff}$ (K) | $logg$ (dex) | $\gamma^a$ | [Fe/H] (dex) | EW(Li)$^b$ (mÅ) | EW(Li) error flag$^c$ | $\gamma$ | $logg$ | RV | Li | H$\alpha$ | [Fe/H] | Randich$^d$ | Cantat-Gaudin$^d$ | Final$^e$ | NMs with Li$^f$ |
|---|---|---|---|---|---|---|---|---|---|---|---|---|---|---|---|---|---|---|
| | | | | | | | | | | | Membership | | | | Gaia studies | | | |
| 2535 | 10341190-6410471 | 65.0 ± 0.6 | 7140 ± 163 | 4.05 ± 0.21 | 1.023 ± 0.001 | -0.13 ± 0.16 | 18 ± 6 | 1 | N | Y | ... | ... | ... | ... | N | ... | n | ... |
| 53926 | 10455214-6429113 | 0.2 ± 0.2 | 4923 ± 56 | ... | 0.993 ± 0.005 | 0.01 ± 0.06 | <41 | 3 | Y | ... | Y | N | N | Y | ... | ... | n | NG |
| 37850 | 10414523-6428035 | 17.8 ± 0.2 | 4377 ± 333 | ... | 0.926 ± 0.003 | 0.03 ± 0.13 | 108 ± 2 | 1 | Y | ... | Y | Y | Y | Y | ... | Y | Y | ... |
| 53927 | 10455224-6424083 | 14.2 ± 0.3 | 5519 ± 67 | ... | 0.993 ± 0.006 | ... | 73 ± 5 | 1 | Y | ... | Y | N | N | ... | ... | ... | n | NG |
| 37851 | 10414548-6449241 | 34.6 ± 0.2 | 5065 ± 148 | ... | 1.023 ± 0.003 | ... | 112 ± 2 | 1 | N | ... | ... | ... | ... | ... | ... | ... | n | Li-rich G |
| 2551 | 10414561-6534337 | 45.3 ± 0.6 | 5133 ± 19 | 2.89 ± 0.03 | ... | -0.03 ± 0.01 | <14 | 3 | ... | N | ... | ... | ... | ... | ... | ... | n | ... |
| 37852 | 10414575-6533516 | 5.8 ± 0.2 | 4580 ± 76 | ... | 1.026 ± 0.003 | ... | <70 | 3 | N | ... | ... | ... | ... | ... | ... | ... | n | G |
| 37853 | 10414623-6448404 | -5.0 ± 0.2 | 4763 ± 44 | 2.56 ± 0.03 | 1.025 ± 0.003 | -0.06 ± 0.03 | <26 | 3 | N | N | ... | ... | ... | ... | ... | ... | n | G |
| 53676 | 10414763-6410027 | 3.0 ± 0.2 | 4575 ± 89 | 2.44 ± 0.14 | 1.013 ± 0.003 | 0.06 ± 0.02 | <38 | 3 | N | N | ... | ... | ... | ... | ... | ... | n | G |
| 37763 | 10402962-6401378 | -6.8 ± 0.2 | 5014 ± 115 | ... | 1.012 ± 0.002 | -0.04 ± 0.06 | <20 | 3 | N | ... | ... | ... | ... | ... | ... | ... | n | G |
| 53677 | 10414892-6412443 | -27.0 ± 0.2 | 5060 ± 3 | ... | 1.032 ± 0.002 | 0.05 ± 0.05 | <21 | 3 | N | ... | ... | ... | ... | ... | ... | ... | n | G |
| 53586 | 10402966-6413174 | 11.3 ± 0.2 | 5829 ± 98 | ... | 1.002 ± 0.003 | ... | 85 ± 5 | 1 | Y | ... | Y | N | N | ... | ... | ... | n | NG? |
| 53678 | 10414902-6407440 | 24.9 ± 0.3 | 4926 ± 184 | 3.58 ± 0.19 | 0.989 ± 0.005 | 0.03 ± 0.02 | <25 | 3 | Y | Y | Y | N | N | Y | ... | ... | n | NG |
| 53928 | 10455299-6433370 | 21.5 ± 0.2 | 5641 ± 46 | 4.28 ± 0.12 | 0.987 ± 0.003 | 0.04 ± 0.13 | 29 ± 5 | 1 | Y | Y | Y | N | N | Y | ... | ... | n | NG |
| 37854 | 10414966-6532330 | 29.1 ± 0.3 | 5142 ± 94 | ... | 1.008 ± 0.006 | ... | <22 | 3 | Y | ... | Y | N | N | ... | ... | ... | n | NG? |
| 37764 | 10403019-6442169 | 17.7 ± 0.2 | 3874 ± 119 | ... | 0.847 ± 0.002 | -0.13 ± 0.13 | ... | ... | Y | ... | Y | ... | N | Y | Y | Y | n | ... |
| 53680 | 10414967-6418033 | -7.6 ± 0.2 | 4626 ± 55 | 2.33 ± 0.11 | 1.031 ± 0.001 | -0.03 ± 0.05 | <47 | 3 | N | N | ... | ... | ... | ... | ... | ... | n | G |
| 53681 | 10414970-6412173 | -30.5 ± 0.4 | 6660 ± 66 | ... | 1.013 ± 0.003 | -0.17 ± 0.06 | <8 | 3 | N | ... | ... | ... | ... | ... | ... | ... | n | ... |
| 37855 | 10415005-6450121 | 66.2 ± 0.2 | 4669 ± 52 | ... | 1.031 ± 0.004 | -0.28 ± 0.08 | <11 | 3 | N | ... | ... | ... | ... | ... | ... | ... | n | G |
| 2577 | 10455398-6419569 | 12.1 ± 0.6 | 4718 ± 26 | 2.58 ± 0.05 | 1.017 ± 0.002 | 0.05 ± 0.04 | 37 ± 9 | 1 | N | N | ... | ... | ... | ... | ... | ... | n | G |
| 53682 | 10415032-6407201 | 6.5 ± 0.9 | 6568 ± 117 | ... | 1.010 ± 0.007 | 0.06 ± 0.09 | <33 | 3 | Y | ... | Y | N | N | Y | ... | ... | n | NG? |
| 53683 | 10415042-6416332 | -11.0 ± 0.2 | 4573 ± 47 | 2.37 ± 0.12 | 1.019 ± 0.004 | -0.02 ± 0.02 | <45 | 3 | N | N | ... | ... | ... | ... | ... | ... | n | G |
| 38122 | 10455504-6323031 | 4.5 ± 0.2 | 4925 ± 146 | ... | 1.022 ± 0.002 | -0.07 ± 0.12 | <14 | 3 | N | ... | ... | ... | ... | ... | ... | ... | n | G |
| 53587 | 10403087-6422359 | -18.7 ± 0.2 | 5741 ± 90 | ... | 1.001 ± 0.001 | -0.37 ± 0.03 | 52 ± 3 | 1 | Y | ... | N | N | N | N | ... | ... | n | NG? |
| 53819 | 10443170-6439027 | 22.4 ± 0.2 | 4745 ± 91 | 2.60 ± 0.11 | 1.014 ± 0.002 | 0.04 ± 0.05 | <35 | 3 | N | N | ... | ... | ... | ... | ... | ... | n | G |
| 37856 | 10415086-6431443 | 20.2 ± 0.2 | 4500 ± 87 | ... | 1.024 ± 0.003 | -0.12 ± 0.09 | <26 | 3 | N | ... | ... | ... | ... | ... | ... | ... | n | G |
| 53929 | 10455624-6428243 | -39.3 ± 0.2 | 5068 ± 53 | 3.34 ± 0.05 | 1.004 ± 0.005 | 0.01 ± 0.01 | <27 | 3 | Y | N | N | N | N | Y | ... | ... | n | NG? |
| 53820 | 10443181-6419471 | 52.5 ± 0.2 | 4533 ± 140 | 2.39 ± 0.14 | 1.011 ± 0.004 | 0.01 ± 0.05 | <40 | 3 | N | N | ... | ... | ... | ... | ... | ... | n | G |
| 37857 | 10415098-6538179 | 35.6 ± 0.2 | 4640 ± 129 | 2.45 ± 0.17 | 1.017 ± 0.007 | 0.00 ± 0.11 | <24 | 3 | N | N | ... | ... | ... | ... | ... | ... | n | G |
| 53930 | 10455697-6432103 | -22.7 ± 0.2 | 5788 ± 102 | ... | 1.000 ± 0.003 | -0.23 ± 0.09 | 42 ± 3 | 1 | Y | ... | N | N | N | Y | ... | ... | n | NG |
| 38123 | 10455785-6325426 | -33.9 ± 0.2 | 4985 ± 111 | ... | 1.019 ± 0.002 | -0.04 ± 0.09 | 346 ± 4 | 1 | N | ... | ... | ... | ... | ... | ... | ... | n | Li-rich G |
| 53684 | 10415106-6413256 | -2.9 ± 0.2 | 4236 ± 293 | ... | 1.041 ± 0.002 | 0.08 ± 0.16 | <40 | 3 | N | ... | ... | ... | ... | ... | ... | ... | n | G |
| 53931 | 10455787-6433271 | -13.0 ± 0.5 | 6931 ± 69 | ... | ... | -0.22 ± 0.07 | ... | ... | ... | ... | ... | ... | ... | ... | ... | ... | n | ... |
| 38124 | 10455842-6317019 | -59.2 ± 0.2 | 4578 ± 111 | 2.38 ± 0.17 | 1.021 ± 0.003 | 0.02 ± 0.01 | <37 | 3 | N | N | ... | ... | ... | ... | ... | ... | n | G |
| 37765 | 10403102-6536169 | 8.7 ± 0.2 | 4998 ± 105 | ... | 1.020 ± 0.004 | ... | ... | ... | N | ... | ... | ... | ... | ... | ... | ... | n | G |
| 37895 | 10422288-6529064 | 57.6 ± 0.3 | 4828 ± 146 | 2.59 ± 0.14 | 1.018 ± 0.010 | -0.07 ± 0.07 | <18 | 3 | N | N | ... | ... | ... | ... | ... | ... | n | G |
| 38125 | 10455861-6405009 | 3.9 ± 0.2 | 4419 ± 182 | ... | 1.038 ± 0.002 | -0.08 ± 0.03 | <34 | 3 | N | ... | ... | ... | ... | ... | ... | ... | n | G |
| 37896 | 10422321-6538316 | 8.5 ± 0.2 | 4575 ± 102 | ... | 1.019 ± 0.005 | -0.15 ± 0.15 | <45 | 3 | N | ... | ... | ... | ... | ... | ... | ... | n | G |
| 53932 | 10455870-6422227 | 38.0 ± 0.2 | 4563 ± 128 | ... | 1.018 ± 0.002 | 0.06 ± 0.10 | <58 | 3 | N | ... | ... | ... | ... | ... | ... | ... | n | G |
| 37897 | 10422397-6436149 | 35.1 ± 0.2 | 4610 ± 82 | ... | 1.061 ± 0.006 | ... | ... | ... | N | ... | ... | ... | ... | ... | ... | ... | n | G |
| 37766 | 10403196-6456301 | 4.0 ± 0.2 | 4624 ± 45 | 2.29 ± 0.10 | 1.025 ± 0.002 | -0.08 ± 0.05 | <29 | 3 | N | N | ... | ... | ... | ... | ... | ... | n | G |
| 53716 | 10422405-6410046 | 39.6 ± 0.2 | 5594 ± 51 | ... | 0.996 ± 0.003 | 0.11 ± 0.04 | 35 ± 7 | 1 | Y | ... | Y | N | N | Y | ... | ... | n | NG |
| 38126 | 10455898-6324534 | 17.4 ± 0.2 | 4789 ± 126 | ... | 1.020 ± 0.002 | -0.09 ± 0.16 | <20 | 3 | N | ... | ... | ... | ... | ... | ... | ... | n | G |
| 37767 | 10403204-6535379 | 31.0 ± 0.2 | 5029 ± 126 | ... | 1.010 ± 0.004 | -0.08 ± 0.15 | <25 | 3 | N | ... | ... | ... | ... | ... | ... | ... | n | G |
| 53833 | 10444099-6422212 | 7.0 ± 0.2 | 4965 ± 154 | ... | 0.970 ± 0.003 | 0.01 ± 0.08 | <31 | 3 | Y | ... | Y | N | Y | Y | ... | ... | n | NG |
| 53717 | 10422413-6418098 | -18.1 ± 0.2 | 4768 ± 29 | 2.62 ± 0.17 | 1.025 ± 0.003 | -0.04 ± 0.01 | <24 | 3 | N | N | ... | ... | ... | ... | ... | ... | n | G |
| 38127 | 10455951-6318421 | -17.1 ± 3.2 | ... | ... | ... | ... | ... | ... | ... | ... | ... | ... | ... | ... | ... | ... | n | ... |
| 37768 | 10403237-6537161 | 47.2 ± 0.3 | 4971 ± 197 | ... | 1.019 ± 0.007 | -0.11 ± 0.14 | <28 | 3 | N | ... | ... | ... | ... | ... | ... | ... | n | G |
| 37898 | 10422420-6544217 | 37.5 ± 0.3 | 4312 ± 230 | 4.45 ± 0.11 | 0.901 ± 0.007 | -0.19 ± 0.10 | <15 | 3 | Y | Y | Y | N | N | Y | ... | ... | n | NG |
| 53933 | 10460043-6434524 | -4.3 ± 0.3 | 6215 ± 128 | ... | 0.997 ± 0.003 | 0.00 ± 0.04 | 94 ± 3 | 1 | Y | ... | Y | N | N | Y | ... | ... | n | NG |
| 38128 | 10460095-6358062 | -42.8 ± 0.2 | 4476 ± 144 | ... | 1.033 ± 0.003 | -0.13 ± 0.14 | <23 | 3 | N | ... | ... | ... | ... | ... | ... | ... | n | G |
| 53834 | 10444113-6419198 | 13.3 ± 0.4 | 6291 ± 129 | ... | 0.996 ± 0.004 | -0.14 ± 0.01 | <15 | 3 | Y | ... | Y | N | N | Y | ... | ... | n | NG |
| 38191 | 10465935-6333052 | 4.1 ± 0.2 | 4580 ± 42 | 2.38 ± 0.13 | 1.021 ± 0.002 | -0.03 ± 0.01 | <43 | 3 | N | N | ... | ... | ... | ... | ... | ... | n | G |
| 38192 | 10465976-6415242 | 19.2 ± 0.2 | 4514 ± 174 | ... | 1.018 ± 0.002 | 0.12 ± 0.05 | <56 | 3 | N | ... | ... | ... | ... | ... | ... | ... | n | G |
| 37899 | 10422592-6536071 | -21.1 ± 0.2 | 5054 ± 124 | ... | 1.015 ± 0.004 | -0.14 ± 0.17 | <26 | 3 | N | ... | ... | ... | ... | ... | ... | ... | n | G |
| 38193 | 10470184-6424359 | -0.9 ± 0.2 | 4556 ± 38 | 2.25 ± 0.11 | 1.032 ± 0.005 | -0.02 ± 0.07 | <50 | 3 | N | N | ... | ... | ... | ... | ... | ... | n | G |
| 37900 | 10422773-6530344 | 11.5 ± 0.3 | 4513 ± 158 | ... | 1.026 ± 0.006 | -0.08 ± 0.07 | ... | ... | N | ... | ... | ... | ... | ... | ... | ... | n | ... |
| 38194 | 10470213-6317123 | -3.8 ± 0.2 | 5204 ± 73 | ... | 1.019 ± 0.002 | 0.03 ± 0.01 | <32 | 3 | N | ... | ... | ... | ... | ... | ... | ... | n | ... |
| 38195 | 10470216-6335548 | 113.6 ± 0.2 | 4899 ± 185 | ... | 1.008 ± 0.002 | -0.15 ± 0.21 | <20 | 3 | Y | ... | N | N | N | Y | ... | ... | n | NG? |

| ID | CNAME | RV (km s$^{-1}$) | T$_{\text{eff}}$ (K) | logg (dex) | $\gamma^a$ | [Fe/H] (dex) | EW(Li)$^b$ (mÅ) | EW(Li) error flag$^c$ | Membership $\gamma$ | logg | RV | Li | H$\alpha$ | [Fe/H] | Gaia studies Randich$^d$ | Cantat-Gaudin$^d$ | Final$^e$ | NMs with Li$^f$ |
|---|---|---|---|---|---|---|---|---|---|---|---|---|---|---|---|---|---|---|
| 37901 | 10422798-6540353 | -35.1 ± 0.2 | 4807 ± 244 | … | 1.013 ± 0.003 | -0.10 ± 0.19 | <23 | 3 | N | … | … | … | … | … | … | … | n | G |
| 38196 | 10470267-6332541 | 43.5 ± 0.2 | 4696 ± 104 | 2.64 ± 0.11 | 1.007 ± 0.002 | 0.01 ± 0.05 | <46 | 3 | Y | N | N | N | N | Y | … | … | n | NG? |
| 2583 | 10470309-6416241 | -4.5 ± 0.6 | 4965 ± 8 | 2.67 ± 0.08 | 1.022 ± 0.003 | -0.13 ± 0.01 | 27 ± 8 | 1 | N | N | … | … | … | … | … | … | n | G |
| 37902 | 10422815-6436128 | 15.7 ± 0.3 | 3539 ± 11 | 4.62 ± 0.17 | 0.813 ± 0.008 | -0.22 ± 0.14 | … | … | Y | Y | Y | … | N | Y | Y | Y | n | … |
| 38197 | 10470362-6412484 | 52.7 ± 1.5 | … | … | … | … | … | … | … | … | … | … | … | … | … | … | n | … |
| 38198 | 10470465-6434298 | 19.4 ± 0.2 | 4530 ± 156 | … | 1.021 ± 0.003 | 0.12 ± 0.03 | <52 | 3 | N | … | … | … | … | … | … | … | n | G |
| 37903 | 10422859-6525332 | 34.7 ± 0.2 | 4569 ± 107 | 2.40 ± 0.19 | 1.017 ± 0.006 | -0.01 ± 0.08 | <34 | 3 | N | N | … | … | … | … | … | … | n | G |
| 38199 | 10470485-6424271 | -43.1 ± 0.2 | 5001 ± 75 | 2.95 ± 0.20 | 1.018 ± 0.005 | -0.04 ± 0.06 | <29 | 3 | N | N | … | … | … | … | … | … | n | G |
| 38200 | 10470579-6330397 | -17.2 ± 0.2 | 4718 ± 37 | … | 1.022 ± 0.002 | -0.09 ± 0.12 | <29 | 3 | N | … | … | … | … | … | … | … | n | G |
| 37904 | 10422968-6540055 | 74.9 ± 0.2 | 5001 ± 65 | … | 1.017 ± 0.006 | -0.13 ± 0.15 | <24 | 3 | N | … | … | … | … | … | … | … | n | G |
| 38201 | 10470595-6326492 | 34.6 ± 3.4 | … | … | … | … | … | … | … | … | … | … | … | … | Y | … | n | … |
| 37905 | 10423054-6524579 | 38.2 ± 0.2 | 4648 ± 1 | 2.41 ± 0.01 | 1.016 ± 0.006 | -0.04 ± 0.01 | <35 | 3 | N | N | … | … | … | … | … | … | n | G |
| 2584 | 10470613-6325568 | -1.7 ± 0.6 | 4697 ± 23 | 2.45 ± 0.11 | 1.021 ± 0.002 | 0.02 ± 0.04 | 35 ± 10 | 1 | N | N | … | … | … | … | … | … | n | G |
| 53718 | 10423083-6416175 | 51.0 ± 0.2 | 4148 ± 293 | … | 1.067 ± 0.002 | -0.24 ± 0.08 | 20 ± 8 | 1 | N | … | … | … | … | … | … | … | n | G |
| 38202 | 10470798-6320085 | -23.3 ± 0.2 | 4674 ± 5 | 2.45 ± 0.06 | 1.022 ± 0.002 | -0.02 ± 0.03 | <37 | 3 | N | N | … | … | … | … | … | … | n | G |
| 53719 | 10423125-6412184 | 8.1 ± 0.3 | 6124 ± 91 | 4.08 ± 0.09 | 0.998 ± 0.005 | -0.05 ± 0.02 | 68 ± 4 | 1 | Y | Y | Y | N | N | Y | … | … | n | NG |
| 38203 | 10470806-6334542 | 32.0 ± 0.2 | 4615 ± 59 | 2.35 ± 0.09 | 1.016 ± 0.002 | -0.08 ± 0.07 | <24 | 3 | N | N | … | … | … | … | … | … | n | G |
| 37906 | 10423143-6449462 | -34.7 ± 0.2 | 4996 ± 75 | … | 1.018 ± 0.003 | -0.05 ± 0.07 | <18 | 3 | N | … | … | … | … | … | … | … | n | G |
| 38204 | 10470862-6322437 | 5.2 ± 0.2 | 4822 ± 9 | 2.68 ± 0.01 | 1.014 ± 0.002 | 0.03 ± 0.07 | <47 | 3 | N | N | … | … | … | … | … | … | n | G |
| 37949 | 10430328-6536231 | -12.5 ± 0.2 | 4716 ± 30 | 2.53 ± 0.19 | 1.024 ± 0.003 | -0.08 ± 0.08 | <23 | 3 | N | N | … | … | … | … | … | … | n | G |
| 38205 | 10470865-6417021 | -17.2 ± 0.2 | 4751 ± 23 | 2.56 ± 0.10 | 1.023 ± 0.002 | -0.06 ± 0.02 | 172 ± 2 | 1 | N | N | … | … | … | … | … | … | n | Li-rich G |
| 38052 | 10444258-6445289 | 10.4 ± 0.2 | 4922 ± 217 | … | 1.011 ± 0.004 | -0.09 ± 0.16 | <29 | 3 | N | … | … | … | … | … | … | … | n | G |
| 38206 | 10471010-6401228 | -2.0 ± 0.2 | 4659 ± 3 | 2.42 ± 0.11 | 1.022 ± 0.002 | -0.09 ± 0.04 | <21 | 3 | N | N | … | … | … | … | … | … | n | G |
| 38207 | 10471270-6424109 | 23.4 ± 0.3 | 6594 ± 124 | … | 0.991 ± 0.010 | 0.63 ± 0.10 | … | … | Y | … | Y | … | N | N | N | … | n | … |
| 53848 | 10445057-6431915 | -15.6 ± 0.2 | 6328 ± 95 | … | 0.995 ± 0.005 | -0.16 ± 0.08 | <31 | 3 | Y | … | N | N | N | Y | … | … | n | NG |
| 38208 | 10471286-6424497 | 20.0 ± 0.2 | 4565 ± 126 | … | 1.022 ± 0.006 | 0.01 ± 0.08 | <44 | 3 | N | … | … | … | … | … | … | … | n | G |
| 16 | 10430362-6403474 | 21.2 ± 0.6 | 5258 ± 88 | 3.76 ± 0.06 | 1.001 ± 0.002 | 0.00 ± 0.04 | 22 ± 5 | 1 | Y | Y | Y | N | N | Y | … | … | n | NG? |
| 38209 | 10471317-6438250 | 17.7 ± 0.2 | 4856 ± 147 | 2.61 ± 0.20 | 1.021 ± 0.002 | -0.07 ± 0.15 | <18 | 3 | N | N | … | … | … | … | … | … | n | G |
| 2555 | 10430377-6402194 | 1.9 ± 0.6 | 4925 ± 28 | 2.69 ± 0.07 | 1.021 ± 0.002 | -0.01 ± 0.09 | 20 ± 3 | 1 | N | N | … | … | … | … | … | … | n | G |
| 38210 | 10471318-6504127 | -11.8 ± 0.2 | 5064 ± 22 | … | 1.017 ± 0.003 | 0.02 ± 0.06 | <22 | 3 | N | … | … | … | … | … | … | … | n | G |
| 38211 | 10471400-6408542 | 8.8 ± 0.2 | 4549 ± 87 | 2.34 ± 0.17 | 1.021 ± 0.002 | -0.04 ± 0.01 | 30 ± 2 | 1 | N | N | … | … | … | … | … | … | n | G |
| 38212 | 10471572-6400381 | -32.2 ± 0.2 | 4539 ± 139 | … | 1.026 ± 0.003 | 0.11 ± 0.01 | <46 | 3 | N | … | … | … | … | … | … | … | n | G |
| 37950 | 10430403-6541439 | 0.7 ± 0.2 | 4745 ± 18 | 2.71 ± 0.05 | 1.008 ± 0.005 | -0.07 ± 0.05 | <35 | 3 | Y | Y | Y | N | N | Y | … | … | n | NG? |
| 38301 | 10483698-6502188 | -29.4 ± 0.2 | 4588 ± 64 | 2.47 ± 0.18 | 1.022 ± 0.004 | 0.09 ± 0.04 | <52 | 3 | N | N | … | … | … | … | … | … | n | G |
| 38057 | 10445086-6450031 | -44.0 ± 0.2 | 4581 ± 81 | … | 1.023 ± 0.002 | -0.06 ± 0.05 | <33 | 3 | N | … | … | … | … | … | … | … | n | G |
| 38302 | 10483757-6510411 | -23.6 ± 0.2 | 4494 ± 109 | … | 1.039 ± 0.003 | -0.09 ± 0.06 | <28 | 3 | N | … | … | … | … | … | … | … | n | G |
| 38303 | 10483788-6455483 | 41.1 ± 0.2 | 4680 ± 129 | … | 1.017 ± 0.003 | -0.12 ± 0.12 | <31 | 3 | N | … | … | … | … | … | … | … | n | G |
| 37951 | 10430628-6540307 | 88.8 ± 0.3 | 4673 ± 7 | 2.12 ± 0.18 | 1.034 ± 0.008 | -0.35 ± 0.07 | … | … | N | N | … | … | … | … | … | … | n | G |
| 38304 | 10483936-6327542 | 51.3 ± 0.2 | 4654 ± 93 | … | 1.030 ± 0.003 | … | 374 ± 4 | 1 | N | … | … | … | … | … | … | … | n | Li-rich G |
| 53849 | 10445126-6430237 | 82.4 ± 0.3 | 5706 ± 127 | … | 0.992 ± 0.007 | -0.59 ± 0.11 | <15 | 3 | Y | … | N | N | N | N | … | … | n | NG |
| 37952 | 10430669-6543595 | 67.1 ± 0.3 | 4666 ± 87 | 2.44 ± 0.16 | 1.019 ± 0.007 | -0.08 ± 0.07 | … | … | N | N | … | … | … | … | … | … | n | G |
| 38305 | 10483957-6323073 | 145.0 ± 0.2 | 4389 ± 228 | … | 1.043 ± 0.006 | -0.36 ± 0.21 | … | … | N | … | … | … | … | … | … | … | n | G |
| 2556 | 10430672-6535134 | 23.4 ± 0.6 | 4752 ± 15 | 2.77 ± 0.08 | 1.009 ± 0.002 | 0.26 ± 0.04 | 50 ± 16 | 1 | Y | N | Y | N | Y | N | … | … | n | NG? |
| 38306 | 10484233-6406100 | 21.3 ± 0.2 | 4525 ± 137 | … | 1.034 ± 0.003 | -0.09 ± 0.10 | <23 | 3 | N | … | … | … | … | … | … | … | n | G |
| 38307 | 10484293-6444071 | -4.8 ± 0.2 | 5122 ± 6 | … | 1.015 ± 0.001 | 0.03 ± 0.02 | <22 | 3 | N | … | … | … | … | … | … | … | n | G |
| 38308 | 10484296-6324481 | 31.0 ± 0.2 | 4530 ± 153 | … | 1.026 ± 0.003 | 0.12 ± 0.03 | <54 | 3 | N | … | … | … | … | … | … | … | n | G |
| 38309 | 10484378-6428500 | 19.1 ± 0.2 | 4616 ± 102 | … | 1.030 ± 0.005 | … | <31 | 3 | N | … | … | … | … | … | … | … | n | G |
| 38310 | 10484452-6348240 | 14.3 ± 0.3 | 3487 ± 64 | … | 0.845 ± 0.015 | -0.22 ± 0.15 | … | … | Y | … | Y | … | N | Y | Y | … | n | … |
| 38311 | 10484598-6431535 | 9.9 ± 0.2 | 4615 ± 57 | 2.50 ± 0.19 | 1.025 ± 0.003 | 0.06 ± 0.01 | <48 | 3 | N | N | … | … | … | … | … | … | n | G |
| 37953 | 10430727-6456318 | -11.0 ± 0.2 | 4626 ± 147 | … | 1.028 ± 0.003 | -0.08 ± 0.06 | 447 ± 6 | 1 | N | … | … | … | … | … | … | … | n | Li-rich G |
| 38312 | 10484696-6349513 | -3.5 ± 0.2 | 4975 ± 147 | … | 1.022 ± 0.002 | -0.08 ± 0.14 | <19 | 3 | N | … | … | … | … | … | … | … | n | G |
| 38313 | 10484718-6350422 | 5.4 ± 0.2 | 4583 ± 104 | … | 1.026 ± 0.002 | 0.02 ± 0.06 | <43 | 3 | N | … | … | … | … | … | … | … | n | G |
| 40 | 10484720-6403081 | -1.3 ± 0.6 | 6844 ± 4 | 4.51 ± 0.22 | … | -0.08 ± 0.02 | <25 | 3 | … | Y | Y | N | N | Y | N | … | n | NG |
| 37954 | 10430767-6541535 | -16.3 ± 0.2 | 4783 ± 82 | 2.73 ± 0.05 | 1.009 ± 0.003 | -0.06 ± 0.09 | <29 | 3 | Y | N | N | N | N | Y | … | … | n | NG? |
| 38314 | 10484812-6458476 | 72.9 ± 0.2 | 4785 ± 142 | … | 1.023 ± 0.003 | … | <23 | 3 | N | … | … | … | … | … | … | … | n | G |
| 37955 | 10430781-6403518 | 15.8 ± 0.2 | 4893 ± 90 | … | 1.020 ± 0.002 | -0.07 ± 0.12 | <21 | 3 | N | … | … | … | … | … | … | … | n | G |
| 38315 | 10484864-6347560 | 2.6 ± 0.2 | 4911 ± 163 | … | 1.018 ± 0.002 | -0.06 ± 0.11 | 27 ± 4 | 1 | N | … | … | … | … | … | … | … | n | G |
| 38316 | 10484889-6516406 | -1.9 ± 0.2 | 4620 ± 42 | 2.39 ± 0.13 | 1.021 ± 0.005 | -0.08 ± 0.03 | <31 | 3 | N | N | … | … | … | … | … | … | n | G |
| 38317 | 10484891-6354029 | 7.3 ± 1.8 | 5974 ± 247 | … | 0.952 ± 0.021 | 0.56 ± 0.14 | … | … | Y | … | Y | … | N | N | … | … | n | … |

**Table C.6.** continued.

| ID | CNAME | RV (km s$^{-1}$) | $T_{\rm eff}$ (K) | $logg$ (dex) | $\gamma^a$ | [Fe/H] (dex) | EW(Li)$^b$ (mÅ) | EW(Li) error flag$^c$ | \multicolumn{6}{c}{Membership} | \multicolumn{2}{c}{Gaia studies} | Final$^e$ | NMs with Li$^f$ |
| | | | | | | | | | $\gamma$ | $logg$ | RV | Li | H$\alpha$ | [Fe/H] | Randich$^d$ | Cantat-Gaudin$^d$ | | |
|---|---|---|---|---|---|---|---|---|---|---|---|---|---|---|---|---|---|---|
| 53729 | 10430807-6425389 | 9.8 ± 0.2 | 6695 ± 47 | … | 0.998 ± 0.003 | 0.04 ± 0.04 | 45 ± 4 | 1 | Y | … | Y | N | N | Y | … | … | n | NG |
| 53747 | 10433396-6428314 | -16.7 ± 0.2 | 4891 ± 132 | … | 1.021 ± 0.003 | -0.08 ± 0.14 | 52 ± 3 | 1 | N | … | … | … | … | … | … | … | n | G |
| 38318 | 10485054-6512026 | -17.9 ± 0.2 | 4982 ± 225 | … | 1.016 ± 0.003 | -0.14 ± 0.18 | <29 | 3 | N | … | … | … | … | … | … | … | n | G |
| 38319 | 10485165-6357419 | 17.1 ± 0.3 | 3341 ± 67 | … | 0.864 ± 0.013 | -0.24 ± 0.13 | … | … | Y | … | Y | … | N | Y | Y | Y | n | … |
| 38320 | 10485180-6405150 | 57.6 ± 0.2 | 3766 ± 84 | … | 0.830 ± 0.005 | -0.17 ± 0.14 | … | … | Y | … | N | … | … | Y | N | … | n | … |
| 38321 | 10485215-6506161 | 23.6 ± 0.2 | 4700 ± 99 | … | 1.019 ± 0.005 | … | <55 | 3 | N | … | … | … | … | … | … | … | n | G |
| 38322 | 10485377-6456006 | 1.3 ± 0.2 | 4673 ± 28 | 2.46 ± 0.02 | 1.017 ± 0.003 | -0.05 ± 0.01 | <33 | 3 | N | N | … | … | … | … | … | … | n | G |
| 37974 | 10433438-6547169 | -14.5 ± 0.2 | 4597 ± 85 | 2.39 ± 0.13 | 1.019 ± 0.005 | -0.01 ± 0.02 | 17 ± … | 3 | N | N | … | … | … | … | … | … | n | G |
| 38323 | 10485383-6434483 | 38.6 ± 0.2 | 4563 ± 163 | … | 1.021 ± 0.004 | -0.13 ± 0.09 | <31 | 3 | N | … | … | … | … | … | … | … | n | G |
| 38060 | 10445892-6351294 | 46.2 ± 0.2 | 4712 ± 114 | 2.54 ± 0.16 | 1.018 ± 0.001 | 0.02 ± 0.06 | 53 ± 6 | 1 | N | N | … | … | … | … | … | … | n | G |
| 38409 | 10500351-6401496 | 32.7 ± 0.2 | 4507 ± 145 | 2.20 ± 0.19 | 1.023 ± 0.002 | -0.03 ± 0.04 | <44 | 3 | N | N | … | … | … | … | … | … | n | G |
| 53864 | 10445897-6418261 | 11.8 ± 0.2 | 4560 ± 113 | 2.74 ± 0.11 | 0.999 ± 0.004 | 0.11 ± 0.04 | <57 | 3 | Y | N | Y | N | N | Y | … | … | n | NG? |
| 38410 | 10500371-6347560 | 29.1 ± 0.2 | 5044 ± 186 | … | 0.968 ± 0.004 | 0.05 ± 0.08 | <26 | 3 | Y | … | Y | N | N | Y | … | … | n | NG |
| 2559 | 10433491-6452243 | -10.3 ± 0.6 | 4641 ± 25 | 2.45 ± 0.08 | 1.018 ± 0.002 | 0.10 ± 0.04 | 33 ± 7 | 1 | N | N | … | … | … | … | … | … | n | G |
| 38411 | 10500394-6339263 | -17.3 ± 0.2 | 5027 ± 133 | … | 1.029 ± 0.002 | -0.12 ± 0.14 | <23 | 3 | N | … | … | … | … | … | … | … | n | G |
| 38412 | 10500455-6513085 | -34.9 ± 0.2 | 4652 ± 109 | … | 1.023 ± 0.005 | -0.09 ± 0.08 | <24 | 3 | N | … | … | … | … | … | … | … | n | G |
| 37975 | 10433519-6357071 | 13.1 ± 0.2 | 4657 ± 66 | 2.38 ± 0.16 | 1.027 ± 0.003 | -0.04 ± 0.11 | <34 | 3 | N | N | … | … | … | … | … | … | n | G |
| 38413 | 10500469-6349097 | 3.2 ± 0.2 | 5376 ± 48 | … | 1.017 ± 0.003 | 0.01 ± 0.06 | 95 ± 2 | 1 | N | … | … | … | … | … | … | … | n | … |
| 37976 | 10433545-6538008 | 28.1 ± 0.3 | 5109 ± 123 | … | 1.017 ± 0.007 | … | <13 | 3 | N | … | … | … | … | … | … | … | n | G |
| 2571 | 10445962-6502191 | 25.3 ± 0.6 | 5795 ± 62 | 4.39 ± 0.12 | … | 0.07 ± 0.12 | <194 | 3 | … | Y | Y | Y | Y | Y | N | Y | n | NG |
| 38414 | 10500665-6346030 | -1.2 ± 0.2 | 4929 ± 127 | 2.97 ± 0.19 | 1.010 ± 0.004 | -0.04 ± 0.07 | <22 | 3 | N | N | … | … | … | … | … | … | n | G |
| 38415 | 10500695-6351117 | -4.6 ± 0.2 | 4623 ± 37 | 2.33 ± 0.16 | 1.035 ± 0.003 | -0.05 ± 0.01 | 43 ± 3 | 1 | N | N | … | … | … | … | … | … | n | G |
| 38416 | 10500701-6340453 | 35.4 ± 0.2 | 4691 ± 175 | … | 1.025 ± 0.004 | -0.04 ± 0.18 | <24 | 3 | N | … | … | … | … | … | … | … | n | G |
| 53748 | 10433584-6435338 | 37.3 ± 0.2 | 4252 ± 259 | … | 1.028 ± 0.004 | 0.08 ± 0.16 | <57 | 3 | N | … | … | … | … | … | … | … | n | G |
| 38417 | 10500782-6418549 | -29.0 ± 0.2 | 4615 ± 154 | … | 1.033 ± 0.003 | -0.01 ± 0.11 | <42 | 3 | N | … | … | … | … | … | … | … | n | G |
| 38418 | 10500886-6436060 | -6.5 ± 0.2 | 4679 ± 34 | 2.46 ± 0.07 | 1.022 ± 0.003 | -0.07 ± 0.02 | <27 | 3 | N | N | … | … | … | … | … | … | n | G |
| 38061 | 10450008-6346403 | 8.3 ± 0.2 | 4736 ± 82 | 2.58 ± 0.01 | 1.020 ± 0.003 | 0.03 ± 0.03 | … | … | N | N | … | … | … | … | … | … | n | G |
| 38419 | 10500963-6511404 | -37.8 ± 0.2 | 4695 ± 59 | … | 1.024 ± 0.003 | -0.10 ± 0.10 | 22 ± 10 | 1 | N | … | … | … | … | … | … | … | n | G |
| 37977 | 10433640-6449556 | 31.3 ± 0.2 | 4533 ± 155 | 2.48 ± 0.13 | 1.009 ± 0.002 | 0.12 ± 0.04 | <43 | 3 | Y | N | Y | N | N | Y | … | … | n | NG? |
| 38420 | 10500991-6334103 | -1.9 ± 0.3 | 4949 ± 152 | … | 1.014 ± 0.005 | -0.04 ± 0.07 | <18 | 3 | N | … | … | … | … | … | … | … | n | G |
| 38421 | 10500998-6350251 | 15.4 ± 0.3 | 4634 ± 51 | 2.42 ± 0.08 | 1.018 ± 0.008 | 0.01 ± 0.11 | <31 | 3 | N | N | … | … | … | … | … | … | n | G |
| 37978 | 10433660-6502214 | 9.0 ± 2.6 | … | … | … | … | … | … | … | … | … | … | … | … | … | … | n | … |
| 2599 | 10501173-6332406 | -14.7 ± 0.6 | 4871 ± 64 | 2.77 ± 0.09 | 1.017 ± 0.002 | 0.24 ± 0.05 | 37 ± 11 | 1 | N | N | … | … | … | … | … | … | n | … |
| 38422 | 10501179-6342011 | 27.0 ± 0.2 | 4505 ± 201 | … | 1.014 ± 0.005 | 0.14 ± 0.04 | <53 | 3 | N | … | … | … | … | … | … | … | n | G |
| 38423 | 10501199-6424342 | 38.1 ± 0.2 | 4650 ± 108 | … | 1.029 ± 0.003 | … | <43 | 3 | N | … | … | … | … | … | … | … | n | G |
| 38062 | 10450017-6501537 | -25.8 ± 0.2 | 5019 ± 191 | … | 1.021 ± 0.002 | -0.10 ± 0.13 | <18 | 3 | N | … | … | … | … | … | … | … | n | G |
| 38424 | 10501219-6433377 | 26.5 ± 0.2 | 4574 ± 75 | 2.31 ± 0.16 | 1.027 ± 0.003 | -0.03 ± 0.01 | <59 | 3 | N | N | … | … | … | … | … | … | n | G |
| 37979 | 10433716-6539232 | 37.5 ± 0.3 | 5032 ± 158 | 3.53 ± 0.12 | 0.998 ± 0.006 | -0.11 ± 0.17 | <40 | 3 | Y | Y | Y | N | N | Y | … | … | n | NG |
| 38425 | 10501350-6516592 | -18.1 ± 0.2 | 4877 ± 15 | 2.71 ± 0.04 | 1.017 ± 0.003 | 0.00 ± 0.02 | … | … | N | N | … | … | … | … | … | … | n | G |
| 38426 | 10501398-6341270 | -1.1 ± 0.2 | 5921 ± 203 | 4.20 ± 0.10 | 0.997 ± 0.001 | 0.25 ± 0.08 | … | … | Y | Y | Y | … | N | N | … | … | n | … |
| 38427 | 10501517-6330505 | -7.2 ± 0.2 | 4939 ± 178 | … | 1.024 ± 0.004 | -0.07 ± 0.07 | <35 | 3 | N | … | … | … | … | … | … | … | n | G |
| 53873 | 10450771-6430461 | -15.1 ± 0.2 | 4577 ± 86 | … | 1.030 ± 0.001 | … | <47 | 3 | N | … | … | … | … | … | … | … | n | G |
| 53749 | 10433732-6427022 | -10.6 ± 0.2 | 4920 ± 151 | … | 1.023 ± 0.002 | -0.06 ± 0.10 | <15 | 3 | N | … | … | … | … | … | … | … | n | G |
| 38428 | 10501544-6330098 | 14.9 ± 0.2 | 4889 ± 159 | 2.70 ± 0.12 | 1.014 ± 0.003 | -0.03 ± 0.06 | <37 | 3 | N | N | … | … | … | … | … | … | n | G |
| 38429 | 10501583-6352425 | 31.3 ± 0.2 | 5114 ± 26 | … | 1.031 ± 0.003 | 0.05 ± 0.05 | <24 | 3 | N | … | … | … | … | … | … | … | n | G |
| 53874 | 10450779-6425430 | 67.4 ± 0.2 | 4345 ± 270 | … | 1.027 ± 0.002 | 0.11 ± 0.12 | <51 | 3 | N | … | … | … | … | … | … | … | n | G |
| 38430 | 10501591-6401093 | -7.8 ± 1.1 | 4951 ± 172 | … | 1.019 ± 0.017 | … | … | … | N | … | … | … | … | … | … | … | n | G |
| 53765 | 10434864-6419323 | -2.7 ± 0.2 | 5874 ± 86 | 4.16 ± 0.05 | 0.994 ± 0.004 | -0.12 ± 0.02 | 39 ± 3 | 1 | Y | Y | Y | N | N | Y | … | … | n | NG |
| 38431 | 10501661-6336275 | 5.2 ± 0.3 | 5179 ± 97 | … | 1.009 ± 0.007 | … | <21 | 3 | Y | … | Y | N | N | … | … | … | n | NG? |
| 38518 | 10510610-6328389 | -31.7 ± 0.2 | 4848 ± 74 | 2.76 ± 0.13 | 1.014 ± 0.005 | 0.00 ± 0.02 | <22 | 3 | N | N | … | … | … | … | … | … | n | G |
| 37995 | 10434875-6346481 | 17.5 ± 0.4 | 3335 ± 58 | 4.67 ± 0.10 | 0.848 ± 0.015 | -0.28 ± 0.14 | … | … | Y | Y | Y | … | N | N | Y | Y | n | … |
| 38519 | 10510674-6334494 | 40.4 ± 0.2 | 4196 ± 161 | 4.57 ± 0.02 | 0.875 ± 0.005 | 0.03 ± 0.22 | … | … | Y | Y | Y | … | … | Y | N | … | n | … |
| 38520 | 10510694-6343350 | 10.1 ± 0.2 | 4758 ± 72 | 2.59 ± 0.09 | 1.013 ± 0.003 | 0.01 ± 0.10 | <27 | 3 | N | N | … | … | … | … | … | … | n | G |
| 38521 | 10510705-6340038 | -20.3 ± 0.2 | 4622 ± 87 | 2.40 ± 0.14 | 1.021 ± 0.005 | -0.01 ± 0.02 | <41 | 3 | N | N | … | … | … | … | … | … | n | G |
| 38522 | 10510727-6423167 | 20.5 ± 0.2 | 4538 ± 152 | … | 1.017 ± 0.002 | 0.15 ± 0.05 | <56 | 3 | N | … | … | … | … | … | … | … | n | G |
| 38523 | 10510795-6344110 | -13.1 ± 0.2 | 4729 ± 9 | 2.51 ± 0.18 | 1.025 ± 0.002 | -0.11 ± 0.10 | <22 | 3 | N | N | … | … | … | … | … | … | n | G |
| 38524 | 10510834-6411319 | -2.1 ± 0.2 | 4844 ± 137 | 2.65 ± 0.14 | 1.018 ± 0.003 | -0.03 ± 0.05 | 41 ± 2 | 1 | N | N | … | … | … | … | … | … | n | G |
| 53766 | 10434933-6433596 | -31.2 ± 0.2 | 4800 ± 149 | … | 1.024 ± 0.001 | -0.08 ± 0.16 | <23 | 3 | N | … | … | … | … | … | … | … | n | G |
| 38525 | 10510937-6328265 | 3.9 ± 0.2 | 5098 ± 79 | 3.04 ± 0.08 | 1.026 ± 0.005 | 0.08 ± 0.06 | 89 ± 3 | 1 | N | N | … | … | … | … | … | … | n | Li-rich G |

| ID | CNAME | RV (km s$^{-1}$) | $T_{\rm eff}$ (K) | $\log g$ (dex) | $\gamma^a$ | [Fe/H] (dex) | EW(Li)$^b$ (mÅ) | EW(Li) error flag$^c$ | $\gamma$ | $\log g$ | Membership RV | Li | H$\alpha$ | [Fe/H] | Gaia studies Randich$^d$ | Cantat-Gaudin$^d$ | Final$^e$ | NMs with Li$^f$ |
|---|---|---|---|---|---|---|---|---|---|---|---|---|---|---|---|---|---|---|
| 37996 | 10434933-6534485 | -15.9 ± 0.2 | 4692 ± 53 | 2.53 ± 0.12 | 1.016 ± 0.005 | -0.05 ± 0.01 | <34 | 3 | N | N | … | … | … | … | … | … | n | G |
| 38526 | 10510967-6334037 | 63.8 ± 0.2 | 4493 ± 179 | … | 1.030 ± 0.005 | -0.01 ± 0.02 | <45 | 3 | N | … | … | … | … | … | … | … | n | G |
| 38527 | 10511009-6424244 | -13.6 ± 0.2 | 4740 ± 215 | … | 1.019 ± 0.002 | -0.11 ± 0.13 | <39 | 3 | N | … | … | … | … | … | … | … | n | G |
| 53767 | 10434937-6435297 | 0.9 ± 0.2 | 4907 ± 21 | 2.76 ± 0.07 | 1.017 ± 0.003 | -0.04 ± 0.07 | <21 | 3 | N | N | … | … | … | … | … | … | n | G |
| 38528 | 10511222-6410311 | 4.2 ± 0.2 | 4761 ± 68 | 2.61 ± 0.08 | 1.014 ± 0.003 | 0.04 ± 0.04 | <30 | 3 | N | N | … | … | … | … | … | … | n | G |
| 38529 | 10511293-6343157 | -32.1 ± 0.2 | 5021 ± 120 | … | 1.021 ± 0.003 | … | <21 | 3 | N | … | … | … | … | … | … | … | n | G |
| 38530 | 10511322-6341296 | -5.1 ± 0.2 | 4958 ± 100 | … | 1.015 ± 0.002 | -0.06 ± 0.10 | <19 | 3 | N | … | … | … | … | … | … | … | n | G |
| 38531 | 10511760-6341056 | 49.8 ± 0.2 | 4974 ± 184 | … | 1.007 ± 0.005 | -0.12 ± 0.17 | <24 | 3 | Y | … | N | N | N | Y | … | … | n | NG? |
| 38532 | 10511761-6331076 | 3.0 ± 0.2 | 4507 ± 61 | 2.14 ± 0.13 | 1.035 ± 0.002 | -0.06 ± 0.07 | <45 | 3 | N | N | … | … | … | … | … | … | n | G |
| 38533 | 10511767-6339166 | 47.1 ± 0.2 | 4961 ± 321 | … | 1.007 ± 0.004 | -0.29 ± 0.28 | <10 | 3 | Y | … | N | N | N | N | … | … | n | … |
| 38070 | 10450850-6344313 | -12.2 ± 0.2 | 4711 ± 123 | 2.54 ± 0.18 | 1.028 ± 0.003 | -0.06 ± 0.03 | <25 | 3 | N | N | … | … | … | … | … | … | n | G |
| 37997 | 10435021-6445022 | 17.6 ± 0.2 | 3394 ± 133 | 4.67 ± 0.13 | 0.849 ± 0.010 | -0.28 ± 0.13 | … | … | Y | Y | N | N | N | N | Y | Y | n | … |
| 38534 | 10511789-6411527 | 3.6 ± 0.4 | 6151 ± 55 | … | 0.987 ± 0.009 | -0.10 ± 0.13 | <16 | 3 | Y | … | Y | N | N | Y | N | … | n | NG |
| 38535 | 10511877-6350043 | -21.7 ± 0.2 | 4847 ± 155 | … | 1.006 ± 0.002 | … | 92 ± 9 | 1 | Y | … | N | Y | N | … | … | … | n | NG? |
| 38536 | 10511896-6350474 | 53.5 ± 0.2 | 4554 ± 114 | 2.37 ± 0.18 | 1.020 ± 0.003 | 0.05 ± 0.01 | <46 | 3 | N | N | … | … | … | … | … | … | n | G |
| 38537 | 10511995-6343120 | -17.1 ± 0.2 | 4602 ± 61 | 2.34 ± 0.18 | 1.027 ± 0.003 | -0.06 ± 0.01 | <34 | 3 | N | N | … | … | … | … | … | … | n | G |
| 38538 | 10512052-6337393 | -2.8 ± 0.2 | 4683 ± 41 | 2.55 ± 0.15 | 1.016 ± 0.004 | -0.07 ± 0.04 | <42 | 3 | N | N | … | … | … | … | … | … | n | G |
| 38009 | 10435961-6400453 | 31.5 ± 251.4 | … | … | … | … | … | … | … | … | … | … | … | … | … | … | n | … |
| 38539 | 10512191-6411500 | -28.5 ± 0.2 | 5095 ± 306 | 2.91 ± 0.18 | 1.029 ± 0.003 | 0.04 ± 0.08 | <41 | 3 | N | N | … | … | … | … | … | … | n | G |
| 38010 | 10435971-6401047 | 12.1 ± 6.3 | … | … | … | … | … | … | … | … | … | … | … | … | … | … | n | … |
| 38540 | 10512211-6349464 | 39.0 ± 0.2 | 4607 ± 40 | 2.39 ± 0.15 | 1.025 ± 0.003 | -0.01 ± 0.04 | <42 | 3 | N | N | … | … | … | … | … | … | n | G |
| 38541 | 10512394-6413419 | 30.2 ± 0.2 | 4321 ± 126 | 4.57 ± 0.11 | 0.900 ± 0.003 | 0.01 ± 0.15 | <26 | 3 | Y | Y | Y | N | N | Y | … | … | n | NG |
| 2611 | 10522844-6416489 | 11.4 ± 0.6 | 5009 ± 25 | 2.60 ± 0.09 | 1.017 ± 0.001 | -0.25 ± 0.03 | 20 ± 5 | 1 | N | N | … | … | … | … | … | … | n | G |
| 38078 | 10451644-6329248 | -32.9 ± 0.2 | 4655 ± 32 | 2.41 ± 0.11 | 1.021 ± 0.002 | -0.07 ± 0.04 | <30 | 3 | N | N | … | … | … | … | … | … | n | G |
| 53784 | 10440019-6419373 | 54.6 ± 0.2 | 4542 ± 121 | 2.28 ± 0.18 | 1.020 ± 0.005 | -0.05 ± 0.04 | <41 | 3 | N | N | … | … | … | … | … | … | n | G |
| 38627 | 10523038-6339191 | 79.7 ± 0.3 | 5063 ± 25 | 2.99 ± 0.15 | 1.015 ± 0.007 | -0.11 ± 0.13 | <18 | 3 | N | N | … | … | … | … | … | … | n | G |
| 38079 | 10451668-6418497 | 8.2 ± 0.2 | 4931 ± 92 | 2.91 ± 0.18 | 1.011 ± 0.001 | 0.01 ± 0.06 | 33 ± 1 | 1 | N | N | … | … | … | … | … | … | n | G |
| 38628 | 10523136-6411389 | -3.4 ± 0.2 | 4993 ± 163 | … | 1.019 ± 0.002 | -0.15 ± 0.19 | <12 | 3 | N | … | … | … | … | … | … | … | n | G |
| 38629 | 10523298-6332269 | -5.5 ± 0.2 | 4582 ± 38 | 2.39 ± 0.11 | 1.015 ± 0.004 | -0.05 ± 0.02 | 37 | 3 | N | N | … | … | … | … | … | … | n | G |
| 38630 | 10523313-6343305 | -29.6 ± 0.2 | 4794 ± 6 | 2.60 ± 0.06 | 1.022 ± 0.002 | -0.02 ± 0.01 | <36 | 3 | N | N | … | … | … | … | … | … | n | G |
| 53785 | 10440089-6427030 | 56.0 ± 0.2 | 4636 ± 80 | … | 0.990 ± 0.006 | 0.05 ± 0.05 | <37 | 3 | Y | … | N | N | N | Y | … | … | n | NG |
| 38631 | 10523375-6335544 | 3.1 ± 0.2 | 4484 ± 199 | … | 1.012 ± 0.005 | 0.15 ± 0.02 | <56 | 3 | N | … | … | … | … | … | … | … | n | G |
| 38632 | 10523594-6426029 | -10.2 ± 0.2 | 4720 ± 13 | 2.50 ± 0.16 | 1.031 ± 0.002 | -0.04 ± 0.01 | <24 | 3 | N | N | … | … | … | … | … | … | n | G |
| 38633 | 10523654-6340504 | -27.5 ± 0.2 | 5004 ± 109 | … | 1.009 ± 0.004 | -0.07 ± 0.11 | <32 | 3 | Y | … | N | N | N | Y | … | … | n | NG? |
| 38011 | 10440108-6538066 | 4.7 ± 0.2 | 4659 ± 120 | … | 1.018 ± 0.004 | -0.10 ± 0.17 | <33 | 3 | N | … | … | … | … | … | … | … | n | G |
| 38634 | 10523726-6347280 | -18.7 ± 0.2 | 5136 ± 24 | … | 1.017 ± 0.003 | -0.05 ± 0.07 | <23 | 3 | N | … | … | … | … | … | … | … | n | G |
| 38635 | 10523762-6347511 | 19.0 ± 0.2 | 4646 ± 81 | 2.49 ± 0.14 | 1.016 ± 0.004 | 0.05 ± 0.09 | <50 | 3 | N | N | … | … | … | … | … | … | n | G |
| 38636 | 10523908-6347073 | -10.0 ± 0.2 | 4547 ± 80 | 2.29 ± 0.20 | 1.026 ± 0.005 | -0.07 ± 0.03 | <30 | 3 | N | N | … | … | … | … | … | … | n | G |
| 38637 | 10523990-6417427 | 4.1 ± 0.2 | 4658 ± 166 | … | 1.018 ± 0.002 | 0.04 ± 0.05 | <33 | 3 | N | … | … | … | … | … | … | … | n | G |
| 38638 | 10524180-6336229 | -29.1 ± 0.2 | 5031 ± 44 | 3.05 ± 0.17 | 1.017 ± 0.003 | 0.03 ± 0.02 | <30 | 3 | N | N | … | … | … | … | … | … | n | G |
| 38012 | 10440169-6349502 | 6.8 ± 0.2 | 4966 ± 50 | 2.99 ± 0.13 | 1.001 ± 0.001 | 0.01 ± 0.03 | <29 | 3 | N | N | … | … | … | … | … | … | n | G |
| 38080 | 10451743-6456434 | 17.8 ± 0.3 | 3411 ± 106 | 4.63 ± 0.15 | 0.845 ± 0.007 | -0.25 ± 0.13 | … | … | Y | Y | Y | N | … | Y | Y | … | n | … |
| 38639 | 10524253-6338194 | -35.6 ± 0.2 | 4718 ± 18 | 2.53 ± 0.09 | 1.023 ± 0.004 | -0.02 ± 0.03 | … | … | N | N | … | … | … | … | … | … | n | G |
| 38640 | 10524280-6346255 | 32.3 ± 0.2 | 4604 ± 69 | 2.33 ± 0.12 | 1.022 ± 0.004 | -0.06 ± 0.04 | <39 | 3 | N | N | … | … | … | … | … | … | n | G |
| 53887 | 10451757-6425500 | 389.8 ± 1.7 | 3509 ± 30 | … | 0.895 ± 0.018 | … | … | … | Y | … | N | … | … | … | … | … | n | … |
| 38641 | 10524325-6414001 | 21.7 ± 0.2 | 4624 ± 102 | … | 1.027 ± 0.002 | … | <40 | 3 | N | … | … | … | … | … | … | … | n | G |
| 38642 | 10524372-6428498 | 8.7 ± 0.2 | 5121 ± 17 | … | 1.019 ± 0.002 | 0.04 ± 0.07 | 38 ± 5 | 1 | N | … | … | … | … | … | … | … | n | G |
| 38023 | 10441075-6539477 | 25.3 ± 0.3 | 4914 ± 154 | 2.30 ± 0.18 | 1.033 ± 0.010 | -0.23 ± 0.03 | <54 | 3 | N | N | … | … | … | … | … | … | n | G |
| 38643 | 10524624-6338259 | 25.6 ± 0.2 | 4704 ± 136 | 2.48 ± 0.20 | 1.017 ± 0.004 | -0.07 ± 0.16 | <36 | 3 | N | N | … | … | … | … | … | … | n | G |
| 38644 | 10524652-6337304 | 0.3 ± 0.2 | 4980 ± 139 | … | 1.019 ± 0.004 | -0.06 ± 0.12 | <18 | 3 | N | … | … | … | … | … | … | … | n | G |
| 38645 | 10524669-6430231 | -17.8 ± 0.2 | 4871 ± 44 | 2.77 ± 0.01 | 1.012 ± 0.003 | 0.00 ± 0.07 | <27 | 3 | N | N | … | … | … | … | … | … | n | G |
| 38085 | 10452464-6330053 | -45.2 ± 0.2 | 4529 ± 126 | … | 1.032 ± 0.002 | 0.04 ± 0.10 | <51 | 3 | N | … | … | … | … | … | … | … | n | G |
| 38024 | 10441157-6441494 | -0.2 ± 0.2 | 4731 ± 33 | 2.52 ± 0.18 | 1.031 ± 0.004 | -0.08 ± 0.06 | <28 | 3 | N | N | … | … | … | … | … | … | n | G |
| 38025 | 10441160-6541186 | 95.7 ± 0.3 | 4598 ± 95 | … | 1.042 ± 0.007 | … | <44 | 3 | N | … | … | … | … | … | … | … | n | G |
| 38086 | 10452481-6502523 | 30.6 ± 0.2 | 4604 ± 168 | … | 1.022 ± 0.002 | 0.01 ± 0.13 | <37 | 3 | N | … | … | … | … | … | … | … | n | G |
| 53791 | 10441179-6435203 | -5.2 ± 0.2 | 6118 ± 105 | 3.87 ± 0.04 | 1.004 ± 0.003 | -0.13 ± 0.08 | 51 ± 3 | 1 | Y | Y | Y | N | N | Y | … | … | n | NG? |
| 53792 | 10441202-6438146 | -1.9 ± 0.2 | 4980 ± 162 | … | 1.016 ± 0.003 | -0.05 ± 0.10 | <31 | 3 | N | … | … | … | … | … | … | … | n | G |
| 38087 | 10452515-6326437 | 36.0 ± 0.2 | 4640 ± 98 | … | 1.021 ± 0.002 | … | <47 | 3 | N | … | … | … | … | … | … | … | n | G |
| 2566 | 10441221-6403452 | 49.6 ± 0.6 | 4824 ± 53 | 2.75 ± 0.07 | 1.013 ± 0.002 | -0.42 ± 0.02 | <12 | 3 | N | N | … | … | … | … | … | … | n | G |

**Table C.6.** continued.

| ID | CNAME | RV (km s$^{-1}$) | $T_{\text{eff}}$ (K) | $\log g$ (dex) | $\gamma^a$ | [Fe/H] (dex) | EW(Li)$^b$ (mÅ) | EW(Li) error flag$^c$ | \multicolumn{6}{c}{Membership} | \multicolumn{2}{c}{Gaia studies} | Final$^e$ | NMs with Li$^f$ |
|---|---|---|---|---|---|---|---|---|---|---|---|---|---|---|---|---|---|---|
| | | | | | | | | | $\gamma$ | $\log g$ | RV | Li | H$\alpha$ | [Fe/H] | Randich$^d$ | Cantat-Gaudin$^d$ | | |
| 38026 | 10441225-6542341 | -5.9 ± 0.2 | 4697 ± 50 | 2.55 ± 0.10 | 1.011 ± 0.006 | 0.00 ± 0.08 | <41 | 3 | N | N | … | … | … | … | … | … | n | G |
| 38088 | 10452558-6419430 | 16.8 ± 0.6 | 3393 ± 70 | 4.66 ± 0.11 | 0.839 ± 0.010 | -0.26 ± 0.13 | <100 | 3 | Y | Y | Y | Y | Y | Y | Y | Y | Y | … |
| 38089 | 10452560-6333465 | 12.5 ± 0.2 | 4381 ± 208 | … | 1.026 ± 0.002 | 0.15 ± 0.11 | <43 | 3 | N | … | … | … | … | … | … | … | n | G |
| 53901 | 10452567-6428589 | -42.6 ± 0.2 | 4821 ± 14 | … | 0.945 ± 0.005 | -0.07 ± 0.07 | <21 | 3 | Y | … | N | N | N | Y | … | … | n | NG |
| 37415 | 10291158-6447022 | 11.8 ± 0.2 | 4818 ± 121 | 2.64 ± 0.19 | 1.017 ± 0.003 | -0.07 ± 0.14 | <25 | 3 | N | N | … | … | … | … | … | … | n | G |
| 37416 | 10291295-6441290 | 4.5 ± 0.3 | 5022 ± 167 | … | 1.015 ± 0.005 | -0.07 ± 0.15 | 23 ± 7 | 1 | N | … | … | … | … | … | … | … | n | G |
| 37435 | 10293809-6440219 | -2.4 ± 0.3 | 4803 ± 183 | 3.03 ± 0.07 | 0.999 ± 0.007 | -0.04 ± 0.07 | <45 | 3 | Y | N | Y | N | Y | Y | … | … | n | NG? |
| 37436 | 10293880-6441042 | 12.0 ± 0.2 | 4754 ± 71 | 2.55 ± 0.12 | 1.020 ± 0.005 | -0.02 ± 0.07 | … | … | N | N | … | … | … | … | … | … | n | G |
| 37437 | 10293907-6329214 | 54.8 ± 0.2 | 4636 ± 112 | 2.45 ± 0.18 | 1.021 ± 0.006 | -0.03 ± 0.02 | <51 | 3 | N | N | … | … | … | … | … | … | n | G |
| 37438 | 10293996-6442405 | -8.4 ± 0.3 | 5009 ± 180 | … | 1.004 ± 0.006 | -0.18 ± 0.24 | <11 | 3 | Y | … | Y | N | N | Y | … | … | n | NG? |
| 37439 | 10294018-6445202 | 35.9 ± 0.2 | 4849 ± 28 | … | 0.996 ± 0.004 | 0.01 ± 0.06 | <30 | 3 | Y | … | Y | N | N | Y | … | … | n | NG |
| 37440 | 10294461-6443591 | 98.5 ± 0.3 | 4894 ± 111 | … | 1.028 ± 0.007 | -0.19 ± 0.16 | … | … | N | … | … | … | … | … | … | … | n | G |
| 37441 | 10294474-6450565 | 9.9 ± 0.2 | 4920 ± 139 | … | 1.017 ± 0.003 | -0.07 ± 0.12 | <35 | 3 | N | … | … | … | … | … | … | … | n | G |
| 37442 | 10294476-6447131 | 34.2 ± 0.3 | 3592 ± 88 | … | 0.820 ± 0.009 | -0.23 ± 0.14 | <100 | 3 | Y | … | Y | Y | Y | Y | N | … | n | … |
| 37443 | 10294501-6444545 | 44.3 ± 0.2 | 4714 ± 28 | 2.50 ± 0.02 | 1.018 ± 0.005 | -0.01 ± 0.03 | … | … | N | N | … | … | … | … | … | … | n | G |
| 37444 | 10294658-6439059 | 11.7 ± 0.3 | 4915 ± 231 | 3.13 ± 0.10 | 1.002 ± 0.006 | -0.06 ± 0.11 | <32 | 3 | Y | N | Y | N | N | Y | … | … | n | NG? |
| 37445 | 10294687-6328143 | 0.5 ± 0.3 | 3886 ± 97 | 4.54 ± 0.13 | 0.832 ± 0.006 | -0.17 ± 0.01 | <23 | 3 | Y | Y | Y | Y | Y | Y | N | … | n | … |
| 37446 | 10294874-6328287 | 50.4 ± 0.2 | 4876 ± 61 | 2.64 ± 0.14 | 1.020 ± 0.002 | -0.12 ± 0.08 | <22 | 3 | N | N | … | … | … | … | … | … | n | G |
| 37447 | 10294889-6445569 | 84.6 ± 0.2 | 4963 ± 163 | … | 1.012 ± 0.005 | -0.29 ± 0.22 | <18 | 3 | N | … | … | … | … | … | … | … | n | G |
| 37448 | 10294993-6444195 | 1.3 ± 0.2 | 4651 ± 1 | 2.43 ± 0.05 | 1.019 ± 0.005 | -0.05 ± 0.02 | <30 | 3 | N | N | … | … | … | … | … | … | n | G |
| 37449 | 10295051-6440356 | 17.3 ± 0.2 | 4867 ± 42 | 2.70 ± 0.15 | 1.017 ± 0.004 | -0.08 ± 0.15 | <25 | 3 | N | N | … | … | … | … | … | … | n | G |
| 37450 | 10295079-6322543 | 80.4 ± 0.2 | 4741 ± 6 | … | 1.025 ± 0.006 | -0.19 ± 0.11 | <23 | 3 | N | … | … | … | … | … | … | … | n | G |
| 37451 | 10295080-6332485 | 3.6 ± 0.2 | 4586 ± 3 | 2.23 ± 0.17 | 1.027 ± 0.002 | -0.15 ± 0.13 | <29 | 3 | N | N | … | … | … | … | … | … | n | G |
| 37452 | 10295255-6446442 | 16.9 ± 0.2 | 4804 ± 134 | 2.60 ± 0.16 | 1.018 ± 0.004 | -0.04 ± 0.11 | <26 | 3 | N | N | … | … | … | … | … | … | n | G |
| 37453 | 10295430-6442574 | -8.7 ± 0.2 | 4569 ± 17 | 2.30 ± 0.02 | 1.016 ± 0.007 | -0.06 ± 0.07 | <17 | 3 | N | N | … | … | … | … | … | … | n | G |
| 37454 | 10295498-6433449 | 48.3 ± 0.2 | 5009 ± 151 | 3.27 ± 0.20 | 1.006 ± 0.005 | -0.02 ± 0.04 | <18 | 3 | Y | N | N | N | N | Y | … | … | n | NG? |
| 37455 | 10295587-6438197 | -33.9 ± 0.2 | 5072 ± 75 | … | 1.012 ± 0.005 | … | 16 ± 3 | 1 | N | … | … | … | … | … | … | … | n | G |
| 37456 | 10295654-6441313 | 8.6 ± 0.2 | 4705 ± 46 | … | 1.022 ± 0.004 | -0.09 ± 0.12 | <24 | 3 | N | … | … | … | … | … | … | … | n | G |
| 37457 | 10295685-6328376 | 66.8 ± 0.2 | 4789 ± 10 | 2.60 ± 0.14 | 1.020 ± 0.006 | -0.14 ± 0.09 | <34 | 3 | N | N | … | … | … | … | … | … | n | G |
| 37458 | 10295726-6328035 | 106.2 ± 0.2 | 4887 ± 169 | 2.86 ± 0.20 | 1.009 ± 0.003 | -0.15 ± 0.21 | <21 | 3 | Y | N | N | N | N | Y | … | … | n | NG? |
| 2527 | 10303798-6330550 | 32.2 ± 0.6 | 4681 ± 31 | 2.69 ± 0.11 | 1.013 ± 0.003 | -0.14 ± 0.02 | 29 ± 9 | 1 | N | N | … | … | … | … | … | … | n | G |
| 37503 | 10303994-6446526 | -27.1 ± 0.2 | 4591 ± 69 | 2.46 ± 0.18 | 1.018 ± 0.003 | 0.05 ± 0.09 | <49 | 3 | N | N | … | … | … | … | … | … | n | G |
| 37504 | 10304227-6435447 | 44.2 ± 0.2 | 4748 ± 32 | 2.53 ± 0.11 | 1.020 ± 0.004 | -0.07 ± 0.05 | … | … | N | N | … | … | … | … | … | … | n | G |
| 2528 | 10304321-6327465 | 12.4 ± 0.6 | 4951 ± 7 | 2.60 ± 0.01 | 1.019 ± 0.003 | -0.11 ± 0.01 | 29 ± 8 | 1 | N | N | … | … | … | … | … | … | n | G |
| 37505 | 10304352-6449146 | -40.7 ± 0.2 | 5084 ± 70 | … | 1.015 ± 0.004 | … | <3 | 3 | N | … | … | … | … | … | … | … | n | G |
| 37506 | 10304458-6448390 | 1.5 ± 0.2 | 5117 ± 54 | … | 1.018 ± 0.002 | -0.03 ± 0.06 | <22 | 3 | N | … | … | … | … | … | … | … | n | G |
| 2529 | 10304475-6443306 | 7.7 ± 0.6 | 4639 ± 1 | 2.39 ± 0.03 | 1.029 ± 0.003 | -0.07 ± 0.02 | 40 ± 5 | 2 | N | N | … | … | … | … | … | … | n | G |
| 37507 | 10304529-6332317 | 39.1 ± 0.8 | … | … | … | … | … | … | … | … | … | … | … | … | … | … | n | … |
| 37508 | 10304562-6446359 | -28.3 ± 0.2 | 4737 ± 9 | 2.56 ± 0.10 | 1.023 ± 0.005 | -0.04 ± 0.09 | <31 | 3 | N | N | … | … | … | … | … | … | n | G |
| 37509 | 10304703-6317438 | 11.8 ± 0.2 | 4607 ± 104 | … | 1.044 ± 0.003 | … | <36 | 3 | N | … | … | … | … | … | … | … | n | G |
| 37510 | 10304712-6321199 | 37.1 ± 0.2 | 4524 ± 164 | … | 1.022 ± 0.003 | 0.11 ± 0.06 | <62 | 3 | N | … | … | … | … | … | … | … | n | G |
| 37511 | 10304826-6444189 | 10.1 ± 0.2 | 4608 ± 91 | … | 1.018 ± 0.005 | … | <64 | 3 | N | … | … | … | … | … | … | … | n | G |
| 37512 | 10304919-6327482 | 17.0 ± 0.2 | 4679 ± 8 | 2.51 ± 0.04 | 1.012 ± 0.005 | -0.04 ± 0.07 | <34 | 3 | N | N | … | … | … | … | … | … | n | G |
| 37513 | 10305228-6325004 | 47.1 ± 0.2 | 4999 ± 155 | … | 1.026 ± 0.005 | -0.15 ± 0.21 | <33 | 3 | N | … | … | … | … | … | … | … | n | G |
| 37514 | 10305254-6337225 | -26.1 ± 0.2 | 5051 ± 146 | … | 1.016 ± 0.003 | -0.05 ± 0.08 | … | … | N | … | … | … | … | … | … | … | n | G |
| 37515 | 10305370-6318098 | 454.7 ± 0.2 | 3508 ± 107 | … | … | … | … | … | … | … | … | … | … | … | … | Y | n | … |
| 37516 | 10305513-6436025 | 38.2 ± 0.2 | 4625 ± 109 | 2.51 ± 0.20 | 1.021 ± 0.004 | 0.07 ± 0.05 | <39 | 3 | N | N | … | … | … | … | … | … | n | G |
| 37517 | 10305524-6332169 | 0.2 ± 0.2 | 4959 ± 169 | … | 1.018 ± 0.003 | -0.12 ± 0.19 | <17 | 3 | N | … | … | … | … | … | … | … | n | G |
| 37518 | 10305555-6438013 | -27.0 ± 0.2 | 5109 ± 118 | … | 1.011 ± 0.003 | -0.26 ± 0.17 | <21 | 3 | N | … | … | … | … | … | … | … | n | G |
| 37519 | 10305592-6439475 | 28.6 ± 0.2 | 4807 ± 98 | … | 0.991 ± 0.006 | 0.04 ± 0.05 | <57 | 3 | Y | … | Y | N | N | Y | … | … | n | NG |
| 37634 | 10354745-6418449 | 18.2 ± 0.2 | 4539 ± 191 | … | 0.860 ± 0.002 | … | 100 ± 3 | 1 | Y | … | Y | Y | Y | … | … | Y | Y | … |
| 37520 | 10305964-6437156 | -2.7 ± 0.3 | 5224 ± 42 | … | 0.987 ± 0.006 | 0.07 ± 0.06 | 164 ± 7 | 1 | Y | … | Y | N | Y | Y | … | … | n | … |
| 37521 | 10310016-6438258 | -7.9 ± 0.3 | 4722 ± 228 | … | 0.956 ± 0.006 | -0.05 ± 0.05 | … | … | Y | … | Y | … | N | Y | … | … | n | … |
| 37522 | 10310037-6316227 | -33.7 ± 0.2 | 4827 ± 136 | 2.64 ± 0.20 | 1.019 ± 0.003 | -0.07 ± 0.11 | <31 | 3 | N | N | … | … | … | … | … | … | n | G |
| 37523 | 10310213-6441518 | 17.6 ± 0.3 | 4690 ± 66 | 2.49 ± 0.02 | 1.013 ± 0.008 | -0.12 ± 0.10 | <43 | 3 | N | N | … | … | … | … | … | … | n | G |
| 37547 | 10312628-6333372 | -40.6 ± 0.2 | 4776 ± 28 | 2.62 ± 0.17 | 1.023 ± 0.003 | -0.06 ± 0.05 | <21 | 3 | N | N | … | … | … | … | … | … | n | G |
| 37548 | 10313037-6318460 | 14.1 ± 0.2 | 4610 ± 40 | 2.35 ± 0.13 | 1.030 ± 0.003 | -0.01 ± 0.10 | <34 | 3 | N | N | … | … | … | … | … | … | n | G |
| 37549 | 10313087-6321203 | 13.7 ± 0.3 | 3602 ± 95 | … | 0.825 ± 0.007 | -0.23 ± 0.14 | … | … | Y | … | Y | … | N | Y | N | … | n | … |
| 37550 | 10313346-6323528 | 15.4 ± 0.2 | 4460 ± 219 | … | 1.039 ± 0.002 | -0.01 ± 0.02 | <42 | 3 | N | … | … | … | … | … | … | … | n | G |

| ID | CNAME | RV (km s$^{-1}$) | T$_{\rm eff}$ (K) | logg (dex) | $\gamma^a$ | [Fe/H] (dex) | EW(Li)$^b$ (mÅ) | EW(Li) error flag$^c$ | $\gamma$ | logg | Membership RV | Li | H$\alpha$ | [Fe/H] | Gaia studies Randich$^d$ | Cantat-Gaudin$^d$ | Final$^e$ | NMs with Li$^f$ |
|---|---|---|---|---|---|---|---|---|---|---|---|---|---|---|---|---|---|---|
| 37551 | 10313481-6323426 | 45.3 ± 0.2 | 4636 ± 98 | … | 1.022 ± 0.007 | … | 439 ± 9 | 1 | N | … | … | … | … | … | … | … | n | Li-rich G |
| 37552 | 10313497-6320069 | -18.0 ± 0.2 | 4946 ± 118 | … | 1.017 ± 0.003 | -0.06 ± 0.11 | 47 ± 2 | 1 | N | … | … | … | … | … | … | … | n | G |
| 37553 | 10313597-6325120 | 17.0 ± 0.2 | 4727 ± 73 | 2.56 ± 0.01 | 1.017 ± 0.004 | 0.00 ± 0.01 | <35 | 3 | N | N | … | … | … | … | … | … | n | G |
| 2531 | 10313722-6327243 | -17.1 ± 0.6 | 5252 ± 20 | 3.11 ± 0.08 | 1.014 ± 0.003 | 0.07 ± 0.03 | 18 ± 2 | 1 | N | N | … | … | … | … | … | … | n | … |
| 37554 | 10313743-6317374 | 9.0 ± 0.2 | 4517 ± 129 | … | 1.036 ± 0.002 | -0.04 ± 0.02 | 108 ± 15 | 1 | N | … | … | … | … | … | … | … | n | G |
| 37635 | 10354926-6337286 | 58.8 ± 0.3 | 4600 ± 130 | 2.35 ± 0.14 | 1.018 ± 0.008 | -0.04 ± 0.10 | … | … | N | N | … | … | … | … | … | … | n | G |
| 37555 | 10313880-6324371 | 26.4 ± 0.2 | 4386 ± 132 | 2.05 ± 0.16 | 1.028 ± 0.004 | 0.23 ± 0.09 | <56 | 3 | N | N | … | … | … | … | … | … | n | G |
| 37556 | 10314034-6330133 | 18.5 ± 0.2 | 4744 ± 38 | 2.57 ± 0.07 | 1.018 ± 0.003 | 0.04 ± 0.06 | <34 | 3 | N | N | … | … | … | … | … | … | n | G |
| 37557 | 10314450-6335141 | 18.8 ± 0.2 | 4929 ± 140 | … | 1.014 ± 0.003 | -0.06 ± 0.10 | … | … | N | … | … | … | … | … | … | … | n | G |
| 2532 | 10314490-6329221 | … | 6544 ± 137 | 4.00 ± 0.16 | … | -0.09 ± 0.14 | <62 | 3 | … | Y | … | N | N | N | … | Y | n | NG |
| 37558 | 10314515-6317241 | -63.1 ± 0.2 | 4974 ± 182 | … | 1.019 ± 0.002 | -0.22 ± 0.08 | <16 | 3 | N | … | … | … | … | … | … | … | n | G |
| 37559 | 10314827-6338170 | 46.0 ± 0.2 | 4932 ± 191 | … | 1.016 ± 0.003 | -0.14 ± 0.18 | <10 | 3 | N | … | … | … | … | … | … | … | n | G |
| 37560 | 10320246-6321166 | -4.1 ± 0.4 | 4968 ± 193 | … | 0.931 ± 0.010 | … | <28 | 3 | Y | … | Y | N | Y | … | … | … | n | NG |
| 37561 | 10320402-6318473 | 17.5 ± 0.2 | 4713 ± 19 | 2.56 ± 0.07 | 1.016 ± 0.002 | 0.05 ± 0.05 | <34 | 3 | N | N | … | … | … | … | … | … | n | G |
| 37562 | 10320502-6319541 | -14.3 ± 0.2 | 4450 ± 172 | … | 1.028 ± 0.002 | 0.16 ± 0.03 | <50 | 3 | N | … | … | … | … | … | … | … | n | G |
| 37563 | 10320620-6320487 | -5.1 ± 0.2 | 4646 ± 34 | 2.49 ± 0.10 | 1.015 ± 0.002 | -0.01 ± 0.01 | <35 | 3 | N | N | … | … | … | … | … | … | n | G |
| 37564 | 10320908-6334533 | 34.7 ± 1.7 | … | … | … | … | … | … | … | … | … | … | … | … | N | … | n | … |
| 37565 | 10321426-6328591 | -2.5 ± 0.2 | 4965 ± 44 | 2.96 ± 0.11 | 1.012 ± 0.002 | 0.02 ± 0.03 | <35 | 3 | N | N | … | … | … | … | … | … | n | G |
| 37566 | 10321459-6321446 | 20.4 ± 0.2 | 4654 ± 32 | 2.47 ± 0.09 | 1.018 ± 0.003 | 0.00 ± 0.02 | … | … | N | N | … | … | … | … | … | … | n | G |
| 2533 | 10321965-6322325 | -24.8 ± 0.6 | 4701 ± 22 | 2.56 ± 0.12 | 1.022 ± 0.002 | 0.01 ± 0.02 | 47 ± 14 | 1 | N | N | … | … | … | … | … | … | n | G |
| 37567 | 10322350-6336064 | 0.0 ± 0.4 | 4864 ± 138 | … | 0.888 ± 0.005 | … | … | … | Y | … | Y | … | N | … | … | … | n | … |
| 37640 | 10355870-6335189 | -28.2 ± 0.2 | 4531 ± 159 | … | 1.033 ± 0.003 | 0.13 ± 0.01 | … | … | N | … | … | … | … | … | … | … | n | G |
| 37599 | 10342188-6411081 | 3.3 ± 0.2 | 4973 ± 170 | … | 1.019 ± 0.002 | -0.08 ± 0.17 | <15 | 3 | N | … | … | … | … | … | … | … | n | G |
| 37641 | 10360003-6340187 | 1.8 ± 0.2 | 4643 ± 5 | 2.35 ± 0.13 | 1.036 ± 0.002 | -0.05 ± 0.02 | 5 ± … | 1 | N | N | … | … | … | … | … | … | n | G |
| 37600 | 10342254-6414394 | -2.8 ± 0.2 | 4961 ± 152 | … | 1.018 ± 0.002 | -0.07 ± 0.11 | <12 | 3 | N | … | … | … | … | … | … | … | n | G |
| 37601 | 10342277-6418095 | 21.0 ± 0.3 | 3423 ± 89 | 4.60 ± 0.19 | 0.846 ± 0.006 | -0.26 ± 0.13 | … | … | Y | Y | Y | … | N | Y | Y | … | n | … |
| 37368 | 10274578-6440597 | -29.8 ± 0.2 | 5123 ± 72 | 3.09 ± 0.09 | 1.015 ± 0.004 | -0.03 ± 0.04 | <31 | 3 | N | N | … | … | … | … | … | … | n | G |
| 37602 | 10342301-6410383 | 6.4 ± 2.7 | … | … | … | … | … | … | … | … | … | … | … | … | … | … | n | … |
| 53553 | 10394403-6404417 | 60.0 ± 0.2 | … | … | … | … | … | … | … | … | … | … | … | … | … | … | n | … |
| 37603 | 10342410-6413107 | -37.2 ± 0.2 | 4505 ± 177 | … | 1.027 ± 0.002 | -0.01 ± 0.06 | <45 | 3 | N | … | … | … | … | … | … | … | n | G |
| 37369 | 10274791-6449143 | 50.7 ± 0.2 | 4852 ± 259 | … | 1.018 ± 0.004 | -0.20 ± 0.20 | <15 | 3 | N | … | … | … | … | … | … | … | n | G |
| 53558 | 10395161-6405166 | -9.4 ± 0.3 | 6047 ± 65 | … | 1.009 ± 0.006 | -0.07 ± 0.01 | <15 | 3 | Y | … | Y | N | N | Y | … | … | n | NG? |
| 37370 | 10275016-6445420 | 36.7 ± 0.2 | 4669 ± 52 | 2.49 ± 0.14 | 1.024 ± 0.003 | 0.06 ± 0.07 | <43 | 3 | N | N | … | … | … | … | … | … | n | G |
| 37371 | 10275065-6436526 | -40.7 ± 0.2 | 5034 ± 151 | … | 1.016 ± 0.004 | -0.21 ± 0.18 | <17 | 3 | N | … | … | … | … | … | … | … | n | G |
| 2537 | 10343957-6407258 | -34.4 ± 0.6 | 4731 ± 35 | 2.56 ± 0.09 | 1.016 ± 0.003 | -0.02 ± 0.02 | <14 | 3 | N | N | … | … | … | … | … | … | n | G |
| 37736 | 10395194-6354529 | 14.6 ± 0.2 | 5016 ± 54 | 3.10 ± 0.12 | 1.010 ± 0.001 | 0.02 ± 0.04 | <34 | 3 | Y | N | Y | N | N | Y | … | … | n | NG? |
| 37372 | 10275188-6446569 | 98.0 ± 0.2 | 4858 ± 152 | … | 1.027 ± 0.003 | -0.20 ± 0.17 | <26 | 3 | N | … | … | … | … | … | … | … | n | G |
| 37606 | 10344018-6414379 | 32.5 ± 0.2 | 4536 ± 89 | … | 1.025 ± 0.004 | -0.14 ± 0.14 | <33 | 3 | N | … | … | … | … | … | … | … | n | G |
| 37373 | 10275658-6446566 | -0.4 ± 0.2 | 4874 ± 210 | … | 1.028 ± 0.004 | -0.05 ± 0.09 | <22 | 3 | N | … | … | … | … | … | … | … | n | G |
| 53559 | 10395292-6407027 | 40.2 ± 0.2 | 4150 ± 245 | … | 1.039 ± 0.003 | -0.07 ± 0.09 | <39 | 3 | N | … | … | … | … | … | … | … | n | G |
| 37374 | 10275818-6450376 | 3.8 ± 0.2 | 4653 ± 109 | … | 1.019 ± 0.006 | -0.13 ± 0.14 | … | … | N | … | … | … | … | … | … | … | n | G |
| 37737 | 10395335-6442104 | 20.7 ± 0.2 | 4502 ± 169 | … | 1.023 ± 0.002 | 0.10 ± 0.08 | <51 | 3 | N | … | … | … | … | … | … | … | n | G |
| 37375 | 10275861-6436569 | 17.2 ± 0.2 | 4906 ± 73 | 2.65 ± 0.15 | 1.024 ± 0.004 | -0.06 ± 0.13 | <23 | 3 | N | N | … | … | … | … | … | … | n | G |
| 37738 | 10395359-6414369 | -1.1 ± 0.2 | 5077 ± 88 | … | 1.015 ± 0.001 | -0.03 ± 0.07 | 24 ± 1 | 1 | N | … | … | … | … | … | … | … | n | G |
| 37376 | 10280157-6437544 | 12.3 ± 0.2 | 4715 ± 85 | 2.53 ± 0.07 | 1.014 ± 0.005 | 0.07 ± 0.04 | <45 | 3 | N | N | … | … | … | … | … | … | n | G |
| 37740 | 10395570-6449563 | 53.7 ± 0.2 | 4585 ± 69 | 2.42 ± 0.12 | 1.012 ± 0.005 | -0.02 ± 0.04 | <42 | 3 | N | N | … | … | … | … | … | … | n | G |
| 37739 | 10395360-6425537 | 9.0 ± 0.2 | 4606 ± 91 | … | 1.023 ± 0.003 | … | <65 | 3 | N | … | … | … | … | … | … | … | n | G |
| 37741 | 10395585-6433457 | 6.4 ± 0.2 | 5023 ± 165 | … | 1.010 ± 0.003 | -0.10 ± 0.19 | <18 | 3 | N | … | … | … | … | … | … | … | n | G |
| 37607 | 10344544-6403506 | 5.5 ± 0.2 | 4705 ± 48 | 2.49 ± 0.03 | 1.018 ± 0.003 | -0.01 ± 0.05 | <41 | 3 | N | N | … | … | … | … | … | … | n | G |
| 37385 | 10281886-6431375 | 17.3 ± 0.3 | 3319 ± 13 | … | 0.858 ± 0.013 | -0.26 ± 0.14 | <100 | 3 | Y | … | Y | Y | Y | Y | Y | … | Y | … |
| 37742 | 10395597-6359300 | 14.9 ± 0.2 | 5120 ± 122 | … | 0.975 ± 0.001 | … | 331 ± 8 | 1 | Y | … | Y | Y | Y | … | Y | Y | Y | … |
| 37751 | 10400586-6437525 | 2.2 ± 0.2 | 4838 ± 91 | 2.67 ± 0.20 | 1.021 ± 0.004 | -0.07 ± 0.12 | <24 | 3 | N | N | … | … | … | … | … | … | n | G |
| 53696 | 10415767-6409342 | -29.8 ± 0.2 | 4109 ± 37 | 1.31 ± 0.07 | 1.065 ± 0.002 | -0.13 ± 0.09 | <24 | 3 | N | N | … | … | … | … | … | … | n | G |
| 6 | 10415772-6409355 | -31.1 ± 0.6 | 3926 ± 99 | 1.21 ± 0.12 | … | -0.09 ± 0.13 | <39 | 3 | … | N | … | … | … | … | … | … | n | … |
| 37386 | 10281895-64450450 | 6.0 ± 0.2 | 4600 ± 63 | 2.34 ± 0.20 | 1.029 ± 0.006 | -0.09 ± 0.04 | <33 | 3 | N | N | … | … | … | … | … | … | n | G |
| 37863 | 10415815-6524592 | 52.3 ± 0.3 | 4681 ± 148 | … | 1.016 ± 0.011 | -0.10 ± 0.10 | <51 | 3 | N | … | … | … | … | … | … | … | n | G |
| 37752 | 10400653-6450374 | 55.8 ± 0.2 | 4502 ± 163 | … | 1.025 ± 0.002 | -0.05 ± 0.02 | <48 | 3 | N | … | … | … | … | … | … | … | n | G |
| 37876 | 10420605-6432480 | -29.3 ± 0.2 | 4709 ± 31 | 2.57 ± 0.09 | 1.012 ± 0.004 | -0.11 ± 0.10 | <40 | 3 | N | N | … | … | … | … | … | … | n | G |
| 53704 | 10420609-6407060 | 2.7 ± 0.6 | 6777 ± 141 | 3.89 ± 0.28 | 1.012 ± 0.007 | -0.29 ± 0.14 | <19 | 3 | N | Y | … | … | … | … | … | … | n | … |

**Table C.6.** continued.

| ID | CNAME | RV (km s$^{-1}$) | $T_{\rm eff}$ (K) | $logg$ (dex) | $\gamma^a$ | [Fe/H] (dex) | EW(Li)$^b$ (mÅ) | EW(Li) error flag$^c$ | Membership $\gamma$ | $logg$ | RV | Li | H$\alpha$ | [Fe/H] | Gaia studies Randich$^d$ | Cantat-Gaudin$^d$ | Final$^e$ | NMs with Li$^f$ |
|---|---|---|---|---|---|---|---|---|---|---|---|---|---|---|---|---|---|---|
| 37613 | 10350175-6405092 | 32.8 ± 0.2 | 4546 ± 150 | ... | 1.024 ± 0.002 | 0.08 ± 0.08 | 389 ± 5 | 1 | N | ... | ... | ... | ... | ... | ... | ... | n | Li-rich G |
| 53711 | 10421229-6417103 | 67.2 ± 0.6 | ... | ... | ... | ... | ... | ... | ... | ... | ... | ... | ... | ... | ... | ... | n | ... |
| 37877 | 10420646-6452427 | 12.2 ± 0.2 | 4143 ± 183 | 4.56 ± 0.02 | 0.864 ± 0.003 | -0.02 ± 0.17 | ... | ... | Y | Y | Y | ... | N | Y | N | ... | n | ... |
| 37883 | 10421277-6532246 | 66.3 ± 0.3 | 5142 ± 98 | ... | 0.992 ± 0.006 | ... | <10 | 3 | Y | ... | N | N | N | ... | ... | ... | n | NG |
| 53712 | 10421334-6413446 | 650.4 ± 15.1 | ... | ... | ... | ... | ... | ... | ... | ... | ... | ... | ... | ... | ... | ... | n | ... |
| 38002 | 10435316-6452006 | 17.5 ± 0.2 | 3784 ± 67 | ... | 0.838 ± 0.003 | -0.16 ± 0.13 | ... | ... | Y | ... | Y | ... | N | Y | Y | Y | n | ... |
| 37878 | 10420708-6446079 | 17.4 ± 0.2 | 5397 ± 42 | 4.06 ± 0.18 | 0.990 ± 0.001 | 0.01 ± 0.06 | 122 ± 1 | 1 | Y | Y | Y | Y | Y | Y | ... | Y | Y | ... |
| 38003 | 10435320-6534032 | -60.2 ± 0.3 | 5141 ± 75 | ... | 1.013 ± 0.002 | -0.34 ± 0.05 | <12 | 3 | N | ... | ... | ... | ... | ... | ... | ... | n | G |
| 53705 | 10420720-6420356 | 3.5 ± 0.2 | 4833 ± 89 | 2.73 ± 0.12 | 1.011 ± 0.003 | 0.07 ± 0.06 | <44 | 3 | N | N | ... | ... | ... | ... | ... | ... | n | G |
| 37907 | 10423152-6437352 | 6.4 ± 0.2 | 4750 ± 73 | 2.55 ± 0.13 | 1.022 ± 0.003 | 0.02 ± 0.07 | 80 ± 5 | 1 | N | N | ... | ... | ... | ... | ... | ... | n | G |
| 37387 | 10282358-6443065 | 48.9 ± 0.2 | 4747 ± 54 | 2.71 ± 0.02 | 1.008 ± 0.005 | -0.11 ± 0.06 | <51 | 3 | Y | N | N | N | N | Y | ... | ... | n | NG? |
| 37879 | 10420774-6541491 | 9.7 ± 0.2 | 4874 ± 149 | ... | 1.021 ± 0.005 | -0.03 ± 0.10 | <24 | 3 | N | ... | ... | ... | ... | ... | ... | ... | n | G |
| 37388 | 10282591-6449054 | 86.3 ± 0.2 | 4636 ± 102 | 2.66 ± 0.03 | 1.005 ± 0.006 | 0.04 ± 0.01 | <39 | 3 | Y | N | N | N | N | Y | ... | ... | n | NG? |
| 53720 | 10423190-6420023 | 5.7 ± 0.3 | 6579 ± 77 | ... | 0.998 ± 0.004 | 0.08 ± 0.06 | 30 ± 6 | 1 | Y | ... | Y | N | N | Y | ... | ... | n | NG |
| 37908 | 10423221-6544479 | -10.9 ± 0.2 | 4859 ± 47 | 2.75 ± 0.02 | 1.011 ± 0.003 | -0.01 ± 0.02 | 14 ± 2 | 1 | N | N | ... | ... | ... | ... | ... | ... | n | G |
| 53706 | 10420833-6419144 | 24.3 ± 0.2 | 4394 ± 243 | ... | 1.036 ± 0.002 | -0.13 ± 0.12 | <42 | 3 | N | ... | ... | ... | ... | ... | ... | ... | n | G |
| 37909 | 10423239-6535368 | 27.8 ± 0.2 | 4643 ± 98 | ... | 1.020 ± 0.004 | ... | <57 | 3 | N | ... | ... | ... | ... | ... | ... | ... | n | G |
| 53565 | 10400808-6409564 | -21.4 ± 0.2 | 7122 ± 33 | ... | 1.032 ± 0.002 | -0.06 ± 0.18 | <3 | 3 | N | ... | ... | ... | ... | ... | ... | ... | n | ... |
| 37614 | 10350470-6401475 | 33.3 ± 0.2 | 5122 ± 95 | ... | 1.012 ± 0.003 | ... | <33 | 3 | N | ... | ... | ... | ... | ... | ... | ... | n | G |
| 37880 | 10420837-6537156 | 5.8 ± 0.2 | 4924 ± 45 | 2.72 ± 0.10 | 1.023 ± 0.002 | -0.01 ± 0.06 | <24 | 3 | N | N | ... | ... | ... | ... | ... | ... | n | G |
| 37910 | 10423265-6524249 | -22.1 ± 0.2 | 5083 ± 102 | ... | 1.017 ± 0.005 | -0.02 ± 0.09 | <34 | 3 | N | ... | ... | ... | ... | ... | ... | ... | n | G |
| 53777 | 10435612-6431227 | -17.9 ± 0.2 | 5069 ± 79 | ... | 1.006 ± 0.002 | -0.22 ± 0.19 | <13 | 3 | Y | ... | N | N | N | Y | ... | ... | n | NG? |
| 37389 | 10282968-6444210 | 73.5 ± 0.2 | 4738 ± 63 | 2.57 ± 0.15 | 1.018 ± 0.005 | -0.09 ± 0.02 | <34 | 3 | N | N | ... | ... | ... | ... | ... | ... | n | G |
| 37881 | 10420884-6545149 | -24.8 ± 0.2 | 5041 ± 124 | ... | 1.019 ± 0.004 | -0.23 ± 0.19 | <16 | 3 | N | ... | ... | ... | ... | ... | ... | ... | n | G |
| 53721 | 10423320-6413486 | 17.5 ± 0.2 | 5143 ± 56 | 3.72 ± 0.10 | 0.998 ± 0.002 | -0.07 ± 0.17 | <32 | 3 | Y | Y | Y | N | N | Y | ... | ... | n | NG |
| 37390 | 10282971-6449464 | 13.0 ± 0.2 | 4790 ± 152 | 2.62 ± 0.20 | 1.014 ± 0.002 | -0.07 ± 0.15 | <22 | 3 | N | N | ... | ... | ... | ... | ... | ... | n | G |
| 37882 | 10420886-6538406 | 35.9 ± 0.2 | 4501 ± 195 | ... | 1.021 ± 0.003 | 0.14 ± 0.03 | <62 | 3 | N | ... | ... | ... | ... | ... | ... | ... | n | G |
| 53722 | 10423323-6409160 | -20.9 ± 0.3 | 6207 ± 66 | 3.85 ± 0.18 | 1.006 ± 0.006 | -0.19 ± 0.10 | ... | ... | Y | Y | N | ... | ... | Y | ... | ... | n | ... |
| 37753 | 10400939-6442244 | 41.9 ± 0.2 | 4779 ± 44 | 2.62 ± 0.07 | 1.014 ± 0.002 | -0.01 ± 0.04 | <39 | 3 | N | N | ... | ... | ... | ... | ... | ... | n | G |
| 37615 | 10350595-6403433 | 52.1 ± 0.2 | 4607 ± 136 | ... | 1.022 ± 0.002 | 0.05 ± 0.07 | <44 | 3 | N | ... | ... | ... | ... | ... | ... | ... | n | G |
| 53707 | 10420953-6411287 | 35.4 ± 0.2 | 4571 ± 135 | 2.39 ± 0.18 | 1.019 ± 0.005 | 0.11 ± 0.10 | <43 | 3 | N | N | ... | ... | ... | ... | ... | ... | n | G |
| 37911 | 10423327-6533239 | 16.6 ± 0.2 | 5087 ± 40 | 3.36 ± 0.20 | 1.008 ± 0.003 | -0.02 ± 0.05 | 21 ± 4 | 1 | Y | N | Y | N | N | Y | ... | ... | n | NG? |
| 37754 | 10400965-6456569 | -17.7 ± 0.2 | 4618 ± 33 | 2.42 ± 0.12 | 1.024 ± 0.002 | 0.03 ± 0.02 | <38 | 3 | N | N | ... | ... | ... | ... | ... | ... | n | G |
| 37391 | 10282989-6447428 | 28.2 ± 0.2 | 4626 ± 171 | ... | 1.023 ± 0.003 | -0.13 ± 0.17 | <22 | 3 | N | ... | ... | ... | ... | ... | ... | ... | n | G |
| 53778 | 10435632-6434018 | 14.3 ± 0.2 | 4989 ± 115 | ... | 1.013 ± 0.001 | -0.06 ± 0.09 | <18 | 3 | N | ... | ... | ... | ... | ... | ... | ... | n | G |
| 53708 | 10420975-6406023 | 14.1 ± 0.2 | 5765 ± 19 | ... | 0.998 ± 0.005 | 0.09 ± 0.08 | 13 ± 4 | 1 | Y | ... | Y | N | N | Y | ... | ... | n | NG |
| 37912 | 10423424-6541569 | 97.0 ± 0.2 | 4595 ± 54 | 2.39 ± 0.15 | 1.029 ± 0.003 | 0.05 ± 0.10 | <42 | 3 | N | N | ... | ... | ... | ... | ... | ... | n | G |
| 38006 | 10435642-6406521 | -18.1 ± 0.2 | 4648 ± 79 | 2.56 ± 0.17 | 1.020 ± 0.004 | 0.11 ± 0.09 | <54 | 3 | N | N | ... | ... | ... | ... | ... | ... | n | G |
| 37392 | 10283024-6440223 | -16.1 ± 0.2 | 5061 ± 92 | ... | 1.016 ± 0.006 | -0.07 ± 0.07 | ... | ... | N | ... | ... | ... | ... | ... | ... | ... | n | G |
| 53709 | 10421117-6414404 | -7.3 ± 0.2 | 5024 ± 42 | ... | 1.024 ± 0.004 | -0.01 ± 0.01 | <29 | 3 | N | ... | ... | ... | ... | ... | ... | ... | n | G |
| 53723 | 10423470-6407599 | 128.9 ± 0.2 | 3902 ± 166 | ... | 1.059 ± 0.005 | -0.26 ± 0.06 | 35 ± 15 | 1 | N | ... | ... | ... | ... | ... | ... | ... | n | G |
| 37393 | 10283239-6451050 | -10.9 ± 0.2 | 4634 ± 43 | 2.35 ± 0.11 | 1.028 ± 0.004 | -0.04 ± 0.03 | <33 | 3 | N | N | ... | ... | ... | ... | ... | ... | n | G |
| 53710 | 10421124-6414076 | -19.9 ± 0.5 | 7241 ± 91 | ... | 1.007 ± 0.004 | ... | <17 | 3 | Y | ... | N | N | N | ... | ... | ... | n | NG? |
| 38007 | 10435648-6358344 | -656.6 ± 220.5 | ... | ... | ... | ... | ... | ... | ... | ... | ... | ... | ... | ... | ... | ... | n | ... |
| 37913 | 10423472-6449492 | 17.6 ± 0.3 | 3480 ± 69 | 4.63 ± 0.14 | 0.827 ± 0.008 | -0.25 ± 0.13 | ... | ... | Y | Y | Y | ... | N | Y | ... | Y | n | ... |
| 2562 | 10435700-6447397 | 22.0 ± 0.6 | 6070 ± 121 | 3.79 ± 0.18 | ... | -0.04 ± 0.11 | <9 | 3 | ... | Y | Y | N | Y | Y | N | ... | n | ... |
| 53567 | 10401016-6417458 | -18.6 ± 0.2 | 4719 ± 29 | 2.49 ± 0.06 | 1.022 ± 0.005 | -0.01 ± 0.02 | <47 | 3 | N | N | ... | ... | ... | ... | ... | ... | n | G |
| 37646 | 10361825-6414564 | 18.4 ± 0.2 | 3706 ± 17 | ... | 0.828 ± 0.003 | -0.19 ± 0.12 | ... | ... | Y | ... | Y | ... | N | Y | Y | Y | n | ... |
| 37914 | 10423509-6431531 | 10.9 ± 0.2 | 4620 ± 42 | 2.30 ± 0.08 | 1.029 ± 0.003 | -0.03 ± 0.02 | <30 | 3 | N | N | ... | ... | ... | ... | ... | ... | n | G |
| 37616 | 10350776-6414151 | -31.7 ± 0.2 | 4994 ± 210 | ... | 1.012 ± 0.002 | -0.15 ± 0.20 | ... | ... | N | ... | ... | ... | ... | ... | ... | ... | n | G |
| 37915 | 10423544-6540236 | 25.7 ± 0.3 | 4987 ± 159 | ... | 1.012 ± 0.007 | -0.10 ± 0.19 | <33 | 3 | N | ... | ... | ... | ... | ... | ... | ... | n | G |
| 37407 | 10285783-6443116 | 146.1 ± 0.3 | ... | ... | ... | ... | ... | ... | ... | ... | ... | ... | ... | ... | ... | ... | n | ... |
| 37916 | 10423573-6535083 | -27.2 ± 0.2 | 4980 ± 89 | ... | 1.020 ± 0.004 | -0.08 ± 0.15 | <23 | 3 | N | ... | ... | ... | ... | ... | ... | ... | n | G |
| 37617 | 10350970-6421082 | 11.2 ± 0.2 | 4584 ± 36 | ... | 1.035 ± 0.004 | -0.08 ± 0.05 | 111 ± 3 | 1 | N | ... | ... | ... | ... | ... | ... | ... | n | G |
| 37408 | 10285795-6440571 | 61.1 ± 0.3 | 5472 ± 89 | ... | 1.008 ± 0.007 | ... | <24 | 3 | Y | ... | N | N | N | ... | ... | ... | n | NG? |
| 38049 | 10443345-6417313 | -41.2 ± 0.2 | 4942 ± 141 | ... | 1.022 ± 0.002 | -0.09 ± 0.13 | <18 | 3 | N | ... | ... | ... | ... | ... | ... | ... | n | G |
| 53724 | 10423589-6410521 | 12.8 ± 0.2 | 4646 ± 84 | 2.47 ± 0.16 | 1.018 ± 0.002 | -0.01 ± 0.06 | <35 | 3 | N | N | ... | ... | ... | ... | ... | ... | n | G |
| 53571 | 10401818-6412467 | 19.8 ± 0.3 | 6370 ± 74 | 4.21 ± 0.16 | 0.999 ± 0.005 | 0.17 ± 0.05 | <9 | 3 | Y | Y | Y | N | N | N | ... | ... | n | ... |
| 37409 | 10285969-6446271 | 40.9 ± 0.3 | 4947 ± 140 | ... | 1.015 ± 0.006 | -0.13 ± 0.16 | <24 | 3 | N | ... | ... | ... | ... | ... | ... | ... | n | G |

| ID | CNAME | RV (km s$^{-1}$) | $T_{\text{eff}}$ (K) | $logg$ (dex) | $\gamma^a$ | [Fe/H] (dex) | EW(Li)$^b$ (mÅ) | EW(Li) error flag$^c$ | Membership $\gamma$ | $logg$ | RV | Li | H$\alpha$ | [Fe/H] | Gaia studies Randich$^d$ | Cantat-Gaudin$^d$ | Final$^e$ | NMs with Li$^f$ |
|---|---|---|---|---|---|---|---|---|---|---|---|---|---|---|---|---|---|---|
| 37649 | 10362753-6329374 | -39.5 ± 0.2 | 5096 ± 62 | … | 1.016 ± 0.003 | -0.07 ± 0.06 | <13 | 3 | N | … | … | … | … | … | … | … | n | G |
| 37625 | 10352296-6400224 | 2.8 ± 0.2 | 4429 ± 183 | … | 1.033 ± 0.002 | 0.05 ± 0.03 | <42 | 3 | N | … | … | … | … | … | … | … | n | G |
| 37917 | 10423611-6539468 | 14.4 ± 0.2 | 4585 ± 161 | 2.36 ± 0.20 | 1.025 ± 0.005 | 0.03 ± 0.01 | <42 | 3 | N | N | … | … | … | … | … | … | n | G |
| 53572 | 10401826-6405284 | 23.5 ± 0.2 | 4614 ± 167 | 2.32 ± 0.15 | 1.021 ± 0.006 | 0.03 ± 0.09 | <38 | 3 | N | N | … | … | … | … | … | … | n | G |
| 37918 | 10423714-6453309 | 17.7 ± 0.6 | … | … | … | … | … | … | … | … | … | … | … | … | … | Y | n | … |
| 37410 | 10290148-6441549 | 45.5 ± 0.2 | 4493 ± 202 | … | 1.025 ± 0.004 | 0.17 ± 0.02 | <60 | 3 | N | … | … | … | … | … | … | … | n | G |
| 38013 | 10440306-6443397 | -24.6 ± 0.2 | 4701 ± 100 | 2.46 ± 0.19 | 1.028 ± 0.003 | -0.03 ± 0.05 | <38 | 3 | N | N | … | … | … | … | … | … | n | G |
| 8 | 10423748-6411137 | 37.9 ± 0.6 | 4750 ± 6 | 2.18 ± 0.02 | … | -0.06 ± 0.02 | <14 | 3 | … | N | … | … | … | … | … | … | n | … |
| 37411 | 10290326-6434532 | 46.7 ± 0.3 | 5726 ± 126 | 4.42 ± 0.15 | 0.982 ± 0.005 | -0.04 ± 0.07 | <29 | 3 | Y | Y | N | N | N | Y | … | … | n | NG |
| 37919 | 10423776-6526573 | 14.2 ± 0.2 | 4670 ± 50 | 2.56 ± 0.17 | 1.023 ± 0.004 | 0.07 ± 0.06 | <53 | 3 | N | N | … | … | … | … | … | … | n | G |
| 37759 | 10401855-6448299 | 6.2 ± 0.2 | 4920 ± 251 | … | 1.012 ± 0.002 | -0.10 ± 0.19 | <21 | 3 | N | … | … | … | … | … | … | … | n | G |
| 37412 | 10290605-6440379 | 37.2 ± 0.3 | 5019 ± 118 | … | 0.993 ± 0.006 | -0.02 ± 0.05 | <24 | 3 | Y | … | Y | N | N | Y | … | … | n | NG |
| 53825 | 10443600-6436124 | 17.6 ± 0.2 | 4745 ± 116 | 2.55 ± 0.18 | 1.019 ± 0.001 | -0.01 ± 0.13 | <31 | 3 | N | N | … | … | … | … | … | … | n | G |
| 37413 | 10290605-6441006 | 37.0 ± 0.2 | 4844 ± 81 | 3.13 ± 0.14 | 0.998 ± 0.006 | 0.00 ± 0.11 | <34 | 3 | Y | N | Y | N | N | Y | … | … | n | NG? |
| 37928 | 10424822-6541233 | -15.6 ± 0.2 | 4749 ± 8 | 2.68 ± 0.10 | 1.008 ± 0.006 | -0.04 ± 0.02 | <44 | 3 | Y | N | N | N | N | Y | … | … | n | NG? |
| 53826 | 10443604-6426569 | 8.6 ± 0.3 | 6063 ± 52 | 4.03 ± 0.04 | 0.998 ± 0.005 | -0.21 ± 0.05 | 60 ± 5 | 1 | Y | Y | N | N | N | Y | … | … | n | NG |
| 53573 | 10401884-6419163 | 13.2 ± 0.2 | 6328 ± 75 | 3.95 ± 0.07 | 1.004 ± 0.001 | -0.11 ± 0.11 | 80 ± 2 | 1 | Y | Y | Y | N | N | Y | … | … | n | NG? |
| 53725 | 10424835-6412555 | 0.8 ± 0.3 | 6794 ± 82 | … | 1.006 ± 0.004 | 0.07 ± 0.07 | <8 | 3 | Y | … | Y | N | N | Y | … | … | n | … |
| 38014 | 10440351-6349493 | 80.7 ± 0.2 | 4607 ± 91 | … | 1.022 ± 0.004 | … | <56 | 3 | N | … | … | … | … | … | … | … | n | G |
| 37650 | 10363085-6333152 | 52.9 ± 0.2 | 4545 ± 107 | 2.27 ± 0.17 | 1.029 ± 0.004 | 0.00 ± 0.04 | <42 | 3 | N | N | … | … | … | … | … | … | n | G |
| 53574 | 10401957-6421506 | -4.6 ± 0.2 | 5026 ± 120 | … | 1.022 ± 0.004 | … | <13 | 3 | N | … | … | … | … | … | … | … | n | G |
| 53827 | 10443706-6421256 | -14.0 ± 0.2 | 4590 ± 101 | 2.29 ± 0.16 | 1.027 ± 0.004 | -0.07 ± 0.01 | <54 | 3 | N | N | … | … | … | … | … | … | n | G |
| 37651 | 10363282-6405556 | 6.4 ± 0.2 | 4830 ± 99 | … | 1.020 ± 0.004 | -0.07 ± 0.13 | <13 | 3 | N | … | … | … | … | … | … | … | n | G |
| 53575 | 10401970-6410258 | 114.7 ± 0.3 | 4739 ± 210 | 2.44 ± 0.03 | 1.018 ± 0.009 | -0.21 ± 0.23 | <40 | 3 | N | N | … | … | … | … | … | … | n | G |
| 37414 | 10290964-6452578 | 32.3 ± 0.2 | 4592 ± 93 | 2.50 ± 0.13 | 1.009 ± 0.005 | 0.09 ± 0.04 | <55 | 3 | Y | N | Y | N | N | Y | … | … | n | NG? |
| 53828 | 10443707-6420439 | -21.0 ± 0.5 | 6951 ± 110 | 4.26 ± 0.20 | 1.006 ± 0.005 | -0.01 ± 0.10 | <22 | 3 | Y | Y | N | N | N | Y | … | … | n | NG? |
| 37929 | 10424968-6529325 | -7.4 ± 0.2 | 4676 ± 75 | 2.49 ± 0.17 | 1.019 ± 0.003 | -0.08 ± 0.06 | <37 | 3 | N | N | … | … | … | … | … | … | n | G |
| 37930 | 10425010-6537128 | 21.3 ± 0.2 | 4594 ± 49 | 2.30 ± 0.09 | 1.022 ± 0.004 | -0.06 ± 0.02 | <31 | 3 | N | N | … | … | … | … | … | … | n | G |
| 37760 | 10402052-6450063 | -9.8 ± 0.2 | 5123 ± 98 | … | 1.023 ± 0.003 | … | <32 | 3 | N | … | … | … | … | … | … | … | n | G |
| 37931 | 10425025-6526561 | 51.1 ± 0.2 | 4556 ± 142 | … | 1.023 ± 0.006 | 0.07 ± 0.09 | <52 | 3 | N | … | … | … | … | … | … | … | n | G |
| 38015 | 10440430-6358590 | 162.9 ± 24.6 | … | … | … | … | … | … | … | … | … | … | … | … | … | … | n | … |
| 37652 | 10363534-6409042 | 27.5 ± 0.2 | 5006 ± 181 | … | 1.010 ± 0.005 | -0.24 ± 0.28 | <15 | 3 | N | … | … | … | … | … | … | … | n | G |
| 53829 | 10443779-6433473 | -37.1 ± 0.3 | … | … | … | … | … | … | … | … | … | … | … | … | … | … | n | … |
| 37653 | 10363555-6330099 | -3.0 ± 0.2 | 5230 ± 121 | … | 1.032 ± 0.002 | 0.03 ± 0.08 | <25 | 3 | N | … | … | … | … | … | … | … | n | … |
| 11 | 10425084-6400148 | -3.0 ± 0.6 | 5610 ± 44 | 3.97 ± 0.03 | … | -0.28 ± 0.01 | <7 | 3 | … | Y | Y | N | N | Y | … | … | n | … |
| 53786 | 10440446-6419148 | -29.9 ± 0.3 | 6585 ± 53 | … | 1.002 ± 0.005 | 0.82 ± 0.05 | <9 | 3 | Y | … | N | N | N | N | … | … | n | … |
| 37654 | 10363617-6340195 | 43.1 ± 0.2 | 4501 ± 198 | … | 1.019 ± 0.003 | 0.08 ± 0.11 | <68 | 3 | N | … | … | … | … | … | … | … | n | G |
| 37932 | 10425119-6535064 | -2.5 ± 0.3 | 4680 ± 29 | 2.47 ± 0.03 | 1.014 ± 0.007 | -0.03 ± 0.05 | <32 | 3 | N | N | … | … | … | … | … | … | n | G |
| 53879 | 10450997-6429597 | 11.9 ± 0.2 | 5804 ± 75 | … | 1.000 ± 0.004 | … | <21 | 3 | Y | … | Y | N | N | … | … | … | n | NG? |
| 37655 | 10363670-6328140 | -22.3 ± 0.2 | 5049 ± 130 | … | 1.018 ± 0.002 | -0.11 ± 0.13 | <19 | 3 | N | … | … | … | … | … | … | … | n | G |
| 37933 | 10425137-6524456 | 52.2 ± 0.3 | 4881 ± 253 | … | 1.043 ± 0.009 | -0.17 ± 0.17 | … | … | N | … | … | … | … | … | … | … | n | … |
| 37761 | 10402161-6410170 | -20.8 ± 0.2 | 4665 ± 14 | 2.46 ± 0.05 | 1.016 ± 0.005 | -0.04 ± 0.10 | 204 ± 3 | 1 | N | N | … | … | … | … | … | … | n | Li-rich G |
| 53835 | 10444272-6416342 | 55.9 ± 0.2 | 3993 ± 63 | … | 1.033 ± 0.003 | -0.13 ± 0.17 | <30 | 3 | N | … | … | … | … | … | … | … | n | … |
| 53576 | 10402174-6418215 | 22.7 ± 0.2 | 4151 ± 361 | … | 1.034 ± 0.007 | -0.16 ± 0.01 | 395 ± 6 | 1 | N | … | … | … | … | … | … | … | n | Li-rich G |
| 37934 | 10425203-6536216 | 5.1 ± 0.2 | 4776 ± 53 | 2.61 ± 0.05 | 1.021 ± 0.003 | -0.02 ± 0.02 | <29 | 3 | N | N | … | … | … | … | … | … | n | G |
| 53591 | 10403252-6404038 | -1.6 ± 0.2 | 4780 ± 188 | 2.57 ± 0.14 | 1.013 ± 0.003 | 0.00 ± 0.10 | <25 | 3 | N | N | … | … | … | … | … | … | n | G |
| 37769 | 10403368-6357105 | -15.6 ± 0.2 | 4630 ± 51 | 2.46 ± 0.12 | 1.019 ± 0.003 | 0.00 ± 0.09 | <37 | 3 | N | N | … | … | … | … | … | … | n | G |
| 37935 | 10425332-6525029 | -11.6 ± 0.2 | 4958 ± 169 | … | 1.014 ± 0.006 | -0.07 ± 0.12 | <32 | 3 | N | … | … | … | … | … | … | … | n | G |
| 38053 | 10444403-6423219 | -19.3 ± 0.5 | 3615 ± 28 | … | 0.827 ± 0.018 | … | … | … | Y | … | N | … | … | … | N | … | n | … |
| 12 | 10425333-6406020 | -15.3 ± 0.6 | 5018 ± 25 | 2.78 ± 0.06 | … | -0.02 ± 0.01 | <14 | 3 | … | N | … | … | … | … | … | … | n | … |
| 2564 | 10440727-6354568 | 9.5 ± 0.6 | 4625 ± 15 | 2.27 ± 0.09 | 1.030 ± 0.001 | 0.00 ± 0.04 | 40 ± 10 | 1 | N | N | … | … | … | … | … | … | n | G |
| 37936 | 10425337-6536533 | -28.9 ± 0.2 | 4969 ± 68 | … | 1.019 ± 0.006 | -0.05 ± 0.10 | <17 | 3 | N | … | … | … | … | … | … | … | n | G |
| 37770 | 10403596-6431004 | 23.7 ± 0.2 | 4306 ± 280 | … | 1.044 ± 0.003 | -0.03 ± 0.01 | <44 | 3 | N | … | … | … | … | … | … | … | n | G |
| 37664 | 10370820-6327577 | -15.1 ± 0.2 | 5137 ± 17 | … | 1.016 ± 0.002 | -0.03 ± 0.04 | <18 | 3 | N | … | … | … | … | … | … | … | n | G |
| 37937 | 10425402-6529386 | 0.2 ± 0.2 | 4742 ± 79 | 2.57 ± 0.21 | 1.020 ± 0.005 | 0.02 ± 0.03 | <49 | 3 | N | N | … | … | … | … | … | … | n | G |
| 37956 | 10430847-6531390 | 13.2 ± 0.3 | 4549 ± 83 | 2.36 ± 0.13 | 1.014 ± 0.007 | -0.03 ± 0.02 | <42 | 3 | N | N | … | … | … | … | … | … | n | … |
| 37957 | 10430910-6546530 | -13.5 ± 0.2 | 4931 ± 219 | … | 1.019 ± 0.004 | -0.32 ± 0.26 | … | … | N | … | … | … | … | … | … | … | n | G |
| 38074 | 10451355-6459366 | 16.4 ± 0.2 | 4773 ± 32 | 2.61 ± 0.03 | 1.014 ± 0.004 | 0.04 ± 0.04 | <40 | 3 | N | N | … | … | … | … | … | … | n | G |
| 38019 | 10440805-6544010 | -6.7 ± 0.2 | 4946 ± 115 | … | 1.021 ± 0.004 | -0.03 ± 0.06 | <41 | 3 | N | … | … | … | … | … | … | … | n | G |

Article number, page 110 of 264**Table C.6.** continued.

| ID | CNAME | RV (km s$^{-1}$) | $T_{\text{eff}}$ (K) | logg (dex) | $\gamma^a$ | [Fe/H] (dex) | EW(Li)$^b$ (mÅ) | EW(Li) error flag$^c$ | $\gamma$ | logg | RV | Li | H$\alpha$ | [Fe/H] | Randich$^d$ | Cantat-Gaudin$^d$ | Final$^e$ | NMs with Li$^f$ |
|---|---|---|---|---|---|---|---|---|---|---|---|---|---|---|---|---|---|---|
| | | | | | | | | | | | Membership | | | | Gaia studies | | | |
| 53836 | 10444474-6430056 | -20.3 ± 0.2 | 5025 ± 120 | … | 1.020 ± 0.001 | … | <17 | 3 | N | … | … | … | … | … | … | … | n | G |
| 37771 | 10403665-6357440 | 6.4 ± 0.2 | 4664 ± 1 | 2.45 ± 0.01 | 1.019 ± 0.004 | 0.00 ± 0.08 | <34 | 3 | N | N | … | … | … | … | … | … | n | G |
| 37665 | 10371253-6338140 | -28.3 ± 0.2 | 4987 ± 96 | … | 1.023 ± 0.002 | -0.05 ± 0.09 | <25 | 3 | N | … | … | … | … | … | … | … | n | G |
| 53730 | 10431151-6422057 | 55.2 ± 0.2 | 4546 ± 110 | … | 1.035 ± 0.005 | -0.15 ± 0.02 | <36 | 3 | N | … | … | … | … | … | … | … | n | G |
| 53837 | 10444484-6424370 | 32.6 ± 0.2 | 3948 ± 17 | … | 1.038 ± 0.002 | 0.02 ± 0.18 | <33 | 3 | N | … | … | … | … | … | … | … | n | … |
| 37666 | 10371302-6341249 | 5.3 ± 0.2 | 5090 ± 38 | … | 1.014 ± 0.003 | -0.01 ± 0.04 | <18 | 3 | N | … | … | … | … | … | … | … | n | G |
| 37667 | 10371730-6340492 | 7.7 ± 0.3 | 5029 ± 123 | … | 1.017 ± 0.007 | … | <34 | 3 | N | … | … | … | … | … | … | … | n | G |
| 37668 | 10372290-6327007 | 44.9 ± 0.2 | 4494 ± 174 | … | 1.026 ± 0.003 | 0.11 ± 0.08 | <55 | 3 | N | … | … | … | … | … | … | … | n | G |
| 37958 | 10431167-6547236 | 31.9 ± 0.2 | 4507 ± 192 | … | 1.020 ± 0.004 | 0.12 ± 0.04 | <64 | 3 | N | … | … | … | … | … | … | … | n | G |
| 37669 | 10372376-6330393 | -6.7 ± 0.2 | 5182 ± 62 | … | 1.016 ± 0.002 | -0.02 ± 0.06 | <15 | 3 | N | … | … | … | … | … | … | … | n | G |
| 53731 | 10431237-6428262 | 6.5 ± 0.2 | 6146 ± 93 | 3.86 ± 0.07 | 1.006 ± 0.003 | -0.21 ± 0.08 | 41 ± 2 | 1 | Y | Y | Y | N | N | Y | … | … | n | NG? |
| 37772 | 10403795-6442324 | -16.8 ± 0.2 | 4943 ± 121 | … | 1.020 ± 0.002 | -0.06 ± 0.11 | <23 | 3 | N | … | … | … | … | … | … | … | n | G |
| 37670 | 10372449-6341335 | 5.8 ± 0.2 | 4873 ± 110 | … | 1.019 ± 0.001 | -0.04 ± 0.11 | 47 ± 6 | 1 | N | … | … | … | … | … | … | … | n | G |
| 37630 | 10354306-6411101 | 60.2 ± 0.2 | 4508 ± 184 | … | 1.018 ± 0.003 | 0.12 ± 0.06 | <44 | 3 | N | … | … | … | … | … | … | … | n | G |
| 38020 | 10440829-6442183 | 18.7 ± 0.3 | 3522 ± 12 | … | 0.843 ± 0.005 | -0.25 ± 0.13 | … | … | Y | … | Y | … | N | Y | Y | … | n | … |
| 53838 | 10444531-6433185 | 96.1 ± 0.2 | … | … | … | … | … | … | … | … | … | … | … | … | … | … | n | … |
| 37671 | 10372824-6326488 | -20.3 ± 0.2 | 4497 ± 191 | … | 1.024 ± 0.004 | 0.09 ± 0.04 | <58 | 3 | N | … | … | … | … | … | … | … | n | G |
| 2565 | 10440835-6401440 | -486.6 ± 0.6 | 7427 ± 348 | 4.07 ± 0.23 | … | -0.17 ± 0.19 | <18 | 3 | … | Y | N | N | N | Y | N | … | n | NG |
| 53593 | 10403819-6415476 | -29.9 ± 0.2 | 4557 ± 117 | … | 1.031 ± 0.001 | 0.00 ± 0.01 | <41 | 3 | N | … | … | … | … | … | … | … | n | G |
| 38075 | 10451425-6409595 | 31.8 ± 0.2 | 4621 ± 91 | … | 1.027 ± 0.002 | … | <52 | 3 | N | … | … | … | … | … | … | … | n | G |
| 37672 | 10373035-6349152 | 36.9 ± 0.2 | 4607 ± 132 | … | 1.019 ± 0.002 | 0.06 ± 0.05 | <54 | 3 | N | … | … | … | … | … | … | … | n | G |
| 37959 | 10431273-6545454 | 9.2 ± 0.2 | 4655 ± 4 | 2.42 ± 0.02 | 1.015 ± 0.005 | -0.07 ± 0.04 | <35 | 3 | N | N | … | … | … | … | … | … | n | G |
| 2543 | 10403825-6501459 | 19.2 ± 0.6 | 4718 ± 9 | 2.38 ± 0.09 | … | -0.25 ± 0.03 | <24 | 3 | … | N | … | … | … | … | … | … | n | … |
| 38139 | 10461291-6324029 | 11.5 ± 0.2 | 4589 ± 99 | 2.41 ± 0.16 | 1.017 ± 0.002 | 0.02 ± 0.05 | … | … | N | N | … | … | … | … | … | … | n | G |
| 38076 | 10451427-6349275 | 2.2 ± 0.2 | 5106 ± 36 | … | 1.019 ± 0.003 | 0.01 ± 0.06 | <19 | 3 | N | … | … | … | … | … | … | … | n | G |
| 37673 | 10373958-6342576 | 28.7 ± 0.2 | 4577 ± 106 | 2.25 ± 0.12 | 1.020 ± 0.004 | -0.08 ± 0.06 | <21 | 3 | N | N | … | … | … | … | … | … | n | … |
| 53842 | 10444744-6424288 | 8.9 ± 0.2 | 6060 ± 115 | 4.05 ± 0.04 | 0.999 ± 0.003 | -0.08 ± 0.03 | 53 ± 7 | 1 | Y | Y | Y | N | N | Y | … | … | n | NG |
| 37698 | 10382807-6342382 | -12.1 ± 0.2 | 5153 ± 14 | … | 1.015 ± 0.002 | -0.05 ± 0.06 | <21 | 3 | N | … | … | … | … | … | … | … | n | G |
| 38140 | 10461304-6426079 | 18.8 ± 0.5 | 3389 ± 152 | 4.68 ± 0.07 | 0.848 ± 0.012 | -0.27 ± 0.13 | <100 | 3 | Y | Y | Y | Y | Y | Y | Y | Y | Y | … |
| 2557 | 10431327-6538057 | -9.8 ± 0.6 | 4977 ± 13 | 2.76 ± 0.10 | … | 0.00 ± 0.03 | <16 | 3 | … | N | … | … | … | … | … | … | n | … |
| 53888 | 10451770-6431416 | 25.2 ± 0.2 | 4270 ± 288 | … | 1.032 ± 0.002 | -0.03 ± 0.01 | <41 | 3 | N | … | … | … | … | … | … | … | n | G |
| 37773 | 10403887-6408273 | 7.4 ± 0.2 | 4815 ± 34 | 2.71 ± 0.01 | 1.011 ± 0.003 | 0.01 ± 0.03 | <40 | 3 | N | N | … | … | … | … | … | … | n | G |
| 53843 | 10444744-6430158 | -38.8 ± 0.2 | 4570 ± 83 | … | 1.021 ± 0.006 | -0.01 ± 0.21 | <44 | 3 | N | … | … | … | … | … | … | … | n | G |
| 37699 | 10382889-6336559 | -8.2 ± 0.2 | 4565 ± 61 | 2.33 ± 0.19 | 1.029 ± 0.003 | -0.04 ± 0.10 | <35 | 3 | N | N | … | … | … | … | … | … | n | G |
| 38081 | 10451776-6401574 | -0.5 ± 0.2 | 4922 ± 209 | … | 1.015 ± 0.002 | -0.07 ± 0.13 | <18 | 3 | N | … | … | … | … | … | … | … | n | G |
| 38141 | 10461326-6420195 | 5.6 ± 0.4 | 3545 ± 35 | 4.67 ± 0.07 | 0.796 ± 0.012 | -0.22 ± 0.15 | … | … | Y | Y | Y | … | N | Y | N | … | n | … |
| 37700 | 10382900-6337537 | -12.6 ± 0.2 | 4904 ± 111 | … | 1.021 ± 0.002 | -0.07 ± 0.12 | <21 | 3 | N | … | … | … | … | … | … | … | n | G |
| 38157 | 10462608-6315431 | 5.6 ± 2.1 | … | … | … | … | … | … | … | … | … | … | … | … | … | … | n | … |
| 53889 | 10451781-6418575 | 37.3 ± 0.2 | 4701 ± 144 | 2.68 ± 0.14 | 1.006 ± 0.003 | 0.04 ± 0.03 | <46 | 3 | Y | N | Y | N | N | Y | … | … | n | NG? |
| 38158 | 10462634-6426239 | 6.7 ± 0.2 | 4730 ± 62 | 2.56 ± 0.19 | 1.022 ± 0.003 | -0.06 ± 0.09 | <28 | 3 | N | N | … | … | … | … | … | … | n | G |
| 37701 | 10382959-6411399 | 17.6 ± 0.2 | 4719 ± 9 | 2.69 ± 0.01 | 1.009 ± 0.002 | 0.06 ± 0.05 | <42 | 3 | Y | N | Y | N | N | Y | … | … | n | NG? |
| 53890 | 10451782-6421560 | 71.8 ± 0.2 | 5833 ± 49 | 4.17 ± 0.17 | 0.996 ± 0.003 | -0.03 ± 0.07 | 32 ± 7 | 1 | Y | Y | N | N | N | Y | … | … | n | NG |
| 53732 | 10431415-6419460 | 17.8 ± 0.2 | 4696 ± 117 | … | 1.020 ± 0.006 | -0.02 ± 0.07 | 15 ± 10 | 1 | N | … | … | … | … | … | … | … | n | G |
| 38159 | 10462689-6423328 | 10.3 ± 0.2 | 4674 ± 23 | 2.42 ± 0.04 | 1.021 ± 0.004 | -0.05 ± 0.02 | 34 ± 11 | 1 | N | N | … | … | … | … | … | … | n | G |
| 37702 | 10383475-6357148 | -2.9 ± 0.2 | 5266 ± 148 | 3.22 ± 0.15 | 1.037 ± 0.002 | 0.05 ± 0.08 | … | … | N | N | … | … | … | … | … | … | n | … |
| 53789 | 10440902-6423411 | 44.0 ± 0.2 | 4581 ± 80 | 2.80 ± 0.15 | 0.998 ± 0.002 | 0.10 ± 0.06 | <57 | 3 | Y | N | N | N | N | Y | … | … | n | NG? |
| 38055 | 10444782-6342583 | -44.6 ± 0.2 | 4763 ± 101 | 2.61 ± 0.20 | 1.017 ± 0.003 | -0.10 ± 0.10 | <28 | 3 | N | N | … | … | … | … | … | … | n | G |
| 53891 | 10451783-6418210 | 4.9 ± 0.2 | 4390 ± 206 | … | 1.040 ± 0.003 | 0.00 ± 0.01 | <47 | 3 | N | … | … | … | … | … | … | … | n | G |
| 38160 | 10462699-6334383 | 0.9 ± 0.2 | 4656 ± 1 | 2.42 ± 0.06 | 1.022 ± 0.002 | -0.04 ± 0.01 | <34 | 3 | N | N | … | … | … | … | … | … | n | G |
| 37960 | 10431466-6452592 | 10.6 ± 0.2 | 4575 ± 86 | … | 1.036 ± 0.002 | … | <38 | 3 | N | … | … | … | … | … | … | … | n | G |
| 37703 | 10384104-6336301 | 2.5 ± 0.2 | 4636 ± 2 | 2.41 ± 0.06 | 1.026 ± 0.003 | 0.02 ± 0.04 | <35 | 3 | N | N | … | … | … | … | … | … | n | G |
| 53892 | 10451786-6430374 | -27.8 ± 0.2 | 5132 ± 3 | … | 1.008 ± 0.003 | -0.17 ± 0.16 | <24 | 3 | Y | … | N | N | N | Y | … | … | n | NG? |
| 53602 | 10404759-6407218 | 0.8 ± 0.2 | 5727 ± 53 | 4.05 ± 0.09 | 0.995 ± 0.002 | 0.10 ± 0.05 | <15 | 3 | Y | Y | Y | N | N | Y | … | … | n | NG |
| 38027 | 10441279-6540127 | -29.9 ± 0.2 | 5128 ± 39 | … | 1.022 ± 0.003 | -0.06 ± 0.09 | <24 | 3 | N | … | … | … | … | … | … | … | n | G |
| 38162 | 10462779-6319127 | 5.1 ± 0.2 | 4678 ± 23 | 2.51 ± 0.05 | 1.017 ± 0.003 | 0.05 ± 0.04 | <42 | 3 | N | N | … | … | … | … | … | … | n | G |
| 53603 | 10404769-6404580 | 44.7 ± 0.3 | 4003 ± 57 | … | 1.060 ± 0.012 | -0.19 ± 0.14 | <28 | 3 | N | … | … | … | … | … | … | … | n | G |
| 33 | 10462898-6407577 | 17.8 ± 0.6 | 6271 ± 115 | 3.94 ± 0.15 | … | -0.02 ± 0.12 | 112 ± 21 | 2 | … | Y | Y | Y | Y | Y | Y | … | Y | … |
| 37704 | 10384642-6400031 | 13.5 ± 0.2 | 4570 ± 103 | 2.46 ± 0.20 | 1.017 ± 0.002 | 0.09 ± 0.07 | <56 | 3 | N | N | … | … | … | … | … | … | n | G |
| 37780 | 10404781-6409037 | 8.2 ± 0.2 | 4634 ± 45 | 2.39 ± 0.06 | 1.021 ± 0.002 | -0.04 ± 0.03 | 37 ± 3 | 1 | N | N | … | … | … | … | … | … | n | G |

A&A proofs: manuscript no. output

| ID | CNAME | RV (km s$^{-1}$) | $T_{\text{eff}}$ (K) | $\log g$ (dex) | $\gamma^a$ | [Fe/H] (dex) | EW(Li)$^b$ (mÅ) | EW(Li) error flag$^c$ | $\gamma$ | $\log g$ | Membership RV | Li | H$\alpha$ | [Fe/H] | Gaia studies Randich$^d$ | Cantat-Gaudin$^d$ | Final$^e$ | NMs with Li$^f$ |
|---|---|---|---|---|---|---|---|---|---|---|---|---|---|---|---|---|---|---|
| 38163 | 10462950-6335519 | 64.8 ± 0.2 | 4409 ± 206 | … | 1.056 ± 0.003 | -0.20 ± 0.09 | 151 ± 5 | 1 | N | … | … | … | … | … | … | … | n | G |
| 53844 | 10444816-6416320 | 6.4 ± 0.2 | 5873 ± 19 | 3.90 ± 0.11 | 1.000 ± 0.002 | -0.38 ± 0.29 | 41 ± 3 | 1 | Y | Y | Y | N | N | N | … | … | n | NG |
| 37705 | 10384756-6338385 | 8.5 ± 0.2 | 4704 ± 68 | 2.49 ± 0.03 | 1.024 ± 0.003 | -0.04 ± 0.03 | <25 | 3 | N | N | … | … | … | … | … | … | n | G |
| 37706 | 10384930-6410267 | -27.9 ± 0.2 | 4938 ± 164 | … | 1.023 ± 0.003 | -0.07 ± 0.12 | <30 | 3 | N | … | … | … | … | … | … | … | n | G |
| 53893 | 10451817-6427378 | -19.2 ± 0.3 | 6502 ± 95 | 3.75 ± 0.19 | 1.013 ± 0.005 | -0.16 ± 0.08 | … | … | N | Y | … | … | … | … | … | … | n | … |
| 2579 | 10463154-6329392 | 44.3 ± 0.6 | 4755 ± 51 | 2.33 ± 0.08 | 1.021 ± 0.002 | 0.02 ± 0.04 | 27 ± 5 | 1 | N | N | … | … | … | … | … | … | n | G |
| 38164 | 10463272-6425533 | -12.8 ± 0.2 | 5160 ± 10 | 3.21 ± 0.12 | 1.012 ± 0.003 | 0.03 ± 0.04 | … | … | N | N | … | … | … | … | … | … | n | G |
| 53733 | 10431547-6432102 | -12.7 ± 0.2 | 5009 ± 151 | … | 1.015 ± 0.004 | -0.11 ± 0.15 | 19 ± 4 | 1 | N | … | … | … | … | … | … | … | n | G |
| 37707 | 10385285-6335490 | -28.2 ± 0.2 | 5119 ± 34 | … | 1.017 ± 0.003 | -0.08 ± 0.07 | <26 | 3 | N | … | … | … | … | … | … | … | n | G |
| 53604 | 10404908-6411105 | 35.3 ± 1.8 | … | … | … | … | … | … | … | … | … | … | … | … | … | … | n | … |
| 38165 | 10463279-6318249 | -0.9 ± 0.4 | 4546 ± 182 | … | 1.010 ± 0.018 | -0.31 ± 0.17 | <65 | 3 | N | … | … | … | … | … | … | … | n | G |
| 37708 | 10385399-6359079 | -3.3 ± 0.2 | 5293 ± 158 | … | 1.026 ± 0.002 | -0.05 ± 0.01 | <23 | 3 | N | … | … | … | … | … | … | … | n | … |
| 38082 | 10451854-6426540 | 17.0 ± 0.3 | 3461 ± 82 | … | 0.833 ± 0.006 | -0.26 ± 0.14 | <100 | 3 | Y | … | Y | Y | Y | Y | Y | Y | Y | … |
| 53951 | 10463295-6429421 | 41.5 ± 0.2 | 6037 ± 214 | 4.17 ± 0.01 | 0.998 ± 0.005 | 0.18 ± 0.09 | <11 | 3 | Y | Y | Y | N | N | N | … | … | n | NG |
| 38028 | 10441312-6400406 | 574.9 ± 838.5 | … | … | … | … | … | … | … | … | … | … | … | … | … | … | n | … |
| 38083 | 10451855-6332265 | 17.7 ± 0.2 | 4068 ± 92 | 4.47 ± 0.12 | 0.876 ± 0.002 | -0.06 ± 0.10 | <17 | 3 | Y | Y | Y | Y | Y | Y | Y | Y | Y | … |
| 38166 | 10463370-6415146 | 167.0 ± 326.9 | … | … | … | … | … | … | … | … | … | … | … | … | … | … | n | … |
| 53738 | 10432316-6433295 | -36.5 ± 0.2 | 6089 ± 154 | 4.05 ± 0.05 | 1.000 ± 0.003 | -0.49 ± 0.13 | <3 | 3 | Y | Y | N | N | N | N | … | … | n | … |
| 37709 | 10385625-6344310 | -26.1 ± 0.2 | 4746 ± 11 | 2.57 ± 0.11 | 1.023 ± 0.003 | -0.05 ± 0.01 | <30 | 3 | N | N | … | … | … | … | … | … | n | G |
| 38167 | 10463483-6414214 | 45.6 ± 0.2 | 4570 ± 149 | … | 1.024 ± 0.002 | 0.06 ± 0.06 | <59 | 3 | N | … | … | … | … | … | … | … | n | G |
| 37710 | 10385663-6352432 | -10.0 ± 0.2 | 4569 ± 261 | … | 1.024 ± 0.002 | -0.03 ± 0.07 | <33 | 3 | N | … | … | … | … | … | … | … | n | G |
| 53793 | 10441319-6421575 | 47.6 ± 0.2 | 4735 ± 58 | … | 0.992 ± 0.004 | 0.05 ± 0.08 | <33 | 3 | Y | … | N | N | N | Y | … | … | n | NG |
| 34 | 10463540-6403400 | … | … | … | … | … | … | … | … | … | … | … | … | … | … | … | n | … |
| 38168 | 10463663-6406101 | 450.2 ± 3.7 | … | … | … | … | … | … | … | … | … | … | … | … | … | … | n | … |
| 37711 | 10385761-6336059 | 7.4 ± 0.2 | 4827 ± 14 | 2.71 ± 0.01 | 1.013 ± 0.003 | 0.05 ± 0.03 | <32 | 3 | N | N | … | … | … | … | … | … | n | G |
| 38169 | 10463869-6405397 | -18.3 ± 0.2 | 4691 ± 25 | 2.46 ± 0.05 | 1.029 ± 0.002 | 0.00 ± 0.07 | <45 | 3 | N | N | … | … | … | … | … | … | n | G |
| 37781 | 10404971-6452371 | 44.9 ± 0.2 | 4978 ± 186 | … | 1.013 ± 0.003 | -0.22 ± 0.25 | … | … | N | … | … | … | … | … | … | … | n | G |
| 38056 | 10444881-6413181 | -11.1 ± 0.2 | 5029 ± 77 | … | 1.016 ± 0.001 | -0.07 ± 0.06 | <24 | 3 | N | … | … | … | … | … | … | … | n | G |
| 2580 | 10463918-6411459 | -18.1 ± 0.6 | 4918 ± 20 | 2.58 ± 0.10 | … | -0.26 ± 0.01 | <13 | 3 | … | N | … | … | … | … | … | … | n | … |
| 53605 | 10404980-6415138 | 17.4 ± 0.2 | 4479 ± 221 | … | 1.039 ± 0.005 | -0.05 ± 0.04 | <38 | 3 | N | … | … | … | … | … | … | … | n | G |
| 35 | 10463922-6407079 | -7.4 ± 0.6 | 6252 ± 129 | 3.87 ± 0.22 | … | 0.12 ± 0.13 | 72 ± 33 | 2 | … | Y | Y | N | N | Y | N | … | n | NG |
| 37712 | 10385866-6341319 | -60.3 ± 0.2 | 4734 ± 15 | 2.43 ± 0.05 | 1.023 ± 0.002 | -0.09 ± 0.07 | <23 | 3 | N | N | … | … | … | … | … | … | n | G |
| 53606 | 10404994-6422226 | 14.5 ± 0.2 | 4522 ± 115 | … | 1.051 ± 0.003 | … | <20 | 3 | N | … | … | … | … | … | … | … | n | G |
| 37966 | 10432424-6546519 | 2.9 ± 0.2 | 4590 ± 70 | 2.48 ± 0.15 | 1.019 ± 0.004 | 0.14 ± 0.09 | <70 | 3 | N | N | … | … | … | … | … | … | n | G |
| 37713 | 10385868-6341107 | 30.4 ± 0.2 | 4636 ± 98 | … | 1.025 ± 0.001 | … | <50 | 3 | N | … | … | … | … | … | … | … | n | G |
| 38170 | 10463932-6424090 | -8.3 ± 0.2 | 4401 ± 200 | … | 1.055 ± 0.003 | -0.15 ± 0.09 | <23 | 3 | N | … | … | … | … | … | … | … | n | G |
| 53739 | 10432490-6429373 | -28.3 ± 0.3 | 5802 ± 61 | … | 1.000 ± 0.005 | -0.27 ± 0.05 | … | … | Y | … | N | N | N | Y | … | … | n | … |
| 37714 | 10390431-6334067 | -4.7 ± 0.2 | 4999 ± 7 | 2.75 ± 0.06 | 1.035 ± 0.004 | 0.04 ± 0.04 | … | … | N | N | … | … | … | … | … | … | n | G |
| 38213 | 10471578-6404393 | 12.1 ± 0.2 | 4720 ± 44 | 2.57 ± 0.09 | 1.020 ± 0.002 | -0.01 ± 0.07 | <28 | 3 | N | N | … | … | … | … | … | … | n | G |
| 38214 | 10471610-6413139 | 19.5 ± 0.2 | 4670 ± 49 | 2.48 ± 0.15 | 1.026 ± 0.002 | 0.06 ± 0.07 | <52 | 3 | N | N | … | … | … | … | … | … | n | G |
| 53850 | 10445206-6416099 | 52.2 ± 0.3 | 5339 ± 26 | 3.86 ± 0.18 | 0.995 ± 0.007 | 0.03 ± 0.01 | 16 ± 6 | 1 | Y | Y | N | N | N | Y | … | … | n | NG |
| 19 | 10432523-6413062 | -17.3 ± 0.6 | 4139 ± 74 | 1.77 ± 0.33 | … | -0.18 ± 0.13 | 135 ± 13 | 2 | … | N | … | … | … | … | N | … | n | … |
| 53607 | 10405038-6403522 | 16.5 ± 0.2 | 4494 ± 192 | … | 1.018 ± 0.002 | 0.06 ± 0.10 | <49 | 3 | N | … | … | … | … | … | … | … | n | G |
| 38215 | 10471770-6413231 | 19.2 ± 0.2 | 4790 ± 16 | 2.62 ± 0.04 | 1.016 ± 0.002 | 0.03 ± 0.05 | <42 | 3 | N | N | … | … | … | … | … | … | n | G |
| 53794 | 10441387-6426460 | -11.2 ± 0.2 | 5753 ± 82 | 4.06 ± 0.15 | 0.997 ± 0.004 | -0.25 ± 0.13 | <16 | 3 | Y | Y | Y | N | N | Y | … | … | n | NG |
| 53608 | 10405084-6418491 | -25.2 ± 0.2 | 5729 ± 16 | … | 1.008 ± 0.002 | -0.35 ± 0.03 | <11 | 3 | Y | … | N | N | N | N | … | … | n | NG? |
| 38216 | 10471863-6442119 | -30.0 ± 0.2 | 5131 ± 35 | 3.11 ± 0.18 | 1.019 ± 0.003 | -0.01 ± 0.01 | <28 | 3 | N | N | … | … | … | … | … | … | n | G |
| 37715 | 10390734-6354136 | -16.4 ± 0.2 | 4734 ± 24 | 2.52 ± 0.05 | 1.020 ± 0.003 | -0.01 ± 0.01 | <39 | 3 | N | N | … | … | … | … | … | … | n | G |
| 38217 | 10471901-6400463 | 30.1 ± 0.2 | 3480 ± 3 | … | 0.880 ± 0.007 | -0.24 ± 0.14 | … | … | Y | … | Y | … | … | Y | N | Y | n | … |
| 53609 | 10405092-6413124 | 126.6 ± 0.2 | 3875 ± 81 | … | 1.037 ± 0.004 | … | 117 ± 20 | 1 | N | … | … | … | … | … | … | … | n | G |
| 38029 | 10441394-6446273 | 17.3 ± 0.2 | 5142 ± 100 | … | 0.975 ± 0.001 | … | 267 ± 8 | 1 | Y | … | Y | Y | Y | Y | … | … | Y | … |
| 53740 | 10432597-6423277 | 376.8 ± 9.6 | … | … | … | … | … | … | … | … | … | … | … | … | … | … | n | … |
| 38218 | 10472089-6441345 | -16.2 ± 0.2 | 4913 ± 226 | … | 1.024 ± 0.002 | -0.06 ± 0.11 | 50 ± 6 | 1 | N | … | … | … | … | … | … | … | n | G |
| 37782 | 10405137-6442479 | 500.1 ± 0.2 | 4599 ± 32 | … | … | 0.04 ± 0.02 | … | … | … | … | … | … | … | … | … | … | n | … |
| 38058 | 10445244-6449066 | 10.4 ± 0.2 | 4718 ± 78 | 2.46 ± 0.02 | 1.024 ± 0.004 | -0.05 ± 0.02 | <24 | 3 | N | N | … | … | … | … | … | … | n | G |
| 38030 | 10441445-6349346 | 33.6 ± 0.2 | 4816 ± 345 | … | 1.016 ± 0.003 | -0.31 ± 0.23 | <28 | 3 | N | … | … | … | … | … | … | … | n | G |
| 37967 | 10432615-6536554 | 1.8 ± 0.3 | 4661 ± 26 | 2.51 ± 0.04 | 1.011 ± 0.007 | -0.12 ± 0.06 | 15 ± 4 | 1 | N | N | … | … | … | … | … | … | n | G |
| 38219 | 10472161-6403002 | -9.0 ± 0.2 | 4419 ± 176 | … | 1.057 ± 0.002 | -0.10 ± 0.10 | … | … | N | … | … | … | … | … | … | … | n | G |
| 53544 | 10393593-6414479 | -10.9 ± 4.1 | … | … | … | … | … | … | … | … | … | … | … | … | … | … | n | … |

**Table C.6.** continued.

| ID | CNAME | RV (km s$^{-1}$) | $T_{\text{eff}}$ (K) | $logg$ (dex) | $\gamma^a$ | [Fe/H] (dex) | EW(Li)$^b$ (mÅ) | EW(Li) error flag$^c$ | $\gamma$ | $logg$ | Membership RV | Li | H$\alpha$ | [Fe/H] | Gaia studies Randich$^d$ | Cantat-Gaudin$^d$ | Final$^e$ | NMs with Li$^f$ |
|---|---|---|---|---|---|---|---|---|---|---|---|---|---|---|---|---|---|---|
| 53896 | 10452139-6432226 | 4.6 ± 0.2 | 5584 ± 72 | ... | 0.995 ± 0.004 | -0.07 ± 0.02 | 42 ± 3 | 1 | Y | ... | Y | N | N | Y | ... | ... | n | NG |
| 38220 | 10472484-6401325 | 9.1 ± 0.2 | 4758 ± 12 | 2.61 ± 0.18 | 1.021 ± 0.003 | -0.08 ± 0.14 | <30 | 3 | N | N | ... | ... | ... | ... | ... | ... | n | G |
| 37728 | 10393612-6413201 | 6.2 ± 0.2 | 5045 ± 119 | ... | 1.014 ± 0.003 | 0.01 ± 0.01 | <30 | 3 | N | ... | ... | ... | ... | ... | ... | ... | n | G |
| 2567 | 10441461-6457559 | 13.7 ± 0.6 | 4917 ± 34 | 2.63 ± 0.08 | 1.021 ± 0.003 | -0.02 ± 0.02 | 22 ± 3 | 1 | N | N | ... | ... | ... | ... | ... | ... | n | G |
| 20 | 10432640-6406387 | 51.9 ± 0.6 | 6483 ± 374 | 3.94 ± 0.20 | ... | 0.01 ± 0.04 | <61 | 3 | ... | Y | N | N | N | N | N | ... | n | NG |
| 37 | 10472500-6400378 | -54.3 ± 0.4 | ... | ... | ... | ... | ... | ... | ... | ... | ... | ... | ... | ... | ... | ... | n | ... |
| 53545 | 10393690-6412363 | -6.4 ± 0.2 | 4232 ± 254 | ... | 1.061 ± 0.003 | -0.10 ± 0.16 | 62 ± 3 | 1 | N | ... | ... | ... | ... | ... | ... | ... | n | G |
| 53851 | 10445269-6432357 | 43.1 ± 0.4 | 5921 ± 277 | 4.30 ± 0.01 | 0.992 ± 0.007 | 0.31 ± 0.16 | 68 ± 12 | 1 | Y | Y | Y | N | N | N | ... | ... | n | NG |
| 53631 | 10410398-6412427 | -2.2 ± 0.2 | 6293 ± 128 | 3.91 ± 0.11 | 1.005 ± 0.001 | -0.13 ± 0.04 | 23 ± 1 | 1 | Y | Y | Y | N | N | Y | ... | ... | n | NG? |
| 53795 | 10441466-6428483 | 14.2 ± 0.2 | 4598 ± 52 | 4.42 ± 0.04 | 0.946 ± 0.003 | -0.01 ± 0.07 | <18 | 3 | Y | Y | Y | N | N | Y | ... | ... | n | NG |
| 2585 | 10472547-6331564 | 24.4 ± 0.6 | 5112 ± 32 | 3.96 ± 0.13 | ... | 0.14 ± 0.03 | 27 ± 6 | 2 | ... | Y | Y | N | Y | N | ... | ... | n | NG |
| 38221 | 10472668-6357592 | 16.6 ± 0.3 | 3479 ± 74 | 4.61 ± 0.19 | 0.834 ± 0.007 | -0.23 ± 0.15 | ... | ... | Y | Y | Y | ... | N | Y | Y | Y | n | ... |
| 53546 | 10393796-6419037 | -3.0 ± 0.2 | 4944 ± 161 | ... | 1.020 ± 0.004 | -0.10 ± 0.17 | <29 | 3 | N | ... | ... | ... | ... | ... | ... | ... | n | G |
| 37803 | 10410408-6429541 | 15.7 ± 0.2 | 4671 ± 27 | 2.46 ± 0.05 | 1.019 ± 0.003 | -0.06 ± 0.08 | <27 | 3 | N | N | ... | ... | ... | ... | ... | ... | n | G |
| 38222 | 10472776-6316054 | -18.3 ± 0.2 | 4532 ± 113 | 2.32 ± 0.20 | 1.023 ± 0.003 | -0.03 ± 0.02 | <40 | 3 | N | N | ... | ... | ... | ... | ... | ... | n | G |
| 38223 | 10472819-6505373 | -0.8 ± 0.2 | 4590 ± 92 | 2.38 ± 0.13 | 1.018 ± 0.004 | 0.02 ± 0.01 | <41 | 3 | N | N | ... | ... | ... | ... | ... | ... | n | G |
| 37804 | 10410411-6529309 | 37.3 ± 0.2 | 4730 ± 341 | ... | 0.950 ± 0.004 | -0.10 ± 0.01 | <19 | 3 | Y | ... | Y | N | N | Y | ... | ... | n | NG |
| 38224 | 10472899-6438543 | -1.2 ± 0.2 | 5052 ± 33 | ... | 1.019 ± 0.001 | -0.04 ± 0.07 | ... | ... | N | ... | ... | ... | ... | ... | ... | ... | n | G |
| 53547 | 10393872-6419539 | 26.1 ± 0.2 | 5030 ± 70 | 3.14 ± 0.08 | 1.007 ± 0.002 | -0.03 ± 0.06 | <34 | 3 | Y | N | Y | N | N | Y | ... | ... | n | NG? |
| 37805 | 10410522-6444231 | -6.9 ± 0.2 | 4873 ± 179 | ... | 1.040 ± 0.002 | -0.07 ± 0.08 | <19 | 3 | N | ... | ... | ... | ... | ... | ... | ... | n | G |
| 53798 | 10441697-6424579 | -2.3 ± 0.2 | 4595 ± 90 | 2.32 ± 0.13 | 1.020 ± 0.005 | -0.08 ± 0.02 | <38 | 3 | N | N | ... | ... | ... | ... | ... | ... | n | ... |
| 37968 | 10432723-6451196 | 41.2 ± 0.2 | 4671 ± 88 | 2.43 ± 0.17 | 1.021 ± 0.002 | -0.06 ± 0.11 | <37 | 3 | N | N | ... | ... | ... | ... | ... | ... | n | G |
| 38225 | 10472962-6500575 | -10.4 ± 0.2 | 4717 ± 157 | ... | 1.022 ± 0.002 | 0.00 ± 0.13 | <30 | 3 | N | ... | ... | ... | ... | ... | ... | ... | n | G |
| 53741 | 10432724-6419165 | -10.9 ± 0.4 | 6047 ± 21 | 3.88 ± 0.02 | 1.002 ± 0.005 | -0.33 ± 0.16 | <53 | 3 | Y | Y | Y | N | N | N | ... | ... | n | NG? |
| 37729 | 10393979-6352555 | 2.9 ± 0.2 | 5146 ± 17 | ... | 1.011 ± 0.002 | -0.07 ± 0.09 | <17 | 3 | N | ... | ... | ... | ... | ... | ... | ... | n | G |
| 38226 | 10473011-6436160 | 8.5 ± 0.2 | 4581 ± 76 | ... | 1.032 ± 0.003 | ... | ... | ... | N | ... | ... | ... | ... | ... | ... | ... | n | G |
| 37730 | 10393999-6402565 | 2.4 ± 0.2 | 4544 ± 35 | ... | 1.036 ± 0.003 | -0.09 ± 0.01 | <31 | 3 | N | ... | ... | ... | ... | ... | ... | ... | n | G |
| 38227 | 10473020-6319239 | -2.4 ± 0.2 | 4931 ± 137 | ... | 1.021 ± 0.002 | -0.08 ± 0.14 | <25 | 3 | N | ... | ... | ... | ... | ... | ... | ... | n | G |
| 53632 | 10410563-6407540 | -47.1 ± 0.2 | 4560 ± 148 | ... | 1.030 ± 0.003 | -0.11 ± 0.11 | <23 | 3 | N | ... | ... | ... | ... | ... | ... | ... | n | G |
| 38228 | 10473091-6405581 | -18.6 ± 0.2 | 4837 ± 110 | 2.65 ± 0.14 | 1.018 ± 0.001 | -0.05 ± 0.07 | <30 | 3 | N | N | ... | ... | ... | ... | ... | ... | n | G |
| 38229 | 10473135-6318527 | -2.6 ± 0.2 | 4912 ± 72 | 2.66 ± 0.15 | 1.036 ± 0.002 | 0.00 ± 0.01 | <26 | 3 | N | N | ... | ... | ... | ... | ... | ... | n | G |
| 37806 | 10410576-6531204 | 3.8 ± 0.2 | 4869 ± 223 | ... | 1.020 ± 0.005 | -0.10 ± 0.17 | <31 | 3 | N | ... | ... | ... | ... | ... | ... | ... | n | G |
| 37980 | 10433732-6537301 | 116.8 ± 0.3 | 5176 ± 96 | ... | 1.007 ± 0.007 | ... | <37 | 3 | Y | ... | N | N | N | N | ... | ... | n | NG? |
| 38230 | 10473303-6422579 | 0.2 ± 0.6 | 4180 ± 102 | ... | 0.898 ± 0.019 | 0.26 ± 0.13 | ... | ... | Y | ... | Y | N | N | N | N | ... | n | ... |
| 53548 | 10394034-6408581 | 9.6 ± 0.2 | 5736 ± 132 | ... | 1.000 ± 0.004 | -0.32 ± 0.24 | 59 ± 3 | 1 | Y | ... | Y | N | N | N | ... | ... | n | NG |
| 53799 | 10441742-6420461 | -34.1 ± 0.2 | 4931 ± 51 | 2.84 ± 0.05 | 1.018 ± 0.003 | 0.02 ± 0.08 | <24 | 3 | N | N | ... | ... | ... | ... | ... | ... | n | G |
| 37981 | 10433733-6533235 | -16.9 ± 0.2 | 5156 ± 3 | 3.50 ± 0.13 | 1.004 ± 0.003 | -0.11 ± 0.08 | 30 ± 4 | 1 | Y | N | N | N | N | Y | ... | ... | n | NG? |
| 38231 | 10473331-6401516 | -29.1 ± 0.2 | 4524 ± 116 | ... | 1.031 ± 0.002 | 0.11 ± 0.03 | <46 | 3 | N | ... | ... | ... | ... | ... | ... | ... | n | G |
| 53633 | 10410710-6417138 | 40.4 ± 0.2 | 5683 ± 141 | ... | 0.996 ± 0.005 | 0.15 ± 0.02 | 32 ± 12 | 1 | Y | ... | Y | N | N | Y | ... | ... | n | NG |
| 53800 | 10441744-6422364 | -45.2 ± 0.3 | 5895 ± 172 | 4.04 ± 0.18 | 0.995 ± 0.005 | -0.37 ± 0.04 | 40 ± 3 | 1 | Y | Y | N | N | N | N | ... | ... | n | NG |
| 38232 | 10473466-6402070 | 1.2 ± 0.2 | 4997 ± 142 | ... | 1.009 ± 0.001 | -0.07 ± 0.15 | <14 | 3 | Y | ... | ... | ... | ... | Y | ... | ... | n | NG? |
| 53549 | 10394171-6420557 | -5.7 ± 0.2 | 5119 ± 104 | ... | 0.999 ± 0.003 | ... | <41 | 3 | Y | ... | Y | N | N | ... | ... | ... | n | NG |
| 37807 | 10410712-6450183 | -0.4 ± 0.2 | 4551 ± 134 | ... | 1.025 ± 0.003 | 0.05 ± 0.02 | 41 ± 5 | 1 | N | ... | ... | ... | ... | ... | ... | ... | n | G |
| 2560 | 10433789-6544597 | 9.8 ± 0.6 | 7092 ± 142 | 4.03 ± 0.20 | ... | -0.14 ± 0.15 | <6 | 3 | ... | Y | Y | N | N | N | N | ... | n | ... |
| 37731 | 10394210-6441110 | -2.9 ± 0.2 | 4602 ± 12 | 2.41 ± 0.09 | 1.021 ± 0.004 | -0.01 ± 0.04 | <40 | 3 | N | N | ... | ... | ... | ... | ... | ... | n | G |
| 53634 | 10410866-6408597 | 10.0 ± 0.3 | 5980 ± 42 | 4.09 ± 0.15 | 1.000 ± 0.005 | 0.12 ± 0.02 | 27 ± 4 | 1 | Y | Y | Y | N | N | Y | ... | ... | n | NG? |
| 38233 | 10473643-6338538 | 41.7 ± 0.2 | 4890 ± 127 | 2.63 ± 0.17 | 1.026 ± 0.003 | -0.06 ± 0.06 | <23 | 3 | N | N | ... | ... | ... | ... | ... | ... | n | G |
| 53750 | 10433805-6428591 | 3.1 ± 0.3 | 6830 ± 73 | ... | 0.990 ± 0.004 | 0.39 ± 0.06 | <26 | 3 | Y | ... | Y | N | N | N | ... | ... | n | NG |
| 53550 | 10394220-6416057 | 56.0 ± 0.2 | 4682 ± 218 | ... | 1.014 ± 0.002 | -0.20 ± 0.18 | <21 | 3 | N | ... | ... | ... | ... | ... | ... | ... | n | G |
| 37808 | 10410898-6534238 | 48.5 ± 0.3 | 4864 ± 189 | ... | 1.020 ± 0.006 | ... | <26 | 3 | N | ... | ... | ... | ... | ... | ... | ... | n | G |
| 38234 | 10473659-6504069 | -28.1 ± 0.2 | 5118 ± 32 | ... | 1.035 ± 0.003 | -0.07 ± 0.01 | <20 | 3 | N | ... | ... | ... | ... | ... | ... | ... | n | G |
| 37982 | 10433844-6352382 | 6.5 ± 0.2 | 4793 ± 27 | 2.65 ± 0.09 | 1.018 ± 0.004 | -0.04 ± 0.05 | ... | ... | N | N | ... | ... | ... | ... | ... | ... | n | G |
| 53551 | 10394224-6414480 | 18.1 ± 0.2 | 4615 ± 122 | 2.69 ± 0.09 | 1.002 ± 0.008 | 0.01 ± 0.02 | 29 ± 5 | 1 | Y | N | Y | N | N | Y | ... | ... | n | NG? |
| 53635 | 10410903-6417138 | 10.4 ± 0.2 | 5016 ± 158 | ... | 1.007 ± 0.001 | -0.28 ± 0.23 | 5 ± 2 | 1 | Y | ... | Y | N | N | Y | ... | ... | n | ... |
| 38256 | 10475849-6431480 | 0.1 ± 0.2 | 4928 ± 233 | ... | 1.008 ± 0.003 | -0.05 ± 0.09 | 39 ± 5 | 1 | Y | ... | Y | N | N | Y | ... | ... | n | NG? |
| 53751 | 10433845-6426480 | -16.6 ± 0.2 | 4715 ± 10 | 2.47 ± 0.09 | 1.030 ± 0.005 | -0.01 ± 0.03 | <30 | 3 | N | N | ... | ... | ... | ... | ... | ... | n | G |
| 37809 | 10410905-6540058 | 23.5 ± 0.2 | 4763 ± 117 | ... | 1.021 ± 0.003 | -0.08 ± 0.14 | <29 | 3 | N | ... | ... | ... | ... | ... | ... | ... | n | G |
| 38257 | 10475881-6320512 | 12.8 ± 0.2 | 4532 ± 153 | ... | 1.023 ± 0.003 | 0.11 ± 0.04 | <44 | 3 | N | ... | ... | ... | ... | ... | ... | ... | n | G |
| 53552 | 10394320-6411059 | -21.7 ± 0.3 | 4513 ± 175 | 2.67 ± 0.12 | 0.996 ± 0.013 | -0.12 ± 0.06 | 40 ± 14 | 1 | Y | N | N | N | N | Y | ... | ... | n | NG? |

| ID | CNAME | RV (km s$^{-1}$) | T$_{eff}$ (K) | logg (dex) | $\gamma^a$ | [Fe/H] (dex) | EW(Li)$^b$ (mÅ) | EW(Li) error flag$^c$ | $\gamma$ | logg | Membership RV | Li | H$\alpha$ | [Fe/H] | Gaia studies Randich$^d$ | Cantat-Gaudin$^d$ | Final$^e$ | NMs with Li$^f$ |
|---|---|---|---|---|---|---|---|---|---|---|---|---|---|---|---|---|---|---|
| 38258 | 10480104-6318303 | -25.5 ± 0.2 | 4729 ± 221 | … | 1.027 ± 0.004 | -0.06 ± 0.07 | <32 | 3 | N | … | … | … | … | … | … | … | n | G |
| 38259 | 10480223-6320515 | 43.7 ± 0.2 | 4572 ± 189 | … | 1.014 ± 0.003 | -0.05 ± 0.06 | <31 | 3 | N | … | … | … | … | … | … | … | n | G |
| 53636 | 10410917-6412029 | 1.6 ± 0.3 | 6588 ± 106 | 4.23 ± 0.10 | 1.001 ± 0.002 | -0.23 ± 0.01 | 57 ± 9 | 1 | Y | Y | Y | N | N | Y | … | … | n | NG? |
| 38561 | 10513625-6346379 | 17.8 ± 0.2 | 4636 ± 102 | … | 1.026 ± 0.003 | … | <59 | 3 | N | … | … | … | … | … | … | … | n | G |
| 38260 | 10480312-6432063 | 3.3 ± 0.2 | 4639 ± 99 | 2.55 ± 0.13 | 1.011 ± 0.003 | 0.04 ± 0.07 | <40 | 3 | N | N | … | … | … | … | … | … | n | G |
| 38562 | 10513640-6341128 | -18.2 ± 0.2 | 5018 ± 150 | … | 1.015 ± 0.003 | -0.08 ± 0.08 | <18 | 3 | N | … | … | … | … | … | … | … | n | G |
| 38261 | 10480456-6402217 | -4.5 ± 0.2 | 4574 ± 113 | … | 1.025 ± 0.002 | 0.01 ± 0.03 | 48 ± 13 | 1 | N | … | … | … | … | … | … | … | n | G |
| 38262 | 10480474-6440326 | 106.5 ± 0.2 | 4958 ± 147 | … | 1.015 ± 0.003 | -0.15 ± 0.21 | <20 | 3 | N | … | … | … | … | … | … | … | n | G |
| 53637 | 10410934-6413566 | -11.1 ± 0.2 | 4502 ± 133 | … | 1.041 ± 0.004 | 0.01 ± 0.09 | 42 ± 4 | 1 | N | … | … | … | … | … | … | … | n | G |
| 38090 | 10452669-6422361 | -42.8 ± 0.2 | 4886 ± 228 | … | 1.017 ± 0.003 | -0.17 ± 0.24 | <16 | 3 | N | … | … | … | … | … | … | … | n | G |
| 37983 | 10433920-6535469 | 31.2 ± 0.2 | 4638 ± 112 | 2.47 ± 0.14 | 1.013 ± 0.005 | 0.02 ± 0.08 | <36 | 3 | N | N | … | … | … | … | … | … | n | G |
| 38563 | 10513742-6422576 | 43.8 ± 0.2 | 4514 ± 178 | … | 1.026 ± 0.003 | 0.11 ± 0.06 | <41 | 3 | N | … | … | … | … | … | … | … | n | G |
| 38263 | 10480512-6335257 | -10.2 ± 12.0 | … | … | … | … | … | … | … | … | … | … | … | … | … | … | n | … |
| 2547 | 10410978-6536045 | 22.7 ± 0.6 | 5025 ± 5 | 3.47 ± 0.01 | … | 0.03 ± 0.01 | <13 | 3 | … | N | … | … | … | … | … | … | n | NG? |
| 38585 | 10515438-6343095 | 95.8 ± 0.3 | 4879 ± 170 | … | 1.010 ± 0.005 | -0.46 ± 0.21 | <16 | 3 | N | … | … | … | … | … | … | … | n | G |
| 38264 | 10480598-6321172 | 22.5 ± 0.2 | 4932 ± 87 | 2.87 ± 0.20 | 1.016 ± 0.006 | -0.06 ± 0.08 | <38 | 3 | N | N | … | … | … | … | … | … | n | G |
| 38265 | 10480621-6405129 | 18.1 ± 0.2 | 4473 ± 191 | 2.16 ± 0.15 | 1.019 ± 0.002 | 0.16 ± 0.05 | <58 | 3 | N | N | … | … | … | … | … | … | n | G |
| 53854 | 10445567-6418528 | 20.9 ± 0.2 | 4960 ± 244 | 3.67 ± 0.07 | 0.992 ± 0.005 | -0.01 ± 0.04 | <35 | 3 | Y | Y | Y | N | Y | Y | … | … | n | NG |
| 53638 | 10411017-6415501 | -27.5 ± 0.2 | 4631 ± 101 | … | 1.035 ± 0.002 | -0.15 ± 0.17 | <17 | 3 | N | … | … | … | … | … | … | … | n | G |
| 38266 | 10480628-6358591 | 15.6 ± 0.2 | 4498 ± 224 | … | 1.031 ± 0.002 | 0.06 ± 0.13 | <39 | 3 | N | … | … | … | … | … | … | … | n | G |
| 38586 | 10515514-6413096 | -23.4 ± 0.2 | 4865 ± 25 | 2.73 ± 0.06 | 1.019 ± 0.002 | -0.01 ± 0.01 | <29 | 3 | N | N | … | … | … | … | … | … | n | G |
| 53639 | 10411032-6413358 | 61.1 ± 0.3 | 5098 ± 126 | … | 0.980 ± 0.008 | -0.06 ± 0.10 | 34 ± 6 | 1 | Y | … | N | N | N | Y | … | … | n | NG |
| 38267 | 10480715-6437093 | 26.8 ± 0.2 | 4530 ± 160 | … | 1.022 ± 0.002 | -0.05 ± 0.09 | <34 | 3 | N | … | … | … | … | … | … | … | n | G |
| 38268 | 10480789-6333349 | -3.6 ± 0.2 | 4780 ± 259 | … | 1.018 ± 0.002 | -0.14 ± 0.23 | <14 | 3 | N | … | … | … | … | … | … | … | n | G |
| 38587 | 10515599-6334345 | -13.9 ± 0.2 | 5028 ± 156 | … | 1.022 ± 0.003 | -0.13 ± 0.14 | <23 | 3 | N | … | … | … | … | … | … | … | n | G |
| 53640 | 10411076-6423553 | -12.1 ± 0.4 | 6673 ± 72 | 3.98 ± 0.14 | 1.015 ± 0.005 | 0.40 ± 0.06 | … | … | N | Y | … | … | … | … | … | … | n | … |
| 37984 | 10433991-6533599 | -52.4 ± 0.2 | 4679 ± 13 | … | 1.036 ± 0.005 | -0.09 ± 0.07 | <16 | 3 | N | … | … | … | … | … | … | … | n | G |
| 53801 | 10441819-6427484 | 100.1 ± 0.2 | 4541 ± 140 | … | 1.032 ± 0.006 | … | <45 | 3 | N | … | … | … | … | … | … | … | n | G |
| 38059 | 10445609-6346445 | 0.0 ± 0.2 | 4476 ± 140 | 2.23 ± 0.19 | 1.025 ± 0.002 | 0.11 ± 0.05 | <37 | 3 | N | N | … | … | … | … | … | … | n | G |
| 38269 | 10480835-6320328 | -33.8 ± 0.2 | 4893 ± 128 | … | 1.016 ± 0.005 | -0.04 ± 0.11 | <27 | 3 | N | … | … | … | … | … | … | … | n | G |
| 38588 | 10515721-6427348 | -41.6 ± 0.2 | 4889 ± 36 | 2.69 ± 0.13 | 1.024 ± 0.003 | -0.03 ± 0.04 | <26 | 3 | N | N | … | … | … | … | … | … | n | G |
| 37822 | 10412053-6536007 | -22.5 ± 0.2 | 4824 ± 29 | 2.71 ± 0.08 | 1.013 ± 0.003 | -0.02 ± 0.03 | <29 | 3 | N | N | … | … | … | … | … | … | n | G |
| 38270 | 10480990-6329094 | 32.0 ± 0.2 | 4400 ± 242 | … | 1.021 ± 0.003 | 0.14 ± 0.08 | … | … | N | … | … | … | … | … | … | … | n | G |
| 38271 | 10481037-6351075 | -2.3 ± 0.2 | 4815 ± 112 | 2.63 ± 0.15 | 1.021 ± 0.003 | -0.01 ± 0.03 | <31 | 3 | N | N | … | … | … | … | … | … | n | G |
| 53902 | 10452739-6426204 | 27.9 ± 0.2 | 4435 ± 186 | … | 1.027 ± 0.003 | -0.13 ± 0.12 | <29 | 3 | N | … | … | … | … | … | … | … | n | G |
| 38589 | 10515726-6339227 | -9.8 ± 0.2 | 4846 ± 13 | 2.65 ± 0.10 | 1.031 ± 0.002 | 0.02 ± 0.06 | 21 ± 3 | 1 | N | N | … | … | … | … | … | … | n | G |
| 37823 | 10412212-6443104 | 48.9 ± 0.2 | 4708 ± 104 | … | 1.014 ± 0.002 | -0.17 ± 0.14 | <25 | 3 | N | … | … | … | … | … | … | … | n | G |
| 38272 | 10481092-6350118 | -25.5 ± 0.2 | 5085 ± 27 | 3.10 ± 0.14 | 1.016 ± 0.003 | 0.01 ± 0.01 | <27 | 3 | N | N | … | … | … | … | … | … | n | G |
| 38590 | 10515734-6336454 | -23.2 ± 0.3 | 5024 ± 113 | … | 1.024 ± 0.006 | … | <26 | 3 | N | … | … | … | … | … | … | … | n | G |
| 2549 | 10412220-6447128 | -16.2 ± 0.6 | 5033 ± 22 | 2.99 ± 0.07 | 1.017 ± 0.002 | 0.19 ± 0.01 | 28 ± 5 | 1 | N | N | … | … | … | … | … | … | n | G |
| 53903 | 10452783-6421325 | -3.3 ± 0.2 | 5810 ± 118 | 4.14 ± 0.01 | 0.993 ± 0.002 | 0.07 ± 0.04 | 94 ± 8 | 1 | Y | Y | Y | N | Y | Y | … | … | n | NG |
| 2588 | 10481093-6513390 | -9.8 ± 0.6 | 6788 ± 313 | 3.99 ± 0.19 | … | -0.05 ± 0.17 | <46 | 3 | … | Y | Y | N | N | Y | N | … | n | NG |
| 38591 | 10515745-6329450 | 12.1 ± 0.2 | 4502 ± 130 | 2.24 ± 0.20 | 1.026 ± 0.005 | -0.04 ± 0.04 | <40 | 3 | N | N | … | … | … | … | … | … | n | G |
| 53652 | 10412228-6414103 | 63.6 ± 0.2 | 4013 ± 108 | … | 1.118 ± 0.004 | -0.11 ± 0.16 | <5 | 3 | N | … | … | … | … | … | … | … | n | G |
| 53855 | 10445622-6417154 | 13.2 ± 0.3 | 6633 ± 50 | … | 1.007 ± 0.003 | 0.07 ± 0.04 | <7 | 3 | Y | … | Y | N | Y | Y | … | … | n | … |
| 38273 | 10481106-6321113 | 11.0 ± 0.3 | 5133 ± 17 | 3.60 ± 0.09 | 1.001 ± 0.006 | -0.09 ± 0.10 | … | … | Y | Y | Y | N | Y | Y | … | … | n | … |
| 37824 | 10412292-6450149 | -14.9 ± 0.2 | 5017 ± 129 | … | 1.007 ± 0.002 | -0.23 ± 0.24 | 5 ± 1 | 1 | Y | … | N | N | N | Y | … | … | n | … |
| 2561 | 10434029-6444223 | -16.5 ± 0.6 | 4970 ± 37 | 2.59 ± 0.10 | 1.020 ± 0.003 | -0.26 ± 0.01 | 20 ± 4 | 1 | N | N | … | … | … | … | … | … | n | G |
| 38274 | 10481233-6351045 | -1.5 ± 0.2 | 5792 ± 13 | 4.01 ± 0.18 | 0.998 ± 0.002 | -0.32 ± 0.11 | 88 ± 2 | 1 | Y | Y | Y | N | Y | N | … | … | n | NG |
| 2608 | 10515818-6340266 | 9.1 ± 0.6 | 4682 ± 21 | 2.45 ± 0.06 | 1.018 ± 0.002 | 0.25 ± 0.09 | 53 ± 16 | 1 | N | N | … | … | … | … | … | … | n | G |
| 37825 | 10412304-6537112 | 3.9 ± 0.3 | 4250 ± 377 | … | 0.894 ± 0.006 | 0.05 ± 0.19 | 67 ± 10 | 1 | Y | … | Y | Y | Y | Y | … | … | Y | … |
| 53810 | 10442320-6431116 | -17.3 ± 0.2 | 4631 ± 55 | 2.34 ± 0.16 | 1.033 ± 0.001 | -0.05 ± 0.08 | 69 ± 9 | 1 | N | N | … | … | … | … | … | … | n | G |
| 38275 | 10481236-6501289 | 47.1 ± 0.2 | 4842 ± 60 | 2.61 ± 0.10 | 1.018 ± 0.003 | -0.08 ± 0.09 | <24 | 3 | N | N | … | … | … | … | … | … | n | G |
| 38592 | 10515939-6408586 | -23.1 ± 0.2 | 4770 ± 28 | 2.59 ± 0.06 | 1.024 ± 0.002 | -0.03 ± 0.02 | <33 | 3 | N | N | … | … | … | … | … | … | n | G |
| 53653 | 10412366-6401566 | 50.3 ± 0.2 | 4605 ± 69 | … | 0.988 ± 0.004 | 0.05 ± 0.13 | <48 | 3 | Y | … | N | N | N | Y | … | … | n | NG |
| 38276 | 10481419-6351121 | -22.5 ± 0.2 | 4567 ± 42 | … | 1.028 ± 0.002 | -0.15 ± 0.12 | <25 | 3 | N | … | … | … | … | … | … | … | n | G |
| 38039 | 10442327-6418128 | -31.4 ± 0.2 | 4748 ± 203 | … | 1.021 ± 0.001 | -0.14 ± 0.17 | 24 ± 3 | 1 | N | … | … | … | … | … | … | … | n | G |
| 53856 | 10445656-6425005 | 87.1 ± 0.2 | 4469 ± 158 | … | 1.040 ± 0.003 | -0.27 ± 0.18 | <30 | 3 | N | … | … | … | … | … | … | … | n | G |
| 38593 | 10515977-6340549 | 66.4 ± 0.2 | 4639 ± 101 | 2.50 ± 0.12 | 1.011 ± 0.005 | -0.01 ± 0.10 | <48 | 3 | N | N | … | … | … | … | … | … | n | G |

**Table C.6.** continued.

| ID | CNAME | RV (km s$^{-1}$) | $T_{\text{eff}}$ (K) | $\log g$ (dex) | $\gamma^a$ | [Fe/H] (dex) | EW(Li)$^b$ (mÅ) | EW(Li) error flag$^c$ | Membership $\gamma$ | $\log g$ | RV | Li | H$\alpha$ | [Fe/H] | Gaia studies Randich$^d$ | Cantat-Gaudin$^d$ | Final$^e$ | NMs with Li$^f$ |
|---|---|---|---|---|---|---|---|---|---|---|---|---|---|---|---|---|---|---|
| 37826 | 10412390-6443564 | 24.0 ± 1.9 | 3410 ± 191 | … | 0.877 ± 0.014 | -0.21 ± 0.14 | … | … | Y | … | Y | … | … | Y | N | Y | n | … |
| 38277 | 10481459-6501140 | 11.1 ± 0.2 | 4580 ± 86 | 2.47 ± 0.18 | 1.019 ± 0.002 | 0.12 ± 0.03 | 51 ± 2 | 1 | N | N | … | … | … | … | … | … | n | G |
| 53752 | 10434105-6431064 | -13.2 ± 0.3 | 5823 ± 74 | … | 1.001 ± 0.006 | … | 81 ± 5 | 1 | Y | … | N | N | N | … | … | … | n | NG? |
| 53857 | 10445659-6423534 | 466.8 ± 3.8 | … | … | … | … | … | … | … | … | … | … | … | … | … | … | n | … |
| 38594 | 10520183-6409210 | 0.1 ± 0.2 | 5087 ± 66 | 3.17 ± 0.17 | 1.011 ± 0.002 | -0.01 ± 0.02 | <32 | 3 | N | N | … | … | … | … | … | … | n | G |
| 2573 | 10452826-6413450 | 22.0 ± 0.6 | 6426 ± 148 | 3.98 ± 0.16 | … | -0.09 ± 0.15 | <77 | 3 | … | Y | Y | Y | N | Y | Y | Y | n | NG |
| 37827 | 10412419-6440100 | -13.4 ± 0.2 | 5139 ± 6 | … | 1.016 ± 0.001 | -0.08 ± 0.08 | … | … | N | … | … | … | … | … | … | … | n | G |
| 38278 | 10481504-6433243 | -28.2 ± 0.2 | 5034 ± 156 | … | 1.015 ± 0.003 | -0.10 ± 0.11 | 27 ± 2 | 1 | N | … | … | … | … | … | … | … | n | G |
| 53858 | 10445659-6426460 | -4.3 ± 0.2 | 5961 ± 179 | 4.06 ± 0.08 | 1.001 ± 0.003 | 0.05 ± 0.07 | 47 ± 5 | 1 | Y | Y | Y | N | N | Y | … | … | n | NG? |
| 38595 | 10520184-6347405 | -5.8 ± 0.2 | 4539 ± 161 | 2.31 ± 0.20 | 1.022 ± 0.003 | 0.02 ± 0.01 | <47 | 3 | N | N | … | … | … | … | … | … | n | G |
| 53654 | 10412434-6422212 | -0.3 ± 0.3 | 6361 ± 120 | 3.92 ± 0.01 | 1.008 ± 0.003 | -0.02 ± 0.03 | <21 | 3 | Y | Y | Y | N | N | Y | … | … | n | NG? |
| 38324 | 10485483-6349051 | 6.1 ± 0.2 | 4865 ± 80 | 2.71 ± 0.04 | 1.017 ± 0.002 | 0.03 ± 0.07 | <41 | 3 | N | N | … | … | … | … | … | … | n | G |
| 38596 | 10520270-6341101 | 0.6 ± 0.3 | 5043 ± 139 | … | 1.006 ± 0.005 | -0.09 ± 0.17 | <29 | 3 | Y | … | Y | N | N | Y | … | … | n | NG? |
| 38091 | 10452846-6400164 | 44.9 ± 0.2 | 5041 ± 106 | … | 1.030 ± 0.003 | -0.01 ± 0.07 | 31 ± 6 | 1 | N | … | … | … | … | … | … | … | n | G |
| 37828 | 10412493-6536163 | 40.2 ± 0.2 | 5042 ± 131 | … | 1.015 ± 0.003 | -0.24 ± 0.17 | <23 | 3 | N | … | … | … | … | … | … | … | n | G |
| 38325 | 10485504-6407090 | 5.4 ± 0.2 | 4610 ± 63 | … | 1.027 ± 0.002 | 0.05 ± 0.01 | <43 | 3 | N | … | … | … | … | … | … | … | n | G |
| 38597 | 10520313-6351196 | -2.2 ± 0.2 | 5005 ± 29 | … | 1.023 ± 0.002 | -0.02 ± 0.04 | <21 | 3 | N | … | … | … | … | … | … | … | n | G |
| 37829 | 10412510-6445155 | -29.5 ± 0.2 | 4566 ± 53 | … | 1.033 ± 0.002 | -0.14 ± 0.11 | <25 | 3 | N | … | … | … | … | … | … | … | n | G |
| 37988 | 10434332-6536247 | 8.1 ± 0.2 | 4684 ± 9 | 2.47 ± 0.09 | 1.023 ± 0.004 | -0.02 ± 0.05 | 17 ± 3 | 1 | N | N | … | … | … | … | … | … | n | G |
| 2590 | 10485573-6403551 | 27.1 ± 0.6 | 4682 ± 36 | 2.37 ± 0.10 | 1.016 ± 0.002 | 0.21 ± 0.06 | 58 ± 19 | 1 | N | N | … | … | … | … | … | … | n | G |
| 53907 | 10453089-6422177 | 9.5 ± 0.2 | 4514 ± 139 | … | 1.030 ± 0.002 | -0.01 ± 0.02 | 160 ± 2 | 1 | N | … | … | … | … | … | … | … | n | Li-rich G |
| 53655 | 10412538-6417355 | -18.7 ± 2.3 | … | … | … | … | … | … | … | … | … | … | … | … | … | … | n | … |
| 38326 | 10485644-6512475 | -15.1 ± 0.2 | 4669 ± 10 | 2.42 ± 0.09 | 1.023 ± 0.003 | -0.06 ± 0.03 | <32 | 3 | N | N | … | … | … | … | … | … | n | G |
| 41 | 10485800-6410030 | … | … | … | … | … | … | … | … | … | … | … | … | … | … | … | n | … |
| 38096 | 10453120-6323103 | -27.0 ± 0.2 | 4864 ± 160 | … | 1.022 ± 0.002 | -0.06 ± 0.10 | … | … | N | … | … | … | … | … | … | … | n | G |
| 38040 | 10442361-6400138 | 453.2 ± 309.3 | … | … | … | … | … | … | … | … | … | … | … | … | … | … | n | … |
| 38598 | 10520652-6430162 | 17.5 ± 0.3 | 3418 ± 102 | 4.69 ± 0.06 | 0.832 ± 0.009 | -0.26 ± 0.13 | … | … | Y | Y | Y | … | N | Y | Y | … | n | … |
| 53656 | 10412543-6420179 | 28.2 ± 0.2 | 4620 ± 147 | … | 1.019 ± 0.005 | 0.03 ± 0.07 | <40 | 3 | N | … | … | … | … | … | … | … | n | G |
| 38327 | 10485814-6508409 | -18.5 ± 0.2 | 4992 ± 42 | 2.98 ± 0.17 | 1.018 ± 0.002 | 0.02 ± 0.01 | <30 | 3 | N | N | … | … | … | … | … | … | n | G |
| 38599 | 10520659-6345388 | 98.4 ± 0.2 | 4665 ± 72 | … | 1.023 ± 0.006 | -0.13 ± 0.13 | <38 | 3 | N | … | … | … | … | … | … | … | n | G |
| 53657 | 10412613-6405142 | 15.2 ± 0.2 | 4541 ± 138 | 2.36 ± 0.12 | 1.012 ± 0.003 | 0.03 ± 0.06 | <38 | 3 | N | N | … | … | … | … | … | … | n | G |
| 38328 | 10485850-6355295 | 35.6 ± 0.2 | 4784 ± 31 | 2.63 ± 0.07 | 1.015 ± 0.001 | 0.01 ± 0.05 | <34 | 3 | N | N | … | … | … | … | … | … | n | G |
| 38600 | 10520688-6335047 | -23.9 ± 0.2 | 4577 ± 44 | 2.38 ± 0.11 | 1.014 ± 0.005 | -0.06 ± 0.03 | <35 | 3 | N | N | … | … | … | … | … | … | n | G |
| 38601 | 10520741-6347231 | -7.1 ± 0.2 | 4662 ± 16 | 2.41 ± 0.07 | 1.025 ± 0.004 | -0.05 ± 0.02 | <34 | 3 | N | N | … | … | … | … | … | … | n | G |
| 37830 | 10412613-6433038 | -25.1 ± 0.2 | 4567 ± 46 | … | 1.031 ± 0.004 | -0.12 ± 0.08 | <23 | 3 | N | … | … | … | … | … | … | … | n | G |
| 38329 | 10485896-6444467 | -25.2 ± 0.2 | 4670 ± 11 | 2.45 ± 0.07 | 1.021 ± 0.002 | -0.03 ± 0.02 | <33 | 3 | N | N | … | … | … | … | … | … | n | G |
| 2574 | 10453134-6451549 | -18.5 ± 0.6 | 4973 ± 39 | 2.70 ± 0.09 | 1.018 ± 0.005 | -0.13 ± 0.05 | <15 | 3 | N | N | … | … | … | … | … | … | n | G |
| 53757 | 10434380-6425561 | 58.9 ± 0.2 | 4458 ± 144 | … | 1.046 ± 0.004 | -0.12 ± 0.07 | <22 | 3 | N | … | … | … | … | … | … | … | n | G |
| 38602 | 10520804-6344380 | 32.7 ± 0.2 | 4651 ± 99 | … | 1.011 ± 0.006 | … | <42 | 3 | N | … | … | … | … | … | … | … | n | G |
| 53658 | 10412727-6401377 | -9.3 ± 0.8 | 7117 ± 135 | … | 1.017 ± 0.007 | … | <23 | 3 | N | … | … | … | … | … | … | … | n | … |
| 38330 | 10485944-6502042 | -55.9 ± 0.2 | 4634 ± 66 | 2.45 ± 0.17 | 1.025 ± 0.004 | 0.00 ± 0.01 | <32 | 3 | N | N | … | … | … | … | … | … | n | G |
| 53865 | 10450063-6419115 | -2.7 ± 0.3 | 5667 ± 151 | … | 1.002 ± 0.006 | -0.34 ± 0.04 | 72 ± 12 | 1 | Y | … | Y | N | N | N | … | … | n | NG? |
| 38603 | 10520839-6427334 | 36.7 ± 0.2 | 4635 ± 98 | … | 1.026 ± 0.003 | … | <66 | 3 | N | … | … | … | … | … | … | … | n | G |
| 37831 | 10412736-6526482 | 18.0 ± 0.2 | 4743 ± 147 | … | 1.014 ± 0.003 | -0.10 ± 0.18 | … | … | N | … | … | … | … | … | … | … | n | G |
| 38331 | 10485955-6455554 | -0.2 ± 0.2 | 5105 ± 71 | 3.56 ± 0.12 | 1.001 ± 0.002 | -0.03 ± 0.10 | <23 | 3 | Y | Y | Y | N | N | Y | … | … | n | NG? |
| 38332 | 10490159-6432429 | -1.3 ± 0.2 | 5067 ± 81 | … | 1.020 ± 0.003 | -0.10 ± 0.11 | … | … | N | … | … | … | … | … | … | … | n | G |
| 38097 | 10453138-6356211 | 1.2 ± 0.2 | 4844 ± 69 | 2.67 ± 0.09 | 1.019 ± 0.003 | 0.01 ± 0.07 | <38 | 3 | N | N | … | … | … | … | … | … | n | G |
| 38604 | 10520916-6346333 | 30.6 ± 0.2 | 4828 ± 121 | 2.75 ± 0.02 | 1.008 ± 0.004 | -0.05 ± 0.11 | <42 | 3 | Y | N | Y | N | N | Y | … | … | n | NG? |
| 53659 | 10412767-6406438 | -8.9 ± 0.2 | 4611 ± 88 | 2.39 ± 0.12 | 1.019 ± 0.005 | -0.03 ± 0.02 | <41 | 3 | N | N | … | … | … | … | … | … | n | G |
| 38333 | 10490216-6444524 | -9.3 ± 0.2 | 4621 ± 74 | 2.37 ± 0.15 | 1.024 ± 0.004 | -0.06 ± 0.03 | 40 ± 5 | 1 | N | N | … | … | … | … | … | … | n | G |
| 38334 | 10490326-6431043 | 43.7 ± 0.2 | 5045 ± 90 | 3.26 ± 0.07 | 1.004 ± 0.001 | -0.04 ± 0.07 | <29 | 3 | Y | N | N | N | N | Y | … | … | n | NG? |
| 38335 | 10490510-6407540 | -20.2 ± 0.2 | 4564 ± 103 | … | 1.033 ± 0.002 | -0.06 ± 0.01 | 156 ± 2 | 1 | N | … | … | … | … | … | … | … | n | Li-rich G |
| 2609 | 10521047-6342542 | 22.4 ± 0.6 | 6066 ± 24 | 4.12 ± 0.08 | … | 0.39 ± 0.02 | <5 | 3 | … | Y | Y | N | N | Y | N | … | n | … |
| 38646 | 10524763-6336552 | 24.7 ± 0.3 | 4830 ± 95 | 2.98 ± 0.19 | 1.000 ± 0.007 | -0.02 ± 0.07 | <14 | 3 | Y | N | Y | N | N | Y | … | … | n | NG? |
| 37832 | 10412786-6445379 | 5.4 ± 0.2 | 5080 ± 107 | … | 1.026 ± 0.003 | -0.02 ± 0.05 | 82 ± 6 | 1 | N | … | … | … | … | … | … | … | n | Li-rich G |
| 38336 | 10490589-6511575 | -23.0 ± 0.2 | 5062 ± 129 | … | 1.014 ± 0.003 | -0.05 ± 0.07 | <26 | 3 | N | … | … | … | … | … | … | … | n | G |
| 38647 | 10524959-6342435 | -32.9 ± 0.4 | 4722 ± 120 | … | 1.010 ± 0.014 | … | … | … | Y | … | N | … | … | … | … | … | n | … |
| 37833 | 10412798-6426452 | -6.4 ± 0.2 | 4798 ± 123 | 2.61 ± 0.12 | 1.015 ± 0.002 | -0.02 ± 0.06 | <38 | 3 | N | N | … | … | … | … | … | … | n | G |
| 37989 | 10434432-6540297 | 96.1 ± 0.2 | 4702 ± 21 | 2.51 ± 0.11 | 1.019 ± 0.005 | -0.07 ± 0.03 | 25 ± 4 | 1 | N | N | … | … | … | … | … | … | n | G |

| ID | CNAME | RV (km s$^{-1}$) | $T_{\text{eff}}$ (K) | $logg$ (dex) | $\gamma^a$ | [Fe/H] (dex) | EW(Li)$^b$ (mÅ) | EW(Li) error flag$^c$ | Membership $\gamma$ | $logg$ | RV | Li | H$\alpha$ | [Fe/H] | Gaia studies Randich$^d$ | Cantat-Gaudin$^d$ | Final$^e$ | NMs with Li$^f$ |
|---|---|---|---|---|---|---|---|---|---|---|---|---|---|---|---|---|---|---|
| 53811 | 10442417-6424337 | 11.8 ± 0.4 | 3718 ± 20 | ... | 0.799 ± 0.018 | ... | ... | ... | Y | ... | Y | ... | N | ... | ... | ... | n | ... |
| 38337 | 10490773-6357173 | -27.9 ± 0.2 | 4878 ± 44 | 2.70 ± 0.09 | 1.021 ± 0.003 | 0.00 ± 0.01 | <41 | 3 | N | N | ... | ... | ... | ... | ... | ... | n | G |
| 38098 | 10453159-6445466 | 42.4 ± 0.3 | 3648 ± 88 | 4.58 ± 0.18 | 0.809 ± 0.007 | -0.21 ± 0.14 | ... | ... | Y | Y | Y | ... | ... | Y | N | ... | n | ... |
| 53660 | 10412946-6416158 | -23.9 ± 0.2 | 4427 ± 178 | ... | 1.046 ± 0.002 | -0.11 ± 0.03 | <29 | 3 | N | ... | ... | ... | ... | ... | ... | ... | n | G |
| 37834 | 10412951-6445049 | 40.8 ± 0.4 | 5020 ± 172 | 3.37 ± 0.16 | 1.002 ± 0.003 | -0.10 ± 0.17 | <22 | 3 | Y | N | Y | N | N | Y | ... | ... | n | NG? |
| 53758 | 10434458-6421368 | -14.9 ± 0.2 | 5900 ± 159 | ... | 1.005 ± 0.004 | 0.03 ± 0.06 | <8 | 3 | Y | ... | N | N | N | Y | ... | ... | n | ... |
| 38648 | 10525100-6338244 | 8.8 ± 0.3 | 4927 ± 182 | 3.14 ± 0.09 | 1.003 ± 0.006 | -0.06 ± 0.13 | <34 | 3 | Y | N | Y | N | N | Y | ... | ... | n | NG? |
| 38338 | 10490853-6505366 | 24.3 ± 0.2 | 4938 ± 184 | ... | 1.014 ± 0.004 | -0.07 ± 0.11 | <22 | 3 | N | ... | ... | ... | ... | ... | ... | ... | n | G |
| 53812 | 10442424-6433321 | -3.7 ± 0.3 | ... | ... | ... | ... | ... | ... | ... | ... | ... | ... | ... | ... | ... | ... | n | ... |
| 2550 | 10412952-6535235 | 6.4 ± 0.6 | 4658 ± 28 | 2.70 ± 0.09 | 1.017 ± 0.002 | 0.10 ± 0.03 | 31 ± 6 | 1 | N | N | ... | ... | ... | ... | ... | ... | n | G |
| 2613 | 10525142-6421139 | 9.3 ± 0.6 | 4676 ± 14 | 2.49 ± 0.02 | 1.021 ± 0.002 | 0.04 ± 0.01 | 31 ± 3 | 2 | N | N | ... | ... | ... | ... | ... | ... | n | G |
| 38339 | 10491024-6511330 | -13.6 ± 0.2 | 4780 ± 159 | ... | 1.021 ± 0.003 | -0.10 ± 0.18 | <13 | 3 | N | ... | ... | ... | ... | ... | ... | ... | n | G |
| 4 | 10415110-6413269 | -9.0 ± 0.6 | 3915 ± 75 | 1.18 ± 0.13 | ... | -0.05 ± 0.12 | <17 | 3 | ... | N | ... | ... | ... | ... | ... | ... | n | ... |
| 37990 | 10434493-6538067 | 76.9 ± 0.2 | 4645 ± 13 | 2.33 ± 0.12 | 1.027 ± 0.006 | -0.11 ± 0.05 | <37 | 3 | N | N | ... | ... | ... | ... | ... | ... | n | G |
| 38340 | 10491026-6458130 | -27.0 ± 0.2 | 4841 ± 200 | ... | 1.021 ± 0.003 | -0.10 ± 0.17 | <26 | 3 | N | ... | ... | ... | ... | ... | ... | ... | n | G |
| 38649 | 10525223-6420349 | -0.3 ± 0.2 | 4782 ± 51 | 2.58 ± 0.13 | 1.028 ± 0.002 | 0.01 ± 0.01 | <28 | 3 | N | N | ... | ... | ... | ... | ... | ... | n | G |
| 53813 | 10442433-6425266 | 24.5 ± 0.2 | 4459 ± 161 | ... | 1.042 ± 0.003 | -0.17 ± 0.16 | 65 ± 8 | 1 | N | ... | ... | ... | ... | ... | ... | ... | n | G |
| 27 | 10450118-6402008 | -2.7 ± 0.6 | 4879 ± 8 | 2.70 ± 0.05 | ... | 0.06 ± 0.03 | <22 | 3 | ... | N | ... | ... | ... | ... | ... | ... | n | ... |
| 53685 | 10415131-6419410 | 165.0 ± 0.3 | 5296 ± 253 | ... | 1.039 ± 0.021 | 0.14 ± 0.15 | <64 | 3 | N | ... | ... | ... | ... | ... | ... | ... | n | ... |
| 53759 | 10434499-6434140 | 4.4 ± 0.4 | 6567 ± 62 | 4.05 ± 0.13 | 1.006 ± 0.003 | -0.09 ± 0.05 | <9 | 3 | Y | Y | Y | N | N | Y | ... | ... | n | ... |
| 53908 | 10453181-6431238 | -9.6 ± 0.2 | 6416 ± 102 | 3.90 ± 0.21 | 1.008 ± 0.006 | -0.08 ± 0.08 | <45 | 3 | Y | Y | Y | N | N | Y | ... | ... | n | NG? |
| 38341 | 10491121-6517522 | -2.0 ± 0.2 | 4589 ± 68 | 2.48 ± 0.16 | 1.018 ± 0.004 | 0.11 ± 0.04 | <52 | 3 | N | N | ... | ... | ... | ... | ... | ... | n | G |
| 38342 | 10491231-6350561 | 2.1 ± 0.2 | 4904 ± 70 | 2.74 ± 0.10 | 1.017 ± 0.002 | -0.02 ± 0.05 | <30 | 3 | N | N | ... | ... | ... | ... | ... | ... | n | G |
| 38650 | 10525393-6418072 | -32.2 ± 0.2 | 4933 ± 131 | ... | 1.020 ± 0.002 | -0.10 ± 0.12 | <22 | 3 | N | ... | ... | ... | ... | ... | ... | ... | n | G |
| 53686 | 10415158-6410286 | 36.6 ± 0.2 | 5278 ± 175 | ... | 0.993 ± 0.004 | -0.16 ± 0.18 | 16 ± 6 | 1 | Y | ... | Y | N | N | Y | ... | ... | n | NG |
| 53760 | 10434541-6431177 | -23.8 ± 0.2 | 4795 ± 63 | 2.62 ± 0.18 | 1.022 ± 0.001 | -0.06 ± 0.09 | <28 | 3 | N | N | ... | ... | ... | ... | ... | ... | n | G |
| 37991 | 10434545-6347036 | -8.7 ± 0.2 | 4570 ± 73 | ... | 1.027 ± 0.003 | -0.08 ± 0.05 | <32 | 3 | N | ... | ... | ... | ... | ... | ... | ... | n | G |
| 38343 | 10491318-6356156 | 0.5 ± 0.2 | 4677 ± 14 | 2.44 ± 0.02 | 1.022 ± 0.003 | -0.01 ± 0.06 | <32 | 3 | N | N | ... | ... | ... | ... | ... | ... | n | G |
| 38063 | 10450119-6402009 | -2.9 ± 0.2 | 4808 ± 119 | 2.60 ± 0.20 | 1.025 ± 0.003 | -0.04 ± 0.07 | <32 | 3 | N | N | ... | ... | ... | ... | ... | ... | n | G |
| 38651 | 10525555-6338568 | 39.7 ± 0.2 | 4578 ± 79 | 2.60 ± 0.01 | 1.005 ± 0.005 | 0.00 ± 0.08 | <59 | 3 | Y | N | Y | N | N | Y | ... | ... | n | NG? |
| 53687 | 10415161-6413532 | -24.0 ± 0.4 | 6343 ± 120 | 3.97 ± 0.03 | 1.007 ± 0.004 | -0.04 ± 0.10 | 31 ± 4 | 1 | Y | Y | N | N | N | Y | ... | ... | n | NG? |
| 38344 | 10491470-6433036 | -15.6 ± 0.2 | 4736 ± 17 | 2.53 ± 0.05 | 1.020 ± 0.002 | 0.00 ± 0.01 | <39 | 3 | N | N | ... | ... | ... | ... | ... | ... | n | G |
| 2614 | 10525690-6427429 | -9.2 ± 0.6 | 5074 ± 3 | 2.84 ± 0.03 | ... | 0.04 ± 0.02 | <16 | 3 | ... | N | ... | ... | ... | ... | ... | ... | n | ... |
| 37858 | 10415170-6440507 | -18.5 ± 0.2 | 4681 ± 76 | ... | 1.024 ± 0.002 | -0.09 ± 0.08 | <24 | 3 | N | ... | ... | ... | ... | ... | ... | ... | n | G |
| 38652 | 10525795-6428162 | 23.0 ± 0.2 | 5068 ± 131 | 3.47 ± 0.19 | 1.002 ± 0.003 | -0.09 ± 0.18 | <25 | 3 | Y | N | Y | N | N | Y | ... | ... | n | NG? |
| 53815 | 10442598-6433278 | 54.9 ± 0.2 | 4339 ± 281 | ... | 1.034 ± 0.001 | 0.07 ± 0.09 | <48 | 3 | N | ... | ... | ... | ... | ... | ... | ... | n | G |
| 38345 | 10491471-6509145 | 4.1 ± 0.2 | 4649 ± 22 | 2.49 ± 0.10 | 1.018 ± 0.005 | -0.01 ± 0.02 | <38 | 3 | N | N | ... | ... | ... | ... | ... | ... | n | G |
| 53688 | 10415172-6410122 | -15.8 ± 0.2 | 4500 ± 134 | ... | 1.054 ± 0.003 | -0.02 ± 0.05 | 54 ± 8 | 1 | N | ... | ... | ... | ... | ... | ... | ... | n | G |
| 53866 | 10450140-6422420 | 5.7 ± 0.2 | 5104 ± 90 | 3.42 ± 0.02 | 1.002 ± 0.004 | -0.04 ± 0.11 | 9 ± 8 | 1 | Y | N | Y | N | N | Y | ... | ... | n | ... |
| 38653 | 10530043-6422423 | 10.0 ± 0.2 | 4854 ± 30 | 2.63 ± 0.11 | 1.023 ± 0.003 | -0.04 ± 0.08 | <28 | 3 | N | N | ... | ... | ... | ... | ... | ... | n | G |
| 53689 | 10415212-6415116 | 2.6 ± 0.5 | 6353 ± 54 | ... | ... | 0.05 ± 0.04 | ... | ... | ... | ... | ... | ... | ... | ... | ... | ... | n | ... |
| 38654 | 10530897-6428372 | -7.0 ± 0.2 | 4881 ± 81 | 2.68 ± 0.11 | 1.021 ± 0.003 | -0.01 ± 0.03 | <36 | 3 | N | N | ... | ... | ... | ... | ... | ... | n | G |
| 38368 | 10493634-6515397 | -8.4 ± 0.2 | 4715 ± 26 | 2.51 ± 0.18 | 1.020 ± 0.003 | -0.12 ± 0.12 | <24 | 3 | N | N | ... | ... | ... | ... | ... | ... | n | G |
| 37859 | 10415228-6500122 | 63.7 ± 0.2 | 4657 ± 23 | 2.30 ± 0.12 | 1.029 ± 0.006 | -0.18 ± 0.05 | <30 | 3 | N | N | ... | ... | ... | ... | ... | ... | n | G |
| 38655 | 10531144-6421510 | -44.0 ± 0.2 | 4833 ± 165 | 2.54 ± 0.20 | 1.023 ± 0.003 | -0.09 ± 0.14 | <28 | 3 | N | N | ... | ... | ... | ... | ... | ... | n | G |
| 38369 | 10493683-6401312 | 57.3 ± 0.2 | 4405 ± 223 | ... | 1.046 ± 0.002 | -0.04 ± 0.02 | <55 | 3 | N | ... | ... | ... | ... | ... | ... | ... | n | G |
| 38043 | 10442631-6505260 | -37.2 ± 0.2 | 4696 ± 67 | 2.45 ± 0.06 | 1.026 ± 0.003 | -0.07 ± 0.01 | 23 ± 3 | 1 | N | N | ... | ... | ... | ... | ... | ... | n | G |
| 37860 | 10415295-6457066 | 24.8 ± 0.2 | 4695 ± 108 | 2.48 ± 0.16 | 1.021 ± 0.002 | 0.02 ± 0.09 | <29 | 3 | N | N | ... | ... | ... | ... | ... | ... | n | G |
| 53768 | 10435071-6433151 | 16.9 ± 0.2 | 6019 ± 211 | 4.05 ± 0.13 | 0.997 ± 0.002 | -0.24 ± 0.04 | 49 ± 2 | 1 | Y | Y | Y | N | N | Y | ... | ... | n | NG |
| 38370 | 10493728-6511019 | 15.4 ± 0.2 | 4631 ± 101 | 2.54 ± 0.18 | 1.017 ± 0.003 | 0.09 ± 0.02 | <40 | 3 | N | N | ... | ... | ... | ... | ... | ... | n | G |
| 38656 | 10531275-6427560 | -16.9 ± 0.2 | 5095 ± 15 | ... | 1.018 ± 0.004 | -0.03 ± 0.08 | <17 | 3 | N | ... | ... | ... | ... | ... | ... | ... | n | G |
| 38064 | 10450182-6444398 | 12.8 ± 0.4 | 3546 ± 73 | 4.70 ± 0.10 | 0.738 ± 0.010 | -0.14 ± 0.11 | ... | ... | Y | Y | Y | ... | N | Y | N | ... | n | ... |
| 37861 | 10415338-6537317 | 0.9 ± 0.3 | 4595 ± 371 | 4.59 ± 0.04 | 0.911 ± 0.007 | 0.02 ± 0.12 | ... | ... | Y | Y | Y | Li | N | Y | ... | ... | n | ... |
| 38371 | 10493808-6337565 | 44.1 ± 0.2 | 4750 ± 75 | ... | 1.024 ± 0.005 | ... | <39 | 3 | N | ... | ... | ... | ... | ... | ... | ... | n | G |
| 38657 | 10531347-6415301 | 86.5 ± 0.3 | 4773 ± 83 | 2.58 ± 0.03 | 1.021 ± 0.008 | -0.06 ± 0.02 | <37 | 3 | N | N | ... | ... | ... | ... | ... | ... | n | G |
| 38044 | 10442661-6359150 | 45.3 ± 0.2 | 4629 ± 94 | ... | 1.019 ± 0.002 | 0.04 ± 0.08 | <38 | 3 | N | ... | ... | ... | ... | ... | ... | ... | n | G |
| 37862 | 10415444-6527401 | -1.0 ± 0.2 | 4702 ± 37 | 2.46 ± 0.10 | 1.023 ± 0.004 | -0.06 ± 0.11 | <34 | 3 | N | N | ... | ... | ... | ... | ... | ... | n | G |
| 38372 | 10493843-6342194 | 14.0 ± 0.2 | 4567 ± 83 | 2.32 ± 0.15 | 1.024 ± 0.003 | -0.03 ± 0.01 | <39 | 3 | N | N | ... | ... | ... | ... | ... | ... | n | G |
| 38658 | 10531426-6430227 | -1.6 ± 0.2 | 4761 ± 63 | 2.59 ± 0.10 | 1.024 ± 0.003 | -0.06 ± 0.03 | <21 | 3 | N | N | ... | ... | ... | ... | ... | ... | n | G |

**Table C.6.** continued.

| ID | CNAME | RV (km s$^{-1}$) | $T_{\text{eff}}$ (K) | $\log g$ (dex) | $\gamma^a$ | [Fe/H] (dex) | EW(Li)$^b$ (mÅ) | EW(Li) error flag$^c$ | Membership $\gamma$ | Membership $\log g$ | Membership RV | Membership Li | Membership H$\alpha$ | Membership [Fe/H] | Gaia studies Randich$^d$ | Gaia studies Cantat-Gaudin$^d$ | Final$^e$ | NMs with Li$^f$ |
|---|---|---|---|---|---|---|---|---|---|---|---|---|---|---|---|---|---|---|
| 28 | 10453911-6407211 | 7.6 ± 0.6 | 4478 ± 14 | 1.80 ± 0.02 | … | -0.37 ± 0.02 | <22 | 3 | … | N | … | … | … | … | … | … | n | … |
| 53690 | 10415467-6407307 | 5.1 ± 0.2 | 4824 ± 176 | … | 1.015 ± 0.002 | -0.03 ± 0.09 | <33 | 3 | N | … | … | … | … | … | … | … | n | G |
| 53769 | 10435089-6424496 | 4.8 ± 0.3 | 5912 ± 56 | 4.14 ± 0.08 | 0.993 ± 0.004 | -0.04 ± 0.04 | 47 ± 4 | 1 | Y | Y | Y | N | N | Y | … | … | n | NG |
| 38373 | 10494021-6446036 | -20.9 ± 0.2 | 4648 ± 41 | 2.34 ± 0.19 | 1.033 ± 0.003 | -0.08 ± 0.06 | <35 | 3 | N | N | … | … | … | … | … | … | n | G |
| 38374 | 10494038-6344239 | 7.3 ± 0.2 | 4762 ± 64 | 2.63 ± 0.07 | 1.012 ± 0.004 | 0.01 ± 0.06 | <34 | 3 | N | N | … | … | … | … | … | … | n | G |
| 38659 | 10531476-6416443 | 25.3 ± 0.2 | 4628 ± 79 | 2.46 ± 0.15 | 1.015 ± 0.005 | -0.04 ± 0.07 | <37 | 3 | N | N | … | … | … | … | … | … | n | G |
| 38045 | 10442672-6421062 | 42.3 ± 0.2 | 4472 ± 209 | … | 1.019 ± 0.002 | -0.08 ± 0.07 | 34 ± 1 | 1 | N | … | … | … | … | … | … | … | n | G |
| 53691 | 10415472-6419308 | -501.0 ± 9.9 | … | … | … | … | … | … | … | … | … | … | … | … | … | … | n | … |
| 38375 | 10494053-6351501 | 14.6 ± 0.2 | 5081 ± 107 | 3.65 ± 0.09 | 0.998 ± 0.001 | 0.00 ± 0.03 | 32 ± 4 | 1 | Y | Y | Y | N | N | Y | … | … | n | NG |
| 38660 | 10531624-6418369 | -1.6 ± 0.2 | 4921 ± 151 | … | 1.016 ± 0.005 | -0.05 ± 0.09 | <39 | 3 | N | … | … | … | … | … | … | … | n | G |
| 5 | 10415473-6407319 | 3.9 ± 0.6 | 5045 ± 64 | 2.97 ± 0.04 | … | -0.01 ± 0.12 | <23 | 3 | … | N | … | … | … | … | … | … | n | … |
| 29 | 10453953-6412335 | -29.3 ± 0.6 | 4851 ± 9 | 2.58 ± 0.05 | … | 0.01 ± 0.08 | 27 ± 3 | 1 | N | N | … | … | … | … | … | … | n | G |
| 38376 | 10494070-6352553 | -23.9 ± 0.2 | 4578 ± 74 | 2.53 ± 0.11 | 1.011 ± 0.001 | 0.12 ± 0.01 | <46 | 3 | N | N | … | … | … | … | … | … | n | G |
| 38661 | 10531757-6417115 | -1.9 ± 0.3 | 4957 ± 58 | 2.89 ± 0.20 | 1.023 ± 0.005 | 0.01 ± 0.07 | <41 | 3 | N | N | … | … | … | … | … | … | n | G |
| 53692 | 10415502-6408254 | -20.9 ± 0.2 | 4457 ± 198 | … | 1.048 ± 0.001 | -0.04 ± 0.02 | <46 | 3 | N | … | … | … | … | … | … | … | n | G |
| 38377 | 10494127-6430308 | -8.4 ± 0.2 | 4888 ± 9 | 2.77 ± 0.01 | 1.014 ± 0.002 | 0.03 ± 0.02 | <41 | 3 | N | N | … | … | … | … | … | … | n | G |
| 38378 | 10494164-6401022 | -3.1 ± 0.2 | 4920 ± 253 | 3.10 ± 0.18 | 1.004 ± 0.004 | -0.08 ± 0.14 | <28 | 3 | Y | N | Y | N | N | Y | … | … | n | NG? |
| 37998 | 10435141-6530214 | 40.0 ± 0.2 | 5012 ± 182 | 3.34 ± 0.20 | 1.003 ± 0.004 | -0.11 ± 0.18 | <21 | 3 | Y | N | Y | N | N | Y | … | … | n | NG? |
| 38662 | 10531883-6424024 | 1.7 ± 0.2 | 4725 ± 36 | 2.62 ± 0.17 | 1.019 ± 0.002 | -0.03 ± 0.03 | <38 | 3 | N | N | … | … | … | … | … | … | n | G |
| 53693 | 10415504-6411122 | -25.8 ± 0.2 | 5022 ± 54 | … | 1.007 ± 0.003 | -0.17 ± 0.08 | <30 | 3 | Y | … | N | N | N | Y | … | … | n | NG? |
| 38379 | 10494621-6338122 | -11.3 ± 0.3 | 5221 ± 125 | 3.48 ± 0.16 | 1.011 ± 0.006 | 0.02 ± 0.08 | … | … | N | N | … | … | … | … | … | … | n | … |
| 53694 | 10415678-6416489 | 7.7 ± 0.3 | 5937 ± 76 | 4.04 ± 0.01 | 0.997 ± 0.007 | -0.14 ± 0.08 | <15 | 3 | Y | Y | Y | N | N | Y | … | … | n | NG |
| 37999 | 10435159-6536409 | 4.2 ± 0.2 | 5068 ± 98 | … | 1.019 ± 0.003 | -0.07 ± 0.14 | <15 | 3 | N | … | … | … | … | … | … | … | n | G |
| 38380 | 10494678-6332230 | -17.3 ± 0.2 | 5112 ± 25 | … | 1.013 ± 0.003 | 0.01 ± 0.03 | 28 ± 3 | 1 | N | … | … | … | … | … | … | … | n | G |
| 38046 | 10442739-6536334 | 83.7 ± 0.3 | 4721 ± 8 | 2.41 ± 0.20 | 1.030 ± 0.007 | -0.13 ± 0.15 | <43 | 3 | N | N | … | … | … | … | … | … | n | G |
| 38105 | 10454028-6329254 | 12.1 ± 0.2 | 4947 ± 80 | 2.77 ± 0.11 | 1.020 ± 0.003 | -0.02 ± 0.04 | <32 | 3 | N | N | … | … | … | … | … | … | n | G |
| 53695 | 10415720-6418055 | 57.2 ± 0.2 | 3901 ± 46 | … | 1.055 ± 0.003 | -0.22 ± 0.16 | <4 | 3 | N | … | … | … | … | … | … | … | n | G |
| 2594 | 10494767-6353192 | -0.2 ± 0.6 | 5037 ± 29 | 3.00 ± 0.07 | 1.012 ± 0.002 | 0.15 ± 0.03 | 35 ± 8 | 1 | N | N | … | … | … | … | … | … | n | G |
| 38381 | 10494835-6405460 | -26.2 ± 0.2 | 4472 ± 189 | … | 1.028 ± 0.002 | 0.08 ± 0.08 | <52 | 3 | N | … | … | … | … | … | … | … | n | G |
| 2595 | 10494839-6446284 | 18.0 ± 0.6 | 5327 ± 123 | 4.34 ± 0.31 | … | -0.12 ± 0.06 | 280 ± 11 | 2 | … | Y | Y | Y | Y | Y | Y | Y | Y | … |
| 38047 | 10442745-6451262 | -25.6 ± 0.2 | 4955 ± 10 | 2.81 ± 0.04 | 1.017 ± 0.003 | 0.00 ± 0.06 | <38 | 3 | N | N | … | … | … | … | … | … | n | G |
| 38663 | 10532612-6418444 | 10.4 ± 0.2 | 4641 ± 98 | … | 1.020 ± 0.003 | … | <62 | 3 | N | … | … | … | … | … | … | … | n | G |
| 38382 | 10494845-6457153 | 9.8 ± 0.2 | 4581 ± 83 | 2.49 ± 0.14 | 1.020 ± 0.003 | 0.10 ± 0.06 | <41 | 3 | N | N | … | … | … | … | … | … | n | G |
| 38000 | 10435197-6543338 | 60.5 ± 0.2 | 5035 ± 52 | 3.53 ± 0.14 | 0.995 ± 0.006 | 0.01 ± 0.08 | <30 | 3 | Y | Y | N | N | N | Y | … | … | n | NG |
| 38383 | 10494892-6345066 | -23.1 ± 0.2 | 5024 ± 155 | … | 1.016 ± 0.004 | -0.18 ± 0.24 | <15 | 3 | N | … | … | … | … | … | … | … | n | G |
| 38106 | 10454043-6348073 | -8.0 ± 0.2 | 5002 ± 108 | … | 1.022 ± 0.003 | -0.04 ± 0.05 | 31 ± 6 | 1 | N | … | … | … | … | … | … | … | n | G |
| 38384 | 10494993-6511071 | -2.0 ± 0.2 | 5114 ± 61 | … | 1.011 ± 0.003 | -0.06 ± 0.09 | <29 | 3 | N | … | … | … | … | … | … | … | n | G |
| 38385 | 10494994-6443007 | 35.3 ± 0.2 | 4604 ± 128 | … | 1.024 ± 0.004 | 0.07 ± 0.03 | 46 ± 5 | 1 | N | … | … | … | … | … | … | … | n | G |
| 38001 | 10435215-6535143 | 71.2 ± 0.2 | 4523 ± 111 | 2.26 | 1.022 ± 0.004 | -0.10 ± 0.02 | <36 | 3 | N | N | … | … | … | … | … | … | n | G |
| 53916 | 10454064-6420248 | 9.1 ± 0.3 | 6115 ± 112 | 4.07 | 0.998 ± 0.004 | -0.07 ± 0.02 | 28 ± 7 | 1 | Y | Y | Y | N | N | Y | … | … | n | NG |
| 38107 | 10454105-6413235 | -23.5 ± 0.2 | 5135 ± 114 | … | 1.010 ± 0.002 | … | … | … | N | … | … | … | … | … | … | … | n | G |
| 38386 | 10495006-6356304 | 19.0 ± 0.2 | 3635 ± 76 | 4.57 ± 0.19 | 0.810 ± 0.005 | -0.21 ± 0.14 | … | … | Y | Y | Y | … | N | Y | Y | … | n | … |
| 53816 | 10442803-6430129 | 432.6 ± 10.8 | … | … | … | … | … | … | … | … | … | … | … | … | … | … | n | … |
| 38067 | 10450544-6402392 | 11.5 ± 0.2 | 4485 ± 207 | … | 1.021 ± 0.003 | 0.12 ± 0.04 | 128 ± 6 | 1 | N | … | … | … | … | … | … | … | n | G |
| 38387 | 10495160-6440168 | 18.7 ± 0.2 | 4745 ± 28 | 2.61 ± 0.02 | 1.016 ± 0.004 | 0.06 ± 0.02 | <36 | 3 | N | N | … | … | … | … | … | … | n | G |
| 53770 | 10435284-6425583 | 11.0 ± 0.2 | 5769 ± 48 | … | 0.996 ± 0.004 | -0.10 ± 0.06 | 29 ± 5 | 1 | Y | … | Y | N | N | Y | … | … | n | NG |
| 38068 | 10450552-6447395 | -19.1 ± 0.2 | 4727 ± 7 | 2.54 ± 0.14 | 1.020 ± 0.003 | -0.08 ± 0.06 | <25 | 3 | N | N | … | … | … | … | … | … | n | G |
| 38388 | 10495228-6513209 | -39.5 ± 0.2 | 4626 ± 52 | 2.38 ± 0.13 | 1.020 ± 0.003 | -0.07 ± 0.04 | <31 | 3 | N | N | … | … | … | … | … | … | n | G |
| 38432 | 10501672-6333390 | 12.9 ± 0.2 | 4662 ± 172 | 2.54 ± 0.16 | 1.010 ± 0.004 | -0.07 ± 0.18 | <25 | 3 | Y | N | Y | N | Y | Y | … | … | n | NG? |
| 38433 | 10501751-6344106 | 59.6 ± 0.2 | 4707 ± 106 | … | 1.009 ± 0.006 | … | <52 | 3 | Y | … | N | N | N | … | … | … | n | NG? |
| 38069 | 10450570-6344234 | -17.3 ± 0.2 | 4648 ± 9 | 2.42 ± 0.12 | 1.026 ± 0.002 | -0.05 ± 0.08 | … | … | N | N | … | … | … | … | … | … | n | G |
| 38434 | 10501849-6349065 | 9.0 ± 0.2 | 4879 ± 168 | … | 1.022 ± 0.006 | -0.05 ± 0.09 | <24 | 3 | N | … | … | … | … | … | … | … | n | G |
| 38435 | 10501862-6331082 | 11.0 ± 0.2 | 4622 ± 89 | 2.45 ± 0.14 | 1.017 ± 0.003 | 0.03 ± 0.04 | <36 | 3 | N | N | … | … | … | … | … | … | n | … |
| 2600 | 10502012-6342055 | -14.4 ± 0.6 | 4765 ± 43 | 2.64 ± 0.09 | 1.020 ± 0.001 | 0.15 ± 0.04 | 39 ± 12 | 1 | N | N | … | … | … | … | … | … | n | G |
| 38436 | 10502046-6407302 | 33.4 ± 0.2 | 4638 ± 98 | … | 1.017 ± 0.003 | … | … | … | N | … | … | … | … | … | … | … | n | G |
| 38115 | 10454594-6318207 | 25.6 ± 0.2 | 5002 ± 64 | 3.02 ± 0.19 | 1.006 ± 0.006 | -0.04 ± 0.09 | <30 | 3 | Y | N | Y | N | N | Y | … | … | n | NG? |
| 38437 | 10502058-6401221 | 14.3 ± 0.2 | 4575 ± 49 | 2.36 ± 0.15 | 1.029 ± 0.003 | 0.00 ± 0.05 | <46 | 3 | N | N | … | … | … | … | … | … | n | G |
| 38438 | 10502062-6354317 | -23.0 ± 0.2 | 5077 ± 83 | … | 1.004 ± 0.002 | -0.16 ± 0.16 | 26 ± 2 | 1 | Y | … | N | N | N | Y | … | … | n | NG? |
| 38439 | 10502078-6406290 | -8.8 ± 0.2 | 4667 ± 15 | 2.40 ± 0.09 | 1.026 ± 0.002 | -0.07 ± 0.01 | <33 | 3 | N | N | … | … | … | … | … | … | n | G |

| ID | CNAME | RV (km s$^{-1}$) | T$_{\text{eff}}$ (K) | logg (dex) | $\gamma^a$ | [Fe/H] (dex) | EW(Li)$^b$ (mÅ) | EW(Li) error flag$^c$ | $\gamma$ | logg | Membership RV | Li | H$\alpha$ | [Fe/H] | Gaia studies Randich$^d$ | Cantat-Gaudin$^d$ | Final$^e$ | NMs with Li$^f$ |
|---|---|---|---|---|---|---|---|---|---|---|---|---|---|---|---|---|---|---|
| 38440 | 10502088-6443232 | 42.2 ± 0.2 | 4715 ± 76 | 2.47 ± 0.19 | 1.022 ± 0.004 | -0.12 ± 0.11 | <26 | 3 | N | N | ... | ... | ... | ... | ... | ... | n | G |
| 53876 | 10450922-6419033 | 27.8 ± 0.2 | 5412 ± 72 | 4.28 ± 0.06 | 0.979 ± 0.002 | 0.15 ± 0.02 | <24 | 3 | Y | Y | Y | N | N | Y | ... | ... | n | NG |
| 38441 | 10502115-6504390 | -7.7 ± 0.2 | 5004 ± 181 | ... | 1.017 ± 0.002 | -0.17 ± 0.21 | <12 | 3 | N | ... | ... | ... | ... | ... | ... | ... | n | G |
| 38116 | 10454623-6421391 | 73.5 ± 0.2 | 5103 ± 125 | ... | 1.010 ± 0.003 | ... | <45 | 3 | Y | ... | N | N | N | ... | ... | ... | n | NG? |
| 38442 | 10502125-6503384 | 15.7 ± 0.2 | 4539 ± 152 | 2.32 ± 0.18 | 1.023 ± 0.004 | 0.13 ± 0.06 | <60 | 3 | N | N | ... | ... | ... | ... | ... | ... | n | G |
| 38443 | 10502151-6358575 | 46.6 ± 0.3 | 4979 ± 218 | ... | 0.993 ± 0.012 | 0.13 ± 0.12 | <21 | 3 | Y | ... | N | N | N | Y | ... | ... | n | NG |
| 38048 | 10443290-6403509 | 3.2 ± 0.2 | 4607 ± 44 | 2.38 ± 0.09 | 1.020 ± 0.002 | -0.01 ± 0.10 | <34 | 3 | N | N | ... | ... | ... | ... | ... | ... | n | G |
| 53877 | 10450938-6434073 | 1.0 ± 0.2 | 4455 ± 183 | ... | 1.038 ± 0.002 | -0.10 ± 0.06 | 64 ± 7 | 1 | N | ... | ... | ... | ... | ... | ... | ... | n | G |
| 38444 | 10502171-6331129 | 29.7 ± 0.3 | 3446 ± 69 | 4.70 ± 0.10 | 0.758 ± 0.006 | -0.27 ± 0.12 | ... | ... | Y | Y | Y | ... | ... | Y | N | Y | n | ... |
| 38445 | 10502280-6350401 | 25.4 ± 0.2 | 4540 ± 174 | ... | 1.014 ± 0.003 | 0.10 ± 0.07 | <52 | 3 | N | ... | ... | ... | ... | ... | ... | ... | n | G |
| 38446 | 10502319-6333584 | 26.9 ± 0.3 | 4819 ± 181 | 2.59 ± 0.20 | 1.022 ± 0.007 | -0.05 ± 0.07 | <33 | 3 | N | N | ... | ... | ... | ... | ... | ... | n | G |
| 53919 | 10454668-6428550 | -7.5 ± 0.6 | ... | ... | ... | ... | ... | ... | ... | ... | ... | ... | ... | ... | ... | ... | n | ... |
| 38447 | 10502364-6443418 | 17.7 ± 0.2 | 4550 ± 123 | 2.34 ± 0.18 | 1.022 ± 0.006 | 0.10 ± 0.09 | 25 5 | 1 | N | N | ... | ... | ... | ... | ... | ... | n | G |
| 38448 | 10502385-6333307 | 15.0 ± 0.2 | 4930 ± 198 | ... | 1.012 ± 0.004 | -0.07 ± 0.13 | <27 | 3 | N | ... | ... | ... | ... | ... | ... | ... | n | G |
| 53821 | 10443313-6435109 | 12.6 ± 0.2 | 5099 ± 83 | 3.63 ± 0.03 | 0.998 ± 0.004 | 0.01 ± 0.05 | 81 ± 4 | 1 | Y | Y | Y | N | N | Y | ... | ... | n | NG |
| 38449 | 10502504-6422434 | -29.5 ± 0.2 | 4396 ± 246 | ... | 1.043 ± 0.002 | 0.12 ± 0.03 | <59 | 3 | N | ... | ... | ... | ... | ... | ... | ... | n | G |
| 53920 | 10454709-6429125 | 22.6 ± 0.2 | 5669 ± 134 | ... | 1.007 ± 0.006 | ... | 148 ± 6 | 1 | Y | ... | Y | Y | N | ... | ... | ... | n | NG? |
| 38450 | 10502511-6347539 | 63.3 ± 0.3 | 4730 ± 24 | ... | 0.998 ± 0.006 | -0.07 ± 0.03 | <27 | 3 | Y | ... | N | N | N | Y | ... | ... | n | NG |
| 38117 | 10454711-6327183 | -27.6 ± 0.2 | 4773 ± 48 | 2.57 ± 0.09 | 1.027 ± 0.002 | -0.04 ± 0.01 | <24 | 3 | N | N | ... | ... | ... | ... | ... | ... | n | G |
| 38451 | 10502778-6332518 | 10.1 ± 0.2 | 4675 ± 114 | 2.49 ± 0.12 | 1.013 ± 0.005 | 0.01 ± 0.08 | <34 | 3 | N | N | ... | ... | ... | ... | ... | ... | n | ... |
| 38452 | 10502820-6355184 | 8.6 ± 0.2 | 4539 ± 139 | ... | 1.025 ± 0.003 | 0.13 ± 0.04 | ... | ... | N | ... | ... | ... | ... | ... | ... | ... | n | G |
| 38453 | 10502831-6347407 | -0.7 ± 0.2 | 5055 ± 117 | ... | 1.017 ± 0.002 | -0.04 ± 0.09 | <28 | 3 | N | ... | ... | ... | ... | ... | ... | ... | n | G |
| 38478 | 10504100-6404144 | 14.2 ± 0.2 | 4635 ± 102 | ... | 1.020 ± 0.003 | ... | <48 | 3 | N | ... | ... | ... | ... | ... | ... | ... | n | G |
| 2575 | 10454736-6457074 | -3.6 ± 0.6 | 4789 ± 56 | 2.66 ± 0.11 | 1.023 ± 0.002 | 0.04 ± 0.03 | 24 ± 5 | 1 | N | N | ... | ... | ... | ... | ... | ... | n | G |
| 38479 | 10504172-6350097 | 26.2 ± 0.3 | 4834 ± 7 | 2.45 ± 0.06 | 1.020 ± 0.006 | -0.26 ± 0.09 | ... | ... | N | N | ... | ... | ... | ... | ... | ... | n | G |
| 38118 | 10454775-6501004 | 14.5 ± 0.2 | 4473 ± 215 | ... | 1.024 ± 0.002 | 0.12 ± 0.07 | 181 ± 3 | 1 | N | ... | ... | ... | ... | ... | ... | ... | n | Li-rich G |
| 2601 | 10504176-6420115 | 7.7 ± 0.6 | 4946 ± 2 | 2.72 ± 0.03 | ... | 0.01 ± 0.01 | <19 | 3 | ... | N | ... | ... | ... | ... | ... | ... | n | ... |
| 38480 | 10504231-6347374 | -2.5 ± 0.2 | 5104 ± 17 | ... | 1.020 ± 0.002 | 0.01 ± 0.02 | <25 | 3 | N | ... | ... | ... | ... | ... | ... | ... | n | G |
| 38119 | 10454789-6452071 | 8.0 ± 0.2 | 4679 ± 77 | 2.53 ± 0.16 | 1.018 ± 0.002 | 0.00 ± 0.06 | <39 | 3 | N | N | ... | ... | ... | ... | ... | ... | n | G |
| 38481 | 10504393-6423540 | 57.1 ± 0.2 | 4527 ± 181 | ... | 1.040 ± 0.003 | 0.08 ± 0.08 | <42 | 3 | N | ... | ... | ... | ... | ... | ... | ... | n | G |
| 53878 | 10450991-6426467 | 7.7 ± 0.7 | 6881 ± 77 | ... | ... | 0.29 ± 0.07 | ... | ... | ... | ... | ... | ... | ... | ... | ... | ... | n | ... |
| 38129 | 10460113-6449358 | 2.1 ± 0.2 | 4844 ± 89 | 2.67 ± 0.13 | 1.019 ± 0.003 | -0.03 ± 0.04 | 25 ± 6 | 1 | N | N | ... | ... | ... | ... | ... | ... | n | G |
| 38482 | 10504398-6420102 | -42.8 ± 0.2 | 4949 ± 130 | ... | 1.024 ± 0.003 | -0.03 ± 0.06 | <21 | 3 | N | ... | ... | ... | ... | ... | ... | ... | n | G |
| 38130 | 10460129-6424492 | 50.0 ± 0.2 | 4841 ± 192 | ... | 1.017 ± 0.002 | -0.21 ± 0.21 | 23 ± 4 | 1 | N | ... | ... | ... | ... | ... | ... | ... | n | G |
| 38483 | 10504455-6346053 | -58.5 ± 0.2 | 4933 ± 136 | ... | 1.017 ± 0.003 | -0.13 ± 0.08 | <17 | 3 | N | ... | ... | ... | ... | ... | ... | ... | n | G |
| 53934 | 10460148-6431594 | 47.8 ± 0.2 | 5591 ± 29 | ... | 0.996 ± 0.005 | -0.26 ± 0.06 | <20 | 3 | Y | ... | N | N | N | Y | ... | ... | n | NG |
| 38484 | 10504459-6334499 | -1.5 ± 0.2 | 4784 ± 88 | 2.82 ± 0.06 | 1.005 ± 0.002 | 0.06 ± 0.04 | <36 | 3 | Y | Y | Y | N | N | Y | ... | ... | n | NG? |
| 38485 | 10504513-6330482 | 19.8 ± 0.2 | 4800 ± 12 | 2.63 ± 0.09 | 1.025 ± 0.004 | 0.03 ± 0.03 | <21 | 3 | N | N | ... | ... | ... | ... | ... | ... | n | G |
| 38131 | 10460226-6414235 | 0.6 ± 0.2 | 4986 ± 196 | ... | 1.013 ± 0.001 | -0.08 ± 0.13 | 32 ± 1 | 1 | N | ... | ... | ... | ... | ... | ... | ... | n | G |
| 38486 | 10504599-6342017 | -9.2 ± 0.2 | 4543 ± 64 | ... | 1.037 ± 0.003 | -0.09 ± 0.05 | <28 | 3 | N | ... | ... | ... | ... | ... | ... | ... | n | G |
| 38132 | 10460323-6454324 | -3.0 ± 0.2 | 4951 ± 67 | 2.90 ± 0.20 | 1.023 ± 0.004 | 0.02 ± 0.03 | <24 | 3 | N | N | ... | ... | ... | ... | ... | ... | n | G |
| 53935 | 10460442-6425219 | -10.0 ± 0.3 | 5857 ± 50 | 3.90 ± 0.12 | 1.001 ± 0.005 | -0.32 ± 0.02 | 62 ± 7 | 1 | Y | Y | Y | N | N | N | ... | ... | n | NG? |
| 38487 | 10504702-6419278 | 21.0 ± 0.2 | 4516 ± 204 | ... | 1.022 ± 0.002 | 0.13 ± 0.13 | 43 ± 2 | 1 | N | ... | ... | ... | ... | ... | ... | ... | n | G |
| 53936 | 10460461-6424565 | -16.8 ± 0.2 | 3585 ± 122 | ... | ... | ... | <100 | 3 | ... | ... | ... | ... | ... | ... | ... | ... | n | NG |
| 53937 | 10460470-6430073 | 41.8 ± 0.2 | 4523 ± 168 | 2.41 ± 0.20 | 1.011 ± 0.003 | 0.08 ± 0.10 | <51 | 3 | N | N | ... | ... | ... | ... | ... | ... | n | G |
| 38488 | 10504784-6331140 | 5.1 ± 0.2 | 4718 ± 44 | 2.64 ± 0.03 | 1.010 ± 0.004 | 0.02 ± 0.09 | <24 | 3 | Y | N | Y | N | N | Y | ... | ... | n | NG? |
| 38489 | 10504919-6344099 | 60.6 ± 0.2 | 4568 ± 109 | 2.32 ± 0.16 | 1.021 ± 0.005 | -0.03 ± 0.02 | <33 | 3 | N | N | ... | ... | ... | ... | ... | ... | n | G |
| 38490 | 10504991-6430131 | 15.0 ± 0.2 | 4292 ± 558 | ... | 0.910 ± 0.003 | -0.04 ± 0.12 | 225 ± 4 | 1 | Y | ... | Y | Y | Y | Y | Y | ... | Y | ... |
| 38133 | 10460594-6411305 | -23.9 ± 0.2 | 4551 ± 113 | ... | 1.041 ± 0.002 | ... | <32 | 3 | N | ... | ... | ... | ... | ... | ... | ... | n | G |
| 38491 | 10505032-6513005 | -2.7 ± 0.2 | 5119 ± 7 | ... | 1.019 ± 0.001 | -0.02 ± 0.06 | <31 | 3 | N | ... | ... | ... | ... | ... | ... | ... | n | G |
| 38134 | 10460612-6321591 | -10.6 ± 0.2 | 4569 ± 32 | ... | 1.042 ± 0.002 | -0.06 ± 0.02 | <39 | 3 | N | ... | ... | ... | ... | ... | ... | ... | n | G |
| 38135 | 10460621-6404471 | 12.7 ± 0.2 | 4576 ± 114 | 2.21 ± 0.20 | 1.025 ± 0.002 | -0.12 ± 0.13 | <33 | 3 | N | N | ... | ... | ... | ... | ... | ... | n | G |
| 38492 | 10505099-6340119 | 52.6 ± 0.2 | 4634 ± 86 | 2.37 ± 0.13 | 1.020 ± 0.004 | -0.07 ± 0.05 | <53 | 3 | N | N | ... | ... | ... | ... | ... | ... | n | G |
| 53939 | 10460642-6433316 | 8.9 ± 0.4 | 6697 ± 49 | ... | ... | 0.21 ± 0.04 | ... | ... | ... | ... | ... | ... | ... | ... | ... | ... | n | ... |
| 38493 | 10505118-6347256 | -3.1 ± 0.2 | 4776 ± 172 | ... | 1.025 ± 0.003 | -0.10 ± 0.17 | <36 | 3 | N | ... | ... | ... | ... | ... | ... | ... | n | G |
| 53940 | 10460668-6422254 | 17.7 ± 0.2 | 3936 ± 13 | ... | 1.049 ± 0.002 | -0.09 ± 0.16 | <22 | 3 | N | ... | ... | ... | ... | ... | ... | ... | n | G |
| 38494 | 10505167-6335573 | -10.6 ± 0.2 | 5000 ± 183 | ... | 1.011 ± 0.004 | -0.16 ± 0.17 | 34 ± 6 | 1 | N | ... | ... | ... | ... | ... | ... | ... | n | G |
| 53941 | 10460806-6427343 | -136.3 ± 2.1 | ... | ... | ... | ... | ... | ... | ... | ... | ... | ... | ... | ... | ... | ... | n | ... |
| 38495 | 10505201-6337398 | -5.7 ± 0.3 | 4696 ± 23 | 2.49 ± 0.14 | 1.026 ± 0.006 | -0.06 ± 0.01 | <20 | 3 | N | N | ... | ... | ... | ... | ... | ... | n | G |

**Table C.6.** continued.

| ID | CNAME | RV (km s$^{-1}$) | $T_{\rm eff}$ (K) | $logg$ (dex) | $\gamma^a$ | [Fe/H] (dex) | EW(Li)$^b$ (mÅ) | EW(Li) error flag$^c$ | $\gamma$ | $logg$ | Membership RV | Li | H$\alpha$ | [Fe/H] | Gaia studies Randich$^d$ | Cantat-Gaudin$^d$ | Final$^e$ | NMs with Li$^f$ |
|---|---|---|---|---|---|---|---|---|---|---|---|---|---|---|---|---|---|---|
| 38136 | 10461002-6447094 | -20.0 ± 0.2 | 4631 ± 58 | 2.33 ± 0.19 | 1.029 ± 0.002 | -0.10 ± 0.07 | <22 | 3 | N | N | ... | ... | ... | ... | ... | ... | n | G |
| 38496 | 10505234-6339251 | 22.5 ± 0.2 | 4730 ± 9 | 2.82 ± 0.18 | 1.003 ± 0.006 | 0.04 ± 0.07 | <58 | 3 | Y | N | Y | N | N | Y | ... | ... | n | NG? |
| 38497 | 10505262-6348045 | 18.0 ± 0.2 | 5037 ± 36 | 2.93 ± 0.12 | 1.024 ± 0.003 | 0.05 ± 0.07 | ... | ... | N | N | ... | ... | ... | ... | ... | ... | n | G |
| 38137 | 10461015-6316517 | 2.8 ± 0.2 | 4497 ± 160 | ... | 1.019 ± 0.003 | 0.11 ± 0.07 | <48 | 3 | N | ... | ... | ... | ... | ... | ... | ... | n | G |
| 38498 | 10505391-6425015 | -1.5 ± 0.2 | 4981 ± 61 | ... | 1.019 ± 0.003 | -0.05 ± 0.08 | <24 | 3 | N | ... | ... | ... | ... | ... | ... | ... | n | G |
| 38138 | 10461115-6353447 | -29.7 ± 0.2 | 4540 ± 91 | ... | 1.041 ± 0.004 | -0.02 ± 0.04 | <23 | 3 | N | ... | ... | ... | ... | ... | ... | ... | n | G |
| 38499 | 10505396-6335367 | -10.6 ± 0.2 | 4802 ± 143 | ... | 1.019 ± 0.003 | -0.09 ± 0.14 | <25 | 3 | N | ... | ... | ... | ... | ... | ... | ... | n | G |
| 38542 | 10512412-6412538 | 2.8 ± 0.2 | 4991 ± 173 | ... | 1.009 ± 0.002 | -0.15 ± 0.19 | ... | ... | Y | ... | Y | ... | N | Y | ... | ... | n | ... |
| 53943 | 10461139-6424042 | 21.4 ± 0.2 | 4486 ± 174 | ... | 1.023 ± 0.003 | -0.05 ± 0.02 | 49 ± 11 | 1 | N | ... | ... | ... | ... | ... | ... | ... | n | G |
| 38543 | 10512417-6333526 | -1.5 ± 0.3 | 4917 ± 157 | ... | 1.008 ± 0.006 | -0.15 ± 0.19 | <23 | 3 | Y | ... | Y | N | N | Y | ... | ... | n | NG? |
| 38544 | 10512455-6433055 | 17.9 ± 0.2 | 6049 ± 210 | 3.93 ± 0.10 | 1.004 ± 0.001 | -0.13 ± 0.02 | 86 ± 1 | 1 | Y | Y | Y | N | N | Y | Y | ... | n | NG? |
| 38545 | 10512479-6335288 | -34.8 ± 0.2 | 4943 ± 141 | ... | 1.025 ± 0.003 | -0.05 ± 0.10 | 48 ± 8 | 1 | N | ... | ... | ... | ... | ... | ... | ... | n | G |
| 38546 | 10512791-6339571 | 69.4 ± 0.3 | 4888 ± 225 | ... | 1.022 ± 0.006 | -0.19 ± 0.09 | <44 | 3 | N | ... | ... | ... | ... | ... | ... | ... | n | G |
| 38547 | 10512800-6340212 | 14.1 ± 0.2 | 4704 ± 93 | 2.50 ± 0.13 | 1.020 ± 0.003 | 0.02 ± 0.08 | <39 | 3 | N | N | ... | ... | ... | ... | ... | ... | n | G |
| 38548 | 10512817-6417388 | 11.1 ± 0.2 | 4529 ± 160 | ... | 1.020 ± 0.003 | 0.10 ± 0.09 | <42 | 3 | N | ... | ... | ... | ... | ... | ... | ... | n | G |
| 38549 | 10512826-6350220 | 8.4 ± 0.3 | 4882 ± 165 | 3.27 ± 0.11 | 0.995 ± 0.008 | -0.11 ± 0.18 | <22 | 3 | Y | N | Y | N | N | Y | ... | ... | n | NG? |
| 38550 | 10512934-6332553 | -6.8 ± 0.3 | 5076 ± 93 | ... | 1.010 ± 0.006 | -0.07 ± 0.15 | ... | ... | N | ... | ... | ... | ... | ... | ... | ... | n | G |
| 38551 | 10512936-6334084 | 52.1 ± 0.2 | 4591 ± 61 | 2.41 ± 0.13 | 1.017 ± 0.003 | -0.01 ± 0.03 | <41 | 3 | N | N | ... | ... | ... | ... | ... | ... | n | G |
| 38552 | 10512957-6345116 | 85.1 ± 0.2 | 4886 ± 315 | ... | 1.000 ± 0.005 | -0.16 ± 0.29 | ... | ... | Y | ... | N | ... | Y | ... | ... | ... | n | ... |
| 38553 | 10513048-6426077 | -3.7 ± 0.2 | 4687 ± 45 | 2.46 ± 0.02 | 1.022 ± 0.003 | -0.04 ± 0.02 | <34 | 3 | N | N | ... | ... | ... | ... | ... | ... | n | G |
| 2605 | 10513100-6344320 | 0.2 ± 0.6 | 5072 ± 16 | 3.10 ± 0.06 | ... | 0.12 ± 0.06 | <14 | 3 | ... | N | ... | ... | ... | ... | ... | ... | n | ... |
| 38554 | 10513122-6348580 | 38.0 ± 0.2 | 4681 ± 49 | 2.69 ± 0.04 | 1.007 ± 0.005 | -0.02 ± 0.12 | <33 | 3 | Y | N | Y | N | N | Y | ... | ... | n | NG? |
| 38555 | 10513208-6345095 | 38.7 ± 0.2 | 4695 ± 105 | 2.57 ± 0.07 | 1.010 ± 0.004 | -0.05 ± 0.11 | <50 | 3 | Y | N | Y | N | N | Y | ... | ... | n | NG? |
| 38556 | 10513215-6333381 | 25.7 ± 0.2 | 4657 ± 93 | ... | 1.025 ± 0.004 | ... | ... | ... | N | ... | ... | ... | ... | ... | ... | ... | n | G |
| 38557 | 10513215-6344243 | 5.8 ± 0.2 | 4925 ± 36 | 2.71 ± 0.04 | 1.020 ± 0.002 | 0.00 ± 0.04 | <36 | 3 | N | N | ... | ... | ... | ... | ... | ... | n | G |
| 38558 | 10513265-6330417 | 42.5 ± 0.2 | 4930 ± 131 | ... | 1.015 ± 0.006 | -0.12 ± 0.10 | <27 | 3 | N | ... | ... | ... | ... | ... | ... | ... | n | G |
| 38559 | 10513345-6339446 | -3.7 ± 0.2 | 4897 ± 245 | ... | 1.016 ± 0.003 | -0.12 ± 0.19 | 31 ± 2 | 1 | N | ... | ... | ... | ... | ... | ... | ... | n | G |
| 38560 | 10513524-6334569 | 20.1 ± 0.2 | 4802 ± 166 | 2.64 ± 0.16 | 1.014 ± 0.004 | -0.02 ± 0.05 | <47 | 3 | N | N | ... | ... | ... | ... | ... | ... | n | G |
| 37395 | 10283396-6453161 | -28.1 ± 0.3 | 5043 ± 162 | ... | 1.017 ± 0.006 | -0.10 ± 0.16 | <28 | 3 | N | ... | ... | ... | ... | ... | ... | ... | n | G |
| 37396 | 10283431-6446196 | 51.3 ± 0.2 | 4579 ± 210 | ... | 1.018 ± 0.006 | -0.07 ± 0.10 | <34 | 3 | N | ... | ... | ... | ... | ... | ... | ... | n | G |
| 37397 | 10283561-6441425 | 14.6 ± 0.2 | 4735 ± 16 | ... | 1.022 ± 0.004 | -0.08 ± 0.14 | <21 | 3 | N | ... | ... | ... | ... | ... | ... | ... | n | G |
| 37398 | 10283584-6448499 | 57.7 ± 0.2 | 4599 ± 58 | 2.25 ± 0.16 | 1.028 ± 0.005 | -0.11 ± 0.07 | <28 | 3 | N | N | ... | ... | ... | ... | ... | ... | n | G |
| 2520 | 10283748-6445080 | 8.1 ± 0.6 | 4969 ± 19 | 2.78 ± 0.05 | 1.015 ± 0.002 | 0.07 ± 0.02 | 29 ± 7 | 1 | N | N | ... | ... | ... | ... | ... | ... | n | G |
| 37399 | 10283782-6452074 | 23.3 ± 0.2 | 4607 ± 91 | ... | 1.025 ± 0.004 | ... | <41 | 3 | N | ... | ... | ... | ... | ... | ... | ... | n | G |
| 37627 | 10353025-6359152 | -9.9 ± 0.4 | 4987 ± 190 | ... | 0.961 ± 0.006 | ... | 247 ± 6 | 1 | Y | ... | Y | Y | Y | ... | N | ... | n | ... |
| 37400 | 10284172-6432171 | 32.7 ± 0.3 | 5157 ± 25 | ... | 1.002 ± 0.007 | -0.18 ± 0.15 | <41 | 3 | Y | ... | Y | N | N | Y | ... | ... | n | NG? |
| 37401 | 10284204-6444036 | 17.8 ± 0.2 | 5032 ± 124 | ... | 1.005 ± 0.002 | -0.06 ± 0.13 | <18 | 3 | Y | ... | Y | N | N | Y | ... | ... | n | NG? |
| 2522 | 10291466-6439127 | -18.8 ± 0.6 | 5136 ± 16 | 3.35 ± 0.03 | 1.000 ± 0.003 | -0.55 ± 0.01 | 10 ± 1 | 1 | Y | N | N | N | N | N | ... | ... | n | NG? |
| 37417 | 10291473-6443309 | -5.9 ± 0.2 | 5114 ± 61 | ... | 1.014 ± 0.004 | -0.05 ± 0.10 | <14 | 3 | N | ... | ... | ... | ... | ... | ... | ... | n | G |
| 37418 | 10291762-6443022 | 115.6 ± 0.2 | 4783 ± 193 | ... | 1.029 ± 0.005 | -0.08 ± 0.09 | <21 | 3 | N | ... | ... | ... | ... | ... | ... | ... | n | G |
| 37419 | 10291899-6440575 | 1.0 ± 0.3 | 5358 ± 119 | ... | 1.012 ± 0.004 | -0.16 ± 0.19 | <29 | 3 | N | ... | ... | ... | ... | ... | ... | ... | n | ... |
| 37420 | 10291904-6438330 | 14.1 ± 0.2 | 4758 ± 62 | 2.80 ± 0.12 | 1.004 ± 0.004 | 0.00 ± 0.10 | <35 | 3 | Y | N | Y | N | N | Y | ... | ... | n | NG? |
| 37421 | 10292187-6445211 | -3.6 ± 0.2 | 5157 ± 28 | ... | 1.010 ± 0.003 | -0.24 ± 0.06 | <31 | 3 | Y | ... | Y | N | N | Y | ... | ... | n | NG? |
| 37422 | 10292391-6445084 | 18.7 ± 0.2 | 5169 ± 39 | ... | 1.004 ± 0.003 | -0.15 ± 0.10 | <17 | 3 | Y | ... | Y | N | N | Y | ... | ... | n | NG? |
| 37423 | 10292478-6437344 | 24.2 ± 0.2 | 4801 ± 35 | ... | 1.001 ± 0.005 | 0.05 ± 0.08 | <38 | 3 | Y | ... | Y | N | N | Y | ... | ... | n | NG? |
| 37424 | 10292538-6435050 | -38.1 ± 0.2 | 5109 ± 51 | 3.34 ± 0.19 | 1.008 ± 0.004 | -0.07 ± 0.06 | <23 | 3 | Y | N | N | N | N | Y | ... | ... | n | NG? |
| 37425 | 10292546-6444328 | 17.2 ± 0.2 | 5283 ± 179 | ... | 1.005 ± 0.003 | -0.30 ± 0.19 | <13 | 3 | Y | ... | Y | N | N | N | ... | ... | n | NG? |
| 37631 | 10354549-6402418 | -9.2 ± 0.2 | 4896 ± 181 | ... | 1.024 ± 0.003 | -0.06 ± 0.11 | <20 | 3 | N | ... | ... | ... | ... | ... | ... | ... | n | G |
| 37426 | 10292706-6451183 | -7.4 ± 0.2 | 4607 ± 80 | 2.34 ± 0.12 | 1.022 ± 0.004 | -0.06 ± 0.10 | <31 | 3 | N | N | ... | ... | ... | ... | ... | ... | n | G |
| 37427 | 10292727-6442423 | -7.0 ± 0.2 | 5016 ± 120 | 3.30 ± 0.04 | 1.001 ± 0.004 | 0.01 ± 0.08 | 44 ± 8 | 1 | Y | N | Y | N | N | Y | ... | ... | n | NG? |
| 37428 | 10292741-6438152 | 17.5 ± 0.2 | 4813 ± 11 | 2.65 ± 0.03 | 1.015 ± 0.003 | 0.01 ± 0.05 | <39 | 3 | N | N | ... | ... | ... | ... | ... | ... | n | G |
| 37429 | 10292749-6448522 | -28.3 ± 0.3 | 5150 ± 25 | 3.54 ± 0.20 | 1.005 ± 0.005 | -0.14 ± 0.12 | <19 | 3 | Y | Y | N | N | N | Y | ... | ... | n | NG? |
| 37430 | 10292996-6452568 | -3.4 ± 0.3 | 5214 ± 107 | 3.53 ± 0.14 | 1.001 ± 0.006 | -0.25 ± 0.08 | <11 | 3 | Y | Y | Y | N | N | Y | ... | ... | n | NG? |
| 37431 | 10293454-6441075 | 7.3 ± 0.2 | 4857 ± 23 | 2.67 ± 0.13 | 1.017 ± 0.003 | 0.04 ± 0.01 | <39 | 3 | N | N | ... | ... | ... | ... | ... | ... | n | G |
| 37432 | 10293494-6438457 | 9.5 ± 0.2 | 5139 ± 94 | ... | 1.013 ± 0.003 | ... | ... | ... | N | ... | ... | ... | ... | ... | ... | ... | n | G |
| 37433 | 10293610-6434336 | 22.2 ± 0.3 | 5008 ± 133 | 4.05 ± 0.15 | 0.979 ± 0.005 | -0.06 ± 0.09 | <26 | 3 | Y | Y | Y | N | N | Y | ... | ... | n | NG |
| 37434 | 10293804-6443363 | -7.9 ± 0.2 | 4952 ± 106 | 3.01 ± 0.14 | 1.008 ± 0.006 | -0.05 ± 0.12 | <31 | 3 | Y | N | Y | N | N | Y | ... | ... | n | NG? |
| 37459 | 10295773-6333342 | 38.8 ± 0.2 | 4816 ± 156 | ... | 1.014 ± 0.003 | -0.18 ± 0.18 | ... | ... | N | ... | ... | ... | ... | ... | ... | ... | n | G |
| 37632 | 10354665-6336583 | 50.5 ± 0.3 | 5376 ± 188 | ... | 1.027 ± 0.009 | ... | <19 | 3 | N | ... | ... | ... | ... | ... | ... | ... | n | ... |

| ID | CNAME | RV (km s$^{-1}$) | $T_{\rm eff}$ (K) | logg (dex) | $\gamma^a$ | [Fe/H] (dex) | EW(Li)$^b$ (mÅ) | EW(Li) error flag$^c$ | $\gamma$ | logg | Membership RV | Li | H$\alpha$ | [Fe/H] | Gaia studies Randich$^d$ | Cantat-Gaudin$^d$ | Final$^e$ | NMs with Li$^f$ |
|---|---|---|---|---|---|---|---|---|---|---|---|---|---|---|---|---|---|---|
| 37460 | 10295793-6332510 | 35.8 ± 0.2 | 4568 ± 188 | ... | 1.020 ± 0.004 | 0.05 ± 0.08 | <44 | 3 | N | ... | ... | ... | ... | ... | ... | ... | n | G |
| 37461 | 10295838-6330483 | -6.4 ± 0.2 | 4575 ± 79 | 2.24 ± 0.16 | 1.039 ± 0.003 | 0.00 ± 0.02 | 38 ± 7 | 1 | N | N | ... | ... | ... | ... | ... | ... | n | G |
| 37462 | 10300060-6328396 | 8.1 ± 0.2 | 4572 ± 120 | ... | 1.019 ± 0.006 | 0.06 ± 0.12 | <51 | 3 | N | ... | ... | ... | ... | ... | ... | ... | n | G |
| 37633 | 10354685-6339087 | -14.2 ± 0.2 | 5169 ± 39 | ... | 1.019 ± 0.002 | -0.05 ± 0.03 | <22 | 3 | N | ... | ... | ... | ... | ... | ... | ... | n | G |
| 37463 | 10300119-6319388 | -17.6 ± 0.2 | 4631 ± 47 | 2.37 ± 0.08 | 1.019 ± 0.004 | -0.07 ± 0.01 | <34 | 3 | N | N | ... | ... | ... | ... | ... | ... | n | G |
| 37464 | 10300183-6319551 | -13.3 ± 0.2 | 4866 ± 159 | ... | 1.009 ± 0.003 | -0.05 ± 0.12 | <26 | 3 | Y | ... | N | N | N | Y | ... | ... | n | NG? |
| 37465 | 10300194-6321203 | -10.2 ± 0.2 | 4612 ± 42 | 2.37 ± 0.13 | 1.023 ± 0.002 | -0.06 ± 0.09 | 363 ± 3 | 1 | N | N | ... | ... | ... | ... | ... | ... | n | Li-rich G |
| 37466 | 10300279-6329261 | 21.7 ± 0.2 | 4583 ± 169 | ... | 1.024 ± 0.003 | 0.06 ± 0.05 | <47 | 3 | N | ... | ... | ... | ... | ... | ... | ... | n | G |
| 37467 | 10300296-6322510 | 74.8 ± 0.2 | 4722 ± 145 | ... | 1.017 ± 0.002 | -0.30 ± 0.17 | ... | ... | N | ... | ... | ... | ... | ... | ... | ... | n | G |
| 37636 | 10355135-6333196 | 38.1 ± 0.2 | 4862 ± 144 | ... | 1.018 ± 0.004 | -0.17 ± 0.14 | <39 | 3 | N | ... | ... | ... | ... | ... | ... | ... | n | G |
| 37468 | 10300314-6324515 | -8.3 ± 0.2 | 4539 ± 143 | ... | 1.023 ± 0.004 | 0.09 ± 0.08 | <50 | 3 | N | ... | ... | ... | ... | ... | ... | ... | n | G |
| 37637 | 10355140-6401051 | 10.5 ± 0.2 | 5141 ± 94 | ... | 1.023 ± 0.006 | ... | ... | ... | N | ... | ... | ... | ... | ... | ... | ... | n | G |
| 37469 | 10300337-6320311 | 30.2 ± 0.3 | 5156 ± 96 | ... | 1.011 ± 0.005 | -0.23 ± 0.25 | 24 ± 3 | 1 | N | ... | ... | ... | ... | ... | ... | ... | n | G |
| 37470 | 10300344-6332217 | 21.4 ± 0.2 | 5000 ± 188 | ... | 1.013 ± 0.003 | -0.27 ± 0.23 | <20 | 3 | N | ... | ... | ... | ... | ... | ... | ... | n | G |
| 37471 | 10300405-6326311 | -11.2 ± 0.2 | 4905 ± 83 | 2.71 ± 0.13 | 1.021 ± 0.002 | -0.03 ± 0.06 | <28 | 3 | N | N | ... | ... | ... | ... | ... | ... | n | G |
| 37472 | 10300558-6438560 | -4.2 ± 0.2 | 4700 ± 10 | 2.60 ± 0.08 | 1.013 ± 0.004 | 0.05 ± 0.05 | ... | ... | N | N | ... | ... | ... | ... | ... | ... | n | G |
| 37473 | 10300568-6329508 | 16.6 ± 0.2 | 5034 ± 48 | ... | 1.015 ± 0.004 | -0.04 ± 0.08 | ... | ... | N | ... | ... | ... | ... | ... | ... | ... | n | G |
| 37474 | 10300624-6319532 | 16.5 ± 0.3 | 4371 ± 316 | 4.59 ± 0.05 | 0.884 ± 0.007 | -0.07 ± 0.05 | ... | ... | Y | Y | Y | ... | N | Y | ... | ... | n | ... |
| 37475 | 10300706-6448455 | 66.7 ± 0.3 | 5028 ± 136 | ... | 1.007 ± 0.005 | -0.65 ± 0.21 | ... | ... | Y | ... | N | ... | ... | N | ... | ... | n | ... |
| 37476 | 10300758-6330222 | -59.0 ± 0.2 | 4722 ± 211 | ... | 1.025 ± 0.003 | -0.29 ± 0.06 | <25 | 3 | N | ... | ... | ... | ... | ... | ... | ... | n | G |
| 37477 | 10300818-6322216 | 48.5 ± 0.2 | 4789 ± 188 | ... | 1.017 ± 0.005 | -0.36 ± 0.18 | <47 | 3 | N | ... | ... | ... | ... | ... | ... | ... | n | G |
| 37478 | 10300834-6323023 | 120.9 ± 0.2 | 4907 ± 232 | ... | 1.017 ± 0.003 | -0.23 ± 0.24 | <21 | 3 | N | ... | ... | ... | ... | ... | ... | ... | n | G |
| 37479 | 10300907-6446329 | 72.1 ± 0.2 | 4706 ± 40 | 2.50 ± 0.10 | 1.018 ± 0.005 | -0.04 ± 0.08 | <37 | 3 | N | N | ... | ... | ... | ... | ... | ... | n | G |
| 37480 | 10301032-6439502 | 30.5 ± 0.2 | 4573 ± 181 | ... | 1.025 ± 0.004 | 0.05 ± 0.07 | <51 | 3 | N | ... | ... | ... | ... | ... | ... | ... | n | G |
| 37638 | 10355294-6408027 | 8.2 ± 0.2 | 4913 ± 208 | ... | 1.026 ± 0.003 | -0.02 ± 0.07 | 61 ± 2 | 1 | N | ... | ... | ... | ... | ... | ... | ... | n | G |
| 37481 | 10301069-6434064 | 27.5 ± 0.2 | 5098 ± 64 | 3.58 ± 0.06 | 0.999 ± 0.003 | -0.07 ± 0.11 | <29 | 3 | Y | Y | Y | N | N | Y | ... | ... | n | NG |
| 37482 | 10301116-6325277 | 21.0 ± 0.2 | 4591 ± 91 | ... | 1.026 ± 0.004 | -0.12 ± 0.12 | <28 | 3 | N | ... | ... | ... | ... | ... | ... | ... | n | G |
| 37524 | 10310473-6325126 | 7.6 ± 0.2 | 4562 ± 114 | ... | 1.021 ± 0.005 | 0.01 ± 0.05 | <28 | 3 | N | ... | ... | ... | ... | ... | ... | ... | n | G |
| 37525 | 10310733-6443290 | 47.6 ± 0.2 | 4596 ± 53 | 2.42 ± 0.12 | 1.017 ± 0.006 | 0.02 ± 0.03 | <45 | 3 | N | N | ... | ... | ... | ... | ... | ... | n | G |
| 37526 | 10310811-6321493 | 9.1 ± 0.3 | 3493 ± 49 | ... | 0.843 ± 0.009 | -0.25 ± 0.13 | ... | ... | Y | ... | Y | ... | N | Y | N | ... | n | ... |
| 37527 | 10310853-6325469 | -4.8 ± 0.2 | 4976 ± 148 | ... | 1.011 ± 0.004 | -0.08 ± 0.15 | <23 | 3 | N | ... | ... | ... | ... | ... | ... | ... | n | G |
| 37528 | 10310953-6329119 | -9.0 ± 0.2 | 4576 ± 125 | ... | 1.028 ± 0.002 | -0.12 ± 0.05 | <25 | 3 | N | ... | ... | ... | ... | ... | ... | ... | n | G |
| 37529 | 10311090-6444421 | 15.9 ± 0.2 | 5138 ± 113 | ... | 1.046 ± 0.002 | ... | <21 | 3 | N | ... | ... | ... | ... | ... | ... | ... | n | G |
| 37530 | 10311130-6332526 | -1.0 ± 0.2 | 4728 ± 18 | 2.62 ± 0.18 | 1.019 ± 0.004 | -0.06 ± 0.07 | <31 | 3 | N | N | ... | ... | ... | ... | ... | ... | n | G |
| 37531 | 10311145-6441201 | 52.0 ± 0.2 | 4608 ± 125 | 2.32 ± 0.18 | 1.019 ± 0.004 | -0.11 ± 0.11 | <38 | 3 | N | N | ... | ... | ... | ... | ... | ... | n | G |
| 37642 | 10360033-6337232 | 20.2 ± 0.2 | 4616 ± 153 | ... | 1.013 ± 0.003 | 0.03 ± 0.10 | <51 | 3 | N | ... | ... | ... | ... | ... | ... | ... | n | G |
| 37532 | 10311172-6335305 | 35.2 ± 0.2 | 4569 ± 90 | 2.47 ± 0.06 | 1.009 ± 0.003 | -0.04 ± 0.07 | <27 | 3 | Y | N | Y | N | N | Y | ... | ... | n | NG? |
| 37533 | 10311190-6322002 | 15.6 ± 0.2 | 4533 ± 137 | ... | 1.041 ± 0.002 | -0.06 ± 0.01 | <40 | 3 | N | ... | ... | ... | ... | ... | ... | ... | n | G |
| 37534 | 10311243-6333561 | -5.1 ± 0.2 | 4529 ± 81 | 2.26 ± 0.13 | 1.022 ± 0.002 | -0.04 ± 0.03 | <37 | 3 | N | N | ... | ... | ... | ... | ... | ... | n | G |
| 37535 | 10311376-6330181 | 71.5 ± 0.2 | 4779 ± 4 | 2.57 ± 0.16 | 1.028 ± 0.004 | -0.10 ± 0.02 | <28 | 3 | N | N | ... | ... | ... | ... | ... | ... | n | G |
| 37536 | 10311407-6328242 | -35.2 ± 0.2 | 5005 ± 195 | ... | 1.015 ± 0.004 | -0.14 ± 0.19 | <12 | 3 | N | ... | ... | ... | ... | ... | ... | ... | n | G |
| 37537 | 10311426-6319310 | 59.1 ± 0.2 | 4639 ± 98 | ... | 1.020 ± 0.006 | ... | ... | ... | N | ... | ... | ... | ... | ... | ... | ... | n | G |
| 37538 | 10311535-6441007 | 50.5 ± 0.2 | 4563 ± 80 | ... | 1.037 ± 0.007 | -0.06 ± 0.04 | <34 | 3 | N | ... | ... | ... | ... | ... | ... | ... | n | G |
| 37539 | 10311637-6336032 | 17.0 ± 0.2 | 4610 ± 28 | 2.32 ± 0.06 | 1.022 ± 0.003 | -0.05 ± 0.10 | <37 | 3 | N | N | ... | ... | ... | ... | ... | ... | n | G |
| 37540 | 10311765-6317569 | 28.7 ± 0.5 | ... | ... | ... | ... | ... | ... | ... | ... | ... | ... | ... | ... | N | ... | n | ... |
| 37541 | 10311778-6324488 | 31.1 ± 0.2 | 4747 ± 79 | 2.57 ± 0.13 | 1.016 ± 0.002 | -0.02 ± 0.09 | <35 | 3 | N | N | ... | ... | ... | ... | ... | ... | n | G |
| 2530 | 10312182-6329281 | -16.9 ± 0.6 | 4654 ± 15 | 2.35 ± 0.05 | 1.021 ± 0.002 | -0.22 ± 0.02 | 18 ± 1 | 1 | N | N | ... | ... | ... | ... | ... | ... | n | G |
| 37717 | 10390884-6342119 | 35.4 ± 0.2 | 4575 ± 178 | ... | 1.020 ± 0.004 | 0.06 ± 0.06 | <39 | 3 | N | ... | ... | ... | ... | ... | ... | ... | n | G |
| 37542 | 10312190-6320169 | -33.7 ± 0.2 | 4647 ± 105 | 2.50 ± 0.15 | 1.013 ± 0.002 | 0.01 ± 0.08 | <33 | 3 | N | N | ... | ... | ... | ... | ... | ... | n | G |
| 37543 | 10312441-6334351 | -25.5 ± 0.2 | 4577 ± 97 | ... | 1.028 ± 0.003 | 0.06 ± 0.11 | <42 | 3 | N | ... | ... | ... | ... | ... | ... | ... | n | G |
| 37718 | 10391188-6456046 | 21.7 ± 1.2 | ... | ... | ... | ... | ... | ... | ... | ... | ... | ... | ... | ... | ... | Y | n | ... |
| 37544 | 10312472-6322205 | -18.8 ± 0.2 | 4720 ± 41 | 2.53 ± 0.04 | 1.014 ± 0.003 | -0.03 ± 0.01 | <35 | 3 | N | N | ... | ... | ... | ... | ... | ... | n | G |
| 53535 | 10391352-6408077 | -7.7 ± 0.2 | 4531 ± 64 | 2.29 ± 0.15 | 1.030 ± 0.005 | 0.01 ± 0.06 | <46 | 3 | N | N | ... | ... | ... | ... | ... | ... | n | G |
| 37545 | 10312578-6325319 | 34.5 ± 0.3 | 5138 ± 99 | ... | 1.007 ± 0.006 | ... | <24 | 3 | Y | ... | Y | N | N | ... | ... | ... | n | NG? |
| 37546 | 10312626-6325050 | 84.4 ± 0.2 | 4655 ± 144 | ... | 1.021 ± 0.006 | -0.13 ± 0.07 | <26 | 3 | N | ... | ... | ... | ... | ... | ... | ... | n | G |
| 37568 | 10322511-6335092 | 63.9 ± 0.2 | 4623 ± 109 | 2.37 ± 0.13 | 1.021 ± 0.004 | -0.02 ± 0.10 | <33 | 3 | N | N | ... | ... | ... | ... | ... | ... | n | G |
| 53536 | 10391511-6414151 | 43.4 ± 0.2 | 5235 ± 74 | ... | 0.991 ± 0.005 | -0.02 ± 0.06 | <9 | 3 | Y | ... | N | N | N | Y | ... | ... | n | ... |
| 37569 | 10323205-6324012 | 13.3 ± 0.2 | 4607 ± 100 | 2.53 ± 0.10 | 1.011 ± 0.003 | 0.13 ± 0.01 | 383 ± 5 | 1 | N | N | ... | ... | ... | ... | ... | ... | n | Li-rich G |
| 37570 | 10323805-6328537 | -29.9 ± 0.2 | 5004 ± 189 | ... | 1.017 ± 0.003 | -0.23 ± 0.16 | <16 | 3 | N | ... | ... | ... | ... | ... | ... | ... | n | G |

**Table C.6.** continued.

| ID | CNAME | RV (km s$^{-1}$) | $T_{\text{eff}}$ (K) | logg (dex) | $\gamma^a$ | [Fe/H] (dex) | EW(Li)$^b$ (mÅ) | EW(Li) error flag$^c$ | $\gamma$ | logg | RV | Li | H$\alpha$ | [Fe/H] | Randich$^d$ | Cantat-Gaudin$^d$ | Final$^e$ | NMs with Li$^f$ |
|---|---|---|---|---|---|---|---|---|---|---|---|---|---|---|---|---|---|---|
| 53537 | 10391559-6415562 | -8.7 ± 0.2 | 4549 ± 106 | … | 1.036 ± 0.003 | … | <25 | 3 | N | … | … | … | … | … | … | … | n | G |
| 37571 | 10324286-6324454 | -28.1 ± 0.2 | 4562 ± 98 | 2.21 ± 0.19 | 1.033 ± 0.003 | -0.06 ± 0.05 | 6 ± 2 | 1 | N | N | … | … | … | … | … | … | n | G |
| 53538 | 10391585-6417528 | 12.9 ± 0.3 | 5556 ± 137 | … | 0.995 ± 0.006 | -0.13 ± 0.14 | <31 | 3 | Y | … | Y | N | N | Y | … | … | n | NG |
| 37572 | 10325373-6407524 | 35.2 ± 0.2 | 4638 ± 98 | … | 1.021 ± 0.002 | … | <55 | 3 | N | … | … | … | … | … | … | … | n | G |
| 37573 | 10325421-6324124 | -22.6 ± 0.2 | 4901 ± 112 | … | 1.024 ± 0.004 | -0.05 ± 0.09 | … | … | N | … | … | … | … | … | … | … | n | G |
| 37719 | 10391730-6458464 | -26.6 ± 0.2 | 5020 ± 193 | … | 1.017 ± 0.003 | -0.10 ± 0.16 | <34 | 3 | N | … | … | … | … | … | … | … | n | G |
| 37574 | 10325587-6409075 | 15.6 ± 0.2 | 5010 ± 168 | … | 1.019 ± 0.004 | -0.11 ± 0.17 | … | … | N | … | … | … | … | … | … | … | n | G |
| 37575 | 10330081-6409181 | -1.6 ± 0.2 | 6265 ± 27 | 4.14 ± 0.06 | 1.000 ± 0.002 | 0.26 ± 0.02 | <12 | 3 | Y | N | Y | N | N | N | N | … | n | NG |
| 37720 | 10391989-6359466 | -9.9 ± 0.2 | 4587 ± 81 | 2.46 ± 0.18 | 1.023 ± 0.003 | 0.07 ± 0.02 | <56 | 3 | N | N | … | … | … | … | … | … | n | G |
| 37576 | 10330100-6332239 | -1.6 ± 0.2 | 5635 ± 222 | 3.97 ± 0.05 | 0.998 ± 0.002 | 0.14 ± 0.02 | 26 ± 5 | 1 | Y | Y | Y | N | N | Y | … | … | n | NG |
| 37577 | 10330502-6409458 | -40.0 ± 0.2 | 4722 ± 164 | … | 1.019 ± 0.004 | -0.13 ± 0.16 | … | … | N | … | … | … | … | … | … | … | n | G |
| 37721 | 10392058-6356202 | 7.3 ± 0.2 | 4635 ± 193 | … | 1.045 ± 0.003 | 0.03 ± 0.06 | <46 | 3 | N | … | … | … | … | … | … | … | n | G |
| 37578 | 10331385-6407223 | 0.8 ± 0.2 | 4978 ± 162 | … | 1.017 ± 0.002 | -0.09 ± 0.13 | <15 | 3 | N | … | … | … | … | … | … | … | n | G |
| 37579 | 10331779-6406481 | 12.6 ± 0.2 | 4547 ± 320 | … | 0.941 ± 0.003 | 0.04 ± 0.14 | … | … | Y | … | Y | … | N | Y | … | … | n | … |
| 37580 | 10332294-6413160 | 9.5 ± 0.2 | 4865 ± 11 | 2.76 ± 0.06 | 1.010 ± 0.002 | 0.02 ± 0.05 | <36 | 3 | N | N | … | … | … | … | … | … | n | G |
| 53539 | 10392129-6414540 | 9.0 ± 0.6 | 7339 ± 95 | … | 1.004 ± 0.005 | … | … | … | Y | … | Y | … | N | Y | … | … | n | … |
| 37581 | 10333241-6410584 | 48.0 ± 0.2 | 4379 ± 239 | … | 1.043 ± 0.003 | -0.10 ± 0.11 | <29 | 3 | N | … | … | … | … | … | … | … | n | G |
| 37582 | 10334051-6403161 | 7.0 ± 0.2 | 4441 ± 147 | … | 1.039 ± 0.002 | -0.22 ± 0.22 | <22 | 3 | N | … | … | … | … | … | … | … | n | G |
| 2534 | 10334181-6413457 | 17.1 ± 0.6 | 5306 ± 29 | 4.21 ± 0.01 | 0.978 ± 0.002 | -0.11 ± 0.02 | 223 ± 1 | 2 | Y | Y | Y | Y | Y | Y | Y | … | Y | … |
| 37583 | 10334213-6413338 | 9.6 ± 0.2 | 4508 ± 96 | … | 1.026 ± 0.002 | -0.09 ± 0.09 | <27 | 3 | N | … | … | … | … | … | … | … | n | G |
| 53554 | 10394465-6415308 | 28.7 ± 0.2 | 4713 ± 102 | … | 1.025 ± 0.003 | … | <40 | 3 | N | … | … | … | … | … | … | … | n | G |
| 37584 | 10334766-6413334 | 34.6 ± 0.2 | 5007 ± 191 | … | 1.017 ± 0.003 | -0.13 ± 0.25 | <22 | 3 | N | … | … | … | … | … | … | … | n | G |
| 53555 | 10394476-6405156 | 137.7 ± 0.2 | 4261 ± 294 | … | 1.019 ± 0.008 | -0.02 ± 0.17 | 50 ± 6 | 1 | N | … | … | … | … | … | … | … | n | G |
| 37585 | 10334842-6358099 | 10.1 ± 0.3 | 6608 ± 125 | 4.18 ± 0.19 | 1.001 ± 0.002 | -0.11 ± 0.16 | … | … | Y | Y | Y | … | N | Y | N | … | n | … |
| 37586 | 10334849-6413230 | 16.6 ± 0.2 | 4518 ± 169 | … | 1.013 ± 0.002 | 0.11 ± 0.07 | <62 | 3 | N | … | … | … | … | … | … | … | n | G |
| 53556 | 10394507-6419088 | 51.7 ± 1.0 | … | … | … | … | … | … | … | … | … | … | … | … | … | … | n | … |
| 37587 | 10335040-6407302 | 28.0 ± 0.2 | 4508 ± 153 | … | 1.049 ± 0.003 | -0.08 ± 0.06 | 80 ± 7 | 1 | N | … | … | … | … | … | … | … | n | G |
| 37588 | 10335054-6358386 | -38.0 ± 0.2 | 4946 ± 109 | … | 1.021 ± 0.002 | -0.03 ± 0.06 | <33 | 3 | N | … | … | … | … | … | … | … | n | G |
| 37732 | 10394543-6451485 | 33.3 ± 0.2 | 4638 ± 57 | 2.36 ± 0.15 | 1.024 ± 0.002 | -0.06 ± 0.10 | <32 | 3 | N | N | … | … | … | … | … | … | n | G |
| 53663 | 10413047-6410253 | -17.9 ± 0.3 | 7295 ± 26 | … | … | … | … | … | … | … | … | … | … | … | … | … | n | … |
| 37589 | 10335064-6404437 | -28.1 ± 0.2 | 4755 ± 199 | … | 1.020 ± 0.002 | -0.06 ± 0.07 | <29 | 3 | N | … | … | … | … | … | … | … | n | G |
| 37733 | 10394555-6400268 | 11.0 ± 0.2 | 4876 ± 209 | … | 1.014 ± 0.002 | -0.08 ± 0.13 | <23 | 3 | N | … | … | … | … | … | … | … | n | G |
| 53664 | 10413107-6416168 | -5.6 ± 0.4 | 6293 ± 225 | … | 1.021 ± 0.009 | -0.26 ± 0.20 | <23 | 3 | N | … | … | … | … | … | … | … | n | … |
| 37590 | 10335193-6411525 | -21.5 ± 0.2 | 5340 ± 66 | … | 1.009 ± 0.003 | 0.03 ± 0.02 | 74 ± 2 | 1 | Y | … | N | N | Y | Y | … | … | n | NG? |
| 53665 | 10413301-6412163 | -12.1 ± 0.3 | 5575 ± 113 | … | 0.991 ± 0.008 | … | <17 | 3 | Y | … | N | N | N | … | … | … | n | NG |
| 37604 | 10342598-6416276 | 58.8 ± 0.2 | 4534 ± 151 | … | 1.023 ± 0.002 | 0.11 ± 0.05 | <48 | 3 | N | … | … | … | … | … | … | … | n | G |
| 37836 | 10413315-6527271 | 61.3 ± 0.3 | 4643 ± 125 | … | 1.023 ± 0.009 | -0.03 ± 0.10 | <51 | 3 | N | … | … | … | … | … | … | … | n | G |
| 37643 | 10360907-6407101 | -2.3 ± 0.2 | 4711 ± 38 | 2.52 ± 0.15 | 1.021 ± 0.002 | -0.07 ± 0.05 | 50 ± 6 | 1 | N | N | … | … | … | … | … | … | n | G |
| 37734 | 10394660-6442217 | 30.5 ± 0.2 | 4571 ± 185 | … | 1.018 ± 0.002 | 0.06 ± 0.06 | <39 | 3 | N | … | … | … | … | … | … | … | n | G |
| 37837 | 10413326-6427153 | -38.9 ± 0.2 | 4785 ± 60 | 2.67 ± 0.14 | 1.017 ± 0.002 | -0.04 ± 0.02 | <33 | 3 | N | N | … | … | … | … | … | … | n | G |
| 53666 | 10413334-6409173 | -6.6 ± 0.7 | 3635 ± 25 | … | 0.827 ± 0.019 | … | … | … | Y | … | Y | … | N | … | … | … | n | … |
| 37838 | 10413381-6541427 | 71.3 ± 0.2 | 4582 ± 104 | 2.32 ± 0.15 | 1.019 ± 0.004 | -0.10 ± 0.02 | <30 | 3 | N | N | … | … | … | … | … | … | n | G |
| 53667 | 10413417-6419046 | 32.3 ± 0.3 | 4646 ± 13 | 2.50 ± 0.01 | 1.016 ± 0.015 | 0.11 ± 0.20 | … | … | N | N | … | … | … | … | … | … | n | G |
| 37839 | 10413443-6535553 | 26.0 ± 0.2 | 4833 ± 68 | 2.63 ± 0.17 | 1.020 ± 0.003 | -0.07 ± 0.13 | <26 | 3 | N | N | … | … | … | … | … | … | n | G |
| 37605 | 10343083-6406546 | 87.4 ± 0.2 | 4434 ± 153 | … | 1.047 ± 0.003 | -0.18 ± 0.05 | <26 | 3 | N | … | … | … | … | … | … | … | n | G |
| 37840 | 10413481-6448582 | 7.4 ± 0.2 | 4658 ± 203 | … | 1.058 ± 0.003 | 0.00 ± 0.04 | … | … | N | … | … | … | … | … | … | … | n | G |
| 37841 | 10413526-6434208 | 0.3 ± 0.2 | 4375 ± 236 | … | 1.026 ± 0.004 | 0.07 ± 0.07 | <49 | 3 | N | … | … | … | … | … | … | … | n | G |
| 37842 | 10413560-6537028 | 41.0 ± 0.2 | 4629 ± 127 | 2.40 ± 0.19 | 1.017 ± 0.005 | -0.03 ± 0.17 | 24 ± 4 | 1 | N | N | … | … | … | … | … | … | n | G |
| 53557 | 10394891-6407357 | 0.8 ± 0.2 | 7342 ± 13 | … | … | … | … | … | … | … | … | … | … | … | … | … | n | … |
| 53668 | 10413631-6415512 | 19.1 ± 0.2 | 4628 ± 102 | … | 1.020 ± 0.004 | … | <72 | 3 | N | … | … | … | … | … | … | … | n | G |
| 53669 | 10413661-6406563 | -15.6 ± 0.3 | 7015 ± 49 | … | … | … | … | … | … | … | … | … | … | … | … | … | n | … |
| 37843 | 10413678-6530174 | -30.2 ± 0.3 | 4725 ± 12 | 2.68 ± 0.08 | 1.014 ± 0.007 | 0.08 ± 0.06 | <44 | 3 | N | N | … | … | … | … | … | … | n | … |
| 53670 | 10413756-6403443 | 42.6 ± 0.2 | 4115 ± 336 | … | 1.061 ± 0.003 | -0.23 ± 0.02 | <23 | 3 | N | … | … | … | … | … | … | … | n | G |
| 37735 | 10394942-6444250 | 17.6 ± 0.2 | 3431 ± 75 | … | 0.848 ± 0.006 | -0.25 ± 0.13 | <100 | 3 | Y | … | Y | Y | Y | Y | Y | Y | Y | … |
| 53671 | 10413787-6413407 | 50.6 ± 0.2 | 4399 ± 87 | 1.89 ± 0.14 | 1.059 ± 0.011 | -0.03 ± 0.13 | <38 | 3 | N | N | … | … | … | … | … | … | n | G |
| 37647 | 10361987-6328500 | 3.7 ± 0.2 | 4748 ± 83 | 2.59 ± 0.11 | 1.015 ± 0.003 | 0.00 ± 0.02 | <32 | 3 | N | N | … | … | … | … | … | … | n | G |
| 37844 | 10413832-6451269 | -8.7 ± 0.2 | 4514 ± 153 | 2.25 ± 0.20 | 1.026 ± 0.002 | 0.09 ± 0.10 | <47 | 3 | N | N | … | … | … | … | … | … | n | G |
| 37743 | 10395638-6400039 | -34.0 ± 0.2 | 4707 ± 118 | … | 1.025 ± 0.003 | -0.03 ± 0.06 | <34 | 3 | N | … | … | … | … | … | … | … | n | G |
| 37845 | 10413906-6457433 | 10.6 ± 0.2 | 4475 ± 186 | … | 1.034 ± 0.002 | -0.05 ± 0.05 | <32 | 3 | N | … | … | … | … | … | … | … | n | G |

**Table C.6.** continued.

| ID | CNAME | RV (km s$^{-1}$) | $T_{\text{eff}}$ (K) | $logg$ (dex) | $\gamma^a$ | [Fe/H] (dex) | EW(Li)$^b$ (mÅ) | EW(Li) error flag$^c$ | $\gamma$ | $logg$ | Membership RV | Li | H$\alpha$ | [Fe/H] | Gaia studies Randich$^d$ | Cantat-Gaudin$^d$ | Final$^e$ | NMs with Li$^f$ |
|---|---|---|---|---|---|---|---|---|---|---|---|---|---|---|---|---|---|---|
| 53672 | 10414012-6413298 | 63.3 ± 0.2 | 4823 ± 42 | … | 1.029 ± 0.006 | -0.19 ± 0.18 | <32 | 3 | N | … | … | … | … | … | … | … | n | G |
| 2536 | 10343769-6410068 | 3.8 ± 0.6 | 4658 ± 24 | 2.60 ± 0.05 | 1.025 ± 0.003 | 0.11 ± 0.04 | 45 ± 15 | 1 | N | N | … | … | … | … | … | … | n | G |
| 53673 | 10414036-6422459 | -15.0 ± 0.2 | 4764 ± 31 | 2.56 ± 0.11 | 1.023 ± 0.003 | -0.04 ± 0.08 | <27 | 3 | N | N | … | … | … | … | … | … | n | G |
| 37864 | 10415931-6532194 | 51.1 ± 0.2 | 4733 ± 20 | … | 0.995 ± 0.005 | -0.01 ± 0.06 | <48 | 3 | Y | … | N | N | N | Y | … | … | n | NG |
| 37744 | 10395722-6445157 | 19.1 ± 0.2 | 4671 ± 77 | 2.50 ± 0.13 | 1.017 ± 0.003 | 0.03 ± 0.08 | <38 | 3 | N | N | … | … | … | … | … | … | n | G |
| 37865 | 10415951-6535035 | 9.9 ± 0.3 | 5159 ± 101 | … | 1.015 ± 0.007 | … | <33 | 3 | N | … | … | … | … | … | … | … | n | G |
| 37866 | 10415962-6534307 | 36.0 ± 0.2 | 4623 ± 198 | … | 1.020 ± 0.004 | 0.02 ± 0.07 | … | … | N | … | … | … | … | … | … | … | n | G |
| 37867 | 10415988-6545419 | -14.6 ± 0.2 | 4472 ± 187 | … | 1.046 ± 0.002 | -0.08 ± 0.07 | <31 | 3 | N | … | … | … | … | … | … | … | n | G |
| 37648 | 10362283-6344137 | -4.3 ± 0.2 | 4660 ± 12 | 2.38 ± 0.09 | 1.030 ± 0.002 | -0.05 ± 0.01 | <43 | 3 | N | N | … | … | … | … | … | … | n | G |
| 37745 | 10395898-6500018 | 40.1 ± 0.5 | 4047 ± 78 | … | 0.857 ± 0.018 | 0.00 ± 0.17 | <51 | 3 | Y | … | Y | Y | N | Y | N | … | n | NG |
| 37868 | 10415994-6532469 | 27.9 ± 0.2 | 4317 ± 165 | 1.80 ± 0.12 | 1.032 ± 0.004 | 0.20 ± 0.03 | <58 | 3 | N | N | … | … | … | … | … | … | n | G |
| 53697 | 10420066-6421333 | 28.1 ± 0.2 | 4518 ± 203 | … | 1.017 ± 0.003 | 0.09 ± 0.13 | 383 ± 5 | 1 | N | … | … | … | … | … | … | … | n | Li-rich G |
| 37869 | 10420135-6455174 | 4.2 ± 0.2 | 4583 ± 100 | … | 1.027 ± 0.004 | 0.08 ± 0.05 | <43 | 3 | N | … | … | … | … | … | … | … | n | G |
| 53698 | 10420142-6415582 | -11.6 ± 0.2 | 4499 ± 109 | 2.25 ± 0.20 | 1.026 ± 0.003 | -0.02 ± 0.01 | <35 | 3 | N | N | … | … | … | … | … | … | n | G |
| 37608 | 10344817-6409411 | 5.5 ± 0.2 | 4621 ± 59 | 2.36 ± 0.10 | 1.020 ± 0.002 | -0.06 ± 0.01 | <34 | 3 | N | N | … | … | … | … | … | … | n | G |
| 53699 | 10420179-6408491 | -29.7 ± 0.2 | 5079 ± 93 | … | 0.982 ± 0.006 | … | <13 | 3 | Y | … | N | N | N | … | … | … | n | NG |
| 37746 | 10395956-6440535 | 44.9 ± 0.2 | 4641 ± 98 | … | 1.022 ± 0.003 | … | 86 ± 3 | 1 | N | … | … | … | … | … | … | … | n | G |
| 37870 | 10420179-6445306 | 0.3 ± 0.2 | 5006 ± 164 | … | 1.019 ± 0.002 | -0.10 ± 0.12 | <23 | 3 | N | … | … | … | … | … | … | … | n | G |
| 53700 | 10420210-6418194 | -7.5 ± 0.3 | 6553 ± 44 | … | 1.032 ± 0.003 | 0.32 ± 0.03 | <8 | 3 | N | … | … | … | … | … | … | … | n | … |
| 37656 | 10364005-6341516 | 9.8 ± 0.2 | 4803 ± 1 | 2.64 ± 0.03 | 1.016 ± 0.002 | 0.04 ± 0.04 | <31 | 3 | N | N | … | … | … | … | … | … | n | G |
| 53560 | 10400054-6421144 | -28.3 ± 0.3 | … | … | … | … | … | … | N | … | … | … | … | … | … | … | n | … |
| 53761 | 10434606-6425447 | -11.9 ± 0.3 | 5857 ± 42 | 4.05 ± 0.09 | 0.997 ± 0.006 | -0.02 ± 0.03 | 51 ± 5 | 1 | Y | Y | N | N | N | Y | … | … | n | NG |
| 53701 | 10420281-6414357 | 138.5 ± 0.4 | 4250 ± 161 | … | 1.076 ± 0.020 | -0.27 ± 0.19 | <47 | 3 | N | … | … | … | … | … | … | … | n | G |
| 37871 | 10420312-6531495 | 29.1 ± 0.2 | 4766 ± 208 | 2.77 ± 0.20 | 1.004 ± 0.005 | -0.03 ± 0.11 | <36 | 3 | Y | N | Y | N | Y | Y | … | … | n | NG? |
| 37657 | 10364174-6345349 | 22.0 ± 0.2 | 4471 ± 216 | … | 1.019 ± 0.003 | 0.15 ± 0.02 | <58 | 3 | N | … | … | … | … | … | … | … | n | G |
| 2542 | 10400080-6449212 | 21.1 ± 0.6 | 4627 ± 9 | 2.49 ± 0.10 | 1.020 ± 0.002 | 0.19 ± 0.04 | 51 ± 17 | 1 | N | N | … | … | … | … | … | … | n | G |
| 37755 | 10401052-6404091 | 11.0 ± 0.2 | 4883 ± 203 | … | 1.014 ± 0.002 | -0.07 ± 0.13 | 21 ± 8 | 1 | N | … | … | … | … | … | … | … | n | G |
| 37872 | 10420335-6442037 | -3.0 ± 0.2 | 4495 ± 104 | … | 1.038 ± 0.003 | -0.10 ± 0.01 | <26 | 3 | N | … | … | … | … | … | … | … | n | G |
| 37992 | 10434654-6538392 | 13.0 ± 0.2 | 4595 ± 85 | … | 1.031 ± 0.005 | 0.00 ± 0.10 | 68 ± 4 | 1 | N | … | … | … | … | … | … | … | n | G |
| 37873 | 10420373-6432398 | -4.9 ± 0.2 | 4587 ± 29 | 2.29 ± 0.15 | 1.037 ± 0.003 | -0.03 ± 0.01 | <31 | 3 | N | N | … | … | … | … | … | … | n | G |
| 37756 | 10401132-6442546 | 15.4 ± 0.2 | 4578 ± 175 | … | 1.021 ± 0.002 | 0.05 ± 0.07 | … | … | N | … | … | … | … | … | … | … | n | G |
| 37993 | 10434657-6540198 | -28.5 ± 0.2 | 4982 ± 65 | 2.99 ± 0.19 | 1.016 ± 0.004 | 0.00 ± 0.04 | 43 ± 6 | 1 | N | N | … | … | … | … | … | … | n | G |
| 37658 | 10364223-6327239 | 35.0 ± 0.2 | 4547 ± 113 | … | 1.036 ± 0.006 | … | <32 | 3 | N | … | … | … | … | … | … | … | n | G |
| 53568 | 10401248-6402561 | 12.9 ± 0.3 | 6209 ± 77 | 4.18 ± 0.20 | 0.995 ± 0.005 | 0.06 ± 0.02 | 34 ± 4 | 1 | Y | Y | Y | N | N | Y | … | … | n | NG |
| 53762 | 10434680-6435072 | 82.9 ± 0.5 | 6440 ± 131 | … | 0.992 ± 0.007 | -0.20 ± 0.11 | <16 | 3 | Y | … | N | N | N | Y | … | … | n | NG |
| 37874 | 10420401-6453326 | 36.7 ± 0.2 | 4405 ± 199 | … | 1.036 ± 0.003 | -0.22 ± 0.21 | <23 | 3 | N | … | … | … | … | … | … | … | n | G |
| 53703 | 10420473-6412535 | 32.6 ± 0.3 | 6541 ± 99 | … | 0.991 ± 0.007 | 0.42 ± 0.07 | <16 | 3 | Y | … | Y | N | N | N | … | … | n | NG |
| 37618 | 10351340-6418401 | 90.6 ± 0.2 | 5073 ± 208 | … | 1.013 ± 0.004 | -0.19 ± 0.19 | <20 | 3 | N | … | … | … | … | … | … | … | n | G |
| 37875 | 10420539-6459515 | -10.0 ± 0.2 | 4666 ± 31 | 2.49 ± 0.05 | 1.020 ± 0.002 | 0.04 ± 0.01 | <36 | 3 | N | N | … | … | … | … | … | … | n | G |
| 37884 | 10421596-6534068 | -2.9 ± 0.2 | 4566 ± 27 | 2.29 ± 0.10 | 1.026 ± 0.003 | -0.03 ± 0.04 | <42 | 3 | N | N | … | … | … | … | … | … | n | G |
| 53763 | 10434780-6425282 | 135.5 ± 0.2 | 3992 ± 12 | … | 1.030 ± 0.003 | -0.04 ± 0.15 | <3 | 3 | N | … | … | … | … | … | … | … | n | G |
| 37619 | 10351416-6420567 | 9.6 ± 0.2 | 4517 ± 128 | … | 1.036 ± 0.003 | -0.01 ± 0.04 | 136 ± 4 | 1 | N | … | … | … | … | … | … | … | n | G |
| 37659 | 10364505-6340281 | 2.9 ± 0.2 | 4560 ± 135 | … | 1.024 ± 0.002 | 0.05 ± 0.09 | <52 | 3 | N | … | … | … | … | … | … | … | n | G |
| 53764 | 10434787-6432343 | -16.2 ± 0.3 | 7060 ± 32 | … | 1.007 ± 0.002 | … | <2 | 3 | Y | … | N | N | N | … | … | … | n | … |
| 37620 | 10351721-6411483 | 0.8 ± 0.2 | 4633 ± 34 | 2.42 ± 0.12 | 1.019 ± 0.002 | -0.07 ± 0.04 | <32 | 3 | N | N | … | … | … | … | … | … | n | G |
| 37885 | 10421667-6545472 | 85.0 ± 0.3 | 4868 ± 219 | 2.77 ± 0.14 | 1.014 ± 0.008 | 0.00 ± 0.06 | <60 | 3 | N | N | … | … | … | … | … | … | n | G |
| 53713 | 10421684-6404433 | -5.8 ± 0.3 | 5117 ± 52 | … | 0.982 ± 0.007 | -0.02 ± 0.03 | <19 | 3 | Y | … | Y | N | Y | Y | … | … | n | NG |
| 37377 | 10280516-6437166 | 73.6 ± 0.3 | 5345 ± 128 | … | 1.019 ± 0.006 | -0.25 ± 0.01 | 13 ± 3 | 1 | N | … | … | … | … | … | … | … | n | … |
| 37994 | 10434816-6400035 | -404.4 ± 107.9 | … | … | … | … | … | … | … | … | … | … | … | … | … | … | n | … |
| 53569 | 10401459-6414290 | -1.4 ± 0.2 | 4966 ± 172 | … | 1.020 ± 0.001 | -0.10 ± 0.12 | <24 | 3 | N | … | … | … | … | … | … | … | n | G |
| 37886 | 10421685-6528316 | -37.3 ± 0.2 | 4678 ± 11 | 2.50 ± 0.12 | 1.020 ± 0.003 | -0.05 ± 0.02 | <36 | 3 | N | N | … | … | … | … | … | … | n | G |
| 53771 | 10435329-6435393 | 32.4 ± 0.2 | 4571 ± 183 | … | 1.017 ± 0.001 | 0.03 ± 0.11 | <51 | 3 | N | … | … | … | … | … | … | … | n | G |
| 7 | 10421769-6407073 | … | … | … | … | … | … | … | … | … | … | … | … | … | … | … | n | … |
| 38004 | 10435331-6347560 | 10.8 ± 0.2 | 5309 ± 155 | 3.48 ± 0.20 | 1.023 ± 0.004 | 0.09 ± 0.08 | <31 | 3 | N | N | … | … | … | … | … | … | n | G |
| 37674 | 10374019-6344312 | 22.3 ± 0.2 | 5525 ± 34 | … | 1.003 ± 0.001 | -0.01 ± 0.13 | <14 | 3 | Y | … | Y | N | N | Y | … | … | n | NG? |
| 37621 | 10351999-6404021 | 18.6 ± 0.3 | 3401 ± 22 | 4.62 ± 0.18 | 0.840 ± 0.007 | -0.26 ± 0.13 | <100 | 3 | Y | Y | Y | Y | Y | Y | Y | Y | Y | … |
| 53570 | 10401600-6408284 | 8.5 ± 0.2 | 6034 ± 82 | 4.05 ± 0.01 | 0.999 ± 0.002 | 0.07 ± 0.02 | 19 ± 5 | 1 | Y | Y | Y | N | N | Y | … | … | n | NG |
| 53772 | 10435340-6424245 | 2.4 ± 0.2 | 4919 ± 44 | 2.83 ± 0.15 | 1.016 ± 0.002 | -0.03 ± 0.08 | <38 | 3 | N | N | … | … | … | … | … | … | n | G |
| 37675 | 10374033-6400419 | 14.5 ± 0.2 | 4741 ± 14 | 2.61 ± 0.18 | 1.021 ± 0.003 | -0.05 ± 0.09 | <25 | 3 | N | N | … | … | … | … | … | … | n | G |

**Table C.6.** continued.

| ID | CNAME | RV (km s$^{-1}$) | $T_{\text{eff}}$ (K) | $\log g$ (dex) | $\gamma^a$ | [Fe/H] (dex) | EW(Li)$^b$ (mÅ) | EW(Li) error flag$^c$ | $\gamma$ | Membership $\log g$ | RV | Li | H$\alpha$ | [Fe/H] | Gaia studies Randich$^d$ | Cantat-Gaudin$^d$ | Final$^e$ | NMs with Li$^f$ |
|---|---|---|---|---|---|---|---|---|---|---|---|---|---|---|---|---|---|---|
| 37622 | 10352026-6357374 | 7.9 ± 0.2 | 4807 ± 221 | ... | 1.033 ± 0.003 | -0.26 ± 0.05 | <31 | 3 | N | ... | ... | ... | ... | ... | ... | ... | n | G |
| 37676 | 10374089-6340048 | 7.9 ± 0.2 | 4509 ± 137 | ... | 1.031 ± 0.002 | -0.03 ± 0.02 | <37 | 3 | N | ... | ... | ... | ... | ... | ... | ... | n | G |
| 37623 | 10352140-6404278 | -10.1 ± 0.2 | 5057 ± 214 | ... | 1.003 ± 0.006 | -0.19 ± 0.32 | ... | ... | Y | ... | Y | ... | ... | Y | ... | ... | n | ... |
| 21 | 10435375-6407514 | 1.5 ± 0.4 | 7094 ± 154 | 4.12 ± 0.15 | ... | -0.13 ± 0.16 | <16 | 3 | ... | Y | Y | N | N | Y | N | ... | n | NG |
| 37378 | 10280884-6450111 | 40.4 ± 0.2 | 5022 ± 162 | ... | 1.009 ± 0.004 | -0.26 ± 0.28 | <25 | 3 | Y | ... | Y | N | N | Y | ... | ... | n | NG? |
| 37677 | 10374111-6345036 | 1.7 ± 0.2 | 4567 ± 132 | ... | 1.030 ± 0.001 | -0.08 ± 0.07 | <34 | 3 | N | ... | ... | ... | ... | ... | ... | ... | n | G |
| 37757 | 10401628-6440432 | -40.8 ± 0.2 | 5059 ± 104 | ... | 1.015 ± 0.002 | -0.19 ± 0.16 | <17 | 3 | N | ... | ... | ... | ... | ... | ... | ... | n | G |
| 37887 | 10421855-6433449 | -10.0 ± 0.2 | 4895 ± 188 | ... | 1.016 ± 0.004 | -0.10 ± 0.17 | <23 | 3 | N | ... | ... | ... | ... | ... | ... | ... | n | G |
| 53773 | 10435383-6417424 | -150.1 ± 26.1 | ... | ... | ... | ... | ... | ... | ... | ... | ... | ... | ... | ... | ... | ... | n | ... |
| 37624 | 10352169-6420320 | -13.6 ± 0.2 | 5073 ± 114 | ... | 1.019 ± 0.004 | -0.06 ± 0.10 | 6 ± 2 | 1 | N | ... | ... | ... | ... | ... | ... | ... | n | G |
| 37758 | 10401639-6448069 | 12.3 ± 0.2 | 4728 ± 1 | 2.56 ± 0.08 | 1.019 ± 0.002 | 0.02 ± 0.05 | <40 | 3 | N | N | ... | ... | ... | ... | ... | ... | n | G |
| 37678 | 10374260-6406333 | 5.6 ± 0.2 | 5116 ± 41 | 3.51 ± 0.20 | 1.004 ± 0.001 | -0.13 ± 0.13 | <28 | 3 | Y | Y | Y | N | N | Y | ... | ... | n | NG? |
| 37888 | 10421947-6427439 | 16.7 ± 0.2 | 4785 ± 188 | 2.71 ± 0.06 | 1.007 ± 0.002 | -0.07 ± 0.16 | <24 | 3 | Y | Y | Y | N | N | Y | ... | ... | n | NG? |
| 37679 | 10374562-6339097 | 41.6 ± 0.2 | 4736 ± 52 | 2.79 ± 0.15 | 1.006 ± 0.002 | 0.09 ± 0.03 | <50 | 3 | Y | N | Y | N | N | Y | ... | ... | n | NG? |
| 37680 | 10374817-6407187 | -6.8 ± 0.2 | 4753 ± 311 | ... | 1.023 ± 0.003 | -0.19 ± 0.21 | <11 | 3 | N | ... | ... | ... | ... | ... | ... | ... | n | G |
| 53774 | 10435390-6418176 | -159.1 ± 3.6 | ... | ... | ... | ... | ... | ... | ... | ... | ... | ... | ... | ... | ... | ... | n | ... |
| 37379 | 10281195-6445267 | 18.7 ± 0.2 | 4748 ± 85 | 2.54 ± 0.18 | 1.023 ± 0.004 | -0.02 ± 0.12 | ... | ... | N | N | ... | ... | ... | ... | ... | ... | n | G |
| 37681 | 10374942-6400505 | 17.7 ± 0.2 | 3893 ± 165 | ... | 0.836 ± 0.004 | -0.15 ± 0.11 | ... | ... | Y | ... | Y | ... | N | Y | Y | Y | n | ... |
| 37682 | 10375564-6356153 | -5.4 ± 0.2 | 4455 ± 173 | ... | 1.031 ± 0.004 | -0.01 ± 0.06 | <39 | 3 | N | ... | ... | ... | ... | ... | ... | ... | n | G |
| 53714 | 10421989-6407049 | 11.0 ± 0.2 | 4512 ± 166 | ... | 1.024 ± 0.003 | 0.14 ± 0.03 | <60 | 3 | N | ... | ... | ... | ... | ... | ... | ... | n | G |
| 38005 | 10435417-6450481 | -23.5 ± 0.2 | 5124 ± 98 | ... | 1.013 ± 0.003 | ... | ... | ... | N | ... | ... | ... | ... | ... | ... | ... | n | ... |
| 37683 | 10375777-6341148 | 25.6 ± 0.2 | 4492 ± 208 | ... | 1.023 ± 0.004 | 0.10 ± 0.08 | 174 ± 5 | 1 | N | ... | ... | ... | ... | ... | ... | ... | n | Li-rich G |
| 37889 | 10422009-6424368 | 16.3 ± 0.2 | 4589 ± 157 | ... | 1.016 ± 0.004 | 0.08 ± 0.03 | <49 | 3 | N | ... | ... | ... | ... | ... | ... | ... | n | G |
| 37380 | 10281301-6444155 | 42.8 ± 0.3 | 4861 ± 118 | ... | 1.015 ± 0.005 | -0.11 ± 0.04 | 16 ± 3 | 1 | N | ... | ... | ... | ... | ... | ... | ... | n | G |
| 53577 | 10402274-6404405 | 0.9 ± 0.2 | 4559 ± 33 | 2.21 ± 0.18 | 1.030 ± 0.004 | -0.12 ± 0.08 | <38 | 3 | N | N | ... | ... | ... | ... | ... | ... | n | G |
| 37684 | 10375821-6329471 | -17.1 ± 0.2 | 4718 ± 55 | 2.52 ± 0.01 | 1.024 ± 0.002 | 0.01 ± 0.07 | <40 | 3 | N | N | ... | ... | ... | ... | ... | ... | n | G |
| 37890 | 10422034-6545258 | 29.3 ± 0.2 | 4549 ± 148 | 2.53 ± 0.11 | 1.007 ± 0.007 | 0.08 ± 0.08 | <56 | 3 | Y | N | Y | N | N | Y | ... | ... | n | NG? |
| 37381 | 10281309-6450563 | 6.9 ± 0.2 | 4655 ± 30 | 2.46 ± 0.10 | 1.023 ± 0.003 | 0.01 ± 0.01 | <36 | 3 | N | N | ... | ... | ... | ... | ... | ... | n | G |
| 53578 | 10402306-6413091 | 23.0 ± 0.2 | 6155 ± 87 | 4.07 ± 0.02 | 1.000 ± 0.002 | -0.01 ± 0.05 | 9 ± 3 | 1 | Y | Y | Y | N | N | Y | ... | ... | n | ... |
| 53775 | 10435437-6427327 | 4.7 ± 0.2 | 6079 ± 144 | 4.12 ± 0.17 | 0.996 ± 0.001 | -0.06 ± 0.08 | 58 ± 2 | 1 | Y | Y | Y | N | N | Y | ... | ... | n | NG |
| 37685 | 10375853-6331455 | 28.9 ± 0.2 | 4536 ± 123 | 2.20 ± 0.19 | 1.026 ± 0.003 | -0.07 ± 0.07 | <37 | 3 | N | N | ... | ... | ... | ... | ... | ... | n | G |
| 37891 | 10422107-6438578 | -31.5 ± 0.2 | 4739 ± 12 | 2.55 ± 0.12 | 1.019 ± 0.005 | -0.10 ± 0.09 | <24 | 3 | N | N | ... | ... | ... | ... | ... | ... | n | G |
| 37382 | 10281374-6446279 | -20.5 ± 0.2 | 5126 ± 98 | ... | 1.002 ± 0.004 | ... | ... | ... | Y | ... | N | ... | ... | ... | ... | ... | n | ... |
| 37686 | 10380704-6403515 | 31.5 ± 0.2 | 4892 ± 111 | ... | 1.018 ± 0.002 | -0.06 ± 0.17 | <35 | 3 | N | ... | ... | ... | ... | ... | ... | ... | n | G |
| 53715 | 10422121-6419595 | 133.6 ± 0.2 | 3476 ± 88 | ... | ... | ... | ... | ... | ... | ... | ... | ... | ... | ... | ... | ... | n | ... |
| 37626 | 10352866-6406108 | 49.3 ± 1.2 | ... | ... | ... | ... | ... | ... | ... | ... | ... | ... | ... | ... | ... | ... | n | ... |
| 53579 | 10402367-6407103 | 32.5 ± 0.2 | 4102 ± 168 | ... | 1.056 ± 0.002 | -0.20 ± 0.05 | 107 ± 2 | 1 | N | ... | ... | ... | ... | ... | ... | ... | n | G |
| 37687 | 10380869-6345393 | -5.9 ± 0.2 | 4898 ± 215 | ... | 1.019 ± 0.002 | -0.05 ± 0.08 | 298 ± 2 | 1 | N | ... | ... | ... | ... | ... | ... | ... | n | Li-rich G |
| 37892 | 10422131-6542388 | -17.2 ± 0.3 | 5038 ± 139 | ... | 1.010 ± 0.006 | -0.13 ± 0.12 | <23 | 3 | Y | ... | N | N | N | Y | ... | ... | n | NG? |
| 37383 | 10281487-6447302 | 38.3 ± 0.2 | 4615 ± 121 | 2.52 ± 0.19 | 1.014 ± 0.005 | 0.07 ± 0.06 | <54 | 3 | N | N | ... | ... | ... | ... | ... | ... | n | G |
| 37688 | 10381177-6344558 | 1.9 ± 0.2 | 4918 ± 173 | ... | 1.015 ± 0.002 | 0.00 ± 0.04 | 73 ± 11 | 1 | N | ... | ... | ... | ... | ... | ... | ... | n | Li-rich G |
| 37893 | 10422212-6452416 | -2.3 ± 0.2 | 4800 ± 134 | ... | 1.026 ± 0.003 | -0.12 ± 0.20 | <19 | 3 | N | ... | ... | ... | ... | ... | ... | ... | n | G |
| 37689 | 10381217-6333375 | -15.5 ± 0.2 | 5013 ± 100 | ... | 1.016 ± 0.002 | -0.09 ± 0.10 | <36 | 3 | N | ... | ... | ... | ... | ... | ... | ... | n | G |
| 53817 | 10442881-6426132 | 19.9 ± 0.2 | 4674 ± 46 | 2.78 ± 0.04 | 1.004 ± 0.001 | 0.09 ± 0.03 | <56 | 3 | Y | N | Y | N | N | Y | ... | ... | n | NG? |
| 37762 | 10402434-6447339 | -24.5 ± 0.2 | 4729 ± 30 | 2.49 ± 0.10 | 1.023 ± 0.002 | -0.04 ± 0.07 | <35 | 3 | N | N | ... | ... | ... | ... | ... | ... | n | G |
| 37690 | 10381276-6406144 | 1.0 ± 0.2 | 4706 ± 38 | 2.56 ± 0.09 | 1.017 ± 0.002 | 0.04 ± 0.05 | <43 | 3 | N | N | ... | ... | ... | ... | ... | ... | n | G |
| 37894 | 10422229-6456051 | -4.8 ± 0.2 | 4892 ± 15 | 2.69 ± 0.16 | 1.026 ± 0.002 | -0.06 ± 0.09 | <16 | 3 | N | N | ... | ... | ... | ... | ... | ... | n | G |
| 53580 | 10402474-6414560 | -29.5 ± 0.2 | 5167 ± 42 | ... | 1.012 ± 0.004 | -0.15 ± 0.09 | 25 ± 4 | 1 | N | ... | ... | ... | ... | ... | ... | ... | n | G |
| 37691 | 10381925-6335521 | -36.1 ± 0.2 | 4943 ± 138 | ... | 1.024 ± 0.002 | -0.05 ± 0.08 | <25 | 3 | N | ... | ... | ... | ... | ... | ... | ... | n | G |
| 9 | 10423830-6408320 | 6.4 ± 0.4 | 6305 ± 127 | 3.97 ± 0.15 | ... | 0.01 ± 0.12 | <101 | 3 | ... | Y | Y | Y | Y | Y | N | ... | n | NG |
| 53581 | 10402480-6406013 | 119.8 ± 0.2 | 3857 ± 177 | ... | 1.058 ± 0.007 | -0.20 ± 0.14 | <21 | 3 | N | ... | ... | ... | ... | ... | ... | ... | n | G |
| 37384 | 10281668-6448565 | 24.0 ± 0.2 | 5144 ± 106 | ... | 1.014 ± 0.005 | ... | <27 | 3 | N | ... | ... | ... | ... | ... | ... | ... | n | G |
| 37692 | 10382031-6339391 | -15.1 ± 0.2 | 4987 ± 22 | 2.82 ± 0.02 | 1.017 ± 0.002 | 0.02 ± 0.07 | <28 | 3 | N | N | ... | ... | ... | ... | ... | ... | n | G |
| 37920 | 10423929-6528025 | -5.9 ± 0.2 | 4795 ± 231 | ... | 1.004 ± 0.004 | -0.13 ± 0.23 | ... | ... | N | ... | ... | ... | ... | ... | ... | ... | n | G |
| 37693 | 10382077-6334581 | 2.0 ± 0.2 | 4862 ± 57 | 2.71 ± 0.14 | 1.020 ± 0.002 | -0.03 ± 0.07 | <33 | 3 | N | N | ... | ... | ... | ... | ... | ... | n | G |
| 37694 | 10382127-6338497 | -2.5 ± 0.2 | 4652 ± 14 | 2.45 ± 0.06 | 1.021 ± 0.003 | 0.01 ± 0.05 | <33 | 3 | N | N | ... | ... | ... | ... | ... | ... | n | G |
| 53582 | 10402600-6417246 | 50.3 ± 0.2 | 4823 ± 30 | 2.68 ± 0.10 | 1.013 ± 0.001 | 0.00 ± 0.04 | <35 | 3 | N | N | ... | ... | ... | ... | ... | ... | n | G |
| 37695 | 10382549-6331391 | 11.0 ± 0.2 | 4573 ± 91 | 2.32 ± 0.15 | 1.028 ± 0.003 | 0.01 ± 0.03 | <41 | 3 | N | N | ... | ... | ... | ... | ... | ... | n | G |
| 37921 | 10424008-6433060 | 20.8 ± 0.2 | 4976 ± 171 | ... | 1.021 ± 0.003 | -0.08 ± 0.16 | <21 | 3 | N | ... | ... | ... | ... | ... | ... | ... | n | G |

| ID | CNAME | RV (km s$^{-1}$) | $T_{\text{eff}}$ (K) | logg (dex) | $\gamma^a$ | [Fe/H] (dex) | EW(Li)$^b$ (mÅ) | EW(Li) error flag$^c$ | $\gamma$ | logg | Membership RV | Li | H$\alpha$ | [Fe/H] | Gaia studies Randich$^d$ | Cantat-Gaudin$^d$ | Final$^e$ | NMs with Li$^f$ |
|---|---|---|---|---|---|---|---|---|---|---|---|---|---|---|---|---|---|---|
| 53776 | 10435537-6432000 | 18.6 ± 0.2 | 5901 ± 54 | 4.17 ± 0.12 | 0.996 ± 0.004 | 0.13 ± 0.02 | 36 ± 5 | 1 | Y | Y | Y | N | N | Y | … | … | n | NG |
| 53870 | 10450676-6436015 | 3.5 ± 0.5 | 6160 ± 104 | … | 1.015 ± 0.005 | -0.59 ± 0.10 | <23 | 3 | N | … | … | … | … | … | … | … | n | … |
| 53583 | 10402621-6415023 | 10.0 ± 0.2 | 5015 ± 170 | … | 1.019 ± 0.003 | -0.17 ± 0.22 | <23 | 3 | N | … | … | … | … | … | … | … | n | G |
| 37394 | 10283341-6446039 | 182.9 ± 0.3 | 4937 ± 243 | … | 1.026 ± 0.006 | -0.17 ± 0.25 | <12 | 3 | N | … | … | … | … | … | … | … | n | G |
| 37696 | 10382635-6336553 | -19.2 ± 0.2 | 5045 ± 12 | 3.00 ± 0.16 | 1.021 ± 0.002 | 0.00 ± 0.06 | <21 | 3 | N | N | … | … | … | … | … | … | n | G |
| 37922 | 10424076-6533565 | -8.6 ± 0.2 | 4962 ± 30 | … | 1.029 ± 0.005 | -0.04 ± 0.07 | <30 | 3 | N | … | … | … | … | … | … | … | n | G |
| 37697 | 10382792-6352085 | -14.3 ± 0.2 | 4992 ± 123 | … | 1.019 ± 0.004 | -0.03 ± 0.05 | <27 | 3 | N | … | … | … | … | … | … | … | n | G |
| 53779 | 10435783-6435544 | -9.2 ± 0.2 | 5862 ± 2 | 3.78 ± 0.20 | 1.004 ± 0.003 | -0.43 ± 0.24 | <27 | 3 | Y | Y | Y | N | N | N | … | … | n | NG? |
| 53871 | 10450677-6436324 | 44.9 ± 0.2 | 4721 ± 282 | … | 1.036 ± 0.002 | -0.14 ± 0.20 | <27 | 3 | N | … | … | … | … | … | … | … | n | G |
| 22 | 10435795-6416421 | -4.3 ± 0.6 | 6208 ± 144 | 3.88 ± 0.12 | … | -0.34 ± 0.12 | <50 | 3 | … | Y | Y | N | Y | Y | N | … | n | NG |
| 53584 | 10402787-6404025 | 5.2 ± 0.2 | 4531 ± 105 | … | 1.049 ± 0.001 | … | <26 | 3 | N | … | … | … | … | … | … | … | n | G |
| 37716 | 10390844-6340020 | 11.9 ± 0.2 | 4924 ± 134 | … | 1.018 ± 0.001 | -0.05 ± 0.09 | <23 | 3 | N | … | … | … | … | … | … | … | n | G |
| 2553 | 10424167-6543415 | -5.3 ± 0.6 | 5044 ± 21 | 3.23 ± 0.07 | 1.010 ± 0.002 | -0.08 ± 0.01 | <13 | 3 | N | N | … | … | … | … | … | … | n | G |
| 25 | 10442965-6408582 | 7.1 ± 0.6 | 4799 ± 11 | 2.50 ± 0.03 | … | 0.06 ± 0.04 | <20 | 3 | … | N | … | … | … | … | … | … | n | … |
| 10 | 10424171-6421047 | 15.1 ± 0.6 | 5814 ± 73 | 4.35 ± 0.14 | … | 0.01 ± 0.12 | <183 | 3 | … | Y | Y | Y | Y | Y | N | … | n | … |
| 53780 | 10435826-6433251 | 19.4 ± 0.2 | 6201 ± 53 | 4.33 ± 0.13 | 0.993 ± 0.003 | -0.03 ± 0.04 | 62 ± 2 | 1 | Y | Y | Y | N | N | Y | … | … | n | NG |
| 38120 | 10454944-6500491 | 91.2 ± 0.2 | 3580 ± 59 | … | 0.815 ± 0.005 | -0.23 ± 0.14 | … | … | Y | … | N | … | … | Y | … | … | n | … |
| 53585 | 10402835-6422132 | -1.1 ± 0.2 | 6309 ± 53 | 3.81 ± 0.09 | 1.009 ± 0.001 | -0.23 ± 0.18 | <1 | 3 | Y | Y | Y | N | N | Y | … | … | n | … |
| 38503 | 10505504-6334099 | 9.9 ± 0.3 | 4712 ± 102 | … | 1.031 ± 0.009 | … | <53 | 3 | N | … | … | … | … | … | … | … | n | G |
| 38008 | 10435833-6356349 | -48.2 ± 323.3 | … | … | … | … | … | … | … | … | … | … | … | … | … | … | n | … |
| 38504 | 10505504-6348446 | 13.6 ± 0.2 | 4696 ± 44 | 2.55 ± 0.09 | 1.015 ± 0.003 | 0.04 ± 0.05 | <37 | 3 | N | N | … | … | … | … | … | … | n | G |
| 37923 | 10424310-6348284 | -16.0 ± 0.2 | 4529 ± 149 | … | 1.028 ± 0.003 | 0.10 ± 0.02 | <49 | 3 | N | … | … | … | … | … | … | … | n | G |
| 53781 | 10435841-6421335 | -22.4 ± 0.3 | 5761 ± 85 | … | 0.999 ± 0.006 | … | 39 ± 8 | 1 | Y | … | N | N | N | … | … | … | n | NG |
| 38505 | 10505543-6350121 | 47.7 ± 0.3 | 4815 ± 100 | … | 1.038 ± 0.006 | -0.20 ± 0.13 | <18 | 3 | N | … | … | … | … | … | … | … | n | G |
| 53922 | 10454992-6426249 | -45.2 ± 0.3 | 5763 ± 58 | 4.18 ± 0.01 | 0.991 ± 0.005 | -0.51 ± 0.17 | … | … | Y | Y | N | … | … | N | … | … | n | … |
| 37924 | 10424389-6529195 | 7.4 ± 0.2 | 4574 ± 52 | 2.38 ± 0.13 | 1.021 ± 0.003 | 0.00 ± 0.04 | <47 | 3 | N | N | … | … | … | … | … | … | n | G |
| 2603 | 10505597-6427028 | -10.5 ± 0.6 | 4876 ± 94 | 1.92 ± 0.12 | 1.024 ± 0.002 | -0.38 ± 0.03 | 29 ± 6 | 1 | N | N | … | … | … | … | … | … | n | … |
| 53872 | 10450766-6434484 | -6.6 ± 0.4 | 6097 ± 69 | … | 1.014 ± 0.004 | -0.47 ± 0.06 | … | … | N | … | … | … | … | … | … | … | n | … |
| 53595 | 10404020-6408346 | -0.5 ± 0.2 | 4684 ± 48 | … | 1.030 ± 0.003 | -0.15 ± 0.18 | <28 | 3 | N | … | … | … | … | … | … | … | n | G |
| 38506 | 10505635-6511137 | 0.6 ± 0.2 | 5184 ± 40 | … | 1.015 ± 0.002 | -0.05 ± 0.07 | <36 | 3 | N | … | … | … | … | … | … | … | n | G |
| 37925 | 10424474-6534166 | -22.9 ± 0.2 | 5027 ± 183 | … | 1.024 ± 0.005 | -0.09 ± 0.14 | <16 | 3 | N | … | … | … | … | … | … | … | n | G |
| 38071 | 10451011-6453364 | 13.7 ± 0.2 | 4515 ± 153 | … | 1.022 ± 0.004 | 0.18 ± 0.01 | <70 | 3 | N | … | … | … | … | … | … | … | n | G |
| 37774 | 10404051-6536011 | -5.7 ± 0.2 | 5109 ± 47 | … | 1.016 ± 0.004 | -0.13 ± 0.16 | <6 | 3 | N | … | … | … | … | … | … | … | n | G |
| 38507 | 10505688-6338597 | -18.3 ± 0.2 | 4571 ± 49 | 2.37 ± 0.12 | 1.019 ± 0.004 | -0.03 ± 0.02 | <25 | 3 | N | N | … | … | … | … | … | … | n | G |
| 37926 | 10424493-6535362 | -12.0 ± 0.3 | 5012 ± 184 | … | 1.013 ± 0.006 | -0.16 ± 0.21 | <27 | 3 | N | … | … | … | … | … | … | … | n | G |
| 38050 | 10443397-6455554 | 17.5 ± 0.2 | 5010 ± 125 | … | 1.008 ± 0.003 | -0.05 ± 0.11 | <27 | 3 | Y | … | Y | N | N | Y | … | … | n | NG? |
| 2604 | 10505858-6343000 | -2.0 ± 0.6 | 4568 ± 3 | 1.91 ± 0.10 | 1.034 ± 0.001 | -0.25 ± 0.04 | 21 ± 3 | 1 | N | N | … | … | … | … | … | … | n | G |
| 37927 | 10424587-6431014 | 28.8 ± 0.2 | 4376 ± 231 | … | 1.024 ± 0.005 | 0.12 ± 0.02 | <53 | 3 | N | … | … | … | … | … | … | … | n | G |
| 53782 | 10435885-6432066 | 63.6 ± 0.2 | 5082 ± 11 | … | 1.008 ± 0.005 | -0.35 ± 0.09 | <12 | 3 | Y | … | N | N | N | N | … | … | n | NG? |
| 38508 | 10505889-6350123 | 10.2 ± 0.2 | 4677 ± 68 | 2.49 ± 0.15 | 1.021 ± 0.006 | 0.03 ± 0.08 | <32 | 3 | N | N | … | … | … | … | … | … | n | G |
| 53880 | 10451081-6436592 | 31.5 ± 0.2 | 5606 ± 19 | 4.23 ± 0.18 | 0.990 ± 0.005 | 0.04 ± 0.11 | <28 | 3 | Y | Y | Y | N | N | Y | … | … | n | NG |
| 53783 | 10435901-6425574 | -366.5 ± 19.7 | … | … | … | … | … | … | … | … | … | … | … | … | … | Y | n | … |
| 38509 | 10505973-6340510 | 86.2 ± 0.2 | 4638 ± 131 | 2.42 ± 0.10 | 1.011 ± 0.005 | -0.07 ± 0.09 | <38 | 3 | N | N | … | … | … | … | … | … | n | G |
| 53923 | 10455080-6421352 | 16.8 ± 0.2 | 4663 ± 149 | … | 1.023 ± 0.006 | -0.04 ± 0.19 | <41 | 3 | N | … | … | … | … | … | … | … | n | G |
| 38510 | 10505995-6510405 | 0.6 ± 0.2 | 4749 ± 34 | 2.54 ± 0.06 | 1.018 ± 0.002 | -0.03 ± 0.08 | <28 | 3 | N | N | … | … | … | … | … | … | n | G |
| 37775 | 10404225-6428459 | 14.7 ± 0.2 | 4685 ± 89 | … | 1.019 ± 0.002 | -0.10 ± 0.18 | <15 | 3 | N | … | … | … | … | … | … | … | n | G |
| 38511 | 10510036-6335181 | -3.9 ± 0.2 | 4873 ± 27 | 2.72 ± 0.11 | 1.020 ± 0.002 | 0.05 ± 0.10 | 20 ± 11 | 1 | N | N | … | … | … | … | … | … | n | G |
| 2554 | 10424692-6353186 | 8.9 ± 0.6 | 5169 ± 17 | 2.88 ± 0.01 | 1.013 ± 0.002 | 0.12 ± 0.03 | 25 ± 3 | 1 | N | N | … | … | … | … | … | … | n | G |
| 53822 | 10443442-6433261 | 126.9 ± 0.2 | 3885 ± 173 | … | 1.052 ± 0.003 | -0.16 ± 0.05 | <15 | 3 | N | … | … | … | … | … | … | … | n | G |
| 38512 | 10510060-6332271 | -11.4 ± 0.3 | 4907 ± 256 | … | 1.014 ± 0.007 | -0.09 ± 0.11 | <30 | 3 | N | … | … | … | … | … | … | … | n | G |
| 53924 | 10455122-6421031 | -28.3 ± 0.2 | 3926 ± 110 | … | 1.052 ± 0.002 | -0.14 ± 0.09 | 72 ± 14 | 1 | N | … | … | … | … | … | … | … | n | G |
| 53597 | 10404304-6405364 | 43.9 ± 0.2 | 4796 ± 30 | … | 0.994 ± 0.002 | 0.07 ± 0.07 | <46 | 3 | Y | … | N | N | N | Y | … | … | n | NG |
| 38513 | 10510140-6345165 | 16.5 ± 0.2 | 4625 ± 102 | … | 1.002 ± 0.006 | … | <68 | 3 | Y | … | Y | N | N | … | … | … | n | NG? |
| 37938 | 10425608-6355510 | 18.1 ± 0.2 | 5135 ± 41 | … | 0.985 ± 0.001 | -0.02 ± 0.05 | 280 ± 10 | 1 | Y | … | Y | Y | Y | Y | Y | Y | Y | … |
| 38072 | 10451116-6425105 | -29.5 ± 0.2 | 4849 ± 182 | … | 1.020 ± 0.002 | -0.06 ± 0.11 | 26 ± 2 | 1 | N | … | … | … | … | … | … | … | n | G |
| 53598 | 10404309-6411465 | -8.6 ± 0.5 | 6155 ± 89 | … | 1.018 ± 0.005 | -0.70 ± 0.07 | <19 | 3 | N | … | … | … | … | … | … | … | n | … |
| 38514 | 10510188-6428340 | -1.4 ± 0.2 | 4628 ± 83 | 2.50 ± 0.20 | 1.024 ± 0.002 | 0.04 ± 0.02 | <36 | 3 | N | N | … | … | … | … | … | … | n | G |
| 37939 | 10425636-6535225 | 9.7 ± 0.2 | 4694 ± 75 | 2.55 ± 0.16 | 1.017 ± 0.003 | -0.06 ± 0.04 | <33 | 3 | N | N | … | … | … | … | … | … | n | G |
| 38121 | 10455198-6449060 | 15.1 ± 0.2 | 4544 ± 118 | 2.28 ± 0.19 | 1.019 ± 0.004 | -0.08 ± 0.09 | <36 | 3 | N | N | … | … | … | … | … | … | n | G |

**Table C.6.** continued.

| ID | CNAME | RV (km s$^{-1}$) | $T_{\rm eff}$ (K) | $logg$ (dex) | $\gamma^a$ | [Fe/H] (dex) | EW(Li)$^b$ (mÅ) | EW(Li) error flag$^c$ | $\gamma$ | $logg$ | Membership RV | Li | H$\alpha$ | [Fe/H] | Gaia studies Randich$^d$ | Cantat-Gaudin$^d$ | Final$^e$ | NMs with Li$^f$ |
|---|---|---|---|---|---|---|---|---|---|---|---|---|---|---|---|---|---|---|
| 53881 | 10451117-6421001 | 41.6 ± 0.2 | 4565 ± 184 | … | 1.026 ± 0.004 | 0.07 ± 0.04 | 54 ± 4 | 1 | N | … | … | … | … | … | … | … | n | G |
| 38515 | 10510393-6348387 | -14.3 ± 0.2 | 5045 ± 15 | … | 1.020 ± 0.004 | -0.02 ± 0.08 | <24 | 3 | N | … | … | … | … | … | … | … | n | G |
| 37940 | 10425650-6359314 | 6.0 ± 0.2 | 4549 ± 108 | 2.21 ± 0.20 | 1.040 ± 0.003 | -0.04 ± 0.03 | <31 | 3 | N | N | … | … | … | … | … | … | n | G |
| 37776 | 10404352-6456157 | 22.2 ± 0.2 | 4592 ± 88 | 2.34 ± 0.11 | 1.013 ± 0.003 | -0.09 ± 0.07 | <41 | 3 | N | N | … | … | … | … | … | … | n | … |
| 38516 | 10510430-6415129 | 0.4 ± 0.2 | 4843 ± 217 | … | 1.024 ± 0.003 | -0.07 ± 0.13 | … | … | N | … | … | … | … | … | … | … | n | G |
| 38517 | 10510539-6330288 | 37.8 ± 0.2 | 4644 ± 73 | 2.46 ± 0.18 | 1.025 ± 0.003 | 0.03 ± 0.08 | <43 | 3 | N | N | … | … | … | … | … | … | n | G |
| 37777 | 10404411-6542536 | 41.4 ± 0.3 | 4717 ± 93 | 2.69 ± 0.03 | 1.007 ± 0.008 | -0.08 ± 0.02 | <37 | 3 | Y | N | Y | N | N | Y | … | … | n | NG? |
| 53925 | 10455212-6433556 | -16.9 ± 0.5 | 6862 ± 139 | … | 0.991 ± 0.007 | 0.06 ± 0.12 | 101 ± 8 | 1 | Y | … | N | Y | N | Y | … | … | n | NG |
| 53823 | 10443489-6416478 | 0.3 ± 0.2 | 5580 ± 98 | 4.17 ± 0.11 | 0.988 ± 0.003 | -0.09 ± 0.03 | <22 | 3 | Y | Y | Y | N | N | Y | … | … | n | NG |
| 38564 | 10513748-6335141 | 3.8 ± 0.2 | 4545 ± 133 | … | 1.016 ± 0.003 | 0.08 ± 0.08 | <53 | 3 | N | … | … | … | … | … | … | … | n | G |
| 37941 | 10425703-6529136 | 7.7 ± 0.3 | 4964 ± 87 | … | 1.025 ± 0.006 | -0.09 ± 0.14 | 30 ± 6 | 1 | N | … | … | … | … | … | … | … | n | G |
| 38142 | 10461329-6421136 | 65.8 ± 0.2 | 4851 ± 114 | … | 1.018 ± 0.002 | -0.17 ± 0.13 | <33 | 3 | N | … | … | … | … | … | … | … | n | G |
| 53824 | 10443494-6432145 | 34.3 ± 1.5 | 3711 ± 18 | … | 0.890 ± 0.018 | … | … | … | Y | … | Y | … | … | … | … | Y | n | … |
| 38073 | 10451139-6348201 | -13.4 ± 0.2 | 4802 ± 221 | … | 1.024 ± 0.002 | -0.10 ± 0.17 | <28 | 3 | N | … | … | … | … | … | … | … | n | G |
| 38565 | 10513826-6336407 | 40.3 ± 0.2 | 5064 ± 221 | … | 1.017 ± 0.003 | -0.32 ± 0.26 | <27 | 3 | N | … | … | … | … | … | … | … | n | G |
| 38566 | 10513847-6335341 | 0.9 ± 0.2 | 4473 ± 216 | … | 1.023 ± 0.005 | 0.20 ± 0.04 | 313 ± 3 | 1 | N | … | … | … | … | … | … | … | n | Li-rich G |
| 13 | 10425723-6402265 | 33.3 ± 0.6 | 4702 ± 3 | 2.74 ± 0.09 | 1.018 ± 0.003 | 0.10 ± 0.01 | 31 ± 2 | 1 | N | N | … | … | … | … | … | … | n | G |
| 38143 | 10461410-6336286 | 1.9 ± 0.2 | 4651 ± 7 | 2.38 ± 0.11 | 1.024 ± 0.003 | -0.10 ± 0.06 | <25 | 3 | N | N | … | … | … | … | … | … | n | G |
| 53882 | 10451144-6427481 | 100.5 ± 0.2 | 3875 ± 84 | … | 1.057 ± 0.004 | -0.11 ± 0.16 | <14 | 3 | N | … | … | … | … | … | … | … | n | G |
| 38567 | 10513866-6339195 | 18.0 ± 0.2 | 4920 ± 113 | … | 1.022 ± 0.004 | -0.07 ± 0.11 | <17 | 3 | N | … | … | … | … | … | … | … | n | G |
| 37942 | 10425830-6546276 | 75.1 ± 0.3 | 4726 ± 65 | 2.49 ± 0.16 | 1.016 ± 0.006 | -0.16 ± 0.13 | … | … | N | N | … | … | … | … | … | … | n | G |
| 53599 | 10404503-6412059 | 39.8 ± 0.2 | 4722 ± 116 | 2.52 ± 0.20 | 1.020 ± 0.004 | -0.03 ± 0.15 | <27 | 3 | N | N | … | … | … | … | … | … | n | G |
| 38144 | 10461429-6325086 | -13.1 ± 0.2 | 5011 ± 182 | … | 1.018 ± 0.002 | -0.12 ± 0.13 | <20 | 3 | N | … | … | … | … | … | … | … | n | G |
| 38568 | 10513911-6342161 | 17.2 ± 0.2 | 4699 ± 99 | … | 1.015 ± 0.005 | … | <45 | 3 | N | … | … | … | … | … | … | … | n | G |
| 38569 | 10514029-6420243 | -14.9 ± 0.2 | 4729 ± 10 | 2.57 ± 0.09 | 1.018 ± 0.003 | -0.04 ± 0.08 | 56 ± 5 | 1 | N | N | … | … | … | … | … | … | n | G |
| 37778 | 10404512-6541296 | 21.0 ± 0.2 | 4681 ± 32 | 2.55 ± 0.09 | 1.012 ± 0.004 | 0.03 ± 0.09 | <34 | 3 | N | N | … | … | … | … | … | … | n | G |
| 37943 | 10430024-6530179 | 16.7 ± 0.2 | 5534 ± 91 | … | 0.992 ± 0.001 | 0.03 ± 0.04 | 188 ± 3 | 1 | Y | … | Y | Y | Y | Y | Y | Y | Y | … |
| 30 | 10461480-6402570 | 19.1 ± 0.6 | 5796 ± 68 | 4.38 ± 0.13 | … | 0.07 ± 0.11 | 221 ± 43 | 2 | … | Y | Y | Y | Y | Y | Y | … | Y | … |
| 38570 | 10514198-6348266 | 37.6 ± 0.2 | 4789 ± 84 | 2.61 ± 0.12 | 1.018 ± 0.002 | -0.10 ± 0.08 | <36 | 3 | N | N | … | … | … | … | … | … | n | G |
| 53726 | 10430032-6425520 | 38.1 ± 0.2 | 6302 ± 77 | 4.33 ± 0.17 | 0.995 ± 0.004 | -0.02 ± 0.06 | 72 ± 6 | 1 | Y | Y | Y | N | N | Y | … | … | n | NG |
| 31 | 10461503-6408551 | -22.1 ± 0.6 | 4946 ± 2 | 2.76 ± 0.02 | … | -0.01 ± 0.02 | < | 3 | … | N | … | … | … | … | … | … | n | … |
| 53600 | 10404598-6416265 | 36.8 ± 0.3 | 6494 ± 79 | … | 0.988 ± 0.004 | 0.01 ± 0.06 | … | … | Y | … | Y | … | … | Y | … | … | n | … |
| 38571 | 10514208-6347281 | 95.6 ± 0.2 | 4754 ± 79 | 2.63 ± 0.10 | 1.016 ± 0.006 | -0.13 ± 0.06 | <10 | 3 | N | N | … | … | … | … | … | … | n | G |
| 38145 | 10461508-6318406 | -14.6 ± 0.2 | 4954 ± 186 | … | 1.027 ± 0.003 | -0.08 ± 0.09 | <26 | 3 | N | … | … | … | … | … | … | … | n | G |
| 37779 | 10404668-6358074 | -25.3 ± 0.2 | 4751 ± 3 | 2.52 ± 0.07 | 1.020 ± 0.002 | -0.05 ± 0.05 | <32 | 3 | N | N | … | … | … | … | … | … | n | G |
| 53727 | 10430064-6427268 | 1.3 ± 0.2 | 4955 ± 143 | … | 1.013 ± 0.005 | -0.10 ± 0.15 | 19 ± 4 | 1 | N | … | … | … | … | … | … | … | n | G |
| 32 | 10461515-6414366 | 14.0 ± 0.6 | 4925 ± 91 | 3.31 ± 0.21 | … | 0.10 ± 0.01 | <24 | 3 | … | N | … | … | … | … | … | … | n | … |
| 37783 | 10405148-6456017 | -4.8 ± 0.2 | 4568 ± 56 | 2.34 ± 0.16 | 1.026 ± 0.002 | -0.03 ± 0.02 | <33 | 3 | N | N | … | … | … | … | … | … | n | G |
| 38572 | 10514321-6331580 | 8.9 ± 0.2 | 5143 ± 106 | … | 1.015 ± 0.004 | … | <23 | 3 | N | … | … | … | … | … | … | … | n | G |
| 37944 | 10430090-6533478 | 94.6 ± 0.2 | 4848 ± 49 | … | 1.027 ± 0.005 | -0.21 ± 0.12 | <36 | 3 | N | … | … | … | … | … | … | … | n | G |
| 38146 | 10461524-6426410 | -16.7 ± 0.2 | 4425 ± 176 | … | 1.042 ± 0.003 | -0.09 ± 0.11 | <32 | 3 | N | … | … | … | … | … | … | … | n | G |
| 37784 | 10405171-6538348 | 65.1 ± 0.3 | 4680 ± 144 | … | 1.020 ± 0.010 | -0.10 ± 0.03 | 59 ± 6 | 1 | N | … | … | … | … | … | … | … | n | G |
| 2606 | 10514339-6336283 | -39.9 ± 0.6 | 5019 ± 7 | 2.66 ± 0.07 | 1.024 ± 0.002 | -0.22 ± 0.05 | <16 | 3 | N | N | … | … | … | … | … | … | n | G |
| 38147 | 10461567-6320125 | -28.1 ± 0.2 | 5128 ± 115 | … | 1.017 ± 0.003 | … | 58 ± 3 | 1 | N | … | … | … | … | … | … | … | n | … |
| 53610 | 10405174-6412018 | 17.4 ± 0.2 | 5975 ± 33 | 3.96 ± 0.16 | 0.999 ± 0.004 | -0.22 ± 0.10 | 24 ± 2 | 1 | Y | Y | Y | N | N | Y | … | … | n | NG |
| 53787 | 10440608-6431436 | -9.9 ± 0.2 | 4728 ± 9 | 2.55 ± 0.11 | 1.020 ± 0.004 | -0.05 ± 0.02 | <35 | 3 | N | N | … | … | … | … | … | … | n | G |
| 38573 | 10514339-6349420 | 2.5 ± 0.2 | 4812 ± 113 | 2.68 ± 0.16 | 1.012 ± 0.003 | -0.07 ± 0.12 | … | … | N | N | … | … | … | … | … | … | n | G |
| 14 | 10430154-6414557 | 27.1 ± 0.6 | 4809 ± 6 | 2.52 ± 0.01 | … | 0.10 ± 0.01 | <49 | 3 | … | N | … | … | … | … | … | … | n | … |
| 53944 | 10461617-6429456 | 43.3 ± 0.3 | 5121 ± 85 | … | 1.010 ± 0.007 | -0.22 ± 0.20 | 21 ± 4 | 1 | Y | … | N | N | N | Y | … | … | n | NG? |
| 53611 | 10405220-6415283 | 20.8 ± 0.2 | 5053 ± 72 | 3.48 ± 0.18 | 0.996 ± 0.002 | -0.03 ± 0.06 | 28 ± 6 | 1 | Y | N | Y | N | N | Y | … | … | n | NG? |
| 38574 | 10514343-6331461 | 1.8 ± 0.2 | 5008 ± 175 | … | 1.013 ± 0.005 | -0.16 ± 0.20 | <14 | 3 | N | … | … | … | … | … | … | … | n | G |
| 15 | 10430175-6411523 | -15.9 ± 0.6 | 6828 ± 246 | 3.99 ± 0.15 | … | -0.04 ± 0.11 | <63 | 3 | … | Y | N | Y | N | Y | N | … | n | NG |
| 53945 | 10461630-6428110 | 17.7 ± 0.2 | 4698 ± 41 | 2.54 ± 0.12 | 1.019 ± 0.004 | 0.04 ± 0.07 | <41 | 3 | N | N | … | … | … | … | … | … | n | G |
| 53612 | 10405234-6424071 | -3.1 ± 7.1 | … | … | … | … | … | … | … | … | … | … | … | … | … | … | n | … |
| 38575 | 10514446-6409096 | 19.0 ± 0.2 | 4594 ± 151 | … | 1.027 ± 0.002 | 0.06 ± 0.06 | <43 | 3 | N | … | … | … | … | … | … | … | n | G |
| 38148 | 10461683-6416524 | 18.7 ± 0.2 | 4586 ± 107 | 2.40 ± 0.19 | 1.025 ± 0.002 | 0.05 ± 0.03 | 28 ± 12 | 1 | N | N | … | … | … | … | … | … | n | G |
| 53788 | 10440631-6428265 | 129.4 ± 1.7 | … | … | … | … | … | … | … | … | … | … | … | … | … | … | n | … |
| 38576 | 10514450-6330166 | -27.7 ± 0.2 | 4715 ± 89 | 2.49 ± 0.02 | 1.026 ± 0.003 | -0.03 ± 0.04 | <35 | 3 | N | N | … | … | … | … | … | … | n | G |
| 37945 | 10430184-6532198 | 42.0 ± 0.3 | 4891 ± 9 | 2.61 ± 0.09 | 1.023 ± 0.007 | -0.13 ± 0.08 | <40 | 3 | N | N | … | … | … | … | … | … | n | G |

| ID | CNAME | RV (km s$^{-1}$) | $T_{\text{eff}}$ (K) | $\log g$ (dex) | $\gamma^a$ | [Fe/H] (dex) | EW(Li)$^b$ (mÅ) | EW(Li) error flag$^c$ | $\gamma$ | $\log g$ | Membership RV | Li | H$\alpha$ | [Fe/H] | Gaia studies Randich$^d$ | Cantat-Gaudin$^d$ | Final$^e$ | NMs with Li$^f$ |
|---|---|---|---|---|---|---|---|---|---|---|---|---|---|---|---|---|---|---|
| 38149 | 10461798-6447476 | 30.9 ± 0.9 | 4334 ± 172 | . . . | 0.993 ± 0.022 | -0.18 ± 0.20 | . . . | . . . | Y | . . . | Y | . . . | . . . | Y | . . . | . . . | n | . . . |
| 37785 | 10405270-6539406 | -13.5 ± 0.2 | 4656 ± 46 | 2.43 ± 0.15 | 1.029 ± 0.004 | 0.01 ± 0.04 | <29 | 3 | N | N | . . . | . . . | . . . | . . . | . . . | . . . | n | G |
| 38577 | 10514497-6348407 | 73.9 ± 0.2 | 4616 ± 91 | . . . | 1.026 ± 0.007 | . . . | . . . | . . . | N | . . . | . . . | . . . | . . . | . . . | . . . | . . . | n | G |
| 37946 | 10430196-6526099 | -16.6 ± 0.2 | 5051 ± 141 | . . . | 1.018 ± 0.003 | -0.03 ± 0.05 | <27 | 3 | N | . . . | . . . | . . . | . . . | . . . | . . . | . . . | n | G |
| 38150 | 10461822-6321323 | -26.8 ± 0.2 | 4837 ± 65 | 2.64 ± 0.09 | 1.022 ± 0.002 | 0.01 ± 0.01 | <25 | 3 | N | N | . . . | . . . | . . . | . . . | . . . | . . . | n | G |
| 37786 | 10405294-6532100 | 52.5 ± 0.3 | 4656 ± 103 | 2.46 ± 0.15 | 1.016 ± 0.006 | -0.05 ± 0.03 | 51 ± 5 | 1 | N | N | . . . | . . . | . . . | . . . | . . . | . . . | n | G |
| 38578 | 10514585-6417337 | 64.2 ± 0.2 | 4546 ± 137 | . . . | 1.020 ± 0.003 | 0.02 ± 0.05 | <36 | 3 | N | . . . | . . . | . . . | . . . | . . . | . . . | . . . | n | G |
| 37947 | 10430218-6528234 | -29.5 ± 0.2 | 4941 ± 187 | . . . | 1.021 ± 0.005 | -0.09 ± 0.15 | 16 ± 3 | 1 | N | . . . | . . . | . . . | . . . | . . . | . . . | . . . | n | G |
| 38151 | 10461837-6455232 | 15.5 ± 0.2 | 5117 ± 54 | . . . | 1.010 ± 0.002 | -0.04 ± 0.07 | <28 | 3 | N | . . . | . . . | . . . | . . . | . . . | . . . | . . . | n | G |
| 53613 | 10405352-6414447 | 8.5 ± 0.2 | 4568 ± 103 | 2.51 ± 0.17 | 1.012 ± 0.004 | 0.07 ± 0.03 | <50 | 3 | N | N | . . . | . . . | . . . | . . . | . . . | . . . | n | G |
| 38579 | 10514784-6332163 | -5.5 ± 0.2 | 4742 ± 41 | 2.54 ± 0.09 | 1.025 ± 0.002 | 0.03 ± 0.04 | <41 | 3 | N | N | . . . | . . . | . . . | . . . | . . . | . . . | n | G |
| 53728 | 10430249-6428355 | 13.5 ± 1.1 | 5563 ± 148 | . . . | 0.996 ± 0.009 | -0.53 ± 0.13 | <42 | 3 | Y | . . . | Y | N | N | N | . . . | . . . | n | NG |
| 38152 | 10461840-6337061 | -40.5 ± 0.2 | 5002 ± 92 | . . . | 1.019 ± 0.003 | -0.07 ± 0.07 | <16 | 3 | N | . . . | . . . | . . . | . . . | . . . | . . . | . . . | n | G |
| 53830 | 10443912-6428256 | -22.7 ± 0.2 | 4690 ± 12 | 2.46 ± 0.05 | 1.021 ± 0.001 | -0.05 ± 0.01 | <35 | 3 | N | N | . . . | . . . | . . . | . . . | . . . | . . . | n | G |
| 53883 | 10451456-6426281 | 4.4 ± 0.2 | 4919 ± 108 | 2.95 ± 0.18 | 1.011 ± 0.002 | 0.00 ± 0.05 | 32 ± 2 | 1 | N | N | . . . | . . . | . . . | . . . | . . . | . . . | n | G |
| 37787 | 10405367-6428545 | 40.1 ± 0.2 | 4754 ± 50 | 2.59 ± 0.16 | 1.018 ± 0.002 | -0.11 ± 0.06 | . . . | . . . | N | N | . . . | . . . | . . . | . . . | . . . | . . . | n | G |
| 2607 | 10514893-6421110 | 15.0 ± 0.6 | 4708 ± 21 | 2.73 ± 0.08 | 1.014 ± 0.002 | -0.33 ± 0.02 | 15 ± 1 | 1 | N | N | . . . | . . . | . . . | . . . | . . . | . . . | n | G |
| 53946 | 10461918-6427521 | -34.5 ± 0.2 | 5083 ± 16 | . . . | 1.013 ± 0.005 | -0.12 ± 0.08 | <26 | 3 | N | . . . | . . . | . . . | . . . | . . . | . . . | . . . | n | G |
| 37788 | 10405411-6538178 | 47.9 ± 0.2 | 4674 ± 151 | . . . | 1.024 ± 0.006 | -0.17 ± 0.15 | <25 | 3 | N | . . . | . . . | . . . | . . . | . . . | . . . | . . . | n | G |
| 38016 | 10440679-6531225 | 30.8 ± 0.3 | 4735 ± 78 | . . . | 1.031 ± 0.007 | . . . | <47 | 3 | N | . . . | . . . | . . . | . . . | . . . | . . . | . . . | n | G |
| 38580 | 10514914-6333234 | 17.0 ± 0.2 | 4883 ± 116 | 2.70 ± 0.12 | 1.015 ± 0.005 | -0.06 ± 0.08 | <25 | 3 | N | N | . . . | . . . | . . . | . . . | . . . | . . . | n | G |
| 37948 | 10430311-6546399 | 19.5 ± 0.2 | 4629 ± 49 | 2.07 ± 0.04 | 1.022 ± 0.003 | -0.28 ± 0.09 | <21 | 3 | N | N | . . . | . . . | . . . | . . . | . . . | . . . | n | G |
| 53947 | 10461978-6424384 | 65.4 ± 0.2 | 4557 ± 111 | . . . | 1.033 ± 0.005 | -0.22 ± 0.11 | 24 ± 4 | 1 | N | . . . | . . . | . . . | . . . | . . . | . . . | . . . | n | G |
| 38581 | 10515151-6337459 | 1.6 ± 0.2 | 4819 ± 128 | 2.64 ± 0.16 | 1.016 ± 0.004 | -0.06 ± 0.09 | <22 | 3 | N | N | . . . | . . . | . . . | . . . | . . . | . . . | n | G |
| 53948 | 10462076-6432420 | -27.6 ± 0.2 | 4734 ± 9 | 2.50 ± 0.05 | 1.020 ± 0.002 | -0.04 ± 0.02 | . . . | . . . | N | N | . . . | . . . | . . . | . . . | . . . | . . . | n | G |
| 38051 | 10443978-6412191 | -26.0 ± 0.2 | 4793 ± 56 | 2.59 ± 0.01 | 1.022 ± 0.001 | 0.01 ± 0.04 | <31 | 3 | N | N | . . . | . . . | . . . | . . . | . . . | . . . | n | G |
| 53884 | 10451483-6428313 | 28.2 ± 0.2 | 4569 ± 82 | . . . | 1.024 ± 0.002 | . . . | <57 | 3 | N | . . . | . . . | . . . | . . . | . . . | . . . | . . . | n | G |
| 2545 | 10405443-6447593 | -23.5 ± 0.6 | 4998 ± 17 | 2.65 ± 0.06 | 1.024 ± 0.002 | 0.08 ± 0.02 | 23 ± 3 | 1 | N | N | . . . | . . . | . . . | . . . | . . . | . . . | n | G |
| 2563 | 10440681-6359351 | 17.5 ± 0.6 | 5792 ± 11 | 4.47 ± 0.10 | . . . | 0.03 ± 0.01 | 182 ± 11 | 2 | . . . | Y | Y | Y | Y | Y | Y | Y | Y | . . . |
| 38582 | 10515284-6343254 | 19.8 ± 0.2 | 4637 ± 102 | . . . | 1.024 ± 0.003 | . . . | <54 | 3 | N | . . . | . . . | . . . | . . . | . . . | . . . | . . . | n | G |
| 17 | 10431551-6404238 | 0.5 ± 0.4 | 7127 ± 154 | 4.00 ± 0.18 | . . . | -0.09 ± 0.14 | <13 | 3 | . . . | Y | Y | N | N | Y | N | . . . | n | NG |
| 53949 | 10462091-6427338 | 8.6 ± 0.2 | 4807 ± 195 | 2.65 ± 0.20 | 1.014 ± 0.004 | -0.04 ± 0.08 | <31 | 3 | N | N | . . . | . . . | . . . | . . . | . . . | . . . | n | G |
| 53614 | 10405507-6418495 | 24.2 ± 0.2 | 4724 ± 84 | 2.55 ± 0.13 | 1.015 ± 0.004 | -0.03 ± 0.11 | <21 | 3 | N | N | . . . | . . . | . . . | . . . | . . . | . . . | n | G |
| 38583 | 10515321-6334114 | -16.6 ± 0.2 | 5012 ± 157 | . . . | 1.011 ± 0.004 | -0.21 ± 0.18 | <16 | 3 | N | . . . | . . . | . . . | . . . | . . . | . . . | . . . | n | G |
| 38153 | 10462183-6418218 | -20.2 ± 0.2 | 5019 ± 191 | . . . | 1.024 ± 0.003 | -0.10 ± 0.14 | <19 | 3 | N | . . . | . . . | . . . | . . . | . . . | . . . | . . . | n | G |
| 37789 | 10405523-6539415 | 238.8 ± 0.2 | 4813 ± 242 | . . . | 1.024 ± 0.002 | -0.62 ± 0.16 | <10 | 3 | N | . . . | . . . | . . . | . . . | . . . | . . . | . . . | n | G |
| 38584 | 10515411-6348132 | 1.8 ± 0.2 | 4715 ± 44 | 2.69 ± 0.07 | 1.007 ± 0.006 | 0.01 ± 0.03 | <37 | 3 | Y | N | Y | N | N | Y | . . . | . . . | n | NG? |
| 38154 | 10462188-6333391 | 14.7 ± 0.2 | 4592 ± 99 | 2.40 ± 0.18 | 1.022 ± 0.002 | 0.02 ± 0.05 | <45 | 3 | N | N | . . . | . . . | . . . | . . . | . . . | . . . | n | G |
| 53831 | 10443998-6428536 | 50.8 ± 0.2 | 4293 ± 245 | . . . | 1.041 ± 0.003 | -0.11 ± 0.08 | <39 | 3 | N | . . . | . . . | . . . | . . . | . . . | . . . | . . . | n | G |
| 38017 | 10440696-6532370 | 21.3 ± 0.2 | 4580 ± 170 | 2.51 ± 0.20 | 1.012 ± 0.007 | 0.12 ± 0.03 | <44 | 3 | N | N | . . . | . . . | . . . | . . . | . . . | . . . | n | G |
| 38605 | 10521090-6340247 | 48.9 ± 0.2 | 4693 ± 52 | 2.56 ± 0.19 | 1.010 ± 0.006 | 0.02 ± 0.02 | <33 | 3 | N | N | . . . | . . . | . . . | . . . | . . . | . . . | n | G |
| 37961 | 10431627-6546495 | 37.1 ± 0.2 | 4505 ± 191 | . . . | 1.020 ± 0.005 | 0.12 ± 0.04 | <50 | 3 | N | . . . | . . . | . . . | . . . | . . . | . . . | . . . | n | G |
| 38155 | 10462396-6338446 | -35.0 ± 0.2 | 4948 ± 94 | . . . | 1.019 ± 0.003 | -0.06 ± 0.10 | <18 | 3 | N | . . . | . . . | . . . | . . . | . . . | . . . | . . . | n | G |
| 38077 | 10451506-6448220 | 19.3 ± 0.8 | 3437 ± 115 | 4.64 ± 0.13 | 0.839 ± 0.008 | -0.26 ± 0.14 | . . . | . . . | Y | Y | Y | . . . | N | Y | Y | Y | n | . . . |
| 37790 | 10405531-6457543 | 2.9 ± 0.3 | 3591 ± 32 | . . . | 0.852 ± 0.005 | -0.21 ± 0.14 | <100 | 3 | Y | . . . | Y | Y | Y | Y | N | . . . | n | . . . |
| 38606 | 10521162-6430324 | 20.1 ± 0.2 | 4503 ± 193 | 2.51 ± 0.13 | 1.006 ± 0.003 | 0.12 ± 0.06 | . . . | . . . | Y | N | Y | . . . | N | Y | . . . | . . . | n | . . . |
| 53734 | 10431691-6425339 | 44.0 ± 0.2 | 4602 ± 74 | 2.30 ± 0.15 | 1.023 ± 0.002 | -0.09 ± 0.08 | <29 | 3 | N | N | . . . | . . . | . . . | . . . | . . . | . . . | n | G |
| 38156 | 10462511-6332016 | -16.3 ± 0.2 | 4678 ± 78 | 2.49 ± 0.17 | 1.017 ± 0.002 | -0.10 ± 0.08 | <29 | 3 | N | N | . . . | . . . | . . . | . . . | . . . | . . . | n | G |
| 37791 | 10405540-6406479 | -17.9 ± 0.2 | 5025 ± 32 | 3.03 ± 0.18 | 1.019 ± 0.001 | 0.01 ± 0.01 | 29 ± 2 | 1 | N | N | . . . | . . . | . . . | . . . | . . . | . . . | n | G |
| 38018 | 10440702-6346071 | 41.8 ± 0.3 | 4573 ± 125 | . . . | 1.031 ± 0.006 | -0.54 ± 0.08 | <36 | 3 | N | . . . | . . . | . . . | . . . | . . . | . . . | . . . | n | G |
| 38607 | 10521167-6346598 | 37.5 ± 0.2 | 4563 ± 131 | 2.56 ± 0.09 | 1.003 ± 0.007 | 0.10 ± 0.05 | <54 | 3 | Y | N | Y | N | N | Y | . . . | . . . | n | NG? |
| 37962 | 10431694-6544450 | -67.5 ± 0.2 | 4756 ± 136 | 2.36 ± 0.16 | 1.024 ± 0.002 | -0.24 ± 0.02 | <19 | 3 | N | N | . . . | . . . | . . . | . . . | . . . | . . . | n | G |
| 38171 | 10463939-6411559 | 12.1 ± 0.2 | 4900 ± 175 | . . . | 1.020 ± 0.002 | -0.08 ± 0.14 | <21 | 3 | N | . . . | . . . | . . . | . . . | . . . | Y | . . . | n | G |
| 37792 | 10405564-6543083 | 20.0 ± 0.2 | 4286 ± 436 | . . . | 0.894 ± 0.004 | -0.04 ± 0.14 | 207 ± 6 | 1 | Y | . . . | Y | Y | Y | Y | Y | . . . | Y | . . . |
| 38608 | 10521204-6413416 | 1.2 ± 0.2 | 4843 ± 153 | 2.66 ± 0.12 | 1.016 ± 0.002 | 0.00 ± 0.03 | 18 ± 4 | 1 | N | N | . . . | . . . | . . . | . . . | . . . | . . . | n | G |
| 38172 | 10463950-6327318 | -4.9 ± 0.2 | 4892 ± 36 | 2.63 ± 0.04 | 1.023 ± 0.002 | 0.01 ± 0.06 | <30 | 3 | N | N | . . . | . . . | . . . | . . . | . . . | . . . | n | G |
| 53832 | 10444046-6425530 | -31.6 ± 0.2 | 5060 ± 35 | . . . | 1.009 ± 0.003 | -0.10 ± 0.05 | <16 | 3 | Y | . . . | N | N | N | Y | . . . | . . . | n | NG? |
| 37793 | 10405597-6529318 | 14.9 ± 0.2 | 4989 ± 158 | . . . | 1.009 ± 0.005 | -0.09 ± 0.17 | <14 | 3 | Y | . . . | Y | N | N | Y | . . . | . . . | n | NG? |
| 38609 | 10521359-6338043 | -16.2 ± 0.2 | 4997 ± 195 | . . . | 1.020 ± 0.003 | -0.23 ± 0.26 | <25 | 3 | N | . . . | . . . | . . . | . . . | . . . | . . . | . . . | n | G |

**Table C.6.** continued.

| ID | CNAME | RV (km s$^{-1}$) | $T_{\text{eff}}$ (K) | $logg$ (dex) | $\gamma^a$ | [Fe/H] (dex) | EW(Li)$^b$ (mÅ) | EW(Li) error flag$^c$ | $\gamma$ | $logg$ | RV | Li | H$\alpha$ | [Fe/H] | Randich$^d$ | Cantat-Gaudin$^d$ | Final$^e$ | NMs with Li$^f$ |
|---|---|---|---|---|---|---|---|---|---|---|---|---|---|---|---|---|---|---|
| 18 | 10431715-6405105 | ... | ... | ... | ... | ... | ... | ... | ... | ... | ... | ... | ... | ... | ... | ... | n | ... |
| 53952 | 10463951-6428145 | 62.0 ± 0.2 | 4516 ± 38 | ... | 1.040 ± 0.004 | -0.29 ± 0.15 | 48 ± 10 | 1 | N | ... | ... | ... | ... | ... | ... | ... | n | G |
| 37794 | 10405648-6529422 | 31.0 ± 0.2 | 5212 ± 136 | ... | 1.022 ± 0.002 | ... | 121 ± 2 | 1 | N | ... | ... | ... | ... | ... | ... | ... | n | ... |
| 38610 | 10521421-6422093 | -20.4 ± 0.2 | 4763 ± 42 | 2.53 ± 0.13 | 1.029 ± 0.002 | -0.07 ± 0.05 | 25 ± 8 | 1 | N | N | ... | ... | ... | ... | ... | ... | n | G |
| 38173 | 10463952-6427570 | -46.8 ± 0.2 | 4967 ± 104 | ... | 1.022 ± 0.003 | -0.04 ± 0.07 | <37 | 3 | N | ... | ... | ... | ... | ... | ... | ... | n | G |
| 2569 | 10444061-6456332 | -35.4 ± 0.6 | 4939 ± 29 | 2.80 ± 0.07 | 1.020 ± 0.003 | -0.02 ± 0.08 | 31 ± 6 | 1 | N | N | ... | ... | ... | ... | ... | ... | n | G |
| 37795 | 10405692-6358225 | 0.7 ± 0.2 | 5089 ± 39 | ... | 1.009 ± 0.003 | -0.03 ± 0.07 | <23 | 3 | Y | ... | Y | N | N | Y | ... | ... | n | NG? |
| 38611 | 10521437-6344586 | -5.4 ± 0.2 | 4647 ± 58 | 2.43 ± 0.15 | 1.020 ± 0.003 | -0.03 ± 0.12 | <33 | 3 | N | N | ... | ... | ... | ... | ... | ... | n | G |
| 38174 | 10464015-6426145 | -11.8 ± 0.2 | 5036 ± 56 | ... | 1.017 ± 0.002 | -0.01 ± 0.03 | ... | ... | N | ... | ... | ... | ... | ... | ... | ... | n | G |
| 53615 | 10405714-6405452 | -30.4 ± 0.2 | 4685 ± 150 | ... | 1.018 ± 0.003 | -0.08 ± 0.08 | <28 | 3 | N | ... | ... | ... | ... | ... | ... | ... | n | G |
| 38612 | 10521465-6342068 | 17.7 ± 0.2 | 5067 ± 90 | 3.54 ± 0.14 | 1.001 ± 0.002 | -0.08 ± 0.12 | <29 | 3 | Y | Y | Y | N | N | Y | ... | ... | n | NG? |
| 37963 | 10431831-6457003 | 10.0 ± 0.2 | 4508 ± 166 | ... | 1.030 ± 0.002 | -0.12 ± 0.09 | <32 | 3 | N | ... | ... | ... | ... | ... | ... | ... | n | G |
| 37796 | 10405733-6429280 | -7.6 ± 0.2 | 4690 ± 68 | 2.49 ± 0.19 | 1.025 ± 0.002 | -0.07 ± 0.04 | <25 | 3 | N | N | ... | ... | ... | ... | ... | ... | n | G |
| 53885 | 10451538-6427077 | 10.4 ± 0.3 | 6566 ± 57 | ... | 1.001 ± 0.003 | -0.27 ± 0.05 | <27 | 3 | Y | ... | Y | N | N | Y | ... | ... | n | NG? |
| 38613 | 10521521-6427257 | -45.8 ± 0.2 | 5080 ± 73 | ... | 1.021 ± 0.002 | -0.19 ± 0.14 | <11 | 3 | N | ... | ... | ... | ... | ... | ... | ... | n | G |
| 38175 | 10464079-6410532 | 2.4 ± 0.2 | 4664 ± 66 | 2.53 ± 0.15 | 1.018 ± 0.001 | 0.02 ± 0.04 | <37 | 3 | N | N | ... | ... | ... | ... | ... | ... | n | G |
| 38614 | 10521576-6345338 | 30.8 ± 0.2 | 4576 ± 165 | ... | 1.021 ± 0.005 | 0.06 ± 0.06 | 78 ± 7 | 1 | N | ... | ... | ... | ... | ... | ... | ... | n | G |
| 38176 | 10464166-6318034 | -7.4 ± 0.2 | 4583 ± 40 | 2.38 ± 0.13 | 1.023 ± 0.003 | 0.00 ± 0.04 | <55 | 3 | N | N | ... | ... | ... | ... | ... | ... | n | G |
| 26 | 10444548-6421013 | -44.2 ± 0.6 | 4751 ± 13 | 2.52 ± 0.13 | 1.018 ± 0.004 | -0.17 ± 0.11 | 33 ± 5 | 1 | N | N | ... | ... | ... | ... | ... | ... | n | G |
| 37810 | 10411097-6529503 | 12.5 ± 0.2 | 4809 ± 7 | 2.65 ± 0.02 | 1.015 ± 0.002 | 0.05 ± 0.04 | <45 | 3 | N | N | ... | ... | ... | ... | ... | ... | n | ... |
| 38615 | 10521651-6347237 | 42.1 ± 0.2 | 4717 ± 92 | ... | 1.019 ± 0.002 | -0.15 ± 0.10 | <38 | 3 | N | ... | ... | ... | ... | ... | ... | ... | n | G |
| 2581 | 10464337-6331244 | 0.4 ± 0.6 | 5004 ± 105 | 2.00 ± 0.03 | 1.039 ± 0.001 | 0.07 ± 0.17 | 33 ± 9 | 1 | N | N | ... | ... | ... | ... | ... | ... | n | G |
| 37811 | 10411183-6438439 | 12.2 ± 0.2 | 4917 ± 208 | ... | 1.018 ± 0.004 | -0.10 ± 0.16 | ... | ... | N | ... | ... | ... | ... | ... | ... | ... | n | G |
| 38616 | 10521826-6410521 | 17.4 ± 0.2 | 3586 ± 68 | ... | 0.816 ± 0.004 | -0.22 ± 0.14 | ... | ... | Y | ... | Y | ... | N | Y | Y | Y | n | ... |
| 37964 | 10431939-6349279 | 19.9 ± 0.2 | 4871 ± 78 | 2.68 ± 0.07 | 1.015 ± 0.003 | -0.02 ± 0.07 | <38 | 3 | N | N | ... | ... | ... | ... | ... | ... | n | G |
| 38177 | 10464405-6424504 | 17.4 ± 0.3 | 3347 ± 56 | ... | 0.858 ± 0.011 | -0.28 ± 0.14 | ... | ... | Y | ... | Y | ... | N | N | Y | Y | n | ... |
| 53839 | 10444551-6425171 | -13.6 ± 0.3 | 5725 ± 119 | ... | 1.001 ± 0.004 | -0.49 ± 0.25 | 38 ± 3 | 1 | Y | ... | N | N | N | N | ... | ... | n | NG? |
| 37812 | 10411237-6450101 | -9.2 ± 0.2 | 5099 ± 143 | ... | 1.019 ± 0.004 | ... | ... | ... | N | ... | ... | ... | ... | ... | ... | ... | n | ... |
| 38617 | 10521909-6425253 | 23.1 ± 0.2 | 4523 ± 163 | ... | 1.018 ± 0.003 | 0.11 ± 0.08 | <48 | 3 | N | ... | ... | ... | ... | ... | ... | ... | n | G |
| 38178 | 10464538-6456301 | -9.8 ± 0.2 | 4645 ± 83 | 2.29 ± 0.06 | 1.046 ± 0.006 | -0.04 ± 0.05 | <43 | 3 | N | N | ... | ... | ... | ... | ... | ... | n | G |
| 53642 | 10411264-6402379 | -17.8 ± 0.2 | 4701 ± 11 | 2.51 ± 0.13 | 1.018 ± 0.002 | -0.10 ± 0.09 | <25 | 3 | N | N | ... | ... | ... | ... | ... | ... | n | G |
| 38618 | 10521934-6336175 | -18.2 ± 0.2 | 4871 ± 242 | ... | 1.022 ± 0.004 | -0.16 ± 0.19 | <35 | 3 | N | ... | ... | ... | ... | ... | ... | ... | n | G |
| 53640 | 10444569-6426020 | -1.7 ± 0.2 | 5234 ± 140 | 3.64 ± 0.01 | 0.997 ± 0.004 | -0.02 ± 0.06 | <26 | 3 | Y | Y | Y | N | N | Y | ... | ... | n | NG |
| 38179 | 10464548-6421467 | 46.3 ± 0.2 | 4623 ± 84 | 2.37 ± 0.16 | 1.023 ± 0.002 | -0.03 ± 0.09 | <32 | 3 | N | N | ... | ... | ... | ... | ... | ... | n | G |
| 53643 | 10411297-6417150 | -37.3 ± 0.4 | 6031 ± 139 | ... | 1.019 ± 0.005 | -0.32 ± 0.10 | <11 | 3 | N | ... | ... | ... | ... | ... | ... | ... | n | ... |
| 38619 | 10521985-6342571 | 1.1 ± 0.2 | 4845 ± 50 | 3.21 ± 0.18 | 0.997 ± 0.004 | 0.06 ± 0.03 | <39 | 3 | Y | N | Y | N | Y | Y | ... | ... | n | NG? |
| 53735 | 10431992-6433058 | 6.1 ± 0.2 | 4455 ± 170 | ... | 1.030 ± 0.002 | 0.01 ± 0.05 | <45 | 3 | N | ... | ... | ... | ... | ... | ... | ... | n | G |
| 37813 | 10411331-6456209 | -32.6 ± 0.2 | 4455 ± 143 | ... | 1.038 ± 0.003 | -0.07 ± 0.02 | <26 | 3 | N | ... | ... | ... | ... | ... | ... | ... | n | G |
| 38180 | 10464781-6415459 | 17.5 ± 0.5 | 3319 ± 85 | ... | 0.867 ± 0.014 | -0.23 ± 0.13 | ... | ... | Y | ... | Y | ... | N | Y | Y | Y | n | ... |
| 38021 | 10441009-6400255 | -43.5 ± 0.2 | 4814 ± 132 | ... | 1.026 ± 0.002 | -0.08 ± 0.09 | <33 | 3 | N | ... | ... | ... | ... | ... | ... | ... | n | G |
| 38620 | 10522084-6423351 | 30.9 ± 0.2 | 4388 ± 111 | 1.89 ± 0.11 | 1.043 ± 0.002 | -0.05 ± 0.16 | 97 ± 2 | 1 | N | N | ... | ... | ... | ... | ... | ... | n | ... |
| 53736 | 10432020-6421250 | 11.8 ± 0.2 | 4939 ± 87 | 2.83 ± 0.04 | 1.011 ± 0.003 | 0.00 ± 0.02 | <35 | 3 | N | N | ... | ... | ... | ... | ... | ... | n | G |
| 37814 | 10411404-6430559 | 49.1 ± 0.2 | 4913 ± 243 | ... | 1.010 ± 0.004 | -0.23 ± 0.23 | ... | ... | Y | ... | N | ... | N | Y | ... | ... | n | ... |
| 38181 | 10464803-6316007 | 88.0 ± 0.9 | ... | ... | ... | ... | ... | ... | ... | ... | ... | ... | ... | ... | ... | ... | n | ... |
| 38621 | 10522130-6419316 | 64.2 ± 0.2 | 4675 ± 60 | 2.45 ± 0.18 | 1.023 ± 0.003 | -0.10 ± 0.08 | <42 | 3 | N | N | ... | ... | ... | ... | ... | ... | n | G |
| 53644 | 10411559-6407101 | -8.1 ± 0.2 | 3538 ± 107 | ... | ... | ... | <100 | 3 | ... | ... | ... | ... | ... | ... | ... | ... | n | NG |
| 53737 | 10432040-6430185 | 11.8 ± 0.2 | 4757 ± 171 | 2.90 ± 0.15 | 0.998 ± 0.005 | -0.08 ± 0.21 | <42 | 3 | Y | N | Y | N | Y | Y | ... | ... | n | NG? |
| 38182 | 10464806-6422560 | 43.4 ± 0.3 | 4868 ± 57 | 2.89 ± 0.17 | 1.006 ± 0.006 | -0.01 ± 0.04 | <37 | 3 | Y | N | N | N | N | Y | ... | ... | n | NG? |
| 53645 | 10411559-6422398 | -20.9 ± 0.4 | 7236 ± 52 | ... | ... | ... | ... | ... | ... | ... | ... | ... | ... | ... | ... | ... | n | ... |
| 38622 | 10522171-6419514 | -27.9 ± 0.2 | 4932 ± 170 | ... | 1.023 ± 0.004 | -0.08 ± 0.12 | <36 | 3 | N | ... | ... | ... | ... | ... | ... | ... | n | G |
| 53646 | 10411595-6406080 | -13.7 ± 0.3 | 5948 ± 185 | 3.99 ± 0.12 | 1.004 ± 0.005 | 0.06 ± 0.18 | 97 ± 4 | 1 | Y | Y | N | N | N | Y | ... | ... | n | NG? |
| 38183 | 10464812-6417262 | 12.9 ± 0.2 | 5143 ± 106 | ... | 1.019 ± 0.003 | ... | <17 | 3 | N | ... | ... | ... | ... | ... | ... | ... | n | G |
| 53894 | 10451925-6438248 | 12.9 ± 0.2 | 4869 ± 182 | ... | 1.021 ± 0.004 | -0.08 ± 0.14 | <25 | 3 | N | ... | ... | ... | ... | ... | ... | ... | n | G |
| 38623 | 10522300-6410527 | -65.8 ± 0.2 | 4564 ± 170 | ... | 1.038 ± 0.003 | -0.23 ± 0.07 | <24 | 3 | N | ... | ... | ... | ... | ... | ... | ... | n | G |
| 37965 | 10432088-6448276 | 11.8 ± 0.2 | 5056 ± 84 | ... | 1.013 ± 0.003 | -0.02 ± 0.06 | <26 | 3 | N | ... | ... | ... | ... | ... | ... | ... | n | G |
| 37815 | 10411621-6530378 | 45.9 ± 0.2 | 4594 ± 68 | 2.48 ± 0.14 | 1.018 ± 0.004 | 0.13 ± 0.01 | <51 | 3 | N | N | ... | ... | ... | ... | ... | ... | n | G |
| 38184 | 10464861-6455374 | -35.7 ± 0.3 | 4773 ± 65 | 2.58 ± 0.06 | 1.021 ± 0.006 | -0.06 ± 0.03 | 30 ± 4 | 1 | N | N | ... | ... | ... | ... | ... | ... | n | G |
| 2610 | 10522362-6332566 | 18.8 ± 0.6 | 6197 ± 13 | 4.07 ± 0.06 | ... | 0.15 ± 0.02 | <5 | 3 | ... | Y | Y | N | N | Y | Y | ... | n | ... |
| 38022 | 10441030-6502495 | -5.7 ± 0.2 | 5116 ± 52 | ... | 1.017 ± 0.003 | -0.03 ± 0.05 | <19 | 3 | N | ... | ... | ... | ... | ... | ... | ... | n | G |

| ID | CNAME | RV (km s$^{-1}$) | $T_{\text{eff}}$ (K) | logg (dex) | $\gamma^a$ | [Fe/H] (dex) | EW(Li)$^b$ (mÅ) | EW(Li) error flag$^c$ | Membership $\gamma$ | logg | RV | Li | H$\alpha$ | [Fe/H] | Gaia studies Randich$^d$ | Cantat-Gaudin$^d$ | Final$^e$ | NMs with Li$^f$ |
|---|---|---|---|---|---|---|---|---|---|---|---|---|---|---|---|---|---|---|
| 38054 | 10444612-6348068 | 33.0 ± 0.2 | 4849 ± 34 | 2.79 ± 0.08 | 1.008 ± 0.002 | 0.00 ± 0.08 | <40 | 3 | Y | N | Y | N | N | Y | … | … | n | NG? |
| 37816 | 10411631-6403484 | 35.4 ± 0.2 | 4971 ± 157 | 3.11 ± 0.17 | 1.006 ± 0.002 | -0.09 ± 0.16 | <29 | 3 | Y | N | Y | N | N | Y | … | … | n | NG? |
| 2582 | 10465181-6334157 | 18.2 ± 0.6 | 4857 ± 128 | 4.63 ± 0.08 | 0.953 ± 0.002 | -0.07 ± 0.06 | 303 ± 10 | 2 | Y | Y | Y | Y | Y | Y | Y | Y | Y | … |
| 38624 | 10522402-6345307 | 21.8 ± 0.3 | 4849 ± 107 | … | 1.048 ± 0.007 | -0.01 ± 0.08 | … | … | N | … | … | … | … | … | … | … | n | G |
| 2548 | 10411636-6454552 | -17.4 ± 0.6 | 4832 ± 76 | 2.59 ± 0.05 | 1.018 ± 0.002 | 0.17 ± 0.06 | 32 ± 8 | 1 | N | … | … | … | … | … | … | N | n | G |
| 38185 | 10465243-6318487 | 23.4 ± 0.2 | 5146 ± 18 | … | 1.010 ± 0.004 | -0.02 ± 0.07 | <40 | 3 | Y | … | Y | N | N | Y | … | … | n | NG? |
| 38625 | 10522636-6345595 | 55.8 ± 0.2 | 4453 ± 166 | 2.13 ± 0.18 | 1.032 ± 0.005 | 0.12 ± 0.05 | 39 ± 10 | 1 | N | N | … | … | … | … | … | … | n | G |
| 37969 | 10432831-6541093 | -1.2 ± 0.3 | 5000 ± 162 | … | 1.011 ± 0.006 | -0.10 ± 0.15 | <28 | 3 | N | … | … | … | … | … | … | … | n | G |
| 53790 | 10441037-6428285 | 22.0 ± 0.2 | 5868 ± 19 | 4.30 ± 0.06 | 0.987 ± 0.003 | 0.05 ± 0.09 | 67 ± 9 | 1 | Y | Y | Y | N | N | Y | … | … | n | NG |
| 37817 | 10411646-6539201 | 54.5 ± 0.2 | 4547 ± 136 | 2.35 ± 0.20 | 1.021 ± 0.006 | 0.06 ± 0.07 | <43 | 3 | N | N | … | … | … | … | … | … | n | G |
| 38186 | 10465283-6316095 | 26.5 ± 0.3 | 4751 ± 71 | 2.73 ± 0.03 | 1.006 ± 0.006 | -0.04 ± 0.14 | <42 | 3 | Y | N | Y | N | N | Y | … | … | n | NG? |
| 53742 | 10411652-6414202 | 56.5 ± 0.3 | 4178 ± 236 | … | 1.029 ± 0.011 | -0.02 ± 0.13 | <56 | 3 | N | … | … | … | … | … | … | … | n | G |
| 38187 | 10465479-6418424 | 14.2 ± 0.2 | 4413 ± 206 | … | 1.041 ± 0.003 | 0.03 ± 0.04 | <46 | 3 | N | … | … | … | … | … | … | … | n | G |
| 38626 | 10522781-6344247 | 15.6 ± 0.2 | 4826 ± 107 | 2.69 ± 0.08 | 1.021 ± 0.004 | -0.04 ± 0.08 | … | … | N | N | … | … | … | … | … | … | n | G |
| 53742 | 10432954-6431271 | 74.5 ± 0.2 | 4551 ± 122 | 2.35 ± 0.14 | 1.018 ± 0.006 | 0.13 ± 0.03 | <62 | 3 | N | N | … | … | … | … | … | … | n | G |
| 37818 | 10411657-6528067 | -15.4 ± 0.3 | 5073 ± 56 | … | 1.008 ± 0.006 | -0.09 ± 0.05 | <12 | 3 | Y | … | N | N | N | Y | … | … | n | NG? |
| 38188 | 10465523-6316091 | 14.3 ± 2.2 | … | … | … | … | … | … | … | … | … | … | … | … | … | … | n | … |
| 38664 | 10533186-6422463 | 15.1 ± 0.2 | 4569 ± 188 | … | 1.021 ± 0.003 | 0.07 ± 0.05 | <35 | 3 | N | … | … | … | … | … | … | … | n | G |
| 38665 | 10533260-6414030 | 41.7 ± 0.2 | 4630 ± 90 | … | 1.022 ± 0.003 | 0.05 ± 0.08 | <45 | 3 | N | … | … | … | … | … | … | … | n | G |
| 53841 | 10444689-6435506 | 117.6 ± 0.2 | 3969 ± 241 | … | 1.062 ± 0.003 | -0.26 ± 0.07 | <17 | 3 | N | … | … | … | … | … | … | … | n | G |
| 37819 | 10411666-6443212 | -13.5 ± 0.2 | 5028 ± 135 | … | 1.014 ± 0.003 | -0.09 ± 0.10 | 32 ± 5 | 1 | N | … | … | … | … | … | … | … | n | G |
| 38189 | 10465667-6335450 | 30.8 ± 0.2 | 4418 ± 212 | 2.06 ± 0.20 | 1.027 ± 0.002 | 0.18 ± 0.01 | <67 | 3 | N | N | … | … | … | … | … | … | n | G |
| 38666 | 10533270-6421313 | 13.4 ± 0.2 | 4511 ± 183 | … | 1.018 ± 0.003 | 0.11 ± 0.07 | <51 | 3 | N | … | … | … | … | … | … | … | n | G |
| 53895 | 10451994-6430386 | 43.4 ± 0.3 | 5748 ± 150 | 4.25 ± 0.15 | 0.993 ± 0.006 | 0.06 ± 0.02 | <34 | 3 | Y | Y | N | N | N | Y | … | … | n | NG |
| 37970 | 10433075-6453093 | 63.7 ± 0.2 | 4169 ± 307 | … | 1.055 ± 0.002 | -0.23 ± 0.05 | <20 | 3 | N | … | … | … | … | … | … | … | n | G |
| 53648 | 10411734-6415361 | 58.2 ± 0.2 | 4848 ± 197 | … | 1.002 ± 0.003 | 0.17 ± 0.15 | <45 | 3 | Y | … | N | N | N | N | … | … | n | NG? |
| 38190 | 10465921-6330366 | -7.4 ± 0.2 | 4730 ± 58 | 2.47 ± 0.12 | 1.025 ± 0.001 | -0.06 ± 0.03 | <27 | 3 | N | N | … | … | … | … | … | … | n | G |
| 38235 | 10473724-6502279 | -3.8 ± 0.2 | 4774 ± 56 | 2.62 ± 0.09 | 1.015 ± 0.004 | -0.02 ± 0.03 | <35 | 3 | N | N | … | … | … | … | … | … | n | G |
| 2572 | 10452008-6355271 | 46.6 ± 0.6 | 4966 ± 40 | 3.32 ± 0.05 | 1.006 ± 0.002 | 0.07 ± 0.02 | <22 | 3 | Y | N | N | N | N | Y | … | … | n | NG? |
| 38667 | 10533611-6424096 | 10.3 ± 0.2 | 4517 ± 93 | 2.30 ± 0.14 | 1.022 ± 0.006 | 0.11 ± 0.03 | <50 | 3 | N | N | … | … | … | … | … | … | n | G |
| 37820 | 10411858-6502059 | -40.2 ± 0.2 | 4724 ± 315 | … | 1.022 ± 0.005 | -0.15 ± 0.25 | <22 | 3 | N | … | … | … | … | … | … | … | n | G |
| 37971 | 10433153-6452432 | 3.4 ± 0.2 | 5365 ± 20 | … | 1.007 ± 0.003 | 0.02 ± 0.12 | 54 ± 2 | 1 | Y | … | Y | N | N | Y | … | … | n | NG? |
| 38236 | 10473831-6446066 | -42.7 ± 0.2 | 4771 ± 15 | 2.65 ± 0.13 | 1.017 ± 0.003 | -0.06 ± 0.06 | <33 | 3 | N | N | … | … | … | … | … | … | n | G |
| 38668 | 10533694-6419061 | -26.6 ± 0.2 | 4690 ± 44 | 2.55 ± 0.10 | 1.017 ± 0.002 | 0.06 ± 0.05 | <46 | 3 | N | N | … | … | … | … | … | … | n | G |
| 53650 | 10411861-6416574 | -21.1 ± 0.2 | 4742 ± 9 | 2.51 ± 0.03 | 1.022 ± 0.002 | -0.01 ± 0.02 | <34 | 3 | N | N | … | … | … | … | … | … | n | G |
| 38084 | 10452012-6324520 | 9.1 ± 0.2 | 4862 ± 14 | 2.75 ± 0.09 | 1.010 ± 0.002 | 0.05 ± 0.07 | <34 | 3 | Y | N | Y | N | N | Y | … | … | n | NG? |
| 38237 | 10473861-6449165 | 30.2 ± 0.2 | 4198 ± 259 | … | 1.047 ± 0.005 | -0.19 ± 0.07 | 64 ± 11 | 1 | N | … | … | … | … | … | … | … | n | G |
| 37821 | 10411904-6444484 | 27.7 ± 0.2 | 4570 ± 101 | 2.44 ± 0.14 | 1.015 ± 0.002 | 0.15 ± 0.01 | <67 | 3 | N | N | … | … | … | … | … | … | n | G |
| 2558 | 10433192-6347311 | -14.1 ± 0.6 | 5052 ± 22 | 2.76 ± 0.04 | … | 0.02 ± 0.02 | <17 | 3 | … | N | … | … | … | … | … | … | n | … |
| 53651 | 10412002-6422130 | -26.8 ± 0.2 | 5935 ± 208 | … | 1.008 ± 0.003 | -0.27 ± 0.07 | … | … | Y | … | N | … | … | Y | … | … | n | … |
| 38238 | 10474092-6326524 | -0.2 ± 0.2 | 4956 ± 167 | … | 1.021 ± 0.005 | -0.07 ± 0.10 | <24 | 3 | N | … | … | … | … | … | … | … | n | G |
| 53661 | 10412974-6406550 | 1.6 ± 0.3 | 6473 ± 95 | … | 0.999 ± 0.005 | -0.12 ± 0.08 | … | … | Y | … | Y | N | N | Y | … | … | n | … |
| 38 | 10474139-6417057 | 18.1 ± 0.6 | 4167 ± 91 | 4.54 ± 0.11 | … | -0.16 ± 0.11 | <48 | 3 | … | Y | Y | Y | Y | Y | Y | … | Y | … |
| 38669 | 10533995-6421143 | 11.9 ± 0.2 | 5214 ± 1 | 4.11 ± 0.08 | 0.979 ± 0.002 | 0.03 ± 0.01 | 20 ± 2 | 1 | Y | Y | Y | N | Y | Y | … | … | n | NG |
| 53743 | 10433204-6431399 | 42.7 ± 0.2 | 4617 ± 102 | … | 1.040 ± 0.003 | … | <30 | 3 | N | … | … | … | … | … | … | … | n | G |
| 53662 | 10412983-6412265 | 75.3 ± 0.2 | 4504 ± 202 | … | 1.038 ± 0.003 | -0.49 ± 0.02 | 45 ± 2 | 1 | N | … | … | … | … | … | … | … | n | G |
| 38239 | 10474288-6407417 | 14.1 ± 0.2 | 4507 ± 168 | … | 1.027 ± 0.003 | 0.15 ± 0.03 | <48 | 3 | N | … | … | … | … | … | … | … | n | G |
| 37835 | 10413030-6459584 | 71.1 ± 0.2 | 4530 ± 167 | … | 1.024 ± 0.004 | -0.21 ± 0.12 | … | … | N | … | … | … | … | … | … | … | n | G |
| 38240 | 10474415-6316374 | -20.5 ± 1.3 | … | … | … | … | … | … | … | … | … | … | … | … | N | … | n | … |
| 38670 | 10534712-6419213 | 18.6 ± 0.2 | 4704 ± 30 | 2.52 ± 0.05 | 1.017 ± 0.003 | 0.06 ± 0.05 | <44 | 3 | N | N | … | … | … | … | … | … | n | G |
| 37972 | 10433231-6355581 | -19.4 ± 0.2 | 4649 ± 62 | 2.43 ± 0.12 | 1.022 ± 0.003 | 0.00 ± 0.03 | <34 | 3 | N | N | … | … | … | … | … | … | n | G |
| 53845 | 10444913-6432203 | 20.0 ± 0.2 | 4946 ± 192 | 3.56 ± 0.19 | 0.989 ± 0.005 | -0.03 ± 0.06 | <36 | 3 | Y | Y | Y | N | N | Y | … | … | n | NG |
| 38241 | 10474513-6404055 | 115.0 ± 0.2 | 4444 ± 169 | … | 1.052 ± 0.003 | -0.24 ± 0.12 | 16 ± 2 | 1 | N | … | … | … | … | … | … | … | n | G |
| 38031 | 10441549-6533497 | 37.3 ± 0.2 | 4798 ± 40 | … | 0.998 ± 0.003 | 0.07 ± 0.08 | <41 | 3 | Y | … | Y | N | N | Y | … | … | n | NG |
| 2586 | 10474621-6332108 | 6.4 ± 0.6 | 5045 ± 24 | 2.92 ± 0.04 | 1.015 ± 0.002 | 0.18 ± 0.02 | 39 ± 11 | 1 | N | N | … | … | … | … | … | … | n | G |
| 53846 | 10444927-6421298 | 47.7 ± 0.2 | 4397 ± 182 | 1.83 ± 0.19 | 1.048 ± 0.002 | -0.09 ± 0.09 | <31 | 3 | N | N | … | … | … | … | … | … | n | G |
| 2587 | 10474712-6434113 | -2.0 ± 0.6 | 6160 ± 35 | 4.27 ± 0.05 | … | 0.15 ± 0.03 | 52 ± 6 | 2 | … | Y | Y | N | N | Y | N | … | n | NG |
| 53744 | 10433258-6419142 | -25.3 ± 0.2 | 4509 ± 139 | … | 1.025 ± 0.002 | -0.07 ± 0.05 | <38 | 3 | N | … | … | … | … | … | … | … | n | G |
| 38242 | 10474743-6448555 | -38.6 ± 0.2 | 4654 ± 105 | … | 1.030 ± 0.002 | -0.12 ± 0.12 | 180 ± 4 | 1 | N | … | … | … | … | … | … | … | n | Li-rich G |

**Table C.6.** continued.

| ID | CNAME | RV (km s$^{-1}$) | $T_{\text{eff}}$ (K) | $logg$ (dex) | $\gamma^a$ | [Fe/H] (dex) | EW(Li)$^b$ (mÅ) | EW(Li) error flag$^c$ | Membership $\gamma$ | Membership $logg$ | Gaia studies RV | Gaia studies Li | Gaia studies H$\alpha$ | Gaia studies [Fe/H] | Randich$^d$ | Cantat-Gaudin$^d$ | Final$^e$ | NMs with Li$^f$ |
|---|---|---|---|---|---|---|---|---|---|---|---|---|---|---|---|---|---|---|
| 53745 | 10433288-6424152 | 19.0 ± 0.3 | 6686 ± 108 | ... | 0.993 ± 0.006 | 0.12 ± 0.09 | ... | ... | Y | ... | Y | ... | N | Y | ... | ... | n | ... |
| 38243 | 10474935-6418082 | -1.1 ± 0.2 | 4579 ± 83 | 2.47 ± 0.15 | 1.017 ± 0.003 | 0.15 ± 0.02 | <53 | 3 | N | N | ... | ... | ... | ... | ... | ... | n | G |
| 38032 | 10441557-6538485 | 84.0 ± 0.2 | 5674 ± 19 | ... | 0.998 ± 0.002 | -0.24 ± 0.08 | ... | ... | Y | ... | N | ... | ... | Y | ... | ... | n | ... |
| 38244 | 10475032-6416040 | 9.8 ± 0.2 | 4631 ± 33 | 2.52 ± 0.17 | 1.025 ± 0.003 | 0.06 ± 0.10 | <57 | 3 | N | N | ... | ... | ... | ... | ... | ... | n | G |
| 53796 | 10441558-6434017 | 13.2 ± 0.2 | 4462 ± 251 | ... | 1.021 ± 0.001 | -0.01 ± 0.10 | <34 | 3 | N | ... | ... | ... | ... | ... | ... | ... | n | G |
| 38245 | 10475043-6504499 | -19.2 ± 0.2 | 4596 ± 4 | 2.33 ± 0.19 | 1.027 ± 0.003 | -0.12 ± 0.08 | <30 | 3 | N | N | ... | ... | ... | ... | ... | ... | n | G |
| 38033 | 10441578-6531072 | -13.6 ± 0.3 | 5002 ± 200 | ... | 1.019 ± 0.005 | -0.12 ± 0.14 | <37 | 3 | N | ... | ... | ... | ... | ... | ... | ... | n | G |
| 38246 | 10475148-6359120 | 5.6 ± 0.2 | 4941 ± 197 | ... | 1.022 ± 0.004 | -0.09 ± 0.18 | <24 | 3 | N | ... | ... | ... | ... | ... | ... | ... | n | G |
| 38247 | 10475343-6334542 | 20.4 ± 0.2 | 4577 ± 76 | ... | 1.026 ± 0.002 | ... | ... | ... | N | ... | ... | ... | ... | ... | ... | ... | n | ... |
| 53797 | 10441582-6438475 | -32.9 ± 0.4 | 6405 ± 114 | 4.14 ± 0.24 | 1.002 ± 0.007 | 0.02 ± 0.09 | 86 ± 5 | 1 | Y | Y | N | N | N | Y | ... | ... | n | NG? |
| 53897 | 10452228-6426201 | 27.9 ± 0.2 | 6439 ± 15 | 4.33 ± 0.03 | 0.997 ± 0.001 | 0.19 ± 0.01 | <12 | 3 | Y | Y | Y | N | N | N | ... | ... | n | NG |
| 38248 | 10475408-6435483 | 26.3 ± 0.2 | 4944 ± 220 | ... | 1.008 ± 0.003 | -0.10 ± 0.19 | <25 | 3 | Y | ... | Y | N | N | Y | ... | ... | n | NG? |
| 53898 | 10452238-6423395 | 30.3 ± 0.2 | 4597 ± 150 | 2.44 ± 0.19 | 1.013 ± 0.002 | 0.03 ± 0.11 | <36 | 3 | N | N | ... | ... | ... | ... | ... | ... | n | G |
| 37973 | 10433364-6532454 | 31.4 ± 0.2 | 4636 ± 98 | ... | 1.018 ± 0.004 | ... | <71 | 3 | N | ... | ... | ... | ... | ... | ... | ... | n | G |
| 38249 | 10475418-6354582 | -21.1 ± 0.2 | 4657 ± 4 | 2.40 ± 0.12 | 1.026 ± 0.003 | -0.08 ± 0.03 | 24 ± 2 | 1 | N | N | ... | ... | ... | ... | ... | ... | n | G |
| 53746 | 10433378-6430013 | 3.5 ± 0.8 | 6765 ± 86 | ... | ... | 0.21 ± 0.07 | ... | ... | ... | ... | ... | ... | ... | ... | ... | ... | n | ... |
| 38250 | 10475533-6446270 | 28.0 ± 0.2 | 4580 ± 172 | ... | 1.015 ± 0.002 | 0.05 ± 0.07 | <46 | 3 | N | ... | ... | ... | ... | ... | ... | ... | n | G |
| 38251 | 10475575-6502317 | 63.7 ± 0.2 | 4468 ± 134 | ... | 1.028 ± 0.004 | -0.15 ± 0.09 | <35 | 3 | N | ... | ... | ... | ... | ... | ... | ... | n | G |
| 37985 | 10434156-6457119 | 8.5 ± 0.2 | 4625 ± 55 | 2.43 ± 0.10 | 1.015 ± 0.002 | -0.03 ± 0.02 | ... | ... | N | N | ... | ... | ... | ... | ... | ... | n | G |
| 38252 | 10475613-6433526 | 11.3 ± 0.2 | 4654 ± 98 | ... | 1.026 ± 0.003 | ... | <39 | 3 | N | ... | ... | ... | ... | ... | ... | ... | n | G |
| 38034 | 10441675-6446580 | 25.7 ± 0.2 | 4606 ± 82 | ... | 1.037 ± 0.003 | ... | <50 | 3 | N | ... | ... | ... | ... | ... | ... | ... | n | G |
| 38253 | 10475684-6319433 | 17.3 ± 0.2 | 4780 ± 82 | 2.62 ± 0.08 | 1.014 ± 0.002 | 0.00 ± 0.03 | 17 ± 2 | 1 | N | N | ... | ... | ... | ... | ... | ... | n | G |
| 37986 | 10434187-6540393 | -25.9 ± 0.3 | 4484 ± 174 | ... | 1.018 ± 0.007 | -0.11 ± 0.11 | <37 | 3 | N | ... | ... | ... | ... | ... | ... | ... | n | G |
| 53847 | 10445010-6436059 | 42.6 ± 0.3 | 3581 ± 80 | 4.63 ± 0.14 | 0.800 ± 0.015 | -0.22 ± 0.15 | <100 | 3 | Y | Y | Y | Y | N | Y | ... | ... | n | NG |
| 53899 | 10452270-6432166 | 12.1 ± 0.2 | 6085 ± 99 | 4.05 ± 0.11 | 0.998 ± 0.003 | -0.12 ± 0.06 | <16 | 3 | Y | Y | Y | N | N | Y | ... | ... | n | NG |
| 38254 | 10475747-6334188 | -1.2 ± 0.2 | 4738 ± 51 | 2.53 ± 0.05 | 1.014 ± 0.002 | -0.04 ± 0.01 | 68 ± 2 | 1 | N | N | ... | ... | ... | ... | ... | ... | n | G |
| 53900 | 10452294-6419473 | -3.7 ± 0.3 | 6340 ± 93 | ... | 1.021 ± 0.005 | -0.25 ± 0.08 | <11 | 3 | N | ... | ... | ... | ... | ... | ... | ... | n | ... |
| 38255 | 10475755-6505083 | -6.8 ± 0.2 | 4863 ± 226 | ... | 1.026 ± 0.004 | -0.07 ± 0.12 | <20 | 3 | N | ... | ... | ... | ... | ... | ... | ... | n | G |
| 53753 | 10434194-6430250 | 8.7 ± 0.2 | 5156 ± 106 | ... | 0.999 ± 0.004 | ... | 15 ± 4 | 1 | Y | ... | Y | N | N | ... | ... | ... | n | NG |
| 38279 | 10481525-6322134 | 7.1 ± 0.2 | 5066 ± 72 | ... | 1.015 ± 0.003 | -0.02 ± 0.06 | 11 ± 2 | 1 | N | ... | ... | ... | ... | ... | ... | ... | n | G |
| 53754 | 10434202-6420474 | 60.2 ± 0.2 | 4637 ± 98 | ... | 1.020 ± 0.003 | ... | 82 ± 9 | 1 | N | ... | ... | ... | ... | ... | ... | ... | n | G |
| 38035 | 10441890-6453211 | 19.3 ± 0.3 | 3592 ± 88 | ... | 0.833 ± 0.008 | -0.23 ± 0.14 | ... | ... | Y | ... | Y | ... | N | Y | Y | ... | n | ... |
| 38280 | 10481570-6324162 | 376.2 ± 5.7 | ... | ... | ... | ... | ... | ... | ... | ... | ... | ... | ... | ... | ... | ... | n | ... |
| 53802 | 10441906-6432096 | 16.2 ± 0.3 | 5069 ± 138 | ... | 0.965 ± 0.006 | ... | <28 | 3 | Y | ... | Y | N | Y | ... | ... | ... | n | NG |
| 38281 | 10481574-6428207 | 39.0 ± 0.2 | 4761 ± 64 | 2.56 ± 0.14 | 1.023 ± 0.003 | -0.01 ± 0.07 | ... | ... | N | N | ... | ... | ... | ... | ... | ... | n | G |
| 38282 | 10481727-6326049 | 10.0 ± 0.2 | 4697 ± 20 | 2.50 ± 0.07 | 1.019 ± 0.001 | -0.04 ± 0.07 | <29 | 3 | N | N | ... | ... | ... | ... | ... | ... | n | G |
| 53755 | 10434224-6426551 | -14.9 ± 0.2 | 4878 ± 146 | ... | 1.016 ± 0.005 | -0.07 ± 0.12 | <26 | 3 | N | ... | ... | ... | ... | ... | ... | ... | n | G |
| 38283 | 10481755-6320555 | 31.9 ± 0.2 | 4790 ± 71 | 2.71 ± 0.05 | 1.008 ± 0.004 | 0.03 ± 0.06 | 42 ± 14 | 1 | Y | N | Y | N | N | Y | ... | ... | n | NG? |
| 53756 | 10434228-6436318 | -9.2 ± 0.2 | 5342 ± 65 | ... | 0.989 ± 0.005 | -0.17 ± 0.09 | 12 ± 3 | 1 | Y | ... | Y | N | N | Y | ... | ... | n | NG |
| 38284 | 10481778-6443341 | 11.3 ± 0.2 | 4861 ± 14 | 2.68 ± 0.02 | 1.016 ± 0.002 | 0.01 ± 0.03 | <33 | 3 | N | N | ... | ... | ... | ... | ... | ... | n | G |
| 53757 | 10441946-6424593 | -17.2 ± 0.2 | 6603 ± 76 | 4.18 ± 0.15 | 1.005 ± 0.005 | 0.35 ± 0.06 | <23 | 3 | Y | Y | N | N | N | N | ... | ... | n | NG? |
| 39 | 10481856-6409537 | 19.2 ± 0.6 | 5793 ± 60 | 4.39 ± 0.13 | ... | 0.06 ± 0.11 | 179 ± 11 | 2 | ... | Y | Y | Y | Y | Y | Y | ... | Y | ... |
| 38285 | 10481898-6500282 | -52.7 ± 0.2 | 5068 ± 14 | 3.06 ± 0.20 | 1.016 ± 0.002 | -0.11 ± 0.05 | <24 | 3 | N | N | ... | ... | ... | ... | ... | ... | n | G |
| 38286 | 10481936-6408188 | 38.1 ± 1.7 | ... | ... | ... | ... | ... | ... | ... | ... | ... | ... | ... | ... | ... | ... | n | ... |
| 38092 | 10452847-6452114 | 59.6 ± 0.2 | 4637 ± 98 | ... | 1.020 ± 0.003 | ... | 52 ± 2 | 1 | N | ... | ... | ... | ... | ... | ... | ... | n | G |
| 38287 | 10482067-6406443 | 4.2 ± 0.2 | 4648 ± 83 | 2.56 ± 0.17 | 1.017 ± 0.002 | 0.08 ± 0.04 | <45 | 3 | N | N | ... | ... | ... | ... | ... | ... | n | G |
| 53804 | 10442005-6432239 | 2.4 ± 0.3 | 6250 ± 107 | ... | 0.988 ± 0.007 | 0.19 ± 0.07 | 52 ± 5 | 1 | Y | ... | Y | N | N | N | ... | ... | n | NG |
| 38288 | 10482113-6445250 | 55.7 ± 0.2 | 4540 ± 110 | 2.28 ± 0.18 | 1.025 ± 0.002 | 0.03 ± 0.09 | <49 | 3 | N | N | ... | ... | ... | ... | ... | ... | n | G |
| 53805 | 10442013-6435277 | -29.4 ± 0.2 | 4586 ± 39 | 2.36 ± 0.14 | 1.029 ± 0.001 | 0.00 ± 0.04 | <42 | 3 | N | N | ... | ... | ... | ... | ... | ... | n | G |
| 38289 | 10482250-6402442 | 59.6 ± 0.2 | 4496 ± 201 | ... | 1.025 ± 0.002 | 0.13 ± 0.04 | 207 ± 2 | 1 | N | ... | ... | ... | ... | ... | ... | ... | n | Li-rich G |
| 38290 | 10482295-6327043 | 10.1 ± 0.2 | 4911 ± 98 | ... | 1.015 ± 0.005 | -0.02 ± 0.05 | ... | ... | N | ... | ... | ... | ... | ... | ... | ... | n | G |
| 2589 | 10482491-6353483 | -15.5 ± 0.6 | 5039 ± 31 | 3.05 ± 0.03 | 1.016 ± 0.001 | 0.06 ± 0.03 | 23 ± 3 | 1 | N | N | ... | ... | ... | ... | ... | ... | n | G |
| 53852 | 10445461-6417039 | 48.3 ± 0.2 | 4461 ± 154 | ... | 1.033 ± 0.005 | -0.10 ± 0.08 | <40 | 3 | N | ... | ... | ... | ... | ... | ... | ... | n | G |
| 38291 | 10482514-6331047 | 3.1 ± 0.2 | 4486 ± 178 | ... | 1.020 ± 0.002 | 0.12 ± 0.06 | <61 | 3 | N | ... | ... | ... | ... | ... | ... | ... | n | G |
| 37987 | 10434319-6527479 | 6.8 ± 0.2 | 4619 ± 128 | ... | 1.017 ± 0.004 | -0.14 ± 0.13 | <22 | 3 | N | ... | ... | ... | ... | ... | ... | ... | n | ... |
| 38292 | 10482588-6444437 | 26.6 ± 0.2 | 4765 ± 54 | 2.61 ± 0.06 | 1.013 ± 0.002 | 0.02 ± 0.08 | <31 | 3 | N | N | ... | ... | ... | ... | ... | ... | n | G |
| 38093 | 10452881-6447079 | 31.1 ± 0.2 | 4731 ± 99 | 2.54 ± 0.15 | 1.014 ± 0.004 | -0.05 ± 0.15 | <30 | 3 | N | N | ... | ... | ... | ... | ... | ... | n | G |
| 38293 | 10482591-6442009 | 30.7 ± 0.2 | 4945 ± 233 | ... | 1.019 ± 0.003 | -0.12 ± 0.27 | <25 | 3 | N | ... | ... | ... | ... | ... | ... | ... | n | G |
| 38294 | 10483096-6355266 | -2.5 ± 0.2 | 4706 ± 21 | 2.52 ± 0.05 | 1.021 ± 0.003 | 0.03 ± 0.01 | <34 | 3 | N | N | ... | ... | ... | ... | ... | ... | n | G |

**Table C.6.** continued.

| ID | CNAME | RV (km s$^{-1}$) | $T_{\rm eff}$ (K) | $logg$ (dex) | $\gamma^a$ | [Fe/H] (dex) | EW(Li)$^b$ (mÅ) | EW(Li) error flag$^c$ | Membership $\gamma$ | Membership $logg$ | Membership RV | Membership Li | Membership H$\alpha$ | Membership [Fe/H] | Gaia studies Randich$^d$ | Gaia studies Cantat-Gaudin$^d$ | Final$^e$ | NMs with Li$^f$ |
|---|---|---|---|---|---|---|---|---|---|---|---|---|---|---|---|---|---|---|
| 2570 | 10445489-6354018 | 47.3 ± 0.6 | 4695 ± 23 | 2.09 ± 0.08 | 1.017 ± 0.002 | -0.75 ± 0.03 | 19 ± 4 | 1 | N | N | ... | ... | ... | ... | ... | ... | n | G |
| 38295 | 10483152-6406465 | 16.5 ± 0.2 | 3864 ± 119 | ... | 0.841 ± 0.002 | -0.16 ± 0.11 | ... | ... | Y | ... | Y | ... | N | Y | Y | Y | n | ... |
| 38296 | 10483222-6326122 | -15.7 ± 0.2 | 4714 ± 45 | 2.52 ± 0.06 | 1.017 ± 0.002 | 0.01 ± 0.09 | <36 | 3 | N | N | ... | ... | ... | ... | ... | ... | n | G |
| 38297 | 10483226-6355327 | -11.6 ± 0.2 | 4718 ± 163 | 2.01 ± 0.01 | 1.025 ± 0.002 | -0.34 ± 0.01 | <27 | 3 | N | N | ... | ... | ... | ... | ... | ... | n | G |
| 38094 | 10452908-6355406 | -9.7 ± 0.2 | 4866 ± 110 | 2.66 ± 0.13 | 1.020 ± 0.003 | -0.03 ± 0.05 | <30 | 3 | N | N | ... | ... | ... | ... | ... | ... | n | G |
| 53853 | 10445504-6432267 | -21.5 ± 0.3 | 6831 ± 66 | ... | 0.999 ± 0.003 | 0.00 ± 0.06 | 58 ± 3 | 1 | Y | ... | N | Y | N | Y | ... | ... | n | NG |
| 38298 | 10483242-6518427 | 35.6 ± 0.2 | 4913 ± 176 | ... | 1.018 ± 0.004 | -0.11 ± 0.14 | <20 | 3 | N | ... | ... | ... | ... | ... | ... | ... | n | G |
| 38299 | 10483452-6439373 | -50.2 ± 0.2 | 4878 ± 228 | ... | 1.020 ± 0.002 | -0.09 ± 0.16 | <20 | 3 | N | ... | ... | ... | ... | ... | ... | ... | n | G |
| 38300 | 10483634-6350521 | -4.6 ± 0.2 | 4859 ± 55 | 2.65 ± 0.12 | 1.023 ± 0.001 | -0.03 ± 0.05 | <31 | 3 | N | N | ... | ... | ... | ... | ... | ... | n | G |
| 38346 | 10491631-6514027 | 3.0 ± 0.2 | 4664 ± 12 | 2.43 ± 0.06 | 1.020 ± 0.003 | -0.06 ± 0.02 | <28 | 3 | N | N | ... | ... | ... | ... | ... | ... | n | G |
| 38095 | 10452984-6444543 | 16.5 ± 0.4 | 3266 ± 88 | ... | 0.874 ± 0.016 | -0.26 ± 0.14 | 383 ± 13 | 1 | Y | ... | Y | Y | Y | Y | Y | ... | Y | ... |
| 38347 | 10491841-6516495 | -43.6 ± 0.2 | 5012 ± 174 | ... | 1.013 ± 0.002 | -0.17 ± 0.17 | <14 | 3 | N | ... | ... | ... | ... | ... | ... | ... | n | G |
| 53904 | 10452987-6424393 | 13.0 ± 0.2 | 4517 ± 59 | ... | 1.040 ± 0.004 | -0.13 ± 0.06 | <28 | 3 | N | ... | ... | ... | ... | ... | ... | ... | n | G |
| 38348 | 10491936-6517526 | 18.9 ± 0.2 | 4654 ± 90 | ... | 1.023 ± 0.004 | -0.13 ± 0.15 | <20 | 3 | N | ... | ... | ... | ... | ... | ... | ... | n | G |
| 38349 | 10492013-6339363 | 8.7 ± 0.2 | 4288 ± 261 | ... | 1.055 ± 0.003 | -0.17 ± 0.04 | <27 | 3 | N | ... | ... | ... | ... | ... | ... | ... | n | G |
| 53905 | 10453002-6430351 | 17.0 ± 0.2 | 4414 ± 169 | ... | 1.035 ± 0.003 | -0.03 ± 0.06 | <37 | 3 | N | ... | ... | ... | ... | ... | ... | ... | n | G |
| 24 | 10442506-6415397 | -10.4 ± 0.6 | 4936 ± 38 | 2.67 ± 0.10 | 1.020 ± 0.004 | -0.09 ± 0.04 | <24 | 3 | N | N | ... | ... | ... | ... | ... | ... | n | G |
| 38350 | 10492123-6405170 | 107.0 ± 0.2 | 4981 ± 183 | ... | 1.019 ± 0.001 | -0.16 ± 0.21 | <25 | 3 | N | ... | ... | ... | ... | ... | ... | ... | n | G |
| 2591 | 10492243-6502120 | -22.0 ± 0.6 | 5039 ± 20 | 2.85 ± 0.06 | ... | -0.08 ± 0.03 | <13 | 3 | ... | N | ... | ... | ... | ... | ... | ... | n | ... |
| 38351 | 10492437-6448099 | 43.7 ± 0.2 | 4338 ± 258 | ... | 1.034 ± 0.002 | 0.16 ± 0.09 | <45 | 3 | N | ... | ... | ... | ... | ... | ... | ... | n | G |
| 53906 | 10453041-6430511 | -29.9 ± 3.5 | ... | ... | ... | ... | ... | ... | ... | ... | ... | ... | ... | ... | ... | ... | n | ... |
| 38352 | 10492478-6504100 | -47.8 ± 0.2 | 4694 ± 4 | 2.50 ± 0.15 | 1.023 ± 0.004 | -0.09 ± 0.07 | <39 | 3 | N | N | ... | ... | ... | ... | ... | ... | n | G |
| 38041 | 10442535-6346418 | 9.7 ± 0.2 | 4804 ± 129 | 2.66 ± 0.15 | 1.014 ± 0.003 | -0.05 ± 0.05 | <32 | 3 | N | N | ... | ... | ... | ... | ... | ... | n | G |
| 38353 | 10492522-6358423 | 46.6 ± 0.2 | 4278 ± 268 | ... | 1.042 ± 0.003 | -0.16 ± 0.12 | <35 | 3 | N | ... | ... | ... | ... | ... | ... | ... | n | G |
| 53859 | 10445729-6423477 | 3.3 ± 0.3 | 3543 ± 140 | 4.65 ± 0.14 | 0.796 ± 0.013 | -0.23 ± 0.14 | ... | ... | Y | Y | Y | ... | N | Y | ... | ... | n | ... |
| 38354 | 10492528-6437399 | -0.9 ± 0.2 | 5015 ± 10 | 3.01 ± 0.13 | 1.014 ± 0.003 | 0.02 ± 0.01 | <30 | 3 | N | N | ... | ... | ... | ... | ... | ... | n | G |
| 38355 | 10492584-6516598 | 54.7 ± 0.2 | 4756 ± 19 | 2.54 ± 0.11 | 1.020 ± 0.003 | -0.09 ± 0.05 | <19 | 3 | N | N | ... | ... | ... | ... | ... | ... | n | G |
| 38356 | 10492624-6439000 | 17.3 ± 0.2 | 4321 ± 485 | ... | 0.901 ± 0.002 | -0.01 ± 0.12 | 182 ± 9 | 1 | Y | ... | Y | Y | Y | Y | Y | ... | Y | ... |
| 53860 | 10445748-6423196 | 48.4 ± 0.2 | 4346 ± 181 | ... | 1.043 ± 0.003 | -0.16 ± 0.03 | 92 ± 11 | 1 | N | ... | ... | ... | ... | ... | ... | ... | n | G |
| 38357 | 10492722-6408137 | 14.2 ± 0.2 | 4843 ± 83 | 2.64 ± 0.12 | 1.018 ± 0.001 | -0.03 ± 0.09 | <27 | 3 | N | N | ... | ... | ... | ... | ... | ... | n | G |
| 38358 | 10492771-6344444 | 43.8 ± 1.0 | ... | ... | ... | ... | ... | ... | ... | ... | ... | ... | ... | ... | N | ... | n | ... |
| 53814 | 10442581-6424570 | 18.6 ± 0.2 | 4609 ± 138 | ... | 1.014 ± 0.001 | 0.03 ± 0.12 | <42 | 3 | N | ... | ... | ... | ... | ... | ... | ... | n | G |
| 2592 | 10492808-6401074 | 53.8 ± 0.6 | 4711 ± 28 | 2.46 ± 0.05 | 1.018 ± 0.002 | 0.07 ± 0.02 | 38 ± 11 | 1 | N | N | ... | ... | ... | ... | ... | ... | n | G |
| 38359 | 10492887-6434239 | 4.5 ± 0.2 | 4826 ± 97 | 2.64 ± 0.14 | 1.020 ± 0.003 | -0.02 ± 0.04 | <28 | 3 | N | N | ... | ... | ... | ... | ... | ... | n | G |
| 53861 | 10445763-6437199 | -7.4 ± 0.2 | 4848 ± 157 | 2.68 ± 0.03 | 1.024 ± 0.004 | -0.01 ± 0.03 | <27 | 3 | N | N | ... | ... | ... | ... | ... | ... | n | G |
| 38042 | 10442591-6358124 | 662.4 ± 372.2 | ... | ... | ... | ... | ... | ... | ... | ... | ... | ... | ... | ... | ... | ... | n | ... |
| 38360 | 10492942-6405234 | -26.8 ± 0.2 | 5116 ± 37 | ... | 1.017 ± 0.003 | -0.08 ± 0.07 | <21 | 3 | N | ... | ... | ... | ... | ... | ... | ... | n | G |
| 38361 | 10493081-6438397 | -15.5 ± 0.2 | 4908 ± 168 | ... | 1.022 ± 0.001 | -0.07 ± 0.13 | <23 | 3 | N | ... | ... | ... | ... | ... | ... | ... | n | G |
| 53862 | 10445774-6432152 | 16.2 ± 0.2 | 3936 ± 89 | ... | 1.051 ± 0.002 | -0.14 ± 0.10 | <28 | 3 | N | ... | ... | ... | ... | ... | ... | ... | n | G |
| 38362 | 10493096-6406353 | -2.3 ± 0.2 | 4592 ± 89 | ... | 1.029 ± 0.002 | 0.06 ± 0.01 | <43 | 3 | N | ... | ... | ... | ... | ... | ... | ... | n | G |
| 53909 | 10453461-6429141 | 24.2 ± 0.2 | 4707 ± 32 | 2.45 ± 0.13 | 1.020 ± 0.004 | -0.10 ± 0.12 | <34 | 3 | N | N | ... | ... | ... | ... | ... | ... | n | G |
| 38363 | 10493139-6434389 | 10.9 ± 0.2 | 4585 ± 76 | ... | 1.026 ± 0.004 | ... | ... | ... | N | ... | ... | ... | ... | ... | ... | ... | n | G |
| 38364 | 10493309-6402534 | 57.0 ± 0.2 | 4662 ± 79 | ... | 1.026 ± 0.003 | -0.09 ± 0.06 | <30 | 3 | N | ... | ... | ... | ... | ... | ... | ... | n | G |
| 38099 | 10453520-6327169 | -14.4 ± 0.2 | 5020 ± 172 | ... | 1.022 ± 0.003 | -0.16 ± 0.21 | <22 | 3 | N | ... | ... | ... | ... | ... | ... | ... | n | G |
| 38365 | 10493386-6506007 | -41.9 ± 0.2 | 4494 ± 171 | 2.23 ± 0.18 | 1.026 ± 0.004 | 0.09 ± 0.09 | <59 | 3 | N | N | ... | ... | ... | ... | ... | ... | n | G |
| 38366 | 10493503-6457131 | -0.9 ± 0.2 | 4758 ± 37 | 2.58 ± 0.08 | 1.021 ± 0.004 | -0.06 ± 0.03 | <29 | 3 | N | N | ... | ... | ... | ... | ... | ... | n | G |
| 53910 | 10453544-6427031 | 17.5 ± 0.2 | 4824 ± 77 | 2.63 ± 0.15 | 1.021 ± 0.003 | -0.04 ± 0.07 | <33 | 3 | N | N | ... | ... | ... | ... | ... | ... | n | G |
| 38367 | 10493529-6335293 | 40.1 ± 0.2 | 4553 ± 125 | 2.65 ± 0.06 | 1.001 ± 0.005 | 0.15 ± 0.09 | <60 | 3 | Y | N | Y | Y | N | Y | ... | ... | n | NG? |
| 38389 | 10495255-6403061 | 14.5 ± 0.2 | 4435 ± 187 | 2.10 ± 0.18 | 1.028 ± 0.003 | 0.18 ± 0.01 | ... | ... | N | N | ... | ... | ... | ... | ... | ... | n | G |
| 53863 | 10445829-6418594 | 26.8 ± 0.2 | 3950 ± 21 | ... | 1.033 ± 0.003 | -0.07 ± 0.15 | 124 ± 3 | 1 | N | ... | ... | ... | ... | ... | ... | ... | n | ... |
| 38390 | 10495292-6505383 | -38.7 ± 0.2 | 4860 ± 112 | 2.74 ± 0.19 | 1.020 ± 0.003 | -0.02 ± 0.06 | <30 | 3 | N | N | ... | ... | ... | ... | ... | ... | n | G |
| 53911 | 10453564-6422495 | -13.5 ± 0.5 | 6437 ± 59 | ... | ... | -0.26 ± 0.05 | ... | ... | ... | ... | ... | ... | ... | ... | ... | ... | n | ... |
| 38391 | 10495370-6445091 | 73.7 ± 0.2 | 4409 ± 199 | ... | 1.045 ± 0.004 | -0.11 ± 0.01 | <29 | 3 | N | ... | ... | ... | ... | ... | ... | ... | n | G |
| 38392 | 10495417-6431056 | -6.0 ± 0.2 | 4581 ± 38 | 2.38 ± 0.11 | 1.023 ± 0.004 | 0.01 ± 0.07 | <41 | 3 | N | N | ... | ... | ... | ... | ... | ... | n | G |
| 38393 | 10495445-6511510 | -60.6 ± 0.2 | 4732 ± 8 | 2.42 ± 0.11 | 1.030 ± 0.003 | -0.05 ± 0.08 | 76 ± 3 | 1 | N | N | ... | ... | ... | ... | ... | ... | n | G |
| 38394 | 10495499-6511118 | -6.0 ± 0.2 | 5089 ± 52 | ... | 1.013 ± 0.004 | -0.07 ± 0.11 | ... | ... | N | ... | ... | ... | ... | ... | ... | ... | n | G |
| 2596 | 10495515-6357402 | -40.2 ± 0.4 | 7121 ± 185 | 3.98 ± 0.19 | ... | -0.14 ± 0.17 | <6 | 3 | ... | Y | N | N | N | Y | N | ... | n | ... |
| 38395 | 10495621-6348452 | 59.3 ± 0.2 | 4772 ± 77 | 2.61 ± 0.01 | 1.016 ± 0.006 | -0.06 ± 0.01 | <43 | 3 | N | N | ... | ... | ... | ... | ... | ... | n | G |
| 53868 | 10450285-6423175 | 92.7 ± 0.2 | 3955 ± 64 | ... | 1.057 ± 0.003 | -0.15 ± 0.09 | 74 ± 3 | 1 | N | ... | ... | ... | ... | ... | ... | ... | n | G |

**Table C.6.** continued.

| ID | CNAME | $RV$ (km s$^{-1}$) | $T_{\text{eff}}$ (K) | $logg$ (dex) | $\gamma^a$ | [Fe/H] (dex) | $EW(Li)^b$ (mÅ) | $EW(Li)$ error flag$^c$ | \multicolumn{6}{c}{Membership} | \multicolumn{2}{c}{Gaia studies} | Final$^e$ | NMs with Li$^f$ |
|---|---|---|---|---|---|---|---|---|---|---|---|---|---|---|---|---|---|---|
| | | | | | | | | | $\gamma$ | $logg$ | RV | Li | H$\alpha$ | [Fe/H] | Randich$^d$ | Cantat-Gaudin$^d$ | | |
| 38396 | 10495631-6344354 | 14.9 ± 0.2 | 4722 ± 4 | 2.67 ± 0.05 | 1.009 ± 0.004 | 0.08 ± 0.08 | <37 | 3 | Y | N | Y | N | N | Y | ... | ... | n | NG? |
| 38397 | 10495661-6352528 | 44.8 ± 0.2 | 4618 ± 206 | ... | 1.018 ± 0.006 | -0.15 ± 0.17 | <32 | 3 | N | ... | ... | ... | ... | ... | ... | ... | n | G |
| 53917 | 10454218-6436509 | 67.6 ± 0.2 | 4399 ± 235 | ... | 1.025 ± 0.005 | 0.12 ± 0.09 | <69 | 3 | N | ... | ... | ... | ... | ... | ... | ... | n | G |
| 2597 | 10495667-6359128 | 4.6 ± 0.6 | 4883 ± 66 | 2.48 ± 0.12 | 1.023 ± 0.002 | 0.04 ± 0.02 | 36 ± 9 | 1 | N | N | ... | ... | ... | ... | ... | ... | n | ... |
| 38108 | 10454219-6457446 | 8.5 ± 0.2 | 4684 ± 40 | 2.66 ± 0.17 | 1.006 ± 0.004 | -0.05 ± 0.02 | <44 | 3 | Y | N | Y | N | N | Y | ... | ... | n | NG? |
| 2598 | 10495685-6348191 | 18.1 ± 0.6 | 4714 ± 103 | 4.46 ± 0.17 | ... | -0.03 ± 0.05 | 305 ± 17 | 2 | ... | Y | Y | Y | Y | Y | Y | ... | Y | ... |
| 38065 | 10450352-6424086 | 0.1 ± 0.2 | 4782 ± 124 | 2.66 ± 0.09 | 1.016 ± 0.003 | 0.08 ± 0.08 | <38 | 3 | N | N | ... | ... | ... | ... | ... | ... | n | G |
| 38109 | 10454226-6349198 | -3.8 ± 0.2 | 4577 ± 86 | ... | 1.027 ± 0.003 | ... | <51 | 3 | N | ... | ... | ... | ... | ... | ... | ... | n | G |
| 38398 | 10495719-6341212 | 13.8 ± 0.2 | 4789 ± 118 | 2.55 ± 0.09 | 1.017 ± 0.002 | 0.03 ± 0.10 | 335 ± 3 | 1 | N | N | ... | ... | ... | ... | ... | ... | n | Li-rich G |
| 38399 | 10495745-6515473 | -27.1 ± 0.2 | 4985 ± 125 | 3.03 ± 0.17 | 1.010 ± 0.003 | -0.06 ± 0.10 | <25 | 3 | Y | N | N | N | N | Y | ... | ... | n | NG? |
| 38110 | 10454301-6411013 | 3.6 ± 0.2 | 4831 ± 99 | ... | 1.024 ± 0.002 | -0.06 ± 0.10 | <33 | 3 | N | ... | ... | ... | ... | ... | ... | ... | n | G |
| 38400 | 10495828-6345473 | 34.6 ± 0.2 | 4899 ± 81 | 2.83 ± 0.05 | 1.008 ± 0.006 | -0.02 ± 0.03 | <4 | 3 | Y | N | Y | N | N | Y | ... | ... | n | ... |
| 38066 | 10450372-6422118 | 26.2 ± 0.2 | 4727 ± 57 | 2.56 ± 0.08 | 1.014 ± 0.002 | 0.03 ± 0.08 | <41 | 3 | N | N | ... | ... | ... | ... | ... | ... | n | G |
| 38401 | 10495890-6346392 | 12.8 ± 0.2 | 5055 ± 86 | ... | 1.016 ± 0.001 | -0.05 ± 0.11 | <32 | 3 | N | ... | ... | ... | ... | ... | ... | ... | n | G |
| 38402 | 10495931-6337032 | 76.2 ± 0.2 | 4821 ± 20 | 2.78 ± 0.12 | 1.007 ± 0.005 | -0.04 ± 0.05 | <31 | 3 | Y | N | N | N | N | Y | ... | ... | n | NG? |
| 38403 | 10495937-6345553 | -14.8 ± 0.2 | 4934 ± 219 | ... | 1.021 ± 0.003 | -0.09 ± 0.13 | 477 ± 5 | 1 | N | ... | ... | ... | ... | ... | ... | ... | n | Li-rich G |
| 38404 | 10495999-6434153 | -17.8 ± 0.2 | 4921 ± 30 | 2.77 ± 0.05 | 1.022 ± 0.003 | 0.00 ± 0.02 | <34 | 3 | N | N | ... | ... | ... | ... | ... | ... | n | G |
| 38405 | 10500061-6459125 | 37.3 ± 0.2 | 4510 ± 148 | 2.28 ± 0.19 | 1.018 ± 0.004 | 0.04 ± 0.08 | <42 | 3 | N | N | ... | ... | ... | ... | ... | ... | n | G |
| 53869 | 10450391-6428050 | 382.6 ± 3.9 | ... | ... | ... | ... | ... | ... | ... | ... | ... | ... | ... | ... | ... | ... | n | ... |
| 38111 | 10454445-6318290 | 3.0 ± 0.2 | 4990 ± 123 | ... | 1.016 ± 0.005 | -0.09 ± 0.14 | ... | ... | N | ... | ... | ... | ... | ... | ... | ... | n | G |
| 38406 | 10500136-6347217 | 53.6 ± 0.2 | 4616 ± 165 | 2.40 ± 0.17 | 1.017 ± 0.005 | 0.01 ± 0.05 | 13 ± 3 | 1 | N | N | ... | ... | ... | ... | ... | ... | n | G |
| 38407 | 10500266-6337434 | -16.6 ± 0.3 | 5131 ± 94 | ... | 1.000 ± 0.006 | ... | <23 | 3 | Y | ... | N | N | N | ... | ... | ... | n | NG |
| 38112 | 10454482-6458402 | 20.0 ± 0.2 | 4304 ± 260 | ... | 1.044 ± 0.003 | -0.20 ± 0.05 | <26 | 3 | N | ... | ... | ... | ... | ... | ... | ... | n | G |
| 38408 | 10500348-6339037 | -3.3 ± 0.2 | 4699 ± 67 | 2.44 ± 0.12 | 1.031 ± 0.004 | 0.04 ± 0.03 | 150 ± 3 | 1 | N | N | ... | ... | ... | ... | ... | ... | n | Li-rich G |
| 53918 | 10454493-6435034 | -29.9 ± 0.2 | 4949 ± 18 | ... | 1.019 ± 0.001 | 0.00 ± 0.05 | <32 | 3 | N | ... | ... | ... | ... | ... | ... | ... | n | G |
| 38454 | 10502835-6503295 | -26.0 ± 0.2 | 4675 ± 74 | 2.49 ± 0.16 | 1.014 ± 0.003 | -0.11 ± 0.11 | <28 | 3 | N | N | ... | ... | ... | ... | ... | ... | n | G |
| 38455 | 10502863-6444176 | -13.8 ± 0.2 | 5280 ± 152 | ... | 1.014 ± 0.003 | -0.03 ± 0.05 | 55 ± 9 | 1 | N | ... | ... | ... | ... | ... | ... | ... | n | ... |
| 38456 | 10503071-6343037 | 33.8 ± 0.2 | 5022 ± 292 | ... | 1.013 ± 0.004 | -0.25 ± 0.19 | <12 | 3 | N | ... | ... | ... | ... | ... | ... | ... | n | G |
| 38457 | 10503148-6344589 | -1.2 ± 0.2 | 5144 ± 9 | ... | 1.016 ± 0.002 | -0.02 ± 0.06 | <26 | 3 | N | ... | ... | ... | ... | ... | ... | ... | n | G |
| 38113 | 10454523-6348339 | -16.3 ± 0.2 | 4930 ± 133 | ... | 1.025 ± 0.003 | -0.04 ± 0.08 | <27 | 3 | N | ... | ... | ... | ... | ... | ... | ... | n | G |
| 38114 | 10454534-6417216 | -1.6 ± 0.2 | 4473 ± 139 | 2.20 ± 0.17 | 1.023 ± 0.002 | 0.14 ± 0.06 | <53 | 3 | N | N | ... | ... | ... | ... | ... | ... | n | G |
| 38458 | 10503152-6359592 | -58.0 ± 0.2 | 4531 ± 85 | 2.18 ± 0.20 | 1.035 ± 0.002 | -0.10 ± 0.09 | <37 | 3 | N | N | ... | ... | ... | ... | ... | ... | n | G |
| 2576 | 10454789-6500411 | -27.2 ± 0.4 | 7223 ± 373 | 3.95 ± 0.32 | ... | -0.27 ± 0.25 | <10 | 3 | ... | Y | N | N | N | Y | N | ... | n | NG |
| 38459 | 10503200-6332481 | 4.3 ± 0.2 | 5055 ± 139 | ... | 1.015 ± 0.001 | -0.06 ± 0.12 | <23 | 3 | N | ... | ... | ... | ... | ... | ... | ... | n | G |
| 38460 | 10503249-6333502 | 15.4 ± 0.2 | 4404 ± 238 | ... | 1.029 ± 0.002 | 0.13 ± 0.09 | <41 | 3 | N | ... | ... | ... | ... | ... | ... | ... | n | G |
| 53921 | 10454817-6430481 | 41.9 ± 0.2 | 4648 ± 42 | 2.40 ± 0.17 | 1.026 ± 0.002 | -0.05 ± 0.09 | <35 | 3 | N | N | ... | ... | ... | ... | ... | ... | n | G |
| 38461 | 10503279-6400355 | 3.5 ± 0.2 | 5044 ± 100 | ... | 1.016 ± 0.002 | -0.03 ± 0.07 | ... | ... | N | ... | ... | ... | ... | ... | ... | ... | n | G |
| 38462 | 10503309-6331510 | -9.5 ± 0.2 | 4736 ± 110 | 2.58 ± 0.12 | 1.017 ± 0.004 | 0.04 ± 0.01 | <45 | 3 | N | N | ... | ... | ... | ... | ... | ... | n | G |
| 38463 | 10503394-6344386 | 6.6 ± 0.2 | 4504 ± 172 | ... | 1.014 ± 0.006 | 0.16 ± 0.01 | <66 | 3 | N | ... | ... | ... | ... | ... | ... | ... | n | G |
| 38464 | 10503413-6345352 | -6.7 ± 0.2 | 4905 ± 173 | ... | 1.018 ± 0.004 | -0.06 ± 0.10 | <21 | 3 | N | ... | ... | ... | ... | ... | ... | ... | n | G |
| 38465 | 10503506-6505395 | -28.8 ± 0.2 | 4987 ± 219 | ... | 1.017 ± 0.003 | -0.14 ± 0.18 | <18 | 3 | N | ... | ... | ... | ... | ... | ... | ... | n | G |
| 38466 | 10503594-6347284 | 4.8 ± 0.2 | 4704 ± 33 | 2.57 ± 0.07 | 1.013 ± 0.004 | 0.05 ± 0.04 | <40 | 3 | N | N | ... | ... | ... | ... | ... | ... | n | G |
| 38467 | 10503595-6510448 | -12.2 ± 0.2 | 4710 ± 51 | 2.48 ± 0.03 | 1.028 ± 0.003 | 0.00 ± 0.05 | 19 ± 7 | 1 | N | N | ... | ... | ... | ... | ... | ... | n | G |
| 38468 | 10503609-6333422 | 48.5 ± 0.2 | 4468 ± 197 | 2.24 ± 0.07 | 1.008 ± 0.005 | 0.08 ± 0.09 | <52 | 3 | Y | N | N | N | N | Y | ... | ... | n | NG? |
| 38469 | 10503631-6512237 | -34.1 ± 0.2 | 4708 ± 61 | 2.49 ± 0.15 | 1.023 ± 0.003 | -0.05 ± 0.06 | 286 ± 4 | 1 | N | N | ... | ... | ... | ... | ... | ... | n | Li-rich G |
| 38470 | 10503658-6345472 | 22.1 ± 0.3 | 4721 ± 247 | ... | 0.896 ± 0.005 | ... | <11 | 3 | Y | ... | Y | N | Y | ... | ... | ... | n | NG |
| 38471 | 10503676-6423294 | -9.4 ± 0.2 | 4645 ± 8 | 2.37 ± 0.16 | 1.031 ± 0.002 | -0.08 ± 0.03 | 45 ± 2 | 1 | N | N | ... | ... | ... | ... | ... | ... | n | G |
| 38472 | 10503694-6511570 | -19.2 ± 0.2 | 5104 ± 45 | 3.49 ± 0.15 | 1.004 ± 0.003 | -0.11 ± 0.07 | <28 | 3 | Y | N | N | N | N | Y | ... | ... | n | NG? |
| 38473 | 10503790-6343297 | -12.8 ± 0.2 | 4920 ± 150 | ... | 1.023 ± 0.002 | -0.04 ± 0.07 | 15 | 3 | N | ... | ... | ... | ... | ... | ... | ... | n | G |
| 38474 | 10503897-6507043 | 22.4 ± 0.2 | 5117 ± 58 | 3.55 ± 0.12 | 1.002 ± 0.004 | -0.02 ± 0.08 | <15 | 3 | Y | Y | Y | N | N | Y | ... | ... | n | NG? |
| 38475 | 10503907-6422237 | 7.3 ± 0.2 | 4573 ± 91 | 2.47 ± 0.16 | 1.017 ± 0.002 | 0.16 ± 0.01 | <67 | 3 | N | N | ... | ... | ... | ... | ... | ... | n | G |
| 38476 | 10503939-6439159 | 74.5 ± 0.2 | 4556 ± 79 | ... | 1.069 ± 0.004 | -0.22 ± 0.09 | ... | ... | N | ... | ... | ... | ... | ... | ... | ... | n | G |
| 38477 | 10504080-6341117 | -11.0 ± 0.2 | 5156 ± 9 | ... | 1.014 ± 0.003 | 0.03 ± 0.02 | <26 | 3 | N | ... | ... | ... | ... | ... | ... | ... | n | G |
| 38500 | 10505457-6418101 | 19.8 ± 0.2 | 4631 ± 102 | ... | 1.021 ± 0.005 | ... | <62 | 3 | N | ... | ... | ... | ... | ... | ... | ... | n | G |
| 38501 | 10505463-6344229 | -22.9 ± 0.2 | 5059 ± 104 | ... | 1.011 ± 0.004 | -0.14 ± 0.11 | ... | ... | N | ... | ... | ... | ... | ... | ... | ... | n | G |
| 38502 | 10505480-6336577 | -41.6 ± 0.2 | 5027 ± 58 | ... | 1.017 ± 0.003 | -0.08 ± 0.09 | <17 | 3 | N | ... | ... | ... | ... | ... | ... | ... | n | G |

**Notes.** [a] Empirical gravity indicator defined by Damiani et al. (2014). [b] The values of $EW$(Li) for this cluster are corrected (subtracted adjacent Fe (6707.43 Å) line). [c] Flags for the errors of the corrected $EW$(Li) values, as follows: 1=$EW$(Li) corrected by blends contribution using models; 2=$EW$(Li) measured separately (Li line resolved - UVES only); and 3=Upper limit (no error for $EW$(Li) is given). [d] Randich et al. (2018), Cantat-Gaudin et al. (2018) [e] The letters "Y" and "N" indicate if the star is a cluster member or not. [f] 'Li-rich G', 'G' and 'NG' indicate "Li-rich G" "giant" and "non-giant" Li field contaminants, respectively.

**Table C.7.** IC 4665

| ID | CNAME | RV (km s⁻¹) | $T_{\rm eff}$ (K) | logg (dex) | $\gamma^a$ | [Fe/H] (dex) | $EW({\rm Li})^b$ (mÅ) | EW(Li) error flag$^c$ | $\gamma$ | logg | RV | Li | H$\alpha$ | [Fe/H] | Randich$^d$ | Cantat-Gaudin$^d$ | Final$^e$ | NMs with Li$^f$ |
|---|---|---|---|---|---|---|---|---|---|---|---|---|---|---|---|---|---|---|
| 38671 | 17442333+0602275 | -25.1 ± 1.0 | … | … | … | … | … | … | … | … | … | … | N | … | N | … | n | … |
| 38672 | 17442342+0555437 | -5.7 ± 0.3 | 5084 ± 172 | … | 0.977 ± 0.011 | -0.05 ± 0.09 | <25 | 3 | Y | … | Y | N | N | Y | … | … | n | NG |
| 38673 | 17442485+0556585 | 0.3 ± 0.3 | 4519 ± 4 | 4.53 ± 0.06 | 0.938 ± 0.006 | -0.05 ± 0.07 | <22 | 3 | Y | Y | Y | Y | N | Y | … | … | n | NG |
| 2615 | 17442695+0554052 | -13.7 ± 0.6 | 5341 ± 180 | 4.35 ± 0.26 | … | -0.01 ± 0.08 | 291 ± 13 | 2 | … | Y | Y | Y | Y | Y | Y | … | Y | … |
| 38674 | 17442698+0551456 | -16.1 ± 0.4 | 3647 ± 23 | … | 0.829 ± 0.010 | -0.19 ± 0.13 | … | … | Y | … | Y | Y | Y | Y | N | Y | n | … |
| 38764 | 17451652+0517499 | -32.0 ± 0.2 | 4471 ± 187 | … | 1.033 ± 0.003 | 0.09 ± 0.06 | <57 | 3 | N | … | … | … | … | … | … | … | n | G |
| 38765 | 17451670+0553564 | -8.9 ± 0.3 | 3643 ± 82 | 4.57 ± 0.20 | 0.812 ± 0.012 | -0.21 ± 0.14 | … | … | Y | Y | Y | … | N | Y | N | … | n | … |
| 38766 | 17451721+0519145 | -1.4 ± 0.2 | 5072 ± 118 | 3.53 ± 0.24 | 0.997 ± 0.006 | -0.18 ± 0.10 | <34 | 3 | Y | Y | Y | N | N | Y | … | … | n | NG |
| 38767 | 17451779+0558003 | -27.0 ± 0.2 | 5357 ± 170 | 3.96 ± 0.17 | 0.994 ± 0.003 | 0.04 ± 0.08 | <35 | 3 | Y | Y | Y | N | N | Y | … | … | n | NG |
| 38768 | 17451835+0515570 | -15.0 ± 0.2 | 4975 ± 198 | … | 1.007 ± 0.002 | -0.22 ± 0.27 | 9 ± 3 | 1 | Y | … | Y | N | N | Y | … | … | n | … |
| 38769 | 17451872+0557341 | -50.3 ± 0.2 | 5361 ± 204 | 4.17 ± 0.09 | 0.983 ± 0.005 | 0.09 ± 0.13 | <33 | 3 | Y | Y | N | Y | Y | Y | … | … | n | NG |
| 38770 | 17451905+0548491 | 32.9 ± 0.3 | 3607 ± 83 | … | 0.822 ± 0.011 | -0.26 ± 0.15 | <100 | 3 | Y | … | N | Y | N | Y | … | … | n | NG |
| 38771 | 17451906+0603394 | 64.2 ± 0.2 | 4643 ± 98 | … | 1.019 ± 0.006 | … | <35 | 3 | N | … | … | … | … | … | … | … | n | G |
| 38772 | 17451922+0528376 | 40.2 ± 0.2 | 4686 ± 4 | 2.59 ± 0.10 | 1.007 ± 0.004 | -0.11 ± 0.09 | <33 | 3 | Y | N | N | N | N | Y | … | … | n | NG? |
| 2623 | 17451940+0547401 | -12.6 ± 0.6 | 5309 ± 106 | 4.47 ± 0.09 | 0.993 ± 0.004 | -0.06 ± 0.01 | 235 ± 9 | 2 | Y | Y | Y | Y | Y | Y | Y | Y | Y | … |
| 38773 | 17451940+0550436 | -34.5 ± 0.2 | 5092 ± 98 | … | 1.015 ± 0.009 | … | <23 | 3 | N | … | … | … | … | … | … | … | n | G |
| 38774 | 17451961+0539392 | -16.3 ± 0.4 | 3528 ± 61 | … | 0.851 ± 0.009 | -0.23 ± 0.15 | … | … | Y | … | Y | Y | Y | Y | N | … | n | … |
| 38775 | 17451964+0536547 | 5.1 ± 0.2 | 5130 ± 30 | 4.30 ± 0.44 | … | -0.09 ± 0.04 | … | … | … | Y | Y | Y | Y | Y | … | Y | n | … |
| 38776 | 17451998+0546295 | -13.5 ± 0.2 | 4255 ± 262 | 4.42 ± 0.18 | 0.896 ± 0.004 | 0.03 ± 0.15 | <15 | 3 | Y | Y | Y | Y | Y | Y | … | Y | Y | … |
| 38777 | 17452017+0557346 | -19.6 ± 0.3 | 4694 ± 264 | … | 1.048 ± 0.007 | -0.07 ± 0.07 | <26 | 3 | N | … | … | … | … | … | … | … | n | G |
| 38778 | 17452041+0528130 | -4.3 ± 0.9 | 3979 ± 240 | 4.60 ± 0.06 | 0.811 ± 0.016 | -0.35 ± 0.30 | … | … | Y | Y | Y | … | Y | N | N | … | n | … |
| 38779 | 17452068+0521225 | 115.1 ± 0.3 | 5232 ± 197 | … | 1.013 ± 0.005 | -0.41 ± 0.15 | … | … | N | … | … | … | … | … | … | … | n | … |
| 38780 | 17452085+0531282 | 6.2 ± 0.3 | 5028 ± 192 | … | 1.009 ± 0.005 | -0.25 ± 0.16 | <34 | 3 | Y | … | Y | N | N | Y | … | … | n | NG? |
| 38781 | 17452087+0554277 | -36.7 ± 0.2 | 4999 ± 127 | … | 1.021 ± 0.004 | -0.28 ± 0.15 | … | … | N | … | … | … | … | … | … | … | n | G |
| 38782 | 17452140+0536265 | -35.5 ± 0.3 | 4287 ± 283 | 4.44 ± 0.06 | 0.889 ± 0.007 | -0.37 ± 0.21 | <15 | 3 | Y | Y | Y | Y | Y | N | … | … | Y | … |
| 38783 | 17452184+0529191 | -74.8 ± 0.2 | 5033 ± 149 | … | 1.019 ± 0.005 | -0.37 ± 0.08 | <10 | 3 | N | … | … | … | … | … | … | … | n | G |
| 38784 | 17452204+0558257 | 30.3 ± 0.2 | 4965 ± 82 | 2.99 ± 0.19 | 1.010 ± 0.004 | -0.16 ± 0.16 | <10 | 3 | Y | N | N | N | N | Y | … | … | n | NG? |
| 38785 | 17452238+0558599 | -10.5 ± 0.3 | 5006 ± 156 | … | 1.002 ± 0.006 | -0.15 ± 0.22 | <24 | 3 | Y | … | Y | N | N | Y | … | … | n | NG? |
| 38786 | 17452291+0548517 | -23.7 ± 0.2 | 4565 ± 51 | 2.35 ± 0.17 | 1.024 ± 0.004 | -0.03 ± 0.04 | <38 | 3 | N | N | … | … | … | … | … | … | n | G |
| 38874 | 17455394+0559234 | -5.2 ± 0.3 | 3888 ± 177 | 4.60 ± 0.06 | 0.807 ± 0.011 | -0.14 ± 0.12 | … | … | Y | Y | Y | … | Y | Y | N | … | n | … |
| 38875 | 17455487+0533252 | 17.8 ± 0.2 | 4564 ± 88 | 4.40 ± 0.11 | 0.944 ± 0.004 | 0.00 ± 0.10 | <19 | 3 | Y | Y | N | N | N | Y | … | … | n | NG |
| 38876 | 17455592+0521315 | -59.2 ± 0.2 | 4655 ± 52 | 2.41 ± 0.14 | 1.023 ± 0.005 | -0.05 ± 0.09 | <33 | 3 | N | N | … | … | … | … | … | … | n | G |
| 2630 | 17455598+0602542 | -14.6 ± 0.6 | 5653 ± 98 | 4.49 ± 0.14 | 0.996 ± 0.003 | 0.02 ± 0.15 | 232 ± 2 | 2 | Y | Y | Y | Y | Y | Y | Y | Y | Y | … |
| 38877 | 17455616+0608460 | 6.1 ± 0.2 | 4896 ± 233 | 3.23 ± 0.03 | 0.998 ± 0.005 | -0.07 ± 0.15 | <19 | 3 | Y | Y | Y | N | Y | Y | … | … | n | NG? |
| 38878 | 17455640+0608291 | -37.2 ± 0.3 | 4926 ± 216 | … | 1.020 ± 0.006 | -0.38 ± 0.25 | <19 | 3 | N | … | … | … | … | … | … | … | n | G |
| 38879 | 17455673+0552240 | 11.0 ± 0.3 | 3996 ± 157 | 4.51 ± 0.07 | 0.846 ± 0.008 | -0.26 ± 0.19 | <40 | 3 | Y | Y | Y | N | Y | Y | N | … | n | NG |
| 38880 | 17455679+0522334 | -24.7 ± 0.2 | 5760 ± 81 | … | 1.000 ± 0.002 | … | 135 ± 3 | 1 | Y | … | Y | Y | N | … | N | Y | n | NG? |
| 2631 | 17455679+0601044 | -35.4 ± 0.6 | 6285 ± 75 | 4.15 ± 0.02 | … | 0.17 ± 0.03 | <14 | 3 | … | Y | N | N | N | N | N | … | n | NG |
| 38881 | 17455682+0529055 | 43.9 ± 0.2 | 4851 ± 57 | 2.72 ± 0.10 | 1.009 ± 0.004 | -0.15 ± 0.13 | <24 | 3 | Y | N | N | N | N | Y | … | … | n | NG? |
| 38882 | 17455685+0603582 | -9.5 ± 0.2 | 4687 ± 208 | … | 1.024 ± 0.003 | -0.34 ± 0.09 | <15 | 3 | N | … | … | … | … | … | … | … | n | … |
| 38884 | 17455842+0521010 | -44.0 ± 0.2 | 4720 ± 43 | 2.87 ± 0.11 | 1.004 ± 0.005 | 0.02 ± 0.10 | <50 | 3 | Y | N | N | Y | Y | Y | … | … | n | NG? |
| 38885 | 17455849+0538087 | -30.1 ± 0.2 | 5159 ± 44 | … | 0.977 ± 0.004 | 0.09 ± 0.09 | <36 | 3 | Y | … | Y | N | Y | Y | … | … | n | NG |
| 38886 | 17455931+0541010 | -44.9 ± 0.3 | 3899 ± 113 | 4.58 ± 0.08 | 0.825 ± 0.012 | -0.08 ± 0.07 | … | … | Y | Y | N | … | … | Y | N | … | n | … |
| 38887 | 17455945+0550392 | 52.7 ± 0.2 | 5059 ± 64 | … | 1.016 ± 0.005 | -0.20 ± 0.13 | <9 | 3 | N | … | … | … | … | … | … | … | n | G |
| 38888 | 17455960+0554007 | -0.3 ± 0.2 | 4411 ± 213 | … | 1.027 ± 0.003 | -0.18 ± 0.12 | <20 | 3 | N | … | … | … | … | … | … | … | n | … |
| 38889 | 17455975+0604227 | -78.4 ± 0.3 | 3293 ± 95 | … | 0.868 ± 0.009 | … | <100 | 3 | Y | … | N | Y | N | … | N | … | n | NG |
| 38890 | 17455991+0608556 | -246.2 ± 0.3 | 4562 ± 198 | … | 1.044 ± 0.007 | -0.46 ± 0.08 | … | … | N | … | … | … | … | … | … | … | n | G |
| 38891 | 17460061+0536037 | 36.0 ± 0.3 | 4918 ± 6 | … | 1.001 ± 0.008 | -0.02 ± 0.12 | <19 | 3 | Y | … | N | N | N | Y | … | … | n | NG? |
| 38892 | 17460089+0537064 | -52.5 ± 0.2 | 4703 ± 28 | 2.43 ± 0.05 | 1.015 ± 0.005 | -0.07 ± 0.04 | <29 | 3 | Y | N | … | … | … | … | … | … | n | G |
| 38893 | 17460097+0600085 | 27.7 ± 0.3 | 4077 ± 27 | 4.61 ± 0.07 | 0.866 ± 0.008 | -0.23 ± 0.12 | <32 | 3 | Y | Y | N | Y | Y | Y | Y | … | n | … |
| 38894 | 17460173+0537166 | 13.0 ± 0.3 | 5645 ± 28 | … | 0.993 ± 0.005 | 0.09 ± 0.13 | <17 | 3 | Y | … | Y | N | Y | Y | … | … | n | NG |
| 38895 | 17460197+0550254 | -24.6 ± 0.3 | 4071 ± 255 | 4.47 ± 0.08 | 0.852 ± 0.008 | -0.15 ± 0.04 | … | … | Y | Y | Y | … | N | Y | N | … | n | … |
| 38987 | 17464065+0528541 | 53.7 ± 0.2 | 4910 ± 118 | … | 1.023 ± 0.006 | -0.26 ± 0.22 | <36 | 3 | Y | … | Y | N | N | Y | … | … | n | G |
| 38988 | 17464101+0521133 | -84.3 ± 0.2 | 5114 ± 228 | … | 1.016 ± 0.002 | -0.45 ± 0.07 | <2 | 3 | N | … | … | … | … | … | … | … | n | G |
| 38989 | 17464104+0527071 | -164.6 ± 0.2 | 4660 ± 190 | 2.39 ± 0.18 | 1.025 ± 0.004 | -0.02 ± 0.03 | … | … | N | N | … | … | … | … | … | … | n | G |
| 38990 | 17464176+0526162 | 1.8 ± 0.2 | 5028 ± 197 | … | 1.013 ± 0.002 | -0.27 ± 0.23 | <18 | 3 | N | … | … | … | … | … | … | … | n | G |
| 38991 | 17464182+0558289 | -3.2 ± 0.4 | 3995 ± 174 | 4.46 ± 0.16 | 0.854 ± 0.015 | 0.02 ± 0.20 | <50 | 3 | Y | Y | Y | N | N | Y | N | … | n | NG |
| 38992 | 17464221+0533420 | -81.1 ± 0.3 | 4794 ± 103 | 4.38 ± 0.13 | 0.962 ± 0.009 | -0.17 ± 0.03 | <18 | 3 | Y | Y | N | N | Y | Y | … | … | n | NG |
| 38993 | 17464224+0606421 | -63.2 ± 0.2 | 5236 ± 62 | … | 0.997 ± 0.003 | … | … | … | Y | … | N | … | … | … | … | … | n | … |

**Table C.7.** continued.

| ID | CNAME | RV (km s$^{-1}$) | $T_{\text{eff}}$ (K) | logg (dex) | $\gamma^a$ | [Fe/H] (dex) | EW(Li)$^b$ (mÅ) | EW(Li) error flag$^c$ | $\gamma$ | logg | RV | Li | H$\alpha$ | [Fe/H] | Randich$^d$ | Cantat-Gaudin$^d$ | Final$^e$ | NMs with Li$^f$ |
|---|---|---|---|---|---|---|---|---|---|---|---|---|---|---|---|---|---|---|
| 38994 | 17464228+0601506 | -13.5 ± 0.3 | 3797 ± 81 | … | 0.823 ± 0.012 | -0.17 ± 0.14 | <100 | 3 | Y | … | Y | Y | Y | Y | Y | … | Y | … |
| 38995 | 17464238+0525514 | -28.9 ± 0.2 | 4994 ± 141 | 4.12 ± 0.04 | 0.973 ± 0.004 | -0.03 ± 0.01 | <14 | 3 | Y | Y | Y | N | N | Y | … | … | n | NG |
| 38996 | 17464256+0528003 | 43.8 ± 0.2 | 4742 ± 106 | … | 1.019 ± 0.006 | -0.23 ± 0.16 | <16 | 3 | N | … | … | … | … | … | … | … | n | G |
| 38997 | 17464332+0535140 | 31.5 ± 0.3 | 3985 ± 158 | … | 0.871 ± 0.007 | -0.22 ± 0.09 | … | … | Y | … | N | … | … | Y | N | … | n | … |
| 38998 | 17464426+0528335 | 9.2 ± 0.2 | 6451 ± 36 | … | 0.999 ± 0.002 | 0.07 ± 0.03 | <17 | 3 | Y | … | Y | N | N | Y | N | … | n | NG |
| 38999 | 17464511+0537339 | 18.7 ± 0.2 | 6002 ± 125 | 4.20 ± 0.07 | 0.996 ± 0.002 | 0.17 ± 0.03 | <21 | 3 | Y | Y | N | N | N | N | N | … | n | NG |
| 39000 | 17464514+0526582 | 121.8 ± 0.2 | 4975 ± 150 | … | 1.012 ± 0.003 | -0.17 ± 0.21 | <16 | 3 | N | … | … | … | … | … | … | … | n | G |
| 39001 | 17464517+0526046 | 10.7 ± 0.2 | 5328 ± 47 | … | 0.998 ± 0.002 | -0.05 ± 0.14 | <18 | 3 | Y | … | Y | N | N | Y | … | … | n | NG |
| 39002 | 17464549+0554555 | -79.1 ± 0.5 | 3541 ± 2 | … | 0.845 ± 0.014 | -0.27 ± 0.13 | … | … | Y | … | N | … | … | Y | N | … | n | … |
| 39003 | 17464554+0603563 | -17.9 ± 0.2 | 5127 ± 309 | … | 1.009 ± 0.004 | -0.59 ± 0.23 | <10 | 3 | Y | … | Y | N | N | N | … | … | n | NG? |
| 39004 | 17464657+0535068 | 45.6 ± 0.2 | 4984 ± 142 | … | 1.011 ± 0.004 | -0.21 ± 0.19 | <17 | 3 | N | … | … | … | … | … | … | … | n | G |
| 39005 | 17464801+0609193 | -14.4 ± 0.3 | 5068 ± 130 | … | 1.004 ± 0.009 | … | <42 | 3 | Y | … | Y | Y | N | … | … | … | n | NG? |
| 39006 | 17464824+0554522 | -17.8 ± 0.3 | 3996 ± 165 | 4.49 ± 0.10 | 0.849 ± 0.010 | -0.24 ± 0.14 | … | … | Y | Y | Y | … | Y | Y | N | … | n | … |
| 39007 | 17464839+0610015 | -16.2 ± 0.4 | 4525 ± 256 | … | 0.946 ± 0.014 | -0.13 ± 0.13 | <31 | 3 | Y | … | Y | Y | Y | Y | … | … | Y | … |
| 39008 | 17464874+0543006 | -4.3 ± 0.2 | 4870 ± 134 | 2.70 ± 0.12 | 1.011 ± 0.006 | -0.16 ± 0.11 | <16 | 3 | N | N | … | … | … | … | … | … | n | G |
| 39009 | 17464898+0522082 | 96.0 ± 0.2 | 5028 ± 154 | … | 1.017 ± 0.003 | -0.14 ± 0.19 | <22 | 3 | N | … | … | … | … | … | … | … | n | G |
| 39010 | 17464911+0526065 | -57.2 ± 0.3 | 5061 ± 42 | 3.64 ± 0.08 | 0.997 ± 0.009 | -0.22 ± 0.11 | <24 | 3 | Y | Y | N | N | N | Y | … | … | n | NG |
| 39102 | 17472822+0555189 | 17.7 ± 0.2 | 4825 ± 27 | 2.66 ± 0.02 | 1.017 ± 0.006 | 0.03 ± 0.04 | <40 | 3 | N | N | … | … | … | … | … | … | n | … |
| 39103 | 17472823+0608349 | -115.0 ± 0.2 | … | … | … | … | … | … | … | … | … | … | … | … | … | … | n | … |
| 39104 | 17472851+0603472 | 14.6 ± 0.2 | 4507 ± 102 | 2.26 ± 0.17 | 1.022 ± 0.002 | 0.01 ± 0.05 | 44 ± 2 | 1 | N | N | … | … | … | … | … | … | n | G |
| 39105 | 17472874+0523066 | -65.3 ± 0.2 | 4867 ± 12 | 3.11 ± 0.13 | 1.001 ± 0.002 | 0.02 ± 0.03 | 20 ± 3 | 1 | Y | N | N | N | N | Y | … | … | n | NG? |
| 39106 | 17472890+0521037 | 32.0 ± 0.2 | 4958 ± 205 | … | 0.970 ± 0.004 | -0.10 ± 0.03 | <15 | 3 | Y | … | N | N | N | Y | … | … | n | NG |
| 39107 | 17472923+0554503 | -37.1 ± 0.3 | 3973 ± 203 | … | 0.854 ± 0.008 | 0.03 ± 0.16 | <18 | 3 | Y | … | Y | Y | N | Y | N | … | n | NG |
| 39108 | 17472926+0603594 | -32.4 ± 0.2 | 5022 ± 120 | … | 1.025 ± 0.002 | … | <11 | 3 | N | … | … | … | … | … | … | … | n | G |
| 39109 | 17472970+0537100 | 10.2 ± 0.2 | 4037 ± 75 | 4.55 ± 0.01 | 0.859 ± 0.004 | -0.12 ± 0.06 | <6 | 3 | Y | Y | Y | N | Y | Y | N | … | n | … |
| 39111 | 17473004+0531365 | -12.2 ± 0.2 | 4988 ± 143 | … | 1.018 ± 0.003 | -0.16 ± 0.19 | <18 | 3 | N | … | … | … | … | … | … | … | n | G |
| 39112 | 17473004+0558496 | -45.2 ± 0.2 | 4732 ± 76 | 2.68 ± 0.06 | 1.008 ± 0.002 | -0.08 ± 0.10 | <31 | 3 | Y | N | N | N | N | Y | … | … | n | NG? |
| 39113 | 17473011+0527266 | -14.2 ± 0.2 | 4946 ± 157 | 3.24 ± 0.27 | 1.000 ± 0.004 | -0.03 ± 0.08 | <34 | 3 | Y | N | Y | N | N | Y | … | … | n | NG? |
| 39114 | 17473030+0540513 | -12.4 ± 0.2 | 4067 ± 154 | 4.56 ± 0.01 | 0.853 ± 0.004 | 0.19 ± 0.31 | <25 | 3 | Y | Y | Y | Y | N | N | Y | … | n | NG |
| 39115 | 17473104+0522166 | -5.5 ± 0.2 | 6053 ± 195 | 4.20 ± 0.01 | 0.996 ± 0.002 | 0.04 ± 0.03 | 63 ± 2 | 1 | Y | Y | Y | N | N | Y | N | … | n | NG |
| 39116 | 17473130+0602117 | 10.8 ± 0.2 | 4799 ± 249 | … | 1.016 ± 0.003 | -0.40 ± 0.07 | <12 | 3 | N | … | … | … | … | … | … | … | n | G |
| 39117 | 17473147+0600060 | -56.1 ± 0.2 | 4905 ± 260 | … | 1.008 ± 0.003 | -0.26 ± 0.14 | <12 | 3 | Y | … | N | N | N | Y | … | … | n | NG? |
| 39118 | 17473152+0601519 | -50.7 ± 0.2 | 5891 ± 144 | 4.19 ± 0.02 | 0.993 ± 0.002 | -0.05 ± 0.04 | 65 ± 4 | 1 | Y | Y | N | N | N | Y | … | … | n | NG |
| 39119 | 17473178+0531054 | -42.7 ± 0.2 | 4890 ± 103 | … | 0.978 ± 0.005 | 0.00 ± 0.07 | <16 | 3 | Y | … | N | N | N | Y | … | … | n | NG |
| 39120 | 17473192+0516594 | 142.7 ± 0.2 | 5289 ± 162 | … | 1.006 ± 0.004 | -0.32 ± 0.23 | <6 | 3 | Y | … | N | N | N | N | … | … | n | … |
| 39121 | 17473210+0527405 | 23.0 ± 0.2 | 4628 ± 136 | 2.43 ± 0.19 | 1.021 ± 0.002 | 0.05 ± 0.07 | 21 ± 4 | 1 | N | N | … | … | … | … | … | … | n | G |
| 39122 | 17473220+0524060 | -30.6 ± 1.3 | … | … | … | … | … | … | … | … | … | … | … | N | … | … | n | … |
| 2642 | 17473252+0552316 | -14.3 ± 0.6 | 6186 ± 38 | 4.28 ± 0.03 | … | 0.15 ± 0.01 | 65 ± 8 | 2 | … | Y | Y | Y | N | Y | Y | … | n | NG |
| 39123 | 17473274+0518502 | -76.3 ± 0.2 | 5123 ± 80 | 3.91 ± 0.06 | 0.985 ± 0.002 | -0.30 ± 0.04 | <16 | 3 | Y | Y | N | N | N | Y | … | … | n | NG |
| 39124 | 17473278+0542267 | 53.2 ± 0.2 | 4848 ± 143 | … | 1.023 ± 0.003 | -0.26 ± 0.18 | <16 | 3 | N | … | … | … | … | … | … | … | n | G |
| 38916 | 17461124+0555309 | -44.8 ± 0.2 | 4940 ± 171 | … | 1.014 ± 0.005 | -0.11 ± 0.12 | <29 | 3 | N | … | … | … | … | … | … | … | n | … |
| 38917 | 17461159+0605417 | -15.9 ± 0.9 | 3437 ± 108 | 4.70 ± 0.04 | 0.793 ± 0.011 | -0.23 ± 0.16 | … | … | Y | Y | … | Y | Y | Y | Y | Y | … | … |
| 38918 | 17461198+0541258 | -14.4 ± 0.3 | 4985 ± 178 | … | 0.983 ± 0.005 | 0.00 ± 0.08 | 250 ± 6 | 1 | Y | … | Y | Y | Y | Y | … | Y | Y | … |
| 38941 | 17461999+0522152 | -83.9 ± 0.3 | 4759 ± 152 | … | 0.915 ± 0.007 | … | <37 | 3 | Y | … | N | N | N | … | … | … | n | NG |
| 38942 | 17462084+0535457 | 19.4 ± 0.2 | 4922 ± 8 | … | 1.001 ± 0.006 | 0.01 ± 0.07 | <43 | 3 | Y | … | N | N | N | Y | … | … | n | NG? |
| 38943 | 17462149+0600336 | 42.4 ± 0.3 | 4720 ± 259 | … | 0.922 ± 0.008 | -0.10 ± 0.03 | <15 | 3 | Y | … | N | N | N | Y | … | … | n | NG |
| 38944 | 17462261+0531159 | -25.3 ± 0.3 | 6847 ± 37 | 4.22 ± 0.07 | 1.005 ± 0.002 | -0.09 ± 0.03 | <5 | 3 | Y | Y | Y | N | N | Y | N | … | n | NG? |
| 38945 | 17462287+0551312 | -40.4 ± 0.3 | 4275 ± 304 | … | 0.901 ± 0.009 | -0.32 ± 0.24 | … | … | Y | … | Y | … | N | N | … | … | n | … |
| 38946 | 17462366+0525344 | -27.2 ± 0.2 | 5476 ± 222 | … | 0.995 ± 0.005 | 0.22 ± 0.04 | <23 | 3 | Y | … | Y | N | N | Y | … | … | n | NG |
| 38958 | 17462680+0605304 | 10.7 ± 0.4 | 3583 ± 78 | 4.70 ± 0.01 | 0.781 ± 0.005 | -0.24 ± 0.14 | … | … | Y | Y | Y | N | N | Y | N | … | n | … |
| 38947 | 17462383+0602344 | -13.8 ± 0.2 | 4080 ± 202 | 4.45 ± 0.19 | 0.871 ± 0.008 | 0.01 ± 0.16 | <14 | 3 | Y | Y | Y | Y | Y | Y | Y | Y | Y | … |
| 38959 | 17462804+0533207 | -1.0 ± 0.3 | 6056 ± 28 | 4.10 ± 0.15 | 1.000 ± 0.003 | 0.04 ± 0.06 | 90 ± 3 | 1 | Y | Y | Y | N | N | Y | N | … | n | NG? |
| 38948 | 17462417+0530188 | -15.5 ± 0.2 | 5110 ± 28 | … | 1.015 ± 0.005 | -0.17 ± 0.20 | <27 | 3 | N | … | … | … | … | … | … | … | n | G |
| 38960 | 17462833+0603129 | -40.7 ± 0.2 | 6299 ± 41 | … | 1.006 ± 0.003 | 0.19 ± 0.03 | 60 ± 8 | 1 | Y | … | N | N | N | N | … | … | n | NG? |
| 2635 | 17462473+0517214 | -13.6 ± 0.6 | 5960 ± 99 | 4.37 ± 0.14 | … | 0.00 ± 0.12 | 141 ± 16 | 2 | … | Y | Y | Y | Y | Y | Y | Y | … | … |
| 38949 | 17462477+0535381 | -7.3 ± 0.2 | 5144 ± 26 | … | 0.985 ± 0.003 | -0.09 ± 0.03 | 263 ± 9 | 1 | Y | … | Y | Y | Y | N | … | Y | … | … |
| 38961 | 17462841+0540180 | 36.7 ± 0.2 | 4553 ± 109 | 2.27 ± 0.13 | 1.013 ± 0.006 | -0.12 ± 0.10 | <37 | 3 | N | N | … | … | … | … | … | … | n | G |
| 38950 | 17462540+0603444 | 5.2 ± 0.3 | 4268 ± 263 | 4.65 ± 0.18 | 0.863 ± 0.008 | -0.04 ± 0.09 | <31 | 3 | Y | Y | Y | Y | N | Y | … | … | n | NG |
| 38962 | 17462850+0517546 | -17.7 ± 0.2 | 4683 ± 109 | 2.51 ± 0.20 | 1.019 ± 0.005 | -0.07 ± 0.07 | <21 | 3 | N | N | … | … | … | … | … | … | n | … |

**Table C.7.** continued.

| ID | CNAME | $RV$ (km s$^{-1}$) | $T_{\text{eff}}$ (K) | $\log g$ (dex) | $\gamma^a$ | [Fe/H] (dex) | $EW(\text{Li})^b$ (mÅ) | $EW(\text{Li})$ error flag$^c$ | $\gamma$ | $\log g$ | Membership RV | Li | H$\alpha$ | [Fe/H] | Gaia studies Randich$^d$ | Cantat-Gaudin$^d$ | Final$^e$ | NMs with Li$^f$ |
|---|---|---|---|---|---|---|---|---|---|---|---|---|---|---|---|---|---|---|
| 38951 | 17462563+0553066 | 26.2 ± 0.3 | 5025 ± 148 | ... | 0.999 ± 0.008 | -0.19 ± 0.17 | <29 | 3 | Y | ... | N | N | N | Y | ... | ... | n | NG |
| 38963 | 17462856+0554473 | 62.1 ± 0.2 | 4545 ± 46 | ... | 1.038 ± 0.007 | -0.18 ± 0.03 | 25 ± 4 | 1 | N | ... | ... | ... | ... | ... | ... | ... | n | G |
| 38952 | 17462572+0531092 | -5.0 ± 0.2 | 5701 ± 17 | 4.35 ± 0.17 | ... | -0.13 ± 0.17 | 123 ± 3 | ... | ... | Y | Y | Y | Y | Y | ... | Y | Y | ... |
| 39011 | 17464952+0526170 | 37.4 ± 0.2 | 4764 ± 97 | 2.55 ± 0.20 | 1.013 ± 0.005 | -0.30 ± 0.05 | <18 | 3 | N | N | ... | ... | ... | ... | ... | ... | n | G |
| 38953 | 17462576+0518353 | -24.8 ± 0.3 | 4583 ± 13 | 4.45 ± 0.14 | 0.951 ± 0.007 | -0.04 ± 0.01 | 12 ± 6 | 1 | Y | Y | Y | N | N | Y | ... | ... | n | NG |
| 38954 | 17462591+0558248 | 25.3 ± 0.3 | 4253 ± 296 | 4.47 ± 0.05 | 0.879 ± 0.009 | -0.05 ± 0.08 | ... | ... | Y | Y | N | ... | ... | Y | ... | ... | n | ... |
| 39012 | 17464999+0604572 | 35.5 ± 0.2 | 4601 ± 75 | 2.28 ± 0.15 | 1.019 ± 0.004 | -0.13 ± 0.13 | <25 | 3 | N | N | ... | ... | ... | ... | ... | ... | n | ... |
| 38955 | 17462597+0557290 | 57.3 ± 0.3 | 4932 ± 150 | ... | 1.011 ± 0.009 | -0.32 ± 0.28 | <16 | 3 | N | ... | ... | ... | ... | ... | ... | ... | n | G |
| 39013 | 17465032+0533038 | -37.1 ± 0.2 | 5293 ± 119 | 3.78 ± 0.18 | 0.998 ± 0.005 | 0.12 ± 0.01 | 27 ± 10 | 1 | Y | Y | Y | N | N | Y | ... | ... | n | NG |
| 38956 | 17462645+0553236 | -6.4 ± 0.2 | 6156 ± 111 | 4.00 ± 0.12 | 1.000 ± 0.003 | -0.08 ± 0.01 | 45 ± 2 | 1 | Y | Y | Y | N | N | Y | N | ... | n | NG? |
| 39014 | 17465034+0525389 | 16.4 ± 0.5 | 4071 ± 314 | ... | 0.883 ± 0.016 | -0.14 ± 0.03 | ... | ... | Y | ... | N | ... | ... | Y | N | ... | n | ... |
| 38957 | 17462649+0558401 | 10.5 ± 0.3 | 4452 ± 188 | 4.55 ± 0.02 | 0.912 ± 0.005 | -0.04 ± 0.01 | <11 | 3 | Y | Y | Y | Y | N | Y | ... | ... | n | NG |
| 39015 | 17465047+0528073 | -47.6 ± 0.2 | 5811 ± 71 | ... | 1.002 ± 0.002 | 0.10 ± 0.02 | <15 | 3 | Y | ... | N | N | N | Y | ... | ... | n | NG? |
| 39016 | 17465080+0542256 | -18.8 ± 0.3 | 5214 ± 75 | ... | 0.985 ± 0.006 | 0.01 ± 0.08 | <31 | 3 | Y | ... | Y | N | Y | Y | ... | ... | n | NG |
| 39017 | 17465108+0552451 | 15.9 ± 0.5 | 4071 ± 262 | ... | 0.888 ± 0.014 | -0.22 ± 0.07 | <28 | 3 | Y | ... | N | N | N | Y | N | ... | n | NG |
| 39018 | 17465117+0530410 | 12.1 ± 0.3 | 5050 ± 3 | 3.20 ± 0.05 | 1.005 ± 0.007 | -0.02 ± 0.09 | ... | ... | Y | N | Y | ... | N | Y | ... | ... | n | ... |
| 39019 | 17465142+0549383 | 5.8 ± 0.3 | 4465 ± 209 | 4.45 ± 0.17 | 0.925 ± 0.009 | -0.15 ± 0.12 | <24 | 3 | Y | Y | Y | Y | Y | Y | ... | ... | Y | ... |
| 2616 | 17442711+0547196 | -13.1 ± 0.6 | 5537 ± 109 | 4.41 ± 0.10 | ... | 0.00 ± 0.01 | 211 ± 16 | 2 | ... | Y | Y | Y | Y | Y | Y | Y | Y | ... |
| 39020 | 17465174+0517028 | -42.5 ± 0.2 | 5717 ± 28 | 4.06 ± 0.19 | 0.996 ± 0.003 | -0.37 ± 0.03 | <4 | 3 | Y | Y | N | N | N | N | ... | ... | n | ... |
| 38675 | 17442847+0523464 | 25.4 ± 0.3 | 5230 ± 35 | 3.60 ± 0.02 | 0.997 ± 0.007 | 0.04 ± 0.10 | ... | ... | Y | Y | N | ... | ... | Y | ... | ... | n | ... |
| 39021 | 17465203+0559079 | 17.0 ± 0.2 | 6114 ± 73 | 3.87 ± 0.06 | 1.001 ± 0.002 | -0.31 ± 0.15 | <2 | 3 | Y | N | N | N | N | Y | ... | ... | n | NG? |
| 39022 | 17465220+0532369 | 51.4 ± 0.2 | 4897 ± 106 | 2.90 ± 0.13 | 1.007 ± 0.006 | -0.21 ± 0.18 | <20 | 3 | Y | N | N | N | N | Y | ... | ... | n | NG? |
| 39023 | 17465272+0520361 | -25.6 ± 0.2 | 4718 ± 171 | 1.99 ± 0.05 | 1.023 ± 0.002 | -0.36 ± 0.09 | <2 | 3 | N | N | ... | ... | ... | ... | ... | ... | n | G |
| 38676 | 17443019+0542317 | -29.1 ± 0.4 | 3926 ± 50 | ... | 0.871 ± 0.017 | 0.00 ± 0.19 | <80 | 3 | Y | ... | Y | Y | N | Y | N | ... | n | NG |
| 38677 | 17443040+0529030 | -11.2 ± 0.3 | 5611 ± 8 | 4.11 ± 0.04 | 0.988 ± 0.005 | 0.06 ± 0.05 | 33 ± 4 | 1 | Y | Y | Y | N | Y | Y | ... | ... | n | NG |
| 39024 | 17465301+0532470 | -80.8 ± 0.2 | 4986 ± 115 | ... | 0.951 ± 0.004 | ... | <12 | 3 | Y | ... | N | N | N | ... | ... | ... | n | NG |
| 38678 | 17443177+0518564 | 11.2 ± 0.2 | 4895 ± 57 | 4.36 ± 0.09 | 0.966 ± 0.004 | 0.03 ± 0.03 | <25 | 3 | Y | Y | Y | N | Y | Y | ... | ... | n | NG |
| 39025 | 17465415+0545327 | -29.5 ± 0.4 | 4330 ± 138 | ... | 0.866 ± 0.017 | ... | ... | ... | Y | ... | Y | ... | N | ... | ... | ... | n | ... |
| 38679 | 17465184+0555058 | -13.6 ± 0.3 | 3945 ± 240 | 4.51 ± 0.07 | 0.820 ± 0.011 | -0.15 ± 0.14 | ... | ... | Y | Y | Y | Y | Y | Y | N | ... | n | ... |
| 39026 | 17465420+0600543 | 17.6 ± 0.8 | ... | ... | ... | ... | ... | ... | ... | ... | ... | ... | ... | ... | ... | ... | n | ... |
| 38680 | 17443216+0529099 | -13.4 ± 0.2 | 5968 ± 110 | 4.16 ± 0.06 | 0.994 ± 0.004 | 0.05 ± 0.09 | 39 ± 6 | 1 | Y | Y | Y | N | N | Y | ... | ... | n | NG |
| 39027 | 17465499+0537452 | 107.9 ± 0.2 | 4432 ± 169 | ... | 1.034 ± 0.005 | -0.25 ± 0.15 | <18 | 3 | N | ... | ... | ... | ... | ... | ... | ... | n | G |
| 38681 | 17443272+0522466 | -79.5 ± 0.2 | 5021 ± 231 | ... | 1.008 ± 0.002 | -0.30 ± 0.09 | 10 ± 2 | 1 | Y | ... | N | N | Y | Y | ... | ... | n | NG? |
| 39028 | 17465546+0556081 | -57.8 ± 0.4 | 3913 ± 115 | 4.52 ± 0.16 | 0.839 ± 0.008 | -0.18 ± 0.03 | ... | ... | Y | Y | N | ... | ... | Y | N | ... | n | ... |
| 38682 | 17443418+0518051 | 49.1 ± 0.2 | 4596 ± 2 | 2.21 ± 0.20 | 1.026 ± 0.005 | -0.27 ± 0.11 | <31 | 3 | N | N | ... | ... | ... | ... | ... | ... | n | G |
| 39029 | 17465562+0530294 | -0.1 ± 0.2 | 3864 ± 192 | 4.71 ± 0.16 | 0.773 ± 0.009 | -0.18 ± 0.10 | <49 | 3 | Y | Y | Y | Y | N | Y | N | ... | n | NG |
| 2617 | 17443565+0557208 | -14.1 ± 0.6 | 6109 ± 36 | 3.80 ± 0.02 | ... | -0.30 ± 0.02 | 36 ± 4 | 2 | ... | Y | Y | N | N | Y | Y | ... | n | NG |
| 38683 | 17443585+0539160 | -15.7 ± 0.3 | 5179 ± 153 | ... | 0.967 ± 0.013 | 0.06 ± 0.10 | <35 | 3 | Y | ... | Y | N | Y | Y | ... | ... | n | NG |
| 39030 | 17465576+0604118 | 24.0 ± 0.8 | ... | ... | ... | ... | ... | ... | ... | ... | ... | ... | ... | ... | ... | ... | n | ... |
| 39031 | 17465585+0523343 | -54.1 ± 0.2 | 6351 ± 25 | 3.98 ± 0.05 | 1.005 ± 0.001 | 0.08 ± 0.02 | 22 ± 1 | 1 | Y | Y | N | N | N | Y | N | ... | n | NG? |
| 38684 | 17443619+0554141 | 8.3 ± 0.2 | 4371 ± 245 | 4.67 ± 0.12 | 0.885 ± 0.009 | -0.14 ± 0.06 | ... | ... | Y | Y | Y | ... | Y | Y | ... | ... | n | ... |
| 39032 | 17465627+0530397 | -87.6 ± 0.2 | 5492 ± 256 | 3.82 ± 0.17 | 0.998 ± 0.002 | 0.06 ± 0.11 | <16 | 3 | Y | Y | N | N | N | Y | ... | ... | n | NG |
| 2618 | 17443667+0537161 | 1.3 ± 0.6 | 6291 ± 112 | 3.89 ± 0.13 | 1.008 ± 0.003 | -0.01 ± 0.11 | 50 ± 1 | 2 | Y | Y | N | N | N | Y | N | ... | n | NG? |
| 39033 | 17465642+0547446 | -14.4 ± 0.3 | 3973 ± 223 | 4.46 ± 0.16 | 0.842 ± 0.007 | 0.02 ± 0.18 | <28 | 3 | Y | Y | Y | Y | Y | Y | Y | Y | Y | ... |
| 38685 | 17443691+0536555 | -43.0 ± 0.4 | 5238 ± 296 | ... | 1.006 ± 0.010 | -0.55 ± 0.13 | <20 | 3 | Y | ... | N | N | N | N | ... | ... | n | NG? |
| 39034 | 17465709+0607550 | 20.5 ± 0.8 | ... | ... | ... | ... | ... | ... | ... | ... | ... | ... | ... | N | ... | ... | n | ... |
| 39058 | 17470874+0608026 | 100.5 ± 0.2 | 4992 ± 189 | ... | 1.011 ± 0.004 | -0.20 ± 0.25 | <13 | 3 | N | ... | ... | ... | ... | ... | ... | ... | n | G |
| 38686 | 17443784+0525421 | -67.0 ± 0.2 | 4702 ± 42 | 2.67 ± 0.13 | 1.015 ± 0.005 | 0.00 ± 0.03 | <38 | 3 | N | N | ... | ... | ... | ... | ... | ... | n | G |
| 39059 | 17470932+0541020 | -32.5 ± 0.5 | 3889 ± 42 | ... | 0.830 ± 0.017 | 0.30 ± 0.15 | <49 | 3 | Y | ... | Y | Y | N | N | N | ... | n | NG |
| 38687 | 17443819+0516522 | 41.9 ± 0.3 | 5464 ± 149 | ... | 1.011 ± 0.005 | ... | ... | ... | N | ... | ... | ... | ... | ... | ... | ... | n | ... |
| 38688 | 17443820+0536180 | -119.8 ± 0.4 | 5023 ± 124 | ... | 1.030 ± 0.014 | ... | <39 | 3 | N | ... | ... | ... | ... | ... | ... | ... | n | G |
| 38689 | 17443887+0542306 | 83.4 ± 0.2 | 4990 ± 94 | ... | 1.011 ± 0.004 | -0.38 ± 0.19 | 12 ± 3 | 1 | N | ... | ... | ... | ... | ... | ... | ... | n | G |
| 39060 | 17470994+0551383 | -39.0 ± 0.3 | 3528 ± 63 | ... | 0.839 ± 0.013 | -0.27 ± 0.13 | ... | ... | Y | ... | ... | ... | ... | Y | N | ... | n | ... |
| 38690 | 17443893+0539341 | 44.4 ± 0.3 | 5769 ± 132 | ... | 0.992 ± 0.005 | 0.01 ± 0.08 | <25 | 3 | Y | ... | N | N | N | Y | ... | ... | n | NG |
| 39061 | 17470996+0543141 | -31.9 ± 0.3 | 3645 ± 19 | ... | 0.829 ± 0.015 | ... | ... | ... | Y | ... | Y | ... | N | ... | N | ... | n | ... |
| 38691 | 17443897+0536053 | -132.8 ± 0.3 | ... | ... | ... | ... | ... | ... | ... | ... | ... | ... | ... | ... | ... | ... | n | ... |
| 39062 | 17470997+0529064 | -25.3 ± 0.3 | 3665 ± 124 | 4.57 ± 0.20 | 0.821 ± 0.008 | -0.21 ± 0.12 | <100 | 3 | Y | Y | Y | Y | N | Y | N | ... | n | NG |
| 2619 | 17444099+0541558 | -17.3 ± 0.6 | 7029 ± 142 | 4.12 ± 0.14 | ... | -0.12 ± 0.16 | <24 | 3 | ... | Y | Y | N | N | Y | N | ... | n | NG |
| 39063 | 17471055+0541189 | -9.6 ± 0.2 | 5030 ± 123 | ... | 1.017 ± 0.003 | ... | <16 | 3 | N | ... | ... | ... | ... | ... | ... | ... | n | G |

| ID | CNAME | RV (km s$^{-1}$) | $T_{eff}$ (K) | logg (dex) | $\gamma^a$ | [Fe/H] (dex) | EW(Li)$^b$ (mÅ) | EW(Li) error flag$^c$ | $\gamma$ | logg | Membership RV | Li | H$\alpha$ | [Fe/H] | Gaia studies Randich$^d$ | Cantat-Gaudin$^d$ | Final$^e$ | NMs with Li$^f$ |
|---|---|---|---|---|---|---|---|---|---|---|---|---|---|---|---|---|---|---|
| 38692 | 17444130+0542524 | 23.9 ± 0.2 | 5111 ± 69 | ... | 1.011 ± 0.004 | -0.17 ± 0.14 | ... | ... | N | ... | ... | ... | ... | ... | ... | ... | n | G |
| 39064 | 17471101+0526105 | 17.0 ± 0.2 | 4590 ± 88 | 2.26 ± 0.17 | 1.023 ± 0.003 | -0.12 ± 0.09 | <25 | 3 | N | N | ... | ... | ... | ... | ... | ... | n | ... |
| 38693 | 17444177+0530043 | -15.7 ± 0.2 | 5266 ± 130 | 4.21 ± 0.02 | 0.981 ± 0.005 | 0.09 ± 0.06 | <18 | 3 | Y | Y | Y | N | N | Y | ... | ... | n | NG |
| 39065 | 17471103+0539560 | 63.5 ± 0.2 | 5000 ± 14 | 3.24 ± 0.16 | 1.001 ± 0.004 | -0.08 ± 0.10 | 4 ± 3 | 1 | Y | N | N | N | N | Y | ... | ... | n | ... |
| 38694 | 17444213+0527167 | 25.5 ± 0.2 | 5751 ± 18 | ... | 1.007 ± 0.003 | -0.15 ± 0.07 | 27 ± 2 | 1 | Y | ... | N | N | N | Y | ... | ... | n | NG? |
| 39066 | 17471123+0544442 | -6.6 ± 0.3 | 4349 ± 239 | ... | 0.873 ± 0.011 | -0.06 ± 0.10 | ... | ... | Y | ... | Y | ... | Y | Y | ... | ... | n | ... |
| 2638 | 17471166+0541502 | -33.9 ± 0.6 | 5525 ± 22 | 3.94 ± 0.05 | ... | 0.13 ± 0.01 | 36 ± 6 | 2 | ... | Y | Y | N | N | Y | ... | ... | n | NG |
| 38718 | 17445666+0534379 | -19.5 ± 0.2 | 5094 ± 76 | ... | 1.020 ± 0.003 | ... | <16 | 3 | N | ... | ... | ... | ... | ... | ... | ... | n | G |
| 39067 | 17471230+0531160 | -47.7 ± 0.2 | 5018 ± 106 | 3.52 ± 0.17 | 0.998 ± 0.003 | 0.08 ± 0.02 | <38 | 3 | Y | Y | N | N | N | Y | ... | ... | n | NG |
| 2621 | 17445810+0551329 | -13.2 ± 0.6 | 5512 ± 124 | 4.49 ± 0.04 | 0.993 ± 0.004 | -0.04 ± 0.07 | 215 ± 6 | 2 | Y | Y | Y | Y | Y | Y | Y | Y | Y | ... |
| 38719 | 17445859+0514317 | -74.7 ± 0.2 | 5516 ± 203 | 3.97 ± 0.02 | 0.994 ± 0.003 | -0.08 ± 0.08 | <10 | 3 | Y | Y | N | N | N | Y | ... | ... | n | NG |
| 38720 | 17445883+0529395 | -12.5 ± 0.2 | 5176 ± 70 | 3.75 ± 0.15 | 1.000 ± 0.004 | -0.14 ± 0.09 | <19 | 3 | Y | Y | N | N | N | Y | ... | ... | n | NG? |
| 39068 | 17471253+0542149 | 169.8 ± 14.8 | 3412 ± 145 | ... | 0.927 ± 0.015 | -0.27 ± 0.14 | ... | ... | Y | ... | N | ... | ... | Y | ... | ... | n | ... |
| 39069 | 17471262+0558244 | 8.5 ± 0.2 | 4628 ± 55 | 4.34 ± 0.11 | 0.948 ± 0.005 | 0.03 ± 0.11 | 20 ± 3 | 1 | Y | Y | N | N | N | Y | ... | ... | n | NG |
| 38721 | 17450031+0542442 | -57.9 ± 0.2 | 4977 ± 152 | ... | 1.006 ± 0.004 | -0.24 ± 0.15 | ... | ... | Y | ... | N | ... | ... | Y | ... | ... | n | ... |
| 39070 | 17471363+0549428 | 19.4 ± 0.3 | 5176 ± 61 | 4.49 ± 0.24 | 0.970 ± 0.008 | -0.16 ± 0.10 | <35 | 3 | Y | Y | N | N | N | Y | ... | ... | n | NG |
| 38722 | 17450096+0537407 | -4.9 ± 0.2 | 4282 ± 276 | ... | 0.902 ± 0.004 | -0.25 ± 0.09 | 9 ± 3 | 1 | Y | ... | Y | N | Y | Y | ... | ... | n | ... |
| 39071 | 17471408+0548498 | -56.9 ± 0.5 | 4587 ± 447 | ... | 0.991 ± 0.016 | -0.25 ± 0.17 | <55 | 3 | Y | ... | N | Y | Y | Y | ... | ... | n | ... |
| 38723 | 17450156+0556223 | -68.2 ± 0.3 | 4048 ± 78 | 4.56 ± 0.03 | 0.862 ± 0.010 | 0.09 ± 0.27 | ... | ... | Y | Y | N | ... | ... | Y | N | ... | n | ... |
| 39072 | 17471507+0558585 | -10.3 ± 0.2 | 4723 ± 6 | 2.65 ± 0.11 | 1.007 ± 0.003 | -0.03 ± 0.03 | <34 | 3 | Y | N | Y | N | Y | Y | ... | ... | n | NG? |
| 38724 | 17450157+0539316 | -18.5 ± 0.3 | 5281 ± 203 | ... | 1.002 ± 0.004 | -0.43 ± 0.14 | 15 ± 3 | 1 | Y | ... | Y | Y | Y | N | ... | ... | n | ... |
| 38725 | 17450217+0535374 | 28.9 ± 0.2 | 4873 ± 254 | ... | 1.017 ± 0.005 | -0.17 ± 0.21 | <24 | 3 | N | ... | ... | ... | ... | ... | ... | ... | n | G |
| 2639 | 17471544+0556358 | -13.3 ± 0.6 | 5733 ± 118 | 4.42 ± 0.12 | 0.995 ± 0.002 | 0.06 ± 0.12 | 228 ± 33 | 1 | Y | Y | Y | Y | Y | Y | Y | Y | Y | ... |
| 38726 | 17450296+0523592 | 51.4 ± 0.2 | 4986 ± 97 | ... | 1.010 ± 0.003 | -0.26 ± 0.21 | <13 | 3 | N | ... | ... | ... | ... | ... | ... | ... | n | G |
| 39073 | 17471559+0604380 | 27.9 ± 0.2 | 5837 ± 15 | 4.10 ± 0.13 | 0.996 ± 0.002 | -0.21 ± 0.11 | 16 ± 2 | 1 | Y | Y | N | N | N | Y | ... | ... | n | NG |
| 38727 | 17450327+0540388 | -68.5 ± 0.2 | 4388 ± 249 | 4.37 ± 0.01 | 0.908 ± 0.006 | 0.09 ± 0.21 | <17 | 3 | Y | Y | N | Y | N | Y | ... | ... | n | NG |
| 38728 | 17450439+0559410 | -88.4 ± 0.3 | 4883 ± 128 | ... | 0.928 ± 0.009 | ... | <23 | 3 | Y | ... | N | N | Y | Y | ... | ... | n | NG |
| 39074 | 17471628+0529495 | -11.8 ± 0.2 | 4332 ± 275 | ... | 0.921 ± 0.004 | -0.01 ± 0.10 | <18 | 3 | Y | ... | Y | Y | Y | Y | ... | ... | n | Y |
| 38729 | 17450471+0524488 | 2.5 ± 0.2 | 4561 ± 138 | 2.24 ± 0.20 | 1.021 ± 0.002 | -0.13 ± 0.08 | <21 | 3 | N | N | ... | ... | ... | ... | ... | ... | n | ... |
| 39075 | 17471641+0555373 | -40.5 ± 0.2 | 4583 ± 80 | 2.42 ± 0.20 | 1.020 ± 0.003 | -0.02 ± 0.06 | <28 | 3 | N | N | ... | ... | ... | ... | ... | ... | n | G |
| 39076 | 17471654+0523500 | 15.7 ± 0.2 | 4547 ± 87 | 1.85 ± 0.10 | 1.024 ± 0.002 | -0.41 ± 0.09 | <18 | 3 | N | N | ... | ... | ... | ... | ... | ... | n | G |
| 38731 | 17450500+0513517 | -35.0 ± 0.2 | 5688 ± 117 | ... | 1.002 ± 0.003 | 0.18 ± 0.04 | 23 ± 2 | 1 | Y | ... | Y | N | N | N | ... | ... | n | NG? |
| 39077 | 17471657+0556322 | -20.6 ± 0.2 | 5839 ± 78 | ... | 0.999 ± 0.003 | -0.05 ± 0.06 | 13 ± 5 | 1 | Y | ... | Y | N | N | Y | ... | ... | n | NG |
| 38732 | 17450514+0531171 | 18.1 ± 0.2 | 4696 ± 121 | 2.13 ± 0.03 | 1.020 ± 0.004 | -0.30 ± 0.05 | <15 | 3 | N | N | ... | ... | ... | ... | ... | ... | n | G |
| 39078 | 17471671+0520379 | -30.1 ± 0.2 | 5046 ± 68 | ... | 1.017 ± 0.003 | -0.54 ± 0.11 | <4 | 3 | N | ... | ... | ... | ... | ... | ... | ... | n | ... |
| 38733 | 17450570+0536105 | 41.3 ± 0.2 | 4596 ± 220 | 4.58 ± 0.04 | 0.925 ± 0.006 | -0.14 ± 0.07 | 18 ± 5 | 1 | Y | Y | N | N | N | Y | ... | ... | n | NG |
| 38734 | 17450599+0520492 | 1.7 ± 0.2 | 4780 ± 5 | ... | 0.999 ± 0.004 | 0.09 ± 0.01 | <37 | 3 | Y | ... | Y | N | Y | Y | ... | ... | n | NG |
| 39125 | 17473297+0559293 | 117.4 ± 0.2 | 5066 ± 160 | ... | 1.014 ± 0.004 | -0.47 ± 0.12 | <27 | 3 | N | ... | ... | ... | ... | ... | ... | ... | n | G |
| 38735 | 17450613+0551064 | 23.6 ± 0.3 | 4189 ± 94 | 4.69 ± 0.17 | 0.869 ± 0.010 | 0.06 ± 0.21 | <18 | 3 | Y | Y | N | Y | N | Y | N | ... | n | NG |
| 39126 | 17473324+0605467 | 5.7 ± 0.2 | 5005 ± 171 | ... | 1.000 ± 0.004 | -0.31 ± 0.21 | ... | ... | Y | ... | Y | ... | N | Y | ... | ... | n | ... |
| 38736 | 17450746+0550594 | -47.7 ± 0.3 | 4996 ± 144 | ... | 1.021 ± 0.006 | -0.32 ± 0.23 | <35 | 3 | N | ... | ... | ... | ... | ... | ... | ... | n | G |
| 38737 | 17450789+0530219 | -24.4 ± 0.2 | 5048 ± 146 | 4.10 ± 0.08 | 0.976 ± 0.003 | 0.05 ± 0.05 | <21 | 3 | Y | Y | N | N | N | Y | ... | ... | n | NG |
| 2643 | 17473331+0531460 | -13.5 ± 0.6 | 5781 ± 58 | 4.38 ± 0.13 | 0.988 ± 0.002 | 0.11 ± 0.11 | 234 ± 6 | 2 | Y | Y | Y | Y | Y | Y | Y | Y | Y | ... |
| 38738 | 17450816+0604037 | 37.4 ± 0.3 | 6085 ± 114 | 4.10 ± 0.11 | 0.997 ± 0.003 | -0.16 ± 0.11 | 31 ± 6 | 1 | Y | Y | N | N | N | Y | N | ... | n | NG |
| 39127 | 17473349+0533004 | 23.4 ± 0.3 | 4274 ± 153 | ... | 0.918 ± 0.007 | 0.04 ± 0.21 | ... | ... | Y | ... | N | ... | ... | Y | ... | ... | n | ... |
| 38739 | 17450838+0525327 | 19.3 ± 0.3 | 5268 ± 192 | ... | 1.011 ± 0.006 | -0.38 ± 0.19 | <28 | 3 | N | ... | ... | ... | ... | ... | ... | ... | n | ... |
| 38740 | 17450852+0555406 | -27.1 ± 0.2 | 5863 ± 99 | ... | 1.006 ± 0.002 | -0.38 ± 0.02 | 42 ± 2 | 1 | Y | ... | Y | N | N | N | ... | ... | n | NG? |
| 39128 | 17473388+0609461 | 39.7 ± 0.2 | 4726 ± 2 | ... | 1.022 ± 0.005 | -0.14 ± 0.13 | <18 | 3 | N | ... | ... | ... | ... | ... | ... | ... | n | G |
| 38787 | 17452363+0518410 | -11.0 ± 0.2 | 6111 ± 118 | 4.06 ± 0.11 | 0.998 ± 0.002 | -0.09 ± 0.05 | 33 ± 3 | 1 | Y | Y | Y | N | N | Y | N | ... | n | NG |
| 39129 | 17473417+0607458 | 77.8 ± 0.3 | 4888 ± 69 | ... | 1.015 ± 0.006 | -0.27 ± 0.22 | <23 | 3 | N | ... | ... | ... | ... | ... | ... | ... | n | G |
| 38788 | 17452389+0558395 | 28.7 ± 0.2 | 5625 ± 52 | ... | 0.998 ± 0.003 | -0.23 ± 0.13 | ... | ... | Y | ... | N | ... | ... | Y | ... | ... | n | ... |
| 39130 | 17473461+0519290 | -87.3 ± 0.2 | 4955 ± 224 | ... | 1.014 ± 0.003 | -0.37 ± 0.07 | <14 | 3 | N | ... | ... | ... | ... | ... | ... | ... | n | G |
| 38789 | 17452412+0526078 | 34.6 ± 0.2 | 5017 ± 85 | ... | 1.018 ± 0.004 | -0.37 ± 0.11 | <19 | 3 | N | ... | ... | ... | ... | ... | ... | ... | n | ... |
| 39131 | 17473490+0550489 | -11.6 ± 0.2 | 5557 ± 32 | ... | 0.993 ± 0.004 | 0.04 ± 0.02 | 32 ± 2 | 1 | Y | ... | Y | N | Y | Y | ... | ... | n | NG |
| 38790 | 17452455+0514481 | -5.6 ± 0.2 | 5977 ± 117 | 4.25 ± 0.03 | 0.993 ± 0.003 | 0.10 ± 0.02 | 35 ± 8 | 1 | Y | Y | N | N | N | Y | ... | ... | n | NG |
| 39132 | 17473595+0515451 | -129.5 ± 0.2 | 4934 ± 182 | 2.87 ± 0.11 | 1.006 ± 0.003 | -0.44 ± 0.13 | <14 | 3 | Y | N | N | N | N | N | ... | ... | n | NG? |
| 38791 | 17452482+0538131 | -68.7 ± 0.2 | 5020 ± 128 | ... | 1.022 ± 0.004 | -0.27 ± 0.01 | ... | ... | N | ... | ... | ... | ... | ... | ... | ... | n | G |
| 39133 | 17473770+0527058 | 53.7 ± 0.2 | 4680 ± 93 | ... | 1.016 ± 0.004 | -0.21 ± 0.12 | 26 ± 8 | 1 | N | ... | ... | ... | ... | ... | ... | ... | n | G |
| 2624 | 17452508+0551388 | -12.4 ± 0.6 | 5349 ± 165 | 4.42 ± 0.18 | 0.998 ± 0.004 | -0.05 ± 0.03 | 301 ± 11 | 2 | Y | Y | Y | Y | Y | Y | Y | Y | Y | ... |

**Table C.7.** continued.

| ID | CNAME | RV (km s$^{-1}$) | $T_{\text{eff}}$ (K) | logg (dex) | $\gamma^a$ | [Fe/H] (dex) | EW(Li)$^b$ (mÅ) | EW(Li) error flag$^c$ | $\gamma$ | logg | Membership RV | Li | H$\alpha$ | [Fe/H] | Randich$^d$ | Cantat-Gaudin$^d$ | Final$^e$ | NMs with Li$^f$ |
|---|---|---|---|---|---|---|---|---|---|---|---|---|---|---|---|---|---|---|
| 39134 | 17473807+0531398 | 1.9 ± 0.2 | 4887 ± 244 | … | 1.015 ± 0.004 | -0.34 ± 0.21 | <9 | 3 | N | … | … | … | … | … | … | … | n | G |
| 39135 | 17473832+0518104 | 58.1 ± 0.2 | 4742 ± 168 | … | 1.016 ± 0.003 | -0.25 ± 0.19 | <9 | 3 | N | … | … | … | … | … | … | … | n | … |
| 38793 | 17452549+0534438 | 8.3 ± 0.3 | 4754 ± 94 | 2.35 ± 0.14 | 1.017 ± 0.007 | -0.29 ± 0.15 | <33 | 3 | N | N | … | … | … | … | … | … | n | G |
| 39136 | 17473861+0548452 | -11.2 ± 0.3 | 5529 ± 183 | … | 1.003 ± 0.005 | -0.44 ± 0.10 | … | … | Y | … | Y | … | N | N | … | … | n | … |
| 38794 | 17452557+0545574 | -23.3 ± 0.2 | 5016 ± 112 | … | 1.012 ± 0.004 | … | <21 | 3 | N | … | … | … | … | … | … | … | n | G |
| 39137 | 17473864+0539286 | -20.0 ± 0.3 | 4687 ± 318 | … | 0.951 ± 0.006 | -0.14 ± 0.07 | <9 | 3 | Y | … | Y | N | N | Y | … | … | n | … |
| 38795 | 17452576+0523107 | -81.1 ± 0.2 | 5643 ± 186 | 3.94 ± 0.10 | 0.999 ± 0.003 | -0.12 ± 0.14 | <8 | 3 | Y | Y | N | N | N | Y | … | … | n | … |
| 39138 | 17473908+0556317 | -31.3 ± 0.2 | 5692 ± 152 | 3.99 ± 0.08 | 0.998 ± 0.002 | 0.19 ± 0.05 | <17 | 3 | Y | Y | Y | N | N | N | … | … | n | NG |
| 38796 | 17452610+0515305 | -60.4 ± 0.2 | 5708 ± 87 | … | 1.003 ± 0.002 | … | 46 ± 4 | 1 | Y | … | N | N | N | N | … | … | n | NG? |
| 39139 | 17473910+0523387 | -52.6 ± 0.4 | 4106 ± 505 | … | 0.839 ± 0.016 | -0.18 ± 0.15 | … | … | Y | … | N | … | … | Y | N | … | n | … |
| 38797 | 17452701+0521369 | -58.8 ± 0.2 | 4878 ± 206 | … | 1.015 ± 0.004 | -0.36 ± 0.17 | … | … | N | … | … | … | … | … | … | … | n | G |
| 39140 | 17473919+0531195 | 3.0 ± 0.2 | 4942 ± 114 | 3.14 ± 0.03 | 1.003 ± 0.002 | -0.07 ± 0.13 | <28 | 3 | Y | Y | Y | N | Y | Y | … | … | n | NG? |
| 38798 | 17452747+0524087 | -13.4 ± 0.2 | 4656 ± 12 | 4.18 ± 0.11 | 0.964 ± 0.005 | 0.02 ± 0.10 | 80 ± 3 | 1 | Y | Y | Y | Y | Y | Y | … | Y | Y | … |
| 39141 | 17473922+0519144 | 7.3 ± 0.2 | 6047 ± 61 | 4.15 ± 0.01 | 0.996 ± 0.002 | 0.00 ± 0.05 | 35 ± 2 | 1 | Y | Y | Y | N | N | Y | N | … | n | NG |
| 38799 | 17452791+0525251 | 57.2 ± 0.3 | 5768 ± 5 | … | 1.002 ± 0.003 | -0.27 ± 0.02 | <7 | 3 | Y | … | N | N | N | Y | … | … | n | … |
| 39142 | 17473947+0527101 | 5.0 ± 0.2 | 6289 ± 112 | 4.13 ± 0.17 | 0.998 ± 0.002 | -0.05 ± 0.04 | <11 | 3 | Y | Y | Y | N | N | Y | N | … | n | NG |
| 38800 | 17452814+0519197 | -28.5 ± 0.3 | … | … | … | … | … | … | … | … | … | … | … | N | … | … | n | … |
| 39143 | 17473995+0518315 | -154.0 ± 0.2 | 4582 ± 301 | … | 1.032 ± 0.004 | -0.04 ± 0.01 | <30 | 3 | N | … | … | … | … | … | … | … | n | G |
| 38801 | 17452869+0524053 | -8.7 ± 0.2 | 6050 ± 136 | 4.07 ± 0.18 | 0.996 ± 0.003 | -0.32 ± 0.16 | <16 | 3 | Y | Y | Y | N | N | N | N | … | n | NG |
| 39144 | 17474089+0523590 | -29.3 ± 0.2 | 4089 ± 167 | 4.47 ± 0.13 | 0.871 ± 0.003 | 0.12 ± 0.27 | … | … | Y | Y | Y | … | N | Y | N | … | n | … |
| 2625 | 17452906+0601462 | -8.8 ± 0.6 | 6475 ± 127 | 4.01 ± 0.15 | … | -0.06 ± 0.15 | 36 ± 23 | 2 | … | Y | Y | N | N | Y | Y | … | n | NG |
| 39145 | 17474137+0551096 | -5.6 ± 0.3 | 4006 ± 200 | 4.53 ± 0.03 | 0.837 ± 0.011 | -0.20 ± 0.04 | <40 | 3 | Y | Y | Y | Y | Y | Y | N | … | n | NG |
| 38802 | 17452921+0519076 | 3.1 ± 0.2 | 4777 ± 19 | … | 0.993 ± 0.006 | 0.09 ± 0.03 | <41 | 3 | Y | … | Y | N | Y | Y | … | … | n | NG |
| 39146 | 17474159+0558125 | -83.8 ± 0.2 | 4579 ± 150 | 2.57 ± 0.15 | 1.009 ± 0.002 | 0.04 ± 0.01 | <46 | 3 | Y | N | N | Y | Y | Y | … | … | n | NG? |
| 39147 | 17474164+0608450 | 25.3 ± 0.2 | 5029 ± 142 | … | 1.012 ± 0.003 | -0.14 ± 0.21 | <13 | 3 | N | … | … | … | … | … | … | … | n | G |
| 38803 | 17452949+0546435 | -34.5 ± 0.2 | 5022 ± 110 | … | 1.004 ± 0.003 | -0.06 ± 0.11 | <30 | 3 | Y | … | Y | N | N | Y | … | … | n | NG? |
| 39168 | 17474798+0546505 | 24.8 ± 0.2 | 3578 ± 88 | … | 0.838 ± 0.005 | -0.23 ± 0.14 | … | … | Y | … | N | … | … | Y | … | … | n | … |
| 38804 | 17452986+0554198 | 57.5 ± 0.2 | 5107 ± 120 | … | 1.013 ± 0.003 | -0.57 ± 0.18 | <3 | 3 | N | … | … | … | … | … | … | … | n | G |
| 38805 | 17453005+0548490 | 25.3 ± 0.2 | 3762 ± 102 | … | 0.831 ± 0.011 | -0.17 ± 0.14 | <100 | 3 | Y | … | N | Y | Y | Y | N | … | n | … |
| 39169 | 17474873+0546010 | 3.4 ± 0.3 | 4069 ± 284 | 4.46 ± 0.09 | 0.848 ± 0.010 | 0.02 ± 0.15 | <26 | 3 | Y | Y | Y | Y | Y | Y | N | … | Y?$^g$ | … |
| 38828 | 17453802+0521373 | 32.4 ± 0.2 | 5088 ± 162 | … | 1.011 ± 0.005 | -0.22 ± 0.20 | <23 | 3 | N | … | … | … | … | … | … | … | n | G |
| 2645 | 17474907+0532396 | -11.1 ± 0.6 | 4925 ± 39 | 4.53 ± 0.03 | … | -0.04 ± 0.05 | <5 | 3 | … | Y | Y | N | N | Y | … | … | n | … |
| 38829 | 17453820+0523431 | -3.7 ± 0.2 | 5856 ± 233 | 4.36 ± 0.19 | 0.985 ± 0.004 | 0.28 ± 0.11 | 21 ± 3 | 1 | Y | Y | Y | N | Y | N | … | … | n | NG |
| 39170 | 17474915+0545333 | -101.4 ± 0.3 | 4635 ± 252 | … | 0.920 ± 0.012 | -0.28 ± 0.11 | <17 | 3 | Y | … | N | N | N | Y | … | … | n | NG? |
| 38830 | 17453825+0543446 | -31.5 ± 0.3 | 4655 ± 46 | 2.46 ± 0.10 | 1.015 ± 0.006 | -0.05 ± 0.08 | <18 | 3 | N | N | … | … | … | … | … | … | n | G |
| 39171 | 17474970+0527342 | -48.3 ± 0.3 | 4358 ± 355 | 4.39 ± 0.04 | 0.895 ± 0.007 | 0.09 ± 0.22 | <39 | 3 | Y | Y | N | Y | Y | Y | … | … | n | NG |
| 38831 | 17453839+0559589 | 39.7 ± 0.2 | 5123 ± 95 | … | 1.006 ± 0.004 | … | <26 | 3 | Y | … | N | N | N | … | … | … | n | NG? |
| 39172 | 17474978+0550246 | -6.9 ± 0.2 | 5087 ± 61 | … | 0.980 ± 0.005 | -0.12 ± 0.02 | <13 | 3 | Y | … | Y | N | N | Y | … | … | n | NG |
| 2628 | 17453881+0536232 | -13.3 ± 0.6 | 6270 ± 93 | 3.92 ± 0.14 | … | -0.03 ± 0.12 | 106 ± 13 | 2 | … | Y | Y | Y | Y | Y | Y | Y | Y | … |
| 39173 | 17475067+0554156 | -36.6 ± 0.2 | 6424 ± 40 | … | 1.005 ± 0.002 | -0.77 ± 0.05 | <8 | 3 | Y | … | Y | N | N | N | … | … | n | … |
| 38832 | 17453903+0607557 | -5.6 ± 0.2 | 4855 ± 204 | … | 1.019 ± 0.004 | -0.09 ± 0.15 | <16 | 3 | N | … | … | … | … | … | … | … | n | G |
| 39174 | 17475117+0529599 | 8.3 ± 0.2 | 4927 ± 161 | … | 1.014 ± 0.004 | -0.16 ± 0.20 | <11 | 3 | N | … | … | … | … | … | … | … | n | G |
| 38833 | 17453916+0557562 | -54.3 ± 0.3 | 4303 ± 226 | 4.59 ± 0.04 | 0.884 ± 0.008 | -0.04 ± 0.18 | … | … | Y | Y | N | … | … | Y | … | … | n | … |
| 39175 | 17475221+0555454 | -69.3 ± 0.3 | 3983 ± 129 | 4.51 ± 0.09 | 0.850 ± 0.008 | -0.28 ± 0.20 | … | … | Y | Y | N | … | … | Y | N | … | n | … |
| 38834 | 17453923+0523445 | 18.7 ± 0.2 | 4460 ± 78 | 1.71 ± 0.04 | 1.026 ± 0.005 | -0.41 ± 0.03 | <15 | 3 | N | N | … | … | … | … | … | … | n | G |
| 2646 | 17475250+0513595 | -35.5 ± 0.6 | 6317 ± 131 | 3.88 ± 0.12 | … | -0.01 ± 0.12 | 64 ± 14 | 2 | … | Y | Y | Y | Y | Y | N | … | n | NG |
| 38835 | 17453924+0609580 | -27.8 ± 0.3 | 5078 ± 4 | … | 0.989 ± 0.006 | -0.07 ± 0.05 | <28 | 3 | Y | … | Y | N | N | Y | … | … | n | NG |
| 39176 | 17475266+0525456 | -31.6 ± 0.7 | 4092 ± 89 | … | 0.954 ± 0.021 | -0.60 ± 0.30 | … | … | Y | … | Y | … | N | N | … | … | n | … |
| 38836 | 17453932+0528191 | -57.3 ± 0.3 | 4052 ± 91 | 4.46 ± 0.15 | 0.876 ± 0.009 | -0.23 ± 0.13 | <19 | 3 | Y | Y | N | Y | N | Y | N | … | n | NG |
| 38837 | 17453967+0530530 | -13.3 ± 0.2 | 3512 ± 149 | … | 0.835 ± 0.007 | -0.14 ± 0.11 | … | … | Y | … | Y | … | Y | Y | Y | … | n | … |
| 39177 | 17475329+0520505 | 3.1 ± 0.2 | 5187 ± 10 | 4.09 ± 0.04 | 0.980 ± 0.003 | -0.03 ± 0.02 | <18 | 3 | Y | Y | Y | N | Y | Y | … | … | n | NG |
| 38838 | 17453973+0546363 | -34.1 ± 0.2 | 5219 ± 129 | … | 0.988 ± 0.002 | … | <21 | 3 | Y | … | Y | N | N | … | … | … | n | NG |
| 39178 | 17475362+0528092 | -13.6 ± 0.2 | 4518 ± 178 | … | 1.025 ± 0.003 | 0.02 ± 0.14 | <44 | 3 | N | … | … | … | … | … | … | … | n | G |
| 38839 | 17454034+0557260 | -32.5 ± 0.3 | 3985 ± 185 | 4.45 ± 0.20 | 0.855 ± 0.011 | -0.12 ± 0.13 | … | … | Y | Y | Y | … | N | Y | N | … | n | … |
| 39179 | 17475478+0534070 | -86.2 ± 0.3 | 4629 ± 222 | … | 0.959 ± 0.006 | 0.02 ± 0.01 | <23 | 3 | Y | … | N | N | N | Y | … | … | n | NG |
| 38840 | 17454087+0526223 | -1.5 ± 0.2 | 6362 ± 26 | … | 0.999 ± 0.003 | -0.18 ± 0.11 | <7 | 3 | Y | … | Y | N | N | Y | N | … | n | … |
| 39180 | 17475541+0558499 | -91.8 ± 0.3 | 4726 ± 260 | … | 0.895 ± 0.011 | … | … | … | Y | … | N | … | … | … | … | … | n | … |
| 38841 | 17454102+0554235 | -11.3 ± 0.3 | 3928 ± 132 | … | 0.853 ± 0.009 | -0.08 ± 0.08 | <21 | 3 | Y | … | Y | Y | Y | Y | Y | … | Y | … |
| 39181 | 17475548+0551545 | -53.8 ± 0.3 | 4800 ± 57 | 2.80 ± 0.06 | 1.006 ± 0.009 | -0.08 ± 0.09 | <20 | 3 | Y | N | N | N | N | Y | … | … | n | NG? |

| ID | CNAME | RV (km s$^{-1}$) | $T_{\rm eff}$ (K) | log g (dex) | $\gamma^a$ | [Fe/H] (dex) | EW(Li)$^b$ (mÅ) | EW(Li) error flag$^c$ | $\gamma$ | logg | Membership RV | Li | H$\alpha$ | [Fe/H] | Gaia studies Randich$^d$ | Cantat-Gaudin$^d$ | Final$^e$ | NMs with Li$^f$ |
|---|---|---|---|---|---|---|---|---|---|---|---|---|---|---|---|---|---|---|
| 38842 | 17454121+0514139 | 4.3 ± 0.2 | 4798 ± 173 | 2.49 ± 0.14 | 1.011 ± 0.006 | -0.30 ± 0.13 | 9 ± 3 | 1 | N | N | ... | ... | ... | ... | ... | ... | n | G |
| 39182 | 17475645+0556242 | -54.6 ± 0.2 | 5090 ± 14 | 3.49 ± 0.05 | 1.001 ± 0.002 | -0.13 ± 0.04 | <21 | 3 | Y | N | N | N | N | Y | ... | ... | n | NG? |
| 2629 | 17454123+0533444 | -10.4 ± 0.6 | 6572 ± 106 | 4.10 ± 0.13 | 0.999 ± 0.002 | -0.03 ± 0.11 | 74 ± 5 | 2 | Y | Y | Y | Y | Y | Y | N | ... | n | ... |
| 39183 | 17475678+0519594 | -8.7 ± 0.2 | 5266 ± 127 | 4.22 ± 0.03 | 0.979 ± 0.005 | 0.09 ± 0.09 | <20 | 3 | Y | Y | Y | N | N | Y | ... | ... | n | NG |
| 39184 | 17475738+0527506 | -21.2 ± 0.3 | 5002 ± 226 | ... | 1.012 ± 0.005 | -0.45 ± 0.20 | <26 | 3 | N | ... | ... | ... | ... | ... | ... | ... | n | G |
| 38843 | 17454145+0602391 | 3.6 ± 0.5 | 4250 ± 388 | ... | 0.912 ± 0.014 | -0.14 ± 0.10 | <31 | 3 | Y | ... | Y | Y | Y | Y | ... | ... | Y | ... |
| 38844 | 17454149+0530336 | 12.3 ± 0.3 | 4976 ± 201 | 3.16 ± 0.17 | 1.003 ± 0.007 | -0.25 ± 0.20 | <12 | 3 | Y | N | Y | N | N | Y | ... | ... | n | NG? |
| 39185 | 17475761+0517072 | -76.9 ± 0.2 | 5107 ± 75 | ... | 1.011 ± 0.004 | -0.44 ± 0.06 | <6 | 3 | N | ... | ... | ... | ... | ... | ... | ... | n | G |
| 38845 | 17454172+0547342 | -126.2 ± 0.4 | 4974 ± 200 | ... | 0.989 ± 0.015 | -0.33 ± 0.16 | ... | ... | Y | ... | N | ... | N | ... | ... | ... | n | ... |
| 39186 | 17475784+0540033 | 8.2 ± 0.3 | 4390 ± 195 | 4.43 ± 0.17 | 0.918 ± 0.006 | 0.04 ± 0.17 | <24 | 3 | Y | Y | Y | Y | N | Y | ... | ... | n | NG |
| 38846 | 17454235+0606281 | -38.8 ± 0.4 | 3935 ± 205 | ... | 0.845 ± 0.016 | -0.17 ± 0.14 | ... | ... | Y | ... | Y | ... | Y | Y | N | ... | n | ... |
| 2647 | 17475820+0559576 | -20.9 ± 0.6 | 5908 ± 11 | 3.96 ± 0.12 | ... | 0.04 ± 0.01 | 58 ± 4 | 2 | ... | Y | Y | N | Y | Y | ... | ... | n | NG |
| 38847 | 17454241+0555411 | -0.3 ± 0.2 | 4976 ± 216 | ... | 1.009 ± 0.004 | -0.22 ± 0.26 | ... | ... | Y | ... | Y | ... | Y | Y | ... | ... | n | ... |
| 39187 | 17475954+0540044 | -33.8 ± 0.2 | 3555 ± 102 | 4.61 ± 0.17 | 0.805 ± 0.006 | -0.26 ± 0.14 | ... | ... | Y | Y | Y | ... | N | Y | N | ... | n | ... |
| 38848 | 17454252+0608478 | -114.6 ± 0.3 | 4962 ± 39 | ... | 1.010 ± 0.006 | -0.10 ± 0.07 | <34 | 3 | N | ... | ... | ... | ... | ... | ... | ... | n | G |
| 39188 | 17475956+0537094 | -0.1 ± 0.3 | 4041 ± 98 | 4.55 ± 0.01 | 0.857 ± 0.006 | -0.14 ± 0.03 | <28 | 3 | Y | Y | Y | Y | N | Y | ... | ... | n | NG |
| 38849 | 17454370+0547214 | -31.1 ± 0.3 | 4897 ± 177 | ... | 1.028 ± 0.009 | -0.21 ± 0.19 | <12 | 3 | N | ... | ... | ... | ... | ... | ... | ... | n | G |
| 38896 | 17460207+0539448 | -19.4 ± 0.3 | ... | ... | ... | ... | ... | ... | ... | ... | ... | ... | N | ... | ... | ... | n | ... |
| 38897 | 17460208+0526008 | 20.7 ± 0.3 | 4572 ± 51 | 2.32 ± 0.12 | 1.022 ± 0.005 | -0.03 ± 0.01 | ... | ... | N | N | ... | ... | ... | ... | ... | ... | n | ... |
| 38898 | 17460270+0526599 | -17.0 ± 0.2 | 4621 ± 53 | 4.32 ± 0.19 | 0.962 ± 0.005 | -0.03 ± 0.01 | <19 | 3 | Y | Y | Y | N | N | Y | ... | ... | n | NG |
| 38899 | 17460277+0531399 | 52.7 ± 0.4 | 4502 ± 291 | 4.62 ± 0.10 | 0.903 ± 0.014 | -0.14 ± 0.04 | <33 | 3 | Y | Y | N | Y | N | Y | ... | ... | n | NG |
| 38900 | 17460330+0520363 | -31.6 ± 0.2 | 4784 ± 157 | ... | 0.974 ± 0.006 | 0.07 ± 0.15 | ... | ... | Y | ... | Y | ... | Y | Y | ... | ... | n | ... |
| 38901 | 17460351+0549426 | -62.8 ± 0.6 | ... | ... | ... | ... | ... | ... | ... | ... | ... | ... | ... | ... | ... | ... | n | ... |
| 38902 | 17460364+0521009 | -28.4 ± 0.5 | 5049 ± 138 | 4.28 ± 0.15 | 0.966 ± 0.014 | -0.32 ± 0.06 | <25 | 3 | Y | Y | Y | N | N | N | ... | ... | n | NG |
| 38903 | 17460366+0532394 | -71.4 ± 0.2 | 4962 ± 48 | 3.44 ± 0.18 | 0.997 ± 0.004 | 0.05 ± 0.03 | <28 | 3 | Y | N | N | N | N | Y | ... | ... | n | NG? |
| 38904 | 17460371+0557020 | -82.5 ± 0.2 | 4798 ± 202 | ... | 1.025 ± 0.004 | -0.08 ± 0.10 | <24 | 3 | N | ... | ... | ... | ... | ... | ... | ... | n | G |
| 38905 | 17460389+0520444 | -19.5 ± 0.2 | 6215 ± 68 | 4.16 ± 0.15 | 0.998 ± 0.005 | 0.28 ± 0.04 | 57 ± 4 | 1 | Y | Y | Y | N | N | N | N | ... | n | NG |
| 38906 | 17460424+0554020 | -60.2 ± 0.2 | 5511 ± 186 | 4.02 ± 0.01 | 0.992 ± 0.004 | -0.06 ± 0.09 | <18 | 3 | Y | Y | N | N | N | Y | ... | ... | n | NG |
| 38907 | 17460480+0529010 | -126.1 ± 0.2 | 4978 ± 170 | ... | 1.014 ± 0.005 | -0.37 ± 0.11 | <8 | 3 | N | ... | ... | ... | ... | ... | ... | ... | n | G |
| 38908 | 17460484+0516171 | 37.4 ± 0.3 | 4912 ± 51 | 2.82 ± 0.11 | 1.009 ± 0.008 | -0.08 ± 0.01 | <30 | 3 | Y | N | Y | N | N | Y | ... | ... | n | NG? |
| 38909 | 17460577+0541546 | -109.3 ± 0.4 | 3724 ± 43 | 4.53 ± 0.17 | 0.812 ± 0.011 | -0.20 ± 0.13 | ... | ... | Y | Y | N | ... | N | Y | ... | ... | n | ... |
| 2632 | 17460709+0555486 | -16.8 ± 0.6 | 6080 ± 42 | 4.43 ± 0.06 | 0.999 ± 0.002 | 0.10 ± 0.01 | 175 ± 2 | 2 | Y | Y | Y | Y | Y | Y | ... | ... | Y | ... |
| 38910 | 17460778+0526097 | 29.1 ± 0.3 | 4128 ± 167 | 4.51 ± 0.06 | 0.872 ± 0.009 | -0.18 ± 0.04 | ... | ... | Y | Y | N | ... | N | Y | N | ... | n | ... |
| 38911 | 17460796+0609482 | -83.7 ± 0.2 | 5056 ± 154 | ... | 1.016 ± 0.006 | -0.29 ± 0.02 | ... | ... | N | ... | ... | ... | ... | ... | ... | ... | n | G |
| 38912 | 17460981+0525340 | 66.5 ± 0.3 | 4891 ± 1 | 2.70 ± 0.06 | 1.022 ± 0.007 | -0.05 ± 0.04 | 99 ± 12 | 1 | N | N | ... | ... | ... | ... | ... | ... | n | Li-rich G |
| 38913 | 17461052+0524417 | 4.1 ± 0.2 | 4881 ± 212 | ... | 1.018 ± 0.004 | -0.12 ± 0.21 | <26 | 3 | N | ... | ... | ... | ... | ... | ... | ... | n | G |
| 38914 | 17461081+0557072 | -8.8 ± 0.3 | 3453 ± 63 | 4.60 ± 0.19 | 0.824 ± 0.009 | -0.25 ± 0.15 | ... | ... | Y | Y | Y | ... | N | Y | N | ... | n | ... |
| 38915 | 17461112+0555482 | -17.1 ± 0.2 | 5054 ± 60 | ... | 1.007 ± 0.002 | -0.07 ± 0.09 | <23 | 3 | Y | ... | Y | N | N | Y | ... | ... | n | NG? |
| 38853 | 17454634+0532288 | 0.5 ± 0.2 | 5062 ± 55 | ... | 1.000 ± 0.003 | -0.25 ± 0.16 | <13 | 3 | Y | ... | Y | N | N | Y | ... | ... | n | NG? |
| 38854 | 17454646+0543549 | -20.8 ± 0.3 | 3682 ± 73 | ... | 0.830 ± 0.008 | -0.19 ± 0.13 | ... | ... | Y | ... | Y | ... | Y | Y | N | ... | n | ... |
| 38855 | 17454768+0525441 | 113.3 ± 0.3 | 4855 ± 108 | ... | 1.015 ± 0.005 | -0.20 ± 0.18 | ... | ... | N | ... | ... | ... | ... | ... | ... | ... | n | G |
| 38856 | 17454777+0552575 | -22.6 ± 0.3 | 4995 ± 111 | 4.59 ± 0.13 | 0.957 ± 0.008 | -0.19 ± 0.09 | <15 | 3 | Y | Y | Y | N | N | Y | ... | ... | n | NG |
| 38857 | 17454838+0531231 | -3.8 ± 0.2 | 4821 ± 174 | 2.44 ± 0.20 | 1.021 ± 0.004 | -0.24 ± 0.10 | <14 | 3 | N | N | ... | ... | ... | ... | ... | ... | n | G |
| 38858 | 17454857+0549044 | 33.2 ± 0.4 | 5209 ± 70 | 3.56 ± 0.02 | 1.002 ± 0.015 | -0.14 ± 0.11 | <30 | 3 | Y | Y | N | N | N | Y | ... | ... | n | NG? |
| 38859 | 17454867+0600413 | -54.4 ± 0.2 | 4872 ± 127 | ... | 0.930 ± 0.009 | ... | <33 | 3 | Y | ... | N | N | Y | ... | ... | ... | n | NG |
| 38860 | 17454929+0540482 | 16.5 ± 0.2 | 5783 ± 97 | 4.31 ± 0.05 | 0.989 ± 0.003 | 0.10 ± 0.06 | <36 | 3 | Y | Y | N | N | N | Y | ... | ... | n | NG |
| 38861 | 17454943+0601272 | 17.4 ± 0.2 | 5983 ± 59 | 4.12 ± 0.12 | 0.998 ± 0.003 | 0.03 ± 0.08 | 44 ± 2 | 1 | Y | Y | N | N | N | Y | ... | ... | n | NG |
| 38862 | 17454961+0525508 | 24.6 ± 0.3 | 5342 ± 61 | 4.02 ± 0.19 | 0.991 ± 0.005 | -0.10 ± 0.04 | 12 ± 5 | 1 | Y | Y | N | N | N | Y | ... | ... | n | NG |
| 38863 | 17454987+0607516 | -76.6 ± 0.3 | 3966 ± 151 | ... | 0.873 ± 0.007 | -0.29 ± 0.21 | ... | ... | Y | ... | N | ... | N | Y | N | ... | n | ... |
| 38864 | 17455032+0607308 | 35.4 ± 0.2 | 4662 ± 30 | 2.28 ± 0.06 | 1.010 ± 0.006 | -0.42 ± 0.10 | <24 | 3 | Y | N | N | N | N | N | ... | ... | n | NG? |
| 38865 | 17455051+0603468 | 50.1 ± 0.3 | 4777 ± 187 | 2.70 ± 0.04 | 1.005 ± 0.007 | -0.19 ± 0.19 | <38 | 3 | Y | N | N | N | N | Y | ... | ... | n | NG? |
| 38866 | 17455125+0525437 | 25.6 ± 0.2 | 5072 ± 71 | ... | 0.998 ± 0.005 | -0.19 ± 0.12 | <26 | 3 | Y | ... | N | N | N | Y | ... | ... | n | NG |
| 38867 | 17455152+0531537 | -39.1 ± 0.2 | 4894 ± 153 | ... | 1.025 ± 0.004 | -0.16 ± 0.14 | <7 | 3 | N | ... | ... | ... | ... | ... | ... | ... | n | G |
| 38868 | 17455198+0559245 | -23.2 ± 0.3 | 4097 ± 216 | 4.57 ± 0.07 | 0.846 ± 0.011 | -0.04 ± 0.17 | ... | ... | Y | Y | Y | ... | Y | Y | N | ... | n | ... |
| 38869 | 17455219+0522487 | -154.9 ± 0.3 | 4793 ± 51 | ... | 1.037 ± 0.05 | -0.87 ± 0.10 | <11 | 3 | N | ... | ... | ... | ... | ... | ... | ... | n | G |
| 38870 | 17455263+0535364 | -70.0 ± 0.2 | 4614 ± 47 | 2.44 ± 0.13 | 1.025 ± 0.004 | 0.06 ± 0.04 | <45 | 3 | N | N | ... | ... | ... | ... | ... | ... | n | G |
| 38871 | 17455281+0517052 | -13.5 ± 0.3 | 6176 ± 76 | 3.92 ± 0.07 | 1.005 ± 0.005 | -0.11 ± 0.07 | 60 ± 3 | 1 | Y | Y | Y | N | N | Y | Y | ... | n | NG? |
| 38872 | 17455294+0546267 | 27.6 ± 0.2 | 4285 ± 201 | 4.57 ± 0.03 | 0.885 ± 0.007 | 0.16 ± 0.30 | <35 | 3 | Y | Y | N | Y | N | N | ... | ... | n | NG |
| 38873 | 17455308+0536293 | -13.7 ± 0.3 | 5247 ± 170 | 3.69 ± 0.13 | 0.994 ± 0.005 | 0.01 ± 0.08 | 19 ± 4 | 1 | Y | Y | Y | Y | Y | Y | ... | Y | Y | ... |

**Table C.7.** continued.

| ID | CNAME | RV (km s$^{-1}$) | $T_{\rm eff}$ (K) | $\log g$ (dex) | $\gamma^a$ | [Fe/H] (dex) | $EW({\rm Li})^b$ (mÅ) | $EW({\rm Li})$ error flag$^c$ | $\gamma$ | $\log g$ | Membership RV | Li | H$\alpha$ | [Fe/H] | Gaia studies Randich$^d$ | Cantat-Gaudin$^d$ | Final$^e$ | NMs with Li$^f$ |
|---|---|---|---|---|---|---|---|---|---|---|---|---|---|---|---|---|---|---|
| 38919 | 17461212+0528362 | 33.7 ± 0.3 | 4990 ± 22 | 2.80 ± 0.09 | 1.015 ± 0.007 | -0.10 ± 0.16 | 29 ± 8 | 1 | N | N | … | … | … | … | … | … | n | G |
| 38920 | 17461249+0604266 | 124.2 ± 0.3 | 4851 ± 6 | 2.65 ± 0.15 | 1.017 ± 0.007 | -0.32 ± 0.01 | 19 ± 4 | 1 | N | N | … | … | … | … | … | … | n | G |
| 38921 | 17461263+0538547 | -6.9 ± 0.2 | 4513 ± 81 | … | 1.030 ± 0.006 | -0.21 ± 0.23 | <14 | 3 | N | … | … | … | … | … | … | … | n | G |
| 38922 | 17461267+0606091 | 17.4 ± 0.2 | 6171 ± 7 | 4.02 ± 0.20 | 0.997 ± 0.002 | -0.32 ± 0.22 | 42 ± 6 | 1 | Y | Y | N | N | N | Y | N | … | n | NG |
| 38923 | 17461292+0559296 | -14.7 ± 0.4 | 3443 ± 91 | … | 0.839 ± 0.014 | -0.22 ± 0.13 | <100 | 3 | Y | … | Y | Y | Y | Y | Y | Y | Y | … |
| 38924 | 17461301+0609515 | -36.0 ± 0.3 | 4466 ± 342 | 4.57 ± 0.04 | 0.897 ± 0.013 | -0.11 ± 0.05 | <32 | 3 | Y | Y | Y | Y | Y | Y | … | … | Y | … |
| 38925 | 17461351+0554301 | -29.7 ± 0.3 | 3609 ± 117 | … | 0.850 ± 0.011 | -0.24 ± 0.14 | … | … | Y | … | Y | … | N | Y | N | … | n | … |
| 38926 | 17461438+0550362 | -12.2 ± 0.3 | 3873 ± 62 | 4.57 ± 0.10 | 0.830 ± 0.007 | -0.21 ± 0.07 | … | … | Y | Y | Y | … | N | Y | Y | … | n | … |
| 38927 | 17461465+0542155 | -23.3 ± 0.2 | 5206 ± 115 | … | 0.989 ± 0.002 | … | <16 | 3 | Y | … | Y | N | N | … | … | … | n | NG |
| 38928 | 17461481+0557449 | -31.3 ± 0.3 | 5086 ± 44 | … | 0.985 ± 0.007 | 0.00 ± 0.03 | <27 | 3 | Y | … | Y | N | N | Y | … | … | n | NG |
| 38929 | 17461523+0601186 | 7.7 ± 0.4 | 3945 ± 184 | … | 0.869 ± 0.011 | -0.08 ± 0.11 | <34 | 3 | Y | … | Y | Y | Y | Y | N | … | Y?$^g$ | … |
| 38930 | 17461625+0523204 | -77.3 ± 0.3 | 4214 ± 419 | … | 0.883 ± 0.018 | 0.04 ± 0.23 | <24 | 3 | Y | … | N | Y | N | Y | … | … | n | NG |
| 38931 | 17461699+0526548 | 26.3 ± 0.2 | 4879 ± 187 | 3.17 ± 0.05 | 0.997 ± 0.004 | -0.13 ± 0.13 | <20 | 3 | Y | N | N | N | N | Y | … | … | n | NG? |
| 38932 | 17461708+0606160 | 32.3 ± 0.2 | 4672 ± 70 | 2.41 ± 0.09 | 1.021 ± 0.004 | 0.03 ± 0.10 | <38 | 3 | N | N | … | … | … | … | … | … | n | G |
| 38933 | 17461730+0556008 | -50.8 ± 0.2 | 4743 ± 43 | 2.89 ± 0.16 | 1.003 ± 0.004 | -0.02 ± 0.05 | <31 | 3 | Y | N | N | N | N | Y | … | … | n | NG? |
| 38934 | 17461743+0607294 | -22.4 ± 0.2 | 4264 ± 276 | 4.51 ± 0.01 | 0.878 ± 0.007 | 0.06 ± 0.20 | <13 | 3 | Y | Y | Y | Y | N | Y | … | … | n | NG |
| 38935 | 17461779+0545090 | 29.3 ± 0.2 | 4533 ± 211 | … | 1.015 ± 0.006 | -0.12 ± 0.09 | <20 | 3 | N | … | … | … | … | … | … | … | n | G |
| 38936 | 17461801+0523216 | -5.9 ± 0.3 | 3468 ± 131 | … | 0.825 ± 0.013 | -0.25 ± 0.14 | … | … | Y | … | Y | … | Y | Y | N | Y | n | … |
| 2633 | 17461845+0528170 | -36.5 ± 0.6 | 6058 ± 17 | 4.15 ± 0.06 | … | 0.10 ± 0.02 | 58 ± 5 | 2 | … | Y | Y | N | N | Y | N | … | n | NG |
| 38937 | 17461847+0552140 | -69.1 ± 0.3 | 5666 ± 60 | … | 1.004 ± 0.006 | -0.33 ± 0.06 | … | … | Y | … | N | … | N | … | … | … | n | … |
| 38938 | 17461870+0554506 | 125.8 ± 0.3 | 4980 ± 158 | … | 1.003 ± 0.007 | -0.61 ± 0.23 | <19 | 3 | Y | … | N | N | N | N | … | … | n | NG? |
| 38939 | 17461935+0601418 | -20.0 ± 0.2 | 5074 ± 94 | … | 1.024 ± 0.005 | … | <7 | 3 | N | … | … | … | … | … | … | … | n | G |
| 2634 | 17461964+0534066 | -27.0 ± 0.6 | 7055 ± 131 | 4.10 ± 0.13 | … | -0.05 ± 0.11 | <15 | 3 | … | Y | Y | N | N | Y | N | … | n | NG |
| 38940 | 17461982+0550231 | 30.9 ± 0.3 | 4771 ± 95 | 2.38 ± 0.03 | 1.012 ± 0.009 | -0.17 ± 0.09 | <24 | 3 | N | N | … | … | … | … | … | … | n | G |
| 38964 | 17462860+0528177 | 43.7 ± 0.3 | 4925 ± 212 | … | 1.009 ± 0.006 | -0.27 ± 0.23 | <16 | 3 | Y | … | N | N | N | Y | … | … | n | NG? |
| 38965 | 17462880+0608062 | 6.2 ± 0.3 | 5172 ± 13 | … | 0.988 ± 0.010 | -0.01 ± 0.01 | … | … | Y | … | Y | … | N | Y | … | … | n | … |
| 38966 | 17462913+0531200 | -2.0 ± 0.2 | 4864 ± 136 | 2.68 ± 0.16 | 1.011 ± 0.004 | -0.19 ± 0.16 | <16 | 3 | N | N | … | … | … | … | … | … | n | G |
| 38967 | 17462918+0555295 | -32.3 ± 0.2 | … | … | … | … | … | … | … | … | … | … | … | N | … | … | n | … |
| 38968 | 17462948+0528457 | -51.8 ± 0.2 | 5012 ± 125 | … | 1.010 ± 0.005 | -0.19 ± 0.12 | <34 | 3 | Y | … | N | N | N | Y | … | … | n | NG? |
| 38969 | 17462973+0524124 | 101.3 ± 0.2 | 4687 ± 100 | 2.63 ± 0.14 | 1.007 ± 0.004 | -0.05 ± 0.05 | <41 | 3 | Y | N | N | Y | N | Y | … | … | n | NG? |
| 38970 | 17463026+0530321 | -3.6 ± 0.3 | 4603 ± 110 | 4.38 ± 0.06 | 0.945 ± 0.008 | -0.34 ± 0.01 | … | … | Y | Y | Y | … | Y | N | … | … | n | … |
| 38971 | 17463111+0606003 | 38.8 ± 0.2 | 5605 ± 50 | … | 0.999 ± 0.002 | 0.02 ± 0.10 | 63 ± 3 | 1 | Y | … | N | N | N | Y | … | … | n | NG |
| 2636 | 17463149+0605280 | -14.5 ± 0.6 | 6029 ± 100 | 4.20 ± 0.13 | 0.998 ± 0.003 | 0.02 ± 0.12 | 157 ± 14 | 2 | Y | Y | Y | Y | Y | Y | Y | Y | Y | … |
| 38972 | 17463169+0549124 | 5.0 ± 0.3 | 4373 ± 212 | 4.46 ± 0.10 | 0.910 ± 0.011 | -0.11 ± 0.01 | <16 | 3 | Y | Y | Y | Y | N | Y | … | … | n | NG |
| 38973 | 17463235+0522264 | -67.8 ± 0.3 | 4642 ± 206 | … | 0.955 ± 0.007 | -0.02 ± 0.04 | <28 | 3 | Y | … | N | N | N | Y | … | … | n | NG |
| 38974 | 17463260+0534414 | 89.3 ± 0.2 | 5037 ± 70 | … | 1.018 ± 0.005 | -0.20 ± 0.16 | <8 | 3 | N | … | … | … | … | … | … | … | n | G |
| 38975 | 17463397+0528512 | -62.6 ± 0.2 | 6241 ± 54 | 4.01 ± 0.12 | 1.002 ± 0.005 | -0.27 ± 0.04 | 51 ± 5 | 1 | Y | Y | N | N | N | Y | N | … | n | NG? |
| 38976 | 17463397+0540541 | -18.9 ± 0.2 | 4644 ± 47 | 2.42 ± 0.10 | 1.016 ± 0.005 | -0.08 ± 0.05 | … | … | N | N | … | … | … | … | … | … | n | G |
| 38977 | 17463399+0539521 | 97.9 ± 0.2 | 4708 ± 61 | … | 1.024 ± 0.003 | -0.21 ± 0.15 | <16 | 3 | N | … | … | … | … | … | … | … | n | G |
| 38978 | 17463516+0526288 | -40.6 ± 0.5 | … | … | … | … | … | … | … | … | … | … | … | … | N | … | n | … |
| 38979 | 17463553+0531076 | -14.5 ± 0.3 | 3719 ± 14 | 4.54 ± 0.20 | 0.819 ± 0.010 | -0.19 ± 0.13 | … | … | Y | Y | Y | … | Y | Y | Y | Y | n | … |
| 38980 | 17463593+0534160 | -39.4 ± 0.2 | 5614 ± 44 | 4.07 ± 0.11 | 0.992 ± 0.002 | -0.01 ± 0.09 | <13 | 3 | Y | Y | Y | N | N | Y | … | … | n | NG |
| 38981 | 17463638+0604294 | -49.4 ± 0.3 | 3973 ± 211 | 4.47 ± 0.14 | 0.841 ± 0.013 | -0.10 ± 0.06 | <45 | 3 | Y | Y | N | Y | N | Y | N | … | n | NG |
| 38982 | 17463764+0603521 | 5.5 ± 0.3 | 5774 ± 162 | … | 0.988 ± 0.007 | 0.24 ± 0.07 | 55 ± 9 | 1 | Y | … | Y | N | N | N | … | … | n | NG |
| 38983 | 17463796+0542264 | 70.8 ± 0.2 | 5039 ± 197 | … | 1.009 ± 0.003 | -0.54 ± 0.17 | … | … | Y | … | N | … | N | N | … | … | n | … |
| 38984 | 17463893+0602129 | 15.4 ± 3.8 | 3925 ± 63 | … | … | -0.20 ± 0.18 | … | … | … | … | … | … | … | … | N | Y | n | … |
| 38985 | 17463998+0605144 | -46.8 ± 0.3 | 4931 ± 66 | 4.39 ± 0.15 | 0.961 ± 0.007 | -0.14 ± 0.02 | … | … | Y | Y | N | … | N | Y | … | … | n | … |
| 38986 | 17464015+0602453 | 0.5 ± 0.4 | 3593 ± 94 | 4.66 ± 0.07 | 0.789 ± 0.011 | -0.24 ± 0.14 | … | … | Y | Y | Y | … | N | Y | N | … | n | … |
| 39035 | 17465739+0529578 | 18.6 ± 0.2 | 5025 ± 279 | … | 1.017 ± 0.005 | -0.30 ± 0.26 | 24 ± 7 | 1 | N | … | … | … | … | … | … | … | n | G |
| 39036 | 17465785+0559100 | 45.9 ± 0.3 | 4789 ± 219 | 4.42 ± 0.17 | 0.956 ± 0.007 | -0.08 ± 0.01 | <33 | 3 | Y | Y | N | N | N | Y | … | … | n | NG |
| 39037 | 17465891+0555524 | 54.2 ± 0.2 | 4634 ± 97 | … | 1.022 ± 0.005 | -0.35 ± 0.10 | … | … | N | … | … | … | … | … | … | … | n | G |
| 39038 | 17465904+0545448 | -46.8 ± 0.3 | 3324 ± 41 | … | 0.888 ± 0.013 | -0.25 ± 0.14 | <100 | 3 | Y | … | N | Y | N | Y | N | … | n | NG |
| 39039 | 17465939+0520079 | -28.3 ± 0.3 | 4603 ± 38 | 4.17 ± 0.19 | 0.962 ± 0.009 | 0.12 ± 0.14 | <19 | 3 | Y | Y | Y | N | N | Y | … | … | n | NG |
| 39040 | 17465956+0525489 | -2.1 ± 0.3 | 4804 ± 254 | 4.54 ± 0.15 | 0.928 ± 0.007 | -0.21 ± 0.22 | <28 | 3 | Y | Y | Y | N | N | Y | … | … | n | NG |
| 39041 | 17465984+0605309 | 4.7 ± 0.2 | 5046 ± 88 | … | 1.015 ± 0.005 | -0.16 ± 0.18 | <8 | 3 | N | … | … | … | … | … | … | … | n | G |
| 39042 | 17465999+0602249 | 19.8 ± 0.2 | 4728 ± 84 | 2.51 ± 0.20 | 1.019 ± 0.002 | -0.08 ± 0.19 | <18 | 3 | N | N | … | … | … | … | … | … | n | G |
| 39043 | 17470023+0530290 | -16.1 ± 0.2 | 4772 ± 29 | … | 0.993 ± 0.004 | 0.07 ± 0.02 | <41 | 3 | Y | … | Y | Y | Y | Y | … | … | Y | … |
| 39044 | 17470039+0606443 | -74.5 ± 0.4 | 3740 ± 108 | 4.58 ± 0.15 | 0.804 ± 0.016 | -0.20 ± 0.13 | … | … | Y | Y | N | … | … | Y | N | … | n | … |
| 39045 | 17470090+0523258 | -45.1 ± 0.2 | 5252 ± 108 | … | 1.002 ± 0.003 | -0.16 ± 0.03 | 33 ± 7 | 1 | Y | … | N | N | N | Y | … | … | n | NG? |

| ID | CNAME | RV (km s$^{-1}$) | $T_{\rm eff}$ (K) | logg (dex) | $\gamma^a$ | [Fe/H] (dex) | EW(Li)$^b$ (mÅ) | EW(Li) error flag$^c$ | \multicolumn{5}{c|}{Membership} | \multicolumn{2}{c|}{Gaia studies} | Final$^e$ | NMs with Li$^f$ |
| | | | | | | | | | $\gamma$ | logg | RV | Li | H$\alpha$ | [Fe/H] | Randich$^d$ | Cantat-Gaudin$^d$ | | |
|---|---|---|---|---|---|---|---|---|---|---|---|---|---|---|---|---|---|---|
| 39046 | 17470103+0606402 | -37.8 ± 0.2 | 6306 ± 26 | 4.16 ± 0.06 | 1.000 ± 0.002 | 0.11 ± 0.02 | 37 ± 5 | 1 | Y | Y | Y | N | N | Y | N | ... | n | NG |
| 39047 | 17470337+0531556 | -27.0 ± 0.2 | 5237 ± 107 | ... | 0.986 ± 0.002 | -0.11 ± 0.04 | <15 | 3 | Y | ... | Y | N | N | Y | ... | ... | n | NG |
| 39048 | 17470360+0602084 | -9.9 ± 0.2 | 4537 ± 105 | ... | 1.031 ± 0.003 | ... | <30 | 3 | N | ... | ... | ... | ... | ... | ... | ... | n | G |
| 39049 | 17470474+0529162 | 39.9 ± 0.2 | 5920 ± 81 | 4.18 ± 0.11 | 0.996 ± 0.002 | 0.04 ± 0.04 | 13 ± 2 | 1 | Y | Y | N | N | Y | Y | ... | ... | n | NG |
| 39050 | 17470499+0558392 | -29.7 ± 0.2 | 5000 ± 180 | ... | 1.018 ± 0.004 | -0.28 ± 0.19 | <19 | 3 | N | ... | ... | ... | ... | ... | ... | ... | n | G |
| 39051 | 17470529+0543313 | -14.4 ± 0.2 | 4888 ± 215 | ... | 1.022 ± 0.004 | -0.11 ± 0.14 | <18 | 3 | N | ... | ... | ... | ... | ... | ... | ... | n | G |
| 2637 | 17470638+0526026 | 7.7 ± 0.6 | 6387 ± 214 | 4.07 ± 0.25 | ... | -0.01 ± 0.10 | 31 ± 6 | 2 | ... | Y | Y | N | N | Y | N | ... | n | NG |
| 39052 | 17470646+0528041 | 1.9 ± 0.2 | 4572 ± 74 | 4.56 ± 0.09 | 0.933 ± 0.004 | -0.07 ± 0.06 | <38 | 3 | Y | Y | Y | Y | Y | Y | ... | ... | Y | ... |
| 39053 | 17470756+0600257 | 44.2 ± 0.2 | 4701 ± 50 | ... | 1.014 ± 0.003 | -0.21 ± 0.18 | <12 | 3 | N | ... | ... | ... | ... | ... | ... | ... | n | G |
| 39054 | 17470781+0533592 | -0.5 ± 0.2 | 5239 ± 27 | ... | 0.988 ± 0.004 | -0.16 ± 0.05 | <23 | 3 | Y | ... | Y | N | N | Y | ... | ... | n | NG |
| 39055 | 17470791+0604144 | -52.1 ± 0.2 | 5109 ± 88 | ... | 0.971 ± 0.005 | ... | <15 | 3 | Y | ... | N | N | Y | ... | ... | ... | n | NG |
| 39056 | 17470826+0556525 | -47.3 ± 0.2 | 4582 ± 120 | ... | 1.016 ± 0.002 | 0.01 ± 0.04 | <34 | 3 | N | ... | ... | ... | ... | ... | ... | ... | n | G |
| 39057 | 17470852+0537376 | -111.1 ± 0.2 | 4803 ± 323 | ... | 1.017 ± 0.005 | -0.21 ± 0.16 | <28 | 3 | N | ... | ... | ... | ... | ... | ... | ... | n | G |
| 39079 | 17471689+0608413 | -55.9 ± 0.2 | 4834 ± 95 | ... | 1.011 ± 0.004 | -0.06 ± 0.12 | <29 | 3 | N | ... | ... | ... | ... | ... | ... | ... | n | G |
| 39080 | 17471720+0531241 | 42.0 ± 0.2 | 4471 ± 125 | ... | 1.025 ± 0.002 | -0.21 ± 0.13 | ... | ... | N | ... | ... | ... | ... | ... | ... | ... | n | G |
| 39081 | 17471776+0556229 | -5.7 ± 0.2 | 4947 ± 207 | ... | 1.005 ± 0.004 | -0.15 ± 0.21 | <13 | 3 | Y | ... | Y | N | N | Y | ... | ... | n | NG? |
| 2641 | 17471834+0530062 | -45.9 ± 0.6 | 6229 ± 132 | 3.98 ± 0.17 | 1.004 ± 0.005 | -0.18 ± 0.17 | 58 ± 5 | 2 | Y | Y | N | N | Y | Y | N | ... | n | NG? |
| 39082 | 17471842+0526438 | 13.3 ± 0.2 | 6007 ± 13 | 4.05 ± 0.01 | 0.997 ± 0.001 | -0.21 ± 0.14 | 50 ± 3 | 1 | Y | Y | Y | N | N | Y | N | ... | n | NG |
| 39083 | 17471868+0520485 | 2.1 ± 0.2 | 6073 ± 2 | 4.18 ± 0.08 | 0.993 ± 0.003 | 0.15 ± 0.01 | 24 ± 7 | 1 | Y | Y | Y | N | N | Y | N | ... | n | NG |
| 39084 | 17471920+0531105 | -2.7 ± 0.2 | 5575 ± 128 | 4.21 ± 0.11 | 0.990 ± 0.002 | -0.06 ± 0.15 | <22 | 3 | Y | Y | Y | N | N | Y | ... | ... | n | NG |
| 39085 | 17471950+0530414 | -22.2 ± 0.4 | 3496 ± 55 | 4.66 ± 0.11 | 0.817 ± 0.015 | -0.25 ± 0.13 | ... | ... | Y | Y | Y | ... | Y | Y | N | ... | n | ... |
| 39086 | 17472001+0546537 | -13.4 ± 0.3 | 3564 ± 53 | ... | 0.819 ± 0.009 | -0.21 ± 0.14 | ... | ... | Y | ... | Y | ... | Y | Y | Y | Y | n | ... |
| 39087 | 17472070+0604000 | -17.1 ± 0.2 | 6002 ± 124 | 4.05 ± 0.17 | 0.996 ± 0.003 | -0.23 ± 0.08 | <16 | 3 | Y | Y | Y | N | N | Y | N | ... | n | NG |
| 39088 | 17472159+0548584 | 10.4 ± 0.3 | 6458 ± 78 | 4.07 ± 0.03 | 1.001 ± 0.003 | -0.22 ± 0.34 | ... | ... | Y | Y | Y | ... | N | Y | N | ... | n | ... |
| 39089 | 17472202+0605308 | -33.7 ± 0.2 | 5279 ± 121 | ... | 0.989 ± 0.004 | 0.02 ± 0.03 | <19 | 3 | Y | ... | Y | N | N | Y | ... | ... | n | NG |
| 39090 | 17472218+0602277 | 179.6 ± 0.2 | 5132 ± 60 | 3.04 ± 0.09 | 1.021 ± 0.002 | -0.01 ± 0.01 | 14 ± 6 | 1 | N | N | ... | ... | ... | ... | ... | ... | n | G |
| 39091 | 17472326+0610267 | -75.1 ± 0.2 | 5037 ± 143 | ... | ... | ... | ... | ... | ... | ... | ... | ... | ... | ... | ... | ... | n | ... |
| 38695 | 17444266+0554454 | -79.8 ± 0.2 | 4739 ± 113 | ... | 1.023 ± 0.005 | -0.28 ± 0.14 | <11 | 3 | N | ... | ... | ... | ... | ... | ... | ... | n | G |
| 39092 | 17472407+0530574 | 36.1 ± 0.2 | 4543 ± 119 | ... | 1.052 ± 0.003 | ... | <8 | 3 | N | ... | ... | ... | ... | ... | ... | ... | n | G |
| 38696 | 17444331+0547537 | -30.7 ± 0.4 | 5269 ± 309 | ... | 1.030 ± 0.012 | -0.51 ± 0.23 | <29 | 3 | N | ... | ... | ... | ... | ... | ... | ... | n | ... |
| 39093 | 17472470+0515193 | -35.5 ± 0.2 | 5167 ± 56 | 4.06 ± 0.10 | 0.983 ± 0.003 | 0.09 ± 0.15 | <18 | 3 | Y | Y | Y | N | Y | Y | ... | ... | n | NG |
| 38697 | 17444351+0544201 | -11.9 ± 0.2 | 5783 ± 107 | 4.15 ± 0.03 | 0.993 ± 0.003 | 0.11 ± 0.01 | <15 | 3 | Y | Y | Y | N | N | Y | ... | ... | n | NG |
| 39094 | 17472494+0533106 | -95.1 ± 0.2 | 5639 ± 132 | 3.94 ± 0.03 | 0.997 ± 0.002 | -0.02 ± 0.10 | <16 | 3 | Y | Y | N | N | N | Y | ... | ... | n | NG |
| 2620 | 17444501+0525346 | -30.2 ± 0.6 | 6389 ± 73 | 3.86 ± 0.03 | ... | -0.13 ± 0.05 | <11 | 3 | ... | Y | Y | N | N | Y | N | ... | n | NG |
| 39095 | 17472530+0535389 | -46.2 ± 0.2 | 4686 ± 7 | 2.49 ± 0.17 | 1.021 ± 0.005 | -0.12 ± 0.10 | ... | ... | N | N | ... | ... | ... | ... | ... | ... | n | G |
| 39096 | 17472576+0559334 | 50.2 ± 0.2 | 4673 ± 45 | ... | 1.016 ± 0.003 | -0.29 ± 0.07 | <18 | 3 | N | ... | ... | ... | ... | ... | ... | ... | n | G |
| 38698 | 17444501+0602115 | -31.3 ± 0.3 | 3625 ± 21 | ... | 0.827 ± 0.012 | -0.21 ± 0.12 | ... | ... | Y | ... | Y | ... | N | Y | N | ... | n | ... |
| 39097 | 17472603+0537280 | -33.6 ± 0.3 | 5239 ± 68 | ... | 0.992 ± 0.004 | -0.12 ± 0.03 | <35 | 3 | Y | ... | Y | N | N | Y | ... | ... | n | NG |
| 38699 | 17444611+0538404 | 44.5 ± 0.2 | 4956 ± 202 | 4.34 ± 0.10 | 0.966 ± 0.004 | -0.19 ± 0.03 | <8 | 3 | Y | Y | N | N | Y | Y | ... | ... | n | ... |
| 39098 | 17472613+0544284 | -13.6 ± 0.3 | 3505 ± 21 | ... | 0.841 ± 0.013 | -0.22 ± 0.14 | ... | ... | Y | ... | Y | ... | Y | Y | Y | Y | n | ... |
| 38700 | 17444642+0533555 | -24.5 ± 0.3 | 4831 ± 138 | ... | 0.946 ± 0.007 | ... | <24 | 3 | Y | ... | Y | N | Y | N | ... | ... | n | NG |
| 39099 | 17472689+0555192 | -26.1 ± 0.2 | 4892 ± 172 | ... | 0.930 ± 0.005 | ... | <8 | 3 | Y | ... | Y | N | N | ... | ... | ... | n | ... |
| 38701 | 17444715+0548020 | 41.4 ± 0.3 | 5085 ± 108 | ... | 0.994 ± 0.007 | -0.38 ± 0.08 | <40 | 3 | Y | ... | N | N | N | N | ... | ... | n | NG |
| 39100 | 17472700+0513467 | -86.1 ± 0.2 | 5590 ± 272 | 3.92 ± 0.14 | 1.001 ± 0.002 | -0.16 ± 0.22 | <15 | 3 | Y | Y | N | N | N | Y | ... | ... | n | NG? |
| 38702 | 17444882+0529193 | -13.7 ± 0.2 | 5225 ± 43 | 4.07 ± 0.10 | 0.986 ± 0.004 | -0.12 ± 0.04 | <8 | 3 | Y | Y | Y | N | N | Y | ... | ... | n | ... |
| 39101 | 17472707+0604256 | 20.4 ± 0.2 | 5824 ± 188 | 4.14 ± 0.13 | 0.997 ± 0.002 | 0.27 ± 0.09 | 66 ± 2 | 1 | Y | Y | N | N | N | N | ... | ... | n | NG |
| 38703 | 17444951+0522105 | 79.7 ± 0.3 | 5114 ± 302 | ... | 1.009 ± 0.005 | -0.48 ± 0.20 | <25 | 3 | Y | ... | N | N | N | N | ... | ... | n | NG? |
| 39148 | 17474192+0601113 | 28.8 ± 0.2 | ... | ... | ... | ... | ... | ... | ... | ... | ... | ... | ... | ... | ... | ... | n | NG |
| 38704 | 17445068+0527176 | 22.7 ± 0.3 | 4730 ± 112 | 4.61 ± 0.07 | 0.944 ± 0.004 | -0.10 ± 0.03 | <6 | 3 | Y | Y | N | N | N | Y | ... | ... | n | ... |
| 38705 | 17445249+0548354 | -10.6 ± 0.2 | 4488 ± 120 | ... | 1.026 ± 0.004 | -0.20 ± 0.18 | 35 ± 6 | 1 | N | ... | ... | ... | ... | ... | ... | ... | n | ... |
| 39149 | 17474230+0547598 | -26.1 ± 0.3 | 4246 ± 302 | 4.49 ± 0.02 | 0.875 ± 0.010 | 0.00 ± 0.12 | ... | ... | Y | Y | Y | ... | Y | Y | ... | ... | n | ... |
| 38706 | 17445275+0527104 | -70.4 ± 0.2 | 5040 ± 177 | ... | 1.002 ± 0.004 | -0.39 ± 0.10 | ... | ... | Y | ... | N | ... | ... | N | ... | ... | n | ... |
| 39150 | 17474232+0533113 | -103.1 ± 0.2 | 4702 ± 130 | ... | 0.949 ± 0.005 | 0.09 ± 0.02 | <22 | 3 | Y | ... | N | N | N | Y | ... | ... | n | NG |
| 38707 | 17445329+0547051 | 30.7 ± 0.2 | 5035 ± 154 | ... | 1.003 ± 0.003 | -0.11 ± 0.18 | <21 | 3 | Y | ... | N | N | N | Y | ... | ... | n | NG? |
| 39151 | 17474306+0553372 | -74.4 ± 0.3 | 4288 ± 259 | 4.50 ± 0.01 | 0.885 ± 0.005 | -0.08 ± 0.15 | <26 | 3 | Y | Y | N | Y | Y | Y | ... | ... | n | ... |
| 38708 | 17445399+0514548 | -16.3 ± 0.3 | 5777 ± 75 | ... | 1.000 ± 0.004 | ... | 61 ± 3 | 1 | Y | ... | Y | N | N | ... | ... | ... | n | NG? |
| 2644 | 17474322+0540474 | 21.1 ± 0.6 | 7099 ± 162 | 4.03 ± 0.21 | ... | -0.13 ± 0.14 | <7 | 3 | ... | Y | N | N | N | Y | N | ... | n | ... |
| 38709 | 17445421+0539489 | -18.9 ± 0.3 | 4843 ± 226 | ... | 0.973 ± 0.005 | -0.07 ± 0.03 | ... | ... | Y | ... | Y | ... | Y | Y | ... | ... | n | ... |
| 39152 | 17474362+0551253 | -57.9 ± 0.2 | 6266 ± 33 | 4.11 ± 0.07 | 0.999 ± 0.002 | -0.28 ± 0.03 | ... | ... | Y | Y | N | ... | ... | Y | N | ... | n | ... |

**Table C.7.** continued.

| ID | CNAME | RV (km s$^{-1}$) | $T_{\rm eff}$ (K) | logg (dex) | $\gamma^a$ | [Fe/H] (dex) | EW(Li)$^b$ (mÅ) | EW(Li) error flag$^c$ | $\gamma$ | logg | \multicolumn{4}{c|}{Membership} | \multicolumn{2}{c|}{Gaia studies} | Final$^e$ | NMs with Li$^f$ |
| | | | | | | | | | | | RV | Li | H$\alpha$ | [Fe/H] | Randich$^d$ | Cantat-Gaudin$^d$ | | |
|---|---|---|---|---|---|---|---|---|---|---|---|---|---|---|---|---|---|---|
| 38710 | 17445460+0513165 | 12.8 ± 0.3 | 5094 ± 54 | … | 1.006 ± 0.005 | -0.30 ± 0.21 | <7 | 3 | Y | … | Y | N | N | Y | … | … | n | … |
| 39153 | 17474403+0521592 | -12.7 ± 0.2 | 5518 ± 220 | 4.12 ± 0.16 | 0.993 ± 0.002 | 0.11 ± 0.02 | 220 ± 2 | 1 | Y | Y | Y | Y | Y | Y | Y | Y | Y | … |
| 38711 | 17445460+0539152 | -13.5 ± 0.2 | 6163 ± 37 | 3.86 ± 0.04 | 1.005 ± 0.002 | -0.07 ± 0.01 | <10 | 3 | Y | Y | Y | N | N | Y | Y | … | n | NG? |
| 39154 | 17474442+0530005 | -24.9 ± 0.2 | 5643 ± 87 | 3.82 ± 0.02 | 1.000 ± 0.001 | 0.04 ± 0.01 | 19 ± 3 | 1 | Y | Y | Y | N | N | Y | … | … | n | NG |
| 38712 | 17445499+0526447 | 18.5 ± 0.3 | 4140 ± 118 | 4.65 ± 0.10 | 0.862 ± 0.007 | 0.02 ± 0.18 | <27 | 3 | Y | Y | N | Y | N | Y | … | … | n | NG |
| 39155 | 17474500+0600493 | 75.9 ± 0.2 | 4750 ± 130 | … | 1.023 ± 0.004 | -0.43 ± 0.08 | <14 | 3 | N | … | … | … | … | … | … | … | n | G |
| 38713 | 17445500+0523478 | -30.2 ± 0.2 | 5017 ± 122 | … | 1.007 ± 0.005 | -0.07 ± 0.10 | <29 | 3 | Y | … | Y | N | N | Y | … | … | n | NG? |
| 39156 | 17474532+0603222 | -34.0 ± 0.2 | 5132 ± 7 | … | 0.984 ± 0.005 | -0.08 ± 0.01 | <22 | 3 | Y | … | Y | N | Y | Y | … | … | n | NG |
| 38714 | 17445512+0518222 | -13.4 ± 0.2 | 5119 ± 7 | 3.65 ± 0.01 | 0.997 ± 0.005 | -0.15 ± 0.15 | 6 ± 3 | 1 | Y | Y | Y | N | N | Y | … | … | n | … |
| 39157 | 17474618+0544254 | -9.7 ± 0.2 | 4622 ± 28 | 2.31 ± 0.06 | 1.016 ± 0.003 | -0.13 ± 0.10 | <24 | 3 | N | N | … | … | … | … | … | … | n | … |
| 38715 | 17445512+0538520 | 19.3 ± 0.2 | 4613 ± 14 | … | 0.931 ± 0.006 | -0.12 ± 0.01 | <25 | 3 | Y | … | N | N | N | Y | … | … | n | NG |
| 39158 | 17474619+0526142 | -47.5 ± 0.3 | 4412 ± 305 | … | 0.925 ± 0.005 | 0.12 ± 0.25 | … | … | Y | … | N | … | … | Y | … | … | n | … |
| 38716 | 17445522+0546182 | 61.2 ± 0.2 | 4992 ± 81 | … | 1.013 ± 0.004 | -0.22 ± 0.13 | <19 | 3 | N | … | … | … | … | … | … | … | n | G |
| 38717 | 17445628+0535087 | 23.2 ± 0.3 | 4648 ± 310 | … | 1.017 ± 0.008 | -0.16 ± 0.32 | <32 | 3 | N | … | … | … | … | … | … | … | n | G |
| 39159 | 17474646+0600418 | 57.8 ± 0.2 | 4612 ± 47 | 1.98 ± 0.16 | 1.028 ± 0.005 | -0.44 ± 0.02 | <23 | 3 | N | N | … | … | … | … | … | … | n | G |
| 38741 | 17450873+0606009 | -170.7 ± 0.3 | 5118 ± 2 | 2.68 ± 0.09 | 1.012 ± 0.008 | -0.70 ± 0.03 | … | … | N | N | … | … | … | … | … | … | n | G |
| 39160 | 17474654+0547018 | 19.9 ± 0.2 | 4885 ± 142 | 2.81 ± 0.09 | 1.007 ± 0.005 | -0.18 ± 0.20 | <22 | 3 | Y | N | N | N | Y | Y | … | … | n | NG? |
| 38742 | 17450967+0526528 | 30.0 ± 0.2 | 5101 ± 104 | … | 1.007 ± 0.004 | … | <13 | 3 | Y | … | N | N | … | … | … | … | n | NG? |
| 39161 | 17474657+0528436 | -160.7 ± 0.3 | … | … | … | … | … | … | … | … | … | … | … | … | … | … | n | … |
| 38743 | 17451007+0549215 | -54.2 ± 0.2 | 4799 ± 104 | … | 0.965 ± 0.004 | -0.03 ± 0.07 | <17 | 3 | Y | … | N | N | N | Y | … | … | n | NG |
| 39162 | 17474677+0554069 | -11.2 ± 0.2 | 5047 ± 108 | … | 1.008 ± 0.003 | -0.07 ± 0.08 | <24 | 3 | Y | … | Y | N | N | Y | … | … | n | NG? |
| 38744 | 17451017+0558109 | -2.7 ± 0.2 | 5681 ± 20 | 4.15 ± 0.08 | 0.987 ± 0.005 | 0.13 ± 0.07 | <12 | 3 | Y | Y | Y | N | N | Y | … | … | n | NG |
| 39163 | 17474682+0522576 | 10.7 ± 0.2 | 4845 ± 205 | … | 1.014 ± 0.004 | -0.39 ± 0.16 | <33 | 3 | N | … | … | … | … | … | … | … | n | G |
| 38745 | 17451048+0546174 | -10.6 ± 0.2 | 4720 ± 72 | … | 0.996 ± 0.005 | 0.01 ± 0.01 | <40 | 3 | Y | … | Y | Y | N | Y | … | … | n | NG |
| 39164 | 17474683+0559435 | 28.6 ± 0.2 | 4373 ± 121 | 4.57 ± 0.06 | 0.905 ± 0.005 | -0.14 ± 0.05 | … | … | Y | Y | N | … | Y | Y | … | … | n | … |
| 38746 | 17451081+0526322 | 1.6 ± 0.2 | 5727 ± 142 | 4.13 ± 0.05 | 0.991 ± 0.002 | 0.21 ± 0.03 | <27 | 3 | Y | Y | Y | N | N | N | … | … | n | NG |
| 39165 | 17474709+0515204 | -13.7 ± 0.2 | 6113 ± 21 | 4.09 ± 0.13 | 0.996 ± 0.001 | -0.06 ± 0.10 | 43 ± 1 | 1 | Y | Y | Y | N | N | Y | Y | … | n | NG |
| 38747 | 17451150+0515212 | 35.4 ± 0.2 | 5640 ± 114 | 4.19 ± 0.08 | 0.990 ± 0.003 | 0.07 ± 0.06 | <9 | 3 | Y | Y | N | N | N | Y | … | … | n | … |
| 39166 | 17474726+0544011 | 40.2 ± 0.2 | 4794 ± 70 | 2.57 ± 0.19 | 1.018 ± 0.003 | -0.19 ± 0.12 | <29 | 3 | N | N | … | … | … | … | … | … | n | G |
| 38748 | 17451152+0532052 | -1.1 ± 0.3 | 3714 ± 39 | 4.58 ± 0.17 | 0.814 ± 0.007 | -0.19 ± 0.12 | … | … | Y | Y | Y | … | Y | Y | N | … | n | … |
| 39167 | 17474729+0554556 | -5.6 ± 0.2 | 5042 ± 121 | … | 1.012 ± 0.005 | -0.26 ± 0.19 | <16 | 3 | N | … | … | … | … | … | … | … | n | G |
| 38749 | 17451204+0531523 | -1.2 ± 0.2 | 3900 ± 70 | 4.64 ± 0.10 | 0.822 ± 0.006 | -0.13 ± 0.04 | <30 | 3 | Y | Y | Y | Y | Y | Y | N | … | Y?$^g$ | … |
| 39189 | 17475991+0515058 | -8.6 ± 0.2 | 4561 ± 90 | 1.89 ± 0.15 | 1.029 ± 0.004 | -0.38 ± 0.07 | <29 | 3 | N | N | … | … | … | … | … | … | n | G |
| 2622 | 17451225+0514597 | -31.5 ± 0.6 | 6432 ± 125 | 4.00 ± 0.11 | … | -0.35 ± 0.08 | <27 | 3 | … | Y | Y | Y | N | N | N | … | n | NG |
| 39190 | 17480040+0517043 | -105.8 ± 0.2 | 5545 ± 237 | 3.93 ± 0.17 | 0.993 ± 0.003 | 0.10 ± 0.22 | <26 | 3 | Y | Y | N | Y | N | Y | … | … | n | NG |
| 39191 | 17480072+0559218 | 14.1 ± 0.2 | 5013 ± 128 | 3.38 ± 0.03 | 0.999 ± 0.005 | -0.09 ± 0.18 | <18 | 3 | Y | N | Y | N | N | Y | … | … | n | NG? |
| 38751 | 17451294+0549505 | -92.0 ± 0.2 | 4655 ± 150 | 2.73 ± 0.07 | 1.006 ± 0.004 | 0.07 ± 0.08 | <47 | 3 | Y | N | N | Y | N | Y | … | … | n | NG? |
| 39192 | 17480091+0538324 | 52.1 ± 0.4 | 4283 ± 111 | … | 0.878 ± 0.010 | -0.41 ± 0.21 | <29 | 3 | Y | … | N | Y | N | N | … | … | n | NG |
| 38752 | 17451333+0517350 | -29.7 ± 0.2 | 4993 ± 144 | … | 1.024 ± 0.004 | -0.14 ± 0.14 | <9 | 3 | N | … | … | … | … | … | … | … | n | G |
| 39193 | 17480117+0607377 | -23.2 ± 0.2 | 4704 ± 64 | 2.49 ± 0.08 | 1.012 ± 0.002 | -0.08 ± 0.09 | <27 | 3 | N | N | … | … | … | … | … | … | n | G |
| 38753 | 17451333+0531577 | -22.3 ± 0.2 | 4852 ± 40 | 4.12 ± 0.10 | 0.969 ± 0.005 | 0.05 ± 0.02 | <19 | 3 | Y | Y | Y | N | N | Y | … | … | n | NG |
| 39194 | 17480124+0555023 | -38.2 ± 0.2 | 4970 ± 145 | … | 1.020 ± 0.005 | -0.16 ± 0.17 | 18 ± 5 | 1 | N | … | … | … | … | … | … | … | n | G |
| 38754 | 17451342+0528065 | -30.6 ± 0.2 | 4559 ± 101 | 4.49 ± 0.02 | 0.934 ± 0.006 | 0.02 ± 0.15 | <27 | 3 | Y | Y | Y | Y | N | Y | … | … | n | NG |
| 38755 | 17451379+0536386 | 13.4 ± 0.4 | 4286 ± 275 | 4.61 ± 0.06 | 0.873 ± 0.011 | -0.26 ± 0.08 | <41 | 3 | Y | Y | Y | Y | Y | Y | … | … | Y | … |
| 39195 | 17480154+0606520 | 22.0 ± 0.3 | 5164 ± 121 | … | 1.019 ± 0.006 | … | 15 ± 5 | 1 | N | … | … | … | … | … | … | … | n | G |
| 38756 | 17451385+0526372 | -24.8 ± 0.6 | … | … | … | … | … | … | … | … | … | … | N | … | … | … | n | … |
| 39196 | 17480166+0605412 | -36.7 ± 0.2 | 6194 ± 27 | 3.84 ± 0.06 | 1.005 ± 0.002 | -0.10 ± 0.09 | 16 ± 6 | 1 | Y | Y | Y | N | N | Y | N | … | n | NG? |
| 38757 | 17451415+0531059 | -27.4 ± 0.2 | 4748 ± 13 | 4.14 ± 0.15 | 0.965 ± 0.005 | 0.09 ± 0.06 | <39 | 3 | Y | Y | Y | N | N | Y | … | … | n | NG |
| 39197 | 17480170+0553262 | -14.5 ± 0.3 | 5591 ± 307 | … | 1.006 ± 0.007 | 0.24 ± 0.07 | <34 | 3 | Y | … | Y | N | N | N | … | … | n | NG? |
| 38758 | 17451435+0518116 | -13.9 ± 0.4 | 5161 ± 93 | … | 0.957 ± 0.004 | … | 281 ± 3 | 1 | Y | … | Y | Y | Y | … | Y | … | Y | … |
| 39198 | 17480254+0520103 | -54.3 ± 0.2 | 5086 ± 45 | 4.32 ± 0.10 | 0.975 ± 0.003 | -0.05 ± 0.08 | <17 | 3 | Y | Y | Y | N | N | Y | … | … | n | NG |
| 38759 | 17451461+0519200 | 55.6 ± 0.3 | 5022 ± 50 | 3.33 ± 0.09 | 0.999 ± 0.005 | -0.24 ± 0.14 | <13 | 3 | Y | N | N | N | Y | Y | … | … | n | NG? |
| 39199 | 17480334+0603219 | -29.4 ± 0.2 | 5792 ± 23 | … | 0.999 ± 0.002 | 0.00 ± 0.05 | 57 ± 2 | 1 | Y | … | Y | N | N | Y | … | … | n | NG |
| 38760 | 17451467+0556230 | 77.2 ± 0.3 | 4526 ± 263 | 4.53 ± 0.05 | 0.915 ± 0.012 | -0.11 ± 0.01 | 28 ± 6 | 1 | Y | Y | N | Y | N | Y | … | … | n | NG |
| 39200 | 17480389+0547403 | -27.7 ± 0.2 | 4878 ± 173 | … | 0.967 ± 0.005 | 0.05 ± 0.14 | <18 | 3 | Y | … | N | Y | Y | Y | … | … | n | NG |
| 38761 | 17451472+0524301 | 44.8 ± 0.3 | 4361 ± 167 | … | 0.924 ± 0.009 | -0.22 ± 0.03 | <18 | 3 | Y | … | N | Y | N | Y | … | … | n | NG |
| 2648 | 17480498+0548421 | -13.6 ± 0.6 | 5239 ± 78 | 4.52 ± 0.12 | … | -0.05 ± 0.12 | 216 ± 33 | 2 | … | Y | Y | Y | Y | Y | Y | Y | Y | … |
| 38762 | 17451497+0549422 | 61.1 ± 0.3 | 5009 ± 114 | … | 1.013 ± 0.008 | -0.61 ± 0.24 | 32 ± 5 | 1 | N | … | … | … | … | … | … | … | n | G |
| 39201 | 17480609+0603176 | -187.7 ± 0.3 | … | … | … | … | … | … | … | … | … | … | … | … | … | … | n | … |

**Table C.7.** continued.

| ID | CNAME | RV (km s$^{-1}$) | $T_{\text{eff}}$ (K) | $\log g$ (dex) | $\gamma^a$ | [Fe/H] (dex) | EW(Li)$^b$ (mÅ) | EW(Li) error flag$^c$ | Membership $\gamma$ | $\log g$ | RV | Li | H$\alpha$ | [Fe/H] | Gaia studies Randich$^d$ | Cantat-Gaudin$^d$ | Final$^e$ | NMs with Li$^f$ |
|---|---|---|---|---|---|---|---|---|---|---|---|---|---|---|---|---|---|---|
| 38763 | 17451553+0526593 | -60.6 ± 0.3 | 5040 ± 114 | ... | 1.024 ± 0.007 | ... | <36 | 3 | N | ... | ... | ... | ... | ... | ... | ... | n | ... |
| 39202 | 17480629+0530567 | 12.0 ± 0.4 | 4282 ± 332 | 4.44 ± 0.18 | 0.891 ± 0.013 | -0.14 ± 0.09 | <25 | 3 | Y | Y | Y | Y | N | Y | ... | ... | n | NG |
| 38806 | 17453011+0536560 | 53.5 ± 0.3 | 5868 ± 13 | 4.08 ± 0.16 | 0.997 ± 0.004 | -0.17 ± 0.19 | 23 ± 7 | 1 | Y | Y | N | N | N | Y | ... | ... | n | NG |
| 39203 | 17480871+0529296 | -19.9 ± 0.4 | 4254 ± 141 | ... | 0.834 ± 0.011 | ... | <41 | 3 | Y | ... | Y | Y | Y | ... | ... | ... | Y | ... |
| 38807 | 17453020+0538426 | 67.7 ± 0.2 | 4904 ± 147 | 2.99 ± 0.04 | 1.004 ± 0.003 | -0.17 ± 0.14 | ... | ... | Y | N | N | ... | N | Y | ... | ... | n | ... |
| 38808 | 17453039+0603473 | 72.5 ± 0.2 | 5661 ± 59 | 4.22 ± 0.14 | 0.991 ± 0.003 | -0.04 ± 0.03 | <17 | 3 | Y | Y | N | N | N | Y | ... | ... | n | NG |
| 38809 | 17453046+0558223 | 23.3 ± 0.2 | 4981 ± 189 | ... | 1.013 ± 0.004 | -0.18 ± 0.19 | <13 | 3 | N | ... | ... | ... | ... | ... | ... | ... | n | G |
| 38810 | 17453070+0517091 | 1.0 ± 0.2 | 5901 ± 175 | 4.21 ± 0.13 | 0.996 ± 0.003 | 0.24 ± 0.08 | 75 ± 5 | 1 | Y | Y | Y | N | N | N | ... | ... | n | NG |
| 38811 | 17453082+0517318 | 35.2 ± 0.2 | 4882 ± 170 | ... | 1.012 ± 0.003 | -0.06 ± 0.15 | <29 | 3 | N | ... | ... | ... | ... | ... | ... | ... | n | G |
| 38812 | 17453088+0607054 | -14.7 ± 0.3 | 4826 ± 135 | ... | 0.948 ± 0.006 | ... | 62 ± 4 | 1 | Y | ... | Y | Y | Y | ... | ... | ... | Y | ... |
| 38813 | 17453137+0607535 | 36.6 ± 0.3 | 5123 ± 9 | 2.98 ± 0.16 | 1.016 ± 0.008 | -0.21 ± 0.14 | <48 | 3 | N | N | ... | ... | ... | ... | ... | ... | n | Li-rich G |
| 38814 | 17453168+0558125 | -40.1 ± 0.3 | 4981 ± 113 | ... | 1.010 ± 0.006 | -0.13 ± 0.15 | <27 | 3 | Y | ... | Y | N | N | Y | ... | ... | n | NG? |
| 38815 | 17453297+0513313 | -9.9 ± 0.3 | 6026 ± 196 | 4.05 ± 0.01 | 1.001 ± 0.003 | 0.11 ± 0.09 | 94 ± 6 | 1 | Y | Y | Y | Y | N | Y | N | ... | n | NG? |
| 38816 | 17453318+0529288 | 48.8 ± 0.2 | 5431 ± 44 | 4.14 ± 0.12 | 0.988 ± 0.005 | 0.13 ± 0.05 | <25 | 3 | Y | Y | N | Y | N | Y | ... | ... | n | NG |
| 38817 | 17453374+0524492 | 49.7 ± 0.3 | 4729 ± 106 | 2.43 ± 0.14 | 1.014 ± 0.006 | -0.19 ± 0.15 | <32 | 3 | N | N | ... | ... | ... | ... | ... | ... | n | G |
| 38818 | 17453396+0552402 | -38.3 ± 0.2 | 5263 ± 118 | ... | 0.996 ± 0.003 | -0.16 ± 0.05 | <16 | 3 | Y | ... | Y | Y | N | Y | ... | ... | n | NG |
| 38819 | 17453402+0604481 | -28.3 ± 0.3 | 4876 ± 151 | ... | 1.036 ± 0.008 | -0.29 ± 0.16 | <27 | 3 | N | ... | ... | ... | ... | ... | ... | ... | n | G |
| 38820 | 17453424+0558157 | -47.6 ± 0.2 | 4780 ± 131 | 2.62 ± 0.06 | 1.023 ± 0.005 | -0.06 ± 0.05 | <9 | 3 | N | N | ... | ... | ... | ... | ... | ... | n | G |
| 38821 | 17453451+0529538 | -14.0 ± 0.3 | 3614 ± 57 | ... | 0.821 ± 0.010 | -0.21 ± 0.14 | <100 | 3 | Y | ... | Y | Y | Y | Y | Y | Y | Y | ... |
| 38822 | 17453524+0527180 | -1.0 ± 0.2 | 4852 ± 142 | 2.73 ± 0.05 | 1.004 ± 0.005 | -0.21 ± 0.25 | <38 | 3 | Y | N | Y | N | N | Y | ... | ... | n | NG? |
| 2626 | 17453550+0519023 | 0.8 ± 0.6 | 6314 ± 135 | 3.87 ± 0.12 | ... | -0.04 ± 0.12 | <12 | 3 | ... | Y | Y | N | N | Y | ... | ... | n | NG |
| 38823 | 17453558+0607071 | 28.5 ± 0.3 | 4699 ± 192 | 4.59 ± 0.05 | 0.937 ± 0.009 | -0.03 ± 0.05 | <23 | 3 | Y | Y | N | N | N | Y | ... | ... | n | NG |
| 38824 | 17453590+0549043 | -65.6 ± 0.3 | 4620 ± 140 | ... | 1.019 ± 0.009 | -0.08 ± 0.13 | <38 | 3 | N | ... | ... | ... | ... | ... | ... | ... | n | G |
| 38825 | 17453737+0530277 | 66.6 ± 0.3 | 5089 ± 124 | ... | 1.010 ± 0.007 | ... | <39 | 3 | N | ... | ... | ... | ... | ... | ... | ... | n | G |
| 38826 | 17453765+0553534 | 74.9 ± 0.2 | 4477 ± 93 | ... | 1.039 ± 0.004 | -0.47 ± 0.06 | <24 | 3 | N | ... | ... | ... | ... | ... | ... | ... | n | G |
| 38827 | 17453783+0545334 | -6.7 ± 0.3 | 4157 ± 175 | 4.54 ± 0.13 | 0.870 ± 0.008 | -0.11 ± 0.06 | ... | ... | Y | Y | Y | ... | N | Y | N | ... | n | ... |
| 38850 | 17454419+0531187 | -54.3 ± 0.2 | 4763 ± 117 | 2.80 ± 0.04 | 1.010 ± 0.004 | 0.11 ± 0.02 | <38 | 3 | N | N | ... | ... | ... | ... | ... | ... | n | G |
| 38851 | 17454558+0556098 | -61.1 ± 0.3 | 4289 ± 294 | 4.53 ± 0.02 | 0.878 ± 0.010 | 0.00 ± 0.16 | <48 | 3 | Y | Y | N | Y | Y | Y | ... | ... | n | NG |
| 38852 | 17454623+0514595 | 8.8 ± 0.3 | 6657 ± 54 | 4.28 ± 0.11 | 1.001 ± 0.003 | 0.01 ± 0.05 | ... | ... | Y | Y | Y | ... | N | Y | N | ... | n | ... |

**Notes.** $^{(a)}$ Empirical gravity indicator defined by Damiani et al. (2014). $^{(b)}$ The values of EW(Li) for this cluster are corrected (subtracted adjacent Fe (6707.43 Å) line). $^{(c)}$ Flags for the errors of the corrected EW(Li) values, as follows: 1=EW(Li) corrected by blends contribution using models; 2=EW(Li) measured separately (Li line resolved - UVES only); and 3=Upper limit (no error for EW(Li) is given). $^{(d)}$ Randich et al. (2018), Cantat-Gaudin et al. (2018) $^{(e)}$ The letters "Y" and "N" indicate if the star is a cluster member or not. $^{(f)}$ 'Li-rich G', 'G' and 'NG' indicate "Li-rich giant", "giant" and "non-giant" Li field contaminants, respectively. $^{(g)}$ For more details about the membership of the stars listed as possible candidates, see the individual notes of Sect. 5 for IC 4665.

**Table C.8.** NGC 2516

| ID | CNAME | RV (km s$^{-1}$) | $T_{\rm eff}$ (K) | $\log g$ (dex) | [Fe/H] (dex) | $EW$(Li)$^a$ (mÅ) | $EW$(Li) error flag$^b$ | Membership RV | Li | $\log g$ | [Fe/H] | Gaia studies Randich$^c$ | Cantat-Gaudin$^c$ | Final$^d$ | NMs with Li$^e$ |
|---|---|---|---|---|---|---|---|---|---|---|---|---|---|---|---|
| 43630 | 07550511-6034396 | 24.4 ± 0.2 | 4223 ± 345 | 4.48 ± 0.13 | 0.11 ± 0.19 | 5 ± 3 | … | Y | Y | Y | Y | Y | Y | Y | … |
| 2908 | 07550546-6036336 | 8.6 ± 0.6 | 5205 ± 116 | 4.38 ± 0.24 | -0.08 ± 0.10 | 26 ± 2 | … | N | … | … | … | … | N | n | NG |
| 2909 | 07550592-6104294 | 23.1 ± 0.6 | 5806 ± 131 | 4.55 ± 0.25 | -0.10 ± 0.10 | 139 ± 1 | … | Y | Y | Y | Y | Y | Y | Y | … |
| 43763 | 07562640-6021143 | 23.6 ± 0.6 | 3615 ± 25 | 4.80 ± 0.20 | … | … | … | Y | … | … | … | Y | … | n | … |
| 43631 | 07550692-6111311 | 23.2 ± 0.2 | 5055 ± 97 | 3.55 ± 0.28 | -0.03 ± 0.19 | 4 ± 3 | … | Y | N | Y | Y | … | … | n | NG |
| 43632 | 07550711-6106356 | 36.7 ± 0.2 | 5024 ± 105 | 3.50 ± 0.23 | -0.21 ± 0.22 | 6 ± 2 | … | N | … | … | … | … | … | n | … |
| 43633 | 07550826-6039579 | -4.7 ± 0.2 | 4896 ± 120 | 2.98 ± 0.26 | 0.00 ± 0.18 | 2 ± 3 | … | N | … | … | … | … | … | n | G |
| 43634 | 07550858-6020088 | 14.7 ± 0.2 | 4860 ± 66 | 3.01 ± 0.29 | 0.20 ± 0.21 | 9 ± 3 | … | N | … | … | … | … | … | n | G |
| 43764 | 07562958-6052125 | 23.3 ± 0.5 | 3586 ± 124 | 4.75 ± 0.14 | -0.06 ± 0.13 | … | … | Y | … | … | … | Y | … | n | … |
| 43635 | 07551059-6043092 | 23.7 ± 0.2 | 4459 ± 361 | 4.64 ± 0.11 | -0.05 ± 0.13 | 9 ± 3 | … | Y | Y | Y | Y | Y | N | Y | … |
| 43765 | 07562978-6100000 | 23.3 ± 0.3 | 3815 ± 220 | 4.81 ± 0.20 | -0.03 ± 0.28 | … | … | Y | … | … | … | Y | … | n | … |
| 43636 | 07551125-6026336 | 24.1 ± 0.5 | 4025 ± 595 | 4.69 ± 0.13 | 0.04 ± 0.20 | … | … | Y | … | … | … | Y | … | n | … |
| 43766 | 07563027-6041370 | 22.2 ± 0.4 | 3617 ± 142 | 4.88 ± 0.15 | -0.06 ± 0.12 | … | … | Y | … | … | … | Y | … | n | … |
| 43638 | 07551178-6026259 | 26.0 ± 0.3 | 4833 ± 347 | 3.81 ± 0.35 | 0.09 ± 0.20 | 221 ± 3 | … | Y | Y | Y | Y | Y | N | Y | … |
| 43767 | 07563059-6106598 | 30.0 ± 1.3 | 3671 ± 579 | 4.14 ± 0.25 | … | … | … | N | … | … | … | Y | … | n | … |
| 43639 | 07551226-6034167 | 33.1 ± 0.2 | 5171 ± 153 | 3.64 ± 0.50 | -0.36 ± 0.19 | 9 ± 2 | … | N | … | … | … | … | … | n | … |
| 43768 | 07563083-6042100 | 40.3 ± 0.2 | 4754 ± 216 | 4.62 ± 0.25 | -0.05 ± 0.14 | 7 ± 2 | … | N | … | … | … | … | … | n | … |
| 43640 | 07551241-6048525 | 23.6 ± 0.2 | 5325 ± 93 | 4.50 ± 0.46 | -0.04 ± 0.15 | 135 ± 3 | … | Y | Y | Y | Y | Y | Y | Y | … |
| 43769 | 07563141-6059331 | 23.5 ± 0.3 | 3878 ± 243 | 4.60 ± 0.10 | -0.06 ± 0.16 | 19 ± 5 | … | Y | Y | Y | Y | Y | Y | Y | … |
| 43641 | 07551340-6041130 | 24.2 ± 0.4 | 3512 ± 127 | 4.43 ± 0.15 | -0.07 ± 0.14 | … | … | Y | … | … | … | Y | … | n | … |
| 43642 | 07551354-6045292 | -15.6 ± 0.6 | 3831 ± 343 | 5.04 ± 0.51 | -0.77 ± 0.79 | … | … | N | … | … | … | N | … | n | … |
| 43770 | 07563195-6028054 | 23.2 ± 0.2 | 6319 ± 178 | 4.38 ± 0.36 | -0.09 ± 0.12 | 79 ± 2 | 1 | Y | Y | Y | Y | Y | Y | Y | … |
| 43643 | 07551361-6020349 | 3.1 ± 1.1 | 3479 ± 27 | 3.99 ± 0.17 | … | … | … | N | … | … | … | N | … | n | … |
| 43644 | 07551451-6055363 | 24.3 ± 0.8 | 3729 ± 194 | 4.75 ± 0.14 | -0.01 ± 0.11 | … | … | Y | … | … | … | Y | … | n | … |
| 43771 | 07563243-6024079 | 25.4 ± 0.3 | 4153 ± 275 | 4.70 ± 0.15 | -0.04 ± 0.15 | 13 ± 4 | … | Y | Y | Y | Y | Y | N | Y | … |
| 43645 | 07551498-6048191 | 25.9 ± 0.3 | 6786 ± 609 | 4.23 ± 0.35 | 0.02 ± 0.19 | … | … | Y | … | … | … | Y | … | n | … |
| 43772 | 07563357-6100558 | 19.9 ± 0.2 | 4732 ± 318 | 4.33 ± 0.31 | -0.23 ± 0.27 | 142 ± 4 | … | N | Y | Y | N | Y | Y | Y | … |
| 2910 | 07551614-6101207 | -13.0 ± 0.6 | 4950 ± 123 | 2.56 ± 0.23 | -0.34 ± 0.10 | 15 ± 2 | … | N | … | … | … | … | … | n | G |
| 43646 | 07551642-6113370 | 24.1 ± 0.3 | 3948 ± 364 | 4.46 ± 0.15 | -0.03 ± 0.19 | … | … | Y | … | … | … | Y | … | n | … |
| 43774 | 07563407-6039060 | 22.2 ± 0.2 | 5908 ± 135 | 4.43 ± 0.32 | -0.13 ± 0.18 | 116 ± 2 | … | Y | Y | Y | Y | Y | Y | Y | … |
| 43775 | 07563410-6030118 | 23.2 ± 0.3 | 3794 ± 278 | 4.49 ± 0.20 | -0.03 ± 0.14 | 75 ± 8 | … | Y | Y | Y | Y | Y | … | Y | … |
| 43647 | 07551675-6038447 | 29.4 ± 2.5 | 3738 ± 358 | … | -0.09 ± 0.13 | … | … | N | … | … | … | Y | … | n | … |
| 43776 | 07563419-6019027 | 11.4 ± 4.7 | 4040 ± 402 | … | … | … | … | N | … | … | … | N | … | n | … |
| 43648 | 07551714-6054141 | 105.0 ± 0.2 | 4851 ± 95 | 2.73 ± 0.25 | 0.00 ± 0.23 | 15 ± 3 | … | N | … | … | … | … | … | n | G |
| 43777 | 07563458-6020521 | 55.3 ± 0.2 | 5459 ± 91 | 4.04 ± 0.36 | -0.21 ± 0.13 | 5 ± 2 | … | N | … | … | … | … | … | n | … |
| 43858 | 07572330-6055520 | 23.9 ± 0.3 | 6924 ± 24 | … | 0.12 ± 0.02 | … | … | Y | … | … | … | Y | … | n | … |
| 43859 | 07572366-6106164 | 24.6 ± 0.2 | 5070 ± 88 | 4.52 ± 0.28 | -0.02 ± 0.13 | 14 ± 3 | … | Y | N | Y | Y | … | … | n | NG |
| 43860 | 07572432-6117295 | 23.5 ± 0.9 | 3550 ± 27 | 5.13 ± 0.20 | … | … | … | Y | … | … | … | Y | … | n | … |
| 43861 | 07572491-6043135 | 32.6 ± 0.2 | 4891 ± 212 | 3.08 ± 0.59 | -0.10 ± 0.19 | 6 ± 2 | … | N | … | … | … | … | … | n | G |
| 43862 | 07572522-6046473 | 23.5 ± 0.2 | 5988 ± 173 | 4.55 ± 0.31 | -0.14 ± 0.16 | 111 ± 2 | … | Y | Y | Y | Y | Y | Y | Y | … |
| 43863 | 07572560-6056510 | 79.7 ± 0.2 | 4686 ± 182 | 3.16 ± 0.42 | 0.13 ± 0.22 | 2 ± 2 | … | N | … | … | … | … | … | n | G |
| 43700 | 07555086-6026294 | 23.3 ± 0.3 | 3967 ± 333 | 4.81 ± 0.26 | -0.02 ± 0.20 | 69 ± 8 | … | Y | Y | Y | Y | Y | Y | Y | … |
| 43864 | 07572567-6117126 | 12.2 ± 0.3 | 4101 ± 246 | 4.85 ± 0.33 | -0.23 ± 0.27 | 11 ± 4 | … | N | … | … | … | N | … | n | NG |
| 43865 | 07572609-6045401 | 24.4 ± 0.2 | 4382 ± 426 | 4.36 ± 0.22 | 0.08 ± 0.19 | 87 ± 4 | … | Y | Y | Y | Y | Y | Y | Y | … |
| 43866 | 07572645-6049131 | -23.3 ± 0.2 | 5917 ± 152 | 4.12 ± 0.30 | -0.29 ± 0.15 | 36 ± 2 | … | N | … | … | … | … | … | n | NG |
| 43701 | 07555144-6058015 | 23.3 ± 0.3 | 4025 ± 135 | 4.48 ± 0.09 | 0.28 ± 0.04 | … | … | Y | … | … | … | Y | … | n | … |
| 43867 | 07572658-6042100 | 22.4 ± 0.5 | 3506 ± 26 | 4.66 ± 0.17 | … | … | … | Y | … | … | … | Y | … | n | … |
| 43868 | 07572674-6055495 | 23.3 ± 0.3 | 4309 ± 420 | 4.67 ± 0.21 | -0.14 ± 0.14 | 13 ± 4 | … | Y | Y | Y | Y | Y | Y | Y | … |
| 43702 | 07555194-6113208 | 23.5 ± 0.4 | 4189 ± 462 | 4.56 ± 0.09 | 0.12 ± 0.22 | 15 ± 5 | … | Y | Y | Y | Y | Y | Y | Y | … |
| 43869 | 07572844-6051295 | 28.0 ± 0.2 | 5791 ± 141 | 4.49 ± 0.37 | -0.04 ± 0.14 | 148 ± 2 | … | N | Y | Y | Y | Y | Y | Y | … |
| 43703 | 07555219-6054036 | 24.5 ± 1.2 | 3845 ± 628 | 4.86 ± 0.28 | 0.04 ± 0.19 | … | … | Y | … | … | … | Y | … | n | … |
| 43870 | 07572926-6046083 | 37.2 ± 4.3 | 4689 ± 1131 | 3.17 ± 0.37 | 0.52 ± 0.05 | … | … | N | … | … | … | Y | … | n | … |
| 43871 | 07572938-6050104 | 21.9 ± 0.2 | 5646 ± 147 | 4.38 ± 0.29 | -0.16 ± 0.20 | 54 ± 2 | … | Y | Y | Y | Y | … | … | Y | … |
| 43872 | 07573082-6047208 | 25.5 ± 0.3 | 3949 ± 353 | 4.43 ± 0.17 | -0.06 ± 0.21 | … | … | Y | … | … | … | Y | … | n | … |
| 43873 | 07573094-6038254 | 24.0 ± 0.5 | 3676 ± 23 | 4.82 ± 0.19 | … | … | … | Y | … | … | … | Y | … | n | … |
| 43874 | 07573119-6112220 | 34.1 ± 0.2 | 5044 ± 86 | 3.62 ± 0.24 | 0.04 ± 0.13 | 20 ± 4 | … | N | … | … | … | … | … | n | NG |
| 43875 | 07573153-6042108 | 24.3 ± 0.2 | 5211 ± 112 | 4.47 ± 0.46 | 0.02 ± 0.13 | 186 ± 3 | … | Y | Y | Y | Y | Y | Y | Y | … |
| 43704 | 07555287-6024532 | 24.3 ± 0.2 | 4460 ± 367 | 4.48 ± 0.20 | -0.05 ± 0.13 | 36 ± 3 | … | Y | Y | Y | Y | Y | Y | Y | … |
| 43876 | 07573278-6033368 | 26.5 ± 1.8 | 3601 ± 19 | 4.54 ± 0.16 | … | … | … | N | … | … | … | Y | … | n | … |

| ID | CNAME | RV (km s$^{-1}$) | $T_{\text{eff}}$ (K) | logg (dex) | [Fe/H] (dex) | $EW(\text{Li})^a$ (mÅ) | $EW(\text{Li})$ error flag$^b$ | Membership | | | | Gaia studies | | Final$^d$ | NMs with Li$^e$ |
| | | | | | | | | RV | Li | logg | [Fe/H] | Randich$^c$ | Cantat-Gaudin$^c$ | | |
|---|---|---|---|---|---|---|---|---|---|---|---|---|---|---|---|
| 43877 | 07573339-6032474 | 22.2 ± 0.6 | 3433 ± 33 | 4.63 ± 0.20 | … | … | … | Y | … | … | … | Y | … | n | … |
| 43878 | 07573354-6050406 | 25.8 ± 3.3 | 3573 ± 21 | 4.93 ± 0.17 | … | … | … | Y | … | … | … | Y | … | n | … |
| 43879 | 07573365-6115183 | 19.6 ± 1.4 | 3877 ± 344 | 4.75 ± 0.17 | 0.10 ± 0.24 | … | … | N | … | … | … | Y | … | n | … |
| 43705 | 07555541-6027338 | -0.9 ± 0.2 | 4970 ± 133 | 2.75 ± 0.41 | -0.16 ± 0.19 | 8 ± 2 | … | N | … | … | … | … | … | n | G |
| 43880 | 07573430-6023282 | 24.3 ± 0.2 | 4452 ± 491 | 4.33 ± 0.22 | 0.12 ± 0.19 | 196 ± 4 | … | Y | Y | Y | Y | Y | N | Y | … |
| 43967 | 07582081-6046180 | 30.3 ± 3.4 | 3503 ± 28 | 4.60 ± 0.18 | … | … | … | N | … | … | … | Y | … | n | … |
| 43706 | 07555570-6041396 | 23.2 ± 0.2 | 5820 ± 64 | 4.44 ± 0.23 | -0.22 ± 0.18 | 108 ± 2 | … | Y | Y | Y | N | Y | … | Y | … |
| 43968 | 07582084-6111460 | 20.1 ± 0.2 | 5803 ± 86 | 4.12 ± 0.23 | -0.40 ± 0.32 | 137 ± 3 | … | N | Y | Y | N | Y | Y | Y | … |
| 43969 | 07582161-6058503 | 23.7 ± 0.2 | 4624 ± 328 | 4.43 ± 0.12 | -0.06 ± 0.14 | 103 ± 3 | … | Y | Y | Y | Y | Y | N | Y | … |
| 43970 | 07582181-6022093 | 16.5 ± 0.3 | 4177 ± 286 | 4.91 ± 0.28 | 0.08 ± 0.19 | 26 ± 5 | … | N | Y | Y | N | Y | … | Y | … |
| 43971 | 07582195-6042333 | 35.8 ± 0.4 | 3426 ± 136 | 4.38 ± 0.18 | -0.08 ± 0.15 | … | … | N | … | … | … | N | … | n | … |
| 43972 | 07582215-6108184 | -5.1 ± 0.2 | 4332 ± 678 | 4.61 ± 0.14 | 0.12 ± 0.21 | … | … | N | … | … | … | N | … | n | … |
| 43973 | 07582232-6111078 | 22.1 ± 0.3 | 3919 ± 211 | 4.62 ± 0.10 | 0.04 ± 0.19 | 21 ± 6 | … | Y | Y | Y | Y | Y | Y | Y | … |
| 43974 | 07582294-6040204 | 24.4 ± 0.3 | 4776 ± 305 | 4.17 ± 0.12 | -0.98 ± 1.00 | 245 ± 4 | … | Y | Y | N | N | Y | Y | Y | … |
| 43975 | 07582339-6054567 | 24.3 ± 0.3 | 4778 ± 286 | 4.12 ± 0.12 | -1.00 ± 1.03 | 233 ± 4 | … | Y | Y | N | N | Y | … | Y | … |
| 43976 | 07582350-6046497 | 24.3 ± 0.2 | 5003 ± 135 | 4.51 ± 0.35 | 0.03 ± 0.13 | 137 ± 4 | … | Y | Y | Y | Y | Y | Y | Y | … |
| 43977 | 07582363-6037220 | 21.7 ± 0.2 | 6147 ± 186 | 4.53 ± 0.43 | -0.16 ± 0.17 | 79 ± 2 | 1 | Y | Y | Y | Y | Y | Y | Y | … |
| 43978 | 07582377-6103445 | 23.9 ± 0.2 | 5840 ± 94 | 4.37 ± 0.26 | -0.23 ± 0.24 | 116 ± 2 | … | Y | Y | Y | N | Y | Y | Y | … |
| 43979 | 07582547-6026305 | 90.2 ± 0.3 | 4288 ± 203 | 4.92 ± 0.26 | 0.01 ± 0.21 | 10 ± 4 | … | N | … | … | … | … | … | n | NG |
| 43980 | 07582713-6112487 | 23.4 ± 0.3 | 4808 ± 186 | 4.49 ± 0.42 | 0.05 ± 0.13 | 100 ± 6 | … | Y | Y | Y | Y | Y | Y | Y | … |
| 43981 | 07583040-6110071 | 28.6 ± 5.4 | 4422 ± 159 | … | 0.44 ± 0.13 | … | … | N | … | … | … | Y | … | n | … |
| 43982 | 07583058-6046298 | 28.1 ± 0.2 | 6853 ± 18 | … | 0.22 ± 0.02 | … | … | N | … | … | … | Y | … | n | … |
| 43983 | 07583073-6046010 | 24.3 ± 0.2 | 4990 ± 138 | 4.51 ± 0.39 | 0.01 ± 0.12 | 134 ± 2 | … | Y | Y | Y | Y | Y | Y | Y | … |
| 2930 | 07583076-6025129 | 50.4 ± 0.6 | 4964 ± 119 | 3.50 ± 0.25 | 0.03 ± 0.10 | <13 | 3 | N | … | … | … | … | … | n | … |
| 43984 | 07583121-6105232 | 23.5 ± 0.4 | 3852 ± 566 | 4.54 ± 0.09 | 0.00 ± 0.20 | … | … | Y | … | … | … | Y | … | n | … |
| 43985 | 07583154-6036588 | 20.2 ± 0.2 | 5361 ± 81 | 3.71 ± 0.24 | -0.08 ± 0.13 | 8 ± 1 | … | N | … | … | … | … | … | n | … |
| 43986 | 07583205-6058268 | 23.5 ± 0.2 | 4328 ± 313 | 4.56 ± 0.10 | 0.10 ± 0.19 | … | … | Y | … | … | … | Y | … | n | … |
| 43987 | 07583259-6059459 | 23.6 ± 0.3 | 4131 ± 240 | 4.87 ± 0.38 | 0.07 ± 0.20 | 11 ± 5 | … | Y | Y | Y | Y | Y | Y | Y | … |
| 43988 | 07583261-6031232 | 21.7 ± 0.9 | 3528 ± 29 | 4.76 ± 0.20 | … | … | … | Y | … | … | … | Y | … | n | … |
| 44075 | 07591796-6045189 | 24.5 ± 0.3 | 3815 ± 331 | 4.74 ± 0.15 | -0.07 ± 0.16 | 271 ± 8 | … | Y | Y | Y | Y | Y | Y | Y | … |
| 44076 | 07591899-6059283 | 21.0 ± 0.2 | 5913 ± 114 | 4.29 ± 0.05 | 0.23 ± 0.14 | 63 ± 2 | … | Y | Y | Y | N | Y | … | Y | … |
| 44077 | 07591973-6034437 | 27.3 ± 0.2 | 6374 ± 215 | 4.59 ± 0.49 | 0.01 ± 0.13 | 68 ± 3 | 1 | N | Y | Y | Y | Y | … | Y | … |
| 44078 | 07592134-6046028 | 29.9 ± 0.2 | 6282 ± 187 | 4.24 ± 0.27 | -0.09 ± 0.13 | 73 ± 2 | 1 | N | Y | Y | Y | Y | N | Y | … |
| 44079 | 07592156-6046453 | 22.7 ± 0.2 | 5175 ± 127 | 4.49 ± 0.31 | 0.00 ± 0.14 | 212 ± 4 | … | Y | Y | Y | Y | Y | Y | Y | … |
| 44080 | 07592249-6037124 | 25.4 ± 0.8 | 3798 ± 313 | 4.77 ± 0.21 | -0.07 ± 0.15 | … | … | Y | … | … | … | Y | … | n | … |
| 44081 | 07592324-6106169 | 63.9 ± 0.2 | 5078 ± 64 | 3.71 ± 0.24 | -0.09 ± 0.14 | 19 ± 3 | … | N | … | … | … | … | … | n | NG |
| 44082 | 07592442-6112110 | 17.2 ± 0.2 | 6267 ± 227 | 4.24 ± 0.16 | 0.05 ± 0.21 | … | … | N | … | … | … | Y | … | n | … |
| 2939 | 07592548-6040508 | 17.1 ± 0.6 | 6022 ± 110 | 4.33 ± 0.22 | 0.08 ± 0.10 | 30 ± 1 | … | N | Y | Y | Y | Y | … | Y | … |
| 44083 | 07592611-6048339 | 23.2 ± 0.2 | 4961 ± 168 | 4.61 ± 0.37 | 0.07 ± 0.14 | 109 ± 4 | … | Y | Y | Y | Y | Y | Y | Y | … |
| 44121 | 07595280-6032498 | 24.2 ± 0.2 | 5084 ± 100 | 4.45 ± 0.38 | -0.01 ± 0.13 | 154 ± 4 | … | Y | Y | Y | Y | Y | … | Y | … |
| 44122 | 07595556-6036385 | 28.3 ± 2.7 | 4014 ± 349 | 4.72 ± 0.14 | -1.26 ± 1.33 | … | … | N | … | … | … | Y | … | n | … |
| 44123 | 07595574-6115296 | 23.5 ± 2.8 | 3432 ± 131 | 4.59 ± 0.16 | 0.00 ± 0.11 | … | … | Y | … | … | … | Y | … | n | … |
| 44124 | 07595583-6102499 | 23.3 ± 0.3 | 3911 ± 219 | 4.71 ± 0.16 | 0.01 ± 0.19 | 42 ± 6 | … | Y | Y | Y | Y | Y | Y | Y | … |
| 44126 | 07595715-6108520 | 71.4 ± 0.2 | 4494 ± 287 | 4.74 ± 0.18 | -0.52 ± 0.21 | 2 ± 2 | … | N | … | … | … | N | … | n | … |
| 44127 | 07595789-6021032 | 23.9 ± 0.5 | 3564 ± 23 | 4.85 ± 0.18 | … | … | … | Y | … | … | … | Y | … | n | … |
| 44128 | 07595837-6024173 | 30.1 ± 0.2 | 4115 ± 531 | 4.44 ± 0.20 | 0.09 ± 0.20 | … | … | N | … | … | … | Y | … | n | … |
| 44152 | 08001867-6050282 | 24.8 ± 0.3 | 4318 ± 382 | 4.58 ± 0.11 | 0.14 ± 0.26 | … | … | Y | … | … | … | Y | … | n | … |
| 44153 | 08001884-6025124 | 24.3 ± 0.2 | 4687 ± 287 | 4.65 ± 0.23 | 0.05 ± 0.13 | 46 ± 3 | … | Y | Y | Y | Y | … | Y | Y | … |
| 44154 | 08001893-6036146 | 24.2 ± 0.2 | 4707 ± 319 | 4.39 ± 0.20 | -0.07 ± 0.14 | 138 ± 3 | … | Y | Y | Y | Y | Y | Y | Y | … |
| 44155 | 08001959-6045204 | 30.1 ± 0.2 | 3934 ± 574 | 4.54 ± 0.09 | 0.11 ± 0.30 | … | … | N | … | … | … | Y | … | n | … |
| 44156 | 08001990-6024319 | -2.0 ± 0.2 | 5378 ± 132 | 3.84 ± 0.12 | 0.04 ± 0.16 | 138 ± 3 | … | N | … | … | … | … | … | n | NG |
| 44157 | 08002059-6042375 | 19.9 ± 0.2 | 5115 ± 217 | 4.35 ± 0.14 | -0.69 ± 0.71 | 204 ± 3 | … | N | Y | Y | N | Y | Y | Y | … |
| 44158 | 08002083-6023163 | 23.8 ± 0.3 | 3786 ± 189 | 4.68 ± 0.12 | -0.01 ± 0.14 | … | … | Y | … | … | … | Y | … | n | … |
| 44159 | 08002198-6103443 | 22.6 ± 0.2 | 4678 ± 257 | 4.77 ± 0.23 | 0.14 ± 0.18 | 61 ± 3 | … | Y | Y | Y | Y | Y | Y | Y | … |
| 44196 | 08010217-6045027 | 24.5 ± 0.2 | 6414 ± 208 | 4.39 ± 0.46 | -0.07 ± 0.13 | 56 ± 2 | 1 | Y | Y | Y | Y | Y | Y | Y | … |
| 44197 | 08010448-6036003 | 42.3 ± 0.2 | 4953 ± 219 | 2.88 ± 0.41 | -0.23 ± 0.23 | 9 ± 3 | … | N | … | … | … | … | … | n | G |
| 44198 | 08010495-6047348 | 23.6 ± 0.2 | 4983 ± 167 | 4.57 ± 0.45 | 0.06 ± 0.12 | 79 ± 4 | … | Y | Y | Y | Y | … | Y | Y | … |
| 44199 | 08010732-6020237 | 12.5 ± 0.6 | 3916 ± 392 | 4.46 ± 0.22 | -0.18 ± 0.29 | … | … | N | … | … | … | N | … | n | … |
| 44200 | 08010953-6027583 | 25.3 ± 1.1 | 3734 ± 351 | 4.50 ± 0.22 | … | … | … | Y | … | … | … | Y | … | n | … |

**Table C.8.** continued.

| ID | CNAME | RV (km s$^{-1}$) | $T_{\rm eff}$ (K) | $\log g$ (dex) | [Fe/H] (dex) | EW(Li)$^a$ (mÅ) | EW(Li) error flag$^b$ | Membership RV | Li | $\log g$ | [Fe/H] | Gaia studies Randich$^c$ | Cantat-Gaudin$^c$ | Final$^d$ | NMs with Li$^e$ |
|---|---|---|---|---|---|---|---|---|---|---|---|---|---|---|---|
| 44201 | 08011107-6034256 | 4.8 ± 0.3 | 6553 ± 353 | 4.16 ± 0.35 | 0.03 ± 0.19 | 56 ± 3 | 1 | N | … | … | … | N | … | n | NG |
| 44202 | 08011270-6023501 | -5.1 ± 0.2 | 4945 ± 141 | 4.30 ± 0.69 | 0.29 ± 0.24 | 12 ± 3 | … | N | … | … | … | … | … | n | NG |
| 44203 | 08011274-6032061 | 24.3 ± 0.8 | 3624 ± 138 | 4.58 ± 0.15 | -0.06 ± 0.12 | … | … | Y | … | … | … | Y | … | n | … |
| 44204 | 08011281-6053531 | 23.4 ± 0.2 | 5928 ± 174 | 4.51 ± 0.35 | -0.13 ± 0.20 | 120 ± 2 | … | Y | Y | Y | Y | Y | Y | Y | … |
| 44205 | 08011376-6044389 | 27.0 ± 0.3 | 4012 ± 307 | 4.77 ± 0.20 | -0.07 ± 0.20 | 18 ± 6 | … | N | Y | Y | Y | Y | … | Y | … |
| 44206 | 08011398-6036125 | -9.0 ± 0.2 | 4971 ± 132 | 3.36 ± 0.21 | 0.10 ± 0.17 | 14 ± 3 | … | N | … | … | … | … | … | n | G |
| 44207 | 08011533-6047395 | 27.1 ± 0.4 | 4852 ± 232 | 4.19 ± 0.13 | 0.07 ± 0.19 | 246 ± 6 | … | N | Y | Y | Y | Y | Y | Y | … |
| 44208 | 08011612-6020192 | 12.8 ± 0.5 | 3989 ± 315 | 4.53 ± 0.14 | -0.24 ± 0.26 | … | … | N | … | … | … | N | … | n | … |
| 44209 | 08011776-6023308 | 223.1 ± 0.2 | 4684 ± 266 | 2.18 ± 0.94 | -0.61 ± 0.36 | … | … | N | … | … | … | … | … | n | G |
| 44210 | 08011939-6030357 | 26.3 ± 0.3 | 3818 ± 181 | 4.67 ± 0.11 | -0.01 ± 0.20 | … | … | N | … | … | … | Y | … | n | … |
| 44211 | 08012174-6025405 | 23.5 ± 0.2 | 5941 ± 124 | 4.37 ± 0.35 | -0.19 ± 0.21 | 111 ± 2 | … | Y | Y | Y | N | Y | Y | Y | … |
| 44212 | 08012289-6053056 | 23.2 ± 0.2 | 4171 ± 352 | 4.62 ± 0.11 | 0.08 ± 0.19 | 67 ± 5 | … | Y | Y | Y | Y | Y | Y | Y | … |
| 44213 | 08012327-6046145 | 27.8 ± 1.0 | 3903 ± 301 | 4.75 ± 0.22 | -0.03 ± 0.13 | … | … | N | … | … | … | Y | … | n | … |
| 44214 | 08012365-6104597 | 25.8 ± 0.2 | 4360 ± 443 | 4.48 ± 0.12 | 0.07 ± 0.19 | 18 ± 3 | … | Y | Y | Y | Y | Y | Y | Y | … |
| 44215 | 08012448-6035158 | 23.5 ± 0.2 | 6282 ± 133 | 4.37 ± 0.39 | 0.06 ± 0.15 | 82 ± 2 | 1 | Y | Y | Y | Y | Y | Y | Y | … |
| 44216 | 08012759-6041408 | 24.0 ± 0.3 | 4313 ± 348 | 4.72 ± 0.15 | -0.11 ± 0.15 | 6 ± 4 | … | Y | Y | Y | Y | Y | Y | Y | … |
| 2953 | 08012920-6027429 | -41.1 ± 0.6 | 5142 ± 118 | 3.64 ± 0.24 | -0.22 ± 0.10 | <14 | 3 | N | … | … | … | … | … | n | NG |
| 44217 | 08013257-6101156 | 23.8 ± 0.3 | 3666 ± 177 | 4.70 ± 0.12 | -0.04 ± 0.14 | … | … | Y | … | … | … | Y | … | n | … |
| 43560 | 07532107-6058131 | 23.4 ± 0.4 | 3824 ± 291 | 4.57 ± 0.10 | -0.01 ± 0.14 | … | … | Y | … | … | … | Y | … | n | … |
| 43561 | 07532163-6102129 | 24.5 ± 0.4 | 3651 ± 192 | 4.47 ± 0.26 | -0.02 ± 0.13 | 5 ± 5 | … | Y | Y | Y | Y | Y | Y | Y | … |
| 43562 | 07532312-6102429 | 23.5 ± 0.2 | 3958 ± 323 | 4.57 ± 0.09 | 0.01 ± 0.20 | 10 ± 3 | … | Y | Y | Y | Y | Y | … | Y | … |
| 43584 | 07542366-6059134 | 22.6 ± 0.3 | 5060 ± 97 | 4.66 ± 0.55 | -0.01 ± 0.13 | 117 ± 4 | … | Y | Y | Y | Y | Y | … | Y | … |
| 43585 | 07542375-6028270 | 48.1 ± 0.2 | 4990 ± 121 | 3.27 ± 0.13 | -0.12 ± 0.22 | 14 ± 3 | … | N | … | … | … | … | … | n | G |
| 43586 | 07542397-6111027 | 61.9 ± 0.2 | 4841 ± 107 | 2.59 ± 0.23 | 0.16 ± 0.26 | 9 ± 3 | … | N | … | … | … | … | … | n | G |
| 43587 | 07542455-6019566 | 24.3 ± 0.3 | 4424 ± 403 | 4.61 ± 0.10 | 0.09 ± 0.19 | 38 ± 5 | … | Y | Y | Y | Y | Y | Y | Y | … |
| 43588 | 07542511-6029238 | 13.1 ± 0.2 | 4887 ± 170 | 2.79 ± 0.56 | -0.25 ± 0.22 | 46 ± 3 | … | N | … | … | … | … | … | n | G |
| 43589 | 07542542-6045146 | 123.7 ± 0.2 | 4711 ± 232 | 2.54 ± 0.35 | 0.01 ± 0.25 | 11 ± 3 | … | N | … | … | … | … | … | n | G |
| 43590 | 07542722-6057451 | 23.7 ± 0.2 | 5548 ± 96 | 4.57 ± 0.43 | 0.03 ± 0.12 | 152 ± 3 | … | Y | Y | Y | Y | Y | Y | Y | … |
| 43591 | 07542890-6046186 | 23.9 ± 0.2 | 4891 ± 195 | 4.56 ± 0.37 | 0.02 ± 0.13 | 130 ± 2 | … | Y | Y | Y | Y | Y | Y | Y | … |
| 43592 | 07543002-6052309 | 40.0 ± 0.2 | 4646 ± 166 | 3.04 ± 0.56 | 0.18 ± 0.20 | 6 ± 3 | … | N | … | … | … | … | … | n | G |
| 43593 | 07543052-6044581 | 19.6 ± 0.2 | 4441 ± 392 | 4.80 ± 0.28 | -0.15 ± 0.19 | 15 ± 3 | … | N | Y | Y | Y | Y | … | Y | … |
| 43594 | 07543141-6039513 | 24.8 ± 0.5 | 3591 ± 125 | 4.59 ± 0.14 | -0.07 ± 0.14 | … | … | Y | … | … | … | Y | … | n | … |
| 43595 | 07543176-6033518 | 51.4 ± 0.2 | 5831 ± 105 | 4.22 ± 0.11 | 0.35 ± 0.17 | 89 ± 4 | … | N | … | … | … | N | … | n | NG |
| 43596 | 07543348-6113265 | 30.1 ± 0.2 | 4686 ± 142 | 4.50 ± 0.34 | 0.21 ± 0.24 | … | … | N | … | … | … | Y | … | n | … |
| 43597 | 07543442-6035456 | 23.9 ± 0.2 | 4735 ± 217 | 4.72 ± 0.23 | 0.00 ± 0.13 | 72 ± 4 | … | Y | Y | Y | Y | Y | Y | Y | … |
| 43598 | 07543520-6101128 | 66.4 ± 0.2 | 5805 ± 105 | 4.39 ± 0.08 | 0.14 ± 0.13 | 37 ± 3 | … | N | … | … | … | … | … | n | NG |
| 43599 | 07543533-6049145 | -14.3 ± 0.2 | 4764 ± 231 | 3.12 ± 0.54 | 0.19 ± 0.23 | 21 ± 3 | … | N | … | … | … | … | … | n | G |
| 43600 | 07543650-6113541 | 3.1 ± 0.3 | 6535 ± 309 | 4.15 ± 0.24 | 0.09 ± 0.20 | 9 ± 2 | 1 | N | … | … | … | N | … | n | … |
| 43601 | 07543785-6102035 | 1.8 ± 0.2 | 5021 ± 133 | 3.27 ± 0.29 | -0.23 ± 0.20 | 7 ± 3 | … | N | … | … | … | … | … | n | G |
| 43602 | 07544020-6105459 | 2.1 ± 0.2 | 6397 ± 205 | 4.05 ± 0.12 | 0.11 ± 0.17 | 2 ± 1 | 1 | N | … | … | … | N | … | n | … |
| 43603 | 07544232-6053534 | 40.9 ± 0.2 | 5738 ± 76 | 4.08 ± 0.22 | -0.11 ± 0.12 | 2 ± 2 | … | N | … | … | … | … | … | n | … |
| 2906 | 07544342-6024437 | 25.1 ± 0.6 | 5549 ± 115 | 4.57 ± 0.24 | 0.01 ± 0.10 | 165 ± 3 | … | Y | Y | Y | Y | … | Y | Y | … |
| 43604 | 07544421-6049473 | 23.4 ± 0.3 | 3749 ± 171 | 4.68 ± 0.11 | -0.02 ± 0.13 | … | … | Y | … | … | … | Y | … | n | … |
| 2907 | 07544443-6108107 | 42.8 ± 0.6 | 5935 ± 113 | 3.76 ± 0.23 | -0.39 ± 0.10 | 34 ± 2 | … | N | … | … | … | … | … | n | NG |
| 44605 | 07544487-6032296 | 245.3 ± 0.3 | 4740 ± 152 | 1.75 ± 0.59 | -1.62 ± 0.26 | … | … | N | … | … | … | … | … | n | G |
| 43649 | 07551741-6038315 | 23.4 ± 0.3 | 3873 ± 399 | 4.56 ± 0.10 | 0.08 ± 0.27 | 148 ± 10 | … | Y | Y | Y | Y | Y | Y | Y | … |
| 43650 | 07551844-6057440 | 23.1 ± 0.3 | 6144 ± 134 | 4.52 ± 0.59 | -0.35 ± 0.29 | 40 ± 2 | 1 | Y | Y | Y | N | Y | Y | Y | … |
| 43899 | 07574613-6023315 | 79.2 ± 0.4 | 3898 ± 343 | 4.56 ± 0.17 | -0.67 ± 0.52 | … | … | N | … | … | … | N | … | n | … |
| 43900 | 07574651-6027229 | 24.3 ± 0.5 | 3745 ± 305 | 4.49 ± 0.21 | … | … | … | Y | … | … | … | Y | … | n | … |
| 43938 | 07580532-6045034 | 24.1 ± 0.2 | 6137 ± 206 | 4.30 ± 0.14 | -0.28 ± 0.27 | 109 ± 2 | 1 | Y | Y | Y | N | Y | Y | Y | … |
| 43651 | 07551956-6044275 | 24.3 ± 0.2 | 4511 ± 270 | 4.63 ± 0.14 | -0.01 ± 0.13 | 19 ± 3 | … | Y | Y | Y | Y | Y | Y | Y | … |
| 43901 | 07574677-6030471 | 88.2 ± 0.2 | 5189 ± 59 | 4.28 ± 0.41 | 0.03 ± 0.14 | 3 ± 3 | … | N | … | … | … | … | … | n | … |
| 43939 | 07580587-6031090 | -7.8 ± 0.3 | 3727 ± 237 | 4.19 ± 0.41 | -0.03 ± 0.10 | … | … | N | … | … | … | N | … | n | … |
| 43923 | 07580010-6058317 | 55.6 ± 0.2 | 4845 ± 62 | 3.29 ± 0.55 | 0.30 ± 0.28 | 19 ± 3 | … | N | … | … | … | … | … | n | G |
| 43940 | 07580587-6032426 | 23.6 ± 0.8 | 3691 ± 334 | 4.54 ± 0.24 | … | … | … | Y | … | … | … | Y | … | n | … |
| 2911 | 07551977-6104200 | 24.2 ± 0.4 | … | … | … | 137 ± 7 | … | Y | … | … | … | Y | Y | n | … |
| 43924 | 07580035-6052122 | 26.1 ± 0.3 | 4098 ± 403 | 4.44 ± 0.15 | 0.11 ± 0.21 | 52 ± 4 | … | N | Y | Y | Y | Y | … | Y | … |
| 2929 | 07580698-6027260 | 75.0 ± 0.6 | 5029 ± 113 | 3.58 ± 0.22 | -0.35 ± 0.09 | <18 | 3 | N | … | … | … | … | … | n | … |
| 43925 | 07580044-6041196 | 18.2 ± 0.3 | 4689 ± 288 | 4.60 ± 0.21 | -0.06 ± 0.13 | 48 ± 3 | … | N | Y | Y | Y | … | Y | Y | … |

**Table C.8.** continued.

| ID | CNAME | RV (km s$^{-1}$) | $T_\mathrm{eff}$ (K) | $logg$ (dex) | [Fe/H] (dex) | $EW$(Li)$^a$ (mÅ) | $EW$(Li) error flag$^b$ | RV | Li | logg | [Fe/H] | Randich$^c$ | Cantat-Gaudin$^c$ | Final$^d$ | NMs with Li$^e$ |
|---|---|---|---|---|---|---|---|---|---|---|---|---|---|---|---|
| 43941 | 07580745-6044334 | 22.0 ± 0.2 | 4424 ± 244 | 4.70 ± 0.20 | -0.11 ± 0.13 | 7 ± 2 | … | Y | Y | Y | Y | Y | Y | Y | … |
| 43652 | 07551991-6057599 | 27.1 ± 0.4 | 3509 ± 136 | 4.62 ± 0.13 | -0.09 ± 0.14 | … | … | N | … | … | … | Y | … | n | … |
| 43926 | 07580073-6108080 | 54.7 ± 0.5 | 3711 ± 578 | 4.37 ± 0.20 | … | … | … | N | … | … | … | N | … | n | … |
| 43942 | 07580887-6044403 | 24.7 ± 0.3 | 6197 ± 231 | 4.76 ± 0.75 | -0.72 ± 0.71 | 42 ± 2 | 1 | Y | Y | Y | N | Y | … | Y | … |
| 43653 | 07552039-6019419 | -4.1 ± 0.3 | 4034 ± 278 | 4.86 ± 0.25 | -0.57 ± 0.35 | … | … | N | … | … | … | N | … | n | … |
| 43927 | 07580085-6103127 | 23.6 ± 0.3 | 4197 ± 402 | 4.53 ± 0.11 | 0.13 ± 0.20 | 18 ± 4 | … | Y | Y | Y | Y | Y | Y | Y | … |
| 43989 | 07583267-6055428 | 23.5 ± 0.4 | 3482 ± 129 | 4.67 ± 0.14 | -0.11 ± 0.13 | … | … | Y | … | … | … | Y | … | n | … |
| 43654 | 07552048-6038025 | 8.2 ± 0.5 | 3678 ± 504 | 4.33 ± 0.24 | … | … | … | N | … | … | … | N | … | n | … |
| 43928 | 07580117-6031234 | 24.5 ± 0.2 | 5022 ± 140 | 4.47 ± 0.50 | 0.06 ± 0.13 | 205 ± 4 | … | Y | Y | Y | Y | Y | Y | Y | … |
| 43990 | 07583277-6048525 | 23.8 ± 0.3 | 3993 ± 312 | 4.64 ± 0.12 | 0.01 ± 0.21 | … | … | Y | … | … | … | Y | … | n | … |
| 43929 | 07580151-6056424 | 162.7 ± 0.2 | 4339 ± 308 | 1.77 ± 0.58 | -0.31 ± 0.39 | 14 ± 3 | … | N | … | … | … | … | … | n | G |
| 43991 | 07583311-6110041 | -0.1 ± 0.3 | 3716 ± 192 | 4.50 ± 0.21 | -0.02 ± 0.13 | … | … | N | … | … | … | N | … | n | … |
| 43655 | 07552105-6034021 | 22.6 ± 0.5 | 3599 ± 156 | 4.76 ± 0.14 | -0.06 ± 0.12 | … | … | Y | … | … | … | Y | … | n | … |
| 43930 | 07580219-6111283 | 24.7 ± 0.2 | 4931 ± 182 | 4.56 ± 0.37 | -0.12 ± 0.16 | 30 ± 4 | … | Y | Y | Y | Y | … | Y | Y | … |
| 43992 | 07583340-6044273 | 23.5 ± 0.2 | 5789 ± 71 | 4.29 ± 0.15 | -0.44 ± 0.36 | 119 ± 2 | … | Y | Y | Y | N | Y | N | Y | … |
| 44168 | 08003236-6113141 | 23.7 ± 0.3 | 4013 ± 406 | 4.51 ± 0.11 | -0.02 ± 0.19 | 12 ± 6 | … | Y | Y | Y | Y | Y | N | Y | … |
| 43656 | 07552115-6026296 | 20.9 ± 0.3 | 3759 ± 201 | 4.50 ± 0.23 | -0.03 ± 0.14 | … | … | N | … | … | … | Y | … | n | … |
| 43931 | 07580295-6057256 | 22.5 ± 0.2 | 4837 ± 186 | 3.16 ± 0.30 | 0.05 ± 0.21 | 3 ± 3 | … | Y | N | N | Y | … | … | n | G |
| 2932 | 07583485-6103121 | 23.5 ± 0.6 | 6043 ± 131 | 4.68 ± 0.23 | -0.08 ± 0.10 | 134 ± 1 | … | Y | Y | Y | Y | Y | Y | Y | … |
| 44657 | 07552202-6050063 | 24.3 ± 0.2 | 5634 ± 109 | 4.32 ± 0.37 | -0.18 ± 0.16 | 129 ± 2 | … | Y | Y | Y | Y | Y | Y | Y | … |
| 43932 | 07580409-6038274 | 24.2 ± 0.8 | 3517 ± 25 | 4.78 ± 0.17 | … | … | … | Y | … | … | … | Y | … | n | … |
| 43993 | 07583572-6046525 | 22.3 ± 0.2 | 5770 ± 147 | 4.38 ± 0.36 | -0.37 ± 0.37 | 138 ± 2 | … | Y | Y | Y | N | Y | Y | Y | … |
| 43658 | 07552206-6105203 | 25.1 ± 0.3 | 5030 ± 103 | 4.65 ± 0.48 | 0.03 ± 0.13 | 76 ± 3 | … | Y | Y | Y | Y | … | Y | Y | … |
| 43933 | 07580444-6052426 | 21.3 ± 0.8 | 3533 ± 28 | 5.06 ± 0.20 | … | … | … | Y | … | … | … | Y | … | n | … |
| 43994 | 07583587-6040407 | -11.0 ± 0.2 | 4806 ± 180 | 4.56 ± 0.35 | -0.06 ± 0.13 | 3 ± 2 | … | N | … | … | … | … | … | n | … |
| 44177 | 08004306-6106116 | 23.7 ± 0.2 | 4602 ± 381 | 4.33 ± 0.16 | -0.12 ± 0.15 | 142 ± 3 | … | Y | Y | Y | Y | Y | Y | Y | … |
| 43934 | 07580458-6040298 | 25.3 ± 0.2 | 4735 ± 218 | 4.71 ± 0.16 | 0.03 ± 0.13 | 82 ± 3 | … | Y | Y | Y | Y | Y | Y | Y | … |
| 43995 | 07583646-6050184 | 24.4 ± 0.2 | 6394 ± 412 | 4.17 ± 0.13 | 0.05 ± 0.22 | 84 ± 2 | 1 | Y | Y | Y | Y | Y | Y | Y | … |
| 43659 | 07552215-6109234 | 24.7 ± 0.2 | 4521 ± 316 | 4.65 ± 0.23 | -0.02 ± 0.13 | 3 ± 3 | … | Y | Y | Y | Y | Y | Y | Y | … |
| 43935 | 07580470-6035037 | 23.4 ± 0.2 | 5881 ± 141 | 4.39 ± 0.30 | -0.25 ± 0.25 | 127 ± 2 | … | Y | Y | Y | N | Y | Y | Y | … |
| 43996 | 07583662-6048063 | 9.7 ± 0.5 | 3640 ± 141 | 4.49 ± 0.14 | -0.06 ± 0.12 | … | … | N | … | … | … | N | … | n | … |
| 43660 | 07552222-6033074 | 24.0 ± 0.2 | 4522 ± 136 | 4.66 ± 0.38 | 0.24 ± 0.22 | 8 ± 3 | … | Y | Y | Y | N | Y | … | Y | … |
| 43936 | 07580515-6046119 | 21.1 ± 0.2 | 4973 ± 203 | 4.57 ± 0.32 | -0.14 ± 0.15 | 54 ± 2 | … | Y | Y | Y | Y | … | … | Y | … |
| 43997 | 07583678-6103548 | 40.0 ± 3.9 | 3965 ± 211 | … | … | … | … | N | … | … | … | N | … | n | … |
| 43937 | 07580525-6047403 | 25.5 ± 0.2 | 6515 ± 200 | 4.46 ± 0.40 | -0.04 ± 0.13 | 49 ± 1 | 1 | Y | Y | Y | Y | Y | Y | Y | … |
| 43998 | 07583693-6104343 | 23.8 ± 0.2 | 5748 ± 108 | 4.42 ± 0.33 | -0.18 ± 0.16 | 116 ± 2 | … | Y | Y | Y | Y | Y | Y | Y | … |
| 43999 | 07583705-6046284 | 24.6 ± 0.2 | 4772 ± 259 | 4.57 ± 0.24 | -0.04 ± 0.13 | 144 ± 3 | … | Y | Y | Y | Y | Y | Y | Y | … |
| 44178 | 08004451-6058165 | 24.0 ± 0.3 | 4056 ± 457 | 4.50 ± 0.11 | 0.08 ± 0.19 | 79 ± 4 | … | Y | Y | Y | Y | Y | … | Y | … |
| 44181 | 08004769-6043368 | 21.7 ± 0.2 | 5554 ± 101 | 4.58 ± 0.32 | 0.02 ± 0.13 | 153 ± 3 | … | Y | Y | Y | Y | Y | Y | Y | … |
| 44000 | 07583730-6035381 | 23.0 ± 0.2 | 4662 ± 304 | 4.69 ± 0.24 | 0.04 ± 0.13 | 20 ± 3 | … | Y | Y | Y | Y | … | Y | Y | … |
| 2913 | 07553236-6023094 | 24.8 ± 0.6 | 5805 ± 198 | 4.49 ± 0.46 | 0.02 ± 0.12 | 129 ± 2 | … | Y | Y | Y | Y | Y | Y | Y | … |
| 44001 | 07583898-6048233 | 24.1 ± 0.3 | 4018 ± 266 | 4.55 ± 0.09 | 0.06 ± 0.23 | 26 ± 5 | … | Y | Y | Y | Y | Y | Y | Y | … |
| 43678 | 07553274-6037074 | 20.9 ± 0.8 | 3926 ± 467 | 4.59 ± 0.09 | 0.06 ± 0.25 | … | … | N | … | … | … | Y | … | n | … |
| 44182 | 08004799-6051564 | 24.0 ± 0.2 | 5419 ± 88 | 4.46 ± 0.44 | 0.04 ± 0.13 | 112 ± 3 | … | Y | Y | Y | Y | Y | Y | Y | … |
| 44002 | 07583905-6045217 | 24.9 ± 0.2 | 5318 ± 113 | 4.47 ± 0.42 | 0.01 ± 0.13 | 121 ± 3 | … | Y | Y | Y | Y | Y | Y | Y | … |
| 44003 | 07583938-6029593 | 24.8 ± 0.2 | 4873 ± 196 | 4.58 ± 0.31 | 0.00 ± 0.12 | 93 ± 3 | … | Y | Y | Y | Y | Y | Y | Y | … |
| 44179 | 08004556-6028586 | 56.0 ± 0.2 | 4664 ± 125 | 2.72 ± 0.31 | 0.01 ± 0.25 | 17 ± 2 | … | N | … | … | … | … | … | n | … |
| 44004 | 07584005-6041429 | 23.1 ± 0.2 | 6314 ± 350 | 4.42 ± 0.32 | -0.07 ± 0.16 | 88 ± 2 | 1 | Y | Y | Y | Y | Y | Y | Y | … |
| 44180 | 08004607-6019473 | 35.8 ± 2.3 | 4330 ± 129 | 4.71 ± 0.52 | -1.25 ± 0.23 | … | … | N | … | … | … | N | … | n | … |
| 43679 | 07553448-6111529 | 25.6 ± 0.3 | 4207 ± 423 | 4.61 ± 0.10 | 0.10 ± 0.19 | 2 ± 3 | … | Y | Y | Y | Y | Y | … | Y | … |
| 44218 | 08013300-6027205 | 22.3 ± 0.6 | 3823 ± 221 | 4.73 ± 0.13 | -0.02 ± 0.11 | … | … | Y | … | … | … | Y | … | n | … |
| 44005 | 07584014-6107216 | 23.1 ± 0.3 | 3998 ± 222 | 4.66 ± 0.13 | 0.06 ± 0.23 | 314 ± 6 | … | Y | Y | Y | Y | Y | Y | Y | … |
| 43680 | 07553485-6041393 | 9.6 ± 0.3 | 4036 ± 408 | 4.62 ± 0.30 | -0.47 ± 0.43 | 6 ± 4 | … | N | … | … | … | N | … | n | … |
| 44219 | 08013340-6034222 | 24.0 ± 0.2 | 4352 ± 297 | 4.89 ± 0.28 | -0.07 ± 0.15 | 44 ± 4 | … | Y | Y | Y | Y | Y | Y | Y | … |
| 44006 | 07584047-6029433 | 13.9 ± 1.8 | 4078 ± 116 | … | … | … | … | N | … | … | … | N | … | n | … |
| 44220 | 08013392-6051381 | 23.8 ± 0.2 | 4405 ± 468 | 4.56 ± 0.09 | 0.07 ± 0.19 | 65 ± 3 | … | Y | Y | Y | Y | Y | Y | Y | … |
| 44007 | 07584125-6058096 | 86.2 ± 0.2 | 4668 ± 99 | 2.59 ± 0.28 | 0.12 ± 0.26 | 11 ± 3 | … | N | … | … | … | … | … | n | G |
| 44221 | 08013470-6043525 | 22.6 ± 0.3 | 3717 ± 137 | 4.63 ± 0.14 | -0.02 ± 0.13 | 19 ± 5 | … | Y | Y | Y | Y | Y | … | Y | … |
| 44008 | 07584149-6050549 | 22.4 ± 0.4 | 3918 ± 255 | 4.55 ± 0.10 | -0.06 ± 0.19 | 54 ± 7 | … | Y | Y | Y | Y | Y | … | Y | … |

**Table C.8.** continued.

| ID | CNAME | RV (km s$^{-1}$) | T$_{eff}$ (K) | logg (dex) | [Fe/H] (dex) | EW(Li)$^a$ (mÅ) | EW(Li) error flag$^b$ | Membership RV | Li | logg | [Fe/H] | Gaia studies Randich$^c$ | Cantat-Gaudin$^c$ | Final$^d$ | NMs with Li$^e$ |
|---|---|---|---|---|---|---|---|---|---|---|---|---|---|---|---|
| 44222 | 08013524-6113067 | 7.5 ± 0.3 | 4231 ± 212 | 4.86 ± 0.35 | -0.21 ± 0.14 | … | … | N | … | … | … | N | … | n | … |
| 2933 | 07584257-6040199 | 25.3 ± 0.6 | 5591 ± 119 | 4.43 ± 0.25 | -0.10 ± 0.11 | 109 ± 4 | … | Y | Y | Y | Y | Y | Y | Y | … |
| 43681 | 07553687-6022470 | 26.3 ± 0.2 | 6702 ± 461 | 3.81 ± 0.21 | 0.04 ± 0.19 | 68 ± 2 | 1 | Y | Y | N | Y | Y | … | Y | … |
| 44223 | 08013638-6045277 | 24.8 ± 0.4 | 3882 ± 253 | 4.54 ± 0.11 | -0.07 ± 0.15 | … | … | Y | … | … | … | Y | … | n | … |
| 44009 | 07584283-6039338 | 21.9 ± 0.3 | 3879 ± 274 | 4.75 ± 0.17 | 0.02 ± 0.19 | … | … | Y | … | … | … | Y | … | n | … |
| 43682 | 07553709-6022374 | 89.0 ± 0.3 | 4619 ± 121 | 2.62 ± 0.37 | 0.10 ± 0.31 | 15 ± 6 | … | N | … | … | … | Y | … | n | G |
| 2954 | 08013658-6059021 | 24.2 ± 0.6 | 5617 ± 120 | 4.39 ± 0.24 | -0.12 ± 0.10 | 141 ± 39 | … | Y | Y | Y | Y | Y | Y | Y | … |
| 44010 | 07584299-6101051 | 53.0 ± 0.2 | 4900 ± 347 | 2.90 ± 0.92 | -0.39 ± 0.26 | … | … | N | … | … | … | … | … | n | G |
| 43683 | 07553731-6109067 | 23.5 ± 0.2 | 6246 ± 166 | 4.45 ± 0.41 | -0.08 ± 0.14 | 107 ± 2 | 1 | Y | Y | Y | Y | Y | N | Y | … |
| 44224 | 08013707-6028527 | 24.8 ± 0.3 | 3848 ± 219 | 4.76 ± 0.13 | 0.06 ± 0.23 | … | … | Y | … | … | … | Y | … | n | … |
| 44033 | 07585360-6054242 | 21.8 ± 0.5 | 3865 ± 333 | 4.61 ± 0.10 | 0.02 ± 0.21 | … | … | Y | … | … | … | Y | … | n | … |
| 44225 | 08013814-6025081 | 44.5 ± 5.7 | 3163 ± 555 | … | … | … | … | N | … | … | … | N | … | n | … |
| 44034 | 07585363-6028088 | 23.6 ± 0.5 | 4650 ± 352 | 4.33 ± 0.17 | -0.06 ± 0.16 | 129 ± 4 | … | Y | Y | Y | Y | Y | … | Y | … |
| 43684 | 07553834-6037305 | 48.8 ± 0.2 | 5811 ± 156 | 4.23 ± 0.09 | 0.26 ± 0.14 | 18 ± 2 | … | N | … | … | … | … | … | n | NG |
| 44226 | 08013837-6101289 | 23.3 ± 0.2 | 3991 ± 321 | 4.54 ± 0.09 | 0.06 ± 0.20 | 1 ± 3 | … | Y | Y | Y | Y | Y | … | Y | … |
| 44035 | 07585364-6056325 | 25.3 ± 0.8 | 3498 ± 24 | 4.35 ± 0.16 | … | … | … | Y | … | … | … | Y | … | n | … |
| 44227 | 08014018-6047068 | 25.5 ± 0.2 | 4647 ± 277 | 4.69 ± 0.26 | -0.01 ± 0.12 | 28 ± 3 | … | Y | Y | Y | Y | Y | Y | Y | … |
| 44036 | 07585511-6118344 | 25.9 ± 1.2 | … | … | … | … | … | Y | … | … | … | Y | … | n | … |
| 44228 | 08014020-6023595 | 7.8 ± 0.2 | 4784 ± 72 | 3.02 ± 0.45 | 0.22 ± 0.22 | 17 ± 4 | … | N | … | … | … | Y | … | n | G |
| 44037 | 07585551-6055595 | 24.6 ± 0.4 | 3904 ± 290 | 4.43 ± 0.17 | 0.05 ± 0.25 | … | … | Y | … | … | … | Y | … | n | … |
| 2955 | 08014071-6106068 | 6.8 ± 0.6 | 5604 ± 113 | 4.44 ± 0.24 | 0.10 ± 0.10 | <11 | 3 | N | … | … | … | … | … | n | … |
| 44038 | 07585563-6025568 | 29.3 ± 2.2 | 3715 ± 41 | … | … | … | … | N | … | … | … | Y | … | n | … |
| 43685 | 07553910-6018305 | 36.8 ± 0.3 | 4052 ± 269 | 4.63 ± 0.10 | -0.28 ± 0.16 | … | … | N | … | … | … | N | … | n | … |
| 44229 | 08014281-6112552 | 30.2 ± 0.4 | 3676 ± 175 | 4.45 ± 0.28 | -0.03 ± 0.14 | … | … | N | … | … | … | Y | … | n | … |
| 44039 | 07585617-6123162 | 26.1 ± 1.0 | … | … | … | … | … | N | … | … | … | Y | … | n | … |
| 43686 | 07553943-6031203 | 22.2 ± 1.0 | 3551 ± 127 | 4.89 ± 0.11 | -0.07 ± 0.12 | … | … | Y | … | … | … | Y | … | n | … |
| 44230 | 08014320-6030039 | 52.2 ± 0.7 | 3669 ± 490 | 4.75 ± 0.79 | -0.68 ± 0.62 | … | … | N | … | … | … | N | … | n | … |
| 44040 | 07585638-6047229 | 19.8 ± 0.3 | 4028 ± 488 | 4.79 ± 0.23 | 0.04 ± 0.22 | … | … | N | … | … | … | N | … | n | … |
| 43687 | 07554003-6101161 | 24.0 ± 0.3 | 3987 ± 308 | 4.60 ± 0.10 | -0.01 ± 0.21 | … | … | Y | … | … | … | Y | … | n | … |
| 44231 | 08014727-6035012 | 23.3 ± 0.6 | 3537 ± 27 | 4.35 ± 0.20 | … | … | … | Y | … | … | … | Y | … | n | … |
| 44041 | 07585658-6052160 | 28.0 ± 0.5 | 3839 ± 578 | 4.74 ± 0.21 | 0.02 ± 0.23 | … | … | N | … | … | … | Y | … | n | … |
| 43688 | 07554040-6104238 | 22.4 ± 0.2 | 5356 ± 201 | 4.44 ± 0.34 | -0.30 ± 0.32 | 155 ± 3 | … | Y | Y | Y | N | Y | Y | Y | … |
| 44232 | 08015578-6044369 | -2.3 ± 0.3 | 4045 ± 312 | 4.50 ± 0.09 | 0.06 ± 0.25 | 21 ± 5 | … | N | … | … | … | N | … | n | NG |
| 44042 | 07585720-6036127 | 24.3 ± 0.4 | 4545 ± 547 | 4.32 ± 0.26 | 0.14 ± 0.19 | 252 ± 4 | … | Y | Y | Y | Y | Y | Y | Y | … |
| 44233 | 08015580-6042558 | 23.6 ± 0.3 | 3947 ± 251 | 4.43 ± 0.14 | -0.02 ± 0.19 | 102 ± 7 | … | Y | Y | Y | Y | Y | Y | Y | … |
| 44043 | 07585752-6037081 | 57.3 ± 0.8 | … | … | … | … | … | N | … | … | … | N | … | n | … |
| 44234 | 08015838-6042508 | 25.3 ± 0.2 | 4057 ± 320 | 4.62 ± 0.11 | 0.06 ± 0.19 | 148 ± 5 | … | Y | Y | Y | Y | Y | Y | Y | … |
| 44044 | 07585754-6043008 | 30.1 ± 0.2 | 4588 ± 482 | 4.22 ± 0.36 | 0.12 ± 0.19 | 226 ± 7 | … | N | Y | Y | Y | Y | N | Y | … |
| 44235 | 08020996-6042421 | 24.7 ± 0.3 | 3971 ± 254 | 4.73 ± 0.21 | 0.02 ± 0.20 | 50 ± 6 | … | Y | Y | Y | Y | Y | Y | Y | … |
| 44045 | 07585760-6040586 | 24.0 ± 0.5 | 3640 ± 209 | 4.66 ± 0.12 | -0.03 ± 0.14 | … | … | Y | … | … | … | Y | … | n | … |
| 43707 | 07555595-6052334 | 31.2 ± 0.3 | 3684 ± 162 | 4.70 ± 0.11 | -0.03 ± 0.14 | … | … | N | … | … | … | Y | … | n | … |
| 44236 | 08021221-6055262 | 23.2 ± 0.2 | 4150 ± 369 | 4.60 ± 0.10 | 0.07 ± 0.19 | 8 ± 3 | … | Y | Y | Y | Y | Y | Y | Y | … |
| 44046 | 07585820-6056328 | 23.7 ± 0.3 | 4105 ± 299 | 4.88 ± 0.39 | 0.06 ± 0.19 | 27 ± 5 | … | Y | Y | Y | Y | Y | Y | Y | … |
| 43708 | 07555597-6056441 | 29.9 ± 0.2 | 4706 ± 288 | 2.65 ± 0.47 | -0.23 ± 0.26 | 11 ± 3 | … | N | … | … | … | … | … | n | … |
| 44237 | 08021438-6057429 | 16.4 ± 0.5 | 3631 ± 494 | 4.70 ± 0.24 | -0.67 ± 0.60 | … | … | N | … | … | … | Y | … | n | … |
| 2936 | 07585971-6110550 | 25.1 ± 0.3 | … | … | … | 187 ± 4 | … | Y | Y | Y | … | Y | … | Y | … |
| 43709 | 07555673-6039322 | 12.4 ± 0.2 | 5674 ± 85 | 4.08 ± 0.16 | 0.06 ± 0.14 | 7 ± 2 | … | N | … | … | … | … | … | n | … |
| 44238 | 08021875-6045186 | 24.3 ± 0.3 | 3954 ± 274 | 4.78 ± 0.18 | -0.43 ± 0.42 | … | … | Y | … | … | … | Y | … | n | … |
| 44047 | 07590019-6040380 | 24.3 ± 0.3 | 4277 ± 294 | 4.90 ± 0.28 | -0.04 ± 0.13 | 10 ± 3 | … | Y | Y | Y | Y | Y | Y | Y | … |
| 44239 | 08022195-6042432 | 13.4 ± 0.3 | 4274 ± 154 | 4.79 ± 0.21 | -0.15 ± 0.14 | 15 ± 4 | … | N | … | … | … | N | … | n | NG |
| 2937 | 07590020-6109585 | 15.9 ± 0.6 | 5592 ± 120 | 4.36 ± 0.22 | -0.49 ± 0.10 | <11 | 3 | N | … | … | … | … | … | n | … |
| 44048 | 07590027-6046423 | 2.5 ± 0.2 | 4826 ± 174 | 2.82 ± 0.25 | -0.11 ± 0.20 | 3 ± 3 | … | N | … | … | … | … | … | n | G |
| 44049 | 07590075-6055489 | -25.7 ± 0.2 | 5700 ± 20 | 4.04 ± 0.27 | -0.32 ± 0.12 | 34 ± 2 | … | N | … | … | … | … | … | n | NG |
| 44250 | 07590124-6044561 | 21.7 ± 0.3 | 3901 ± 168 | 4.81 ± 0.25 | 0.08 ± 0.25 | 15 ± 5 | … | Y | Y | Y | Y | Y | Y | Y | … |
| 43710 | 07555749-6041438 | 23.7 ± 0.4 | 3601 ± 200 | 4.42 ± 0.16 | -0.05 ± 0.13 | … | … | Y | … | … | … | Y | … | n | … |
| 44051 | 07590306-6106245 | 24.4 ± 0.5 | 3753 ± 224 | 4.55 ± 0.19 | -0.02 ± 0.13 | … | … | Y | … | … | … | Y | … | n | … |
| 44052 | 07590447-6033400 | 33.5 ± 0.3 | 5792 ± 147 | 4.46 ± 0.29 | -0.04 ± 0.14 | … | … | N | … | … | … | … | … | n | … |
| 44053 | 07590452-6101141 | 28.1 ± 0.2 | 4689 ± 134 | 2.53 ± 0.28 | -0.02 ± 0.20 | 23 ± 3 | … | N | … | … | … | … | … | n | G |
| 43711 | 07555802-6111369 | 23.8 ± 0.3 | 5350 ± 82 | 4.57 ± 0.38 | 0.31 ± 0.20 | 22 ± 4 | … | Y | N | Y | N | … | … | n | NG |

| ID | CNAME | RV (km s$^{-1}$) | $T_{\text{eff}}$ (K) | logg (dex) | [Fe/H] (dex) | $EW$(Li)$^a$ (mÅ) | $EW$(Li) error flag$^b$ | Membership RV | Li | logg | [Fe/H] | Gaia studies Randich$^c$ | Cantat-Gaudin$^c$ | Final$^d$ | NMs with Li$^e$ |
|---|---|---|---|---|---|---|---|---|---|---|---|---|---|---|---|
| 44054 | 07590471-6037086 | 23.3 ± 0.2 | 4692 ± 275 | 4.65 ± 0.18 | -0.01 ± 0.13 | 65 ± 3 | … | Y | Y | Y | Y | Y | Y | Y | … |
| 43712 | 07555838-6053052 | 23.9 ± 0.2 | 5095 ± 103 | 4.60 ± 0.46 | -0.05 ± 0.16 | 134 ± 4 | … | Y | Y | Y | Y | Y | Y | Y | … |
| 44084 | 07592637-6019164 | 38.6 ± 0.3 | 4694 ± 81 | 2.53 ± 0.28 | 0.28 ± 0.35 | 9 ± 4 | … | N | … | … | … | … | … | n | G |
| 44085 | 07592727-6026370 | 71.2 ± 0.2 | 5016 ± 106 | 3.27 ± 0.14 | -0.10 ± 0.21 | 2 ± 3 | … | N | … | … | … | … | … | n | G |
| 43713 | 07555844-6056126 | 23.5 ± 0.2 | 4563 ± 289 | 4.62 ± 0.20 | -0.04 ± 0.14 | 27 ± 4 | … | Y | Y | Y | Y | Y | Y | Y | … |
| 44086 | 07592893-6104406 | 6.9 ± 0.2 | 5148 ± 58 | 4.52 ± 0.28 | 0.15 ± 0.15 | 11 ± 3 | … | N | … | … | … | … | … | n | NG |
| 44087 | 07592900-6039417 | 23.2 ± 0.9 | 3963 ± 564 | 4.46 ± 0.12 | 0.04 ± 0.19 | … | … | Y | … | … | … | Y | … | n | … |
| 43714 | 07555860-6023482 | 23.9 ± 0.5 | 3576 ± 23 | 4.54 ± 0.18 | … | … | … | Y | … | … | … | Y | … | n | … |
| 43715 | 07555914-6050442 | 23.2 ± 0.2 | 4583 ± 456 | 4.47 ± 0.12 | 0.12 ± 0.19 | 182 ± 3 | … | Y | Y | Y | Y | Y | Y | Y | … |
| 44088 | 07592931-6025476 | -6.6 ± 0.3 | 3819 ± 180 | 4.78 ± 0.15 | -0.10 ± 0.35 | … | … | N | … | … | … | N | … | n | … |
| 44089 | 07592938-6040026 | 40.3 ± 0.3 | 3663 ± 125 | 4.70 ± 0.09 | -0.03 ± 0.14 | … | … | N | … | … | … | N | … | n | … |
| 44090 | 07592983-6028554 | 10.9 ± 0.6 | 3687 ± 488 | 4.79 ± 0.24 | … | … | … | N | … | … | … | N | … | n | … |
| 44091 | 07593093-6025165 | 22.4 ± 0.2 | 5168 ± 59 | 4.49 ± 0.37 | 0.00 ± 0.13 | 36 ± 2 | … | Y | N | Y | Y | … | N | n | NG |
| 43727 | 07560458-6038303 | 9.6 ± 0.4 | 4548 ± 507 | 3.87 ± 1.04 | 0.07 ± 0.20 | 173 ± 8 | … | N | … | … | … | … | … | n | NG |
| 2940 | 07593210-6048450 | 49.6 ± 0.6 | 5490 ± 125 | 4.74 ± 0.28 | -0.14 ± 0.11 | 220 ± 2 | … | N | … | … | … | N | Y | n | NG |
| 43728 | 07560587-6055349 | 26.0 ± 0.4 | 3623 ± 140 | 4.57 ± 0.14 | -0.03 ± 0.14 | … | … | Y | … | … | … | Y | … | n | … |
| 44092 | 07593231-6058460 | 23.8 ± 0.3 | 4807 ± 374 | 4.02 ± 0.23 | 0.16 ± 0.21 | 203 ± 5 | … | Y | Y | N | Y | Y | Y | Y | … |
| 44093 | 07593236-6045025 | 24.4 ± 0.2 | 4573 ± 282 | 4.61 ± 0.18 | -0.06 ± 0.13 | 30 ± 3 | … | Y | Y | Y | Y | Y | Y | Y | … |
| 43729 | 07560698-6031438 | 25.4 ± 0.2 | 4737 ± 178 | 4.59 ± 0.36 | -0.02 ± 0.13 | 37 ± 4 | … | Y | Y | Y | Y | … | Y | Y | … |
| 44094 | 07593261-6104529 | 24.0 ± 1.3 | 3991 ± 881 | 4.63 ± 0.10 | 0.04 ± 0.19 | … | … | Y | … | … | … | Y | … | n | … |
| 44095 | 07593402-6045184 | 23.7 ± 0.2 | 5298 ± 92 | 4.47 ± 0.37 | 0.08 ± 0.13 | 107 ± 3 | … | Y | Y | Y | Y | Y | Y | Y | … |
| 43730 | 07560814-6037597 | 25.4 ± 0.2 | 5175 ± 143 | 4.49 ± 0.37 | -0.06 ± 0.15 | 156 ± 5 | … | Y | Y | Y | Y | Y | … | Y | … |
| 43731 | 07560871-6045266 | 24.0 ± 0.2 | 4201 ± 385 | 4.52 ± 0.11 | 0.13 ± 0.21 | 4 ± 3 | … | Y | Y | Y | Y | Y | Y | Y | … |
| 43732 | 07560885-6049366 | 22.0 ± 0.3 | 4001 ± 312 | 4.59 ± 0.10 | 0.00 ± 0.20 | … | … | Y | … | … | … | Y | … | n | … |
| 43733 | 07560924-6047360 | 22.6 ± 0.6 | 3528 ± 126 | 4.58 ± 0.13 | -0.09 ± 0.13 | … | … | Y | … | … | … | Y | … | n | … |
| 44104 | 07594242-6024492 | -1.7 ± 0.3 | 4543 ± 361 | 4.89 ± 0.26 | -0.43 ± 0.23 | 4 ± 4 | … | N | … | … | … | N | … | n | … |
| 43541 | 07515457-6047568 | 7.5 ± 0.3 | 4406 ± 289 | 4.83 ± 0.25 | -0.33 ± 0.23 | … | … | N | … | … | … | N | … | n | … |
| 43734 | 07561042-6105132 | 23.1 ± 0.5 | 3607 ± 19 | 4.51 ± 0.17 | … | … | … | Y | … | … | … | Y | … | n | … |
| 43542 | 07515966-6047220 | 25.7 ± 0.2 | 4206 ± 337 | 4.50 ± 0.11 | 0.11 ± 0.21 | 42 ± 5 | … | Y | Y | Y | Y | Y | Y | Y | … |
| 43735 | 07561050-6054150 | 23.6 ± 0.3 | 3930 ± 314 | 4.76 ± 0.21 | 0.02 ± 0.19 | … | … | Y | … | … | … | Y | … | n | … |
| 43543 | 07520129-6043233 | 16.4 ± 0.8 | 3890 ± 496 | 4.92 ± 0.59 | -0.97 ± 0.05 | … | … | N | … | … | … | Y | … | n | … |
| 43736 | 07561129-6050580 | 23.6 ± 0.2 | 4520 ± 320 | 4.39 ± 0.41 | -0.08 ± 0.16 | 248 ± 4 | … | Y | Y | Y | Y | Y | Y | Y | … |
| 43544 | 07520389-6050116 | 25.5 ± 0.2 | 4193 ± 233 | 4.88 ± 0.39 | 0.07 ± 0.19 | 10 ± 4 | … | Y | Y | Y | Y | Y | Y | Y | … |
| 43737 | 07561189-6111384 | 2.8 ± 0.2 | 6624 ± 273 | 4.08 ± 0.03 | -0.07 ± 0.28 | 42 ± 2 | 1 | N | … | … | … | N | … | n | NG |
| 43545 | 07521002-6044245 | 25.6 ± 0.3 | 4124 ± 339 | 4.61 ± 0.13 | 0.09 ± 0.19 | 13 ± 4 | … | Y | Y | Y | Y | Y | Y | Y | … |
| 43738 | 07561223-6101165 | 39.2 ± 0.2 | 4763 ± 53 | 2.55 ± 0.23 | 0.08 ± 0.19 | 12 ± 3 | … | N | … | … | … | … | … | n | … |
| 43546 | 07521382-6047151 | 24.1 ± 0.2 | 3929 ± 192 | 4.65 ± 0.11 | 0.04 ± 0.20 | 46 ± 6 | … | Y | Y | Y | Y | Y | N | Y | … |
| 43739 | 07561313-6102284 | 52.4 ± 0.3 | 4086 ± 468 | 4.02 ± 0.50 | 0.06 ± 0.19 | 1 ± 4 | … | N | … | … | … | … | Y | n | … |
| 43547 | 07521911-6039241 | -6.3 ± 0.7 | 3725 ± 491 | 4.70 ± 0.17 | … | … | … | N | … | … | … | N | … | n | … |
| 43740 | 07561327-6108061 | 23.2 ± 0.2 | 4711 ± 270 | 4.75 ± 0.32 | 0.06 ± 0.15 | 11 ± 3 | … | Y | Y | Y | Y | Y | Y | Y | … |
| 44105 | 07594430-6032198 | 19.7 ± 0.3 | 3551 ± 55 | 4.30 ± 0.12 | -0.07 ± 0.12 | … | … | N | … | … | … | Y | … | n | … |
| 43548 | 07522464-6043006 | -2.5 ± 0.6 | 3833 ± 168 | 4.29 ± 0.26 | 0.00 ± 0.19 | … | … | N | … | … | … | N | … | n | … |
| 43549 | 07523060-6049134 | 23.5 ± 0.3 | 3602 ± 122 | 4.59 ± 0.12 | -0.07 ± 0.14 | … | … | Y | … | … | … | Y | … | n | … |
| 43741 | 07561340-6059071 | 23.8 ± 0.3 | 5154 ± 133 | 4.40 ± 0.41 | 0.02 ± 0.12 | 102 ± 5 | … | Y | Y | Y | Y | Y | Y | Y | … |
| 43550 | 07523629-6046405 | 31.4 ± 0.4 | 3899 ± 306 | 4.85 ± 0.21 | 0.00 ± 0.12 | … | … | N | … | … | … | Y | … | n | … |
| 43551 | 07524604-6039537 | 23.8 ± 0.3 | 4106 ± 302 | 4.65 ± 0.16 | 0.08 ± 0.20 | 20 ± 4 | … | Y | Y | Y | Y | Y | Y | Y | … |
| 43742 | 07561399-6111109 | 23.1 ± 0.3 | 4395 ± 362 | 4.75 ± 0.21 | 0.11 ± 0.19 | 9 ± 5 | … | Y | Y | Y | Y | Y | … | Y | … |
| 43552 | 07525073-6054177 | 35.4 ± 2.9 | 3905 ± 428 | 4.65 ± 0.13 | 0.04 ± 0.19 | … | … | N | … | … | … | N | … | n | … |
| 43778 | 07563502-6110084 | 494.9 ± 11.1 | … | … | … | … | … | N | … | … | … | … | … | n | … |
| 44106 | 07594539-6111064 | 23.0 ± 0.6 | 3593 ± 21 | 4.66 ± 0.18 | … | … | … | Y | … | … | … | Y | … | n | … |
| 43553 | 07525994-6053288 | 23.8 ± 0.3 | 3936 ± 180 | 4.76 ± 0.21 | 0.03 ± 0.19 | … | … | Y | … | … | … | Y | … | n | … |
| 43554 | 07530001-6042137 | 6.6 ± 0.3 | 3961 ± 347 | 4.59 ± 0.20 | -0.34 ± 0.18 | … | … | N | … | … | … | N | … | n | … |
| 43555 | 07530057-6048094 | 24.7 ± 0.3 | 3920 ± 218 | 4.68 ± 0.13 | -0.03 ± 0.23 | … | … | Y | … | … | … | Y | … | n | … |
| 43779 | 07563692-6106549 | 21.9 ± 0.3 | 3904 ± 241 | 4.33 ± 0.27 | -0.05 ± 0.24 | … | … | Y | … | … | … | Y | … | n | … |
| 43556 | 07530259-6050259 | 22.4 ± 0.6 | 4235 ± 540 | 4.77 ± 0.25 | 0.10 ± 0.19 | 21 ± 5 | … | Y | Y | Y | Y | Y | Y | Y | … |
| 43557 | 07531031-6046400 | 23.3 ± 0.3 | 3587 ± 133 | 4.46 ± 0.12 | -0.05 ± 0.12 | … | … | Y | … | … | … | Y | … | n | … |
| 44107 | 07594658-6051212 | 24.1 ± 0.5 | 3547 ± 26 | 4.79 ± 0.19 | … | … | … | Y | … | … | … | Y | … | n | … |
| 43558 | 07531177-6041390 | 24.0 ± 0.3 | 3837 ± 215 | 4.56 ± 0.17 | -0.02 ± 0.13 | … | … | Y | … | … | … | Y | … | n | … |
| 43559 | 07531326-6043422 | 4.1 ± 0.3 | 4039 ± 314 | 4.90 ± 0.31 | -0.23 ± 0.23 | 38 ± 5 | … | N | … | … | … | N | … | n | NG |

**Table C.8.** continued.

| ID | CNAME | RV (km s$^{-1}$) | $T_{\rm eff}$ (K) | logg (dex) | [Fe/H] (dex) | EW(Li)$^a$ (mÅ) | EW(Li) error flag$^b$ | Membership RV | Li | logg | [Fe/H] | Gaia studies Randich$^c$ | Cantat-Gaudin$^c$ | Final$^d$ | NMs with Li$^e$ |
|---|---|---|---|---|---|---|---|---|---|---|---|---|---|---|---|
| 43780 | 07563905-6103466 | 22.3 ± 0.3 | 3940 ± 350 | 4.61 ± 0.10 | 0.00 ± 0.20 | … | … | Y | … | … | … | Y | … | n | … |
| 43781 | 07563912-6106466 | 23.0 ± 0.6 | 3562 ± 26 | 4.62 ± 0.20 | … | … | … | Y | … | … | … | Y | … | n | … |
| 44108 | 07594689-6037266 | 89.6 ± 0.2 | 4803 ± 146 | 3.00 ± 0.21 | -0.11 ± 0.17 | 6 ± 3 | … | N | … | … | … | … | … | n | … |
| 43782 | 07563964-6104256 | 20.3 ± 0.5 | 3524 ± 26 | 4.37 ± 0.17 | … | … | … | N | … | … | … | Y | … | n | … |
| 43783 | 07564014-6027412 | 15.0 ± 0.3 | 4079 ± 252 | 4.74 ± 0.16 | -0.19 ± 0.28 | … | … | N | … | … | … | N | … | n | … |
| 44110 | 07594747-6054153 | 24.4 ± 0.2 | 5513 ± 111 | 4.42 ± 0.43 | -0.13 ± 0.13 | 142 ± 2 | … | Y | Y | Y | Y | Y | Y | Y | … |
| 44129 | 07595914-6104543 | 53.3 ± 0.4 | 3892 ± 260 | 4.58 ± 0.10 | 0.01 ± 0.12 | … | … | N | … | … | … | N | … | n | … |
| 43784 | 07564302-6027026 | 34.6 ± 0.2 | 6132 ± 150 | 3.98 ± 0.14 | -0.04 ± 0.18 | 4 ± 1 | … | N | … | … | … | N | … | n | … |
| 44130 | 07595940-6031110 | 10.9 ± 0.2 | 5617 ± 136 | 3.97 ± 0.29 | -0.17 ± 0.18 | 23 ± 2 | … | N | … | … | … | … | … | n | NG |
| 2917 | 07564410-6034523 | 24.3 ± 0.6 | 5608 ± 118 | 4.40 ± 0.25 | -0.06 ± 0.10 | 127 ± 2 | … | Y | Y | Y | Y | Y | Y | Y | … |
| 43785 | 07564410-6103120 | -8.4 ± 0.2 | 6158 ± 200 | 4.13 ± 0.12 | -0.06 ± 0.13 | 20 ± 2 | 1 | N | … | … | … | N | … | n | NG |
| 43786 | 07564523-6116013 | 27.9 ± 1.3 | … | … | … | … | … | N | … | … | … | … | … | n | … |
| 44131 | 07595998-6036307 | 83.9 ± 0.2 | 4912 ± 164 | 2.85 ± 0.32 | -0.24 ± 0.21 | 4 ± 1 | … | N | … | … | … | … | … | n | G |
| 43787 | 07564607-6024132 | 23.8 ± 0.4 | 3893 ± 207 | 4.47 ± 0.15 | 0.00 ± 0.19 | 128 ± 10 | … | Y | Y | Y | Y | Y | N | Y | … |
| 43788 | 07564663-6041273 | 23.3 ± 0.4 | 3586 ± 140 | 4.84 ± 0.17 | -0.08 ± 0.14 | … | … | Y | … | … | … | Y | … | n | … |
| 43789 | 07564709-6043225 | 23.1 ± 0.2 | 4247 ± 398 | 4.50 ± 0.33 | -0.04 ± 0.12 | 109 ± 4 | … | Y | Y | Y | Y | Y | Y | Y | … |
| 2943 | 08000021-6040083 | 22.2 ± 0.6 | 5175 ± 120 | 4.39 ± 0.24 | -0.11 ± 0.10 | 140 ± 3 | … | Y | Y | Y | Y | Y | Y | Y | … |
| 43790 | 07564721-6101236 | 23.1 ± 0.6 | 3575 ± 23 | 4.54 ± 0.18 | … | … | … | Y | … | … | … | Y | … | n | … |
| 43812 | 07570068-6057296 | 24.6 ± 0.2 | 4841 ± 198 | 4.54 ± 0.34 | -0.07 ± 0.14 | 53 ± 3 | … | Y | Y | Y | Y | … | Y | Y | … |
| 44132 | 08000034-6023219 | 114.0 ± 0.3 | 4782 ± 515 | 4.03 ± 0.16 | -0.51 ± 0.81 | 7 ± 2 | … | N | … | … | … | … | … | n | … |
| 43813 | 07570096-6044322 | 24.3 ± 0.6 | 3517 ± 28 | 4.36 ± 0.19 | … | … | … | Y | … | … | … | Y | … | n | … |
| 44133 | 08000049-6052559 | 23.0 ± 0.2 | 4823 ± 224 | 4.63 ± 0.31 | 0.03 ± 0.12 | 93 ± 3 | … | Y | Y | Y | Y | Y | Y | Y | … |
| 2920 | 07570115-6108175 | -20.3 ± 0.6 | 4973 ± 153 | 3.34 ± 0.28 | -0.01 ± 0.12 | <18 | 3 | N | … | … | … | … | … | n | G |
| 43814 | 07570128-6052061 | 23.6 ± 0.3 | 4080 ± 273 | 4.74 ± 0.25 | 0.02 ± 0.19 | 115 ± 5 | … | Y | Y | Y | Y | Y | Y | Y | … |
| 43815 | 07570128-6101101 | 24.5 ± 0.5 | 3540 ± 27 | 4.48 ± 0.19 | … | … | … | Y | … | … | … | Y | … | n | … |
| 43816 | 07570134-6042136 | 24.0 ± 0.2 | 4536 ± 307 | 4.63 ± 0.17 | -0.01 ± 0.13 | 16 ± 3 | … | Y | Y | Y | Y | Y | Y | Y | … |
| 43817 | 07570153-6106129 | 11.4 ± 0.3 | 5174 ± 37 | 3.92 ± 0.15 | -0.08 ± 0.15 | 21 ± 4 | … | N | … | … | … | … | … | n | NG |
| 43818 | 07570160-6103085 | 22.4 ± 0.2 | 5833 ± 105 | 4.32 ± 0.28 | -0.22 ± 0.23 | 106 ± 2 | … | Y | Y | Y | N | Y | Y | Y | … |
| 44134 | 08000128-6102224 | 22.1 ± 0.3 | 5218 ± 38 | 4.48 ± 0.38 | -0.07 ± 0.16 | 11 ± 3 | … | Y | N | Y | Y | … | … | n | NG |
| 43819 | 07570240-6102049 | 25.3 ± 0.2 | 6500 ± 234 | 4.51 ± 0.53 | -0.11 ± 0.13 | 31 ± 2 | 1 | Y | Y | Y | Y | Y | Y | Y | … |
| 43820 | 07570251-6047567 | 24.2 ± 0.3 | 4235 ± 348 | 4.85 ± 0.29 | -0.10 ± 0.16 | 7 ± 3 | … | Y | Y | Y | Y | Y | Y | Y | … |
| 43821 | 07570309-6032524 | 52.3 ± 0.2 | 5952 ± 105 | 4.25 ± 0.19 | -0.17 ± 0.15 | 131 ± 2 | … | N | … | … | … | N | N | n | NG |
| 44135 | 08000214-6030311 | 22.0 ± 0.6 | 3717 ± 315 | 4.54 ± 0.17 | … | … | … | Y | … | … | … | Y | … | n | … |
| 43822 | 07570312-6021415 | 28.1 ± 0.2 | 4911 ± 257 | 3.03 ± 0.45 | -0.17 ± 0.23 | 26 ± 3 | … | N | … | … | … | … | … | n | G |
| 43823 | 07570332-6108161 | 11.4 ± 0.4 | 5586 ± 29 | 4.34 ± 0.21 | -0.06 ± 0.19 | 116 ± 4 | … | N | … | … | … | N | … | n | NG |
| 43824 | 07570418-6052438 | 18.6 ± 0.3 | 5034 ± 112 | 3.34 ± 0.18 | -0.26 ± 0.19 | … | … | N | … | … | … | … | … | n | G |
| 44136 | 08000265-6036554 | 24.0 ± 0.2 | 6031 ± 108 | 4.25 ± 0.15 | -0.18 ± 0.29 | 25 ± 2 | … | Y | Y | Y | Y | Y | Y | Y | … |
| 43825 | 07570537-6059115 | 0.0 ± 0.2 | 4942 ± 192 | 3.48 ± 0.35 | 0.01 ± 0.17 | 16 ± 3 | … | N | … | … | … | … | … | n | G |
| 43826 | 07570701-6102081 | 24.0 ± 0.5 | 3852 ± 260 | 4.54 ± 0.14 | -0.03 ± 0.12 | … | … | Y | … | … | … | Y | … | n | … |
| 43827 | 07570882-6108507 | 21.3 ± 0.2 | 4747 ± 234 | 4.60 ± 0.32 | 0.01 ± 0.12 | 19 ± 3 | … | Y | Y | Y | Y | … | Y | Y | … |
| 44141 | 08000793-6043234 | 21.6 ± 0.3 | 4089 ± 329 | 4.62 ± 0.22 | -0.38 ± 0.29 | … | … | Y | … | … | … | Y | … | n | … |
| 43828 | 07570886-6106462 | 23.3 ± 0.2 | 4782 ± 237 | 4.50 ± 0.36 | 0.03 ± 0.13 | 74 ± 3 | … | Y | Y | Y | Y | Y | Y | Y | … |
| 43829 | 07570892-6129186 | 23.3 ± 1.0 | … | … | … | … | … | Y | … | … | … | Y | … | n | … |
| 43830 | 07570938-6048027 | 23.7 ± 0.2 | 6293 ± 184 | 4.48 ± 0.48 | -0.15 ± 0.13 | 98 ± 2 | 1 | Y | Y | Y | Y | Y | Y | Y | … |
| 43831 | 07570960-6056067 | 23.8 ± 0.4 | 3845 ± 385 | 4.58 ± 0.10 | -0.08 ± 0.17 | … | … | Y | … | … | … | Y | … | n | … |
| 44142 | 08000867-6051320 | 25.9 ± 0.3 | 4039 ± 316 | 4.64 ± 0.16 | 0.07 ± 0.20 | … | … | Y | … | … | … | Y | … | n | … |
| 43832 | 07570970-6032134 | 23.8 ± 0.3 | 3977 ± 376 | 4.86 ± 0.22 | -0.52 ± 0.36 | … | … | Y | … | … | … | Y | … | n | … |
| 43833 | 07570982-6024219 | 61.4 ± 0.2 | 5027 ± 97 | 4.55 ± 0.35 | -0.05 ± 0.13 | 10 ± 2 | … | N | … | … | … | … | … | n | NG |
| 43834 | 07570998-6044103 | 24.7 ± 0.2 | 5838 ± 140 | 4.39 ± 0.32 | -0.25 ± 0.23 | 136 ± 2 | … | Y | Y | Y | N | Y | Y | Y | … |
| 2923 | 07573608-6048128 | 23.2 ± 0.6 | 5601 ± 115 | 4.52 ± 0.25 | -0.02 ± 0.09 | 155 ± 2 | … | Y | Y | Y | Y | Y | Y | Y | … |
| 43881 | 07573634-6121293 | 23.2 ± 0.3 | 3970 ± 280 | 4.78 ± 0.24 | -0.02 ± 0.20 | 20 ± 5 | … | Y | Y | Y | Y | Y | Y | Y | … |
| 2945 | 08000944-6033355 | 23.8 ± 0.6 | 5896 ± 123 | 4.52 ± 0.24 | -0.05 ± 0.09 | 119 ± 2 | … | Y | Y | Y | Y | Y | Y | Y | … |
| 43882 | 07573649-6019060 | 498.4 ± 7.4 | … | … | … | … | … | N | … | … | … | … | … | n | … |
| 43883 | 07573650-6102085 | 5.1 ± 0.3 | 4368 ± 270 | 3.76 ± 0.14 | 0.10 ± 0.04 | … | … | N | … | … | … | N | … | n | … |
| 43884 | 07573664-6042168 | 11.4 ± 0.2 | 5760 ± 135 | 4.35 ± 0.06 | 0.18 ± 0.12 | 11 ± 1 | … | N | … | … | … | … | … | n | NG |
| 43885 | 07573777-6111491 | 26.5 ± 0.5 | 4248 ± 433 | 4.50 ± 0.11 | 0.11 ± 0.21 | … | … | N | … | … | … | Y | … | n | … |
| 43886 | 07573812-6027181 | 3.2 ± 0.3 | 3631 ± 121 | 4.40 ± 0.13 | -0.05 ± 0.12 | … | … | N | … | … | … | N | … | n | … |
| 2946 | 08001019-6107367 | 34.4 ± 0.6 | 4827 ± 113 | 2.73 ± 0.22 | -0.21 ± 0.10 | <16 | 3 | N | … | … | … | … | Y | n | G |
| 43887 | 07573888-6043529 | 23.5 ± 0.2 | 4775 ± 250 | 4.50 ± 0.40 | 0.00 ± 0.14 | 59 ± 4 | … | Y | Y | Y | Y | … | Y | Y | … |

**Table C.8.** continued.

| ID | CNAME | RV (km s$^{-1}$) | $T_{\rm eff}$ (K) | logg (dex) | [Fe/H] (dex) | EW(Li)$^a$ (mÅ) | EW(Li) error flag$^b$ | Membership RV | Li | logg | [Fe/H] | Gaia studies Randich$^c$ | Cantat-Gaudin$^c$ | Final$^d$ | NMs with Li$^e$ |
|---|---|---|---|---|---|---|---|---|---|---|---|---|---|---|---|
| 43888 | 07573891-6038168 | 7.5 ± 0.7 | 3405 ± 37 | 4.85 ± 0.21 | … | … | … | N | … | … | … | N | … | n | … |
| 43889 | 07573908-6026036 | 27.4 ± 0.2 | 4875 ± 216 | 4.41 ± 0.34 | -0.07 ± 0.13 | 8 ± 3 | … | N | … | … | … | … | … | n | … |
| 44143 | 08001140-6101333 | 23.1 ± 0.2 | 4343 ± 329 | 4.66 ± 0.14 | -0.02 ± 0.14 | … | … | Y | … | … | … | Y | … | n | … |
| 43890 | 07573915-6102095 | 24.3 ± 0.2 | 6084 ± 198 | 4.13 ± 0.07 | -0.30 ± 0.35 | 54 ± 2 | 1 | Y | Y | Y | N | Y | Y | Y | … |
| 43891 | 07573937-6028347 | 25.5 ± 0.2 | 4289 ± 383 | 4.33 ± 0.26 | 0.11 ± 0.19 | 9 ± 3 | … | Y | Y | Y | Y | Y | Y | Y | … |
| 43892 | 07573984-6046394 | 23.8 ± 0.3 | 4262 ± 462 | 4.31 ± 0.28 | 0.11 ± 0.19 | 29 ± 4 | … | Y | Y | Y | Y | Y | Y | Y | … |
| 43893 | 07574058-6038301 | 20.8 ± 1.0 | 3463 ± 39 | 4.75 ± 0.24 | … | … | … | N | … | … | … | Y | … | n | … |
| 44144 | 08001200-6042537 | 63.3 ± 0.2 | 4706 ± 109 | 2.44 ± 0.31 | -0.23 ± 0.19 | 18 ± 3 | … | N | … | … | … | … | … | n | … |
| 44160 | 08002579-6054212 | 23.7 ± 0.3 | 3928 ± 302 | 4.65 ± 0.12 | 0.04 ± 0.21 | 84 ± 9 | … | Y | Y | Y | Y | Y | Y | Y | … |
| 43894 | 07574272-6044207 | 24.8 ± 0.2 | 6104 ± 190 | 4.29 ± 0.14 | -0.24 ± 0.23 | 91 ± 2 | 1 | Y | Y | Y | N | Y | Y | Y | … |
| 44161 | 08002682-6042452 | 19.9 ± 0.3 | 4017 ± 253 | 4.76 ± 0.16 | -0.41 ± 0.24 | … | … | N | … | … | … | Y | … | n | … |
| 43895 | 07574418-6045153 | -6.0 ± 0.4 | 3708 ± 472 | 4.89 ± 0.63 | -0.62 ± 0.45 | … | … | N | … | … | … | N | … | n | … |
| 43896 | 07574472-6123326 | 23.8 ± 0.7 | 3742 ± 307 | 4.22 ± 0.23 | … | … | … | Y | … | … | … | Y | … | n | … |
| 44162 | 08002735-6042581 | 24.0 ± 0.3 | 4051 ± 339 | 4.41 ± 0.14 | 0.11 ± 0.21 | 15 ± 4 | … | Y | Y | Y | Y | Y | Y | Y | … |
| 43897 | 07574508-6038277 | 9.8 ± 0.5 | 3936 ± 473 | 4.99 ± 0.17 | -0.41 ± 0.16 | … | … | N | … | … | … | N | … | n | … |
| 43898 | 07574552-6100231 | 23.6 ± 0.2 | 5113 ± 107 | 4.54 ± 0.52 | 0.05 ± 0.12 | 136 ± 4 | … | Y | Y | Y | Y | Y | Y | Y | … |
| 2947 | 08002823-6021348 | 51.7 ± 0.6 | 5465 ± 119 | 4.31 ± 0.22 | -0.38 ± 0.10 | <27 | 3 | N | … | … | … | … | … | n | NG |
| 44163 | 08002876-6107450 | 26.2 ± 0.7 | 4031 ± 590 | 4.89 ± 0.31 | 0.11 ± 0.22 | … | … | N | … | … | … | Y | … | n | … |
| 44164 | 08002884-6100483 | 23.8 ± 0.2 | 5973 ± 178 | 4.43 ± 0.28 | -0.20 ± 0.22 | 125 ± 2 | … | Y | Y | Y | N | Y | Y | Y | … |
| 44165 | 08002907-6055428 | 23.2 ± 0.2 | 4803 ± 228 | 4.72 ± 0.34 | 0.11 ± 0.15 | 73 ± 3 | … | Y | Y | Y | Y | Y | Y | Y | … |
| 44166 | 08002967-6042288 | 25.4 ± 0.4 | 3953 ± 223 | 4.84 ± 0.28 | -0.04 ± 0.20 | … | … | Y | … | … | … | Y | … | n | … |
| 44167 | 08003110-6027487 | 23.7 ± 0.2 | 6189 ± 116 | 4.38 ± 0.40 | -0.11 ± 0.20 | … | … | Y | … | … | … | Y | … | n | … |
| 2948 | 08003117-6108595 | -8.5 ± 0.6 | 5483 ± 122 | 4.47 ± 0.24 | 0.06 ± 0.10 | <15 | 3 | N | … | … | … | … | … | n | NG |
| 43838 | 07571065-6111160 | 23.2 ± 0.3 | 4106 ± 401 | 4.38 ± 0.20 | 0.11 ± 0.21 | 110 ± 6 | … | Y | Y | Y | Y | Y | Y | Y | … |
| 43839 | 07571082-6049067 | 23.9 ± 0.5 | 3668 ± 186 | 4.60 ± 0.16 | -0.03 ± 0.14 | … | … | Y | … | … | … | Y | … | n | … |
| 43840 | 07571084-6050501 | 24.0 ± 0.2 | 6069 ± 204 | 4.33 ± 0.19 | -0.19 ± 0.21 | 121 ± 2 | 1 | Y | Y | Y | N | Y | Y | Y | … |
| 43841 | 07571090-6033392 | 59.7 ± 0.2 | 4834 ± 198 | 2.76 ± 0.50 | -0.08 ± 0.19 | 17 ± 4 | … | N | … | … | … | … | … | n | G |
| 43842 | 07571111-6048156 | 24.1 ± 0.2 | 5085 ± 96 | 4.50 ± 0.50 | 0.03 ± 0.12 | 168 ± 3 | … | Y | Y | Y | Y | Y | Y | Y | … |
| 43843 | 07571215-6033090 | 23.2 ± 0.2 | 4752 ± 605 | 4.02 ± 0.07 | 0.16 ± 0.20 | 226 ± 4 | … | Y | Y | N | Y | Y | Y | Y | … |
| 43844 | 07571254-6102130 | 23.5 ± 0.5 | 3612 ± 20 | 4.73 ± 0.18 | … | … | … | Y | … | … | … | Y | … | n | … |
| 43845 | 07571300-6055159 | 23.2 ± 0.4 | 3601 ± 16 | 4.73 ± 0.14 | … | … | … | Y | … | … | … | Y | … | n | … |
| 43846 | 07571418-6040524 | 30.2 ± 0.2 | 5946 ± 147 | 4.19 ± 0.21 | -0.26 ± 0.24 | 136 ± 1 | … | N | Y | Y | N | Y | … | Y | … |
| 43847 | 07571474-6041418 | 24.9 ± 0.2 | 5072 ± 101 | 4.57 ± 0.39 | 0.13 ± 0.16 | 121 ± 3 | … | Y | Y | Y | Y | Y | Y | Y | … |
| 43848 | 07571524-6030065 | 101.6 ± 0.2 | 4450 ± 130 | 2.39 ± 0.36 | 0.14 ± 0.29 | 15 ± 4 | … | N | … | … | … | … | … | n | G |
| 43849 | 07571540-6038534 | 24.3 ± 0.2 | 4674 ± 404 | 4.35 ± 0.22 | 0.12 ± 0.19 | 248 ± 4 | … | Y | Y | Y | Y | Y | Y | Y | … |
| 43563 | 07532554-6037497 | 6.3 ± 0.4 | 3737 ± 221 | 4.75 ± 0.13 | -0.02 ± 0.13 | … | … | N | … | … | … | N | … | n | … |
| 43850 | 07571593-6041269 | 23.0 ± 0.3 | 4201 ± 278 | 4.91 ± 0.20 | -0.19 ± 0.20 | 14 ± 4 | … | Y | Y | Y | Y | Y | Y | Y | … |
| 43564 | 07533218-6108527 | 54.3 ± 1.5 | 3979 ± 391 | 4.61 ± 0.10 | -0.31 ± 0.23 | … | … | N | … | … | … | N | … | n | … |
| 43851 | 07571627-6047136 | 25.7 ± 0.2 | 6511 ± 210 | 4.41 ± 0.39 | -0.12 ± 0.12 | 46 ± 1 | 1 | Y | Y | Y | Y | Y | Y | Y | … |
| 43565 | 07533937-6039585 | 34.7 ± 0.3 | 3802 ± 400 | 4.93 ± 0.47 | -0.63 ± 0.51 | … | … | N | … | … | … | N | … | n | … |
| 43852 | 07571639-6107463 | 23.8 ± 0.3 | 4276 ± 375 | 4.89 ± 0.25 | -0.09 ± 0.14 | 33 ± 5 | … | Y | Y | Y | Y | Y | Y | Y | … |
| 43566 | 07534864-6039002 | 48.2 ± 0.7 | 3610 ± 548 | 4.88 ± 0.19 | … | … | … | N | … | … | … | N | … | n | … |
| 43853 | 07571643-6102445 | 27.0 ± 0.5 | 3765 ± 399 | 3.77 ± 0.19 | … | … | … | N | … | … | … | Y | … | n | … |
| 43567 | 07534951-6111396 | 30.0 ± 0.4 | 4153 ± 484 | 4.18 ± 0.41 | 0.11 ± 0.21 | … | … | N | … | … | … | Y | … | n | … |
| 43854 | 07572011-6107141 | 23.6 ± 0.2 | 5990 ± 194 | 4.38 ± 0.23 | -0.11 ± 0.23 | 76 ± 2 | … | Y | Y | Y | Y | Y | Y | Y | … |
| 43568 | 07540165-6045292 | 21.6 ± 0.5 | 3401 ± 148 | 4.96 ± 0.18 | -0.08 ± 0.15 | … | … | Y | N | N | Y | Y | … | n | … |
| 43855 | 07572080-6055434 | 30.1 ± 0.3 | 3873 ± 681 | 4.63 ± 0.10 | 0.08 ± 0.22 | … | … | N | … | … | … | Y | … | n | … |
| 43569 | 07540422-6110109 | 23.2 ± 0.3 | 3548 ± 128 | 4.54 ± 0.14 | -0.07 ± 0.14 | … | … | Y | … | … | … | Y | … | n | … |
| 43856 | 07572083-6044035 | 26.9 ± 0.2 | 4842 ± 297 | 4.30 ± 0.27 | 0.10 ± 0.23 | 222 ± 4 | … | N | Y | Y | Y | Y | … | Y | … |
| 43570 | 07540602-6101251 | 23.9 ± 0.2 | 4109 ± 396 | 4.43 ± 0.16 | 0.07 ± 0.19 | 6 ± 2 | … | Y | Y | Y | Y | Y | Y | Y | … |
| 43857 | 07572240-6107074 | 22.6 ± 0.2 | 4352 ± 321 | 4.54 ± 0.21 | -0.05 ± 0.15 | 19 ± 3 | … | Y | Y | Y | Y | Y | Y | Y | … |
| 43902 | 07574708-6051389 | 25.4 ± 0.3 | 4170 ± 363 | 4.61 ± 0.10 | 0.10 ± 0.20 | 28 ± 5 | … | Y | Y | Y | Y | Y | Y | Y | … |
| 43572 | 07540943-6107554 | 24.8 ± 0.2 | 4458 ± 176 | 4.78 ± 0.26 | -0.07 ± 0.12 | 5 ± 3 | … | Y | Y | Y | Y | Y | Y | Y | … |
| 43903 | 07574742-6040003 | 24.5 ± 1.1 | … | … | … | … | … | Y | … | … | … | Y | … | n | … |
| 2903 | 07541167-6048001 | 23.8 ± 0.6 | 5518 ± 130 | 4.33 ± 0.25 | -0.11 ± 0.11 | 203 ± 2 | … | Y | Y | Y | Y | Y | Y | Y | … |
| 2926 | 07574792-6056131 | 25.4 ± 0.6 | 5560 ± 121 | 4.53 ± 0.23 | 0.00 ± 0.09 | 158 ± 17 | … | Y | Y | Y | Y | Y | Y | Y | … |
| 43573 | 07541520-6025485 | 5.2 ± 0.7 | 3877 ± 362 | 4.43 ± 0.20 | -0.29 ± 0.32 | … | … | N | … | … | … | N | … | n | … |
| 43904 | 07574834-6031307 | 13.1 ± 0.2 | 6196 ± 226 | 4.16 ± 0.18 | -0.23 ± 0.14 | 70 ± 1 | 1 | N | … | … | … | N | … | n | NG |
| 2904 | 07541553-6058079 | 24.4 ± 0.6 | 5515 ± 129 | 4.54 ± 0.27 | 0.01 ± 0.12 | 188 ± 2 | … | Y | Y | Y | Y | Y | Y | Y | … |

**Table C.8.** continued.

| ID | CNAME | RV (km s$^{-1}$) | $T_{\rm eff}$ (K) | $\log g$ (dex) | [Fe/H] (dex) | $EW$(Li)$^a$ (mÅ) | $EW$(Li) error flag$^b$ | Membership | | | | Gaia studies | | Final$^d$ | NMs with Li$^e$ |
| | | | | | | | | RV | Li | $\log g$ | [Fe/H] | Randich$^c$ | Cantat-Gaudin$^c$ | | |
|---|---|---|---|---|---|---|---|---|---|---|---|---|---|---|---|
| 43905 | 07575101-6024111 | 22.5 ± 0.5 | 3947 ± 422 | 4.35 ± 0.21 | 0.06 ± 0.29 | … | … | Y | … | … | … | Y | … | n | … |
| 43574 | 07541858-6041516 | 23.3 ± 0.3 | 3678 ± 168 | 4.79 ± 0.15 | -0.03 ± 0.14 | … | … | Y | … | … | … | Y | … | n | … |
| 43906 | 07575115-6042107 | 23.5 ± 0.6 | 3823 ± 349 | 4.06 ± 0.23 | … | … | … | Y | … | … | … | Y | … | n | … |
| 43575 | 07541880-6042229 | 23.4 ± 0.2 | 4437 ± 264 | 4.70 ± 0.22 | -0.01 ± 0.15 | 12 ± 3 | … | Y | Y | Y | Y | Y | Y | Y | … |
| 43907 | 07575159-6106547 | 23.6 ± 0.7 | 3899 ± 427 | 4.69 ± 0.17 | -0.08 ± 0.14 | … | … | Y | … | … | … | Y | … | n | … |
| 43576 | 07541976-6049150 | 50.0 ± 0.2 | 5368 ± 74 | 4.47 ± 0.29 | 0.06 ± 0.12 | 4 ± 3 | … | N | … | … | … | … | … | n | … |
| 43908 | 07575188-6104220 | 24.1 ± 0.3 | 4149 ± 309 | 4.77 ± 0.23 | 0.08 ± 0.22 | 1 ± 3 | … | Y | Y | Y | Y | Y | Y | Y | … |
| 43577 | 07542108-6027081 | -4.2 ± 0.3 | 4227 ± 231 | 4.85 ± 0.24 | -0.36 ± 0.18 | 9 ± 4 | … | N | … | … | … | N | … | n | … |
| 2927 | 07575215-6100318 | 24.4 ± 0.6 | 5304 ± 120 | 4.49 ± 0.25 | -0.09 ± 0.10 | 107 ± 2 | … | Y | Y | Y | Y | Y | Y | Y | … |
| 43578 | 07542118-6058130 | 24.5 ± 1.0 | 3870 ± 409 | 4.55 ± 0.09 | -0.07 ± 0.13 | … | … | Y | … | … | … | Y | … | n | … |
| 43909 | 07575281-6030478 | 22.4 ± 0.3 | 3920 ± 177 | 4.73 ± 0.18 | 0.01 ± 0.19 | 75 ± 8 | … | Y | Y | Y | Y | Y | Y | Y | … |
| 43579 | 07542124-6111181 | 2.2 ± 0.2 | 4981 ± 131 | 3.27 ± 0.27 | -0.10 ± 0.17 | 12 ± 4 | … | N | … | … | … | … | … | n | G |
| 43910 | 07575299-6019360 | 65.8 ± 0.6 | 3565 ± 30 | 4.46 ± 0.23 | … | … | … | N | … | … | … | N | … | n | … |
| 43580 | 07542147-6022136 | 28.4 ± 0.3 | 4074 ± 327 | 4.70 ± 0.48 | -0.18 ± 0.14 | 26 ± 6 | … | N | Y | Y | Y | Y | … | Y | … |
| 43911 | 07575479-6051450 | 20.4 ± 0.2 | 5002 ± 75 | 2.83 ± 0.28 | -0.12 ± 0.18 | … | … | N | … | … | … | … | … | n | G |
| 43581 | 07542312-6041099 | 28.7 ± 1.7 | 3899 ± 507 | 5.17 ± 0.61 | -0.04 ± 0.15 | … | … | N | … | … | … | Y | … | n | … |
| 43912 | 07575482-6050269 | 23.5 ± 0.3 | 3910 ± 208 | 4.65 ± 0.12 | 0.00 ± 0.20 | 121 ± 11 | … | Y | Y | Y | Y | Y | Y | Y | … |
| 2905 | 07542322-6034452 | 89.8 ± 0.6 | 5899 ± 118 | 4.19 ± 0.24 | 0.14 ± 0.09 | 32 ± 2 | … | N | … | … | … | … | … | n | NG |
| 43913 | 07575542-6048272 | 24.5 ± 0.2 | 5823 ± 104 | 4.24 ± 0.17 | -0.55 ± 0.50 | 106 ± 2 | … | Y | Y | Y | N | Y | Y | Y | … |
| 43582 | 07542355-6114383 | 24.6 ± 2.8 | 3699 ± 266 | 4.43 ± 1.38 | -0.08 ± 0.17 | … | … | Y | … | … | … | Y | … | n | … |
| 43914 | 07575561-6029074 | 51.4 ± 0.2 | 4660 ± 129 | 2.84 ± 0.39 | 0.11 ± 0.26 | 14 ± 3 | … | N | … | … | … | … | … | n | … |
| 43915 | 07575562-6036060 | 24.5 ± 0.3 | 4199 ± 432 | 4.46 ± 0.15 | 0.09 ± 0.19 | 68 ± 5 | … | Y | Y | Y | Y | Y | … | Y | … |
| 43583 | 07542364-6051540 | 37.9 ± 0.2 | 4465 ± 380 | 4.60 ± 0.27 | -0.26 ± 0.26 | 23 ± 3 | … | N | … | … | … | N | Y | n | NG |
| 43916 | 07575611-6054591 | 24.3 ± 0.2 | 5137 ± 91 | 4.41 ± 0.39 | 0.03 ± 0.12 | 204 ± 3 | … | Y | Y | Y | Y | Y | Y | Y | … |
| 43606 | 07544660-6112211 | 41.2 ± 0.2 | 4861 ± 156 | 3.29 ± 0.41 | 0.21 ± 0.25 | 13 ± 3 | … | N | … | … | … | … | … | n | G |
| 43607 | 07544740-6039592 | 124.0 ± 0.2 | 4890 ± 156 | 3.04 ± 0.51 | -0.27 ± 0.25 | 13 ± 3 | … | N | … | … | … | … | … | n | G |
| 43917 | 07575767-6053390 | 24.3 ± 0.2 | 4946 ± 463 | 4.12 ± 0.12 | -0.64 ± 0.68 | 235 ± 4 | … | Y | Y | N | N | Y | Y | Y | … |
| 43608 | 07544751-6057590 | -9.2 ± 0.3 | 4210 ± 447 | 4.41 ± 0.20 | 0.10 ± 0.20 | 4 ± 4 | … | N | … | … | … | N | Y | n | … |
| 43918 | 07575805-6045446 | 24.7 ± 0.2 | 4514 ± 330 | 4.56 ± 0.08 | -0.17 ± 0.20 | 42 ± 3 | … | Y | Y | Y | Y | Y | … | Y | … |
| 43609 | 07544819-6058285 | 24.3 ± 0.2 | 4611 ± 265 | 4.65 ± 0.20 | -0.04 ± 0.13 | 10 ± 2 | … | Y | Y | Y | Y | Y | Y | Y | … |
| 43919 | 07575851-6033517 | 106.0 ± 0.5 | 3690 ± 589 | 4.36 ± 0.25 | … | … | … | N | … | … | … | … | … | n | … |
| 43610 | 07545020-6050407 | 24.1 ± 0.2 | 6085 ± 257 | 4.42 ± 0.18 | -0.25 ± 0.33 | 102 ± 2 | 1 | Y | Y | Y | N | Y | Y | Y | … |
| 43920 | 07575861-6024107 | 25.5 ± 0.2 | 4853 ± 199 | 4.66 ± 0.34 | 0.02 ± 0.13 | 58 ± 3 | … | Y | Y | Y | Y | … | N | Y | … |
| 43611 | 07545144-6048109 | 24.3 ± 0.2 | 4623 ± 290 | 4.68 ± 0.23 | 0.03 ± 0.13 | 42 ± 3 | … | Y | Y | Y | Y | Y | Y | Y | … |
| 43921 | 07575863-6103390 | 23.8 ± 0.2 | 4904 ± 197 | 4.43 ± 0.39 | 0.02 ± 0.12 | 143 ± 4 | … | Y | Y | Y | Y | Y | Y | Y | … |
| 43612 | 07545162-6106074 | 37.6 ± 0.2 | 5807 ± 140 | 4.49 ± 0.20 | 0.08 ± 0.14 | 10 ± 2 | … | N | … | … | … | … | … | n | NG |
| 43922 | 07575932-6056535 | 26.7 ± 0.2 | 5967 ± 141 | 4.30 ± 0.12 | -0.34 ± 0.28 | 92 ± 2 | … | N | Y | Y | N | Y | Y | Y | … |
| 43943 | 07580948-6059412 | 25.3 ± 0.2 | 4613 ± 253 | 4.75 ± 0.21 | 0.03 ± 0.13 | 61 ± 4 | … | Y | Y | Y | Y | Y | Y | Y | … |
| 43613 | 07545163-6025304 | 23.9 ± 0.3 | 4069 ± 486 | 4.44 ± 0.16 | 0.11 ± 0.20 | … | … | Y | … | … | … | Y | … | n | … |
| 43944 | 07581100-6025028 | 24.8 ± 0.3 | 6194 ± 211 | 4.74 ± 0.85 | -0.77 ± 0.61 | 31 ± 2 | 1 | Y | Y | Y | N | Y | Y | Y | … |
| 43614 | 07545228-6020112 | 69.4 ± 0.2 | 5010 ± 120 | 2.61 ± 0.46 | -0.25 ± 0.23 | 12 ± 2 | … | N | … | … | … | … | … | n | G |
| 43945 | 07581123-6049599 | 32.2 ± 0.2 | 5375 ± 104 | 4.41 ± 0.33 | -0.10 ± 0.13 | 172 ± 3 | … | N | Y | Y | Y | Y | Y | Y | … |
| 43615 | 07545314-6019437 | 76.2 ± 0.2 | 4809 ± 144 | 2.61 ± 0.27 | 0.07 ± 0.27 | 18 ± 2 | … | N | … | … | … | … | … | n | G |
| 43946 | 07581182-6023594 | 65.3 ± 1.8 | … | … | … | … | … | N | … | … | … | … | … | n | … |
| 43616 | 07545378-6046173 | 95.7 ± 0.2 | 4606 ± 172 | 2.71 ± 0.44 | 0.02 ± 0.23 | 21 ± 3 | … | N | … | … | … | … | … | n | … |
| 43947 | 07581251-6038058 | 23.0 ± 0.4 | 4002 ± 307 | 4.79 ± 0.25 | 0.00 ± 0.20 | … | … | Y | … | … | … | Y | … | n | … |
| 43617 | 07545588-6054583 | 42.2 ± 0.3 | 4110 ± 259 | 4.83 ± 0.27 | -0.52 ± 0.27 | 8 ± 3 | … | N | … | … | … | N | … | n | … |
| 43948 | 07581274-6101519 | 23.1 ± 0.2 | 5326 ± 102 | 4.52 ± 0.36 | -0.15 ± 0.20 | 96 ± 2 | … | Y | Y | Y | Y | Y | Y | Y | … |
| 43618 | 07545617-6027432 | 8.7 ± 0.3 | 3637 ± 273 | 4.54 ± 0.09 | -0.11 ± 0.15 | … | … | N | … | … | … | N | … | n | … |
| 43949 | 07581287-6038365 | 23.9 ± 0.2 | 6467 ± 301 | 4.42 ± 0.30 | -0.01 ± 0.15 | 53 ± 1 | 1 | Y | Y | Y | Y | Y | Y | Y | … |
| 43619 | 07545637-6105118 | 22.1 ± 0.6 | 3733 ± 335 | 4.77 ± 0.20 | -0.05 ± 0.14 | … | … | Y | … | … | … | Y | … | n | … |
| 43950 | 07581383-6059348 | -2.0 ± 0.2 | 4977 ± 101 | 2.78 ± 0.41 | -0.13 ± 0.19 | 12 ± 2 | … | N | … | … | … | … | … | n | … |
| 43620 | 07545737-6038255 | 32.4 ± 0.2 | 5055 ± 92 | 3.44 ± 0.38 | -0.18 ± 0.18 | 8 ± 3 | … | N | … | … | … | … | … | n | G |
| 43951 | 07581385-6049512 | 25.5 ± 0.2 | 4035 ± 244 | 4.64 ± 0.12 | -0.22 ± 0.22 | 17 ± 4 | … | Y | Y | Y | N | Y | … | Y | … |
| 43621 | 07545756-6055475 | 46.1 ± 0.2 | 4711 ± 77 | 4.55 ± 0.28 | 0.15 ± 0.18 | 4 ± 3 | … | N | … | … | … | … | … | n | … |
| 43952 | 07581395-6110387 | 61.4 ± 0.2 | 5767 ± 149 | 4.01 ± 0.21 | 0.04 ± 0.14 | 14 ± 2 | … | N | … | … | … | … | … | n | NG |
| 43953 | 07581415-6106009 | 7.8 ± 0.2 | 4867 ± 190 | 4.59 ± 0.39 | 0.05 ± 0.13 | 9 ± 4 | … | N | … | … | … | … | … | n | … |
| 43622 | 07545769-6059373 | 17.0 ± 0.3 | 4093 ± 235 | 4.72 ± 0.17 | -0.17 ± 0.24 | 23 ± 5 | … | N | Y | Y | Y | Y | … | Y | … |
| 43954 | 07581528-6059334 | 24.4 ± 0.4 | 3637 ± 184 | 4.55 ± 0.16 | -0.01 ± 0.11 | 199 ± 11 | … | Y | Y | Y | Y | Y | Y | Y | … |

| ID | CNAME | RV (km s$^{-1}$) | $T_{\text{eff}}$ (K) | logg (dex) | [Fe/H] (dex) | EW(Li)$^a$ (mÅ) | EW(Li) error flag$^b$ | Membership | | | | Gaia studies | | Final$^d$ | NMs with Li$^e$ |
|---|---|---|---|---|---|---|---|---|---|---|---|---|---|---|---|
| | | | | | | | | RV | Li | logg | [Fe/H] | Randich$^c$ | Cantat-Gaudin$^c$ | | |
| 43623 | 07545954-6104227 | 25.8 ± 0.3 | 3782 ± 292 | 4.57 ± 0.15 | -0.03 ± 0.14 | ... | ... | Y | ... | ... | ... | Y | ... | n | ... |
| 43955 | 07581683-6054526 | 22.7 ± 0.3 | 5433 ± 90 | 4.46 ± 0.37 | 0.03 ± 0.13 | 128 ± 3 | ... | Y | Y | Y | Y | Y | Y | Y | ... |
| 43956 | 07581719-6044368 | 23.9 ± 0.4 | 3538 ± 126 | 4.58 ± 0.16 | -0.07 ± 0.14 | ... | ... | Y | ... | ... | ... | Y | ... | n | ... |
| 43624 | 07550080-6019235 | 8.0 ± 0.6 | 3646 ± 25 | 4.19 ± 0.21 | ... | ... | ... | N | ... | ... | ... | N | ... | n | ... |
| 44145 | 08001303-6041460 | 25.4 ± 0.2 | 4374 ± 318 | 4.52 ± 0.12 | 0.11 ± 0.19 | 10 ± 2 | ... | Y | Y | Y | Y | Y | Y | Y | ... |
| 43957 | 07581783-6129451 | 60.2 ± 1.2 | 4404 ± 79 | 5.11 ± 0.26 | -0.34 ± 0.09 | ... | ... | N | ... | ... | ... | N | ... | n | ... |
| 43625 | 07550144-6027506 | 24.7 ± 0.3 | 3971 ± 328 | 4.50 ± 0.11 | -0.07 ± 0.23 | 47 ± 5 | ... | Y | Y | Y | Y | Y | N | Y | ... |
| 43958 | 07581811-6020196 | 23.7 ± 0.2 | 4776 ± 216 | 4.69 ± 0.20 | 0.06 ± 0.14 | 30 ± 3 | ... | Y | Y | Y | Y | ... | Y | Y | ... |
| 43626 | 07550223-6051257 | 89.0 ± 0.3 | 4004 ± 385 | 4.50 ± 0.33 | -0.46 ± 0.28 | ... | ... | N | ... | ... | ... | N | ... | n | ... |
| 43627 | 07550360-6028452 | 25.4 ± 0.2 | 4282 ± 371 | 4.66 ± 0.23 | -0.03 ± 0.14 | 20 ± 4 | ... | Y | Y | Y | Y | Y | Y | Y | ... |
| 43959 | 07581813-6059052 | 46.3 ± 0.2 | 5853 ± 130 | 4.24 ± 0.24 | -0.31 ± 0.14 | 12 ± 2 | ... | N | ... | ... | ... | ... | ... | n | NG |
| 43628 | 07550440-6048392 | 23.9 ± 0.3 | 4949 ± 169 | 4.37 ± 0.16 | 0.14 ± 0.19 | 175 ± 5 | ... | Y | Y | Y | Y | Y | ... | Y | ... |
| 43960 | 07581825-6121091 | 24.1 ± 0.4 | 3869 ± 247 | 4.50 ± 0.13 | 0.03 ± 0.19 | ... | ... | Y | ... | ... | ... | Y | ... | n | ... |
| 43629 | 07550485-6043438 | 39.3 ± 0.2 | 4504 ± 170 | 2.42 ± 0.40 | 0.07 ± 0.26 | 24 ± 3 | ... | N | ... | ... | ... | ... | ... | n | G |
| 44146 | 08001450-6045154 | 22.0 ± 0.2 | 6131 ± 197 | 4.32 ± 0.22 | -0.21 ± 0.12 | 95 ± 1 | ... | Y | Y | Y | N | Y | Y | Y | ... |
| 43961 | 07581853-6019148 | 62.0 ± 11.7 | 4715 ± 204 | 3.32 ± 0.43 | 0.03 ± 0.15 | ... | ... | N | ... | ... | ... | N | ... | n | ... |
| 43661 | 07552347-6041512 | 23.4 ± 2.5 | ... | ... | ... | ... | ... | Y | ... | ... | ... | Y | ... | n | ... |
| 43962 | 07581885-6117219 | 99.7 ± 0.4 | 4225 ± 269 | 4.62 ± 0.33 | -0.11 ± 0.12 | 45 ± 10 | ... | N | ... | ... | ... | ... | ... | n | NG |
| 43963 | 07581918-6123523 | 24.4 ± 0.3 | 4387 ± 173 | 4.78 ± 0.23 | 0.01 ± 0.15 | 29 ± 5 | ... | Y | Y | Y | Y | Y | Y | Y | ... |
| 43964 | 07581986-6041581 | 25.2 ± 0.2 | 6166 ± 132 | 4.41 ± 0.53 | -0.32 ± 0.27 | 45 ± 2 | 1 | Y | N | N | N | Y | N | n | NG |
| 43662 | 07552448-6039006 | 62.7 ± 0.2 | 4875 ± 153 | 3.02 ± 0.24 | -0.13 ± 0.20 | 11 ± 2 | ... | N | ... | ... | ... | ... | ... | n | G |
| 43965 | 07581994-6043402 | 24.0 ± 0.2 | 4684 ± 250 | 4.63 ± 0.28 | 0.01 ± 0.13 | 20 ± 3 | ... | Y | Y | Y | Y | ... | Y | Y | ... |
| 43663 | 07552460-6054451 | 25.6 ± 0.2 | 4767 ± 228 | 2.70 ± 0.33 | -0.16 ± 0.23 | 15 ± 3 | ... | Y | N | N | Y | ... | ... | n | G |
| 43966 | 07582031-6041397 | 25.2 ± 0.2 | 6019 ± 146 | 4.35 ± 0.26 | -0.11 ± 0.14 | 132 ± 2 | ... | Y | Y | Y | Y | Y | Y | Y | ... |
| 44147 | 08001614-6030498 | 44.3 ± 0.2 | 4445 ± 187 | 2.43 ± 0.30 | 0.27 ± 0.30 | 11 ± 3 | ... | N | ... | ... | ... | ... | ... | n | ... |
| 44011 | 07584326-6055257 | 26.2 ± 0.3 | 5099 ± 410 | 4.04 ± 0.26 | -0.73 ± 1.02 | 204 ± 4 | ... | N | Y | Y | N | Y | Y | Y | ... |
| 44148 | 08001626-6022431 | 51.4 ± 0.2 | 5983 ± 145 | 4.41 ± 0.12 | 0.30 ± 0.16 | 39 ± 2 | ... | N | ... | ... | ... | ... | ... | n | NG |
| 43664 | 07552562-6033260 | 16.6 ± 0.2 | 5095 ± 122 | 3.47 ± 0.29 | -0.21 ± 0.17 | 3 ± 2 | ... | N | ... | ... | ... | ... | ... | n | G |
| 2934 | 07584349-6051425 | 68.7 ± 0.6 | 5444 ± 122 | 3.82 ± 0.24 | -0.24 ± 0.10 | <16 | 3 | N | ... | ... | ... | ... | ... | n | NG |
| 44149 | 08001647-6046151 | 24.7 ± 0.2 | 4310 ± 314 | 4.59 ± 0.20 | -0.77 ± 0.78 | 41 ± 4 | ... | Y | Y | Y | N | Y | Y | Y | ... |
| 43665 | 07552568-6036207 | 35.7 ± 1.4 | 3896 ± 521 | 4.73 ± 0.10 | 0.00 ± 0.12 | ... | ... | N | ... | ... | ... | N | ... | n | ... |
| 44012 | 07584375-6032568 | 19.4 ± 0.2 | 5547 ± 150 | 4.16 ± 0.21 | -0.50 ± 0.47 | 145 ± 2 | ... | N | Y | Y | N | Y | ... | Y | ... |
| 43666 | 07552681-6044348 | 24.9 ± 0.2 | 4925 ± 187 | 4.58 ± 0.35 | 0.00 ± 0.12 | 147 ± 4 | ... | Y | Y | Y | Y | Y | ... | Y | ... |
| 44013 | 07584415-6045374 | 22.1 ± 0.3 | 3813 ± 203 | 4.72 ± 0.12 | -0.02 ± 0.13 | ... | ... | Y | ... | ... | ... | Y | ... | n | ... |
| 44150 | 08001727-6051271 | -7.5 ± 0.2 | 5444 ± 185 | 4.34 ± 0.31 | 0.15 ± 0.14 | 8 ± 2 | ... | N | ... | ... | ... | ... | ... | n | ... |
| 43667 | 07552745-6033307 | 23.2 ± 0.3 | 3853 ± 198 | 4.58 ± 0.13 | 0.04 ± 0.20 | ... | ... | Y | ... | ... | ... | Y | ... | n | ... |
| 44014 | 07584416-6056011 | 24.1 ± 0.2 | 5198 ± 151 | 4.47 ± 0.44 | -0.02 ± 0.14 | 136 ± 4 | ... | Y | Y | Y | Y | Y | Y | Y | ... |
| 43668 | 07552751-6048332 | 23.5 ± 0.3 | 3703 ± 176 | 4.55 ± 0.18 | -0.02 ± 0.11 | ... | ... | Y | ... | ... | ... | Y | ... | n | ... |
| 44015 | 07584438-6033596 | 22.2 ± 0.6 | 3500 ± 39 | 5.11 ± 0.24 | ... | ... | ... | Y | ... | ... | ... | Y | ... | n | ... |
| 44151 | 08001818-6048173 | 24.0 ± 0.4 | 3673 ± 174 | 4.73 ± 0.12 | -0.03 ± 0.14 | ... | ... | Y | ... | ... | ... | Y | ... | n | ... |
| 43669 | 07552752-6019047 | 639.2 ± 339.4 | ... | ... | ... | ... | ... | N | ... | ... | ... | ... | ... | n | ... |
| 44016 | 07584489-6102357 | 24.3 ± 0.4 | 3754 ± 313 | 4.42 ± 0.22 | -0.04 ± 0.15 | ... | ... | Y | ... | ... | ... | Y | ... | n | ... |
| 44169 | 08003318-6031235 | 25.3 ± 0.2 | 6222 ± 198 | 4.25 ± 0.31 | -0.17 ± 0.23 | 70 ± 2 | 1 | Y | Y | Y | Y | Y | Y | Y | ... |
| 44017 | 07584632-6051457 | 73.1 ± 0.2 | 4770 ± 126 | 2.38 ± 0.33 | -0.16 ± 0.18 | 16 ± 3 | ... | N | ... | ... | ... | ... | ... | n | G |
| 43670 | 07552770-6048138 | 24.8 ± 0.4 | 3930 ± 356 | 4.59 ± 0.10 | 0.05 ± 0.21 | ... | ... | Y | ... | ... | ... | Y | ... | n | ... |
| 44018 | 07584646-6105588 | 26.6 ± 0.2 | 6716 ± 278 | 4.22 ± 0.39 | -0.08 ± 0.22 | 17 ± 2 | 1 | N | Y | Y | Y | Y | Y | Y | ... |
| 43671 | 07552817-6110296 | 22.9 ± 0.5 | 3573 ± 123 | 4.70 ± 0.16 | -0.07 ± 0.14 | ... | ... | Y | ... | ... | ... | Y | ... | n | ... |
| 44019 | 07584804-6054144 | 24.6 ± 0.3 | 4864 ± 495 | 3.84 ± 0.46 | -0.60 ± 0.87 | 209 ± 4 | ... | Y | Y | N | N | Y | Y | Y | ... |
| 44020 | 07584833-6106149 | 23.4 ± 0.2 | 6291 ± 246 | 4.35 ± 0.32 | -0.08 ± 0.16 | 85 ± 2 | 1 | Y | Y | Y | Y | Y | Y | Y | ... |
| 43672 | 07552858-6024560 | 17.8 ± 0.4 | 4100 ± 396 | 4.51 ± 0.10 | -0.94 ± 0.85 | 16 ± 6 | ... | N | Y | N | N | Y | ... | Y | ... |
| 44021 | 07584866-6055248 | 25.9 ± 0.2 | 6462 ± 440 | 4.54 ± 0.42 | -0.11 ± 0.30 | 41 ± 2 | 1 | Y | Y | Y | Y | Y | Y | Y | ... |
| 43673 | 07552876-6028198 | -3.4 ± 0.2 | 4244 ± 179 | 4.71 ± 0.18 | -0.05 ± 0.13 | 6 ± 2 | ... | N | ... | ... | ... | N | ... | n | ... |
| 44170 | 08003437-6042225 | 23.6 ± 0.2 | 5957 ± 129 | 4.42 ± 0.33 | -0.13 ± 0.15 | 116 ± 1 | ... | Y | Y | Y | Y | Y | Y | Y | ... |
| 44022 | 07584886-6047466 | 24.1 ± 0.5 | 3615 ± 230 | 4.60 ± 0.16 | -0.05 ± 0.12 | ... | ... | Y | ... | ... | ... | Y | ... | n | ... |
| 43674 | 07552900-6104407 | 23.1 ± 0.3 | 3901 ± 208 | 4.58 ± 0.10 | 0.07 ± 0.24 | 32 ± 6 | ... | Y | Y | Y | Y | Y | Y | Y | ... |
| 44023 | 07584895-6104049 | 58.7 ± 0.2 | 4725 ± 153 | 2.68 ± 0.33 | -0.06 ± 0.19 | ... | ... | N | ... | ... | ... | ... | ... | n | G |
| 44024 | 07584939-6028213 | 23.4 ± 0.3 | 3918 ± 249 | 4.60 ± 0.10 | 0.08 ± 0.24 | ... | ... | Y | ... | ... | ... | Y | ... | n | ... |
| 43675 | 07553025-6107468 | 78.4 ± 0.2 | 4674 ± 249 | 2.38 ± 0.44 | -0.43 ± 0.21 | 8 ± 2 | ... | N | ... | ... | ... | ... | ... | n | G |
| 44025 | 07585063-6106507 | 25.8 ± 0.2 | 5751 ± 82 | 4.50 ± 0.43 | -0.18 ± 0.15 | 126 ± 2 | ... | Y | Y | Y | Y | Y | Y | Y | ... |

**Table C.8.** continued.

| ID | CNAME | RV (km s$^{-1}$) | $T_{\rm eff}$ (K) | $\log g$ (dex) | [Fe/H] (dex) | $EW$(Li)$^a$ (mÅ) | $EW$(Li) error flag$^b$ | Membership | | | | Gaia studies | | Final$^d$ | NMs with Li$^e$ |
|---|---|---|---|---|---|---|---|---|---|---|---|---|---|---|---|
| | | | | | | | | RV | Li | $\log g$ | [Fe/H] | Randich$^c$ | Cantat-Gaudin$^c$ | | |
| 44026 | 07585090-6111444 | 23.3 ± 0.8 | 3541 ± 31 | 4.79 ± 0.22 | … | … | … | Y | … | … | … | … | … | n | … |
| 43676 | 07553120-6056508 | 23.9 ± 0.2 | 6442 ± 251 | 4.36 ± 0.31 | 0.05 ± 0.16 | 18 ± 1 | 1 | Y | Y | Y | Y | Y | Y | Y | … |
| 44027 | 07585108-6023472 | 32.7 ± 0.3 | 3596 ± 125 | 4.69 ± 0.13 | -0.07 ± 0.14 | … | … | N | … | … | … | N | … | n | … |
| 44028 | 07585176-6113408 | 21.0 ± 4.5 | … | … | … | … | … | N | … | … | … | Y | … | n | … |
| 44029 | 07585217-6123592 | 28.6 ± 2.2 | … | … | … | … | … | N | … | … | … | … | … | n | … |
| 44171 | 08003552-6048264 | 24.3 ± 0.2 | 4935 ± 184 | 4.54 ± 0.35 | -0.01 ± 0.12 | 124 ± 4 | … | Y | Y | Y | Y | Y | Y | Y | … |
| 44030 | 07585273-6102021 | 23.9 ± 0.5 | 3568 ± 22 | 4.93 ± 0.18 | … | … | … | Y | … | … | … | Y | … | n | … |
| 44172 | 08003560-6101214 | 25.8 ± 0.3 | 4576 ± 342 | 4.30 ± 0.33 | -0.72 ± 0.74 | 240 ± 4 | … | Y | Y | Y | N | Y | … | Y | … |
| 44031 | 07585290-6031150 | 102.3 ± 0.2 | 4692 ± 85 | 2.68 ± 0.32 | -0.05 ± 0.22 | 17 ± 3 | … | N | … | … | … | … | … | n | G |
| 44173 | 08003647-6105272 | 3.1 ± 0.3 | 5165 ± 77 | 3.26 ± 0.30 | -0.28 ± 0.15 | 21 ± 4 | … | N | … | … | … | Y | … | n | G |
| 43677 | 07553173-6034206 | 23.0 ± 0.4 | 3832 ± 197 | 4.45 ± 0.25 | 0.01 ± 0.13 | … | … | Y | … | … | … | Y | … | n | … |
| 44032 | 07585309-6037057 | 23.9 ± 0.3 | 3783 ± 200 | 4.65 ± 0.12 | -0.01 ± 0.13 | … | … | Y | … | … | … | Y | … | n | … |
| 44055 | 07590501-6056361 | 26.5 ± 0.2 | 7088 ± 515 | 4.11 ± 0.20 | 0.14 ± 0.15 | 51 ± 2 | 1 | N | Y | Y | Y | Y | Y | Y | … |
| 44174 | 08003704-6103078 | 23.5 ± 0.3 | 3804 ± 184 | 4.71 ± 0.12 | -0.02 ± 0.13 | 83 ± 6 | … | Y | Y | Y | Y | Y | … | Y | … |
| 43689 | 07554378-6037522 | 24.9 ± 0.3 | 6027 ± 202 | 4.18 ± 0.06 | -0.40 ± 0.30 | 82 ± 2 | … | Y | Y | Y | N | Y | Y | Y | … |
| 44056 | 07590677-6051533 | 23.3 ± 0.2 | 4454 ± 308 | 4.68 ± 0.17 | -0.02 ± 0.13 | 88 ± 4 | … | Y | Y | Y | Y | Y | Y | Y | … |
| 44057 | 07590678-6103172 | 61.9 ± 0.2 | 4437 ± 285 | 2.56 ± 0.58 | -0.01 ± 0.29 | 13 ± 3 | … | N | … | … | … | … | … | n | G |
| 43690 | 07554408-6021336 | 23.5 ± 0.4 | 3539 ± 26 | 4.42 ± 0.18 | … | … | … | Y | … | … | … | Y | … | n | … |
| 44058 | 07590681-6101230 | 22.3 ± 0.4 | 3649 ± 413 | 4.73 ± 0.19 | … | … | … | Y | … | … | … | Y | … | n | … |
| 44175 | 08003775-6036331 | 24.0 ± 0.2 | 4973 ± 261 | 4.39 ± 0.28 | -0.02 ± 0.14 | 208 ± 5 | … | Y | Y | Y | Y | Y | N | Y | … |
| 44059 | 07590793-6111287 | 26.2 ± 0.2 | 5004 ± 77 | 2.81 ± 0.30 | 0.03 ± 0.18 | 8 ± 4 | … | N | … | … | … | … | … | n | G |
| 44060 | 07590857-6032596 | 25.6 ± 0.3 | 3724 ± 184 | 4.60 ± 0.14 | -0.02 ± 0.13 | 33 ± 9 | … | Y | Y | Y | Y | Y | Y | Y | … |
| 44061 | 07590959-6050515 | 23.1 ± 0.7 | 3879 ± 595 | 4.43 ± 0.17 | 0.00 ± 0.20 | … | … | Y | … | … | … | Y | … | n | … |
| 43691 | 07554488-6044129 | 24.3 ± 0.6 | 3766 ± 321 | 4.85 ± 0.29 | 0.01 ± 0.11 | … | … | Y | … | … | … | Y | … | n | … |
| 44176 | 08003869-6106244 | 22.4 ± 0.9 | 3475 ± 32 | 4.60 ± 0.21 | … | … | … | Y | … | … | … | Y | … | n | … |
| 44062 | 07590961-6025210 | 21.0 ± 0.3 | 4629 ± 338 | 4.63 ± 0.34 | -0.37 ± 0.23 | 20 ± 4 | … | N | … | … | … | … | … | n | NG |
| 43692 | 07554498-6039236 | 57.5 ± 0.2 | 5036 ± 162 | 3.66 ± 0.40 | -0.43 ± 0.17 | 13 ± 3 | … | N | … | … | … | … | … | n | NG |
| 44063 | 07591010-6119178 | 23.5 ± 2.4 | … | … | … | … | … | Y | … | … | … | … | … | n | … |
| 44064 | 07591012-6112330 | 22.5 ± 0.3 | 4423 ± 393 | 4.92 ± 0.31 | -0.11 ± 0.15 | 3 ± 4 | … | Y | Y | Y | Y | Y | … | Y | … |
| 44065 | 07591034-6102557 | 23.7 ± 0.3 | 4074 ± 267 | 4.86 ± 0.31 | 0.02 ± 0.20 | 20 ± 5 | … | Y | Y | Y | Y | Y | … | Y | … |
| 43693 | 07554638-6022590 | 24.3 ± 0.9 | 3697 ± 356 | 4.32 ± 0.21 | … | … | … | Y | … | … | … | Y | … | n | … |
| 43694 | 07554732-6059044 | 62.5 ± 0.2 | 4785 ± 137 | 2.58 ± 0.30 | -0.05 ± 0.20 | 3 ± 2 | … | N | … | … | … | … | … | n | G |
| 43695 | 07554739-6021259 | 14.2 ± 0.3 | 4466 ± 260 | 4.31 ± 0.37 | 0.14 ± 0.17 | … | … | N | … | … | … | N | … | n | … |
| 44066 | 07591146-6049056 | 24.5 ± 0.2 | 4954 ± 192 | 4.38 ± 0.44 | 0.08 ± 0.13 | 95 ± 4 | … | Y | Y | Y | Y | Y | Y | Y | … |
| 43696 | 07554768-6023434 | 23.9 ± 0.7 | 3532 ± 28 | 4.73 ± 0.20 | … | … | … | Y | … | … | … | Y | … | n | … |
| 44067 | 07591276-6026144 | 24.1 ± 0.3 | 3864 ± 281 | 4.63 ± 0.11 | -0.01 ± 0.19 | 67 ± 7 | … | Y | Y | Y | Y | Y | Y | Y | … |
| 44183 | 08005054-6020039 | 12.4 ± 0.3 | 4524 ± 316 | 4.76 ± 0.18 | -0.31 ± 0.21 | … | … | N | … | … | … | N | … | n | … |
| 43697 | 07554811-6107145 | 7.6 ± 0.2 | 5043 ± 67 | 4.49 ± 0.47 | 0.04 ± 0.13 | 11 ± 2 | … | N | … | … | … | … | … | n | NG |
| 44068 | 07591289-6119100 | 24.0 ± 1.0 | … | … | … | … | … | Y | … | … | … | … | … | n | … |
| 44184 | 08005057-6059216 | 22.0 ± 0.3 | 3791 ± 243 | 4.50 ± 0.18 | 0.01 ± 0.13 | … | … | Y | … | … | … | Y | … | n | … |
| 44069 | 07591341-6040239 | 22.1 ± 0.2 | 5722 ± 165 | 4.48 ± 0.34 | -0.13 ± 0.19 | 153 ± 3 | … | Y | Y | Y | Y | Y | Y | Y | … |
| 44185 | 08005092-6055548 | -0.5 ± 0.2 | 4332 ± 241 | 4.67 ± 0.30 | -0.04 ± 0.13 | 12 ± 3 | … | N | … | … | … | N | … | n | NG |
| 43698 | 07554835-6107229 | -5.0 ± 0.2 | 4549 ± 87 | 4.84 ± 0.30 | 0.04 ± 0.15 | … | … | N | … | … | … | N | … | n | … |
| 44070 | 07591373-6054292 | 42.0 ± 0.2 | 6017 ± 139 | 4.28 ± 0.21 | -0.15 ± 0.13 | 56 ± 2 | … | N | … | … | … | N | … | n | NG |
| 43699 | 07554907-6023479 | 20.1 ± 1.5 | 4174 ± 376 | 4.66 ± 0.16 | 0.19 ± 0.26 | … | … | N | … | … | … | Y | … | n | … |
| 44071 | 07591559-6041200 | 24.1 ± 0.3 | 4061 ± 327 | 4.60 ± 0.12 | 0.08 ± 0.19 | 26 ± 5 | … | Y | Y | Y | Y | Y | N | Y | … |
| 44186 | 08005131-6046322 | 24.8 ± 0.3 | 3927 ± 220 | 4.59 ± 0.09 | 0.03 ± 0.19 | 26 ± 5 | … | Y | Y | Y | Y | Y | Y | Y | … |
| 44072 | 07591661-6121062 | 17.2 ± 1.8 | … | … | … | … | … | N | … | … | … | … | … | n | … |
| 44187 | 08005189-6046598 | 24.2 ± 0.2 | 5749 ± 76 | 4.42 ± 0.33 | -0.08 ± 0.13 | 132 ± 2 | … | Y | Y | Y | Y | Y | Y | Y | … |
| 44073 | 07591703-6040341 | 97.5 ± 0.2 | 4878 ± 166 | 3.07 ± 0.33 | -0.24 ± 0.28 | … | … | N | … | … | … | … | … | n | G |
| 44074 | 07591704-6034133 | 40.8 ± 0.2 | 5991 ± 134 | 4.16 ± 0.12 | 0.09 ± 0.14 | 111 ± 1 | … | N | … | … | … | N | … | n | NG |
| 43716 | 07560100-6037242 | 24.5 ± 0.2 | 5115 ± 96 | 4.49 ± 0.48 | 0.01 ± 0.12 | 88 ± 4 | … | Y | Y | Y | Y | Y | Y | Y | … |
| 44188 | 08005479-6024058 | 20.0 ± 0.4 | 3757 ± 262 | 4.58 ± 0.16 | 0.01 ± 0.13 | … | … | N | … | … | … | Y | … | n | … |
| 43717 | 07560125-6057268 | 23.8 ± 0.3 | 6050 ± 206 | 4.10 ± 0.11 | -0.43 ± 0.42 | 139 ± 4 | … | Y | Y | Y | N | Y | Y | Y | … |
| 44189 | 08005481-6104556 | 43.3 ± 0.2 | 4609 ± 176 | 4.74 ± 0.26 | 0.11 ± 0.16 | … | … | N | … | … | … | Y | … | n | … |
| 43718 | 07560174-6037388 | 23.0 ± 0.4 | 3724 ± 255 | 4.60 ± 0.14 | -0.05 ± 0.12 | … | … | Y | … | … | … | Y | … | n | … |
| 43719 | 07560181-6045067 | 43.5 ± 0.2 | 4679 ± 138 | 2.84 ± 0.38 | 0.20 ± 0.25 | 2 ± 3 | … | N | … | … | … | … | … | n | G |
| 44190 | 08005558-6045409 | 22.3 ± 0.3 | 4826 ± 303 | 3.98 ± 0.35 | -0.62 ± 0.64 | 235 ± 4 | … | Y | Y | N | N | Y | Y | Y | … |
| 44097 | 07593662-6111475 | 21.3 ± 0.2 | 5659 ± 61 | 4.34 ± 0.18 | 0.20 ± 0.13 | … | … | Y | … | … | … | … | … | n | … |

**Table C.8.** continued.

| ID | CNAME | RV (km s$^{-1}$) | $T_{\rm eff}$ (K) | $\log g$ (dex) | [Fe/H] (dex) | $EW$(Li)$^a$ (mÅ) | $EW$(Li) error flag$^b$ | RV | Li | Membership logg | [Fe/H] | Randich$^c$ | Gaia studies Cantat-Gaudin$^c$ | Final$^d$ | NMs with Li$^e$ |
|---|---|---|---|---|---|---|---|---|---|---|---|---|---|---|---|
| 43720 | 07560191-6024375 | 22.2 ± 0.2 | 4890 ± 189 | 2.97 ± 0.35 | -0.06 ± 0.20 | 17 ± 3 | ... | Y | N | N | Y | ... | ... | n | G |
| 44191 | 08005606-6117247 | 1.1 ± 0.4 | 3572 ± 22 | 4.45 ± 0.16 | ... | ... | ... | N | ... | ... | ... | N | ... | n | ... |
| 43721 | 07560192-6109023 | 23.7 ± 1.0 | 3781 ± 288 | 4.61 ± 0.13 | -0.08 ± 0.17 | ... | ... | Y | ... | ... | ... | Y | ... | n | ... |
| 44192 | 08005722-6048575 | 20.8 ± 0.3 | 3956 ± 289 | 4.64 ± 0.23 | 0.09 ± 0.28 | 28 ± 7 | ... | N | Y | Y | Y | Y | ... | Y | ... |
| 43722 | 07560223-6020505 | 23.6 ± 0.3 | 4253 ± 470 | 4.26 ± 0.32 | 0.12 ± 0.22 | 116 ± 6 | ... | Y | Y | Y | Y | Y | ... | Y | ... |
| 44098 | 07593776-6019161 | -7.0 ± 0.2 | 4959 ± 124 | 3.04 ± 0.31 | -0.10 ± 0.15 | 11 ± 3 | ... | N | ... | ... | ... | ... | ... | n | G |
| 44193 | 08005781-6033329 | 24.4 ± 0.3 | 3919 ± 243 | 4.53 ± 0.11 | 0.02 ± 0.19 | ... | ... | Y | ... | ... | ... | Y | ... | n | ... |
| 43723 | 07560226-6038459 | 23.8 ± 0.2 | 5565 ± 91 | 4.47 ± 0.39 | 0.04 ± 0.15 | 155 ± 3 | ... | Y | Y | Y | Y | Y | Y | Y | ... |
| 44194 | 08005862-6020078 | 16.4 ± 0.2 | 5306 ± 112 | 4.58 ± 0.40 | 0.12 ± 0.13 | 5 ± 3 | ... | N | ... | ... | ... | ... | ... | n | ... |
| 44099 | 07593812-6112078 | 70.7 ± 0.2 | 4702 ± 201 | 2.97 ± 0.34 | 0.04 ± 0.24 | ... | ... | N | ... | ... | ... | ... | ... | n | G |
| 44195 | 08005917-6106242 | 24.3 ± 2.5 | 3844 ± 461 | 4.76 ± 0.10 | 0.02 ± 0.11 | ... | ... | Y | ... | ... | ... | Y | ... | n | ... |
| 44100 | 07593818-6108146 | 66.7 ± 0.2 | 4782 ± 198 | 2.30 ± 0.60 | -0.19 ± 0.17 | 7 ± 2 | ... | N | ... | ... | ... | ... | ... | n | G |
| 44101 | 07593913-6045373 | 19.4 ± 3.9 | 3454 ± 33 | 4.61 ± 0.20 | ... | ... | ... | N | ... | ... | ... | Y | ... | n | ... |
| 44240 | 08022910-6044507 | 10.7 ± 0.3 | 3654 ± 189 | 4.63 ± 0.14 | -0.03 ± 0.14 | 42 ± 8 | ... | N | ... | ... | ... | N | ... | n | NG |
| 43724 | 07560266-6028101 | 131.5 ± 0.2 | 4790 ± 129 | 2.64 ± 0.30 | -0.16 ± 0.23 | 15 ± 2 | ... | N | ... | ... | ... | ... | ... | n | G |
| 44241 | 08023471-6046046 | 23.4 ± 0.4 | 3435 ± 12 | 4.44 ± 0.14 | -0.11 ± 0.13 | ... | ... | Y | ... | ... | ... | Y | ... | n | ... |
| 43725 | 07560307-6102051 | -1.4 ± 0.2 | 4081 ± 316 | 4.92 ± 0.35 | -0.43 ± 0.33 | 10 ± 3 | ... | N | ... | ... | ... | N | ... | n | NG |
| 44242 | 08023533-6040197 | 23.5 ± 0.3 | 3997 ± 305 | 4.76 ± 0.28 | 0.00 ± 0.20 | 16 ± 5 | ... | Y | Y | Y | Y | Y | Y | Y | ... |
| 44102 | 07594001-6042311 | 28.3 ± 0.2 | 4794 ± 373 | 4.05 ± 0.14 | -0.36 ± 0.36 | 215 ± 4 | ... | N | Y | Y | N | Y | ... | Y | ... |
| 44243 | 08023898-6057231 | 24.8 ± 0.6 | 3485 ± 29 | 4.54 ± 0.18 | ... | ... | ... | Y | ... | ... | ... | Y | ... | n | ... |
| 44244 | 08024318-6053051 | 24.0 ± 0.5 | 3696 ± 726 | 4.74 ± 0.16 | -0.95 ± 0.06 | ... | ... | Y | ... | ... | ... | Y | ... | n | ... |
| 2942 | 07594121-6109251 | -19.9 ± 0.6 | 5861 ± 134 | 3.99 ± 0.20 | -0.06 ± 0.14 | 61 ± 2 | 1 | N | ... | ... | ... | N | ... | n | NG |
| 44245 | 08024838-6047096 | 24.9 ± 0.4 | 3638 ± 140 | 4.73 ± 0.13 | -0.06 ± 0.12 | 250 ± 9 | ... | Y | Y | Y | Y | Y | N | Y | ... |
| 44246 | 08024877-6052010 | 24.8 ± 0.6 | 3759 ± 216 | 4.70 ± 0.11 | -0.03 ± 0.14 | ... | ... | Y | ... | ... | ... | Y | ... | n | ... |
| 44247 | 08025517-6039245 | 25.4 ± 0.3 | 4034 ± 431 | 4.61 ± 0.13 | 0.06 ± 0.20 | 28 ± 6 | ... | Y | Y | Y | Y | Y | Y | Y | ... |
| 44248 | 08031131-6038165 | 475.7 ± 7.0 | ... | ... | ... | ... | ... | N | ... | ... | ... | ... | ... | n | ... |
| 43726 | 07560365-6020523 | -3.5 ± 0.2 | 5904 ± 72 | 3.99 ± 0.10 | 0.04 ± 0.13 | 45 ± 2 | ... | N | ... | ... | ... | ... | ... | n | NG |
| 44249 | 08031783-6049582 | 23.5 ± 0.3 | 4030 ± 343 | 4.94 ± 0.35 | -0.20 ± 0.33 | 27 ± 5 | ... | Y | Y | Y | N | Y | Y | Y | ... |
| 44103 | 07594214-6046000 | 23.5 ± 0.2 | 4979 ± 147 | 4.82 ± 0.39 | 0.02 ± 0.12 | 8 ± 3 | ... | Y | N | Y | Y | ... | ... | n | NG |
| 44250 | 08033219-6044262 | 23.3 ± 1.0 | ... | ... | ... | ... | ... | Y | ... | ... | ... | Y | ... | n | ... |
| 43743 | 07561423-6058156 | 25.4 ± 0.3 | 4090 ± 284 | 4.92 ± 0.42 | -0.13 ± 0.21 | 8 ± 3 | ... | Y | Y | Y | Y | Y | Y | Y | ... |
| 44111 | 07594768-6101212 | 12.3 ± 0.2 | 5014 ± 109 | 3.47 ± 0.14 | -0.16 ± 0.16 | 13 ± 3 | ... | N | ... | ... | ... | ... | ... | n | G |
| 44251 | 08034358-6051252 | 32.7 ± 0.4 | 3374 ± 38 | 3.95 ± 0.24 | ... | ... | ... | N | ... | ... | ... | N | ... | n | ... |
| 43744 | 07561430-6109320 | 42.3 ± 1.2 | 4465 ± 312 | 2.92 ± 1.12 | 0.15 ± 0.22 | ... | ... | N | ... | ... | ... | N | ... | n | ... |
| 44112 | 07594824-6105200 | 23.9 ± 0.2 | 5721 ± 129 | 4.50 ± 0.31 | -0.07 ± 0.14 | 114 ± 2 | ... | Y | Y | Y | Y | Y | Y | Y | ... |
| 43745 | 07561448-6054314 | 21.4 ± 0.5 | 3960 ± 427 | 4.52 ± 0.11 | 0.08 ± 0.29 | ... | ... | Y | ... | ... | ... | Y | ... | n | ... |
| 43746 | 07561491-6038481 | 24.6 ± 0.4 | 3602 ± 19 | 4.45 ± 0.16 | ... | ... | ... | Y | ... | ... | ... | Y | ... | n | ... |
| 43747 | 07561521-6023267 | 26.0 ± 1.4 | 3398 ± 36 | 4.93 ± 0.22 | ... | ... | ... | N | ... | ... | ... | Y | ... | n | ... |
| 43748 | 07561522-6038219 | 28.0 ± 1.6 | 3849 ± 285 | 4.55 ± 0.10 | -0.15 ± 0.25 | ... | ... | N | ... | ... | ... | Y | ... | n | ... |
| 43749 | 07561544-6053084 | 22.9 ± 0.4 | 3873 ± 205 | 4.68 ± 0.14 | -0.01 ± 0.20 | ... | ... | Y | ... | ... | ... | Y | ... | n | ... |
| 43750 | 07561556-6047204 | 23.3 ± 0.3 | 3972 ± 241 | 4.77 ± 0.23 | -0.01 ± 0.19 | 9 ± 4 | ... | Y | Y | Y | Y | Y | ... | Y | ... |
| 44113 | 07595015-6043342 | 23.4 ± 0.2 | 4927 ± 206 | 4.44 ± 0.25 | -0.04 ± 0.13 | 117 ± 3 | ... | Y | Y | Y | Y | Y | ... | Y | ... |
| 43751 | 07561707-6022188 | 24.8 ± 0.3 | 4197 ± 484 | 4.34 ± 0.27 | 0.15 ± 0.23 | 28 ± 5 | ... | Y | Y | Y | Y | Y | ... | Y | ... |
| 44114 | 07595031-6044149 | 24.6 ± 0.2 | 5122 ± 89 | 4.54 ± 0.37 | 0.01 ± 0.12 | 89 ± 3 | ... | Y | Y | Y | Y | Y | Y | Y | ... |
| 43752 | 07561890-6055298 | 6.9 ± 0.3 | 4147 ± 340 | 4.77 ± 0.28 | -0.14 ± 0.17 | 10 ± 4 | ... | N | ... | ... | ... | N | ... | n | NG |
| 43753 | 07561895-6056535 | 24.3 ± 0.2 | 6269 ± 151 | 4.33 ± 0.26 | -0.03 ± 0.12 | 84 ± 2 | 1 | Y | Y | Y | Y | Y | ... | Y | ... |
| 2915 | 07562154-6058089 | 25.9 ± 0.6 | 5118 ± 126 | 4.50 ± 0.23 | -0.08 ± 0.09 | 98 ± 5 | ... | Y | Y | Y | Y | Y | ... | Y | ... |
| 44115 | 07595144-6108016 | 22.1 ± 0.3 | 3655 ± 168 | 4.54 ± 0.21 | -0.05 ± 0.12 | ... | ... | Y | ... | ... | ... | Y | ... | n | ... |
| 43754 | 07562200-6056340 | 21.7 ± 0.2 | 5568 ± 59 | 4.26 ± 0.38 | -0.37 ± 0.27 | 131 ± 3 | ... | Y | Y | Y | N | Y | N | Y | ... |
| 43755 | 07562272-6051423 | 24.7 ± 0.3 | 6690 ± 513 | 3.98 ± 0.09 | 0.10 ± 0.20 | 29 ± 1 | 1 | Y | Y | Y | Y | Y | Y | Y | ... |
| 44116 | 07595171-6022095 | 24.1 ± 0.2 | 6284 ± 271 | 4.37 ± 0.23 | -0.07 ± 0.14 | 92 ± 2 | 1 | Y | Y | Y | Y | Y | Y | Y | ... |
| 43756 | 07562277-6112359 | 507.1 ± 4.5 | ... | ... | ... | ... | ... | N | ... | ... | ... | ... | ... | n | ... |
| 44117 | 07595202-6029134 | 16.9 ± 0.9 | 3781 ± 262 | 4.82 ± 0.25 | -0.09 ± 0.14 | 47 ± 7 | ... | N | Y | Y | Y | Y | N | Y | ... |
| 43757 | 07562434-6053561 | 50.5 ± 0.3 | 4876 ± 181 | 2.95 ± 0.47 | -0.13 ± 0.20 | 10 ± 3 | ... | N | ... | ... | ... | ... | ... | n | G |
| 44118 | 07595210-6028433 | 3.2 ± 1.1 | 3533 ± 26 | 4.87 ± 0.18 | ... | ... | ... | N | ... | ... | ... | N | ... | n | ... |
| 43758 | 07562439-6112084 | 7.6 ± 0.5 | 4758 ± 280 | 3.22 ± 0.20 | 0.00 ± 0.25 | ... | ... | N | ... | ... | ... | ... | ... | n | G |
| 44119 | 07595247-6102008 | 24.3 ± 0.3 | 3990 ± 184 | 4.83 ± 0.28 | -0.02 ± 0.21 | 14 ± 5 | ... | Y | Y | Y | Y | Y | Y | Y | ... |
| 43759 | 07562494-6121393 | 23.3 ± 2.4 | 5049 ± 238 | 3.93 ± 0.55 | -0.32 ± 0.20 | ... | ... | Y | ... | ... | ... | Y | ... | n | ... |
| 44120 | 07595262-6041159 | 24.0 ± 0.2 | 6353 ± 299 | 4.66 ± 0.51 | -0.07 ± 0.36 | 103 ± 2 | 1 | Y | Y | Y | Y | Y | Y | Y | ... |

**Table C.8.** continued.

| ID | CNAME | RV (km s$^{-1}$) | $T_{\text{eff}}$ (K) | logg (dex) | [Fe/H] (dex) | EW(Li)$^a$ (mÅ) | EW(Li) error flag$^b$ | Membership RV | Li | logg | [Fe/H] | Gaia studies Randich$^c$ | Cantat-Gaudin$^c$ | Final$^d$ | NMs with Li$^e$ |
|---|---|---|---|---|---|---|---|---|---|---|---|---|---|---|---|
| 43760 | 07562550-6045198 | 50.5 ± 0.2 | 5112 ± 101 | 2.91 ± 0.38 | -0.01 ± 0.16 | 8 ± 2 | … | N | … | … | … | … | … | n | G |
| 44137 | 08000304-6037243 | 25.8 ± 0.2 | 6863 ± 726 | 4.33 ± 0.28 | -0.08 ± 0.32 | 66 ± 1 | 1 | Y | Y | Y | Y | Y | Y | Y | … |
| 43761 | 07562557-6102378 | 23.4 ± 0.3 | 3733 ± 226 | 4.65 ± 0.12 | -0.02 ± 0.13 | … | … | Y | … | … | … | Y | … | n | … |
| 43762 | 07562639-6034245 | 23.4 ± 1.8 | … | … | … | … | … | Y | … | … | … | Y | … | n | … |
| 43791 | 07564760-6104373 | 23.9 ± 0.2 | 4738 ± 240 | 4.71 ± 0.27 | -0.02 ± 0.12 | 22 ± 3 | … | Y | Y | Y | Y | … | Y | Y | … |
| 43792 | 07564817-6046311 | 25.1 ± 0.2 | 4201 ± 367 | 4.20 ± 0.40 | 0.09 ± 0.20 | … | … | Y | … | … | … | Y | … | n | … |
| 43793 | 07564825-6051086 | 21.5 ± 0.4 | 3727 ± 219 | 4.66 ± 0.11 | -0.02 ± 0.11 | … | … | Y | … | … | … | Y | … | n | … |
| 2918 | 07564930-6030483 | 60.7 ± 0.6 | 5332 ± 141 | 3.90 ± 0.29 | -0.15 ± 0.12 | <35 | 3 | N | … | … | … | … | … | n | NG |
| 44138 | 08000488-6045194 | 29.1 ± 0.3 | 6507 ± 27 | … | 0.15 ± 0.02 | … | … | N | … | … | … | Y | … | n | … |
| 43794 | 07564935-6059527 | 26.2 ± 0.2 | 6340 ± 163 | 4.09 ± 0.17 | -0.24 ± 0.12 | 2 ± 1 | 1 | N | … | … | … | … | … | n | … |
| 43795 | 07564984-6041278 | 19.6 ± 0.2 | 4464 ± 136 | 4.66 ± 0.29 | 0.10 ± 0.17 | 2 ± 3 | … | N | Y | Y | Y | Y | … | Y | … |
| 43796 | 07565000-6039458 | 23.5 ± 0.2 | 5282 ± 157 | 4.54 ± 0.41 | 0.01 ± 0.15 | 117 ± 4 | … | Y | Y | Y | Y | Y | Y | Y | … |
| 43797 | 07565055-6057370 | 48.7 ± 0.8 | 3471 ± 33 | 3.95 ± 0.21 | … | … | … | N | … | … | … | N | … | n | … |
| 44139 | 08000541-6022527 | 25.5 ± 0.2 | 4682 ± 291 | 4.63 ± 0.30 | -0.10 ± 0.16 | 26 ± 3 | … | Y | Y | Y | Y | … | Y | Y | … |
| 43798 | 07565111-6111276 | 49.4 ± 0.2 | 4579 ± 176 | 2.93 ± 0.47 | 0.30 ± 0.20 | 22 ± 4 | … | N | … | … | … | … | … | n | … |
| 43799 | 07565124-6030259 | -460.8 ± 790.0 | … | … | … | … | … | N | … | … | … | … | … | n | … |
| 43800 | 07565168-6036139 | 33.8 ± 3.4 | 4084 ± 527 | 4.67 ± 0.13 | 0.01 ± 0.11 | … | … | N | … | … | … | Y | … | n | … |
| 44140 | 08000635-6042333 | -5.3 ± 0.2 | 4918 ± 149 | 2.86 ± 0.28 | -0.11 ± 0.15 | 14 ± 2 | … | N | … | … | … | … | … | n | G |
| 43801 | 07565270-6047409 | 95.0 ± 0.2 | 4895 ± 51 | 2.61 ± 0.29 | 0.05 ± 0.22 | 18 ± 2 | … | N | … | … | … | … | … | n | G |
| 43802 | 07565408-6050056 | 23.4 ± 0.2 | 4802 ± 237 | 4.56 ± 0.22 | -0.06 ± 0.15 | 148 ± 3 | … | Y | Y | Y | Y | Y | Y | Y | … |
| 43803 | 07565502-6028512 | -16.1 ± 0.4 | 4061 ± 263 | 4.61 ± 0.11 | -0.58 ± 0.37 | … | … | N | … | … | … | N | … | n | … |
| 43804 | 07565520-6058118 | 22.3 ± 0.3 | 3965 ± 341 | 4.57 ± 0.09 | 0.03 ± 0.27 | 53 ± 6 | … | Y | Y | Y | Y | Y | Y | Y | … |
| 43805 | 07565554-6039300 | 28.7 ± 2.0 | 3851 ± 335 | 4.93 ± 0.38 | 0.00 ± 0.11 | … | … | N | … | … | … | Y | … | n | … |
| 43806 | 07565600-6019424 | 24.8 ± 0.2 | 5191 ± 61 | 4.48 ± 0.44 | 0.10 ± 0.13 | 159 ± 4 | … | Y | Y | Y | Y | Y | Y | Y | … |
| 43807 | 07565636-6100575 | -10.2 ± 0.2 | 4292 ± 169 | 4.67 ± 0.17 | 0.19 ± 0.33 | … | … | N | … | … | … | N | … | n | … |
| 43808 | 07565692-6123217 | 19.7 ± 0.7 | 4170 ± 163 | 5.00 ± 0.15 | -0.44 ± 0.46 | … | … | N | … | … | … | Y | … | n | … |
| 43809 | 07565746-6017559 | 25.3 ± 0.2 | 6172 ± 131 | 4.18 ± 0.11 | -0.33 ± 0.37 | 115 ± 3 | 1 | Y | Y | Y | N | Y | Y | Y | … |
| 43810 | 07565956-6049567 | 24.5 ± 0.4 | 4110 ± 475 | 4.25 ± 0.38 | 0.13 ± 0.26 | 52 ± 5 | … | Y | Y | Y | Y | Y | Y | Y | … |
| 43811 | 07565986-6106322 | 24.2 ± 0.3 | 3970 ± 390 | 4.47 ± 0.14 | -0.07 ± 0.22 | 52 ± 6 | … | Y | Y | Y | Y | Y | Y | Y | … |
| 43835 | 07571030-6044527 | 71.3 ± 0.2 | 4797 ± 206 | 2.79 ± 0.29 | -0.18 ± 0.23 | 12 ± 3 | … | N | … | … | … | … | … | n | G |
| 43836 | 07571035-6101134 | 22.0 ± 1.0 | 3564 ± 24 | 4.79 ± 0.19 | … | … | … | Y | … | … | … | Y | … | n | … |
| 43837 | 07571050-6039362 | 27.6 ± 0.2 | 5304 ± 141 | 4.48 ± 0.31 | -0.12 ± 0.14 | 113 ± 2 | … | N | Y | Y | Y | Y | … | Y | … |

**Notes.** $^{(a)}$ The values of EW(Li) for this cluster are corrected (subtracted adjacent Fe (6707.43 Å) line). $^{(b)}$ Flags for the errors of the corrected EW(Li) values, as follows: 1=EW(Li) corrected by blends contribution using models; and 3=Upper limit (no error for EW(Li) is given). $^{(c)}$ Randich et al. (2018), Cantat-Gaudin et al. (2018). $^{(d)}$ The letters "Y" and "N" indicate if the star is a cluster member or not. $^{(e)}$ 'Li-rich G', 'G' and 'NG' indicate "Li-rich gaint", "giant" and "non-giant" Li field contaminants, respectively.

**Table C.9.** NGC 6705

| ID | CNAME | RV (km s$^{-1}$) | $T_{\rm eff}$ (K) | $\log g$ (dex) | [Fe/H] (dex) | EW(Li)$^a$ (mÅ) | EW(Li) error flag$^b$ | Membership RV | Li | $\log g$ | [Fe/H] | Gaia study Cantat-Gaudin$^c$ | Final$^d$ | NMs with Li$^e$ |
|---|---|---|---|---|---|---|---|---|---|---|---|---|---|---|
| 3230 | 18502831-0615122 | 34.5 ± 0.6 | 4557 ± 125 | 1.83 ± 0.24 | 0.17 ± 0.12 | 96 ± 3 | ... | Y | Y | Y | Y | ... | Y | ... |
| 3231 | 18503724-0614364 | 36.5 ± 0.6 | 4860 ± 116 | 2.43 ± 0.24 | 0.02 ± 0.10 | 109 ± 1 | ... | Y | Y | Y | Y | ... | Y | ... |
| 3232 | 18504563-0612038 | 35.1 ± 0.6 | 4674 ± 126 | 2.09 ± 0.23 | 0.06 ± 0.10 | 79 ± 2 | ... | Y | Y | Y | Y | ... | Y | ... |
| 3233 | 18504737-0617184 | 32.9 ± 0.6 | 4274 ± 121 | 1.58 ± 0.23 | 0.04 ± 0.10 | 65 ± 1 | ... | Y | Y | Y | Y | ... | Y | ... |
| 3234 | 18505494-0616182 | 35.8 ± 0.6 | 4667 ± 118 | 2.22 ± 0.24 | 0.07 ± 0.10 | 60 ± 2 | ... | Y | Y | Y | Y | ... | Y | ... |
| 3236 | 18505551-0618317 | 27.4 ± 0.6 | ... | ... | ... | ... | ... | N | ... | ... | ... | ... | n | ... |
| 3237 | 18505581-0618148 | 36.1 ± 0.6 | 4473 ± 128 | 1.89 ± 0.24 | 0.16 ± 0.11 | 102 ± 2 | ... | Y | Y | Y | Y | ... | Y | ... |
| 3238 | 18505755-0613461 | 31.6 ± 0.6 | 4860 ± 120 | 2.58 ± 0.23 | 0.02 ± 0.11 | 102 ± 1 | ... | Y | Y | Y | Y | ... | Y | ... |
| 3239 | 18505884-0614409 | 33.3 ± 0.5 | ... | ... | ... | 67 ± 36 | 1 | Y | N | ... | ... | Y | n | ... |
| 3240 | 18505944-0612435 | 35.5 ± 0.6 | 4888 ± 120 | 2.45 ± 0.23 | -0.02 ± 0.09 | 69 ± 1 | ... | Y | Y | Y | Y | ... | Y | ... |
| 3241 | 18505976-0616255 | 34.1 ± 1.0 | ... | ... | ... | ... | ... | Y | ... | ... | ... | ... | n | ... |
| 3243 | 18505998-0617359 | 19.9 ± 0.1 | ... | ... | ... | ... | ... | N | ... | ... | ... | ... | n | ... |
| 3244 | 18510023-0616594 | 36.3 ± 0.6 | 4289 ± 153 | 1.47 ± 0.35 | 0.11 ± 0.11 | 140 ± 3 | ... | Y | Y | Y | Y | ... | Y | ... |
| 3246 | 18510032-0617183 | 36.5 ± 0.6 | 4857 ± 120 | 2.31 ± 0.24 | 0.07 ± 0.10 | 59 ± 1 | ... | Y | Y | Y | Y | ... | Y | ... |
| 3247 | 18510093-0614564 | 43.0 ± 0.6 | 4709 ± 121 | 2.20 ± 0.24 | -0.02 ± 0.10 | 18 ± 1 | ... | N | ... | ... | ... | ... | n | G |
| 3248 | 18510200-0617265 | 33.2 ± 0.6 | 4283 ± 121 | 1.62 ± 0.23 | 0.04 ± 0.10 | 15 ± 2 | ... | Y | Y | Y | Y | ... | Y | ... |
| 3249 | 18510289-0615301 | 34.0 ± 0.6 | 4728 ± 117 | 2.27 ± 0.23 | 0.03 ± 0.09 | 79 ± 4 | ... | Y | Y | Y | Y | ... | Y | ... |
| 3250 | 18510304-0613595 | 38.0 ± 0.6 | ... | ... | ... | ... | ... | Y | ... | ... | ... | ... | n | ... |
| 3251 | 18510341-0616202 | 37.3 ± 0.6 | 4781 ± 119 | 2.21 ± 0.23 | 0.01 ± 0.10 | 60 ± 1 | ... | Y | Y | Y | Y | ... | Y | ... |
| 3252 | 18510358-0616112 | 36.6 ± 0.6 | 4851 ± 119 | 2.29 ± 0.21 | 0.10 ± 0.09 | 57 ± 1 | ... | Y | Y | Y | Y | ... | Y | ... |
| 3253 | 18510399-0620414 | 34.0 ± 0.6 | 4634 ± 112 | 2.08 ± 0.24 | 0.10 ± 0.10 | 60 ± 1 | ... | Y | Y | Y | Y | ... | Y | ... |
| 3257 | 18510662-0612442 | 33.2 ± 0.6 | 4751 ± 120 | 2.17 ± 0.23 | 0.13 ± 0.10 | 71 ± 2 | ... | Y | Y | Y | Y | ... | Y | ... |
| 3260 | 18510763-0614195 | 24.8 ± 0.9 | ... | ... | ... | ... | ... | N | ... | ... | ... | ... | n | ... |
| 3261 | 18510786-0617119 | 34.7 ± 0.6 | 4751 ± 126 | 2.29 ± 0.26 | 0.14 ± 0.11 | 67 ± 1 | ... | Y | Y | Y | Y | ... | Y | ... |
| 3262 | 18510812-0616017 | 88.3 ± 0.2 | ... | ... | ... | ... | ... | N | ... | ... | ... | ... | n | ... |
| 3263 | 18510833-0616532 | 34.3 ± 0.6 | 4834 ± 133 | 2.17 ± 0.31 | 0.17 ± 0.10 | 50 ± 3 | ... | Y | Y | Y | Y | ... | Y | ... |
| 3264 | 18510837-0617495 | -70.9 ± 0.6 | 4153 ± 112 | 1.39 ± 0.23 | -0.20 ± 0.09 | 3 ± 0 | ... | N | ... | ... | ... | ... | n | G |
| 3266 | 18511013-0615486 | 38.5 ± 0.6 | 4420 ± 121 | 1.81 ± 0.23 | 0.10 ± 0.10 | 11 ± 1 | ... | Y | Y | Y | Y | ... | Y | ... |
| 3267 | 18511048-0615470 | 34.2 ± 0.6 | 4704 ± 126 | 2.12 ± 0.23 | 0.00 ± 0.09 | 71 ± 1 | ... | Y | Y | Y | Y | ... | Y | ... |
| 3268 | 18511116-0614340 | 33.4 ± 0.6 | 4310 ± 133 | 1.64 ± 0.24 | 0.04 ± 0.10 | 165 ± 1 | ... | Y | Y | Y | Y | ... | Y | ... |
| 3270 | 18511452-0616551 | 36.3 ± 0.6 | 4796 ± 117 | 2.37 ± 0.23 | 0.10 ± 0.11 | 65 ± 1 | ... | Y | Y | Y | Y | ... | Y | ... |
| 3271 | 18511534-0618359 | 34.7 ± 0.6 | 4781 ± 128 | 2.25 ± 0.24 | 0.18 ± 0.11 | 73 ± 1 | ... | Y | Y | Y | Y | ... | Y | ... |
| 3272 | 18511571-0618146 | 36.0 ± 0.6 | 4644 ± 116 | 2.10 ± 0.24 | 0.13 ± 0.10 | 73 ± 2 | ... | Y | Y | Y | Y | ... | Y | ... |
| 3273 | 18511654-0612082 | 51.7 ± 0.4 | ... | ... | ... | ... | ... | N | ... | ... | ... | ... | n | ... |
| 3274 | 18512283-0621589 | 4.1 ± 0.6 | 4248 ± 127 | 1.76 ± 0.22 | 0.10 ± 0.10 | 27 ± 1 | ... | N | ... | ... | ... | ... | n | G |
| 3275 | 18512662-0614537 | 34.8 ± 0.6 | 4331 ± 118 | 1.80 ± 0.24 | 0.08 ± 0.10 | 138 ± 1 | ... | Y | Y | Y | Y | ... | Y | ... |
| 3276 | 18513636-0617499 | 0.3 ± 0.6 | 3920 ± 150 | 1.33 ± 0.30 | -0.13 ± 0.14 | 62 ± 1 | ... | N | ... | ... | ... | ... | n | G |
| 3277 | 18514034-0617128 | 33.3 ± 0.6 | 4670 ± 122 | 1.98 ± 0.23 | 0.06 ± 0.10 | 76 ± 1 | ... | Y | Y | Y | Y | ... | Y | ... |
| 3278 | 18514130-0620125 | 34.3 ± 0.6 | 4637 ± 116 | 2.10 ± 0.23 | 0.12 ± 0.10 | 35 ± 1 | ... | Y | Y | Y | Y | ... | Y | ... |
| 49919 | 18501559-0615288 | 30.9 ± 0.3 | 5845 ± 375 | 4.45 ± 0.35 | 0.30 ± 0.30 | ... | ... | N | ... | ... | ... | ... | n | ... |
| 49920 | 18502001-0611319 | 107.4 ± 1.1 | 6166 ± 333 | 4.04 ± 0.58 | -0.08 ± 0.47 | ... | ... | N | ... | ... | ... | ... | n | ... |
| 49921 | 18502037-0622456 | 23.4 ± 1.6 | 4942 ± 190 | 4.38 ± 0.32 | 0.06 ± 0.15 | ... | ... | N | ... | ... | ... | ... | n | ... |
| 49922 | 18502227-0619460 | 39.2 ± 0.3 | 5955 ± 270 | 4.32 ± 0.07 | 0.29 ± 0.15 | 39 ± 35 | 1 | Y | Y | Y | Y | N | Y | ... |
| 49923 | 18502321-0614292 | -43.3 ± 0.3 | 5744 ± 39 | 4.12 ± 0.34 | 0.16 ± 0.14 | ... | ... | N | ... | ... | ... | ... | n | ... |
| 49924 | 18502321-0615267 | -5.4 ± 0.3 | 6321 ± 196 | 4.08 ± 0.07 | 0.26 ± 0.24 | ... | ... | N | ... | ... | ... | ... | n | ... |
| 49925 | 18502379-0615209 | 29.7 ± 0.2 | 5789 ± 62 | 3.59 ± 0.76 | -0.88 ± 0.44 | ... | ... | N | ... | ... | ... | ... | n | ... |
| 49926 | 18502494-0613044 | 1.1 ± 0.3 | 6061 ± 112 | 4.45 ± 0.24 | 0.14 ± 0.13 | 53 ± 26 | 1 | N | ... | ... | ... | ... | n | NG |
| 49927 | 18502524-0612246 | -2.3 ± 0.3 | 5854 ± 164 | 4.39 ± 0.09 | 0.08 ± 0.14 | ... | ... | N | ... | ... | ... | ... | n | ... |
| 49928 | 18502541-0617094 | 35.4 ± 0.6 | 6343 ± 250 | 4.76 ± 0.54 | 0.08 ± 0.32 | 70 ± 45 | 1 | Y | Y | Y | Y | Y | Y | ... |
| 49929 | 18502627-0617135 | -79.4 ± 0.4 | 5898 ± 217 | 4.67 ± 0.54 | -0.09 ± 0.20 | 59 ± 48 | 1 | N | ... | ... | ... | ... | n | NG |
| 49931 | 18502640-0610143 | 5.8 ± 0.9 | 5578 ± 369 | 3.56 ± 0.53 | -0.38 ± 0.72 | ... | ... | N | ... | ... | ... | ... | n | ... |
| 49932 | 18502655-0613260 | 34.7 ± 0.6 | 6695 ± 505 | 4.43 ± 0.38 | 0.23 ± 0.21 | ... | ... | Y | ... | ... | ... | ... | n | ... |
| 49933 | 18502655-0615001 | 37.7 ± 0.4 | 6834 ± 539 | 4.38 ± 0.36 | 0.22 ± 0.28 | 53 ± 43 | 1 | Y | Y | Y | Y | Y | Y | ... |
| 49934 | 18502696-0620216 | -67.5 ± 0.3 | 6172 ± 40 | 3.98 ± 0.35 | -0.10 ± 0.27 | ... | ... | N | ... | ... | ... | ... | n | ... |
| 49935 | 18502710-0620184 | 28.6 ± 0.3 | 6254 ± 186 | 4.32 ± 0.16 | -0.06 ± 0.14 | ... | ... | N | ... | ... | ... | ... | n | ... |
| 49936 | 18502750-0619540 | 1.9 ± 0.5 | 5693 ± 229 | 4.47 ± 0.29 | 0.10 ± 0.27 | ... | ... | N | ... | ... | ... | ... | n | ... |
| 49937 | 18502753-0610321 | 113.8 ± 0.9 | 6213 ± 280 | 3.81 ± 0.67 | 0.30 ± 0.44 | ... | ... | N | ... | ... | ... | ... | n | ... |
| 49938 | 18502781-0611432 | 35.5 ± 0.4 | 5829 ± 81 | 4.36 ± 0.26 | -0.33 ± 0.41 | 100 ± 36 | 1 | Y | Y | Y | N | Y | Y | ... |
| 49939 | 18502784-0611244 | -12.0 ± 0.1 | 5927 ± 180 | 4.25 ± 0.12 | 0.20 ± 0.13 | 98 ± 58 | ... | N | ... | ... | ... | ... | n | NG |

Gutiérrez Albarrán et al.: Calibrating the lithium–age relation I.

**Table C.9.** continued.

| ID | CNAME | RV (km s$^{-1}$) | $T_{\rm eff}$ (K) | $\log g$ (dex) | [Fe/H] (dex) | EW(Li)$^a$ (mÅ) | EW(Li) error flag$^b$ | Membership RV | Li | $\log g$ | [Fe/H] | Gaia study Cantat-Gaudin$^c$ | Final$^d$ | NMs with Li$^e$ |
|---|---|---|---|---|---|---|---|---|---|---|---|---|---|---|
| 49941 | 18502827-0611031 | -12.1 ± 0.3 | 5968 ± 160 | 4.45 ± 0.21 | 0.35 ± 0.16 | 105 ± 41 | 1 | N | … | … | … | … | n | NG |
| 49942 | 18502834-0621053 | -13.2 ± 0.3 | 5275 ± 90 | 4.47 ± 0.27 | 0.08 ± 0.14 | … | … | N | … | … | … | … | n | … |
| 49943 | 18502885-0610475 | -24.4 ± 0.3 | 6108 ± 154 | 4.27 ± 0.39 | -0.02 ± 0.14 | 97 ± 29 | 1 | N | … | … | … | N | n | NG |
| 49944 | 18502900-0611392 | 33.8 ± 1.8 | … | … | … | … | … | Y | … | … | … | … | n | … |
| 49945 | 18502900-0618340 | 3.5 ± 0.3 | 6263 ± 33 | 3.55 ± 0.35 | 0.06 ± 0.26 | … | … | N | … | … | … | … | n | … |
| 49946 | 18502921-0620209 | 13.3 ± 0.2 | 5546 ± 50 | 4.57 ± 0.37 | 0.01 ± 0.13 | … | … | N | … | … | … | … | n | … |
| 49948 | 18502964-0612020 | 17.5 ± 0.3 | 6512 ± 230 | 3.92 ± 0.47 | 0.16 ± 0.19 | … | … | N | … | … | … | … | n | … |
| 49949 | 18503054-0608235 | 41.5 ± 0.4 | 6130 ± 223 | 4.54 ± 0.19 | 0.33 ± 0.21 | … | … | N | … | … | … | … | n | … |
| 49950 | 18503056-0616418 | -59.3 ± 0.6 | 5581 ± 89 | 4.57 ± 0.84 | -0.61 ± 0.29 | … | … | N | … | … | … | … | n | … |
| 49952 | 18503106-0617453 | 73.8 ± 0.3 | 6185 ± 260 | 4.28 ± 0.25 | -0.64 ± 0.32 | <21 | 3 | N | … | … | … | … | n | NG |
| 49954 | 18503156-0616365 | 24.2 ± 0.3 | 6192 ± 160 | 4.39 ± 0.27 | -0.06 ± 0.13 | 125 ± 26 | 1 | N | … | … | … | … | n | NG |
| 49955 | 18503215-0621368 | 26.8 ± 0.3 | 5877 ± 156 | 3.97 ± 0.58 | 0.18 ± 0.15 | <59 | 3 | N | … | … | … | … | n | NG |
| 49956 | 18503218-0616449 | 38.8 ± 0.8 | 6566 ± 292 | 4.41 ± 0.49 | 0.10 ± 0.28 | <60 | 3 | Y | Y | Y | Y | Y | Y | … |
| 49957 | 18503228-0621470 | 55.2 ± 0.3 | 6269 ± 166 | 4.15 ± 0.17 | 0.28 ± 0.21 | 149 ± 24 | 1 | N | … | … | … | … | n | NG |
| 49960 | 18503230-0617563 | 18.6 ± 0.6 | 7376 ± 71 | … | … | … | … | N | … | … | … | … | n | … |
| 49961 | 18503325-0618119 | 66.1 ± 0.3 | 5876 ± 239 | 4.62 ± 0.18 | 0.04 ± 0.17 | … | … | N | … | … | … | … | n | … |
| 49962 | 18503326-0616346 | 16.7 ± 0.3 | 6148 ± 235 | 4.33 ± 0.15 | 0.31 ± 0.21 | 78 ± 27 | 1 | N | … | … | … | … | n | NG |
| 49964 | 18503353-0609058 | 21.0 ± 0.3 | 6210 ± 159 | 4.14 ± 0.19 | -0.05 ± 0.22 | … | … | N | … | … | … | … | n | … |
| 49965 | 18503362-0619129 | 46.1 ± 0.4 | 4828 ± 308 | 3.24 ± 0.43 | -0.13 ± 0.16 | <104 | 3 | N | … | … | … | … | n | Li-rich G |
| 49966 | 18503371-0623084 | 23.8 ± 0.5 | 6088 ± 109 | 4.10 ± 0.13 | 0.10 ± 0.14 | 82 ± 59 | 1 | N | … | … | … | … | n | NG |
| 49967 | 18503383-0618439 | 33.3 ± 0.3 | 6052 ± 154 | 4.44 ± 0.18 | 0.03 ± 0.13 | 126 ± 33 | 1 | Y | Y | Y | Y | N | Y | … |
| 49968 | 18503386-0609534 | 69.0 ± 0.3 | 6207 ± 159 | 4.31 ± 0.22 | 0.14 ± 0.17 | 46 ± 37 | 1 | N | … | … | … | Y | n | NG |
| 49969 | 18503396-0617572 | 19.8 ± 0.3 | 6094 ± 249 | 4.19 ± 0.17 | 0.40 ± 0.24 | … | … | N | … | … | … | … | n | … |
| 49971 | 18503421-0621204 | 2.9 ± 0.1 | 5898 ± 151 | 4.53 ± 0.30 | -0.04 ± 0.13 | 106 ± 23 | … | N | … | … | … | … | n | NG |
| 49972 | 18503432-0619130 | 34.9 ± 0.3 | 5947 ± 375 | 4.58 ± 0.33 | 0.16 ± 0.25 | 111 ± 55 | 1 | Y | Y | Y | Y | N | Y | … |
| 49973 | 18503447-0621521 | -18.2 ± 0.3 | 5896 ± 236 | 4.00 ± 0.24 | 0.23 ± 0.18 | … | … | N | … | … | … | … | n | … |
| 49974 | 18503479-0621024 | -34.2 ± 0.3 | 6339 ± 173 | 4.34 ± 0.29 | 0.16 ± 0.19 | … | … | N | … | … | … | … | n | … |
| 49975 | 18503481-0614102 | 51.4 ± 0.4 | 6326 ± 247 | 4.05 ± 0.04 | 0.17 ± 0.19 | … | … | N | … | … | … | … | n | … |
| 49976 | 18503484-0615562 | 2.5 ± 0.3 | 6432 ± 64 | 4.01 ± 0.14 | 0.04 ± 0.13 | 43 ± 0 | … | N | … | … | … | … | n | NG |
| 49978 | 18503507-0616527 | 35.0 ± 0.5 | 6365 ± 215 | 4.40 ± 0.29 | 0.17 ± 0.16 | 135 ± 59 | 1 | Y | Y | Y | Y | Y | Y | … |
| 49979 | 18503539-0619590 | 40.2 ± 0.3 | 6558 ± 347 | 4.32 ± 0.20 | 0.15 ± 0.34 | 125 ± 47 | 1 | N | … | … | … | N | n | NG |
| 49980 | 18503553-0617448 | 17.9 ± 0.2 | 5652 ± 147 | 3.91 ± 0.65 | -0.35 ± 0.59 | … | … | N | … | … | … | … | n | … |
| 49982 | 18503584-0623368 | 21.0 ± 0.4 | 5524 ± 133 | 3.97 ± 0.13 | -0.16 ± 0.17 | <39 | 3 | N | … | … | … | … | n | NG |
| 49983 | 18503594-0616379 | 39.7 ± 0.4 | 6813 ± 409 | 4.51 ± 0.40 | 0.29 ± 0.27 | 108 ± 43 | 1 | Y | Y | Y | Y | Y | Y | … |
| 49984 | 18503601-0623426 | -42.3 ± 0.3 | 6012 ± 207 | 4.51 ± 0.31 | 0.27 ± 0.16 | 40 ± 29 | 1 | N | … | … | … | … | n | NG |
| 49985 | 18503603-0619495 | 37.5 ± 0.9 | 6787 ± 563 | 3.79 ± 0.21 | 0.12 ± 0.19 | 36 ± 27 | 1 | Y | Y | Y | Y | Y | Y | … |
| 49987 | 18503616-0617504 | 21.9 ± 0.5 | 5905 ± 177 | 4.52 ± 0.44 | 0.20 ± 0.13 | 84 ± 82 | 1 | N | … | … | … | … | n | NG |
| 49988 | 18503618-0614116 | 36.1 ± 0.7 | 7458 ± 746 | 4.11 ± 0.18 | 0.03 ± 0.22 | … | … | Y | … | … | … | … | n | … |
| 49989 | 18503637-0621530 | -21.7 ± 0.2 | 5718 ± 102 | 4.22 ± 0.14 | 0.29 ± 0.16 | … | … | N | … | … | … | … | n | … |
| 49990 | 18503641-0613435 | 40.3 ± 0.6 | 6806 ± 446 | 4.15 ± 0.21 | 0.20 ± 0.23 | … | … | N | … | … | … | … | n | … |
| 49992 | 18503672-0613088 | -4.7 ± 0.3 | 6098 ± 192 | 3.79 ± 0.17 | -0.22 ± 0.13 | … | … | N | … | … | … | … | n | … |
| 49994 | 18503678-0623512 | -40.6 ± 0.3 | 6173 ± 54 | 4.03 ± 0.16 | -0.40 ± 0.15 | … | … | N | … | … | … | … | n | … |
| 49996 | 18503711-0620330 | 74.6 ± 0.3 | 6193 ± 158 | 4.12 ± 0.11 | 0.18 ± 0.16 | … | … | N | … | … | … | … | n | … |
| 49998 | 18503724-0616210 | -15.8 ± 0.1 | 6312 ± 158 | 4.23 ± 0.55 | 0.01 ± 0.23 | … | … | N | … | … | … | … | n | … |
| 49999 | 18503732-0622247 | 61.5 ± 0.3 | 5974 ± 190 | 4.30 ± 0.15 | -0.21 ± 0.20 | 65 ± 30 | 1 | N | … | … | … | … | n | NG |
| 50000 | 18503742-0613550 | 14.8 ± 0.5 | 7386 ± 58 | … | … | … | … | N | … | … | … | … | n | … |
| 50001 | 18503745-0623371 | 45.7 ± 0.3 | 5969 ± 82 | 3.85 ± 0.30 | 0.34 ± 0.18 | … | … | N | … | … | … | … | n | … |
| 50002 | 18503776-0611078 | -61.5 ± 0.3 | 5311 ± 101 | 3.88 ± 0.39 | 0.19 ± 0.24 | <29 | 3 | N | … | … | … | … | n | NG |
| 50003 | 18503789-0612111 | 18.6 ± 0.3 | 6047 ± 184 | 4.37 ± 0.12 | 0.23 ± 0.17 | 82 ± 22 | 1 | N | … | … | … | … | n | NG |
| 50004 | 18503793-0623438 | -14.5 ± 0.3 | 6108 ± 224 | 3.96 ± 0.20 | 0.32 ± 0.16 | 51 ± 25 | 1 | N | … | … | … | … | n | NG |
| 50005 | 18503822-0615566 | 8.2 ± 0.4 | 5848 ± 165 | 3.98 ± 0.11 | 0.09 ± 0.38 | <46 | 3 | N | … | … | … | … | n | NG |
| 50008 | 18503865-0610028 | 39.6 ± 0.3 | 6107 ± 68 | 4.35 ± 0.40 | -0.11 ± 0.15 | … | … | Y | … | … | … | … | n | … |
| 50009 | 18503883-0624000 | 78.0 ± 0.7 | 6604 ± 328 | 3.97 ± 0.01 | 0.25 ± 0.24 | … | … | N | … | … | … | … | n | … |
| 50010 | 18503890-0616304 | -25.1 ± 0.3 | 6010 ± 179 | 4.35 ± 0.19 | 0.17 ± 0.14 | … | … | N | … | … | … | … | n | … |
| 50013 | 18503921-0623197 | 8.6 ± 0.3 | 6194 ± 250 | 4.30 ± 0.08 | 0.24 ± 0.21 | 69 ± 21 | 1 | N | … | … | … | … | n | NG |
| 50014 | 18503924-0611317 | 43.2 ± 0.3 | 6107 ± 156 | 4.14 ± 0.12 | 0.29 ± 0.15 | … | … | N | … | … | … | … | n | … |
| 50015 | 18503924-0614105 | 32.6 ± 0.3 | 5718 ± 68 | 4.36 ± 0.29 | 0.07 ± 0.16 | … | … | Y | … | … | … | … | n | … |
| 50016 | 18503941-0616320 | 70.2 ± 0.3 | 6404 ± 241 | 4.13 ± 0.07 | 0.26 ± 0.21 | 122 ± 33 | 1 | N | … | … | … | N | n | NG |
| 50017 | 18503953-0616420 | -11.0 ± 0.3 | 5912 ± 101 | 4.27 ± 0.06 | 0.39 ± 0.21 | 32 ± 31 | 1 | N | … | … | … | … | n | NG |

| ID | CNAME | RV (km s$^{-1}$) | $T_{\text{eff}}$ (K) | logg (dex) | [Fe/H] (dex) | EW(Li)$^a$ (mÅ) | EW(Li) error flag$^b$ | Membership RV | Li | logg | [Fe/H] | Gaia study Cantat-Gaudin$^c$ | Final$^d$ | NMs with Li$^e$ |
|---|---|---|---|---|---|---|---|---|---|---|---|---|---|---|
| 50018 | 18503958-0621149 | 0.4 ± 0.3 | 5532 ± 44 | 4.20 ± 0.34 | -0.24 ± 0.13 | … | … | N | … | … | … | … | n | … |
| 50019 | 18503978-0615502 | 34.3 ± 0.9 | 6289 ± 219 | 4.69 ± 0.75 | -0.07 ± 0.36 | … | … | Y | … | … | … | … | n | … |
| 50022 | 18504001-0608388 | 33.8 ± 0.6 | 6767 ± 472 | 4.49 ± 0.44 | 0.27 ± 0.23 | 99 ± 48 | 1 | Y | Y | Y | Y | N | Y | … |
| 50023 | 18504004-0619125 | -17.0 ± 0.3 | 5568 ± 69 | 3.96 ± 0.24 | -0.14 ± 0.14 | … | … | N | … | … | … | … | n | … |
| 50025 | 18504028-0611013 | -48.3 ± 0.4 | 5613 ± 238 | 4.20 ± 0.97 | -0.21 ± 0.15 | … | … | N | … | … | … | … | n | … |
| 50027 | 18504057-0622357 | 34.7 ± 0.4 | 5407 ± 118 | 3.94 ± 0.14 | -0.81 ± 0.87 | <30 | 3 | Y | N | N | N | … | n | NG |
| 50028 | 18504098-0610258 | 34.8 ± 0.4 | 6869 ± 507 | 4.43 ± 0.31 | 0.11 ± 0.21 | … | … | Y | … | … | … | … | n | … |
| 50029 | 18504118-0624576 | -13.3 ± 0.3 | 6318 ± 255 | 4.19 ± 0.06 | 0.43 ± 0.26 | … | … | N | … | … | … | … | n | … |
| 50030 | 18504122-0609271 | 35.3 ± 0.6 | 6209 ± 158 | 4.92 ± 0.60 | -0.13 ± 0.26 | 63 ± 46 | 1 | Y | Y | Y | Y | N | Y | … |
| 50031 | 18504130-0614061 | 4.8 ± 0.1 | 5536 ± 177 | 3.97 ± 0.23 | 0.03 ± 0.16 | … | … | N | … | … | … | … | n | … |
| 50032 | 18504146-0620333 | 13.3 ± 0.3 | 5718 ± 122 | 4.28 ± 0.18 | 0.22 ± 0.13 | … | … | N | … | … | … | … | n | … |
| 50033 | 18504147-0611134 | 8.0 ± 0.3 | 6027 ± 198 | 4.45 ± 0.23 | -0.19 ± 0.14 | <35 | 3 | N | … | … | … | … | n | NG |
| 50034 | 18504148-0615122 | 27.7 ± 0.3 | 5775 ± 70 | 4.30 ± 0.20 | -0.36 ± 0.24 | 32 ± 26 | 1 | N | … | … | … | … | n | NG |
| 50035 | 18504157-0611257 | 8.2 ± 0.4 | 5902 ± 124 | 4.41 ± 0.26 | 0.04 ± 0.12 | 91 ± 28 | 1 | N | … | … | … | … | n | NG |
| 50036 | 18504163-0622579 | 17.7 ± 0.9 | 7140 ± 953 | 3.83 ± 0.19 | 0.08 ± 0.13 | 47 ± 28 | 1 | N | … | … | … | … | n | NG |
| 50037 | 18504171-0614536 | 18.5 ± 0.3 | 5944 ± 63 | 4.16 ± 0.11 | -0.02 ± 0.22 | … | … | N | … | … | … | … | n | … |
| 50038 | 18504183-0621252 | -3.0 ± 0.3 | 5615 ± 109 | 4.08 ± 0.30 | 0.30 ± 0.18 | … | … | N | … | … | … | … | n | … |
| 50039 | 18504183-0623464 | 82.3 ± 0.3 | 6171 ± 299 | 4.25 ± 0.21 | -0.14 ± 0.63 | … | … | N | … | … | … | … | n | … |
| 50040 | 18504188-0622567 | 14.9 ± 0.2 | 6485 ± 246 | 4.34 ± 0.24 | -0.02 ± 0.29 | 122 ± 32 | 1 | N | … | … | … | Y | n | NG |
| 50043 | 18504222-0607490 | 28.1 ± 0.4 | 6758 ± 463 | 4.34 ± 0.32 | 0.30 ± 0.45 | … | … | N | … | … | … | … | n | … |
| 50044 | 18504230-0620521 | 53.8 ± 0.8 | 6385 ± 266 | 4.55 ± 0.46 | -0.14 ± 0.31 | … | … | N | … | … | … | … | n | … |
| 50047 | 18504252-0612377 | 8.8 ± 0.7 | 5450 ± 31 | 3.93 ± 0.14 | 0.19 ± 0.16 | … | … | N | … | … | … | … | n | … |
| 50048 | 18504263-0623450 | -8.1 ± 0.3 | 6239 ± 112 | 4.48 ± 0.32 | 0.10 ± 0.16 | … | … | N | … | … | … | … | n | … |
| 50049 | 18504288-0619126 | 34.8 ± 0.1 | 6655 ± 148 | 4.71 ± 0.75 | 0.16 ± 0.19 | 55 ± 0 | … | Y | Y | Y | Y | Y | Y | … |
| 50050 | 18504289-0612200 | 30.7 ± 0.3 | 5869 ± 255 | 4.25 ± 0.11 | 0.09 ± 0.14 | 54 ± 27 | 1 | N | … | … | … | … | n | NG |
| 50052 | 18504305-0618035 | -0.1 ± 0.1 | 5990 ± 29 | 4.28 ± 0.12 | 0.09 ± 0.15 | 46 ± 30 | 1 | N | … | … | … | … | n | NG |
| 50053 | 18504307-0616199 | -57.3 ± 0.3 | 5651 ± 65 | 4.32 ± 0.42 | 0.26 ± 0.22 | … | … | N | … | … | … | … | n | … |
| 50054 | 18504320-0609533 | -2.2 ± 0.3 | 5992 ± 157 | 4.32 ± 0.26 | 0.32 ± 0.17 | 70 ± 26 | 1 | N | … | … | … | … | n | NG |
| 50055 | 18504323-0623114 | 29.9 ± 0.9 | 6752 ± 501 | 4.32 ± 0.27 | 0.06 ± 0.20 | … | … | N | … | … | … | … | n | … |
| 50056 | 18504331-0613473 | 26.5 ± 10.0 | 7273 ± 72 | … | … | … | … | N | … | … | … | … | n | … |
| 50058 | 18504350-0613598 | 13.1 ± 0.3 | 6791 ± 138 | 4.22 ± 0.29 | 0.08 ± 0.23 | … | … | N | … | … | … | … | n | … |
| 50059 | 18504366-0613229 | 18.1 ± 0.5 | 7395 ± 751 | 4.18 ± 0.16 | 0.14 ± 0.15 | … | … | N | … | … | … | … | n | … |
| 50060 | 18504367-0616536 | 15.3 ± 0.1 | 5329 ± 102 | 4.50 ± 0.40 | -0.10 ± 0.12 | … | … | N | … | … | … | … | n | … |
| 50061 | 18504372-0609528 | 25.7 ± 0.3 | 7178 ± 813 | 4.05 ± 0.22 | 0.14 ± 0.12 | … | … | N | … | … | … | … | n | … |
| 50062 | 18504385-0617067 | 2.6 ± 0.3 | 5980 ± 83 | 4.19 ± 0.02 | 0.28 ± 0.16 | 92 ± 29 | 1 | N | … | … | … | … | n | NG |
| 50063 | 18504403-0623330 | 20.5 ± 1.5 | 7468 ± 955 | 4.15 ± 0.17 | -0.01 ± 0.36 | … | … | N | … | … | … | … | n | … |
| 50064 | 18504417-0623267 | 50.6 ± 0.5 | 5840 ± 28 | 4.40 ± 0.11 | 0.19 ± 0.14 | 157 ± 98 | 1 | N | … | … | … | … | n | NG |
| 50066 | 18504446-0617314 | -17.5 ± 0.5 | 6629 ± 491 | 4.11 ± 0.26 | 0.26 ± 0.24 | … | … | N | … | … | … | … | n | … |
| 50067 | 18504448-0619067 | 61.7 ± 0.2 | 6290 ± 322 | 4.31 ± 0.14 | 0.19 ± 0.20 | 72 ± 37 | 1 | N | … | … | … | … | n | NG |
| 50068 | 18504450-0610083 | 48.6 ± 0.3 | 6249 ± 179 | 4.31 ± 0.16 | 0.22 ± 0.20 | … | … | N | … | … | … | … | n | … |
| 50069 | 18504488-0623067 | 4.7 ± 0.3 | 6138 ± 204 | 4.60 ± 0.29 | 0.16 ± 0.14 | 57 ± 22 | 1 | N | … | … | … | … | n | NG |
| 50070 | 18504489-0616552 | 35.2 ± 0.7 | 6470 ± 201 | 4.51 ± 0.43 | -0.05 ± 0.20 | 44 ± 42 | 1 | Y | Y | Y | Y | Y | Y | … |
| 50071 | 18504514-0613091 | 32.3 ± 0.6 | 6330 ± 217 | 4.91 ± 0.61 | 0.03 ± 0.16 | 53 ± 41 | 1 | Y | Y | Y | Y | Y | Y | … |
| 50072 | 18504514-0617285 | 23.4 ± 0.1 | 5651 ± 118 | 3.87 ± 0.20 | 0.01 ± 0.15 | 97 ± 32 | 1 | N | … | … | … | … | n | NG |
| 50073 | 18504534-0616164 | 38.3 ± 0.8 | 6319 ± 233 | 4.99 ± 0.82 | -0.03 ± 0.29 | 60 ± 49 | 1 | Y | Y | Y | Y | Y | Y | … |
| 50075 | 18504556-0619011 | 65.5 ± 0.6 | 6350 ± 314 | 4.21 ± 0.08 | 0.19 ± 0.30 | 47 ± 27 | 1 | N | … | … | … | … | n | NG |
| 50076 | 18504557-0612549 | 12.8 ± 0.3 | 5463 ± 126 | 4.43 ± 0.41 | -0.17 ± 0.22 | 206 ± 33 | 1 | N | … | … | … | … | n | NG |
| 50077 | 18504561-0615222 | 21.3 ± 0.7 | 6549 ± 349 | 4.28 ± 0.49 | 0.08 ± 0.23 | … | … | N | … | … | … | … | n | … |
| 50078 | 18504588-0624404 | 52.0 ± 0.4 | 6156 ± 305 | 4.66 ± 0.26 | 0.30 ± 0.24 | 88 ± 38 | 1 | N | … | … | … | … | n | NG |
| 50081 | 18504600-0619166 | 35.8 ± 0.5 | … | … | … | … | … | Y | … | … | … | … | n | … |
| 50082 | 18504615-0621404 | 38.2 ± 1.2 | 6866 ± 571 | 4.08 ± 0.05 | 0.06 ± 0.29 | … | … | Y | … | … | … | … | n | … |
| 50084 | 18504637-0617528 | 37.2 ± 0.6 | 6304 ± 93 | 4.21 ± 0.19 | 0.03 ± 0.13 | … | … | Y | … | … | … | … | n | … |
| 50085 | 18504647-0612289 | 39.0 ± 0.6 | 6661 ± 485 | 4.35 ± 0.47 | 0.13 ± 0.24 | … | … | Y | … | … | … | … | n | … |
| 50088 | 18504664-0627066 | 53.3 ± 0.8 | 4726 ± 363 | 3.35 ± 0.44 | -0.54 ± 0.37 | … | … | N | … | … | … | … | n | G |
| 50089 | 18504668-0614524 | 9.5 ± 0.3 | 5962 ± 138 | 4.36 ± 0.16 | 0.06 ± 0.12 | 58 ± 18 | 1 | N | … | … | … | … | n | NG |
| 50091 | 18504686-0616311 | 90.5 ± 0.5 | 4789 ± 161 | 2.85 ± 0.54 | -0.47 ± 0.29 | … | … | N | … | … | … | … | n | G |
| 50092 | 18504700-0615497 | 37.1 ± 0.3 | 5988 ± 138 | 4.39 ± 0.11 | 0.09 ± 0.12 | 69 ± 36 | 1 | Y | Y | Y | Y | … | Y | … |
| 50093 | 18504713-0615000 | 38.6 ± 0.7 | 6799 ± 219 | 4.68 ± 0.67 | 0.12 ± 0.19 | 67 ± 52 | 1 | Y | Y | Y | Y | Y | Y | … |
| 50095 | 18504734-0612247 | -1.4 ± 0.2 | 5918 ± 104 | 4.19 ± 0.15 | 0.39 ± 0.21 | 80 ± 30 | 1 | N | … | … | … | … | n | NG |

**Table C.9.** continued.

| ID | CNAME | RV (km s$^{-1}$) | $T_{\rm eff}$ (K) | $\log g$ (dex) | [Fe/H] (dex) | EW(Li)$^a$ (mÅ) | EW(Li) error flag$^b$ | Membership RV | Li | $\log g$ | [Fe/H] | Gaia study Cantat-Gaudin$^c$ | Final$^d$ | NMs with Li$^e$ |
|---|---|---|---|---|---|---|---|---|---|---|---|---|---|---|
| 50096 | 18504736-0612079 | 4.0 ± 0.2 | 6138 ± 316 | 4.65 ± 0.33 | 0.28 ± 0.21 | 93 ± 30 | 1 | N | … | … | … | N | n | NG |
| 50097 | 18504746-0620024 | 29.0 ± 10.0 | … | … | … | … | … | N | … | … | … | … | n | … |
| 50098 | 18504754-0609413 | 18.7 ± 0.3 | 6077 ± 37 | 4.15 ± 0.11 | 0.00 ± 0.16 | … | … | N | … | … | … | … | n | … |
| 50099 | 18504754-0621367 | -27.0 ± 0.3 | 6326 ± 175 | 4.28 ± 0.33 | -0.20 ± 0.15 | … | … | N | … | … | … | … | n | … |
| 50100 | 18504758-0619038 | 9.9 ± 0.4 | 7210 ± 397 | 4.25 ± 0.11 | 0.12 ± 0.14 | … | … | N | … | … | … | … | n | … |
| 50101 | 18504765-0618319 | -10.4 ± 0.4 | 6035 ± 73 | 4.55 ± 0.39 | -0.13 ± 0.16 | 60 ± 39 | 1 | N | … | … | … | … | n | NG |
| 50103 | 18504768-0614257 | -18.8 ± 0.3 | 6217 ± 294 | 4.28 ± 0.14 | 0.30 ± 0.20 | … | … | N | … | … | … | … | n | … |
| 50105 | 18504779-0610337 | -32.3 ± 0.3 | 5856 ± 69 | 4.30 ± 0.54 | -0.02 ± 0.15 | 90 ± 31 | 1 | N | … | … | … | … | n | NG |
| 50106 | 18504786-0613072 | 114.8 ± 0.4 | 5900 ± 16 | 4.18 ± 0.13 | 0.76 ± 0.25 | … | … | N | … | … | … | … | n | … |
| 50110 | 18504822-0622198 | 38.1 ± 0.8 | 7020 ± 658 | 4.15 ± 0.21 | 0.12 ± 0.14 | … | … | Y | … | … | … | … | n | … |
| 50111 | 18504837-0608102 | 119.0 ± 1.3 | 7316 ± 119 | … | … | … | … | N | … | … | … | … | n | … |
| 50112 | 18504857-0619229 | 35.8 ± 0.4 | 6448 ± 223 | 4.59 ± 0.47 | 0.06 ± 0.18 | … | … | Y | … | … | … | … | n | … |
| 50114 | 18504878-0611487 | 38.8 ± 0.9 | 6189 ± 70 | 4.15 ± 0.10 | 0.08 ± 0.35 | … | … | Y | … | … | … | … | n | … |
| 50115 | 18504884-0622571 | 37.4 ± 0.4 | 6769 ± 131 | 4.39 ± 0.32 | 0.09 ± 0.23 | 56 ± 29 | 1 | Y | Y | Y | Y | Y | Y | … |
| 50116 | 18504891-0615170 | 41.5 ± 0.4 | 5927 ± 221 | 4.37 ± 0.27 | 0.21 ± 0.15 | 94 ± 61 | 1 | N | … | … | … | N | n | NG |
| 50117 | 18504892-0618091 | -0.4 ± 0.3 | 6770 ± 405 | 4.14 ± 0.11 | 0.15 ± 0.19 | 61 ± 17 | 1 | N | … | … | … | Y | n | NG |
| 50118 | 18504896-0614324 | 48.9 ± 0.3 | 6161 ± 194 | 4.53 ± 0.42 | 0.03 ± 0.16 | 92 ± 30 | 1 | N | … | … | … | … | n | NG |
| 50119 | 18504906-0613019 | 38.7 ± 0.7 | 6938 ± 67 | … | 0.42 ± 0.06 | … | … | Y | … | … | … | … | n | … |
| 50120 | 18504935-0618072 | 60.0 ± 0.4 | 6102 ± 147 | 4.31 ± 0.13 | 0.03 ± 0.42 | … | … | N | … | … | … | … | n | … |
| 50121 | 18504954-0614585 | 34.6 ± 0.2 | 6536 ± 155 | 4.19 ± 0.25 | 0.01 ± 0.15 | … | … | Y | … | … | … | … | n | … |
| 50122 | 18504957-0614493 | 30.7 ± 0.3 | 6748 ± 487 | 4.15 ± 0.16 | 0.12 ± 0.19 | … | … | N | … | … | … | … | n | … |
| 50123 | 18504957-0619025 | 39.3 ± 0.3 | 5971 ± 116 | 4.41 ± 0.24 | 0.12 ± 0.12 | … | … | Y | … | … | … | … | n | … |
| 50124 | 18504966-0618175 | 48.7 ± 0.2 | 5796 ± 285 | 3.79 ± 0.34 | -0.63 ± 0.37 | 79 ± 46 | 1 | N | … | … | … | … | n | NG |
| 50125 | 18504988-0612326 | 43.8 ± 0.3 | 6096 ± 186 | 4.45 ± 0.24 | 0.00 ± 0.19 | 53 ± 25 | 1 | N | … | … | … | … | n | NG |
| 50127 | 18504999-0614162 | 41.6 ± 0.7 | 6363 ± 200 | 3.85 ± 0.24 | 0.14 ± 0.26 | 42 ± 29 | 1 | N | … | … | … | N | n | NG |
| 50129 | 18505011-0615467 | -9.4 ± 0.3 | 6090 ± 112 | 4.36 ± 0.24 | 0.20 ± 0.15 | 100 ± 43 | 1 | N | … | … | … | … | n | NG |
| 50130 | 18505019-0621276 | 36.6 ± 0.7 | 4535 ± 592 | 4.54 ± 1.00 | -0.17 ± 0.25 | … | … | Y | … | … | … | … | n | … |
| 50132 | 18505027-0615531 | 34.4 ± 0.6 | 5189 ± 431 | 4.62 ± 0.36 | 0.38 ± 0.51 | … | … | Y | … | … | … | … | n | … |
| 50133 | 18505028-0625020 | -23.4 ± 0.3 | 5324 ± 114 | 4.32 ± 0.45 | 0.01 ± 0.12 | … | … | N | … | … | … | … | n | … |
| 50134 | 18505040-0612599 | 6.6 ± 0.3 | 5677 ± 25 | 4.20 ± 0.14 | 0.13 ± 0.16 | … | … | N | … | … | … | … | n | … |
| 50135 | 18505045-0617310 | 29.8 ± 0.3 | 4893 ± 265 | 4.44 ± 0.49 | -0.37 ± 0.27 | … | … | N | … | … | … | … | n | … |
| 50137 | 18505051-0620596 | -40.3 ± 0.4 | 5749 ± 116 | 4.44 ± 0.35 | -0.15 ± 0.28 | … | … | N | … | … | … | … | n | … |
| 50138 | 18505058-0624481 | 33.1 ± 0.7 | … | … | … | … | … | Y | … | … | … | … | n | … |
| 50139 | 18505060-0616585 | 36.5 ± 0.5 | … | … | … | … | … | Y | … | … | … | … | n | … |
| 50140 | 18505062-0619132 | 33.0 ± 0.3 | 5955 ± 113 | 4.65 ± 0.41 | -0.01 ± 0.15 | 134 ± 34 | 1 | Y | Y | Y | Y | N | Y | … |
| 50141 | 18505088-0620418 | 39.2 ± 0.6 | 6993 ± 394 | 4.79 ± 0.68 | 0.12 ± 0.20 | … | … | Y | … | … | … | … | n | … |
| 50142 | 18505091-0611028 | -78.8 ± 0.3 | 5762 ± 150 | 4.18 ± 0.30 | -0.19 ± 0.17 | 63 ± 32 | 1 | N | … | … | … | … | n | NG |
| 50144 | 18505108-0616323 | 59.8 ± 0.3 | 6021 ± 114 | 4.24 ± 0.13 | 0.02 ± 0.17 | … | … | N | … | … | … | … | n | … |
| 50145 | 18505112-0618562 | 10.5 ± 0.3 | 5939 ± 239 | 4.07 ± 0.02 | 0.49 ± 0.24 | … | 1 | N | … | … | … | N | n | … |
| 50147 | 18505141-0609550 | 65.0 ± 0.3 | 6055 ± 208 | 4.06 ± 0.09 | -0.27 ± 0.14 | 43 ± 25 | 1 | N | … | … | … | … | n | NG |
| 50148 | 18505144-0608453 | -17.1 ± 0.2 | 5897 ± 69 | 4.28 ± 0.14 | 0.25 ± 0.14 | 55 ± 19 | 1 | N | … | … | … | … | n | NG |
| 50149 | 18505148-0607339 | 35.8 ± 0.9 | 6596 ± 459 | 3.85 ± 0.24 | 0.28 ± 0.27 | … | … | Y | … | … | … | … | n | … |
| 50150 | 18505152-0619079 | 43.3 ± 0.3 | 5881 ± 72 | 4.20 ± 0.11 | 0.00 ± 0.20 | … | … | N | … | … | … | … | n | … |
| 50151 | 18505156-0624541 | -35.6 ± 0.3 | 5867 ± 216 | 4.33 ± 0.33 | 0.28 ± 0.16 | … | … | N | … | … | … | … | n | … |
| 50152 | 18505161-0612138 | 59.9 ± 0.3 | 5943 ± 294 | 4.25 ± 0.17 | 0.35 ± 0.18 | … | … | N | … | … | … | … | n | … |
| 50153 | 18505186-0618382 | 38.9 ± 0.6 | … | … | … | … | … | Y | … | … | … | … | n | … |
| 50154 | 18505187-0618147 | 39.0 ± 0.4 | 6738 ± 460 | 4.43 ± 0.34 | 0.13 ± 0.19 | 62 ± 38 | 1 | Y | Y | Y | Y | Y | Y | … |
| 50155 | 18505195-0624106 | 45.5 ± 0.3 | 5172 ± 96 | 2.49 ± 0.54 | -1.13 ± 0.21 | … | … | N | … | … | … | … | n | G |
| 50157 | 18505221-0608524 | 37.2 ± 0.9 | 6627 ± 96 | … | -0.02 ± 0.08 | 43 ± 29 | 1 | Y | Y | … | Y | Y | Y | … |
| 50158 | 18505223-0618572 | 34.4 ± 0.5 | 5184 ± 154 | 4.07 ± 0.10 | -0.02 ± 0.24 | … | … | Y | … | … | … | … | n | … |
| 50160 | 18505238-0616072 | 35.5 ± 0.2 | 6329 ± 214 | 4.52 ± 0.53 | 0.19 ± 0.24 | 70 ± 45 | 1 | Y | Y | Y | Y | … | Y | … |
| 50163 | 18505244-0625355 | 24.0 ± 0.5 | 7161 ± 647 | 4.18 ± 0.15 | 0.14 ± 0.17 | … | 1 | N | … | … | … | … | n | … |
| 50165 | 18505252-0617468 | 52.2 ± 0.3 | 6532 ± 254 | 4.39 ± 0.35 | 0.09 ± 0.16 | … | … | N | … | … | … | … | n | … |
| 50167 | 18505260-0619465 | 39.8 ± 0.3 | … | … | … | … | … | Y | … | … | … | … | n | … |
| 50168 | 18505266-0611041 | 81.2 ± 0.3 | 5961 ± 98 | 3.79 ± 0.19 | -0.28 ± 0.20 | … | … | N | … | … | … | … | n | … |
| 50169 | 18505270-0621406 | 47.1 ± 1.5 | 7228 ± 412 | 4.05 ± 0.23 | -0.09 ± 0.19 | 34 ± 30 | 1 | N | … | … | … | N | n | NG |
| 50170 | 18505280-0622538 | 37.7 ± 0.6 | 6685 ± 486 | 4.48 ± 0.69 | 0.18 ± 0.22 | 89 ± 70 | 1 | Y | Y | Y | Y | Y | Y | … |
| 50171 | 18505285-0609071 | 36.0 ± 0.5 | 6892 ± 693 | 4.47 ± 0.69 | 0.21 ± 0.28 | … | … | Y | … | … | … | … | n | … |
| 50172 | 18505292-0617209 | 35.9 ± 0.7 | 6489 ± 246 | 4.37 ± 0.69 | 0.05 ± 0.27 | 54 ± 43 | 1 | Y | Y | Y | Y | Y | Y | … |

| ID | CNAME | RV (km s$^{-1}$) | $T_{\text{eff}}$ (K) | logg (dex) | [Fe/H] (dex) | EW(Li)$^a$ (mÅ) | EW(Li) error flag$^b$ | Membership RV | Li | logg | [Fe/H] | Gaia study Cantat-Gaudin$^c$ | Final$^d$ | NMs with Li$^e$ |
|---|---|---|---|---|---|---|---|---|---|---|---|---|---|---|
| 50173 | 18505295-0624586 | -18.8 ± 0.3 | 5800 ± 150 | 4.21 ± 0.34 | 0.18 ± 0.15 | … | … | N | … | … | … | … | n | … |
| 50179 | 18505318-0620134 | 40.1 ± 0.6 | 7441 ± 408 | 4.12 ± 0.17 | 0.02 ± 0.20 | 55 ± 35 | 1 | N | … | … | … | N | n | NG |
| 50182 | 18505344-0624142 | 34.8 ± 0.7 | 6543 ± 380 | 4.38 ± 0.57 | 0.11 ± 0.32 | … | … | Y | … | … | … | … | n | … |
| 50183 | 18505345-0614142 | 41.0 ± 0.4 | 7423 ± 562 | 4.17 ± 0.15 | 0.04 ± 0.21 | … | … | N | … | … | … | … | n | … |
| 50185 | 18505350-0613072 | 63.6 ± 2.7 | 5805 ± 694 | 3.95 ± 0.94 | -0.88 ± 0.97 | … | … | N | … | … | … | … | n | … |
| 50186 | 18505351-0615405 | 29.1 ± 0.7 | 5911 ± 742 | 5.03 ± 0.20 | 0.15 ± 0.60 | … | … | N | … | … | … | … | n | … |
| 50187 | 18505352-0616599 | 33.3 ± 0.4 | 5837 ± 204 | 4.53 ± 0.15 | 0.17 ± 0.15 | 151 ± 78 | 1 | Y | Y | Y | Y | … | Y | … |
| 50188 | 18505356-0617572 | 15.0 ± 0.3 | 5888 ± 75 | 4.29 ± 0.06 | 0.25 ± 0.13 | 53 ± 32 | 1 | N | … | … | … | … | n | NG |
| 50189 | 18505357-0616379 | 35.9 ± 0.8 | 7480 ± 82 | … | … | … | … | Y | … | … | … | … | n | … |
| 50190 | 18505366-0620312 | -6.6 ± 0.6 | 6095 ± 241 | 3.86 ± 0.50 | 0.16 ± 0.27 | … | … | N | … | … | … | … | n | … |
| 50191 | 18505369-0606495 | 16.6 ± 0.3 | 6775 ± 412 | 4.52 ± 0.41 | 0.24 ± 0.32 | … | … | N | … | … | … | … | n | … |
| 50192 | 18505375-0615439 | 37.5 ± 0.4 | 6664 ± 463 | 4.28 ± 0.49 | 0.17 ± 0.21 | … | … | Y | … | … | … | … | n | … |
| 50195 | 18505399-0620484 | -44.8 ± 0.3 | 5669 ± 305 | 4.51 ± 0.44 | 0.36 ± 0.26 | … | … | N | … | … | … | … | n | … |
| 50197 | 18505417-0615222 | 41.7 ± 0.6 | 6491 ± 264 | 4.90 ± 0.73 | 0.01 ± 0.13 | … | … | N | … | … | … | … | n | … |
| 50198 | 18505430-0617155 | 33.1 ± 0.7 | … | … | … | … | … | Y | … | … | … | … | n | … |
| 50200 | 18505433-0611552 | 23.9 ± 0.1 | 5972 ± 213 | 4.00 ± 0.31 | -0.43 ± 0.18 | 47 ± 43 | 1 | N | … | … | … | … | n | NG |
| 50201 | 18505433-0624187 | 42.5 ± 0.3 | 5257 ± 144 | 3.69 ± 0.25 | 0.14 ± 0.28 | 26 ± 21 | 1 | N | N | N | Y | Y | n | NG |
| 50203 | 18505442-0622228 | -10.0 ± 0.3 | 6229 ± 133 | 4.27 ± 0.13 | 0.22 ± 0.19 | … | … | N | … | … | … | … | n | … |
| 50204 | 18505445-0616368 | 36.4 ± 0.5 | 5445 ± 34 | 4.63 ± 0.44 | 0.01 ± 0.13 | 137 ± 70 | 1 | Y | Y | Y | Y | Y | Y | … |
| 50205 | 18505454-0618066 | 52.8 ± 0.4 | 4919 ± 313 | 3.07 ± 0.37 | -0.45 ± 0.63 | 51 ± 39 | 1 | N | … | … | … | … | n | G |
| 50209 | 18505488-0615215 | 43.2 ± 0.7 | 4863 ± 259 | 3.83 ± 0.20 | 0.02 ± 0.38 | … | … | N | … | … | … | … | n | … |
| 50210 | 18505490-0624241 | -17.6 ± 0.4 | 6035 ± 243 | 4.04 ± 0.12 | -0.21 ± 0.18 | … | … | N | … | … | … | … | n | … |
| 50211 | 18505492-0620326 | 34.5 ± 0.4 | 7272 ± 311 | 4.10 ± 0.16 | 0.11 ± 0.11 | 36 ± 32 | 1 | Y | Y | Y | Y | Y | Y | … |
| 50213 | 18505503-0617434 | 35.9 ± 0.5 | 6048 ± 103 | 4.58 ± 0.36 | -0.10 ± 0.19 | 134 ± 44 | 1 | Y | Y | Y | Y | Y | Y | … |
| 50215 | 18505533-0619141 | -0.3 ± 0.3 | 5929 ± 163 | 4.36 ± 0.30 | -0.13 ± 0.16 | 61 ± 30 | 1 | N | … | … | … | … | n | NG |
| 50216 | 18505536-0619280 | 33.8 ± 0.8 | 6791 ± 671 | 4.45 ± 0.40 | 0.17 ± 0.20 | <48 | 3 | Y | Y | Y | Y | Y | Y | … |
| 50217 | 18505543-0609579 | -3.4 ± 2.2 | 5870 ± 163 | 3.52 ± 0.49 | 0.16 ± 0.10 | … | … | N | … | … | … | … | n | … |
| 50218 | 18505547-0610542 | 35.4 ± 0.6 | 6313 ± 207 | 4.88 ± 0.72 | 0.02 ± 0.16 | 68 ± 47 | 1 | Y | Y | Y | Y | Y | Y | … |
| 50219 | 18505548-0615294 | -4.8 ± 0.3 | 5840 ± 39 | 4.38 ± 0.13 | 0.14 ± 0.12 | <43 | 3 | N | … | … | … | … | n | NG |
| 50220 | 18505556-0617345 | 55.0 ± 0.2 | 4966 ± 146 | 3.44 ± 0.12 | -0.08 ± 0.18 | … | … | N | … | … | … | … | n | G |
| 50227 | 18505593-0610225 | 43.2 ± 0.3 | 6771 ± 242 | 4.36 ± 0.26 | 0.30 ± 0.28 | 81 ± 30 | 1 | N | … | … | … | … | n | NG |
| 50228 | 18505594-0619386 | 35.8 ± 0.5 | 6866 ± 502 | 4.60 ± 0.49 | 0.24 ± 0.29 | … | … | Y | … | … | … | … | n | … |
| 50229 | 18505596-0626246 | 6.1 ± 0.3 | 6098 ± 122 | 4.38 ± 0.24 | 0.22 ± 0.15 | 66 ± 38 | 1 | N | … | … | … | N | n | NG |
| 50231 | 18505607-0619194 | 24.4 ± 0.3 | 5664 ± 66 | 4.37 ± 0.49 | 0.17 ± 0.13 | … | … | N | … | … | … | … | n | … |
| 50232 | 18505608-0616556 | 36.4 ± 0.6 | … | … | … | … | … | Y | … | … | … | … | n | … |
| 50233 | 18505613-0617443 | 0.6 ± 0.5 | 5563 ± 64 | 4.11 ± 0.21 | 0.06 ± 0.15 | … | … | N | … | … | … | … | n | … |
| 50235 | 18505621-0619329 | 29.7 ± 0.1 | 6946 ± 148 | 3.96 ± 0.29 | 0.12 ± 0.30 | 77 ± 58 | 1 | Y | Y | Y | Y | Y | Y | … |
| 50237 | 18505626-0617340 | 52.2 ± 0.5 | 4793 ± 293 | 2.85 ± 0.86 | -0.82 ± 0.66 | … | … | N | … | … | … | … | n | G |
| 50238 | 18505627-0620182 | 12.1 ± 0.4 | 5984 ± 79 | 4.11 ± 0.51 | 0.07 ± 0.14 | 137 ± 68 | 1 | N | … | … | … | … | n | NG |
| 50240 | 18505635-0618459 | 35.9 ± 0.7 | 6503 ± 393 | 4.07 ± 0.01 | 0.10 ± 0.24 | … | … | Y | … | … | … | … | n | … |
| 50241 | 18505637-0618193 | 33.8 ± 0.4 | 6090 ± 245 | 4.61 ± 0.23 | 0.17 ± 0.25 | 133 ± 53 | 1 | Y | Y | Y | Y | … | Y | … |
| 50245 | 18505654-0612414 | 37.3 ± 1.6 | 6841 ± 51 | 3.96 ± 0.17 | 0.08 ± 0.23 | … | … | Y | … | … | … | … | n | … |
| 50247 | 18505672-0616163 | 1.5 ± 1.0 | 5732 ± 76 | 4.42 ± 0.16 | -0.19 ± 0.15 | … | … | N | … | … | … | … | n | … |
| 50248 | 18505672-0620578 | 23.4 ± 0.3 | 5580 ± 59 | 3.73 ± 0.41 | 0.13 ± 0.25 | 71 ± 30 | 1 | N | … | … | … | … | n | NG |
| 50249 | 18505677-0612361 | -38.6 ± 0.3 | 4918 ± 103 | 3.35 ± 0.59 | 0.15 ± 0.14 | … | … | N | … | … | … | … | n | G |
| 50250 | 18505680-0606421 | 0.2 ± 0.3 | 6113 ± 231 | 4.44 ± 0.17 | 0.26 ± 0.18 | 77 ± 35 | 1 | N | … | … | … | … | n | NG |
| 50252 | 18505683-0609319 | 20.9 ± 0.2 | 5611 ± 105 | 4.16 ± 0.20 | 0.28 ± 0.17 | 75 ± 25 | 1 | N | … | … | … | … | n | NG |
| 50253 | 18505686-0619107 | 37.5 ± 0.4 | 7267 ± 752 | 4.18 ± 0.15 | 0.14 ± 0.17 | 53 ± 36 | 1 | Y | Y | Y | Y | N | Y | … |
| 50255 | 18505701-0615044 | 29.9 ± 0.2 | 6324 ± 298 | 4.43 ± 0.28 | -0.30 ± 0.57 | 134 ± 34 | 1 | Y | Y | Y | N | Y | Y | … |
| 50256 | 18505705-0618335 | 34.5 ± 0.5 | 4523 ± 207 | 4.00 ± 0.26 | -0.24 ± 0.28 | … | … | Y | … | … | … | … | n | … |
| 50259 | 18505737-0621132 | 80.4 ± 0.3 | 6121 ± 185 | 4.33 ± 0.12 | 0.18 ± 0.15 | … | … | N | … | … | … | … | n | … |
| 50260 | 18505740-0608561 | 17.2 ± 0.3 | 6684 ± 493 | 4.37 ± 0.50 | 0.21 ± 0.28 | … | … | N | … | … | … | … | n | … |
| 50261 | 18505741-0617373 | 31.1 ± 0.5 | 6289 ± 217 | 4.61 ± 0.42 | -0.15 ± 0.32 | 99 ± 54 | 1 | Y | Y | Y | N | Y | Y | … |
| 50263 | 18505745-0615391 | 0.0 ± 0.2 | 5785 ± 20 | 3.92 ± 0.21 | -0.01 ± 0.15 | 49 ± 22 | 1 | N | … | … | … | … | n | NG |
| 50264 | 18505758-0624040 | -8.9 ± 0.3 | 6473 ± 182 | 4.31 ± 0.29 | -0.06 ± 0.13 | 93 ± 29 | 1 | N | … | … | … | N | n | NG |
| 50265 | 18505760-0618351 | 34.8 ± 0.2 | 5934 ± 202 | 4.42 ± 0.16 | -0.03 ± 0.36 | 138 ± 57 | 1 | Y | Y | Y | Y | Y | Y | … |
| 50266 | 18505764-0617479 | 7.1 ± 0.1 | 5967 ± 151 | 4.30 ± 0.48 | -0.19 ± 0.19 | … | … | N | … | … | … | … | n | … |
| 50267 | 18505771-0618323 | 57.5 ± 0.5 | 5628 ± 180 | 4.04 ± 0.69 | 0.22 ± 0.25 | … | … | N | … | … | … | … | n | … |
| 50268 | 18505774-0624289 | 21.0 ± 0.3 | 5978 ± 77 | 4.39 ± 0.09 | 0.21 ± 0.13 | <29 | 3 | N | … | … | … | … | n | NG |

**Table C.9.** continued.

| ID | CNAME | RV (km s$^{-1}$) | $T_{\text{eff}}$ (K) | $logg$ (dex) | [Fe/H] (dex) | EW(Li)$^a$ (mÅ) | EW(Li) error flag$^b$ | Membership RV | Li | $logg$ | [Fe/H] | Gaia study Cantat-Gaudin$^c$ | Final$^d$ | NMs with Li$^e$ |
|---|---|---|---|---|---|---|---|---|---|---|---|---|---|---|
| 50270 | 18505782-0617583 | 57.4 ± 0.8 | … | … | … | … | … | N | … | … | … | … | n | … |
| 50271 | 18505789-0615301 | 21.5 ± 10.0 | 7178 ± 102 | … | -2.56 ± 0.76 | … | … | N | … | … | … | … | n | … |
| 50272 | 18505792-0617217 | 57.1 ± 0.8 | 4886 ± 196 | 4.73 ± 0.27 | -0.30 ± 0.42 | <106 | 3 | N | … | … | … | … | n | NG |
| 50273 | 18505793-0624311 | 2.2 ± 0.1 | 6400 ± 150 | 4.18 ± 0.20 | 0.11 ± 0.16 | 115 ± 22 | 1 | N | … | … | … | … | n | NG |
| 50277 | 18505802-0608316 | 16.2 ± 10.0 | 6071 ± 114 | 4.13 ± 0.38 | -0.41 ± 0.15 | 54 ± 27 | 1 | N | … | … | … | … | n | NG |
| 50278 | 18505803-0615555 | 35.7 ± 0.6 | 7350 ± 67 | … | … | … | … | Y | … | … | … | … | n | … |
| 50280 | 18505805-0618086 | 36.6 ± 0.9 | 6797 ± 151 | 4.40 ± 0.47 | 0.04 ± 0.19 | 97 ± 60 | 1 | Y | Y | Y | Y | Y | Y | … |
| 50281 | 18505806-0617433 | 34.7 ± 0.4 | 7125 ± 555 | 4.03 ± 0.22 | 0.01 ± 0.33 | … | … | Y | … | … | … | … | n | … |
| 50284 | 18505823-0620477 | 35.2 ± 0.9 | 6646 ± 285 | 4.15 ± 0.21 | 0.11 ± 0.21 | … | … | Y | … | … | … | … | n | … |
| 50286 | 18505829-0626286 | 35.5 ± 1.6 | 5110 ± 213 | 3.85 ± 0.58 | -0.83 ± 0.21 | … | … | Y | … | … | … | … | n | … |
| 50288 | 18505836-0620226 | -2.6 ± 0.5 | 5106 ± 97 | 4.19 ± 0.37 | 0.02 ± 0.14 | … | … | N | … | … | … | … | n | … |
| 50290 | 18505841-0617093 | 69.2 ± 0.3 | 6079 ± 177 | 3.83 ± 0.31 | 0.10 ± 0.12 | … | … | N | … | … | … | … | n | … |
| 50293 | 18505855-0611380 | 24.3 ± 0.2 | 6310 ± 195 | 4.47 ± 0.49 | 0.05 ± 0.25 | … | … | N | … | … | … | … | n | … |
| 50294 | 18505858-0613545 | 21.0 ± 0.3 | 7093 ± 66 | 4.64 ± 0.14 | -0.57 ± 0.09 | … | … | N | … | … | … | … | n | … |
| 50295 | 18505861-0610093 | 46.3 ± 0.3 | 6017 ± 206 | 4.46 ± 0.30 | -0.08 ± 0.14 | … | … | N | … | … | … | … | n | … |
| 50296 | 18505861-0612334 | -4.8 ± 0.3 | 5986 ± 152 | 4.43 ± 0.25 | 0.11 ± 0.16 | 83 ± 31 | 1 | N | … | … | … | … | n | NG |
| 50297 | 18505863-0619244 | 120.5 ± 0.4 | 5058 ± 218 | 3.65 ± 0.49 | -0.07 ± 0.26 | … | … | N | … | … | … | … | n | … |
| 50300 | 18505879-0619068 | -49.6 ± 0.4 | 5149 ± 56 | 4.27 ± 0.72 | 0.05 ± 0.19 | … | … | N | … | … | … | … | n | … |
| 50301 | 18505881-0620296 | 45.7 ± 1.2 | 6416 ± 159 | 4.55 ± 0.53 | 0.18 ± 0.23 | … | … | N | … | … | … | … | n | … |
| 50304 | 18505897-0621212 | 32.2 ± 0.4 | 7176 ± 645 | 4.16 ± 0.17 | 0.16 ± 0.19 | 74 ± 53 | 1 | Y | Y | Y | Y | … | Y | … |
| 50305 | 18505898-0609391 | 64.0 ± 0.3 | 6115 ± 199 | 4.29 ± 0.11 | -0.02 ± 0.12 | … | … | N | … | … | … | … | n | … |
| 50307 | 18505909-0613102 | 13.2 ± 0.3 | 6305 ± 237 | 4.34 ± 0.23 | 0.13 ± 0.19 | 110 ± 31 | 1 | N | … | … | … | … | n | NG |
| 50308 | 18505912-0618123 | 38.4 ± 0.4 | … | … | … | … | … | Y | … | … | … | … | n | … |
| 50310 | 18505918-0611039 | 46.2 ± 0.1 | 6066 ± 210 | 4.26 ± 0.17 | 0.05 ± 0.13 | 54 ± 50 | 1 | N | … | … | … | … | n | NG |
| 50311 | 18505924-0607494 | -12.2 ± 0.3 | 5764 ± 27 | 4.60 ± 0.24 | 0.06 ± 0.15 | … | … | N | … | … | … | … | n | … |
| 50312 | 18505926-0614301 | 77.5 ± 1.1 | 5962 ± 128 | 4.02 ± 0.20 | 0.00 ± 0.16 | … | … | N | … | … | … | … | n | … |
| 50314 | 18505930-0616476 | 7.8 ± 0.5 | 5919 ± 182 | 4.23 ± 0.32 | 0.17 ± 0.25 | 100 ± 88 | 1 | N | … | … | … | … | n | NG |
| 50317 | 18505940-0610029 | 57.2 ± 0.6 | 6416 ± 229 | 4.07 ± 0.15 | 0.18 ± 0.65 | … | … | N | … | … | … | … | n | … |
| 50321 | 18505952-0618561 | 55.0 ± 0.2 | 6254 ± 142 | 4.40 ± 0.31 | -0.17 ± 0.13 | 43 ± 14 | 1 | N | … | … | … | … | n | NG |
| 50322 | 18505962-0614475 | 2.7 ± 0.1 | 6712 ± 638 | 4.53 ± 0.41 | 0.36 ± 0.29 | 96 ± 23 | 1 | N | … | … | … | … | n | NG |
| 50324 | 18505969-0610506 | -16.3 ± 0.3 | 5575 ± 88 | 4.28 ± 0.15 | 0.29 ± 0.17 | … | … | N | … | … | … | … | n | … |
| 50326 | 18505971-0624046 | 51.1 ± 0.3 | 5963 ± 98 | 4.15 ± 0.21 | 0.27 ± 0.17 | … | … | N | … | … | … | … | n | … |
| 50327 | 18505972-0625058 | -49.1 ± 0.5 | 5678 ± 110 | 4.13 ± 0.06 | -0.03 ± 0.27 | … | … | N | … | … | … | … | n | … |
| 50328 | 18505976-0617477 | -24.3 ± 0.3 | 4332 ± 244 | 2.28 ± 0.33 | 0.40 ± 0.28 | … | … | N | … | … | … | … | n | … |
| 50329 | 18505979-0616068 | 75.7 ± 1.0 | 6305 ± 205 | 4.31 ± 0.28 | 0.15 ± 0.22 | … | … | N | … | … | … | … | n | … |
| 50331 | 18505981-0616167 | 35.4 ± 0.8 | 7445 ± 75 | … | … | … | … | Y | … | … | … | … | n | … |
| 50333 | 18505993-0606488 | 39.3 ± 0.3 | 4973 ± 161 | 3.71 ± 0.23 | 0.17 ± 0.21 | … | … | Y | … | … | … | … | n | … |
| 50334 | 18510000-0615529 | 38.9 ± 0.5 | 5099 ± 109 | 4.69 ± 0.36 | 0.04 ± 0.14 | … | … | Y | … | … | … | … | n | … |
| 50335 | 18510003-0614220 | 34.6 ± 0.2 | 7210 ± 396 | 4.05 ± 0.23 | -0.09 ± 0.19 | … | … | Y | … | … | … | … | n | … |
| 50336 | 18510004-0615080 | 36.8 ± 0.5 | 6738 ± 514 | 3.89 ± 0.28 | 0.22 ± 0.25 | 27 ± 13 | 1 | Y | Y | Y | Y | … | Y | … |
| 50337 | 18510005-0610534 | 36.5 ± 0.5 | 6272 ± 123 | 5.03 ± 0.74 | -0.03 ± 0.14 | … | … | Y | … | … | … | … | n | … |
| 50338 | 18510007-0611197 | 4.3 ± 0.3 | 6377 ± 181 | 4.22 ± 0.33 | -0.13 ± 0.14 | … | … | N | … | … | … | … | n | … |
| 50339 | 18510010-0613556 | 35.3 ± 0.5 | 5440 ± 160 | 4.42 ± 0.81 | 0.06 ± 0.16 | … | … | Y | … | … | … | … | n | … |
| 50341 | 18510015-0615342 | 10.0 ± 0.2 | 6935 ± 42 | 4.62 ± 0.52 | -0.02 ± 0.21 | … | … | N | … | … | … | … | n | … |
| 50342 | 18510015-0618579 | 38.1 ± 0.4 | 6718 ± 318 | 4.31 ± 0.42 | 0.08 ± 0.21 | 98 ± 57 | 1 | Y | Y | Y | Y | Y | Y | … |
| 50343 | 18510016-0615023 | 39.5 ± 0.3 | 6615 ± 185 | 4.08 ± 0.12 | 0.05 ± 0.19 | … | … | Y | … | … | … | … | n | … |
| 50344 | 18510018-0626059 | -42.7 ± 0.3 | 5136 ± 140 | 4.19 ± 0.15 | 0.09 ± 0.13 | <42 | 3 | N | … | … | … | … | n | NG |
| 50346 | 18510021-0614005 | 27.2 ± 0.6 | 6399 ± 306 | 3.83 ± 0.19 | 0.13 ± 0.19 | … | … | N | … | … | … | … | n | … |
| 50348 | 18510027-0621387 | -20.6 ± 0.3 | 6137 ± 143 | 4.39 ± 0.45 | -0.14 ± 0.13 | 60 ± 34 | 1 | N | … | … | … | … | n | NG |
| 50349 | 18510030-0618504 | 31.5 ± 4.6 | 6922 ± 441 | 4.96 ± 0.85 | 0.46 ± 0.39 | … | … | Y | … | … | … | … | n | … |
| 50350 | 18510033-0610285 | 15.3 ± 0.3 | 6138 ± 147 | 4.12 ± 0.06 | 0.21 ± 0.14 | … | … | N | … | … | … | … | n | … |
| 50351 | 18510037-0611145 | -9.9 ± 0.7 | 6323 ± 126 | 3.97 ± 0.39 | -1.23 ± 0.23 | … | … | N | … | … | … | … | n | … |
| 50352 | 18510039-0616416 | 35.9 ± 0.7 | 6965 ± 606 | 4.30 ± 0.19 | 0.09 ± 0.19 | … | … | Y | … | … | … | … | n | … |
| 50353 | 18510051-0614343 | 33.1 ± 0.3 | 7248 ± 53 | … | … | 20 ± 18 | 1 | Y | Y | … | … | Y | Y | … |
| 50354 | 18510055-0622579 | 36.3 ± 0.5 | … | … | … | … | … | Y | … | … | … | … | n | … |
| 50357 | 18510066-0614100 | 17.8 ± 0.5 | 5703 ± 93 | 4.47 ± 0.12 | 0.16 ± 0.16 | … | … | N | … | … | … | … | n | … |
| 50358 | 18510068-0615020 | 21.3 ± 1.6 | 4990 ± 139 | 4.65 ± 0.11 | -0.41 ± 0.24 | … | … | N | … | … | … | … | n | … |
| 50359 | 18510068-0617574 | 40.2 ± 0.4 | 6374 ± 173 | 4.22 ± 0.43 | 0.13 ± 0.19 | 49 ± 34 | 1 | Y | Y | Y | Y | Y | Y | … |
| 50360 | 18510068-0618517 | 10.3 ± 0.5 | 6235 ± 244 | 4.51 ± 0.54 | 0.44 ± 0.31 | … | … | N | … | … | … | … | n | … |

| ID | CNAME | RV (km s$^{-1}$) | T$_{\rm eff}$ (K) | logg (dex) | [Fe/H] (dex) | EW(Li)$^a$ (mÅ) | EW(Li) error flag$^b$ | Membership | | | | Gaia study Cantat-Gaudin$^c$ | Final$^d$ | NMs with Li$^e$ |
| | | | | | | | | RV | Li | logg | [Fe/H] | | | |
|---|---|---|---|---|---|---|---|---|---|---|---|---|---|---|
| 50361 | 18510069-0626218 | 33.8 ± 0.4 | 5960 ± 59 | 4.15 ± 0.15 | -0.07 ± 0.18 | 124 ± 37 | 1 | Y | Y | Y | Y | N | Y | … |
| 50362 | 18510070-0614070 | 36.9 ± 0.6 | 5236 ± 253 | 4.66 ± 0.33 | -0.18 ± 0.30 | <80 | 3 | Y | Y | Y | N | … | Y | … |
| 50364 | 18510073-0622333 | 4.8 ± 3.0 | 6158 ± 35 | … | -1.69 ± 1.61 | … | … | N | … | … | … | … | n | … |
| 50366 | 18510081-0615368 | 63.4 ± 0.4 | 6003 ± 112 | 4.32 ± 0.24 | 0.13 ± 0.13 | … | … | N | … | … | … | … | n | … |
| 50372 | 18510105-0617110 | 31.3 ± 10.0 | 6067 ± 95 | … | … | … | … | Y | … | … | … | … | n | … |
| 50378 | 18510134-0607348 | 11.5 ± 0.3 | 5200 ± 163 | 3.75 ± 0.52 | -0.75 ± 0.24 | … | … | N | … | … | … | … | n | … |
| 50379 | 18510134-0621224 | 11.0 ± 0.9 | 4788 ± 747 | 4.48 ± 0.82 | -0.12 ± 0.53 | … | … | N | … | … | … | … | n | … |
| 50380 | 18510136-0619082 | 39.3 ± 0.9 | 5632 ± 96 | 4.00 ± 0.60 | 0.08 ± 0.32 | … | … | Y | … | … | … | … | n | … |
| 50382 | 18510153-0617283 | 35.9 ± 0.5 | 6150 ± 215 | 4.63 ± 0.33 | -0.09 ± 0.27 | 109 ± 55 | 1 | Y | Y | Y | Y | Y | Y | … |
| 50383 | 18510162-0619168 | 34.5 ± 0.4 | 5940 ± 289 | 4.78 ± 0.39 | 0.04 ± 0.34 | 112 ± 53 | 1 | Y | Y | Y | Y | Y | Y | … |
| 50384 | 18510164-0614156 | 35.3 ± 0.5 | 5555 ± 298 | 5.05 ± 0.56 | 0.06 ± 0.15 | 138 ± 81 | 1 | Y | Y | Y | Y | N | Y | … |
| 50385 | 18510164-0620098 | 36.5 ± 0.3 | 6636 ± 373 | 4.42 ± 0.44 | 0.12 ± 0.19 | 84 ± 24 | 1 | Y | Y | Y | Y | Y | Y | … |
| 50387 | 18510170-0617077 | 33.6 ± 0.4 | 5746 ± 60 | 4.62 ± 0.32 | 0.05 ± 0.12 | 80 ± 55 | 1 | Y | Y | Y | Y | … | Y | … |
| 50390 | 18510179-0609245 | -22.3 ± 0.3 | 5774 ± 81 | 4.09 ± 0.22 | 0.32 ± 0.19 | 66 ± 28 | 1 | N | … | … | … | … | n | NG |
| 50391 | 18510180-0613461 | 33.9 ± 0.7 | 6849 ± 650 | 4.58 ± 0.80 | 0.23 ± 0.31 | … | … | Y | … | … | … | … | n | … |
| 50392 | 18510180-0615554 | 15.7 ± 0.2 | 5811 ± 62 | 4.31 ± 0.31 | -0.03 ± 0.12 | … | … | N | … | … | … | … | n | … |
| 50393 | 18510202-0614371 | 36.9 ± 0.4 | 6373 ± 193 | 4.42 ± 0.32 | 0.09 ± 0.14 | … | … | Y | … | … | … | … | n | … |
| 50394 | 18510208-0615133 | 32.9 ± 0.5 | 5934 ± 139 | 5.01 ± 0.56 | 0.09 ± 0.14 | 135 ± 69 | 1 | Y | Y | Y | Y | N | Y | … |
| 50395 | 18510212-0620493 | 90.9 ± 0.8 | 5532 ± 70 | 4.28 ± 0.10 | -0.04 ± 0.15 | … | … | N | … | … | … | … | n | … |
| 50397 | 18510221-0613303 | -15.6 ± 0.3 | 5985 ± 141 | 4.12 ± 0.49 | -0.47 ± 0.17 | … | … | N | … | … | … | … | n | … |
| 50398 | 18510221-0615206 | 35.9 ± 3.0 | 6205 ± 233 | 2.62 ± 1.58 | -1.06 ± 0.25 | … | … | Y | … | … | … | … | n | … |
| 50400 | 18510229-0613161 | 34.7 ± 0.2 | 6393 ± 138 | 4.20 ± 0.31 | 0.11 ± 0.14 | 46 ± 41 | 1 | Y | Y | Y | Y | Y | Y | … |
| 50402 | 18510234-0614430 | 33.8 ± 0.9 | 7070 ± 846 | 3.89 ± 0.28 | 0.13 ± 0.16 | … | … | Y | … | … | … | … | n | … |
| 50403 | 18510235-0622567 | 39.0 ± 0.4 | 7120 ± 600 | 4.18 ± 0.15 | 0.20 ± 0.13 | … | … | Y | … | … | … | … | n | … |
| 50406 | 18510243-0607506 | 37.8 ± 0.9 | 6766 ± 79 | … | 0.31 ± 0.06 | … | … | Y | … | … | … | … | n | … |
| 50408 | 18510248-0615320 | 26.3 ± 4.9 | … | … | … | … | … | N | … | … | … | … | n | … |
| 50411 | 18510274-0626391 | 3.8 ± 0.3 | 6217 ± 147 | 4.28 ± 0.31 | 0.00 ± 0.13 | … | … | N | … | … | … | … | n | … |
| 50420 | 18510306-0616282 | 36.0 ± 0.4 | 6378 ± 133 | 4.50 ± 0.54 | 0.06 ± 0.20 | 68 ± 36 | 1 | Y | Y | Y | Y | … | Y | … |
| 50422 | 18510319-0611014 | -1.6 ± 0.1 | 5819 ± 154 | 4.04 ± 0.21 | -0.40 ± 0.13 | … | … | N | … | … | … | … | n | … |
| 50424 | 18510322-0625388 | -1.9 ± 0.1 | 5099 ± 61 | 4.64 ± 0.22 | -0.11 ± 0.13 | … | … | N | … | … | … | … | n | … |
| 50425 | 18510330-0608167 | 37.7 ± 0.4 | 7394 ± 60 | … | … | … | … | Y | … | … | … | … | n | … |
| 50426 | 18510333-0617552 | 51.3 ± 0.3 | 6842 ± 547 | 4.38 ± 0.34 | 0.20 ± 0.19 | 61 ± 25 | 1 | N | … | … | … | Y | n | NG |
| 50427 | 18510348-0616494 | 144.5 ± 0.2 | 4818 ± 152 | 2.74 ± 0.34 | -0.03 ± 0.23 | … | … | N | … | … | … | … | n | G |
| 50429 | 18510355-0618135 | -54.1 ± 0.4 | 5588 ± 192 | 4.09 ± 0.54 | 0.52 ± 0.29 | … | … | N | … | … | … | … | n | … |
| 50431 | 18510361-0626456 | 37.4 ± 0.8 | 6909 ± 375 | 4.11 ± 0.20 | 0.08 ± 0.31 | … | … | Y | … | … | … | … | n | … |
| 50434 | 18510372-0609321 | 35.1 ± 0.3 | 6471 ± 234 | 4.59 ± 0.50 | 0.08 ± 0.15 | 45 ± 19 | 1 | Y | Y | Y | Y | … | Y | … |
| 50437 | 18510391-0619588 | 34.1 ± 0.6 | 6514 ± 227 | 4.84 ± 0.77 | -0.16 ± 0.23 | … | … | Y | … | … | … | … | n | … |
| 50439 | 18510393-0626377 | 43.4 ± 0.6 | 6813 ± 243 | 4.59 ± 0.58 | -0.01 ± 0.22 | … | … | N | … | … | … | … | n | … |
| 50440 | 18510394-0618482 | 39.6 ± 0.5 | 6756 ± 197 | 4.15 ± 0.21 | 0.16 ± 0.19 | 50 ± 32 | 1 | Y | Y | Y | Y | Y | Y | … |
| 50441 | 18510400-0613360 | 35.0 ± 10.0 | 7321 ± 95 | … | … | … | … | Y | … | … | … | … | n | … |
| 50442 | 18510401-0615387 | 54.6 ± 1.4 | 4645 ± 247 | 4.48 ± 0.15 | -0.75 ± 0.64 | … | … | N | … | … | … | … | n | … |
| 50448 | 18510417-0618147 | 38.9 ± 1.8 | 7003 ± 190 | 4.05 ± 0.23 | -0.03 ± 0.20 | … | … | Y | … | … | … | … | n | … |
| 50449 | 18510417-0619206 | 30.6 ± 0.5 | 5861 ± 161 | 4.62 ± 0.24 | 0.03 ± 0.22 | 128 ± 92 | 1 | N | … | … | … | N | n | NG |
| 50450 | 18510427-0612144 | 38.7 ± 0.5 | 6871 ± 421 | 4.66 ± 0.65 | 0.00 ± 0.20 | 27 ± 18 | 1 | Y | Y | Y | Y | Y | Y | … |
| 50451 | 18510433-0614539 | 81.7 ± 0.4 | 4178 ± 468 | 2.57 ± 1.11 | 0.12 ± 0.87 | … | … | N | … | … | … | … | n | G |
| 50452 | 18510438-0614236 | 12.0 ± 2.4 | 6097 ± 77 | 4.63 ± 0.37 | 0.22 ± 0.22 | … | … | N | … | … | … | … | n | … |
| 50453 | 18510445-0606018 | 8.9 ± 0.8 | 6574 ± 411 | 4.65 ± 0.65 | 0.48 ± 0.58 | <56 | 3 | N | … | … | … | … | n | NG |
| 50457 | 18510460-0618285 | 35.2 ± 0.7 | … | … | … | … | … | Y | … | … | … | … | n | … |
| 50460 | 18510465-0617295 | 37.1 ± 0.6 | 6670 ± 549 | 4.35 ± 0.29 | 0.30 ± 0.33 | … | … | Y | … | … | … | … | n | … |
| 50464 | 18510485-0614460 | 85.1 ± 0.3 | 4926 ± 79 | 2.63 ± 0.35 | 0.04 ± 0.19 | … | … | N | … | … | … | … | n | G |
| 50471 | 18510517-0614197 | 38.2 ± 0.3 | 6738 ± 416 | 4.21 ± 0.37 | 0.16 ± 0.33 | 75 ± 50 | 1 | Y | Y | Y | Y | Y | Y | … |
| 50474 | 18510539-0621392 | 38.0 ± 0.7 | 6892 ± 390 | 4.43 ± 0.27 | 0.24 ± 0.20 | … | … | Y | … | … | … | … | n | … |
| 50476 | 18510541-0620584 | 38.8 ± 0.3 | 6023 ± 132 | 4.22 ± 0.26 | -0.17 ± 0.19 | 49 ± 30 | 1 | Y | Y | Y | N | … | Y | … |
| 50479 | 18510552-0613529 | 28.6 ± 2.0 | 7090 ± 291 | 3.96 ± 0.17 | 0.00 ± 0.16 | … | … | N | … | … | … | … | n | … |
| 50480 | 18510556-0615353 | 34.3 ± 0.9 | 6727 ± 528 | 4.13 ± 0.34 | 0.13 ± 0.19 | … | … | Y | … | … | … | … | n | … |
| 50481 | 18510558-0626204 | -6.4 ± 0.3 | 6144 ± 219 | 4.56 ± 0.20 | 0.36 ± 0.21 | … | … | N | … | … | … | … | n | … |
| 50483 | 18510568-0622427 | 15.8 ± 0.3 | 6133 ± 137 | 4.59 ± 0.38 | -0.07 ± 0.13 | 73 ± 19 | 1 | N | … | … | … | … | n | NG |
| 50487 | 18510582-0608139 | 53.0 ± 0.3 | 6482 ± 241 | 4.74 ± 0.68 | 0.06 ± 0.14 | <34 | 3 | N | … | … | … | … | n | NG |
| 50488 | 18510582-0615491 | 5.0 ± 0.3 | 5688 ± 49 | 4.39 ± 0.33 | 0.00 ± 0.14 | … | … | N | … | … | … | … | n | … |

**Table C.9.** continued.

| ID | CNAME | RV (km s$^{-1}$) | $T_{eff}$ (K) | logg (dex) | [Fe/H] (dex) | EW(Li)$^a$ (mÅ) | EW(Li) error flag$^b$ | Membership RV | Li | logg | [Fe/H] | Gaia study Cantat-Gaudin$^c$ | Final$^d$ | NMs with Li$^e$ |
|---|---|---|---|---|---|---|---|---|---|---|---|---|---|---|
| 50490 | 18510590-0612198 | -8.2 ± 0.3 | 6036 ± 173 | 4.28 ± 0.05 | 0.29 ± 0.16 | 60 ± 30 | 1 | N | … | … | … | N | n | NG |
| 50494 | 18510605-0613315 | 34.3 ± 10.0 | 7118 ± 92 | … | … | … | … | Y | … | … | … | … | n | … |
| 50496 | 18510611-0607538 | 20.7 ± 0.5 | 6691 ± 416 | 4.46 ± 0.43 | 0.10 ± 0.21 | … | … | N | … | … | … | … | n | … |
| 50497 | 18510628-0614524 | 56.0 ± 0.5 | 4096 ± 533 | 2.19 ± 1.13 | -0.29 ± 0.50 | … | … | N | … | … | … | … | n | G |
| 50498 | 18510630-0615318 | 37.1 ± 0.9 | 7428 ± 501 | 4.05 ± 0.22 | -0.06 ± 0.18 | … | … | Y | … | … | … | … | n | … |
| 50499 | 18510630-0619580 | 10.0 ± 10.0 | 7174 ± 78 | … | … | … | … | N | … | … | … | … | n | … |
| 50500 | 18510632-0618158 | 35.6 ± 0.6 | 6576 ± 140 | 4.38 ± 0.50 | -0.02 ± 0.31 | 79 ± 51 | 1 | Y | Y | Y | Y | Y | Y | … |
| 50502 | 18510636-0611016 | 32.4 ± 0.2 | 6255 ± 167 | 3.90 ± 0.31 | 0.36 ± 0.18 | 111 ± 42 | 1 | Y | Y | Y | N | … | Y | … |
| 50504 | 18510646-0625316 | 68.5 ± 0.3 | 6201 ± 124 | 4.27 ± 0.20 | 0.16 ± 0.18 | … | … | N | … | … | … | … | n | … |
| 50505 | 18510650-0610467 | 35.4 ± 0.5 | 6116 ± 91 | 4.81 ± 0.56 | -0.09 ± 0.18 | 107 ± 49 | 1 | Y | Y | Y | Y | N | Y | … |
| 50506 | 18510650-0619381 | 35.3 ± 1.2 | 7397 ± 109 | … | … | 32 ± 23 | 1 | Y | Y | … | … | Y | Y | … |
| 50508 | 18510658-0607402 | 10.4 ± 0.7 | 7337 ± 93 | … | … | … | … | N | … | … | … | … | n | … |
| 50509 | 18510661-0624375 | 32.8 ± 0.7 | 7160 ± 560 | 4.15 ± 0.21 | 0.11 ± 0.15 | 40 ± 31 | 1 | Y | Y | Y | Y | N | Y | … |
| 50510 | 18510673-0610010 | 35.1 ± 0.6 | 6371 ± 211 | 4.57 ± 0.40 | -0.10 ± 0.35 | … | … | Y | … | … | … | … | n | … |
| 50511 | 18510673-0613144 | -11.1 ± 0.7 | 5574 ± 378 | 4.84 ± 1.06 | -0.06 ± 0.17 | … | … | N | … | … | … | … | n | … |
| 50514 | 18510680-0617043 | 31.5 ± 0.4 | 6554 ± 240 | 4.52 ± 0.37 | -0.02 ± 0.19 | <34 | 3 | Y | Y | Y | Y | … | Y | … |
| 50515 | 18510687-0613424 | 38.8 ± 0.5 | 5781 ± 110 | 4.66 ± 0.56 | 0.40 ± 0.15 | … | … | Y | … | … | … | … | n | … |
| 50517 | 18510695-0624151 | 55.8 ± 0.3 | 6706 ± 442 | 4.41 ± 0.44 | 0.09 ± 0.22 | 53 ± 29 | 1 | N | … | … | … | … | n | NG |
| 50518 | 18510700-0615069 | 39.9 ± 0.4 | 7127 ± 670 | 4.17 ± 0.17 | 0.06 ± 0.35 | 62 ± 35 | 1 | Y | Y | Y | Y | … | Y | … |
| 50520 | 18510705-0613358 | -56.2 ± 1.6 | 6402 ± 171 | 5.66 ± 0.50 | -0.30 ± 0.16 | … | … | N | … | … | … | … | n | … |
| 50521 | 18510706-0609582 | 34.5 ± 0.3 | 6478 ± 248 | 4.60 ± 0.53 | 0.12 ± 0.22 | 56 ± 40 | 1 | Y | Y | Y | Y | Y | Y | … |
| 50522 | 18510708-0615442 | -2.7 ± 0.3 | 5845 ± 148 | 4.46 ± 0.11 | 0.27 ± 0.14 | … | … | N | … | … | … | … | n | … |
| 50524 | 18510729-0613388 | 35.2 ± 0.3 | 6689 ± 395 | 4.20 ± 0.22 | 0.15 ± 0.19 | 75 ± 23 | 1 | Y | Y | Y | Y | N | Y | … |
| 50525 | 18510731-0612538 | 25.4 ± 10.0 | 7422 ± 88 | … | … | … | … | N | … | … | … | … | n | … |
| 50528 | 18510741-0614559 | 35.7 ± 0.3 | … | … | … | … | … | Y | … | … | … | … | n | … |
| 50529 | 18510745-0614133 | 34.2 ± 0.9 | 6776 ± 481 | 4.03 ± 0.24 | 0.12 ± 0.20 | 31 ± 24 | 1 | Y | Y | Y | Y | Y | Y | … |
| 50530 | 18510747-0615353 | 34.9 ± 0.4 | 5885 ± 242 | 5.08 ± 0.55 | 0.11 ± 0.33 | 148 ± 68 | 1 | Y | Y | Y | Y | … | Y | … |
| 50531 | 18510750-0613339 | 35.6 ± 0.6 | 6718 ± 422 | 4.65 ± 0.63 | 0.18 ± 0.21 | 72 ± 49 | 1 | Y | Y | Y | Y | N | Y | … |
| 50532 | 18510756-0624184 | 44.0 ± 0.4 | 6756 ± 495 | 4.34 ± 0.36 | 0.14 ± 0.19 | <47 | 3 | Y | Y | Y | Y | Y | Y | … |
| 50534 | 18510768-0608490 | 9.6 ± 0.4 | 6042 ± 115 | 3.99 ± 0.01 | -0.22 ± 0.31 | 39 ± 30 | 1 | N | … | … | … | … | n | NG |
| 50537 | 18510784-0613183 | 37.7 ± 0.7 | 6414 ± 228 | 4.84 ± 0.73 | -0.37 ± 0.41 | 29 ± 23 | 1 | Y | Y | Y | N | Y | Y | … |
| 50538 | 18510786-0616058 | 15.2 ± 0.3 | 6275 ± 168 | 4.57 ± 0.53 | 0.22 ± 0.18 | 103 ± 40 | 1 | N | … | … | … | … | n | NG |
| 50541 | 18510795-0628519 | 44.2 ± 2.1 | … | … | … | … | … | N | … | … | … | … | n | … |
| 50542 | 18510799-0612152 | 40.5 ± 0.6 | 5406 ± 204 | 4.21 ± 0.70 | -0.14 ± 0.16 | 113 ± 44 | 1 | N | … | … | … | N | n | NG |
| 50543 | 18510799-0617342 | 37.6 ± 1.1 | 6871 ± 78 | 3.96 ± 0.17 | -0.05 ± 0.30 | 35 ± 25 | 1 | Y | Y | Y | Y | … | Y | … |
| 50544 | 18510801-0615160 | 37.6 ± 0.8 | 6336 ± 257 | 4.48 ± 0.36 | 0.06 ± 0.19 | 67 ± 59 | 1 | Y | Y | Y | Y | N | Y | … |
| 50545 | 18510802-0607249 | 41.9 ± 0.4 | 6450 ± 163 | 3.81 ± 0.33 | -0.01 ± 0.12 | 86 ± 35 | 1 | N | … | … | … | … | n | NG |
| 50546 | 18510810-0614192 | 34.5 ± 0.8 | 6574 ± 531 | 3.95 ± 0.08 | 0.23 ± 0.28 | … | … | Y | … | … | … | … | n | … |
| 50547 | 18510812-0625438 | 45.4 ± 0.5 | 5937 ± 149 | 4.37 ± 0.10 | 0.09 ± 0.13 | 34 ± 27 | 1 | N | … | … | … | … | n | NG |
| 50550 | 18510838-0623244 | -22.5 ± 0.3 | 5990 ± 141 | 4.55 ± 0.28 | 0.02 ± 0.13 | 66 ± 32 | 1 | N | … | … | … | … | n | NG |
| 50551 | 18510848-0618256 | 34.9 ± 0.4 | 6458 ± 310 | 4.39 ± 0.36 | 0.08 ± 0.17 | 75 ± 34 | 1 | Y | Y | Y | Y | … | Y | … |
| 50552 | 18510851-0610102 | 37.8 ± 0.7 | 6527 ± 302 | 3.78 ± 0.21 | 0.06 ± 0.22 | … | … | Y | … | … | … | … | n | … |
| 50554 | 18510855-0614530 | -9.4 ± 0.5 | 5025 ± 212 | 4.85 ± 0.56 | -0.15 ± 0.18 | … | … | N | … | … | … | … | n | … |
| 50555 | 18510876-0623526 | 41.5 ± 0.3 | 5764 ± 178 | 4.06 ± 0.14 | 0.31 ± 0.16 | <32 | 3 | N | … | … | … | … | n | NG |
| 50556 | 18510878-0612260 | 37.2 ± 2.8 | 5900 ± 253 | 5.05 ± 0.72 | -1.29 ± 0.23 | … | … | Y | … | … | … | … | n | … |
| 50557 | 18510883-0608436 | 35.6 ± 0.3 | 6998 ± 71 | 4.61 ± 0.13 | 0.24 ± 0.07 | 53 ± 31 | 1 | Y | Y | Y | Y | Y | Y | … |
| 50558 | 18510883-0615498 | 31.9 ± 0.5 | … | … | … | … | … | Y | … | … | … | … | n | … |
| 50559 | 18510886-0616218 | 33.8 ± 0.8 | 7428 ± 75 | … | … | … | … | Y | … | … | … | … | n | … |
| 50560 | 18510889-0617166 | -40.2 ± 0.8 | 5713 ± 76 | 4.34 ± 0.06 | 0.19 ± 0.24 | … | … | N | … | … | … | … | n | … |
| 50562 | 18510898-0621313 | 47.4 ± 10.0 | … | … | … | … | … | N | … | … | … | … | n | … |
| 50563 | 18510902-0613132 | 95.8 ± 0.6 | 4750 ± 228 | 2.72 ± 0.44 | -0.15 ± 0.22 | … | … | N | … | … | … | … | n | G |
| 50564 | 18510904-0610150 | 11.9 ± 0.2 | 7473 ± 655 | 4.19 ± 0.16 | 0.12 ± 0.14 | 67 ± 20 | 1 | N | … | … | … | Y | n | NG |
| 50565 | 18510905-0608527 | 32.0 ± 1.0 | 6568 ± 325 | 4.53 ± 0.39 | 0.11 ± 0.22 | … | … | Y | … | … | … | … | n | … |
| 50567 | 18510915-0611254 | 33.3 ± 0.4 | … | … | … | … | … | Y | … | … | … | … | n | … |
| 50568 | 18510917-0613278 | 35.3 ± 0.4 | 6991 ± 455 | 4.63 ± 0.53 | 0.12 ± 0.20 | 50 ± 43 | 1 | Y | Y | Y | Y | Y | Y | … |
| 50569 | 18510921-0614505 | 34.3 ± 10.0 | 7476 ± 99 | … | … | … | … | Y | … | … | … | … | n | … |
| 50570 | 18510924-0610595 | 37.8 ± 0.6 | 7167 ± 808 | 4.12 ± 0.20 | 0.04 ± 0.36 | 67 ± 55 | 1 | Y | Y | Y | Y | … | Y | … |
| 50572 | 18510930-0623513 | 20.0 ± 2.8 | … | … | … | … | … | N | … | … | … | … | n | … |
| 50573 | 18510934-0615150 | 41.9 ± 10.0 | … | … | … | … | … | N | … | … | … | … | n | … |

**Table C.9.** continued.

| ID | CNAME | RV (km s$^{-1}$) | $T_{\rm eff}$ (K) | $logg$ (dex) | [Fe/H] (dex) | EW(Li)$^a$ (mÅ) | EW(Li) error flag$^b$ | Membership RV | Li | $logg$ | [Fe/H] | Gaia study Cantat-Gaudin$^c$ | Final$^d$ | NMs with Li$^e$ |
|---|---|---|---|---|---|---|---|---|---|---|---|---|---|---|
| 50574 | 18510936-0617562 | 33.4 ± 0.7 | 6402 ± 208 | 4.76 ± 0.51 | 0.18 ± 0.18 | 68 ± 30 | … | Y | Y | Y | Y | … | Y | … |
| 50575 | 18510937-0626377 | 11.4 ± 0.4 | 5838 ± 399 | 4.35 ± 0.30 | 0.11 ± 0.27 | … | … | N | … | … | … | … | n | … |
| 50577 | 18510950-0618366 | 30.9 ± 0.5 | 6904 ± 417 | 4.56 ± 0.46 | 0.22 ± 0.21 | 81 ± 61 | 1 | Y | Y | Y | Y | Y | Y | … |
| 50579 | 18510953-0617014 | 36.8 ± 0.4 | 6390 ± 198 | 4.20 ± 0.25 | 0.01 ± 0.19 | 40 ± 33 | 1 | Y | Y | Y | Y | N | Y | … |
| 50581 | 18510968-0617225 | 14.1 ± 0.3 | 5664 ± 232 | 4.46 ± 0.17 | 0.37 ± 0.17 | … | … | N | … | … | … | … | n | … |
| 50582 | 18510974-0628515 | 19.6 ± 8.7 | … | … | … | … | … | N | … | … | … | … | n | … |
| 50583 | 18510979-0612334 | 53.6 ± 2.4 | 5127 ± 289 | … | -0.42 ± 0.30 | … | … | N | … | … | … | … | n | Li-rich G |
| 50584 | 18510982-0617447 | 102.2 ± 0.6 | 5937 ± 324 | 4.19 ± 0.07 | 0.36 ± 0.24 | <166 | 3 | N | … | … | … | … | n | NG |
| 50587 | 18510990-0617003 | 38.5 ± 0.8 | 6768 ± 317 | 4.35 ± 0.33 | 0.11 ± 0.20 | 58 ± 56 | 1 | Y | Y | Y | Y | … | Y | … |
| 50589 | 18511001-0622205 | 41.8 ± 0.9 | 6773 ± 549 | 3.75 ± 0.16 | 0.10 ± 0.20 | 50 ± 41 | 1 | Y | Y | Y | Y | Y | Y | … |
| 50590 | 18511002-0611503 | 53.3 ± 1.0 | 5071 ± 134 | 3.01 ± 0.44 | -0.47 ± 0.11 | … | … | N | … | … | … | … | n | Li-rich G |
| 50591 | 18511002-0616414 | 35.6 ± 0.4 | 6430 ± 237 | 4.51 ± 0.35 | 0.03 ± 0.18 | <38 | 3 | Y | Y | Y | Y | … | Y | … |
| 50592 | 18511003-0615403 | 38.0 ± 0.7 | … | … | … | … | … | Y | … | … | … | … | n | … |
| 50593 | 18511012-0616141 | 32.7 ± 0.3 | 6245 ± 143 | 4.25 ± 0.16 | 0.12 ± 0.17 | … | … | Y | … | … | … | … | n | … |
| 50595 | 18511018-0614226 | 38.2 ± 0.5 | 7271 ± 79 | … | … | 26 ± 14 | 1 | Y | Y | … | … | … | Y | … |
| 50597 | 18511030-0626321 | -0.5 ± 0.3 | 5976 ± 166 | 4.37 ± 0.23 | 0.06 ± 0.13 | … | … | N | … | … | … | … | n | … |
| 50600 | 18511042-0614077 | 91.0 ± 0.3 | 6026 ± 102 | 4.45 ± 0.15 | 0.29 ± 0.16 | … | … | N | … | … | … | … | n | … |
| 50602 | 18511055-0621199 | 24.0 ± 0.3 | 6390 ± 222 | 4.19 ± 0.11 | 0.07 ± 0.14 | … | … | N | … | … | … | … | n | … |
| 50603 | 18511060-0619206 | 29.8 ± 0.1 | 7312 ± 362 | 4.09 ± 0.23 | -0.08 ± 0.19 | 54 ± 36 | 1 | Y | Y | Y | Y | Y | Y | … |
| 50604 | 18511063-0618531 | 35.2 ± 10.0 | 4653 ± 1423 | 4.09 ± 0.51 | -1.87 ± 0.36 | … | … | Y | … | … | … | … | n | … |
| 50605 | 18511066-0610200 | 20.2 ± 0.4 | 6777 ± 500 | 4.30 ± 0.27 | 0.06 ± 0.30 | … | … | N | … | … | … | … | n | … |
| 50606 | 18511078-0611282 | -4.4 ± 0.4 | 7442 ± 55 | … | … | … | … | N | … | … | … | … | n | … |
| 50607 | 18511083-0611454 | 36.9 ± 0.9 | 6351 ± 153 | 3.74 ± 0.25 | 0.06 ± 0.34 | … | … | Y | … | … | … | … | n | … |
| 50608 | 18511084-0613266 | 38.0 ± 0.7 | 6541 ± 548 | 4.67 ± 0.57 | 0.24 ± 0.21 | … | … | Y | … | … | … | … | n | … |
| 50610 | 18511086-0619334 | -39.1 ± 10.0 | 6176 ± 16 | 3.87 ± 0.04 | -0.16 ± 0.12 | … | … | N | … | … | … | … | n | … |
| 50611 | 18511087-0617168 | 34.7 ± 0.4 | 5153 ± 33 | 4.44 ± 0.55 | 0.25 ± 0.21 | 126 ± 74 | 1 | Y | Y | Y | Y | … | Y | … |
| 50614 | 18511102-0608370 | 33.1 ± 0.6 | 6613 ± 96 | 4.57 ± 0.63 | 0.10 ± 0.19 | 117 ± 98 | 1 | Y | Y | Y | Y | … | Y | … |
| 50615 | 18511102-0617320 | 38.1 ± 0.5 | 5575 ± 253 | 4.20 ± 0.25 | 0.09 ± 0.16 | 150 ± 73 | 1 | Y | Y | Y | Y | … | Y | … |
| 50616 | 18511109-0616052 | 93.8 ± 0.5 | 5526 ± 91 | 3.98 ± 0.23 | -0.22 ± 0.18 | … | … | N | … | … | … | … | n | … |
| 50617 | 18511109-0625449 | -64.9 ± 0.3 | 5643 ± 124 | 4.28 ± 0.28 | -0.12 ± 0.22 | … | … | N | … | … | … | … | n | … |
| 50619 | 18511122-0607436 | 6.9 ± 0.1 | 5883 ± 155 | 4.45 ± 0.24 | 0.18 ± 0.16 | 34 ± 26 | 1 | N | … | … | … | … | n | NG |
| 50621 | 18511131-0617491 | 13.3 ± 0.2 | 5770 ± 46 | 4.74 ± 0.25 | -0.02 ± 0.18 | 64 ± 0 | … | N | … | … | … | … | n | NG |
| 50622 | 18511133-0615162 | 38.2 ± 0.6 | 6837 ± 267 | 4.49 ± 0.48 | -0.02 ± 0.23 | … | … | Y | … | … | … | … | n | … |
| 50624 | 18511135-0613028 | 1.5 ± 0.3 | 6235 ± 109 | 4.14 ± 0.18 | -0.18 ± 0.19 | 42 ± 22 | 1 | N | … | … | … | … | n | NG |
| 50626 | 18511146-0618403 | -34.9 ± 0.2 | 4892 ± 138 | 3.11 ± 0.31 | -0.03 ± 0.17 | … | … | N | … | … | … | … | n | G |
| 50627 | 18511148-0614238 | 34.4 ± 0.5 | 5959 ± 389 | 4.90 ± 0.52 | 0.09 ± 0.28 | 113 ± 81 | 1 | Y | Y | Y | Y | Y | Y | … |
| 50628 | 18511152-0614157 | 37.5 ± 0.8 | 5914 ± 303 | 3.41 ± 0.40 | -0.07 ± 0.58 | … | … | Y | … | … | … | … | n | … |
| 50629 | 18511152-0622253 | 37.2 ± 1.0 | 6507 ± 221 | 4.95 ± 0.88 | -0.80 ± 0.73 | 48 ± 39 | 1 | Y | Y | Y | N | Y | Y | … |
| 50630 | 18511154-0612214 | -3.1 ± 0.3 | 5885 ± 159 | 4.34 ± 0.09 | 0.35 ± 0.17 | 97 ± 29 | 1 | N | … | … | … | … | n | NG |
| 50631 | 18511154-0623277 | 3.5 ± 0.4 | 5551 ± 92 | 4.22 ± 0.19 | 0.20 ± 0.12 | <123 | 3 | N | … | … | … | … | n | NG |
| 50633 | 18511164-0614425 | 35.7 ± 0.5 | 6625 ± 349 | 4.12 ± 0.08 | 0.18 ± 0.21 | 80 ± 44 | 1 | Y | Y | Y | Y | Y | Y | … |
| 50636 | 18511175-0621098 | 113.0 ± 0.8 | 6235 ± 242 | 4.19 ± 0.23 | 0.31 ± 0.16 | … | … | N | … | … | … | … | n | … |
| 50638 | 18511194-0617294 | 41.1 ± 0.4 | 6521 ± 239 | 4.38 ± 0.33 | 0.19 ± 0.19 | <53 | 3 | Y | Y | Y | Y | Y | Y | … |
| 50641 | 18511198-0615194 | 48.8 ± 0.4 | 5969 ± 252 | 4.26 ± 0.26 | 0.41 ± 0.20 | <120 | 3 | N | … | … | … | … | n | NG |
| 50642 | 18511208-0614493 | 32.5 ± 1.1 | 7405 ± 890 | 4.12 ± 0.18 | -0.16 ± 0.48 | … | … | Y | … | … | … | … | n | … |
| 50643 | 18511220-0614233 | 24.0 ± 0.4 | 5970 ± 173 | 4.26 ± 0.41 | 0.27 ± 0.13 | <66 | 3 | N | … | … | … | … | n | NG |
| 50648 | 18511230-0620556 | -18.7 ± 0.3 | 5921 ± 87 | 4.05 ± 0.16 | 0.27 ± 0.17 | … | … | N | … | … | … | … | n | … |
| 50649 | 18511233-0617012 | 29.8 ± 0.3 | 6093 ± 151 | 4.21 ± 0.17 | 0.20 ± 0.17 | 30 ± 20 | 1 | N | … | … | … | … | n | NG |
| 50651 | 18511242-0622102 | 87.0 ± 0.9 | 6747 ± 472 | 4.40 ± 0.48 | 0.20 ± 0.20 | 84 ± 79 | 1 | N | … | … | … | N | n | NG |
| 50653 | 18511259-0617119 | 23.2 ± 0.5 | 6671 ± 472 | 4.18 ± 0.39 | 0.19 ± 0.24 | … | … | N | … | … | … | … | n | … |
| 50655 | 18511272-0615413 | 31.3 ± 0.7 | 5923 ± 407 | 4.85 ± 0.40 | 0.31 ± 0.33 | <135 | 3 | Y | Y | Y | Y | N | Y | … |
| 50656 | 18511272-0625467 | 39.4 ± 0.6 | 6239 ± 217 | 4.76 ± 0.81 | -0.01 ± 0.18 | … | … | Y | … | … | … | … | n | … |
| 50657 | 18511276-0614451 | 18.6 ± 0.4 | 6426 ± 237 | 4.20 ± 0.28 | 0.12 ± 0.19 | … | … | N | … | … | … | … | n | … |
| 50658 | 18511284-0615337 | 35.9 ± 0.4 | 6047 ± 118 | 4.74 ± 0.55 | -0.01 ± 0.15 | 102 ± 40 | 1 | Y | Y | Y | Y | … | Y | … |
| 50659 | 18511287-0614307 | 37.8 ± 0.5 | 6686 ± 488 | 3.79 ± 0.21 | 0.20 ± 0.25 | … | … | Y | … | … | … | … | n | … |
| 50661 | 18511293-0615206 | 60.9 ± 1.5 | 5560 ± 61 | 4.04 ± 0.35 | -0.60 ± 0.30 | … | … | N | … | … | … | … | n | … |
| 50662 | 18511295-0616550 | 36.7 ± 1.2 | 6793 ± 364 | 3.94 ± 0.17 | 0.17 ± 0.33 | 37 ± 30 | 1 | Y | Y | Y | Y | … | Y | … |
| 50663 | 18511297-0610466 | -22.4 ± 0.3 | 5380 ± 129 | 4.40 ± 0.57 | 0.05 ± 0.13 | 41 ± 25 | 1 | N | … | … | … | … | n | NG |
| 50664 | 18511301-0615524 | 33.3 ± 0.7 | 5355 ± 136 | 4.56 ± 0.16 | 0.02 ± 0.26 | … | … | Y | … | … | … | … | n | … |

**Table C.9.** continued.

| ID | CNAME | RV (km s$^{-1}$) | $T_{\text{eff}}$ (K) | log g (dex) | [Fe/H] (dex) | EW(Li)$^a$ (mÅ) | EW(Li) error flag$^b$ | Membership RV | Li | log g | [Fe/H] | Gaia study Cantat-Gaudin$^c$ | Final$^d$ | NMs with Li$^e$ |
|---|---|---|---|---|---|---|---|---|---|---|---|---|---|---|
| 50667 | 18511318-0628238 | 48.9 ± 16.2 | … | … | … | … | … | N | … | … | … | … | n | … |
| 50668 | 18511320-0620387 | 41.4 ± 0.4 | 6580 ± 320 | 4.26 ± 0.27 | 0.16 ± 0.20 | 43 ± 27 | 1 | N | … | … | … | N | n | NG |
| 50669 | 18511324-0612410 | 22.6 ± 1.6 | 6146 ± 1219 | 3.72 ± 0.41 | -0.30 ± 0.13 | … | … | N | … | … | … | … | n | … |
| 50671 | 18511331-0614374 | 41.6 ± 0.7 | 6876 ± 581 | 4.92 ± 0.91 | 0.21 ± 0.25 | 69 ± 47 | 1 | Y | Y | Y | Y | Y | Y | … |
| 50673 | 18511351-0620373 | 8.2 ± 0.3 | 5905 ± 100 | 3.93 ± 0.42 | 0.11 ± 0.19 | 33 ± 26 | 1 | N | … | … | … | … | n | NG |
| 50674 | 18511355-0612375 | 34.9 ± 1.0 | 6514 ± 354 | 4.14 ± 0.08 | 0.16 ± 0.20 | … | … | Y | … | … | … | … | n | … |
| 50675 | 18511356-0615363 | 37.3 ± 0.6 | 6323 ± 220 | 4.52 ± 0.29 | 0.29 ± 0.22 | 85 ± 35 | 1 | Y | Y | Y | Y | Y | Y | … |
| 50676 | 18511366-0618194 | 37.1 ± 0.9 | 6804 ± 445 | 4.12 ± 0.20 | 0.14 ± 0.25 | … | … | Y | … | … | … | … | n | … |
| 50679 | 18511404-0610577 | 37.5 ± 0.5 | 6060 ± 98 | 4.95 ± 0.79 | -0.10 ± 0.18 | 51 ± 33 | 1 | Y | Y | Y | Y | N | Y | … |
| 50680 | 18511406-0613078 | 34.2 ± 0.9 | 5217 ± 158 | 2.89 ± 0.46 | -0.27 ± 0.12 | … | … | Y | … | … | … | … | n | … |
| 50682 | 18511422-0609452 | -31.2 ± 0.3 | 6038 ± 184 | 4.35 ± 0.20 | 0.35 ± 0.17 | 56 ± 41 | 1 | N | … | … | … | … | n | NG |
| 50683 | 18511427-0615506 | 37.4 ± 1.0 | 6495 ± 95 | 5.07 ± 0.90 | -0.79 ± 0.72 | … | … | Y | … | … | … | … | n | … |
| 50684 | 18511428-0615151 | 33.7 ± 0.5 | 6285 ± 201 | 4.86 ± 0.67 | 0.07 ± 0.19 | 55 ± 34 | 1 | Y | Y | Y | Y | Y | Y | … |
| 50685 | 18511433-0614259 | -43.7 ± 0.3 | 5696 ± 233 | 4.48 ± 0.44 | 0.37 ± 0.23 | … | … | N | … | … | … | … | n | … |
| 50687 | 18511436-0621598 | -1.9 ± 0.3 | 5969 ± 118 | 3.72 ± 0.39 | 0.05 ± 0.15 | … | … | N | … | … | … | … | n | … |
| 50689 | 18511445-0619430 | 35.4 ± 0.4 | 6516 ± 190 | 4.69 ± 0.53 | -0.01 ± 0.12 | 38 ± 27 | 1 | Y | Y | Y | Y | N | Y | … |
| 50690 | 18511454-0615157 | -16.0 ± 0.9 | 5864 ± 473 | 4.86 ± 0.41 | 0.23 ± 1.00 | … | … | N | … | … | … | … | n | … |
| 50691 | 18511459-0616271 | 35.7 ± 0.4 | 6725 ± 463 | 4.40 ± 0.42 | 0.20 ± 0.26 | 93 ± 44 | 1 | Y | Y | Y | Y | Y | Y | … |
| 50694 | 18511472-0618529 | -9.6 ± 0.3 | 6098 ± 192 | 4.44 ± 0.15 | 0.23 ± 0.26 | 74 ± 35 | 1 | N | … | … | … | … | n | NG |
| 50695 | 18511473-0624077 | -6.9 ± 0.3 | 6371 ± 210 | 4.34 ± 0.23 | 0.45 ± 0.26 | … | … | N | … | … | … | … | n | … |
| 50696 | 18511478-0606035 | 4.0 ± 3.5 | 5604 ± 105 | 5.04 ± 0.31 | -0.02 ± 0.11 | … | … | N | … | … | … | … | n | … |
| 50697 | 18511478-0616356 | 35.6 ± 0.5 | 5674 ± 213 | 4.58 ± 0.44 | 0.25 ± 0.15 | <193 | 3 | Y | Y | Y | Y | … | Y | … |
| 50698 | 18511481-0612570 | 21.8 ± 0.2 | 5877 ± 39 | 4.34 ± 0.09 | 0.34 ± 0.16 | <57 | 3 | N | … | … | … | … | n | NG |
| 50700 | 18511493-0624371 | -26.0 ± 0.3 | 5969 ± 235 | 4.42 ± 0.11 | 0.33 ± 0.20 | 74 ± 31 | 1 | N | … | … | … | … | n | NG |
| 50701 | 18511497-0622202 | 38.0 ± 0.3 | 6203 ± 69 | 4.33 ± 0.20 | 0.07 ± 0.14 | … | … | Y | … | … | … | … | n | … |
| 50702 | 18511501-0611126 | 38.4 ± 0.5 | … | … | … | … | … | Y | … | … | … | … | n | … |
| 50704 | 18511517-0615541 | 35.1 ± 10.0 | 7150 ± 162 | 4.04 ± 0.19 | 0.03 ± 0.15 | … | … | Y | … | … | … | … | n | … |
| 50705 | 18511517-0617119 | 46.2 ± 0.2 | 5998 ± 176 | 4.20 ± 0.16 | -0.17 ± 0.29 | 36 ± 0 | … | N | … | … | … | … | n | NG |
| 50707 | 18511522-0613561 | 47.0 ± 0.3 | 6382 ± 257 | 4.46 ± 0.28 | 0.29 ± 0.21 | 39 ± 35 | 1 | N | … | … | … | … | n | NG |
| 50709 | 18511529-0620559 | 42.8 ± 0.1 | 5978 ± 45 | 4.15 ± 0.07 | 0.17 ± 0.14 | … | … | N | … | … | … | … | n | … |
| 50710 | 18511542-0620506 | 50.7 ± 0.3 | 6019 ± 227 | 4.52 ± 0.39 | -0.10 ± 0.16 | … | … | N | … | … | … | … | n | … |
| 50712 | 18511550-0616126 | 32.2 ± 0.1 | 6144 ± 43 | 4.14 ± 0.19 | 0.12 ± 0.25 | 61 ± 34 | 1 | Y | Y | Y | Y | … | Y | … |
| 50714 | 18511576-0625177 | 5.1 ± 0.3 | 6162 ± 281 | 4.20 ± 0.20 | -0.12 ± 0.12 | 52 ± 41 | 1 | N | … | … | … | … | n | NG |
| 50716 | 18511595-0626251 | -8.7 ± 0.3 | 5768 ± 246 | 4.02 ± 0.10 | 0.30 ± 0.14 | … | … | N | … | … | … | … | n | … |
| 50718 | 18511605-0616402 | -1.7 ± 0.3 | 5469 ± 130 | 4.57 ± 0.18 | 0.33 ± 0.18 | … | … | N | … | … | … | … | n | … |
| 50719 | 18511607-0620524 | 32.7 ± 0.4 | 6098 ± 51 | 4.68 ± 0.40 | 0.10 ± 0.15 | 130 ± 64 | 1 | Y | Y | Y | Y | … | Y | … |
| 50721 | 18511623-0617415 | 33.7 ± 0.4 | 6780 ± 483 | 4.37 ± 0.35 | 0.21 ± 0.25 | 40 ± 33 | 1 | Y | Y | Y | Y | … | Y | … |
| 50722 | 18511626-0615122 | 34.4 ± 0.7 | 6404 ± 218 | 5.29 ± 1.25 | -0.44 ± 0.57 | 58 ± 32 | 1 | Y | Y | Y? | N | Y | Y | … |
| 50723 | 18511628-0616549 | 35.6 ± 1.0 | 7085 ± 96 | … | … | … | … | Y | … | … | … | … | n | … |
| 50725 | 18511631-0614261 | 35.4 ± 2.1 | … | … | … | … | … | Y | … | … | … | … | n | … |
| 50726 | 18511652-0617079 | -1.6 ± 0.4 | 5660 ± 85 | 4.75 ± 0.30 | -0.04 ± 0.23 | … | … | N | … | … | … | … | n | … |
| 50727 | 18511658-0610353 | -6.0 ± 0.3 | 6080 ± 165 | 4.53 ± 0.22 | 0.27 ± 0.16 | 68 ± 38 | 1 | N | … | … | … | … | n | NG |
| 50728 | 18511661-0624096 | -17.9 ± 4.0 | 5047 ± 51 | … | -2.53 ± 0.18 | … | … | N | … | … | … | … | n | Li-rich G |
| 50729 | 18511664-0620203 | 33.1 ± 0.6 | 6241 ± 121 | 3.54 ± 0.55 | -0.49 ± 0.41 | <63 | 3 | Y | Y | Y? | N | … | Y | … |
| 50730 | 18511668-0622033 | 39.3 ± 0.2 | 6024 ± 118 | 4.27 ± 0.21 | -0.11 ± 0.14 | … | … | Y | … | … | … | … | n | … |
| 50731 | 18511681-0615094 | 21.1 ± 1.2 | 6407 ± 359 | 4.74 ± 0.83 | 0.02 ± 0.24 | … | … | N | … | … | … | … | n | … |
| 50732 | 18511688-0617302 | -11.9 ± 0.2 | 5918 ± 242 | 4.38 ± 0.11 | 0.27 ± 0.20 | 76 ± 25 | 1 | N | … | … | … | … | n | NG |
| 50733 | 18511696-0615340 | 34.7 ± 0.6 | … | … | … | 27 ± 16 | 1 | Y | N | … | … | Y | n | … |
| 50734 | 18511697-0617556 | -46.7 ± 0.6 | 5052 ± 159 | 3.88 ± 0.41 | -0.15 ± 0.15 | … | … | N | … | … | … | … | n | … |
| 50735 | 18511703-0619533 | 70.8 ± 0.4 | 6236 ± 261 | 4.42 ± 0.23 | 0.26 ± 0.17 | … | … | N | … | … | … | … | n | … |
| 50737 | 18511718-0616252 | 32.4 ± 0.7 | 5368 ± 205 | 4.76 ± 0.52 | -0.41 ± 0.46 | 192 ± 89 | 1 | Y | Y | Y | N | … | Y | … |
| 50738 | 18511719-0622231 | 26.1 ± 10.0 | … | … | … | … | … | N | … | … | … | … | n | … |
| 50739 | 18511726-0623247 | -52.9 ± 1.1 | 4747 ± 707 | 3.95 ± 0.91 | -0.03 ± 0.27 | … | … | N | … | … | … | … | n | … |
| 50740 | 18511730-0615394 | 37.9 ± 0.9 | 7459 ± 884 | 4.17 ± 0.17 | -0.06 ± 0.35 | … | … | Y | … | … | … | … | n | … |
| 50741 | 18511731-0618154 | 13.1 ± 0.3 | 5790 ± 181 | 4.14 ± 0.12 | 0.04 ± 0.21 | … | … | N | … | … | … | … | n | … |
| 50742 | 18511755-0613173 | 35.8 ± 0.9 | 4911 ± 309 | 4.44 ± 0.19 | 0.13 ± 0.23 | … | … | Y | … | … | … | … | n | … |
| 50743 | 18511758-0615177 | 37.0 ± 0.1 | 5764 ± 54 | 4.23 ± 0.18 | 0.25 ± 0.21 | 68 ± 37 | 1 | Y | Y | Y | Y | … | Y | … |
| 50744 | 18511792-0612426 | 35.7 ± 0.8 | 7096 ± 492 | 4.17 ± 0.17 | -0.06 ± 0.35 | … | … | Y | … | … | … | … | n | … |
| 50745 | 18511795-0618107 | 22.4 ± 0.3 | 6526 ± 259 | 4.38 ± 0.33 | 0.14 ± 0.18 | 79 ± 40 | 1 | N | … | … | … | Y | n | NG |

| ID | CNAME | RV (km s$^{-1}$) | $T_{\text{eff}}$ (K) | logg (dex) | [Fe/H] (dex) | EW(Li)$^a$ (mÅ) | EW(Li) error flag$^b$ | Membership | | | | Gaia study Cantat-Gaudin$^c$ | Final$^d$ | NMs with Li$^e$ |
|---|---|---|---|---|---|---|---|---|---|---|---|---|---|---|
| | | | | | | | | RV | Li | logg | [Fe/H] | | | |
| 50746 | 18511801-0615587 | 40.1 ± 0.7 | 6828 ± 390 | 4.09 ± 0.20 | -0.02 ± 0.22 | … | … | N | … | … | … | … | n | … |
| 50747 | 18511808-0617223 | -3.6 ± 0.3 | 6487 ± 227 | 4.25 ± 0.30 | 0.29 ± 0.21 | … | … | N | … | … | … | … | n | … |
| 50748 | 18511815-0622444 | 56.3 ± 0.8 | 6939 ± 434 | 4.16 ± 0.17 | 0.04 ± 0.19 | … | … | N | … | … | … | … | n | … |
| 50749 | 18511828-0615038 | 39.1 ± 0.7 | 6965 ± 671 | 4.40 ± 0.37 | 0.13 ± 0.19 | 28 ± 16 | 1 | Y | Y | Y | Y | Y | Y | … |
| 50751 | 18511836-0619458 | 36.1 ± 0.1 | 6289 ± 198 | 4.60 ± 0.63 | 0.12 ± 0.20 | 54 ± 27 | 1 | Y | Y | Y | Y | … | Y | … |
| 50752 | 18511845-0611382 | 130.7 ± 1.4 | 5430 ± 92 | 3.62 ± 0.46 | -0.28 ± 0.10 | … | … | N | … | … | … | … | n | … |
| 50754 | 18511870-0621100 | 39.3 ± 0.3 | 5276 ± 201 | 4.06 ± 0.19 | -0.29 ± 0.18 | … | … | Y | … | … | … | … | n | … |
| 50755 | 18511878-0624399 | 61.3 ± 0.8 | 5819 ± 140 | 3.76 ± 0.27 | 0.37 ± 0.15 | … | … | N | … | … | … | … | n | … |
| 50756 | 18511883-0615030 | 41.0 ± 0.3 | 7030 ± 527 | 4.13 ± 0.19 | 0.11 ± 0.17 | 65 ± 25 | 1 | Y | Y | Y | Y | Y | Y | … |
| 50757 | 18511885-0620062 | 92.5 ± 0.3 | 5904 ± 105 | 4.01 ± 0.36 | -0.41 ± 0.15 | 69 ± 26 | 1 | N | … | … | … | … | n | NG |
| 50758 | 18511904-0615497 | -27.6 ± 0.3 | 6580 ± 322 | 4.19 ± 0.23 | 0.25 ± 0.28 | … | … | N | … | … | … | … | n | … |
| 50760 | 18511907-0614540 | -22.6 ± 0.3 | 5796 ± 132 | 4.09 ± 0.14 | 0.15 ± 0.12 | 33 ± 31 | 1 | N | … | … | … | … | n | NG |
| 50762 | 18511921-0616562 | 37.5 ± 0.4 | 6745 ± 481 | 4.39 ± 0.44 | 0.27 ± 0.36 | … | … | Y | … | … | … | … | n | … |
| 50763 | 18511924-0626319 | 16.4 ± 2.8 | 4747 ± 352 | 4.33 ± 0.41 | -0.56 ± 0.70 | … | … | N | … | … | … | … | n | … |
| 50764 | 18511939-0622131 | 25.8 ± 0.6 | 6368 ± 255 | 4.82 ± 0.59 | 0.13 ± 0.27 | <35 | 3 | N | … | … | … | … | n | NG |
| 50765 | 18511956-0617123 | 34.0 ± 0.5 | … | … | … | … | … | Y | … | … | … | … | n | … |
| 50766 | 18511959-0619441 | 39.6 ± 1.7 | 6773 ± 478 | 4.86 ± 0.86 | -0.03 ± 0.51 | … | … | Y | … | … | … | … | n | … |
| 50767 | 18511967-0608115 | 34.9 ± 0.6 | 6707 ± 433 | 4.36 ± 0.44 | 0.21 ± 0.21 | 66 ± 39 | 1 | Y | Y | Y | Y | Y | Y | … |
| 50769 | 18511976-0626347 | -41.6 ± 0.9 | 4984 ± 416 | 4.08 ± 0.78 | -0.50 ± 0.17 | … | … | N | … | … | … | … | n | … |
| 50770 | 18511980-0617373 | 34.1 ± 0.5 | 6312 ± 123 | 4.06 ± 0.14 | 0.03 ± 0.26 | 30 ± 25 | 1 | Y | Y | Y | Y | Y | Y | … |
| 50771 | 18511982-0623306 | -49.3 ± 0.3 | 5999 ± 110 | 3.82 ± 0.15 | 0.08 ± 0.13 | … | … | N | … | … | … | … | n | … |
| 50772 | 18511989-0614326 | 21.3 ± 0.3 | 5838 ± 229 | 4.59 ± 0.42 | -0.02 ± 0.20 | 159 ± 51 | 1 | N | … | … | … | … | n | NG |
| 50774 | 18511996-0617506 | 36.8 ± 0.6 | 7374 ± 511 | 4.16 ± 0.15 | 0.04 ± 0.21 | … | … | Y | … | … | … | … | n | … |
| 50775 | 18512003-0622024 | -5.4 ± 0.3 | 5981 ± 114 | 4.21 ± 0.20 | -0.03 ± 0.12 | … | … | N | … | … | … | … | n | … |
| 50776 | 18512008-0623066 | 62.0 ± 1.4 | 5348 ± 118 | 3.98 ± 0.09 | -0.25 ± 0.28 | … | … | N | … | … | … | … | n | … |
| 50777 | 18512009-0614221 | 0.8 ± 0.1 | 6161 ± 136 | 4.34 ± 0.14 | -0.04 ± 0.20 | … | … | N | … | … | … | … | n | … |
| 50779 | 18512011-0617103 | 29.0 ± 0.5 | 5664 ± 48 | 4.79 ± 0.37 | -0.17 ± 0.29 | 186 ± 69 | 1 | N | … | … | … | … | N | n | NG |
| 50780 | 18512021-0614294 | 73.9 ± 0.8 | 6181 ± 253 | 4.38 ± 0.17 | 0.23 ± 0.34 | <92 | 3 | N | … | … | … | … | n | NG |
| 50782 | 18512039-0613160 | 55.4 ± 0.3 | 5441 ± 129 | 3.95 ± 0.18 | 0.02 ± 0.21 | <46 | 3 | N | … | … | … | … | n | NG |
| 50783 | 18512052-0614375 | 25.3 ± 0.4 | 6205 ± 69 | 4.47 ± 0.65 | -0.18 ± 0.21 | 94 ± 49 | 1 | N | … | … | … | … | n | NG |
| 50784 | 18512082-0620040 | 35.5 ± 0.7 | 6262 ± 105 | 4.92 ± 0.68 | 0.29 ± 0.32 | 39 ± 49 | 1 | Y | Y | Y | Y | Y | Y | … |
| 50785 | 18512089-0617421 | 37.6 ± 0.4 | … | … | … | … | … | Y | … | … | … | … | n | … |
| 50786 | 18512094-0618085 | -56.9 ± 0.3 | 6049 ± 139 | 4.35 ± 0.31 | 0.00 ± 0.13 | … | … | N | … | … | … | … | n | … |
| 50787 | 18512100-0623192 | -40.8 ± 0.3 | 5814 ± 101 | 4.20 ± 0.11 | -0.52 ± 0.17 | … | … | N | … | … | … | … | n | … |
| 50788 | 18512111-0617151 | 13.6 ± 0.3 | 5927 ± 187 | 4.52 ± 0.20 | 0.12 ± 0.16 | 88 ± 29 | 1 | N | … | … | … | … | n | NG |
| 50789 | 18512127-0612152 | 34.8 ± 0.4 | 6306 ± 138 | 4.45 ± 0.45 | 0.15 ± 0.18 | 74 ± 54 | 1 | Y | Y | Y | Y | Y | Y | … |
| 50790 | 18512139-0610389 | 30.8 ± 0.8 | 7486 ± 86 | … | … | … | … | N | … | … | … | … | n | … |
| 50792 | 18512139-0621404 | -19.1 ± 0.3 | 5944 ± 108 | 4.28 ± 0.12 | 0.34 ± 0.16 | … | … | N | … | … | … | … | n | … |
| 50797 | 18512173-0623074 | 13.7 ± 0.4 | 5787 ± 81 | 4.25 ± 0.25 | 0.02 ± 0.27 | 103 ± 37 | 1 | N | … | … | … | … | n | NG |
| 50799 | 18512200-0614489 | -3.9 ± 0.3 | 5940 ± 191 | 4.32 ± 0.07 | 0.34 ± 0.16 | 110 ± 59 | 1 | N | … | … | … | … | n | NG |
| 50800 | 18512203-0609346 | 55.0 ± 0.1 | 6148 ± 269 | 4.31 ± 0.25 | -0.20 ± 0.35 | <45 | 3 | N | … | … | … | … | n | NG |
| 50801 | 18512213-0613284 | 34.5 ± 0.3 | 7206 ± 558 | 4.17 ± 0.16 | 0.14 ± 0.15 | 52 ± 33 | 1 | Y | Y | Y | Y | Y | Y | … |
| 50803 | 18512228-0615544 | -7.0 ± 0.4 | 5624 ± 111 | 4.14 ± 0.33 | -0.24 ± 0.19 | <40 | 3 | N | … | … | … | Y | n | NG |
| 50804 | 18512242-0616061 | 32.5 ± 0.3 | 6640 ± 450 | 4.26 ± 0.34 | 0.25 ± 0.27 | … | … | Y | … | … | … | … | n | … |
| 50805 | 18512251-0616256 | -16.0 ± 0.3 | 5642 ± 199 | 3.93 ± 0.24 | -0.10 ± 0.17 | <41 | 3 | N | … | … | … | … | n | NG |
| 50806 | 18512281-0615282 | 32.1 ± 1.1 | 6289 ± 284 | 4.90 ± 0.85 | -0.06 ± 0.24 | 116 ± 67 | 1 | Y | Y | Y | Y | Y | Y | … |
| 50807 | 18512303-0611459 | -6.5 ± 0.3 | 6493 ± 277 | 4.31 ± 0.38 | 0.10 ± 0.21 | … | … | N | … | … | … | … | n | … |
| 50808 | 18512307-0620045 | 38.4 ± 1.9 | … | … | … | … | … | Y | … | … | … | … | n | … |
| 50810 | 18512323-0625076 | 35.8 ± 1.6 | 7122 ± 699 | 4.15 ± 0.21 | 0.09 ± 0.24 | … | … | Y | … | … | … | … | n | … |
| 50811 | 18512363-0620112 | 15.0 ± 0.3 | 6130 ± 135 | 4.38 ± 0.22 | 0.14 ± 0.13 | 64 ± 32 | 1 | N | … | … | … | … | n | NG |
| 50812 | 18512373-0613465 | 34.9 ± 0.7 | 6585 ± 361 | 3.79 ± 0.21 | 0.20 ± 0.32 | 34 ± 18 | 1 | Y | Y | Y | Y | Y | Y | … |
| 50813 | 18512384-0625366 | 8.2 ± 0.1 | 5515 ± 176 | 4.45 ± 0.37 | -0.12 ± 0.20 | 169 ± 28 | 1 | N | … | … | … | … | n | NG |
| 50815 | 18512397-0612014 | 36.1 ± 0.7 | 6833 ± 536 | 4.04 ± 0.24 | 0.21 ± 0.26 | <24 | 3 | Y | Y | Y | Y | Y | Y | … |
| 50817 | 18512418-0621503 | 45.0 ± 0.5 | … | … | … | 40 ± 29 | 1 | N | … | … | … | N | n | … |
| 50818 | 18512431-0611092 | 34.4 ± 0.5 | 7375 ± 74 | … | … | 43 ± 29 | 1 | Y | Y | … | … | Y | Y | … |
| 50819 | 18512436-0616031 | 33.2 ± 0.3 | 7281 ± 459 | 4.09 ± 0.15 | 0.07 ± 0.14 | 66 ± 31 | 1 | Y | Y | Y | Y | Y | Y | … |
| 50820 | 18512440-0615017 | 32.9 ± 0.3 | 6160 ± 175 | 4.36 ± 0.16 | -0.01 ± 0.25 | 94 ± 44 | 1 | Y | Y | Y | Y | … | Y | … |
| 50823 | 18512475-0617471 | 37.8 ± 0.6 | 7041 ± 526 | 4.12 ± 0.18 | 0.14 ± 0.22 | … | … | Y | … | … | … | … | n | … |
| 50824 | 18512477-0616482 | 36.4 ± 0.5 | 6653 ± 305 | 4.43 ± 0.32 | 0.18 ± 0.21 | … | … | Y | … | … | … | … | n | … |

**Table C.9.** continued.

| ID | CNAME | RV (km s$^{-1}$) | $T_{eff}$ (K) | logg (dex) | [Fe/H] (dex) | EW(Li)$^a$ (mÅ) | EW(Li) error flag$^b$ | Membership RV | Li | logg | [Fe/H] | Gaia study Cantat-Gaudin$^c$ | Final$^d$ | NMs with Li$^e$ |
|---|---|---|---|---|---|---|---|---|---|---|---|---|---|---|
| 50825 | 18512477-0619247 | 48.7 ± 0.4 | 6298 ± 198 | 4.29 ± 0.12 | 0.24 ± 0.23 | … | … | N | … | … | … | … | n | … |
| 50826 | 18512493-0623288 | -34.0 ± 0.3 | 5832 ± 217 | 4.30 ± 0.11 | -0.08 ± 0.24 | … | … | N | … | … | … | … | n | … |
| 50828 | 18512555-0621524 | 75.6 ± 0.3 | 6197 ± 223 | 4.37 ± 0.25 | 0.12 ± 0.22 | <60 | 3 | N | … | … | … | … | n | NG |
| 50829 | 18512556-0618415 | 37.1 ± 0.4 | 6515 ± 233 | 4.49 ± 0.40 | 0.18 ± 0.19 | 40 ± 32 | 1 | Y | Y | Y | Y | Y | Y | … |
| 50830 | 18512566-0616223 | 24.8 ± 0.2 | 6492 ± 293 | 4.05 ± 0.25 | 0.15 ± 0.19 | … | … | N | … | … | … | … | n | … |
| 50832 | 18512581-0611426 | -32.7 ± 0.3 | 5930 ± 122 | 4.24 ± 0.33 | 0.10 ± 0.13 | <60 | 3 | N | … | … | … | … | n | NG |
| 50833 | 18512604-0610504 | 40.1 ± 0.2 | 6718 ± 457 | 4.21 ± 0.22 | 0.22 ± 0.29 | 65 ± 29 | 1 | Y | Y | Y | Y | Y | Y | … |
| 50835 | 18512631-0618575 | 2.1 ± 0.3 | 5296 ± 133 | 4.31 ± 0.25 | -0.19 ± 0.13 | … | … | N | … | … | … | … | n | … |
| 50837 | 18512655-0611425 | 34.2 ± 0.3 | 5256 ± 51 | 4.62 ± 0.43 | -0.10 ± 0.13 | … | … | Y | … | … | … | … | n | … |
| 50838 | 18512663-0619306 | -29.6 ± 0.3 | 5629 ± 106 | 4.02 ± 0.19 | -0.24 ± 0.18 | … | … | N | … | … | … | … | n | … |
| 50839 | 18512708-0616020 | 36.6 ± 1.5 | 6660 ± 95 | … | 0.56 ± 0.07 | … | … | Y | … | … | … | … | n | … |
| 50840 | 18512722-0617417 | 19.5 ± 0.4 | 6215 ± 125 | 3.97 ± 0.13 | 0.11 ± 0.14 | <73 | 3 | N | … | … | … | … | n | NG |
| 50841 | 18512724-0619494 | 33.1 ± 0.3 | 6442 ± 169 | 4.53 ± 0.48 | 0.07 ± 0.14 | 87 ± 45 | 1 | Y | Y | Y | Y | … | Y | … |
| 50842 | 18512773-0622494 | -17.2 ± 0.3 | 5711 ± 273 | 3.96 ± 0.10 | 0.17 ± 0.14 | 42 ± 38 | 1 | N | … | … | … | … | n | NG |
| 50843 | 18512790-0618027 | 36.8 ± 0.3 | 6085 ± 66 | 4.68 ± 0.45 | -0.09 ± 0.13 | … | … | Y | … | … | … | … | n | … |
| 50844 | 18512794-0610014 | 15.6 ± 0.8 | 6480 ± 347 | 4.43 ± 0.38 | 0.25 ± 0.24 | … | … | N | … | … | … | … | n | … |
| 50845 | 18512794-0616404 | 13.4 ± 0.3 | 6073 ± 154 | 4.40 ± 0.16 | 0.04 ± 0.17 | 63 ± 30 | 1 | N | … | … | … | … | n | NG |
| 50849 | 18512850-0617229 | 111.4 ± 0.2 | 4491 ± 121 | 2.34 ± 0.29 | 0.22 ± 0.17 | … | … | N | … | … | … | … | n | G |
| 50850 | 18512851-0608213 | -23.1 ± 0.2 | 6052 ± 201 | 4.32 ± 0.17 | 0.11 ± 0.13 | 50 ± 26 | 1 | N | … | … | … | … | n | NG |
| 50852 | 18512873-0624333 | 92.8 ± 0.2 | 6352 ± 235 | 4.32 ± 0.23 | 0.25 ± 0.16 | 74 ± 43 | … | N | … | … | … | … | n | NG |
| 50854 | 18512917-0615089 | 40.9 ± 0.7 | 6890 ± 380 | 4.21 ± 0.12 | -0.01 ± 0.20 | 37 ± 34 | 1 | Y | Y | Y | Y | Y | Y | … |
| 50855 | 18512933-0611377 | -18.3 ± 1.0 | 6553 ± 354 | 3.79 ± 0.21 | 0.19 ± 0.24 | … | … | N | … | … | … | … | n | … |
| 50856 | 18512945-0626465 | 33.6 ± 8.9 | … | … | … | … | … | Y | … | … | … | … | n | … |
| 50857 | 18512965-0621256 | 26.2 ± 0.3 | 6533 ± 185 | 4.40 ± 0.33 | -0.16 ± 0.16 | 76 ± 22 | 1 | N | … | … | … | … | n | NG |
| 50858 | 18512967-0617038 | 35.5 ± 0.6 | 6717 ± 423 | 4.15 ± 0.12 | 0.13 ± 0.19 | 14 ± 9 | 1 | Y | Y | Y | Y | Y | Y | … |
| 50859 | 18512970-0608324 | 34.0 ± 0.7 | 7186 ± 583 | 4.20 ± 0.13 | 0.10 ± 0.24 | … | … | Y | … | … | … | … | n | … |
| 50863 | 18513015-0617465 | 34.9 ± 0.9 | 5259 ± 280 | 4.67 ± 0.40 | -0.25 ± 0.20 | … | … | Y | … | … | … | … | n | … |
| 50864 | 18513021-0610390 | 35.0 ± 0.3 | 5589 ± 75 | 4.15 ± 0.40 | -0.47 ± 0.14 | … | … | Y | … | … | … | … | n | … |
| 50865 | 18513023-0622138 | 11.0 ± 0.3 | 6396 ± 180 | 4.22 ± 0.33 | 0.08 ± 0.24 | 79 ± 29 | 1 | N | … | … | … | … | n | NG |
| 50867 | 18513092-0613003 | 33.6 ± 0.8 | 6964 ± 540 | 5.10 ± 1.00 | 0.13 ± 0.19 | … | … | Y | … | … | … | … | n | … |
| 50868 | 18513103-0609074 | -13.7 ± 0.3 | 6229 ± 144 | 4.33 ± 0.34 | -0.04 ± 0.17 | … | … | N | … | … | … | … | n | … |
| 50870 | 18513133-0614106 | 37.1 ± 0.9 | 6519 ± 262 | 5.23 ± 1.02 | -0.15 ± 0.39 | … | … | Y | … | … | … | … | n | … |
| 50871 | 18513145-0610278 | -27.7 ± 0.3 | 6093 ± 241 | 3.98 ± 0.45 | -0.03 ± 0.31 | … | … | N | … | … | … | … | n | … |
| 50874 | 18513159-0620582 | 34.0 ± 0.4 | 6713 ± 512 | 4.24 ± 0.44 | 0.31 ± 0.55 | 66 ± 44 | … | Y | Y | Y | Y | Y | Y | … |
| 50875 | 18513161-0612399 | -32.3 ± 0.3 | 5931 ± 294 | 3.99 ± 0.17 | 0.31 ± 0.17 | … | … | N | … | … | … | … | n | … |
| 50876 | 18513165-0613421 | 34.6 ± 0.5 | … | … | … | … | … | Y | … | … | … | … | n | … |
| 50877 | 18513166-0618032 | -36.2 ± 0.3 | 5932 ± 215 | 4.18 ± 0.13 | 0.27 ± 0.18 | 68 ± 28 | 1 | N | … | … | … | … | n | NG |
| 50878 | 18513170-0616480 | 34.3 ± 0.9 | 6889 ± 595 | 4.04 ± 0.24 | 0.18 ± 0.21 | <21 | 3 | Y | Y | Y | Y | Y | Y | … |
| 50879 | 18513191-0612425 | 10.5 ± 0.3 | 6537 ± 259 | 3.69 ± 0.32 | 0.17 ± 0.20 | 108 ± 25 | 1 | N | … | … | … | … | n | NG |
| 50882 | 18513237-0613175 | 36.5 ± 0.3 | 7067 ± 706 | 4.03 ± 0.22 | 0.00 ± 0.34 | 65 ± 29 | 1 | Y | Y | Y | Y | Y | Y | … |
| 50883 | 18513258-0612330 | 39.0 ± 0.7 | 6265 ± 67 | 3.79 ± 0.21 | 0.10 ± 0.21 | … | … | Y | … | … | … | … | n | … |
| 50884 | 18513260-0621245 | 39.7 ± 0.3 | 5932 ± 191 | 4.56 ± 0.18 | 0.27 ± 0.19 | <41 | 3 | Y | Y | Y | Y | … | Y | … |
| 50885 | 18513275-0618047 | 48.9 ± 0.4 | 5450 ± 151 | 4.47 ± 0.17 | 0.11 ± 0.17 | <118 | 3 | N | … | … | … | … | n | NG |
| 50886 | 18513296-0623088 | -49.6 ± 1.1 | 5197 ± 291 | 3.39 ± 0.91 | -0.09 ± 0.15 | … | … | N | … | … | … | … | n | Li-rich G |
| 50887 | 18513303-0612581 | 14.9 ± 0.3 | 6330 ± 254 | 4.01 ± 0.17 | -0.16 ± 0.12 | <56 | 3 | N | … | … | … | … | n | NG |
| 50888 | 18513303-0616136 | 21.2 ± 0.2 | 6326 ± 119 | 4.54 ± 0.46 | 0.15 ± 0.17 | … | … | N | … | … | … | … | n | … |
| 50889 | 18513330-0617114 | 7.3 ± 0.5 | 6652 ± 428 | 4.17 ± 0.37 | 0.15 ± 0.22 | <19 | 3 | N | … | … | … | Y | n | NG |
| 50890 | 18513354-0609395 | -26.0 ± 0.4 | 6533 ± 257 | 4.15 ± 0.19 | -0.06 ± 0.37 | … | … | N | … | … | … | … | n | … |
| 50891 | 18513362-0615298 | -34.0 ± 0.3 | 5905 ± 140 | 4.31 ± 0.20 | 0.08 ± 0.15 | 76 ± 46 | 1 | N | … | … | … | … | n | NG |
| 50893 | 18513376-0615178 | 99.8 ± 1.5 | 5128 ± 90 | 2.47 ± 0.44 | -0.27 ± 0.10 | … | … | N | … | … | … | … | n | Li-rich G |
| 50894 | 18513389-0621125 | 65.7 ± 0.3 | 6351 ± 203 | 4.53 ± 0.35 | 0.20 ± 0.19 | … | … | N | … | … | … | … | n | … |
| 50895 | 18513389-0621234 | -58.5 ± 0.3 | 5568 ± 73 | 3.95 ± 0.12 | 0.00 ± 0.16 | 69 ± 37 | 1 | N | … | … | … | … | n | NG |
| 50896 | 18513412-0620079 | -72.5 ± 0.4 | 5706 ± 137 | 4.20 ± 0.44 | -0.67 ± 0.25 | … | … | N | … | … | … | … | n | … |
| 50897 | 18513425-0613169 | -3.4 ± 0.4 | 6771 ± 472 | 3.99 ± 0.05 | 0.20 ± 0.24 | … | … | N | … | … | … | … | n | … |
| 50898 | 18513447-0617066 | 22.2 ± 0.3 | 6204 ± 160 | 4.16 ± 0.10 | 0.00 ± 0.13 | … | … | N | … | … | … | … | n | … |
| 50899 | 18513451-0610356 | 33.9 ± 0.9 | 7403 ± 87 | … | … | … | … | Y | … | … | … | … | n | … |
| 50900 | 18513463-0613176 | 61.2 ± 0.2 | 6226 ± 117 | 4.55 ± 0.37 | 0.03 ± 0.16 | … | … | N | … | … | … | … | n | … |
| 50903 | 18513540-0622395 | 43.5 ± 0.5 | 6375 ± 110 | 4.59 ± 0.85 | -0.38 ± 0.28 | 85 ± 38 | 1 | N | … | … | … | … | n | NG |
| 50904 | 18513564-0613051 | -1.9 ± 0.3 | 6111 ± 72 | 4.31 ± 0.27 | -0.05 ± 0.12 | 69 ± 21 | 1 | N | … | … | … | … | n | NG |

**Table C.9.** continued.

| ID | CNAME | RV (km s$^{-1}$) | $T_{\text{eff}}$ (K) | $\log g$ (dex) | [Fe/H] (dex) | EW(Li)$^a$ (mÅ) | EW(Li) error flag$^b$ | Membership RV | Li | $\log g$ | [Fe/H] | Gaia study Cantat-Gaudin$^c$ | Final$^d$ | NMs with Li$^e$ |
|---|---|---|---|---|---|---|---|---|---|---|---|---|---|---|
| 50906 | 18513575-0609069 | 8.1 ± 0.5 | 6594 ± 333 | 4.23 ± 0.25 | 0.09 ± 0.22 | … | … | N | … | … | … | … | n | … |
| 50907 | 18513588-0614595 | -73.6 ± 0.2 | 5177 ± 49 | 4.14 ± 0.09 | 0.08 ± 0.18 | … | … | N | … | … | … | … | n | … |
| 50908 | 18513590-0623006 | 64.7 ± 0.3 | 4952 ± 156 | 3.44 ± 0.26 | 0.06 ± 0.15 | <39 | 3 | N | … | … | … | … | n | G |
| 50910 | 18513609-0612534 | 1.0 ± 1.1 | 6835 ± 233 | 4.20 ± 0.13 | 0.20 ± 0.26 | … | … | N | … | … | … | … | n | … |
| 50911 | 18513615-0612479 | 48.9 ± 0.3 | 5854 ± 45 | 4.36 ± 0.19 | -0.43 ± 0.15 | 20 ± 18 | 1 | N | … | … | … | … | n | NG |
| 50912 | 18513636-0616190 | 24.9 ± 0.1 | 6301 ± 564 | 4.42 ± 0.23 | 0.06 ± 0.85 | <54 | 3 | N | … | … | … | Y | n | NG |
| 50913 | 18513672-0616308 | -1.0 ± 0.3 | 5898 ± 251 | 4.32 ± 0.12 | 0.04 ± 0.22 | 70 ± 32 | 1 | N | … | … | … | … | n | NG |
| 50914 | 18513697-0619375 | 31.1 ± 0.3 | 6767 ± 136 | 4.41 ± 0.36 | 0.10 ± 0.22 | 44 ± 42 | 1 | Y | Y | Y | Y | … | Y | … |
| 50916 | 18513760-0617547 | 36.6 ± 0.9 | … | … | … | … | … | Y | … | … | … | … | n | … |
| 50918 | 18513837-0619229 | -10.0 ± 0.3 | 5526 ± 157 | 3.76 ± 0.19 | 0.15 ± 0.19 | 57 ± 29 | 1 | N | … | … | … | … | n | NG |
| 50919 | 18513858-0616012 | 18.8 ± 0.3 | 6234 ± 160 | 4.42 ± 0.23 | 0.21 ± 0.20 | 34 ± 32 | 1 | N | … | … | … | … | n | NG |
| 50921 | 18513861-0619307 | -72.5 ± 0.3 | 5666 ± 264 | 4.18 ± 0.26 | 0.10 ± 0.24 | … | … | N | … | … | … | … | n | … |
| 50922 | 18513864-0616468 | 40.4 ± 2.2 | 6754 ± 122 | 3.96 ± 0.17 | 0.17 ± 0.22 | 25 ± 24 | 1 | Y | Y | Y | Y | Y | Y | … |
| 50923 | 18513879-0620558 | 44.1 ± 3.7 | 6314 ± 167 | … | 0.40 ± 0.11 | <55 | 3 | N | … | … | … | … | n | … |
| 50924 | 18513906-0616192 | 17.8 ± 0.7 | 6504 ± 224 | 4.10 ± 0.69 | 0.11 ± 0.22 | … | … | N | … | … | … | … | n | … |
| 50925 | 18513911-0620544 | -45.2 ± 0.3 | 6031 ± 107 | 3.95 ± 0.22 | -0.35 ± 0.56 | … | … | N | … | … | … | … | n | … |
| 50926 | 18513933-0614416 | 28.0 ± 0.4 | 6146 ± 250 | 3.99 ± 0.18 | 0.26 ± 0.15 | … | … | N | … | … | … | … | n | … |
| 50927 | 18513993-0612168 | -13.9 ± 0.4 | 6460 ± 290 | 4.63 ± 0.59 | 0.06 ± 0.17 | … | … | N | … | … | … | … | n | … |
| 50928 | 18514000-0610405 | 34.2 ± 0.1 | … | … | … | … | … | Y | … | … | … | … | n | … |
| 50930 | 18514062-0618328 | 11.2 ± 0.3 | 6082 ± 172 | 4.29 ± 0.16 | 0.25 ± 0.16 | … | … | N | … | … | … | … | n | … |
| 50931 | 18514103-0618406 | 29.5 ± 0.7 | 6549 ± 307 | 4.52 ± 0.44 | 0.02 ± 0.13 | <45 | 3 | N | … | … | … | N | n | NG |
| 50932 | 18514165-0623559 | 111.6 ± 4.9 | … | … | … | … | … | N | … | … | … | … | n | … |
| 50933 | 18514193-0616574 | 0.7 ± 0.1 | 5573 ± 223 | 4.21 ± 0.17 | 0.05 ± 0.19 | <34 | 3 | N | … | … | … | … | n | NG |
| 50934 | 18514208-0618480 | 34.6 ± 0.6 | 6527 ± 303 | 4.56 ± 0.76 | 0.18 ± 0.25 | … | … | Y | … | … | … | … | n | … |
| 50935 | 18514395-0614538 | 18.9 ± 0.3 | 6315 ± 249 | 4.48 ± 0.50 | -0.50 ± 0.24 | … | … | N | … | … | … | … | n | … |
| 50936 | 18514499-0616300 | 34.9 ± 1.2 | 6488 ± 377 | 4.38 ± 0.33 | 0.16 ± 0.19 | … | … | Y | … | … | … | … | n | … |
| 50937 | 18514666-0616457 | 2.2 ± 0.6 | 6297 ± 183 | 4.81 ± 0.55 | 0.43 ± 0.25 | <62 | 3 | N | … | … | … | … | n | NG |
| 50938 | 18514869-0621075 | 25.5 ± 1.1 | 5533 ± 174 | 3.69 ± 0.52 | -0.33 ± 0.61 | … | … | N | … | … | … | … | n | … |
| 50939 | 18514946-0620231 | 115.6 ± 2.3 | 5357 ± 142 | 3.73 ± 0.67 | -0.44 ± 0.15 | … | … | N | … | … | … | … | n | … |

**Notes.** [a] The values of EW(Li) for this cluster are corrected (subtracted adjacent Fe (6707.43 Å) line). [b] Flags for the errors of the corrected EW(Li) values, as follows: 1=EW(Li) corrected by blends contribution using models; and 3=Upper limit (no error for EW(Li) is given). [c] Cantat-Gaudin et al. (2018). [d] The letters "Y" and "N" indicate if the star is a cluster member or not. [e] 'Li-rich G', 'G' and 'NG' indicate "Li-rich giant", "giant" and "non-giant" Li field contaminants, respectively.

**Table C.10.** NGC 4815

| ID | CNAME | RV (km s$^{-1}$) | $T_{\text{eff}}$ (K) | $\log g$ (dex) | [Fe/H] (dex) | EW(Li)$^a$ (mÅ) | EW(Li) error flag$^b$ | Membership RV | Li | $\log g$ | [Fe/H] | Gaia study Cantat-Gaudin$^c$ | Final$^d$ | NMs with Li$^e$ |
|---|---|---|---|---|---|---|---|---|---|---|---|---|---|---|
| 45601 | 12575787-6458270 | -26.6 ± 0.3 | 6015 ± 237 | 4.57 ± 0.46 | 0.21 ± 0.19 | 70 ± 34 | 1 | Y | Y | Y | Y | … | Y | … |
| 45700 | 12575818-6459323 | -29.4 ± 0.6 | 5577 ± 151 | 3.71 ± 0.49 | -0.22 ± 0.14 | <135 | 3 | Y | N | N | Y | … | n | NG |
| 45602 | 12575822-6454316 | -12.7 ± 0.3 | 6259 ± 131 | 4.50 ± 0.23 | 0.37 ± 0.20 | 104 ± 38 | 1 | N | … | … | … | … | n | NG |
| 45603 | 12575878-6458030 | -30.2 ± 0.1 | … | … | … | … | … | Y | … | … | … | Y | n | … |
| 45604 | 12580025-6500457 | -35.2 ± 1.2 | … | … | … | … | … | Y | … | … | … | … | n | … |
| 45605 | 12580070-6458074 | -17.6 ± 5.8 | … | … | … | … | … | Y | … | … | … | … | n | … |
| 45606 | 12580098-6454462 | -15.8 ± 1.6 | … | … | … | … | … | N | … | … | … | Y | n | … |
| 45607 | 12580134-6455432 | -55.0 ± 8.6 | … | … | … | … | … | N | … | … | … | … | n | … |
| 45701 | 12580138-6454441 | -56.7 ± 2.4 | 6844 ± 201 | 4.36 ± 0.38 | 0.50 ± 0.17 | … | … | N | … | … | … | … | n | … |
| 45608 | 12580145-6454028 | -26.4 ± 3.1 | … | … | … | … | … | Y | … | … | … | N | n | … |
| 45702 | 12580148-6451284 | -16.2 ± 0.9 | 6777 ± 450 | 4.23 ± 0.20 | 0.11 ± 0.28 | <56 | 3 | Y | Y | Y | Y | … | Y | … |
| 45609 | 12580171-6456192 | -1.2 ± 4.0 | … | … | … | … | … | N | … | … | … | N | n | … |
| 45610 | 12580189-6452492 | -15.8 ± 0.6 | 6426 ± 178 | 4.63 ± 0.15 | 0.13 ± 0.47 | <39 | 3 | N | … | … | … | … | n | NG |
| 45703 | 12580234-6458157 | -30.2 ± 0.6 | 6362 ± 205 | 3.93 ± 0.36 | 0.02 ± 0.23 | … | … | Y | … | … | … | … | n | … |
| 3060 | 12580262-6456492 | -15.8 ± 0.6 | 4931 ± 114 | 2.51 ± 0.23 | 0.04 ± 0.10 | … | … | N | … | … | … | … | n | … |
| 45611 | 12580281-6454586 | -28.6 ± 11.0 | … | … | … | … | … | Y | … | … | … | Y | n | … |
| 45612 | 12580453-6458376 | -20.0 ± 4.7 | … | … | … | … | … | Y | … | … | … | Y | n | … |
| 45705 | 12580478-6458281 | -22.8 ± 1.1 | 6341 ± 210 | 4.47 ± 0.51 | -0.44 ± 0.21 | … | … | Y | … | … | … | Y | n | … |
| 45706 | 12580515-6453338 | 27.9 ± 0.3 | 5363 ± 112 | 4.39 ± 0.20 | 0.18 ± 0.14 | <79 | 3 | N | … | … | … | … | n | NG |
| 45613 | 12580541-6457475 | -25.5 ± 2.6 | … | … | … | … | … | Y | … | … | … | Y | n | … |
| 45614 | 12580543-6459389 | -34.4 ± 0.2 | … | … | … | … | … | Y | … | … | … | Y | n | … |
| 45543 | 12571382-6453464 | -15.2 ± 0.3 | … | … | … | … | … | N | … | … | … | … | n | … |
| 45657 | 12571414-6453147 | -23.1 ± 0.3 | 6025 ± 270 | 4.32 ± 0.48 | -0.27 ± 0.15 | … | … | Y | … | … | … | … | n | … |
| 45544 | 12571436-6454050 | -2.0 ± 0.7 | … | … | … | … | … | N | … | … | … | Y | n | … |
| 45545 | 12571478-6456218 | -26.4 ± 2.9 | … | … | … | … | … | Y | … | … | … | Y | n | … |
| 45546 | 12571533-6457039 | -27.0 ± 1.0 | … | … | … | … | … | Y | … | … | … | N | n | … |
| 45547 | 12571591-6456469 | -15.1 ± 2.7 | … | … | … | … | … | N | … | … | … | … | n | … |
| 45548 | 12571606-6455355 | -26.5 ± 2.3 | … | … | … | … | … | Y | … | … | … | Y | n | … |
| 45658 | 12571691-6455285 | -0.5 ± 0.3 | 5703 ± 241 | 4.33 ± 0.25 | 0.36 ± 0.16 | … | … | N | … | … | … | … | n | … |
| 45550 | 12571790-6455163 | -48.1 ± 0.2 | … | … | … | … | … | N | … | … | … | … | n | … |
| 45551 | 12572000-6457587 | -2.3 ± 0.5 | … | … | … | … | … | N | … | … | … | N | n | … |
| 45552 | 12572017-6458209 | -24.5 ± 0.7 | 7142 ± 129 | … | … | … | … | … | … | … | … | N | n | … |
| 45659 | 12572031-6501055 | -3.8 ± 1.5 | 7026 ± 664 | 4.15 ± 0.21 | 0.15 ± 0.16 | <80 | 3 | N | … | … | … | … | n | NG |
| 45553 | 12572130-6459192 | -29.1 ± 1.9 | … | … | … | … | … | Y | … | … | … | Y | n | … |
| 45660 | 12572141-6458001 | 0.0 ± 0.4 | 5352 ± 109 | 4.32 ± 0.23 | 0.17 ± 0.15 | <111 | 3 | N | … | … | … | … | n | NG |
| 45661 | 12572191-6500513 | -37.8 ± 1.6 | 6538 ± 343 | 3.82 ± 0.43 | -0.32 ± 0.55 | … | … | Y | … | … | … | Y | n | … |
| 45662 | 12572194-6452312 | -13.4 ± 1.1 | 5723 ± 168 | 4.51 ± 0.77 | -0.05 ± 0.18 | <159 | 3 | N | … | … | … | … | n | NG |
| 45663 | 12572267-6453583 | -32.0 ± 1.1 | 6820 ± 495 | 4.03 ± 0.22 | 0.17 ± 0.40 | … | … | Y | … | … | … | N | n | … |
| 45664 | 12572276-6455294 | -0.5 ± 1.2 | 5863 ± 119 | 4.51 ± 0.25 | 0.08 ± 0.26 | 230 ± 134 | 1 | N | … | … | … | … | n | NG |
| 45665 | 12572286-6454288 | -24.6 ± 0.4 | 5948 ± 146 | 4.00 ± 0.36 | 0.02 ± 0.22 | <82 | 3 | Y | Y | Y | Y | N | Y | … |
| 45554 | 12572345-6501287 | -33.9 ± 0.3 | 6025 ± 306 | 4.29 ± 0.49 | 0.24 ± 0.22 | 52 ± 34 | 1 | Y | Y | Y | Y | … | Y | … |
| 45555 | 12572360-6455563 | -10.1 ± 0.2 | … | … | … | … | … | N | … | … | … | … | n | … |
| 45556 | 12572370-6459189 | -15.1 ± 0.1 | 6123 ± 77 | 4.05 ± 0.04 | … | … | … | N | … | … | … | … | n | … |
| 45666 | 12572387-6452366 | -31.8 ± 0.4 | 5444 ± 234 | 4.80 ± 0.42 | 0.14 ± 0.20 | <67 | 3 | Y | Y | Y | Y | … | Y | … |
| 3049 | 12572442-6455173 | -29.5 ± 0.6 | 4323 ± 129 | 1.77 ± 0.26 | -0.13 ± 0.11 | 26 ± 2 | … | Y | N | Y | Y | … | Y | … |
| 45557 | 12572534-6458431 | -28.8 ± 2.0 | … | … | … | … | … | Y | … | … | … | Y | n | … |
| 45667 | 12572592-6500340 | 1.3 ± 0.4 | 4990 ± 90 | 4.42 ± 0.61 | -0.04 ± 0.14 | … | … | N | … | … | … | … | n | … |
| 45668 | 12572603-6454596 | -44.6 ± 0.5 | 6293 ± 302 | 4.28 ± 0.39 | -0.04 ± 0.43 | 91 ± 65 | 1 | N | … | … | … | N | n | NG |
| 45670 | 12572808-6452433 | -51.5 ± 1.4 | 4385 ± 631 | 1.51 ± 0.55 | -1.07 ± 1.06 | … | … | N | … | … | … | … | n | G |
| 45558 | 12572875-6456171 | -25.7 ± 1.1 | … | … | … | … | … | Y | … | … | … | Y | n | … |
| 45671 | 12572889-6457049 | -31.3 ± 1.2 | 6278 ± 223 | 3.93 ± 0.12 | -0.31 ± 0.38 | … | … | Y | … | … | … | … | n | … |
| 45672 | 12572909-6458203 | -30.2 ± 0.9 | 6764 ± 504 | 4.06 ± 0.13 | 0.16 ± 0.23 | <70 | 3 | Y | Y | Y | Y | … | Y | … |
| 45673 | 12573016-6459257 | -42.5 ± 0.6 | 6033 ± 249 | 3.97 ± 0.49 | 0.06 ± 0.33 | … | … | N | … | … | … | … | n | … |
| 45559 | 12573029-6500401 | -18.2 ± 0.8 | … | … | … | … | … | Y | … | … | … | … | n | … |
| 45674 | 12573034-6458482 | -3.7 ± 0.3 | 5974 ± 146 | 4.38 ± 0.28 | 0.38 ± 0.17 | … | … | N | … | … | … | … | n | … |
| 45560 | 12573036-6455287 | -22.3 ± 10.2 | … | … | … | … | … | Y | … | … | … | … | n | … |
| 45561 | 12573076-6452300 | -30.2 ± 0.1 | … | … | … | … | … | Y | … | … | … | … | n | … |
| 45562 | 12573153-6451432 | -30.4 ± 3.1 | … | … | … | … | … | Y | … | … | … | Y | n | … |
| 45675 | 12573199-6455397 | -48.9 ± 1.1 | 5890 ± 343 | 4.18 ± 0.44 | -0.55 ± 1.00 | … | … | N | … | … | … | … | n | … |

**Table C.10.** continued.

| ID | CNAME | RV (km s$^{-1}$) | $T_{\text{eff}}$ (K) | $logg$ (dex) | [Fe/H] (dex) | $EW$(Li)$^a$ (mÅ) | $EW$(Li) error flag$^b$ | Membership RV | Li | $logg$ | [Fe/H] | Gaia study Cantat-Gaudin$^c$ | Final$^d$ | NMs with Li$^e$ |
|---|---|---|---|---|---|---|---|---|---|---|---|---|---|---|
| 3050 | 12573217-6455167 | -22.2 ± 0.6 | 5020 ± 121 | 2.56 ± 0.23 | 0.02 ± 0.10 | 31 ± 1 | … | Y | Y | Y | Y | … | Y | … |
| 45563 | 12573261-6500022 | 13.6 ± 15.8 | … | … | … | … | … | N | … | … | … | Y | n | … |
| 45677 | 12573361-6459543 | 12.3 ± 0.7 | 6522 ± 363 | 4.19 ± 0.32 | -0.02 ± 0.23 | … | … | N | … | … | … | … | n | … |
| 45678 | 12573460-6453231 | -19.2 ± 1.4 | 7239 ± 671 | 4.16 ± 0.17 | 0.01 ± 0.34 | <68 | 3 | Y | Y | Y | Y | Y | Y | … |
| 45687 | 12574313-6455591 | -27.6 ± 1.0 | 6147 ± 200 | 4.27 ± 0.37 | -0.25 ± 0.27 | 75 ± 60 | 1 | Y | Y | Y | Y | Y | Y | … |
| 45577 | 12574321-6457132 | -26.7 ± 9.1 | … | … | … | … | … | Y | … | … | … | Y | n | … |
| 3052 | 12574328-6457386 | -30.4 ± 0.6 | 4900 ± 122 | 2.27 ± 0.22 | -0.04 ± 0.10 | <5 | 3 | Y | Y | Y | Y | … | Y | … |
| 3053 | 12574341-6458045 | -49.0 ± 0.6 | 4992 ± 118 | 2.81 ± 0.24 | -0.03 ± 0.10 | <7 | 3 | N | … | … | … | … | n | G |
| 45578 | 12574416-6454479 | -22.1 ± 4.0 | … | … | … | … | … | Y | … | … | … | Y | n | … |
| 45579 | 12574450-6459019 | 11.7 ± 11.4 | … | … | … | … | … | N | … | … | … | Y | n | … |
| 3054 | 12574457-6459398 | -19.3 ± 0.6 | 4749 ± 112 | 2.97 ± 0.23 | 0.03 ± 0.10 | 54 ± 3 | … | Y | Y | Y | Y | … | Y | … |
| 45580 | 12574504-6458120 | -22.8 ± 4.5 | … | … | … | … | … | Y | … | … | … | Y | n | … |
| 45581 | 12574537-6456598 | -27.1 ± 1.5 | … | … | … | … | … | Y | … | … | … | Y | n | … |
| 45688 | 12574586-6453177 | -5.3 ± 0.4 | 6300 ± 313 | 4.71 ± 0.46 | 0.44 ± 0.27 | 74 ± 57 | 1 | N | … | … | … | … | n | NG |
| 45582 | 12574638-6500023 | -29.5 ± 1.5 | … | … | … | … | … | Y | … | … | … | Y | n | … |
| 45583 | 12574674-6500244 | -29.3 ± 2.0 | … | … | … | … | … | Y | … | … | … | Y | n | … |
| 45584 | 12574702-6452353 | -22.6 ± 5.1 | … | … | … | … | … | Y | … | … | … | Y | n | … |
| 45689 | 12574705-6451338 | -34.9 ± 0.9 | 6643 ± 570 | 3.89 ± 0.06 | 0.12 ± 0.43 | <81 | 3 | Y | Y | Y | Y | N | Y | … |
| 45585 | 12574715-6459412 | -22.7 ± 5.1 | … | … | … | … | … | Y | … | … | … | Y | n | … |
| 45690 | 12574734-6458105 | -29.4 ± 0.5 | 6458 ± 249 | 4.79 ± 0.71 | -0.04 ± 0.21 | … | … | Y | … | … | … | Y | n | … |
| 45691 | 12574845-6454129 | 5.7 ± 0.5 | 5849 ± 145 | 4.31 ± 0.09 | 0.35 ± 0.18 | <167 | 3 | N | … | … | … | … | n | NG |
| 45586 | 12574851-6453102 | -35.2 ± 1.7 | … | … | … | … | … | Y | … | … | … | Y | n | … |
| 45587 | 12574864-6454568 | -30.5 ± 0.2 | … | … | … | … | … | Y | … | … | … | … | n | … |
| 3055 | 12574905-6458511 | -70.4 ± 0.6 | 4800 ± 124 | 2.67 ± 0.24 | -0.22 ± 0.10 | <11 | 3 | N | … | … | … | … | n | G |
| 45588 | 12574917-6454103 | -24.6 ± 3.4 | … | … | … | … | … | Y | … | … | … | Y | n | … |
| 45589 | 12574925-6455166 | -27.0 ± 4.3 | … | … | … | … | … | Y | … | … | … | … | n | … |
| 45590 | 12574953-6458417 | -6.6 ± 14.8 | … | … | … | … | … | N | … | … | … | Y | n | … |
| 45591 | 12574984-6450388 | -47.3 ± 0.1 | … | … | … | … | … | N | … | … | … | … | n | … |
| 45615 | 12580615-6459128 | -31.4 ± 4.4 | … | … | … | … | … | Y | … | … | … | Y | n | … |
| 45616 | 12580711-6457187 | -53.4 ± 1.8 | … | … | … | … | … | N | … | … | … | … | n | … |
| 45707 | 12580780-6455324 | -36.2 ± 3.8 | 6555 ± 764 | 4.40 ± 0.36 | 0.06 ± 0.40 | <134 | 3 | Y | N | Y | Y | … | n | NG |
| 45617 | 12580783-6500084 | -28.9 ± 0.7 | … | … | … | … | … | Y | … | … | … | Y | n | … |
| 45618 | 12580818-6456344 | -20.5 ± 5.1 | … | … | … | … | … | Y | … | … | … | Y | n | … |
| 45708 | 12580826-6454525 | -41.8 ± 1.5 | 6700 ± 850 | 5.13 ± 1.10 | -0.09 ± 0.32 | <118 | 3 | N | … | … | … | … | n | NG |
| 45619 | 12580852-6452146 | -1.7 ± 0.4 | … | … | … | … | … | N | … | … | … | … | n | … |
| 45620 | 12580876-6455050 | -26.2 ± 1.1 | … | … | … | … | … | Y | … | … | … | Y | n | … |
| 45709 | 12580914-6458021 | -46.9 ± 0.3 | 5913 ± 193 | 4.05 ± 0.23 | -0.47 ± 0.12 | … | … | N | … | … | … | … | n | … |
| 45710 | 12581017-6459230 | -19.9 ± 0.7 | 7124 ± 136 | … | … | … | … | Y | … | … | … | Y | n | … |
| 45711 | 12581050-6455173 | -34.0 ± 2.1 | 6351 ± 394 | 4.18 ± 0.35 | -0.19 ± 0.40 | <107 | 3 | Y | Y | Y | Y | Y | Y | … |
| 45621 | 12581070-6458130 | 3.5 ± 0.3 | 6197 ± 132 | 4.47 ± 0.08 | 0.27 ± 0.18 | 57 ± 25 | 1 | N | … | … | … | … | n | NG |
| 45622 | 12581088-6454441 | 10.3 ± 0.3 | 5713 ± 26 | 4.34 ± 0.08 | -0.11 ± 0.17 | … | … | N | … | … | … | … | n | … |
| 45623 | 12581158-6457101 | -25.0 ± 4.9 | … | … | … | … | … | Y | … | … | … | Y | n | … |
| 45624 | 12581188-6456404 | 18.3 ± 0.2 | 5667 ± 111 | 4.02 ± 0.05 | 0.14 ± 0.22 | … | … | N | … | … | … | … | n | … |
| 45625 | 12581226-6500253 | -27.9 ± 9.7 | … | … | … | … | … | Y | … | … | … | Y | n | … |
| 45626 | 12581275-6457413 | -25.3 ± 3.7 | … | … | … | … | … | Y | … | … | … | … | n | … |
| 45627 | 12581279-6454041 | 5.7 ± 0.1 | … | … | … | … | … | N | … | … | … | … | n | … |
| 45712 | 12581293-6456546 | -15.6 ± 0.6 | 6786 ± 352 | 4.27 ± 0.17 | -0.03 ± 0.24 | 68 ± 53 | 1 | Y | Y | Y | Y | Y | Y | … |
| 45713 | 12581382-6459370 | -15.1 ± 0.9 | 7181 ± 175 | … | … | … | … | N | … | … | … | … | n | … |
| 45723 | 12582518-6458259 | -32.3 ± 1.8 | 7065 ± 790 | 4.01 ± 0.22 | 0.10 ± 0.14 | <99 | 3 | Y | Y | Y | Y | … | Y | … |
| 45724 | 12582561-6454180 | 3.6 ± 0.3 | 5651 ± 206 | 4.16 ± 0.22 | 0.33 ± 0.18 | … | … | N | … | … | … | … | n | … |
| 45638 | 12582664-6455372 | -31.3 ± 0.2 | … | … | … | … | … | Y | … | … | … | … | n | … |
| 45639 | 12582728-6457368 | -14.9 ± 0.2 | 5469 ± 141 | 3.03 ± 0.19 | -0.50 ± 0.19 | … | … | N | … | … | … | … | n | … |
| 45726 | 12582784-6455512 | -17.1 ± 1.7 | 6630 ± 305 | 4.14 ± 0.11 | -0.01 ± 0.21 | <151 | 3 | Y | N | Y | Y | N | n | NG |
| 45727 | 12582875-6453420 | -6.4 ± 1.8 | 7077 ± 339 | … | … | <130 | 3 | N | … | … | … | … | n | … |
| 45728 | 12582905-6458018 | -32.9 ± 2.3 | 7469 ± 188 | … | … | … | … | Y | … | … | … | Y | n | … |
| 45640 | 12582952-6453081 | -66.7 ± 0.2 | … | … | … | … | … | N | … | … | … | … | n | … |
| 45641 | 12582989-6453544 | -32.1 ± 0.2 | 5619 ± 140 | 3.48 ± 0.15 | -0.29 ± 0.20 | … | … | Y | … | … | … | … | n | … |
| 45729 | 12583077-6452290 | -3.3 ± 0.5 | 5793 ± 733 | 4.06 ± 0.56 | 0.24 ± 0.63 | … | … | N | … | … | … | … | n | … |
| 45730 | 12583363-6455595 | -23.9 ± 2.7 | 7192 ± 215 | … | … | … | … | Y | … | … | … | Y | n | … |

**Table C.10.** continued.

| ID | CNAME | $RV$ (km s$^{-1}$) | $T_{\rm eff}$ (K) | $logg$ (dex) | [Fe/H] (dex) | $EW$(Li)$^a$ (mÅ) | $EW$(Li) error flag$^b$ | Membership RV | Membership Li | Membership $logg$ | Membership [Fe/H] | Gaia study Cantat-Gaudin$^c$ | Final$^d$ | NMs with Li$^e$ |
|---|---|---|---|---|---|---|---|---|---|---|---|---|---|---|
| 45731 | 12583412-6501089 | -46.2 ± 0.3 | 4980 ± 201 | 4.21 ± 0.60 | 0.06 ± 0.16 | ... | ... | N | ... | ... | ... | ... | n | ... |
| 45733 | 12583456-6453419 | -24.2 ± 1.6 | 6673 ± 724 | 5.04 ± 1.04 | 0.29 ± 0.59 | <59 | 3 | Y | Y | N | Y | ... | n | NG |
| 45734 | 12583587-6454459 | -17.0 ± 0.3 | 6296 ± 140 | 4.21 ± 0.34 | -0.16 ± 0.14 | 32 ± 25 | 1 | Y | Y | Y | Y | ... | Y | ... |
| 45735 | 12583604-6459404 | -0.9 ± 0.8 | 6176 ± 341 | 4.95 ± 0.87 | -0.06 ± 0.51 | <41 | 3 | N | ... | ... | ... | ... | n | NG |
| 45737 | 12584045-6455543 | 35.2 ± 0.3 | 4946 ± 493 | 4.24 ± 0.64 | -0.37 ± 0.29 | <65 | 3 | N | ... | ... | ... | ... | n | NG |
| 45738 | 12584092-6453499 | -2.4 ± 1.1 | 6232 ± 102 | 3.79 ± 0.74 | 0.10 ± 0.21 | ... | ... | N | ... | ... | ... | ... | n | ... |
| 45646 | 12584097-6452410 | -72.2 ± 10.9 | ... | ... | ... | ... | ... | N | ... | ... | ... | Y | n | ... |
| 45647 | 12584497-6453010 | -61.1 ± 0.3 | ... | ... | ... | ... | ... | N | ... | ... | ... | ... | n | ... |
| 45648 | 12584516-6457467 | -40.9 ± 0.3 | 6107 ± 123 | 4.48 ± 0.44 | 0.11 ± 0.22 | ... | ... | N | ... | ... | ... | ... | n | ... |
| 45740 | 12585692-6457513 | 51.5 ± 0.4 | 5344 ± 104 | 4.36 ± 0.28 | -0.43 ± 0.18 | ... | ... | N | ... | ... | ... | ... | n | ... |
| 45649 | 12585776-6456268 | -7.2 ± 0.3 | 5654 ± 47 | 4.51 ± 0.54 | 0.12 ± 0.16 | ... | ... | N | ... | ... | ... | ... | n | ... |
| 45650 | 12590203-6458225 | -6.9 ± 0.5 | ... | ... | ... | ... | ... | N | ... | ... | ... | ... | n | ... |
| 45741 | 12590285-6457037 | 21.8 ± 0.4 | 5466 ± 270 | 3.72 ± 0.31 | 0.13 ± 0.17 | ... | ... | N | ... | ... | ... | ... | n | ... |
| 45651 | 12565137-6457506 | -36.2 ± 0.6 | 5926 ± 61 | 4.21 ± 0.43 | -0.19 ± 0.22 | 95 ± 77 | 1 | Y | Y | Y | Y | ... | Y | ... |
| 45652 | 12565617-6455220 | -69.0 ± 0.5 | 6181 ± 368 | 4.20 ± 0.27 | -0.17 ± 0.29 | 55 ± 53 | 1 | N | ... | ... | ... | ... | n | NG |
| 45538 | 12570063-6457046 | -27.0 ± 2.9 | ... | ... | ... | ... | ... | Y | ... | ... | ... | Y | n | ... |
| 45539 | 12570169-6457338 | -32.6 ± 0.7 | ... | ... | ... | ... | ... | Y | ... | ... | ... | Y | n | ... |
| 45653 | 12570258-6458007 | -14.3 ± 1.0 | 6693 ± 530 | 3.94 ± 0.08 | 0.23 ± 0.20 | ... | ... | N | ... | ... | ... | ... | n | ... |
| 45540 | 12570931-6457537 | -24.7 ± 0.4 | 6966 ± 112 | 4.40 ± 0.21 | 0.33 ± 0.10 | ... | ... | Y | ... | ... | ... | Y | n | ... |
| 45654 | 12570959-6456443 | 235.6 ± 0.5 | 6126 ± 544 | 4.05 ± 0.18 | -1.29 ± 0.62 | ... | ... | N | ... | ... | ... | ... | n | ... |
| 45655 | 12571134-6452552 | -25.7 ± 0.5 | 5424 ± 113 | 4.04 ± 0.03 | -0.44 ± 0.16 | ... | ... | Y | ... | ... | ... | ... | n | ... |
| 45541 | 12571169-6458368 | -20.0 ± 0.4 | ... | ... | ... | ... | ... | Y | ... | ... | ... | ... | n | ... |
| 45656 | 12571226-6456406 | -61.3 ± 0.5 | 6156 ± 354 | 4.54 ± 0.50 | -0.21 ± 0.35 | ... | ... | N | ... | ... | ... | ... | n | ... |
| 45542 | 12571227-6453266 | -31.5 ± 0.3 | ... | ... | ... | ... | ... | Y | ... | ... | ... | ... | n | ... |
| 3048 | 12571312-6456090 | -9.1 ± 0.6 | 4745 ± 116 | 2.53 ± 0.23 | -0.09 ± 0.10 | <8 | 3 | N | ... | ... | ... | ... | n | G |
| 45679 | 12573581-6452226 | -65.1 ± 2.0 | 6362 ± 35 | 5.37 ± 0.30 | 0.16 ± 0.15 | ... | ... | N | ... | ... | ... | ... | n | ... |
| 45565 | 12573585-6454490 | -19.2 ± 3.0 | ... | ... | ... | ... | ... | Y | ... | ... | ... | Y | n | ... |
| 45566 | 12573654-6457120 | -28.8 ± 4.9 | ... | ... | ... | ... | ... | Y | ... | ... | ... | Y | n | ... |
| 45567 | 12573677-6454092 | -16.2 ± 0.3 | 5948 ± 153 | 4.57 ± 0.28 | 0.26 ± 0.17 | ... | ... | Y | ... | ... | ... | Y | n | ... |
| 45568 | 12573693-6458021 | -29.1 ± 0.5 | ... | ... | ... | ... | ... | Y | ... | ... | ... | Y | n | ... |
| 45569 | 12573749-6451315 | -31.6 ± 0.6 | ... | ... | ... | ... | ... | Y | ... | ... | ... | Y | n | ... |
| 45680 | 12573771-6455515 | -34.2 ± 1.7 | 7375 ± 163 | ... | ... | ... | ... | ... | ... | ... | ... | Y | n | ... |
| 45681 | 12573831-6452279 | -46.0 ± 0.3 | 6183 ± 363 | 4.48 ± 0.19 | 0.04 ± 0.42 | <74 | 3 | N | ... | ... | ... | ... | n | NG |
| 45570 | 12573849-6459052 | -25.7 ± 1.0 | ... | ... | ... | ... | ... | Y | ... | ... | ... | Y | n | ... |
| 45682 | 12573867-6454519 | -20.5 ± 0.4 | 6351 ± 255 | 4.28 ± 0.32 | -0.20 ± 0.22 | <46 | 3 | Y | Y | Y | Y | ... | Y | ... |
| 45573 | 12574030-6453585 | -85.3 ± 3.4 | ... | ... | ... | ... | ... | N | ... | ... | ... | Y | n | ... |
| 45574 | 12574067-6456520 | -31.0 ± 0.6 | ... | ... | ... | ... | ... | Y | ... | ... | ... | Y | n | ... |
| 3051 | 12574080-6455572 | -29.3 ± 0.6 | ... | ... | ... | ... | ... | Y | ... | ... | ... | ... | n | ... |
| 45683 | 12574131-6452292 | -18.2 ± 0.6 | 6071 ± 389 | 4.22 ± 0.07 | 0.11 ± 0.70 | <76 | 3 | Y | Y | Y | Y | ... | Y | ... |
| 45575 | 12574187-6457570 | -28.0 ± 0.2 | ... | ... | ... | ... | ... | Y | ... | ... | ... | Y | n | ... |
| 45684 | 12574256-6500242 | -38.0 ± 0.4 | 6339 ± 348 | 4.35 ± 0.23 | 0.17 ± 0.51 | ... | ... | Y | ... | ... | ... | ... | n | ... |
| 45685 | 12574261-6458421 | -32.1 ± 1.0 | 6252 ± 355 | 4.31 ± 0.22 | 0.03 ± 0.34 | 91 ± 80 | 1 | Y | Y | Y | Y | ... | Y | ... |
| 45576 | 12574263-6459521 | -26.9 ± 4.3 | ... | ... | ... | ... | ... | Y | ... | ... | ... | Y | n | ... |
| 45686 | 12574279-6454449 | -41.4 ± 0.4 | 5918 ± 381 | 4.42 ± 0.21 | -0.08 ± 0.42 | 90 ± 87 | 1 | N | ... | ... | ... | ... | n | NG |
| 45693 | 12575070-6459567 | -29.2 ± 0.3 | 6273 ± 229 | 4.72 ± 0.70 | 0.04 ± 0.23 | 50 ± 26 | 1 | Y | Y | Y | Y | ... | Y | ... |
| 45694 | 12575122-6457266 | -26.5 ± 1.0 | 7125 ± 617 | 4.15 ± 0.18 | -0.11 ± 0.45 | ... | ... | Y | ... | ... | ... | ... | n | ... |
| 45593 | 12575223-6459227 | -27.4 ± 1.5 | ... | ... | ... | ... | ... | Y | ... | ... | ... | Y | n | ... |
| 45695 | 12575267-6452173 | -19.1 ± 2.6 | 6361 ± 171 | 4.88 ± 0.75 | 0.48 ± 0.42 | <74 | 3 | Y | Y | Y | N | ... | Y | ... |
| 3056 | 12575308-6457182 | -20.7 ± 0.6 | ... | ... | ... | ... | ... | Y | ... | ... | ... | ... | n | ... |
| 45696 | 12575485-6500533 | 66.5 ± 0.3 | 5259 ± 57 | 4.62 ± 0.28 | 0.23 ± 0.26 | ... | ... | N | ... | ... | ... | ... | n | ... |
| 3057 | 12575511-6458483 | -29.3 ± 0.6 | 4923 ± 121 | 2.48 ± 0.23 | 0.00 ± 0.10 | 26 ± 1 | ... | Y | Y | Y | Y | ... | Y | ... |
| 45594 | 12575520-6456193 | -33.1 ± 8.7 | ... | ... | ... | ... | ... | Y | ... | ... | ... | Y | n | ... |
| 3058 | 12575529-6456536 | -29.5 ± 0.6 | 5029 ± 115 | 2.63 ± 0.23 | -0.03 ± 0.10 | 26 ± 1 | ... | Y | Y | Y | Y | ... | Y | ... |
| 45595 | 12575529-6458262 | -38.3 ± 0.2 | ... | ... | ... | ... | ... | N | ... | ... | ... | Y | n | ... |
| 3059 | 12575531-6500412 | -43.9 ± 0.6 | 4826 ± 111 | 2.78 ± 0.23 | -0.14 ± 0.11 | 36 ± 1 | ... | N | ... | ... | ... | ... | n | G |
| 45697 | 12575535-6500327 | -22.5 ± 0.7 | 6641 ± 123 | 4.13 ± 0.26 | 0.69 ± 0.10 | ... | ... | Y | ... | ... | ... | ... | n | ... |
| 45596 | 12575613-6454355 | -22.4 ± 1.9 | ... | ... | ... | ... | ... | Y | ... | ... | ... | ... | n | ... |
| 45597 | 12575632-6500005 | -24.7 ± 0.3 | 6911 ± 112 | 4.33 ± 0.21 | 0.39 ± 0.10 | 49 ± 45 | 1 | Y | Y | Y | Y | Y | Y | ... |
| 45698 | 12575647-6459119 | -28.1 ± 1.1 | 6156 ± 252 | 4.31 ± 0.60 | -0.08 ± 0.20 | ... | ... | Y | ... | ... | ... | ... | n | ... |

**Table C.10.** continued.

| ID | CNAME | $RV$ (km s$^{-1}$) | $T_{\rm eff}$ (K) | $logg$ (dex) | [Fe/H] (dex) | $EW$(Li)$^a$ (mÅ) | $EW$(Li) error flag$^b$ | Membership RV | Li | $logg$ | [Fe/H] | Gaia study Cantat-Gaudin$^c$ | Final$^d$ | NMs with Li$^e$ |
|---|---|---|---|---|---|---|---|---|---|---|---|---|---|---|
| 45598 | 12575661-6451542 | -29.6 ± 2.5 | … | … | … | … | … | Y | … | … | … | … | n | … |
| 45599 | 12575692-6459171 | -34.5 ± 0.2 | … | … | … | … | … | Y | … | … | … | … | n | … |
| 45699 | 12575730-6500494 | 82.3 ± 0.3 | 5833 ± 534 | 4.53 ± 0.17 | 0.12 ± 0.31 | … | … | N | … | … | … | … | n | … |
| 45600 | 12575773-6457324 | -37.7 ± 6.7 | … | … | … | … | … | Y | … | … | … | … | n | … |
| 45714 | 12581393-6452266 | -49.1 ± 1.7 | 7139 ± 161 | … | … | … | … | N | … | … | … | … | n | … |
| 45629 | 12581428-6455087 | -30.3 ± 8.7 | … | … | … | … | … | Y | … | … | … | Y | n | … |
| 45715 | 12581524-6456492 | -27.4 ± 2.1 | 6827 ± 174 | 4.04 ± 0.32 | 0.52 ± 0.15 | <66 | 3 | Y | Y | Y | N | … | Y | … |
| 45716 | 12581536-6501250 | -21.3 ± 1.7 | 6832 ± 192 | 4.89 ± 0.39 | 0.16 ± 0.17 | … | … | Y | … | … | … | … | n | … |
| 45717 | 12581635-6456266 | -0.4 ± 0.5 | 5650 ± 40 | 4.55 ± 0.18 | 0.13 ± 0.13 | <92 | 3 | N | … | … | … | … | n | NG |
| 45630 | 12581638-6457252 | -44.7 ± 0.2 | 5779 ± 132 | 3.34 ± 0.02 | -0.08 ± 0.21 | … | … | N | … | … | … | … | n | … |
| 45632 | 12581840-6456314 | -4.7 ± 7.9 | … | … | … | … | … | N | … | … | … | Y | n | … |
| 45718 | 12581897-6500052 | -25.6 ± 1.7 | 5941 ± 115 | 4.10 ± 0.38 | -0.50 ± 0.36 | … | … | Y | … | … | … | … | n | … |
| 3061 | 12581939-6453533 | -34.1 ± 0.6 | 3959 ± 129 | 1.44 ± 0.26 | -0.02 ± 0.12 | <9 | 3 | Y | Y | Y | Y | … | Y | … |
| 45633 | 12581951-6450480 | -23.4 ± 2.7 | … | … | … | … | … | Y | … | … | … | … | n | … |
| 45720 | 12582063-6454149 | -12.7 ± 0.3 | 5598 ± 134 | 4.61 ± 0.48 | 0.37 ± 0.19 | <89 | 3 | N | … | … | … | … | n | NG |
| 45721 | 12582069-6458378 | -29.6 ± 0.7 | 6494 ± 219 | 4.06 ± 0.23 | 0.09 ± 0.21 | <66 | 3 | Y | Y | Y | Y | Y | Y | … |
| 45634 | 12582136-6455151 | -28.6 ± 7.3 | … | … | … | … | … | Y | … | … | … | Y | n | … |
| 45635 | 12582178-6458437 | -36.6 ± 3.7 | … | … | … | … | … | Y | … | … | … | … | n | … |
| 45722 | 12582401-6459136 | -33.8 ± 2.0 | 7462 ± 180 | … | … | … | … | Y | … | … | … | … | n | … |
| 45636 | 12582449-6458526 | -24.0 ± 5.8 | … | … | … | … | … | Y | … | … | … | Y | n | … |
| 45643 | 12583958-6458495 | -27.4 ± 0.3 | … | … | … | … | … | Y | … | … | … | … | n | … |
| 45644 | 12583994-6456464 | 12.8 ± 0.1 | … | … | … | … | … | N | … | … | … | … | n | … |
| 45645 | 12584037-6457416 | -159.4 ± 22.8 | … | … | … | … | … | N | … | … | … | Y | n | … |

**Notes.** $^{(a)}$ The values of $EW$(Li) for this cluster are corrected (subtracted adjacent Fe (6707.43 Å) line). $^{(b)}$ Flags for the errors of the corrected $EW$(Li) values, as follows: 1=$EW$(Li) corrected by blends contribution using models; and 3=Upper limit (no error for $EW$(Li) is given). $^{(c)}$ Cantat-Gaudin et al. (2018). $^{(d)}$ The letters "Y" and "N" indicate if the star is a cluster member or not. $^{(e)}$ 'Li-rich G', 'G' and 'NG' indicate "Li-rich giant", "giant" and "non-giant" Li field contaminants, respectively.

**Table C.11.** NGC 6633

| ID | CNAME | RV (km s$^{-1}$) | $T_{\rm eff}$ (K) | logg (dex) | [Fe/H] (dex) | EW(Li)$^a$ (mÅ) | EW(Li) error flag$^b$ | Membership RV | Li | logg | [Fe/H] | Gaia studies Randich$^c$ | Cantat-Gaudin$^c$ | Final$^d$ | NMs with Li$^e$ |
|---|---|---|---|---|---|---|---|---|---|---|---|---|---|---|---|
| 320 | 18274431+0631059 | -28.8 ± 0.6 | 7166 ± 2000 | 4.85 ± 1.00 | … | … | … | Y | … | … | … | … | … | n | … |
| 3150 | 18263213+0626485 | 12.4 ± 0.6 | 6077 ± 129 | 4.22 ± 0.29 | 0.07 ± 0.106 | … | … | N | … | … | … | N | … | n | … |
| 3151 | 18263242+0626496 | -0.8 ± 0.6 | 6089 ± 171 | 4.32 ± 0.33 | 0.00 ± 0.118 | … | … | Y | … | … | … | N | … | n | … |
| 3152 | 18263318+0621466 | -15.0 ± 0.6 | 6095 ± 116 | 4.24 ± 0.24 | -0.21 ± 0.1 | 39 ± 1 | … | Y | Y | Y | Y | N | … | Y?$^g$ | … |
| 3154 | 18263376+0631350 | 17.1 ± 0.6 | 4983 ± 118 | 2.63 ± 0.22 | -0.09 ± 0.096 | 4 ± 0.3 | 1 | N | … | … | … | N | … | n | G |
| 3156 | 18263404+0637467 | -28.9 ± 0.6 | 5721 ± 115 | 4.37 ± 0.22 | -0.09 ± 0.093 | 64 ± 2 | … | Y | Y | Y | Y | Y | … | Y | … |
| 3158 | 18263798+0623454 | 22.1 ± 0.6 | … | … | … | … | … | N | … | … | … | N | … | n | … |
| 3159 | 18263966+0629501 | -27.8 ± 0.6 | 5576 ± 112 | 4.36 ± 0.23 | -0.08 ± 0.092 | 44 ± 1 | 1 | Y | Y | Y | Y | Y | … | Y | … |
| 3160 | 18264024+0618536 | 7.3 ± 0.6 | 5515 ± 117 | 3.68 ± 0.22 | -0.29 ± 0.094 | <4 | 3 | N | .. | … | … | N | … | n | … |
| 3161 | 18264026+0637500 | -31.7 ± 0.6 | 6455 ± 2000 | 5.00 ± 1.00 | … | 39 ± 21 | 1 | Y | Y | N | … | N | … | n | … |
| 3162 | 18264060+0643352 | -28.9 ± 0.6 | 6600 ± 50 | 4.90 ± 0.20 | 0.00 ± 0.1 | 48 ± 13 | 1 | Y | Y | Y? | Y | Y | Y | Y | … |
| 3163 | 18264117+0614522 | 19.3 ± 0.6 | 6019 ± 115 | 4.02 ± 0.24 | -0.61 ± 0.095 | 37 ± 1 | … | N | .. | … | … | N | … | n | NG |
| 3164 | 18264301+0632354 | -55.7 ± 0.6 | 5674 ± 116 | 4.34 ± 0.23 | -0.35 ± 0.102 | 32 ± 1 | … | N | .. | … | … | N | … | n | NG |
| 3165 | 18264491+0624181 | -26.0 ± 0.1 | 7450 ± 50 | 1.50 ± 0.20 | -0.20 ± 0.1 | … | … | Y | … | … | … | N | Y | n | … |
| 3166 | 18264580+0645386 | 14.6 ± 0.6 | 6564 ± 120 | 3.87 ± 0.23 | -0.22 ± 0.102 | <6 | 3 | N | .. | … | … | N | … | n | … |
| 3167 | 18264594+0636077 | -17.0 ± 0.6 | 5988 ± 131 | 3.97 ± 0.27 | 0.33 ± 0.113 | … | … | Y | … | … | … | N | … | n | … |
| 3169 | 18264786+0614598 | -14.1 ± 0.6 | 7250 ± 50 | 4.50 ± 0.20 | -0.10 ± 0.1 | … | … | Y | … | … | … | N | … | n | … |
| 3170 | 18265055+0626594 | 29.1 ± 0.6 | 5522 ± 126 | 4.27 ± 0.23 | -0.06 ± 0.097 | <6 | 3 | N | .. | … | … | N | … | n | … |
| 3171 | 18265061+0644331 | 20.4 ± 0.6 | 6040 ± 121 | 4.28 ± 0.23 | -0.04 ± 0.098 | 57 ± 1 | … | N | .. | … | … | N | … | n | NG |
| 3174 | 18265503+0620566 | -27.5 ± 0.6 | 7250 ± 50 | 3.70 ± 0.20 | … | … | … | Y | … | … | … | Y | Y | n | … |
| 3175 | 18265521+0633543 | 23.0 ± 0.6 | 6748 ± 143 | 4.54 ± 0.29 | -0.38 ± 0.184 | 85 ± 31 | 1 | N | .. | … | … | N | … | n | NG |
| 3176 | 18265539+0627575 | 1.4 ± 0.1 | 7300 ± 50 | 4.50 ± 0.20 | 0.00 ± 0.1 | … | … | Y | … | … | … | N | … | n | … |
| 3177 | 18265591+0635559 | -18.5 ± 0.1 | 5459 ± 2000 | 4.85 ± 1.00 | … | 41 ± 13 | … | Y | Y | Y | … | Y | … | Y | … |
| 3178 | 18270066+0641148 | 21.6 ± 0.6 | 6494 ± 125 | 4.10 ± 0.25 | -0.04 ± 0.096 | <7 | 3 | N | .. | … | … | N | … | n | … |
| 3179 | 18270297+0616243 | 26.2 ± 0.6 | 5758 ± 126 | 4.33 ± 0.27 | -0.12 ± 0.095 | … | … | N | … | … | … | N | … | n | … |
| 3181 | 18270509+0632141 | -42.5 ± 0.6 | 5239 ± 116 | 2.86 ± 0.26 | -0.07 ± 0.105 | 12 ± 1 | … | Y | N | Y | Y | N | … | n | … |
| 3182 | 18270721+0618284 | -35.8 ± 0.6 | 6057 ± 116 | 4.18 ± 0.24 | -0.15 ± 0.102 | 46 ± 8 | … | Y | Y | Y | Y | N | … | Y?$^g$ | … |
| 3183 | 18270752+0627556 | -27.6 ± 0.6 | 5738 ± 113 | 4.43 ± 0.23 | -0.06 ± 0.091 | 67 ± 2 | … | Y | Y | Y | Y | Y | Y | Y | … |
| 3185 | 18271445+0621373 | 9.0 ± 0.6 | 6230 ± 115 | 4.33 ± 0.23 | 0.08 ± 0.095 | 39 ± 3 | 1 | N | .. | … | … | N | … | n | NG |
| 3187 | 18271687+0633533 | -53.0 ± 0.6 | 6500 ± 2000 | 5.00 ± 1.00 | … | … | … | N | … | … | … | N | … | n | … |
| 3188 | 18271823+0626146 | 36.3 ± 0.6 | 6065 ± 119 | 3.87 ± 0.23 | 0.32 ± 0.096 | <2 | 3 | N | .. | … | … | N | … | n | … |
| 3192 | 18272787+0620520 | -30.8 ± 0.6 | 6325 ± 173 | 3.81 ± 0.45 | 0.22 ± 0.119 | <5 | 3 | Y | N | N | Y | Y | … | n | … |
| 3193 | 18273098+0618484 | -92.1 ± 0.6 | 5360 ± 119 | 4.47 ± 0.23 | -0.10 ± 0.095 | <5 | 3 | N | .. | … | … | … | … | n | … |
| 3194 | 18273258+0619524 | 11.8 ± 0.6 | 5491 ± 124 | 4.39 ± 0.23 | 0.01 ± 0.099 | <9 | 3 | N | .. | … | … | N | … | n | … |
| 3195 | 18273275+0640456 | -0.6 ± 0.6 | … | … | … | … | … | Y | … | … | … | N | … | n | … |
| 3198 | 18273949+0628080 | -8.0 ± 0.6 | 5846 ± 118 | 4.26 ± 0.22 | -0.40 ± 0.099 | <4 | 3 | Y | N | N | N | N | … | n | … |
| 3199 | 18274143+0630541 | -45.0 ± 0.6 | 5027 ± 123 | 4.50 ± 0.23 | -0.27 ± 0.103 | <6 | 3 | Y | N | Y | Y | N | … | n | … |
| 3200 | 18274148+0623448 | -35.5 ± 0.6 | 6007 ± 114 | 3.95 ± 0.23 | 0.29 ± 0.093 | <4 | 3 | Y | N | N | Y | N | … | n | … |
| 3201 | 18274267+0639082 | -27.1 ± 0.6 | 6153 ± 125 | 3.84 ± 0.24 | 0.04 ± 0.105 | 124 ± 1 | … | Y | Y | Y | Y | N | … | Y | … |
| 3202 | 18274416+0640040 | -31.0 ± 0.6 | 5594 ± 118 | 4.19 ± 0.22 | 0.42 ± 0.096 | <6 | 3 | Y | N | N | N | Y | … | n | … |
| 3205 | 18274725+0640462 | -28.3 ± 0.6 | 5227 ± 114 | 4.45 ± 0.23 | -0.06 ± 0.103 | 20 ± 1 | 1 | Y | Y | Y | Y | Y | … | Y | … |
| 3206 | 18274752+0629267 | 23.8 ± 0.6 | 5619 ± 115 | 3.93 ± 0.23 | 0.17 ± 0.1 | 5 ± 1 | … | N | … | … | … | N | … | n | … |
| 3208 | 18275112+0622485 | -29.2 ± 0.6 | … | … | … | … | … | Y | … | … | … | Y | … | n | … |
| 3209 | 18275473+0636003 | -29.2 ± 0.6 | 4991 ± 114 | 2.61 ± 0.22 | 0.00 ± 0.096 | 7 ± 0.3 | 1 | Y | Y | Y | Y | Y | Y | Y | … |
| 3210 | 18275733+0646499 | -21.9 ± 0.6 | 6127 ± 117 | 4.20 ± 0.23 | 0.06 ± 0.094 | 42 ± 3 | 1 | Y | Y | Y | Y | N | … | Y?$^g$ | … |
| 3211 | 18275771+0642479 | -47.5 ± 0.6 | 5426 ± 108 | 4.36 ± 0.21 | 0.20 ± 0.094 | <6 | 3 | N | .. | … | … | N | … | n | … |
| 3212 | 18280018+0654514 | -28.7 ± 0.6 | 5049 ± 113 | 2.80 ± 0.23 | 0.02 ± 0.094 | 32 ± 1 | … | Y | Y | Y | Y | Y | … | Y | … |
| 3214 | 18280063+0623517 | -10.1 ± 0.6 | … | … | … | … | … | Y | … | … | … | N | … | n | … |
| 3215 | 18280145+0642302 | -45.7 ± 0.6 | 5864 ± 114 | 4.06 ± 0.23 | 0.29 ± 0.095 | 112 ± 6 | 1 | Y | Y | Y | Y | N | … | Y?$^g$ | … |
| 3216 | 18280276+0650238 | 18.3 ± 0.6 | … | … | … | … | … | N | … | … | … | N | … | n | … |
| 3217 | 18280330+0639516 | -29.9 ± 0.6 | 6373 ± 124 | 4.15 ± 0.22 | -0.18 ± 0.096 | 9 ± 1 | … | Y | Y | Y | Y | Y | … | Y | … |
| 3219 | 18280866+0638090 | -27.0 ± 0.6 | 5125 ± 123 | 4.52 ± 0.24 | -0.10 ± 0.107 | 33 ± 2 | 1 | Y | Y | Y | Y | Y | Y | Y | … |
| 3222 | 18281667+0642150 | 26.5 ± 0.6 | 5366 ± 118 | 4.41 ± 0.23 | 0.11 ± 0.1 | <6 | 3 | N | .. | … | … | N | … | n | … |
| 3223 | 18281702+0636284 | -78.3 ± 0.6 | 5181 ± 113 | 4.46 ± 0.23 | 0.20 ± 0.097 | <6 | 3 | N | .. | … | … | N | … | n | … |
| 3224 | 18282047+0635248 | 15.4 ± 0.6 | 4765 ± 127 | 2.56 ± 0.28 | 0.03 ± 0.1 | … | … | N | … | … | … | N | … | n | … |
| 3225 | 18282204+0650074 | -53.6 ± 0.6 | 6391 ± 125 | 4.33 ± 0.23 | 0.02 ± 0.096 | 55 ± 1 | … | N | .. | … | … | N | … | n | NG |
| 3226 | 18282297+0642293 | -29.6 ± 0.6 | 5074 ± 112 | 2.84 ± 0.23 | -0.01 ± 0.092 | 20 ± 1 | … | Y | Y | Y | Y | Y | … | Y | … |
| 3227 | 18283246+0644404 | -27.3 ± 0.6 | 5902 ± 118 | 4.34 ± 0.23 | 0.00 ± 0.094 | 81 ± 1 | 1 | Y | Y | Y | Y | Y | Y | Y | … |
| 3228 | 18283247+0644249 | 8.5 ± 0.6 | 6057 ± 121 | 4.17 ± 0.23 | 0.06 ± 0.1 | 39 ± 3 | … | N | .. | … | … | N | … | n | NG |

| ID | CNAME | RV (km s$^{-1}$) | $T_{\rm eff}$ (K) | logg (dex) | [Fe/H] (dex) | EW(Li)$^a$ (mÅ) | EW(Li) error flag$^b$ | Membership RV | Li | logg | [Fe/H] | Gaia studies Randich$^c$ | Cantat-Gaudin$^c$ | Final$^d$ | NMs with Li$^e$ |
|---|---|---|---|---|---|---|---|---|---|---|---|---|---|---|---|
| 48399 | 18260669+0617594 | -23.6 ± 0.3 | 5057 ± 77 | 3.00 ± 0.31 | -0.16 ± 0.15 | … | … | Y | … | … | … | … | … | n | G |
| 48400 | 18260712+0620257 | -11.4 ± 0.6 | 5163 ± 166 | 4.34 ± 0.19 | -0.80 ± 0.88 | <51 | 3 | Y | Y | Y | N | … | … | Y | … |
| 48401 | 18260752+0623519 | 21.8 ± 0.3 | 4576 ± 261 | 2.59 ± 0.40 | 0.25 ± 0.31 | … | … | N | … | … | … | … | … | n | G |
| 48402 | 18260780+0621598 | 32.2 ± 0.4 | 5145 ± 64 | 4.42 ± 0.28 | -0.17 ± 0.13 | … | … | N | … | … | … | … | … | n | … |
| 48403 | 18260782+0618088 | 38.9 ± 0.7 | 4564 ± 74 | 4.90 ± 0.35 | -0.15 ± 0.3 | … | … | N | … | … | … | … | … | n | … |
| 48404 | 18260795+0617545 | -38.3 ± 0.3 | 4898 ± 185 | 2.81 ± 0.28 | -0.08 ± 0.16 | … | … | Y | … | … | … | … | … | n | G |
| 48405 | 18260802+0625009 | 44.6 ± 0.3 | 4982 ± 135 | 3.45 ± 0.14 | -0.20 ± 0.2 | … | … | N | … | … | … | … | … | n | G |
| 48406 | 18260834+0624119 | -30.1 ± 0.3 | 4652 ± 253 | 4.61 ± 0.36 | -0.36 ± 0.26 | <37 | 3 | Y | Y | Y | N | Y | … | Y | … |
| 48407 | 18260877+0625084 | 46.3 ± 0.4 | 4869 ± 56 | 3.12 ± 0.39 | 0.26 ± 0.2 | <156 | 3 | N | .. | … | … | … | … | n | Li-rich G |
| 48408 | 18260879+0626352 | 160.6 ± 0.3 | 5055 ± 69 | 2.95 ± 0.49 | -0.26 ± 0.2 | … | … | N | … | … | … | … | … | n | G |
| 48409 | 18260881+0621359 | -8.9 ± 0.4 | 4893 ± 176 | 2.95 ± 0.25 | -0.11 ± 0.16 | … | … | Y | … | … | … | … | … | n | G |
| 48410 | 18260891+0622258 | -32.6 ± 1.1 | 4790 ± 86 | 3.20 ± 0.37 | -0.36 ± 0.09 | … | … | Y | … | … | … | N | … | n | G |
| 48411 | 18260905+0626381 | 53.9 ± 0.5 | 5107 ± 176 | 3.40 ± 0.53 | -0.45 ± 0.17 | … | … | N | … | … | … | … | … | n | Li-rich G |
| 48412 | 18260963+0620448 | -28.0 ± 0.4 | 4023 ± 279 | 4.80 ± 0.25 | -0.41 ± 0.27 | <40 | 3 | Y | Y | Y | N | Y | Y | Y | … |
| 48413 | 18260973+0621338 | -80.5 ± 0.2 | 5871 ± 144 | 4.30 ± 0.23 | 0.10 ± 0.2 | 95 ± 15 | 1 | N | .. | … | … | … | … | n | NG |
| 48414 | 18260983+0624542 | -102.9 ± 0.3 | 4552 ± 183 | 4.39 ± 0.70 | 0.05 ± 0.15 | … | … | N | … | … | … | … | … | n | … |
| 48415 | 18261013+0618023 | 10.4 ± 0.3 | 4500 ± 125 | 4.40 ± 0.21 | 0.27 ± 0.27 | … | … | N | … | … | … | … | … | n | … |
| 48416 | 18261028+0625511 | -23.9 ± 0.5 | 4072 ± 333 | 4.86 ± 0.36 | -0.08 ± 0.21 | … | … | Y | … | … | … | N | … | n | … |
| 48417 | 18261053+0620394 | 40.3 ± 0.3 | 4662 ± 195 | 2.94 ± 0.45 | -0.14 ± 0.16 | … | … | N | … | … | … | … | … | n | G |
| 48418 | 18261090+0626547 | -17.9 ± 0.3 | 5599 ± 104 | 4.37 ± 0.27 | -0.10 ± 0.13 | … | … | Y | … | … | … | … | … | n | … |
| 48419 | 18261114+0619560 | 43.3 ± 0.3 | 4894 ± 114 | 3.34 ± 0.38 | 0.03 ± 0.13 | <35 | 3 | N | .. | … | … | … | … | n | G |
| 48420 | 18261120+0622483 | 0.0 ± 0.3 | 4998 ± 92 | 3.39 ± 0.47 | 0.04 ± 0.13 | … | … | Y | … | … | … | … | … | n | G |
| 48421 | 18261126+0619464 | 24.6 ± 0.2 | 4585 ± 156 | 2.65 ± 0.35 | 0.27 ± 0.27 | <31 | 3 | N | .. | … | … | … | … | n | G |
| 48422 | 18261176+0618404 | -29.0 ± 0.2 | 5012 ± 78 | 3.46 ± 0.20 | 0.08 ± 0.16 | … | … | Y | … | … | … | … | … | n | G |
| 48423 | 18261259+0624221 | 93.4 ± 0.2 | 4888 ± 156 | 2.42 ± 0.72 | -0.36 ± 0.26 | … | … | N | … | … | … | … | … | n | G |
| 48424 | 18261265+0622234 | -44.6 ± 0.3 | 5088 ± 63 | 4.43 ± 0.58 | 0.12 ± 0.15 | … | … | Y | … | … | … | N | … | n | … |
| 48425 | 18261269+0628111 | -40.3 ± 0.2 | 5721 ± 125 | 4.35 ± 0.36 | 0.00 ± 0.18 | … | … | Y | … | … | … | N | … | n | … |
| 48426 | 18261297+0627482 | 0.9 ± 0.3 | 5527 ± 91 | 4.64 ± 0.59 | -0.06 ± 0.15 | 155 ± 70 | 1 | Y?$^f$ | N | Y | Y | … | … | n | NG |
| 48427 | 18261303+0626402 | -26.9 ± 0.3 | 4407 ± 337 | 4.69 ± 0.18 | -0.20 ± 0.27 | <28 | 3 | Y | Y | Y | Y | Y | … | Y | … |
| 48428 | 18261308+0627042 | 98.1 ± 0.3 | 4909 ± 126 | 3.09 ± 0.42 | -0.13 ± 0.18 | … | … | N | … | … | … | … | … | n | G |
| 48429 | 18261312+0622509 | 7.7 ± 0.2 | 4379 ± 233 | 2.43 ± 0.44 | 0.14 ± 0.26 | <31 | 3 | N | .. | … | … | … | … | n | G |
| 48430 | 18261318+0622236 | -85.9 ± 0.3 | 4665 ± 347 | 2.37 ± 0.53 | -0.57 ± 0.17 | … | … | N | … | … | … | … | … | n | G |
| 48431 | 18261324+0619438 | -176.5 ± 0.4 | 4763 ± 140 | 2.05 ± 0.45 | -1.10 ± 0.19 | … | … | N | … | … | … | … | … | n | G |
| 48432 | 18261346+0625205 | -61.2 ± 0.8 | 4849 ± 87 | 4.00 ± 0.25 | -0.18 ± 0.1 | … | … | N | … | … | … | N | … | n | … |
| 48433 | 18261369+0623089 | 55.1 ± 0.8 | 4992 ± 175 | 3.05 ± 2.30 | -0.60 ± 0.29 | … | … | N | … | … | … | N | … | n | Li-rich G |
| 48434 | 18261445+0625464 | 113.5 ± 0.3 | 4768 ± 255 | 2.84 ± 0.39 | -0.21 ± 0.27 | … | … | N | … | … | … | … | … | n | G |
| 48435 | 18261458+0623223 | -180.5 ± 3.8 | 4934 ± 216 | 3.70 ± 0.87 | -1.28 ± 0.28 | … | … | N | … | … | … | … | … | n | … |
| 48436 | 18261461+0627564 | -6.5 ± 0.9 | 4989 ± 238 | 2.63 ± 0.26 | -0.43 ± 0.2 | … | … | Y | … | … | … | … | … | n | G |
| 48437 | 18261466+0622488 | 42.6 ± 0.4 | 4428 ± 284 | 4.09 ± 0.51 | -0.29 ± 0.26 | <73 | 3 | N | .. | … | … | … | … | n | NG |
| 48438 | 18261472+0626371 | 24.6 ± 0.3 | 6053 ± 189 | 4.04 ± 0.14 | 0.18 ± 0.13 | … | … | N | … | … | … | … | … | n | … |
| 48439 | 18261480+0626093 | -17.8 ± 0.3 | 4957 ± 359 | 3.33 ± 0.76 | -0.87 ± 0.17 | … | … | Y | … | … | … | … | … | n | G |
| 48440 | 18261525+0625520 | -54.4 ± 1.5 | … | … | … | … | … | N | … | … | … | … | … | n | … |
| 48441 | 18261583+0626587 | 137.1 ± 0.3 | 4835 ± 201 | 3.09 ± 0.21 | -0.26 ± 0.22 | … | … | N | … | … | … | … | … | n | G |
| 48442 | 18261611+0625459 | -61.2 ± 0.2 | 5716 ± 114 | 4.08 ± 0.30 | -0.01 ± 0.15 | … | … | N | … | … | … | … | … | n | … |
| 48443 | 18261643+0623498 | 35.2 ± 0.4 | 5248 ± 80 | 3.75 ± 0.47 | -0.05 ± 0.15 | … | … | N | … | … | … | … | … | n | … |
| 48444 | 18261650+0620177 | 0.1 ± 0.5 | 4309 ± 458 | 4.41 ± 0.33 | -0.37 ± 0.33 | … | … | Y | … | … | … | … | … | n | … |
| 48445 | 18261670+0621520 | 52.3 ± 0.3 | 5947 ± 213 | 4.31 ± 0.27 | 0.23 ± 0.2 | … | … | N | … | … | … | … | … | n | … |
| 48446 | 18261673+0620132 | 76.2 ± 0.2 | 4701 ± 122 | 2.66 ± 0.33 | 0.09 ± 0.27 | … | … | N | … | … | … | … | … | n | G |
| 48447 | 18261709+0618489 | 5.7 ± 0.3 | 4637 ± 70 | 2.72 ± 0.39 | 0.09 ± 0.2 | … | … | N | … | … | … | … | … | n | G |
| 48448 | 18261731+0617562 | -1.5 ± 0.3 | 4086 ± 272 | 4.76 ± 0.27 | -0.17 ± 0.16 | <67 | 3 | Y | N | N | Y | N | … | n | NG |
| 48449 | 18261782+0623259 | -25.3 ± 0.7 | 4763 ± 216 | 3.21 ± 0.58 | -0.49 ± 0.18 | … | … | Y | … | … | … | N | … | n | G |
| 48450 | 18261803+0625324 | -25.1 ± 0.3 | 5700 ± 192 | 4.06 ± 0.15 | -0.41 ± 0.65 | … | … | Y | … | … | … | … | … | n | … |
| 48451 | 18261806+0626480 | -42.8 ± 0.2 | 5687 ± 265 | 4.17 ± 0.15 | 0.24 ± 0.13 | … | … | Y | … | … | … | … | … | n | … |
| 48452 | 18261845+0621077 | -130.7 ± 0.9 | 4669 ± 690 | 3.34 ± 1.13 | -0.24 ± 0.19 | <57 | 3 | N | .. | … | … | … | … | n | G |
| 48453 | 18261853+0621131 | 65.7 ± 0.2 | 5989 ± 178 | 3.95 ± 0.26 | 0.30 ± 0.14 | 76 ± 29 | 1 | N | … | … | … | … | … | n | NG |
| 48454 | 18261855+0620540 | -30.5 ± 0.3 | 4692 ± 267 | 2.81 ± 0.50 | -0.12 ± 0.25 | … | … | Y | … | … | … | … | … | n | G |
| 48455 | 18261865+0617576 | -148.2 ± 0.3 | 4689 ± 131 | 1.40 ± 0.46 | -1.68 ± 0.43 | … | … | N | … | … | … | … | … | n | G |
| 48456 | 18261878+0625333 | 17.3 ± 0.3 | 5147 ± 189 | 3.06 ± 0.88 | -0.45 ± 0.23 | … | … | N | … | … | … | … | … | n | G |
| 48457 | 18261883+0625488 | -40.3 ± 0.3 | 4449 ± 220 | 2.46 ± 0.44 | 0.28 ± 0.33 | … | … | Y | … | … | … | … | … | n | G |

**Table C.11.** continued.

| ID | CNAME | RV (km s$^{-1}$) | $T_{\text{eff}}$ (K) | logg (dex) | [Fe/H] (dex) | EW(Li)$^a$ (mÅ) | EW(Li) error flag$^b$ | Membership RV | Li | logg | [Fe/H] | Gaia studies Randich$^c$ | Cantat-Gaudin$^c$ | Final$^d$ | NMs with Li$^e$ |
|---|---|---|---|---|---|---|---|---|---|---|---|---|---|---|---|
| 48458 | 18261895+0627501 | 120.3 ± 0.2 | 4907 ± 183 | 2.98 ± 0.36 | -0.03 ± 0.24 | ... | ... | N | ... | ... | ... | ... | ... | n | G |
| 48459 | 18261896+0622536 | 13.6 ± 0.3 | 5153 ± 120 | 3.24 ± 0.14 | -0.27 ± 0.2 | ... | ... | N | ... | ... | ... | ... | ... | n | G |
| 48460 | 18261962+0620434 | -3.7 ± 0.4 | 4466 ± 386 | 4.24 ± 0.74 | 0.00 ± 0.27 | ... | ... | Y | ... | ... | ... | ... | ... | n | ... |
| 48461 | 18261984+0623433 | -125.5 ± 0.3 | 4636 ± 296 | 2.21 ± 0.48 | -0.62 ± 0.28 | ... | ... | N | ... | ... | ... | ... | ... | n | G |
| 48462 | 18261997+0624559 | -24.0 ± 0.3 | 4689 ± 216 | 2.67 ± 0.40 | -0.06 ± 0.23 | ... | ... | Y | ... | ... | ... | ... | ... | n | G |
| 48463 | 18262012+0618471 | -27.2 ± 0.3 | 3916 ± 332 | 4.58 ± 0.09 | -0.31 ± 0.27 | <102 | 3 | Y | Y | Y | Y | Y | ... | Y | ... |
| 48464 | 18262030+0619328 | 2.5 ± 0.3 | 4846 ± 176 | 2.64 ± 0.36 | -0.02 ± 0.15 | ... | ... | Y | ... | ... | ... | ... | ... | n | G |
| 48465 | 18262037+0620554 | 26.6 ± 0.3 | 4295 ± 200 | 4.86 ± 0.24 | -0.06 ± 0.19 | <35 | 3 | N | .. | ... | ... | N | ... | n | NG |
| 48466 | 18262044+0618371 | -65.8 ± 0.2 | 4921 ± 43 | 4.31 ± 0.62 | 0.19 ± 0.2 | ... | ... | N | ... | ... | ... | ... | ... | n | ... |
| 48467 | 18262067+0624250 | 21.8 ± 0.4 | 4946 ± 130 | 3.33 ± 0.32 | -0.41 ± 0.23 | ... | ... | N | ... | ... | ... | ... | ... | n | G |
| 48468 | 18262132+0618174 | 24.6 ± 0.3 | 6107 ± 189 | 4.30 ± 0.25 | -0.34 ± 0.16 | 33 ± 16 | 1 | N | .. | ... | ... | ... | ... | n | NG |
| 48469 | 18262138+0619317 | -15.5 ± 0.3 | 5866 ± 209 | 4.16 ± 0.29 | -0.15 ± 0.15 | ... | ... | Y | ... | ... | ... | ... | ... | n | ... |
| 48470 | 18262149+0620439 | -38.6 ± 0.3 | 5799 ± 69 | 4.34 ± 0.24 | 0.22 ± 0.13 | ... | ... | Y | ... | ... | ... | ... | ... | n | ... |
| 48471 | 18262152+0621270 | -111.1 ± 0.5 | 3728 ± 301 | 4.13 ± 0.19 | ... | ... | ... | N | ... | ... | ... | ... | ... | n | ... |
| 48472 | 18262203+0624068 | 9.6 ± 0.3 | 4624 ± 154 | 2.79 ± 0.43 | 0.26 ± 0.32 | ... | ... | N | ... | ... | ... | ... | ... | n | G |
| 48473 | 18262230+0626024 | 42.5 ± 0.2 | 4798 ± 222 | 2.86 ± 0.42 | -0.28 ± 0.22 | ... | ... | N | ... | ... | ... | ... | ... | n | G |
| 48474 | 18262269+0627312 | -45.3 ± 0.3 | 3907 ± 368 | 4.67 ± 0.28 | -0.69 ± 0.58 | <56 | 3 | Y | Y | N | N | N | ... | n | NG |
| 48475 | 18262282+0621399 | -65.2 ± 0.3 | 4687 ± 323 | 2.43 ± 0.42 | -0.34 ± 0.13 | ... | ... | N | ... | ... | ... | ... | ... | n | G |
| 48476 | 18262319+0627267 | -11.3 ± 0.3 | 4661 ± 305 | 2.83 ± 0.43 | -0.19 ± 0.2 | ... | ... | Y | ... | ... | ... | ... | ... | n | G |
| 48477 | 18262335+0626314 | 13.2 ± 0.2 | 4621 ± 131 | 2.71 ± 0.42 | 0.22 ± 0.23 | ... | ... | N | ... | ... | ... | ... | ... | n | G |
| 48478 | 18262343+0619477 | 50.1 ± 0.5 | 4319 ± 185 | 4.32 ± 0.62 | -0.09 ± 0.39 | ... | ... | N | ... | ... | ... | N | ... | n | ... |
| 48479 | 18262345+0620532 | -68.0 ± 0.3 | 4118 ± 260 | 1.97 ± 0.65 | 0.00 ± 0.37 | ... | ... | N | ... | ... | ... | ... | ... | n | ... |
| 48480 | 18262349+0627351 | -46.0 ± 0.3 | 4222 ± 260 | 4.95 ± 0.37 | -0.36 ± 0.25 | ... | ... | Y | ... | ... | ... | N | ... | n | ... |
| 48481 | 18262363+0623207 | 39.1 ± 0.2 | 4772 ± 138 | 2.85 ± 0.28 | -0.04 ± 0.16 | ... | ... | N | ... | ... | ... | ... | ... | n | G |
| 48482 | 18262384+0622578 | 5.0 ± 0.6 | 4175 ± 217 | 4.13 ± 0.49 | -0.19 ± 0.17 | ... | ... | N | ... | ... | ... | ... | ... | n | ... |
| 48483 | 18262418+0620145 | 25.5 ± 0.2 | 4954 ± 127 | 3.40 ± 0.22 | 0.03 ± 0.15 | ... | ... | N | ... | ... | ... | ... | ... | n | G |
| 48484 | 18262419+0623484 | 49.9 ± 0.2 | 5514 ± 225 | 3.97 ± 0.16 | 0.22 ± 0.15 | <37 | 3 | N | .. | ... | ... | ... | ... | n | NG |
| 48485 | 18262429+0626399 | -52.3 ± 0.3 | 4183 ± 264 | 4.95 ± 0.36 | -0.35 ± 0.21 | <45 | 3 | N | .. | ... | ... | N | ... | n | NG |
| 48486 | 18262433+0618449 | 46.5 ± 0.3 | 4689 ± 311 | 4.49 ± 0.21 | -0.31 ± 0.19 | ... | ... | N | ... | ... | ... | ... | ... | n | ... |
| 48487 | 18262486+0621425 | 15.0 ± 0.2 | 5024 ± 115 | 4.54 ± 0.31 | 0.17 ± 0.16 | <34 | 3 | N | .. | ... | ... | ... | ... | n | NG |
| 48488 | 18262488+0617101 | 32.3 ± 0.5 | 4983 ± 150 | 3.18 ± 0.79 | -0.22 ± 0.21 | ... | ... | N | ... | ... | ... | ... | ... | n | G |
| 48489 | 18262492+0622351 | -11.8 ± 0.4 | 6115 ± 201 | 4.25 ± 0.23 | -0.27 ± 0.28 | 125 ± 50 | 1 | Y | Y | Y | Y | ... | ... | Y | ... |
| 48490 | 18262526+0621371 | -41.7 ± 0.3 | 5110 ± 70 | 3.65 ± 0.20 | -0.09 ± 0.14 | ... | ... | Y | ... | ... | ... | ... | ... | n | ... |
| 48491 | 18262554+0618047 | 1.6 ± 0.3 | 5243 ± 75 | 3.92 ± 0.16 | 0.09 ± 0.13 | 37 ± 31 | 1 | Y?$^f$ | Y | Y | Y | N | ... | n | NG |
| 48492 | 18262559+0621336 | 3.9 ± 0.3 | 5196 ± 257 | 3.73 ± 0.55 | -0.48 ± 0.23 | ... | ... | N | ... | ... | ... | ... | ... | n | ... |
| 48493 | 18262569+0615456 | 57.7 ± 0.8 | 4484 ± 73 | 4.87 ± 0.18 | -0.03 ± 0.09 | ... | ... | N | ... | ... | ... | N | ... | n | ... |
| 48494 | 18262572+0630228 | -55.2 ± 0.2 | 5024 ± 96 | 3.43 ± 0.27 | -0.04 ± 0.13 | ... | ... | N | ... | ... | ... | ... | ... | n | G |
| 48495 | 18262577+0630370 | -27.7 ± 0.4 | 3969 ± 260 | 4.64 ± 0.11 | 0.01 ± 0.2 | ... | 3 | Y | ... | ... | ... | Y | Y | n | ... |
| 48496 | 18262586+0632394 | 32.8 ± 0.4 | 6205 ± 129 | 4.01 ± 0.18 | -0.16 ± 0.15 | ... | ... | N | ... | ... | ... | ... | ... | n | ... |
| 48497 | 18262604+0630582 | -9.2 ± 0.6 | 5400 ± 51 | 4.21 ± 0.36 | 0.39 ± 0.17 | ... | ... | Y | ... | ... | ... | ... | ... | n | ... |
| 48498 | 18262606+0628420 | 24.0 ± 0.3 | 5380 ± 28 | 4.72 ± 0.36 | -0.03 ± 0.14 | ... | ... | N | ... | ... | ... | ... | ... | n | ... |
| 48499 | 18262609+0613417 | 30.0 ± 0.3 | 4459 ± 252 | 4.83 ± 0.35 | -0.34 ± 0.24 | ... | ... | N | ... | ... | ... | N | ... | n | ... |
| 48500 | 18262615+0632066 | 14.9 ± 2.0 | ... | ... | ... | ... | ... | N | ... | ... | ... | ... | ... | n | ... |
| 48501 | 18262625+0630144 | -24.5 ± 0.3 | 3519 ± 21 | 3.97 ± 0.14 | ... | ... | ... | Y | ... | ... | ... | N | ... | n | ... |
| 48502 | 18262625+0644373 | -7.6 ± 0.5 | 4908 ± 198 | 2.56 ± 0.70 | -0.43 ± 0.29 | ... | ... | Y | ... | ... | ... | ... | ... | n | G |
| 48503 | 18262633+0622087 | 131.7 ± 1.0 | 5117 ± 455 | 3.63 ± 0.72 | 0.01 ± 0.33 | ... | ... | N | ... | ... | ... | ... | ... | n | ... |
| 48504 | 18262633+0638335 | 32.3 ± 0.6 | 5172 ± 166 | 4.45 ± 0.22 | 0.08 ± 0.17 | ... | ... | N | ... | ... | ... | ... | ... | n | ... |
| 48505 | 18262635+0641320 | -11.1 ± 1.0 | 4588 ± 103 | 4.43 ± 0.27 | -0.11 ± 0.11 | ... | ... | Y | ... | ... | ... | ... | ... | n | ... |
| 48506 | 18262639+0626566 | 21.7 ± 0.2 | 4949 ± 158 | 3.55 ± 0.33 | 0.04 ± 0.16 | ... | ... | N | .. | ... | ... | ... | ... | n | ... |
| 48507 | 18262640+0617354 | 7.0 ± 0.6 | 6585 ± 290 | 4.04 ± 0.24 | 0.13 ± 0.19 | 31 ± 26 | 1 | N | ... | ... | ... | ... | ... | n | NG |
| 48508 | 18262642+0635517 | -34.2 ± 0.3 | 4707 ± 140 | 2.44 ± 0.39 | -0.34 ± 0.13 | 45 ± 34 | 1 | Y | Y | Y | Y | ... | ... | Y | ... |
| 48509 | 18262648+0645511 | -15.0 ± 0.3 | 5863 ± 193 | 4.46 ± 0.18 | 0.23 ± 0.15 | <32 | 3 | Y | Y | Y | Y | ... | ... | Y | ... |
| 48510 | 18262659+0620559 | -33.7 ± 0.3 | 5648 ± 220 | 4.41 ± 0.19 | 0.23 ± 0.14 | ... | ... | Y | ... | ... | ... | ... | ... | n | ... |
| 48511 | 18262665+0644076 | -11.1 ± 0.7 | 4042 ± 363 | 4.69 ± 0.23 | -0.82 ± 0.58 | <152 | 3 | Y | N | Y | N | ... | ... | n | NG |
| 48512 | 18262669+0633452 | 36.2 ± 1.3 | 4485 ± 110 | 4.86 ± 0.31 | -0.07 ± 0.14 | ... | ... | N | ... | ... | ... | N | ... | n | ... |
| 48513 | 18262693+0613455 | -3.7 ± 0.3 | 4744 ± 253 | 3.27 ± 0.35 | 0.26 ± 0.28 | ... | ... | Y | ... | ... | ... | ... | ... | n | G |
| 48514 | 18262697+0646116 | 27.9 ± 0.3 | 4814 ± 71 | 2.94 ± 0.43 | 0.11 ± 0.24 | ... | ... | N | ... | ... | ... | ... | ... | n | G |
| 48515 | 18262704+0632456 | -88.8 ± 2.0 | 4865 ± 110 | ... | -1.37 ± 0.18 | ... | ... | N | ... | ... | ... | ... | ... | n | ... |
| 48516 | 18262709+0637326 | 14.5 ± 0.4 | 5787 ± 169 | 4.30 ± 0.25 | 0.22 ± 0.17 | ... | ... | N | ... | ... | ... | ... | ... | n | ... |

| ID | CNAME | RV (km s$^{-1}$) | $T_{\rm eff}$ (K) | log g (dex) | [Fe/H] (dex) | EW(Li)$^a$ (mÅ) | EW(Li) error flag$^b$ | Membership RV | Li | log g | [Fe/H] | Gaia studies Randich$^c$ | Cantat-Gaudin$^c$ | Final$^d$ | NMs with Li$^e$ |
|---|---|---|---|---|---|---|---|---|---|---|---|---|---|---|---|
| 48517 | 18262715+0621046 | -9.6 ± 0.3 | 3758 ± 230 | 4.50 ± 0.24 | -0.03 ± 0.14 | … | … | Y | … | … | … | N | … | n | … |
| 48518 | 18262719+0630267 | -19.4 ± 0.5 | 4696 ± 157 | 2.27 ± 0.49 | -0.20 ± 0.37 | … | … | Y | … | … | … | … | … | n | G |
| 48519 | 18262722+0645205 | 28.2 ± 0.5 | 5048 ± 142 | 4.75 ± 0.42 | -0.02 ± 0.14 | … | … | N | … | … | … | … | … | n | … |
| 48520 | 18262729+0628028 | -42.9 ± 0.3 | 4947 ± 133 | 3.09 ± 0.42 | -0.21 ± 0.15 | … | … | Y | … | … | … | … | … | n | G |
| 48521 | 18262730+0631296 | -20.5 ± 0.2 | 6010 ± 173 | 4.35 ± 0.15 | 0.29 ± 0.15 | 61 ± 24 | 1 | Y | Y | Y | Y | … | … | Y | … |
| 48522 | 18262737+0640353 | 23.1 ± 0.3 | 4658 ± 276 | 2.73 ± 0.44 | -0.26 ± 0.24 | … | … | N | … | … | … | … | … | n | G |
| 48523 | 18262738+0617198 | -30.5 ± 0.3 | 5069 ± 83 | 4.50 ± 0.38 | 0.09 ± 0.15 | <69 | 3 | Y | Y | Y | Y | … | … | Y | … |
| 48524 | 18262739+0645167 | 74.6 ± 0.4 | 4404 ± 193 | 4.60 ± 0.28 | 0.06 ± 0.23 | … | … | N | … | … | … | … | … | n | … |
| 48525 | 18262747+0634096 | 49.2 ± 0.3 | 4995 ± 150 | 3.47 ± 0.37 | 0.00 ± 0.2 | … | … | N | … | … | … | … | … | n | Li-rich G |
| 48526 | 18262749+0636123 | -56.6 ± 0.3 | 3852 ± 291 | 4.60 ± 0.13 | -0.06 ± 0.12 | <23 | 3 | N | .. | … | … | N | … | n | NG |
| 48527 | 18262758+0618261 | 28.5 ± 0.2 | 4661 ± 158 | 2.86 ± 0.43 | 0.21 ± 0.26 | … | … | N | … | … | … | … | … | n | G |
| 48528 | 18262758+0644124 | 29.1 ± 0.5 | 3962 ± 248 | 4.62 ± 0.11 | -0.06 ± 0.25 | <133 | 3 | N | .. | … | … | … | … | n | NG |
| 48529 | 18262764+0641274 | -14.9 ± 0.3 | 5665 ± 361 | 4.60 ± 0.25 | -0.01 ± 0.18 | … | … | Y | … | … | … | … | … | n | … |
| 48530 | 18262770+0628503 | 8.3 ± 0.2 | 4719 ± 255 | 4.55 ± 0.30 | 0.09 ± 0.16 | <32 | 3 | N | .. | … | … | … | … | n | NG |
| 48531 | 18262789+0632038 | -91.9 ± 0.3 | 5687 ± 113 | 4.26 ± 0.22 | -0.08 ± 0.18 | … | … | N | … | … | … | … | … | n | … |
| 48532 | 18262789+0646116 | -18.1 ± 0.3 | 5627 ± 70 | 3.71 ± 0.57 | -0.31 ± 0.15 | … | … | Y | … | … | … | … | … | n | … |
| 48533 | 18262790+0621331 | -5.6 ± 0.3 | 5332 ± 131 | 4.21 ± 0.42 | 0.07 ± 0.14 | … | … | Y | … | … | … | … | … | n | … |
| 48534 | 18262792+0628454 | 17.7 ± 0.3 | 5839 ± 63 | 4.41 ± 0.11 | 0.25 ± 0.16 | … | … | N | … | … | … | … | … | n | … |
| 48535 | 18262798+0613556 | 11.0 ± 1.0 | 4778 ± 100 | 4.86 ± 0.27 | 0.53 ± 0.1 | … | … | N | … | … | … | N | … | n | … |
| 48536 | 18262800+0630534 | -3.0 ± 0.4 | 4618 ± 314 | 4.62 ± 0.50 | -0.09 ± 0.13 | <79 | 3 | Y | N | N | Y | … | … | n | NG |
| 48537 | 18262801+0620013 | -33.8 ± 0.7 | 4570 ± 126 | 5.11 ± 0.21 | -0.59 ± 0.15 | … | … | Y | … | … | … | N | … | n | … |
| 48538 | 18262802+0644011 | 73.0 ± 0.3 | 4669 ± 129 | 2.50 ± 0.23 | -0.20 ± 0.18 | … | … | N | … | … | … | … | … | n | G |
| 48539 | 18262806+0632194 | -15.0 ± 0.3 | 5904 ± 286 | 4.16 ± 0.06 | 0.26 ± 0.15 | 58 ± 28 | 1 | Y | Y | Y | Y | … | … | Y | … |
| 48540 | 18262808+0613430 | 57.1 ± 2.7 | … | … | … | … | … | N | … | … | … | … | … | n | … |
| 48541 | 18262837+0624127 | -24.4 ± 0.3 | 4940 ± 143 | 2.96 ± 0.43 | -0.35 ± 0.17 | … | … | Y | … | … | … | … | … | n | G |
| 48542 | 18262839+0630313 | -75.3 ± 0.3 | 4951 ± 97 | 3.77 ± 0.18 | -0.40 ± 0.14 | <37 | 3 | N | .. | … | … | … | … | n | NG |
| 48543 | 18262841+0644311 | -1.3 ± 0.3 | 4691 ± 65 | 4.36 ± 0.60 | 0.38 ± 0.25 | … | … | Y | … | … | … | … | … | n | … |
| 48544 | 18262845+0628337 | -52.3 ± 0.3 | 4921 ± 253 | 3.33 ± 0.44 | -0.47 ± 0.22 | … | … | N | … | … | … | … | … | n | G |
| 48545 | 18262849+0625411 | -130.5 ± 0.3 | 4893 ± 186 | 2.37 ± 0.33 | -0.69 ± 0.24 | … | … | N | … | … | … | … | … | n | G |
| 48546 | 18262852+0616295 | 25.7 ± 0.4 | 4224 ± 217 | 4.60 ± 0.36 | -0.23 ± 0.14 | … | … | N | … | … | … | … | … | n | … |
| 48547 | 18262852+0617507 | -9.2 ± 0.3 | 5821 ± 53 | 4.13 ± 0.13 | 0.08 ± 0.12 | … | … | Y | … | … | … | … | … | n | … |
| 48548 | 18262852+0631557 | -34.4 ± 0.3 | 5666 ± 349 | 4.11 ± 0.41 | 0.41 ± 0.22 | 60 ± 50 | 1 | Y | Y | Y | N | … | … | Y | … |
| 48549 | 18262881+0636019 | 2.8 ± 1.0 | … | … | … | … | … | Y | … | … | … | N | … | n | … |
| 48550 | 18262884+0634190 | -83.7 ± 0.3 | 4404 ± 296 | 2.28 ± 0.66 | -0.20 ± 0.34 | … | … | N | … | … | … | … | … | n | G |
| 48551 | 18262887+0617332 | 31.2 ± 0.2 | 4736 ± 288 | 4.45 ± 0.28 | 0.14 ± 0.15 | … | … | N | … | … | … | … | … | n | … |
| 48552 | 18262888+0616463 | -36.3 ± 0.4 | 4931 ± 456 | 3.70 ± 0.40 | -0.78 ± 0.68 | <39 | 3 | Y | Y | Y | N | … | … | Y | … |
| 48553 | 18262892+0630286 | -2.1 ± 0.6 | 4848 ± 71 | 5.01 ± 0.57 | -0.04 ± 0.24 | … | … | Y | … | … | … | N | … | n | … |
| 48554 | 18262896+0614254 | -14.7 ± 0.7 | 4066 ± 367 | 4.18 ± 0.37 | 0.56 ± 0.58 | … | … | Y | … | … | … | N | … | n | … |
| 48555 | 18262898+0620191 | -58.7 ± 0.4 | 4143 ± 191 | 4.67 ± 0.20 | -0.27 ± 0.32 | … | … | N | … | … | … | N | … | n | … |
| 48556 | 18262934+0614314 | -389.5 ± 0.7 | 5446 ± 140 | 2.44 ± 0.69 | -1.66 ± 0.37 | … | … | N | … | … | … | … | … | n | … |
| 48557 | 18262943+0642425 | -30.4 ± 0.3 | 5114 ± 97 | 3.57 ± 0.42 | 0.24 ± 0.25 | … | … | Y | … | … | … | … | … | n | … |
| 48558 | 18262957+0623139 | 55.2 ± 0.3 | 5093 ± 86 | 4.16 ± 0.50 | 0.07 ± 0.14 | … | … | N | … | … | … | … | … | n | … |
| 48559 | 18262964+0641508 | -49.2 ± 0.3 | 5266 ± 70 | 3.62 ± 0.37 | -0.11 ± 0.16 | … | … | N | … | … | … | … | … | n | … |
| 48560 | 18262985+0626034 | 3.7 ± 0.3 | 4698 ± 105 | 3.10 ± 0.58 | 0.26 ± 0.24 | <53 | 3 | N | .. | … | … | … | … | n | G |
| 48561 | 18262995+0634176 | -43.9 ± 0.3 | 5828 ± 82 | 3.97 ± 0.23 | 0.00 ± 0.15 | … | … | Y | … | … | … | … | … | n | … |
| 48562 | 18262995+0644358 | 23.6 ± 0.3 | 5175 ± 109 | 3.57 ± 0.29 | 0.03 ± 0.22 | … | … | N | … | … | … | N | … | n | … |
| 48563 | 18263012+0616570 | -61.0 ± 0.5 | 3881 ± 322 | 4.93 ± 0.23 | 0.06 ± 0.24 | … | … | N | … | … | … | N | … | n | … |
| 48564 | 18263022+0620187 | 13.3 ± 0.3 | 4696 ± 156 | 2.92 ± 0.47 | 0.10 ± 0.27 | <54 | 3 | N | .. | … | … | … | … | n | G |
| 48565 | 18263024+0613557 | 14.8 ± 0.3 | 5038 ± 203 | 3.44 ± 0.63 | -0.19 ± 0.42 | … | … | N | … | … | … | … | … | n | G |
| 48566 | 18263027+0630335 | -14.4 ± 0.3 | 6213 ± 363 | 4.20 ± 0.09 | 0.28 ± 0.32 | … | … | Y | … | … | … | … | … | n | … |
| 48567 | 18263035+0629059 | -50.0 ± 0.2 | 5221 ± 57 | 4.33 ± 0.32 | 0.09 ± 0.12 | … | … | N | … | … | … | … | … | n | … |
| 48568 | 18263037+0637569 | 12.5 ± 0.3 | 5117 ± 84 | 3.69 ± 0.28 | -0.04 ± 0.13 | … | … | N | … | … | … | … | … | n | … |
| 48569 | 18263042+0625164 | -29.5 ± 0.3 | 4741 ± 157 | 3.20 ± 0.59 | 0.25 ± 0.28 | … | … | Y | … | … | … | … | … | n | G |
| 48570 | 18263043+0646125 | -8.2 ± 0.3 | 6335 ± 226 | 4.31 ± 0.31 | 0.10 ± 0.21 | 48 ± 20 | 1 | Y | Y | Y | Y | … | … | Y | … |
| 48571 | 18263059+0627576 | 39.3 ± 0.3 | 5822 ± 104 | 4.31 ± 0.19 | -0.03 ± 0.12 | … | … | N | … | … | … | … | … | n | … |
| 48572 | 18263063+0642230 | 14.1 ± 0.4 | 4792 ± 317 | 3.35 ± 0.28 | -0.05 ± 0.27 | … | … | N | … | … | … | … | … | n | G |
| 48573 | 18263066+0624106 | 73.2 ± 0.2 | 4949 ± 113 | 3.31 ± 0.53 | -0.35 ± 0.2 | <27 | 3 | N | .. | … | … | … | … | n | G |
| 48574 | 18263073+0615270 | -167.8 ± 0.4 | 4611 ± 461 | 2.93 ± 0.72 | -0.15 ± 0.4 | <106 | 3 | N | .. | … | … | … | … | n | G |
| 48575 | 18263096+0642428 | -8.9 ± 0.4 | 4476 ± 495 | 4.05 ± 0.46 | -0.51 ± 0.39 | <58 | 3 | Y | N | N | N | … | … | n | NG |

**Table C.11.** continued.

| ID | CNAME | RV (km s$^{-1}$) | $T_{\rm eff}$ (K) | logg (dex) | [Fe/H] (dex) | EW(Li)$^a$ (mÅ) | EW(Li) error flag$^b$ | Membership RV | Li | logg | [Fe/H] | Gaia studies Randich$^c$ | Cantat-Gaudin$^c$ | Final$^d$ | NMs with Li$^e$ |
|---|---|---|---|---|---|---|---|---|---|---|---|---|---|---|---|
| 48576 | 18263112+0620205 | 24.6 ± 0.4 | 3839 ± 467 | 3.65 ± 0.07 | -1.20 ± 1.17 | 44 ± 42 | 1 | N | .. | … | … | … | … | n | NG |
| 48577 | 18263129+0638274 | 5.4 ± 0.4 | 6448 ± 195 | 4.16 ± 0.16 | 0.10 ± 0.16 | … | … | N | … | … | … | … | … | n | … |
| 48578 | 18263130+0622504 | -63.2 ± 0.3 | 4522 ± 169 | 2.02 ± 0.34 | -0.36 ± 0.19 | … | … | N | … | … | … | … | … | n | G |
| 48579 | 18263136+0623283 | 81.3 ± 0.3 | 4821 ± 107 | 2.74 ± 0.37 | -0.17 ± 0.23 | … | … | N | … | … | … | … | … | n | G |
| 48580 | 18263136+0639151 | 8.6 ± 0.3 | 4936 ± 119 | 5.15 ± 0.44 | 0.19 ± 0.18 | … | … | N | … | … | … | … | … | n | … |
| 48581 | 18263139+0614542 | -20.5 ± 1.4 | … | … | … | … | … | Y | … | … | … | N | … | n | … |
| 48582 | 18263139+0645256 | 64.2 ± 0.3 | 5422 ± 183 | 3.95 ± 0.19 | 0.03 ± 0.14 | … | … | N | … | … | … | … | … | n | … |
| 48583 | 18263149+0641517 | 72.5 ± 0.2 | 4765 ± 249 | 2.83 ± 0.30 | -0.36 ± 0.28 | … | … | N | … | … | … | … | … | n | G |
| 48584 | 18263154+0615430 | 37.0 ± 0.5 | 4595 ± 285 | 4.10 ± 0.54 | 0.18 ± 0.39 | … | … | N | … | … | … | … | … | n | … |
| 48585 | 18263154+0644023 | 15.2 ± 0.3 | 6309 ± 126 | 4.24 ± 0.28 | -0.09 ± 0.15 | … | … | N | … | … | … | … | … | n | … |
| 48586 | 18263157+0637178 | -8.8 ± 0.9 | 4338 ± 58 | 5.12 ± 0.23 | -0.49 ± 0.06 | … | … | Y | … | … | … | N | … | n | … |
| 48587 | 18263168+0618261 | 86.8 ± 0.3 | 4802 ± 75 | 3.34 ± 0.41 | 0.15 ± 0.25 | <46 | 3 | N | .. | … | … | … | … | n | G |
| 48588 | 18263178+0624323 | 25.2 ± 0.2 | 4461 ± 188 | 2.72 ± 0.33 | 0.37 ± 0.26 | … | … | N | … | … | … | … | … | n | G |
| 48589 | 18263182+0639213 | 17.4 ± 0.3 | 5182 ± 47 | 3.66 ± 0.19 | -0.08 ± 0.14 | … | … | N | … | … | … | … | … | n | … |
| 48590 | 18263190+0624466 | 93.1 ± 0.2 | 4710 ± 264 | 2.51 ± 0.41 | -0.31 ± 0.26 | … | … | N | … | … | … | … | … | n | G |
| 48592 | 18263195+0622211 | -8.7 ± 0.2 | 4775 ± 195 | 4.66 ± 0.32 | -0.10 ± 0.13 | <18 | 3 | Y | Y | Y | Y | N | … | Y?$^g$ | … |
| 48593 | 18263202+0619447 | -25.1 ± 0.2 | 5090 ± 42 | 3.28 ± 0.34 | -0.13 ± 0.2 | … | … | Y | … | … | … | … | … | n | G |
| 48594 | 18263255+0641547 | -22.2 ± 0.3 | 4635 ± 242 | 4.68 ± 0.44 | 0.07 ± 0.19 | … | … | Y | … | … | … | … | … | n | … |
| 48595 | 18263256+0628364 | 45.8 ± 0.3 | 4644 ± 180 | 4.64 ± 0.35 | -0.05 ± 0.12 | … | … | N | … | … | … | … | … | n | … |
| 48596 | 18263262+0617471 | 74.4 ± 0.2 | 4586 ± 161 | 4.70 ± 0.23 | 0.07 ± 0.16 | … | … | N | … | … | … | … | … | n | … |
| 48597 | 18263270+0634346 | -34.6 ± 0.3 | 5110 ± 38 | 4.47 ± 0.23 | 0.23 ± 0.21 | … | … | Y | … | … | … | … | … | n | … |
| 48598 | 18263272+0624130 | 130.4 ± 0.3 | 4898 ± 186 | 3.46 ± 0.13 | -0.15 ± 0.17 | … | … | N | … | … | … | … | … | n | G |
| 48599 | 18263272+0628392 | 25.1 ± 0.3 | 4511 ± 213 | 4.58 ± 0.09 | -0.34 ± 0.53 | … | … | N | … | … | … | N | … | n | … |
| 48600 | 18263275+0618534 | -63.4 ± 1.0 | 4471 ± 447 | 4.04 ± 0.53 | -0.03 ± 0.21 | … | … | N | … | … | … | N | … | n | … |
| 48601 | 18263295+0636394 | 1.2 ± 0.4 | 4811 ± 419 | 4.16 ± 0.41 | -0.81 ± 0.3 | <37 | 3 | Y?$^f$ | Y | Y | N | … | … | n | NG |
| 48602 | 18263300+0615228 | 12.8 ± 0.3 | 5877 ± 147 | 4.26 ± 0.32 | 0.32 ± 0.16 | 95 ± 40 | 1 | N | .. | … | … | … | … | n | NG |
| 48603 | 18263304+0633395 | -65.5 ± 0.3 | 4333 ± 299 | 2.22 ± 0.41 | 0.14 ± 0.24 | … | … | N | … | … | … | … | … | n | G |
| 48604 | 18263316+0626113 | 131.1 ± 0.3 | 4711 ± 87 | 3.26 ± 0.65 | 0.24 ± 0.26 | … | … | N | … | … | … | … | … | n | G |
| 48605 | 18263326+0629438 | -50.3 ± 0.3 | 6182 ± 94 | 4.07 ± 0.01 | -0.24 ± 0.4 | … | … | N | … | … | … | … | … | n | … |
| 48606 | 18263344+0644463 | 14.1 ± 0.3 | 5262 ± 142 | 4.86 ± 0.59 | 0.01 ± 0.13 | … | … | N | … | … | … | … | … | n | … |
| 48607 | 18263346+0614414 | 56.0 ± 0.3 | 5966 ± 118 | 4.13 ± 0.21 | 0.34 ± 0.23 | … | … | N | … | … | … | … | … | n | … |
| 48608 | 18263354+0618505 | -27.5 ± 0.4 | 3478 ± 133 | 4.47 ± 0.13 | -0.05 ± 0.13 | <114 | 3 | Y | N | Y | Y | … | … | n | NG |
| 48609 | 18263360+0643226 | -3.8 ± 0.4 | 5051 ± 289 | 3.41 ± 0.65 | -0.49 ± 0.25 | <41 | 3 | Y | Y | Y | N | … | … | Y | … |
| 48610 | 18263382+0627084 | -170.0 ± 0.3 | 4686 ± 149 | 1.93 ± 0.49 | -1.05 ± 0.28 | … | … | N | … | … | … | … | … | n | G |
| 48611 | 18263384+0616019 | -5.3 ± 0.2 | 4853 ± 62 | 2.95 ± 0.48 | 0.24 ± 0.21 | … | … | Y | … | … | … | … | … | n | G |
| 48612 | 18263391+0619514 | -20.8 ± 0.3 | 4453 ± 296 | 4.46 ± 0.45 | 0.07 ± 0.15 | … | … | Y | … | … | … | … | … | n | … |
| 48613 | 18263401+0628302 | -27.9 ± 0.3 | 3613 ± 126 | 4.65 ± 0.10 | -0.05 ± 0.12 | <95 | 3 | Y | Y | Y | Y | Y | … | Y | … |
| 48614 | 18263404+0625161 | -62.7 ± 0.2 | 4395 ± 230 | 2.14 ± 0.45 | -0.21 ± 0.19 | … | … | N | … | … | … | … | … | n | G |
| 48615 | 18263409+0634356 | 11.9 ± 0.3 | 5257 ± 93 | 4.96 ± 0.54 | -0.14 ± 0.23 | … | … | N | … | … | … | … | … | n | … |
| 48616 | 18263416+0645326 | -22.4 ± 0.3 | 5933 ± 283 | 4.41 ± 0.13 | 0.16 ± 0.17 | … | … | Y | … | … | … | … | … | n | … |
| 48617 | 18263426+0623335 | -25.5 ± 0.2 | 4793 ± 261 | 2.78 ± 0.49 | -0.21 ± 0.18 | … | … | Y | … | … | … | … | … | n | G |
| 48618 | 18263435+0618424 | -4.2 ± 0.2 | 4482 ± 138 | 4.70 ± 0.36 | 0.22 ± 0.24 | … | … | Y | … | … | … | N | … | n | … |
| 48619 | 18263438+0616274 | -29.2 ± 0.2 | 4677 ± 109 | 2.95 ± 0.43 | 0.27 ± 0.24 | … | … | Y | … | … | … | … | … | n | G |
| 48620 | 18263466+0636418 | -12.5 ± 0.3 | 5232 ± 82 | 4.27 ± 0.33 | 0.17 ± 0.17 | … | … | Y | … | … | … | … | … | n | … |
| 48621 | 18263469+0629397 | 99.6 ± 0.3 | 5111 ± 222 | 3.26 ± 0.74 | -0.54 ± 0.18 | … | … | N | … | … | … | … | … | n | G |
| 48622 | 18263474+0614010 | 158.5 ± 0.4 | 5173 ± 209 | 3.43 ± 0.67 | -0.69 ± 0.24 | … | … | N | … | … | … | … | … | n | G |
| 48623 | 18263478+0635476 | -15.4 ± 0.4 | 4586 ± 346 | 4.49 ± 0.56 | 0.10 ± 0.15 | <58 | 3 | Y | N | Y | Y | … | … | n | NG |
| 48624 | 18263482+0629170 | 34.7 ± 0.2 | 5791 ± 161 | 4.08 ± 0.13 | 0.00 ± 0.14 | 70 ± 19 | 1 | N | .. | … | … | … | … | n | NG |
| 48625 | 18263483+0641123 | -42.7 ± 0.2 | 5302 ± 110 | 3.73 ± 0.15 | -0.08 ± 0.12 | … | … | Y | … | … | … | … | … | n | … |
| 48626 | 18263503+0641255 | -21.2 ± 0.4 | 4331 ± 321 | 4.46 ± 0.29 | -0.13 ± 0.28 | <84 | 3 | Y | N | Y | Y | … | … | n | NG |
| 48627 | 18263506+0629422 | -17.3 ± 0.4 | 5111 ± 143 | 2.43 ± 0.98 | -0.56 ± 0.35 | … | … | Y | … | … | … | … | … | n | G |
| 48628 | 18263527+0628559 | -37.0 ± 0.3 | 5710 ± 49 | 4.22 ± 0.11 | 0.20 ± 0.14 | … | … | Y | … | … | … | … | … | n | … |
| 48629 | 18263552+0624179 | 0.3 ± 0.4 | 3562 ± 133 | 4.33 ± 0.11 | -0.07 ± 0.14 | … | … | Y | … | … | … | N | … | n | … |
| 48630 | 18263554+0629381 | 18.8 ± 0.3 | 4713 ± 170 | 4.29 ± 0.48 | 0.06 ± 0.14 | … | … | N | … | … | … | … | … | n | … |
| 48631 | 18263562+0619455 | -34.8 ± 0.2 | 6294 ± 217 | 4.34 ± 0.34 | -0.13 ± 0.15 | 65 ± 14 | 1 | Y | Y | Y | Y | … | … | Y | … |
| 48632 | 18263570+0636575 | -21.8 ± 0.5 | 4607 ± 379 | 2.58 ± 0.64 | -0.33 ± 0.25 | … | … | Y | … | … | … | … | … | n | G |
| 48633 | 18263570+0638261 | 5.4 ± 0.3 | 6245 ± 146 | 4.16 ± 0.19 | -0.27 ± 0.14 | … | … | N | … | … | … | … | … | n | … |
| 48634 | 18263586+0616027 | 61.3 ± 0.2 | 4887 ± 111 | 3.09 ± 0.39 | 0.13 ± 0.19 | … | … | N | … | … | … | … | … | n | G |
| 48635 | 18263588+0639244 | -4.7 ± 0.4 | 4758 ± 289 | 3.20 ± 0.39 | -0.01 ± 0.36 | <79 | 3 | Y | N | N | Y | … | … | n | G |

**Table C.11.** continued.

| ID | CNAME | RV (km s$^{-1}$) | $T_{\text{eff}}$ (K) | logg (dex) | [Fe/H] (dex) | EW(Li)$^a$ (mÅ) | EW(Li) error flag$^b$ | Membership RV | Li | logg | [Fe/H] | Gaia studies Randich$^c$ | Cantat-Gaudin$^c$ | Final$^d$ | NMs with Li$^e$ |
|---|---|---|---|---|---|---|---|---|---|---|---|---|---|---|---|
| 48636 | 18263608+0643294 | 26.6 ± 0.3 | 4814 ± 239 | 3.23 ± 0.38 | 0.01 ± 0.24 | … | … | N | … | … | … | … | … | n | G |
| 48637 | 18263612+0615383 | 45.8 ± 0.3 | 4613 ± 141 | 2.16 ± 0.67 | -0.14 ± 0.27 | <74 | 3 | N | .. | … | … | … | … | n | G |
| 48638 | 18263616+0642094 | 32.2 ± 0.5 | 5032 ± 109 | 3.30 ± 0.43 | -0.08 ± 0.12 | … | … | N | … | … | … | … | … | n | G |
| 48639 | 18263617+0637398 | 42.7 ± 0.4 | 4623 ± 367 | 4.41 ± 0.27 | -0.02 ± 0.14 | … | … | N | … | … | … | … | … | n | … |
| 48640 | 18263630+0636317 | 7.6 ± 0.3 | 5667 ± 95 | 3.93 ± 0.34 | 0.42 ± 0.21 | … | … | N | … | … | … | … | … | n | … |
| 48641 | 18263638+0629022 | -144.0 ± 1.5 | 4306 ± 87 | 5.12 ± 0.38 | -0.25 ± 0.11 | … | … | N | … | … | … | … | … | n | … |
| 48642 | 18263638+0639011 | -16.5 ± 0.3 | 5898 ± 81 | 4.20 ± 0.21 | 0.21 ± 0.13 | … | … | Y | … | … | … | … | … | n | … |
| 48643 | 18263653+0629308 | -10.3 ± 0.2 | 4772 ± 223 | 2.74 ± 0.37 | -0.13 ± 0.19 | … | … | Y | … | … | … | … | … | n | G |
| 48644 | 18263660+0617546 | -52.1 ± 0.3 | 5161 ± 196 | 3.65 ± 0.61 | -0.84 ± 0.23 | … | … | N | … | … | … | … | … | n | … |
| 48645 | 18263688+0626296 | 14.0 ± 0.4 | 4540 ± 249 | 4.83 ± 0.25 | -0.09 ± 0.16 | <47 | 3 | N | .. | … | … | N | … | n | NG |
| 48646 | 18263695+0621261 | 75.4 ± 0.3 | 5457 ± 201 | 3.94 ± 0.17 | 0.30 ± 0.18 | <35 | 3 | N | .. | … | … | … | … | n | NG |
| 48647 | 18263703+0620447 | -90.9 ± 0.2 | 4827 ± 193 | 2.79 ± 0.24 | -0.42 ± 0.12 | … | … | N | … | … | … | … | … | n | G |
| 48648 | 18263707+0635003 | 10.0 ± 0.3 | 4790 ± 247 | 3.30 ± 0.54 | 0.23 ± 0.26 | … | … | N | … | … | … | … | … | n | G |
| 48649 | 18263713+0641399 | -36.8 ± 0.3 | 5226 ± 53 | 4.34 ± 0.32 | 0.12 ± 0.16 | … | … | Y | … | … | … | … | … | n | … |
| 48650 | 18263719+0639171 | -32.0 ± 0.3 | 4822 ± 411 | 1.85 ± 0.40 | -1.20 ± 0.21 | … | … | Y | … | … | … | … | … | n | G |
| 48651 | 18263728+0618384 | 10.8 ± 0.3 | 4774 ± 242 | 3.03 ± 0.28 | 0.07 ± 0.22 | … | … | N | … | … | … | … | … | n | G |
| 48652 | 18263728+0645481 | 8.3 ± 0.4 | 4508 ± 331 | 4.45 ± 0.41 | 0.00 ± 0.18 | … | … | N | … | … | … | … | … | n | … |
| 48653 | 18263732+0629373 | -13.7 ± 0.4 | 4907 ± 191 | 4.18 ± 0.32 | 0.31 ± 0.43 | … | … | Y | … | … | … | … | … | n | … |
| 48654 | 18263735+0643252 | 19.0 ± 0.3 | 4955 ± 113 | 3.12 ± 0.33 | -0.34 ± 0.18 | … | … | N | … | … | … | … | … | n | G |
| 48655 | 18263747+0642236 | -6.6 ± 0.3 | 5689 ± 262 | 4.52 ± 0.37 | -0.09 ± 0.21 | … | … | Y | … | … | … | … | … | n | … |
| 48656 | 18263748+0627073 | -6.4 ± 0.3 | 4426 ± 144 | 4.75 ± 0.26 | -0.05 ± 0.12 | <50 | 3 | Y | N | N | Y | N | … | n | NG |
| 48657 | 18263754+0615024 | 29.1 ± 0.3 | 4831 ± 186 | 2.88 ± 0.45 | -0.20 ± 0.2 | <45 | 3 | N | .. | … | … | … | … | n | G |
| 48658 | 18263757+0633234 | -29.2 ± 0.3 | 3828 ± 397 | 4.53 ± 0.11 | -0.04 ± 0.14 | <55 | 3 | Y | Y | Y | Y | Y | … | Y | … |
| 48659 | 18263772+0614507 | 58.2 ± 0.3 | 5158 ± 80 | 3.66 ± 0.27 | -0.28 ± 0.15 | … | … | N | … | … | … | … | … | n | … |
| 48660 | 18263783+0618386 | 70.5 ± 0.4 | 4317 ± 414 | 4.42 ± 0.56 | -0.17 ± 0.17 | … | … | N | … | … | … | … | … | n | … |
| 48661 | 18263784+0629580 | 6.7 ± 0.3 | 4828 ± 205 | 3.11 ± 0.23 | -0.27 ± 0.24 | … | … | N | … | … | … | … | … | n | G |
| 48662 | 18263785+0615373 | -58.0 ± 0.7 | 5100 ± 503 | 4.88 ± 0.35 | -0.25 ± 0.55 | … | … | N | … | … | … | … | … | n | … |
| 48663 | 18263791+0627013 | 5.7 ± 0.3 | 4304 ± 245 | 4.62 ± 0.27 | 0.08 ± 0.17 | … | … | N | … | … | … | N | … | n | … |
| 48664 | 18263813+0638525 | -16.9 ± 0.3 | 4881 ± 96 | 4.04 ± 0.31 | 0.17 ± 0.18 | … | … | Y | … | … | … | … | … | n | … |
| 48665 | 18263829+0635048 | 7.4 ± 0.3 | 5792 ± 71 | 4.00 ± 0.17 | 0.30 ± 0.15 | … | … | N | … | … | … | … | … | n | … |
| 48666 | 18263831+0614597 | -2.6 ± 0.3 | 5700 ± 54 | 4.38 ± 0.43 | -0.14 ± 0.14 | … | … | Y | … | … | … | … | … | n | … |
| 48667 | 18263832+0617014 | 55.7 ± 0.5 | 4403 ± 84 | 4.43 ± 0.46 | 0.14 ± 0.4 | … | … | N | … | … | … | N | … | n | … |
| 48668 | 18263839+0625491 | 55.9 ± 0.3 | 5454 ± 84 | 4.56 ± 0.28 | 0.05 ± 0.12 | … | … | N | … | … | … | … | … | n | … |
| 48669 | 18263842+0615541 | 79.6 ± 0.3 | 4756 ± 135 | 3.20 ± 0.30 | 0.12 ± 0.23 | <31 | 3 | N | .. | … | … | … | … | n | G |
| 48670 | 18263842+0616039 | -21.3 ± 0.5 | 3576 ± 26 | 4.37 ± 0.21 | … | … | … | Y | … | … | … | N | … | n | … |
| 48671 | 18263851+0618292 | 41.1 ± 0.3 | 4783 ± 219 | 4.72 ± 0.27 | -0.14 ± 0.17 | … | … | N | … | … | … | … | … | n | … |
| 48672 | 18263853+0631533 | -56.4 ± 0.3 | 5629 ± 90 | 4.10 ± 0.43 | -0.08 ± 0.24 | … | … | N | … | … | … | … | … | n | … |
| 48673 | 18263862+0630387 | 15.4 ± 0.8 | 7034 ± 710 | 4.03 ± 0.22 | 0.00 ± 0.33 | … | … | N | … | … | … | … | … | n | … |
| 48674 | 18263873+0633274 | -42.7 ± 0.3 | 5968 ± 483 | 4.34 ± 0.15 | 0.26 ± 0.21 | … | … | Y | … | … | … | … | … | n | … |
| 48675 | 18263883+0637014 | 7.5 ± 0.3 | 5787 ± 192 | 4.30 ± 0.31 | 0.17 ± 0.13 | … | … | N | … | … | … | … | … | n | … |
| 48676 | 18263886+0636134 | -4.9 ± 0.4 | 4017 ± 341 | 4.42 ± 0.18 | -0.07 ± 0.31 | <109 | 3 | Y | N | N | Y | N | … | n | NG |
| 48677 | 18263888+0621179 | 53.6 ± 0.2 | 4549 ± 129 | 2.62 ± 0.30 | 0.18 ± 0.19 | … | … | N | … | … | … | … | … | n | G |
| 48678 | 18263892+0618079 | 14.4 ± 0.3 | 4719 ± 283 | 4.84 ± 0.24 | -0.30 ± 0.2 | … | … | N | … | … | … | … | … | n | … |
| 48680 | 18263901+0643471 | -29.8 ± 0.3 | 4985 ± 133 | 3.50 ± 0.19 | 0.01 ± 0.12 | … | … | Y | … | … | … | … | … | n | … |
| 48681 | 18263909+0625000 | 45.5 ± 0.7 | 4216 ± 207 | 3.75 ± 0.47 | -0.50 ± 0.49 | … | … | N | … | … | … | … | … | n | … |
| 48682 | 18263917+0620105 | 88.8 ± 0.3 | 4903 ± 227 | 3.63 ± 0.47 | -0.29 ± 0.14 | … | … | N | … | … | … | … | … | n | … |
| 48683 | 18263923+0625394 | 20.0 ± 0.3 | 6372 ± 179 | 4.13 ± 0.21 | -0.17 ± 0.22 | <34 | 3 | N | .. | … | … | … | … | n | NG |
| 48684 | 18263942+0637459 | -13.7 ± 0.2 | 4832 ± 158 | 3.29 ± 0.55 | 0.38 ± 0.28 | … | … | Y | … | … | … | … | … | n | G |
| 48685 | 18263971+0623408 | -22.1 ± 0.2 | 4538 ± 125 | 2.67 ± 0.46 | 0.18 ± 0.2 | … | … | Y | … | … | … | … | … | n | G |
| 48686 | 18263974+0622089 | 106.1 ± 0.3 | 5075 ± 91 | 3.49 ± 0.14 | -0.12 ± 0.22 | … | … | N | … | … | … | … | … | n | G |
| 48687 | 18263985+0637317 | -50.1 ± 0.4 | 4611 ± 304 | 3.96 ± 0.85 | 0.07 ± 0.17 | <109 | 3 | N | .. | … | … | … | … | n | NG |
| 48688 | 18263986+0630046 | -48.0 ± 0.2 | 4745 ± 101 | 3.30 ± 0.61 | 0.41 ± 0.27 | 66 ± 27 | … | N | .. | … | … | … | … | n | G |
| 48689 | 18263999+0617574 | 37.1 ± 0.3 | 4669 ± 96 | 2.50 ± 0.24 | 0.20 ± 0.22 | … | … | N | … | … | … | … | … | n | … |
| 48690 | 18264005+0637273 | 92.8 ± 0.4 | 5209 ± 197 | 3.63 ± 0.49 | -0.60 ± 0.26 | … | … | N | … | … | … | … | … | n | … |
| 48691 | 18264019+0639076 | -23.1 ± 0.3 | 5082 ± 80 | 3.14 ± 0.26 | -0.41 ± 0.23 | … | … | Y | … | … | … | … | … | n | G |
| 48692 | 18264027+0618114 | 21.0 ± 0.3 | 5773 ± 132 | 4.22 ± 0.19 | 0.30 ± 0.16 | <46 | 3 | N | .. | … | … | … | … | n | NG |
| 48693 | 18264029+0616075 | 2.7 ± 0.2 | 5925 ± 150 | 4.34 ± 0.18 | 0.03 ± 0.12 | … | … | Y | … | … | … | … | … | n | … |
| 48694 | 18264035+0644415 | -60.5 ± 0.3 | 4767 ± 283 | 2.73 ± 0.64 | -0.33 ± 0.26 | … | … | N | … | … | … | … | … | n | G |
| 48695 | 18264038+0627500 | -15.0 ± 0.3 | 5040 ± 99 | 3.49 ± 0.40 | 0.01 ± 0.17 | … | … | Y | … | … | … | … | … | n | G |

**Table C.11.** continued.

| ID | CNAME | RV (km s$^{-1}$) | $T_{\text{eff}}$ (K) | logg (dex) | [Fe/H] (dex) | EW(Li)$^a$ (mÅ) | EW(Li) error flag$^b$ | Membership RV | Li | logg | [Fe/H] | Gaia studies Randich$^c$ | Cantat-Gaudin$^c$ | Final$^d$ | NMs with Li$^e$ |
|---|---|---|---|---|---|---|---|---|---|---|---|---|---|---|---|
| 48696 | 18264048+0622042 | 37.7 ± 0.3 | 5074 ± 96 | 4.60 ± 0.36 | 0.12 ± 0.14 | … | … | N | … | … | … | … | … | n | … |
| 48697 | 18264052+0614434 | -9.7 ± 0.3 | 4850 ± 196 | 3.40 ± 0.46 | 0.15 ± 0.19 | … | … | Y | … | … | … | … | … | n | G |
| 48698 | 18264062+0644336 | 12.8 ± 0.5 | 5868 ± 308 | 4.19 ± 0.08 | -0.25 ± 0.59 | <56 | 3 | N | .. | … | … | … | … | n | NG |
| 48699 | 18264069+0631114 | -3.5 ± 0.3 | 5445 ± 77 | 4.04 ± 0.29 | 0.23 ± 0.15 | 36 ± 32 | 1 | Y | Y | Y | Y | … | … | Y | … |
| 48700 | 18264071+0615158 | 18.9 ± 0.3 | 5261 ± 231 | 4.42 ± 0.28 | -0.40 ± 0.16 | 73 ± 45 | 1 | N | … | … | … | … | … | n | NG |
| 48702 | 18264083+0644487 | -19.2 ± 0.3 | 5226 ± 153 | 4.61 ± 0.38 | 0.02 ± 0.13 | <32 | 3 | Y | Y | Y | Y | … | … | Y | … |
| 48703 | 18264101+0625374 | 21.3 ± 0.2 | 4723 ± 103 | 2.59 ± 0.27 | 0.21 ± 0.23 | … | … | N | … | … | … | … | … | n | G |
| 48704 | 18264104+0646150 | 56.5 ± 0.3 | 5375 ± 114 | 3.97 ± 0.29 | -0.25 ± 0.3 | <40 | 3 | N | .. | … | … | … | … | n | NG |
| 48705 | 18264108+0632307 | -109.8 ± 0.6 | 4838 ± 554 | 2.82 ± 0.21 | -1.13 ± 1.6 | … | … | N | … | … | … | … | … | n | … |
| 48706 | 18264118+0644169 | 46.1 ± 0.3 | 4942 ± 161 | 3.26 ± 0.56 | -0.22 ± 0.21 | … | … | N | … | … | … | … | … | n | G |
| 48707 | 18264123+0616344 | 49.1 ± 0.6 | 4720 ± 95 | 2.66 ± 0.27 | -0.55 ± 0.2 | … | … | N | … | … | … | … | … | n | G |
| 48708 | 18264125+0616285 | -11.0 ± 0.3 | 4700 ± 227 | 2.99 ± 0.44 | -0.07 ± 0.14 | … | … | Y | … | … | … | … | … | n | G |
| 48709 | 18264133+0622469 | -88.4 ± 0.4 | 4047 ± 505 | 4.36 ± 0.07 | -0.06 ± 0.29 | … | … | N | … | … | … | … | … | n | … |
| 48710 | 18264134+0639130 | -55.7 ± 0.3 | 4448 ± 157 | 2.22 ± 0.31 | -0.09 ± 0.15 | … | … | N | … | … | … | … | … | n | G |
| 48711 | 18264140+0645197 | 10.1 ± 0.3 | 5489 ± 85 | 4.48 ± 0.29 | -0.05 ± 0.12 | … | … | N | … | … | … | … | … | n | … |
| 48712 | 18264148+0630278 | -19.9 ± 0.3 | 5312 ± 86 | 3.77 ± 0.32 | 0.34 ± 0.26 | 105 ± 37 | 1 | Y | N | N | N | … | … | n | NG |
| 48713 | 18264152+0634573 | -40.3 ± 0.3 | 4481 ± 289 | 3.95 ± 0.36 | -0.12 ± 0.12 | … | … | Y | … | … | … | … | … | n | … |
| 48714 | 18264159+0615521 | -34.8 ± 0.2 | 4962 ± 105 | 3.57 ± 0.26 | -0.08 ± 0.13 | <19 | 3 | Y | Y | Y? | Y | … | … | Y | … |
| 48715 | 18264162+0635092 | -10.1 ± 0.3 | 4181 ± 288 | 4.86 ± 0.40 | -0.18 ± 0.18 | <59 | 3 | Y | N | N | Y | N | … | n | NG |
| 48716 | 18264166+0644304 | -43.8 ± 0.4 | 3465 ± 129 | 4.43 ± 0.16 | -0.12 ± 0.13 | <116 | 3 | Y | Y | Y | Y | … | … | Y | … |
| 48717 | 18264186+0619270 | 88.9 ± 0.3 | 4910 ± 194 | 3.49 ± 0.38 | 0.01 ± 0.19 | … | … | N | … | … | … | … | … | n | G |
| 48718 | 18264186+0628481 | 72.7 ± 1.4 | … | … | … | … | … | N | … | … | … | … | … | n | … |
| 48719 | 18264204+0626342 | -4.4 ± 0.2 | 4916 ± 131 | 3.41 ± 0.10 | -0.04 ± 0.14 | … | … | Y | … | … | … | … | … | n | G |
| 48720 | 18264221+0637084 | 1.4 ± 0.3 | 4877 ± 177 | 3.10 ± 0.26 | -0.13 ± 0.15 | … | … | Y | … | … | … | … | … | n | G |
| 48721 | 18264226+0631459 | 14.9 ± 0.3 | 6160 ± 90 | 4.10 ± 0.06 | 0.09 ± 0.15 | … | … | N | … | … | … | … | … | n | … |
| 48722 | 18264229+0614150 | -49.9 ± 1.5 | 5179 ± 362 | 4.13 ± 0.89 | -0.46 ± 0.91 | … | … | N | … | … | … | … | … | n | … |
| 48723 | 18264244+0642027 | 79.8 ± 0.3 | 4795 ± 160 | 2.75 ± 0.25 | 0.00 ± 0.14 | … | … | N | … | … | … | … | … | n | G |
| 48724 | 18264249+0646098 | -20.9 ± 0.3 | 4938 ± 171 | 2.90 ± 0.34 | -0.28 ± 0.17 | … | … | Y | … | … | … | … | … | n | G |
| 48725 | 18264250+0626245 | 7.6 ± 0.5 | 4007 ± 352 | … | 0.03 ± 0.21 | <87 | 3 | N | … | … | … | N | … | n | NG |
| 48726 | 18264253+0632120 | -22.1 ± 0.4 | 5172 ± 266 | 4.45 ± 1.01 | 0.00 ± 0.16 | <61 | 3 | Y | Y | Y | Y | N | … | Y?$^g$ | … |
| 48727 | 18264269+0634427 | 103.8 ± 0.2 | 4540 ± 124 | 2.03 ± 0.31 | -0.20 ± 0.26 | … | … | N | … | … | … | … | … | n | G |
| 48728 | 18264273+0617014 | 80.5 ± 0.7 | 4556 ± 404 | 3.39 ± 0.56 | 0.20 ± 0.21 | … | … | N | … | … | … | … | … | n | G |
| 48729 | 18264281+0632036 | -33.7 ± 0.4 | 4910 ± 139 | 3.00 ± 0.53 | -0.25 ± 0.21 | … | … | Y | … | … | … | … | … | n | G |
| 48730 | 18264288+0617548 | 30.1 ± 0.3 | 5966 ± 33 | 4.26 ± 0.09 | -0.06 ± 0.16 | <40 | 3 | N | .. | … | … | … | … | n | NG |
| 48731 | 18264290+0619184 | -22.1 ± 0.2 | 4947 ± 131 | 3.47 ± 0.21 | -0.23 ± 0.16 | … | … | Y | … | … | … | … | … | n | G |
| 48732 | 18264295+0616489 | 55.8 ± 0.3 | 4636 ± 82 | 2.75 ± 0.37 | 0.16 ± 0.28 | … | … | N | … | … | … | … | … | n | G |
| 48733 | 18264296+0627090 | -0.2 ± 0.5 | 4442 ± 105 | 4.25 ± 0.16 | -0.05 ± 0.16 | … | … | Y | … | … | … | N | … | n | … |
| 48734 | 18264301+0615102 | 2.8 ± 0.4 | 3763 ± 397 | 3.97 ± 0.14 | … | … | … | Y | … | … | … | N | … | n | … |
| 48735 | 18264307+0619291 | -31.8 ± 0.3 | 6053 ± 85 | 4.19 ± 0.25 | 0.10 ± 0.13 | … | … | Y | … | … | … | … | … | n | … |
| 48736 | 18264307+0633300 | 13.6 ± 0.6 | 5067 ± 275 | 3.63 ± 0.31 | 0.21 ± 0.18 | … | … | N | … | … | … | … | … | n | … |
| 48737 | 18264318+0620091 | 54.8 ± 1.1 | 4668 ± 127 | 2.99 ± 0.51 | -0.39 ± 0.16 | … | … | N | … | … | … | … | … | n | … |
| 48738 | 18264323+0618393 | 11.2 ± 0.6 | 3847 ± 315 | 4.56 ± 0.19 | -0.03 ± 0.14 | <72 | 3 | N | .. | … | … | N | … | n | NG |
| 48739 | 18264331+0628310 | 50.5 ± 0.2 | 4956 ± 151 | 2.62 ± 0.78 | -0.34 ± 0.25 | … | … | N | … | … | … | … | … | n | G |
| 48740 | 18264334+0635065 | 2.3 ± 0.3 | 4695 ± 330 | 2.66 ± 0.51 | -0.25 ± 0.22 | … | … | Y | … | … | … | … | … | n | G |
| 48741 | 18264336+0626009 | 7.9 ± 0.4 | 4654 ± 226 | 4.29 ± 0.39 | -0.09 ± 0.13 | … | … | N | … | … | … | … | … | n | … |
| 48742 | 18264347+0615263 | -50.0 ± 0.3 | 5392 ± 220 | 4.63 ± 0.43 | 0.04 ± 0.18 | … | … | N | … | … | … | … | … | n | … |
| 48743 | 18264350+0633393 | 4.6 ± 0.3 | 5744 ± 63 | 4.07 ± 0.46 | 0.04 ± 0.14 | 77 ± 30 | 1 | N | .. | … | … | … | … | n | NG |
| 48744 | 18264351+0643347 | -53.5 ± 0.3 | 4607 ± 304 | 2.45 ± 0.45 | -0.31 ± 0.16 | … | … | N | … | … | … | … | … | n | G |
| 48745 | 18264354+0644349 | -31.6 ± 1.0 | 4438 ± 95 | 3.62 ± 0.94 | -0.28 ± 0.45 | … | … | Y | … | … | … | … | … | n | … |
| 48746 | 18264370+0627055 | 29.7 ± 0.4 | 4316 ± 238 | 4.69 ± 0.14 | 0.07 ± 0.18 | … | … | N | … | … | … | … | … | n | … |
| 48747 | 18264371+0635507 | 20.4 ± 0.3 | 4777 ± 210 | 4.72 ± 0.30 | 0.23 ± 0.22 | … | … | N | … | … | … | N | … | n | … |
| 48748 | 18264384+0641500 | 71.5 ± 0.2 | 5794 ± 103 | 4.00 ± 0.19 | 0.08 ± 0.13 | … | … | N | … | … | … | … | … | n | … |
| 48749 | 18264407+0621195 | 16.8 ± 0.3 | 4369 ± 192 | 2.11 ± 0.50 | -0.09 ± 0.31 | … | … | N | … | … | … | … | … | n | G |
| 48750 | 18264411+0635337 | 4.1 ± 0.3 | 5057 ± 99 | 4.61 ± 0.24 | -0.18 ± 0.13 | … | … | N | … | … | … | … | … | n | … |
| 48751 | 18264412+0630531 | 20.0 ± 0.3 | 4416 ± 196 | 4.68 ± 0.19 | -0.06 ± 0.13 | … | … | N | … | … | … | N | … | n | … |
| 48752 | 18264414+0620179 | -50.2 ± 0.3 | 4748 ± 104 | 2.65 ± 0.27 | 0.10 ± 0.19 | … | … | N | … | … | … | … | … | n | G |
| 48753 | 18264416+0622108 | -29.8 ± 0.3 | 5077 ± 61 | 4.50 ± 0.49 | -0.08 ± 0.13 | 38 ± 27 | 1 | Y | Y | Y | Y | … | … | Y | … |
| 48754 | 18264441+0624261 | -16.8 ± 0.3 | 6511 ± 222 | 3.93 ± 0.06 | -0.03 ± 0.23 | 75 ± 21 | 1 | Y | Y | Y | Y | … | … | Y | … |
| 48755 | 18264452+0625459 | -19.1 ± 0.3 | 4565 ± 121 | 2.65 ± 0.38 | 0.10 ± 0.22 | … | … | Y | … | … | … | … | … | n | G |

| ID | CNAME | RV (km s$^{-1}$) | $T_{\text{eff}}$ (K) | logg (dex) | [Fe/H] (dex) | EW(Li)$^a$ (mÅ) | EW(Li) error flag$^b$ | Membership RV | Li | logg | [Fe/H] | Gaia studies Randich$^c$ | Cantat-Gaudin$^c$ | Final$^d$ | NMs with Li$^e$ |
|---|---|---|---|---|---|---|---|---|---|---|---|---|---|---|---|
| 48756 | 18264460+0622336 | 66.8 ± 0.3 | 4576 ± 256 | 2.52 ± 0.30 | -0.06 ± 0.15 | … | … | N | … | … | … | … | … | n | G |
| 48757 | 18264470+0623334 | -29.5 ± 0.2 | 4786 ± 56 | 2.93 ± 0.38 | 0.15 ± 0.18 | … | … | Y | … | … | … | … | … | n | G |
| 48758 | 18264490+0640023 | 17.8 ± 0.3 | 4532 ± 274 | 4.75 ± 0.30 | 0.08 ± 0.17 | … | … | N | … | … | … | … | … | n | … |
| 48759 | 18264499+0639447 | 11.3 ± 0.3 | 5309 ± 126 | 3.74 ± 0.10 | 0.03 ± 0.12 | … | … | N | … | … | … | … | … | n | … |
| 48760 | 18264505+0641573 | -13.2 ± 0.3 | 5618 ± 135 | 4.07 ± 0.31 | -0.54 ± 0.19 | … | … | Y | … | … | … | … | … | n | … |
| 48761 | 18264515+0636254 | 28.8 ± 0.3 | 5484 ± 70 | 4.01 ± 0.22 | -0.30 ± 0.14 | … | … | N | … | … | … | … | … | n | … |
| 48762 | 18264518+0633439 | -56.3 ± 0.3 | 5146 ± 145 | 4.52 ± 0.58 | -0.18 ± 0.19 | <29 | 3 | N | .. | … | … | N | … | n | NG |
| 48763 | 18264518+0644112 | -34.7 ± 0.3 | 5233 ± 369 | 4.17 ± 0.51 | -0.66 ± 0.8 | <33 | 3 | Y | Y | Y | N | … | … | Y | … |
| 48764 | 18264532+0637421 | -5.3 ± 0.3 | 6295 ± 135 | 3.99 ± 0.13 | 0.10 ± 0.17 | … | … | Y | … | … | … | … | … | n | … |
| 48765 | 18264534+0621521 | 38.8 ± 0.2 | 4526 ± 138 | 2.46 ± 0.32 | 0.25 ± 0.29 | <15 | 3 | N | .. | … | … | … | … | n | G |
| 48766 | 18264534+0637008 | 57.8 ± 0.3 | 4892 ± 193 | 3.15 ± 0.35 | -0.11 ± 0.28 | … | … | N | … | … | … | … | … | n | G |
| 48767 | 18264542+0633292 | 2.6 ± 0.3 | 4692 ± 256 | 2.92 ± 0.49 | -0.03 ± 0.27 | … | … | Y | … | … | … | … | … | n | G |
| 48768 | 18264543+0616326 | -17.9 ± 0.4 | 4726 ± 174 | 4.31 ± 0.99 | -0.03 ± 0.19 | <95 | 3 | Y | N | Y | Y | N | … | n | NG |
| 48769 | 18264554+0625454 | -27.0 ± 0.5 | 4238 ± 336 | 3.90 ± 0.19 | -0.01 ± 0.25 | … | … | Y | … | … | … | Y | … | n | … |
| 48770 | 18264562+0630554 | -19.5 ± 0.4 | 5050 ± 86 | 3.26 ± 0.52 | -0.15 ± 0.23 | … | … | Y | … | … | … | … | … | n | G |
| 48771 | 18264566+0645593 | -11.2 ± 0.2 | 5239 ± 132 | 3.72 ± 0.44 | 0.46 ± 0.32 | … | … | Y | … | … | … | … | … | n | … |
| 48772 | 18264568+0613586 | 43.8 ± 0.3 | 4663 ± 184 | 3.21 ± 0.67 | 0.31 ± 0.33 | … | … | N | … | … | … | … | … | n | G |
| 48773 | 18264570+0618514 | 103.5 ± 0.3 | 4900 ± 174 | 3.29 ± 0.36 | -0.10 ± 0.19 | … | … | N | … | … | … | … | … | n | G |
| 48774 | 18264585+0616078 | -0.2 ± 0.3 | 5005 ± 77 | 3.27 ± 0.29 | 0.00 ± 0.18 | … | … | Y | … | … | … | … | … | n | G |
| 48775 | 18264585+0635006 | 77.6 ± 0.3 | 5036 ± 131 | 3.23 ± 0.15 | -0.13 ± 0.14 | … | … | N | … | … | … | … | … | n | G |
| 48776 | 18264589+0616380 | 28.1 ± 0.3 | 5613 ± 31 | 4.39 ± 0.16 | 0.10 ± 0.15 | … | … | N | … | … | … | … | … | n | … |
| 48777 | 18264622+0613560 | 81.8 ± 0.3 | 4827 ± 186 | 2.75 ± 0.40 | -0.11 ± 0.26 | … | … | N | … | … | … | … | … | n | G |
| 48778 | 18264642+0615400 | 43.5 ± 0.3 | 5426 ± 185 | 4.20 ± 0.08 | -0.20 ± 0.13 | … | … | N | … | … | … | … | … | n | … |
| 48779 | 18264648+0638045 | -19.4 ± 0.2 | 5906 ± 146 | 4.28 ± 0.07 | 0.15 ± 0.13 | 42 ± 24 | 1 | Y | Y | Y | Y | … | … | Y | … |
| 48780 | 18264659+0620098 | -6.6 ± 0.3 | 4287 ± 251 | 2.41 ± 0.55 | 0.43 ± 0.3 | … | … | Y | … | … | … | … | … | n | G |
| 48781 | 18264676+0617107 | -47.1 ± 0.3 | 4599 ± 88 | 2.52 ± 0.24 | 0.22 ± 0.18 | … | … | Y | … | … | … | … | … | n | G |
| 48782 | 18264679+0639417 | 96.1 ± 0.3 | 4831 ± 215 | 3.35 ± 0.21 | -0.23 ± 0.24 | … | … | N | … | … | … | … | … | n | G |
| 48783 | 18264706+0615158 | 30.6 ± 0.3 | 6053 ± 113 | 4.34 ± 0.46 | -0.16 ± 0.14 | 73 ± 50 | 1 | N | .. | … | … | … | … | n | NG |
| 48784 | 18264713+0617591 | 52.5 ± 0.3 | 5119 ± 113 | 3.68 ± 0.47 | -0.31 ± 0.13 | … | … | N | … | … | … | … | … | n | … |
| 48785 | 18264717+0639567 | 41.6 ± 0.3 | 4953 ± 149 | 3.37 ± 0.39 | 0.06 ± 0.2 | … | … | N | … | … | … | … | … | n | G |
| 48786 | 18264720+0625376 | 34.7 ± 0.2 | 5788 ± 97 | 4.57 ± 0.28 | 0.15 ± 0.12 | … | … | N | … | … | … | … | … | n | … |
| 48787 | 18264725+0628425 | -29.2 ± 0.3 | 4980 ± 93 | 3.22 ± 0.21 | 0.00 ± 0.15 | … | … | Y | … | … | … | … | … | n | G |
| 48788 | 18264736+0614263 | 60.5 ± 0.2 | 4475 ± 192 | 2.34 ± 0.38 | -0.02 ± 0.2 | … | … | N | … | … | … | … | … | n | … |
| 48789 | 18264755+0638549 | -11.1 ± 0.3 | 4992 ± 113 | 3.42 ± 0.52 | -0.06 ± 0.22 | … | … | Y | … | … | … | … | … | n | G |
| 48790 | 18264760+0626343 | -27.4 ± 0.3 | 4021 ± 257 | 4.89 ± 0.34 | -0.26 ± 0.25 | <26 | 3 | Y | Y | Y | Y | Y | Y | Y | … |
| 48791 | 18264762+0630162 | -9.3 ± 0.3 | 6249 ± 82 | 4.16 ± 0.31 | 0.01 ± 0.14 | … | … | Y | … | … | … | … | … | n | … |
| 48792 | 18264763+0633199 | -2.1 ± 0.3 | 4702 ± 275 | 2.70 ± 0.51 | -0.26 ± 0.22 | … | … | Y | … | … | … | … | … | n | G |
| 48793 | 18264771+0618408 | -1.9 ± 0.3 | 5174 ± 80 | 4.02 ± 0.21 | 0.19 ± 0.19 | … | … | Y | … | … | … | … | … | n | … |
| 48794 | 18264780+0617381 | -6.0 ± 0.2 | 4509 ± 189 | 2.94 ± 0.50 | 0.40 ± 0.29 | … | … | Y | … | … | … | … | … | n | G |
| 48795 | 18264783+0633089 | -28.3 ± 0.2 | 4469 ± 136 | 4.67 ± 0.30 | -0.04 ± 0.13 | … | … | Y | … | … | … | Y | Y | n | … |
| 48796 | 18264784+0631544 | 55.1 ± 0.3 | 4680 ± 105 | 3.14 ± 0.47 | 0.32 ± 0.33 | <48 | 3 | N | .. | … | … | … | … | n | G |
| 48797 | 18264791+0646135 | 71.2 ± 0.3 | 4503 ± 294 | 2.49 ± 0.43 | -0.19 ± 0.32 | … | … | N | … | … | … | … | … | n | G |
| 48798 | 18264806+0618134 | -86.2 ± 0.3 | 4699 ± 184 | 2.75 ± 0.33 | 0.10 ± 0.3 | … | … | N | … | … | … | … | … | n | G |
| 48799 | 18264824+0615075 | -17.2 ± 0.2 | 5048 ± 74 | 3.29 ± 0.36 | -0.06 ± 0.19 | … | … | Y | … | … | … | … | … | n | G |
| 48800 | 18264833+0620223 | -22.4 ± 0.3 | 4565 ± 119 | 4.51 ± 0.26 | 0.23 ± 0.2 | … | … | Y | … | … | … | N | … | n | … |
| 48801 | 18264847+0629208 | -32.9 ± 0.4 | 4060 ± 261 | 4.31 ± 0.21 | -0.01 ± 0.19 | <118 | 3 | Y | N | Y | Y | N | … | n | NG |
| 48802 | 18264851+0637495 | -46.1 ± 0.2 | 4979 ± 140 | 2.90 ± 0.42 | -0.13 ± 0.16 | … | … | Y | … | … | … | … | … | n | G |
| 48803 | 18264862+0634422 | -14.6 ± 1.3 | 3642 ± 10 | 5.27 ± 0.09 | … | … | … | Y | … | … | … | N | … | n | … |
| 48804 | 18264879+0619528 | 26.1 ± 0.3 | 4978 ± 95 | 3.23 ± 0.24 | 0.05 ± 0.15 | … | … | N | … | … | … | … | … | n | G |
| 48805 | 18264885+0637591 | -8.8 ± 0.3 | 4919 ± 148 | 3.06 ± 0.45 | -0.45 ± 0.34 | … | … | Y | … | … | … | … | … | n | G |
| 48806 | 18264888+0619038 | -12.9 ± 0.6 | 3753 ± 447 | 4.23 ± 0.22 | … | … | … | Y | … | … | … | N | … | n | … |
| 48807 | 18264902+0631518 | -80.2 ± 0.2 | 4704 ± 282 | 4.11 ± 0.04 | -0.51 ± 0.61 | <85 | 3 | N | .. | … | … | … | … | n | NG |
| 48808 | 18264911+0619224 | -3.3 ± 0.3 | 4668 ± 341 | 4.71 ± 0.33 | -0.17 ± 0.23 | <24 | 3 | Y | Y | Y | Y | … | … | Y | … |
| 48809 | 18264921+0617594 | 13.1 ± 0.3 | 5153 ± 96 | 3.66 ± 0.16 | -0.08 ± 0.23 | … | … | N | … | … | … | … | … | n | … |
| 48810 | 18264922+0630200 | -34.9 ± 0.3 | 4712 ± 300 | 2.47 ± 0.62 | -0.52 ± 0.31 | … | … | Y | … | … | … | … | … | n | G |
| 48811 | 18264927+0634528 | -24.5 ± 0.6 | 4583 ± 446 | 3.68 ± 1.32 | 0.21 ± 0.15 | … | … | Y | … | … | … | … | … | n | … |
| 48812 | 18264929+0616570 | 46.9 ± 0.3 | 4910 ± 201 | 3.45 ± 0.44 | -0.18 ± 0.29 | … | … | N | … | … | … | … | … | n | G |
| 48813 | 18264936+0627550 | -25.1 ± 0.3 | 5987 ± 160 | 3.98 ± 0.15 | 0.21 ± 0.15 | 46 ± 16 | 1 | Y | Y | Y | Y | … | … | Y | … |
| 48814 | 18264936+0635520 | -22.0 ± 0.4 | 4676 ± 198 | 4.66 ± 0.14 | -0.20 ± 0.16 | <71 | 3 | Y | N | Y | Y | … | … | n | NG |

**Table C.11.** continued.

| ID | CNAME | RV (km s$^{-1}$) | $T_{\text{eff}}$ (K) | logg (dex) | [Fe/H] (dex) | EW(Li)$^a$ (mÅ) | EW(Li) error flag$^b$ | Membership RV | Li | logg | [Fe/H] | Gaia studies Randich$^c$ | Cantat-Gaudin$^c$ | Final$^d$ | NMs with Li$^e$ |
|---|---|---|---|---|---|---|---|---|---|---|---|---|---|---|---|
| 48815 | 18264939+0622360 | 1.4 ± 0.3 | 5815 ± 52 | 4.38 ± 0.50 | 0.09 ± 0.13 | 46 ± 29 | 1 | Y?$^f$ | Y | Y | Y | … | … | n | NG |
| 48816 | 18264939+0638370 | -29.6 ± 0.3 | 5001 ± 166 | 4.55 ± 0.25 | -0.25 ± 0.15 | … | … | Y | … | … | … | … | … | n | … |
| 48817 | 18264941+0628507 | 46.5 ± 1.3 | 4478 ± 127 | 4.55 ± 0.38 | -0.21 ± 0.17 | … | … | N | … | … | … | … | … | n | … |
| 48818 | 18264953+0614249 | 37.3 ± 0.3 | 4888 ± 115 | 2.85 ± 0.33 | 0.27 ± 0.22 | … | … | N | … | … | … | … | … | n | G |
| 48819 | 18264964+0636056 | -48.2 ± 0.3 | 4756 ± 226 | 2.78 ± 0.25 | -0.08 ± 0.17 | … | … | N | … | … | … | … | … | n | G |
| 48820 | 18264971+0616238 | 32.1 ± 0.2 | 5077 ± 108 | 4.56 ± 0.33 | -0.12 ± 0.13 | … | … | N | … | … | … | … | … | n | … |
| 48821 | 18264990+0634320 | -17.3 ± 0.4 | 5396 ± 147 | 4.32 ± 0.08 | 0.39 ± 0.31 | … | … | Y | … | … | … | … | … | n | … |
| 48822 | 18264992+0638223 | 23.3 ± 0.3 | 6261 ± 151 | 4.36 ± 0.31 | -0.22 ± 0.16 | … | … | N | … | … | … | … | … | n | … |
| 48823 | 18264995+0632223 | 20.9 ± 0.3 | 4770 ± 289 | 3.98 ± 0.73 | 0.32 ± 0.23 | … | … | N | … | … | … | … | … | n | … |
| 48824 | 18265000+0645567 | -37.0 ± 0.3 | 4990 ± 81 | 4.59 ± 0.36 | 0.08 ± 0.14 | … | … | Y | … | … | … | … | … | n | … |
| 48825 | 18265013+0634368 | 101.7 ± 0.2 | 4779 ± 112 | 2.97 ± 0.48 | 0.14 ± 0.23 | … | … | N | … | … | … | … | … | n | G |
| 48826 | 18265014+0645015 | 0.1 ± 0.3 | 4898 ± 145 | 3.30 ± 0.22 | -0.07 ± 0.18 | … | … | Y | … | … | … | … | … | n | G |
| 48827 | 18265020+0642441 | -47.3 ± 0.3 | 6105 ± 212 | 3.84 ± 0.05 | -0.18 ± 0.18 | … | … | Y | … | … | … | … | … | n | … |
| 48828 | 18265025+0632471 | -79.3 ± 0.3 | 5546 ± 49 | 4.26 ± 0.39 | -0.51 ± 0.13 | … | … | N | … | … | … | … | … | n | … |
| 48829 | 18265025+0640378 | -16.1 ± 0.2 | 6114 ± 161 | 4.06 ± 0.22 | -0.09 ± 0.18 | … | … | Y | … | … | … | … | … | n | … |
| 48830 | 18265028+0630207 | 24.7 ± 0.3 | 5181 ± 162 | 4.88 ± 0.29 | -0.23 ± 0.24 | … | … | N | … | … | … | … | … | n | … |
| 48831 | 18265032+0618124 | 109.1 ± 0.3 | 4816 ± 228 | 2.69 ± 0.42 | -0.40 ± 0.3 | … | … | N | … | … | … | … | … | n | G |
| 48832 | 18265049+0620538 | -44.6 ± 0.3 | 4842 ± 156 | 2.33 ± 0.39 | -0.13 ± 0.17 | … | … | Y | … | … | … | … | … | n | G |
| 48833 | 18265069+0613490 | -1.1 ± 0.5 | 4331 ± 232 | 4.65 ± 0.24 | -0.03 ± 0.18 | … | … | Y | … | … | … | … | … | n | … |
| 48834 | 18265074+0618117 | 10.8 ± 0.3 | 5687 ± 96 | 4.34 ± 0.27 | -0.05 ± 0.16 | … | … | N | … | … | … | … | … | n | … |
| 48835 | 18265078+0630165 | 6.8 ± 0.3 | 4625 ± 210 | 4.58 ± 0.17 | 0.15 ± 0.16 | … | … | N | … | … | … | … | … | n | … |
| 48836 | 18265094+0631406 | -9.8 ± 0.3 | 5605 ± 31 | 3.99 ± 0.09 | 0.15 ± 0.16 | 73 ± 30 | 1 | Y | Y | Y | Y | … | … | Y | … |
| 48837 | 18265119+0614410 | -25.1 ± 0.3 | 4201 ± 282 | 4.35 ± 0.28 | 0.11 ± 0.2 | <23 | 3 | Y | N | Y | Y | N | … | n | NG |
| 48838 | 18265145+0640196 | -38.2 ± 0.3 | 5947 ± 135 | 4.02 ± 0.14 | 0.18 ± 0.14 | 53 ± 30 | 1 | Y | Y | Y | Y | … | … | Y | … |
| 48839 | 18265148+0631329 | 48.0 ± 0.3 | 4733 ± 186 | 2.71 ± 0.34 | -0.23 ± 0.21 | … | … | N | … | … | … | … | … | n | G |
| 48840 | 18265160+0617391 | 21.0 ± 0.3 | 5058 ± 86 | 4.74 ± 0.73 | 0.28 ± 0.23 | <72 | 3 | N | .. | … | … | … | … | n | NG |
| 48841 | 18265164+0642526 | 52.6 ± 5.3 | 4296 ± 644 | 4.60 ± 0.13 | -1.14 ± 1.33 | <73 | 3 | N | .. | … | … | … | … | n | NG |
| 48842 | 18265170+0635587 | 12.6 ± 0.3 | 5584 ± 115 | 4.57 ± 0.47 | 0.21 ± 0.13 | 49 ± 34 | 1 | N | .. | … | … | … | … | n | NG |
| 48843 | 18265194+0646039 | 23.5 ± 0.3 | 5928 ± 120 | 4.14 ± 0.33 | -0.35 ± 0.13 | … | … | N | … | … | … | … | … | n | … |
| 48844 | 18265201+0633532 | -14.9 ± 0.5 | 4300 ± 136 | 5.07 ± 0.54 | -0.35 ± 0.49 | <116 | 3 | Y | N | N | Y | … | … | n | NG |
| 48845 | 18265204+0634183 | 66.2 ± 0.3 | 5766 ± 189 | 4.25 ± 0.27 | 0.34 ± 0.15 | … | … | N | … | … | … | … | … | n | … |
| 48846 | 18265208+0614224 | -99.5 ± 0.3 | 4693 ± 222 | 4.52 ± 0.34 | -0.04 ± 0.25 | … | … | N | … | … | … | … | … | n | … |
| 48847 | 18265208+0631085 | 43.3 ± 0.2 | 4422 ± 168 | 2.42 ± 0.49 | 0.02 ± 0.26 | … | … | N | … | … | … | … | … | n | G |
| 48848 | 18265210+0637138 | -7.6 ± 0.3 | 5692 ± 455 | 4.60 ± 0.26 | 0.36 ± 0.21 | … | … | Y | … | … | … | … | … | n | … |
| 48849 | 18265211+0615122 | -43.2 ± 0.3 | 5340 ± 112 | 4.38 ± 0.53 | 0.07 ± 0.21 | … | … | Y | … | … | … | … | … | n | … |
| 48850 | 18265215+0627446 | -16.8 ± 0.3 | 5033 ± 74 | 3.55 ± 0.17 | -0.15 ± 0.16 | … | … | Y | … | … | … | … | … | n | … |
| 48851 | 18265219+0626433 | 41.4 ± 0.2 | 4752 ± 95 | 2.87 ± 0.34 | 0.22 ± 0.24 | … | … | N | … | … | … | … | … | n | G |
| 48852 | 18265219+0634138 | 21.4 ± 0.4 | 4804 ± 418 | 3.18 ± 0.92 | -0.61 ± 0.4 | … | … | N | … | … | … | … | … | n | G |
| 48853 | 18265231+0624050 | 48.0 ± 0.5 | 5682 ± 67 | 3.75 ± 0.67 | -0.23 ± 0.15 | 104 ± 94 | 1 | N | .. | … | … | … | … | n | NG |
| 48854 | 18265248+0627259 | 32.4 ± 0.3 | 4903 ± 202 | 3.18 ± 0.43 | -0.73 ± 0.63 | 261 ± 41 | 1 | N | .. | … | … | … | … | n | Li-rich G |
| 48855 | 18265259+0639162 | -22.1 ± 0.4 | 4800 ± 275 | 2.84 ± 0.47 | -0.15 ± 0.27 | … | … | Y | … | … | … | … | … | n | G |
| 48856 | 18265261+0644151 | 4.7 ± 0.3 | 5539 ± 171 | 3.81 ± 0.20 | 0.22 ± 0.19 | <35 | 3 | N | .. | … | … | … | … | n | NG |
| 48857 | 18265272+0637097 | -27.7 ± 1.0 | 4613 ± 79 | 3.97 ± 0.25 | -0.45 ± 0.09 | … | … | Y | … | … | … | … | … | n | … |
| 48858 | 18265277+0615284 | -5.5 ± 0.2 | 4877 ± 226 | 4.49 ± 0.29 | -0.09 ± 0.15 | … | … | Y | … | … | … | … | … | n | … |
| 48859 | 18265314+0613509 | -89.9 ± 0.3 | 5741 ± 271 | 4.32 ± 0.28 | 0.10 ± 0.19 | … | … | N | … | … | … | … | … | n | … |
| 48860 | 18265318+0634038 | 82.8 ± 0.5 | 4817 ± 255 | 2.99 ± 0.62 | 0.12 ± 0.55 | <97 | 3 | N | .. | … | … | … | … | n | G |
| 48861 | 18265336+0643263 | 23.2 ± 0.3 | 4703 ± 317 | 2.99 ± 0.24 | 0.04 ± 0.26 | … | … | N | … | … | … | … | … | n | G |
| 48862 | 18265338+0628174 | 16.9 ± 0.3 | 4736 ± 283 | 2.88 ± 0.48 | -0.06 ± 0.27 | … | … | N | … | … | … | … | … | n | G |
| 48863 | 18265364+0633211 | 23.5 ± 0.9 | 5166 ± 153 | 4.47 ± 0.19 | -0.39 ± 0.34 | <172 | 3 | N | .. | … | … | … | … | n | NG |
| 48864 | 18265365+0631484 | 42.4 ± 0.3 | 3945 ± 252 | 4.56 ± 0.10 | 0.04 ± 0.19 | <59 | 3 | N | .. | … | … | … | … | n | NG |
| 48865 | 18265406+0616000 | -29.5 ± 0.3 | 5482 ± 247 | 4.27 ± 0.39 | 0.05 ± 0.12 | … | … | Y | … | … | … | … | … | n | … |
| 48866 | 18265406+0633423 | -3.1 ± 0.2 | 4822 ± 207 | 4.43 ± 0.66 | 0.25 ± 0.22 | … | … | Y | … | … | … | N | … | n | … |
| 48867 | 18265413+0635072 | -31.2 ± 0.2 | 4393 ± 225 | 2.50 ± 0.65 | -0.09 ± 0.29 | … | … | Y | … | … | … | … | … | n | … |
| 48868 | 18265434+0632375 | -0.1 ± 0.5 | 6559 ± 251 | 4.28 ± 0.35 | -0.03 ± 0.35 | … | … | Y | … | … | … | … | … | n | … |
| 48869 | 18265434+0633275 | 35.4 ± 0.3 | 4989 ± 126 | 3.58 ± 0.37 | 0.08 ± 0.14 | … | … | N | … | … | … | … | … | n | … |
| 48870 | 18265437+0629563 | -22.1 ± 0.3 | 6275 ± 170 | 4.09 ± 0.09 | 0.08 ± 0.23 | … | … | Y | … | … | … | … | … | n | … |
| 48871 | 18265454+0614555 | 89.5 ± 0.3 | 4929 ± 136 | 3.03 ± 0.45 | -0.16 ± 0.2 | … | … | N | … | … | … | … | … | n | G |
| 48872 | 18265466+0616270 | 79.1 ± 0.3 | 4969 ± 124 | 3.06 ± 0.39 | -0.27 ± 0.19 | … | … | N | … | … | … | … | … | n | G |
| 48873 | 18265477+0615281 | 70.1 ± 0.2 | 4726 ± 172 | 2.52 ± 0.33 | -0.07 ± 0.18 | … | … | N | … | … | … | … | … | n | G |

| ID | CNAME | RV (km s$^{-1}$) | $T_{\text{eff}}$ (K) | logg (dex) | [Fe/H] (dex) | EW(Li)$^a$ (mÅ) | EW(Li) error flag$^b$ | Membership RV | Li | logg | [Fe/H] | Gaia studies Randich$^c$ | Cantat-Gaudin$^c$ | Final$^d$ | NMs with Li$^e$ |
|---|---|---|---|---|---|---|---|---|---|---|---|---|---|---|---|
| 48874 | 18265491+0637274 | 21.0 ± 0.3 | 6059 ± 101 | 4.35 ± 0.32 | -0.08 ± 0.13 | 67 ± 46 | 1 | N | .. | … | … | … | … | n | NG |
| 48875 | 18265509+0637115 | 54.4 ± 0.4 | 4549 ± 283 | 4.38 ± 0.52 | 0.23 ± 0.24 | <130 | 3 | N | .. | … | … | … | … | n | NG |
| 48876 | 18265512+0636577 | 54.6 ± 0.5 | 3962 ± 341 | 4.46 ± 0.13 | -0.55 ± 0.44 | … | … | N | … | … | … | … | … | n | … |
| 48877 | 18265527+0614307 | -58.6 ± 0.3 | 5055 ± 109 | 3.51 ± 0.21 | -0.44 ± 0.2 | <16 | 3 | N | .. | … | … | … | … | n | NG |
| 48878 | 18265527+0636161 | -41.8 ± 0.3 | 5533 ± 96 | 4.17 ± 0.38 | -0.26 ± 0.23 | … | … | Y | … | … | … | … | … | n | … |
| 48879 | 18265546+0642393 | -54.3 ± 0.3 | 5801 ± 123 | 4.35 ± 0.24 | -0.45 ± 0.19 | … | … | N | … | … | … | … | … | n | … |
| 48880 | 18265547+0630427 | -27.6 ± 0.2 | 4988 ± 89 | 4.58 ± 0.44 | -0.02 ± 0.13 | <25 | 3 | Y | Y | Y | Y | Y | Y | Y | … |
| 48881 | 18265551+0632465 | -29.2 ± 0.3 | 5078 ± 86 | 3.36 ± 0.12 | -0.19 ± 0.15 | … | … | Y | … | … | … | … | … | n | … |
| 48882 | 18265565+0633338 | 17.0 ± 0.2 | 4905 ± 149 | 2.92 ± 0.30 | 0.02 ± 0.21 | … | … | N | … | … | … | … | … | n | G |
| 48883 | 18265569+0623126 | -99.4 ± 0.4 | 4687 ± 651 | 2.42 ± 0.12 | -0.06 ± 0.03 | … | … | N | … | … | … | … | … | n | … |
| 48884 | 18265573+0615402 | 5.8 ± 0.2 | 5009 ± 74 | 3.59 ± 0.40 | 0.11 ± 0.16 | … | … | N | … | … | … | N | … | n | … |
| 48885 | 18265579+0640429 | -65.1 ± 0.3 | 5214 ± 217 | 4.59 ± 0.47 | -0.36 ± 0.42 | … | … | N | … | … | … | … | … | n | … |
| 48886 | 18265579+0643132 | -41.1 ± 0.3 | 5801 ± 191 | 4.08 ± 0.15 | 0.09 ± 0.16 | 68 ± 30 | 1 | Y | Y | Y | Y | … | … | Y | … |
| 48887 | 18265612+0624183 | -9.3 ± 0.3 | 4564 ± 88 | 2.26 ± 0.30 | -0.05 ± 0.12 | <62 | 3 | Y | N | Y | Y | … | … | n | G |
| 48888 | 18265629+0613451 | -47.4 ± 0.3 | 5488 ± 226 | 4.07 ± 0.27 | 0.22 ± 0.19 | … | … | N | … | … | … | N | … | n | … |
| 48889 | 18265630+0632279 | -21.2 ± 0.3 | 4780 ± 161 | 4.65 ± 0.18 | 0.08 ± 0.17 | … | … | Y | … | … | … | … | … | n | … |
| 48890 | 18265636+0626557 | -18.8 ± 0.3 | 4979 ± 120 | 4.72 ± 0.48 | 0.20 ± 0.17 | … | … | Y | … | … | … | … | … | n | … |
| 48891 | 18265640+0615084 | -12.3 ± 0.3 | 4738 ± 70 | 2.60 ± 0.26 | 0.09 ± 0.23 | … | … | Y | … | … | … | … | … | n | G |
| 48892 | 18265644+0640269 | -11.2 ± 0.3 | 5752 ± 79 | 4.25 ± 0.29 | -0.46 ± 0.19 | … | … | Y | … | … | … | … | … | n | … |
| 48893 | 18265661+0624188 | 54.4 ± 0.2 | 5401 ± 160 | 3.87 ± 0.35 | -0.23 ± 0.18 | … | … | N | … | … | … | … | … | n | … |
| 48894 | 18265663+0624446 | 87.9 ± 0.2 | 4733 ± 105 | 2.79 ± 0.33 | 0.14 ± 0.25 | … | … | N | … | … | … | … | … | n | G |
| 48895 | 18265707+0638131 | -10.9 ± 0.5 | 3995 ± 149 | 5.13 ± 0.34 | 0.30 ± 0.15 | … | … | Y | … | … | … | … | … | n | … |
| 48896 | 18265715+0635174 | 24.0 ± 0.5 | 5286 ± 114 | 3.77 ± 0.24 | -0.45 ± 0.29 | … | … | N | … | … | … | … | … | n | … |
| 48897 | 18265721+0639304 | -20.3 ± 0.2 | 4781 ± 198 | 2.90 ± 0.42 | 0.04 ± 0.23 | … | … | Y | … | … | … | … | … | n | G |
| 48898 | 18265732+0628403 | 3.9 ± 1.9 | 6176 ± 301 | 3.66 ± 0.67 | -0.89 ± 1.19 | <50 | 3 | N | .. | … | … | … | … | n | NG |
| 48899 | 18265753+0615249 | 110.5 ± 0.4 | 5115 ± 121 | 3.49 ± 0.17 | -0.21 ± 0.22 | … | … | N | … | … | … | … | … | n | G |
| 48900 | 18265753+0643407 | -36.8 ± 0.3 | 5264 ± 73 | 4.48 ± 0.32 | -0.12 ± 0.13 | … | … | Y | … | … | … | … | … | n | … |
| 48901 | 18265756+0640377 | 147.0 ± 0.3 | 4831 ± 185 | 2.56 ± 0.49 | -0.32 ± 0.23 | … | … | N | … | … | … | … | … | n | … |
| 48902 | 18265790+0640465 | -25.2 ± 0.2 | 5379 ± 55 | 4.05 ± 0.28 | 0.01 ± 0.13 | … | … | Y | … | … | … | … | … | n | … |
| 48903 | 18265803+0628281 | -5.8 ± 0.3 | 5478 ± 217 | 3.70 ± 0.37 | 0.05 ± 0.13 | <38 | 3 | Y | Y | Y | Y | … | … | Y | … |
| 48904 | 18265808+0631575 | 14.3 ± 0.2 | 4569 ± 114 | 2.51 ± 0.28 | 0.03 ± 0.17 | <43 | 3 | N | .. | … | … | … | … | n | G |
| 48905 | 18265811+0644105 | -23.1 ± 0.3 | 4526 ± 245 | 2.65 ± 0.44 | 0.01 ± 0.19 | … | … | Y | … | … | … | … | … | n | G |
| 48906 | 18265827+0643454 | -64.6 ± 0.5 | 4895 ± 202 | 2.94 ± 0.46 | -0.63 ± 0.59 | … | … | N | … | … | … | … | … | n | G |
| 48907 | 18265839+0638047 | -51.4 ± 0.3 | 5095 ± 77 | 3.43 ± 0.40 | 0.00 ± 0.21 | … | … | N | … | … | … | … | … | n | G |
| 48909 | 18265874+0614232 | -34.2 ± 0.3 | 5959 ± 189 | 4.24 ± 0.06 | 0.09 ± 0.13 | … | … | Y | … | … | … | … | … | n | … |
| 48910 | 18265877+0623395 | -15.8 ± 0.2 | 5713 ± 327 | 4.35 ± 0.19 | 0.37 ± 0.19 | … | … | Y | … | … | … | … | … | n | … |
| 48911 | 18265890+0633228 | 13.5 ± 0.3 | 5052 ± 119 | 3.36 ± 0.50 | -0.32 ± 0.19 | 102 ± 40 | 1 | N | .. | … | … | … | … | n | Li-rich G |
| 48912 | 18265903+0635319 | 22.3 ± 0.3 | 4362 ± 217 | 4.81 ± 0.27 | -0.16 ± 0.13 | … | … | N | … | … | … | … | … | n | … |
| 48913 | 18265912+0631507 | 24.8 ± 0.2 | 4187 ± 273 | 1.88 ± 0.59 | -0.03 ± 0.3 | 134 ± 29 | 1 | N | .. | … | … | … | … | n | G |
| 48914 | 18265912+0633463 | -0.8 ± 0.2 | 4336 ± 166 | 4.67 ± 0.29 | 0.07 ± 0.18 | <29 | 3 | Y | N | N | Y | N | … | n | NG |
| 48915 | 18265916+0615041 | 57.0 ± 1.3 | 4534 ± 115 | 4.06 ± 0.28 | -0.13 ± 0.12 | … | … | N | … | … | … | N | … | n | … |
| 48916 | 18265917+0614067 | -68.5 ± 0.3 | 5469 ± 77 | 4.44 ± 0.56 | 0.22 ± 0.25 | … | … | N | … | … | … | … | … | n | … |
| 48917 | 18265919+0613542 | 31.3 ± 0.3 | 5863 ± 40 | 4.49 ± 0.38 | 0.00 ± 0.12 | 57 ± 31 | 1 | N | .. | … | … | … | … | n | NG |
| 48918 | 18265921+0618092 | 105.1 ± 0.3 | 5099 ± 200 | 3.06 ± 0.26 | -0.15 ± 0.17 | … | … | N | … | … | … | … | … | n | G |
| 48919 | 18265932+0638346 | -145.7 ± 0.3 | 4905 ± 89 | 1.77 ± 0.69 | -1.18 ± 0.35 | … | … | N | … | … | … | … | … | n | G |
| 48920 | 18265941+0628000 | 17.7 ± 0.2 | 4880 ± 155 | 2.90 ± 0.30 | -0.05 ± 0.18 | … | … | N | … | … | … | … | … | n | G |
| 48921 | 18265943+0640416 | -30.4 ± 0.3 | 4490 ± 268 | 4.87 ± 0.36 | -0.49 ± 0.31 | … | … | Y | … | … | … | … | … | n | … |
| 48922 | 18265945+0632188 | -4.7 ± 0.3 | 4751 ± 209 | 2.91 ± 0.22 | 0.10 ± 0.26 | … | … | Y | … | … | … | … | … | n | G |
| 48923 | 18265958+0633398 | 27.9 ± 0.2 | 4853 ± 159 | 2.81 ± 0.40 | -0.42 ± 0.26 | … | … | N | … | … | … | … | … | n | G |
| 48924 | 18265963+0638291 | -27.7 ± 0.2 | 5205 ± 102 | 4.40 ± 0.51 | -0.02 ± 0.13 | 18 ± 9 | 1 | Y | Y | Y | Y | Y | Y | Y | … |
| 48925 | 18265967+0639199 | -19.0 ± 0.3 | 5398 ± 261 | 4.54 ± 0.15 | 0.14 ± 0.15 | … | … | Y | … | … | … | … | … | n | … |
| 48926 | 18265986+0633071 | -30.0 ± 0.3 | 4992 ± 111 | 4.71 ± 0.34 | 0.04 ± 0.13 | … | … | Y | … | … | … | Y | … | n | … |
| 48927 | 18265988+0623253 | 52.1 ± 0.6 | 5411 ± 238 | 3.61 ± 0.20 | 0.07 ± 0.15 | … | … | N | … | … | … | … | … | n | … |
| 48928 | 18265999+0636597 | 41.0 ± 0.4 | 5177 ± 148 | 4.03 ± 0.72 | -0.47 ± 0.23 | … | … | N | … | … | … | … | … | n | G |
| 48929 | 18270002+0634196 | -19.2 ± 0.3 | 4611 ± 368 | 2.49 ± 0.62 | -0.53 ± 0.26 | … | … | Y | … | … | … | … | … | n | G |
| 48930 | 18270010+0614163 | 2.8 ± 0.2 | 5351 ± 97 | 3.86 ± 0.23 | 0.38 ± 0.23 | … | … | Y | … | … | … | … | … | n | … |
| 48931 | 18270032+0640317 | 32.4 ± 0.3 | 4773 ± 346 | 2.62 ± 0.47 | -0.27 ± 0.34 | … | … | N | … | … | … | … | … | n | G |
| 48932 | 18270032+0644446 | -23.0 ± 0.3 | 4949 ± 141 | 3.17 ± 0.48 | -0.21 ± 0.2 | <47 | 3 | Y | Y | Y | Y | … | … | Y | … |
| 48933 | 18270040+0613382 | -133.9 ± 0.3 | 4920 ± 57 | 2.71 ± 0.24 | -0.17 ± 0.19 | … | … | N | … | … | … | … | … | n | G |

**Table C.11.** continued.

| ID | CNAME | RV (km s$^{-1}$) | $T_{\rm eff}$ (K) | $\log g$ (dex) | [Fe/H] (dex) | EW(Li)$^a$ (mÅ) | EW(Li) error flag$^b$ | Membership RV | Li | $\log g$ | [Fe/H] | Gaia studies Randich$^c$ | Cantat-Gaudin$^c$ | Final$^d$ | NMs with Li$^e$ |
|---|---|---|---|---|---|---|---|---|---|---|---|---|---|---|---|
| 48934 | 18270054+0617398 | 0.4 ± 0.4 | 4636 ± 266 | 4.85 ± 0.32 | -0.15 ± 0.14 | … | … | Y | … | … | … | … | … | n | … |
| 48935 | 18270054+0644262 | -85.0 ± 0.3 | 4991 ± 133 | 3.29 ± 0.51 | -0.35 ± 0.12 | … | … | N | … | … | … | … | … | n | G |
| 48936 | 18270055+0638417 | -4.1 ± 0.2 | 4163 ± 256 | 4.82 ± 0.20 | 0.07 ± 0.18 | … | … | Y | … | … | … | N | … | n | … |
| 48937 | 18270057+0637095 | -248.7 ± 0.2 | 4232 ± 361 | 1.81 ± 0.60 | -0.34 ± 0.19 | … | … | N | … | … | … | … | … | n | G |
| 48938 | 18270071+0639374 | 75.0 ± 0.3 | 5540 ± 67 | 4.08 ± 0.25 | -0.32 ± 0.14 | … | … | N | … | … | … | … | … | n | … |
| 48939 | 18270087+0633395 | 25.1 ± 0.2 | 4053 ± 562 | 4.71 ± 0.20 | -0.04 ± 0.23 | <26 | 3 | N | .. | … | … | N | … | n | NG |
| 48940 | 18270091+0613449 | 74.8 ± 0.5 | 4619 ± 421 | 3.49 ± 0.21 | -0.24 ± 0.25 | … | … | N | … | … | … | … | … | n | G |
| 48941 | 18270096+0642583 | 50.3 ± 0.2 | 4821 ± 207 | 2.47 ± 0.65 | -0.25 ± 0.3 | … | … | N | … | … | … | … | … | n | G |
| 48942 | 18270099+0635439 | -4.1 ± 0.3 | 5982 ± 311 | 4.36 ± 0.34 | 0.33 ± 0.17 | … | … | Y | … | … | … | … | … | n | … |
| 48943 | 18270104+0627442 | 28.5 ± 0.2 | 6200 ± 66 | 4.04 ± 0.26 | -0.20 ± 0.16 | 34 ± 13 | … | N | .. | … | … | … | … | n | NG |
| 48944 | 18270109+0636321 | 101.4 ± 0.3 | 4902 ± 149 | 2.87 ± 0.33 | -0.13 ± 0.18 | … | … | N | … | … | … | … | … | n | G |
| 48945 | 18270110+0631416 | -45.9 ± 0.2 | 5375 ± 66 | 4.40 ± 0.40 | -0.20 ± 0.13 | … | … | Y | … | … | … | … | … | n | … |
| 48946 | 18270128+0633271 | 11.0 ± 0.3 | 6430 ± 159 | 3.93 ± 0.23 | -0.11 ± 0.12 | … | … | N | … | … | … | … | … | n | … |
| 48947 | 18270135+0630399 | 55.3 ± 0.2 | 4649 ± 136 | 2.86 ± 0.45 | 0.06 ± 0.2 | … | … | N | … | … | … | … | … | n | G |
| 48948 | 18270166+0626140 | -34.8 ± 0.2 | 4945 ± 137 | 4.61 ± 0.37 | 0.10 ± 0.15 | <16 | 3 | Y | Y | Y | Y | N | … | Y?$^g$ | … |
| 48949 | 18270181+0629408 | -35.6 ± 0.2 | 5758 ± 174 | 4.11 ± 0.04 | 0.33 ± 0.16 | … | … | Y | … | … | … | … | … | n | … |
| 48950 | 18270192+0621263 | 9.7 ± 0.2 | 4765 ± 159 | 3.02 ± 0.24 | 0.07 ± 0.17 | … | … | N | … | … | … | … | … | n | G |
| 48951 | 18270192+0622375 | 42.0 ± 0.3 | 4889 ± 193 | 3.36 ± 0.14 | -0.10 ± 0.23 | … | … | N | … | … | … | … | … | n | G |
| 48952 | 18270209+0629584 | -37.7 ± 0.2 | 4848 ± 203 | 2.86 ± 0.45 | -0.29 ± 0.22 | … | … | Y | … | Y | … | … | … | n | G |
| 48953 | 18270216+0623359 | 56.0 ± 0.2 | 4747 ± 70 | 2.96 ± 0.42 | 0.16 ± 0.2 | <32 | 3 | N | .. | … | … | … | … | n | G |
| 48954 | 18270228+0636186 | -317.4 ± 0.6 | 4419 ± 325 | 2.64 ± 1.03 | -1.37 ± 0.2 | … | … | N | … | … | … | … | … | n | G |
| 48955 | 18270231+0617251 | 38.1 ± 0.3 | 5048 ± 184 | 3.16 ± 0.40 | -0.14 ± 0.22 | … | … | N | … | … | … | … | … | n | G |
| 48956 | 18270238+0645545 | -1.2 ± 0.2 | 4938 ± 145 | 2.58 ± 0.58 | -0.36 ± 0.25 | … | … | Y | … | … | … | … | … | n | G |
| 48957 | 18270245+0623535 | 30.1 ± 0.2 | 4569 ± 128 | 2.87 ± 0.53 | 0.27 ± 0.24 | … | … | N | … | … | … | … | … | n | G |
| 48958 | 18270252+0622084 | 30.1 ± 0.2 | 4433 ± 129 | 2.29 ± 0.45 | -0.09 ± 0.22 | … | … | N | … | … | … | … | … | n | G |
| 48959 | 18270255+0643462 | -47.0 ± 0.2 | 5646 ± 309 | 4.35 ± 0.18 | 0.37 ± 0.17 | … | … | Y | … | … | … | … | … | n | … |
| 48960 | 18270257+0645361 | 17.5 ± 0.3 | 4957 ± 329 | 3.20 ± 0.97 | -0.55 ± 0.28 | <74 | 3 | N | .. | … | … | … | … | n | Li-rich G |
| 48961 | 18270259+0617200 | 14.4 ± 0.2 | 4579 ± 169 | 2.83 ± 0.60 | 0.22 ± 0.29 | … | … | N | … | … | … | … | … | n | G |
| 48962 | 18270261+0623109 | -24.2 ± 0.3 | 6678 ± 403 | 4.05 ± 0.12 | 0.19 ± 0.19 | … | … | Y | … | … | … | … | … | n | … |
| 48963 | 18270266+0645032 | -22.4 ± 0.2 | 4839 ± 123 | 3.22 ± 0.34 | 0.15 ± 0.17 | … | … | Y | … | … | … | … | … | n | G |
| 48964 | 18270270+0627140 | -20.3 ± 0.2 | 4969 ± 88 | 3.24 ± 0.21 | -0.33 ± 0.17 | … | … | Y | … | … | … | … | … | n | G |
| 48965 | 18270282+0640286 | -70.0 ± 0.3 | 5805 ± 149 | 4.18 ± 0.12 | -0.47 ± 0.12 | … | … | N | … | … | … | … | … | n | … |
| 48966 | 18270285+0628023 | 88.9 ± 0.2 | 4743 ± 161 | 2.64 ± 0.33 | -0.06 ± 0.22 | … | … | N | … | … | … | … | … | n | G |
| 48967 | 18270297+0620046 | 3.7 ± 0.3 | 5046 ± 105 | 3.39 ± 0.28 | -0.01 ± 0.16 | <46 | 3 | N | .. | … | … | … | … | n | G |
| 48968 | 18270303+0639394 | 39.5 ± 0.3 | 5014 ± 206 | 3.59 ± 0.45 | -0.38 ± 0.19 | … | … | N | … | … | … | … | … | n | … |
| 48969 | 18270312+0614518 | -163.5 ± 0.4 | 4935 ± 113 | 2.43 ± 0.40 | -0.22 ± 0.39 | … | … | N | … | … | … | … | … | n | G |
| 48970 | 18270312+0641206 | -55.6 ± 0.2 | 4856 ± 147 | 3.27 ± 0.63 | 0.38 ± 0.32 | <34 | 3 | N | .. | … | … | … | … | n | G |
| 48971 | 18270318+0633222 | 62.1 ± 0.3 | 4902 ± 147 | 3.25 ± 0.35 | -0.11 ± 0.19 | … | … | N | … | … | … | … | … | n | G |
| 48972 | 18270329+0629481 | -26.0 ± 0.2 | 4701 ± 229 | 2.57 ± 0.38 | -0.27 ± 0.25 | … | … | Y | … | … | … | … | … | n | G |
| 48973 | 18270344+0622556 | 98.2 ± 0.3 | 5174 ± 211 | 3.52 ± 0.72 | -0.63 ± 0.26 | … | … | N | … | … | … | … | … | n | … |
| 48974 | 18270350+0637412 | -57.1 ± 0.3 | 5803 ± 122 | 4.24 ± 0.15 | -0.06 ± 0.21 | … | … | N | … | … | … | … | … | n | … |
| 48975 | 18270359+0617399 | 21.3 ± 0.2 | 4681 ± 130 | 3.10 ± 0.52 | 0.19 ± 0.19 | … | … | N | … | … | … | … | … | n | … |
| 48976 | 18270372+0619102 | 17.9 ± 0.3 | 4713 ± 325 | 3.16 ± 0.14 | -0.06 ± 0.16 | <75 | 3 | N | .. | … | … | … | … | n | G |
| 48977 | 18270385+0621235 | -88.2 ± 0.3 | 5004 ± 91 | 3.16 ± 0.20 | -0.45 ± 0.12 | … | … | N | … | … | … | … | … | n | G |
| 48978 | 18270386+0631133 | -11.0 ± 0.2 | 5785 ± 80 | 4.21 ± 0.24 | -0.19 ± 0.14 | … | … | Y | … | … | … | … | … | n | … |
| 48979 | 18270396+0639543 | 9.3 ± 0.8 | 5667 ± 791 | 3.70 ± 0.41 | -0.46 ± 0.63 | … | … | N | … | … | … | N | … | n | … |
| 48980 | 18270415+0634181 | -63.2 ± 0.3 | 4548 ± 202 | 2.17 ± 0.52 | 0.03 ± 0.44 | … | … | N | … | … | … | … | … | n | G |
| 48981 | 18270426+0636415 | -9.0 ± 0.2 | 4524 ± 111 | 2.35 ± 0.41 | -0.09 ± 0.2 | <54 | 3 | Y | Y | Y | Y | … | … | Y | … |
| 48982 | 18270435+0622264 | 1.1 ± 0.3 | 4966 ± 144 | 3.01 ± 0.31 | -0.07 ± 0.16 | … | … | Y | … | … | … | … | … | n | G |
| 48983 | 18270444+0630508 | -39.2 ± 0.6 | 4348 ± 82 | 4.08 ± 0.51 | -0.66 ± 0.21 | … | … | Y | … | … | … | … | … | n | … |
| 48984 | 18270448+0639192 | 63.9 ± 0.6 | 5109 ± 106 | 3.67 ± 0.70 | 0.11 ± 0.45 | … | … | N | … | … | … | … | … | n | … |
| 48985 | 18270451+0640061 | 8.0 ± 0.2 | 4186 ± 229 | 2.17 ± 0.54 | 0.06 ± 0.26 | … | … | N | … | … | … | … | … | n | G |
| 48986 | 18270454+0639415 | -11.1 ± 0.6 | 4922 ± 269 | 4.09 ± 0.19 | -0.71 ± 0.44 | … | 3 | Y | … | … | … | … | … | n | … |
| 48987 | 18270460+0619291 | 34.9 ± 0.2 | 5524 ± 121 | 4.29 ± 0.23 | 0.25 ± 0.14 | … | … | N | … | … | … | … | … | n | … |
| 48988 | 18270481+0638365 | -2.5 ± 0.3 | 4861 ± 169 | 3.02 ± 0.46 | 0.05 ± 0.33 | … | … | Y | … | … | … | … | … | n | G |
| 48989 | 18270494+0616539 | -56.9 ± 0.2 | 4879 ± 125 | 3.62 ± 0.65 | 0.21 ± 0.15 | … | … | N | … | … | … | N | … | n | … |
| 48990 | 18270504+0640443 | -40.8 ± 0.8 | 4926 ± 610 | 3.37 ± 1.30 | -0.16 ± 0.29 | … | … | Y | … | … | … | … | … | n | G |
| 48991 | 18270506+0642212 | 29.0 ± 0.5 | 3970 ± 294 | 4.68 ± 0.17 | -0.55 ± 0.38 | … | … | N | … | … | … | … | … | n | … |
| 48992 | 18270512+0634213 | 3.1 ± 0.4 | 6133 ± 72 | 4.17 ± 0.27 | -0.14 ± 0.15 | … | … | N | … | … | … | … | … | n | … |

**Table C.11.** continued.

| ID | CNAME | RV (km s$^{-1}$) | $T_{\rm eff}$ (K) | logg (dex) | [Fe/H] (dex) | EW(Li)$^a$ (mÅ) | EW(Li) error flag$^b$ | Membership RV | Li | logg | [Fe/H] | Gaia studies Randich$^c$ | Cantat-Gaudin$^c$ | Final$^d$ | NMs with Li$^e$ |
|---|---|---|---|---|---|---|---|---|---|---|---|---|---|---|---|
| 48993 | 18270513+0628356 | -13.7 ± 0.3 | 3895 ± 352 | 4.56 ± 0.28 | -0.51 ± 0.42 | <73 | 3 | Y | N | Y | N | N | … | n | NG |
| 48994 | 18270513+0639228 | 37.0 ± 0.3 | 5217 ± 152 | 2.86 ± 0.44 | -1.11 ± 0.19 | … | … | N | … | … | … | … | … | n | … |
| 48995 | 18270514+0637095 | -64.0 ± 0.2 | 4952 ± 82 | 3.59 ± 0.46 | 0.14 ± 0.21 | … | … | N | … | … | … | … | … | n | … |
| 48996 | 18270515+0631303 | -79.4 ± 0.2 | 5924 ± 212 | 4.14 ± 0.22 | -0.18 ± 0.19 | 46 ± 14 | 1 | N | .. | … | … | … | … | n | NG |
| 48997 | 18270526+0624165 | 4.3 ± 0.3 | 5855 ± 25 | 4.00 ± 0.18 | -0.37 ± 0.13 | … | … | N | … | … | … | … | … | n | … |
| 48998 | 18270527+0622285 | -15.5 ± 0.2 | 4428 ± 275 | 4.77 ± 0.26 | -0.26 ± 0.16 | … | … | Y | … | … | … | N | … | n | … |
| 48999 | 18270527+0624251 | 4.2 ± 0.2 | 5199 ± 44 | 4.32 ± 0.44 | 0.07 ± 0.14 | … | … | N | … | … | … | … | … | n | … |
| 49000 | 18270527+0627554 | -27.2 ± 0.2 | 4685 ± 216 | 4.60 ± 0.38 | -0.01 ± 0.13 | … | … | Y | … | … | … | Y | Y | n | … |
| 49001 | 18270539+0643436 | -16.4 ± 0.2 | 5593 ± 134 | 4.41 ± 0.49 | 0.05 ± 0.14 | 113 ± 26 | 1 | Y | N | Y | Y | … | … | n | NG |
| 49002 | 18270557+0633319 | -65.3 ± 0.5 | 3717 ± 168 | 4.51 ± 0.21 | -0.06 ± 0.12 | <135 | 3 | N | .. | … | … | N | … | n | NG |
| 49003 | 18270574+0629265 | 19.4 ± 0.2 | 4922 ± 169 | 2.79 ± 0.47 | -0.18 ± 0.19 | … | … | N | … | … | … | … | … | n | G |
| 49004 | 18270575+0639314 | 59.8 ± 0.5 | 5248 ± 128 | 4.46 ± 0.36 | -0.43 ± 0.17 | … | … | N | … | … | … | … | … | n | … |
| 49005 | 18270587+0625093 | -25.2 ± 0.3 | 5062 ± 137 | 2.83 ± 0.43 | -0.26 ± 0.21 | … | … | Y | … | … | … | … | … | n | Li-rich G |
| 49006 | 18270592+0639220 | 26.3 ± 0.3 | 4651 ± 98 | 2.11 ± 0.40 | -0.09 ± 0.13 | … | … | N | … | … | … | … | … | n | G |
| 49007 | 18270598+0618270 | 38.0 ± 0.3 | 5842 ± 33 | 4.19 ± 0.12 | -0.51 ± 0.18 | 31 ± 28 | 1 | N | .. | … | … | … | … | n | NG |
| 49008 | 18270614+0639459 | -11.9 ± 0.3 | 5921 ± 196 | 4.05 ± 0.18 | 0.00 ± 0.13 | … | … | Y | … | … | … | … | … | n | … |
| 49009 | 18270620+0633424 | -39.2 ± 0.3 | 4863 ± 128 | 4.52 ± 0.32 | 0.08 ± 0.18 | … | … | Y | … | … | … | … | … | n | … |
| 49010 | 18270624+0627247 | -31.5 ± 0.4 | 4091 ± 320 | 4.93 ± 0.17 | -0.17 ± 0.18 | … | … | Y | … | … | … | N | … | n | … |
| 49011 | 18270627+0624277 | 48.3 ± 0.2 | 4701 ± 88 | 3.17 ± 0.58 | 0.25 ± 0.22 | <18 | 3 | N | .. | … | … | … | … | n | G |
| 49012 | 18270638+0638474 | -16.8 ± 0.3 | 4797 ± 237 | 3.10 ± 0.36 | -0.03 ± 0.24 | … | … | Y | … | … | … | … | … | n | G |
| 49013 | 18270643+0624124 | 56.8 ± 0.2 | 4834 ± 232 | 3.12 ± 0.29 | -0.14 ± 0.21 | … | … | N | … | … | … | … | … | n | G |
| 49014 | 18270646+0643051 | -30.2 ± 0.3 | 4552 ± 258 | 4.59 ± 0.22 | -0.17 ± 0.16 | … | … | Y | … | … | … | … | … | n | … |
| 49015 | 18270651+0633145 | 86.8 ± 0.2 | 4911 ± 143 | 2.72 ± 0.41 | -0.21 ± 0.18 | … | … | N | … | … | … | … | … | n | G |
| 49016 | 18270656+0645309 | -33.5 ± 0.3 | 4805 ± 199 | 4.27 ± 0.69 | -0.23 ± 0.14 | … | … | Y | … | … | … | … | … | n | … |
| 49017 | 18270660+0622592 | 25.8 ± 0.3 | 4678 ± 300 | 2.67 ± 0.44 | -0.04 ± 0.27 | … | … | N | … | … | … | … | … | n | G |
| 49018 | 18270671+0625275 | 11.3 ± 0.3 | 4735 ± 310 | 2.89 ± 0.54 | -0.34 ± 0.26 | … | … | N | … | … | … | … | … | n | G |
| 49019 | 18270672+0629492 | -42.6 ± 0.3 | 4554 ± 339 | 2.16 ± 0.44 | -0.37 ± 0.27 | … | … | Y | … | … | … | … | … | n | G |
| 49020 | 18270674+0639135 | 36.4 ± 0.4 | 5726 ± 78 | 4.38 ± 0.36 | -0.02 ± 0.13 | … | … | N | … | … | … | … | … | n | … |
| 49021 | 18270686+0631251 | -60.9 ± 0.3 | 4194 ± 319 | 4.44 ± 0.36 | -0.15 ± 0.22 | … | … | N | … | … | … | … | … | n | … |
| 49022 | 18270695+0620549 | -18.4 ± 0.2 | 5452 ± 188 | 4.32 ± 0.12 | 0.36 ± 0.2 | … | … | Y | … | … | … | … | … | n | … |
| 49023 | 18270709+0637573 | -24.9 ± 0.7 | 3928 ± 188 | 3.58 ± 0.21 | 0.21 ± 0.15 | … | … | Y | … | … | … | N | … | n | … |
| 49024 | 18270712+0630182 | 95.5 ± 0.2 | 5004 ± 50 | 2.71 ± 0.56 | -0.41 ± 0.18 | … | … | N | … | … | … | … | … | n | G |
| 49025 | 18270715+0619119 | 21.0 ± 0.2 | 5883 ± 15 | 4.22 ± 0.09 | 0.26 ± 0.17 | … | … | N | … | … | … | … | … | n | … |
| 49026 | 18270740+0638558 | -0.3 ± 0.3 | 4629 ± 56 | 4.72 ± 0.44 | 0.33 ± 0.3 | … | … | Y | … | … | … | N | … | n | … |
| 49028 | 18270742+0619185 | 134.7 ± 0.3 | 4988 ± 255 | 3.40 ± 0.59 | -0.45 ± 0.28 | … | … | N | … | … | … | … | … | n | G |
| 49029 | 18270743+0644234 | -46.8 ± 0.4 | 4901 ± 174 | 3.06 ± 0.48 | -0.05 ± 0.25 | … | … | Y | … | … | … | … | … | n | G |
| 49030 | 18270751+0626399 | 55.7 ± 0.3 | 4758 ± 297 | 2.98 ± 0.43 | -0.24 ± 0.23 | … | … | N | … | … | … | … | … | n | G |
| 49031 | 18270752+0622558 | 17.9 ± 0.3 | 4414 ± 194 | 4.74 ± 0.17 | -0.05 ± 0.15 | … | … | Y | … | … | … | … | … | n | … |
| 49032 | 18270766+0619461 | -10.3 ± 0.3 | 5115 ± 74 | 2.93 ± 0.73 | -0.43 ± 0.24 | … | … | Y | … | … | … | … | … | n | G |
| 49033 | 18270766+0637019 | 54.7 ± 0.2 | 5787 ± 149 | 4.24 ± 0.10 | 0.25 ± 0.13 | 62 ± 25 | 1 | N | .. | … | … | … | … | n | NG |
| 49034 | 18270776+0629098 | 15.5 ± 0.3 | 4961 ± 112 | 3.14 ± 0.38 | -0.29 ± 0.22 | … | … | N | … | … | … | … | … | n | G |
| 49035 | 18270779+0645568 | 61.2 ± 0.3 | 4925 ± 157 | 3.08 ± 0.47 | -0.34 ± 0.2 | … | … | N | … | … | … | … | … | n | G |
| 49036 | 18270780+0643335 | -47.9 ± 0.3 | 5528 ± 273 | 4.57 ± 0.39 | 0.14 ± 0.21 | … | … | N | … | … | … | … | … | n | … |
| 49037 | 18270784+0623588 | -68.1 ± 0.2 | 6039 ± 213 | 4.18 ± 0.26 | -0.15 ± 0.18 | 49 ± 22 | 1 | N | .. | … | … | … | … | n | NG |
| 49038 | 18270800+0627134 | 6.5 ± 0.2 | 4592 ± 146 | 2.93 ± 0.40 | 0.33 ± 0.26 | … | … | N | … | … | … | … | … | n | G |
| 49039 | 18270812+0627585 | -60.3 ± 0.7 | 5582 ± 867 | 4.28 ± 0.15 | 0.18 ± 0.32 | … | … | N | … | … | … | N | … | n | … |
| 49040 | 18270815+0643497 | -24.5 ± 0.2 | 6078 ± 11 | 4.18 ± 0.17 | -0.06 ± 0.16 | 54 ± 18 | … | Y | Y | Y | Y | N | … | Y?$^g$ | … |
| 49041 | 18270822+0641288 | -25.5 ± 0.2 | 4477 ± 167 | 2.55 ± 0.40 | 0.19 ± 0.26 | … | … | Y | … | … | … | … | … | n | G |
| 49042 | 18270825+0642005 | -17.1 ± 0.2 | 4634 ± 245 | 2.71 ± 0.46 | -0.15 ± 0.22 | … | … | Y | … | … | … | … | … | n | … |
| 49043 | 18270844+0631152 | -12.0 ± 0.3 | 5270 ± 139 | 4.42 ± 0.37 | 0.11 ± 0.15 | … | … | Y | … | … | … | … | … | n | … |
| 49044 | 18270855+0634025 | -39.5 ± 0.2 | 4730 ± 273 | 2.97 ± 0.44 | -0.26 ± 0.2 | … | … | Y | … | … | … | … | … | n | G |
| 49045 | 18270869+0635302 | -57.3 ± 0.3 | 4903 ± 63 | 4.50 ± 0.46 | 0.08 ± 0.14 | … | … | N | … | … | … | … | … | n | … |
| 49046 | 18270871+0628502 | 53.1 ± 0.2 | 4817 ± 51 | 2.94 ± 0.45 | 0.17 ± 0.18 | … | … | N | … | … | … | … | … | n | G |
| 49048 | 18270876+0637027 | -41.3 ± 0.4 | 4564 ± 82 | 4.84 ± 0.15 | 0.37 ± 0.16 | … | … | Y | … | … | … | N | … | n | … |
| 49049 | 18270889+0627497 | -5.5 ± 0.2 | 4910 ± 162 | 2.60 ± 0.48 | -0.34 ± 0.23 | … | … | Y | … | … | … | … | … | n | G |
| 49050 | 18270890+0619186 | -8.1 ± 0.3 | 3852 ± 304 | 4.67 ± 0.11 | -0.03 ± 0.14 | <53 | 3 | Y | N | N | Y | N | … | n | NG |
| 49051 | 18270891+0619028 | -13.1 ± 0.3 | 4517 ± 126 | 2.37 ± 0.34 | 0.01 ± 0.16 | … | … | Y | … | … | … | … | … | n | G |
| 49052 | 18270907+0629126 | 2.8 ± 0.3 | 4490 ± 183 | 4.43 ± 0.41 | 0.20 ± 0.21 | … | … | Y | … | … | … | N | … | n | … |
| 49053 | 18270921+0620077 | 61.2 ± 0.3 | 5892 ± 51 | 4.36 ± 0.30 | -0.19 ± 0.16 | … | … | N | … | … | … | … | … | n | … |

**Table C.11.** continued.

| ID | CNAME | RV (km s$^{-1}$) | $T_{\rm eff}$ (K) | logg (dex) | [Fe/H] (dex) | EW(Li)$^a$ (mÅ) | EW(Li) error flag$^b$ | Membership RV | Li | logg | [Fe/H] | Gaia studies Randich$^c$ | Cantat-Gaudin$^c$ | Final$^d$ | NMs with Li$^e$ |
|---|---|---|---|---|---|---|---|---|---|---|---|---|---|---|---|
| 49054 | 18270921+0629406 | 86.8 ± 0.2 | 4973 ± 109 | 3.00 ± 0.30 | -0.27 ± 0.21 | … | … | N | … | … | … | … | … | n | G |
| 49055 | 18270938+0629306 | -70.0 ± 0.3 | 3995 ± 357 | 4.34 ± 0.25 | 0.09 ± 0.38 | <102 | 3 | N | .. | … | … | N | … | n | NG |
| 49056 | 18270943+0641273 | -41.2 ± 0.3 | 4497 ± 278 | 4.55 ± 0.38 | -0.15 ± 0.12 | … | … | Y | … | … | … | … | … | n | … |
| 49057 | 18270947+0618343 | 8.6 ± 0.3 | 4909 ± 162 | 3.06 ± 0.28 | -0.19 ± 0.17 | … | … | N | … | … | … | … | … | n | G |
| 49058 | 18270953+0645236 | -65.3 ± 0.3 | 4555 ± 230 | 4.76 ± 0.30 | -0.42 ± 0.21 | … | … | N | … | … | … | … | … | n | … |
| 49059 | 18270956+0639451 | -25.3 ± 0.3 | 5084 ± 40 | 3.35 ± 0.41 | 0.28 ± 0.25 | … | … | Y | … | … | … | … | … | n | G |
| 49060 | 18270975+0623372 | -23.1 ± 0.3 | 4077 ± 238 | 4.64 ± 0.23 | -0.43 ± 0.31 | <40 | 3 | Y | N | Y | N | N | … | n | NG |
| 49061 | 18270975+0637070 | -1.3 ± 0.3 | 4128 ± 234 | 4.94 ± 0.35 | -0.25 ± 0.24 | <39 | 3 | Y | N | N | Y | N | … | n | NG |
| 49062 | 18270976+0619111 | -42.7 ± 0.2 | 4948 ± 115 | 3.42 ± 0.43 | 0.00 ± 0.14 | … | … | Y | … | … | … | … | … | n | G |
| 49063 | 18270976+0620125 | -81.8 ± 0.3 | 4214 ± 371 | 4.65 ± 0.29 | -0.04 ± 0.17 | <88 | 3 | N | .. | … | … | … | … | n | NG |
| 49064 | 18270980+0644083 | 15.0 ± 0.3 | 5859 ± 52 | 4.25 ± 0.11 | 0.04 ± 0.13 | … | … | N | … | … | … | … | … | n | … |
| 49065 | 18270994+0623338 | -3.6 ± 0.3 | 4549 ± 176 | 4.39 ± 0.77 | 0.33 ± 0.19 | … | … | Y | … | … | … | … | … | n | … |
| 49066 | 18271036+0623128 | -15.6 ± 0.3 | 6357 ± 150 | 4.19 ± 0.35 | -0.10 ± 0.12 | … | … | Y | … | … | … | … | … | n | … |
| 49067 | 18271048+0640322 | -39.0 ± 0.3 | 4927 ± 109 | 2.79 ± 0.34 | -0.05 ± 0.22 | … | … | Y | … | … | … | … | … | n | G |
| 49068 | 18271049+0629393 | -103.9 ± 0.2 | 4257 ± 371 | 4.63 ± 0.24 | -0.24 ± 0.18 | … | … | N | … | … | … | … | … | n | … |
| 49069 | 18271050+0623021 | -93.6 ± 54.7 | … | … | … | … | … | N | … | … | … | … | … | n | … |
| 49070 | 18271056+0621060 | 202.3 ± 96.7 | … | … | … | … | … | N | … | … | … | … | … | n | … |
| 49071 | 18271059+0620500 | -47.0 ± 0.3 | 4322 ± 269 | 4.82 ± 0.29 | -0.27 ± 0.16 | <26 | 3 | Y | N | Y | Y | N | … | n | NG |
| 49073 | 18271088+0619467 | 71.5 ± 0.2 | 4938 ± 173 | 2.88 ± 0.94 | -0.53 ± 0.22 | … | … | N | … | … | … | … | … | n | G |
| 49074 | 18271090+0640592 | -59.5 ± 0.2 | 4916 ± 94 | 4.32 ± 0.43 | 0.21 ± 0.21 | … | … | N | … | … | … | … | … | n | … |
| 49075 | 18271094+0626198 | -120.9 ± 0.3 | 4728 ± 209 | 3.30 ± 0.41 | -0.23 ± 0.13 | … | … | N | … | … | … | … | … | n | G |
| 49076 | 18271105+0643389 | -32.3 ± 0.3 | 5770 ± 59 | 4.28 ± 0.33 | -0.15 ± 0.13 | 40 ± 21 | 1 | Y | Y | Y | Y | … | … | Y | … |
| 49077 | 18271106+0622126 | -32.9 ± 0.3 | 4971 ± 92 | 4.18 ± 0.68 | 0.24 ± 0.2 | … | … | Y | … | … | … | … | … | n | … |
| 49078 | 18271109+0640125 | 77.1 ± 0.3 | 5372 ± 40 | 3.77 ± 0.49 | -0.24 ± 0.13 | … | … | N | … | … | … | … | … | n | … |
| 49079 | 18271112+0637427 | -28.9 ± 0.2 | 6288 ± 111 | 4.34 ± 0.33 | -0.13 ± 0.17 | 59 ± 9 | … | Y | Y | Y | Y | Y | Y | Y | … |
| 49080 | 18271124+0632204 | -9.1 ± 0.3 | 4776 ± 152 | 2.92 ± 0.37 | 0.01 ± 0.19 | … | … | Y | … | … | … | … | … | n | G |
| 49081 | 18271147+0630108 | -35.9 ± 0.3 | 4374 ± 187 | 4.69 ± 0.32 | 0.18 ± 0.24 | … | … | Y | … | … | … | N | … | n | … |
| 49082 | 18271157+0638520 | 113.8 ± 0.3 | 5194 ± 86 | 3.31 ± 0.68 | -0.40 ± 0.13 | … | … | N | … | … | … | … | … | n | Li-rich G |
| 49083 | 18271158+0620235 | -28.1 ± 0.3 | 5876 ± 197 | 4.56 ± 0.23 | 0.33 ± 0.16 | … | … | Y | … | … | … | … | … | n | … |
| 49084 | 18271164+0631060 | -29.0 ± 0.2 | 4516 ± 280 | 4.59 ± 0.31 | -0.13 ± 0.15 | <20 | 3 | Y | Y | Y | Y | Y | Y | Y | … |
| 49085 | 18271174+0627073 | 14.2 ± 0.3 | 3868 ± 230 | 4.59 ± 0.15 | -0.02 ± 0.2 | … | … | N | … | … | … | N | … | n | … |
| 49086 | 18271175+0628017 | -31.5 ± 0.2 | 4852 ± 186 | 3.09 ± 0.39 | -0.25 ± 0.19 | … | … | Y | … | … | … | … | … | n | G |
| 49087 | 18271179+0636578 | -108.5 ± 0.3 | 4846 ± 200 | 2.68 ± 0.48 | -0.13 ± 0.37 | … | … | N | … | … | … | … | … | n | G |
| 49088 | 18271182+0642579 | 23.4 ± 0.3 | 6170 ± 191 | 4.15 ± 0.10 | -0.03 ± 0.14 | 67 ± 35 | 1 | N | .. | … | … | … | … | n | NG |
| 49089 | 18271192+0632594 | -119.5 ± 0.2 | 4557 ± 261 | 2.38 ± 0.51 | -0.16 ± 0.22 | 72 ± 28 | 1 | N | .. | … | … | … | … | n | … |
| 49090 | 18271207+0618495 | -39.0 ± 0.3 | 5381 ± 179 | 4.52 ± 0.36 | 0.13 ± 0.15 | … | … | Y | … | … | … | … | … | n | … |
| 49091 | 18271215+0625123 | 32.6 ± 0.2 | 4583 ± 200 | 2.54 ± 0.31 | 0.02 ± 0.2 | … | … | N | … | … | … | … | … | n | … |
| 49092 | 18271224+0624044 | -10.6 ± 0.2 | 5752 ± 215 | 4.53 ± 0.27 | 0.34 ± 0.16 | … | … | Y | … | … | … | … | … | n | … |
| 49093 | 18271234+0636291 | 25.6 ± 0.3 | 5409 ± 78 | 4.28 ± 0.36 | -0.25 ± 0.18 | … | … | N | … | … | … | … | … | n | … |
| 49094 | 18271249+0645324 | -81.7 ± 0.4 | 4176 ± 418 | 4.78 ± 0.26 | -0.11 ± 0.28 | <118 | 3 | N | .. | … | … | … | … | n | NG |
| 49095 | 18271250+0618435 | -27.4 ± 0.3 | 5131 ± 72 | 4.44 ± 0.40 | -0.14 ± 0.14 | … | … | Y | … | … | … | … | … | n | … |
| 49096 | 18271277+0641147 | -67.7 ± 0.2 | 5261 ± 54 | 4.41 ± 0.40 | 0.15 ± 0.23 | … | … | N | … | … | … | … | … | n | … |
| 49097 | 18271279+0621396 | -8.0 ± 7.4 | … | … | … | … | … | Y | … | … | … | … | … | n | … |
| 49098 | 18271295+0627315 | 19.2 ± 0.3 | 5777 ± 234 | 4.42 ± 0.24 | 0.04 ± 0.16 | … | … | N | … | … | … | … | … | n | … |
| 49099 | 18271295+0643135 | -41.4 ± 0.2 | 4846 ± 122 | 3.24 ± 0.39 | 0.06 ± 0.14 | … | … | Y | … | … | … | … | … | n | G |
| 49100 | 18271302+0644127 | -31.4 ± 0.3 | 3699 ± 157 | 4.52 ± 0.21 | -0.03 ± 0.14 | <54 | 3 | Y | Y | Y | Y | Y | Y | Y | … |
| 49101 | 18271317+0632143 | -39.2 ± 0.4 | 4768 ± 143 | 4.69 ± 0.18 | -0.18 ± 0.21 | <72 | 3 | Y | N | Y | Y | … | … | n | NG |
| 49102 | 18271317+0642036 | -67.6 ± 0.4 | 4283 ± 308 | 4.55 ± 0.30 | -0.26 ± 0.21 | <103 | 3 | N | .. | … | … | … | … | n | NG |
| 49103 | 18271324+0622353 | 51.2 ± 0.2 | 4802 ± 141 | 3.10 ± 0.26 | 0.10 ± 0.18 | … | … | N | … | … | … | … | … | n | G |
| 49104 | 18271331+0621586 | 55.2 ± 0.3 | 5077 ± 93 | 3.80 ± 0.26 | 0.00 ± 0.13 | … | … | N | … | … | … | … | … | n | … |
| 49105 | 18271331+0625159 | -15.9 ± 0.2 | 6073 ± 64 | 3.71 ± 0.13 | -0.40 ± 0.16 | 64 ± 11 | … | Y | Y | Y | N | … | … | Y | … |
| 49106 | 18271338+0642357 | -42.5 ± 0.3 | 6317 ± 196 | 4.18 ± 0.29 | 0.11 ± 0.17 | … | … | Y | … | … | … | … | … | n | … |
| 49107 | 18271339+0625595 | 38.8 ± 0.2 | 4496 ± 143 | 2.36 ± 0.35 | -0.12 ± 0.2 | … | … | N | … | … | … | … | … | n | G |
| 49108 | 18271346+0622250 | 12.0 ± 0.3 | 6878 ± 554 | 4.47 ± 0.44 | -0.06 ± 0.28 | … | … | N | … | … | … | … | … | n | … |
| 49109 | 18271351+0643174 | -27.4 ± 0.3 | 5998 ± 150 | 4.17 ± 0.12 | -0.12 ± 0.12 | 52 ± 26 | 1 | Y | Y | Y | Y | … | … | Y | … |
| 49110 | 18271372+0632336 | -59.4 ± 0.3 | 4686 ± 199 | 4.21 ± 0.63 | 0.36 ± 0.25 | … | … | N | … | … | … | … | … | n | … |
| 49111 | 18271379+0618071 | -38.7 ± 0.3 | 4688 ± 272 | 2.66 ± 0.45 | -0.20 ± 0.2 | … | … | Y | … | … | … | … | … | n | G |
| 49112 | 18271388+0638402 | -197.9 ± 0.3 | 4908 ± 179 | 2.35 ± 0.47 | -0.43 ± 0.38 | … | … | N | … | … | … | … | … | n | G |
| 49113 | 18271398+0639317 | -288.6 ± 0.3 | 4238 ± 228 | 1.64 ± 1.28 | -0.82 ± 0.45 | … | … | N | … | … | … | … | … | n | G |

| ID | CNAME | RV (km s$^{-1}$) | $T_{\text{eff}}$ (K) | logg (dex) | [Fe/H] (dex) | EW(Li)$^a$ (mÅ) | EW(Li) error flag$^b$ | Membership RV | Li | logg | [Fe/H] | Gaia studies Randich$^c$ | Cantat-Gaudin$^c$ | Final$^d$ | NMs with Li$^e$ |
|---|---|---|---|---|---|---|---|---|---|---|---|---|---|---|---|
| 49114 | 18271425+0619006 | -44.1 ± 0.3 | 4690 ± 317 | 2.78 ± 0.70 | -0.46 ± 0.3 | … | … | Y | … | … | … | … | … | n | G |
| 49115 | 18271426+0637384 | -25.3 ± 0.3 | 4827 ± 214 | 4.82 ± 0.47 | 0.01 ± 0.13 | … | … | Y | … | … | … | … | … | n | … |
| 49116 | 18271449+0622436 | 72.8 ± 0.2 | 4868 ± 108 | 2.98 ± 0.23 | -0.03 ± 0.16 | … | … | N | … | … | … | … | … | n | G |
| 49117 | 18271468+0640311 | 35.9 ± 0.2 | 4523 ± 115 | 2.61 ± 0.45 | 0.14 ± 0.2 | <27 | 3 | N | .. | … | … | … | … | n | G |
| 49118 | 18271472+0623407 | -103.6 ± 0.3 | 4506 ± 251 | 2.56 ± 0.43 | -0.11 ± 0.16 | … | … | N | … | … | … | … | … | n | G |
| 49119 | 18271477+0619295 | 2.9 ± 0.2 | 5672 ± 193 | 4.04 ± 0.16 | 0.34 ± 0.16 | … | … | Y | … | … | … | … | … | n | … |
| 49120 | 18271483+0625075 | -1.0 ± 0.2 | 5281 ± 105 | 4.29 ± 0.43 | -0.01 ± 0.12 | … | … | Y | … | … | … | … | … | n | … |
| 49121 | 18271484+0634311 | -29.7 ± 0.3 | 5958 ± 130 | 4.19 ± 0.38 | -0.47 ± 0.18 | … | … | Y | … | … | … | … | … | n | … |
| 49122 | 18271488+0643454 | 30.1 ± 0.2 | 4919 ± 118 | 2.75 ± 0.28 | 0.04 ± 0.16 | … | … | N | … | … | … | … | … | n | G |
| 49123 | 18271492+0625404 | 49.3 ± 0.3 | 4792 ± 302 | 2.74 ± 1.02 | -0.65 ± 0.33 | … | … | N | … | … | … | … | … | n | G |
| 49124 | 18271505+0631462 | -2.8 ± 0.3 | 5101 ± 68 | 3.83 ± 0.43 | 0.11 ± 0.14 | … | … | Y | … | … | … | … | … | n | … |
| 49125 | 18271514+0645404 | -19.2 ± 0.3 | 5743 ± 98 | 4.06 ± 0.23 | 0.26 ± 0.13 | … | … | Y | … | … | … | … | … | n | … |
| 49126 | 18271521+0627061 | 3.8 ± 0.2 | 4826 ± 178 | 4.51 ± 0.30 | -0.06 ± 0.13 | <21 | 3 | N | .. | … | … | … | … | n | NG |
| 49127 | 18271523+0638285 | 2.0 ± 0.3 | 4966 ± 122 | 3.48 ± 0.11 | -0.25 ± 0.21 | … | … | Y | … | … | … | … | … | n | G |
| 49128 | 18271526+0625564 | 0.8 ± 0.2 | 4923 ± 132 | 3.31 ± 0.38 | 0.16 ± 0.19 | … | … | Y | … | … | … | … | … | n | G |
| 49129 | 18271528+0622578 | -4.7 ± 0.3 | 4360 ± 172 | 4.71 ± 0.26 | 0.18 ± 0.22 | … | … | Y | … | … | … | N | … | n | … |
| 49130 | 18271529+0635183 | 77.3 ± 0.2 | 4827 ± 176 | 2.94 ± 0.43 | -0.15 ± 0.2 | … | … | N | … | … | … | … | … | n | G |
| 49131 | 18271534+0645247 | 35.9 ± 0.3 | 5336 ± 204 | 4.66 ± 0.21 | -0.08 ± 0.14 | … | … | N | … | … | … | … | … | n | … |
| 49132 | 18271537+0642458 | 31.4 ± 0.2 | 5061 ± 144 | 2.95 ± 0.34 | -0.12 ± 0.19 | 48 ± 21 | 1 | N | .. | … | … | … | … | n | Li-rich G |
| 49133 | 18271538+0637189 | -1.5 ± 0.2 | 4538 ± 117 | 2.57 ± 0.35 | 0.19 ± 0.22 | … | … | Y | … | … | … | … | … | n | … |
| 49134 | 18271539+0623484 | 49.7 ± 0.3 | 4745 ± 57 | 2.85 ± 0.27 | 0.08 ± 0.21 | … | … | N | … | … | … | … | … | n | G |
| 49135 | 18271542+0638234 | 65.6 ± 0.2 | 4743 ± 137 | 2.62 ± 0.29 | 0.02 ± 0.22 | … | … | N | … | … | … | … | … | n | G |
| 49136 | 18271548+0640513 | -28.1 ± 0.2 | 4380 ± 198 | 4.72 ± 0.31 | -0.07 ± 0.16 | <17 | 3 | Y | Y | Y | Y | Y | Y | Y | … |
| 49137 | 18271591+0623347 | -25.8 ± 0.2 | 5793 ± 91 | 4.25 ± 0.13 | 0.26 ± 0.13 | … | … | Y | … | … | … | … | … | n | … |
| 49138 | 18271599+0636150 | 14.9 ± 0.3 | 4866 ± 211 | 2.65 ± 0.57 | -0.32 ± 0.21 | … | … | N | … | … | … | … | … | n | G |
| 49139 | 18271608+0643108 | -21.3 ± 0.3 | 4955 ± 53 | 4.40 ± 0.70 | 0.25 ± 0.2 | … | … | Y | … | … | … | … | … | n | … |
| 49140 | 18271618+0639086 | 28.1 ± 0.3 | 4453 ± 96 | 4.74 ± 0.48 | 0.34 ± 0.4 | <97 | 3 | N | .. | … | … | … | … | n | NG |
| 49141 | 18271637+0633171 | -9.8 ± 0.2 | 4411 ± 174 | 2.45 ± 0.43 | 0.10 ± 0.21 | <41 | 3 | Y | N | N | Y | … | … | n | G |
| 49142 | 18271644+0642266 | -87.9 ± 0.2 | 4310 ± 332 | 4.52 ± 0.29 | -0.01 ± 0.2 | … | … | N | … | … | … | N | … | n | … |
| 49143 | 18271648+0633539 | -1.6 ± 0.3 | 6163 ± 106 | 4.30 ± 0.31 | -0.20 ± 0.12 | … | … | Y | … | … | … | … | … | n | … |
| 49144 | 18271696+0637434 | -26.2 ± 0.7 | 4860 ± 62 | 3.12 ± 0.89 | -0.25 ± 0.17 | … | … | Y | … | … | … | … | … | n | G |
| 49145 | 18271705+0632208 | -45.6 ± 0.3 | 5272 ± 129 | 4.51 ± 0.42 | 0.05 ± 0.13 | … | … | Y | … | … | … | … | … | n | … |
| 49146 | 18271717+0636178 | -19.2 ± 0.3 | 5109 ± 76 | 4.35 ± 0.67 | 0.14 ± 0.23 | <59 | 3 | Y | Y | Y | Y | … | … | Y | … |
| 49147 | 18271722+0637235 | -169.1 ± 0.2 | 4971 ± 98 | 2.23 ± 0.43 | -0.55 ± 0.14 | … | … | N | … | … | … | … | … | n | G |
| 49148 | 18271731+0634545 | -47.9 ± 0.2 | 5741 ± 112 | 4.28 ± 0.11 | 0.21 ± 0.14 | … | … | N | … | … | … | … | … | n | … |
| 49149 | 18271732+0621039 | -86.1 ± 0.3 | 4903 ± 145 | 2.59 ± 0.34 | -0.30 ± 0.17 | … | … | N | … | … | … | … | … | n | G |
| 49150 | 18271733+0619334 | -1.4 ± 0.3 | 5120 ± 89 | 3.72 ± 0.14 | -0.23 ± 0.15 | … | … | Y | … | … | … | N | … | n | … |
| 49151 | 18271738+0636202 | -19.2 ± 0.3 | 4164 ± 171 | 4.67 ± 0.12 | -0.01 ± 0.19 | <66 | 3 | Y | N | Y | Y | N | … | n | NG |
| 49152 | 18271757+0621557 | -24.8 ± 0.3 | 5575 ± 144 | 4.17 ± 0.35 | -0.20 ± 0.13 | … | … | Y | … | … | … | … | … | n | … |
| 49153 | 18271760+0624323 | -59.0 ± 0.3 | 5088 ± 37 | 4.41 ± 0.60 | 0.19 ± 0.21 | … | … | N | … | … | … | … | … | n | … |
| 49154 | 18271764+0644071 | 4.5 ± 0.2 | 4646 ± 110 | 2.80 ± 0.45 | 0.04 ± 0.19 | 32 ± 25 | 1 | N | .. | … | … | … | … | n | G |
| 49155 | 18271764+0630367 | -2.1 ± 0.2 | 5892 ± 75 | 4.11 ± 0.09 | 0.13 ± 0.14 | 74 ± 22 | 1 | Y | Y | Y | Y | … | … | Y | … |
| 49156 | 18271780+0644416 | -60.5 ± 0.2 | 4887 ± 126 | 4.43 ± 0.36 | -0.12 ± 0.13 | … | … | N | … | … | … | N | … | n | … |
| 49157 | 18271785+0628083 | 58.0 ± 0.3 | 5102 ± 197 | 3.29 ± 0.95 | -0.48 ± 0.31 | … | … | N | … | … | … | … | … | n | G |
| 49158 | 18271788+0623591 | -40.7 ± 0.3 | 4913 ± 168 | 2.69 ± 0.57 | -0.14 ± 0.19 | … | … | Y | … | … | … | … | … | n | G |
| 49159 | 18271789+0639010 | -602.2 ± 48.4 | … | … | … | <127 | 3 | N | .. | … | … | … | … | n | … |
| 49160 | 18271792+0637147 | 60.2 ± 0.3 | 4955 ± 141 | 3.56 ± 0.41 | 0.15 ± 0.26 | … | … | N | … | … | … | … | … | n | … |
| 49161 | 18271805+0621453 | -58.8 ± 0.3 | 5786 ± 152 | 4.44 ± 0.27 | 0.23 ± 0.14 | … | … | N | … | … | … | … | … | n | … |
| 49162 | 18271816+0632543 | 9.1 ± 0.2 | 4678 ± 113 | 2.89 ± 0.39 | 0.27 ± 0.22 | … | … | N | … | … | … | … | … | n | G |
| 49163 | 18271828+0642517 | -11.4 ± 0.5 | 4323 ± 315 | 4.90 ± 0.37 | -0.58 ± 0.44 | <91 | 3 | Y | N | N | N | N | … | n | NG |
| 49164 | 18271830+0621098 | -41.1 ± 0.3 | 4280 ± 327 | 4.85 ± 0.25 | 0.13 ± 0.17 | <92 | 3 | Y | N | Y | Y | … | … | n | NG |
| 49165 | 18271845+0634354 | 43.4 ± 0.3 | 5074 ± 255 | 3.64 ± 0.65 | -0.56 ± 0.26 | … | … | N | … | … | … | … | … | n | … |
| 49166 | 18271850+0640219 | 8.2 ± 0.3 | 5813 ± 68 | 4.18 ± 0.04 | -0.17 ± 0.14 | 52 ± 32 | 1 | N | .. | … | … | … | … | n | NG |
| 49167 | 18271856+0624156 | -6.0 ± 0.2 | 4792 ± 211 | 4.43 ± 0.28 | -0.35 ± 0.2 | <22 | 3 | Y | Y | Y | Y | … | … | Y | … |
| 49168 | 18271871+0630074 | -22.7 ± 0.2 | 4746 ± 70 | 2.54 ± 0.24 | 0.11 ± 0.22 | … | … | Y | … | … | … | … | … | n | G |
| 49169 | 18271873+0639393 | 54.2 ± 0.2 | 4602 ± 231 | 2.68 ± 0.50 | 0.19 ± 0.24 | … | … | N | … | … | … | … | … | n | … |
| 49170 | 18271891+0644334 | -3.3 ± 0.2 | 4682 ± 273 | 2.81 ± 0.42 | -0.10 ± 0.24 | … | … | Y | … | … | … | … | … | n | G |
| 49171 | 18271892+0622430 | -50.0 ± 0.3 | 4615 ± 84 | 2.38 ± 0.28 | 0.07 ± 0.19 | … | … | N | … | … | … | … | … | n | G |
| 49172 | 18271897+0642447 | 23.5 ± 0.3 | 5035 ± 100 | 3.42 ± 0.13 | 0.05 ± 0.14 | … | … | N | … | … | … | … | … | n | G |

**Table C.11.** continued.

| ID | CNAME | RV (km s$^{-1}$) | $T_{\rm eff}$ (K) | logg (dex) | [Fe/H] (dex) | EW(Li)$^a$ (mÅ) | EW(Li) error flag$^b$ | Membership RV | Li | logg | [Fe/H] | Gaia studies Randich$^c$ | Cantat-Gaudin$^c$ | Final$^d$ | NMs with Li$^e$ |
|---|---|---|---|---|---|---|---|---|---|---|---|---|---|---|---|
| 49173 | 18271914+0625093 | 21.8 ± 0.2 | 4974 ± 131 | 3.43 ± 0.13 | 0.01 ± 0.16 | … | … | N | … | … | … | … | … | n | G |
| 49174 | 18271923+0645053 | 27.1 ± 0.3 | 4900 ± 130 | 3.25 ± 0.21 | 0.12 ± 0.2 | … | … | N | … | … | … | … | … | n | G |
| 49175 | 18271943+0631227 | -4.3 ± 0.3 | 6088 ± 148 | 4.30 ± 0.24 | 0.13 ± 0.13 | 91 ± 28 | 1 | Y | Y | Y | Y | … | … | Y | … |
| 49176 | 18271944+0624496 | 30.3 ± 0.2 | 4504 ± 189 | 2.65 ± 0.41 | 0.22 ± 0.2 | … | … | N | … | … | … | … | … | n | G |
| 49177 | 18271957+0631358 | -4.2 ± 2.2 | … | … | … | … | … | Y | … | … | … | … | … | n | … |
| 49178 | 18271962+0620322 | 42.2 ± 0.3 | 5014 ± 277 | 3.57 ± 0.64 | -0.51 ± 0.22 | … | … | N | … | … | … | … | … | n | … |
| 49179 | 18271966+0637366 | 64.3 ± 0.2 | 4600 ± 93 | 2.48 ± 0.33 | 0.00 ± 0.17 | <37 | 3 | N | .. | … | … | … | … | n | G |
| 49180 | 18271967+0636082 | -65.3 ± 0.3 | 5641 ± 175 | 4.59 ± 0.28 | -0.17 ± 0.22 | … | … | N | … | … | … | … | … | n | … |
| 49181 | 18271972+0621069 | 71.6 ± 0.3 | 4842 ± 150 | 4.69 ± 0.25 | 0.08 ± 0.19 | <79 | 3 | N | .. | … | … | … | … | n | NG |
| 49182 | 18271979+0638188 | -5.6 ± 0.3 | 5732 ± 127 | 4.29 ± 0.48 | -0.03 ± 0.16 | … | … | Y | … | … | … | … | … | n | … |
| 49183 | 18271979+0645152 | 11.1 ± 0.3 | 4723 ± 286 | 3.21 ± 0.29 | 0.02 ± 0.2 | … | … | N | … | … | … | … | … | n | G |
| 49184 | 18271986+0645423 | 51.1 ± 0.3 | 4258 ± 327 | 2.43 ± 0.43 | 0.05 ± 0.22 | … | … | N | … | … | … | … | … | n | G |
| 49185 | 18271996+0637206 | -67.7 ± 0.4 | 5180 ± 286 | 4.16 ± 0.64 | 0.19 ± 0.18 | … | … | N | … | … | … | … | … | n | … |
| 49186 | 18271999+0620036 | -61.2 ± 0.3 | 5354 ± 137 | 4.52 ± 0.32 | 0.21 ± 0.18 | … | … | N | … | … | … | … | … | n | … |
| 49187 | 18272006+0631333 | 40.9 ± 0.3 | 4829 ± 247 | 4.46 ± 0.31 | -0.40 ± 0.19 | … | … | N | … | … | … | … | … | n | … |
| 49188 | 18272013+0625216 | -24.0 ± 0.2 | 4851 ± 165 | 3.22 ± 0.32 | -0.03 ± 0.15 | … | … | Y | … | … | … | … | … | n | G |
| 49189 | 18272017+0639352 | 5.4 ± 0.2 | 4691 ± 198 | 2.94 ± 0.39 | 0.11 ± 0.18 | … | … | N | … | … | … | … | … | n | G |
| 49190 | 18272021+0634093 | 20.9 ± 0.2 | 4951 ± 134 | 2.81 ± 0.47 | -0.21 ± 0.2 | … | … | N | … | … | … | … | … | n | G |
| 49191 | 18272030+0642518 | 59.8 ± 0.3 | 4647 ± 105 | 2.70 ± 0.31 | 0.14 ± 0.22 | … | … | N | … | … | … | … | … | n | G |
| 49192 | 18272035+0634487 | 13.3 ± 0.8 | 4366 ± 200 | 4.64 ± 0.13 | 0.05 ± 0.3 | … | … | N | … | … | … | … | … | n | … |
| 49193 | 18272066+0625054 | -17.9 ± 0.2 | 4480 ± 125 | 4.70 ± 0.35 | -0.11 ± 0.15 | <25 | 3 | Y | Y | Y | Y | N | … | Y?$^g$ | … |
| 49194 | 18272084+0635223 | -59.3 ± 0.3 | 5770 ± 153 | 4.15 ± 0.06 | 0.15 ± 0.15 | 99 ± 33 | 1 | N | .. | … | … | … | … | n | NG |
| 49195 | 18272088+0630510 | 10.5 ± 0.2 | 4350 ± 236 | 2.42 ± 0.46 | 0.42 ± 0.34 | … | … | N | … | … | … | … | … | n | G |
| 49196 | 18272092+0624510 | -27.7 ± 0.3 | 6097 ± 207 | 4.25 ± 0.13 | 0.11 ± 0.15 | 44 ± 27 | 1 | Y | Y | Y | Y | … | … | Y | … |
| 49197 | 18272097+0634160 | 9.3 ± 0.3 | 5012 ± 176 | 3.44 ± 0.53 | -0.49 ± 0.25 | … | … | N | … | … | … | … | … | n | G |
| 49199 | 18272123+0643560 | 20.8 ± 0.2 | 4772 ± 226 | 2.90 ± 0.44 | -0.19 ± 0.23 | … | … | N | … | … | … | … | … | n | … |
| 49200 | 18272129+0630192 | -23.0 ± 0.3 | 4973 ± 102 | 4.37 ± 0.77 | 0.31 ± 0.3 | … | … | Y | … | … | … | … | … | n | … |
| 49201 | 18272133+0620123 | -67.5 ± 0.2 | 5092 ± 65 | 3.69 ± 0.31 | 0.00 ± 0.2 | <29 | 3 | N | .. | … | … | … | … | n | NG |
| 49202 | 18272133+0621459 | 77.0 ± 0.2 | 4923 ± 107 | 2.59 ± 0.63 | -0.44 ± 0.19 | … | … | N | … | … | … | … | … | n | G |
| 49203 | 18272142+0633236 | 20.8 ± 0.4 | 4484 ± 487 | 2.58 ± 0.62 | -0.55 ± 0.48 | <65 | 3 | N | .. | … | … | … | … | n | G |
| 49204 | 18272151+0628012 | -96.7 ± 0.2 | 4379 ± 285 | 4.53 ± 0.66 | 0.19 ± 0.2 | … | … | N | … | … | … | … | … | n | … |
| 49205 | 18272165+0643466 | -2.6 ± 1.4 | … | … | … | … | … | Y | … | … | … | … | … | n | … |
| 49206 | 18272179+0624258 | -26.8 ± 0.3 | 5593 ± 159 | 4.25 ± 0.29 | 0.25 ± 0.16 | … | … | Y | … | … | … | … | … | n | … |
| 49207 | 18272194+0619021 | -27.0 ± 0.2 | 5725 ± 110 | 4.16 ± 0.07 | 0.01 ± 0.13 | … | … | Y | … | … | … | … | … | n | … |
| 49208 | 18272203+0644052 | -5.0 ± 0.3 | 5149 ± 33 | 4.15 ± 0.64 | 0.43 ± 0.34 | … | … | Y | … | … | … | … | … | n | … |
| 49209 | 18272213+0625432 | 33.5 ± 0.2 | 4927 ± 76 | 3.35 ± 0.39 | 0.19 ± 0.21 | … | … | N | … | … | … | N | … | n | G |
| 49210 | 18272216+0640205 | 46.8 ± 0.3 | 5739 ± 446 | 4.58 ± 0.35 | 0.04 ± 0.2 | … | … | N | … | … | … | … | … | n | … |
| 49211 | 18272234+0641320 | -57.4 ± 0.5 | 3750 ± 487 | 4.29 ± 0.21 | … | … | … | N | … | … | … | … | … | n | … |
| 49212 | 18272238+0622108 | -14.9 ± 0.2 | 5807 ± 110 | 4.23 ± 0.17 | 0.02 ± 0.13 | … | … | Y | … | … | … | … | … | n | … |
| 49213 | 18272241+0621126 | -18.1 ± 0.3 | 4013 ± 258 | 4.54 ± 0.10 | 0.04 ± 0.2 | <54 | 3 | Y | N | Y | Y | … | … | n | NG |
| 49214 | 18272243+0633308 | -179.2 ± 0.4 | 4405 ± 339 | 1.81 ± 0.35 | -0.59 ± 0.27 | … | … | N | … | … | … | … | … | n | G |
| 49215 | 18272259+0638498 | 20.2 ± 0.3 | 5556 ± 209 | 4.37 ± 0.32 | -0.07 ± 0.13 | … | … | N | … | … | … | … | … | n | … |
| 49216 | 18272260+0627584 | -8.9 ± 0.2 | 4629 ± 167 | 2.40 ± 0.32 | -0.19 ± 0.16 | … | … | Y | … | … | … | … | … | n | G |
| 49217 | 18272266+0644107 | 45.9 ± 0.2 | 4884 ± 273 | 2.85 ± 0.35 | -0.13 ± 0.27 | … | … | N | … | … | … | … | … | n | G |
| 49218 | 18272267+0640053 | -17.1 ± 0.3 | 4855 ± 262 | 4.50 ± 0.25 | 0.37 ± 0.31 | <93 | 3 | Y | N | Y | N | … | … | n | NG |
| 49219 | 18272278+0635522 | 22.6 ± 0.2 | 5570 ± 150 | 4.31 ± 0.17 | 0.38 ± 0.18 | … | … | N | … | … | … | … | … | n | … |
| 49220 | 18272280+0632384 | 57.3 ± 0.2 | 4532 ± 174 | 2.52 ± 0.35 | 0.03 ± 0.19 | <13 | 3 | N | .. | … | … | … | … | n | G |
| 49221 | 18272292+0630384 | -37.1 ± 0.2 | 5126 ± 35 | 3.71 ± 0.35 | -0.15 ± 0.14 | … | … | Y | … | … | … | … | … | n | … |
| 49222 | 18272302+0622590 | 110.6 ± 0.2 | 4821 ± 151 | 2.98 ± 0.27 | 0.03 ± 0.22 | … | … | N | … | … | … | … | … | n | G |
| 49223 | 18272303+0619289 | 70.0 ± 0.7 | 4568 ± 211 | 4.35 ± 0.13 | 0.01 ± 0.15 | … | … | N | … | … | … | … | … | n | … |
| 49224 | 18272331+0629335 | 90.5 ± 0.2 | 4740 ± 86 | 2.79 ± 0.44 | 0.11 ± 0.24 | … | … | N | … | … | … | … | … | n | G |
| 49225 | 18272333+0625398 | 133.8 ± 0.2 | 4848 ± 100 | 2.80 ± 0.34 | 0.01 ± 0.27 | … | … | N | … | … | … | … | … | n | G |
| 49226 | 18272336+0621483 | 12.0 ± 0.3 | 4976 ± 141 | 3.49 ± 0.40 | -0.32 ± 0.23 | 57 ± 34 | 1 | N | .. | … | … | … | … | n | G |
| 49227 | 18272349+0620188 | 79.0 ± 0.3 | 4979 ± 175 | 3.48 ± 0.68 | -0.47 ± 0.2 | … | … | N | … | … | … | … | … | n | G |
| 49228 | 18272350+0625066 | -1.6 ± 0.2 | 5192 ± 65 | 3.61 ± 0.34 | 0.23 ± 0.21 | … | … | Y | … | … | … | … | … | n | … |
| 49229 | 18272354+0621239 | -45.2 ± 0.2 | 4966 ± 115 | 3.39 ± 0.42 | 0.07 ± 0.17 | … | … | Y | … | … | … | N | … | n | G |
| 49230 | 18272359+0643252 | -110.4 ± 0.3 | 4588 ± 236 | 2.67 ± 0.24 | -0.21 ± 0.23 | … | … | N | … | … | … | … | … | n | G |
| 49231 | 18272382+0629015 | -30.1 ± 0.2 | 5012 ± 97 | 4.52 ± 0.34 | -0.04 ± 0.13 | … | … | Y | … | … | … | … | … | n | … |
| 49232 | 18272383+0628009 | 31.3 ± 16.0 | … | … | … | … | … | N | … | … | … | … | … | n | … |

**Table C.11.** continued.

| ID | CNAME | RV (km s$^{-1}$) | $T_{\rm eff}$ (K) | logg (dex) | [Fe/H] (dex) | EW(Li)$^a$ (mÅ) | EW(Li) error flag$^b$ | Membership RV | Li | logg | [Fe/H] | Gaia studies Randich$^c$ | Cantat-Gaudin$^c$ | Final$^d$ | NMs with Li$^e$ |
|---|---|---|---|---|---|---|---|---|---|---|---|---|---|---|---|
| 49233 | 18272390+0623100 | 20.9 ± 0.2 | 5660 ± 58 | 3.90 ± 0.25 | -0.34 ± 0.16 | 47 ± 9 | 1 | N | .. | … | … | … | … | n | NG |
| 49234 | 18272399+0642154 | 2.4 ± 0.2 | 6236 ± 113 | 3.99 ± 0.06 | -0.11 ± 0.18 | … | … | Y | … | … | … | … | … | n | … |
| 49235 | 18272413+0636162 | 62.5 ± 0.2 | 4969 ± 65 | 3.30 ± 0.37 | 0.21 ± 0.21 | … | … | N | … | … | … | … | … | n | G |
| 49236 | 18272424+0619596 | -11.6 ± 0.3 | 4985 ± 259 | 2.99 ± 0.59 | -0.66 ± 0.29 | … | … | Y | … | … | … | … | … | n | G |
| 49237 | 18272424+0626476 | 22.2 ± 0.2 | 4624 ± 120 | 2.70 ± 0.46 | 0.14 ± 0.23 | … | … | N | … | … | … | … | … | n | … |
| 49238 | 18272426+0637582 | -60.5 ± 0.3 | 5198 ± 129 | 4.42 ± 0.38 | -0.50 ± 0.16 | … | … | N | … | … | … | … | … | n | … |
| 49239 | 18272429+0632029 | 55.0 ± 0.3 | 5077 ± 100 | 3.25 ± 0.14 | -0.09 ± 0.15 | … | … | N | … | … | … | … | … | n | G |
| 49240 | 18272437+0620179 | -201.4 ± 0.3 | 4917 ± 176 | 2.72 ± 0.72 | -0.55 ± 0.3 | … | … | N | … | … | … | … | … | n | G |
| 49241 | 18272439+0621174 | -50.4 ± 0.2 | 5666 ± 76 | 4.05 ± 0.29 | -0.06 ± 0.19 | … | … | N | … | … | … | … | … | n | … |
| 49242 | 18272440+0634348 | -2.0 ± 0.3 | 4835 ± 65 | 3.21 ± 0.50 | 0.28 ± 0.22 | … | … | Y | … | … | … | … | … | n | G |
| 49243 | 18272444+0639287 | -43.4 ± 0.4 | 4998 ± 210 | 3.77 ± 0.50 | -0.24 ± 0.22 | <146 | 3 | Y | N | Y | Y | … | … | n | NG |
| 49244 | 18272450+0626239 | 10.2 ± 0.3 | 4292 ± 30 | 4.86 ± 0.30 | -0.11 ± 0.14 | <48 | 3 | N | .. | … | … | … | … | n | NG |
| 49245 | 18272458+0620136 | -45.2 ± 0.3 | 4395 ± 361 | 4.54 ± 0.49 | -0.08 ± 0.18 | … | … | Y | … | … | … | … | … | n | … |
| 49246 | 18272467+0640164 | -2.4 ± 0.5 | 5155 ± 96 | 4.42 ± 0.47 | 0.23 ± 0.14 | <98 | 3 | Y | N | Y | Y | … | … | n | NG |
| 49247 | 18272478+0643410 | -17.8 ± 0.2 | 6342 ± 171 | 4.30 ± 0.25 | -0.02 ± 0.15 | <71 | 3 | Y | N | N | Y | N | … | n | NG |
| 49248 | 18272486+0635410 | 6.8 ± 0.5 | 3751 ± 310 | 4.44 ± 0.12 | -0.06 ± 0.12 | … | 3 | N | … | … | … | … | … | n | … |
| 49249 | 18272526+0621599 | -76.0 ± 0.3 | 4873 ± 157 | 3.70 ± 0.39 | -0.02 ± 0.14 | <27 | 3 | N | .. | … | … | … | … | n | NG |
| 49250 | 18272531+0621062 | 2.7 ± 0.4 | 4404 ± 359 | 4.32 ± 0.34 | -0.24 ± 0.26 | <72 | 3 | Y?$^f$ | N | Y | Y | … | … | n | NG |
| 49251 | 18272533+0619042 | -55.2 ± 0.2 | 5711 ± 88 | 4.09 ± 0.11 | 0.20 ± 0.14 | … | … | N | … | … | … | … | … | n | … |
| 49252 | 18272549+0620192 | 25.5 ± 0.3 | 5595 ± 31 | 4.01 ± 0.14 | 0.27 ± 0.15 | … | … | N | … | … | … | … | … | n | … |
| 49253 | 18272551+0644399 | -34.2 ± 0.2 | 5994 ± 146 | 4.13 ± 0.13 | 0.10 ± 0.13 | 51 ± 16 | 1 | Y | Y | Y | Y | N | … | Y?$^g$ | … |
| 49254 | 18272553+0641584 | -13.6 ± 0.3 | 4324 ± 204 | 4.60 ± 0.31 | 0.21 ± 0.23 | <40 | 3 | Y | N | Y | Y | N | … | n | NG |
| 49255 | 18272556+0642289 | 65.7 ± 0.3 | 4836 ± 185 | 2.92 ± 0.52 | -0.40 ± 0.21 | … | … | N | … | … | … | … | … | n | G |
| 49256 | 18272557+0619222 | 76.9 ± 0.3 | 5573 ± 46 | 4.18 ± 0.19 | -0.52 ± 0.18 | … | … | N | … | … | … | … | … | n | … |
| 49257 | 18272557+0641075 | 0.8 ± 0.3 | 5346 ± 175 | 4.59 ± 0.37 | 0.13 ± 0.15 | … | … | Y | … | … | … | … | … | n | … |
| 49258 | 18272570+0626491 | 48.4 ± 0.2 | 4582 ± 115 | 2.38 ± 0.25 | 0.16 ± 0.22 | … | … | N | … | … | … | … | … | n | G |
| 49259 | 18272574+0634554 | 60.3 ± 0.3 | 6506 ± 212 | 3.67 ± 0.33 | -0.07 ± 0.38 | … | … | N | … | … | … | … | … | n | … |
| 49260 | 18272590+0637595 | 21.8 ± 0.3 | 4512 ± 267 | 4.79 ± 0.28 | -0.05 ± 0.13 | … | … | N | … | … | … | N | … | n | … |
| 49261 | 18272590+0642497 | 86.3 ± 0.3 | 5022 ± 251 | 3.67 ± 0.51 | -0.46 ± 0.26 | … | … | N | … | … | … | … | … | n | … |
| 49262 | 18272599+0639332 | -74.1 ± 0.2 | 4722 ± 235 | 2.94 ± 0.46 | -0.21 ± 0.13 | … | … | N | … | … | … | … | … | n | G |
| 49263 | 18272604+0620008 | -30.6 ± 0.2 | 4361 ± 231 | 4.64 ± 0.27 | 0.24 ± 0.23 | … | … | Y | … | … | … | Y | … | n | … |
| 49264 | 18272608+0634104 | 72.2 ± 0.5 | 5046 ± 72 | 3.61 ± 0.44 | -0.28 ± 0.14 | … | … | N | … | … | … | … | … | n | … |
| 49265 | 18272618+0621278 | -45.1 ± 0.2 | 4796 ± 121 | 3.29 ± 0.43 | 0.10 ± 0.15 | … | … | Y | … | … | … | … | … | n | G |
| 49266 | 18272619+0639031 | 79.3 ± 0.3 | 5858 ± 67 | 3.74 ± 0.12 | -0.50 ± 0.22 | … | … | N | … | … | … | … | … | n | … |
| 49267 | 18272627+0645159 | 56.9 ± 0.2 | 4874 ± 176 | 2.70 ± 0.52 | -0.30 ± 0.24 | … | … | N | … | … | … | … | … | n | G |
| 49268 | 18272628+0621247 | 51.4 ± 0.3 | 4736 ± 193 | 2.97 ± 0.41 | 0.03 ± 0.2 | … | … | N | … | … | … | … | … | n | G |
| 49269 | 18272630+0636259 | 27.8 ± 0.2 | 5720 ± 28 | 3.97 ± 0.18 | -0.30 ± 0.15 | 57 ± 16 | 1 | N | .. | … | … | … | … | n | NG |
| 49270 | 18272631+0629514 | 33.6 ± 0.2 | 4611 ± 148 | 2.96 ± 0.44 | 0.23 ± 0.23 | … | … | N | … | … | … | … | … | n | G |
| 49271 | 18272637+0635320 | 77.7 ± 0.2 | 4348 ± 201 | 2.26 ± 0.33 | 0.05 ± 0.18 | … | … | N | … | … | … | … | … | n | … |
| 49272 | 18272640+0634181 | -25.0 ± 0.2 | 6157 ± 100 | 4.32 ± 0.39 | -0.32 ± 0.19 | 38 ± 10 | … | Y | Y | Y | Y | N | … | Y?$^g$ | … |
| 49273 | 18272642+0627100 | -104.8 ± 0.3 | 5664 ± 22 | 4.04 ± 0.48 | -0.20 ± 0.21 | … | … | N | … | … | … | … | … | n | … |
| 49274 | 18272649+0637591 | 48.1 ± 0.3 | 4818 ± 290 | 2.67 ± 0.69 | -0.30 ± 0.35 | … | … | N | … | … | … | … | … | n | G |
| 49275 | 18272657+0643399 | -12.4 ± 0.3 | 6258 ± 139 | 4.18 ± 0.31 | -0.20 ± 0.12 | 40 ± 16 | 1 | Y | Y | Y | Y | … | … | Y | … |
| 49276 | 18272665+0634212 | -38.1 ± 1.7 | … | … | … | … | … | Y | … | … | … | N | … | n | … |
| 49277 | 18272678+0629392 | -30.3 ± 0.2 | 4647 ± 242 | 4.54 ± 0.37 | -0.03 ± 0.12 | … | … | Y | … | … | … | … | … | n | … |
| 49278 | 18272684+0627323 | -73.6 ± 0.3 | 4868 ± 181 | 2.71 ± 0.37 | -0.65 ± 0.2 | … | … | N | … | … | … | … | … | n | G |
| 49279 | 18272686+0625055 | 39.7 ± 0.3 | 5738 ± 95 | 4.03 ± 0.17 | -0.03 ± 0.13 | 66 ± 20 | 1 | N | .. | … | … | … | … | n | NG |
| 49280 | 18272689+0644370 | -41.0 ± 0.3 | 5142 ± 52 | 2.91 ± 0.43 | -0.22 ± 0.15 | <55 | 3 | Y | Y | Y | Y | … | … | Y | … |
| 49281 | 18272704+0619116 | 45.9 ± 0.3 | 5485 ± 195 | 4.02 ± 0.29 | -0.18 ± 0.19 | 61 ± 23 | 1 | N | .. | … | … | … | … | n | NG |
| 49282 | 18272705+0620440 | 12.1 ± 0.2 | 5285 ± 69 | 4.38 ± 0.31 | 0.03 ± 0.13 | … | … | N | … | … | … | … | … | n | … |
| 49283 | 18272713+0627220 | 82.7 ± 0.2 | 4753 ± 84 | 2.65 ± 0.29 | -0.05 ± 0.18 | … | … | N | … | … | … | … | … | n | G |
| 49284 | 18272724+0640136 | 78.3 ± 0.3 | 4877 ± 276 | 2.77 ± 1.05 | -0.59 ± 0.21 | … | … | N | … | … | … | … | … | n | G |
| 49285 | 18272747+0623261 | -18.6 ± 0.2 | 4150 ± 368 | 4.37 ± 0.24 | 0.12 ± 0.22 | <23 | 3 | Y | N | Y | Y | N | … | n | NG |
| 49286 | 18272754+0626564 | -15.5 ± 0.2 | 4702 ± 190 | 2.82 ± 0.39 | -0.05 ± 0.18 | … | … | Y | … | … | … | … | … | n | G |
| 49287 | 18272754+0642575 | -55.1 ± 0.3 | 5208 ± 52 | 3.46 ± 0.27 | -0.06 ± 0.21 | … | … | N | … | … | … | … | … | n | … |
| 49288 | 18272761+0625440 | 48.7 ± 0.2 | 4639 ± 105 | 2.68 ± 0.36 | 0.18 ± 0.23 | … | … | N | … | … | … | … | … | n | … |
| 49289 | 18272762+0639413 | -14.1 ± 0.3 | 4742 ± 301 | 2.87 ± 0.55 | -0.14 ± 0.22 | … | … | Y | … | … | … | … | … | n | G |
| 49290 | 18272769+0643418 | -36.0 ± 0.3 | 5760 ± 153 | 3.90 ± 0.19 | 0.00 ± 0.14 | … | … | Y | … | … | … | … | … | n | … |
| 49291 | 18272783+0643341 | 5.5 ± 0.2 | 4468 ± 166 | 2.44 ± 0.45 | 0.00 ± 0.22 | … | … | N | … | … | … | … | … | n | G |

**Table C.11.** continued.

| ID | CNAME | RV (km s$^{-1}$) | $T_{\rm eff}$ (K) | logg (dex) | [Fe/H] (dex) | EW(Li)$^a$ (mÅ) | EW(Li) error flag$^b$ | Membership RV | Li | logg | [Fe/H] | Gaia studies Randich$^c$ | Cantat-Gaudin$^c$ | Final$^d$ | NMs with Li$^e$ |
|---|---|---|---|---|---|---|---|---|---|---|---|---|---|---|---|
| 49293 | 18272784+0635537 | -10.8 ± 0.3 | 5112 ± 77 | 3.62 ± 0.38 | 0.38 ± 0.32 | … | … | Y | … | … | … | … | … | n | … |
| 49294 | 18272796+0633445 | 36.4 ± 0.4 | 4919 ± 132 | 3.01 ± 0.38 | 0.06 ± 0.25 | … | … | N | … | … | … | … | … | n | G |
| 49295 | 18272799+0623259 | -71.7 ± 0.3 | 5145 ± 39 | 4.24 ± 0.49 | -0.25 ± 0.14 | … | … | N | … | … | … | … | … | n | … |
| 49296 | 18272800+0619295 | -18.4 ± 0.2 | 4281 ± 288 | 4.38 ± 0.15 | 0.10 ± 0.22 | 68 ± 22 | 1 | Y | N | Y | Y | N | … | n | NG |
| 49297 | 18272800+0636514 | -29.4 ± 0.2 | 6377 ± 101 | 4.19 ± 0.10 | 0.22 ± 0.18 | … | … | Y | … | … | … | Y | … | n | … |
| 49298 | 18272803+0634048 | -36.4 ± 0.2 | 5621 ± 291 | 4.37 ± 0.18 | 0.36 ± 0.16 | … | … | Y | … | … | … | … | … | n | … |
| 49299 | 18272813+0620289 | 46.5 ± 0.2 | 5728 ± 72 | 4.43 ± 0.20 | -0.23 ± 0.14 | … | … | N | … | … | … | … | … | n | … |
| 49300 | 18272815+0627140 | -44.0 ± 0.2 | 5934 ± 143 | 4.14 ± 0.22 | 0.02 ± 0.2 | 45 ± 20 | 1 | Y | Y | Y | Y | … | … | Y | … |
| 49301 | 18272816+0625491 | 22.3 ± 0.3 | 5855 ± 94 | 4.16 ± 0.11 | 0.07 ± 0.13 | 69 ± 24 | 1 | N | … | … | … | … | … | n | NG |
| 49302 | 18272819+0637559 | -3.8 ± 0.3 | 4805 ± 247 | 2.71 ± 0.47 | -0.09 ± 0.27 | … | … | Y | … | … | … | … | … | n | G |
| 49303 | 18272824+0619378 | -8.7 ± 0.2 | 5023 ± 68 | 4.41 ± 0.38 | -0.20 ± 0.14 | … | … | Y | … | … | … | N | … | n | … |
| 49304 | 18272832+0629237 | -8.8 ± 0.2 | 4347 ± 221 | 2.35 ± 0.54 | -0.03 ± 0.23 | … | … | Y | … | … | … | … | … | n | G |
| 49305 | 18272844+0619570 | 22.5 ± 0.2 | 5444 ± 53 | 3.99 ± 0.19 | -0.03 ± 0.14 | 36 ± 22 | 1 | N | … | … | … | … | … | n | NG |
| 49306 | 18272846+0626499 | 27.7 ± 0.3 | 6532 ± 238 | 4.15 ± 0.12 | 0.11 ± 0.2 | 38 ± 18 | 1 | N | … | … | … | … | … | n | NG |
| 49307 | 18272866+0628144 | -68.2 ± 0.5 | 5035 ± 120 | 4.59 ± 1.07 | -0.18 ± 0.19 | … | … | N | … | … | … | … | … | n | … |
| 49308 | 18272874+0631061 | 41.4 ± 0.2 | 4960 ± 143 | 3.32 ± 0.25 | -0.06 ± 0.16 | … | … | N | … | … | … | … | … | n | G |
| 49309 | 18272875+0621250 | -5.5 ± 0.2 | 5612 ± 118 | 4.36 ± 0.24 | 0.20 ± 0.13 | … | … | Y | … | … | … | N | … | n | … |
| 49310 | 18272877+0640380 | 13.6 ± 0.2 | 5691 ± 84 | 4.04 ± 0.17 | 0.18 ± 0.12 | 93 ± 25 | 1 | N | … | … | … | … | … | n | NG |
| 49311 | 18272910+0630407 | -65.7 ± 0.3 | 6431 ± 234 | 4.11 ± 0.15 | -0.18 ± 0.21 | 82 ± 36 | 1 | N | … | … | … | … | … | n | NG |
| 49312 | 18272918+0638410 | 40.4 ± 0.3 | 5521 ± 211 | 3.82 ± 0.21 | 0.21 ± 0.16 | <61 | 3 | N | … | … | … | … | … | n | NG |
| 49313 | 18272930+0629520 | -7.3 ± 0.2 | 4810 ± 129 | 4.60 ± 0.27 | 0.15 ± 0.18 | <31 | 3 | Y | Y | Y | Y | … | … | Y | … |
| 49314 | 18272931+0637103 | 72.7 ± 0.2 | 4801 ± 201 | 2.74 ± 0.35 | -0.23 ± 0.2 | … | … | N | … | … | … | … | … | n | G |
| 49315 | 18272933+0640179 | -43.0 ± 0.3 | 5243 ± 92 | 3.64 ± 0.37 | 0.01 ± 0.17 | … | … | Y | … | … | … | … | … | n | … |
| 49316 | 18272971+0645296 | -57.9 ± 0.3 | 4581 ± 145 | 2.36 ± 0.36 | -0.14 ± 0.15 | … | … | N | … | … | … | … | … | n | G |
| 49317 | 18272995+0619201 | -65.7 ± 0.2 | 5518 ± 113 | 4.31 ± 0.32 | 0.23 ± 0.14 | … | … | N | … | … | … | … | … | n | … |
| 49318 | 18273010+0619111 | 23.3 ± 0.3 | 5034 ± 94 | 4.32 ± 0.60 | 0.21 ± 0.2 | … | … | N | … | … | … | … | … | n | … |
| 49319 | 18273016+0620145 | 70.9 ± 0.3 | 4972 ± 229 | 3.33 ± 0.49 | -0.19 ± 0.17 | … | … | N | … | … | … | … | … | n | G |
| 49320 | 18273017+0629261 | 34.5 ± 0.2 | 4440 ± 196 | 2.44 ± 0.32 | 0.37 ± 0.28 | … | … | N | … | … | … | … | … | n | … |
| 49321 | 18273023+0635544 | 19.8 ± 0.4 | 4199 ± 277 | 4.42 ± 0.27 | 0.10 ± 0.24 | … | … | N | … | … | … | N | … | n | … |
| 49322 | 18273030+0640047 | -261.5 ± 0.7 | 4616 ± 231 | 1.34 ± 0.52 | -1.00 ± 0.39 | … | … | N | … | … | … | … | … | n | G |
| 49323 | 18273033+0619345 | -47.2 ± 0.3 | 4906 ± 88 | 4.43 ± 0.48 | 0.13 ± 0.18 | … | … | Y | … | … | … | … | … | n | … |
| 49324 | 18273041+0641118 | 39.7 ± 0.4 | 4252 ± 227 | 4.78 ± 0.22 | -0.09 ± 0.18 | <75 | 3 | N | … | … | … | … | … | n | NG |
| 49325 | 18273043+0636345 | 167.2 ± 8.3 | … | … | … | … | … | N | … | … | … | … | … | n | … |
| 49326 | 18273045+0619399 | -25.5 ± 0.3 | 6100 ± 189 | 4.14 ± 0.02 | 0.10 ± 0.13 | … | … | Y | … | … | … | … | … | n | … |
| 49327 | 18273064+0625255 | -58.6 ± 0.2 | 4893 ± 189 | 2.90 ± 0.48 | -0.21 ± 0.17 | … | … | N | … | … | … | … | … | n | G |
| 49328 | 18273088+0636097 | -5.1 ± 0.3 | 5796 ± 188 | 4.20 ± 0.22 | 0.04 ± 0.19 | … | … | Y | … | … | … | … | … | n | … |
| 49329 | 18273092+0628026 | -11.0 ± 0.2 | 4927 ± 162 | 3.18 ± 0.33 | -0.14 ± 0.19 | … | … | Y | … | … | … | … | … | n | G |
| 49330 | 18273096+0625085 | -70.3 ± 0.2 | 4618 ± 182 | 2.52 ± 0.44 | -0.23 ± 0.19 | … | … | N | … | … | … | … | … | n | G |
| 49331 | 18273103+0626057 | 100.5 ± 0.2 | 4957 ± 133 | 3.35 ± 0.25 | -0.01 ± 0.17 | … | … | N | … | … | … | … | … | n | G |
| 49332 | 18273109+0628276 | -28.8 ± 0.2 | 4904 ± 122 | 3.36 ± 0.31 | -0.05 ± 0.13 | … | … | Y | … | … | … | … | … | n | G |
| 49333 | 18273110+0639190 | 16.3 ± 0.3 | 5485 ± 169 | 4.33 ± 0.12 | 0.49 ± 0.33 | … | … | N | … | … | … | … | … | n | … |
| 49334 | 18273110+0642220 | -49.7 ± 0.3 | 5068 ± 72 | 2.97 ± 0.43 | -0.11 ± 0.27 | … | … | N | … | … | … | … | … | n | G |
| 49335 | 18273112+0639345 | 24.9 ± 0.2 | 4739 ± 300 | 2.78 ± 0.50 | -0.39 ± 0.31 | … | … | N | … | … | … | … | … | n | G |
| 49336 | 18273116+0621473 | 44.2 ± 0.3 | 4831 ± 217 | 3.25 ± 0.42 | 0.04 ± 0.22 | … | … | N | … | … | … | … | … | n | G |
| 49337 | 18273119+0631131 | -51.6 ± 0.3 | 4139 ± 348 | 4.57 ± 0.22 | -0.04 ± 0.21 | <108 | 3 | N | … | … | … | N | … | n | NG |
| 49338 | 18273128+0618401 | -14.6 ± 0.5 | 5133 ± 133 | 3.08 ± 0.32 | -0.35 ± 0.23 | … | … | Y | … | … | … | … | … | n | G |
| 49339 | 18273130+0618226 | -45.4 ± 0.3 | 5225 ± 354 | 3.55 ± 0.35 | -0.58 ± 0.7 | … | … | Y | … | … | … | … | … | n | … |
| 49340 | 18273134+0636580 | 35.7 ± 0.3 | 5932 ± 186 | 4.33 ± 0.10 | -0.14 ± 0.17 | … | … | N | … | … | … | … | … | n | … |
| 49341 | 18273136+0629010 | -26.9 ± 0.3 | 4833 ± 137 | 2.69 ± 0.30 | -0.18 ± 0.15 | … | … | Y | … | … | … | … | … | n | G |
| 49342 | 18273138+0636290 | 6.6 ± 0.3 | 5869 ± 30 | 4.11 ± 0.16 | 0.04 ± 0.13 | 70 ± 39 | 1 | N | … | … | … | … | … | n | NG |
| 49343 | 18273148+0627583 | 19.8 ± 0.3 | 4924 ± 201 | 3.48 ± 0.66 | -0.09 ± 0.19 | … | … | N | … | … | … | … | … | n | G |
| 49344 | 18273154+0622007 | -27.2 ± 0.2 | 4647 ± 229 | 4.65 ± 0.39 | -0.16 ± 0.16 | … | … | Y | … | … | … | Y | … | n | … |
| 49345 | 18273162+0639096 | 3.7 ± 0.6 | 5070 ± 73 | 3.00 ± 0.68 | -0.21 ± 0.25 | … | … | N | … | … | … | … | … | n | G |
| 49346 | 18273179+0625456 | -84.0 ± 0.4 | 4054 ± 390 | 4.72 ± 0.19 | -0.44 ± 0.36 | <81 | 3 | N | … | … | … | … | … | n | NG |
| 49347 | 18273186+0638597 | 98.5 ± 0.2 | 4699 ± 118 | 2.55 ± 0.30 | 0.04 ± 0.26 | … | … | N | … | … | … | … | … | n | G |
| 49348 | 18273203+0636522 | -23.8 ± 0.2 | 4346 ± 213 | 4.46 ± 0.24 | -0.71 ± 0.68 | <12 | 3 | Y | Y | Y | N | N | Y | Y?$^g$ | … |
| 49349 | 18273206+0642008 | 1.8 ± 0.2 | 4895 ± 212 | 2.95 ± 0.44 | -0.11 ± 0.22 | 40 ± 16 | 1 | Y?$^f$ | Y | Y | Y | … | … | n | G |
| 49350 | 18273207+0636435 | -10.4 ± 0.5 | 4674 ± 81 | 4.43 ± 0.34 | 0.26 ± 0.17 | … | … | Y | … | … | … | … | … | n | … |
| 49351 | 18273238+0637521 | -65.3 ± 0.2 | 4910 ± 129 | 3.23 ± 0.38 | -0.05 ± 0.14 | … | … | N | … | … | … | … | … | n | G |

| ID | CNAME | RV (km s$^{-1}$) | $T_{\rm eff}$ (K) | logg (dex) | [Fe/H] (dex) | EW(Li)$^a$ (mÅ) | EW(Li) error flag$^b$ | Membership RV | Li | logg | [Fe/H] | Gaia studies Randich$^c$ | Cantat-Gaudin$^c$ | Final$^d$ | NMs with Li$^e$ |
|---|---|---|---|---|---|---|---|---|---|---|---|---|---|---|---|
| 49352 | 18273271+0623396 | 13.1 ± 0.3 | 5011 ± 173 | 3.49 ± 0.39 | -0.46 ± 0.22 | … | … | N | … | … | … | … | … | n | G |
| 49353 | 18273275+0640505 | -16.6 ± 0.3 | 5584 ± 100 | 4.15 ± 0.28 | 0.02 ± 0.12 | … | … | Y | … | … | … | … | … | n | … |
| 49354 | 18273284+0627295 | -21.3 ± 0.3 | 4765 ± 57 | 2.48 ± 0.24 | 0.19 ± 0.28 | … | … | Y | … | … | … | … | … | n | G |
| 49355 | 18273284+0641540 | 0.6 ± 0.3 | 5101 ± 59 | 4.53 ± 0.49 | 0.13 ± 0.14 | … | … | Y | … | … | … | … | … | n | … |
| 49356 | 18273285+0626577 | 72.2 ± 0.3 | 4756 ± 203 | 2.81 ± 0.24 | -0.21 ± 0.19 | … | … | N | … | … | … | … | … | n | G |
| 49357 | 18273287+0639554 | 15.6 ± 0.3 | 5641 ± 138 | 4.63 ± 0.27 | -0.04 ± 0.24 | … | … | N | … | … | … | … | … | n | … |
| 49358 | 18273289+0644129 | 79.9 ± 0.2 | 4540 ± 139 | 2.54 ± 0.46 | 0.02 ± 0.24 | <21 | 3 | N | .. | … | … | … | … | n | G |
| 49359 | 18273315+0636246 | -53.7 ± 0.3 | 5549 ± 125 | 3.98 ± 0.16 | 0.04 ± 0.13 | … | … | N | … | … | … | … | … | n | … |
| 49360 | 18273321+0631500 | 59.8 ± 0.2 | 5709 ± 116 | 4.09 ± 0.26 | -0.11 ± 0.13 | 124 ± 27 | 1 | N | .. | … | … | … | … | n | NG |
| 49361 | 18273327+0631256 | -20.0 ± 0.3 | 4886 ± 178 | 4.46 ± 0.34 | 0.12 ± 0.15 | <55 | 3 | Y | Y | Y | Y | … | … | Y | … |
| 49362 | 18273335+0643008 | 37.2 ± 0.4 | 5061 ± 75 | 3.44 ± 0.49 | 0.01 ± 0.16 | … | … | N | … | … | … | … | … | n | G |
| 49363 | 18273340+0634473 | -13.4 ± 0.2 | 4594 ± 154 | 4.64 ± 0.19 | 0.01 ± 0.13 | … | … | Y | … | … | … | … | … | n | … |
| 49364 | 18273343+0624340 | 87.1 ± 0.2 | 4699 ± 110 | 2.57 ± 0.26 | -0.07 ± 0.18 | … | … | N | … | … | … | … | … | n | G |
| 49365 | 18273346+0620313 | 16.4 ± 0.2 | 5325 ± 47 | 3.92 ± 0.40 | 0.18 ± 0.15 | … | … | N | … | … | … | N | … | n | … |
| 49366 | 18273366+0620183 | 11.0 ± 0.3 | 4994 ± 167 | 3.54 ± 0.40 | -0.16 ± 0.17 | … | … | N | … | … | … | … | … | n | … |
| 49367 | 18273370+0635489 | -51.4 ± 0.3 | 4833 ± 156 | 2.92 ± 0.43 | 0.03 ± 0.27 | … | … | N | … | … | … | … | … | n | G |
| 49368 | 18273375+0618380 | -53.6 ± 0.2 | 4815 ± 118 | 4.50 ± 0.34 | -0.04 ± 0.14 | <27 | 3 | N | .. | … | … | N | … | n | NG |
| 49369 | 18273381+0639214 | -85.7 ± 0.3 | 5401 ± 60 | 3.95 ± 0.18 | -0.16 ± 0.15 | … | … | N | … | … | … | … | … | n | … |
| 49370 | 18273385+0626554 | -18.1 ± 0.3 | 4396 ± 239 | 4.64 ± 0.36 | -0.06 ± 0.13 | … | … | Y | … | … | … | N | … | n | … |
| 49371 | 18273422+0634253 | -34.1 ± 0.3 | 4440 ± 154 | 2.55 ± 0.39 | 0.29 ± 0.24 | … | … | Y | … | … | … | … | … | n | … |
| 49372 | 18273422+0641210 | -70.0 ± 0.2 | 4742 ± 230 | 3.18 ± 0.28 | -0.10 ± 0.16 | <24 | 3 | N | … | … | … | … | … | n | G |
| 49373 | 18273428+0638571 | -3.9 ± 0.2 | 4004 ± 284 | 4.72 ± 0.14 | -0.48 ± 0.32 | <17 | 3 | Y | N | Y | N | N | … | n | NG |
| 49374 | 18273447+0633525 | 51.0 ± 0.3 | 5265 ± 296 | 3.75 ± 0.27 | -0.10 ± 0.18 | … | … | N | … | … | … | … | … | n | … |
| 49375 | 18273456+0638283 | 41.1 ± 0.3 | 4979 ± 162 | 3.43 ± 0.20 | -0.09 ± 0.2 | … | … | N | … | … | … | … | … | n | G |
| 49376 | 18273459+0619090 | -30.3 ± 0.2 | 5627 ± 191 | 4.32 ± 0.25 | 0.25 ± 0.16 | … | … | Y | … | … | … | … | … | n | … |
| 49377 | 18273460+0620222 | -36.2 ± 0.3 | 5615 ± 42 | 4.01 ± 0.17 | -0.15 ± 0.13 | 49 ± 31 | 1 | Y | Y | Y | Y | … | … | Y | … |
| 49378 | 18273461+0622470 | -18.1 ± 0.3 | 4529 ± 237 | 4.58 ± 0.28 | 0.10 ± 0.17 | … | … | Y | … | … | … | … | … | n | … |
| 49379 | 18273476+0643376 | -23.7 ± 0.2 | 5778 ± 44 | 4.15 ± 0.30 | 0.25 ± 0.17 | 49 ± 18 | 1 | Y | Y | Y | Y | … | … | Y | … |
| 49380 | 18273477+0621112 | 67.2 ± 0.3 | 4990 ± 97 | 3.37 ± 0.42 | -0.33 ± 0.17 | … | … | N | … | … | … | … | … | n | G |
| 49381 | 18273488+0640451 | -41.6 ± 0.4 | 4561 ± 209 | 5.01 ± 0.62 | 0.13 ± 0.21 | … | … | Y | … | … | … | N | … | n | … |
| 49382 | 18273489+0619242 | -11.6 ± 0.2 | 5821 ± 45 | 4.30 ± 0.13 | 0.06 ± 0.12 | 33 ± 23 | 1 | Y | Y | Y | Y | … | … | Y | … |
| 49383 | 18273497+0628066 | -19.5 ± 0.2 | 5811 ± 69 | 4.20 ± 0.19 | 0.08 ± 0.12 | … | … | Y | … | … | … | … | … | n | … |
| 49384 | 18273510+0628296 | 1.3 ± 0.3 | 4640 ± 67 | 2.86 ± 0.47 | 0.31 ± 0.3 | <60 | 3 | Y?$^f$ | N | Y | Y | … | … | n | G |
| 49385 | 18273518+0630339 | -31.1 ± 0.3 | 4772 ± 193 | 4.67 ± 0.29 | -0.04 ± 0.12 | <38 | 3 | Y | Y | Y | Y | … | … | Y | … |
| 49386 | 18273529+0626402 | -55.1 ± 0.3 | 4637 ± 170 | 4.38 ± 0.42 | 0.27 ± 0.22 | … | … | N | … | … | … | … | … | n | … |
| 49387 | 18273532+0636157 | -193.6 ± 0.3 | 4980 ± 53 | 2.04 ± 0.47 | -1.60 ± 0.47 | … | … | N | … | … | … | … | … | n | G |
| 49388 | 18273545+0622565 | -188.9 ± 0.3 | 5008 ± 83 | 2.59 ± 0.41 | -0.32 ± 0.17 | … | … | N | … | … | … | … | … | n | G |
| 49389 | 18273557+0631425 | 63.3 ± 0.2 | 4559 ± 172 | 2.47 ± 0.34 | -0.14 ± 0.2 | … | … | N | … | … | … | … | … | n | G |
| 49390 | 18273560+0623036 | 26.8 ± 0.2 | 5704 ± 82 | 4.39 ± 0.27 | -0.15 ± 0.13 | … | … | N | … | … | … | … | … | n | … |
| 49391 | 18273578+0634112 | 6.9 ± 0.3 | 4876 ± 161 | 3.03 ± 0.34 | -0.25 ± 0.19 | <55 | 3 | N | .. | … | … | … | … | n | G |
| 49392 | 18273607+0623384 | -0.9 ± 0.2 | 4518 ± 167 | 2.45 ± 0.32 | 0.24 ± 0.29 | … | … | Y | … | … | … | … | … | n | … |
| 49393 | 18273609+0621561 | -29.3 ± 0.3 | 4757 ± 166 | 4.89 ± 0.28 | -0.06 ± 0.12 | … | … | Y | … | … | … | … | … | n | … |
| 49394 | 18273613+0624424 | 21.4 ± 0.3 | 3618 ± 147 | 4.38 ± 0.11 | -0.06 ± 0.12 | <87 | 3 | N | .. | … | … | N | … | n | NG |
| 49395 | 18273626+0642476 | 1.5 ± 0.2 | 4613 ± 205 | 2.70 ± 0.54 | 0.11 ± 0.24 | … | … | Y | … | … | … | … | … | n | G |
| 49396 | 18273628+0618439 | 32.0 ± 0.3 | 5008 ± 130 | 3.56 ± 0.29 | -0.27 ± 0.19 | … | … | N | … | … | … | … | … | n | … |
| 49397 | 18273630+0632046 | 15.9 ± 0.3 | 6600 ± 366 | 3.81 ± 0.22 | 0.09 ± 0.28 | … | 1 | N | … | … | … | … | … | n | … |
| 49398 | 18273640+0626057 | -54.2 ± 0.5 | 4066 ± 181 | 3.77 ± 0.34 | 0.48 ± 0.14 | … | … | N | … | … | … | N | … | n | … |
| 49399 | 18273641+0644199 | 15.6 ± 0.2 | 4484 ± 173 | 2.72 ± 0.49 | 0.25 ± 0.22 | <27 | 3 | N | .. | … | … | … | … | n | G |
| 49400 | 18273642+0634121 | 1.3 ± 0.2 | 6349 ± 158 | 4.12 ± 0.08 | 0.17 ± 0.18 | 125 ± 21 | 1 | Y?$^f$ | Y | Y | Y | … | … | n | NG |
| 49401 | 18273651+0623058 | 16.5 ± 0.3 | 6374 ± 129 | 4.05 ± 0.15 | 0.00 ± 0.13 | 91 ± 27 | 1 | N | .. | … | … | … | … | n | NG |
| 49402 | 18273654+0625175 | -115.0 ± 0.2 | 4799 ± 174 | 2.81 ± 0.39 | -0.04 ± 0.2 | 56 ± 18 | 1 | N | … | … | … | … | … | n | … |
| 49403 | 18273656+0626124 | -44.1 ± 0.3 | 4777 ± 208 | 3.19 ± 0.46 | 0.00 ± 0.17 | … | … | Y | … | … | … | … | … | n | G |
| 49404 | 18273663+0636442 | 20.2 ± 0.3 | 4464 ± 289 | 4.97 ± 0.35 | -0.43 ± 0.25 | <34 | 3 | N | .. | … | … | N | … | n | NG |
| 49405 | 18273677+0628312 | 0.8 ± 0.3 | 4829 ± 178 | 2.95 ± 0.32 | -0.03 ± 0.17 | … | … | Y | … | … | … | … | … | n | G |
| 49406 | 18273682+0626286 | 13.0 ± 0.3 | 4812 ± 219 | 2.56 ± 0.76 | -0.57 ± 0.35 | … | … | N | … | … | … | … | … | n | G |
| 49407 | 18273683+0621120 | -8.8 ± 0.3 | 5154 ± 47 | 3.77 ± 0.45 | -0.11 ± 0.14 | … | … | Y | … | … | … | … | … | n | … |
| 49408 | 18273684+0632221 | 73.6 ± 0.3 | 4670 ± 70 | 2.56 ± 0.27 | 0.09 ± 0.19 | … | … | N | … | … | … | … | … | n | G |
| 49409 | 18273686+0626236 | -13.4 ± 0.2 | 4619 ± 189 | 2.63 ± 0.36 | -0.09 ± 0.18 | … | … | Y | … | … | … | … | … | n | G |
| 49410 | 18273690+0620587 | -2.9 ± 0.3 | 4171 ± 247 | 4.69 ± 0.19 | -0.23 ± 0.17 | <55 | 3 | Y | N | N | Y | N | … | n | NG |

**Table C.11.** continued.

| ID | CNAME | RV (km s$^{-1}$) | $T_{\rm eff}$ (K) | logg (dex) | [Fe/H] (dex) | EW(Li)$^a$ (mÅ) | EW(Li) error flag$^b$ | Membership RV | Li | logg | [Fe/H] | Gaia studies Randich$^c$ | Cantat-Gaudin$^c$ | Final$^d$ | NMs with Li$^e$ |
|---|---|---|---|---|---|---|---|---|---|---|---|---|---|---|---|
| 49411 | 18273696+0622216 | -49.2 ± 0.2 | 5426 ± 195 | 4.33 ± 0.38 | 0.19 ± 0.13 | … | … | N | … | … | … | … | … | n | … |
| 49412 | 18273696+0627149 | -93.6 ± 0.3 | 4579 ± 382 | 2.50 ± 0.52 | -0.17 ± 0.18 | <73 | 3 | N | .. | … | … | … | … | n | G |
| 49413 | 18273710+0643340 | 22.5 ± 0.2 | 4569 ± 178 | 2.50 ± 0.27 | 0.21 ± 0.24 | … | … | N | … | … | … | … | … | n | G |
| 49414 | 18273713+0638152 | -1.6 ± 0.2 | 4929 ± 157 | 2.73 ± 0.50 | -0.29 ± 0.23 | … | … | Y | … | … | … | … | … | n | G |
| 49415 | 18273719+0633260 | 2.1 ± 0.3 | 5360 ± 150 | 4.33 ± 0.27 | 0.05 ± 0.12 | … | … | Y | … | … | … | … | … | n | … |
| 49416 | 18273737+0627364 | 27.0 ± 0.2 | 4797 ± 249 | 3.24 ± 0.16 | -0.13 ± 0.18 | … | … | N | … | … | … | … | … | n | G |
| 49417 | 18273743+0636340 | 70.6 ± 0.3 | 5971 ± 151 | 4.18 ± 0.09 | -0.05 ± 0.14 | 62 ± 23 | 1 | N | .. | … | … | … | … | n | NG |
| 49418 | 18273750+0637138 | -14.5 ± 0.3 | 4846 ± 172 | 3.31 ± 0.30 | -0.24 ± 0.17 | … | … | Y | … | … | … | … | … | n | G |
| 49419 | 18273756+0622059 | -12.2 ± 0.2 | 5453 ± 182 | 4.22 ± 0.23 | 0.40 ± 0.22 | … | … | Y | … | … | … | … | … | n | … |
| 49420 | 18273764+0620441 | -28.2 ± 0.3 | 3596 ± 14 | 4.34 ± 0.12 | … | … | … | Y | … | … | … | … | N | n | … |
| 49421 | 18273764+0623476 | -9.0 ± 0.3 | 6219 ± 190 | 4.28 ± 0.19 | 0.08 ± 0.26 | … | … | Y | … | … | … | … | … | n | … |
| 49422 | 18273772+0626492 | -7.9 ± 0.3 | 5858 ± 132 | 4.31 ± 0.21 | 0.34 ± 0.17 | 112 ± 34 | 1 | Y | N | Y | N | … | … | n | … |
| 49423 | 18273773+0626574 | -16.4 ± 0.3 | 4844 ± 122 | 4.48 ± 0.43 | 0.10 ± 0.17 | … | … | Y | … | … | … | … | … | n | … |
| 49424 | 18273775+0635072 | 34.4 ± 0.3 | 5972 ± 89 | 4.27 ± 0.37 | -0.33 ± 0.17 | … | … | N | … | … | … | … | … | n | … |
| 49425 | 18273794+0637227 | 32.9 ± 0.2 | 5387 ± 77 | 4.45 ± 0.22 | 0.04 ± 0.13 | … | … | N | … | … | … | … | … | n | … |
| 49426 | 18273795+0619225 | 23.4 ± 0.5 | 6311 ± 253 | 3.88 ± 0.22 | -0.04 ± 0.33 | 155 ± 58 | 1 | N | .. | … | … | … | … | n | NG |
| 49427 | 18273796+0629330 | 24.4 ± 0.3 | 4702 ± 142 | 2.70 ± 0.33 | 0.01 ± 0.18 | … | … | N | … | … | … | … | … | n | G |
| 49428 | 18273798+0621091 | 50.7 ± 0.2 | 4898 ± 159 | 3.48 ± 0.26 | 0.07 ± 0.2 | … | … | N | … | … | … | … | … | n | G |
| 49429 | 18273806+0624496 | 15.5 ± 0.3 | 5063 ± 94 | 4.51 ± 0.38 | -0.24 ± 0.16 | … | … | N | … | … | … | … | … | n | … |
| 49430 | 18273810+0637293 | 23.1 ± 0.7 | 4245 ± 281 | 4.15 ± 0.49 | -0.39 ± 0.25 | … | … | N | … | … | … | … | … | n | … |
| 49431 | 18273837+0619595 | 25.2 ± 0.3 | 5697 ± 73 | 3.85 ± 0.28 | -0.51 ± 0.38 | <13 | 3 | N | .. | … | … | … | … | n | NG |
| 49432 | 18273842+0643318 | -6.4 ± 0.2 | 4392 ± 207 | 2.41 ± 0.59 | -0.04 ± 0.25 | … | … | Y | … | … | … | … | … | n | G |
| 49433 | 18273844+0641484 | 34.6 ± 0.2 | 4533 ± 123 | 2.56 ± 0.27 | 0.03 ± 0.21 | … | … | N | … | … | … | … | … | n | G |
| 49434 | 18273859+0641508 | -22.8 ± 0.3 | 5875 ± 75 | 4.10 ± 0.24 | -0.49 ± 0.15 | … | … | Y | … | … | … | … | … | n | … |
| 49435 | 18273861+0618591 | 18.5 ± 0.5 | 6372 ± 146 | 4.48 ± 0.46 | -0.17 ± 0.17 | … | … | N | … | … | … | … | … | n | … |
| 49436 | 18273863+0628183 | 95.2 ± 0.3 | 4640 ± 98 | 2.66 ± 0.34 | 0.02 ± 0.18 | … | … | N | … | … | … | … | … | n | G |
| 49437 | 18273872+0641595 | -27.5 ± 0.2 | 4747 ± 87 | 2.81 ± 0.37 | 0.20 ± 0.24 | … | … | Y | … | … | … | … | … | n | G |
| 49438 | 18273889+0623338 | 82.9 ± 0.3 | 4817 ± 222 | 2.82 ± 0.35 | -0.05 ± 0.22 | … | … | N | … | … | … | … | … | n | G |
| 49439 | 18273909+0625339 | 31.1 ± 0.8 | 4146 ± 52 | … | … | … | … | N | … | … | … | N | … | n | … |
| 49440 | 18273915+0637485 | 41.6 ± 0.2 | 4818 ± 211 | 2.95 ± 0.32 | -0.15 ± 0.2 | … | … | N | … | … | … | … | … | n | G |
| 49441 | 18273928+0624168 | 112.5 ± 0.2 | 4585 ± 125 | 2.33 ± 0.29 | 0.13 ± 0.27 | … | … | N | … | … | … | … | … | n | G |
| 49442 | 18273946+0620558 | 5.9 ± 0.2 | 4680 ± 124 | 2.58 ± 0.27 | 0.14 ± 0.21 | … | … | N | … | … | … | … | … | n | G |
| 49443 | 18273947+0631278 | 107.3 ± 0.3 | 4645 ± 222 | 2.44 ± 0.27 | -0.12 ± 0.16 | … | … | N | … | … | … | … | … | n | G |
| 49444 | 18273950+0643418 | -17.1 ± 0.2 | 4729 ± 212 | 2.88 ± 0.52 | 0.07 ± 0.2 | … | … | Y | … | … | … | … | … | n | G |
| 49445 | 18273954+0643491 | 48.7 ± 0.3 | 4760 ± 234 | 2.70 ± 0.42 | -0.05 ± 0.22 | … | … | N | … | … | … | … | … | n | G |
| 49446 | 18273956+0634590 | 40.3 ± 0.2 | 5411 ± 118 | 4.20 ± 0.18 | -0.18 ± 0.12 | … | … | N | … | … | … | … | … | n | … |
| 49447 | 18273972+0634303 | 96.7 ± 0.2 | 4683 ± 104 | 2.65 ± 0.31 | 0.21 ± 0.29 | … | … | N | … | … | … | … | … | n | … |
| 49448 | 18273973+0641206 | 11.4 ± 0.3 | 4733 ± 221 | 3.29 ± 0.26 | 0.03 ± 0.18 | <55 | 3 | N | .. | … | … | … | … | n | G |
| 49449 | 18273974+0639451 | 31.2 ± 1.0 | 4595 ± 72 | 5.07 ± 0.18 | -0.22 ± 0.09 | … | … | N | … | … | … | … | … | n | … |
| 49450 | 18273987+0639198 | 56.8 ± 0.3 | 4870 ± 132 | 2.86 ± 0.33 | 0.26 ± 0.28 | … | … | N | … | … | … | … | … | n | G |
| 49451 | 18273992+0639364 | -83.8 ± 0.2 | 4578 ± 200 | 2.53 ± 0.36 | -0.14 ± 0.18 | … | … | N | … | … | … | … | … | n | … |
| 49452 | 18274011+0620199 | -73.9 ± 0.3 | 4938 ± 134 | 3.03 ± 0.24 | -0.22 ± 0.16 | … | … | N | … | … | … | … | … | n | G |
| 49453 | 18274013+0626073 | -15.5 ± 0.2 | 5584 ± 52 | 4.17 ± 0.17 | 0.08 ± 0.14 | … | … | Y | … | … | … | … | … | n | … |
| 49454 | 18274015+0639075 | -39.2 ± 0.3 | 3690 ± 431 | 4.56 ± 0.10 | -0.12 ± 0.13 | <63 | 3 | Y | Y | Y | Y | N | … | Y?$^g$ | … |
| 49455 | 18274018+0621109 | 19.7 ± 0.3 | 4742 ± 258 | 2.99 ± 0.39 | -0.11 ± 0.22 | <94 | 3 | N | .. | … | … | … | … | n | G |
| 49456 | 18274029+0641294 | 12.7 ± 0.3 | 5587 ± 86 | 4.11 ± 0.17 | 0.22 ± 0.15 | … | … | N | … | … | … | … | … | n | … |
| 49457 | 18274031+0625206 | -10.3 ± 0.2 | 4659 ± 183 | 2.63 ± 0.50 | -0.06 ± 0.24 | … | … | Y | … | … | … | … | … | n | … |
| 49458 | 18274048+0636217 | -3.6 ± 0.2 | 4566 ± 209 | 2.09 ± 0.31 | -0.11 ± 0.2 | … | … | Y | … | … | … | … | … | n | G |
| 49459 | 18274051+0626016 | -48.3 ± 0.2 | 4892 ± 198 | 3.29 ± 0.36 | -0.31 ± 0.18 | <37 | 3 | N | .. | … | … | … | … | n | G |
| 49460 | 18274057+0634344 | 11.8 ± 0.9 | 5048 ± 85 | 3.81 ± 0.31 | -0.28 ± 0.1 | … | … | N | … | … | … | … | … | n | … |
| 49461 | 18274070+0639449 | -1.6 ± 0.3 | 5322 ± 118 | 4.32 ± 0.63 | 0.21 ± 0.17 | … | … | Y | … | … | … | … | … | n | … |
| 49462 | 18274078+0623248 | 79.4 ± 0.2 | 4839 ± 174 | 2.90 ± 0.82 | -0.61 ± 0.2 | … | … | N | … | … | … | … | … | n | G |
| 49463 | 18274086+0619183 | 34.3 ± 0.3 | 5054 ± 233 | 3.39 ± 0.80 | -0.53 ± 0.24 | … | … | N | … | … | … | … | … | n | G |
| 49464 | 18274100+0643062 | -25.4 ± 0.3 | 4452 ± 190 | 4.69 ± 0.27 | -0.05 ± 0.16 | <34 | 3 | Y | Y | Y | Y | … | … | Y | … |
| 49465 | 18274104+0621052 | 21.1 ± 0.3 | 6126 ± 113 | 4.44 ± 0.36 | -0.04 ± 0.14 | 47 ± 14 | 1 | N | .. | … | … | … | … | n | NG |
| 49466 | 18274117+0643027 | -25.5 ± 0.3 | 4188 ± 287 | 4.86 ± 0.46 | -0.09 ± 0.28 | <101 | 3 | Y | N | Y | Y | … | … | n | NG |
| 49467 | 18274134+0625310 | 50.6 ± 0.2 | 4389 ± 185 | 2.38 ± 0.44 | 0.02 ± 0.21 | 27 ± 20 | 1 | N | .. | … | … | … | … | n | G |
| 49468 | 18274152+0633229 | -17.7 ± 0.3 | 4449 ± 202 | 4.50 ± 0.38 | 0.25 ± 0.22 | … | … | Y | … | … | … | … | … | n | … |
| 49469 | 18274156+0643115 | -23.8 ± 0.3 | 4822 ± 215 | 3.11 ± 0.55 | -0.21 ± 0.22 | … | … | Y | … | … | … | … | … | n | G |

| ID | CNAME | RV (km s$^{-1}$) | $T_{\rm eff}$ (K) | logg (dex) | [Fe/H] (dex) | EW(Li)$^a$ (mÅ) | EW(Li) error flag$^b$ | Membership RV | Li | logg | [Fe/H] | Gaia studies Randich$^c$ | Cantat-Gaudin$^c$ | Final$^d$ | NMs with Li$^e$ |
|---|---|---|---|---|---|---|---|---|---|---|---|---|---|---|---|
| 49470 | 18274161+0635139 | 26.8 ± 0.2 | 4571 ± 81 | 2.66 ± 0.37 | 0.28 ± 0.25 | ... | ... | N | ... | ... | ... | ... | ... | n | G |
| 49471 | 18274167+0626184 | -29.1 ± 0.4 | 3798 ± 400 | 5.02 ± 0.46 | -0.85 ± 0.81 | <84 | 3 | Y | Y | Y | N | Y | Y | Y | ... |
| 49472 | 18274178+0645098 | 41.6 ± 0.2 | 4551 ± 188 | 2.45 ± 0.27 | 0.04 ± 0.17 | ... | ... | N | ... | ... | ... | ... | ... | n | G |
| 49473 | 18274190+0628204 | 21.7 ± 0.2 | 5678 ± 80 | 4.35 ± 0.09 | 0.08 ± 0.14 | ... | ... | N | ... | ... | ... | ... | ... | n | ... |
| 49474 | 18274212+0630556 | 71.6 ± 0.2 | 4829 ± 70 | 3.27 ± 0.54 | 0.26 ± 0.22 | ... | ... | N | ... | ... | ... | ... | ... | n | G |
| 49475 | 18274228+0639476 | 97.5 ± 0.3 | 4546 ± 302 | 3.05 ± 0.60 | 0.19 ± 0.42 | ... | ... | N | ... | ... | ... | ... | ... | n | G |
| 49476 | 18274236+0633416 | -14.7 ± 0.2 | 5016 ± 137 | 3.17 ± 0.57 | -0.39 ± 0.22 | ... | ... | Y | ... | ... | ... | ... | ... | n | G |
| 49477 | 18274237+0625353 | -28.0 ± 0.2 | 5389 ± 158 | 4.32 ± 0.47 | -0.07 ± 0.13 | ... | ... | Y | ... | ... | ... | ... | N | n | ... |
| 49478 | 18274237+0637573 | 1.0 ± 0.4 | 3978 ± 252 | 4.66 ± 0.13 | -0.22 ± 0.4 | <77 | 3 | Y?$^f$ | Y | Y | Y | N | ... | n | NG |
| 49479 | 18274238+0627029 | 100.4 ± 0.2 | 5169 ± 43 | 3.67 ± 0.26 | 0.07 ± 0.16 | ... | ... | N | ... | ... | ... | ... | ... | n | ... |
| 49480 | 18274258+0641539 | 3.8 ± 0.3 | 5329 ± 157 | 4.28 ± 0.58 | 0.02 ± 0.13 | ... | ... | N | ... | ... | ... | ... | ... | n | ... |
| 49481 | 18274272+0628261 | -10.1 ± 0.4 | 4475 ± 238 | 4.76 ± 0.34 | -0.19 ± 0.17 | ... | ... | Y | ... | ... | ... | ... | ... | n | ... |
| 49482 | 18274280+0623079 | 44.3 ± 0.3 | 4837 ± 179 | 4.83 ± 0.26 | 0.14 ± 0.19 | <46 | 3 | N | .. | ... | ... | ... | ... | n | NG |
| 49483 | 18274289+0639221 | 62.7 ± 0.2 | 4514 ± 194 | 2.50 ± 0.37 | 0.22 ± 0.31 | ... | ... | N | ... | ... | ... | ... | ... | n | G |
| 49484 | 18274292+0635377 | -35.2 ± 0.5 | 4266 ± 295 | 4.70 ± 0.40 | -0.53 ± 0.63 | ... | ... | Y | ... | ... | ... | N | ... | n | ... |
| 49485 | 18274296+0636553 | 34.2 ± 0.2 | 4866 ± 200 | 2.90 ± 0.40 | -0.42 ± 0.28 | ... | ... | N | ... | ... | ... | ... | ... | n | G |
| 49486 | 18274297+0645113 | 49.7 ± 0.9 | 5037 ± 89 | 4.09 ± 0.27 | 0.15 ± 0.08 | ... | ... | N | ... | ... | ... | ... | ... | n | ... |
| 49487 | 18274304+0623546 | 18.1 ± 0.3 | 5323 ± 316 | 4.45 ± 0.93 | 0.41 ± 0.26 | <56 | 3 | N | .. | ... | ... | ... | ... | n | NG |
| 49488 | 18274316+0644125 | -27.7 ± 0.4 | 4123 ± 316 | 4.65 ± 0.22 | -0.53 ± 0.35 | <67 | 3 | Y | N | Y | N | ... | ... | n | NG |
| 49489 | 18274331+0625100 | -33.9 ± 0.3 | 3592 ± 124 | 4.45 ± 0.08 | -0.06 ± 0.13 | 72 ± 48 | 1 | Y | N | Y | Y | N | ... | n | NG |
| 49490 | 18274334+0624179 | 21.7 ± 0.3 | 4918 ± 178 | 2.92 ± 0.50 | -0.34 ± 0.24 | ... | ... | N | ... | ... | ... | ... | ... | n | G |
| 49491 | 18274340+0633314 | -6.7 ± 0.2 | 4998 ± 111 | 4.43 ± 0.27 | 0.11 ± 0.15 | ... | ... | Y | ... | ... | ... | ... | ... | n | ... |
| 49493 | 18274364+0637282 | 10.8 ± 0.3 | 4762 ± 241 | 2.60 ± 0.61 | -0.24 ± 0.21 | ... | ... | N | ... | ... | ... | ... | ... | n | G |
| 49494 | 18274367+0638498 | -52.7 ± 0.4 | 5014 ± 130 | 2.83 ± 0.52 | -0.26 ± 0.19 | ... | ... | N | ... | ... | ... | ... | ... | n | G |
| 49495 | 18274382+0639395 | -32.7 ± 0.3 | 4606 ± 96 | 4.60 ± 0.42 | -0.45 ± 0.6 | ... | ... | Y | ... | ... | ... | N | ... | n | ... |
| 49496 | 18274383+0635125 | 9.9 ± 0.3 | 4810 ± 225 | 2.54 ± 0.60 | -0.21 ± 0.22 | ... | ... | N | ... | ... | ... | ... | ... | n | G |
| 49497 | 18274399+0627020 | -41.7 ± 0.2 | 4273 ± 269 | 4.83 ± 0.32 | -0.16 ± 0.17 | <16 | 3 | Y | N | Y | Y | N | ... | n | NG |
| 49498 | 18274447+0635548 | -12.8 ± 0.2 | 4882 ± 181 | 3.18 ± 0.26 | -0.25 ± 0.2 | ... | ... | Y | ... | ... | ... | ... | ... | n | G |
| 49499 | 18274451+0645529 | -10.6 ± 0.4 | 5713 ± 27 | 3.94 ± 0.25 | -0.79 ± 0.59 | ... | ... | Y | ... | ... | ... | ... | ... | n | ... |
| 49500 | 18274462+0625467 | -41.2 ± 0.2 | 5365 ± 148 | 4.28 ± 0.22 | 0.30 ± 0.17 | ... | ... | Y | ... | ... | ... | ... | ... | n | ... |
| 49501 | 18274469+0634151 | 106.6 ± 0.3 | 4925 ± 128 | 3.57 ± 0.37 | 0.11 ± 0.19 | ... | ... | N | ... | ... | ... | ... | ... | n | ... |
| 49502 | 18274478+0622537 | -31.8 ± 0.2 | 4459 ± 154 | 4.70 ± 0.30 | 0.08 ± 0.18 | <18 | 3 | Y | Y | Y | Y | Y | ... | Y | ... |
| 49503 | 18274483+0631281 | -51.7 ± 0.2 | 4790 ± 120 | 3.13 ± 0.28 | -0.02 ± 0.14 | <29 | 3 | N | .. | ... | ... | ... | ... | n | G |
| 49504 | 18274504+0635336 | 120.6 ± 0.2 | 4929 ± 109 | 2.76 ± 0.49 | -0.41 ± 0.19 | ... | ... | N | ... | ... | ... | ... | ... | n | G |
| 49505 | 18274522+0627295 | 44.8 ± 0.2 | 4635 ± 221 | 2.70 ± 0.47 | -0.08 ± 0.27 | ... | ... | N | ... | ... | ... | ... | ... | n | G |
| 49506 | 18274530+0622552 | 57.3 ± 0.2 | 4712 ± 87 | 3.05 ± 0.51 | 0.32 ± 0.3 | <39 | 3 | N | .. | ... | ... | ... | ... | n | G |
| 49507 | 18274543+0634466 | -29.2 ± 0.3 | 4383 ± 304 | 4.62 ± 0.39 | 0.30 ± 0.21 | ... | ... | Y | ... | ... | ... | ... | ... | n | ... |
| 49508 | 18274544+0637141 | 65.8 ± 0.2 | 4832 ± 175 | 2.56 ± 0.44 | -0.53 ± 0.22 | ... | ... | N | ... | ... | ... | ... | ... | n | G |
| 49509 | 18274549+0636360 | -28.0 ± 0.2 | 5593 ± 110 | 4.28 ± 0.40 | -0.09 ± 0.14 | 49 ± 13 | 1 | Y | Y | Y | Y | Y | Y | Y | ... |
| 49510 | 18274566+0632165 | 50.9 ± 0.3 | 3757 ± 159 | 4.48 ± 0.24 | -0.02 ± 0.13 | ... | ... | N | ... | ... | ... | ... | ... | n | ... |
| 49511 | 18274572+0640267 | 27.8 ± 0.2 | 6244 ± 91 | 4.16 ± 0.07 | 0.12 ± 0.15 | ... | ... | N | ... | ... | ... | ... | ... | n | ... |
| 49512 | 18274582+0623053 | 28.1 ± 0.2 | 4476 ± 168 | 2.53 ± 0.58 | -0.06 ± 0.27 | ... | ... | N | ... | ... | ... | ... | ... | n | G |
| 49513 | 18274617+0634441 | 1.7 ± 0.3 | 4672 ± 209 | 4.72 ± 0.25 | 0.08 ± 0.16 | ... | ... | Y | ... | ... | ... | ... | ... | n | ... |
| 49514 | 18274622+0644485 | 6.0 ± 0.2 | 4752 ± 126 | 2.85 ± 0.44 | 0.13 ± 0.2 | ... | ... | N | ... | ... | ... | ... | ... | n | G |
| 49515 | 18274631+0633054 | 8.4 ± 0.2 | 4933 ± 79 | 3.29 ± 0.36 | 0.28 ± 0.21 | ... | ... | N | ... | ... | ... | ... | ... | n | G |
| 49516 | 18274650+0645393 | -45.5 ± 0.2 | 4512 ± 181 | 2.62 ± 0.38 | 0.20 ± 0.25 | ... | ... | Y | ... | ... | ... | ... | ... | n | G |
| 49517 | 18274663+0630061 | 31.9 ± 0.2 | 4734 ± 129 | 2.94 ± 0.36 | 0.18 ± 0.2 | ... | ... | N | ... | ... | ... | ... | ... | n | G |
| 49518 | 18274666+0635591 | -34.9 ± 0.3 | 4651 ± 53 | 2.68 ± 0.28 | 0.06 ± 0.17 | ... | ... | Y | ... | ... | ... | ... | ... | n | G |
| 49519 | 18274670+0629072 | 84.4 ± 0.2 | 4782 ± 84 | 2.71 ± 0.28 | 0.11 ± 0.23 | ... | ... | N | ... | ... | ... | ... | ... | n | G |
| 49520 | 18274680+0635258 | -2.8 ± 0.2 | 6128 ± 7 | 3.84 ± 0.03 | -0.25 ± 0.34 | ... | ... | Y | ... | ... | ... | N | ... | n | ... |
| 49521 | 18274690+0636396 | -17.7 ± 0.3 | 4535 ± 267 | 4.42 ± 0.56 | 0.40 ± 0.3 | <115 | 3 | Y | N | Y | N | ... | ... | n | NG |
| 49522 | 18274720+0635066 | -39.2 ± 0.3 | 5861 ± 98 | 3.99 ± 0.16 | 0.09 ± 0.14 | ... | ... | Y | ... | ... | ... | ... | ... | n | ... |
| 49523 | 18274727+0636240 | -49.2 ± 0.2 | 4746 ± 203 | 3.03 ± 0.45 | -0.06 ± 0.15 | ... | ... | N | ... | ... | ... | ... | ... | n | G |
| 49524 | 18274739+0637129 | 3.4 ± 0.2 | 4324 ± 206 | 4.54 ± 0.14 | -0.03 ± 0.13 | ... | ... | Y | ... | ... | ... | N | ... | n | ... |
| 49525 | 18274741+0622388 | 0.2 ± 0.3 | 5516 ± 131 | 3.97 ± 0.24 | -0.37 ± 0.17 | <24 | 3 | Y?$^f$ | Y | Y | Y | ... | ... | n | NG |
| 49526 | 18274747+0632037 | 23.3 ± 0.4 | 4187 ± 282 | 4.73 ± 0.16 | -0.42 ± 0.4 | <115 | 3 | Y | .. | ... | ... | N | ... | n | NG |
| 49527 | 18274760+0638249 | -23.1 ± 0.2 | 5030 ± 88 | 3.69 ± 0.43 | 0.26 ± 0.2 | <42 | 3 | Y | Y | Y? | Y | ... | ... | Y | ... |
| 49528 | 18274764+0628186 | -5.4 ± 0.3 | 5097 ± 44 | 3.16 ± 0.50 | -0.31 ± 0.2 | ... | ... | Y | ... | ... | ... | ... | ... | n | G |
| 49529 | 18274770+0624316 | 12.7 ± 0.3 | 4843 ± 218 | 2.64 ± 0.60 | -0.22 ± 0.21 | ... | ... | N | ... | ... | ... | ... | ... | n | G |

**Table C.11.** continued.

| ID | CNAME | RV (km s$^{-1}$) | $T_{\text{eff}}$ (K) | logg (dex) | [Fe/H] (dex) | EW(Li)$^a$ (mÅ) | EW(Li) error flag$^b$ | Membership RV | Li | logg | [Fe/H] | Gaia studies Randich$^c$ | Cantat-Gaudin$^c$ | Final$^d$ | NMs with Li$^e$ |
|---|---|---|---|---|---|---|---|---|---|---|---|---|---|---|---|
| 49530 | 18274776+0635400 | -3.1 ± 0.2 | 4861 ± 166 | 2.90 ± 0.35 | -0.10 ± 0.18 | … | … | Y | … | … | … | … | … | n | G |
| 49531 | 18274776+0645375 | 2.0 ± 0.3 | 4378 ± 336 | 4.67 ± 0.27 | -0.18 ± 0.17 | … | … | Y | … | … | … | … | … | n | … |
| 49532 | 18274784+0632468 | -12.2 ± 0.2 | 5944 ± 174 | 4.18 ± 0.14 | 0.17 ± 0.13 | … | … | Y | … | … | … | … | … | n | … |
| 49533 | 18274795+0642268 | 70.3 ± 0.2 | 5984 ± 132 | 4.17 ± 0.09 | 0.18 ± 0.13 | 104 ± 24 | 1 | N | .. | … | … | … | … | n | NG |
| 49534 | 18274795+0643073 | 62.0 ± 0.2 | 4800 ± 119 | 2.97 ± 0.33 | 0.14 ± 0.21 | … | … | N | … | … | … | … | … | n | G |
| 49535 | 18274802+0630147 | 40.3 ± 0.3 | 4806 ± 164 | 2.55 ± 0.38 | -0.43 ± 0.27 | … | … | N | … | … | … | … | … | n | G |
| 49536 | 18274809+0625330 | -111.8 ± 0.3 | 4774 ± 267 | 2.83 ± 0.53 | -0.31 ± 0.14 | … | … | N | … | … | … | … | … | n | G |
| 49537 | 18274824+0645391 | 6.5 ± 0.3 | 6235 ± 166 | 4.39 ± 0.45 | 0.07 ± 0.16 | … | … | N | … | … | … | … | … | n | … |
| 49538 | 18274852+0631344 | 3.1 ± 0.3 | 5560 ± 184 | 4.45 ± 0.38 | -0.29 ± 0.19 | <18 | 3 | N | .. | … | … | … | … | n | NG |
| 49539 | 18274858+0634381 | -45.4 ± 0.2 | 5956 ± 161 | 4.27 ± 0.16 | 0.30 ± 0.18 | 76 ± 13 | … | Y | Y | Y | Y | N | … | Y?$^g$ | … |
| 49540 | 18274859+0624302 | 21.5 ± 0.3 | 5051 ± 302 | 2.96 ± 0.83 | -0.58 ± 0.19 | … | … | N | … | … | … | … | … | n | G |
| 49541 | 18274871+0626105 | -40.0 ± 0.7 | 5360 ± 341 | 4.35 ± 0.56 | 0.09 ± 0.17 | … | … | Y | … | … | … | N | … | n | … |
| 49542 | 18274880+0622338 | -16.0 ± 0.3 | 4676 ± 229 | 4.57 ± 0.27 | -0.02 ± 0.14 | … | … | Y | … | … | … | … | … | n | … |
| 49543 | 18274881+0628500 | -15.5 ± 0.3 | 5128 ± 127 | 4.47 ± 0.35 | 0.23 ± 0.19 | <60 | 3 | Y | Y | Y | Y | … | … | Y | … |
| 49544 | 18274926+0638189 | 32.5 ± 0.2 | 4675 ± 299 | 2.84 ± 0.40 | -0.05 ± 0.24 | … | … | N | … | … | … | … | … | n | G |
| 49545 | 18274931+0643372 | 18.1 ± 0.2 | 5775 ± 142 | 4.29 ± 0.11 | 0.24 ± 0.13 | … | … | N | … | … | … | … | … | n | … |
| 49546 | 18274935+0634023 | 7.7 ± 0.3 | 6119 ± 76 | 4.26 ± 0.16 | -0.03 ± 0.13 | … | … | N | … | … | … | … | … | n | … |
| 49547 | 18274944+0625040 | 39.1 ± 0.2 | 4904 ± 177 | 2.75 ± 0.49 | -0.36 ± 0.22 | … | … | N | … | … | … | … | … | n | G |
| 49548 | 18274964+0629212 | 6.8 ± 0.4 | 4349 ± 197 | 4.59 ± 0.17 | 0.02 ± 0.15 | … | … | N | … | … | … | N | … | n | … |
| 49549 | 18274971+0633520 | 58.3 ± 0.2 | 4833 ± 179 | 2.99 ± 0.40 | -0.35 ± 0.21 | … | … | N | … | … | … | … | … | n | G |
| 49550 | 18274990+0625245 | -3.0 ± 0.2 | 6199 ± 97 | 4.09 ± 0.20 | -0.08 ± 0.12 | … | … | Y | … | … | … | … | … | n | … |
| 49551 | 18274995+0622548 | -30.3 ± 0.2 | 4224 ± 252 | 2.15 ± 0.48 | 0.00 ± 0.2 | <34 | 3 | Y | N | Y | Y | … | … | n | G |
| 49552 | 18274995+0630580 | 45.6 ± 0.3 | 5762 ± 190 | 3.97 ± 0.23 | -0.08 ± 0.14 | <62 | 3 | N | .. | … | … | … | … | n | NG |
| 49553 | 18274995+0635457 | -11.4 ± 0.4 | 3706 ± 440 | 4.28 ± 0.11 | -0.63 ± 0.47 | … | … | Y | … | … | … | … | … | n | … |
| 49554 | 18274998+0629304 | -20.0 ± 0.3 | 4853 ± 158 | 2.98 ± 0.18 | -0.31 ± 0.14 | … | … | Y | … | … | … | … | … | n | G |
| 49555 | 18275065+0630556 | 2.5 ± 0.2 | 5562 ± 72 | 4.19 ± 0.27 | 0.10 ± 0.15 | … | … | Y | … | … | … | … | … | n | … |
| 49556 | 18275069+0623038 | -0.7 ± 0.2 | 5848 ± 138 | 4.04 ± 0.02 | 0.35 ± 0.16 | 82 ± 26 | 1 | Y | Y | Y | N | … | … | Y | … |
| 49557 | 18275076+0624423 | 33.7 ± 0.3 | 4960 ± 143 | 2.81 ± 0.46 | -0.38 ± 0.24 | … | … | N | … | … | … | … | … | n | G |
| 49558 | 18275085+0628584 | -16.1 ± 0.2 | 4704 ± 90 | 2.52 ± 0.25 | 0.16 ± 0.25 | … | … | Y | … | … | … | … | … | n | G |
| 49559 | 18275090+0641031 | 37.4 ± 0.3 | 4488 ± 148 | 2.60 ± 0.32 | 0.12 ± 0.13 | … | … | N | … | … | … | … | … | n | G |
| 49560 | 18275114+0639515 | 4.5 ± 0.2 | 4923 ± 154 | 2.75 ± 0.42 | -0.13 ± 0.18 | … | … | N | … | … | … | … | … | n | G |
| 49561 | 18275133+0643477 | 5.8 ± 0.4 | 5431 ± 280 | 3.88 ± 0.41 | 0.08 ± 0.15 | <84 | 3 | N | .. | … | … | … | … | n | NG |
| 49562 | 18275172+0638302 | 24.7 ± 0.3 | 4772 ± 166 | 3.57 ± 0.53 | 0.12 ± 0.18 | … | … | N | … | … | … | … | … | n | … |
| 49563 | 18275180+0640487 | -33.3 ± 0.4 | 5722 ± 96 | 4.08 ± 0.17 | 0.20 ± 0.13 | … | … | Y | … | … | … | … | … | n | … |
| 49564 | 18275187+0624499 | -24.7 ± 0.3 | 6543 ± 235 | 3.88 ± 0.06 | 0.13 ± 0.2 | 61 ± 21 | 1 | Y | Y | Y | Y | … | … | Y | … |
| 49565 | 18275187+0642487 | 46.3 ± 0.2 | 4922 ± 148 | 3.31 ± 0.23 | 0.10 ± 0.16 | … | … | N | … | … | … | … | … | n | G |
| 49566 | 18275189+0629321 | -3.6 ± 0.2 | 4858 ± 149 | 3.11 ± 0.33 | 0.18 ± 0.23 | … | … | Y | … | … | … | … | … | n | G |
| 49567 | 18275190+0624166 | -17.3 ± 0.2 | 5145 ± 41 | 4.61 ± 0.31 | 0.17 ± 0.17 | … | … | Y | … | … | … | … | … | n | … |
| 49568 | 18275205+0633413 | -17.0 ± 0.2 | 5639 ± 64 | 4.16 ± 0.25 | 0.13 ± 0.15 | … | … | Y | … | … | … | … | … | n | … |
| 49569 | 18275213+0641591 | 20.9 ± 0.3 | 6384 ± 204 | 4.31 ± 0.30 | 0.14 ± 0.23 | … | … | N | … | … | … | … | … | n | … |
| 49570 | 18275227+0631479 | -63.7 ± 0.3 | 4460 ± 182 | 4.48 ± 0.39 | 0.23 ± 0.22 | … | … | N | … | … | … | … | … | n | … |
| 49571 | 18275247+0637162 | 11.0 ± 0.2 | 5729 ± 157 | 4.55 ± 0.19 | -0.05 ± 0.19 | … | … | N | … | … | … | … | … | n | … |
| 49572 | 18275252+0630342 | 65.0 ± 0.2 | 4676 ± 156 | 2.52 ± 0.32 | 0.00 ± 0.15 | … | … | N | … | … | … | … | … | n | G |
| 49573 | 18275259+0628189 | 62.6 ± 0.4 | 4840 ± 304 | 3.58 ± 0.52 | -0.58 ± 0.5 | 48 ± 46 | 1 | N | .. | … | … | … | … | n | NG |
| 49574 | 18275260+0635045 | 12.2 ± 0.4 | 4010 ± 303 | 4.56 ± 0.08 | -0.57 ± 0.44 | <80 | 3 | N | .. | … | … | N | … | n | NG |
| 49575 | 18275262+0638186 | 7.7 ± 0.3 | 4675 ± 112 | 2.46 ± 0.39 | -0.16 ± 0.19 | … | … | N | … | … | … | … | … | n | G |
| 49576 | 18275270+0622319 | 25.5 ± 0.2 | 4717 ± 158 | 2.64 ± 0.30 | 0.10 ± 0.26 | … | … | N | … | … | … | … | … | n | G |
| 49577 | 18275290+0637378 | 26.7 ± 0.3 | 5095 ± 123 | 3.82 ± 0.46 | 0.05 ± 0.12 | … | … | N | … | … | … | … | … | n | … |
| 49578 | 18275296+0632110 | 64.9 ± 0.2 | 4610 ± 136 | 2.46 ± 0.26 | 0.09 ± 0.21 | … | … | N | … | … | … | … | … | n | G |
| 49579 | 18275296+0634467 | 14.2 ± 0.3 | 4935 ± 137 | 3.13 ± 0.36 | -0.20 ± 0.21 | … | … | N | … | … | … | … | … | n | … |
| 49580 | 18275299+0645423 | 30.4 ± 0.3 | 4491 ± 266 | 2.55 ± 0.52 | 0.00 ± 0.29 | … | … | N | … | … | … | … | … | n | G |
| 49581 | 18275301+0627189 | 9.6 ± 0.2 | 4867 ± 85 | 2.88 ± 0.33 | 0.11 ± 0.2 | … | … | N | … | … | … | … | … | n | G |
| 49582 | 18275303+0626583 | 53.6 ± 0.3 | 5721 ± 153 | 4.18 ± 0.21 | 0.06 ± 0.13 | … | … | N | … | … | … | … | … | n | … |
| 49583 | 18275316+0639321 | 1.5 ± 0.3 | 6155 ± 185 | 4.11 ± 0.14 | 0.15 ± 0.25 | … | … | Y | … | … | … | … | … | n | … |
| 49584 | 18275327+0638524 | -38.6 ± 0.3 | 4008 ± 221 | 4.60 ± 0.10 | -0.03 ± 0.22 | … | … | Y | … | … | … | N | … | n | … |
| 49585 | 18275340+0625217 | 41.1 ± 0.2 | 4814 ± 170 | 3.05 ± 0.32 | -0.01 ± 0.2 | … | … | N | … | … | … | … | … | n | G |
| 49586 | 18275351+0639030 | 160.2 ± 0.2 | 4775 ± 137 | 2.61 ± 0.30 | -0.09 ± 0.2 | … | … | N | … | … | … | … | … | n | G |
| 49587 | 18275356+0622295 | -23.3 ± 0.3 | 4362 ± 208 | 4.61 ± 0.26 | -0.07 ± 0.13 | <41 | 3 | Y | N | Y | Y | N | … | n | NG |
| 49588 | 18275362+0636386 | -20.8 ± 0.4 | 5561 ± 34 | 4.07 ± 0.38 | -0.41 ± 0.49 | … | … | Y | … | … | … | … | … | n | … |

**Table C.11.** continued.

| ID | CNAME | RV (km s$^{-1}$) | $T_{\rm eff}$ (K) | logg (dex) | [Fe/H] (dex) | EW(Li)$^a$ (mÅ) | EW(Li) error flag$^b$ | Membership RV | Li | logg | [Fe/H] | Gaia studies Randich$^c$ | Cantat-Gaudin$^c$ | Final$^d$ | NMs with Li$^e$ |
|---|---|---|---|---|---|---|---|---|---|---|---|---|---|---|---|
| 49589 | 18275370+0640057 | 10.7 ± 0.3 | 5937 ± 116 | 4.14 ± 0.10 | 0.42 ± 0.22 | … | … | N | … | … | … | … | … | n | … |
| 49590 | 18275392+0644327 | 6.7 ± 0.3 | 4612 ± 194 | 2.52 ± 0.26 | 0.01 ± 0.26 | … | … | N | … | … | … | … | … | n | G |
| 49591 | 18275395+0624538 | -10.9 ± 0.3 | 4703 ± 218 | 4.87 ± 0.25 | -0.23 ± 0.19 | … | … | Y | … | … | … | … | … | n | … |
| 49592 | 18275404+0640437 | -38.9 ± 0.5 | 5272 ± 336 | 3.47 ± 0.33 | -0.34 ± 0.15 | … | … | Y | … | … | … | N | … | n | … |
| 49593 | 18275406+0634250 | 48.8 ± 0.3 | 5236 ± 128 | 3.60 ± 0.58 | -0.25 ± 0.16 | … | … | N | … | … | … | … | … | n | … |
| 49594 | 18275417+0639405 | 126.2 ± 0.2 | 4587 ± 142 | 2.54 ± 0.41 | -0.15 ± 0.19 | … | … | N | … | … | … | … | … | n | G |
| 49595 | 18275441+0637088 | -111.0 ± 0.3 | 4419 ± 331 | 2.56 ± 0.63 | 0.13 ± 0.31 | … | … | N | … | … | … | … | … | n | G |
| 49596 | 18275446+0640513 | 5.3 ± 0.6 | 6964 ± 603 | 4.50 ± 0.38 | 0.14 ± 0.19 | … | … | N | … | … | … | … | … | n | … |
| 49597 | 18275453+0629104 | 19.3 ± 0.2 | 4225 ± 247 | 4.83 ± 0.32 | -0.19 ± 0.2 | <13 | 3 | N | .. | … | … | N | … | n | NG |
| 49598 | 18275478+0629572 | -24.8 ± 0.3 | 5934 ± 122 | 3.98 ± 0.29 | -0.09 ± 0.16 | 42 ± 35 | 1 | Y | Y | Y | Y | … | … | Y | … |
| 49599 | 18275479+0643304 | -21.2 ± 0.3 | 4922 ± 132 | 4.59 ± 0.39 | 0.15 ± 0.2 | … | … | Y | … | … | … | … | … | n | … |
| 49600 | 18275481+0632186 | -73.0 ± 0.2 | 5454 ± 110 | 4.42 ± 0.46 | 0.02 ± 0.18 | … | … | N | … | … | … | … | … | n | … |
| 49601 | 18275495+0625567 | 102.9 ± 0.4 | 4658 ± 400 | 3.17 ± 0.42 | -0.12 ± 0.27 | … | … | N | … | … | … | … | … | n | G |
| 49602 | 18275504+0643214 | 39.3 ± 0.2 | 4636 ± 182 | 2.56 ± 0.28 | 0.07 ± 0.22 | … | … | N | … | … | … | … | … | n | G |
| 49603 | 18275510+0622377 | 11.9 ± 0.4 | 4019 ± 302 | 4.69 ± 0.16 | -0.14 ± 0.28 | … | … | N | … | … | … | N | … | n | … |
| 49604 | 18275512+0639268 | 76.0 ± 0.3 | 4864 ± 239 | 3.50 ± 0.22 | -0.04 ± 0.18 | … | … | N | … | … | … | … | … | n | … |
| 49605 | 18275543+0635045 | -12.2 ± 0.3 | 4522 ± 224 | 2.58 ± 0.64 | -0.02 ± 0.15 | … | … | Y | … | … | … | … | … | n | G |
| 49606 | 18275567+0630430 | 32.4 ± 0.2 | 4482 ± 175 | 2.25 ± 0.45 | -0.15 ± 0.22 | … | … | N | … | … | … | … | … | n | G |
| 49607 | 18275591+0623310 | -13.3 ± 0.3 | 4844 ± 149 | 2.84 ± 0.34 | -0.10 ± 0.15 | … | … | Y | … | … | … | … | … | n | G |
| 49608 | 18275614+0637051 | -31.6 ± 0.3 | 5954 ± 194 | 4.44 ± 0.38 | -0.06 ± 0.18 | 64 ± 32 | 1 | Y | Y | Y | Y | … | … | Y | … |
| 49609 | 18275625+0644563 | -36.6 ± 0.2 | 4976 ± 121 | 4.50 ± 0.29 | 0.05 ± 0.13 | <26 | 3 | Y | Y | Y | Y | … | … | Y | … |
| 49610 | 18275642+0625496 | -17.1 ± 0.3 | 4900 ± 212 | 3.44 ± 0.31 | -0.18 ± 0.18 | … | … | Y | … | … | … | … | … | n | G |
| 49611 | 18275654+0624389 | 30.9 ± 0.3 | 4938 ± 125 | 3.54 ± 0.41 | 0.18 ± 0.18 | … | … | N | … | … | … | … | … | n | … |
| 49612 | 18275661+0639017 | 23.5 ± 0.3 | 3978 ± 244 | 4.63 ± 0.11 | -0.11 ± 0.41 | <44 | 3 | N | .. | … | … | N | … | n | NG |
| 49613 | 18275666+0628408 | 35.9 ± 0.3 | 4979 ± 136 | 4.42 ± 0.65 | 0.29 ± 0.23 | … | … | N | … | … | … | … | … | n | … |
| 49614 | 18275672+0626219 | -17.0 ± 0.2 | 4841 ± 154 | 3.16 ± 0.36 | 0.12 ± 0.19 | … | … | Y | … | … | … | … | … | n | G |
| 49615 | 18275683+0632574 | 24.7 ± 0.3 | 5124 ± 65 | 4.28 ± 0.59 | 0.26 ± 0.21 | … | … | N | … | … | … | … | … | n | … |
| 49616 | 18275683+0639364 | 28.8 ± 0.2 | 4452 ± 151 | 1.92 ± 0.34 | -0.37 ± 0.16 | … | … | N | … | … | … | … | … | n | G |
| 49617 | 18275693+0638207 | 51.6 ± 0.5 | 5664 ± 103 | 4.32 ± 0.22 | 0.11 ± 0.15 | … | … | N | … | … | … | … | … | n | … |
| 49618 | 18275707+0642127 | -29.8 ± 0.3 | 4293 ± 245 | 4.66 ± 0.11 | -0.64 ± 0.5 | … | … | Y | … | … | … | Y | … | n | … |
| 49619 | 18275713+0632348 | 27.4 ± 0.3 | 5115 ± 66 | 3.98 ± 0.16 | 0.11 ± 0.18 | 65 ± 38 | 1 | N | .. | … | … | … | … | n | NG |
| 49620 | 18275722+0625474 | -7.4 ± 0.2 | 4342 ± 187 | 2.14 ± 0.53 | -0.08 ± 0.27 | … | … | Y | … | … | … | … | … | n | G |
| 49621 | 18275727+0649306 | 73.9 ± 0.5 | 4716 ± 349 | 3.18 ± 0.42 | -0.18 ± 0.19 | … | … | N | … | … | … | … | … | n | G |
| 49622 | 18275733+0622427 | -49.7 ± 0.2 | 5472 ± 212 | 4.19 ± 0.35 | 0.05 ± 0.13 | … | … | N | … | … | … | … | … | n | … |
| 49623 | 18275737+0623492 | -0.1 ± 0.2 | 4517 ± 132 | 4.66 ± 0.26 | 0.02 ± 0.15 | … | … | Y | … | … | … | … | … | n | … |
| 49624 | 18275749+0631445 | 63.3 ± 0.2 | 4748 ± 231 | 2.58 ± 0.32 | -0.31 ± 0.19 | … | … | N | … | … | … | … | … | n | G |
| 49625 | 18275753+0633018 | 25.8 ± 0.3 | 4389 ± 192 | 4.59 ± 0.30 | 0.26 ± 0.26 | … | … | N | … | … | … | … | … | n | … |
| 49626 | 18275773+0626037 | 92.9 ± 0.2 | 4884 ± 177 | 2.79 ± 0.32 | -0.08 ± 0.18 | … | … | N | … | … | … | … | … | n | G |
| 49627 | 18275774+0625193 | -2.0 ± 0.3 | 5879 ± 79 | 4.17 ± 0.44 | -0.17 ± 0.15 | … | … | Y | … | … | … | … | … | n | … |
| 49628 | 18275776+0645437 | 19.1 ± 0.3 | 5924 ± 134 | 4.50 ± 0.24 | 0.06 ± 0.13 | 65 ± 29 | 1 | N | .. | … | … | … | … | n | NG |
| 49629 | 18275786+0635084 | 5.2 ± 0.2 | 5353 ± 52 | 4.28 ± 0.11 | 0.27 ± 0.18 | … | … | N | … | … | … | … | … | n | … |
| 49630 | 18275792+0648334 | 43.5 ± 0.2 | 4687 ± 234 | 2.79 ± 0.30 | -0.17 ± 0.2 | … | … | N | … | … | … | … | … | n | G |
| 49631 | 18275813+0646266 | -40.8 ± 0.3 | 4411 ± 227 | 4.68 ± 0.26 | 0.09 ± 0.16 | … | … | Y | … | … | … | N | … | n | … |
| 49632 | 18275832+0644291 | 31.1 ± 0.2 | 4884 ± 170 | 3.33 ± 0.20 | 0.00 ± 0.16 | … | … | N | … | … | … | … | … | n | G |
| 49633 | 18275833+0647570 | 2.8 ± 0.3 | 4938 ± 136 | 3.51 ± 0.38 | -0.27 ± 0.19 | … | … | Y | … | … | … | … | … | n | … |
| 49634 | 18275855+0644554 | 11.4 ± 0.5 | 4218 ± 149 | 4.88 ± 0.08 | -0.57 ± 0.25 | … | … | N | … | … | … | N | … | n | … |
| 49635 | 18275867+0631290 | 66.1 ± 0.3 | 4753 ± 252 | 2.86 ± 0.60 | -0.39 ± 0.24 | … | … | N | … | … | … | … | … | n | G |
| 49636 | 18275872+0644109 | 1.2 ± 0.2 | 4805 ± 273 | 2.81 ± 0.43 | -0.14 ± 0.22 | … | … | Y | … | … | … | … | … | n | G |
| 49637 | 18275877+0630479 | 41.6 ± 0.3 | 5020 ± 127 | 3.64 ± 0.27 | -0.13 ± 0.2 | … | … | N | … | … | … | … | … | n | … |
| 49638 | 18275880+0638206 | 73.3 ± 0.3 | 4660 ± 190 | 2.65 ± 0.36 | -0.04 ± 0.23 | … | … | N | … | … | … | … | … | n | G |
| 49639 | 18275880+0641213 | -110.2 ± 0.6 | 3961 ± 201 | 3.51 ± 0.24 | -0.38 ± 0.26 | … | … | N | … | … | … | … | … | n | … |
| 49640 | 18275889+0627137 | 5.1 ± 0.2 | 4742 ± 153 | 3.11 ± 0.46 | 0.20 ± 0.22 | … | … | N | … | … | … | … | … | n | G |
| 49641 | 18275896+0629050 | -29.5 ± 0.3 | 4973 ± 197 | 3.16 ± 0.46 | -0.74 ± 0.57 | <38 | 3 | Y | Y | Y | N | … | … | Y | … |
| 49642 | 18275900+0623351 | 17.5 ± 0.3 | 6329 ± 120 | 3.50 ± 0.44 | 0.22 ± 0.26 | … | … | N | … | … | … | … | … | n | … |
| 49643 | 18275904+0635514 | 1.8 ± 0.4 | 4218 ± 314 | 4.46 ± 0.07 | -0.09 ± 0.17 | <166 | 3 | Y?$^f$ | N | Y | Y | … | … | n | NG |
| 49644 | 18275924+0624406 | 6.8 ± 0.2 | 5836 ± 88 | 3.90 ± 0.29 | 0.28 ± 0.13 | … | … | N | … | … | … | … | … | n | … |
| 49645 | 18275924+0634086 | -27.6 ± 0.3 | 3882 ± 338 | 4.83 ± 0.24 | -0.58 ± 0.47 | <40 | 3 | Y | Y | Y | N | Y | Y | Y | … |
| 49646 | 18275924+0641357 | -28.9 ± 0.2 | 5906 ± 129 | 4.18 ± 0.20 | 0.02 ± 0.13 | 56 ± 20 | 1 | Y | Y | Y | Y | … | … | Y | … |
| 49647 | 18275932+0650379 | 0.1 ± 0.2 | 5540 ± 24 | 3.60 ± 0.58 | -0.77 ± 0.63 | … | … | Y | … | … | … | … | … | n | … |

**Table C.11.** continued.

| ID | CNAME | RV (km s$^{-1}$) | $T_{\rm eff}$ (K) | logg (dex) | [Fe/H] (dex) | EW(Li)$^a$ (mÅ) | EW(Li) error flag$^b$ | Membership RV | Li | logg | [Fe/H] | Gaia studies Randich$^c$ | Cantat-Gaudin$^c$ | Final$^d$ | NMs with Li$^e$ |
|---|---|---|---|---|---|---|---|---|---|---|---|---|---|---|---|
| 49648 | 18275942+0650069 | 36.8 ± 0.2 | 6015 ± 61 | 4.20 ± 0.06 | -0.09 ± 0.13 | 37 ± 22 | 1 | N | .. | … | … | … | … | n | NG |
| 49649 | 18275944+0638448 | 19.2 ± 0.3 | 4920 ± 252 | 3.34 ± 0.59 | -0.51 ± 0.27 | … | … | N | … | … | … | … | … | n | G |
| 49650 | 18275950+0648560 | -6.5 ± 0.3 | 4812 ± 221 | 3.26 ± 0.34 | 0.12 ± 0.22 | … | … | Y | … | … | … | … | … | n | G |
| 49652 | 18275964+0636583 | -50.2 ± 0.3 | 4928 ± 201 | 3.36 ± 0.29 | -0.27 ± 0.2 | … | … | N | … | … | … | … | … | n | G |
| 49653 | 18275970+0626170 | 32.8 ± 0.3 | 5122 ± 240 | 2.96 ± 0.74 | -0.55 ± 0.18 | … | … | N | … | … | … | … | … | n | Li-rich G |
| 49654 | 18275981+0650220 | 98.2 ± 0.3 | 4793 ± 188 | 3.56 ± 0.32 | -0.26 ± 0.26 | … | … | N | … | … | … | … | … | n | … |
| 49655 | 18275990+0628046 | -6.5 ± 0.5 | 4201 ± 223 | 4.80 ± 0.46 | -0.12 ± 0.21 | <83 | 3 | Y | N | N | Y | N | … | n | NG |
| 49656 | 18275993+0627301 | 12.9 ± 0.4 | 4779 ± 183 | 4.62 ± 0.22 | 0.04 ± 0.18 | … | … | N | … | … | … | … | … | n | … |
| 49657 | 18280023+0629553 | 73.8 ± 0.2 | 4436 ± 198 | 2.69 ± 0.40 | 0.33 ± 0.26 | … | … | N | … | … | … | … | … | n | … |
| 49658 | 18280069+0636199 | 90.1 ± 0.2 | 4683 ± 157 | 2.71 ± 0.48 | 0.11 ± 0.26 | … | … | N | … | … | … | … | … | n | G |
| 49659 | 18280082+0649521 | -18.1 ± 0.3 | 4106 ± 212 | 4.63 ± 0.17 | -0.45 ± 0.25 | … | … | Y | … | … | … | N | … | n | … |
| 49660 | 18280086+0643269 | 17.6 ± 0.3 | 6445 ± 167 | 4.19 ± 0.19 | -0.19 ± 0.14 | 39 ± 36 | 1 | N | .. | … | … | … | … | n | NG |
| 49661 | 18280087+0624197 | 39.5 ± 0.3 | 4882 ± 185 | 3.34 ± 0.39 | -0.04 ± 0.17 | … | … | N | … | … | … | … | … | n | G |
| 49662 | 18280095+0636162 | 14.8 ± 0.2 | 4226 ± 241 | 2.27 ± 0.51 | 0.20 ± 0.33 | 92 ± 26 | 1 | N | .. | … | … | … | … | n | G |
| 49663 | 18280100+0623144 | -27.8 ± 0.4 | 3546 ± 14 | 4.48 ± 0.10 | … | … | … | Y | … | … | … | … | … | n | … |
| 49664 | 18280117+0624430 | 81.1 ± 0.3 | 4758 ± 268 | 3.18 ± 0.43 | 0.23 ± 0.34 | … | … | N | … | … | … | … | … | n | G |
| 49665 | 18280134+0646231 | -11.0 ± 0.3 | 4771 ± 300 | 2.72 ± 0.50 | -0.36 ± 0.25 | … | … | Y | … | … | … | … | … | n | G |
| 49666 | 18280137+0628140 | -99.6 ± 0.3 | 4564 ± 207 | 2.70 ± 0.37 | 0.10 ± 0.16 | … | … | N | … | … | … | … | … | n | … |
| 49667 | 18280141+0632531 | 24.4 ± 0.3 | 5009 ± 130 | 3.77 ± 0.37 | -0.15 ± 0.24 | … | … | N | … | … | … | … | … | n | … |
| 49668 | 18280141+0646297 | 18.6 ± 0.3 | 5910 ± 63 | 3.94 ± 0.17 | -0.10 ± 0.13 | … | … | N | … | … | … | … | … | n | … |
| 49669 | 18280144+0625264 | -54.4 ± 0.3 | 5177 ± 40 | 3.60 ± 0.46 | -0.17 ± 0.13 | … | … | N | … | … | … | … | … | n | … |
| 49670 | 18280151+0641171 | 18.7 ± 0.3 | 4552 ± 185 | 3.19 ± 0.59 | 0.42 ± 0.32 | … | … | N | … | … | … | … | … | n | G |
| 49671 | 18280160+0624190 | 32.3 ± 0.4 | 3817 ± 216 | 4.71 ± 0.11 | -0.05 ± 0.12 | … | … | N | … | … | … | … | … | n | … |
| 49672 | 18280162+0638105 | -13.3 ± 0.2 | 4855 ± 175 | 2.68 ± 0.37 | -0.12 ± 0.18 | … | … | Y | … | … | … | … | … | n | G |
| 49673 | 18280175+0646535 | -10.0 ± 0.2 | 6194 ± 60 | 4.10 ± 0.28 | -0.11 ± 0.13 | … | … | Y | … | … | … | … | … | n | … |
| 49674 | 18280177+0644054 | 46.4 ± 0.3 | 4869 ± 150 | 2.93 ± 0.39 | 0.39 ± 0.27 | … | … | N | … | … | … | … | … | n | G |
| 49675 | 18280206+0630197 | 30.2 ± 0.3 | 4583 ± 217 | 2.53 ± 0.37 | -0.08 ± 0.17 | … | … | N | … | … | … | … | … | n | G |
| 49676 | 18280221+0638314 | 10.8 ± 0.3 | 6450 ± 228 | 4.25 ± 0.29 | 0.04 ± 0.26 | … | … | N | … | … | … | … | … | n | … |
| 49677 | 18280231+0633525 | 38.7 ± 0.2 | 4559 ± 97 | 2.18 ± 0.31 | -0.30 ± 0.15 | … | … | N | … | … | … | … | … | n | G |
| 49678 | 18280239+0634014 | 15.1 ± 0.3 | 4678 ± 206 | 4.83 ± 0.26 | 0.03 ± 0.13 | … | … | N | … | … | … | … | … | n | … |
| 49679 | 18280240+0640088 | -9.2 ± 0.2 | 4685 ± 119 | 3.03 ± 0.56 | 0.28 ± 0.28 | … | … | Y | … | … | … | … | … | n | G |
| 49680 | 18280276+0632345 | 13.3 ± 0.2 | 4010 ± 265 | 4.79 ± 0.23 | -0.34 ± 0.24 | <25 | 3 | N | .. | … | … | N | … | n | NG |
| 49681 | 18280277+0631355 | -12.3 ± 0.3 | 5235 ± 147 | 3.85 ± 0.26 | 0.18 ± 0.16 | … | … | Y | … | … | … | … | … | n | … |
| 49682 | 18280280+0638255 | -29.8 ± 0.2 | 5790 ± 77 | 4.41 ± 0.48 | -0.28 ± 0.16 | 57 ± 11 | 1 | Y | Y | Y | Y | Y | … | Y | … |
| 49683 | 18280284+0646449 | -21.1 ± 0.2 | 6094 ± 225 | 4.15 ± 0.09 | 0.05 ± 0.12 | 57 ± 17 | 1 | Y | Y | Y | Y | … | … | Y | … |
| 49684 | 18280285+0628430 | -49.1 ± 0.2 | 5776 ± 64 | 4.19 ± 0.24 | 0.08 ± 0.13 | … | … | N | … | … | … | … | … | n | … |
| 49685 | 18280299+0623095 | -15.8 ± 0.3 | 4739 ± 164 | 4.59 ± 0.34 | 0.14 ± 0.16 | <51 | 3 | Y | Y | Y | Y | N | … | Y?$^g$ | … |
| 49686 | 18280302+0630249 | 78.0 ± 0.2 | 4797 ± 280 | 2.49 ± 0.54 | -0.30 ± 0.26 | … | … | N | … | … | … | … | … | n | G |
| 49687 | 18280311+0636278 | 61.3 ± 0.3 | 5937 ± 143 | 4.17 ± 0.08 | 0.02 ± 0.15 | 45 ± 23 | 1 | N | .. | … | … | … | … | n | NG |
| 49688 | 18280317+0631543 | 47.2 ± 0.3 | 4376 ± 224 | 2.32 ± 0.32 | 0.34 ± 0.24 | … | … | N | … | … | … | … | … | n | G |
| 49689 | 18280318+0641270 | 0.2 ± 0.2 | 5545 ± 128 | 4.23 ± 0.17 | 0.01 ± 0.13 | … | … | Y | … | … | … | N | … | n | … |
| 49690 | 18280328+0631150 | -78.4 ± 0.3 | 4925 ± 177 | 3.47 ± 0.39 | -0.17 ± 0.13 | <41 | 3 | N | .. | … | … | … | … | n | G |
| 49691 | 18280348+0623180 | -26.8 ± 0.3 | 5853 ± 204 | 4.30 ± 0.24 | -0.04 ± 0.15 | … | … | Y | … | … | … | … | … | n | … |
| 49692 | 18280369+0645315 | 2.2 ± 0.2 | 6153 ± 97 | 4.26 ± 0.25 | -0.01 ± 0.12 | 73 ± 20 | 1 | Y?$^f$ | Y | Y | Y | … | … | n | NG |
| 49693 | 18280397+0638585 | -56.7 ± 0.2 | 4486 ± 227 | 4.80 ± 0.28 | -0.31 ± 0.17 | <40 | 3 | N | .. | … | … | N | … | n | NG |
| 49694 | 18280403+0623316 | -94.7 ± 0.4 | 4730 ± 283 | 2.45 ± 0.53 | -0.27 ± 0.22 | … | … | N | … | … | … | … | … | n | G |
| 49695 | 18280403+0639183 | -13.8 ± 0.2 | 4740 ± 247 | 2.78 ± 0.39 | -0.13 ± 0.2 | … | … | Y | … | … | … | … | … | n | G |
| 49696 | 18280424+0637258 | -4.5 ± 0.3 | 5106 ± 284 | 3.54 ± 0.56 | -0.56 ± 0.26 | … | … | Y | … | … | … | … | … | n | … |
| 49697 | 18280431+0628418 | 19.1 ± 0.3 | 4155 ± 223 | 4.77 ± 0.19 | -0.04 ± 0.19 | <86 | 3 | N | .. | … | … | N | … | n | NG |
| 49698 | 18280458+0627038 | -5.5 ± 0.3 | 4917 ± 123 | 3.16 ± 0.36 | -0.08 ± 0.13 | … | … | Y | … | … | … | … | … | n | G |
| 49699 | 18280463+0637018 | -10.0 ± 0.3 | 5681 ± 87 | 4.12 ± 0.15 | 0.07 ± 0.12 | 63 ± 36 | 1 | Y | Y | Y | Y | … | … | Y | … |
| 49700 | 18280479+0626432 | 33.3 ± 0.3 | 4850 ± 214 | 2.94 ± 0.56 | -0.29 ± 0.21 | … | … | N | … | … | … | … | … | n | G |
| 49701 | 18280493+0635554 | 7.4 ± 0.3 | 4582 ± 338 | 4.59 ± 0.35 | 0.05 ± 0.16 | … | … | N | … | … | … | N | … | n | … |
| 49702 | 18280494+0623471 | -182.3 ± 0.3 | 5024 ± 190 | 2.12 ± 0.54 | -0.60 ± 0.34 | … | … | N | … | … | … | … | … | n | G |
| 49703 | 18280512+0631340 | 69.1 ± 0.2 | 4899 ± 131 | 3.34 ± 0.30 | 0.11 ± 0.17 | … | … | N | … | … | … | … | … | n | G |
| 49704 | 18280514+0623119 | -58.6 ± 0.3 | 3674 ± 123 | 4.41 ± 0.11 | -0.06 ± 0.12 | <61 | 3 | N | .. | … | … | N | … | n | NG |
| 49705 | 18280528+0624320 | 157.1 ± 0.2 | 4316 ± 196 | 2.24 ± 0.50 | -0.05 ± 0.26 | … | … | N | … | … | … | … | … | n | G |
| 49706 | 18280533+0648467 | 42.1 ± 0.3 | 4596 ± 160 | 2.96 ± 0.49 | 0.16 ± 0.2 | … | … | N | … | … | … | … | … | n | G |
| 49707 | 18280543+0628144 | -54.3 ± 0.3 | 4137 ± 340 | 4.84 ± 0.38 | -0.15 ± 0.24 | <86 | 3 | N | .. | … | … | N | … | n | NG |

**Table C.11.** continued.

| ID | CNAME | RV (km s$^{-1}$) | $T_{\text{eff}}$ (K) | logg (dex) | [Fe/H] (dex) | EW(Li)$^a$ (mÅ) | EW(Li) error flag$^b$ | Membership RV | Li | logg | [Fe/H] | Gaia studies Randich$^c$ | Cantat-Gaudin$^c$ | Final$^d$ | NMs with Li$^e$ |
|---|---|---|---|---|---|---|---|---|---|---|---|---|---|---|---|
| 49708 | 18280547+0632540 | -30.7 ± 0.2 | 4703 ± 126 | 2.96 ± 0.38 | 0.13 ± 0.15 | … | … | Y | … | … | … | … | … | n | G |
| 49709 | 18280549+0643269 | 70.9 ± 0.3 | 4737 ± 95 | 2.80 ± 0.33 | 0.25 ± 0.25 | <43 | 3 | N | .. | … | … | … | … | n | G |
| 49710 | 18280552+0636440 | 0.6 ± 0.3 | 5284 ± 63 | 3.68 ± 0.23 | -0.27 ± 0.14 | … | … | Y | … | … | … | … | … | n | … |
| 49711 | 18280557+0634374 | 2.8 ± 0.3 | 6120 ± 93 | 4.22 ± 0.12 | 0.03 ± 0.12 | … | … | Y | … | … | … | … | … | n | … |
| 49712 | 18280570+0623088 | -37.2 ± 0.3 | 4961 ± 159 | 3.29 ± 0.51 | 0.34 ± 0.23 | … | … | Y | … | … | … | … | … | n | G |
| 49713 | 18280573+0637317 | -30.6 ± 0.3 | 4569 ± 139 | 4.62 ± 0.28 | 0.19 ± 0.19 | … | … | Y | … | … | … | … | … | n | … |
| 49714 | 18280573+0637410 | 43.3 ± 0.3 | 4548 ± 202 | 4.63 ± 0.10 | -0.29 ± 0.2 | … | … | N | … | … | … | … | … | n | … |
| 49715 | 18280606+0636117 | 57.8 ± 0.3 | 5625 ± 133 | 4.16 ± 0.13 | -0.08 ± 0.15 | … | … | N | … | … | … | … | … | n | … |
| 49716 | 18280636+0630383 | -41.1 ± 0.2 | 4400 ± 225 | 2.47 ± 0.60 | -0.03 ± 0.26 | … | … | Y | … | … | … | … | … | n | G |
| 49717 | 18280638+0622535 | 90.9 ± 0.2 | 4803 ± 185 | 2.86 ± 0.29 | -0.45 ± 0.2 | … | … | N | … | … | … | … | … | n | G |
| 49718 | 18280648+0641078 | -160.7 ± 2.2 | 5550 ± 296 | 3.99 ± 0.74 | 0.02 ± 0.22 | … | … | N | … | … | … | … | … | n | … |
| 49719 | 18280652+0626324 | -18.3 ± 0.3 | 5736 ± 135 | 4.18 ± 0.22 | 0.27 ± 0.15 | … | … | Y | … | … | … | … | … | n | … |
| 49720 | 18280652+0635285 | 1.6 ± 0.2 | 5353 ± 137 | 4.18 ± 0.46 | 0.06 ± 0.12 | … | … | Y | … | … | … | … | … | n | … |
| 49721 | 18280664+0622507 | -249.1 ± 0.4 | 4161 ± 422 | 2.47 ± 1.02 | 0.18 ± 0.27 | … | … | N | … | … | … | … | … | n | G |
| 49722 | 18280666+0626492 | -88.2 ± 0.4 | 4708 ± 191 | 4.46 ± 0.28 | 0.00 ± 0.18 | <104 | 3 | N | .. | … | … | … | … | n | NG |
| 49723 | 18280670+0644596 | 24.4 ± 0.2 | 4599 ± 104 | 2.41 ± 0.27 | 0.16 ± 0.27 | … | … | N | … | … | … | … | … | n | … |
| 49724 | 18280672+0638317 | 8.4 ± 0.2 | 4389 ± 177 | 2.53 ± 0.51 | 0.15 ± 0.26 | <22 | 3 | N | … | … | … | … | … | n | … |
| 49725 | 18280678+0650469 | -23.6 ± 0.3 | 5359 ± 162 | 3.69 ± 0.24 | 0.04 ± 0.13 | <35 | 3 | Y | Y | Y | Y | … | … | Y | … |
| 49726 | 18280686+0644438 | -4.3 ± 0.2 | 4808 ± 226 | 2.88 ± 0.55 | -0.43 ± 0.27 | … | … | Y | … | … | … | … | … | n | G |
| 49727 | 18280697+0626379 | -45.2 ± 0.2 | 5588 ± 75 | 4.23 ± 0.31 | 0.03 ± 0.13 | … | … | Y | … | … | … | … | … | n | … |
| 49728 | 18280734+0623423 | -6.9 ± 0.3 | 4895 ± 185 | 2.90 ± 0.35 | -0.14 ± 0.17 | … | … | Y | … | … | … | … | … | n | G |
| 49729 | 18280745+0623486 | -13.7 ± 0.3 | 4180 ± 213 | 4.68 ± 0.34 | 0.01 ± 0.27 | <47 | 3 | Y | N | Y | Y | N | … | n | NG |
| 49730 | 18280753+0646567 | -35.9 ± 0.2 | 5014 ± 84 | 2.62 ± 0.51 | -0.37 ± 0.19 | … | … | Y | … | … | … | … | … | n | G |
| 49731 | 18280754+0639246 | 14.5 ± 0.3 | 7441 ± 226 | 4.01 ± 0.21 | 0.00 ± 0.17 | … | … | N | … | … | … | … | … | n | … |
| 49732 | 18280775+0631131 | 32.0 ± 0.2 | 4907 ± 168 | 3.12 ± 0.20 | -0.09 ± 0.16 | … | … | N | … | … | … | … | … | n | G |
| 49733 | 18280810+0629215 | 45.6 ± 0.3 | 5759 ± 150 | 4.15 ± 0.22 | -0.12 ± 0.16 | 56 ± 22 | 1 | N | .. | … | … | … | … | n | NG |
| 49734 | 18280825+0635172 | 26.7 ± 0.2 | 4743 ± 228 | 2.86 ± 0.32 | 0.06 ± 0.22 | 87 ± 18 | 1 | N | .. | … | … | … | … | n | G |
| 49735 | 18280826+0645348 | -35.6 ± 0.3 | 5575 ± 79 | 4.12 ± 0.25 | -0.07 ± 0.14 | … | … | Y | … | … | … | … | … | n | … |
| 49736 | 18280830+0628552 | -34.3 ± 0.3 | 5878 ± 230 | 4.29 ± 0.08 | 0.18 ± 0.14 | 38 ± 18 | 1 | Y | Y | Y | Y | … | … | Y | … |
| 49737 | 18280846+0622434 | -20.1 ± 0.3 | 4622 ± 248 | 4.66 ± 0.43 | 0.10 ± 0.19 | … | … | Y | … | … | … | … | … | n | … |
| 49738 | 18280856+0632000 | 37.0 ± 0.3 | 5001 ± 101 | 4.49 ± 0.30 | 0.07 ± 0.13 | … | … | N | … | … | … | … | … | n | … |
| 49739 | 18280860+0624591 | -7.4 ± 0.3 | 5455 ± 88 | 4.28 ± 0.18 | 0.06 ± 0.16 | … | … | Y | … | … | … | … | … | n | … |
| 49740 | 18280864+0633296 | -22.9 ± 0.3 | 5338 ± 203 | 4.44 ± 0.37 | 0.09 ± 0.15 | … | … | Y | … | … | … | … | … | n | … |
| 49741 | 18280867+0640178 | -43.5 ± 0.2 | 4510 ± 131 | 2.77 ± 0.42 | 0.34 ± 0.23 | … | … | Y | … | … | … | … | … | n | G |
| 49742 | 18280867+0646067 | 59.1 ± 0.2 | 4638 ± 242 | 2.73 ± 0.31 | -0.02 ± 0.2 | … | … | N | … | … | … | … | … | n | G |
| 49743 | 18280872+0631487 | -6.5 ± 0.2 | 4804 ± 206 | 2.67 ± 0.39 | -0.12 ± 0.18 | … | … | Y | … | … | … | … | … | n | G |
| 49744 | 18280894+0650469 | -0.4 ± 0.3 | 4580 ± 222 | 4.53 ± 0.33 | -0.05 ± 0.12 | … | … | Y | … | … | … | … | … | n | … |
| 49745 | 18280907+0631015 | -32.8 ± 0.3 | 4772 ± 259 | 3.03 ± 0.48 | -0.34 ± 0.24 | … | … | Y | … | … | … | … | … | n | G |
| 49746 | 18280920+0627070 | 3.1 ± 0.3 | 4584 ± 93 | 4.84 ± 0.35 | 0.01 ± 0.14 | … | … | N | … | … | … | … | … | n | … |
| 49747 | 18280930+0645115 | 28.5 ± 0.3 | 4737 ± 286 | 4.34 ± 0.41 | 0.17 ± 0.21 | <43 | 3 | N | .. | … | … | … | … | n | NG |
| 49748 | 18280945+0636420 | -35.9 ± 0.3 | 5727 ± 43 | 4.06 ± 0.26 | -0.33 ± 0.13 | 48 ± 19 | 1 | Y | Y | Y | Y | … | … | Y | … |
| 49749 | 18280957+0630089 | -0.5 ± 0.3 | 4900 ± 158 | 3.59 ± 0.34 | 0.07 ± 0.17 | … | … | Y | … | … | … | … | … | n | … |
| 49750 | 18280957+0632467 | -22.4 ± 0.3 | 4799 ± 71 | 4.65 ± 0.29 | 0.20 ± 0.22 | <81 | 3 | Y | N | Y | Y | … | … | n | NG |
| 49751 | 18280965+0630379 | 22.4 ± 0.2 | 4610 ± 64 | 2.63 ± 0.32 | 0.22 ± 0.22 | … | … | N | … | … | … | … | … | n | G |
| 49753 | 18280984+0650003 | 16.4 ± 0.3 | 4835 ± 95 | 3.63 ± 0.44 | 0.23 ± 0.19 | … | … | N | … | … | … | … | … | n | … |
| 49754 | 18280986+0648241 | 1.6 ± 0.3 | 4086 ± 218 | 4.66 ± 0.14 | 0.07 ± 0.2 | <22 | 3 | Y?$^f$ | Y | Y | Y | N | … | n | NG |
| 49755 | 18280987+0646547 | -46.2 ± 0.2 | 6058 ± 231 | 4.13 ± 0.10 | 0.18 ± 0.15 | 43 ± 16 | 1 | Y | Y | Y | Y | N | … | Y?$^g$ | … |
| 49756 | 18280999+0644537 | -23.7 ± 0.2 | 4576 ± 230 | 4.20 ± 0.67 | 0.30 ± 0.25 | <31 | 3 | Y | Y | Y | Y | … | … | Y | … |
| 49757 | 18281010+0623341 | -55.8 ± 2.2 | 4743 ± 263 | 3.33 ± 0.80 | -0.71 ± 0.27 | … | … | N | … | … | … | … | … | n | … |
| 49758 | 18281027+0649011 | 1.6 ± 0.3 | 3979 ± 327 | 4.82 ± 0.26 | -0.48 ± 0.38 | … | … | Y | … | … | … | N | … | n | … |
| 49759 | 18281032+0643102 | 17.5 ± 0.3 | 5089 ± 264 | 2.45 ± 0.39 | -1.10 ± 0.18 | … | … | N | … | … | … | … | … | n | G |
| 49760 | 18281054+0628581 | -3.6 ± 0.3 | 6285 ± 184 | 4.34 ± 0.38 | 0.00 ± 0.23 | … | … | Y | … | … | … | … | … | n | … |
| 49761 | 18281061+0636027 | -50.0 ± 0.4 | 5101 ± 54 | 4.16 ± 0.62 | -0.17 ± 0.17 | … | … | N | … | … | … | … | … | n | … |
| 49762 | 18281067+0638093 | 43.4 ± 0.3 | 5173 ± 104 | 3.62 ± 0.23 | -0.07 ± 0.15 | … | … | N | … | … | … | … | … | n | … |
| 49763 | 18281069+0628121 | 46.7 ± 0.3 | 4673 ± 278 | 2.99 ± 0.31 | -0.03 ± 0.18 | … | … | N | … | … | … | … | … | n | G |
| 49764 | 18281070+0641544 | -41.2 ± 0.3 | 4754 ± 206 | 2.60 ± 0.31 | -0.18 ± 0.18 | … | … | Y | … | … | … | … | … | n | G |
| 49765 | 18281078+0642217 | -23.6 ± 0.2 | 4804 ± 142 | 4.56 ± 0.41 | 0.03 ± 0.13 | <14 | 3 | Y | Y | Y | Y | N | … | Y?$^g$ | … |
| 49766 | 18281080+0636089 | 78.4 ± 0.2 | 4862 ± 111 | 3.16 ± 0.25 | -0.07 ± 0.15 | … | … | N | … | … | … | … | … | n | G |
| 49767 | 18281090+0646066 | 33.3 ± 0.3 | 4792 ± 217 | 3.19 ± 0.16 | -0.04 ± 0.21 | … | … | N | … | … | … | … | … | n | G |

**Table C.11.** continued.

| ID | CNAME | RV (km s$^{-1}$) | $T_{\rm eff}$ (K) | logg (dex) | [Fe/H] (dex) | EW(Li)$^a$ (mÅ) | EW(Li) error flag$^b$ | Membership RV | Li | logg | [Fe/H] | Gaia studies Randich$^c$ | Cantat-Gaudin$^c$ | Final$^d$ | NMs with Li$^e$ |
|---|---|---|---|---|---|---|---|---|---|---|---|---|---|---|---|
| 49768 | 18281095+0637173 | -89.3 ± 0.3 | 4740 ± 310 | 3.07 ± 0.88 | -0.39 ± 0.31 | … | … | N | … | … | … | … | … | n | G |
| 49769 | 18281103+0628057 | 39.4 ± 0.3 | 4724 ± 55 | 3.04 ± 0.46 | 0.15 ± 0.21 | … | … | N | … | … | … | … | … | n | G |
| 49770 | 18281139+0627396 | 2.3 ± 0.3 | 6185 ± 195 | 4.02 ± 0.22 | -0.10 ± 0.18 | 89 ± 26 | 1 | Y?$^f$ | Y | Y | Y | … | … | n | NG |
| 49771 | 18281148+0648293 | -0.7 ± 0.6 | 4974 ± 135 | 3.59 ± 0.77 | -0.18 ± 0.19 | … | … | Y | … | … | … | … | … | n | … |
| 49772 | 18281153+0638597 | 27.0 ± 0.3 | 4903 ± 129 | 3.39 ± 0.52 | 0.19 ± 0.16 | … | … | N | … | … | … | … | … | n | G |
| 49773 | 18281154+0634366 | -34.5 ± 0.3 | 5288 ± 82 | 4.68 ± 0.49 | 0.13 ± 0.14 | … | … | Y | … | … | … | … | … | n | … |
| 49774 | 18281161+0630433 | 90.4 ± 0.3 | 5062 ± 169 | 3.05 ± 0.58 | -0.41 ± 0.28 | <60 | 3 | N | .. | … | … | … | … | n | Li-rich G |
| 49775 | 18281162+0650028 | -70.2 ± 0.3 | 4088 ± 343 | 4.40 ± 0.24 | 0.02 ± 0.24 | … | … | N | … | … | … | N | … | n | … |
| 49776 | 18281170+0636324 | 66.3 ± 0.3 | 4791 ± 193 | 3.04 ± 0.40 | -0.64 ± 0.33 | 32 ± 15 | 1 | N | .. | … | … | … | … | n | G |
| 49777 | 18281195+0643264 | 7.5 ± 0.3 | 5013 ± 181 | 4.35 ± 0.58 | -0.17 ± 0.17 | … | … | N | … | … | … | … | … | n | … |
| 49778 | 18281197+0648436 | -17.6 ± 0.3 | 4856 ± 298 | 3.16 ± 0.64 | -0.56 ± 0.35 | … | … | Y | … | … | … | … | … | n | G |
| 49779 | 18281229+0633073 | 91.4 ± 0.3 | 4610 ± 188 | 3.20 ± 0.63 | 0.15 ± 0.18 | <72 | 3 | N | .. | … | … | … | … | n | G |
| 49780 | 18281231+0624089 | 119.5 ± 0.2 | 4579 ± 134 | 2.71 ± 0.47 | 0.21 ± 0.31 | … | … | N | … | … | … | … | … | n | G |
| 49781 | 18281243+0644492 | 25.8 ± 0.2 | 4626 ± 240 | 2.80 ± 0.41 | -0.03 ± 0.26 | … | … | N | … | … | … | … | … | n | G |
| 49782 | 18281247+0627405 | -4.6 ± 0.2 | 4841 ± 223 | 4.37 ± 0.47 | 0.05 ± 0.13 | … | … | Y | … | … | … | … | … | n | … |
| 49783 | 18281249+0623259 | -31.4 ± 0.3 | 5186 ± 32 | 4.53 ± 0.24 | -0.08 ± 0.13 | … | … | Y | … | … | … | … | … | n | … |
| 49784 | 18281251+0632529 | 72.8 ± 0.2 | 4564 ± 147 | 2.56 ± 0.34 | 0.21 ± 0.24 | … | … | N | … | … | … | … | … | n | … |
| 49785 | 18281255+0649343 | -59.5 ± 0.4 | 4290 ± 296 | 4.27 ± 0.90 | 0.05 ± 0.21 | … | … | N | … | … | … | N | … | n | … |
| 49786 | 18281270+0634306 | 55.0 ± 0.2 | 4906 ± 126 | 2.93 ± 0.18 | -0.14 ± 0.18 | … | … | N | … | … | … | … | … | n | G |
| 49787 | 18281288+0650425 | 38.1 ± 0.6 | 4418 ± 335 | 4.31 ± 0.48 | -0.56 ± 0.25 | … | … | N | … | … | … | … | … | n | … |
| 49788 | 18281295+0650249 | 4.7 ± 0.3 | 5073 ± 78 | 4.75 ± 0.38 | 0.26 ± 0.24 | <50 | 3 | N | .. | … | … | … | … | n | NG |
| 49789 | 18281296+0641173 | 17.5 ± 0.3 | 3653 ± 15 | 4.68 ± 0.13 | … | … | … | N | … | … | … | N | … | n | … |
| 49790 | 18281306+0650479 | -41.7 ± 0.3 | 5704 ± 211 | 4.31 ± 0.17 | 0.23 ± 0.15 | … | … | Y | … | … | … | … | … | n | … |
| 49791 | 18281307+0639042 | 19.0 ± 3.9 | 4912 ± 158 | 3.67 ± 0.74 | -1.20 ± 0.23 | … | … | N | … | … | … | … | … | n | … |
| 49792 | 18281312+0627187 | 64.3 ± 0.3 | 4836 ± 122 | 3.16 ± 0.41 | -0.05 ± 0.16 | … | … | N | … | … | … | … | … | n | G |
| 49793 | 18281316+0641335 | 4.3 ± 0.3 | 5003 ± 116 | 3.56 ± 0.22 | -0.19 ± 0.19 | … | … | N | … | … | … | … | … | n | … |
| 49794 | 18281330+0649473 | -45.9 ± 0.3 | 4875 ± 119 | 2.63 ± 0.30 | -0.12 ± 0.17 | … | … | Y | … | … | … | … | … | n | G |
| 49795 | 18281332+0643388 | 8.6 ± 0.3 | 5127 ± 42 | 4.37 ± 0.62 | 0.02 ± 0.15 | … | … | N | … | … | … | … | … | n | … |
| 49796 | 18281337+0648158 | -6.7 ± 0.8 | 5739 ± 290 | 3.73 ± 0.25 | -0.02 ± 0.42 | … | … | Y | … | … | … | N | … | n | … |
| 49797 | 18281346+0635304 | 26.5 ± 0.3 | 4840 ± 175 | 3.18 ± 0.25 | 0.03 ± 0.23 | … | … | N | … | … | … | … | … | n | G |
| 49798 | 18281364+0625053 | 75.8 ± 0.2 | 4848 ± 150 | 3.03 ± 0.34 | 0.06 ± 0.21 | … | … | N | … | … | … | … | … | n | G |
| 49799 | 18281391+0624537 | 21.0 ± 0.3 | 5739 ± 108 | 3.97 ± 0.15 | -0.35 ± 0.15 | … | … | N | … | … | … | … | … | n | … |
| 49800 | 18281395+0644317 | -29.2 ± 0.3 | 3795 ± 353 | 2.46 ± 0.16 | -0.04 ± 0.14 | <40 | 3 | Y | Y | Y? | Y | Y | … | Y | … |
| 49801 | 18281398+0636343 | -54.2 ± 0.3 | 4525 ± 314 | 2.18 ± 0.33 | -0.22 ± 0.19 | … | … | N | … | … | … | … | … | n | G |
| 49802 | 18281429+0647554 | 57.5 ± 0.3 | 4696 ± 275 | 2.89 ± 0.52 | -0.20 ± 0.2 | <74 | 3 | N | .. | … | … | … | … | n | G |
| 49803 | 18281437+0635126 | 15.7 ± 0.2 | 4882 ± 192 | 2.82 ± 0.40 | -0.09 ± 0.2 | … | … | N | … | … | … | … | … | n | G |
| 49804 | 18281448+0635536 | 16.7 ± 0.3 | 5523 ± 40 | 4.08 ± 0.20 | 0.19 ± 0.12 | … | … | N | … | … | … | … | … | n | … |
| 49805 | 18281492+0633396 | -14.7 ± 0.3 | 4600 ± 221 | 2.58 ± 0.34 | -0.14 ± 0.17 | … | … | Y | … | … | … | … | … | n | G |
| 49806 | 18281494+0643241 | -4.4 ± 0.2 | 4546 ± 159 | 2.56 ± 0.43 | -0.09 ± 0.21 | … | … | Y | … | … | … | … | … | n | G |
| 49807 | 18281499+0640157 | -11.0 ± 0.3 | 6239 ± 144 | 4.48 ± 0.55 | 0.02 ± 0.14 | … | … | Y | … | … | … | … | … | n | … |
| 49808 | 18281503+0637255 | -148.7 ± 0.3 | 4997 ± 96 | 2.74 ± 0.47 | -0.63 ± 0.27 | … | … | N | … | … | … | … | … | n | G |
| 49809 | 18281508+0642267 | -43.3 ± 0.2 | 5653 ± 143 | 4.32 ± 0.29 | 0.23 ± 0.13 | … | … | Y | … | … | … | … | … | n | … |
| 49810 | 18281514+0640323 | -33.7 ± 0.3 | 5043 ± 241 | 2.86 ± 0.35 | -0.65 ± 0.25 | … | … | Y | … | … | … | … | … | n | G |
| 49811 | 18281515+0649534 | 5.6 ± 0.3 | 5417 ± 334 | 3.78 ± 0.31 | -0.06 ± 0.17 | … | … | N | … | … | … | … | … | n | … |
| 49812 | 18281516+0638470 | -27.9 ± 0.2 | 6445 ± 202 | 4.28 ± 0.35 | -0.23 ± 0.13 | 49 ± 14 | 1 | Y | Y | Y | Y | Y | … | Y | … |
| 49813 | 18281537+0643037 | -58.3 ± 0.3 | 4811 ± 126 | 3.10 ± 0.60 | 0.27 ± 0.26 | … | … | N | … | … | … | … | … | n | G |
| 49814 | 18281539+0622463 | 96.3 ± 0.3 | 4728 ± 301 | 2.61 ± 0.47 | -0.23 ± 0.26 | … | … | N | … | … | … | … | … | n | G |
| 49815 | 18281550+0647328 | -20.9 ± 0.3 | 5926 ± 167 | 4.19 ± 0.23 | -0.19 ± 0.16 | 49 ± 31 | 1 | Y | Y | Y | Y | … | … | Y | … |
| 49816 | 18281556+0641413 | -21.2 ± 0.3 | 6189 ± 73 | 3.76 ± 0.19 | -0.02 ± 0.14 | 62 ± 19 | 1 | Y | Y | Y | Y | … | … | Y | … |
| 49817 | 18281559+0638057 | 31.4 ± 0.3 | 5013 ± 119 | 4.41 ± 0.28 | 0.22 ± 0.19 | … | … | N | … | … | … | … | … | n | … |
| 49818 | 18281582+0626174 | -51.3 ± 0.2 | 4857 ± 155 | 2.85 ± 0.41 | -0.11 ± 0.21 | … | … | N | … | … | … | … | … | n | G |
| 49819 | 18281639+0638473 | 35.3 ± 0.2 | 4363 ± 154 | 2.29 ± 0.48 | 0.01 ± 0.22 | … | … | N | … | … | … | … | … | n | G |
| 49820 | 18281640+0631129 | 37.4 ± 0.2 | 4683 ± 144 | 2.88 ± 0.38 | 0.16 ± 0.22 | … | … | N | … | … | … | … | … | n | … |
| 49821 | 18281640+0645121 | -27.2 ± 0.2 | 4709 ± 158 | 2.82 ± 0.50 | 0.16 ± 0.24 | … | … | Y | … | … | … | … | … | n | G |
| 49822 | 18281656+0638354 | -14.2 ± 0.3 | 4758 ± 221 | 4.34 ± 0.64 | 0.01 ± 0.12 | <72 | 3 | Y | N | Y | Y | … | … | n | NG |
| 49823 | 18281658+0649161 | -36.3 ± 0.2 | 4774 ± 172 | 2.82 ± 0.31 | -0.13 ± 0.17 | … | … | Y | … | … | … | … | … | n | … |
| 49824 | 18281674+0632042 | -64.9 ± 0.3 | 5099 ± 56 | 2.85 ± 0.42 | -0.42 ± 0.16 | … | … | N | … | … | … | … | … | n | G |
| 49825 | 18281691+0628241 | -55.8 ± 0.2 | 5564 ± 226 | 4.24 ± 0.24 | 0.17 ± 0.14 | … | … | N | … | … | … | … | … | n | … |
| 49826 | 18281697+0640049 | -124.0 ± 0.4 | 4969 ± 114 | 2.67 ± 0.39 | -0.32 ± 0.15 | … | … | N | … | … | … | … | … | n | G |

| ID | CNAME | RV (km s$^{-1}$) | $T_{\text{eff}}$ (K) | logg (dex) | [Fe/H] (dex) | EW(Li)$^a$ (mÅ) | EW(Li) error flag$^b$ | Membership RV | Li | logg | [Fe/H] | Gaia studies Randich$^c$ | Cantat-Gaudin$^c$ | Final$^d$ | NMs with Li$^e$ |
|---|---|---|---|---|---|---|---|---|---|---|---|---|---|---|---|
| 49827 | 18281699+0641157 | -74.8 ± 0.2 | 4735 ± 264 | 2.83 ± 0.45 | -0.38 ± 0.13 | … | … | N | … | … | … | … | … | n | G |
| 49828 | 18281709+0627532 | -28.8 ± 0.2 | 4918 ± 155 | 3.21 ± 0.28 | -0.18 ± 0.17 | … | … | Y | … | … | … | … | … | n | G |
| 49829 | 18281710+0644344 | -14.9 ± 0.3 | 5528 ± 77 | 4.25 ± 0.29 | -0.17 ± 0.15 | <20 | 3 | Y | Y | Y | Y | … | … | Y | … |
| 49830 | 18281717+0624060 | -68.0 ± 0.3 | 4777 ± 188 | 2.90 ± 0.51 | -0.47 ± 0.2 | … | … | N | … | … | … | … | … | n | G |
| 49831 | 18281745+0623516 | -181.8 ± 0.3 | 4826 ± 189 | 2.01 ± 0.63 | -1.22 ± 0.23 | … | … | N | … | … | … | … | … | n | G |
| 49832 | 18281777+0625068 | 104.7 ± 0.3 | 5194 ± 116 | 3.34 ± 0.81 | -0.59 ± 0.21 | … | … | N | … | … | … | … | … | n | G |
| 49833 | 18281792+0643065 | 38.0 ± 0.9 | 4503 ± 91 | 4.00 ± 0.22 | -0.15 ± 0.1 | … | … | N | … | … | … | … | … | n | … |
| 49834 | 18281796+0626267 | 30.1 ± 0.4 | 3993 ± 308 | 4.54 ± 0.02 | -0.22 ± 0.34 | … | … | N | … | … | … | … | … | n | … |
| 49835 | 18281814+0644426 | 49.3 ± 0.8 | 4233 ± 286 | 4.47 ± 0.34 | -0.66 ± 0.34 | <71 | 3 | N | … | … | … | … | … | n | NG |
| 49836 | 18281820+0629436 | 7.3 ± 0.4 | 5692 ± 72 | 4.02 ± 0.15 | 0.22 ± 0.14 | … | … | N | … | … | … | … | … | n | … |
| 49837 | 18281823+0627147 | 34.4 ± 0.2 | 4922 ± 284 | 2.91 ± 0.94 | -0.53 ± 0.21 | … | … | N | … | … | … | … | … | n | G |
| 49838 | 18281831+0648596 | -48.9 ± 0.3 | 4699 ± 188 | 2.81 ± 0.36 | -0.06 ± 0.19 | … | … | N | … | … | … | … | … | n | G |
| 49839 | 18281840+0642517 | 61.7 ± 0.2 | 4868 ± 169 | 3.27 ± 0.23 | -0.06 ± 0.16 | … | … | N | … | … | … | … | … | n | G |
| 49840 | 18281842+0641576 | 49.0 ± 0.2 | 4559 ± 131 | 2.55 ± 0.36 | 0.22 ± 0.25 | <22 | 3 | N | .. | … | … | … | … | n | … |
| 49841 | 18281842+0646289 | -26.9 ± 0.2 | 4082 ± 226 | 4.67 ± 0.40 | -0.16 ± 0.24 | … | … | Y | … | … | … | … | Y | n | … |
| 49842 | 18281872+0647059 | 42.5 ± 0.5 | 4553 ± 122 | 4.01 ± 0.01 | -0.18 ± 0.18 | … | … | N | … | … | … | … | … | n | … |
| 49843 | 18281876+0644541 | 34.6 ± 0.3 | 5226 ± 90 | 3.91 ± 0.26 | 0.29 ± 0.22 | … | … | N | … | … | … | … | … | n | … |
| 49844 | 18281889+0642314 | -49.9 ± 0.3 | 5113 ± 50 | 4.61 ± 0.45 | 0.03 ± 0.13 | … | … | N | … | … | … | … | … | n | … |
| 49845 | 18281902+0648138 | -26.8 ± 0.3 | 5156 ± 40 | 3.26 ± 0.30 | -0.16 ± 0.16 | … | … | Y | … | … | … | … | … | n | G |
| 49846 | 18281914+0644313 | 60.5 ± 0.3 | 4649 ± 105 | 2.54 ± 0.31 | -0.12 ± 0.21 | … | … | N | … | … | … | … | … | n | G |
| 49847 | 18281931+0641557 | 6.7 ± 0.2 | 4742 ± 194 | 3.11 ± 0.40 | 0.17 ± 0.2 | … | … | N | … | … | … | … | … | n | G |
| 49848 | 18281957+0642244 | 79.4 ± 0.2 | 4654 ± 175 | 2.73 ± 0.20 | -0.21 ± 0.23 | … | … | N | … | … | … | … | … | n | G |
| 49849 | 18282004+0642358 | -53.4 ± 0.3 | 4419 ± 201 | 4.53 ± 0.37 | 0.25 ± 0.21 | … | … | N | … | … | … | … | … | n | … |
| 49850 | 18282028+0641354 | 0.8 ± 0.3 | 4842 ± 202 | 3.62 ± 0.36 | 0.07 ± 0.15 | … | … | Y | … | … | … | … | … | n | … |
| 49851 | 18282029+0642454 | -42.4 ± 0.2 | 4660 ± 106 | 2.51 ± 0.30 | 0.14 ± 0.22 | … | … | Y | … | … | … | … | … | n | … |
| 49852 | 18282041+0640343 | -49.9 ± 0.4 | 4197 ± 280 | 4.62 ± 0.17 | 0.03 ± 0.19 | … | … | N | … | … | … | N | … | n | … |
| 49853 | 18282069+0641299 | 2.8 ± 0.2 | 5410 ± 209 | 4.38 ± 0.50 | 0.04 ± 0.15 | 24 ± 21 | 1 | Y?$^f$ | Y | Y | Y | … | … | n | NG |
| 49854 | 18282069+0641541 | 1.0 ± 0.6 | 4234 ± 91 | 4.43 ± 0.36 | 0.09 ± 0.13 | … | … | Y | … | … | … | N | … | n | … |
| 49855 | 18282076+0648009 | -27.3 ± 0.5 | 3996 ± 261 | 4.82 ± 0.25 | -0.47 ± 0.28 | <54 | 3 | Y | Y | Y | N | Y | … | Y | … |
| 49856 | 18282137+0648266 | -53.5 ± 0.2 | 5619 ± 205 | 4.30 ± 0.18 | 0.23 ± 0.14 | … | … | N | … | … | … | … | … | n | … |
| 49858 | 18282154+0640399 | -61.4 ± 0.3 | 4841 ± 86 | 4.09 ± 0.79 | 0.23 ± 0.17 | <43 | 3 | N | .. | … | … | … | … | n | NG |
| 49859 | 18282166+0648192 | -52.5 ± 0.3 | 4778 ± 201 | 2.67 ± 0.27 | -0.13 ± 0.23 | … | … | N | … | … | … | … | … | n | G |
| 49860 | 18282195+0642505 | -7.7 ± 0.3 | 5581 ± 22 | 3.95 ± 0.33 | -0.72 ± 0.22 | … | … | Y | … | … | … | … | … | n | … |
| 49861 | 18282198+0646452 | -10.6 ± 0.3 | 5262 ± 161 | 4.17 ± 0.34 | 0.09 ± 0.15 | … | … | Y | … | … | … | … | … | n | … |
| 49862 | 18282200+0643449 | -47.3 ± 0.4 | 4248 ± 257 | 4.60 ± 0.39 | -0.03 ± 0.16 | … | … | Y | … | … | … | Y | … | n | … |
| 49864 | 18282356+0648253 | 77.3 ± 0.2 | 4570 ± 142 | 2.70 ± 0.37 | 0.05 ± 0.2 | … | … | N | … | … | … | … | … | n | G |
| 49865 | 18282364+0648547 | -17.9 ± 0.3 | 5413 ± 75 | 3.88 ± 0.06 | 0.12 ± 0.17 | … | … | Y | … | … | … | … | … | n | … |
| 49866 | 18282414+0648210 | -5.2 ± 0.2 | 4638 ± 231 | 2.77 ± 0.48 | 0.01 ± 0.21 | … | … | Y | … | … | … | … | … | n | G |
| 49867 | 18282417+0645234 | 21.9 ± 0.5 | 4370 ± 192 | 5.14 ± 0.23 | -0.21 ± 0.15 | … | … | N | … | … | … | … | … | n | … |
| 49868 | 18282449+0643079 | 33.7 ± 0.2 | 4403 ± 175 | 2.46 ± 0.38 | 0.18 ± 0.23 | <26 | 3 | N | .. | … | … | … | … | n | G |
| 49869 | 18282462+0644174 | -107.1 ± 0.3 | 4675 ± 55 | 1.41 ± 0.47 | -1.74 ± 0.4 | … | … | N | … | … | … | … | … | n | G |
| 49870 | 18282483+0647002 | 21.5 ± 0.3 | 5052 ± 113 | 4.71 ± 0.43 | -0.07 ± 0.13 | … | … | N | … | … | … | … | … | n | … |
| 49871 | 18282484+0641247 | 14.9 ± 0.2 | 4744 ± 249 | 3.05 ± 0.22 | -0.09 ± 0.18 | … | … | N | … | … | … | … | … | n | G |
| 49872 | 18282587+0644451 | 22.4 ± 0.4 | 4634 ± 353 | 4.51 ± 0.58 | -0.14 ± 0.13 | <80 | 3 | N | .. | … | … | … | … | n | NG |
| 49873 | 18282590+0643217 | -128.5 ± 0.3 | 4845 ± 186 | 2.78 ± 0.31 | -0.29 ± 0.16 | … | … | N | … | … | … | … | … | n | G |
| 49874 | 18282592+0647567 | 44.9 ± 0.2 | 4535 ± 171 | 2.47 ± 0.40 | -0.03 ± 0.23 | <17 | 3 | N | .. | … | … | … | … | n | G |
| 49875 | 18282628+0649552 | 38.7 ± 0.2 | 4835 ± 241 | 3.25 ± 0.27 | -0.14 ± 0.26 | … | … | N | … | … | … | … | … | n | G |
| 49876 | 18282649+0643339 | -63.6 ± 0.2 | 4175 ± 291 | 2.04 ± 0.57 | 0.07 ± 0.2 | … | … | N | … | … | … | … | … | n | G |
| 49877 | 18282655+0642193 | 3.7 ± 0.2 | 4577 ± 191 | 2.86 ± 0.54 | 0.30 ± 0.25 | … | … | N | … | … | … | … | … | n | G |
| 49878 | 18282667+0643183 | -25.7 ± 0.4 | 4459 ± 206 | 4.24 ± 0.24 | -0.57 ± 0.29 | <106 | 3 | Y | N | Y | N | N | … | n | NG |
| 49879 | 18282686+0641388 | -106.1 ± 0.3 | 4873 ± 207 | 2.93 ± 0.35 | -0.21 ± 0.15 | … | … | N | … | … | … | … | … | n | G |
| 49880 | 18282686+0644507 | 9.2 ± 0.3 | 4499 ± 283 | 2.92 ± 0.52 | 0.45 ± 0.44 | … | … | N | … | … | … | … | … | n | G |
| 49881 | 18282703+0643130 | 5.1 ± 0.2 | 4225 ± 302 | 2.20 ± 0.54 | 0.13 ± 0.34 | … | … | N | … | … | … | … | … | n | G |
| 49882 | 18282726+0645308 | 10.7 ± 0.3 | 4874 ± 191 | 2.80 ± 0.48 | -0.21 ± 0.2 | … | … | N | … | … | … | … | … | n | G |
| 49883 | 18282738+0648161 | 90.7 ± 0.3 | 4966 ± 170 | 3.17 ± 0.18 | -0.19 ± 0.37 | … | … | N | … | … | … | … | … | n | G |
| 49884 | 18282762+0646165 | -10.9 ± 0.4 | 4196 ± 282 | 4.65 ± 0.25 | 0.04 ± 0.17 | <57 | 3 | Y | N | Y | Y | N | … | n | NG |
| 49885 | 18282780+0650232 | -0.1 ± 0.3 | 4620 ± 138 | 2.69 ± 0.44 | 0.01 ± 0.19 | <38 | 3 | Y | Y | Y | Y | … | … | Y | … |
| 49886 | 18282843+0646418 | 3.7 ± 0.4 | 3967 ± 369 | 4.35 ± 0.24 | -0.09 ± 0.2 | <66 | 3 | N | .. | … | … | … | … | n | NG |
| 49887 | 18282857+0641038 | -237.4 ± 0.2 | 4535 ± 158 | 2.20 ± 0.47 | -0.27 ± 0.48 | 87 ± 46 | 1 | N | .. | … | … | … | … | n | G |

**Table C.11.** continued.

| ID | CNAME | RV (km s$^{-1}$) | $T_{\rm eff}$ (K) | logg (dex) | [Fe/H] (dex) | EW(Li)$^a$ (mÅ) | EW(Li) error flag$^b$ | Membership RV | Membership Li | Membership logg | Membership [Fe/H] | Gaia studies Randich$^c$ | Gaia studies Cantat-Gaudin$^c$ | Final$^d$ | NMs with Li$^e$ |
|---|---|---|---|---|---|---|---|---|---|---|---|---|---|---|---|
| 49888 | 18282874+0645217 | 30.8 ± 0.2 | 4738 ± 249 | 3.01 ± 0.28 | -0.02 ± 0.15 | … | … | N | … | … | … | … | … | n | G |
| 49889 | 18282877+0644365 | 39.3 ± 0.2 | 4416 ± 177 | 2.46 ± 0.33 | 0.24 ± 0.22 | … | … | N | … | … | … | … | … | n | G |
| 49890 | 18282879+0645037 | -27.3 ± 0.2 | 4906 ± 164 | 3.32 ± 0.19 | -0.22 ± 0.18 | … | … | Y | … | … | … | … | … | n | G |
| 49891 | 18282897+0646162 | -17.8 ± 0.3 | 4908 ± 113 | 4.49 ± 0.25 | 0.14 ± 0.16 | … | … | Y | … | … | … | … | … | n | … |
| 49892 | 18282929+0644312 | -37.1 ± 0.2 | 4642 ± 135 | 2.65 ± 0.45 | 0.06 ± 0.2 | … | … | Y | … | … | … | … | … | n | G |
| 49893 | 18282944+0643449 | -18.5 ± 0.3 | 5857 ± 78 | 4.08 ± 0.21 | 0.30 ± 0.16 | … | … | Y | … | … | … | … | … | n | … |
| 49894 | 18283004+0648066 | 88.8 ± 0.2 | 4702 ± 201 | 2.61 ± 0.37 | -0.30 ± 0.18 | … | … | N | … | … | … | … | … | n | G |
| 49895 | 18283021+0642111 | 7.6 ± 0.3 | 4955 ± 72 | 3.60 ± 0.51 | 0.31 ± 0.25 | … | … | N | … | … | … | … | … | n | … |
| 49896 | 18283045+0643532 | -0.9 ± 0.2 | 4651 ± 294 | 2.49 ± 0.52 | -0.12 ± 0.22 | … | … | Y | … | … | … | … | … | n | G |
| 49897 | 18283128+0642299 | 57.4 ± 0.2 | 4718 ± 115 | 2.65 ± 0.28 | 0.09 ± 0.23 | … | … | N | … | … | … | … | … | n | … |
| 49898 | 18283185+0644093 | -17.2 ± 0.3 | 4799 ± 208 | 3.03 ± 0.15 | -0.02 ± 0.21 | … | … | Y | … | … | … | … | … | n | G |
| 49899 | 18283278+0648284 | 81.7 ± 0.3 | 5235 ± 184 | 3.36 ± 0.62 | -0.46 ± 0.18 | … | … | N | … | … | … | … | … | n | … |
| 49900 | 18283289+0643201 | -10.3 ± 0.2 | 4631 ± 89 | 2.60 ± 0.37 | -0.15 ± 0.18 | … | … | Y | … | … | … | … | … | n | … |
| 49901 | 18283294+0649160 | -27.5 ± 0.2 | 6328 ± 244 | 4.76 ± 0.94 | -0.55 ± 0.48 | … | … | Y | … | … | … | Y | Y | n | … |
| 49902 | 18283310+0646106 | 12.4 ± 0.2 | 4837 ± 245 | 2.62 ± 0.55 | -0.40 ± 0.24 | … | … | N | … | … | … | … | … | n | G |
| 49903 | 18283318+0645420 | -25.6 ± 1.9 | 3309 ± 156 | … | -0.14 ± 0.14 | <127 | 3 | Y | Y | … | Y | Y | … | Y | … |
| 49904 | 18283319+0642537 | 21.9 ± 0.3 | 4962 ± 114 | 3.15 ± 0.20 | -0.32 ± 0.26 | … | … | N | … | … | … | … | … | n | G |
| 49905 | 18283320+0650104 | -29.8 ± 0.3 | 4392 ± 203 | 4.72 ± 0.31 | 0.07 ± 0.15 | … | … | Y | … | … | … | Y | … | n | … |
| 49906 | 18283377+0641325 | 24.5 ± 0.3 | 4766 ± 92 | 3.25 ± 0.52 | 0.23 ± 0.23 | <22 | 3 | N | .. | … | … | … | … | n | G |
| 49907 | 18283412+0644437 | -62.5 ± 0.3 | 4145 ± 296 | 4.85 ± 0.36 | -0.55 ± 0.32 | <41 | 3 | N | .. | … | … | N | … | n | NG |
| 49908 | 18283484+0642003 | 61.2 ± 0.4 | 4968 ± 187 | 3.47 ± 0.37 | 0.02 ± 0.14 | <91 | 3 | N | … | … | … | … | … | n | Li-rich G |
| 49909 | 18283506+0650305 | -27.5 ± 0.2 | 4752 ± 166 | 4.63 ± 0.38 | -0.01 ± 0.12 | <26 | 3 | Y | Y | Y | Y | Y | Y | Y | … |
| 49910 | 18283507+0646116 | -56.2 ± 0.3 | 4609 ± 219 | 2.38 ± 0.32 | -0.28 ± 0.14 | <72 | 3 | N | .. | … | … | … | … | n | G |
| 49911 | 18283613+0645330 | -29.1 ± 0.5 | 4029 ± 100 | 4.53 ± 0.32 | 0.36 ± 0.13 | … | … | Y | … | … | … | Y | … | n | … |
| 49912 | 18283617+0646444 | 19.3 ± 0.2 | 4676 ± 144 | 2.93 ± 0.66 | 0.19 ± 0.25 | … | … | N | … | … | … | … | … | n | G |
| 49913 | 18261497+0622189 | -362.8 ± 712.7 | … | … | … | … | … | N | … | … | … | … | … | n | … |
| 49914 | 18261888+0624028 | -2.2 ± 0.1 | … | … | … | … | … | Y | … | … | … | … | … | n | … |
| 49915 | 18262719+0631371 | -25.1 ± 0.1 | … | … | … | … | … | Y | … | … | … | … | Y | n | … |
| 49916 | 18263184+0616052 | -3.0 ± 0.2 | … | … | … | … | … | Y | … | … | … | … | … | n | … |
| 49917 | 18270204+0624399 | -25.1 ± 0.1 | … | … | … | … | … | Y | … | … | … | … | Y | n | … |
| 49918 | 18273582+0632590 | 14.9 ± 0.1 | … | … | … | … | … | N | … | … | … | … | … | n | … |
| 54821 | 18264217+0627514 | 55.3 ± 0.1 | … | … | … | … | … | N | … | … | … | … | … | n | … |
| 54822 | 18264670+0633257 | 117.6 ± 0.1 | … | … | … | … | … | N | … | … | … | … | … | n | … |
| 54823 | 18264681+0630104 | -25.2 ± 0.1 | … | … | … | … | … | Y | … | … | … | … | Y | n | … |
| 54824 | 18265134+0630042 | 29.3 ± 0.1 | … | … | … | … | … | N | … | … | … | … | … | n | … |
| 54825 | 18265714+0631295 | -14.0 ± 0.1 | … | … | … | … | … | Y | … | … | … | … | … | n | … |
| 54826 | 18265721+0629178 | 83.0 ± 0.1 | … | … | … | … | … | N | … | … | … | … | … | n | … |
| 54827 | 18265728+0638421 | -12.5 ± 0.1 | … | … | … | … | … | Y | … | … | … | … | … | n | … |
| 54828 | 18265971+0637570 | -41.6 ± 0.1 | … | … | … | … | … | Y | … | … | … | … | … | n | … |
| 54829 | 18270004+0630474 | 8.3 ± 0.1 | … | … | … | … | … | N | … | … | … | … | Y | n | … |
| 54830 | 18270028+0627184 | 36.0 ± 0.1 | … | … | … | … | … | N | … | … | … | … | … | n | … |
| 54831 | 18270042+0625020 | -35.4 ± 0.1 | … | … | … | … | … | Y | … | … | … | … | Y | n | … |
| 54832 | 18270156+0633543 | 55.2 ± 0.1 | … | … | … | … | … | N | … | … | … | … | … | n | … |
| 54833 | 18270202+0639105 | -17.8 ± 0.1 | … | … | … | … | … | Y | … | … | … | … | … | n | … |
| 54834 | 18270273+0634514 | 64.3 ± 0.1 | … | … | … | … | … | N | … | … | … | … | … | n | … |
| 54835 | 18270349+0630244 | 29.9 ± 0.1 | … | … | … | … | … | N | … | … | … | … | … | n | … |
| 54836 | 18270705+0623588 | 32.4 ± 0.1 | … | … | … | … | … | N | … | … | … | … | … | n | … |
| 54837 | 18270758+0634499 | -80.1 ± 0.1 | … | … | … | … | … | N | … | … | … | … | … | n | … |
| 54838 | 18270938+0632145 | 28.4 ± 0.1 | … | … | … | … | … | N | … | … | … | … | … | n | … |
| 54839 | 18271051+0623569 | 80.8 ± 0.1 | … | … | … | … | … | N | … | … | … | … | … | n | … |
| 54840 | 18271202+0631280 | 14.5 ± 0.1 | … | … | … | … | … | N | … | … | … | … | … | n | … |
| 54841 | 18271330+0625158 | -15.8 ± 0.1 | … | … | … | … | … | Y | … | … | … | … | … | n | … |
| 54842 | 18271474+0634486 | -3.6 ± 0.1 | … | … | … | … | … | Y | … | … | … | … | … | n | … |
| 54843 | 18271574+0632119 | 21.7 ± 0.1 | … | … | … | … | … | N | … | … | … | … | … | n | … |
| 54844 | 18271594+0634526 | -56.3 ± 0.1 | … | … | … | … | … | N | … | … | … | … | … | n | … |
| 54845 | 18271648+0629552 | -42.4 ± 0.1 | … | … | … | … | … | Y | … | … | … | … | … | n | … |
| 54846 | 18271721+0638371 | -34.3 ± 0.1 | … | … | … | … | … | Y | … | … | … | … | … | n | … |
| 54847 | 18271935+0626075 | -17.0 ± 0.1 | … | … | … | … | … | Y | … | … | … | … | … | n | … |
| 54848 | 18271940+0635593 | 14.3 ± 0.1 | … | … | … | … | … | N | … | … | … | … | … | n | … |

| ID | CNAME | RV (km s$^{-1}$) | $T_{\text{eff}}$ (K) | logg (dex) | [Fe/H] (dex) | EW(Li)$^a$ (mÅ) | EW(Li) error flag$^b$ | Membership RV | Li | logg | [Fe/H] | Gaia studies Randich$^c$ | Cantat-Gaudin$^c$ | Final$^d$ | NMs with Li$^e$ |
|---|---|---|---|---|---|---|---|---|---|---|---|---|---|---|---|
| 54849 | 18272020+0638548 | -28.9 ± 0.1 | … | … | … | … | … | Y | … | … | … | … | … | n | … |
| 54850 | 18272060+0624495 | 26.9 ± 0.1 | … | … | … | … | … | N | … | … | … | … | … | n | … |
| 54851 | 18272331+0633446 | -12.2 ± 0.1 | … | … | … | … | … | Y | … | … | … | … | … | n | … |
| 54852 | 18272499+0625430 | -51.1 ± 0.5 | … | … | … | … | … | N | … | … | … | … | … | n | … |
| 54853 | 18272510+0631584 | 1.9 ± 0.8 | … | … | … | … | … | Y | … | … | … | … | … | n | … |
| 54854 | 18272763+0627228 | 25.2 ± 0.1 | … | … | … | … | … | N | … | … | … | … | … | n | … |
| 54855 | 18272784+0637239 | -32.0 ± 0.1 | … | … | … | … | … | Y | … | … | … | … | … | n | … |
| 54856 | 18272848+0624328 | 10.8 ± 0.1 | … | … | … | … | … | N | … | … | … | … | … | n | … |
| 54857 | 18272903+0632490 | -3.2 ± 0.1 | … | … | … | … | … | Y | … | … | … | … | … | n | … |
| 54858 | 18272976+0625124 | -62.2 ± 0.2 | … | … | … | … | … | N | … | … | … | … | … | n | … |
| 54859 | 18273357+0640300 | 35.4 ± 0.1 | … | … | … | … | … | N | … | … | … | … | … | n | … |
| 54860 | 18273412+0632489 | 17.7 ± 0.1 | … | … | … | … | … | N | … | … | … | … | … | n | … |
| 54861 | 18273428+0634246 | 415.0 ± 1.4 | … | … | … | … | … | N | … | … | … | … | Y | n | … |
| 54862 | 18273555+0629415 | 14.9 ± 0.1 | … | … | … | … | … | N | … | … | … | … | … | n | … |
| 54863 | 18273756+0633008 | -25.6 ± 0.1 | … | … | … | … | … | Y | … | … | … | … | Y | n | … |
| 54864 | 18273852+0627480 | 4.2 ± 0.1 | … | … | … | … | … | N | … | … | … | … | … | n | … |
| 54865 | 18273936+0626558 | -23.7 ± 0.1 | … | … | … | … | … | Y | … | … | … | … | … | n | … |
| 54866 | 18273997+0640116 | 35.8 ± 0.1 | … | … | … | … | … | N | … | … | … | … | … | n | … |
| 54867 | 18274122+0630205 | -18.1 ± 0.1 | … | … | … | … | … | Y | … | … | … | … | … | n | … |
| 54868 | 18274128+0640280 | -27.5 ± 0.1 | … | … | … | … | … | Y | … | … | … | … | Y | n | … |
| 54869 | 18274236+0625354 | -27.6 ± 0.1 | … | … | … | … | … | Y | … | … | … | … | N | n | … |
| 54870 | 18274583+0638597 | 3.9 ± 0.4 | … | … | … | … | … | N | … | … | … | … | … | n | … |
| 54871 | 18274726+0637329 | 9.4 ± 0.1 | … | … | … | … | … | N | … | … | … | … | … | n | … |
| 54872 | 18274897+0633534 | -20.6 ± 0.1 | … | … | … | … | … | Y | … | … | … | … | … | n | … |
| 54873 | 18275003+0627450 | 41.3 ± 0.1 | … | … | … | … | … | N | … | … | … | … | … | n | … |
| 54874 | 18275100+0638032 | -2.2 ± 0.1 | … | … | … | … | … | Y | … | … | … | … | … | n | … |
| 54875 | 18275101+0629591 | 22.5 ± 0.1 | … | … | … | … | … | N | … | … | … | … | … | n | … |
| 54876 | 18275301+0634023 | -26.9 ± 0.1 | … | … | … | … | … | Y | … | … | … | … | … | n | … |
| 54877 | 18275447+0628041 | 3.6 ± 0.1 | … | … | … | … | … | N | … | … | … | … | … | n | … |
| 54878 | 18275832+0632231 | 5.6 ± 0.1 | … | … | … | … | … | N | … | … | … | … | … | n | … |
| 54879 | 18275929+0632480 | 34.0 ± 0.1 | … | … | … | … | … | N | … | … | … | … | … | n | … |
| 54880 | 18275951+0630237 | 32.0 ± 0.1 | … | … | … | … | … | N | … | … | … | … | … | n | … |
| 54881 | 18275992+0633096 | 35.8 ± 0.1 | … | … | … | … | … | N | … | … | … | … | … | n | … |
| 54882 | 18280081+0632155 | 78.4 ± 0.1 | … | … | … | … | … | N | … | … | … | … | … | n | … |
| 54883 | 18280660+0638080 | -16.5 ± 0.7 | … | … | … | … | … | Y | … | … | … | … | … | n | … |
| 54884 | 18281441+0630440 | -50.2 ± 0.1 | … | … | … | … | … | N | … | … | … | … | N | n | … |
| 54885 | 18281493+0634159 | -30.8 ± 0.1 | … | … | … | … | … | Y | … | … | … | … | … | n | … |

**Notes.** $^{(a)}$ The values of EW(Li) for this cluster are corrected (subtracted adjacent Fe (6707.43 Å) line). $^{(b)}$ Flags for the errors of the corrected EW(Li) values, as follows: 1=EW(Li) corrected by blends contribution using models; and 3=Upper limit (no error for EW(Li) is given). $^{(c)}$ Randich et al. (2018), Cantat-Gaudin et al. (2018). $^{(d)}$ The letters "Y" and "N" indicate if the star is a cluster member or not. $^{(e)}$ 'Li-rich G', 'G' and 'NG' indicate "Li-rich giant", "giant" and "non-giant" Li field outliers, respectively. $^{(f)}$ For more details about the membership of these particular stars, see the individual notes of Appendix A for NGC 6633. gFor more details about the membership of the stars listed as possible candidates, see the individual notes of Appendix A for NGC 6633.

**Table C.12.** Trumpler 23

| ID | CNAME | RV (km s$^{-1}$) | $T_{\rm eff}$ (K) | $logg$ (dex) | [Fe/H] (dex) | $EW({\rm Li})^a$ (mÅ) | $EW({\rm Li})$ error flag$^b$ | RV | Li | Membership $logg$ | [Fe/H] | Final$^c$ | Li-rich non-mem$^d$ |
|---|---|---|---|---|---|---|---|---|---|---|---|---|---|
| 53351 | 16004226-5335450 | -68.5 ± 2.9 | ... | ... | ... | ... | ... | N | ... | ... | ... | n | ... |
| 53279 | 16004240-5332325 | -24.6 ± 0.3 | 5890 ± 169 | 4.12 ± 0.18 | 0.37 ± 0.16 | 93 ± 18 | 1 | N | ... | ... | ... | n | NG |
| 53352 | 16004252-5330337 | -32.4 ± 0.5 | ... | ... | ... | ... | ... | N | ... | ... | ... | n | ... |
| 53353 | 16004295-5330566 | -60.4 ± 2.0 | ... | ... | ... | ... | ... | Y | ... | ... | ... | n | ... |
| 3405 | 16004312-5330509 | -61.8 ± 0.6 | 4913 ± 121 | 2.86 ± 0.23 | 0.11 ± 0.10 | 59 ± 1 | ... | Y | Y | Y | Y | Y | ... |
| 53280 | 16004350-5331482 | -60.1 ± 1.0 | 6584 ± 440 | 4.05 ± 0.21 | 0.30 ± 0.42 | ... | ... | Y | ... | ... | ... | n | ... |
| 53354 | 16004351-5333456 | -31.2 ± 0.1 | ... | ... | ... | ... | 1 | N | ... | ... | ... | n | ... |
| 53281 | 16004357-5328513 | -62.1 ± 0.5 | 6722 ± 435 | 4.13 ± 0.02 | 0.33 ± 0.35 | 39 ± 18 | ... | Y | Y | Y | Y | Y | ... |
| 53355 | 16004375-5335538 | -33.9 ± 0.1 | ... | ... | ... | ... | ... | N | ... | ... | ... | n | ... |
| 53283 | 16004452-5329038 | -34.8 ± 0.3 | 6273 ± 204 | 4.27 ± 0.35 | 0.08 ± 0.21 | ... | ... | N | ... | ... | ... | n | ... |
| 53356 | 16004496-5333060 | -63.9 ± 1.7 | ... | ... | ... | ... | ... | Y | ... | ... | ... | n | ... |
| 53358 | 16004544-5333062 | -61.2 ± 3.7 | ... | ... | ... | ... | ... | Y | ... | ... | ... | n | ... |
| 53284 | 16004556-5332159 | -25.0 ± 0.2 | 4542 ± 144 | 4.61 ± 0.43 | 0.15 ± 0.16 | ... | ... | N | ... | ... | ... | n | ... |
| 3407 | 16004569-5329177 | -55.4 ± 0.6 | 5008 ± 123 | 2.98 ± 0.23 | 0.18 ± 0.10 | <5 | 3 | N | ... | ... | ... | n | G |
| 53319 | 16005857-5329161 | -32.2 ± 0.3 | 5882 ± 115 | 4.10 ± 0.15 | 0.16 ± 0.15 | ... | ... | N | ... | ... | ... | n | ... |
| 53320 | 16005947-5328024 | 10.4 ± 0.2 | 4933 ± 166 | 4.64 ± 0.20 | 0.04 ± 0.15 | ... | ... | N | ... | ... | ... | n | ... |
| 53321 | 16005947-5330396 | -54.3 ± 0.5 | 6169 ± 120 | 4.63 ± 0.77 | -0.01 ± 0.31 | ... | ... | N | ... | ... | ... | n | ... |
| 53392 | 16005951-5335225 | -62.6 ± 3.4 | ... | ... | ... | ... | ... | Y | ... | ... | ... | n | ... |
| 53393 | 16005961-5331283 | -67.4 ± 8.4 | ... | ... | ... | ... | ... | N | ... | ... | ... | n | ... |
| 53394 | 16005980-5334552 | -38.1 ± 2.3 | ... | ... | ... | ... | ... | N | ... | ... | ... | n | ... |
| 53395 | 16005986-5329431 | -62.4 ± 0.2 | ... | ... | ... | ... | ... | Y | ... | ... | ... | n | ... |
| 3416 | 16010025-5333101 | -60.2 ± 0.6 | 4884 ± 119 | 2.79 ± 0.22 | 0.10 ± 0.10 | 54 ± 1 | ... | Y | Y | Y | Y | Y | ... |
| 53396 | 16010069-5329384 | -61.2 ± 4.5 | ... | ... | ... | ... | ... | Y | ... | ... | ... | n | ... |
| 53397 | 16010155-5328151 | 458.8 ± 0.1 | ... | ... | ... | ... | ... | N | ... | ... | ... | n | ... |
| 53398 | 16010204-5332060 | -62.1 ± 2.3 | ... | ... | ... | ... | ... | Y | ... | ... | ... | n | ... |
| 53399 | 16010227-5333466 | -64.6 ± 1.3 | ... | ... | ... | ... | ... | Y | ... | ... | ... | n | ... |
| 53323 | 16010238-5331368 | -22.5 ± 0.4 | 6369 ± 193 | 4.56 ± 0.46 | 0.29 ± 0.23 | 91 ± 26 | 1 | N | ... | ... | ... | n | NG |
| 53400 | 16010286-5328087 | -32.9 ± 0.2 | ... | ... | ... | ... | ... | N | ... | ... | ... | n | ... |
| 53401 | 16010322-5332285 | -51.0 ± 1.6 | ... | ... | ... | ... | ... | N | ... | ... | ... | n | ... |
| 53324 | 16010323-5330273 | -61.1 ± 0.4 | 6647 ± 492 | 4.05 ± 0.12 | 0.39 ± 0.43 | ... | ... | Y | ... | ... | ... | n | ... |
| 53410 | 16010922-5330561 | -21.3 ± 3.4 | ... | ... | ... | ... | ... | N | ... | ... | ... | n | ... |
| 53411 | 16010935-5332003 | -61.8 ± 0.7 | ... | ... | ... | ... | ... | Y | ... | ... | ... | n | ... |
| 53267 | 16003312-5332533 | -88.4 ± 0.2 | 6100 ± 311 | 4.24 ± 0.23 | 0.30 ± 0.30 | ... | ... | N | ... | ... | ... | n | ... |
| 53336 | 16003334-5329569 | -28.8 ± 0.1 | 5897 ± 101 | 4.16 ± 0.06 | -0.18 ± 0.15 | ... | ... | N | ... | ... | ... | n | ... |
| 53337 | 16003389-5330469 | -48.0 ± 1.7 | ... | ... | ... | ... | ... | N | ... | ... | ... | n | ... |
| 53268 | 16003500-5334335 | -47.1 ± 0.6 | 6752 ± 62 | 4.15 ± 0.12 | 0.45 ± 0.05 | ... | ... | N | ... | ... | ... | n | ... |
| 53338 | 16003549-5331214 | -38.6 ± 0.5 | ... | ... | ... | ... | ... | N | ... | ... | ... | n | ... |
| 53339 | 16003558-5329192 | -75.3 ± 0.1 | ... | ... | ... | ... | ... | N | ... | ... | ... | n | ... |
| 53340 | 16003652-5330052 | -80.3 ± 0.1 | ... | ... | ... | ... | ... | N | ... | ... | ... | n | ... |
| 53270 | 16003787-5331536 | -54.4 ± 0.6 | 7080 ± 718 | 4.15 ± 0.21 | 0.12 ± 0.14 | ... | ... | N | ... | ... | ... | n | ... |
| 53341 | 16003791-5331140 | -44.1 ± 3.5 | ... | ... | ... | ... | ... | N | ... | ... | ... | n | ... |
| 53271 | 16003799-5334228 | -4.6 ± 0.3 | 6569 ± 39 | 4.46 ± 0.08 | 0.45 ± 0.03 | ... | ... | N | ... | ... | ... | n | ... |
| 53272 | 16003805-5331267 | -68.0 ± 1.2 | 6325 ± 602 | 4.01 ± 0.19 | 0.02 ± 0.24 | ... | ... | N | ... | ... | ... | n | ... |
| 53342 | 16003881-5331400 | -38.3 ± 0.1 | ... | ... | ... | ... | ... | N | ... | ... | ... | n | ... |
| 53285 | 16004569-5332567 | -64.9 ± 0.7 | 7058 ± 488 | 4.15 ± 0.21 | 0.11 ± 0.15 | ... | ... | Y | ... | ... | ... | n | ... |
| 3408 | 16004572-5332095 | -43.3 ± 0.6 | 4900 ± 125 | 2.68 ± 0.26 | 0.15 ± 0.10 | 20 ± 1 | ... | N | ... | ... | ... | n | G |
| 53286 | 16004581-5331557 | -64.2 ± 1.0 | 6766 ± 407 | 4.46 ± 0.36 | 0.21 ± 0.40 | 57 ± 38 | 1 | Y | Y | Y | Y | Y | ... |
| 53287 | 16004583-5330286 | -0.7 ± 0.3 | 5649 ± 43 | 4.29 ± 0.05 | -0.01 ± 0.15 | ... | ... | N | ... | ... | ... | n | ... |
| 53359 | 16004639-5333149 | -62.8 ± 0.5 | ... | ... | ... | ... | ... | Y | ... | ... | ... | n | ... |
| 53360 | 16004658-5334434 | -14.6 ± 0.2 | ... | ... | ... | ... | ... | N | ... | ... | ... | n | ... |
| 53288 | 16004689-5328513 | 16.7 ± 0.3 | 6082 ± 252 | 4.37 ± 0.22 | 0.11 ± 0.21 | 71 ± 28 | 1 | N | ... | ... | ... | n | NG |
| 53361 | 16004693-5334523 | -64.7 ± 0.3 | ... | ... | ... | ... | ... | Y | ... | ... | ... | n | ... |
| 53289 | 16004767-5332334 | 12.2 ± 0.3 | 5497 ± 114 | 4.32 ± 0.10 | -0.09 ± 0.21 | ... | ... | N | ... | ... | ... | n | ... |
| 53362 | 16004787-5334169 | -59.4 ± 2.0 | ... | ... | ... | ... | ... | Y | ... | ... | ... | n | ... |
| 53363 | 16004804-5333020 | -14.7 ± 0.2 | ... | ... | ... | ... | ... | N | ... | ... | ... | n | ... |
| 53295 | 16005175-5332339 | -64.2 ± 0.6 | 7306 ± 783 | 4.15 ± 0.18 | -0.07 ± 0.46 | ... | ... | Y | ... | ... | ... | n | ... |
| 53376 | 16005205-5335459 | -35.5 ± 0.3 | ... | ... | ... | ... | ... | N | ... | ... | ... | n | ... |
| 3414 | 16005220-5333362 | -62.4 ± 0.6 | 4917 ± 116 | 2.63 ± 0.22 | 0.18 ± 0.10 | 12 ± 1 | ... | Y | Y | Y | Y | Y | ... |
| 53377 | 16005229-5333251 | -34.0 ± 0.2 | ... | ... | ... | ... | ... | N | ... | ... | ... | n | ... |

| ID | CNAME | RV (km s$^{-1}$) | $T_{\text{eff}}$ (K) | $\log g$ (dex) | [Fe/H] (dex) | EW(Li)$^a$ (mÅ) | EW(Li) error flag$^b$ | Membership RV | Li | $\log g$ | [Fe/H] | Final$^c$ | Li-rich non-mem$^d$ |
|---|---|---|---|---|---|---|---|---|---|---|---|---|---|
| 53297 | 16005241-5331345 | -63.4 ± 0.6 | 6602 ± 414 | 4.07 ± 0.19 | 0.27 ± 0.38 | … | … | Y | … | … | … | n | … |
| 53378 | 16005288-5331550 | -12.7 ± 1.4 | … | … | … | … | … | N | … | … | … | n | … |
| 53300 | 16005372-5328100 | -37.3 ± 0.3 | 6355 ± 168 | 3.93 ± 0.03 | 0.20 ± 0.21 | 114 ± 25 | 1 | N | … | … | … | n | NG |
| 53301 | 16005380-5336063 | -40.5 ± 0.3 | 5966 ± 155 | 4.26 ± 0.26 | 0.02 ± 0.15 | 50 ± 35 | 1 | N | … | … | … | n | NG |
| 53379 | 16005402-5331231 | 13.9 ± 0.1 | 5257 ± 108 | 3.16 ± 0.05 | -0.10 ± 0.24 | … | … | N | … | … | … | n | … |
| 53302 | 16005412-5328296 | 13.4 ± 0.3 | 6285 ± 177 | 4.27 ± 0.26 | 0.10 ± 0.14 | … | … | N | … | … | … | n | … |
| 53380 | 16005412-5336140 | -64.3 ± 0.1 | … | … | … | … | … | Y | … | … | … | n | … |
| 53303 | 16005429-5329248 | -90.0 ± 0.3 | 5914 ± 206 | 4.04 ± 0.25 | -0.25 ± 0.17 | … | … | N | … | … | … | n | … |
| 53381 | 16005462-5330451 | -62.7 ± 0.2 | … | … | … | … | … | Y | … | … | … | n | … |
| 53304 | 16005469-5334324 | -57.3 ± 0.3 | 5993 ± 156 | 4.18 ± 0.42 | -0.02 ± 0.21 | 39 ± 23 | 1 | Y | Y | Y | Y | Y | … |
| 53305 | 16005470-5334459 | 48.5 ± 0.2 | 4755 ± 142 | 4.66 ± 0.22 | -0.15 ± 0.13 | … | … | N | … | … | … | n | … |
| 53382 | 16005475-5335022 | 78.4 ± 0.1 | … | … | … | … | … | N | … | … | … | n | … |
| 53306 | 16005553-5330246 | -4.4 ± 0.6 | 7410 ± 74 | … | … | … | … | N | … | … | … | n | … |
| 3417 | 16010433-5332336 | -60.5 ± 0.6 | 4776 ± 122 | 2.48 ± 0.24 | 0.17 ± 0.10 | 31 ± 1 | … | Y | Y | Y | Y | Y | … |
| 53402 | 16010444-5330031 | -68.8 ± 2.7 | … | … | … | … | … | N | … | … | … | n | … |
| 53403 | 16010451-5332442 | -50.2 ± 0.1 | … | … | … | … | … | N | … | … | … | n | … |
| 53404 | 16010454-5328438 | -61.2 ± 0.9 | … | … | … | … | … | Y | … | … | … | n | … |
| 53405 | 16010476-5328190 | -63.0 ± 1.9 | … | … | … | … | … | Y | … | … | … | n | … |
| 53406 | 16010489-5329440 | -61.1 ± 0.5 | … | … | … | … | … | Y | … | … | … | n | … |
| 53327 | 16010496-5332496 | 7.5 ± 0.2 | 6614 ± 305 | 4.15 ± 0.23 | -0.04 ± 0.19 | 64 ± 17 | 1 | N | … | … | … | n | NG |
| 53329 | 16010533-5333195 | -60.9 ± 0.3 | 6378 ± 181 | 4.32 ± 0.27 | 0.17 ± 0.18 | 64 ± 21 | 1 | Y | Y | Y | Y | Y | … |
| 53407 | 16010556-5333133 | -60.0 ± 0.5 | … | … | … | … | … | Y | … | … | … | n | … |
| 53330 | 16010614-5329010 | -11.0 ± 0.3 | 5941 ± 136 | 4.26 ± 0.05 | 0.19 ± 0.12 | 78 ± 20 | 1 | N | … | … | … | n | NG |
| 3418 | 16010639-5331056 | -61.9 ± 0.6 | 4848 ± 121 | 2.74 ± 0.24 | 0.14 ± 0.11 | 44 ± 2 | … | Y | Y | Y | Y | Y | … |
| 53331 | 16010669-5328395 | -123.5 ± 0.3 | 5238 ± 115 | 3.99 ± 0.32 | 0.19 ± 0.29 | … | … | N | … | … | … | n | … |
| 53408 | 16010676-5333432 | -37.6 ± 1.1 | … | … | … | … | … | N | … | … | … | n | … |
| 53409 | 16010748-5333272 | -17.9 ± 3.9 | … | … | … | … | … | N | … | … | … | n | … |
| 3419 | 16010770-5329374 | -61.5 ± 0.6 | 4832 ± 115 | 2.64 ± 0.23 | 0.16 ± 0.10 | 62 ± 1 | … | Y | Y | Y | Y | Y | … |
| 53333 | 16010826-5331372 | 5.6 ± 0.2 | 4967 ± 151 | 4.56 ± 0.36 | 0.04 ± 0.14 | … | … | N | … | … | … | n | … |
| 53309 | 16005585-5335245 | -61.3 ± 0.7 | 7062 ± 553 | 4.10 ± 0.19 | -0.07 ± 0.46 | … | … | Y | … | … | … | n | … |
| 53311 | 16005659-5327535 | 25.8 ± 0.3 | 5794 ± 73 | 4.49 ± 0.19 | -0.21 ± 0.16 | … | … | N | … | … | … | n | … |
| 53312 | 16005675-5335029 | -60.4 ± 0.4 | 6527 ± 203 | 4.36 ± 0.43 | 0.25 ± 0.25 | … | … | Y | … | … | … | n | … |
| 53384 | 16005684-5329384 | -22.2 ± 0.2 | … | … | … | … | … | N | … | … | … | n | … |
| 53385 | 16005689-5332277 | -61.0 ± 1.2 | … | … | … | … | … | Y | … | … | … | n | … |
| 53313 | 16005700-5330159 | -47.4 ± 0.3 | 6535 ± 54 | 3.81 ± 0.11 | 0.44 ± 0.04 | 58 ± 20 | … | N | … | … | … | n | NG |
| 53386 | 16005705-5333242 | -62.9 ± 0.1 | 5501 ± 150 | 4.42 ± 0.02 | 0.65 ± 0.29 | … | … | Y | … | … | … | n | … |
| 53387 | 16005715-5332459 | -57.4 ± 2.1 | … | … | … | … | … | Y | … | … | … | n | … |
| 53388 | 16005739-5332261 | -68.8 ± 0.1 | … | … | … | … | … | N | … | … | … | n | … |
| 53315 | 16005774-5334375 | -59.0 ± 0.6 | 6812 ± 307 | 4.06 ± 0.19 | 0.19 ± 0.30 | … | … | Y | … | … | … | n | … |
| 53389 | 16005793-5329383 | -37.2 ± 0.1 | 5319 ± 109 | 4.69 ± 0.05 | -0.19 ± 0.20 | … | … | N | … | … | … | n | … |
| 3415 | 16005798-5331476 | -60.0 ± 0.6 | 4863 ± 118 | 2.69 ± 0.24 | 0.13 ± 0.10 | <7 | 3 | Y | Y | Y | Y | Y | … |
| 53316 | 16005817-5334494 | 7.8 ± 0.3 | 5274 ± 109 | 4.71 ± 0.37 | 0.04 ± 0.12 | … | … | N | … | … | … | n | … |
| 53317 | 16005823-5333282 | -80.7 ± 0.3 | 6253 ± 389 | 4.17 ± 0.39 | 0.14 ± 0.33 | … | … | N | … | … | … | n | … |
| 53318 | 16005827-5330323 | -60.4 ± 0.4 | 7055 ± 666 | 4.05 ± 0.23 | 0.12 ± 0.14 | 79 ± 30 | 1 | Y | Y | Y | Y | Y | … |
| 53390 | 16005835-5329550 | -54.4 ± 0.1 | … | … | … | … | … | N | … | … | … | n | … |
| 53391 | 16005846-5335490 | -56.0 ± 1.6 | … | … | … | … | … | N | … | … | … | n | … |
| 53412 | 16010967-5332235 | -55.2 ± 0.1 | … | … | … | … | … | N | … | … | … | n | … |
| 53334 | 16011001-5331319 | -29.8 ± 0.4 | 6717 ± 294 | 4.00 ± 0.11 | 0.15 ± 0.32 | 41 ± 26 | 1 | N | … | … | … | n | NG |
| 53265 | 16003077-5331541 | -58.0 ± 0.2 | 6039 ± 83 | 4.05 ± 0.26 | 0.04 ± 0.20 | … | … | Y | … | … | … | n | … |
| 53266 | 16003142-5330137 | -35.7 ± 0.2 | 6051 ± 125 | 4.31 ± 0.39 | -0.01 ± 0.25 | 101 ± 20 | 1 | N | … | … | … | n | NG |
| 53335 | 16003266-5329243 | -61.2 ± 1.9 | … | … | … | … | … | Y | … | … | … | n | … |
| 3401 | 16003885-5334507 | -61.2 ± 0.6 | 4509 ± 135 | 2.42 ± 0.24 | 0.08 ± 0.10 | <7 | 3 | Y | Y | Y | Y | Y | … |
| 53274 | 16003891-5330304 | -60.3 ± 0.8 | 6364 ± 181 | 4.64 ± 0.83 | 0.25 ± 0.20 | … | … | Y | … | … | … | n | … |
| 53343 | 16003903-5333026 | -69.2 ± 4.0 | … | … | … | … | … | N | … | … | … | n | … |
| 3402 | 16003935-5332367 | -60.9 ± 0.6 | 4796 ± 123 | 2.57 ± 0.23 | 0.14 ± 0.10 | 35 ± 2 | … | Y | Y | Y | Y | Y | … |
| 53344 | 16003972-5331217 | -59.6 ± 0.1 | … | … | … | … | … | Y | … | … | … | n | … |
| 53345 | 16003979-5329242 | -31.7 ± 0.1 | 5083 ± 93 | 4.55 ± 0.04 | 0.18 ± 0.14 | … | … | N | … | … | … | n | … |
| 53346 | 16003985-5333514 | -55.1 ± 0.1 | … | … | … | … | … | N | … | … | … | n | … |
| 53275 | 16003991-5335351 | 48.8 ± 0.3 | 4479 ± 118 | 4.79 ± 0.25 | -0.03 ± 0.13 | … | … | N | … | … | … | n | … |

**Table C.12.** continued.

| ID | CNAME | RV (km s$^{-1}$) | $T_{\text{eff}}$ (K) | logg (dex) | [Fe/H] (dex) | EW(Li)$^a$ (mÅ) | EW(Li) error flag$^b$ | RV | Li | Membership logg | [Fe/H] | Final$^c$ | Li-rich non-mem$^d$ |
|---|---|---|---|---|---|---|---|---|---|---|---|---|---|
| 3403 | 16004025-5329439 | -56.5 ± 0.6 | 4912 ± 125 | 2.77 ± 0.24 | 0.11 ± 0.10 | 56 ± 1 | … | Y | Y | Y | Y | Y | … |
| 53347 | 16004034-5333240 | -61.2 ± 0.1 | … | … | … | … | … | Y | … | … | … | n | … |
| 3404 | 16004035-5333047 | -68.6 ± 0.6 | 4897 ± 121 | 3.40 ± 0.26 | 0.00 ± 0.13 | … | … | N | … | … | … | n | Li-rich G |
| 53348 | 16004058-5329583 | -66.2 ± 0.1 | … | … | … | … | … | N | … | … | … | n | … |
| 53276 | 16004146-5328081 | -32.9 ± 0.5 | 6509 ± 263 | 4.22 ± 0.40 | 0.32 ± 0.25 | … | … | N | … | … | … | n | … |
| 53277 | 16004161-5333182 | 42.3 ± 0.3 | 5188 ± 51 | 4.56 ± 0.34 | 0.02 ± 0.13 | … | … | N | … | … | … | n | … |
| 53349 | 16004168-5332182 | -62.8 ± 0.5 | … | … | … | … | … | Y | … | … | … | n | … |
| 53278 | 16004174-5333545 | -58.3 ± 0.3 | 6407 ± 186 | 4.24 ± 0.32 | 0.06 ± 0.15 | 75 ± 27 | 1 | Y | Y | Y | Y | Y | … |
| 53350 | 16004196-5335545 | -25.3 ± 0.4 | … | … | … | … | … | N | … | … | … | n | … |
| 53364 | 16004890-5328089 | -16.3 ± 0.2 | … | … | … | … | … | N | … | … | … | n | … |
| 53291 | 16004906-5330204 | -58.1 ± 1.0 | 6809 ± 344 | 4.13 ± 0.19 | 0.25 ± 0.36 | … | … | Y | … | … | … | n | … |
| 53365 | 16004917-5333257 | 42.0 ± 0.6 | … | … | … | … | … | N | … | … | … | n | … |
| 53292 | 16004925-5334087 | -26.8 ± 0.4 | 6824 ± 318 | 4.25 ± 0.05 | 0.29 ± 0.27 | 30 ± 20 | 1 | N | … | … | … | n | NG |
| 53366 | 16004926-5333097 | -91.2 ± 0.1 | … | … | … | … | … | N | … | … | … | n | … |
| 53367 | 16004929-5331023 | -62.5 ± 0.2 | … | … | … | … | … | Y | … | … | … | n | … |
| 53368 | 16004932-5332384 | -24.9 ± 0.1 | 5365 ± 87 | 4.27 ± 0.02 | 0.09 ± 0.18 | … | … | N | … | … | … | n | … |
| 53369 | 16004946-5329113 | -57.4 ± 7.5 | … | … | … | … | … | Y | … | … | … | n | … |
| 53293 | 16004956-5329564 | -40.4 ± 0.3 | 6189 ± 124 | 3.99 ± 0.19 | 0.02 ± 0.23 | … | … | N | … | … | … | n | … |
| 3410 | 16004973-5331459 | -10.3 ± 0.6 | 4449 ± 126 | 2.43 ± 0.24 | 0.21 ± 0.09 | 24 ± 2 | … | N | … | … | … | n | G |
| 53370 | 16005026-5333580 | -3.5 ± 1.1 | … | … | … | … | … | N | … | … | … | n | … |
| 53371 | 16005040-5331031 | -112.4 ± 0.8 | … | … | … | … | … | N | … | … | … | n | … |
| 53372 | 16005062-5331383 | -62.3 ± 0.4 | … | … | … | … | … | Y | … | … | … | n | … |
| 3411 | 16005072-5335536 | -9.0 ± 0.6 | 4878 ± 126 | 2.95 ± 0.24 | 0.19 ± 0.10 | <12 | 3 | N | … | … | … | n | G |
| 53373 | 16005086-5332030 | -62.6 ± 0.1 | 6063 ± 145 | 3.17 ± 0.03 | … | … | … | Y | … | … | … | n | … |
| 53374 | 16005117-5330529 | -66.0 ± 1.2 | … | … | … | … | … | N | … | … | … | n | … |
| 53375 | 16005131-5333440 | -58.8 ± 0.5 | … | … | … | … | … | Y | … | … | … | n | … |
| 53294 | 16005146-5332497 | -61.0 ± 1.1 | 6517 ± 397 | 3.90 ± 0.16 | 0.27 ± 0.28 | … | … | Y | … | … | … | n | … |
| 3412 | 16005168-5332013 | -62.8 ± 0.6 | 4881 ± 117 | 2.60 ± 0.23 | 0.17 ± 0.10 | 16 ± 1 | … | Y | Y | Y | Y | Y | … |
| 53383 | 16005577-5329141 | -86.4 ± 0.1 | … | … | … | … | … | N | … | … | … | n | … |

$^{(a)}$ The values of $EW$(Li) for this cluster are corrected (subtracted adjacent Fe (6707.43 Å) line). $^{(b)}$ Flags for the errors of the corrected $EW$(Li) values, as follows: 1=$EW$(Li) corrected by blends contribution using models; and 3=Upper limit (no error for $EW$(Li) is given). $^{(c)}$ The letters "Y" and "N" indicate if the star is a cluster member or not. $^{(d)}$ "G" and "NG" indicate "giant" and "non-giant" Li field outliers, respectively.

**Table C.13.** Berkeley 81

| ID | CNAME | RV (km s$^{-1}$) | $T_{\text{eff}}$ (K) | $\log g$ (dex) | [Fe/H] (dex) | $EW(\text{Li})^a$ (mÅ) | $EW(\text{Li})$ error flag$^b$ | Membership RV | Li | $\log g$ | [Fe/H] | Gaia study Cantat-Gaudin$^c$ | Final$^d$ | NMs with Li$^e$ |
|---|---|---|---|---|---|---|---|---|---|---|---|---|---|---|
| 3877 | 19012987-0029026 | 24.1 ± 2.9 | … | … | … | … | … | N | … | … | … | … | n | … |
| 3878 | 19012992-0025377 | 32.5 ± 0.2 | 6491 ± 145 | 4.44 ± 0.05 | … | … | … | N | … | … | … | … | n | … |
| 3696 | 19012992-0029545 | 38.7 ± 0.4 | 6164 ± 215 | 4.65 ± 0.38 | 0.29 ± 0.24 | 71 ± 57 | 1 | N | … | … | … | … | n | NG |
| 3879 | 19013051-0023395 | 22.4 ± 0.2 | … | … | … | … | … | N | … | … | … | … | n | … |
| 3697 | 19013052-0023354 | 36.8 ± 0.4 | 6024 ± 200 | 4.74 ± 0.33 | 0.14 ± 0.17 | 112 ± 70 | 1 | N | … | … | … | N | n | NG |
| 3698 | 19013112-0027173 | -33.8 ± 0.3 | 5802 ± 99 | 4.16 ± 0.37 | 0.31 ± 0.18 | … | … | N | … | … | … | … | n | … |
| 3699 | 19013115-0022486 | 17.0 ± 0.3 | 6589 ± 170 | 4.57 ± 0.14 | 0.03 ± 0.28 | 90 ± 30 | … | N | … | … | … | … | n | NG |
| 3700 | 19013127-0029415 | 33.3 ± 0.4 | 6941 ± 122 | … | -1.61 ± 0.35 | … | … | N | … | … | … | N | n | … |
| 3880 | 19013141-0022149 | 6.2 ± 0.1 | 5807 ± 101 | 4.83 ± 0.04 | 0.04 ± 0.13 | … | … | N | … | … | … | … | n | … |
| 3702 | 19013148-0030049 | -51.8 ± 0.3 | 6035 ± 276 | 4.29 ± 0.21 | 0.15 ± 0.26 | … | … | N | … | … | … | … | n | … |
| 3703 | 19013164-0030465 | 4.0 ± 0.3 | 6306 ± 177 | 4.55 ± 0.16 | 0.29 ± 0.27 | … | … | N | … | … | … | … | n | … |
| 3704 | 19013178-0026470 | 68.9 ± 0.9 | 6290 ± 500 | 5.03 ± 0.41 | -0.41 ± 0.38 | … | … | N | … | … | … | … | n | … |
| 3705 | 19013178-0028535 | 44.5 ± 0.3 | 6335 ± 274 | 4.55 ± 0.17 | -0.06 ± 0.40 | 83 ± 34 | … | Y | Y | Y | Y | … | Y | … |
| 3706 | 19013227-0031112 | 87.8 ± 0.5 | 6363 ± 327 | 4.32 ± 0.43 | -0.12 ± 0.22 | … | … | N | … | … | … | … | n | … |
| 3707 | 19013240-0023593 | 47.9 ± 0.7 | 6371 ± 239 | … | 0.05 ± 0.19 | <107 | 3 | Y | Y | Y | Y | … | Y | … |
| 3708 | 19013243-0027190 | -22.4 ± 0.3 | 5475 ± 132 | 4.89 ± 0.50 | 0.08 ± 0.16 | <82 | 3 | N | … | … | … | … | n | NG |
| 3709 | 19013256-0023474 | 29.1 ± 0.5 | 6344 ± 363 | 4.19 ± 0.18 | -0.24 ± 0.23 | … | … | N | … | … | … | … | n | … |
| 3881 | 19013277-0022436 | 27.6 ± 0.5 | … | … | … | … | … | N | … | … | … | … | n | … |
| 3882 | 19013280-0025031 | 14.9 ± 3.0 | … | … | … | … | … | N | … | … | … | … | n | … |
| 3711 | 19013287-0031312 | 44.5 ± 0.4 | 6285 ± 259 | 4.48 ± 0.28 | 0.32 ± 0.25 | <72 | 3 | Y | Y | Y | Y | … | Y | … |
| 3712 | 19013299-0024437 | 17.9 ± 2.1 | … | … | … | … | … | N | … | … | … | … | n | … |
| 3713 | 19013307-0030576 | 49.5 ± 1.4 | 7370 ± 137 | … | … | … | … | Y | … | … | … | N | n | … |
| 3774 | 19014328-0029491 | 57.1 ± 0.5 | 5866 ± 122 | 4.46 ± 0.30 | 0.16 ± 0.15 | <67 | 3 | N | … | … | … | … | n | NG |
| 3775 | 19014347-0027595 | 15.3 ± 0.4 | 5962 ± 67 | 4.57 ± 0.34 | -0.11 ± 0.13 | 84 ± 39 | 1 | N | … | … | … | … | n | NG |
| 3911 | 19014349-0026374 | 49.4 ± 0.1 | 5757 ± 147 | 4.91 ± 0.05 | 0.35 ± 0.18 | … | … | Y | … | … | … | … | n | … |
| 3912 | 19014355-0027432 | -37.2 ± 0.1 | 5785 ± 130 | 4.16 ± 0.08 | -0.14 ± 0.21 | … | … | N | … | … | … | … | n | … |
| 3776 | 19014383-0025585 | 43.9 ± 2.2 | 7425 ± 199 | … | … | … | … | Y | … | … | … | N | n | … |
| 3913 | 19014399-0026493 | 52.0 ± 3.1 | … | … | … | … | … | Y | … | … | … | Y | n | … |
| 3777 | 19014399-0031316 | 15.0 ± 0.3 | 6223 ± 276 | 4.50 ± 0.22 | 0.16 ± 0.16 | 41 ± 36 | 1 | N | … | … | … | … | n | NG |
| 3778 | 19014404-0022572 | -4.9 ± 0.5 | 6422 ± 328 | 5.12 ± 1.09 | 0.31 ± 0.20 | … | … | N | … | … | … | … | n | … |
| 3779 | 19014417-0025115 | 51.9 ± 1.6 | 6695 ± 575 | 4.80 ± 0.77 | 0.30 ± 0.33 | 80 ± 77 | 1 | Y | Y | Y | Y | … | Y | … |
| 3780 | 19014419-0030238 | 6.2 ± 0.3 | 5844 ± 245 | 4.58 ± 0.56 | 0.15 ± 0.20 | 108 ± 34 | 1 | N | … | … | … | … | n | NG |
| 3781 | 19014436-0029370 | -103.3 ± 0.6 | 5921 ± 213 | 4.37 ± 0.06 | -1.14 ± 0.61 | … | … | N | … | … | … | … | n | … |
| 3782 | 19014443-0026202 | 47.1 ± 1.0 | 6688 ± 581 | 4.44 ± 0.62 | 0.27 ± 0.54 | 86 ± 70 | 1 | Y | Y | Y | Y | Y | Y | … |
| 3783 | 19014447-0024227 | 58.6 ± 1.4 | 7089 ± 930 | 3.93 ± 0.18 | 0.17 ± 0.18 | … | … | N | … | … | … | N | n | … |
| 3784 | 19014448-0027471 | -29.7 ± 0.3 | 6108 ± 239 | 4.34 ± 0.16 | 0.37 ± 0.21 | … | … | N | … | … | … | … | n | … |
| 3785 | 19014452-0026570 | -19.0 ± 0.4 | 5003 ± 255 | 4.37 ± 0.14 | 0.02 ± 0.24 | … | … | N | … | … | … | … | n | … |
| 3914 | 19014469-0028446 | 23.6 ± 0.8 | … | … | … | … | … | N | … | … | … | N | n | … |
| 482 | 19014498-0027496 | 48.3 ± 0.6 | 4742 ± 119 | 2.66 ± 0.23 | 0.09 ± 0.10 | 103 ± 4 | … | Y | N | Y | Y | … | Y | … |
| 3915 | 19014511-0027303 | 40.3 ± 5.9 | … | … | … | … | … | N | … | … | … | Y | n | … |
| 483 | 19014525-0023580 | 49.4 ± 0.6 | 4966 ± 121 | 2.81 ± 0.23 | 0.15 ± 0.10 | 19 ± 1 | … | Y | Y | Y | Y | … | Y | … |
| 3786 | 19014532-0026401 | 47.8 ± 0.5 | 5953 ± 285 | 4.94 ± 0.44 | 0.26 ± 0.15 | <93 | 3 | Y | Y | Y? | Y | … | Y | … |
| 3916 | 19014533-0032403 | 49.1 ± 2.2 | … | … | … | … | … | Y | … | … | … | Y | n | … |
| 3787 | 19014577-0024134 | 13.5 ± 0.3 | 6072 ± 168 | 4.73 ± 0.40 | 0.00 ± 0.19 | 74 ± 30 | 1 | N | … | … | … | … | n | NG |
| 3788 | 19014588-0027103 | 12.1 ± 0.5 | 6179 ± 146 | 4.88 ± 0.67 | -0.34 ± 0.26 | <74 | 3 | N | … | … | … | … | n | NG |
| 3789 | 19014589-0027348 | 51.3 ± 1.4 | 6692 ± 85 | … | 0.71 ± 0.07 | … | … | Y | … | … | … | Y | n | … |
| 3858 | 19020010-0023547 | 26.9 ± 0.4 | 5999 ± 322 | 4.11 ± 0.27 | 0.19 ± 0.25 | … | … | N | … | … | … | … | n | … |
| 3859 | 19020043-0025034 | 27.0 ± 0.6 | 5978 ± 195 | 4.56 ± 0.22 | 0.14 ± 0.20 | … | … | N | … | … | … | … | n | … |
| 3860 | 19020052-0025161 | 21.0 ± 0.4 | 5762 ± 156 | 4.57 ± 0.39 | 0.32 ± 0.22 | 153 ± 130 | 1 | N | … | … | … | … | n | NG |
| 3861 | 19020063-0027554 | 21.2 ± 0.4 | 5639 ± 168 | 4.70 ± 0.53 | 0.22 ± 0.26 | <132 | 3 | N | … | … | … | … | n | NG |
| 3862 | 19020106-0027155 | 55.2 ± 1.4 | 6292 ± 177 | 4.30 ± 0.39 | 0.07 ± 0.13 | <101 | 3 | N | … | … | … | … | n | NG |
| 3863 | 19020110-0028129 | 9.6 ± 0.5 | 6339 ± 321 | 4.35 ± 0.24 | 0.36 ± 0.37 | … | … | N | … | … | … | … | n | … |
| 3864 | 19020111-0024423 | -78.2 ± 0.4 | 5581 ± 78 | 4.04 ± 0.54 | -0.50 ± 0.14 | … | … | N | … | … | … | … | n | … |
| 3942 | 19020121-0025128 | 48.1 ± 1.4 | … | … | … | … | … | Y | … | … | … | Y | n | … |
| 3865 | 19020130-0027075 | 97.0 ± 0.4 | 5939 ± 158 | 4.02 ± 0.30 | -0.46 ± 0.15 | 80 ± 59 | 1 | N | … | … | … | … | n | NG |
| 3943 | 19020169-0026580 | -39.7 ± 0.2 | 5793 ± 109 | 3.12 ± 0.10 | -0.09 ± 0.21 | … | … | N | … | … | … | … | n | … |
| 3866 | 19020173-0025579 | 36.6 ± 0.3 | … | … | … | … | … | N | … | … | … | Y | n | … |
| 3867 | 19020179-0025303 | 16.8 ± 0.3 | 6311 ± 142 | 4.19 ± 0.07 | 0.32 ± 0.15 | … | … | N | … | … | … | … | n | … |
| 3868 | 19020258-0026529 | 8.2 ± 0.4 | 6043 ± 183 | 4.35 ± 0.25 | -0.01 ± 0.15 | <65 | 3 | N | … | … | … | … | n | NG |

**Table C.13.** continued.

| ID | CNAME | RV (km s$^{-1}$) | $T_{\text{eff}}$ (K) | logg (dex) | [Fe/H] (dex) | EW(Li)$^a$ (mÅ) | EW(Li) error flag$^b$ | RV | Membership Li | logg | [Fe/H] | Gaia study Cantat-Gaudin$^c$ | Final$^d$ | NMs with Li$^e$ |
|---|---|---|---|---|---|---|---|---|---|---|---|---|---|---|
| 3944 | 19020260-0026030 | -3.5 ± 0.1 | 5835 ± 74 | 4.31 ± 0.01 | 0.09 ± 0.13 | … | … | N | … | … | … | … | n | … |
| 3945 | 19020275-0027274 | 12.9 ± 0.1 | 5783 ± 118 | 3.52 ± 0.05 | -0.07 ± 0.21 | … | … | N | … | … | … | … | n | … |
| 3946 | 19020318-0029117 | 15.4 ± 0.1 | 6051 ± 91 | 4.49 ± 0.02 | … | … | … | N | … | … | … | … | n | … |
| 3869 | 19020325-0025502 | 20.1 ± 0.5 | 6505 ± 131 | 4.55 ± 0.09 | 0.03 ± 0.33 | <47 | 3 | N | … | … | … | … | n | NG |
| 3870 | 19020346-0025431 | 44.9 ± 1.5 | 7109 ± 823 | 4.01 ± 0.26 | 0.18 ± 0.13 | <98 | 3 | Y | Y | Y | Y | N | Y | … |
| 3947 | 19020347-0024442 | 51.4 ± 0.1 | 5969 ± 112 | 3.49 ± 0.06 | 0.21 ± 0.22 | … | … | Y | … | … | … | … | n | … |
| 3925 | 19014775-0026521 | -2.1 ± 0.5 | … | … | … | … | … | N | … | … | … | Y | n | … |
| 3802 | 19014777-0028311 | 21.0 ± 0.5 | 6087 ± 195 | 4.64 ± 0.42 | 0.23 ± 0.16 | <54 | 3 | N | … | … | … | … | n | NG |
| 3803 | 19014802-0022219 | 45.4 ± 1.0 | 7466 ± 113 | … | … | … | … | Y | … | … | … | Y | n | … |
| 3819 | 19015017-0023178 | 20.0 ± 0.4 | 5954 ± 323 | 4.54 ± 0.38 | -0.16 ± 0.43 | 167 ± 60 | 1 | N | … | … | … | … | n | NG |
| 3820 | 19015024-0029114 | 7.7 ± 0.6 | 6791 ± 103 | 4.49 ± 0.20 | 0.39 ± 0.09 | … | … | N | … | … | … | … | n | … |
| 3821 | 19015029-0028468 | 47.8 ± 2.3 | 6502 ± 174 | 5.24 ± 0.43 | 0.66 ± 0.13 | … | … | Y | … | … | … | … | n | … |
| 3822 | 19015034-0023522 | 73.2 ± 0.3 | 6225 ± 509 | 4.75 ± 0.44 | -0.05 ± 0.25 | … | … | N | … | … | … | … | n | … |
| 3823 | 19015047-0025342 | 26.7 ± 0.8 | 6431 ± 295 | 4.67 ± 0.76 | 0.24 ± 0.34 | <111 | 3 | N | … | … | … | N | n | NG |
| 3935 | 19015059-0026088 | 40.0 ± 0.1 | 6105 ± 123 | 3.76 ± 0.08 | … | … | … | N | … | … | … | … | n | … |
| 3824 | 19015059-0031310 | 133.5 ± 0.3 | 5496 ± 241 | 4.09 ± 0.32 | 0.04 ± 0.12 | … | … | N | … | … | … | … | n | … |
| 3825 | 19015069-0030166 | -14.8 ± 0.4 | 5370 ± 183 | 4.53 ± 0.26 | -0.23 ± 0.15 | <117 | 3 | N | … | … | … | … | n | NG |
| 3826 | 19015095-0031428 | -40.2 ± 0.3 | 5704 ± 379 | 4.24 ± 0.08 | 0.11 ± 0.23 | … | … | N | … | … | … | … | n | … |
| 3827 | 19015104-0030434 | -35.8 ± 0.3 | 5801 ± 239 | 4.48 ± 0.18 | 0.32 ± 0.18 | … | … | N | … | … | … | … | n | … |
| 3828 | 19015111-0025269 | 37.4 ± 0.6 | 7137 ± 95 | … | … | <46 | 3 | N | … | … | … | … | n | … |
| 3829 | 19015147-0024234 | 23.1 ± 0.4 | 5942 ± 215 | 4.99 ± 0.59 | 0.33 ± 0.22 | 123 ± 77 | 1 | N | … | … | … | … | n | NG |
| 3830 | 19015153-0026121 | 41.6 ± 1.4 | 7398 ± 110 | … | … | … | … | N | … | … | … | Y | n | … |
| 3831 | 19015186-0026120 | -7.2 ± 0.5 | 5802 ± 302 | 4.22 ± 0.54 | 0.21 ± 0.19 | … | … | N | … | … | … | … | n | … |
| 3832 | 19015192-0027340 | 47.6 ± 2.0 | 6888 ± 122 | … | 0.57 ± 0.11 | … | … | Y | … | … | … | Y | n | … |
| 3833 | 19015195-0023078 | 44.2 ± 0.7 | 6509 ± 313 | 4.70 ± 0.50 | 0.20 ± 0.21 | … | … | Y | … | … | … | … | n | … |
| 3936 | 19015195-0025058 | 67.7 ± 0.2 | 5545 ± 146 | 3.09 ± 0.05 | 0.04 ± 0.25 | … | … | N | … | … | … | … | n | … |
| 3834 | 19015196-0028545 | 58.3 ± 0.4 | 6239 ± 307 | 4.75 ± 0.32 | 0.40 ± 0.28 | 75 ± 67 | 1 | N | … | … | … | … | n | NG |
| 3835 | 19015218-0027052 | 43.6 ± 1.8 | 7173 ± 134 | … | … | … | … | Y | … | … | … | N | n | … |
| 3836 | 19015226-0022264 | 27.6 ± 0.9 | 6660 ± 565 | 4.06 ± 0.18 | 0.04 ± 0.28 | … | … | N | … | … | … | … | n | … |
| 485 | 19015261-0025318 | 48.5 ± 0.6 | 4954 ± 132 | 2.92 ± 0.26 | 0.23 ± 0.12 | 29 ± 2 | … | Y | Y | Y | Y | … | Y | … |
| 3937 | 19015278-0028549 | 44.3 ± 0.4 | 6781 ± 149 | 4.33 ± 0.05 | … | … | … | Y | … | … | … | … | n | … |
| 3837 | 19015300-0022239 | 45.5 ± 0.8 | 6525 ± 303 | 4.51 ± 0.33 | 0.25 ± 0.21 | … | … | Y | … | … | … | … | n | … |
| 3714 | 19013310-0029162 | 28.9 ± 0.6 | 5920 ± 111 | 4.22 ± 0.57 | 0.20 ± 0.16 | 121 ± 119 | 1 | N | … | … | … | … | n | NG |
| 3715 | 19013316-0026018 | -55.0 ± 0.4 | 5456 ± 111 | 3.86 ± 0.24 | -0.35 ± 0.32 | … | … | N | … | … | … | … | n | … |
| 3883 | 19013326-0023376 | 33.1 ± 0.6 | … | … | … | … | … | N | … | … | … | … | n | … |
| 3884 | 19013340-0024088 | 11.4 ± 1.2 | … | … | … | … | … | N | … | … | … | … | n | … |
| 3885 | 19013359-0030550 | -92.3 ± 0.2 | … | … | … | … | … | N | … | … | … | … | n | … |
| 3886 | 19013366-0026312 | -55.5 ± 0.1 | 5871 ± 138 | 4.87 ± 0.02 | 0.19 ± 0.23 | … | … | N | … | … | … | … | n | … |
| 3716 | 19013372-0025163 | 38.2 ± 2.1 | 6676 ± 381 | 4.38 ± 0.36 | 0.15 ± 0.19 | … | … | N | … | … | … | N | n | … |
| 3717 | 19013373-0023487 | 10.7 ± 0.7 | 6675 ± 101 | 4.32 ± 0.20 | 0.26 ± 0.08 | … | … | N | … | … | … | … | n | … |
| 3718 | 19013420-0030383 | -50.1 ± 0.3 | 5399 ± 201 | 4.58 ± 0.25 | 0.03 ± 0.19 | … | … | N | … | … | … | … | n | … |
| 3720 | 19013446-0027369 | 48.7 ± 1.2 | 7288 ± 115 | … | … | … | … | Y | … | … | … | Y | n | … |
| 3721 | 19013466-0023103 | 57.5 ± 0.4 | 6171 ± 296 | 4.68 ± 0.32 | 0.02 ± 0.22 | <50 | 3 | N | … | … | … | … | n | NG |
| 3887 | 19013467-0030595 | -19.0 ± 0.2 | 6093 ± 119 | 3.47 ± 0.03 | … | … | … | N | … | … | … | … | n | … |
| 3722 | 19013485-0028506 | -16.8 ± 1.1 | 6167 ± 18 | 3.82 ± 0.17 | -0.61 ± 0.02 | … | … | N | … | … | … | … | n | … |
| 3723 | 19013495-0028320 | 46.6 ± 0.7 | 7148 ± 810 | 4.00 ± 0.21 | 0.11 ± 0.14 | … | … | Y | … | … | … | … | n | … |
| 3724 | 19013516-0028216 | 7.3 ± 0.5 | 5544 ± 85 | 4.45 ± 0.56 | -0.17 ± 0.16 | <86 | 3 | N | … | … | … | … | n | NG |
| 3725 | 19013532-0026272 | -9.3 ± 0.5 | 5947 ± 177 | 4.81 ± 0.43 | 0.14 ± 0.17 | … | … | N | … | … | … | … | n | … |
| 473 | 19013537-0028186 | 47.5 ± 0.6 | 5167 ± 142 | 3.30 ± 0.30 | 0.27 ± 0.13 | <16 | 3 | Y | Y | Y | Y | … | Y | … |
| 3888 | 19013544-0027120 | 48.1 ± 1.2 | … | … | … | … | … | Y | … | … | … | Y | n | … |
| 3889 | 19013549-0024162 | 6.8 ± 0.1 | 5847 ± 112 | 4.39 ± 0.01 | -0.03 ± 0.20 | … | … | N | … | … | … | … | n | … |
| 3726 | 19013549-0029565 | 76.8 ± 1.5 | 6344 ± 326 | 5.05 ± 0.82 | 0.05 ± 0.39 | <94 | 3 | N | … | … | … | … | n | NG |
| 3890 | 19013565-0027330 | 4.9 ± 0.2 | 5901 ± 150 | 3.58 ± 0.01 | -0.04 ± 0.26 | … | … | N | … | … | … | … | n | … |
| 3727 | 19013567-0026318 | 25.8 ± 0.6 | 6071 ± 73 | 4.79 ± 0.60 | -0.21 ± 0.20 | 79 ± 77 | 1 | N | … | … | … | … | n | NG |
| 3728 | 19013585-0027549 | -9.6 ± 0.7 | 5603 ± 77 | 4.35 ± 0.41 | -0.59 ± 0.24 | … | … | N | … | … | … | … | n | … |
| 3745 | 19013809-0027105 | 48.0 ± 1.0 | 6459 ± 336 | 4.86 ± 0.59 | 0.25 ± 0.24 | <45 | 3 | Y | Y | Y? | Y | N | Y | … |
| 3746 | 19013812-0029237 | 68.0 ± 1.7 | 6467 ± 250 | 5.04 ± 0.30 | -0.30 ± 0.50 | <95 | 3 | N | … | … | … | … | n | NG |
| 3897 | 19013825-0025436 | 46.4 ± 0.3 | … | … | … | … | … | Y | … | … | … | Y | n | … |
| 3747 | 19013840-0022155 | -12.4 ± 0.3 | 5724 ± 19 | 4.60 ± 0.32 | 0.11 ± 0.12 | … | … | N | … | … | … | … | n | … |

**Table C.13.** continued.

| ID | CNAME | RV (km s$^{-1}$) | $T_{\rm eff}$ (K) | logg (dex) | [Fe/H] (dex) | EW(Li)$^a$ (mÅ) | EW(Li) error flag$^b$ | Membership RV | Li | logg | [Fe/H] | Gaia study Cantat-Gaudin$^c$ | Final$^d$ | NMs with Li$^e$ |
|---|---|---|---|---|---|---|---|---|---|---|---|---|---|---|
| 3748 | 19013850-0027550 | 51.1 ± 0.3 | 6431 ± 130 | 4.68 ± 0.30 | 0.11 ± 0.16 | … | … | Y | … | … | … | … | n | … |
| 3749 | 19013856-0030379 | -20.3 ± 0.3 | 5914 ± 124 | 4.52 ± 0.52 | 0.23 ± 0.14 | <41 | 3 | N | … | … | … | … | n | NG |
| 3898 | 19013864-0022455 | -65.9 ± 0.1 | 5765 ± 107 | 4.67 ± 0.06 | -0.22 ± 0.19 | … | … | N | … | … | … | … | n | … |
| 3899 | 19013869-0030208 | -7.6 ± 0.1 | 5841 ± 132 | 4.14 ± 0.06 | -0.31 ± 0.20 | … | … | N | … | … | … | … | n | … |
| 3750 | 19013886-0026309 | 49.6 ± 3.5 | 6584 ± 464 | 4.69 ± 0.65 | 0.20 ± 0.19 | … | … | Y | … | … | … | … | n | … |
| 3900 | 19013891-0027029 | -29.3 ± 0.2 | … | … | … | … | … | N | … | … | … | … | n | … |
| 3901 | 19013900-0027441 | 76.0 ± 0.3 | … | … | … | … | … | N | … | … | … | … | n | … |
| 476 | 19013910-0027114 | 48.8 ± 0.6 | 5012 ± 117 | 2.96 ± 0.24 | 0.12 ± 0.10 | 9 ± 1 | … | Y | Y | Y | Y | … | Y | … |
| 3751 | 19013911-0025473 | -17.0 ± 0.3 | 6254 ± 282 | 4.60 ± 0.11 | 0.18 ± 0.35 | 43 ± 17 | 1 | N | … | … | … | N | n | NG |
| 3902 | 19013925-0027568 | 42.5 ± 1.0 | … | … | … | … | … | N | … | … | … | … | n | … |
| 3752 | 19013935-0028300 | 47.5 ± 1.4 | 7185 ± 722 | 4.12 ± 0.19 | 0.11 ± 0.21 | … | … | Y | … | … | … | N | n | … |
| 3903 | 19013938-0026492 | 44.8 ± 0.1 | … | … | … | … | … | Y | … | … | … | Y | n | … |
| 3754 | 19013967-0027010 | 51.1 ± 3.0 | 6225 ± 293 | … | 0.44 ± 0.19 | … | … | Y | … | … | … | … | n | … |
| 3755 | 19013968-0027220 | 14.6 ± 0.4 | 5267 ± 34 | 4.99 ± 0.72 | 0.15 ± 0.15 | <65 | 3 | N | … | … | … | … | n | NG |
| 3904 | 19013980-0025536 | 45.4 ± 5.0 | … | … | … | … | … | Y | … | … | … | Y | n | … |
| 3905 | 19013982-0030271 | 47.2 ± 1.2 | … | … | … | … | … | Y | … | … | … | Y | n | … |
| 477 | 19013997-0028213 | 48.6 ± 0.6 | 4970 ± 149 | 3.27 ± 0.27 | 0.32 ± 0.10 | <17 | 3 | Y | Y | Y | Y | … | Y | … |
| 478 | 19014004-0028129 | 48.7 ± 0.6 | 4940 ± 127 | 2.77 ± 0.24 | 0.26 ± 0.11 | 12 ± 1 | … | Y | Y | Y | Y | … | Y | … |
| 3756 | 19014025-0026284 | 38.1 ± 0.4 | 6041 ± 134 | 4.54 ± 0.22 | 0.06 ± 0.19 | <45 | 3 | N | … | … | … | … | n | NG |
| 3790 | 19014637-0024287 | 45.7 ± 0.5 | 6171 ± 208 | 4.55 ± 0.31 | 0.27 ± 0.30 | … | … | Y | … | … | … | … | n | … |
| 3917 | 19014640-0025137 | 42.0 ± 5.1 | … | … | … | … | … | N | … | … | … | Y | n | … |
| 3791 | 19014647-0021577 | -1.4 ± 0.4 | 6207 ± 15 | 4.44 ± 0.53 | 0.15 ± 0.22 | … | … | N | … | … | … | … | n | … |
| 3792 | 19014648-0027210 | 49.8 ± 0.7 | 7049 ± 856 | … | … | <52 | 3 | Y | Y | … | … | Y | Y | … |
| 3918 | 19014667-0032351 | -70.7 ± 0.1 | 5365 ± 94 | 4.05 ± 0.05 | -0.01 ± 0.19 | … | … | N | … | … | … | … | n | … |
| 3919 | 19014677-0029522 | 46.4 ± 0.4 | … | … | … | … | … | Y | … | … | … | … | n | … |
| 3793 | 19014687-0031305 | -10.2 ± 0.3 | 6173 ± 283 | 4.54 ± 0.21 | 0.35 ± 0.40 | 117 ± 57 | 1 | N | … | … | … | … | n | NG |
| 3920 | 19014691-0029379 | -19.6 ± 0.2 | 5331 ± 150 | 2.73 ± 0.07 | -0.41 ± 0.24 | … | … | N | … | … | … | … | n | … |
| 3794 | 19014707-0026219 | 37.3 ± 0.6 | 6494 ± 321 | 4.42 ± 0.31 | 0.27 ± 0.21 | … | … | N | … | … | … | … | n | … |
| 3921 | 19014717-0031597 | -25.6 ± 0.1 | … | … | … | … | … | N | … | … | … | … | n | … |
| 3795 | 19014727-0030467 | 81.9 ± 0.9 | 7127 ± 1165 | 4.01 ± 0.22 | -0.03 ± 0.19 | <67 | 3 | N | … | … | … | … | n | NG |
| 3922 | 19014728-0021464 | 8.9 ± 0.1 | 5509 ± 119 | 3.35 ± 0.03 | 0.10 ± 0.20 | … | … | N | … | … | … | … | n | … |
| 3796 | 19014728-0025559 | 46.6 ± 1.7 | 6560 ± 440 | 5.07 ± 1.04 | 0.21 ± 0.19 | <64 | 3 | Y | Y | N | Y | Y | Y | … |
| 3923 | 19014740-0028273 | 45.4 ± 7.0 | … | … | … | … | … | Y | … | … | … | Y | n | … |
| 3797 | 19014741-0025397 | -21.0 ± 0.3 | 5334 ± 208 | 4.36 ± 0.48 | -0.28 ± 0.15 | … | … | N | … | … | … | … | n | … |
| 3798 | 19014742-0023253 | 34.4 ± 0.6 | 5718 ± 113 | 4.50 ± 0.18 | -0.13 ± 0.15 | … | … | N | … | … | … | … | n | … |
| 3799 | 19014742-0026555 | -89.1 ± 0.7 | 5353 ± 33 | 3.06 ± 0.31 | -1.28 ± 0.77 | … | … | N | … | … | … | … | n | … |
| 3924 | 19014762-0028120 | -8.7 ± 0.2 | … | … | … | … | … | N | … | … | … | … | n | … |
| 3800 | 19014766-0029392 | -16.9 ± 0.3 | 5655 ± 43 | 4.63 ± 0.35 | 0.05 ± 0.16 | 66 ± 41 | 1 | N | … | … | … | … | n | NG |
| 484 | 19014769-0025108 | 48.0 ± 0.6 | 4785 ± 120 | 2.59 ± 0.23 | 0.20 ± 0.10 | 15 ± 1 | … | Y | Y | Y | Y | … | Y | … |
| 3801 | 19014772-0027550 | 46.1 ± 1.0 | 6086 ± 103 | 4.18 ± 0.07 | 0.10 ± 0.33 | … | … | Y | … | … | … | … | n | … |
| 3759 | 19014092-0027403 | 3.5 ± 0.4 | 6056 ± 244 | 4.42 ± 0.22 | -0.01 ± 0.22 | 134 ± 89 | 1 | N | … | … | … | N | n | NG |
| 3761 | 19014102-0022479 | 41.7 ± 2.6 | 6876 ± 543 | 3.97 ± 0.24 | 0.25 ± 0.21 | … | … | N | … | … | … | Y | n | … |
| 3907 | 19014103-0027233 | 53.0 ± 3.0 | … | … | … | … | … | N | … | … | … | Y | n | … |
| 479 | 19014127-0026444 | 47.2 ± 0.6 | 4870 ± 128 | 2.98 ± 0.26 | 0.23 ± 0.10 | <13 | 3 | Y | Y | Y | Y | … | Y | … |
| 3763 | 19014128-0023169 | 58.9 ± 0.5 | 6200 ± 325 | 4.31 ± 0.38 | -0.31 ± 0.15 | … | … | N | … | … | … | N | n | … |
| 3764 | 19014134-0029247 | -19.3 ± 0.3 | 6318 ± 197 | 4.41 ± 0.31 | 0.36 ± 0.29 | … | … | N | … | … | … | … | n | … |
| 3765 | 19014135-0030068 | -5.5 ± 0.3 | 5959 ± 163 | 4.72 ± 0.34 | 0.15 ± 0.17 | 57 ± 37 | 1 | N | … | … | … | … | n | NG |
| 3766 | 19014167-0025379 | -54.5 ± 0.3 | 5497 ± 160 | 4.13 ± 0.30 | -0.26 ± 0.22 | … | … | N | … | … | … | … | n | … |
| 3767 | 19014169-0024008 | 56.7 ± 0.4 | 5949 ± 185 | 4.58 ± 0.28 | 0.18 ± 0.13 | 100 ± 72 | 1 | N | … | … | … | … | n | NG |
| 480 | 19014194-0028172 | 48.4 ± 0.6 | 4952 ± 126 | 2.92 ± 0.27 | 0.20 ± 0.10 | 24 ± 2 | … | Y | Y | Y | Y | … | Y | … |
| 3769 | 19014198-0027485 | 12.8 ± 0.3 | 5487 ± 108 | 4.29 ± 0.12 | 0.31 ± 0.24 | … | … | N | … | … | … | … | n | … |
| 3770 | 19014200-0026438 | 50.3 ± 0.9 | 7148 ± 137 | … | … | … | … | Y | … | … | … | N | n | … |
| 3908 | 19014207-0024371 | -29.4 ± 0.3 | … | … | … | … | … | N | … | … | … | … | n | … |
| 481 | 19014228-0027388 | 48.6 ± 0.6 | 4916 ± 126 | 2.83 ± 0.24 | 0.22 ± 0.10 | 18 ± 1 | … | Y | Y | Y | Y | … | Y | … |
| 3772 | 19014235-0028130 | 52.3 ± 0.4 | 6086 ± 174 | 4.13 ± 0.15 | 0.16 ± 0.19 | … | … | Y | … | … | … | … | n | … |
| 3909 | 19014240-0024069 | 458.8 ± 0.2 | … | … | … | … | … | N | … | … | … | … | n | … |
| 3910 | 19014304-0030399 | 33.9 ± 0.2 | 6419 ± 150 | 4.03 ± 0.02 | … | … | … | N | … | … | … | … | n | … |
| 3773 | 19014324-0031277 | -29.2 ± 0.3 | 5397 ± 355 | 4.27 ± 0.45 | -0.54 ± 0.22 | … | … | N | … | … | … | … | n | … |
| 3804 | 19014802-0028536 | 16.8 ± 0.4 | 5905 ± 110 | 4.73 ± 0.34 | 0.16 ± 0.14 | 68 ± 67 | 1 | N | … | … | … | … | n | NG |

**Table C.13.** continued.

| ID | CNAME | RV (km s$^{-1}$) | $T_{\text{eff}}$ (K) | logg (dex) | [Fe/H] (dex) | EW(Li)$^a$ (mÅ) | EW(Li) error flag$^b$ | Membership RV | Li | logg | [Fe/H] | Gaia study Cantat-Gaudin$^c$ | Final$^d$ | NMs with Li$^e$ |
|---|---|---|---|---|---|---|---|---|---|---|---|---|---|---|
| 3805 | 19014802-0031461 | 48.4 ± 1.0 | 7070 ± 137 | … | … | … | … | Y | … | … | … | N | n | … |
| 3806 | 19014804-0024382 | 37.0 ± 0.3 | 5984 ± 199 | 4.58 ± 0.35 | -0.11 ± 0.12 | … | … | N | … | … | … | … | n | … |
| 3926 | 19014805-0031002 | -17.6 ± 0.1 | … | … | … | … | … | N | … | … | … | … | n | … |
| 3927 | 19014817-0022147 | -5.0 ± 0.2 | … | … | … | … | … | N | … | … | … | … | n | … |
| 3928 | 19014826-0026233 | 37.6 ± 0.6 | … | … | … | … | … | N | … | … | … | Y | n | … |
| 3929 | 19014832-0024500 | 56.4 ± 5.8 | … | … | … | … | … | N | … | … | … | Y | n | … |
| 3807 | 19014841-0026391 | 4.4 ± 0.3 | 5923 ± 199 | 4.60 ± 0.24 | -0.20 ± 0.20 | … | … | N | … | … | … | … | n | … |
| 3930 | 19014852-0030393 | -17.7 ± 0.1 | 5685 ± 145 | 4.91 ± 0.05 | 0.43 ± 0.16 | … | … | N | … | … | … | … | n | … |
| 3808 | 19014865-0030469 | -51.5 ± 0.3 | 5288 ± 249 | 4.34 ± 0.35 | 0.15 ± 0.15 | … | … | N | … | … | … | … | n | … |
| 3809 | 19014885-0025181 | 44.4 ± 1.7 | 6890 ± 114 | … | 0.46 ± 0.10 | … | … | Y | … | … | … | N | n | … |
| 3810 | 19014895-0023504 | -20.3 ± 0.3 | 5940 ± 141 | 4.91 ± 0.61 | -0.39 ± 0.69 | … | … | N | … | … | … | … | n | … |
| 3811 | 19014900-0024203 | 66.0 ± 1.2 | 6748 ± 600 | 5.13 ± 1.19 | 0.16 ± 0.20 | <99 | 3 | N | … | … | … | … | n | NG |
| 3812 | 19014904-0024413 | -36.9 ± 0.3 | 5729 ± 22 | 4.29 ± 0.46 | -0.62 ± 0.23 | … | … | N | … | … | … | … | n | … |
| 3931 | 19014910-0023258 | -13.9 ± 0.3 | 5501 ± 150 | 3.94 ± 0.01 | -0.19 ± 0.41 | … | … | N | … | … | … | … | n | … |
| 3813 | 19014929-0030101 | 9.7 ± 0.3 | 6323 ± 193 | 4.52 ± 0.10 | 0.20 ± 0.21 | 74 ± 25 | 1 | N | … | … | … | … | n | NG |
| 3814 | 19014943-0027212 | -55.7 ± 0.7 | 5393 ± 64 | 3.38 ± 0.27 | -0.75 ± 0.28 | <87 | 3 | N | … | … | … | … | n | … |
| 3815 | 19014945-0023099 | 103.0 ± 0.5 | 5963 ± 129 | 4.60 ± 0.33 | -0.72 ± 0.31 | … | … | N | … | … | … | … | n | … |
| 3816 | 19014951-0028219 | -18.1 ± 0.5 | 5371 ± 274 | 4.62 ± 0.21 | 0.10 ± 0.25 | <103 | 3 | N | … | … | … | … | n | NG |
| 3932 | 19014957-0024060 | 40.6 ± 7.0 | … | … | … | … | … | N | … | … | … | … | n | … |
| 3817 | 19014982-0028078 | 43.8 ± 2.1 | 6464 ± 187 | 4.37 ± 0.40 | 0.12 ± 0.14 | … | … | Y | … | … | … | Y | n | … |
| 3818 | 19014984-0024223 | 26.1 ± 0.4 | 6347 ± 90 | 4.43 ± 0.07 | 0.06 ± 0.19 | … | … | N | … | … | … | … | n | … |
| 3933 | 19015012-0030384 | 47.9 ± 0.2 | … | … | … | … | … | Y | … | … | … | Y | n | … |
| 3934 | 19015015-0027206 | 48.6 ± 1.3 | … | … | … | … | … | Y | … | … | … | Y | n | … |
| 3839 | 19015336-0023150 | 8.3 ± 0.3 | 6467 ± 191 | 4.56 ± 0.18 | 0.32 ± 0.29 | … | … | N | … | … | … | … | n | … |
| 3840 | 19015339-0025546 | 21.2 ± 0.4 | 5833 ± 133 | 4.38 ± 0.15 | -0.05 ± 0.12 | <169 | 3 | N | … | … | … | … | n | NG |
| 3841 | 19015405-0027382 | 26.5 ± 0.3 | 5933 ± 128 | 4.53 ± 0.21 | -0.01 ± 0.16 | <74 | 3 | N | … | … | … | … | n | NG |
| 3842 | 19015415-0030008 | 51.3 ± 0.4 | 6089 ± 283 | 4.32 ± 0.20 | -0.44 ± 0.22 | <124 | 3 | Y | N | Y | N | … | n | NG |
| 3843 | 19015432-0022452 | 0.5 ± 0.4 | 5996 ± 202 | 4.53 ± 0.16 | 0.30 ± 0.16 | 48 ± 37 | 1 | N | … | … | … | N | n | NG |
| 3844 | 19015503-0030141 | 17.5 ± 0.3 | 6100 ± 153 | 4.62 ± 0.27 | 0.26 ± 0.18 | <39 | 3 | N | … | … | … | N | n | NG |
| 3938 | 19015511-0026338 | 14.8 ± 0.1 | … | … | … | … | … | N | … | … | … | … | n | … |
| 3845 | 19015539-0024379 | 0.3 ± 0.3 | 5906 ± 268 | 4.47 ± 0.31 | 0.31 ± 0.18 | … | … | N | … | … | … | … | n | … |
| 3939 | 19015561-0027402 | -19.8 ± 0.2 | … | … | … | … | … | N | … | … | … | … | n | … |
| 3846 | 19015561-0028186 | 46.8 ± 1.9 | 7387 ± 1025 | 4.15 ± 0.21 | 0.15 ± 0.16 | … | … | Y | … | … | … | Y | n | … |
| 3847 | 19015594-0026332 | 8.9 ± 0.3 | 5149 ± 97 | 4.79 ± 0.44 | -0.02 ± 0.16 | … | … | N | … | … | … | … | n | … |
| 3848 | 19015603-0031116 | 4.6 ± 0.5 | 5833 ± 452 | 4.75 ± 0.43 | 0.17 ± 0.20 | … | … | N | … | … | … | … | n | … |
| 3940 | 19015615-0025210 | 28.7 ± 0.2 | 5905 ± 110 | 3.86 ± 0.04 | 0.01 ± 0.18 | … | … | N | … | … | … | … | n | … |
| 3849 | 19015657-0026161 | 57.6 ± 0.3 | 5444 ± 45 | 4.65 ± 0.41 | 0.29 ± 0.18 | … | … | N | … | … | … | … | n | … |
| 3850 | 19015659-0028092 | 11.4 ± 0.4 | 5627 ± 194 | 4.58 ± 0.35 | 0.24 ± 0.16 | … | … | N | … | … | … | … | n | … |
| 3851 | 19015661-0030376 | 28.7 ± 0.4 | 6356 ± 231 | 4.17 ± 0.19 | 0.23 ± 0.20 | … | … | N | … | … | … | … | n | … |
| 3852 | 19015662-0026424 | 46.9 ± 2.6 | 7128 ± 221 | 4.40 ± 0.42 | -0.16 ± 0.24 | <73 | 3 | Y | Y | Y | N | Y | Y | … |
| 3853 | 19015673-0030212 | 24.1 ± 0.4 | 5746 ± 118 | 4.54 ± 0.48 | 0.30 ± 0.15 | … | … | N | … | … | … | … | n | … |
| 3854 | 19015717-0026199 | 50.9 ± 0.9 | 6170 ± 153 | 4.33 ± 0.27 | 0.11 ± 0.17 | <84 | 3 | Y | Y | Y | Y | … | Y | … |
| 3941 | 19015720-0025571 | -38.9 ± 0.1 | 5675 ± 123 | 3.40 ± 0.09 | -0.25 ± 0.18 | … | … | N | … | … | … | … | n | … |
| 3855 | 19015748-0025108 | 0.4 ± 0.7 | 6161 ± 108 | 4.06 ± 0.42 | 0.40 ± 0.27 | 133 ± 76 | 1 | N | … | … | … | … | n | NG |
| 3856 | 19015751-0027262 | -22.9 ± 0.3 | 5965 ± 137 | 4.18 ± 0.11 | 0.34 ± 0.15 | <86 | 3 | N | … | … | … | … | n | NG |
| 486 | 19015978-0028183 | 47.0 ± 0.6 | 4694 ± 112 | 2.37 ± 0.27 | -0.34 ± 0.10 | <6 | 3 | Y | Y | Y | N | … | Y | … |
| 3857 | 19015997-0025565 | 24.4 ± 0.3 | 5966 ± 79 | 4.49 ± 0.09 | 0.32 ± 0.15 | … | … | N | … | … | … | … | n | … |
| 3871 | 19012049-0027090 | 22.1 ± 0.1 | … | … | … | … | … | N | … | … | … | … | n | … |
| 3682 | 19012201-0028029 | 29.7 ± 0.5 | 5954 ± 96 | 4.27 ± 0.09 | 0.16 ± 0.24 | <95 | 3 | N | … | … | … | … | n | NG |
| 3683 | 19012248-0028482 | 17.2 ± 0.3 | 5954 ± 150 | 4.97 ± 0.60 | -0.06 ± 0.13 | 46 ± 41 | 1 | N | … | … | … | … | n | NG |
| 3872 | 19012270-0029151 | 30.1 ± 0.2 | … | … | … | … | … | N | … | … | … | … | n | … |
| 3684 | 19012337-0028180 | 6.2 ± 0.5 | 5761 ± 114 | 5.14 ± 0.74 | 0.08 ± 0.15 | 107 ± 82 | 1 | N | … | … | … | … | n | NG |
| 3685 | 19012357-0026230 | -19.9 ± 0.5 | 5881 ± 102 | 4.01 ± 0.38 | -0.35 ± 0.16 | … | … | N | … | … | … | … | n | … |
| 3873 | 19012386-0024211 | -12.9 ± 0.2 | … | … | … | … | … | N | … | … | … | … | n | … |
| 3686 | 19012389-0026282 | 81.4 ± 0.3 | 5468 ± 87 | 4.50 ± 0.36 | -0.12 ± 0.12 | … | … | N | … | … | … | … | n | … |
| 3874 | 19012407-0028403 | 11.8 ± 0.1 | 5579 ± 115 | 3.90 ± 0.07 | -0.04 ± 0.23 | … | … | N | … | … | … | … | n | … |
| 3688 | 19012526-0025487 | 40.1 ± 0.3 | 5946 ± 202 | 4.54 ± 0.23 | 0.03 ± 0.16 | … | … | N | … | … | … | … | n | … |
| 3875 | 19012533-0026479 | 18.4 ± 0.2 | 6469 ± 149 | 4.72 ± 0.02 | … | … | … | N | … | … | … | … | n | … |
| 3689 | 19012771-0029507 | -4.6 ± 0.4 | 5612 ± 180 | 4.68 ± 0.38 | 0.23 ± 0.17 | <94 | 3 | N | … | … | … | … | n | NG |

**Table C.13.** continued.

| ID | CNAME | RV (km s$^{-1}$) | $T_{\text{eff}}$ (K) | $logg$ (dex) | [Fe/H] (dex) | $EW$(Li)$^a$ (mÅ) | $EW$(Li) error flag$^b$ | Membership RV | Li | $logg$ | [Fe/H] | Gaia study Cantat-Gaudin$^c$ | Final$^d$ | NMs with Li$^e$ |
|---|---|---|---|---|---|---|---|---|---|---|---|---|---|---|
| 3690 | 19012871-0030120 | 50.6 ± 0.4 | 6005 ± 200 | 4.71 ± 0.38 | 0.20 ± 0.18 | … | … | Y | … | … | … | … | n | … |
| 3691 | 19012909-0024251 | 34.4 ± 0.9 | 6631 ± 519 | 4.28 ± 0.43 | 0.10 ± 0.19 | … | … | N | … | … | … | … | n | … |
| 3692 | 19012910-0027350 | 56.5 ± 0.5 | 6082 ± 244 | 4.64 ± 0.26 | 0.28 ± 0.19 | 42 ± 32 | 1 | N | … | … | … | … | n | NG |
| 3693 | 19012912-0026442 | 50.3 ± 1.7 | 6446 ± 203 | 5.14 ± 0.48 | -0.02 ± 0.17 | … | … | Y | … | … | … | N | n | … |
| 3694 | 19012930-0029096 | 1.8 ± 0.3 | 6092 ± 339 | 4.29 ± 0.19 | 0.30 ± 0.22 | 101 ± 66 | 1 | N | … | … | … | N | n | NG |
| 3876 | 19012952-0026386 | 44.0 ± 1.9 | … | … | … | … | … | Y | … | … | … | … | n | … |
| 3695 | 19012975-0027170 | 51.3 ± 1.1 | 7045 ± 199 | 3.74 ± 0.35 | -0.16 ± 0.21 | <48 | 3 | Y | Y | Y | N | Y | Y | … |
| 3729 | 19013586-0032086 | 63.8 ± 0.3 | 5991 ± 247 | 4.47 ± 0.30 | -0.11 ± 0.17 | … | … | N | … | … | … | … | n | … |
| 3730 | 19013609-0029295 | -58.7 ± 0.9 | 5704 ± 189 | 4.16 ± 0.56 | -0.42 ± 0.62 | … | … | N | … | … | … | … | n | … |
| 3731 | 19013616-0030052 | -62.6 ± 0.3 | 5609 ± 19 | 4.14 ± 0.57 | -0.21 ± 0.17 | … | … | N | … | … | … | … | n | … |
| 474 | 19013631-0027447 | 47.6 ± 0.6 | 4991 ± 138 | 2.79 ± 0.30 | 0.25 ± 0.11 | <7 | 3 | Y | Y | Y | Y | … | Y | … |
| 3733 | 19013633-0024141 | 50.6 ± 0.8 | 6835 ± 672 | 4.64 ± 0.90 | 0.26 ± 0.24 | <126 | 3 | Y | N | Y | Y | … | n | NG |
| 3734 | 19013649-0026149 | 46.3 ± 0.8 | 7172 ± 123 | … | … | … | … | Y | … | … | … | Y | n | … |
| 3735 | 19013651-0025306 | 39.5 ± 0.4 | 5793 ± 130 | 4.93 ± 0.39 | -0.12 ± 0.16 | … | … | N | … | … | … | … | n | … |
| 475 | 19013651-0027021 | 48.9 ± 0.6 | 4939 ± 127 | 3.14 ± 0.24 | 0.23 ± 0.12 | <13 | 3 | Y | Y | Y | Y | … | Y | … |
| 3891 | 19013653-0031293 | -15.1 ± 1.9 | … | … | … | … | … | N | … | … | … | … | n | … |
| 3892 | 19013670-0029302 | 9.4 ± 0.5 | … | … | … | … | … | N | … | … | … | … | n | … |
| 3736 | 19013682-0025086 | 28.1 ± 0.3 | 6106 ± 344 | 4.57 ± 0.25 | 0.11 ± 0.40 | … | … | N | … | … | … | … | n | … |
| 3737 | 19013691-0031582 | 8.9 ± 0.3 | 6039 ± 117 | 4.79 ± 0.20 | -0.02 ± 0.19 | 62 ± 30 | 1 | N | … | … | … | … | n | NG |
| 3738 | 19013705-0028220 | 52.4 ± 2.5 | 6399 ± 245 | 5.35 ± 0.60 | -0.07 ± 0.20 | … | … | Y | … | … | … | … | n | … |
| 3740 | 19013727-0032112 | 25.8 ± 0.3 | 5777 ± 186 | 4.59 ± 0.60 | 0.35 ± 0.20 | <93 | 3 | N | … | … | … | … | n | NG |
| 3741 | 19013745-0032011 | 53.1 ± 0.4 | 5824 ± 226 | 4.70 ± 0.32 | -0.36 ± 0.13 | … | … | N | … | … | … | … | n | … |
| 3893 | 19013750-0025117 | -4.4 ± 0.2 | … | … | … | … | … | N | … | … | … | … | n | … |
| 3894 | 19013752-0029178 | 5.4 ± 0.2 | 5671 ± 113 | 4.05 ± 0.09 | -0.10 ± 0.17 | … | … | N | … | … | … | … | n | … |
| 3895 | 19013762-0026113 | 53.7 ± 5.6 | … | … | … | … | … | N | … | … | … | Y | n | … |
| 3742 | 19013779-0026552 | -82.6 ± 0.7 | 5497 ± 193 | 3.91 ± 0.41 | -0.24 ± 0.17 | … | … | N | … | … | … | … | n | … |
| 3743 | 19013788-0030417 | -11.3 ± 0.3 | 6072 ± 303 | 4.49 ± 0.38 | 0.00 ± 0.25 | … | … | N | … | … | … | … | n | … |
| 3896 | 19013797-0030114 | 15.3 ± 0.2 | … | … | … | … | … | N | … | … | … | … | n | … |
| 3744 | 19013803-0024516 | 64.8 ± 0.4 | 7042 ± 783 | 3.88 ± 0.24 | 0.10 ± 0.14 | <40 | 3 | N | … | … | … | … | n | NG |
| 3757 | 19014027-0030401 | 35.2 ± 0.3 | 6133 ± 252 | 4.37 ± 0.13 | 0.32 ± 0.27 | 36 ± 20 | 1 | N | … | … | … | … | n | NG |
| 3906 | 19014056-0027196 | 58.1 ± 0.4 | … | … | … | … | … | N | … | … | … | Y | n | … |
| 3758 | 19014066-0029046 | 45.6 ± 1.2 | 7402 ± 1050 | 4.09 ± 0.20 | 0.01 ± 0.33 | 63 ± 53 | 1 | Y | Y | Y | Y | Y | Y | … |
| 3771 | 19014212-0025400 | 48.0 ± 1.9 | … | … | … | … | … | Y | … | … | … | N | n | … |

**Notes.** $^{(a)}$ The values of $EW$(Li) for this cluster are corrected (subtracted adjacent Fe (6707.43 Å) line). $^{(b)}$ Flags for the errors of the corrected $EW$(Li) values, as follows: 1=$EW$(Li) corrected by blends contribution using models; and 3=Upper limit (no error for $EW$(Li) is given). $^{(c)}$ Cantat-Gaudin et al. (2018). $^{(d)}$ The letters "Y" and "N" indicate if the star is a cluster member or not. $^{(e)}$ 'Li-rich G', 'G' and 'NG' indicate "Li-rich giant", "giant" and "non-giant" Li field outliers, respectively.

**Table C.14.** NGC 6005

| ID | CNAME | RV (km s$^{-1}$) | $T_{\rm eff}$ (K) | logg (dex) | [Fe/H] (dex) | EW(Li)$^a$ (mÅ) | EW(Li) error flag$^b$ | Membership RV | Li | logg | [Fe/H] | Gaia study Cantat-Gaudin$^c$ | Final$^d$ | NMs with Li$^e$ |
|---|---|---|---|---|---|---|---|---|---|---|---|---|---|---|
| 46026 | 15555208-5727049 | -26.3 ± 0.1 | … | … | … | … | … | Y | … | … | … | … | n | … |
| 46349 | 15555211-5728196 | -18.5 ± 0.7 | 6709 ± 347 | 4.15 ± 0.21 | 0.29 ± 0.40 | … | … | Y | … | … | … | Y | n | … |
| 46027 | 15555213-5728015 | -28.2 ± 0.2 | 5973 ± 127 | 4.91 ± 0.05 | 0.39 ± 0.25 | … | … | Y | … | … | … | … | n | … |
| 46350 | 15555220-5727353 | -36.7 ± 0.7 | 6036 ± 468 | 4.09 ± 0.14 | 0.12 ± 0.70 | … | … | N | … | … | … | … | n | … |
| 46351 | 15555223-5723302 | 24.5 ± 0.6 | 4452 ± 396 | 3.68 ± 0.22 | -0.89 ± 0.72 | … | … | N | … | … | … | … | n | … |
| 46028 | 15555228-5723149 | -24.7 ± 0.5 | … | … | … | … | … | Y | … | … | … | Y | n | … |
| 46352 | 15555235-5722452 | -90.3 ± 0.6 | 5449 ± 165 | 3.88 ± 0.76 | -0.03 ± 0.43 | … | … | N | … | … | … | … | n | … |
| 46029 | 15555236-5728598 | -20.1 ± 0.1 | … | … | … | … | … | Y | … | … | … | … | n | … |
| 46030 | 15555279-5735124 | -67.8 ± 0.3 | … | … | … | … | … | N | … | … | … | … | n | … |
| 46031 | 15555291-5724283 | -25.6 ± 1.1 | … | … | … | … | … | Y | … | … | … | Y | n | … |
| 46353 | 15555291-5728407 | -119.4 ± 0.4 | 5034 ± 296 | 3.02 ± 0.17 | 0.03 ± 0.28 | … | … | N | … | … | … | … | n | G |
| 46032 | 15555309-5724441 | -25.8 ± 0.3 | … | … | … | … | … | Y | … | … | … | N | n | … |
| 46354 | 15555312-5729152 | -56.1 ± 0.6 | 5946 ± 331 | 4.12 ± 0.30 | -0.04 ± 0.18 | <90 | 3 | N | … | … | … | N | n | NG |
| 46033 | 15555337-5723339 | -2.7 ± 0.1 | … | … | … | … | … | N | … | … | … | … | n | … |
| 46355 | 15555337-5725416 | 2.1 ± 0.6 | 5612 ± 942 | 4.13 ± 0.50 | -0.06 ± 0.31 | … | … | N | … | … | … | … | n | … |
| 46356 | 15555345-5724564 | -9.7 ± 0.3 | 5823 ± 212 | 4.32 ± 0.30 | 0.35 ± 0.16 | … | … | N | … | … | … | … | n | … |
| 46034 | 15555352-5725105 | -29.6 ± 0.4 | 6038 ± 75 | 4.15 ± 0.54 | 0.11 ± 0.26 | 85 ± 56 | … | Y | Y | Y | Y | Y | Y | … |
| 46357 | 15555353-5726232 | -96.0 ± 1.2 | 6041 ± 703 | 3.33 ± 0.76 | -0.46 ± 0.54 | … | … | N | … | … | … | … | n | … |
| 46035 | 15555406-5726174 | -34.5 ± 0.3 | 5501 ± 127 | 2.34 ± 0.01 | -0.37 ± 0.30 | … | … | N | … | … | … | … | n | … |
| 46358 | 15555410-5727232 | -93.7 ± 0.5 | 5590 ± 288 | 4.37 ± 0.45 | 0.06 ± 0.24 | … | … | N | … | … | … | … | n | … |
| 46036 | 15555415-5724382 | -35.2 ± 0.1 | … | … | … | … | … | N | … | … | … | Y | n | … |
| 46359 | 15555429-5724432 | -29.1 ± 0.5 | 5139 ± 52 | 4.31 ± 0.13 | 0.20 ± 0.15 | … | … | Y | … | … | … | … | n | … |
| 46037 | 15555437-5723257 | -10.1 ± 0.1 | … | … | … | … | … | N | … | … | … | Y | n | … |
| 46360 | 15555449-5727021 | -26.6 ± 0.8 | 6026 ± 78 | 4.42 ± 0.08 | 0.33 ± 0.16 | … | … | Y | … | … | … | … | n | … |
| 46419 | 15560438-5724166 | -6.8 ± 0.8 | 6266 ± 197 | 3.92 ± 0.13 | -0.13 ± 0.17 | … | … | N | … | … | … | … | n | … |
| 46074 | 15560455-5727240 | -71.0 ± 0.2 | 5665 ± 127 | 4.08 ± 0.04 | 0.25 ± 0.28 | … | … | N | … | … | … | … | n | … |
| 46420 | 15560485-5725539 | -53.8 ± 0.4 | 5633 ± 112 | 4.87 ± 0.48 | -0.07 ± 0.19 | 108 ± 56 | 1 | N | … | … | … | N | n | NG |
| 46421 | 15560487-5730265 | -53.9 ± 0.5 | 5822 ± 92 | 4.06 ± 0.51 | -0.10 ± 0.38 | <48 | 3 | N | … | … | … | … | n | NG |
| 46422 | 15560495-5725476 | -24.6 ± 0.7 | 5031 ± 359 | 4.57 ± 0.67 | 0.06 ± 0.39 | … | … | Y | … | … | … | … | n | … |
| 46075 | 15560503-5723202 | -15.1 ± 0.2 | … | … | … | … | … | N | … | … | … | N | n | … |
| 46423 | 15560530-5725099 | -74.4 ± 0.3 | 5703 ± 279 | 4.07 ± 0.19 | 0.10 ± 0.17 | … | … | N | … | … | … | … | n | … |
| 46076 | 15560534-5728045 | -60.9 ± 0.3 | … | … | … | … | … | N | … | … | … | … | n | … |
| 46077 | 15560537-5725039 | -25.1 ± 0.7 | … | … | … | … | … | Y | … | … | … | Y | n | … |
| 46424 | 15560537-5728417 | -77.5 ± 0.5 | 5239 ± 74 | 3.88 ± 0.21 | 0.04 ± 0.17 | <114 | 3 | N | … | … | … | … | n | NG |
| 46078 | 15560557-5732021 | -24.9 ± 0.2 | … | … | … | … | … | Y | … | … | … | … | n | … |
| 46425 | 15560562-5727300 | -10.4 ± 0.8 | 5143 ± 43 | 3.80 ± 0.15 | -0.08 ± 0.26 | … | … | N | … | … | … | … | n | … |
| 46079 | 15560569-5725234 | -30.2 ± 0.1 | … | … | … | … | … | Y | … | … | … | Y | n | … |
| 3089 | 15560582-5726193 | -22.8 ± 0.6 | 4947 ± 120 | 3.03 ± 0.23 | 0.15 ± 0.10 | 22 ± 1 | … | Y | Y | Y | Y | Y | Y | … |
| 46080 | 15560584-5730335 | -49.7 ± 0.3 | … | … | … | … | … | N | … | … | … | … | n | … |
| 46426 | 15560595-5722086 | -38.1 ± 0.3 | 6052 ± 232 | 3.95 ± 0.26 | -0.03 ± 0.21 | 54 ± 43 | 1 | N | … | … | … | … | n | NG |
| 46081 | 15560598-5723082 | -40.2 ± 0.2 | 5805 ± 122 | 3.58 ± 0.20 | -0.20 ± 0.24 | … | … | N | … | … | … | … | n | … |
| 46427 | 15560602-5729538 | -15.6 ± 0.4 | 5877 ± 244 | 4.35 ± 0.22 | -0.64 ± 0.16 | <54 | 3 | N | … | … | … | … | n | NG |
| 46428 | 15560618-5723563 | -13.4 ± 0.9 | 6178 ± 754 | 3.24 ± 0.44 | -0.21 ± 0.39 | … | … | N | … | … | … | … | n | … |
| 46429 | 15560623-5727391 | -46.9 ± 0.7 | 5921 ± 79 | 3.61 ± 0.25 | 0.24 ± 0.23 | … | … | N | … | … | … | … | n | … |
| 46082 | 15560624-5728077 | -71.4 ± 0.1 | … | … | … | … | … | N | … | … | … | … | n | … |
| 46430 | 15560625-5723440 | 15.5 ± 0.6 | 5072 ± 179 | 4.82 ± 0.16 | 0.10 ± 0.27 | … | … | N | … | … | … | … | n | … |
| 46431 | 15560659-5729261 | -82.7 ± 0.7 | 5587 ± 152 | 4.29 ± 0.71 | -0.39 ± 0.14 | … | … | N | … | … | … | … | n | … |
| 46432 | 15560676-5726086 | -37.2 ± 0.5 | 5598 ± 346 | 4.84 ± 0.63 | 0.12 ± 0.17 | … | … | N | … | … | … | … | n | … |
| 46134 | 15562903-5716068 | 16.9 ± 0.1 | … | … | … | … | … | N | … | … | … | … | n | … |
| 46135 | 15562931-5733067 | -30.0 ± 0.1 | … | … | … | … | … | Y | … | … | … | … | n | … |
| 3094 | 15562994-5726584 | -70.4 ± 0.6 | 4786 ± 123 | 2.59 ± 0.22 | 0.15 ± 0.10 | 9 ± 1 | … | N | … | … | … | … | n | G |
| 46136 | 15563000-5717058 | -28.9 ± 0.1 | 5501 ± 150 | 4.47 ± 0.12 | 0.00 ± 0.22 | … | … | Y | … | … | … | … | n | … |
| 46137 | 15563015-5726319 | -18.2 ± 0.2 | … | … | … | … | … | Y | … | … | … | … | n | … |
| 46138 | 15563026-5722226 | -61.1 ± 0.3 | … | … | … | … | … | N | … | … | … | … | n | … |
| 46139 | 15563212-5732172 | 45.5 ± 0.1 | 5179 ± 231 | 4.41 ± 0.43 | 0.11 ± 0.25 | … | … | N | … | … | … | … | n | … |
| 46140 | 15563215-5716019 | -80.1 ± 0.1 | 4757 ± 67 | 3.09 ± 0.02 | -0.08 ± 0.14 | … | … | N | … | … | … | … | n | … |
| 46141 | 15563283-5722325 | -70.4 ± 0.1 | … | … | … | … | … | N | … | … | … | … | n | … |
| 46142 | 15563375-5729460 | -66.4 ± 0.1 | 5043 ± 78 | 3.53 ± 0.05 | 0.12 ± 0.20 | … | … | N | … | … | … | … | n | … |
| 46143 | 15563466-5732239 | -35.4 ± 0.1 | … | … | … | … | … | N | … | … | … | … | n | … |

**Table C.14.** continued.

| ID | CNAME | RV (km s$^{-1}$) | $T_{\text{eff}}$ (K) | logg (dex) | [Fe/H] (dex) | EW(Li)$^a$ (mÅ) | EW(Li) error flag$^b$ | Membership RV | Li | logg | [Fe/H] | Gaia study Cantat-Gaudin$^c$ | Final$^d$ | NMs with Li$^e$ |
|---|---|---|---|---|---|---|---|---|---|---|---|---|---|---|
| 46144 | 15563488-5728354 | 31.1 ± 0.1 | … | … | … | … | … | N | … | … | … | … | n | … |
| 46145 | 15563631-5732380 | 15.2 ± 0.2 | … | … | … | … | … | N | … | … | … | … | n | … |
| 46146 | 15563818-5732190 | -43.0 ± 0.9 | … | … | … | … | … | N | … | … | … | … | n | … |
| 46196 | 15552328-5722527 | -12.0 ± 0.5 | 5527 ± 275 | 3.99 ± 0.42 | 0.20 ± 0.28 | <66 | 3 | N | … | … | … | … | n | NG |
| 46197 | 15552353-5723556 | -73.9 ± 0.7 | 5393 ± 262 | 3.92 ± 0.31 | -0.03 ± 0.21 | … | … | N | … | … | … | … | n | … |
| 46198 | 15552354-5726418 | -89.3 ± 0.5 | 5248 ± 43 | 3.93 ± 0.35 | 0.15 ± 0.32 | … | … | N | … | … | … | … | n | … |
| 45953 | 15552368-5731204 | -24.4 ± 0.1 | 5155 ± 84 | 3.74 ± 0.05 | 0.03 ± 0.18 | … | … | Y | … | … | … | Y | n | … |
| 46199 | 15552373-5728351 | -49.3 ± 0.5 | 5802 ± 117 | 3.80 ± 0.40 | 0.12 ± 0.20 | … | … | N | … | … | … | … | n | … |
| 46200 | 15552389-5728056 | -25.8 ± 0.8 | 5382 ± 155 | 4.46 ± 0.13 | -0.08 ± 0.24 | … | … | Y | … | … | … | … | n | … |
| 46201 | 15552408-5725217 | -19.6 ± 0.9 | 5907 ± 378 | 4.01 ± 0.18 | -0.11 ± 0.61 | … | … | Y | … | … | … | … | n | … |
| 46202 | 15552409-5727499 | -43.4 ± 0.7 | 5478 ± 26 | 3.87 ± 0.68 | -0.15 ± 0.17 | … | … | N | … | … | … | … | n | … |
| 46203 | 15552446-5723274 | 15.6 ± 0.4 | 5030 ± 243 | 4.19 ± 0.40 | 0.08 ± 0.13 | … | … | N | … | … | … | … | n | … |
| 46204 | 15552466-5728394 | -39.1 ± 0.4 | 5946 ± 344 | 4.17 ± 0.34 | 0.31 ± 0.16 | … | … | N | … | … | … | … | n | … |
| 45954 | 15552480-5726542 | -28.5 ± 0.3 | … | … | … | … | … | Y | … | … | … | … | n | … |
| 46205 | 15552483-5727523 | -74.9 ± 0.3 | 5977 ± 250 | 4.12 ± 0.22 | 0.00 ± 0.21 | … | … | N | … | … | … | … | n | … |
| 45955 | 15552488-5728325 | -22.1 ± 0.5 | 6913 ± 150 | 4.91 ± 0.05 | … | … | … | Y | … | … | … | … | n | … |
| 46206 | 15552496-5731174 | -4.2 ± 0.8 | 5971 ± 232 | 4.83 ± 0.23 | 0.23 ± 0.36 | … | … | N | … | … | … | … | n | … |
| 46207 | 15552498-5726486 | 4.5 ± 0.4 | 5652 ± 147 | 4.00 ± 0.21 | 0.12 ± 0.17 | 125 ± 64 | 1 | N | … | … | … | N | n | NG |
| 45956 | 15552533-5732007 | 31.0 ± 0.2 | 6017 ± 149 | 4.91 ± 0.05 | … | … | … | N | … | … | … | … | n | … |
| 46208 | 15552549-5729511 | -20.1 ± 0.6 | 5819 ± 443 | 4.63 ± 0.47 | 0.09 ± 0.38 | … | … | Y | … | … | … | … | n | … |
| 46209 | 15552587-5725269 | -79.2 ± 0.4 | 6328 ± 263 | 4.51 ± 0.58 | 0.00 ± 0.30 | … | … | N | … | … | … | … | n | … |
| 45957 | 15552594-5730520 | -5.2 ± 0.1 | … | … | … | … | … | N | … | … | … | … | n | … |
| 45958 | 15552610-5727303 | -11.2 ± 0.3 | 5521 ± 26 | 4.14 ± 0.22 | 0.17 ± 0.17 | … | … | N | … | … | … | … | n | … |
| 46210 | 15552628-5722368 | -10.6 ± 0.6 | 6585 ± 493 | 4.48 ± 0.69 | 0.08 ± 0.21 | 47 ± 32 | 1 | N | … | … | … | … | n | NG |
| 46211 | 15552631-5726136 | -47.2 ± 0.6 | 6176 ± 463 | 3.61 ± 0.70 | -0.25 ± 0.21 | <65 | 3 | N | Y | N | N | … | n | NG |
| 46212 | 15552664-5721599 | -44.7 ± 0.8 | 5606 ± 222 | 4.00 ± 0.05 | -0.04 ± 0.27 | … | … | N | … | … | … | … | n | … |
| 46213 | 15552720-5726370 | -26.8 ± 0.6 | 5308 ± 85 | 3.64 ± 0.77 | 0.06 ± 0.20 | … | … | Y | … | … | … | … | n | … |
| 46272 | 15553989-5720279 | -89.5 ± 1.7 | 6177 ± 410 | 5.68 ± 1.04 | 0.21 ± 0.28 | … | … | N | … | … | … | … | n | … |
| 46273 | 15553992-5728012 | -82.2 ± 1.5 | 6222 ± 170 | 4.67 ± 0.35 | 0.13 ± 0.15 | … | … | N | … | … | … | … | n | … |
| 3079 | 15554015-5726002 | -24.8 ± 0.6 | 4868 ± 120 | 2.67 ± 0.24 | 0.17 ± 0.10 | <6 | 3 | Y | Y | Y | Y | … | Y | … |
| 46274 | 15554021-5725215 | -42.2 ± 0.5 | 5857 ± 195 | 4.27 ± 0.13 | 0.22 ± 0.13 | <61 | 3 | N | … | … | … | … | n | NG |
| 46275 | 15554026-5725499 | -24.5 ± 0.5 | 6249 ± 207 | 4.56 ± 0.41 | 0.05 ± 0.26 | 33 ± 25 | 1 | Y | Y | Y | Y | … | Y | … |
| 46276 | 15554028-5724508 | -143.0 ± 0.9 | 5618 ± 105 | 4.05 ± 0.79 | 0.08 ± 0.29 | … | … | N | … | … | … | … | n | … |
| 46277 | 15554035-5727065 | -23.4 ± 0.5 | 5680 ± 380 | 4.62 ± 0.17 | -0.05 ± 0.25 | … | … | Y | … | … | … | … | n | … |
| 45994 | 15554037-5714451 | -56.7 ± 0.1 | 5671 ± 116 | 4.91 ± 0.05 | 0.23 ± 0.24 | … | … | N | … | … | … | … | n | … |
| 45995 | 15554037-5724298 | -17.2 ± 0.6 | … | … | … | … | … | Y | … | … | … | Y | n | … |
| 46278 | 15554051-5718487 | -112.4 ± 2.4 | 4813 ± 95 | 3.47 ± 0.46 | -1.09 ± 0.13 | … | … | N | … | … | … | … | n | G |
| 46280 | 15554073-5726223 | -77.1 ± 0.5 | 5680 ± 140 | 4.23 ± 0.44 | -0.09 ± 0.24 | <54 | 3 | N | … | … | … | … | n | NG |
| 45996 | 15554079-5726380 | 21.3 ± 0.1 | 5835 ± 104 | 4.91 ± 0.05 | -0.02 ± 0.15 | … | … | N | … | … | … | … | n | … |
| 45997 | 15554080-5724175 | -67.4 ± 0.3 | 5521 ± 117 | 3.18 ± 0.05 | -0.04 ± 0.23 | … | … | N | … | … | … | … | n | … |
| 46281 | 15554082-5729498 | -24.3 ± 0.6 | 5738 ± 142 | 4.25 ± 0.24 | 0.19 ± 0.22 | 106 ± 58 | 1 | Y | Y | Y | Y | Y | Y | … |
| 3080 | 15554101-5727497 | -12.1 ± 0.6 | 4785 ± 120 | 2.88 ± 0.24 | 0.05 ± 0.10 | 50 ± 2 | 1 | N | … | … | … | … | n | G |
| 46282 | 15554103-5728258 | -34.0 ± 0.6 | 5723 ± 20 | 4.52 ± 0.21 | -0.11 ± 0.26 | <55 | 3 | Y | Y | Y | Y | … | Y | … |
| 46283 | 15554105-5729053 | -92.5 ± 2.7 | 6711 ± 690 | 3.83 ± 0.53 | -0.01 ± 0.09 | … | … | N | … | … | … | … | n | … |
| 3081 | 15554127-5726574 | -26.3 ± 0.6 | 4909 ± 117 | 2.80 ± 0.29 | 0.15 ± 0.10 | 22 ± 2 | … | Y | Y | Y | Y | … | Y | … |
| 46284 | 15554131-5723428 | -93.9 ± 2.5 | 5603 ± 406 | 3.87 ± 0.61 | -0.62 ± 0.30 | … | … | N | … | … | … | … | n | … |
| 46285 | 15554141-5727511 | -28.9 ± 0.5 | 5912 ± 236 | 4.07 ± 0.32 | 0.21 ± 0.19 | 75 ± 52 | 1 | Y | Y | Y | Y | … | Y | … |
| 46286 | 15554144-5729166 | -81.7 ± 0.3 | 5761 ± 167 | 4.08 ± 0.53 | 0.08 ± 0.18 | 100 ± 44 | 1 | N | … | … | … | … | n | NG |
| 46287 | 15554154-5722452 | 3.5 ± 3.1 | 5242 ± 320 | 3.22 ± 0.47 | -0.91 ± 0.95 | … | … | N | … | … | … | … | n | … |
| 46288 | 15554155-5727151 | -43.8 ± 0.8 | 5740 ± 84 | 4.26 ± 0.30 | 0.13 ± 0.32 | … | … | N | … | … | … | … | n | … |
| 45933 | 15545889-5731191 | 50.7 ± 0.1 | 4985 ± 112 | 3.95 ± 0.05 | -0.09 ± 0.23 | … | … | N | … | … | … | … | n | … |
| 46162 | 15551286-5727472 | -90.5 ± 1.0 | 5643 ± 256 | 3.94 ± 0.66 | -0.20 ± 0.72 | … | … | N | … | … | … | … | n | … |
| 46163 | 15551317-5730146 | -27.6 ± 0.6 | 5999 ± 129 | 4.27 ± 0.05 | 0.04 ± 0.20 | <99 | 3 | Y | Y | Y | Y | … | Y | … |
| 46164 | 15551379-5728125 | 19.5 ± 0.3 | 5117 ± 96 | 4.24 ± 0.44 | 0.21 ± 0.18 | … | … | N | … | … | … | … | n | … |
| 45939 | 15551388-5733066 | -41.1 ± 0.4 | … | … | … | … | … | N | … | … | … | … | n | … |
| 45940 | 15551406-5726548 | -25.4 ± 0.4 | … | … | … | … | … | Y | … | … | … | … | n | … |
| 45942 | 15551754-5720096 | -28.6 ± 0.1 | … | … | … | … | … | Y | … | … | … | … | n | … |
| 45941 | 15551488-5727230 | -77.8 ± 0.3 | 6505 ± 149 | 4.91 ± 0.05 | … | … | … | N | … | … | … | … | n | … |
| 46177 | 15551772-5731358 | -86.0 ± 0.7 | 5606 ± 142 | 3.12 ± 0.18 | -0.01 ± 0.31 | … | … | N | … | … | … | … | n | … |

**Table C.14.** continued.

| ID | CNAME | RV (km s$^{-1}$) | $T_{\text{eff}}$ (K) | logg (dex) | [Fe/H] (dex) | EW(Li)$^a$ (mÅ) | EW(Li) error flag$^b$ | RV | Membership Li | logg | [Fe/H] | Gaia study Cantat-Gaudin$^c$ | Final$^d$ | NMs with Li$^e$ |
|---|---|---|---|---|---|---|---|---|---|---|---|---|---|---|
| 46165 | 15551498-5727091 | -86.0 ± 0.8 | 5634 ± 231 | 4.05 ± 0.43 | -0.18 ± 0.27 | ... | ... | N | ... | ... | ... | ... | n | ... |
| 46178 | 15551779-5727022 | 48.2 ± 0.8 | 5878 ± 493 | 4.28 ± 0.86 | -0.15 ± 0.47 | ... | ... | N | ... | ... | ... | ... | n | ... |
| 46166 | 15551530-5726509 | -31.6 ± 0.4 | 5953 ± 402 | 4.15 ± 0.19 | 0.42 ± 0.27 | 96 ± 86 | 1 | Y | Y | Y | N | ... | Y | ... |
| 45943 | 15551790-5716441 | -5.5 ± 0.1 | ... | ... | ... | ... | ... | N | ... | ... | ... | ... | n | ... |
| 46167 | 15551541-5718555 | -32.8 ± 1.0 | 6002 ± 395 | 4.65 ± 0.47 | -0.24 ± 0.56 | ... | ... | Y | ... | ... | ... | ... | n | ... |
| 46179 | 15551797-5725300 | -50.4 ± 0.7 | 5902 ± 464 | 3.64 ± 0.26 | 0.06 ± 0.19 | ... | ... | N | ... | ... | ... | ... | n | ... |
| 46168 | 15551544-5726271 | -25.5 ± 1.0 | 5803 ± 173 | 3.87 ± 0.87 | 0.24 ± 0.18 | ... | ... | Y | ... | ... | ... | ... | n | ... |
| 46180 | 15551816-5725501 | -13.9 ± 0.6 | 4527 ± 381 | 4.67 ± 0.16 | 0.19 ± 0.19 | ... | ... | N | ... | ... | ... | ... | n | ... |
| 46169 | 15551556-5728102 | -58.0 ± 0.5 | 5852 ± 364 | 4.41 ± 0.51 | 0.28 ± 0.19 | <73 | 3 | N | ... | ... | ... | ... | n | NG |
| 45959 | 15552720-5727353 | -23.1 ± 2.8 | ... | ... | ... | ... | ... | Y | ... | ... | ... | Y | n | ... |
| 46170 | 15551575-5727378 | -36.3 ± 0.9 | 5928 ± 880 | 4.40 ± 0.54 | -0.16 ± 0.87 | ... | ... | N | ... | ... | ... | ... | n | ... |
| 46214 | 15552720-5728112 | -76.6 ± 0.5 | 4774 ± 284 | 3.49 ± 0.37 | 0.18 ± 0.20 | ... | ... | N | ... | ... | ... | ... | n | G |
| 46171 | 15551609-5730238 | -27.9 ± 0.4 | 4866 ± 121 | 3.84 ± 0.35 | 0.14 ± 0.16 | ... | ... | Y | ... | ... | ... | ... | n | ... |
| 46215 | 15552728-5731286 | -83.1 ± 0.8 | 5225 ± 463 | 4.73 ± 0.14 | 0.29 ± 0.44 | ... | ... | N | ... | ... | ... | ... | n | ... |
| 46172 | 15551628-5722514 | -79.4 ± 0.5 | 5676 ± 193 | 4.14 ± 0.76 | 0.05 ± 0.19 | ... | ... | N | ... | ... | ... | ... | n | ... |
| 46216 | 15552730-5727569 | -10.1 ± 0.3 | 6234 ± 209 | 4.35 ± 0.17 | 0.10 ± 0.17 | 55 ± 28 | 1 | N | ... | ... | ... | ... | n | NG |
| 46173 | 15551634-5729043 | -38.3 ± 0.6 | 5617 ± 112 | 4.52 ± 0.42 | -0.16 ± 0.22 | <81 | 3 | N | ... | ... | ... | ... | n | NG |
| 46217 | 15552753-5723288 | 13.3 ± 0.8 | 5224 ± 63 | 3.93 ± 0.66 | -0.27 ± 0.38 | ... | ... | N | ... | ... | ... | ... | n | ... |
| 45960 | 15552761-5727484 | -40.4 ± 0.1 | ... | ... | ... | ... | ... | N | ... | ... | ... | N | n | ... |
| 46175 | 15551733-5722487 | -70.2 ± 0.4 | 5958 ± 227 | 4.26 ± 0.31 | -0.10 ± 0.21 | ... | ... | N | ... | ... | ... | ... | n | ... |
| 46218 | 15552767-5723514 | -68.5 ± 0.6 | 5342 ± 444 | 3.92 ± 0.13 | 0.06 ± 0.15 | ... | ... | N | ... | ... | ... | ... | n | ... |
| 46176 | 15551743-5723015 | -54.4 ± 1.5 | 5918 ± 216 | 5.13 ± 0.44 | 0.39 ± 0.15 | ... | ... | N | ... | ... | ... | ... | n | ... |
| 46090 | 15560951-5732057 | -77.6 ± 0.1 | 4709 ± 70 | 2.91 ± 0.03 | 0.23 ± 0.16 | ... | ... | N | ... | ... | ... | ... | n | ... |
| 46219 | 15552781-5722476 | 3.2 ± 0.5 | 5536 ± 79 | 3.91 ± 0.32 | -0.13 ± 0.22 | ... | ... | N | ... | ... | ... | N | n | ... |
| 46220 | 15552781-5725392 | -17.2 ± 0.5 | 5726 ± 445 | 4.38 ± 0.59 | 0.21 ± 0.20 | ... | ... | Y | ... | ... | ... | ... | n | ... |
| 46444 | 15560961-5726016 | -104.9 ± 0.6 | 4863 ± 181 | 3.20 ± 0.45 | -0.14 ± 0.19 | ... | ... | N | ... | ... | ... | ... | n | G |
| 46221 | 15552787-5722119 | -25.4 ± 0.7 | 5855 ± 230 | 4.52 ± 0.73 | 0.22 ± 0.21 | ... | ... | Y | ... | ... | ... | ... | n | ... |
| 46445 | 15560962-5723591 | 57.6 ± 0.8 | 4984 ± 297 | 4.00 ± 1.57 | -0.68 ± 0.28 | ... | ... | N | ... | ... | ... | ... | n | ... |
| 46222 | 15552793-5728418 | -96.6 ± 1.1 | 6125 ± 126 | 2.58 ± 1.06 | -0.66 ± 0.33 | ... | ... | N | ... | ... | ... | ... | n | ... |
| 46456 | 15561365-5727545 | -94.7 ± 0.7 | 5565 ± 456 | 3.42 ± 0.41 | -0.40 ± 0.33 | ... | ... | N | ... | ... | ... | ... | n | ... |
| 46223 | 15552861-5730197 | -93.8 ± 0.8 | 5850 ± 372 | 4.32 ± 0.17 | 0.08 ± 0.39 | ... | ... | N | ... | ... | ... | ... | n | ... |
| 46457 | 15561405-5726001 | -22.5 ± 0.7 | 6068 ± 298 | 5.15 ± 0.46 | -0.59 ± 0.49 | ... | ... | Y | ... | ... | ... | ... | n | ... |
| 46105 | 15561408-5725267 | -18.5 ± 0.1 | 6047 ± 122 | 4.18 ± 0.05 | ... | ... | ... | Y | ... | ... | ... | ... | n | ... |
| 46224 | 15552886-5726227 | -24.5 ± 0.5 | 6558 ± 279 | 4.40 ± 0.36 | 0.06 ± 0.44 | 106 ± 41 | 1 | Y | Y | Y | Y | ... | Y | ... |
| 46458 | 15561412-5731495 | -42.7 ± 1.0 | 5821 ± 82 | 4.32 ± 0.74 | 0.22 ± 0.32 | ... | ... | N | ... | ... | ... | ... | n | ... |
| 46225 | 15552909-5722475 | 6.4 ± 0.3 | 4531 ± 379 | 4.21 ± 0.45 | -0.17 ± 0.20 | <41 | 3 | N | ... | ... | ... | ... | n | NG |
| 46106 | 15561428-5728359 | -21.0 ± 0.1 | ... | ... | ... | ... | ... | Y | ... | ... | ... | ... | n | ... |
| 46226 | 15552922-5726009 | -47.6 ± 0.5 | 4912 ± 266 | 4.51 ± 0.44 | 0.24 ± 0.55 | ... | ... | N | ... | ... | ... | ... | n | ... |
| 46107 | 15561494-5725388 | -16.6 ± 0.1 | 5577 ± 99 | 4.54 ± 0.02 | -0.05 ± 0.17 | ... | ... | N | ... | ... | ... | ... | n | ... |
| 45961 | 15552926-5716216 | -67.0 ± 0.1 | 5251 ± 75 | 4.91 ± 0.05 | 0.12 ± 0.20 | ... | ... | N | ... | ... | ... | ... | n | ... |
| 46108 | 15561532-5725030 | -47.2 ± 0.1 | ... | ... | ... | ... | ... | N | ... | ... | ... | ... | n | ... |
| 46227 | 15552932-5723595 | -65.9 ± 1.0 | 5119 ± 289 | 3.40 ± 1.79 | -0.37 ± 0.53 | ... | ... | N | ... | ... | ... | ... | n | G |
| 46459 | 15561543-5720241 | 0.9 ± 0.5 | 5404 ± 137 | 4.09 ± 0.26 | -0.12 ± 0.18 | <79 | 3 | N | ... | ... | ... | ... | n | NG |
| 46228 | 15552943-5723356 | -54.8 ± 0.4 | 5935 ± 255 | 4.53 ± 0.30 | 0.12 ± 0.36 | ... | ... | N | ... | ... | ... | ... | n | ... |
| 46460 | 15561578-5725201 | -7.4 ± 0.5 | 5546 ± 44 | 4.32 ± 0.74 | 0.27 ± 0.28 | ... | ... | N | ... | ... | ... | ... | n | ... |
| 3076 | 15552962-5724408 | -25.0 ± 0.6 | 4974 ± 117 | 3.05 ± 0.22 | 0.16 ± 0.09 | 24 ± 1 | ... | Y | Y | Y | Y | ... | Y | ... |
| 46461 | 15561609-5727484 | 5.2 ± 0.4 | 5706 ± 72 | 4.58 ± 0.28 | 0.27 ± 0.13 | ... | ... | N | ... | ... | ... | ... | n | ... |
| 46462 | 15561630-5724403 | -42.3 ± 0.7 | 5812 ± 649 | 4.26 ± 0.32 | 0.20 ± 0.80 | ... | ... | N | ... | ... | ... | ... | n | ... |
| 45962 | 15552965-5729293 | -25.3 ± 0.1 | 5069 ± 76 | 3.64 ± 0.05 | 0.08 ± 0.16 | ... | ... | Y | ... | ... | ... | Y | n | ... |
| 46109 | 15561641-5726126 | -17.5 ± 0.3 | ... | ... | ... | ... | ... | Y | ... | ... | ... | ... | n | ... |
| 46229 | 15552968-5727537 | -62.4 ± 0.8 | 5690 ± 239 | 3.85 ± 0.09 | -0.19 ± 0.39 | <56 | 3 | N | ... | ... | ... | ... | n | NG |
| 46110 | 15561656-5728097 | -20.5 ± 1.5 | ... | ... | ... | ... | ... | Y | ... | ... | ... | Y | n | ... |
| 45963 | 15552977-5729142 | -22.4 ± 0.1 | 6229 ± 150 | 4.87 ± 0.02 | ... | ... | ... | Y | ... | ... | ... | ... | n | ... |
| 46463 | 15561679-5720291 | 5.9 ± 0.6 | 4943 ± 269 | 4.82 ± 0.18 | -0.37 ± 0.17 | ... | ... | N | ... | ... | ... | ... | n | ... |
| 46230 | 15553004-5729209 | -43.4 ± 0.5 | 4830 ± 478 | 3.12 ± 0.41 | 0.18 ± 0.31 | ... | ... | N | ... | ... | ... | ... | n | G |
| 46111 | 15561685-5728326 | -48.6 ± 0.2 | ... | ... | ... | ... | ... | N | ... | ... | ... | ... | n | ... |
| 46112 | 15561699-5733058 | -32.1 ± 0.1 | 4711 ± 105 | 3.23 ± 0.07 | 0.10 ± 0.23 | ... | ... | Y | ... | ... | ... | ... | n | ... |
| 46231 | 15553060-5730066 | -54.0 ± 1.1 | 5444 ± 296 | 4.50 ± 1.04 | -0.57 ± 0.63 | ... | ... | N | ... | ... | ... | ... | n | ... |
| 46464 | 15561718-5726356 | -105.5 ± 0.6 | 6061 ± 355 | 4.84 ± 0.62 | -0.13 ± 0.22 | ... | ... | N | ... | ... | ... | ... | n | ... |

| ID | CNAME | RV (km s$^{-1}$) | $T_{\rm eff}$ (K) | $logg$ (dex) | [Fe/H] (dex) | EW(Li)$^a$ (mÅ) | EW(Li) error flag$^b$ | Membership RV | Li | $logg$ | [Fe/H] | Gaia study Cantat-Gaudin$^c$ | Final$^d$ | NMs with Li$^e$ |
|---|---|---|---|---|---|---|---|---|---|---|---|---|---|---|
| 46244 | 15553395-5724172 | -84.9 ± 1.0 | 5702 ± 214 | 3.90 ± 0.42 | -0.17 ± 0.25 | … | … | N | … | … | … | … | n | … |
| 46113 | 15561743-5724579 | -47.0 ± 0.3 | … | … | … | … | … | N | … | … | … | … | n | … |
| 46245 | 15553406-5728462 | -13.4 ± 0.5 | 5727 ± 171 | 4.20 ± 0.37 | 0.16 ± 0.16 | … | … | N | … | … | … | … | n | … |
| 46114 | 15561766-5726430 | 4.2 ± 0.3 | … | … | … | … | … | N | … | … | … | … | n | … |
| 46246 | 15553407-5726090 | -24.7 ± 0.4 | 6262 ± 187 | 4.14 ± 0.07 | 0.24 ± 0.26 | 37 ± 36 | 1 | Y | Y | Y | Y | Y | Y | … |
| 46465 | 15561770-5726188 | -125.2 ± 1.4 | 5273 ± 142 | 4.27 ± 0.18 | 0.16 ± 0.24 | … | … | N | … | … | … | … | n | … |
| 46115 | 15561794-5723540 | -6.6 ± 0.4 | … | … | … | … | … | N | … | … | … | … | n | … |
| 46247 | 15553426-5732080 | -52.8 ± 0.4 | 5794 ± 89 | 4.24 ± 0.43 | 0.23 ± 0.20 | <82 | 3 | N | … | … | … | … | n | NG |
| 46466 | 15561867-5731002 | -81.2 ± 1.0 | 5495 ± 171 | 3.76 ± 0.15 | -0.36 ± 0.25 | … | … | N | … | … | … | … | n | … |
| 46248 | 15553457-5726124 | -23.1 ± 0.5 | 5539 ± 136 | 4.18 ± 0.35 | 0.10 ± 0.18 | … | … | Y | … | … | … | … | n | … |
| 3093 | 15561874-5731192 | -7.1 ± 0.6 | 4664 ± 122 | 2.58 ± 0.23 | -0.05 ± 0.10 | 8 ± 1 | … | N | … | … | … | … | n | G |
| 45975 | 15553459-5730013 | -25.0 ± 0.2 | … | … | … | … | … | Y | … | … | … | Y | n | … |
| 46249 | 15553473-5723491 | -1.4 ± 0.6 | 5412 ± 29 | 4.83 ± 0.50 | 0.06 ± 0.15 | … | … | N | … | … | … | … | n | … |
| 45976 | 15553480-5719348 | 44.2 ± 0.1 | 5675 ± 141 | 3.84 ± 0.09 | -0.23 ± 0.14 | … | … | N | … | … | … | … | n | … |
| 46250 | 15553485-5728520 | -65.9 ± 0.4 | 5693 ± 285 | 3.94 ± 0.28 | 0.08 ± 0.14 | <71 | 3 | N | … | … | … | … | n | NG |
| 45977 | 15553502-5715297 | -16.6 ± 0.3 | … | … | … | … | … | N | … | … | … | … | n | … |
| 45978 | 15553507-5728118 | -25.1 ± 0.3 | … | … | … | … | … | Y | … | … | … | Y | n | … |
| 45979 | 15553520-5729380 | -20.4 ± 0.1 | 6709 ± 132 | 4.29 ± 0.02 | … | … | … | Y | … | … | … | … | n | … |
| 46251 | 15553528-5729215 | -40.2 ± 0.6 | 5572 ± 118 | 4.03 ± 0.43 | 0.30 ± 0.15 | … | … | N | … | … | … | … | n | … |
| 46252 | 15553547-5724116 | -47.8 ± 0.6 | 5855 ± 530 | 3.75 ± 0.06 | -0.17 ± 0.49 | … | … | N | … | … | … | … | n | … |
| 45980 | 15553563-5722592 | -6.4 ± 0.2 | … | … | … | … | … | N | … | … | … | … | n | … |
| 45981 | 15553573-5728309 | -29.8 ± 0.4 | 6173 ± 26 | 4.07 ± 0.15 | 0.04 ± 0.17 | 88 ± 57 | … | Y | Y | Y | Y | … | Y | … |
| 45982 | 15553590-5725283 | -22.5 ± 0.1 | 6059 ± 98 | 4.91 ± 0.05 | … | … | … | Y | … | … | … | … | n | … |
| 46253 | 15553611-5725167 | -76.7 ± 0.4 | 6330 ± 372 | 4.07 ± 0.15 | 0.36 ± 0.29 | 104 ± 61 | 1 | N | … | … | … | … | n | NG |
| 46254 | 15553623-5725505 | -24.5 ± 0.7 | 5453 ± 262 | 3.13 ± 0.27 | -0.16 ± 0.15 | … | … | Y | … | … | … | … | n | … |
| 46255 | 15553630-5730007 | -27.8 ± 0.6 | 5718 ± 103 | 4.62 ± 0.33 | 0.20 ± 0.15 | … | … | Y | … | … | … | … | n | … |
| 45983 | 15553642-5727092 | -17.3 ± 1.7 | … | … | … | … | … | Y | … | … | … | Y | n | … |
| 46256 | 15553656-5726134 | -29.0 ± 0.8 | 5511 ± 277 | 4.58 ± 0.10 | -0.12 ± 0.70 | … | … | Y | … | … | … | … | n | … |
| 46257 | 15553670-5724089 | -17.6 ± 0.5 | 6065 ± 201 | 4.07 ± 0.22 | 0.10 ± 0.22 | … | … | Y | … | … | … | … | n | … |
| 3078 | 15553687-5729569 | -23.4 ± 0.6 | 4918 ± 121 | 3.00 ± 0.21 | 0.15 ± 0.10 | 47 ± 2 | … | Y | Y | Y | Y | … | Y | … |
| 46289 | 15554155-5731137 | -18.9 ± 0.3 | 5889 ± 99 | 4.04 ± 0.14 | 0.08 ± 0.12 | <47 | 3 | Y | Y | Y | Y | N | Y | … |
| 46290 | 15554165-5729092 | -20.6 ± 0.4 | 5468 ± 120 | 4.01 ± 0.14 | 0.26 ± 0.16 | … | … | Y | … | … | … | … | n | … |
| 3082 | 15554167-5725533 | -23.7 ± 0.6 | 4912 ± 118 | 2.79 ± 0.23 | 0.16 ± 0.10 | <5 | 3 | Y | Y | Y | Y | … | Y | … |
| 46291 | 15554170-5724159 | -133.3 ± 1.2 | 6134 ± 508 | 4.01 ± 0.09 | -0.06 ± 0.78 | … | … | N | … | … | … | … | n | … |
| 45998 | 15554171-5726144 | -30.1 ± 0.6 | … | … | … | … | … | Y | … | … | … | N | n | … |
| 46292 | 15554175-5729433 | 42.6 ± 0.8 | 5572 ± 26 | 4.93 ± 0.22 | -0.33 ± 0.15 | … | … | N | … | … | … | … | n | … |
| 45999 | 15554194-5726213 | -34.0 ± 0.3 | … | … | … | … | … | N | … | … | … | Y | n | … |
| 46293 | 15554217-5724409 | -132.4 ± 1.2 | 5744 ± 778 | 4.24 ± 0.40 | -0.47 ± 0.30 | … | … | N | … | … | … | … | n | … |
| 46000 | 15554224-5729247 | -30.2 ± 0.1 | … | … | … | … | … | Y | … | … | … | Y | n | … |
| 46294 | 15554242-5724524 | -4.5 ± 0.6 | 4992 ± 360 | 4.33 ± 0.06 | -0.34 ± 0.24 | … | … | N | … | … | … | … | n | … |
| 46295 | 15554274-5725052 | -116.2 ± 1.0 | 6456 ± 159 | 3.96 ± 0.19 | 0.13 ± 0.16 | … | … | N | … | … | … | … | n | … |
| 46296 | 15554277-5719436 | -65.2 ± 1.5 | 5919 ± 407 | 4.11 ± 0.19 | 0.01 ± 0.24 | … | … | N | … | … | … | … | n | … |
| 46297 | 15554285-5724129 | -15.4 ± 0.7 | 5612 ± 201 | 3.80 ± 0.90 | -0.18 ± 0.17 | <135 | 3 | N | … | … | … | … | n | NG |
| 46002 | 15554552-5730063 | -24.9 ± 0.3 | … | … | … | … | … | Y | … | … | … | Y | n | … |
| 46298 | 15554558-5728557 | -22.9 ± 0.5 | 7106 ± 753 | 4.17 ± 0.16 | 0.26 ± 0.15 | 33 ± 17 | 1 | Y | N | N | Y | Y | n | NG |
| 46299 | 15554560-5728164 | -65.3 ± 0.5 | 6011 ± 260 | 4.44 ± 0.28 | 0.34 ± 0.16 | <147 | 3 | N | … | … | … | … | n | NG |
| 46300 | 15554573-5727458 | -25.4 ± 0.4 | 6084 ± 273 | 4.56 ± 0.22 | 0.18 ± 0.22 | <60 | 3 | Y | Y | Y | Y | N | Y | … |
| 46003 | 15554580-5725360 | -25.1 ± 0.2 | … | … | … | … | … | Y | … | … | … | Y | n | … |
| 46301 | 15554592-5724303 | 0.0 ± 0.7 | 5051 ± 348 | 3.54 ± 0.09 | 0.18 ± 0.38 | … | … | N | … | … | … | … | n | … |
| 46004 | 15554596-5734268 | -31.9 ± 0.2 | … | … | … | … | … | Y | … | … | … | … | n | … |
| 46318 | 15554810-5728480 | -9.8 ± 0.3 | 5968 ± 212 | 4.15 ± 0.13 | 0.40 ± 0.21 | 59 ± 47 | 1 | N | … | … | … | … | n | NG |
| 46319 | 15554811-5729405 | -50.5 ± 0.4 | 6274 ± 106 | 4.39 ± 0.35 | 0.19 ± 0.15 | <34 | 3 | N | … | … | … | … | n | NG |
| 46320 | 15554836-5720107 | -33.4 ± 0.7 | 5677 ± 38 | 4.43 ± 0.10 | 0.11 ± 0.47 | … | … | N | … | … | … | … | n | … |
| 46321 | 15554844-5728074 | -24.4 ± 0.6 | 5365 ± 362 | 4.24 ± 0.06 | 0.20 ± 0.37 | … | … | Y | … | … | … | … | n | … |
| 46322 | 15554853-5724577 | -23.6 ± 0.4 | 5966 ± 195 | 4.67 ± 0.39 | 0.14 ± 0.16 | <91 | 3 | Y | Y | Y | Y | … | Y | … |
| 46012 | 15554857-5731111 | -34.9 ± 0.9 | … | … | … | … | … | N | … | … | … | … | n | … |
| 46323 | 15554858-5730355 | -29.8 ± 0.5 | 5502 ± 63 | 4.16 ± 0.53 | -0.20 ± 0.26 | … | … | Y | … | … | … | … | n | … |
| 46013 | 15554865-5728460 | -15.1 ± 0.1 | … | … | … | … | … | N | … | … | … | Y | n | … |
| 46014 | 15554884-5725196 | -22.2 ± 0.1 | … | … | … | … | … | Y | … | … | … | … | n | … |

**Table C.14.** continued.

| ID | CNAME | RV (km s$^{-1}$) | $T_{\text{eff}}$ (K) | logg (dex) | [Fe/H] (dex) | EW(Li)$^a$ (mÅ) | EW(Li) error flag$^b$ | Membership RV | Li | logg | [Fe/H] | Gaia study Cantat-Gaudin$^c$ | Final$^d$ | NMs with Li$^e$ |
|---|---|---|---|---|---|---|---|---|---|---|---|---|---|---|
| 46015 | 15554908-5724520 | -20.1 ± 0.1 | … | … | … | … | … | Y | … | … | … | Y | n | … |
| 46324 | 15554917-5727071 | -20.2 ± 0.7 | 5769 ± 226 | 4.00 ± 0.39 | -0.21 ± 0.39 | <63 | 3 | Y | Y | Y | Y | Y | Y | … |
| 46016 | 15554920-5728418 | -20.4 ± 0.3 | 6045 ± 125 | 4.42 ± 0.05 | … | … | … | Y | … | … | … | … | n | … |
| 46325 | 15554925-5729498 | -23.2 ± 0.7 | 5536 ± 203 | 4.14 ± 0.08 | -0.03 ± 0.16 | … | … | Y | … | … | … | … | n | … |
| 46326 | 15554933-5726355 | -24.7 ± 0.3 | 5270 ± 181 | 4.33 ± 0.28 | 0.07 ± 0.14 | <108 | 3 | Y | N | Y | Y | … | n | NG |
| 46327 | 15554967-5727251 | 15.7 ± 0.5 | 5817 ± 253 | 4.05 ± 0.32 | 0.09 ± 0.34 | <101 | 3 | N | … | … | … | … | n | NG |
| 46017 | 15554973-5727016 | -22.0 ± 0.4 | … | … | … | … | … | Y | … | … | … | N | n | … |
| 46328 | 15554974-5720276 | -14.3 ± 0.5 | 6191 ± 167 | 4.53 ± 0.36 | 0.14 ± 0.20 | 57 ± 35 | 1 | N | … | … | … | … | n | NG |
| 46329 | 15555010-5725189 | -63.6 ± 0.3 | 5782 ± 118 | 3.78 ± 0.44 | 0.04 ± 0.17 | <55 | 3 | N | … | … | … | N | n | NG |
| 46330 | 15555013-5722582 | -25.0 ± 0.4 | 6244 ± 350 | 4.38 ± 0.24 | 0.07 ± 0.45 | 76 ± 41 | 1 | Y | Y | Y | Y | Y | Y | … |
| 46331 | 15555013-5727480 | -24.6 ± 0.5 | 6320 ± 148 | 4.51 ± 0.38 | 0.17 ± 0.18 | … | … | Y | … | … | … | N | n | … |
| 46018 | 15555014-5725482 | -47.9 ± 0.4 | 5855 ± 99 | 4.20 ± 0.73 | 0.12 ± 0.36 | 68 ± 28 | … | … | … | … | … | … | n | NG |
| 46332 | 15555021-5725275 | -65.2 ± 0.3 | 6159 ± 335 | 4.17 ± 0.11 | 0.30 ± 0.26 | … | … | N | … | … | … | … | n | … |
| 46333 | 15555033-5723204 | -69.2 ± 1.1 | 5070 ± 286 | 4.46 ± 0.61 | -0.18 ± 0.25 | … | … | N | … | … | … | … | n | … |
| 46334 | 15555036-5729393 | -38.8 ± 0.5 | 5951 ± 263 | 4.27 ± 0.23 | 0.22 ± 0.24 | … | … | N | … | … | … | … | n | … |
| 46361 | 15555468-5724358 | -45.2 ± 0.3 | 5840 ± 223 | 4.01 ± 0.16 | 0.16 ± 0.20 | … | … | N | … | … | … | … | n | … |
| 46038 | 15555486-5728226 | -31.2 ± 0.2 | 6155 ± 150 | 4.73 ± 0.05 | … | … | … | Y | … | … | … | … | n | … |
| 46362 | 15555508-5719504 | -87.4 ± 0.4 | 5462 ± 136 | 4.07 ± 0.15 | 0.04 ± 0.24 | … | … | N | … | … | … | … | n | … |
| 46039 | 15555514-5727566 | -25.3 ± 0.4 | … | … | … | … | … | Y | … | … | … | … | n | … |
| 46364 | 15555551-5725082 | 14.1 ± 1.0 | 6158 ± 253 | 4.84 ± 0.62 | -0.12 ± 0.23 | … | … | N | … | … | … | … | n | … |
| 46365 | 15555551-5725518 | -25.0 ± 0.4 | 6198 ± 218 | 4.36 ± 0.34 | 0.09 ± 0.18 | 55 ± 46 | 1 | Y | Y | Y | Y | Y | Y | … |
| 46040 | 15555551-5730213 | -37.7 ± 0.1 | … | … | … | … | … | N | … | … | … | … | n | … |
| 46366 | 15555552-5724081 | -32.7 ± 0.6 | 5827 ± 273 | 4.37 ± 0.36 | -0.26 ± 0.18 | … | … | Y | … | … | … | … | n | … |
| 46367 | 15555585-5724568 | -23.4 ± 0.7 | 7164 ± 869 | 4.04 ± 0.24 | 0.11 ± 0.14 | … | … | Y | … | … | … | N | n | … |
| 46368 | 15555599-5726194 | -42.7 ± 0.2 | 5838 ± 155 | 4.46 ± 0.29 | 0.16 ± 0.13 | 57 ± 13 | 1 | N | … | … | … | … | n | NG |
| 46369 | 15555611-5728241 | -20.2 ± 0.5 | 5441 ± 168 | 4.17 ± 0.16 | 0.40 ± 0.27 | … | … | Y | … | … | … | … | n | … |
| 3087 | 15555612-5726539 | -23.7 ± 0.6 | 4936 ± 113 | 2.99 ± 0.23 | 0.15 ± 0.10 | 12 ± 1 | … | Y | Y | Y | Y | … | Y | … |
| 46041 | 15555613-5731449 | -75.0 ± 0.4 | … | … | … | … | … | N | … | … | … | … | n | … |
| 46370 | 15555628-5728522 | -73.5 ± 4.3 | 6053 ± 299 | 2.62 ± 0.64 | -0.36 ± 0.27 | <71 | 3 | N | … | … | … | … | n | … |
| 46371 | 15555645-5725340 | -19.3 ± 0.8 | 6460 ± 349 | 4.43 ± 0.38 | 0.17 ± 0.19 | … | … | Y | … | … | … | N | n | … |
| 46042 | 15555646-5726327 | -0.1 ± 0.1 | … | … | … | … | … | N | … | … | … | … | n | … |
| 46043 | 15555647-5731272 | -33.8 ± 0.3 | 5465 ± 292 | 4.48 ± 0.98 | -0.10 ± 0.36 | 290 ± 90 | … | N | N | Y | Y | … | n | NG |
| 46372 | 15555654-5725195 | -22.1 ± 1.3 | 5052 ± 442 | 3.73 ± 0.19 | -0.78 ± 0.89 | <118 | 3 | Y | N | N | N | N | n | NG |
| 46373 | 15555668-5723474 | -27.2 ± 0.7 | 5789 ± 169 | 4.27 ± 0.45 | -0.49 ± 0.16 | … | … | Y | … | … | … | … | n | … |
| 46374 | 15555680-5724236 | -15.6 ± 0.9 | 6250 ± 277 | 3.59 ± 0.54 | -0.10 ± 0.38 | … | … | N | … | … | … | N | n | … |
| 46375 | 15555683-5729171 | 6.9 ± 0.6 | 5761 ± 326 | 3.50 ± 0.24 | 0.08 ± 0.35 | … | … | N | … | … | … | N | n | … |
| 46376 | 15555684-5730028 | -27.7 ± 0.6 | 5640 ± 356 | 4.07 ± 0.79 | 0.36 ± 0.35 | … | … | Y | … | … | … | … | n | … |
| 46044 | 15555689-5731549 | -36.5 ± 0.1 | … | … | … | … | … | N | … | … | … | … | n | … |
| 46391 | 15555916-5722125 | -34.2 ± 0.4 | 5651 ± 158 | 4.15 ± 0.12 | -0.03 ± 0.18 | … | … | N | … | … | … | … | n | … |
| 46055 | 15555930-5726070 | -37.0 ± 0.1 | … | … | … | … | … | N | … | … | … | Y | n | … |
| 46392 | 15555930-5726395 | -66.4 ± 2.2 | 5922 ± 144 | 3.57 ± 0.83 | -0.28 ± 0.34 | … | … | N | … | … | … | … | n | … |
| 46056 | 15555936-5731558 | -36.8 ± 0.3 | … | … | … | … | … | N | … | … | … | … | n | … |
| 46393 | 15555942-5721551 | -9.8 ± 0.4 | 5959 ± 430 | 4.36 ± 0.23 | 0.06 ± 0.46 | <77 | 3 | N | … | … | … | … | n | NG |
| 46057 | 15555958-5724240 | -25.1 ± 0.1 | … | … | … | … | … | Y | … | … | … | Y | n | … |
| 46058 | 15555982-5725524 | -33.2 ± 0.5 | 6139 ± 114 | 3.41 ± 0.05 | … | … | … | N | … | … | … | … | n | … |
| 46394 | 15560002-5723479 | -32.2 ± 0.3 | 5873 ± 105 | 4.34 ± 0.19 | 0.14 ± 0.13 | … | … | Y | … | … | … | … | n | … |
| 46395 | 15560030-5726114 | -36.7 ± 0.7 | 6417 ± 138 | 4.35 ± 0.64 | -0.07 ± 0.28 | … | … | N | … | … | … | … | n | … |
| 46396 | 15560037-5724113 | -121.0 ± 0.6 | 4960 ± 525 | 3.94 ± 0.83 | 0.13 ± 0.25 | … | … | N | … | … | … | … | n | … |
| 46059 | 15560044-5730000 | -20.9 ± 0.2 | … | … | … | … | … | Y | … | … | … | … | n | … |
| 46060 | 15560048-5719218 | -14.6 ± 0.1 | 5989 ± 78 | 4.25 ± 0.02 | -0.31 ± 0.13 | … | … | N | … | … | … | … | n | … |
| 46397 | 15560067-5727216 | -109.7 ± 0.7 | 5420 ± 386 | 3.56 ± 0.10 | -0.26 ± 0.15 | … | … | N | … | … | … | … | n | … |
| 46398 | 15560068-5726126 | -26.9 ± 0.4 | 5685 ± 38 | 4.32 ± 0.22 | 0.18 ± 0.16 | <56 | 3 | Y | Y | Y | Y | N | Y | … |
| 46061 | 15560086-5727145 | -25.9 ± 1.7 | … | … | … | … | … | Y | … | … | … | Y | n | … |
| 46399 | 15560094-5719450 | -76.5 ± 0.4 | 5209 ± 198 | 4.41 ± 0.27 | 0.01 ± 0.25 | … | … | N | … | … | … | … | n | … |
| 46062 | 15560128-5724584 | -26.0 ± 0.3 | … | … | … | … | … | Y | … | … | … | Y | n | … |
| 46063 | 15560128-5725448 | -26.3 ± 0.8 | … | … | … | … | … | Y | … | … | … | Y | n | … |
| 46400 | 15560139-5725490 | -98.5 ± 0.9 | 5881 ± 462 | 4.05 ± 0.34 | -0.17 ± 0.44 | … | … | N | … | … | … | … | n | … |
| 46401 | 15560143-5724278 | -26.0 ± 0.3 | 6626 ± 416 | 4.19 ± 0.41 | 0.17 ± 0.26 | 68 ± 25 | 1 | Y | Y | Y | Y | … | Y | … |
| 46064 | 15560148-5724122 | -40.4 ± 0.4 | … | … | … | … | … | N | … | … | … | … | n | … |

**Table C.14.** continued.

| ID | CNAME | RV (km s$^{-1}$) | $T_{\text{eff}}$ (K) | logg (dex) | [Fe/H] (dex) | EW(Li)$^a$ (mÅ) | EW(Li) error flag$^b$ | Membership RV | Li | logg | [Fe/H] | Gaia study Cantat-Gaudin$^c$ | Final$^d$ | NMs with Li$^e$ |
|---|---|---|---|---|---|---|---|---|---|---|---|---|---|---|
| 46402 | 15560167-5726064 | -5.2 ± 0.5 | 5988 ± 291 | 4.95 ± 0.61 | 0.31 ± 0.15 | … | … | N | … | … | … | … | n | … |
| 46065 | 15560168-5735189 | -53.7 ± 0.7 | … | … | … | … | … | N | … | … | … | … | n | … |
| 46403 | 15560174-5726418 | -26.1 ± 0.4 | 5713 ± 130 | 4.44 ± 0.25 | 0.09 ± 0.14 | 124 ± 57 | 1 | Y | N | Y | Y | N | n | NG |
| 46083 | 15560682-5726192 | -47.4 ± 0.5 | … | … | … | … | … | N | … | … | … | … | n | … |
| 46433 | 15560690-5728560 | -58.0 ± 0.5 | 5182 ± 35 | 3.89 ± 0.07 | 0.08 ± 0.18 | … | … | N | … | … | … | … | n | … |
| 46434 | 15560700-5731129 | -109.9 ± 1.2 | 5631 ± 46 | 4.00 ± 0.48 | -0.34 ± 0.17 | … | … | N | … | … | … | … | n | … |
| 3090 | 15560707-5722408 | -23.0 ± 0.6 | 4960 ± 118 | 2.89 ± 0.23 | 0.18 ± 0.09 | 47 ± 3 | … | Y | Y | Y | Y | Y | Y | … |
| 46084 | 15560722-5735543 | -80.0 ± 0.2 | … | … | … | … | … | N | … | … | … | … | n | … |
| 46435 | 15560724-5731000 | -65.3 ± 0.7 | 4843 ± 621 | 3.88 ± 0.65 | -0.30 ± 0.58 | … | … | N | … | … | … | … | n | … |
| 46085 | 15560724-5734054 | -23.6 ± 0.1 | … | … | … | … | … | Y | … | … | … | … | n | … |
| 46086 | 15560749-5724484 | -26.5 ± 0.2 | … | … | … | … | … | Y | … | … | … | … | n | … |
| 3091 | 15560777-5722216 | 7.6 ± 0.6 | 4603 ± 125 | 2.92 ± 0.23 | 0.13 ± 0.09 | <6 | 3 | N | … | … | … | … | n | G |
| 46436 | 15560777-5729292 | -25.2 ± 0.5 | 5542 ± 96 | 4.55 ± 0.11 | 0.36 ± 0.15 | … | … | Y | … | … | … | … | n | … |
| 46437 | 15560778-5726566 | 3.3 ± 0.8 | 5936 ± 204 | 4.47 ± 0.96 | 0.10 ± 0.24 | … | … | N | … | … | … | … | n | … |
| 46438 | 15560782-5722086 | -45.4 ± 1.5 | 5577 ± 23 | 3.37 ± 0.70 | -0.05 ± 0.45 | … | … | N | … | … | … | … | n | … |
| 3092 | 15560789-5726360 | -34.7 ± 0.6 | 4640 ± 114 | 2.58 ± 0.24 | -0.05 ± 0.09 | <7 | 3 | N | Y | N | Y | … | n | G |
| 46439 | 15560793-5723108 | 68.1 ± 0.7 | 5029 ± 147 | 3.13 ± 0.27 | -0.16 ± 0.16 | … | … | N | … | … | … | … | n | G |
| 46087 | 15560814-5725322 | 37.8 ± 0.1 | 5781 ± 129 | 4.54 ± 0.06 | 0.06 ± 0.21 | … | … | N | … | … | … | … | n | … |
| 46440 | 15560840-5728221 | -15.5 ± 0.6 | 6095 ± 285 | 4.35 ± 0.12 | 0.33 ± 0.16 | … | … | N | … | … | … | … | n | … |
| 46088 | 15560858-5719557 | -26.4 ± 1.0 | 6118 ± 203 | 4.52 ± 0.66 | 0.12 ± 0.41 | … | … | Y | … | … | … | … | n | … |
| 46441 | 15560884-5721541 | -22.3 ± 1.1 | 5839 ± 334 | 2.46 ± 1.17 | -0.53 ± 0.44 | … | … | Y | … | … | … | … | n | … |
| 46089 | 15560922-5732317 | -19.6 ± 0.2 | … | … | … | … | … | Y | … | … | … | … | n | … |
| 46442 | 15560931-5726300 | -99.2 ± 0.6 | 5082 ± 167 | 3.74 ± 1.21 | 0.36 ± 0.55 | … | … | N | … | … | … | … | n | … |
| 46443 | 15560936-5724359 | -85.7 ± 0.4 | 6102 ± 344 | 4.20 ± 0.07 | 0.12 ± 0.39 | … | … | N | … | … | … | … | n | … |
| 46147 | 15545980-5720172 | -63.5 ± 0.7 | 5879 ± 774 | 4.53 ± 0.21 | 0.04 ± 0.68 | … | … | N | … | … | … | … | n | … |
| 45934 | 15550021-5726521 | -10.7 ± 0.1 | 4371 ± 150 | 2.78 ± 0.25 | 0.32 ± 0.26 | … | … | N | … | … | … | … | n | … |
| 45935 | 15550247-5723227 | -33.2 ± 0.1 | … | … | … | … | … | N | … | … | … | … | n | … |
| 46405 | 15560217-5729098 | -66.0 ± 0.6 | 6219 ± 374 | 4.15 ± 0.23 | 0.08 ± 0.26 | <89 | 3 | N | … | … | … | … | n | NG |
| 46148 | 15550334-5722152 | -61.9 ± 0.5 | 5864 ± 152 | 4.07 ± 0.19 | 0.28 ± 0.16 | … | … | N | … | … | … | … | n | … |
| 46067 | 15560221-5717194 | -35.3 ± 0.1 | … | … | … | … | … | N | … | … | … | … | n | … |
| 46149 | 15550433-5719198 | -23.7 ± 1.1 | 5308 ± 79 | 3.42 ± 0.34 | 0.28 ± 0.07 | … | … | Y | … | … | … | … | n | … |
| 46406 | 15560231-5728555 | -51.8 ± 0.5 | 4999 ± 120 | 4.32 ± 0.87 | 0.10 ± 0.28 | … | … | N | … | … | … | … | n | … |
| 46068 | 15560234-5727084 | -25.8 ± 0.7 | … | … | … | … | … | Y | … | … | … | Y | n | … |
| 46407 | 15560237-5729228 | -92.6 ± 0.6 | 5662 ± 39 | 4.70 ± 0.71 | 0.21 ± 0.22 | … | … | N | … | … | … | … | n | … |
| 45936 | 15550740-5733275 | -4.7 ± 0.1 | … | … | … | … | … | N | … | … | … | … | n | … |
| 46408 | 15560239-5727476 | -14.2 ± 0.6 | 5838 ± 460 | 3.86 ± 0.16 | 0.11 ± 0.29 | … | … | N | … | … | … | … | n | … |
| 46150 | 15550748-5728264 | -22.6 ± 0.4 | 5742 ± 63 | 4.22 ± 0.45 | 0.13 ± 0.14 | <97 | 3 | Y | Y | Y | Y | … | Y | … |
| 46409 | 15560239-5728445 | -13.3 ± 0.8 | 5245 ± 296 | 3.94 ± 0.11 | -0.26 ± 0.34 | … | … | N | … | … | … | … | n | … |
| 45937 | 15550849-5732276 | 67.6 ± 0.6 | … | … | … | … | … | N | … | … | … | … | n | … |
| 46410 | 15560240-5726517 | -19.3 ± 1.0 | 4797 ± 100 | 4.04 ± 0.66 | -0.27 ± 0.41 | … | … | Y | … | … | … | … | n | … |
| 46151 | 15550874-5723035 | -28.0 ± 0.9 | 5777 ± 544 | 4.59 ± 0.74 | -0.34 ± 0.74 | <85 | 3 | Y | Y | N | N | … | n | NG |
| 46411 | 15560286-5728105 | -48.3 ± 0.3 | 5677 ± 116 | 4.00 ± 0.31 | 0.20 ± 0.14 | 83 ± 25 | 1 | N | … | … | … | … | n | NG |
| 46412 | 15560288-5728210 | -46.7 ± 0.6 | 4695 ± 341 | 4.42 ± 0.38 | 0.19 ± 0.37 | … | … | N | … | … | … | … | n | … |
| 46152 | 15550951-5725389 | -7.7 ± 0.5 | 5608 ± 90 | 3.63 ± 0.47 | 0.24 ± 0.27 | … | … | N | … | … | … | … | n | … |
| 46413 | 15560291-5728072 | -53.8 ± 0.4 | 5505 ± 32 | 4.03 ± 0.21 | 0.20 ± 0.14 | … | … | N | … | … | … | … | n | … |
| 46153 | 15550991-5724378 | -59.1 ± 0.8 | 5881 ± 221 | 3.98 ± 0.87 | 0.18 ± 0.35 | … | … | N | … | … | … | … | n | … |
| 46414 | 15560292-5725287 | -50.8 ± 0.7 | 6090 ± 333 | 4.39 ± 0.16 | 0.32 ± 0.34 | … | … | N | … | … | … | … | n | … |
| 46154 | 15551107-5725194 | -2.2 ± 0.3 | 6409 ± 187 | 4.38 ± 0.38 | 0.03 ± 0.14 | 57 ± 24 | 1 | N | … | … | … | Y | n | NG |
| 46415 | 15560295-5724074 | -3.3 ± 1.3 | 4525 ± 597 | 4.04 ± 0.05 | -0.22 ± 0.21 | … | … | N | … | … | … | … | n | … |
| 46155 | 15551111-5732232 | -98.1 ± 1.5 | 6154 ± 271 | 4.53 ± 0.34 | 0.19 ± 0.36 | … | … | N | … | … | … | … | n | … |
| 46069 | 15560310-5723476 | -26.0 ± 0.6 | 5252 ± 63 | 4.33 ± 0.06 | -0.63 ± 0.23 | … | … | Y | … | … | … | … | n | … |
| 45938 | 15551118-5731212 | -25.4 ± 0.1 | … | … | … | … | … | Y | … | … | … | Y | n | … |
| 46070 | 15560352-5728525 | -55.7 ± 0.1 | … | … | … | … | … | N | … | … | … | Y | n | … |
| 46156 | 15551160-5730409 | -88.1 ± 2.1 | 5433 ± 498 | 4.30 ± 0.61 | -0.10 ± 0.27 | … | … | N | … | … | … | … | n | … |
| 46071 | 15560368-5725491 | -27.7 ± 0.2 | … | … | … | … | … | Y | … | … | … | Y | n | … |
| 46157 | 15551160-5731092 | -110.5 ± 0.5 | 4934 ± 604 | 3.79 ± 1.70 | 0.09 ± 0.63 | … | … | N | … | … | … | … | n | … |
| 46072 | 15560400-5733130 | -29.9 ± 0.5 | … | … | … | … | … | Y | … | … | … | … | n | … |
| 46158 | 15551250-5722519 | -25.0 ± 0.9 | 5715 ± 152 | 4.29 ± 0.31 | 0.06 ± 0.34 | … | … | Y | … | … | … | … | n | … |
| 46416 | 15560404-5726221 | -28.2 ± 0.4 | 5458 ± 93 | 4.60 ± 0.65 | -0.11 ± 0.14 | … | … | Y | … | … | … | … | n | … |

**Table C.14.** continued.

| ID | CNAME | RV (km s$^{-1}$) | $T_{\rm eff}$ (K) | logg (dex) | [Fe/H] (dex) | EW(Li)$^a$ (mÅ) | EW(Li) error flag$^b$ | Membership RV | Li | logg | [Fe/H] | Gaia study Cantat-Gaudin$^c$ | Final$^d$ | NMs with Li$^e$ |
|---|---|---|---|---|---|---|---|---|---|---|---|---|---|---|
| 46159 | 15551260-5730471 | -77.3 ± 0.5 | 5526 ± 175 | 4.12 ± 0.38 | -0.24 ± 0.20 | <62 | 3 | N | … | … | … | … | n | NG |
| 46073 | 15560404-5732262 | -61.0 ± 0.3 | 5899 ± 128 | 4.38 ± 0.05 | 0.40 ± 0.26 | … | … | N | … | … | … | … | n | … |
| 46160 | 15551282-5720195 | -53.9 ± 0.5 | 5982 ± 251 | 4.23 ± 0.27 | -0.10 ± 0.22 | <67 | 3 | N | … | … | … | … | n | NG |
| 46417 | 15560417-5729183 | -61.5 ± 1.1 | 5424 ± 57 | 3.97 ± 1.20 | -0.39 ± 0.36 | … | … | N | … | … | … | … | n | … |
| 46161 | 15551285-5726319 | -71.0 ± 2.8 | 5592 ± 320 | 3.40 ± 0.84 | -0.60 ± 0.56 | <60 | 3 | N | … | … | … | N | n | … |
| 46181 | 15551835-5729352 | -25.5 ± 0.7 | 5016 ± 170 | 3.65 ± 1.36 | -0.24 ± 0.36 | … | … | Y | … | … | … | … | n | … |
| 45944 | 15551843-5725003 | -82.4 ± 0.1 | … | … | … | … | … | N | … | … | … | … | n | … |
| 46418 | 15560430-5722306 | 10.6 ± 0.4 | 5184 ± 62 | 3.77 ± 0.58 | 0.07 ± 0.13 | … | … | N | … | … | … | … | n | … |
| 46182 | 15551905-5721577 | -11.3 ± 0.7 | 5943 ± 155 | 3.85 ± 0.25 | -0.14 ± 0.27 | … | … | N | … | … | … | … | n | … |
| 46183 | 15551969-5727071 | -24.4 ± 0.5 | 5705 ± 54 | 4.91 ± 0.61 | 0.04 ± 0.17 | … | … | Y | … | … | … | N | n | … |
| 46091 | 15560987-5728096 | -25.3 ± 0.1 | 5039 ± 85 | 3.49 ± 0.04 | 0.19 ± 0.18 | … | … | Y | … | … | … | Y | n | … |
| 45945 | 15551977-5732310 | -20.4 ± 1.0 | 6398 ± 34 | 3.72 ± 0.06 | 0.21 ± 0.15 | … | … | Y | … | … | … | … | n | … |
| 45946 | 15551986-5724198 | -57.6 ± 0.1 | … | … | … | … | … | N | … | … | … | … | n | … |
| 46092 | 15560998-5731295 | -30.2 ± 0.1 | … | … | … | … | … | Y | … | … | … | Y | n | … |
| 46184 | 15551996-5726108 | 13.0 ± 0.3 | 5794 ± 159 | 4.33 ± 0.34 | 0.07 ± 0.16 | … | … | N | … | … | … | … | n | … |
| 46093 | 15561006-5728430 | -8.8 ± 0.1 | … | … | … | … | … | N | … | … | … | … | n | … |
| 46094 | 15561008-5716143 | -18.2 ± 0.2 | … | … | … | … | … | Y | … | … | … | … | n | … |
| 46185 | 15552050-5727452 | -79.5 ± 0.4 | 5471 ± 245 | 3.57 ± 0.15 | -0.03 ± 0.18 | … | … | N | … | … | … | … | n | … |
| 46095 | 15561035-5724433 | -24.0 ± 0.3 | … | … | … | … | … | Y | … | … | … | … | n | … |
| 46186 | 15552059-5726136 | -34.1 ± 0.6 | 5645 ± 33 | 3.79 ± 0.88 | -0.13 ± 0.23 | … | … | N | … | … | … | … | n | … |
| 46096 | 15561048-5719587 | -63.8 ± 0.1 | … | … | … | … | … | N | … | … | … | … | n | … |
| 45947 | 15552066-5727098 | -32.2 ± 0.3 | … | … | … | … | … | Y | … | … | … | … | n | … |
| 46447 | 15561057-5728221 | -70.5 ± 1.1 | 5302 ± 45 | 4.31 ± 0.57 | -0.76 ± 0.22 | … | … | N | … | … | … | … | n | … |
| 46187 | 15552126-5727325 | -69.7 ± 0.6 | 5742 ± 104 | 3.77 ± 0.58 | 0.11 ± 0.54 | … | … | N | … | … | … | … | n | … |
| 46448 | 15561079-5720239 | -74.3 ± 1.1 | 5228 ± 453 | 3.37 ± 0.90 | 0.00 ± 0.52 | … | … | N | … | … | … | … | n | … |
| 46188 | 15552128-5728073 | -46.7 ± 0.7 | 5619 ± 60 | 4.23 ± 0.70 | -0.04 ± 0.17 | … | … | N | … | … | … | … | n | … |
| 46449 | 15561114-5726067 | -72.4 ± 0.6 | 5346 ± 145 | 4.80 ± 0.58 | 0.26 ± 0.17 | … | … | N | … | … | … | … | n | … |
| 45948 | 15552131-5725048 | -64.7 ± 0.1 | … | … | … | … | … | N | … | … | … | … | n | … |
| 46097 | 15561163-5734469 | -0.2 ± 0.1 | … | … | … | … | … | N | … | … | … | … | n | … |
| 46189 | 15552132-5724204 | 11.8 ± 0.3 | 6123 ± 304 | 4.20 ± 0.13 | 0.39 ± 0.25 | 75 ± 30 | 1 | N | … | … | … | … | n | NG |
| 46098 | 15561166-5726258 | 16.8 ± 0.4 | 6253 ± 135 | 4.91 ± 0.05 | … | … | … | N | … | … | … | … | n | … |
| 46099 | 15561178-5726011 | -39.7 ± 0.4 | … | … | … | … | … | N | … | … | … | … | n | … |
| 46190 | 15552175-5719579 | -42.7 ± 0.9 | 6559 ± 301 | 4.12 ± 0.08 | 0.24 ± 0.34 | … | 3 | N | … | … | … | N | n | … |
| 46100 | 15561180-5733417 | -23.0 ± 0.2 | … | … | … | … | … | Y | … | … | … | … | n | … |
| 46191 | 15552176-5723007 | -29.3 ± 0.3 | 6017 ± 327 | 4.51 ± 0.18 | 0.17 ± 0.38 | … | … | Y | … | … | … | … | n | … |
| 46450 | 15561184-5729482 | -72.6 ± 0.6 | 5177 ± 516 | 3.58 ± 1.12 | 0.14 ± 0.46 | … | … | N | … | … | … | … | n | … |
| 45949 | 15552176-5734510 | -76.8 ± 0.1 | … | … | … | … | … | N | … | … | … | … | n | … |
| 46101 | 15561230-5729055 | -43.6 ± 0.2 | 5469 ± 129 | 3.67 ± 0.02 | 0.01 ± 0.22 | … | … | N | … | … | … | … | n | … |
| 46192 | 15552182-5730524 | -81.6 ± 0.5 | 5626 ± 122 | 4.08 ± 0.21 | 0.05 ± 0.22 | <105 | 3 | N | … | … | … | … | n | NG |
| 46451 | 15561240-5726450 | -32.5 ± 0.4 | 5587 ± 316 | 4.26 ± 0.17 | 0.16 ± 0.21 | … | … | Y | … | … | … | … | n | … |
| 45950 | 15552208-5731389 | -24.1 ± 0.1 | … | … | … | … | … | Y | … | … | … | Y | n | … |
| 46452 | 15561265-5727163 | -59.5 ± 0.3 | 5247 ± 81 | 3.62 ± 0.06 | -0.14 ± 0.20 | <55 | 3 | N | … | … | … | … | n | NG |
| 46193 | 15552248-5729356 | -65.9 ± 0.4 | 5831 ± 359 | 4.14 ± 0.19 | 0.28 ± 0.27 | … | … | N | … | … | … | … | n | … |
| 46453 | 15561287-5718157 | -9.6 ± 0.6 | 4921 ± 278 | 3.44 ± 0.17 | -0.40 ± 0.30 | … | … | N | … | … | … | … | n | G |
| 45951 | 15552275-5733410 | 4.8 ± 0.1 | … | … | … | … | … | N | … | … | … | … | n | … |
| 46102 | 15561292-5726402 | -30.2 ± 0.1 | … | … | … | … | … | Y | … | … | … | Y | n | … |
| 45952 | 15552280-5733156 | -37.5 ± 2.1 | … | … | … | … | … | N | … | … | … | N | n | … |
| 46454 | 15561298-5726512 | -73.7 ± 0.5 | 6045 ± 279 | 4.35 ± 0.32 | -0.15 ± 0.30 | 48 ± 41 | 1 | N | … | … | … | … | n | NG |
| 46194 | 15552282-5727147 | -19.3 ± 0.6 | 5573 ± 70 | 4.20 ± 0.23 | -0.10 ± 0.16 | … | … | Y | … | … | … | … | n | … |
| 46455 | 15561332-5720191 | -81.6 ± 0.3 | 6027 ± 240 | 4.38 ± 0.30 | 0.26 ± 0.22 | … | … | N | … | … | … | … | n | … |
| 46195 | 15552317-5723051 | -7.9 ± 0.4 | 6118 ± 229 | 4.23 ± 0.34 | -0.01 ± 0.24 | 54 ± 45 | 1 | N | … | … | … | … | n | NG |
| 46103 | 15561355-5715317 | -12.3 ± 0.2 | … | … | … | … | … | N | … | … | … | … | n | … |
| 46104 | 15561360-5727489 | 43.1 ± 0.1 | 5401 ± 95 | 4.41 ± 0.05 | -0.04 ± 0.20 | … | … | N | … | … | … | … | n | … |
| 46232 | 15553061-5720224 | -92.6 ± 0.8 | 5156 ± 478 | 3.37 ± 0.48 | -0.32 ± 0.26 | … | … | N | … | … | … | … | n | G |
| 46117 | 15561916-5728398 | -25.1 ± 0.1 | … | … | … | … | … | Y | … | … | … | Y | n | … |
| 45964 | 15553073-5724442 | -35.2 ± 0.1 | … | … | … | … | … | N | … | … | … | … | n | … |
| 46467 | 15561931-5718134 | -68.3 ± 1.3 | 4635 ± 135 | 3.14 ± 0.49 | -0.21 ± 0.16 | … | … | N | … | … | … | … | n | … |
| 46233 | 15553075-5726332 | -43.1 ± 0.3 | 5714 ± 271 | 4.19 ± 0.19 | 0.28 ± 0.15 | … | … | N | … | … | … | … | n | … |
| 46118 | 15561932-5723439 | 9.1 ± 0.2 | … | … | … | … | … | N | … | … | … | … | n | … |

**Table C.14.** continued.

| ID | CNAME | RV (km s$^{-1}$) | $T_{\text{eff}}$ (K) | logg (dex) | [Fe/H] (dex) | EW(Li)$^a$ (mÅ) | EW(Li) error flag$^b$ | Membership RV | Li | logg | [Fe/H] | Gaia study Cantat-Gaudin$^c$ | Final$^d$ | NMs with Li$^e$ |
|---|---|---|---|---|---|---|---|---|---|---|---|---|---|---|
| 46234 | 15553082-5724527 | -26.6 ± 0.4 | 6059 ± 246 | 4.54 ± 0.38 | 0.14 ± 0.22 | 63 ± 36 | 1 | Y | Y | Y | Y | ... | Y | ... |
| 46468 | 15561957-5728048 | -38.4 ± 0.8 | 5677 ± 563 | 4.55 ± 0.32 | -0.85 ± 0.22 | ... | ... | N | ... | ... | ... | ... | n | ... |
| 46469 | 15561980-5724528 | -91.8 ± 3.6 | 6647 ± 859 | 4.27 ± 0.50 | -1.02 ± 0.11 | ... | ... | N | ... | ... | ... | ... | n | ... |
| 45965 | 15553086-5724597 | 16.1 ± 0.4 | 5917 ± 149 | 4.55 ± 0.05 | 0.71 ± 0.23 | ... | ... | N | ... | ... | ... | ... | n | ... |
| 46470 | 15561997-5719079 | -29.1 ± 0.7 | 5218 ± 302 | 4.05 ± 0.90 | -0.29 ± 0.15 | ... | ... | Y | ... | ... | ... | ... | n | ... |
| 46235 | 15553102-5728582 | -83.7 ± 0.6 | 5776 ± 300 | 4.40 ± 0.35 | 0.28 ± 0.44 | <101 | 3 | N | ... | ... | ... | ... | n | NG |
| 46119 | 15562109-5735556 | -14.8 ± 0.1 | ... | ... | ... | ... | ... | N | ... | ... | ... | ... | n | ... |
| 46236 | 15553122-5729360 | -10.4 ± 0.4 | 5950 ± 237 | 3.78 ± 0.47 | 0.33 ± 0.17 | 137 ± 82 | 1 | N | ... | ... | ... | ... | n | NG |
| 46120 | 15562176-5731452 | -25.2 ± 0.3 | ... | ... | ... | ... | ... | Y | ... | ... | ... | ... | n | ... |
| 45966 | 15553129-5732091 | -57.7 ± 0.5 | ... | ... | ... | ... | ... | N | ... | ... | ... | N | n | ... |
| 46471 | 15562180-5724438 | -16.7 ± 0.7 | 5885 ± 682 | 4.45 ± 0.15 | -0.04 ± 0.75 | ... | ... | N | ... | ... | ... | ... | n | ... |
| 46237 | 15553139-5723088 | -118.5 ± 0.5 | 4897 ± 149 | 3.44 ± 0.87 | 0.21 ± 0.21 | ... | ... | N | ... | ... | ... | ... | n | G |
| 46121 | 15562218-5733492 | -30.9 ± 0.3 | ... | ... | ... | ... | ... | Y | ... | ... | ... | ... | n | ... |
| 46238 | 15553156-5729150 | -35.0 ± 0.5 | 5019 ± 211 | 3.66 ± 0.37 | -0.40 ± 0.15 | ... | ... | N | ... | ... | ... | ... | n | ... |
| 46472 | 15562253-5726172 | -8.2 ± 1.2 | 5410 ± 91 | 4.15 ± 0.36 | 0.14 ± 0.09 | ... | ... | N | ... | ... | ... | ... | n | ... |
| 46239 | 15553173-5725454 | -23.9 ± 0.6 | 5061 ± 185 | 4.90 ± 0.20 | 0.08 ± 0.44 | ... | ... | Y | ... | ... | ... | ... | n | ... |
| 46122 | 15562300-5733283 | -37.0 ± 0.1 | 4853 ± 72 | 3.44 ± 0.05 | -0.07 ± 0.15 | ... | ... | N | ... | ... | ... | ... | n | ... |
| 45967 | 15553187-5729419 | -18.6 ± 0.7 | ... | ... | ... | ... | ... | Y | ... | ... | ... | ... | n | ... |
| 46123 | 15562306-5725135 | -12.1 ± 0.1 | 5231 ± 89 | 4.53 ± 0.02 | -0.02 ± 0.19 | ... | ... | N | ... | ... | ... | ... | n | ... |
| 45968 | 15553204-5722474 | -38.1 ± 0.1 | ... | ... | ... | ... | ... | N | ... | ... | ... | ... | n | ... |
| 46124 | 15562315-5728125 | -39.5 ± 0.4 | ... | ... | ... | ... | ... | N | ... | ... | ... | ... | n | ... |
| 46240 | 15553265-5718100 | -25.0 ± 0.7 | 4778 ± 254 | 4.50 ± 1.01 | 0.03 ± 0.30 | ... | ... | Y | ... | ... | ... | ... | n | ... |
| 46125 | 15562397-5718484 | -45.4 ± 0.1 | 6043 ± 142 | 4.41 ± 0.09 | ... | ... | ... | N | ... | ... | ... | ... | n | ... |
| 46241 | 15553290-5728556 | -16.5 ± 0.5 | 5772 ± 133 | 4.16 ± 0.42 | -0.21 ± 0.18 | <54 | 3 | N | ... | ... | ... | ... | n | NG |
| 46126 | 15562468-5731093 | 73.8 ± 0.1 | 5997 ± 125 | 4.80 ± 0.09 | -0.17 ± 0.19 | ... | ... | N | ... | ... | ... | ... | n | ... |
| 45969 | 15553291-5735443 | -41.1 ± 0.1 | 5089 ± 98 | 3.87 ± 0.05 | -0.14 ± 0.20 | ... | ... | N | ... | ... | ... | ... | n | ... |
| 46127 | 15562478-5725550 | -19.1 ± 0.1 | ... | ... | ... | ... | ... | Y | ... | ... | ... | Y | n | ... |
| 3077 | 15553294-5725298 | -33.0 ± 0.6 | 4883 ± 122 | 2.80 ± 0.24 | 0.17 ± 0.10 | 7 ± 1 | ... | N | ... | ... | ... | N | n | ... |
| 46128 | 15562530-5725399 | -53.0 ± 0.1 | 5457 ± 149 | 4.01 ± 0.01 | -0.18 ± 0.16 | ... | ... | N | ... | ... | ... | ... | n | ... |
| 46242 | 15553303-5724508 | -57.3 ± 0.9 | 4889 ± 424 | 4.20 ± 0.22 | 0.09 ± 0.36 | ... | ... | N | ... | ... | ... | ... | n | ... |
| 46129 | 15562550-5728061 | -35.7 ± 0.1 | ... | ... | ... | ... | ... | N | ... | ... | ... | ... | n | ... |
| 45970 | 15553303-5728349 | -32.9 ± 0.2 | 6185 ± 123 | 3.35 ± 0.02 | ... | ... | ... | Y | ... | ... | ... | ... | n | ... |
| 46473 | 15562600-5723087 | -7.5 ± 1.0 | 5527 ± 141 | 4.81 ± 0.27 | 0.19 ± 0.10 | ... | ... | N | ... | ... | ... | ... | n | ... |
| 45971 | 15553304-5727150 | -39.2 ± 0.7 | ... | ... | ... | ... | ... | N | ... | ... | ... | ... | n | ... |
| 46130 | 15562605-5732079 | -77.3 ± 0.3 | ... | ... | ... | ... | ... | N | ... | ... | ... | ... | n | ... |
| 45972 | 15553305-5723250 | -85.4 ± 0.3 | ... | ... | ... | ... | ... | N | ... | ... | ... | ... | n | ... |
| 46131 | 15562665-5720018 | 6.5 ± 0.1 | 5699 ± 130 | 4.91 ± 0.05 | 0.48 ± 0.22 | ... | ... | N | ... | ... | ... | ... | n | ... |
| 46243 | 15553315-5725499 | 57.9 ± 0.4 | 5616 ± 270 | 4.42 ± 0.17 | 0.35 ± 0.20 | <59 | 3 | N | ... | ... | ... | N | n | NG |
| 46132 | 15562753-5723434 | -16.4 ± 0.2 | ... | ... | ... | ... | ... | N | ... | ... | ... | ... | n | ... |
| 45973 | 15553355-5728032 | -21.4 ± 0.1 | ... | ... | ... | ... | ... | Y | ... | ... | ... | ... | n | ... |
| 46133 | 15562774-5722190 | -12.0 ± 0.2 | ... | ... | ... | ... | ... | N | ... | ... | ... | ... | n | ... |
| 45974 | 15553367-5729364 | -54.3 ± 0.2 | 6463 ± 124 | 4.91 ± 0.05 | ... | ... | ... | N | ... | ... | ... | ... | n | ... |
| 45984 | 15553707-5723409 | -19.5 ± 0.2 | 6391 ± 150 | 4.71 ± 0.05 | ... | ... | ... | Y | ... | ... | ... | ... | n | ... |
| 46258 | 15553734-5726109 | -80.6 ± 1.0 | 5382 ± 251 | 3.88 ± 0.42 | 0.14 ± 0.30 | ... | ... | N | ... | ... | ... | ... | n | ... |
| 45985 | 15553743-5723159 | 15.7 ± 0.1 | ... | ... | ... | ... | ... | N | ... | ... | ... | ... | n | ... |
| 46259 | 15553784-5727442 | 5.0 ± 0.5 | 5861 ± 111 | 4.09 ± 0.31 | 0.05 ± 0.13 | <83 | 3 | N | ... | ... | ... | ... | n | NG |
| 46260 | 15553791-5725374 | 62.1 ± 1.3 | 5172 ± 128 | 3.48 ± 2.06 | -0.93 ± 0.23 | ... | ... | N | ... | ... | ... | ... | n | Li-rich G |
| 46261 | 15553791-5726522 | -36.9 ± 0.3 | 5351 ± 127 | 4.01 ± 0.62 | 0.00 ± 0.13 | 199 ± 55 | 1 | N | N | N | Y | ... | n | NG |
| 46262 | 15553799-5728254 | -96.8 ± 0.5 | 5219 ± 214 | 4.27 ± 0.22 | -0.64 ± 0.29 | <57 | 3 | N | ... | ... | ... | ... | n | NG |
| 46263 | 15553801-5718336 | -49.5 ± 1.2 | 5816 ± 571 | 4.54 ± 0.29 | 0.02 ± 0.36 | ... | ... | N | ... | ... | ... | ... | n | ... |
| 46264 | 15553801-5729487 | -33.5 ± 0.7 | 6032 ± 385 | 4.44 ± 0.28 | -0.15 ± 0.40 | ... | ... | N | ... | ... | ... | ... | n | ... |
| 45986 | 15553808-5726083 | -103.9 ± 5.6 | ... | ... | ... | ... | ... | N | ... | ... | ... | ... | n | ... |
| 45987 | 15553826-5724484 | -24.0 ± 2.1 | ... | ... | ... | ... | ... | Y | ... | ... | ... | Y | n | ... |
| 46265 | 15553831-5723116 | -45.6 ± 0.5 | 5818 ± 498 | 5.11 ± 0.41 | 0.26 ± 0.49 | ... | ... | N | ... | ... | ... | ... | n | ... |
| 45988 | 15553835-5734517 | -38.3 ± 0.3 | 6105 ± 138 | 3.40 ± 0.05 | ... | ... | ... | N | ... | ... | ... | ... | n | ... |
| 46266 | 15553839-5727322 | -25.0 ± 0.3 | 6477 ± 219 | 4.19 ± 0.22 | 0.26 ± 0.24 | ... | ... | Y | ... | ... | ... | Y | n | ... |
| 46267 | 15553842-5730392 | -20.6 ± 0.6 | 5617 ± 34 | 3.99 ± 0.65 | 0.53 ± 0.34 | ... | ... | Y | ... | ... | ... | ... | n | ... |
| 45989 | 15553843-5727380 | -25.5 ± 0.3 | ... | ... | ... | ... | ... | Y | ... | ... | ... | Y | n | ... |
| 45990 | 15553848-5723545 | -25.0 ± 1.1 | ... | ... | ... | ... | ... | Y | ... | ... | ... | Y | n | ... |

**Table C.14.** continued.

| ID | CNAME | RV (km s$^{-1}$) | $T_{\rm eff}$ (K) | $logg$ (dex) | [Fe/H] (dex) | EW(Li)[a] (mÅ) | EW(Li) error flag[b] | Membership RV | Li | $logg$ | [Fe/H] | Gaia study Cantat-Gaudin[c] | Final[d] | NMs with Li[e] |
|---|---|---|---|---|---|---|---|---|---|---|---|---|---|---|
| 46268 | 15553854-5724181 | -14.2 ± 0.7 | 6076 ± 415 | 4.20 ± 0.03 | 0.24 ± 0.61 | … | … | N | … | … | … | … | n | … |
| 45992 | 15553879-5725479 | -39.3 ± 0.3 | 6673 ± 150 | 4.91 ± 0.05 | … | … | … | N | … | … | … | … | n | … |
| 45993 | 15553884-5730114 | 10.9 ± 0.3 | 6017 ± 128 | 3.79 ± 0.03 | … | … | … | N | … | … | … | … | n | … |
| 46269 | 15553888-5728572 | -25.5 ± 0.7 | 5950 ± 220 | 4.03 ± 0.25 | -0.17 ± 0.28 | <70 | 3 | Y | Y | Y | Y | … | Y | … |
| 46270 | 15553906-5728096 | -23.3 ± 0.6 | 5167 ± 178 | 4.74 ± 0.17 | 0.16 ± 0.19 | … | … | Y | … | … | … | … | n | … |
| 46271 | 15553951-5729430 | -16.8 ± 0.5 | 5772 ± 79 | 4.06 ± 0.33 | 0.16 ± 0.21 | … | … | N | … | … | … | … | n | … |
| 3083 | 15554609-5723481 | -22.7 ± 0.6 | 4924 ± 125 | 2.83 ± 0.24 | 0.15 ± 0.11 | 15 ± 1 | … | Y | Y | Y | Y | N | Y | … |
| 46302 | 15554632-5724193 | -70.4 ± 0.5 | 5408 ± 47 | 3.80 ± 0.57 | -0.17 ± 0.21 | … | … | N | … | … | … | … | n | … |
| 46005 | 15554637-5731091 | -24.7 ± 0.1 | 5119 ± 74 | 3.68 ± 0.05 | 0.01 ± 0.16 | … | … | Y | … | … | … | Y | n | … |
| 46303 | 15554643-5727030 | 27.9 ± 0.4 | 5593 ± 17 | 4.07 ± 0.30 | 0.21 ± 0.18 | … | … | N | … | … | … | … | n | … |
| 46006 | 15554643-5728009 | -20.1 ± 0.1 | … | … | … | … | … | Y | … | … | … | Y | n | … |
| 46304 | 15554646-5726400 | -28.5 ± 0.7 | 5908 ± 7 | 4.49 ± 0.27 | -0.05 ± 0.23 | … | … | Y | … | … | … | … | n | … |
| 46007 | 15554647-5732044 | -28.8 ± 0.4 | … | … | … | … | … | Y | … | … | … | … | n | … |
| 46008 | 15554653-5728229 | -97.3 ± 0.3 | 6021 ± 149 | 3.82 ± 0.04 | … | … | … | N | … | … | … | … | n | … |
| 46305 | 15554659-5722517 | -61.2 ± 0.4 | 6167 ± 288 | 4.54 ± 0.51 | 0.15 ± 0.23 | <57 | 3 | N | … | … | … | … | n | NG |
| 46306 | 15554662-5728500 | -80.3 ± 0.8 | 5037 ± 435 | 4.33 ± 0.57 | -0.06 ± 0.31 | … | … | N | … | … | … | … | n | … |
| 46307 | 15554669-5725439 | -46.2 ± 0.7 | 5513 ± 57 | 4.51 ± 0.60 | 0.03 ± 0.19 | … | … | N | … | … | … | … | n | … |
| 46308 | 15554677-5724280 | -28.9 ± 0.6 | 5832 ± 362 | 4.52 ± 0.41 | -0.17 ± 0.46 | <48 | 3 | Y | Y | Y | Y | … | Y | … |
| 46309 | 15554698-5725268 | -32.5 ± 0.6 | 4934 ± 340 | 4.33 ± 0.20 | 0.07 ± 0.18 | … | … | Y | … | … | … | … | n | … |
| 46310 | 15554702-5727465 | -123.7 ± 0.9 | 5108 ± 195 | 3.65 ± 0.51 | -0.32 ± 0.36 | … | … | N | … | … | … | … | n | … |
| 46010 | 15554735-5722597 | -36.5 ± 0.8 | … | … | … | … | … | N | … | … | … | N | n | … |
| 46011 | 15554736-5725094 | -26.4 ± 0.4 | … | … | … | … | … | Y | … | … | … | N | n | … |
| 46311 | 15554740-5726381 | -74.1 ± 0.4 | 5564 ± 246 | 3.92 ± 0.04 | 0.18 ± 0.23 | … | … | N | … | … | … | … | n | … |
| 46312 | 15554754-5725467 | -22.9 ± 0.5 | 5699 ± 507 | 4.79 ± 0.56 | 0.30 ± 0.31 | … | … | Y | … | … | … | … | n | … |
| 46313 | 15554758-5728154 | -37.1 ± 0.8 | 5823 ± 45 | 3.96 ± 0.32 | -0.18 ± 0.27 | 112 ± 71 | 1 | N | N | Y | N | … | n | NG |
| 46314 | 15554778-5725353 | -23.9 ± 0.6 | 5682 ± 362 | 4.21 ± 0.74 | 0.19 ± 0.22 | … | … | Y | … | … | … | … | n | … |
| 46315 | 15554783-5727058 | -36.6 ± 0.6 | 5591 ± 90 | 4.08 ± 0.22 | 0.02 ± 0.17 | … | … | N | … | … | … | … | n | … |
| 46316 | 15554786-5727569 | -67.3 ± 0.5 | 5937 ± 292 | 4.19 ± 0.54 | 0.11 ± 0.23 | 64 ± 56 | 1 | N | … | … | … | … | n | NG |
| 46317 | 15554791-5725221 | -58.0 ± 0.4 | 6047 ± 345 | 3.97 ± 0.24 | 0.07 ± 0.16 | 88 ± 45 | 1 | N | … | … | … | … | n | NG |
| 46019 | 15555041-5729053 | -22.2 ± 2.0 | … | … | … | … | … | Y | … | … | … | Y | n | … |
| 46335 | 15555042-5727043 | -101.8 ± 0.4 | 4787 ± 243 | 3.16 ± 0.33 | -0.59 ± 0.20 | … | … | N | … | … | … | … | n | G |
| 3084 | 15555069-5726255 | -28.0 ± 0.6 | 4826 ± 110 | 2.66 ± 0.22 | 0.11 ± 0.10 | 8 ± 1 | … | Y | Y | Y | Y | … | Y | … |
| 46336 | 15555073-5730183 | -14.6 ± 0.6 | 5885 ± 241 | 4.38 ± 0.20 | 0.43 ± 0.30 | … | … | N | … | … | … | … | n | … |
| 46337 | 15555079-5724312 | -37.4 ± 0.9 | 5062 ± 265 | 4.20 ± 0.69 | 0.19 ± 0.30 | … | … | N | … | … | … | … | n | … |
| 46338 | 15555081-5724506 | -35.0 ± 0.7 | 4751 ± 246 | 4.59 ± 0.28 | 0.36 ± 0.38 | … | … | N | … | … | … | … | n | … |
| 46339 | 15555081-5727454 | 2.6 ± 0.7 | 5554 ± 326 | 4.55 ± 1.16 | -0.18 ± 0.37 | … | … | N | … | … | … | … | n | … |
| 46020 | 15555082-5729452 | -64.9 ± 0.6 | … | … | … | … | … | N | … | … | … | … | n | … |
| 46021 | 15555119-5729284 | -26.4 ± 0.3 | … | … | … | … | … | Y | … | … | … | … | n | … |
| 46022 | 15555128-5725272 | 25.4 ± 0.3 | 4571 ± 72 | 4.86 ± 0.20 | -0.20 ± 0.18 | 44 ± 38 | … | N | … | … | … | … | n | NG |
| 46340 | 15555129-5730105 | -51.8 ± 0.7 | 5809 ± 468 | 4.62 ± 0.66 | -0.29 ± 0.44 | … | … | N | … | … | … | … | n | … |
| 46023 | 15555132-5723581 | -15.1 ± 0.1 | … | … | … | … | … | N | … | … | … | Y | n | … |
| 46341 | 15555135-5726020 | -23.9 ± 0.5 | 5348 ± 179 | 4.20 ± 1.07 | 0.22 ± 0.19 | … | … | Y | … | … | … | … | n | … |
| 46342 | 15555143-5729027 | -23.9 ± 0.6 | 6402 ± 312 | 4.17 ± 0.33 | 0.06 ± 0.25 | <38 | 3 | Y | Y | Y | Y | Y | Y | … |
| 46343 | 15555161-5724591 | -27.8 ± 0.3 | 6349 ± 190 | 4.13 ± 0.25 | 0.06 ± 0.19 | … | … | Y | … | … | … | … | n | … |
| 3085 | 15555162-5724181 | -22.9 ± 0.6 | 4880 ± 123 | 2.85 ± 0.25 | 0.21 ± 0.10 | 6 ± 1 | … | Y | Y | Y | Y | … | Y | … |
| 46024 | 15555162-5727226 | -40.5 ± 1.0 | … | … | … | … | … | N | … | … | … | … | n | … |
| 3086 | 15555171-5725390 | -21.3 ± 0.6 | 4893 ± 113 | 2.70 ± 0.24 | 0.14 ± 0.10 | 10 ± 1 | … | Y | Y | Y | Y | … | Y | … |
| 46344 | 15555175-5725248 | -27.0 ± 0.3 | 5980 ± 202 | 4.59 ± 0.23 | -0.04 ± 0.35 | 72 ± 27 | 1 | Y | Y | Y | Y | Y | Y | … |
| 46345 | 15555178-5722295 | -25.1 ± 0.4 | 5958 ± 252 | 4.38 ± 0.37 | 0.00 ± 0.21 | 86 ± 45 | 1 | Y | Y | Y | Y | … | Y | … |
| 46346 | 15555182-5723524 | -43.7 ± 0.3 | 5861 ± 47 | 4.14 ± 0.42 | 0.11 ± 0.13 | 63 ± 60 | 1 | N | … | … | … | … | n | NG |
| 46347 | 15555190-5729024 | -74.0 ± 0.4 | 5544 ± 171 | 4.61 ± 0.44 | 0.15 ± 0.19 | … | … | N | … | … | … | … | n | … |
| 46348 | 15555196-5725080 | -94.4 ± 0.5 | 5206 ± 220 | 3.72 ± 0.18 | -0.04 ± 0.38 | … | … | N | … | … | … | … | n | … |
| 46025 | 15555197-5730065 | -19.5 ± 0.1 | … | … | … | … | … | Y | … | … | … | N | n | … |
| 46377 | 15555700-5726139 | -31.3 ± 0.4 | 5815 ± 157 | 4.42 ± 0.45 | 0.29 ± 0.16 | … | … | Y | … | … | … | … | n | … |
| 46045 | 15555707-5733035 | -63.8 ± 0.3 | 5813 ± 259 | 3.77 ± 0.18 | 0.01 ± 0.24 | … | … | N | … | … | … | … | n | … |
| 46046 | 15555723-5732423 | 49.1 ± 0.1 | 5409 ± 108 | 3.67 ± 0.05 | -0.29 ± 0.19 | … | … | N | … | … | … | … | n | … |
| 46378 | 15555761-5722421 | 8.8 ± 0.3 | 4865 ± 195 | 4.71 ± 0.33 | 0.04 ± 0.13 | … | … | N | … | … | … | … | n | … |
| 46047 | 15555769-5726343 | -26.2 ± 0.3 | … | … | … | … | … | Y | … | … | … | … | n | … |
| 46048 | 15555789-5718128 | -22.5 ± 0.1 | 4425 ± 101 | 2.21 ± 0.18 | 0.01 ± 0.18 | … | … | Y | … | … | … | … | n | … |

**Table C.14.** continued.

| ID | CNAME | RV (km s$^{-1}$) | $T_{\rm eff}$ (K) | $\log g$ (dex) | [Fe/H] (dex) | $EW$(Li)$^a$ (mÅ) | $EW$(Li) error flag$^b$ | Membership RV | Li | $\log g$ | [Fe/H] | Gaia study Cantat-Gaudin$^c$ | Final$^d$ | NMs with Li$^e$ |
|---|---|---|---|---|---|---|---|---|---|---|---|---|---|---|
| 46049 | 15555791-5725589 | -73.3 ± 0.1 | 6237 ± 111 | 4.11 ± 0.01 | … | … | … | N | … | … | … | … | n | … |
| 46379 | 15555823-5725104 | -43.1 ± 0.6 | 5154 ± 110 | 4.16 ± 0.15 | 0.05 ± 0.21 | … | … | N | … | … | … | … | n | … |
| 46050 | 15555824-5719227 | -17.1 ± 0.1 | … | … | … | … | … | Y | … | … | … | … | n | … |
| 46380 | 15555831-5726020 | 9.9 ± 0.5 | 4785 ± 150 | 3.71 ± 0.73 | 0.34 ± 0.42 | … | … | N | … | … | … | … | n | … |
| 46051 | 15555834-5735358 | -21.8 ± 0.3 | … | … | … | … | … | Y | … | … | … | … | n | … |
| 46052 | 15555842-5722592 | -50.7 ± 3.0 | … | … | … | … | … | N | … | … | … | … | n | … |
| 46053 | 15555847-5731401 | -27.0 ± 0.4 | … | … | … | … | … | Y | … | … | … | … | n | … |
| 46381 | 15555850-5723449 | -47.5 ± 0.5 | 5898 ± 373 | 3.38 ± 0.62 | 0.19 ± 0.44 | … | … | N | … | … | … | … | n | … |
| 46382 | 15555871-5731303 | -72.2 ± 0.5 | 5830 ± 423 | 4.51 ± 0.32 | -0.28 ± 0.58 | … | … | N | … | … | … | … | n | … |
| 46383 | 15555874-5722306 | -32.5 ± 0.9 | 5456 ± 109 | 4.08 ± 0.09 | -0.12 ± 0.15 | … | … | Y | … | … | … | … | n | … |
| 46054 | 15555879-5722532 | -7.7 ± 0.2 | 6253 ± 138 | 4.76 ± 0.05 | … | … | … | N | … | … | … | … | n | … |
| 46384 | 15555885-5726145 | -45.4 ± 0.9 | 4685 ± 459 | 3.31 ± 1.86 | -0.73 ± 0.32 | … | … | N | … | … | … | … | n | G |
| 46385 | 15555888-5727294 | -43.1 ± 0.6 | 5880 ± 409 | 4.89 ± 0.40 | 0.12 ± 0.17 | … | … | N | … | … | … | … | n | … |
| 46386 | 15555889-5727251 | -14.4 ± 0.6 | 6769 ± 343 | 4.07 ± 0.07 | -0.01 ± 0.19 | 58 ± 26 | 1 | N | … | … | … | … | n | NG |
| 46387 | 15555906-5720153 | -45.9 ± 0.7 | 5812 ± 95 | 3.97 ± 0.10 | -0.02 ± 0.21 | … | … | N | … | … | … | … | n | … |
| 46388 | 15555906-5724013 | -54.7 ± 0.7 | 5469 ± 268 | 4.79 ± 0.47 | -0.17 ± 0.21 | … | … | N | … | … | … | … | n | … |
| 46389 | 15555910-5730373 | -42.9 ± 0.8 | 5355 ± 286 | 3.12 ± 0.24 | 0.04 ± 0.19 | … | … | N | … | … | … | … | n | … |
| 46390 | 15555915-5724266 | -5.8 ± 0.4 | 5661 ± 270 | 4.47 ± 0.38 | 0.25 ± 0.13 | … | … | N | … | … | … | … | n | … |
| 46404 | 15560175-5731423 | -71.7 ± 0.5 | 5274 ± 80 | 3.94 ± 0.09 | 0.04 ± 0.19 | … | … | N | … | … | … | … | n | … |
| 46066 | 15560193-5731535 | -27.7 ± 0.3 | … | … | … | … | … | Y | … | … | … | … | n | … |
| 3088 | 15560208-5727418 | -24.7 ± 0.6 | 4943 ± 119 | 2.92 ± 0.23 | 0.18 ± 0.09 | 24 ± 1 | … | Y | Y | Y | Y | Y | Y | … |

**Notes.** $^{(a)}$ The values of $EW$(Li) for this cluster are corrected (subtracted adjacent Fe (6707.43 Å) line). $^{(b)}$ Flags for the errors of the corrected $EW$(Li) values, as follows: 1=$EW$(Li) corrected by blends contribution using models; and 3=Upper limit (no error for $EW$(Li) is given). $^{(c)}$ Cantat-Gaudin et al. (2018). $^{(d)}$ The letters "Y" and "N" indicate if the star is a cluster member or not. $^{(e)}$ 'Li-rich G', 'G' and 'NG' indicate "Li-rich giant", "giant" and "non-giant" Li field outliers, respectively.

**Table C.15.** NGC 6802

| ID | CNAME | RV (km s$^{-1}$) | $T_{\rm eff}$ (K) | logg (dex) | [Fe/H] (dex) | EW(Li)$^a$ (mÅ) | EW(Li) error flag$^b$ | Membership RV | Li | logg | [Fe/H] | Gaia study Cantat-Gaudin$^c$ | Final$^d$ | NMs with Li$^e$ |
|---|---|---|---|---|---|---|---|---|---|---|---|---|---|---|
| 51167 | 19302635+2013343 | 16.1 ± 2.0 | 6624 ± 383 | 3.81 ± 0.21 | -0.05 ± 0.29 | ... | ... | Y | ... | ... | ... | ... | n | ... |
| 51059 | 19302638+2017520 | 11.2 ± 0.6 | ... | ... | ... | ... | ... | Y | ... | ... | ... | ... | n | ... |
| 51060 | 19302643+2014385 | 19.3 ± 4.0 | ... | ... | ... | ... | ... | N | ... | ... | ... | ... | n | ... |
| 51168 | 19302647+2011358 | 17.0 ± 1.6 | 6543 ± 102 | 4.16 ± 0.17 | -0.02 ± 0.35 | ... | ... | N | ... | ... | ... | ... | n | ... |
| 51169 | 19302662+2018112 | 14.8 ± 0.8 | 6270 ± 345 | 3.85 ± 0.37 | -0.15 ± 0.17 | <73 | 3 | Y | Y | Y | Y | Y | Y | ... |
| 51061 | 19302700+2016226 | 13.6 ± 2.2 | ... | ... | ... | ... | ... | Y | ... | ... | ... | Y | n | ... |
| 51170 | 19302749+2015434 | 12.2 ± 1.1 | 6814 ± 452 | 4.15 ± 0.21 | 0.20 ± 0.24 | ... | ... | Y | ... | ... | ... | Y | n | ... |
| 51171 | 19302781+2014295 | 12.8 ± 0.5 | 6201 ± 226 | 5.11 ± 0.71 | 0.01 ± 0.15 | 167 ± 83 | 1 | Y | Y | Y | Y | Y | Y | ... |
| 51062 | 19302826+2014367 | 14.3 ± 0.1 | ... | ... | ... | ... | ... | Y | ... | ... | ... | Y | n | ... |
| 51063 | 19302866+2012367 | 7.4 ± 3.6 | ... | ... | ... | ... | ... | N | ... | ... | ... | N | n | ... |
| 51064 | 19302898+2016123 | 65.9 ± 3.4 | ... | ... | ... | ... | ... | N | ... | ... | ... | Y | n | ... |
| 51109 | 19303852+2015228 | 26.9 ± 5.4 | ... | ... | ... | ... | ... | N | ... | ... | ... | Y | n | ... |
| 51201 | 19303856+2015159 | 17.8 ± 1.0 | 6653 ± 558 | 4.79 ± 1.01 | 0.00 ± 0.19 | <51 | 3 | N | Y | Y | Y | ... | Y | ... |
| 51110 | 19303858+2012324 | 10.1 ± 6.9 | ... | ... | ... | ... | ... | Y | ... | ... | ... | ... | n | ... |
| 51111 | 19303872+2013089 | 16.7 ± 0.5 | ... | ... | ... | ... | ... | N | ... | ... | ... | ... | n | ... |
| 51202 | 19303882+2016556 | 11.9 ± 0.5 | 6441 ± 194 | 4.49 ± 0.47 | 0.00 ± 0.17 | ... | ... | Y | ... | ... | ... | ... | n | ... |
| 3291 | 19303884+2014005 | 13.5 ± 0.6 | 5065 ± 119 | 2.84 ± 0.25 | 0.12 ± 0.10 | 14 ± 1 | ... | Y | Y | Y | Y | ... | Y | ... |
| 51203 | 19303884+2018153 | 17.0 ± 0.9 | 6583 ± 359 | 4.30 ± 0.24 | 0.15 ± 0.19 | ... | ... | N | ... | ... | ... | ... | n | ... |
| 51204 | 19303909+2017295 | 21.4 ± 0.3 | 6220 ± 142 | 4.40 ± 0.32 | 0.05 ± 0.15 | ... | ... | N | ... | ... | ... | ... | n | ... |
| 51205 | 19303915+2016454 | 10.8 ± 0.6 | 5793 ± 60 | 4.23 ± 0.36 | -0.21 ± 0.14 | 108 ± 76 | 1 | Y | Y | Y | Y | ... | Y | ... |
| 51112 | 19303922+2015167 | 4.7 ± 2.2 | ... | ... | ... | ... | ... | N | ... | ... | ... | ... | n | ... |
| 51113 | 19303936+2016163 | -13.0 ± 0.2 | ... | ... | ... | ... | ... | N | ... | ... | ... | ... | n | ... |
| 51206 | 19303940+2017171 | 18.1 ± 0.6 | 6072 ± 162 | 4.79 ± 0.52 | -0.20 ± 0.27 | 100 ± 52 | 1 | N | ... | ... | ... | ... | n | NG |
| 3292 | 19303943+2015237 | 10.4 ± 0.6 | 4948 ± 120 | 2.63 ± 0.24 | 0.10 ± 0.11 | 24 ± 1 | ... | Y | Y | Y | Y | ... | Y | ... |
| 51207 | 19303959+2016139 | 15.1 ± 0.5 | 6914 ± 447 | 4.58 ± 0.45 | 0.10 ± 0.24 | ... | ... | Y | ... | ... | ... | ... | n | ... |
| 3293 | 19303970+2013474 | 24.1 ± 0.6 | 5143 ± 127 | 2.33 ± 0.24 | 0.04 ± 0.10 | 31 ± 1 | ... | N | ... | ... | ... | ... | n | Li-rich G |
| 51114 | 19303972+2014560 | 12.1 ± 1.3 | ... | ... | ... | ... | ... | Y | ... | ... | ... | ... | n | ... |
| 51209 | 19303999+2013253 | 16.3 ± 1.0 | 6804 ± 480 | 4.35 ± 0.34 | 0.07 ± 0.21 | ... | ... | Y | ... | ... | ... | ... | n | ... |
| 51115 | 19304006+2017524 | 10.8 ± 8.4 | ... | ... | ... | ... | ... | Y | ... | ... | ... | ... | n | ... |
| 51116 | 19304007+2018391 | 8.1 ± 0.4 | ... | ... | ... | ... | ... | N | ... | ... | ... | ... | n | ... |
| 51214 | 19304370+2013100 | 30.1 ± 0.5 | 7170 ± 704 | 4.10 ± 0.19 | 0.01 ± 0.34 | 42 ± 37 | 1 | N | ... | ... | ... | Y | n | NG |
| 51132 | 19304381+2015026 | 12.7 ± 5.5 | ... | ... | ... | ... | ... | Y | ... | ... | ... | ... | n | ... |
| 51143 | 19304845+2012006 | -24.0 ± 0.2 | ... | ... | ... | ... | ... | N | ... | ... | ... | ... | n | ... |
| 51222 | 19304849+2018587 | 14.6 ± 1.3 | 6325 ± 504 | 4.96 ± 0.72 | -0.23 ± 0.39 | ... | ... | Y | ... | ... | ... | ... | n | ... |
| 51223 | 19304862+2016316 | 11.9 ± 1.3 | 6166 ± 185 | ... | -0.23 ± 0.16 | ... | ... | Y | ... | ... | ... | ... | n | ... |
| 51144 | 19304888+2013182 | 14.0 ± 2.1 | ... | ... | ... | ... | ... | Y | ... | ... | ... | ... | n | ... |
| 51150 | 19305211+2014411 | 19.6 ± 1.0 | ... | ... | ... | ... | ... | N | ... | ... | ... | ... | n | ... |
| 51145 | 19304961+2013488 | 26.1 ± 2.3 | ... | ... | ... | ... | ... | N | ... | ... | ... | ... | n | ... |
| 51228 | 19305232+2013230 | 8.9 ± 3.9 | 7101 ± 253 | ... | ... | ... | ... | N | ... | ... | ... | ... | n | ... |
| 51151 | 19305294+2015282 | 17.3 ± 4.7 | ... | ... | ... | ... | ... | N | ... | ... | ... | ... | n | ... |
| 51146 | 19304985+2012423 | 28.3 ± 1.3 | ... | ... | ... | ... | ... | N | ... | ... | ... | ... | n | ... |
| 51147 | 19304994+2015252 | 6.5 ± 6.6 | ... | ... | ... | ... | ... | N | ... | ... | ... | ... | n | ... |
| 51224 | 19305047+2014362 | 12.8 ± 0.5 | 6096 ± 76 | 5.22 ± 0.44 | -0.18 ± 0.25 | <49 | 3 | Y | Y | N | Y | Y | Y | ... |
| 51225 | 19305049+2012527 | -4.4 ± 0.5 | 5919 ± 721 | 4.04 ± 0.33 | -0.08 ± 0.60 | ... | ... | N | ... | ... | ... | ... | n | ... |
| 51148 | 19305091+2013113 | 23.1 ± 1.6 | 6546 ± 152 | 5.20 ± 0.36 | -0.04 ± 0.13 | ... | ... | N | ... | ... | ... | ... | n | ... |
| 51149 | 19305131+2016301 | 11.1 ± 5.8 | ... | ... | ... | ... | ... | Y | ... | ... | ... | ... | n | ... |
| 51226 | 19305143+2016240 | 12.4 ± 0.5 | 6311 ± 174 | 4.51 ± 0.45 | -0.07 ± 0.16 | ... | ... | Y | ... | ... | ... | ... | n | ... |
| 51227 | 19305159+2017413 | 14.6 ± 1.3 | 6888 ± 564 | 4.74 ± 0.73 | 0.02 ± 0.19 | <33 | 3 | Y | Y | Y | Y | Y | Y | ... |
| 51046 | 19301861+2015280 | -1.4 ± 0.8 | ... | ... | ... | ... | ... | N | ... | ... | ... | ... | n | ... |
| 51048 | 19301909+2013571 | 21.6 ± 0.2 | ... | ... | ... | ... | ... | N | ... | ... | ... | ... | n | ... |
| 51154 | 19301932+2015503 | 21.3 ± 0.7 | 6785 ± 235 | 4.44 ± 0.34 | 0.08 ± 0.22 | ... | ... | N | ... | ... | ... | ... | n | ... |
| 51049 | 19301985+2013424 | 18.4 ± 1.7 | ... | ... | ... | ... | ... | N | ... | ... | ... | ... | n | ... |
| 51050 | 19302001+2013368 | -1.8 ± 0.4 | 6334 ± 227 | 4.23 ± 0.04 | 0.01 ± 0.27 | ... | ... | N | ... | ... | ... | ... | n | ... |
| 51155 | 19302019+2016175 | 14.9 ± 1.9 | 6396 ± 275 | 4.52 ± 0.27 | -0.20 ± 0.25 | ... | ... | Y | ... | ... | ... | ... | n | ... |
| 51051 | 19302064+2016238 | 23.3 ± 6.7 | ... | ... | ... | ... | ... | N | ... | ... | ... | ... | n | ... |
| 51053 | 19302120+2018430 | 62.0 ± 21.0 | ... | ... | ... | ... | ... | N | ... | ... | ... | ... | n | ... |
| 51156 | 19302141+2014181 | 13.7 ± 1.2 | 7110 ± 601 | 4.10 ± 0.19 | -0.16 ± 0.47 | ... | ... | Y | ... | ... | ... | ... | n | ... |
| 51173 | 19302920+2015317 | 13.3 ± 0.8 | 6789 ± 691 | 4.41 ± 0.36 | 0.04 ± 0.19 | ... | ... | Y | ... | ... | ... | ... | n | ... |
| 51065 | 19302968+2016286 | 19.4 ± 4.4 | ... | ... | ... | ... | ... | N | ... | ... | ... | ... | n | ... |

| ID | CNAME | RV (km s$^{-1}$) | $T_{\text{eff}}$ (K) | $logg$ (dex) | [Fe/H] (dex) | EW(Li)$^a$ (mÅ) | EW(Li) error flag$^b$ | Membership RV | Li | logg | [Fe/H] | Gaia study Cantat-Gaudin$^c$ | Final$^d$ | NMs with Li$^e$ |
|---|---|---|---|---|---|---|---|---|---|---|---|---|---|---|
| 51174 | 19302998+2017491 | 61.3 ± 0.3 | 5324 ± 87 | 4.47 ± 0.26 | 0.20 ± 0.14 | … | … | N | … | … | … | … | n | … |
| 51066 | 19303010+2016182 | 20.0 ± 7.0 | … | … | … | … | … | N | … | … | … | … | n | … |
| 51067 | 19303023+2013299 | 12.0 ± 1.0 | … | … | … | … | … | Y | … | … | … | … | n | … |
| 51068 | 19303031+2015283 | 10.8 ± 4.6 | … | … | … | … | … | Y | … | … | … | … | n | … |
| 51175 | 19303038+2018124 | 17.7 ± 0.5 | 6092 ± 226 | 3.99 ± 0.05 | 0.03 ± 0.19 | <71 | 3 | N | Y | Y | Y | … | Y | … |
| 3284 | 19303058+2013163 | 11.6 ± 0.6 | 5086 ± 114 | 2.81 ± 0.23 | 0.11 ± 0.10 | 18 ± 1 | … | Y | Y | Y | Y | … | Y | … |
| 3285 | 19303085+2016274 | 12.3 ± 0.6 | 5031 ± 156 | 2.80 ± 0.33 | 0.08 ± 0.12 | 11 ± 1 | … | Y | Y | Y | Y | … | Y | … |
| 51176 | 19303102+2019194 | -17.7 ± 0.4 | 5903 ± 88 | 3.89 ± 0.25 | -0.45 ± 0.18 | 59 ± 29 | 1 | N | … | … | … | … | n | NG |
| 51177 | 19303118+2018297 | 14.1 ± 0.9 | 6595 ± 312 | 4.16 ± 0.36 | -0.05 ± 0.13 | … | … | Y | … | … | … | … | n | … |
| 51178 | 19303119+2015328 | 2.9 ± 0.5 | 6241 ± 234 | 4.29 ± 0.18 | -0.02 ± 0.14 | 82 ± 56 | 1 | N | … | … | … | … | n | NG |
| 51179 | 19303120+2012585 | -34.9 ± 0.3 | 5864 ± 298 | 4.25 ± 0.30 | 0.08 ± 0.19 | … | … | N | … | … | … | … | n | … |
| 51181 | 19303161+2015557 | 13.1 ± 1.1 | 6287 ± 15 | 3.89 ± 0.15 | 0.07 ± 0.27 | … | … | Y | … | … | … | … | n | … |
| 3286 | 19303184+2014459 | 12.0 ± 0.6 | 5049 ± 119 | 2.92 ± 0.24 | 0.13 ± 0.10 | 40 ± 3 | … | Y | Y | Y | Y | … | Y | … |
| 51069 | 19303192+2014102 | 13.9 ± 4.4 | … | … | … | … | … | Y | … | … | … | … | n | … |
| 51070 | 19303199+2013154 | 18.6 ± 8.2 | … | … | … | … | … | N | … | … | … | … | n | … |
| 51189 | 19303409+2019228 | 14.2 ± 1.0 | 6462 ± 62 | 4.26 ± 0.24 | -0.03 ± 0.24 | … | … | Y | … | … | … | … | n | … |
| 51190 | 19303415+2017082 | 7.7 ± 3.0 | 6284 ± 402 | 4.65 ± 0.49 | 0.37 ± 0.76 | … | … | N | … | … | … | … | n | … |
| 51082 | 19303423+2015280 | 74.3 ± 1.7 | … | … | … | … | … | N | … | … | … | … | n | … |
| 51083 | 19303424+2015519 | 11.8 ± 2.2 | … | … | … | … | … | Y | … | … | … | … | n | … |
| 51084 | 19303429+2018218 | 14.4 ± 6.2 | … | … | … | … | … | Y | … | … | … | … | n | … |
| 51085 | 19303453+2017540 | 12.8 ± 4.9 | … | … | … | … | … | Y | … | … | … | … | n | … |
| 51086 | 19303468+2018370 | 14.0 ± 1.2 | 6843 ± 154 | 4.63 ± 0.32 | -0.04 ± 0.15 | … | … | Y | … | … | … | … | n | … |
| 51087 | 19303473+2011524 | -16.7 ± 0.2 | … | … | … | … | … | N | … | … | … | … | n | … |
| 51088 | 19303482+2016297 | -6.1 ± 0.3 | 5827 ± 275 | 4.46 ± 0.45 | -0.20 ± 0.18 | … | … | N | … | … | … | … | n | … |
| 51191 | 19303484+2011303 | 7.3 ± 1.1 | 7388 ± 132 | … | … | … | … | N | … | … | … | … | n | … |
| 51192 | 19303484+2015035 | 36.2 ± 5.2 | 7076 ± 246 | … | … | … | … | N | … | … | … | … | n | … |
| 51089 | 19303491+2014402 | 29.2 ± 1.2 | … | … | … | … | … | N | … | … | … | … | n | … |
| 51193 | 19303495+2013511 | -45.4 ± 0.4 | 5748 ± 310 | 3.84 ± 0.30 | -0.17 ± 0.19 | … | … | N | … | … | … | … | n | … |
| 51090 | 19303506+2016010 | 10.9 ± 5.9 | … | … | … | … | … | Y | … | … | … | … | n | … |
| 51092 | 19303550+2014324 | 17.0 ± 1.0 | … | … | … | … | … | N | … | … | … | … | n | … |
| 51093 | 19303556+2013438 | 14.1 ± 0.8 | … | … | … | … | … | Y | … | … | … | … | n | … |
| 51094 | 19303583+2014199 | 16.0 ± 2.9 | … | … | … | … | … | Y | … | … | … | … | n | … |
| 51095 | 19303602+2016541 | 76.3 ± 0.2 | … | … | … | … | … | N | … | … | … | … | n | … |
| 3289 | 19303611+2016329 | 12.5 ± 0.6 | 4965 ± 121 | 2.61 ± 0.23 | 0.10 ± 0.10 | 12 ± 1 | … | Y | Y | Y | Y | … | Y | … |
| 51117 | 19304027+2017431 | 14.8 ± 1.3 | … | … | … | … | … | Y | … | … | … | … | n | … |
| 51118 | 19304040+2016169 | 11.1 ± 0.6 | … | … | … | … | … | Y | … | … | … | … | n | … |
| 51119 | 19304050+2017259 | 13.4 ± 8.0 | … | … | … | … | … | Y | … | … | … | … | n | … |
| 51210 | 19304060+2015480 | 43.1 ± 0.6 | 5753 ± 124 | 4.67 ± 0.47 | -0.16 ± 0.40 | <68 | 3 | N | … | … | … | … | n | NG |
| 51120 | 19304061+2016266 | 84.1 ± 1.8 | … | … | … | … | … | N | … | … | … | … | n | … |
| 51121 | 19304102+2013596 | -19.0 ± 0.4 | 6397 ± 122 | 4.31 ± 0.52 | -0.01 ± 0.29 | … | … | N | … | … | … | … | n | … |
| 51122 | 19304126+2017142 | 14.6 ± 3.4 | … | … | … | … | … | Y | … | … | … | … | n | … |
| 51123 | 19304134+2016406 | 23.8 ± 13.0 | … | … | … | … | … | N | … | … | … | … | n | … |
| 51211 | 19304135+2018523 | 12.3 ± 0.9 | 6316 ± 165 | 4.80 ± 0.69 | -0.11 ± 0.14 | … | … | Y | … | … | … | … | n | … |
| 51124 | 19304152+2012503 | 15.5 ± 0.9 | 6969 ± 135 | 4.97 ± 0.27 | 0.08 ± 0.13 | … | … | Y | … | … | … | … | n | … |
| 3294 | 19304170+2015224 | 11.5 ± 0.6 | 5178 ± 123 | 3.13 ± 0.23 | 0.09 ± 0.10 | 29 ± 4 | … | Y | Y | Y | Y | … | Y | … |
| 51125 | 19304171+2014275 | 24.0 ± 6.7 | … | … | … | … | … | N | … | … | … | … | n | … |
| 51126 | 19304196+2017545 | 54.0 ± 0.1 | … | … | … | … | … | N | … | … | … | … | n | … |
| 51212 | 19304207+2019498 | 13.2 ± 1.1 | 6371 ± 266 | 4.74 ± 0.54 | -0.21 ± 0.45 | … | … | Y | … | … | … | … | n | … |
| 51128 | 19304213+2014462 | 20.2 ± 2.3 | … | … | … | … | … | N | … | … | … | … | n | … |
| 51129 | 19304233+2016522 | 13.2 ± 1.6 | … | … | … | … | … | Y | … | … | … | … | n | … |
| 51213 | 19304251+2017216 | 19.4 ± 1.5 | 6483 ± 160 | 4.57 ± 0.43 | 0.06 ± 0.17 | … | … | N | … | … | … | … | n | … |
| 3295 | 19304281+2016107 | 17.4 ± 0.6 | 4766 ± 121 | 2.63 ± 0.25 | -0.10 ± 0.10 | 273 ± 3 | … | N | N | N | Y | … | n | Li-rich G |
| 51130 | 19304314+2014539 | 17.9 ± 5.0 | … | … | … | … | … | R | … | … | … | … | n | … |
| 51196 | 19303670+2014290 | -29.4 ± 0.4 | 6787 ± 461 | 4.08 ± 0.05 | 0.12 ± 0.30 | … | … | N | … | … | … | … | n | … |
| 51097 | 19303688+2014542 | 19.9 ± 0.1 | … | … | … | … | … | N | … | … | … | … | n | … |
| 51098 | 19303688+2015069 | 14.5 ± 1.0 | 7146 ± 107 | … | … | … | … | Y | … | … | … | … | n | … |
| 51099 | 19303719+2018299 | 107.9 ± 2.0 | … | … | … | … | … | N | … | … | … | … | n | … |
| 51100 | 19303721+2014101 | 12.4 ± 2.3 | … | … | … | … | … | Y | … | … | … | … | n | … |
| 51197 | 19303727+2013208 | 48.3 ± 0.5 | 5974 ± 219 | 5.06 ± 0.57 | -0.12 ± 0.20 | <92 | 3 | N | … | … | … | … | n | NG |

**Table C.15.** continued.

| ID | CNAME | RV (km s$^{-1}$) | $T_{\rm eff}$ (K) | $\log g$ (dex) | [Fe/H] (dex) | EW(Li)$^a$ (mÅ) | EW(Li) error flag$^b$ | RV | Li | Membership logg | [Fe/H] | Gaia study Cantat-Gaudin$^c$ | Final$^d$ | NMs with Li$^e$ |
|---|---|---|---|---|---|---|---|---|---|---|---|---|---|---|
| 51101 | 19303731+2017014 | 11.8 ± 4.1 | … | … | … | … | … | Y | … | … | … | … | n | … |
| 51102 | 19303736+2016037 | 14.0 ± 5.4 | … | … | … | … | … | Y | … | … | … | … | n | … |
| 51103 | 19303752+2013486 | 14.9 ± 0.7 | … | … | … | … | … | Y | … | … | … | … | n | … |
| 51104 | 19303761+2012238 | 13.1 ± 1.7 | … | … | … | … | … | Y | … | … | … | … | n | … |
| 51105 | 19303764+2020140 | 9.6 ± 0.4 | 6645 ± 57 | 4.14 ± 0.14 | 1.02 ± 0.06 | 114 ± 35 | … | N | … | … | … | … | n | NG |
| 51198 | 19303769+2011020 | 34.5 ± 1.7 | 7320 ± 271 | … | … | <89 | 3 | N | … | … | … | … | n | … |
| 3290 | 19303773+2016196 | 16.6 ± 0.6 | 4984 ± 117 | 2.89 ± 0.22 | 0.00 ± 0.09 | 46 ± 3 | … | N | Y | Y | Y | … | n | Li-rich G |
| 51199 | 19303801+2015401 | 12.3 ± 0.4 | 6501 ± 191 | 4.92 ± 0.70 | -0.17 ± 0.25 | <54 | 3 | Y | Y | Y | Y | Y | Y | … |
| 51106 | 19303808+2013211 | 11.1 ± 2.7 | … | … | … | … | … | Y | … | … | … | … | n | … |
| 51107 | 19303814+2020033 | 11.2 ± 1.2 | … | … | … | … | … | Y | … | … | … | … | n | … |
| 51108 | 19303827+2011487 | -4.8 ± 1.0 | … | … | … | … | … | N | … | … | … | … | n | … |
| 51200 | 19303828+2013536 | 17.4 ± 1.8 | 7119 ± 895 | 3.79 ± 0.21 | 0.07 ± 0.13 | … | … | N | … | … | … | … | n | … |
| 51215 | 19304392+2017366 | 15.0 ± 0.8 | 6145 ± 317 | 4.19 ± 0.33 | -0.14 ± 0.63 | … | … | Y | … | … | … | … | n | … |
| 51133 | 19304409+2018102 | 14.5 ± 2.2 | … | … | … | … | … | Y | … | … | … | … | n | … |
| 51216 | 19304446+2012089 | -14.9 ± 0.5 | 5545 ± 104 | 3.67 ± 0.66 | -0.45 ± 0.19 | … | … | N | … | … | … | … | n | … |
| 51217 | 19304543+2016289 | 21.1 ± 0.9 | 5827 ± 8 | 4.41 ± 0.28 | -0.46 ± 0.15 | 111 ± 68 | 1 | N | … | … | … | N | n | NG |
| 51134 | 19304556+2015215 | 11.7 ± 0.2 | … | … | … | … | … | Y | … | … | … | … | n | … |
| 51135 | 19304587+2018011 | 18.4 ± 0.9 | … | … | … | … | … | N | … | … | … | … | n | … |
| 51136 | 19304596+2016103 | 12.9 ± 0.3 | … | … | … | … | … | Y | … | … | … | … | n | … |
| 51218 | 19304635+2014320 | -43.7 ± 0.8 | 5946 ± 468 | 3.67 ± 0.08 | -0.21 ± 0.54 | … | … | N | … | … | … | … | n | … |
| 3296 | 19304646+2015140 | 61.8 ± 0.6 | 4709 ± 122 | 2.90 ± 0.24 | 0.00 ± 0.10 | <6 | 3 | N | … | … | … | … | n | G |
| 51137 | 19304662+2013488 | 13.0 ± 0.1 | … | … | … | … | … | Y | … | … | … | … | n | … |
| 51219 | 19304663+2018473 | -43.0 ± 0.3 | 5559 ± 226 | 3.78 ± 0.44 | 0.16 ± 0.13 | <71 | 3 | N | … | … | … | … | n | NG |
| 51220 | 19304691+2016334 | 30.3 ± 1.5 | 7158 ± 695 | 4.13 ± 0.19 | 0.11 ± 0.21 | … | … | N | … | … | … | … | n | … |
| 51138 | 19304763+2013456 | 14.1 ± 2.3 | … | … | … | … | … | Y | … | … | … | … | n | … |
| 51139 | 19304786+2014323 | 14.6 ± 1.0 | … | … | … | … | … | Y | … | … | … | … | n | … |
| 51140 | 19304788+2018250 | 14.6 ± 2.4 | … | … | … | … | … | Y | … | … | … | … | n | … |
| 51141 | 19304796+2015407 | 10.9 ± 0.4 | … | … | … | … | … | Y | … | … | … | … | n | … |
| 51221 | 19304801+2012141 | -25.7 ± 0.7 | 6285 ± 233 | 4.79 ± 0.93 | -0.26 ± 0.53 | <74 | 3 | N | … | … | … | … | n | NG |
| 51142 | 19304841+2018458 | 15.0 ± 1.4 | … | … | … | … | … | Y | … | … | … | … | n | … |
| 51152 | 19305440+2016122 | 12.4 ± 3.9 | … | … | … | … | … | Y | … | … | … | … | n | … |
| 51229 | 19305451+2016037 | -3.5 ± 0.4 | 5976 ± 193 | 4.19 ± 0.26 | -0.32 ± 0.15 | 115 ± 42 | 1 | N | … | … | … | … | n | NG |
| 51157 | 19302237+2013203 | 13.6 ± 0.7 | 6122 ± 140 | 4.33 ± 0.60 | -0.14 ± 0.27 | <150 | 3 | Y | Y | Y | Y | … | Y | … |
| 51158 | 19302283+2013079 | 11.3 ± 0.8 | 7254 ± 674 | 4.19 ± 0.16 | 0.11 ± 0.16 | … | … | Y | … | … | … | … | n | … |
| 51159 | 19302335+2015364 | -345.6 ± 1.0 | 6024 ± 157 | 4.04 ± 0.60 | -1.33 ± 0.96 | … | … | N | … | … | … | … | n | … |
| 51160 | 19302347+2018574 | 28.0 ± 0.7 | 7267 ± 452 | 4.25 ± 0.11 | 0.12 ± 0.14 | 39 ± 28 | 1 | N | … | … | … | … | n | NG |
| 51161 | 19302361+2016464 | 13.1 ± 0.4 | 6364 ± 237 | 4.51 ± 0.45 | 0.00 ± 0.15 | 58 ± 30 | 1 | Y | Y | Y | Y | … | Y | … |
| 51162 | 19302487+2015553 | 21.2 ± 1.3 | 7460 ± 1102 | 4.12 ± 0.20 | 0.05 ± 0.35 | <127 | 3 | N | … | … | … | … | n | NG |
| 51055 | 19302489+2016207 | 16.1 ± 2.8 | … | … | … | … | … | Y | … | … | … | … | n | … |
| 51056 | 19302491+2018001 | 6.1 ± 12.9 | … | … | … | … | … | N | … | … | … | … | n | … |
| 51163 | 19302520+2014465 | 3.5 ± 1.0 | 6296 ± 151 | 4.11 ± 0.56 | 0.03 ± 0.16 | … | … | N | … | … | … | … | n | … |
| 51164 | 19302528+2017024 | 52.9 ± 0.4 | 6317 ± 232 | 4.51 ± 0.35 | -0.05 ± 0.14 | … | … | N | … | … | … | … | n | … |
| 51165 | 19302531+2018425 | 12.1 ± 0.6 | 5845 ± 38 | 4.50 ± 0.37 | -0.02 ± 0.15 | 137 ± 91 | 1 | Y | Y | Y | Y | … | Y | … |
| 51057 | 19302535+2015250 | 14.8 ± 1.2 | … | … | … | … | … | Y | … | … | … | … | n | … |
| 51058 | 19302605+2011448 | 9.9 ± 5.0 | … | … | … | … | … | N | … | … | … | … | n | … |
| 51166 | 19302625+2016218 | 17.6 ± 0.9 | 6164 ± 227 | 4.76 ± 0.48 | -0.08 ± 0.13 | … | … | N | … | … | … | … | n | … |
| 51182 | 19303217+2014363 | 12.3 ± 0.6 | 6269 ± 368 | 4.13 ± 0.11 | -0.06 ± 0.24 | <76 | 3 | Y | Y | Y | Y | … | Y | … |
| 51071 | 19303250+2012348 | 14.0 ± 1.2 | … | … | … | … | … | Y | … | … | … | … | n | … |
| 51072 | 19303254+2015184 | 12.8 ± 4.7 | … | … | … | … | … | Y | … | … | … | … | n | … |
| 51073 | 19303258+2016383 | 15.3 ± 0.7 | … | … | … | … | … | Y | … | … | … | … | n | … |
| 51183 | 19303260+2018533 | 14.5 ± 1.3 | 6327 ± 171 | 4.91 ± 0.88 | -0.14 ± 0.33 | <33 | 3 | Y | Y | Y | Y | Y | Y | … |
| 51074 | 19303261+2012513 | 17.8 ± 3.7 | … | … | … | … | … | N | … | … | … | … | n | … |
| 3287 | 19303274+2014498 | 12.2 ± 0.6 | 4941 ± 119 | 2.62 ± 0.23 | 0.03 ± 0.10 | 12 ± 1 | … | Y | Y | N | Y | … | n | G |
| 51075 | 19303295+2014360 | 21.5 ± 2.0 | … | … | … | … | … | N | … | … | … | … | n | … |
| 51076 | 19303297+2017558 | 20.6 ± 2.9 | … | … | … | … | … | N | … | … | … | … | n | … |
| 51184 | 19303306+2019494 | 9.8 ± 0.9 | 6462 ± 460 | 4.25 ± 0.06 | 0.11 ± 0.23 | … | … | N | … | … | … | … | n | … |
| 3288 | 19303309+2015442 | 11.5 ± 0.6 | 4800 ± 121 | 2.43 ± 0.23 | 0.08 ± 0.10 | 15 ± 1 | … | Y | Y | Y | Y | … | Y | … |
| 51077 | 19303316+2013365 | 12.8 ± 4.3 | … | … | … | … | … | Y | … | … | … | … | n | … |
| 51185 | 19303324+2012072 | 26.1 ± 0.7 | 6533 ± 256 | 4.84 ± 0.74 | -0.02 ± 0.15 | <53 | 3 | N | … | … | … | Y | n | NG |

| ID | CNAME | RV (km s$^{-1}$) | $T_{\text{eff}}$ (K) | logg (dex) | [Fe/H] (dex) | EW(Li)$^a$ (mÅ) | EW(Li) error flag$^b$ | Membership | | | | Gaia study Cantat-Gaudin$^c$ | Final$^d$ | NMs with Li$^e$ |
| | | | | | | | | RV | Li | logg | [Fe/H] | | | |
|---|---|---|---|---|---|---|---|---|---|---|---|---|---|---|
| 51078 | 19303324+2018278 | 18.7 ± 5.5 | … | … | … | … | … | N | … | … | … | … | n | … |
| 51186 | 19303333+2017420 | -31.1 ± 0.5 | 5920 ± 263 | 4.18 ± 0.18 | 0.01 ± 0.25 | … | … | N | … | … | … | … | n | … |
| 51079 | 19303366+2017341 | -5.2 ± 4.6 | … | … | … | … | … | N | … | … | … | … | n | … |
| 51187 | 19303372+2014206 | 10.9 ± 0.7 | 6579 ± 240 | 4.48 ± 0.52 | 0.04 ± 0.21 | 100 ± 45 | 1 | Y | Y | Y | Y | … | Y | … |
| 51188 | 19303379+2016311 | 14.5 ± 1.0 | 6856 ± 257 | 4.34 ± 0.32 | 0.04 ± 0.19 | … | … | Y | … | … | … | … | n | … |
| 51080 | 19303391+2017527 | 14.9 ± 0.7 | … | … | … | … | … | Y | … | … | … | … | n | … |
| 51081 | 19303403+2015042 | 14.9 ± 0.1 | … | … | … | … | … | Y | … | … | … | … | n | … |
| 51194 | 19303633+2016424 | 27.2 ± 0.7 | 6069 ± 63 | 4.42 ± 0.42 | -0.71 ± 0.60 | … | … | N | … | … | … | … | n | … |
| 51096 | 19303646+2012320 | 3.8 ± 1.5 | 6668 ± 129 | … | 0.02 ± 0.11 | … | … | N | … | … | … | … | n | … |
| 51195 | 19303655+2011575 | 20.3 ± 1.3 | 6511 ± 290 | 5.12 ± 0.78 | -0.41 ± 0.23 | … | … | N | … | … | … | … | n | … |

**Notes.** [a] The values of EW(Li) for this cluster are corrected (subtracted adjacent Fe (6707.43 Å) line). [b] Flags for the errors of the corrected EW(Li) values, as follows: 1=EW(Li) corrected by blends contribution using models; and 3=Upper limit (no error for EW(Li) is given). [c] Cantat-Gaudin et al. (2018). [d] The letters "Y" and "N" indicate if the star is a cluster member or not. [e] 'Li-rich G', 'G' and 'NG' indicate "Li-rich giant", "giant" and "non-giant" Li field outliers, respectively.

**Table C.16.** Pismis 18

| ID | CNAME | RV (km s$^{-1}$) | $T_{\text{eff}}$ (K) | $logg$ (dex) | [Fe/H] (dex) | EW(Li)$^a$ (mÅ) | EW(Li) error flag$^b$ | RV | Li | Membership logg | [Fe/H] | Final$^c$ | NMs with Li$^d$ |
|---|---|---|---|---|---|---|---|---|---|---|---|---|---|
| 51230 | 13355159-6200378 | -30.2 ± 0.1 | 6149 ± 61 | 3.67 ± 0.01 | … | … | … | Y | … | … | … | n | … |
| 51281 | 13355267-6155382 | -24.7 ± 0.4 | 6404 ± 212 | 4.03 ± 0.16 | 0.29 ± 0.24 | … | … | Y | … | … | … | n | … |
| 51282 | 13355292-6156473 | -47.9 ± 0.3 | 6721 ± 123 | 4.19 ± 0.04 | 0.09 ± 0.23 | 54 ± 23 | 1 | N | .. | … | … | n | NG |
| 51231 | 13355361-6201377 | -16.4 ± 0.2 | … | … | … | … | … | N | … | … | … | n | … |
| 51283 | 13355412-6202332 | -5.5 ± 0.3 | 6122 ± 198 | 4.39 ± 0.22 | 0.09 ± 0.14 | 74 ± 19 | 1 | N | … | … | … | n | NG |
| 51284 | 13355550-6213529 | -43.0 ± 0.3 | 5802 ± 134 | 4.20 ± 0.16 | -0.39 ± 0.14 | … | … | N | … | … | … | n | … |
| 51285 | 13355695-6213179 | -42.4 ± 0.3 | 5936 ± 325 | 4.52 ± 0.32 | 0.19 ± 0.25 | … | … | N | … | … | … | n | … |
| 51232 | 13355779-6156185 | -6.5 ± 0.3 | … | … | … | … | … | N | … | … | … | n | … |
| 51286 | 13360304-6212080 | -33.6 ± 0.9 | 7201 ± 843 | 4.12 ± 0.20 | 0.04 ± 0.36 | … | … | Y | … | … | … | n | … |
| 51233 | 13360315-6205566 | -32.3 ± 11.4 | … | … | … | … | … | Y | … | … | … | n | … |
| 51234 | 13360526-6212326 | -30.4 ± 0.1 | … | … | … | … | … | Y | … | … | … | n | … |
| 51287 | 13360529-6216116 | -41.1 ± 0.2 | 6015 ± 249 | 4.25 ± 0.20 | -0.05 ± 0.23 | 42 ± 20 | 1 | N | … | … | … | n | NG |
| 51235 | 13360565-6156597 | -18.6 ± 0.1 | … | … | … | … | … | N | … | … | … | n | … |
| 51267 | 13370770-6202480 | -29.7 ± 0.2 | 5650 ± 49 | 4.13 ± 0.22 | 0.04 ± 0.16 | … | … | Y | … | … | … | n | … |
| 51268 | 13370807-6159084 | 2.0 ± 0.2 | 6299 ± 134 | 4.21 ± 0.13 | 0.01 ± 0.16 | … | … | N | .. | … | … | n | … |
| 51269 | 13370830-6204477 | -19.8 ± 0.8 | … | … | … | … | … | N | … | … | … | n | … |
| 51270 | 13370918-6206569 | -26.6 ± 1.2 | … | … | … | … | … | Y | … | … | … | n | … |
| 51346 | 13370997-6204268 | -38.4 ± 0.6 | 6389 ± 229 | 4.53 ± 0.39 | 0.18 ± 0.24 | … | … | N | … | … | … | n | … |
| 51347 | 13371006-6206352 | -31.6 ± 0.4 | 6136 ± 419 | 4.74 ± 0.51 | -0.08 ± 0.40 | 118 ± 34 | 1 | Y | N | N | Y | n | NG |
| 51348 | 13371060-6204071 | -49.6 ± 0.6 | 5714 ± 223 | 3.89 ± 0.22 | -0.19 ± 0.18 | … | … | N | .. | … | … | n | … |
| 51271 | 13371063-6204512 | -23.6 ± 0.3 | … | … | … | … | … | N | … | … | … | n | … |
| 51349 | 13371098-6205371 | -33.7 ± 0.4 | 6499 ± 214 | 4.18 ± 0.21 | -0.01 ± 0.21 | … | … | Y | … | … | … | n | … |
| 51272 | 13371138-6204165 | -1.3 ± 0.2 | … | … | … | … | … | N | … | … | … | n | … |
| 51350 | 13371147-6205141 | -25.7 ± 1.3 | 7176 ± 696 | 4.13 ± 0.20 | 0.06 ± 0.21 | … | … | Y | … | … | … | n | … |
| 51273 | 13371180-6208085 | -23.4 ± 1.1 | … | … | … | … | … | N | … | … | … | n | … |
| 3304 | 13371182-6206030 | -28.3 ± 0.6 | 4898 ± 123 | 2.52 ± 0.23 | 0.11 ± 0.10 | 8 ± 1 | … | Y | Y | Y | Y | Y | … |
| 51274 | 13371184-6205052 | -35.3 ± 0.1 | … | … | … | … | … | Y | … | … | … | n | … |
| 51351 | 13371188-6206233 | -35.0 ± 0.3 | 6219 ± 218 | 3.81 ± 0.17 | -0.25 ± 0.28 | … | … | Y | … | … | … | n | … |
| 51352 | 13371285-6206532 | -52.2 ± 0.5 | 5858 ± 311 | 4.49 ± 0.35 | -0.22 ± 0.38 | <91 | 3 | N | .. | … | … | n | NG |
| 51275 | 13371532-6213078 | -3.8 ± 0.2 | … | … | … | … | … | N | … | … | … | n | … |
| 51353 | 13371768-6204447 | -27.7 ± 0.2 | 6081 ± 201 | 4.36 ± 0.29 | -0.06 ± 0.17 | 89 ± 18 | 1 | Y | Y | Y | Y | Y | … |
| 51276 | 13371831-6216285 | -16.4 ± 0.3 | … | … | … | … | … | N | … | … | … | n | … |
| 51277 | 13371858-6200580 | -49.2 ± 0.9 | … | … | … | … | … | N | .. | … | … | n | … |
| 51354 | 13372120-6202431 | -5.8 ± 0.2 | 5533 ± 201 | 4.48 ± 0.37 | -0.02 ± 0.14 | 52 ± 15 | 1 | N | … | … | … | n | NG |
| 51279 | 13372196-6154365 | -25.6 ± 0.9 | … | … | … | … | … | Y | … | … | … | n | … |
| 51236 | 13360660-6215514 | -2.6 ± 0.1 | … | … | … | … | … | N | … | … | … | n | … |
| 51288 | 13360912-6200130 | -6.1 ± 0.3 | 6342 ± 168 | 4.29 ± 0.30 | 0.02 ± 0.14 | 64 ± 15 | 1 | N | … | … | … | n | NG |
| 51237 | 13361189-6156396 | -7.0 ± 0.1 | … | … | … | … | … | N | … | … | … | n | … |
| 51238 | 13361402-6207025 | -61.2 ± 0.2 | 6232 ± 102 | 4.28 ± 0.07 | -0.44 ± 0.16 | … | … | N | .. | … | … | n | … |
| 51292 | 13361764-6206058 | -33.5 ± 1.0 | 7473 ± 94 | … | … | … | … | Y | … | … | … | n | … |
| 51293 | 13361875-6208205 | -52.4 ± 0.3 | 6369 ± 188 | 4.15 ± 0.13 | 0.26 ± 0.23 | … | … | N | .. | … | … | n | … |
| 51294 | 13362236-6158449 | -8.8 ± 0.2 | 5957 ± 72 | 4.04 ± 0.07 | -0.10 ± 0.13 | 46 ± 20 | 1 | N | … | … | … | n | NG |
| 51239 | 13362331-6200358 | -17.1 ± 0.2 | … | … | … | … | … | N | … | … | … | n | … |
| 51240 | 13362394-6201038 | -20.1 ± 0.3 | … | … | … | … | … | N | … | … | … | n | … |
| 51241 | 13362487-6215210 | -4.2 ± 0.2 | 5878 ± 273 | 4.07 ± 0.04 | -0.11 ± 0.20 | … | … | N | .. | … | … | n | … |
| 51242 | 13362692-6214515 | -4.3 ± 0.3 | 5836 ± 407 | 4.78 ± 0.70 | -0.56 ± 0.53 | … | … | N | … | … | … | n | … |
| 51243 | 13362746-6158018 | -10.1 ± 0.1 | … | … | … | … | … | N | … | … | … | n | … |
| 51244 | 13362763-6207280 | -8.5 ± 0.1 | 5425 ± 93 | 4.71 ± 0.06 | 0.05 ± 0.16 | … | … | N | … | … | … | n | … |
| 51297 | 13363080-6156129 | -30.6 ± 0.4 | 6425 ± 218 | 4.30 ± 0.24 | 0.06 ± 0.17 | … | … | Y | … | … | … | n | … |
| 51298 | 13363222-6156268 | 16.4 ± 0.2 | 6076 ± 164 | 4.26 ± 0.28 | 0.18 ± 0.15 | 64 ± 18 | 1 | N | .. | … | … | n | NG |
| 51245 | 13363246-6201586 | 3.1 ± 0.1 | 5511 ± 43 | 4.20 ± 0.05 | -0.19 ± 0.18 | … | … | N | … | … | … | n | … |
| 51299 | 13363637-6202086 | -20.5 ± 0.3 | 5983 ± 196 | 4.24 ± 0.33 | -0.08 ± 0.18 | 58 ± 33 | 1 | N | … | … | … | n | NG |
| 51300 | 13363956-6155287 | -38.6 ± 0.6 | 6480 ± 367 | 4.15 ± 0.28 | 0.06 ± 0.19 | … | … | N | … | … | … | n | … |
| 51301 | 13363992-6157171 | -13.6 ± 0.3 | 5447 ± 119 | 4.42 ± 0.28 | -0.07 ± 0.12 | … | … | N | … | … | … | n | … |
| 51253 | 13365109-6205550 | -28.7 ± 0.5 | … | … | … | … | … | Y | … | … | … | n | … |
| 51254 | 13365119-6155187 | 4.4 ± 0.1 | 5713 ± 113 | 4.09 ± 0.01 | 0.07 ± 0.19 | … | … | N | .. | … | … | n | … |
| 51255 | 13365119-6212116 | -11.0 ± 0.2 | … | … | … | … | … | N | … | … | … | n | … |
| 51319 | 13365218-6206468 | -30.1 ± 0.5 | 5295 ± 431 | 3.55 ± 0.88 | -0.46 ± 0.31 | 147 ± 86 | 1 | Y | N | N | N | n | NG |
| 51320 | 13365219-6205531 | -32.0 ± 1.9 | 5617 ± 400 | 3.33 ± 0.51 | -0.38 ± 0.61 | … | … | Y | … | … | … | n | … |

| ID | CNAME | RV (km s$^{-1}$) | $T_\mathrm{eff}$ (K) | $\log g$ (dex) | [Fe/H] (dex) | EW(Li)$^a$ (mÅ) | EW(Li) error flag$^b$ | \multicolumn{5}{c}{Membership} | NMs with Li$^d$ |
|---|---|---|---|---|---|---|---|---|---|---|---|---|---|
|  |  |  |  |  |  |  |  | RV | Li | $\log g$ | [Fe/H] | Final$^c$ |  |
| 51321 | 13365226-6207322 | 13.6 ± 0.3 | 5923 ± 286 | 4.21 ± 0.16 | 0.24 ± 0.17 | … | … | N | .. | … | … | n | … |
| 51322 | 13365279-6207090 | -33.5 ± 0.9 | 6300 ± 203 | 4.78 ± 0.85 | 0.34 ± 0.25 | … | … | Y | … | … | … | n | … |
| 51323 | 13365304-6204298 | -31.4 ± 1.0 | 6843 ± 484 | 4.30 ± 0.20 | 0.07 ± 0.19 | … | … | Y | … | … | … | n | … |
| 51256 | 13365306-6204204 | -28.1 ± 4.5 | … | … | … | … | … | Y | … | … | … | n | … |
| 51324 | 13365314-6206371 | -16.4 ± 0.2 | 4864 ± 155 | 4.48 ± 0.33 | 0.07 ± 0.14 | 115 ± 18 | 1 | N | N | Y | Y | n | NG |
| 3299 | 13365321-6210050 | -40.2 ± 0.6 | 4866 ± 115 | 2.85 ± 0.23 | 0.06 ± 0.10 | <6 | 3 | N | … | … | … | n | … |
| 51325 | 13365340-6206124 | -12.0 ± 0.3 | 6033 ± 154 | 4.10 ± 0.08 | 0.12 ± 0.15 | … | … | N | … | … | … | n | … |
| 51258 | 13365432-6205028 | -61.7 ± 0.1 | … | … | … | … | … | N | .. | … | … | n | … |
| 51326 | 13365524-6203262 | -13.9 ± 0.2 | 6200 ± 122 | 4.19 ± 0.12 | 0.08 ± 0.13 | … | … | N | … | … | … | n | … |
| 3300 | 13365597-6205130 | -28.1 ± 0.6 | 4921 ± 120 | 2.55 ± 0.24 | 0.09 ± 0.09 | 5 ± 1 | … | Y | Y | Y | Y | Y | … |
| 51327 | 13365597-6206066 | -29.7 ± 0.5 | 5778 ± 275 | 4.11 ± 0.12 | -0.08 ± 0.15 | 81 ± 34 | 1 | Y | Y | Y | Y | Y | … |
| 51328 | 13365611-6204447 | -22.5 ± 0.3 | 5817 ± 70 | 4.35 ± 0.17 | -0.06 ± 0.18 | … | … | N | … | … | … | n | … |
| 51329 | 13365704-6208323 | -18.4 ± 0.3 | 6361 ± 192 | 4.42 ± 0.41 | 0.04 ± 0.23 | 46 ± 15 | 1 | N | Y | Y | Y | Y | … |
| 51259 | 13365737-6206023 | -30.4 ± 1.1 | … | … | … | … | … | Y | … | … | … | n | … |
| 3301 | 13365882-6205197 | -27.3 ± 0.6 | 5082 ± 122 | 3.00 ± 0.23 | 0.08 ± 0.10 | 36 ± 1 | … | Y | Y | Y | Y | Y | … |
| 51356 | 13372399-6158582 | -10.0 ± 0.2 | 6474 ± 217 | 3.75 ± 0.21 | 0.23 ± 0.23 | … | … | N | … | … | … | n | … |
| 51357 | 13372422-6205072 | -20.7 ± 0.3 | 6167 ± 216 | 4.16 ± 0.22 | 0.30 ± 0.29 | … | … | N | … | … | … | n | … |
| 51358 | 13372583-6157089 | -51.5 ± 0.2 | 6644 ± 428 | 4.10 ± 0.21 | 0.27 ± 0.41 | 65 ± 25 | 1 | N | .. | … | … | n | NG |
| 3305 | 13372623-6204585 | -5.8 ± 0.6 | 4475 ± 151 | 2.31 ± 0.33 | 0.16 ± 0.11 | <6 | 3 | N | N | N | N | n | G |
| 51359 | 13372737-6155008 | -31.5 ± 0.3 | 6293 ± 171 | 4.17 ± 0.28 | -0.02 ± 0.15 | … | … | Y | … | … | … | n | … |
| 51280 | 13372766-6208447 | -31.1 ± 5.0 | … | … | … | … | … | Y | … | … | … | n | … |
| 306 | 13372876-6202527 | -33.8 ± 0.6 | 4910 ± 125 | 2.89 ± 0.24 | 0.21 ± 0.10 | <6 | 3 | Y | Y | Y | Y | Y | … |
| 51360 | 13373248-6208191 | -30.7 ± 0.5 | 6548 ± 332 | 4.04 ± 0.16 | 0.12 ± 0.19 | 84 ± 23 | 1 | Y | Y | Y | Y | Y | … |
| 51361 | 13373403-6207036 | -38.2 ± 0.3 | 5952 ± 158 | 4.35 ± 0.30 | 0.06 ± 0.13 | 66 ± 12 | 1 | N | … | … | … | n | NG |
| 51246 | 13364002-6212345 | -45.0 ± 0.3 | 7060 ± 48 | … | … | … | … | N | … | … | … | n | … |
| 51247 | 13364117-6205166 | -21.1 ± 1.7 | … | … | … | … | … | N | … | … | … | n | … |
| 51302 | 13364277-6204288 | -28.0 ± 0.3 | 5851 ± 168 | 4.35 ± 0.24 | 0.23 ± 0.14 | … | … | Y | … | … | … | n | … |
| 51248 | 13364291-6204081 | -3.3 ± 124.3 | … | … | … | … | … | N | .. | … | … | n | … |
| 51303 | 13364304-6206482 | -47.8 ± 0.4 | 6199 ± 84 | 4.27 ± 0.41 | -0.18 ± 0.16 | 62 ± 23 | 1 | N | .. | … | … | n | NG |
| 51249 | 13364397-6206104 | -30.0 ± 3.9 | … | … | … | … | … | Y | … | … | … | n | … |
| 51304 | 13364410-6202144 | -22.3 ± 0.2 | 5903 ± 190 | 4.33 ± 0.10 | 0.16 ± 0.13 | … | … | N | … | … | … | n | … |
| 51305 | 13364418-6204483 | -34.1 ± 0.7 | … | … | … | … | … | Y | … | … | … | n | … |
| 51306 | 13364430-6205471 | -22.8 ± 0.6 | 7221 ± 766 | 4.11 ± 0.19 | -0.02 ± 0.34 | 82 ± 24 | 1 | N | Y | Y | Y | Y | … |
| 51307 | 13364503-6207088 | -63.4 ± 0.3 | 5659 ± 439 | 3.88 ± 0.24 | 0.16 ± 0.23 | 90 ± 36 | 1 | N | … | … | … | n | NG |
| 51308 | 13364586-6207361 | -46.4 ± 8.8 | 6761 ± 854 | 3.89 ± 0.66 | -0.01 ± 0.09 | … | … | N | .. | … | … | n | … |
| 51250 | 13364687-6205483 | -30.3 ± 0.1 | … | … | … | … | … | Y | … | … | … | n | … |
| 51310 | 13364748-6206251 | -35.3 ± 1.1 | 6734 ± 535 | 3.99 ± 0.19 | 0.16 ± 0.20 | … | … | Y | … | … | … | n | … |
| 3297 | 13364831-6206517 | -26.7 ± 0.6 | 5005 ± 118 | 2.73 ± 0.25 | 0.11 ± 0.09 | 8 ± 1 | … | Y | Y | Y | Y | Y | … |
| 51312 | 13364895-6204476 | -45.6 ± 1.0 | 5826 ± 242 | 3.96 ± 0.86 | -0.54 ± 0.25 | <69 | 3 | N | … | … | … | n | NG |
| 51313 | 13364946-6204100 | -33.0 ± 1.4 | 6806 ± 565 | 4.08 ± 0.26 | 0.10 ± 0.25 | … | … | Y | … | … | … | n | … |
| 51251 | 13364968-6156238 | -8.1 ± 0.1 | … | … | … | … | … | N | … | … | … | n | … |
| 3298 | 13365001-6205376 | -18.3 ± 0.6 | 4978 ± 125 | 2.73 ± 0.23 | 0.08 ± 0.10 | 44 ± 2 | … | N | Y | Y | Y | Y | … |
| 51314 | 13365003-6200530 | -22.8 ± 0.2 | 5439 ± 96 | 4.45 ± 0.42 | -0.09 ± 0.14 | … | … | N | … | … | … | n | … |
| 51315 | 13365005-6200333 | 33.5 ± 0.2 | 5907 ± 175 | 4.35 ± 0.13 | 0.06 ± 0.15 | 47 ± 14 | 1 | N | .. | … | … | n | NG |
| 51316 | 13365027-6206392 | -26.8 ± 2.5 | 6410 ± 711 | 4.33 ± 0.15 | -0.13 ± 0.38 | … | … | Y | … | … | … | n | … |
| 51331 | 13365949-6205340 | -28.2 ± 3.0 | 7361 ± 252 | … | … | … | … | Y | … | … | … | n | … |
| 51332 | 13370097-6212435 | -43.4 ± 0.3 | 5801 ± 61 | 3.94 ± 0.23 | -0.52 ± 0.16 | 72 ± 15 | 1 | N | … | … | … | n | NG |
| 51333 | 13370145-6205201 | -8.6 ± 0.3 | 6301 ± 227 | 4.33 ± 0.22 | 0.09 ± 0.24 | 65 ± 24 | 1 | N | … | … | … | n | NG |
| 51261 | 13370213-6204032 | -0.2 ± 0.1 | … | … | … | … | … | N | .. | … | … | n | … |
| 3302 | 13370214-6206095 | -26.8 ± 0.6 | 4970 ± 121 | 2.69 ± 0.23 | 0.15 ± 0.10 | <5 | 3 | Y | Y | Y | Y | Y | … |
| 51262 | 13370260-6205262 | -51.6 ± 0.1 | … | … | … | … | … | N | .. | … | … | n | … |
| 51335 | 13370289-6204256 | -14.1 ± 1.0 | 5508 ± 710 | 3.41 ± 1.29 | -0.21 ± 0.47 | … | … | N | … | … | … | n | … |
| 51337 | 13370340-6206352 | -52.0 ± 1.1 | 6249 ± 93 | 4.47 ± 0.72 | -0.17 ± 0.32 | … | … | N | … | … | … | n | … |
| 51339 | 13370370-6217577 | -12.0 ± 0.2 | 6028 ± 169 | 4.22 ± 0.19 | 0.04 ± 0.15 | 53 ± 13 | 1 | N | … | … | … | n | NG |
| 51263 | 13370385-6214581 | -30.2 ± 0.1 | … | … | … | … | … | Y | … | … | … | n | … |
| 51340 | 13370391-6206005 | -29.2 ± 0.8 | 6230 ± 77 | 4.55 ± 0.61 | 0.07 ± 0.19 | … | … | Y | … | … | … | n | … |
| 51264 | 13370473-6204579 | -17.6 ± 3.0 | … | … | … | … | … | N | … | … | … | n | … |
| 3303 | 13370523-6206433 | -27.7 ± 0.6 | 4953 ± 117 | 2.66 ± 0.23 | 0.10 ± 0.10 | 26 ± 1 | … | Y | Y | Y | Y | Y | … |
| 51341 | 13370545-6205095 | -24.7 ± 0.4 | 5866 ± 62 | 4.52 ± 0.46 | -0.07 ± 0.22 | 57 ± 46 | 1 | Y | Y | Y | Y | Y | … |

**Table C.16.** continued.

| ID | CNAME | RV (km s$^{-1}$) | $T_{\text{eff}}$ (K) | $logg$ (dex) | [Fe/H] (dex) | EW(Li)$^a$ (mÅ) | EW(Li) error flag$^b$ | Membership RV | Li | $logg$ | [Fe/H] | Final$^c$ | NMs with Li$^d$ |
|---|---|---|---|---|---|---|---|---|---|---|---|---|---|
| 51265 | 13370611-6159537 | -2.2 ± 0.1 | … | … | … | … | … | N | .. | … | … | n | … |
| 51342 | 13370631-6215436 | -30.0 ± 0.2 | 5364 ± 141 | 4.46 ± 0.46 | -0.09 ± 0.12 | 50 ± 19 | 1 | Y | Y | Y | Y | Y | … |
| 51343 | 13370635-6206124 | 5.1 ± 0.5 | 6093 ± 294 | 4.53 ± 0.44 | -0.18 ± 0.30 | 69 ± 54 | 1 | N | .. | … | … | n | NG |
| 51344 | 13370683-6206493 | -9.0 ± 0.3 | 5854 ± 187 | 4.34 ± 0.11 | 0.28 ± 0.16 | 57 ± 47 | 1 | N | … | … | … | n | NG |
| 51266 | 13370693-6205236 | -30.6 ± 1.1 | … | … | … | … | … | Y | … | … | … | n | … |
| 51345 | 13370756-6206350 | 186.5 ± 0.3 | 5781 ± 57 | 3.72 ± 0.26 | -0.68 ± 0.26 | … | … | N | .. | … | … | n | … |

**Notes.** $^{(a)}$ The values of EW(Li) for this cluster are corrected (subtracted adjacent Fe (6707.43 Å) line). $^{(b)}$ Flags for the errors of the corrected EW(Li) values, as follows: 1=EW(Li) corrected by blends contribution using models; and 3=Upper limit (no error for EW(Li) is given). $^{(c)}$ The letters "Y" and "N" indicate if the star is a cluster member or not. $^{(d)}$ 'Li-rich G', 'G' and 'NG' indicate "Li-rich giant", "giant" and "non-giant" Li field outliers, respectively.

**Table C.17.** Trumpler 20

| ID | CNAME | RV (km s$^{-1}$) | $T_{\rm eff}$ (K) | $\log g$ (dex) | [Fe/H] (dex) | $EW$(Li)[a] (mÅ) | $EW$(Li) error flag[b] | Membership RV | Li | $\log g$ | [Fe/H] | Gaia study Cantat-Gaudin[c] | Final[d] | NMs with Li[e] |
|---|---|---|---|---|---|---|---|---|---|---|---|---|---|---|
| 55753 | 12385773-6029244 | -34.5 ± 0.7 | ... | ... | ... | ... | ... | N | ... | ... | ... | ... | n | ... |
| 52778 | 12385783-6035158 | 1.5 ± 0.3 | 5828 ± 242 | 4.53 ± 0.43 | 0.16 ± 0.18 | <33 | 3 | N | ... | ... | ... | ... | n | NG |
| 52779 | 12385787-6039505 | 31.6 ± 0.7 | 5527 ± 272 | 3.86 ± 0.05 | -0.36 ± 0.28 | ... | ... | N | ... | ... | ... | ... | n | ... |
| 55754 | 12385804-6030286 | -39.7 ± 0.1 | 4311 ± 94 | 1.95 ± 0.11 | -0.33 ± 0.18 | ... | ... | Y | ... | ... | ... | ... | n | ... |
| 52780 | 12385804-6033109 | -25.0 ± 1.2 | 6037 ± 285 | 4.11 ± 0.63 | 0.08 ± 0.20 | ... | ... | N | ... | ... | ... | ... | n | ... |
| 3361 | 12385807-6030286 | -40.8 ± 0.6 | 4551 ± 133 | 2.18 ± 0.29 | 0.09 ± 0.11 | <4 | 3 | Y | Y | Y | Y | ... | Y | ... |
| 55755 | 12385814-6036020 | 49.2 ± 0.2 | 5949 ± 129 | 3.52 ± 0.05 | -0.09 ± 0.23 | ... | ... | N | ... | ... | ... | ... | n | ... |
| 55756 | 12385819-6036566 | -39.5 ± 0.3 | ... | ... | ... | ... | ... | Y | ... | ... | ... | ... | n | ... |
| 55757 | 12385857-6037035 | -35.3 ± 0.1 | ... | ... | ... | ... | ... | N | ... | ... | ... | ... | n | ... |
| 55758 | 12385858-6037388 | -33.0 ± 0.1 | ... | ... | ... | ... | ... | N | ... | ... | ... | ... | n | ... |
| 52781 | 12385859-6037035 | -37.3 ± 0.4 | 6677 ± 90 | 4.01 ± 0.10 | 0.13 ± 0.19 | 44 ± 25 | 1 | Y | Y | Y | Y | ... | Y | ... |
| 52782 | 12385861-6037388 | -30.0 ± 0.2 | 6100 ± 174 | 4.15 ± 0.14 | 0.04 ± 0.13 | ... | ... | N | ... | ... | ... | ... | n | ... |
| 52783 | 12385871-6034229 | -52.4 ± 0.8 | 6077 ± 343 | 4.21 ± 0.11 | -0.29 ± 0.13 | <76 | 3 | N | ... | ... | ... | ... | n | NG |
| 52784 | 12385903-6041577 | -38.9 ± 0.5 | 6004 ± 423 | 4.41 ± 0.24 | 0.24 ± 0.28 | ... | ... | Y | ... | ... | ... | ... | n | ... |
| 55759 | 12385914-6035449 | -38.5 ± 0.3 | ... | ... | ... | ... | ... | Y | ... | ... | ... | ... | n | ... |
| 55760 | 12385921-6041203 | -14.6 ± 0.1 | ... | ... | ... | ... | ... | N | ... | ... | ... | ... | n | ... |
| 55761 | 12385937-6037281 | -35.7 ± 1.2 | ... | ... | ... | ... | ... | N | ... | ... | ... | ... | n | ... |
| 55762 | 12385942-6031308 | -37.2 ± 0.2 | ... | ... | ... | ... | ... | Y | ... | ... | ... | ... | n | ... |
| 52785 | 12385949-6038087 | -10.5 ± 0.6 | 5678 ± 113 | 4.44 ± 0.44 | 0.22 ± 0.19 | ... | ... | N | ... | ... | ... | ... | n | ... |
| 55763 | 12385965-6033124 | -14.1 ± 0.1 | 4299 ± 134 | 2.96 ± 0.07 | 0.19 ± 0.24 | ... | ... | N | ... | ... | ... | ... | n | ... |
| 55764 | 12385977-6037320 | -31.4 ± 0.5 | ... | ... | ... | ... | ... | N | ... | ... | ... | ... | n | ... |
| 52786 | 12385980-6034475 | -38.0 ± 0.2 | 6357 ± 53 | 3.47 ± 0.16 | 0.22 ± 0.15 | 94 ± 40 | ... | Y | Y | Y | Y | Y | Y | ... |
| 55765 | 12385982-6041076 | -10.1 ± 0.1 | ... | ... | ... | ... | ... | N | ... | ... | ... | ... | n | ... |
| 55766 | 12390013-6033592 | -31.6 ± 0.3 | ... | ... | ... | ... | ... | N | ... | ... | ... | ... | n | ... |
| 55827 | 12391084-6044037 | -38.9 ± 1.0 | ... | ... | ... | ... | ... | Y | ... | ... | ... | ... | n | ... |
| 55828 | 12391087-6040393 | -47.1 ± 0.2 | ... | ... | ... | ... | ... | N | ... | ... | ... | ... | n | ... |
| 55829 | 12391088-6032128 | -13.5 ± 0.1 | 5859 ± 136 | 4.21 ± 0.03 | 0.03 ± 0.25 | ... | ... | N | ... | ... | ... | ... | n | ... |
| 3366 | 12391113-6036528 | -41.1 ± 0.6 | 4967 ± 115 | 2.93 ± 0.23 | 0.18 ± 0.10 | <9 | 3 | Y | Y | Y | Y | ... | Y | ... |
| 52819 | 12391116-6036059 | -57.1 ± 0.4 | 5591 ± 152 | 4.03 ± 0.77 | -0.17 ± 0.16 | 78 ± 46 | 1 | N | ... | ... | ... | ... | n | NG |
| 52820 | 12391120-6032492 | -20.6 ± 0.4 | 6062 ± 353 | 3.87 ± 0.05 | 0.30 ± 0.28 | <52 | 3 | N | ... | ... | ... | ... | n | NG |
| 55830 | 12391127-6030572 | 17.1 ± 0.2 | 5927 ± 150 | 3.45 ± 0.05 | 0.27 ± 0.27 | ... | ... | N | ... | ... | ... | ... | n | ... |
| 52821 | 12391128-6038019 | 15.9 ± 0.2 | 4795 ± 138 | 3.22 ± 0.23 | -0.27 ± 0.15 | ... | ... | N | ... | ... | ... | ... | n | G |
| 55831 | 12391157-6036401 | -41.5 ± 1.5 | ... | ... | ... | ... | ... | Y | ... | ... | ... | ... | n | ... |
| 52822 | 12391158-6040270 | -38.6 ± 0.4 | 6164 ± 226 | 4.25 ± 0.20 | 0.27 ± 0.18 | <77 | 3 | Y | Y | Y | Y | Y | Y | ... |
| 55832 | 12391160-6037220 | -39.0 ± 0.1 | 5501 ± 134 | 4.30 ± 0.04 | 0.58 ± 0.24 | ... | ... | Y | ... | ... | ... | ... | n | ... |
| 55833 | 12391166-6042243 | -38.7 ± 1.0 | ... | ... | ... | ... | ... | Y | ... | ... | ... | ... | n | ... |
| 55834 | 12391183-6028589 | -9.3 ± 0.2 | ... | ... | ... | ... | ... | N | ... | ... | ... | ... | n | ... |
| 52823 | 12391188-6035588 | -55.9 ± 0.3 | 5989 ± 422 | 4.22 ± 0.16 | 0.12 ± 0.24 | <69 | 3 | N | ... | ... | ... | ... | n | NG |
| 52824 | 12391192-6043336 | -22.4 ± 0.5 | 5550 ± 31 | 3.83 ± 0.29 | -0.04 ± 0.13 | <103 | 3 | N | ... | ... | ... | ... | n | NG |
| 3367 | 12391200-6036322 | -40.9 ± 0.6 | 4939 ± 112 | 2.83 ± 0.23 | 0.15 ± 0.10 | <7 | 3 | Y | Y | Y | Y | ... | Y | ... |
| 52825 | 12391204-6041313 | -39.5 ± 0.7 | 5884 ± 43 | 4.31 ± 0.11 | 0.27 ± 0.25 | ... | ... | Y | ... | ... | ... | ... | n | ... |
| 55835 | 12391207-6033160 | -41.9 ± 1.7 | ... | ... | ... | ... | ... | Y | ... | ... | ... | ... | n | ... |
| 52826 | 12391210-6038007 | -39.3 ± 0.6 | 6727 ± 78 | 3.77 ± 0.15 | 0.22 ± 0.06 | <16 | 3 | Y | Y | Y | Y | Y | Y | ... |
| 55836 | 12391217-6031446 | 9.3 ± 0.1 | 5691 ± 75 | 2.95 ± 0.05 | 0.10 ± 0.15 | ... | ... | N | ... | ... | ... | ... | n | ... |
| 55837 | 12391227-6039121 | -38.8 ± 0.6 | ... | ... | ... | ... | ... | Y | ... | ... | ... | ... | n | ... |
| 55838 | 12391246-6035507 | -32.9 ± 1.5 | ... | ... | ... | ... | ... | N | ... | ... | ... | ... | n | ... |
| 52827 | 12391248-6035506 | -41.0 ± 0.4 | 6758 ± 534 | 4.02 ± 0.22 | 0.06 ± 0.22 | ... | ... | Y | ... | ... | ... | ... | n | ... |
| 52871 | 12391959-6037534 | -9.0 ± 0.3 | 6479 ± 54 | 3.80 ± 0.18 | 0.08 ± 0.16 | ... | ... | N | ... | ... | ... | ... | n | ... |
| 55892 | 12391971-6037571 | 9.5 ± 0.2 | ... | ... | ... | ... | ... | N | ... | ... | ... | ... | n | ... |
| 55893 | 12391975-6033351 | -42.2 ± 1.1 | ... | ... | ... | ... | ... | Y | ... | ... | ... | ... | n | ... |
| 52872 | 12391977-6037065 | -6.9 ± 0.6 | 6100 ± 330 | 4.47 ± 0.30 | 0.26 ± 0.26 | 66 ± 60 | 1 | N | ... | ... | ... | ... | n | NG |
| 52873 | 12391978-6035065 | -39.4 ± 0.5 | 5942 ± 299 | 4.15 ± 0.12 | 0.09 ± 0.23 | <53 | 3 | Y | Y | Y | Y | ... | Y | ... |
| 55894 | 12391980-6037414 | -38.8 ± 1.4 | ... | ... | ... | ... | ... | Y | ... | ... | ... | ... | n | ... |
| 55896 | 12391998-6035147 | 25.5 ± 0.2 | 6265 ± 134 | 4.60 ± 0.03 | ... | ... | ... | N | ... | ... | ... | ... | n | ... |
| 55897 | 12391999-6031562 | -6.7 ± 0.2 | 6161 ± 149 | 3.47 ± 0.02 | ... | ... | ... | N | ... | ... | ... | ... | n | ... |
| 55898 | 12392004-6039296 | -36.2 ± 1.6 | ... | ... | ... | ... | ... | Y | ... | ... | ... | ... | n | ... |
| 52874 | 12392006-6036533 | 104.9 ± 0.3 | 4874 ± 467 | 3.43 ± 0.23 | -0.28 ± 0.31 | ... | ... | N | ... | ... | ... | ... | n | G |
| 52875 | 12392006-6037163 | -65.3 ± 0.4 | 5952 ± 67 | 3.93 ± 0.24 | -0.09 ± 0.14 | ... | ... | N | ... | ... | ... | ... | n | ... |
| 55899 | 12392029-6033383 | 37.5 ± 0.1 | ... | ... | ... | ... | ... | N | ... | ... | ... | ... | n | ... |

**Table C.17.** continued.

| ID | CNAME | RV (km s$^{-1}$) | $T_{\text{eff}}$ (K) | $\log g$ (dex) | [Fe/H] (dex) | EW(Li)$^a$ (mÅ) | EW(Li) error flag$^b$ | Membership RV | Li | $\log g$ | [Fe/H] | Gaia study Cantat-Gaudin$^c$ | Final$^d$ | NMs with Li$^e$ |
|---|---|---|---|---|---|---|---|---|---|---|---|---|---|---|
| 52876 | 12392034-6036316 | -32.1 ± 0.3 | 5470 ± 268 | 4.54 ± 0.42 | -0.07 ± 0.17 | … | … | N | … | … | … | … | n | … |
| 52877 | 12392042-6036275 | -39.2 ± 0.3 | 6793 ± 69 | 4.17 ± 0.13 | 0.29 ± 0.06 | 61 ± 20 | … | Y | Y | Y | Y | Y | Y | … |
| 52878 | 12392058-6030307 | -50.9 ± 1.6 | 6309 ± 830 | 3.89 ± 0.31 | -1.11 ± 0.20 | … | … | N | … | … | … | … | n | … |
| 52879 | 12392075-6031284 | 12.8 ± 0.2 | 5100 ± 123 | 3.31 ± 0.15 | -0.01 ± 0.17 | … | … | N | … | … | … | … | n | G |
| 55900 | 12392098-6036263 | -40.8 ± 0.1 | 4723 ± 124 | 3.21 ± 0.07 | 0.33 ± 0.26 | … | … | Y | … | … | … | … | n | … |
| 55901 | 12392100-6040408 | -43.2 ± 1.3 | … | … | … | … | … | N | … | … | … | … | n | … |
| 52880 | 12392105-6035380 | -47.3 ± 0.5 | 5932 ± 214 | 4.29 ± 0.24 | -0.10 ± 0.23 | <56 | 3 | N | … | … | … | … | n | NG |
| 55902 | 12392116-6047130 | -35.3 ± 0.1 | … | … | … | … | … | N | … | … | … | … | n | … |
| 52881 | 12392137-6040491 | -45.7 ± 0.5 | 6716 ± 69 | 4.09 ± 0.14 | 0.08 ± 0.06 | … | … | N | … | … | … | … | n | … |
| 52883 | 12392152-6037305 | -35.1 ± 0.6 | 6885 ± 633 | 4.14 ± 0.02 | 0.00 ± 0.20 | 59 ± 48 | 1 | N | … | … | … | N | n | NG |
| 55954 | 12392745-6039028 | -33.6 ± 0.9 | … | … | … | … | … | N | … | … | … | … | n | … |
| 52923 | 12392747-6034549 | -29.0 ± 0.5 | 5615 ± 675 | 4.66 ± 0.12 | -0.12 ± 0.48 | … | … | N | … | … | … | … | n | … |
| 52924 | 12392751-6039251 | -16.8 ± 0.3 | 5252 ± 75 | 4.28 ± 0.61 | 0.18 ± 0.15 | <89 | 3 | N | … | … | … | … | n | NG |
| 52925 | 12392756-6037442 | 20.2 ± 0.7 | 5276 ± 127 | 3.41 ± 0.41 | -0.44 ± 0.39 | 112 ± 99 | 1 | N | … | … | … | … | n | … |
| 52926 | 12392760-6035476 | -25.0 ± 0.3 | 5611 ± 182 | 3.94 ± 0.04 | -0.22 ± 0.42 | <77 | 3 | N | … | … | … | … | n | NG |
| 55955 | 12392760-6042158 | -38.9 ± 1.0 | … | … | … | … | … | Y | … | … | … | … | n | … |
| 55956 | 12392769-6044150 | -11.6 ± 0.1 | … | … | … | … | … | N | … | … | … | … | n | … |
| 52927 | 12392774-6037371 | 9.5 ± 0.4 | 5466 ± 286 | 4.14 ± 0.27 | -0.22 ± 0.16 | … | … | N | … | … | … | … | n | … |
| 55957 | 12392785-6039094 | -5.3 ± 0.1 | … | … | … | … | … | N | … | … | … | … | n | … |
| 52928 | 12392786-6031196 | -41.1 ± 0.5 | 5891 ± 69 | 4.25 ± 0.61 | 0.11 ± 0.14 | … | … | Y | … | … | … | … | n | … |
| 55958 | 12392793-6032460 | -35.1 ± 0.1 | 6295 ± 150 | 4.25 ± 0.05 | … | … | … | N | … | … | … | … | n | … |
| 55959 | 12392793-6039031 | -38.9 ± 0.5 | … | … | … | … | … | Y | … | … | … | … | n | … |
| 55960 | 12392794-6035331 | -36.1 ± 1.2 | … | … | … | … | … | Y | … | … | … | … | n | … |
| 55961 | 12392799-6037187 | -76.8 ± 0.1 | … | … | … | … | … | N | … | … | … | … | n | … |
| 52929 | 12392803-6044541 | -16.6 ± 0.5 | 5593 ± 313 | 4.29 ± 0.24 | -0.01 ± 0.23 | … | … | N | … | … | … | … | n | … |
| 55962 | 12392803-6045028 | -41.6 ± 1.0 | … | … | … | … | … | Y | … | … | … | … | n | … |
| 55963 | 12392819-6039598 | -43.6 ± 0.1 | … | … | … | … | … | N | … | … | … | … | n | … |
| 52930 | 12392824-6041134 | -11.1 ± 0.4 | 5688 ± 98 | 4.47 ± 0.19 | -0.12 ± 0.14 | 49 ± 31 | 1 | N | … | … | … | … | n | NG |
| 52931 | 12392833-6036112 | -41.1 ± 0.3 | 5995 ± 287 | 4.36 ± 0.22 | 0.18 ± 0.17 | … | … | Y | … | … | … | … | n | … |
| 55964 | 12392840-6039443 | -36.5 ± 1.0 | … | … | … | … | … | Y | … | … | … | … | n | … |
| 52932 | 12392842-6037090 | -36.5 ± 0.6 | 6403 ± 264 | 4.14 ± 0.11 | 0.06 ± 0.19 | … | … | Y | … | … | … | … | n | … |
| 55965 | 12392843-6044329 | -36.1 ± 0.1 | 5351 ± 72 | 4.47 ± 0.02 | 0.11 ± 0.16 | … | … | Y | … | … | … | … | n | … |
| 52933 | 12392851-6042121 | -16.7 ± 0.5 | 5396 ± 217 | 4.24 ± 0.36 | 0.02 ± 0.12 | … | … | N | … | … | … | … | n | … |
| 55966 | 12392857-6036301 | -35.3 ± 0.4 | 6617 ± 149 | 3.27 ± 0.05 | … | … | … | N | … | … | … | … | n | … |
| 52970 | 12393512-6036217 | -37.9 ± 0.4 | 5767 ± 30 | 4.32 ± 0.23 | -0.09 ± 0.14 | <63 | 3 | Y | Y | Y | Y | … | Y | … |
| 52971 | 12393513-6039523 | -34.4 ± 0.5 | 6379 ± 268 | 4.54 ± 0.76 | 0.01 ± 0.17 | … | … | N | … | … | … | … | n | … |
| 56026 | 12393520-6047344 | -29.5 ± 2.2 | … | … | … | … | … | N | … | … | … | … | n | … |
| 56027 | 12393527-6037066 | -39.3 ± 0.2 | … | … | … | … | … | Y | … | … | … | … | n | … |
| 52972 | 12393530-6040442 | -17.1 ± 0.3 | 6994 ± 245 | 4.21 ± 0.08 | -0.04 ± 0.20 | 34 ± 15 | 1 | N | … | … | … | … | n | NG |
| 56028 | 12393532-6038396 | -38.7 ± 0.1 | 4143 ± 126 | 1.70 ± 0.17 | -0.02 ± 0.24 | … | … | Y | … | … | … | … | n | … |
| 52973 | 12393537-6039386 | -54.7 ± 0.5 | 5635 ± 85 | 4.01 ± 0.63 | -0.26 ± 0.16 | <115 | 3 | N | … | … | … | … | n | NG |
| 56029 | 12393547-6045476 | -39.3 ± 0.1 | … | … | … | … | … | Y | … | … | … | … | n | … |
| 52974 | 12393548-6035036 | -34.0 ± 0.4 | 5407 ± 461 | 4.41 ± 0.71 | -0.25 ± 0.45 | … | … | N | … | … | … | … | n | … |
| 52975 | 12393556-6034269 | -3.0 ± 5.5 | 5668 ± 158 | … | -0.73 ± 1.11 | … | … | N | … | … | … | … | n | … |
| 52977 | 12393570-6039521 | -10.0 ± 0.4 | 5373 ± 153 | 4.32 ± 0.34 | -0.53 ± 0.22 | … | … | N | … | … | … | … | n | … |
| 56030 | 12393571-6037027 | -39.3 ± 1.1 | … | … | … | … | … | Y | … | … | … | … | n | … |
| 56031 | 12393575-6031417 | -28.9 ± 0.1 | 5799 ± 100 | 3.71 ± 0.07 | -0.65 ± 0.16 | … | … | N | … | … | … | … | n | … |
| 52978 | 12393576-6036266 | -25.0 ± 0.3 | 6039 ± 216 | 4.17 ± 0.21 | -0.35 ± 0.32 | 49 ± 23 | … | N | … | … | … | … | n | NG |
| 52979 | 12393581-6031225 | -14.7 ± 0.8 | 5568 ± 575 | 4.78 ± 0.38 | -0.39 ± 0.29 | … | … | N | … | … | … | … | n | … |
| 52980 | 12393589-6038214 | -43.6 ± 0.2 | 4938 ± 81 | 2.97 ± 0.32 | 0.24 ± 0.25 | … | … | N | … | … | … | … | n | G |
| 56032 | 12393609-6034453 | -45.8 ± 0.2 | … | … | … | … | … | N | … | … | … | … | n | … |
| 56033 | 12393612-6036001 | -41.6 ± 0.1 | 5357 ± 177 | 4.91 ± 0.05 | 0.26 ± 0.22 | … | … | Y | … | … | … | … | n | … |
| 52981 | 12393618-6037564 | -74.4 ± 0.6 | 5447 ± 60 | 4.25 ± 0.32 | -0.33 ± 0.20 | … | … | N | … | … | … | … | n | … |
| 56034 | 12393619-6037270 | -41.8 ± 0.7 | … | … | … | … | … | Y | … | … | … | … | n | … |
| 52982 | 12393629-6034526 | -23.6 ± 0.3 | 5851 ± 169 | 4.48 ± 0.32 | 0.21 ± 0.13 | … | … | N | … | … | … | … | n | … |
| 52983 | 12393634-6040502 | -18.9 ± 0.4 | 5749 ± 92 | 4.17 ± 0.39 | 0.14 ± 0.14 | … | … | N | … | … | … | … | n | … |
| 52984 | 12393638-6035181 | -39.3 ± 0.4 | 5921 ± 216 | 4.13 ± 0.18 | 0.04 ± 0.14 | <127 | 3 | Y | N | Y | Y | … | n | NG |
| 56091 | 12394295-6036023 | -41.1 ± 0.1 | 5283 ± 132 | 4.25 ± 0.04 | 0.60 ± 0.24 | … | … | Y | … | … | … | … | n | … |
| 3377 | 12394307-6039193 | -40.2 ± 0.6 | 4880 ± 117 | 2.76 ± 0.23 | 0.12 ± 0.10 | <7 | 3 | Y | Y | Y | Y | … | Y | … |

| ID | CNAME | RV (km s$^{-1}$) | $T_{\rm eff}$ (K) | logg (dex) | [Fe/H] (dex) | EW(Li)$^a$ (mÅ) | EW(Li) error flag$^b$ | Membership RV | Li | logg | [Fe/H] | Gaia study Cantat-Gaudin$^c$ | Final$^d$ | NMs with Li$^e$ |
|---|---|---|---|---|---|---|---|---|---|---|---|---|---|---|
| 56093 | 12394309-6039279 | -39.0 ± 1.7 | … | … | … | … | … | Y | … | … | … | … | n | … |
| 56094 | 12394311-6040351 | -0.7 ± 0.2 | … | … | … | … | … | N | … | … | … | … | n | … |
| 56095 | 12394313-6041431 | -15.1 ± 0.1 | … | … | … | … | … | N | … | … | … | … | n | … |
| 53165 | 12400721-6035267 | -6.1 ± 0.2 | 4832 ± 127 | 2.90 ± 0.46 | 0.01 ± 0.23 | … | … | N | … | … | … | … | n | G |
| 56096 | 12394341-6038489 | -40.6 ± 0.6 | … | … | … | … | … | Y | … | … | … | … | n | … |
| 53166 | 12400735-6039226 | -37.0 ± 0.4 | 5877 ± 222 | 4.36 ± 0.41 | -0.23 ± 0.14 | … | … | Y | … | … | … | … | n | … |
| 56097 | 12394343-6037329 | -40.5 ± 0.5 | … | … | … | … | … | Y | … | … | … | … | n | … |
| 53167 | 12400752-6034473 | -41.7 ± 0.5 | 5982 ± 377 | 4.07 ± 0.09 | -0.21 ± 0.31 | <48 | 3 | Y | N | Y | Y | N | n | NG |
| 53021 | 12394345-6029273 | -37.9 ± 1.0 | 5435 ± 135 | 3.54 ± 0.42 | -0.25 ± 0.20 | … | … | Y | … | … | … | … | n | … |
| 3398 | 12400754-6035445 | -39.5 ± 0.6 | 4435 ± 115 | 2.10 ± 0.23 | 0.06 ± 0.10 | <5 | 3 | Y | Y | Y | Y | … | Y | … |
| 53022 | 12394352-6036458 | -36.7 ± 0.5 | 5316 ± 231 | 4.09 ± 0.32 | -0.07 ± 0.18 | … | … | Y | … | … | … | … | n | … |
| 53168 | 12400756-6038235 | -31.7 ± 0.3 | 5937 ± 363 | 4.30 ± 0.51 | 0.14 ± 0.22 | 49 ± 32 | 1 | N | … | … | … | … | n | NG |
| 56098 | 12394353-6037129 | -40.6 ± 0.6 | … | … | … | … | … | Y | … | … | … | … | n | … |
| 53023 | 12394353-6038447 | -57.2 ± 0.5 | 5728 ± 146 | 4.33 ± 0.22 | -0.52 ± 0.35 | <65 | 3 | N | … | … | … | … | n | NG |
| 56285 | 12400760-6044397 | -39.7 ± 0.5 | … | … | … | … | … | Y | … | … | … | … | n | … |
| 53024 | 12394365-6038071 | -28.9 ± 0.6 | 5731 ± 104 | 3.81 ± 0.49 | -0.01 ± 0.16 | … | … | N | … | … | … | … | n | … |
| 53169 | 12400761-6034111 | -25.1 ± 1.6 | 4970 ± 186 | 4.29 ± 0.57 | 0.23 ± 0.12 | … | … | N | … | … | … | … | n | … |
| 56099 | 12394372-6037433 | -14.0 ± 0.1 | 6275 ± 131 | 4.26 ± 0.02 | … | … | … | N | … | … | … | … | n | … |
| 56286 | 12400761-6039018 | -6.8 ± 0.2 | 5927 ± 138 | 3.77 ± 0.03 | -0.21 ± 0.25 | … | … | N | … | … | … | … | n | … |
| 56100 | 12394373-6034265 | -40.5 ± 0.7 | … | … | … | … | … | Y | … | … | … | … | n | … |
| 53170 | 12400768-6042178 | -37.0 ± 0.7 | 6805 ± 74 | … | 0.56 ± 0.06 | … | … | Y | … | … | … | … | n | … |
| 53025 | 12394381-6036019 | -32.3 ± 0.3 | 6313 ± 50 | 4.54 ± 0.55 | 0.11 ± 0.15 | … | … | N | … | … | … | … | n | … |
| 56287 | 12400775-6043362 | -40.9 ± 0.1 | 5069 ± 115 | 3.64 ± 0.07 | 0.30 ± 0.16 | … | … | Y | … | … | … | … | n | … |
| 3378 | 12394385-6033165 | -40.1 ± 0.6 | 5004 ± 113 | 2.87 ± 0.23 | 0.09 ± 0.10 | <8 | 3 | Y | Y | Y | Y | … | Y | … |
| 56288 | 12400783-6033028 | -48.8 ± 0.5 | … | … | … | … | … | N | … | … | … | … | n | … |
| 53026 | 12394393-6039396 | 12.8 ± 0.5 | 5697 ± 314 | 3.78 ± 0.32 | 0.07 ± 0.20 | <64 | 3 | N | … | … | … | … | n | NG |
| 56289 | 12400783-6041358 | -26.1 ± 0.1 | 6119 ± 101 | 3.61 ± 0.05 | … | … | … | N | … | … | … | … | n | … |
| 53027 | 12394395-6038427 | -13.5 ± 0.3 | 6207 ± 109 | 4.10 ± 0.08 | 0.18 ± 0.20 | … | … | N | … | … | … | … | n | … |
| 53171 | 12400811-6038262 | -42.4 ± 0.6 | 5562 ± 73 | 4.04 ± 0.32 | -0.05 ± 0.13 | <76 | 3 | Y | Y? | N | Y | … | n | NG |
| 56101 | 12394399-6034351 | -36.1 ± 2.0 | … | … | … | … | … | Y | … | … | … | … | n | … |
| 56290 | 12400816-6033308 | -15.5 ± 0.1 | 5185 ± 112 | 3.98 ± 0.06 | 0.09 ± 0.20 | … | … | N | … | … | … | … | n | … |
| 53028 | 12394405-6042203 | 14.5 ± 0.5 | 6242 ± 322 | 3.91 ± 0.28 | -0.35 ± 0.17 | … | … | N | … | … | … | … | n | … |
| 53172 | 12400824-6033110 | -45.2 ± 0.6 | 6244 ± 394 | 4.66 ± 0.49 | -0.06 ± 0.26 | <98 | 3 | N | … | … | … | … | n | NG |
| 56102 | 12394407-6042114 | -7.8 ± 0.1 | … | … | … | … | … | N | … | … | … | … | n | … |
| 56291 | 12400828-6044424 | -21.0 ± 0.3 | … | … | … | … | … | N | … | … | … | … | n | … |
| 56103 | 12394414-6039423 | -41.7 ± 0.8 | … | … | … | … | … | Y | … | … | … | … | n | … |
| 56292 | 12400837-6038476 | -40.0 ± 0.1 | 5145 ± 107 | 4.11 ± 0.05 | 0.27 ± 0.21 | … | … | Y | … | … | … | … | n | … |
| 53173 | 12400843-6038566 | -39.1 ± 0.4 | 5524 ± 229 | 4.01 ± 0.36 | -0.25 ± 0.23 | … | … | Y | … | … | … | … | n | … |
| 3379 | 12394418-6034410 | -39.6 ± 0.6 | 4963 ± 118 | 2.81 ± 0.23 | 0.15 ± 0.10 | <8 | 3 | Y | Y | Y | Y | … | Y | … |
| 53174 | 12400860-6033138 | -30.1 ± 0.6 | 5658 ± 63 | 3.98 ± 0.19 | -0.52 ± 0.22 | 73 ± 61 | 1 | N | Y | N | N | … | n | NG |
| 56151 | 12394953-6042452 | -39.5 ± 1.9 | … | … | … | … | … | Y | … | … | … | … | n | … |
| 56293 | 12400870-6042344 | -48.9 ± 0.2 | 6535 ± 148 | 4.35 ± 0.01 | … | … | … | N | … | … | … | … | n | … |
| 53068 | 12394959-6034140 | -0.9 ± 0.4 | 5935 ± 49 | 4.63 ± 0.39 | 0.10 ± 0.14 | <133 | 3 | N | … | … | … | … | n | NG |
| 53175 | 12400872-6037051 | -23.1 ± 0.4 | 5876 ± 342 | 4.35 ± 0.08 | 0.11 ± 0.17 | … | … | N | … | … | … | … | n | … |
| 56152 | 12394961-6031577 | -22.4 ± 0.2 | … | … | … | … | … | N | … | … | … | … | n | … |
| 53176 | 12400875-6037572 | -39.9 ± 0.4 | 6726 ± 69 | 4.29 ± 0.14 | 0.22 ± 0.06 | 48 ± 41 | … | Y | Y | Y | Y | N | Y | … |
| 53069 | 12394966-6036300 | -50.1 ± 0.8 | 6289 ± 411 | 4.91 ± 0.73 | 0.01 ± 0.26 | <77 | 3 | N | … | … | … | … | n | NG |
| 53177 | 12400883-6036514 | 8.4 ± 0.2 | 5756 ± 12 | 4.09 ± 0.24 | -0.20 ± 0.18 | 48 ± 15 | … | N | … | … | … | … | n | NG |
| 55661 | 12384055-6037442 | 2.3 ± 0.2 | … | … | … | … | … | N | … | … | … | … | n | … |
| 56153 | 12394977-6034466 | -41.9 ± 0.2 | … | … | … | … | … | Y | … | … | … | … | n | … |
| 53178 | 12400888-6035552 | -26.0 ± 0.5 | 6139 ± 206 | 4.14 ± 0.36 | 0.20 ± 0.17 | <119 | 3 | N | … | … | … | … | n | NG |
| 55662 | 12384072-6033538 | -3.9 ± 0.4 | … | … | … | … | … | N | … | … | … | … | n | … |
| 56340 | 12401888-6038215 | -28.5 ± 1.3 | … | … | … | … | … | N | … | … | … | … | n | … |
| 56154 | 12394983-6038069 | -40.4 ± 0.6 | … | … | … | … | … | Y | … | … | … | … | n | … |
| 55663 | 12384073-6032474 | -55.7 ± 0.2 | 5631 ± 149 | 2.39 ± 0.05 | 0.40 ± 0.25 | … | … | N | … | … | … | … | n | … |
| 56341 | 12401889-6044597 | -10.1 ± 0.3 | … | … | … | … | … | N | … | … | … | … | n | … |
| 56155 | 12395007-6038584 | -37.0 ± 0.9 | … | … | … | … | … | Y | … | … | … | … | n | … |
| 56156 | 12395007-6043316 | -23.3 ± 0.3 | … | … | … | … | … | N | … | … | … | … | n | … |
| 56342 | 12401896-6034407 | -40.6 ± 0.7 | … | … | … | … | … | Y | … | … | … | … | n | … |

**Table C.17.** continued.

| ID | CNAME | RV (km s$^{-1}$) | $T_{\text{eff}}$ (K) | logg (dex) | [Fe/H] (dex) | EW(Li)$^a$ (mÅ) | EW(Li) error flag$^b$ | Membership RV | Li | logg | [Fe/H] | Gaia study Cantat-Gaudin$^c$ | Final$^d$ | NMs with Li$^e$ |
|---|---|---|---|---|---|---|---|---|---|---|---|---|---|---|
| 55664 | 12384105-6032513 | -26.4 ± 0.2 | … | … | … | … | … | N | … | … | … | … | n | … |
| 53071 | 12395009-6038584 | -40.6 ± 0.5 | 6641 ± 186 | 3.99 ± 0.04 | 0.13 ± 0.19 | 53 ± 29 | 1 | Y | Y | Y | Y | Y | Y | … |
| 53226 | 12401939-6044267 | -33.5 ± 0.5 | 5651 ± 498 | 4.74 ± 0.22 | 0.19 ± 0.17 | … | … | N | … | … | … | … | n | … |
| 55665 | 12384108-6037590 | -0.5 ± 0.3 | … | … | … | … | … | N | … | … | … | … | n | … |
| 56157 | 12395015-6035087 | -41.2 ± 0.7 | … | … | … | … | … | Y | … | … | … | … | n | … |
| 53227 | 12401957-6036586 | -13.6 ± 0.4 | 4820 ± 258 | 4.31 ± 0.52 | -0.10 ± 0.17 | 242 ± 78 | 1 | N | … | … | … | … | n | NG |
| 55666 | 12384122-6035233 | -20.3 ± 0.4 | … | … | … | … | … | N | … | … | … | … | n | … |
| 56158 | 12395015-6040229 | -30.6 ± 2.8 | … | … | … | … | … | N | … | … | … | … | n | … |
| 56343 | 12401959-6044030 | -6.7 ± 0.2 | 6117 ± 147 | 3.75 ± 0.04 | … | … | … | N | … | … | … | … | n | … |
| 55667 | 12384162-6046283 | -10.1 ± 0.1 | … | … | … | … | … | N | … | … | … | … | n | … |
| 56159 | 12395016-6029334 | -5.3 ± 0.2 | 6741 ± 148 | 4.91 ± 0.05 | … | … | … | N | … | … | … | … | n | … |
| 56344 | 12401990-6030499 | -36.0 ± 0.4 | … | … | … | … | … | N | … | … | … | … | n | … |
| 55668 | 12384172-6028412 | -49.9 ± 0.2 | 6045 ± 133 | 3.51 ± 0.01 | … | … | … | N | … | … | … | … | n | … |
| 53072 | 12395017-6027518 | -34.9 ± 1.0 | 5911 ± 333 | 4.63 ± 0.34 | 0.11 ± 0.16 | … | … | N | … | … | … | … | n | … |
| 56345 | 12401999-6037472 | -36.7 ± 1.7 | … | … | … | … | … | Y | … | … | … | … | n | … |
| 55669 | 12384182-6041484 | 21.6 ± 0.2 | 5601 ± 148 | 3.60 ± 0.05 | 0.01 ± 0.30 | … | … | N | … | … | … | … | n | … |
| 53073 | 12395021-6037168 | 2.4 ± 0.2 | 6208 ± 86 | 4.53 ± 0.15 | 0.26 ± 0.16 | 64 ± 12 | … | N | … | … | … | … | n | NG |
| 53228 | 12402003-6037524 | -37.2 ± 0.9 | 6806 ± 98 | 3.81 ± 0.18 | 0.33 ± 0.08 | … | … | Y | … | … | … | … | n | … |
| 55670 | 12384237-6041516 | -31.7 ± 0.1 | 6849 ± 149 | 3.93 ± 0.05 | … | … | … | N | … | … | … | … | n | … |
| 56160 | 12395028-6036145 | -42.3 ± 0.9 | … | … | … | … | … | Y | … | … | … | … | n | … |
| 53229 | 12402039-6042428 | -5.0 ± 0.6 | 5877 ± 418 | 4.48 ± 0.30 | -0.22 ± 0.43 | … | … | N | … | … | … | … | n | … |
| 55671 | 12384279-6042169 | -4.5 ± 0.2 | … | … | … | … | … | N | … | … | … | … | n | … |
| 56161 | 12395031-6040417 | -37.3 ± 1.2 | … | … | … | … | … | Y | … | … | … | … | n | … |
| 56346 | 12402045-6046071 | -33.8 ± 0.2 | … | … | … | … | … | N | … | … | … | … | n | … |
| 55672 | 12384280-6032372 | -12.8 ± 0.2 | … | … | … | … | … | N | … | … | … | … | n | … |
| 56162 | 12395033-6031389 | -10.1 ± 0.1 | … | … | … | … | … | N | … | … | … | … | n | … |
| 56347 | 12402052-6046462 | -7.8 ± 0.1 | 4587 ± 119 | 3.02 ± 0.07 | 0.27 ± 0.23 | … | … | N | … | … | … | … | n | … |
| 55673 | 12384324-6039256 | -37.6 ± 0.9 | … | … | … | … | … | Y | … | … | … | … | n | … |
| 53074 | 12395047-6036294 | -40.3 ± 0.6 | 5791 ± 22 | 4.56 ± 0.40 | -0.21 ± 0.28 | … | … | Y | … | … | … | … | n | … |
| 56348 | 12402084-6032301 | -68.8 ± 0.2 | 6003 ± 248 | 3.36 ± 0.18 | … | … | … | N | … | … | … | … | n | … |
| 55674 | 12384368-6047015 | 27.7 ± 0.5 | … | … | … | … | … | N | … | … | … | … | n | … |
| 53075 | 12395050-6032014 | 38.8 ± 0.4 | 4962 ± 72 | 4.42 ± 0.58 | 0.01 ± 0.21 | … | … | N | … | … | … | … | n | … |
| 56349 | 12402114-6038565 | -39.5 ± 0.1 | 5031 ± 167 | 3.99 ± 0.07 | 0.31 ± 0.21 | … | … | Y | … | … | … | … | n | … |
| 52753 | 12384372-6039044 | 7.7 ± 0.5 | 6201 ± 217 | 4.17 ± 0.20 | -0.18 ± 0.21 | 64 ± 59 | 1 | N | … | … | … | … | n | NG |
| 56163 | 12395050-6037543 | -39.1 ± 0.5 | … | … | … | … | … | Y | … | … | … | … | n | … |
| 56350 | 12402160-6034199 | -39.6 ± 1.0 | … | … | … | … | … | Y | … | … | … | … | n | … |
| 53076 | 12395050-6042265 | -39.0 ± 0.5 | 6170 ± 337 | 4.23 ± 0.40 | -0.12 ± 0.42 | <59 | 3 | Y | Y | Y | Y | … | Y | … |
| 56351 | 12402161-6041246 | 18.3 ± 0.1 | 5121 ± 127 | 4.57 ± 0.05 | 0.29 ± 0.25 | … | … | N | … | … | … | … | n | … |
| 55676 | 12384388-6045404 | -33.6 ± 0.2 | … | … | … | … | … | N | … | … | … | … | n | … |
| 53077 | 12395051-6037541 | -38.3 ± 0.3 | 6610 ± 264 | 4.01 ± 0.05 | 0.15 ± 0.19 | 86 ± 32 | 1 | Y | Y | Y | Y | Y | Y | … |
| 56352 | 12402174-6034073 | -40.2 ± 0.1 | … | … | … | … | … | Y | … | … | … | … | n | … |
| 55677 | 12384428-6036433 | -42.3 ± 2.2 | … | … | … | … | … | Y | … | … | … | … | n | … |
| 53078 | 12395063-6035445 | -36.1 ± 0.4 | 5629 ± 147 | 4.09 ± 0.41 | -0.29 ± 0.20 | … | … | Y | … | … | … | … | n | … |
| 56353 | 12402208-6032320 | -18.7 ± 0.1 | … | … | … | … | … | N | … | … | … | … | n | … |
| 55678 | 12384428-6037359 | -30.0 ± 0.1 | 5781 ± 133 | 4.03 ± 0.08 | -0.29 ± 0.20 | … | … | N | … | … | … | … | n | … |
| 55679 | 12384442-6040413 | -17.7 ± 0.2 | … | … | … | … | … | N | … | … | … | … | n | … |
| 53079 | 12395066-6036124 | -41.2 ± 0.5 | 6655 ± 63 | 3.78 ± 0.12 | 0.20 ± 0.05 | 88 ± 40 | … | Y | Y | Y | Y | Y | Y | … |
| 53230 | 12402208-6039582 | 91.1 ± 0.4 | 5571 ± 120 | 3.90 ± 0.27 | -0.56 ± 0.24 | … | … | N | … | … | … | … | n | … |
| 55680 | 12384466-6039483 | -4.4 ± 0.1 | 5957 ± 84 | 4.01 ± 0.05 | -0.02 ± 0.15 | … | … | N | … | … | … | … | n | … |
| 56219 | 12395652-6042097 | -29.3 ± 0.2 | … | … | … | … | … | N | … | … | … | … | n | … |
| 56354 | 12402215-6036188 | -38.9 ± 0.1 | 5179 ± 116 | 4.07 ± 0.05 | 0.02 ± 0.23 | … | … | Y | … | … | … | … | n | … |
| 3390 | 12395654-6039012 | -38.2 ± 0.6 | 4932 ± 122 | 2.83 ± 0.25 | 0.16 ± 0.10 | <7 | 3 | Y | Y | Y | Y | … | Y | … |
| 3399 | 12402227-6037419 | -40.6 ± 0.6 | 4936 ± 121 | 2.77 ± 0.23 | 0.12 ± 0.10 | <9 | 3 | Y | Y | Y | Y | … | Y | … |
| 56220 | 12395658-6044562 | -40.6 ± 0.9 | … | … | … | … | … | Y | … | … | … | … | n | … |
| 53231 | 12402277-6041260 | -24.9 ± 0.4 | 5519 ± 227 | 4.37 ± 0.37 | 0.12 ± 0.31 | … | … | N | … | … | … | … | n | … |
| 56221 | 12395666-6047038 | -40.4 ± 0.9 | … | … | … | … | … | Y | … | … | … | … | n | … |
| 56355 | 12402281-6037000 | 39.2 ± 0.1 | 5377 ± 90 | 3.94 ± 0.05 | -0.17 ± 0.19 | … | … | N | … | … | … | … | n | … |
| 56222 | 12395667-6034015 | -36.0 ± 0.5 | … | … | … | … | … | Y | … | … | … | … | n | … |
| 56356 | 12402284-6040334 | -39.4 ± 0.1 | 4407 ± 91 | 2.45 ± 0.05 | 0.00 ± 0.21 | … | … | Y | … | … | … | … | n | … |

**Table C.17.** continued.

| ID | CNAME | RV (km s$^{-1}$) | $T_{\rm eff}$ (K) | $logg$ (dex) | [Fe/H] (dex) | EW(Li)$^a$ (mÅ) | EW(Li) error flag$^b$ | Membership RV | Li | $logg$ | [Fe/H] | Gaia study Cantat-Gaudin$^c$ | Final$^d$ | NMs with Li$^e$ |
|---|---|---|---|---|---|---|---|---|---|---|---|---|---|---|
| 56223 | 12395668-6043229 | -37.8 ± 1.7 | … | … | … | … | … | Y | … | … | … | … | n | … |
| 56423 | 12404901-6035346 | -12.2 ± 0.3 | … | … | … | … | … | N | … | … | … | … | n | … |
| 56224 | 12395682-6031599 | -63.2 ± 0.1 | … | … | … | … | … | N | … | … | … | … | n | … |
| 56424 | 12404909-6033353 | -39.6 ± 0.1 | 5321 ± 109 | 3.94 ± 0.05 | 0.21 ± 0.21 | … | … | Y | … | … | … | … | n | … |
| 53261 | 12404922-6042263 | -47.2 ± 0.6 | 6618 ± 1007 | 4.81 ± 0.20 | -0.02 ± 0.05 | … | … | N | … | … | … | … | n | … |
| 56225 | 12395688-6038336 | -39.2 ± 0.5 | … | … | … | … | … | Y | … | … | … | … | n | … |
| 56425 | 12404956-6043173 | 13.3 ± 0.1 | 5561 ± 122 | 4.73 ± 0.02 | 0.66 ± 0.22 | … | … | N | … | … | … | … | n | … |
| 53119 | 12395698-6041397 | -5.9 ± 0.5 | 5753 ± 51 | 4.73 ± 0.36 | -0.14 ± 0.19 | … | … | N | … | … | … | … | n | … |
| 53262 | 12404958-6032338 | -15.1 ± 0.5 | 5595 ± 233 | 4.37 ± 0.09 | 0.43 ± 0.26 | … | … | N | … | … | … | … | n | … |
| 56226 | 12395699-6034285 | -38.9 ± 0.1 | 4829 ± 124 | 2.96 ± 0.09 | 0.13 ± 0.29 | … | … | Y | … | … | … | … | n | … |
| 56426 | 12404992-6035062 | -5.1 ± 0.1 | … | … | … | … | … | N | … | … | … | … | n | … |
| 3391 | 12395711-6039335 | -40.6 ± 0.6 | 4964 ± 116 | 2.83 ± 0.23 | 0.14 ± 0.09 | … | … | Y | … | … | … | … | n | G |
| 56427 | 12405044-6036083 | 51.7 ± 0.1 | 4559 ± 103 | 3.49 ± 0.06 | -0.04 ± 0.29 | … | … | N | … | … | … | … | n | … |
| 53120 | 12395723-6035207 | 24.2 ± 0.4 | 5232 ± 275 | 4.16 ± 0.31 | -0.10 ± 0.18 | … | … | N | … | … | … | … | n | … |
| 56428 | 12405096-6037009 | -58.2 ± 0.2 | 5897 ± 129 | 2.70 ± 0.05 | 0.16 ± 0.26 | … | … | N | … | … | … | … | n | … |
| 56227 | 12395725-6035594 | -35.2 ± 1.1 | … | … | … | … | … | N | … | … | … | … | n | … |
| 56429 | 12405126-6035449 | -25.4 ± 0.2 | … | … | … | … | … | N | … | … | … | … | n | … |
| 56228 | 12395725-6041162 | -14.2 ± 0.1 | 5961 ± 128 | 4.73 ± 0.06 | 0.10 ± 0.19 | … | … | N | … | … | … | … | n | … |
| 56430 | 12405165-6035191 | -39.5 ± 0.9 | … | … | … | … | … | Y | … | … | … | … | n | … |
| 56431 | 12405255-6036358 | -30.1 ± 0.1 | … | … | … | … | … | N | … | … | … | … | n | … |
| 56229 | 12395730-6034395 | -35.0 ± 1.0 | … | … | … | … | … | N | … | … | … | … | n | … |
| 53263 | 12405329-6044360 | 21.4 ± 1.4 | 5839 ± 219 | 4.22 ± 0.87 | -0.75 ± 0.92 | … | … | N | … | … | … | … | n | … |
| 53121 | 12395734-6035236 | -38.4 ± 0.3 | 6629 ± 46 | 4.09 ± 0.09 | 0.15 ± 0.04 | 67 ± 28 | … | Y | Y | Y | Y | Y | Y | … |
| 56432 | 12405333-6035148 | -35.4 ± 0.1 | … | … | … | … | … | N | … | … | … | … | n | … |
| 56230 | 12395757-6039128 | -26.5 ± 0.2 | … | … | … | … | … | N | … | … | … | … | n | … |
| 56433 | 12405376-6039313 | -24.2 ± 1.6 | … | … | … | … | … | N | … | … | … | … | n | … |
| 56231 | 12395773-6030145 | 34.4 ± 0.2 | 6049 ± 150 | 3.47 ± 0.03 | … | … | … | N | … | … | … | … | n | … |
| 53264 | 12405522-6042139 | -12.4 ± 0.5 | 5321 ± 290 | 4.21 ± 1.04 | 0.40 ± 0.58 | … | … | N | … | … | … | … | n | … |
| 56232 | 12395775-6039498 | -36.9 ± 0.8 | … | … | … | … | … | Y | … | … | … | … | n | … |
| 56233 | 12395783-6041403 | -10.1 ± 0.1 | … | … | … | … | … | N | … | … | … | … | n | … |
| 56234 | 12395795-6033044 | -40.2 ± 1.4 | … | … | … | … | … | Y | … | … | … | … | n | … |
| 56235 | 12395806-6038445 | -19.8 ± 0.2 | … | … | … | … | … | N | … | … | … | … | n | … |
| 53122 | 12395806-6040301 | -65.3 ± 0.3 | 5311 ± 478 | 3.89 ± 0.89 | -0.79 ± 0.88 | … | … | N | … | … | … | … | n | … |
| 55852 | 12391458-6044123 | -39.6 ± 0.2 | … | … | … | … | … | Y | … | … | … | … | n | … |
| 52837 | 12391471-6035010 | -24.4 ± 0.6 | 5939 ± 377 | 3.76 ± 0.30 | -0.47 ± 0.48 | <71 | 3 | N | … | … | … | … | n | NG |
| 55591 | 12381637-6042512 | -16.2 ± 0.1 | 5457 ± 117 | 4.77 ± 0.02 | 0.34 ± 0.20 | … | … | N | … | … | … | … | n | … |
| 52838 | 12391473-6035353 | -49.6 ± 0.7 | 6105 ± 353 | 4.20 ± 0.69 | 0.08 ± 0.41 | <81 | 3 | N | … | … | … | … | n | NG |
| 55592 | 12381718-6045311 | 8.3 ± 0.1 | … | … | … | … | … | N | … | … | … | … | n | … |
| 55866 | 12391674-6037146 | -36.8 ± 0.3 | … | … | … | … | … | Y | … | … | … | … | n | … |
| 55867 | 12391678-6030289 | -41.6 ± 0.6 | … | … | … | … | … | Y | … | … | … | … | n | … |
| 55593 | 12381775-6043594 | -15.6 ± 0.2 | … | … | … | … | … | N | … | … | … | … | n | … |
| 52849 | 12391683-6036119 | 2.1 ± 0.5 | 5850 ± 66 | 4.48 ± 0.16 | 0.32 ± 0.27 | … | … | N | … | … | … | … | n | … |
| 55618 | 12382731-6038142 | -45.1 ± 0.2 | … | … | … | … | … | N | … | … | … | … | n | … |
| 55619 | 12382787-6046010 | -81.7 ± 0.1 | 5023 ± 101 | 3.85 ± 0.03 | -0.09 ± 0.22 | … | … | N | … | … | … | … | n | … |
| 52850 | 12391686-6035154 | 3.9 ± 0.4 | 5811 ± 340 | 4.59 ± 0.33 | 0.41 ± 0.27 | <73 | 3 | N | … | … | … | … | n | NG |
| 55620 | 12382899-6036523 | -48.8 ± 0.2 | … | … | … | … | … | N | … | … | … | … | n | … |
| 55868 | 12391692-6035303 | -40.6 ± 0.1 | 5369 ± 122 | 3.35 ± 0.05 | 0.51 ± 0.26 | … | … | Y | … | … | … | … | n | … |
| 55621 | 12382925-6039103 | -39.2 ± 0.3 | … | … | … | … | … | Y | … | … | … | … | n | … |
| 55869 | 12391711-6043337 | -40.4 ± 0.3 | … | … | … | … | … | Y | … | … | … | … | n | … |
| 55870 | 12391725-6041080 | -41.3 ± 1.2 | … | … | … | … | … | Y | … | … | … | … | n | … |
| 55623 | 12382945-6039022 | -15.1 ± 0.1 | … | … | … | … | … | N | … | … | … | … | n | … |
| 52851 | 12391729-6037359 | -27.1 ± 0.3 | 5378 ± 282 | 4.13 ± 0.11 | -0.01 ± 0.13 | <103 | 3 | N | … | … | … | … | n | NG |
| 55624 | 12382956-6044356 | 1.1 ± 0.3 | … | … | … | … | … | N | … | … | … | … | n | … |
| 55871 | 12391734-6047080 | -12.4 ± 0.2 | … | … | … | … | … | N | … | … | … | … | n | … |
| 52750 | 12382989-6039332 | -19.2 ± 0.9 | 5767 ± 296 | 4.20 ± 0.33 | 0.12 ± 0.17 | … | … | N | … | … | … | … | n | … |
| 52852 | 12391739-6036496 | -13.5 ± 1.1 | 6114 ± 428 | 4.87 ± 0.82 | -0.78 ± 0.66 | <67 | 3 | N | … | … | … | … | n | NG |
| 55625 | 12383026-6037281 | -34.6 ± 0.3 | … | … | … | … | … | N | … | … | … | … | n | … |
| 52853 | 12391741-6038144 | -49.4 ± 0.8 | 6515 ± 193 | 4.01 ± 0.02 | -0.10 ± 0.19 | … | … | N | … | … | … | … | n | … |
| 55626 | 12383139-6033185 | -39.2 ± 0.3 | … | … | … | … | … | Y | … | … | … | … | n | … |

**Table C.17.** continued.

| ID | CNAME | RV (km s$^{-1}$) | $T_{\text{eff}}$ (K) | $\log g$ (dex) | [Fe/H] (dex) | $EW(\text{Li})^a$ (mÅ) | $EW(\text{Li})$ error flag$^b$ | Membership RV | Li | $\log g$ | [Fe/H] | Gaia study Cantat-Gaudin$^c$ | Final$^d$ | NMs with Li$^e$ |
|---|---|---|---|---|---|---|---|---|---|---|---|---|---|---|
| 52854 | 12391762-6035230 | 76.1 ± 0.5 | 5292 ± 663 | 4.18 ± 0.37 | 0.35 ± 0.66 | ... | ... | N | ... | ... | ... | ... | n | ... |
| 52751 | 12383141-6032157 | -0.8 ± 0.7 | 5932 ± 210 | 4.49 ± 0.35 | -0.18 ± 0.30 | <71 | 3 | N | ... | ... | ... | ... | n | NG |
| 55627 | 12383159-6038393 | 6.6 ± 0.1 | 5629 ± 108 | 3.47 ± 0.08 | -0.23 ± 0.19 | ... | ... | N | ... | ... | ... | ... | n | ... |
| 52855 | 12391771-6035086 | 15.6 ± 0.3 | 5389 ± 207 | 4.23 ± 0.43 | 0.57 ± 0.37 | <49 | 3 | N | ... | ... | ... | ... | n | NG |
| 55628 | 12383166-6039058 | -40.4 ± 0.3 | ... | ... | ... | ... | ... | Y | ... | ... | ... | ... | n | ... |
| 55873 | 12391779-6036052 | -40.2 ± 0.4 | ... | ... | ... | ... | ... | Y | ... | ... | ... | ... | n | ... |
| 55629 | 12383199-6030083 | -10.5 ± 0.2 | ... | ... | ... | ... | ... | N | ... | ... | ... | ... | n | ... |
| 55630 | 12383206-6040035 | -17.3 ± 0.2 | 5589 ± 116 | 2.44 ± 0.05 | 0.01 ± 0.26 | ... | ... | N | ... | ... | ... | ... | n | ... |
| 55874 | 12391781-6036397 | -39.3 ± 0.4 | ... | ... | ... | ... | ... | Y | ... | ... | ... | ... | n | ... |
| 55631 | 12383299-6039363 | -40.4 ± 0.5 | ... | ... | ... | ... | ... | Y | ... | ... | ... | ... | n | ... |
| 55875 | 12391784-6031266 | -17.4 ± 0.1 | ... | ... | ... | ... | ... | N | ... | ... | ... | ... | n | ... |
| 55632 | 12383319-6042136 | -39.7 ± 0.1 | 5497 ± 124 | 4.15 ± 0.05 | 0.65 ± 0.24 | ... | ... | Y | ... | ... | ... | ... | n | ... |
| 52856 | 12391786-6039140 | -58.3 ± 0.5 | 5092 ± 375 | 4.37 ± 0.79 | -0.15 ± 0.21 | <84 | 3 | N | ... | ... | ... | ... | n | NG |
| 52857 | 12391794-6038098 | -64.7 ± 0.8 | 5761 ± 102 | 4.80 ± 0.85 | -0.40 ± 0.18 | <98 | 3 | N | ... | ... | ... | ... | n | NG |
| 55634 | 12383408-6038434 | -7.8 ± 0.1 | 5295 ± 130 | 4.26 ± 0.05 | -0.12 ± 0.26 | ... | ... | N | ... | ... | ... | ... | n | ... |
| 55876 | 12391795-6038497 | -41.8 ± 2.0 | ... | ... | ... | ... | ... | Y | ... | ... | ... | ... | n | ... |
| 55635 | 12383433-6036316 | -41.9 ± 1.1 | ... | ... | ... | ... | ... | Y | ... | ... | ... | ... | n | ... |
| 52858 | 12391824-6036261 | 2.7 ± 0.3 | 5789 ± 432 | 4.26 ± 0.21 | 0.33 ± 0.18 | <34 | 3 | N | ... | ... | ... | ... | n | NG |
| 55636 | 12383466-6036574 | -38.2 ± 0.3 | ... | ... | ... | ... | ... | Y | ... | ... | ... | ... | n | ... |
| 52859 | 12391832-6032159 | -29.9 ± 0.5 | 5844 ± 123 | 3.93 ± 0.33 | 0.13 ± 0.18 | 87 ± 87 | 1 | N | ... | ... | ... | ... | n | NG |
| 55637 | 12383512-6030182 | 62.1 ± 0.2 | 5497 ± 142 | 3.33 ± 0.05 | -0.24 ± 0.22 | ... | ... | N | ... | ... | ... | ... | n | ... |
| 55877 | 12391832-6035229 | -50.2 ± 0.1 | ... | ... | ... | ... | ... | N | ... | ... | ... | ... | n | ... |
| 55638 | 12383522-6038274 | -54.8 ± 0.1 | 5245 ± 118 | 4.22 ± 0.03 | 0.47 ± 0.21 | ... | ... | N | ... | ... | ... | ... | n | ... |
| 55878 | 12391833-6041042 | -40.3 ± 0.1 | 5385 ± 88 | 3.72 ± 0.02 | 0.23 ± 0.18 | ... | ... | Y | ... | ... | ... | ... | n | ... |
| 55639 | 12383553-6034155 | -30.5 ± 0.3 | ... | ... | ... | ... | ... | N | ... | ... | ... | ... | n | ... |
| 55903 | 12392162-6044179 | -41.4 ± 0.1 | 5371 ± 142 | 4.24 ± 0.02 | 0.14 ± 0.21 | ... | ... | Y | ... | ... | ... | ... | n | ... |
| 55681 | 12384467-6037540 | -40.2 ± 1.5 | ... | ... | ... | ... | ... | Y | ... | ... | ... | ... | n | ... |
| 52884 | 12392166-6035216 | -30.1 ± 0.4 | 5566 ± 132 | 4.28 ± 0.34 | -0.24 ± 0.17 | <59 | 3 | N | ... | ... | ... | ... | n | NG |
| 55682 | 12384467-6041494 | -36.4 ± 0.4 | ... | ... | ... | ... | ... | Y | ... | ... | ... | ... | n | ... |
| 55904 | 12392167-6033463 | -39.8 ± 0.5 | ... | ... | ... | ... | ... | Y | ... | ... | ... | ... | n | ... |
| 55683 | 12384483-6036022 | -50.6 ± 0.4 | ... | ... | ... | ... | ... | N | ... | ... | ... | ... | n | ... |
| 55905 | 12392175-6037567 | 1.2 ± 0.2 | ... | ... | ... | ... | ... | N | ... | ... | ... | ... | n | ... |
| 55684 | 12384495-6041542 | -39.4 ± 0.3 | ... | ... | ... | ... | ... | Y | ... | ... | ... | ... | n | ... |
| 55685 | 12384500-6037069 | -40.5 ± 1.3 | ... | ... | ... | ... | ... | Y | ... | ... | ... | ... | n | ... |
| 55906 | 12392180-6046293 | -42.9 ± 0.1 | 5109 ± 94 | 2.73 ± 0.06 | -0.25 ± 0.20 | ... | ... | Y | ... | ... | ... | ... | n | ... |
| 55686 | 12384510-6037516 | -15.8 ± 0.1 | 6195 ± 109 | 4.54 ± 0.04 | ... | ... | ... | N | ... | ... | ... | ... | n | ... |
| 55907 | 12392188-6036285 | -38.0 ± 1.1 | ... | ... | ... | ... | ... | Y | ... | ... | ... | ... | n | ... |
| 52754 | 12384530-6037449 | -38.8 ± 0.4 | 5377 ± 285 | 4.14 ± 1.13 | -0.16 ± 0.16 | ... | ... | Y | ... | ... | ... | ... | n | ... |
| 55908 | 12392195-6039077 | -41.1 ± 0.1 | 5517 ± 117 | 4.91 ± 0.05 | 0.20 ± 0.21 | ... | ... | Y | ... | ... | ... | ... | n | ... |
| 55687 | 12384577-6038495 | -42.1 ± 0.1 | 5439 ± 109 | 4.33 ± 0.05 | 0.43 ± 0.21 | ... | ... | Y | ... | ... | ... | ... | n | ... |
| 55688 | 12384588-6038212 | -26.7 ± 0.3 | ... | ... | ... | ... | ... | N | ... | ... | ... | ... | n | ... |
| 55909 | 12392200-6035371 | -45.4 ± 0.8 | ... | ... | ... | ... | ... | N | ... | ... | ... | ... | n | ... |
| 55689 | 12384598-6035204 | -0.6 ± 0.1 | 6537 ± 149 | 4.32 ± 0.02 | ... | ... | ... | N | ... | ... | ... | ... | n | ... |
| 52885 | 12392214-6035104 | -39.1 ± 0.5 | 6115 ± 344 | 4.39 ± 0.19 | -0.02 ± 0.23 | 99 ± 58 | 1 | Y | Y | Y | Y | ... | Y | ... |
| 55690 | 12384603-6042173 | 10.7 ± 0.1 | 6189 ± 149 | 4.21 ± 0.07 | ... | ... | ... | N | ... | ... | ... | ... | n | ... |
| 55910 | 12392221-6045006 | -39.1 ± 1.1 | ... | ... | ... | ... | ... | Y | ... | ... | ... | ... | n | ... |
| 55691 | 12384629-6042037 | 6.7 ± 0.1 | 6365 ± 141 | 4.61 ± 0.05 | ... | ... | ... | N | ... | ... | ... | ... | n | ... |
| 55911 | 12392222-6045587 | -12.7 ± 0.1 | 5477 ± 569 | 4.91 ± 0.05 | 0.28 ± 0.28 | ... | ... | N | ... | ... | ... | ... | n | ... |
| 52755 | 12384634-6036553 | -31.1 ± 0.2 | 5433 ± 71 | 3.61 ± 0.36 | -0.29 ± 0.15 | ... | ... | N | ... | ... | ... | ... | n | ... |
| 55912 | 12392227-6030373 | -15.3 ± 0.2 | ... | ... | ... | ... | ... | N | ... | ... | ... | ... | n | ... |
| 55692 | 12384656-6043599 | -23.0 ± 2.2 | ... | ... | ... | ... | ... | N | ... | ... | ... | ... | n | ... |
| 56117 | 12394553-6040433 | -38.9 ± 0.3 | ... | ... | ... | ... | ... | Y | ... | ... | ... | ... | n | ... |
| 55913 | 12392227-6046371 | 5.5 ± 0.1 | ... | ... | ... | ... | ... | N | ... | ... | ... | ... | n | ... |
| 52756 | 12384675-6038103 | -37.8 ± 0.6 | 6239 ± 320 | 4.43 ± 0.22 | 0.08 ± 0.25 | 64 ± 48 | 1 | Y | Y | Y | Y | Y | Y | ... |
| 56118 | 12394557-6035434 | -41.7 ± 0.2 | ... | ... | ... | ... | ... | Y | ... | ... | ... | ... | n | ... |
| 55914 | 12392231-6041346 | 2.5 ± 0.2 | ... | ... | ... | ... | ... | N | ... | ... | ... | ... | n | ... |
| 55693 | 12384676-6038146 | 11.1 ± 0.1 | 5579 ± 99 | 4.13 ± 0.07 | 0.16 ± 0.15 | ... | ... | N | ... | ... | ... | ... | n | ... |
| 56119 | 12394558-6039591 | -40.7 ± 0.1 | ... | ... | ... | ... | ... | Y | ... | ... | ... | ... | n | ... |
| 55915 | 12392251-6032427 | 2.9 ± 0.2 | ... | ... | ... | ... | ... | N | ... | ... | ... | ... | n | ... |

**Table C.17.** continued.

| ID | CNAME | RV (km s$^{-1}$) | $T_\text{eff}$ (K) | $logg$ (dex) | [Fe/H] (dex) | EW(Li)$^a$ (mÅ) | EW(Li) error flag$^b$ | Membership RV | Li | $logg$ | [Fe/H] | Gaia study Cantat-Gaudin$^c$ | Final$^d$ | NMs with Li$^e$ |
|---|---|---|---|---|---|---|---|---|---|---|---|---|---|---|
| 52757 | 12384692-6027388 | -399.9 ± 6.3 | … | … | … | … | … | N | … | … | … | … | n | … |
| 53045 | 12394676-6035103 | -38.1 ± 0.3 | 6208 ± 183 | 4.23 ± 0.35 | -0.10 ± 0.29 | 43 ± 29 | 1 | Y | Y | Y | Y | … | Y | … |
| 55916 | 12392251-6039254 | -24.2 ± 0.2 | 5779 ± 76 | 3.15 ± 0.01 | -0.28 ± 0.28 | … | … | N | … | … | … | … | n | … |
| 55694 | 12384697-6036204 | -42.6 ± 1.4 | … | … | … | … | … | Y | … | … | … | … | n | … |
| 3383 | 12394688-6033540 | -40.3 ± 0.6 | 4961 ± 116 | 2.90 ± 0.23 | 0.14 ± 0.09 | <8 | 3 | Y | Y | Y | Y | … | Y | … |
| 52886 | 12392258-6034477 | -53.7 ± 3.5 | 6629 ± 178 | … | 0.20 ± 0.14 | <31 | 3 | N | … | … | … | … | n | … |
| 55695 | 12384698-6038378 | -19.4 ± 0.1 | 5901 ± 135 | 3.70 ± 0.05 | 0.01 ± 0.18 | … | … | N | … | … | … | … | n | … |
| 53046 | 12394705-6039183 | -11.5 ± 0.6 | 5604 ± 101 | 4.19 ± 0.43 | -0.74 ± 0.56 | 90 ± 52 | 1 | N | … | … | … | Y | n | NG |
| 52887 | 12392282-6039194 | -39.2 ± 0.4 | 6209 ± 316 | 4.37 ± 0.29 | 0.12 ± 0.21 | 58 ± 56 | 1 | Y | Y | Y | Y | … | Y | … |
| 56133 | 12394706-6031138 | -19.6 ± 0.3 | … | … | … | … | … | N | … | … | … | … | n | … |
| 55696 | 12384720-6035414 | -34.0 ± 0.2 | … | … | … | … | … | N | … | … | … | … | n | … |
| 55917 | 12392289-6042506 | 9.3 ± 0.1 | … | … | … | … | … | N | … | … | … | … | n | … |
| 56134 | 12394708-6034195 | -18.1 ± 0.4 | … | … | … | … | … | N | … | … | … | … | n | … |
| 55697 | 12384725-6042292 | -33.2 ± 1.9 | … | … | … | … | … | N | … | … | … | … | n | … |
| 55918 | 12392303-6045514 | -18.4 ± 0.2 | … | … | … | … | … | N | … | … | … | … | n | … |
| 56135 | 12394709-6037518 | -40.1 ± 0.1 | 5411 ± 294 | 4.09 ± 0.27 | 0.55 ± 0.31 | … | … | Y | … | … | … | … | n | … |
| 55919 | 12392305-6044193 | -41.7 ± 2.0 | … | … | … | … | … | Y | … | … | … | … | n | … |
| 53047 | 12394715-6035172 | -48.5 ± 0.6 | 5999 ± 125 | 4.35 ± 0.22 | -0.13 ± 0.30 | 97 ± 33 | 1 | N | … | … | … | N | n | NG |
| 55920 | 12392309-6044090 | -26.9 ± 0.1 | 6067 ± 150 | 3.95 ± 0.05 | … | … | … | N | … | … | … | … | n | … |
| 55699 | 12384747-6038043 | -31.7 ± 0.3 | … | … | … | … | … | N | … | … | … | … | n | … |
| 3384 | 12394715-6040584 | -38.7 ± 0.6 | 4586 ± 123 | 2.29 ± 0.24 | 0.05 ± 0.10 | 7 ± 1 | … | Y | Y | Y | Y | … | Y | … |
| 52888 | 12392323-6038533 | -18.6 ± 0.2 | 6075 ± 125 | 4.29 ± 0.45 | -0.08 ± 0.16 | 55 ± 13 | … | N | … | … | … | … | n | NG |
| 52758 | 12384753-6037429 | -43.9 ± 0.8 | 5612 ± 188 | 3.99 ± 0.25 | -0.46 ± 0.20 | … | … | N | … | … | … | … | n | … |
| 56136 | 12394727-6033137 | -39.2 ± 0.5 | … | … | … | … | … | Y | … | … | … | … | n | … |
| 55921 | 12392341-6037529 | 18.9 ± 0.9 | … | … | … | … | … | N | … | … | … | … | n | … |
| 55720 | 12385208-6043548 | -47.9 ± 0.1 | 6421 ± 113 | 3.67 ± 0.05 | … | … | … | N | … | … | … | … | n | … |
| 56137 | 12394727-6042297 | -47.2 ± 0.4 | … | … | … | … | … | N | … | … | … | … | n | … |
| 52903 | 12392452-6035361 | -34.1 ± 0.2 | 4462 ± 165 | 2.42 ± 0.32 | 0.29 ± 0.28 | 122 ± 25 | 1 | N | N | Y | Y | Y | Y | … |
| 55721 | 12385210-6036423 | 16.3 ± 0.2 | 6017 ± 336 | 3.26 ± 0.77 | … | … | … | N | … | … | … | … | n | … |
| 56138 | 12394734-6034562 | -36.4 ± 0.9 | … | … | … | … | … | Y | … | … | … | … | n | … |
| 55722 | 12385212-6042106 | -39.1 ± 0.3 | … | … | … | … | … | Y | … | … | … | … | n | … |
| 55929 | 12392455-6040579 | -53.1 ± 0.1 | 6047 ± 122 | 3.51 ± 0.03 | … | … | … | N | … | … | … | … | n | … |
| 53048 | 12394740-6035001 | -19.4 ± 0.4 | 6127 ± 414 | 4.44 ± 0.13 | 0.23 ± 0.39 | … | … | N | … | … | … | … | n | … |
| 55723 | 12385217-6038154 | -1.0 ± 0.1 | 5947 ± 107 | 4.76 ± 0.08 | -0.01 ± 0.14 | … | … | N | … | … | … | … | n | … |
| 56139 | 12394741-6038411 | -39.2 ± 0.1 | 4995 ± 108 | 3.49 ± 0.01 | 0.36 ± 0.23 | … | … | Y | … | … | … | … | n | … |
| 55724 | 12385234-6041533 | -26.2 ± 0.2 | 5579 ± 117 | 3.37 ± 0.07 | -0.61 ± 0.21 | … | … | N | … | … | … | … | n | … |
| 52904 | 12392480-6038245 | -39.5 ± 0.5 | 5515 ± 319 | 4.78 ± 0.53 | -0.03 ± 0.24 | … | … | Y | … | … | … | … | n | … |
| 53049 | 12394742-6036440 | -36.4 ± 0.5 | 6063 ± 379 | 4.42 ± 0.24 | 0.05 ± 0.24 | … | … | Y | … | … | … | … | n | … |
| 55725 | 12385236-6036408 | -35.5 ± 0.4 | … | … | … | … | … | N | … | … | … | … | n | … |
| 52763 | 12385250-6036353 | 2.9 ± 0.2 | 6189 ± 91 | 4.33 ± 0.24 | 0.29 ± 0.21 | … | … | N | … | … | … | … | n | … |
| 55931 | 12392485-6036239 | -40.2 ± 0.1 | 5233 ± 99 | 3.96 ± 0.05 | 0.04 ± 0.20 | … | … | Y | … | … | … | … | n | … |
| 3385 | 12394742-6038411 | -38.4 ± 0.6 | 4924 ± 117 | 2.77 ± 0.23 | 0.08 ± 0.10 | <9 | 3 | Y | Y | Y | Y | … | Y | … |
| 55726 | 12385253-6040091 | -49.9 ± 2.3 | … | … | … | … | … | N | … | … | … | … | n | … |
| 55932 | 12392487-6036469 | -37.4 ± 0.1 | 5259 ± 79 | 3.86 ± 0.05 | -0.04 ± 0.18 | … | … | Y | … | … | … | … | n | … |
| 53050 | 12394776-6040044 | -40.6 ± 0.4 | 6710 ± 58 | 3.95 ± 0.11 | 0.43 ± 0.05 | 106 ± 55 | … | Y | Y | Y | Y | Y | Y | … |
| 52905 | 12392497-6038459 | -45.3 ± 0.3 | 6277 ± 422 | 4.50 ± 0.34 | 0.41 ± 0.32 | … | … | N | … | … | … | … | n | … |
| 55727 | 12385271-6031067 | -14.8 ± 0.2 | … | … | … | … | … | N | … | … | … | … | n | … |
| 56372 | 12402754-6041047 | 13.4 ± 0.9 | … | … | … | … | … | N | … | … | … | … | n | … |
| 56140 | 12394805-6038005 | -39.1 ± 0.2 | … | … | … | … | … | Y | … | … | … | … | n | … |
| 55728 | 12385289-6044553 | -12.8 ± 0.3 | … | … | … | … | … | N | … | … | … | … | n | … |
| 55933 | 12392504-6038133 | -40.9 ± 0.2 | … | … | … | … | … | Y | … | … | … | … | n | … |
| 56373 | 12402795-6036457 | -40.3 ± 0.4 | … | … | … | … | … | Y | … | … | … | … | n | … |
| 52764 | 12385296-6033398 | -38.5 ± 0.3 | 6857 ± 64 | 4.08 ± 0.12 | 0.14 ± 0.06 | 83 ± 41 | … | Y | Y | Y | Y | Y | Y | … |
| 53051 | 12394807-6034052 | 30.2 ± 0.4 | 5257 ± 102 | 4.58 ± 0.20 | -0.03 ± 0.16 | … | … | N | … | … | … | … | n | … |
| 55934 | 12392509-6039331 | -39.0 ± 0.6 | … | … | … | … | … | Y | … | … | … | … | n | … |
| 52765 | 12385298-6033591 | -3.5 ± 0.3 | 6070 ± 264 | 4.33 ± 0.13 | 0.02 ± 0.19 | 52 ± 13 | 1 | N | … | … | … | … | n | NG |
| 56374 | 12402816-6042433 | -26.5 ± 0.6 | … | … | … | … | … | N | … | … | … | … | n | … |
| 53052 | 12394808-6038460 | -37.4 ± 0.2 | 5972 ± 120 | 3.87 ± 0.30 | -0.54 ± 0.19 | … | … | Y | … | … | … | … | n | … |
| 55935 | 12392512-6039458 | -43.9 ± 2.1 | … | … | … | … | … | N | … | … | … | … | n | … |

**Table C.17.** continued.

| ID | CNAME | RV (km s$^{-1}$) | $T_{\text{eff}}$ (K) | $\log g$ (dex) | [Fe/H] (dex) | $EW(\text{Li})^a$ (mÅ) | $EW(\text{Li})$ error flag$^b$ | Membership RV | Li | $\log g$ | [Fe/H] | Gaia study Cantat-Gaudin$^c$ | Final$^d$ | NMs with Li$^e$ |
|---|---|---|---|---|---|---|---|---|---|---|---|---|---|---|
| 55729 | 12385321-6041598 | -8.0 ± 0.1 | 5239 ± 111 | 3.65 ± 0.05 | 0.44 ± 0.17 | … | … | N | … | … | … | … | n | … |
| 56390 | 12403330-6038052 | -39.6 ± 0.1 | 5637 ± 136 | 4.74 ± 0.05 | 0.65 ± 0.25 | … | … | Y | … | … | … | … | n | … |
| 56141 | 12394808-6043235 | -35.9 ± 1.6 | … | … | … | … | … | N | … | … | … | … | n | … |
| 55936 | 12392517-6040077 | -41.9 ± 1.8 | … | … | … | … | … | Y | … | … | … | … | n | … |
| 52766 | 12385331-6040329 | -24.5 ± 0.3 | 5428 ± 59 | 4.19 ± 0.38 | 0.13 ± 0.18 | … | … | N | … | … | … | … | n | … |
| 56391 | 12403391-6030573 | -73.6 ± 0.2 | … | … | … | … | … | N | … | … | … | … | n | … |
| 53053 | 12394809-6044208 | -21.1 ± 0.4 | 5909 ± 274 | 4.36 ± 0.08 | 0.14 ± 0.12 | <85 | 3 | N | … | … | … | … | n | NG |
| 52906 | 12392518-6034222 | 2.5 ± 0.6 | 5134 ± 68 | 3.83 ± 0.34 | -0.25 ± 0.15 | … | … | N | … | … | … | … | n | … |
| 56392 | 12403416-6038006 | -9.0 ± 0.1 | … | … | … | … | … | N | … | … | … | … | n | … |
| 55730 | 12385336-6042534 | -51.2 ± 0.2 | … | … | … | … | … | N | … | … | … | … | n | … |
| 56164 | 12395073-6039166 | -32.3 ± 0.2 | … | … | … | … | … | N | … | … | … | … | n | … |
| 52907 | 12392519-6040077 | -38.3 ± 0.4 | 6856 ± 267 | 3.96 ± 0.08 | 0.06 ± 0.20 | 51 ± 23 | 1 | Y | Y | Y | Y | Y | Y | … |
| 55731 | 12385350-6042140 | -46.8 ± 0.2 | … | … | … | … | … | N | … | … | … | … | n | … |
| 55732 | 12385375-6038431 | -35.2 ± 0.1 | … | … | … | … | … | N | … | … | … | … | n | … |
| 56393 | 12403462-6032351 | -41.7 ± 0.6 | … | … | … | … | … | Y | … | … | … | … | n | … |
| 56165 | 12395077-6042009 | -62.9 ± 0.2 | … | … | … | … | … | N | … | … | … | … | n | … |
| 55937 | 12392526-6031102 | -0.4 ± 0.1 | 6227 ± 133 | 4.88 ± 0.02 | … | … | … | N | … | … | … | … | n | … |
| 55733 | 12385387-6033536 | -15.9 ± 0.2 | 6173 ± 121 | 3.51 ± 0.01 | … | … | … | N | … | … | … | … | n | … |
| 56394 | 12403478-6037102 | -34.9 ± 1.9 | … | … | … | … | … | N | … | … | … | … | n | … |
| 56166 | 12395080-6045174 | -38.5 ± 0.7 | … | … | … | … | … | Y | … | … | … | … | n | … |
| 55938 | 12392526-6039056 | -41.1 ± 0.4 | … | … | … | … | … | Y | … | … | … | … | n | … |
| 55734 | 12385420-6041580 | -40.7 ± 0.3 | … | … | … | … | … | Y | … | … | … | … | n | … |
| 55735 | 12385424-6032177 | -64.8 ± 0.2 | … | … | … | … | … | N | … | … | … | … | n | … |
| 56395 | 12403480-6038227 | -39.7 ± 0.1 | 5087 ± 82 | 3.65 ± 0.05 | 0.07 ± 0.17 | … | … | Y | … | … | … | … | n | … |
| 56167 | 12395092-6043122 | -40.4 ± 0.2 | … | … | … | … | … | Y | … | … | … | … | n | … |
| 53251 | 12403696-6035297 | -42.3 ± 1.5 | 6019 ± 229 | 4.24 ± 0.55 | 0.66 ± 0.13 | … | … | Y | … | … | … | … | n | … |
| 56396 | 12403484-6034342 | -28.7 ± 0.9 | … | … | … | … | … | N | … | … | … | … | n | … |
| 55736 | 12385427-6040099 | -26.9 ± 0.2 | … | … | … | … | … | N | … | … | … | … | n | … |
| 52908 | 12392536-6035030 | -37.7 ± 0.4 | 6744 ± 72 | 3.96 ± 0.14 | 0.16 ± 0.06 | 72 ± 27 | … | Y | Y | Y | Y | … | Y | … |
| 53080 | 12395104-6038239 | 13.5 ± 0.5 | 5968 ± 169 | 4.26 ± 0.12 | -0.05 ± 0.15 | … | … | N | … | … | … | … | n | … |
| 56403 | 12403706-6036329 | -39.0 ± 0.7 | … | … | … | … | … | Y | … | … | … | … | n | … |
| 56397 | 12403547-6043302 | -8.7 ± 1.1 | … | … | … | … | … | N | … | … | … | … | n | … |
| 55737 | 12385444-6038252 | -36.0 ± 1.7 | … | … | … | … | … | N | … | … | … | … | n | … |
| 55939 | 12392538-6037263 | -57.6 ± 3.7 | … | … | … | … | … | N | … | … | … | … | n | … |
| 56168 | 12395106-6045478 | -50.9 ± 0.2 | … | … | … | … | … | N | … | … | … | … | n | … |
| 56404 | 12403707-6040432 | -26.2 ± 0.1 | 5353 ± 92 | 4.22 ± 0.04 | 0.13 ± 0.19 | … | … | N | … | … | … | … | n | … |
| 53246 | 12403552-6040441 | 44.0 ± 0.6 | 5905 ± 197 | 5.63 ± 0.14 | 0.07 ± 0.29 | … | … | N | … | … | … | … | n | … |
| 55738 | 12385448-6032537 | 47.9 ± 0.2 | 6219 ± 116 | 4.85 ± 0.03 | … | … | … | N | … | … | … | … | n | … |
| 55940 | 12392538-6041444 | -39.1 ± 0.4 | … | … | … | … | … | Y | … | … | … | … | n | … |
| 56169 | 12395116-6042378 | -16.1 ± 0.1 | 5703 ± 108 | 3.83 ± 0.04 | -0.26 ± 0.19 | … | … | N | … | … | … | … | n | … |
| 56405 | 12403713-6039293 | 2.8 ± 0.2 | … | … | … | … | … | N | … | … | … | … | n | … |
| 55767 | 12390024-6037340 | -39.1 ± 0.7 | … | … | … | … | … | Y | … | … | … | … | n | … |
| 55941 | 12392540-6029102 | -27.6 ± 0.4 | … | … | … | … | … | N | … | … | … | … | n | … |
| 55942 | 12392547-6042455 | -12.9 ± 0.2 | … | … | … | … | … | N | … | … | … | … | n | … |
| 56170 | 12395117-6043526 | -15.1 ± 0.2 | … | … | … | … | … | N | … | … | … | … | n | … |
| 56406 | 12403729-6035124 | -7.5 ± 0.3 | … | … | … | … | … | N | … | … | … | … | n | … |
| 56399 | 12403581-6038076 | -40.1 ± 0.3 | … | … | … | … | … | Y | … | … | … | … | n | … |
| 52909 | 12392549-6039347 | -19.2 ± 0.6 | 5776 ± 145 | 3.94 ± 0.36 | -0.12 ± 0.30 | … | … | N | … | … | … | … | n | … |
| 55768 | 12390039-6029013 | -28.6 ± 1.1 | … | … | … | … | … | N | … | … | … | … | n | … |
| 56407 | 12403742-6035024 | -16.5 ± 3.9 | … | … | … | … | … | N | … | … | … | … | n | … |
| 56400 | 12403601-6038421 | -38.5 ± 1.1 | … | … | … | … | … | Y | … | … | … | … | n | … |
| 55769 | 12390047-6028304 | -24.2 ± 0.7 | … | … | … | … | … | N | … | … | … | … | n | … |
| 52910 | 12392553-6035470 | -40.5 ± 0.5 | 6356 ± 579 | 3.95 ± 0.18 | 0.21 ± 0.32 | <41 | 3 | Y | Y | Y | Y | N | Y | … |
| 53081 | 12395143-6032477 | -31.3 ± 0.4 | 5577 ± 73 | 4.60 ± 0.65 | -0.18 ± 0.24 | 62 ± 26 | 1 | N | … | … | … | … | n | NG |
| 53247 | 12403601-6044553 | -3.3 ± 0.7 | 5304 ± 216 | 4.06 ± 0.59 | 0.09 ± 0.14 | … | … | N | … | … | … | … | n | … |
| 55770 | 12390051-6043172 | -65.9 ± 0.1 | … | … | … | … | … | N | … | … | … | … | n | … |
| 52911 | 12392554-6036282 | 1.9 ± 0.3 | 5108 ± 193 | 4.33 ± 0.19 | 0.06 ± 0.14 | … | … | N | … | … | … | … | n | … |
| 56172 | 12395151-6044310 | -38.2 ± 1.8 | … | … | … | … | … | Y | … | … | … | … | n | … |
| 53248 | 12403630-6045263 | -9.6 ± 1.0 | 6348 ± 149 | 4.07 ± 0.04 | 0.11 ± 0.82 | … | … | N | … | … | … | … | n | … |

**Table C.17.** continued.

| ID | CNAME | RV (km s$^{-1}$) | $T_{\rm eff}$ (K) | $logg$ (dex) | [Fe/H] (dex) | EW(Li)$^a$ (mÅ) | EW(Li) error flag$^b$ | Membership RV | Li | $logg$ | [Fe/H] | Gaia study Cantat-Gaudin$^c$ | Final$^d$ | NMs with Li$^e$ |
|---|---|---|---|---|---|---|---|---|---|---|---|---|---|---|
| 55771 | 12390053-6038582 | -19.0 ± 0.2 | … | … | … | … | … | N | … | … | … | … | n | … |
| 55943 | 12392572-6039319 | -41.3 ± 0.1 | 5187 ± 110 | 3.88 ± 0.04 | 0.28 ± 0.23 | … | … | Y | … | … | … | … | n | … |
| 56173 | 12395154-6041566 | -29.9 ± 0.1 | … | … | … | … | … | N | … | … | … | … | n | … |
| 53249 | 12403634-6033203 | -40.2 ± 0.8 | 5947 ± 530 | 4.55 ± 0.22 | -0.06 ± 0.66 | … | … | Y | … | … | … | … | n | … |
| 55967 | 12392859-6040533 | -8.6 ± 0.1 | 4295 ± 131 | 2.52 ± 0.07 | 0.12 ± 0.28 | … | … | N | … | … | … | … | n | … |
| 55772 | 12390069-6045394 | -4.9 ± 0.2 | 5627 ± 129 | 2.97 ± 0.05 | 0.06 ± 0.28 | … | … | N | … | … | … | … | n | … |
| 55773 | 12390079-6039263 | 1.3 ± 0.1 | 4599 ± 90 | 2.77 ± 0.07 | 0.20 ± 0.20 | … | … | N | … | … | … | … | n | … |
| 56174 | 12395156-6031138 | -30.2 ± 0.5 | … | … | … | … | … | N | … | … | … | … | n | … |
| 53250 | 12403637-6035153 | 28.1 ± 0.9 | 6206 ± 142 | 3.41 ± 0.15 | 0.06 ± 0.39 | … | … | N | … | … | … | … | n | … |
| 52934 | 12392860-6038054 | -40.3 ± 0.4 | 5773 ± 48 | 4.22 ± 0.58 | 0.05 ± 0.14 | <74 | 3 | Y | Y | Y | Y | … | Y | … |
| 52787 | 12390101-6041284 | -20.7 ± 0.4 | 5145 ± 132 | 4.14 ± 0.55 | -0.02 ± 0.18 | … | … | N | … | … | … | … | n | … |
| 56175 | 12395159-6029550 | -42.8 ± 0.9 | … | … | … | … | … | Y | … | … | … | … | n | … |
| 56401 | 12403664-6043357 | -40.6 ± 0.3 | … | … | … | … | … | Y | … | … | … | … | n | … |
| 52788 | 12390132-6031298 | -62.5 ± 1.0 | 6109 ± 132 | 4.63 ± 0.42 | -0.03 ± 0.53 | … | … | N | … | … | … | … | n | … |
| 55968 | 12392877-6038542 | -10.1 ± 0.1 | … | … | … | … | … | N | … | … | … | … | n | … |
| 53083 | 12395161-6037146 | -40.5 ± 0.4 | 5838 ± 62 | 4.53 ± 0.28 | -0.03 ± 0.16 | … | … | Y | … | … | … | … | n | … |
| 56402 | 12403687-6033462 | -9.0 ± 0.1 | 5061 ± 77 | 4.00 ± 0.05 | -0.01 ± 0.16 | … | … | N | … | … | … | … | n | … |
| 52935 | 12392879-6039518 | 4.1 ± 0.5 | 5544 ± 152 | 4.19 ± 0.63 | 0.16 ± 0.23 | … | … | N | … | … | … | … | n | … |
| 55969 | 12392881-6037323 | -32.9 ± 2.2 | … | … | … | … | … | N | … | … | … | … | n | … |
| 53084 | 12395162-6033361 | -17.1 ± 0.6 | 6492 ± 654 | 4.61 ± 0.19 | 0.53 ± 0.04 | … | … | N | … | … | … | … | n | … |
| 55774 | 12390140-6044594 | 13.0 ± 0.3 | … | … | … | … | … | N | … | … | … | … | n | … |
| 52936 | 12392890-6034379 | -39.1 ± 0.4 | 5534 ± 55 | 4.78 ± 0.15 | 0.31 ± 0.20 | … | … | Y | … | … | … | … | n | … |
| 53085 | 12395165-6040534 | -40.1 ± 0.4 | 6212 ± 206 | 4.48 ± 0.53 | 0.09 ± 0.14 | 106 ± 40 | 1 | Y | Y | Y | Y | N | Y | … |
| 55775 | 12390144-6033211 | -64.6 ± 0.1 | 6155 ± 119 | 4.22 ± 0.14 | … | … | … | N | … | … | … | … | n | … |
| 55970 | 12392894-6045542 | -34.4 ± 0.5 | … | … | … | … | … | N | … | … | … | … | n | … |
| 53086 | 12395176-6037423 | 18.7 ± 0.3 | 5293 ± 132 | 4.48 ± 0.35 | 0.33 ± 0.26 | … | … | N | … | … | … | … | n | … |
| 55776 | 12390147-6038480 | -40.6 ± 0.7 | … | … | … | … | … | Y | … | … | … | … | n | … |
| 52789 | 12390174-6038558 | 6.6 ± 0.4 | 5809 ± 274 | 4.82 ± 0.61 | 0.43 ± 0.26 | <53 | 3 | N | … | … | … | … | n | NG |
| 52937 | 12392912-6035410 | -32.4 ± 0.4 | 5583 ± 506 | 4.56 ± 0.77 | -0.13 ± 0.40 | … | … | N | … | … | … | … | n | … |
| 56176 | 12395189-6039463 | -39.7 ± 0.2 | … | … | … | … | … | Y | … | … | … | … | n | … |
| 55777 | 12390188-6029101 | -24.2 ± 0.6 | … | … | … | … | … | N | … | … | … | … | n | … |
| 55971 | 12392920-6030037 | -37.4 ± 1.2 | … | … | … | … | … | Y | … | … | … | … | n | … |
| 55972 | 12392921-6035483 | -40.0 ± 0.7 | … | … | … | … | … | Y | … | … | … | … | n | … |
| 56177 | 12395189-6041088 | -15.1 ± 0.1 | … | … | … | … | … | N | … | … | … | … | n | … |
| 55778 | 12390195-6040277 | -40.2 ± 2.5 | … | … | … | … | … | Y | … | … | … | … | n | … |
| 55973 | 12392949-6030422 | 15.7 ± 0.1 | … | … | … | … | … | N | … | … | … | … | n | … |
| 56178 | 12395190-6037283 | -5.2 ± 0.2 | 6507 ± 137 | 4.29 ± 0.05 | … | … | … | N | … | … | … | … | n | … |
| 55779 | 12390209-6033392 | -48.8 ± 0.4 | … | … | … | … | … | N | … | … | … | … | n | … |
| 55974 | 12392955-6039109 | -39.4 ± 0.4 | … | … | … | … | … | Y | … | … | … | … | n | … |
| 53087 | 12395191-6041089 | -36.6 ± 0.4 | 6656 ± 139 | 3.92 ± 0.13 | 0.15 ± 0.19 | 73 ± 32 | 1 | Y | Y | Y | Y | Y | Y | … |
| 55780 | 12390214-6037148 | -39.3 ± 0.4 | … | … | … | … | … | Y | … | … | … | … | n | … |
| 55975 | 12392973-6035475 | -24.8 ± 0.1 | 6187 ± 107 | 4.90 ± 0.05 | … | … | … | N | … | … | … | … | n | … |
| 53088 | 12395194-6037283 | -5.6 ± 0.3 | 6546 ± 317 | 4.37 ± 0.34 | -0.01 ± 0.15 | … | … | N | … | … | … | … | n | … |
| 52790 | 12390230-6040300 | 41.6 ± 0.6 | 5156 ± 93 | 4.10 ± 0.41 | -0.27 ± 0.23 | … | … | N | … | … | … | … | n | … |
| 55976 | 12392977-6045334 | -16.4 ± 0.2 | … | … | … | … | … | N | … | … | … | … | n | … |
| 56179 | 12395206-6038283 | -40.9 ± 0.3 | … | … | … | … | … | Y | … | … | … | … | n | … |
| 55781 | 12390267-6030374 | -7.0 ± 0.1 | … | … | … | … | … | N | … | … | … | … | n | … |
| 55977 | 12392979-6035382 | -24.4 ± 0.2 | … | … | … | … | … | N | … | … | … | … | n | … |
| 56193 | 12395339-6043238 | -41.6 ± 0.1 | 5679 ± 124 | 4.91 ± 0.05 | 0.30 ± 0.22 | … | … | Y | … | … | … | … | n | … |
| 55782 | 12390270-6035434 | -68.4 ± 0.1 | … | … | … | … | … | N | … | … | … | … | n | … |
| 53100 | 12395340-6039474 | -43.0 ± 0.6 | 6788 ± 461 | 4.16 ± 0.14 | 0.05 ± 0.19 | 64 ± 31 | 1 | Y | Y | Y | Y | … | Y | … |
| 52938 | 12393007-6035548 | -45.8 ± 0.4 | 4969 ± 377 | 4.42 ± 0.17 | -0.53 ± 0.38 | … | … | N | … | … | … | … | n | … |
| 55783 | 12390290-6037230 | 5.9 ± 0.1 | 4647 ± 129 | 3.07 ± 0.07 | 0.25 ± 0.25 | … | … | N | … | … | … | … | n | … |
| 53101 | 12395342-6039545 | -41.0 ± 0.3 | 6416 ± 351 | 4.67 ± 0.65 | 0.04 ± 0.28 | <33 | 3 | Y | Y | Y | Y | Y | Y | … |
| 55979 | 12393011-6040224 | -20.1 ± 0.1 | … | … | … | … | … | N | … | … | … | … | n | … |
| 52791 | 12390294-6035510 | -1.7 ± 0.2 | 5964 ± 177 | 4.36 ± 0.33 | -0.03 ± 0.17 | 32 ± 12 | … | N | … | … | … | … | n | NG |
| 56194 | 12395346-6031329 | -41.6 ± 1.0 | … | … | … | … | … | Y | … | … | … | … | n | … |
| 55980 | 12393013-6037230 | -41.1 ± 0.1 | … | … | … | … | … | Y | … | … | … | … | n | … |
| 55784 | 12390304-6045218 | -5.1 ± 0.1 | … | … | … | … | … | N | … | … | … | … | n | … |

**Table C.17.** continued.

| ID | CNAME | RV (km s$^{-1}$) | $T_{\rm eff}$ (K) | $logg$ (dex) | [Fe/H] (dex) | $EW$(Li)$^a$ (mÅ) | $EW$(Li) error flag$^b$ | Membership RV | Li | $logg$ | [Fe/H] | Gaia study Cantat-Gaudin$^c$ | Final$^d$ | NMs with Li$^e$ |
|---|---|---|---|---|---|---|---|---|---|---|---|---|---|---|
| 56195 | 12395358-6039045 | -46.0 ± 0.9 | … | … | … | … | … | N | … | … | … | … | n | … |
| 55981 | 12393014-6036117 | -38.7 ± 0.8 | … | … | … | … | … | Y | … | … | … | … | n | … |
| 56196 | 12395360-6041429 | -2.9 ± 0.1 | … | … | … | … | … | N | … | … | … | … | n | … |
| 52792 | 12390322-6040299 | -38.3 ± 0.5 | 6600 ± 76 | 3.77 ± 0.15 | 0.09 ± 0.06 | 56 ± 26 | … | Y | Y | Y | Y | N | Y | … |
| 55982 | 12393015-6036496 | -55.2 ± 0.1 | … | … | … | … | … | N | … | … | … | … | n | … |
| 56197 | 12395394-6034050 | -44.9 ± 0.9 | … | … | … | … | … | N | … | … | … | … | n | … |
| 52801 | 12390531-6041338 | -38.5 ± 0.5 | 5395 ± 208 | 3.59 ± 0.42 | -0.62 ± 0.26 | … | … | Y | … | … | … | … | n | … |
| 52939 | 12393016-6031430 | -21.3 ± 2.6 | 5114 ± 505 | 5.24 ± 0.85 | -0.83 ± 1.14 | … | … | N | … | … | … | … | n | … |
| 53102 | 12395399-6034453 | -33.8 ± 0.6 | 6640 ± 827 | 4.16 ± 0.18 | -0.13 ± 0.20 | … | … | N | … | … | … | … | n | … |
| 55983 | 12393024-6032213 | 25.5 ± 0.1 | … | … | … | … | … | N | … | … | … | … | n | … |
| 56198 | 12395400-6034171 | -40.5 ± 0.1 | … | … | … | … | … | Y | … | … | … | … | n | … |
| 55799 | 12390547-6033525 | -26.7 ± 0.1 | 5655 ± 106 | 3.65 ± 0.06 | -0.18 ± 0.18 | … | … | N | … | … | … | … | n | … |
| 52940 | 12393024-6037097 | -39.4 ± 1.3 | 4484 ± 523 | 2.66 ± 0.83 | -0.70 ± 0.43 | <79 | 3 | Y | N | Y? | N | … | Y | … |
| 56199 | 12395401-6041536 | -19.2 ± 0.6 | … | … | … | … | … | N | … | … | … | … | n | … |
| 55800 | 12390568-6044357 | -41.7 ± 0.4 | … | … | … | … | … | Y | … | … | … | … | n | … |
| 52951 | 12393213-6039558 | -37.0 ± 0.4 | 5557 ± 85 | 4.47 ± 0.33 | -0.18 ± 0.14 | … | … | Y | … | … | … | … | n | … |
| 56200 | 12395401-6041587 | -40.5 ± 1.8 | … | … | … | … | … | Y | … | … | … | … | n | … |
| 55801 | 12390579-6043455 | -58.1 ± 0.1 | 4669 ± 59 | 2.70 ± 0.04 | -0.37 ± 0.14 | … | … | N | … | … | … | … | n | … |
| 52952 | 12393214-6033008 | -67.9 ± 1.1 | 6026 ± 79 | 4.15 ± 0.27 | -0.07 ± 0.33 | … | … | N | … | … | … | … | n | … |
| 53103 | 12395404-6041354 | -28.1 ± 0.4 | 5750 ± 32 | 4.61 ± 0.60 | -0.01 ± 0.21 | … | … | N | … | … | … | … | n | … |
| 52802 | 12390591-6027535 | -40.2 ± 1.6 | 5899 ± 630 | 4.15 ± 0.80 | -0.50 ± 0.70 | … | … | Y | … | … | … | … | n | … |
| 52953 | 12393216-6042172 | -0.5 ± 0.5 | 4929 ± 53 | 4.05 ± 0.80 | -0.40 ± 0.36 | … | … | N | … | … | … | … | n | … |
| 53104 | 12395411-6031264 | -35.8 ± 1.0 | 5628 ± 123 | 5.11 ± 0.19 | 0.24 ± 0.24 | … | … | N | … | … | … | … | n | … |
| 55802 | 12390595-6036391 | -49.2 ± 0.2 | … | … | … | … | … | N | … | … | … | … | n | … |
| 52954 | 12393219-6035373 | -39.3 ± 0.3 | 6441 ± 37 | 3.84 ± 0.08 | 0.31 ± 0.03 | … | … | Y | … | … | … | … | n | … |
| 56201 | 12395416-6044342 | -49.7 ± 0.3 | … | … | … | … | … | N | … | … | … | … | n | … |
| 52803 | 12390597-6042028 | 44.1 ± 0.6 | 5889 ± 421 | 3.91 ± 0.02 | -0.07 ± 0.24 | … | … | N | … | … | … | … | n | … |
| 52955 | 12393231-6037073 | -40.4 ± 0.6 | 6826 ± 227 | 4.01 ± 0.03 | 0.19 ± 0.22 | 63 ± 41 | 1 | Y | Y | Y | Y | Y | Y | … |
| 3387 | 12395424-6038370 | -41.7 ± 0.6 | 4913 ± 109 | 2.80 ± 0.23 | 0.15 ± 0.10 | <6 | 3 | Y | Y | Y | Y | … | Y | … |
| 55803 | 12390610-6042313 | -12.4 ± 0.8 | … | … | … | … | … | N | … | … | … | … | n | … |
| 52956 | 12393237-6036320 | -54.9 ± 0.9 | 5626 ± 348 | 4.14 ± 0.77 | -0.27 ± 0.15 | … | … | N | … | … | … | … | n | … |
| 56202 | 12395431-6043370 | -33.2 ± 0.2 | … | … | … | … | … | N | … | … | … | … | n | … |
| 55804 | 12390616-6042075 | -39.5 ± 1.4 | … | … | … | … | … | Y | … | … | … | … | n | … |
| 52804 | 12390668-6032098 | -18.8 ± 0.5 | 6173 ± 258 | 4.25 ± 0.28 | 0.22 ± 0.28 | <108 | 3 | N | … | … | … | … | n | NG |
| 53105 | 12395434-6038134 | -40.5 ± 0.3 | 6318 ± 168 | 4.59 ± 0.52 | 0.07 ± 0.16 | <40 | 3 | Y | Y | Y | Y | … | Y | … |
| 55998 | 12393246-6036148 | -41.9 ± 1.7 | … | … | … | … | … | Y | … | … | … | … | n | … |
| 53106 | 12395437-6032168 | 4.6 ± 0.4 | 5281 ± 99 | 4.55 ± 0.31 | 0.31 ± 0.27 | <80 | 3 | N | … | … | … | … | n | NG |
| 3364 | 12390709-6038056 | -39.8 ± 0.6 | 5026 ± 120 | 3.14 ± 0.23 | 0.10 ± 0.10 | <9 | 3 | Y | Y | Y | Y | … | Y | … |
| 52957 | 12393256-6034398 | -40.8 ± 0.4 | 6204 ± 407 | 4.29 ± 0.28 | 0.10 ± 0.29 | <45 | 3 | Y | Y | Y | Y | N | Y | … |
| 53107 | 12395448-6037128 | -41.0 ± 0.6 | 6421 ± 233 | 3.86 ± 0.08 | 0.10 ± 0.31 | 52 ± 50 | 1 | Y | Y | Y | Y | … | Y | … |
| 52805 | 12390733-6035150 | -58.7 ± 0.4 | 6276 ± 440 | 4.47 ± 0.31 | -0.04 ± 0.25 | 65 ± 26 | 1 | N | … | … | … | … | n | NG |
| 53108 | 12395455-6036096 | -29.7 ± 0.3 | 6697 ± 87 | 3.90 ± 0.17 | 0.40 ± 0.07 | 67 ± 24 | … | N | Y | Y | Y | Y | Y | … |
| 52806 | 12390737-6039341 | -5.7 ± 0.3 | 6059 ± 87 | 3.91 ± 0.02 | -0.37 ± 0.34 | … | … | N | … | … | … | … | n | … |
| 56000 | 12393272-6040037 | -35.5 ± 0.5 | … | … | … | … | … | N | … | … | … | … | n | … |
| 56203 | 12395461-6044070 | -16.7 ± 0.8 | … | … | … | … | … | N | … | … | … | … | n | … |
| 55806 | 12390746-6028421 | -19.0 ± 3.1 | … | … | … | … | … | N | … | … | … | … | n | … |
| 52807 | 12390754-6043168 | -17.2 ± 0.4 | 7064 ± 494 | 4.04 ± 0.22 | 0.00 ± 0.33 | 57 ± 36 | 1 | N | … | … | … | N | n | NG |
| 56001 | 12393272-6042355 | -39.7 ± 1.4 | … | … | … | … | … | Y | … | … | … | … | n | … |
| 53109 | 12395467-6036273 | -7.0 ± 0.5 | 5624 ± 331 | 4.36 ± 0.64 | -0.11 ± 0.16 | … | … | N | … | … | … | … | n | … |
| 55807 | 12390772-6039005 | -52.5 ± 0.3 | … | … | … | … | … | N | … | … | … | … | n | … |
| 52958 | 12393292-6038311 | 22.2 ± 0.6 | 4687 ± 68 | 2.94 ± 1.00 | -0.66 ± 0.39 | <60 | 3 | N | … | … | … | … | n | G |
| 52808 | 12390793-6036303 | 4.6 ± 0.2 | 4681 ± 123 | 3.10 ± 0.38 | 0.25 ± 0.24 | … | … | N | … | … | … | … | n | G |
| 53110 | 12395469-6033270 | 5.4 ± 0.5 | 5580 ± 211 | 3.65 ± 0.52 | -0.10 ± 0.14 | <59 | 3 | N | … | … | … | … | n | NG |
| 56002 | 12393317-6037309 | -40.4 ± 0.3 | … | … | … | … | … | Y | … | … | … | … | n | … |
| 53111 | 12395481-6042173 | -10.0 ± 0.5 | 5151 ± 232 | 4.52 ± 0.83 | -0.11 ± 0.13 | … | … | N | … | … | … | … | n | … |
| 52959 | 12393317-6039566 | -2.9 ± 0.3 | 6344 ± 81 | 4.17 ± 0.33 | -0.06 ± 0.16 | … | … | N | … | … | … | … | n | … |
| 56236 | 12395809-6034394 | -40.0 ± 0.6 | … | … | … | … | … | Y | … | … | … | … | n | … |
| 55809 | 12390822-6046469 | -56.0 ± 5.9 | … | … | … | … | … | N | … | … | … | … | n | … |
| 56003 | 12393318-6041416 | -34.4 ± 0.4 | … | … | … | … | … | N | … | … | … | … | n | … |

**Table C.17.** continued.

| ID | CNAME | RV (km s$^{-1}$) | $T_{\rm eff}$ (K) | $logg$ (dex) | [Fe/H] (dex) | $EW({\rm Li})^a$ (mÅ) | $EW({\rm Li})$ error flag$^b$ | Membership RV | Membership Li | Membership $logg$ | Membership [Fe/H] | Gaia study Cantat-Gaudin$^c$ | Final$^d$ | NMs with Li$^e$ |
|---|---|---|---|---|---|---|---|---|---|---|---|---|---|---|
| 56237 | 12395814-6041469 | -60.7 ± 0.3 | 6089 ± 316 | 3.10 ± 0.45 | … | … | … | N | … | … | … | … | n | … |
| 52809 | 12390842-6036528 | 29.1 ± 0.4 | 6036 ± 110 | 4.05 ± 0.30 | 0.63 ± 0.17 | … | … | N | … | … | … | … | n | … |
| 56004 | 12393322-6042255 | -35.9 ± 1.0 | … | … | … | … | … | N | … | … | … | … | n | … |
| 53124 | 12395816-6041260 | -14.1 ± 0.3 | 5406 ± 151 | 4.41 ± 0.26 | 0.27 ± 0.19 | … | … | N | … | … | … | … | n | … |
| 55810 | 12390850-6037107 | -31.8 ± 0.2 | 5881 ± 122 | 2.60 ± 0.05 | -0.26 ± 0.24 | … | … | N | … | … | … | … | n | … |
| 52960 | 12393329-6036008 | -22.2 ± 0.3 | 6703 ± 487 | 4.09 ± 0.20 | 0.19 ± 0.26 | … | … | N | … | … | … | … | n | … |
| 56238 | 12395821-6042059 | 49.8 ± 0.1 | 5805 ± 134 | 4.91 ± 0.05 | 0.28 ± 0.21 | … | … | N | … | … | … | … | n | … |
| 55811 | 12390856-6047163 | -39.2 ± 0.1 | 5177 ± 99 | 3.78 ± 0.05 | 0.22 ± 0.20 | … | … | Y | … | … | … | … | n | … |
| 56005 | 12393332-6042455 | -35.7 ± 3.2 | … | … | … | … | … | N | … | … | … | … | n | … |
| 56239 | 12395823-6037133 | -9.9 ± 0.1 | … | … | … | … | … | N | … | … | … | … | n | … |
| 55812 | 12390865-6042186 | -24.9 ± 0.4 | … | … | … | … | … | N | … | … | … | … | n | … |
| 56006 | 12393336-6033524 | -11.0 ± 0.2 | … | … | … | … | … | N | … | … | … | … | n | … |
| 56240 | 12395829-6035111 | -39.5 ± 1.1 | … | … | … | … | … | Y | … | … | … | … | n | … |
| 52961 | 12393355-6038255 | 36.9 ± 0.4 | 5549 ± 42 | 4.25 ± 0.11 | 0.19 ± 0.15 | <51 | 3 | N | … | … | … | … | n | NG |
| 52828 | 12391254-6039243 | -16.0 ± 2.0 | 6314 ± 266 | 4.57 ± 0.60 | -0.05 ± 0.21 | … | … | N | … | … | … | … | n | … |
| 56241 | 12395874-6037039 | -40.1 ± 0.2 | … | … | … | … | … | Y | … | … | … | … | n | … |
| 56242 | 12395881-6035106 | -54.4 ± 0.2 | 6819 ± 150 | 4.91 ± 0.05 | … | … | … | N | … | … | … | … | n | … |
| 52829 | 12391261-6038541 | -6.4 ± 0.5 | 4993 ± 168 | 4.46 ± 0.30 | -0.07 ± 0.15 | … | … | N | … | … | … | … | n | … |
| 56007 | 12393370-6036505 | -28.9 ± 0.1 | 6171 ± 142 | 4.13 ± 0.06 | … | … | … | N | … | … | … | … | n | … |
| 53125 | 12395892-6035325 | -19.2 ± 0.5 | 7039 ± 526 | 4.15 ± 0.21 | 0.11 ± 0.15 | … | … | N | … | … | … | … | n | … |
| 52830 | 12391264-6035414 | 21.8 ± 0.5 | 6134 ± 239 | 4.37 ± 0.29 | 0.08 ± 0.29 | <128 | 3 | N | … | … | … | … | n | NG |
| 52963 | 12393385-6038038 | -16.2 ± 0.2 | 4385 ± 201 | 2.29 ± 0.30 | 0.42 ± 0.29 | … | … | N | … | … | … | … | n | G |
| 55840 | 12391269-6038162 | -37.2 ± 0.4 | … | … | … | … | … | Y | … | … | … | … | n | … |
| 56243 | 12395898-6042431 | -37.6 ± 0.6 | … | … | … | … | … | Y | … | … | … | … | n | … |
| 55841 | 12391277-6033283 | -33.4 ± 1.1 | … | … | … | … | … | N | … | … | … | … | n | … |
| 56035 | 12393651-6030146 | -67.5 ± 0.2 | 6371 ± 138 | 4.60 ± 0.02 | … | … | … | N | … | … | … | … | n | … |
| 56244 | 12395900-6028107 | 0.8 ± 0.1 | 6177 ± 149 | 4.12 ± 0.02 | … | … | … | N | … | … | … | … | n | … |
| 56036 | 12393657-6043196 | -39.6 ± 0.1 | 5627 ± 118 | 4.53 ± 0.05 | 0.40 ± 0.20 | … | … | Y | … | … | … | … | n | … |
| 55842 | 12391286-6036094 | -4.6 ± 0.2 | … | … | … | … | … | N | … | … | … | … | n | … |
| 56245 | 12395901-6039096 | -40.5 ± 0.6 | … | … | … | … | … | Y | … | … | … | … | n | … |
| 55843 | 12391301-6037485 | 2.2 ± 0.1 | … | … | … | … | … | N | … | … | … | … | n | … |
| 56037 | 12393665-6044028 | -21.1 ± 1.8 | … | … | … | … | … | N | … | … | … | … | n | … |
| 56246 | 12395905-6034314 | -44.8 ± 0.3 | … | … | … | … | … | N | … | … | … | … | n | … |
| 55844 | 12391306-6045403 | -43.4 ± 2.2 | … | … | … | … | … | N | … | … | … | … | n | … |
| 52985 | 12393675-6041449 | -40.4 ± 0.5 | 5707 ± 61 | 4.03 ± 0.37 | -0.08 ± 0.17 | … | … | Y | … | … | … | … | n | … |
| 53126 | 12395907-6035479 | -21.0 ± 0.5 | 5834 ± 112 | 4.46 ± 0.43 | -0.25 ± 0.27 | … | … | N | … | … | … | … | n | … |
| 52831 | 12391337-6041382 | -40.3 ± 0.3 | 6378 ± 422 | 4.52 ± 0.44 | 0.29 ± 0.33 | 90 ± 28 | 1 | Y | Y | Y | Y | N | Y | … |
| 52986 | 12393689-6040505 | 12.5 ± 0.6 | 6236 ± 163 | 4.21 ± 0.39 | -0.09 ± 0.14 | … | … | N | … | … | … | … | n | … |
| 53127 | 12395919-6042023 | -36.5 ± 0.5 | 6632 ± 50 | … | -0.04 ± 0.17 | 90 ± 40 | … | Y | Y | … | Y | … | Y | … |
| 52832 | 12391344-6045559 | -38.5 ± 0.6 | 5699 ± 779 | 3.75 ± 0.66 | -0.31 ± 0.89 | … | … | Y | … | … | … | … | n | … |
| 56038 | 12393697-6038300 | -39.3 ± 0.8 | … | … | … | … | … | Y | … | … | … | … | n | … |
| 53128 | 12395940-6039267 | -43.3 ± 0.8 | 5087 ± 473 | 4.05 ± 0.77 | -0.19 ± 0.39 | … | … | N | … | … | … | … | n | … |
| 52833 | 12391356-6035259 | -40.5 ± 0.4 | 6340 ± 479 | 4.42 ± 0.28 | 0.29 ± 0.34 | 52 ± 47 | 1 | Y | Y | Y | Y | … | Y | … |
| 56039 | 12393703-6036540 | -45.3 ± 0.3 | … | … | … | … | … | N | … | … | … | … | n | … |
| 55845 | 12391367-6042041 | -59.2 ± 0.8 | … | … | … | … | … | N | … | … | … | … | n | … |
| 55846 | 12391376-6042224 | 7.4 ± 0.2 | 5491 ± 150 | 3.36 ± 0.07 | -0.28 ± 0.19 | … | … | N | … | … | … | … | n | … |
| 53129 | 12395949-6038313 | -31.3 ± 0.3 | 5805 ± 31 | 4.45 ± 0.13 | 0.21 ± 0.14 | … | … | N | … | … | … | … | n | … |
| 55847 | 12391388-6044273 | -24.5 ± 0.4 | … | … | … | … | … | N | … | … | … | … | n | … |
| 52987 | 12393704-6034430 | 8.4 ± 0.7 | 5743 ± 135 | 4.64 ± 0.19 | -0.02 ± 0.24 | … | … | N | … | … | … | … | n | … |
| 56247 | 12395958-6040069 | -39.4 ± 0.1 | 5371 ± 103 | 4.24 ± 0.04 | 0.54 ± 0.19 | … | … | Y | … | … | … | … | n | … |
| 52988 | 12393710-6036325 | -39.1 ± 0.5 | 5223 ± 119 | 4.60 ± 0.23 | -0.27 ± 0.18 | … | … | Y | … | … | … | … | n | … |
| 55848 | 12391409-6043499 | -34.8 ± 0.2 | 6231 ± 129 | 3.43 ± 0.05 | … | … | … | N | … | … | … | … | n | … |
| 3392 | 12395973-6035072 | -41.1 ± 0.6 | 4876 ± 113 | 2.76 ± 0.22 | 0.10 ± 0.10 | <5 | 3 | Y | Y | Y | Y | … | Y | … |
| 52989 | 12393711-6039553 | 10.1 ± 0.6 | 5452 ± 71 | 4.33 ± 0.25 | -0.14 ± 0.19 | … | … | N | … | … | … | … | n | … |
| 52835 | 12391412-6041170 | -32.4 ± 0.3 | 6250 ± 101 | 4.39 ± 0.90 | 0.15 ± 0.26 | 26 ± 16 | … | N | Y | Y | Y | Y | Y | … |
| 53130 | 12400011-6040387 | -31.2 ± 0.4 | 6686 ± 79 | 4.03 ± 0.16 | 0.19 ± 0.06 | <15 | 3 | N | Y | Y | Y | … | Y | … |
| 56040 | 12393733-6039251 | -39.8 ± 0.9 | … | … | … | … | … | Y | … | … | … | … | n | … |
| 53131 | 12400016-6040123 | -41.0 ± 0.5 | 5667 ± 336 | 4.34 ± 0.23 | -0.31 ± 0.37 | … | … | Y | … | … | … | … | n | … |
| 55849 | 12391422-6037559 | -17.9 ± 0.3 | … | … | … | … | … | N | … | … | … | … | n | … |

**Table C.17.** continued.

| ID | CNAME | RV (km s$^{-1}$) | $T_{\rm eff}$ (K) | logg (dex) | [Fe/H] (dex) | EW(Li)$^a$ (mÅ) | EW(Li) error flag$^b$ | Membership RV | Li | logg | [Fe/H] | Gaia study Cantat-Gaudin$^c$ | Final$^d$ | NMs with Li$^e$ |
|---|---|---|---|---|---|---|---|---|---|---|---|---|---|---|
| 56041 | 12393736-6043522 | 3.1 ± 0.1 | 4493 ± 150 | 2.49 ± 0.07 | 0.19 ± 0.29 | … | … | N | … | … | … | … | n | … |
| 53132 | 12400028-6032466 | 5.5 ± 0.4 | 5267 ± 105 | 4.35 ± 0.26 | -0.04 ± 0.17 | <68 | 3 | N | … | … | … | … | n | NG |
| 55850 | 12391437-6033118 | -38.8 ± 0.1 | … | … | … | … | … | Y | … | … | … | … | n | … |
| 52990 | 12393737-6035498 | 51.5 ± 0.4 | 5166 ± 84 | 4.44 ± 0.67 | 0.18 ± 0.15 | <65 | 3 | N | … | … | … | … | n | NG |
| 53133 | 12400033-6037515 | 23.6 ± 0.5 | 5695 ± 27 | 3.92 ± 0.42 | -0.73 ± 0.23 | 87 ± 51 | 3 | N | … | … | … | … | n | NG |
| 52836 | 12391454-6038121 | -36.8 ± 0.4 | 6642 ± 51 | 3.84 ± 0.10 | 0.35 ± 0.04 | … | … | Y | … | … | … | Y | n | … |
| 53134 | 12400047-6041437 | -41.5 ± 0.4 | 5993 ± 97 | 4.52 ± 0.47 | -0.09 ± 0.19 | 129 ± 39 | 1 | Y | N | Y | Y | Y | n | NG |
| 55851 | 12391457-6040111 | -17.6 ± 0.8 | … | … | … | … | … | N | … | … | … | … | n | … |
| 56042 | 12393738-6041213 | -62.1 ± 0.1 | 6041 ± 143 | 3.77 ± 0.03 | … | … | … | N | … | … | … | … | n | … |
| 56260 | 12400228-6040076 | -18.2 ± 0.1 | 5479 ± 146 | 4.08 ± 0.04 | 0.33 ± 0.22 | … | … | N | … | … | … | … | n | … |
| 3373 | 12393740-6032568 | -40.6 ± 0.6 | 5031 ± 120 | 3.00 ± 0.23 | 0.14 ± 0.10 | <10 | 3 | Y | Y | Y | Y | … | Y | … |
| 53146 | 12400248-6039294 | -40.1 ± 0.4 | 6056 ± 253 | 4.23 ± 0.23 | 0.08 ± 0.34 | <78 | 3 | Y | Y | Y | Y | … | Y | … |
| 56044 | 12393745-6038430 | 0.4 ± 0.3 | … | … | … | … | … | N | … | … | … | … | n | … |
| 53147 | 12400257-6035402 | -50.1 ± 0.5 | 6141 ± 276 | 4.05 ± 0.43 | -0.31 ± 0.19 | … | … | N | … | … | … | … | n | … |
| 52991 | 12393761-6037372 | -30.0 ± 0.2 | 6171 ± 269 | 4.47 ± 0.26 | -0.49 ± 0.75 | … | … | N | … | … | … | … | n | … |
| 3395 | 12400259-6039545 | -39.6 ± 0.6 | 4521 ± 115 | 2.20 ± 0.22 | 0.03 ± 0.09 | <4 | 3 | Y | Y | Y | Y | … | Y | … |
| 53148 | 12400271-6035345 | -38.6 ± 0.3 | 7495 ± 93 | … | … | … | … | Y | … | … | … | … | n | … |
| 52992 | 12393766-6030285 | 18.0 ± 0.7 | 5552 ± 122 | 4.53 ± 0.26 | -0.19 ± 0.16 | … | … | N | … | … | … | … | n | … |
| 56261 | 12400276-6041192 | -38.7 ± 0.1 | 5229 ± 331 | 4.53 ± 0.54 | 0.07 ± 0.28 | … | … | Y | … | … | … | … | n | … |
| 52993 | 12393766-6038189 | -47.9 ± 0.6 | 6400 ± 143 | 4.23 ± 0.48 | -0.24 ± 0.23 | … | … | N | … | … | … | … | n | … |
| 56262 | 12400277-6039022 | -37.2 ± 1.1 | … | … | … | … | … | N | … | … | … | … | n | … |
| 56046 | 12393771-6036022 | -55.9 ± 0.1 | … | … | … | … | … | N | … | … | … | … | n | … |
| 52994 | 12393781-6037171 | -39.6 ± 0.9 | 6463 ± 252 | 4.26 ± 0.34 | 0.04 ± 0.14 | … | … | Y | … | … | … | … | n | … |
| 3396 | 12400278-6041192 | -37.7 ± 0.6 | 4951 ± 117 | 2.94 ± 0.22 | 0.12 ± 0.09 | 13 ± 1 | … | Y | Y | Y | Y | … | Y | … |
| 3374 | 12393781-6039051 | -39.2 ± 0.6 | 4935 ± 128 | 2.79 ± 0.22 | 0.12 ± 0.11 | <9 | 3 | Y | Y | Y | Y | … | Y | … |
| 56060 | 12393945-6042306 | -40.8 ± 0.1 | … | … | … | … | … | Y | … | … | … | … | n | … |
| 56264 | 12400303-6045061 | -40.6 ± 0.1 | 5121 ± 121 | 3.54 ± 0.09 | 0.29 ± 0.22 | … | … | Y | … | … | … | … | n | … |
| 56061 | 12393950-6043063 | -50.1 ± 0.1 | 5017 ± 103 | 3.70 ± 0.09 | -0.05 ± 0.16 | … | … | N | … | … | … | … | n | … |
| 53149 | 12400308-6039585 | 32.0 ± 1.4 | 4859 ± 272 | 3.62 ± 0.25 | -0.98 ± 0.51 | … | … | N | … | … | … | … | n | … |
| 56062 | 12393951-6046029 | -36.5 ± 3.9 | … | … | … | … | … | Y | … | … | … | … | n | … |
| 53150 | 12400314-6037232 | -32.4 ± 0.5 | 5962 ± 324 | 3.83 ± 0.27 | 0.02 ± 0.23 | 57 ± 43 | 1 | N | … | … | … | … | n | NG |
| 53006 | 12393957-6036401 | -63.3 ± 0.4 | 5934 ± 277 | 4.11 ± 0.47 | -0.38 ± 0.23 | <51 | 3 | N | … | … | … | … | n | NG |
| 56265 | 12400317-6035013 | -23.7 ± 0.4 | … | … | … | … | … | N | … | … | … | … | n | … |
| 56063 | 12393960-6038375 | -53.5 ± 0.6 | … | … | … | … | … | N | … | … | … | … | n | … |
| 56266 | 12400351-6033050 | -29.5 ± 0.1 | … | … | … | … | … | N | … | … | … | … | n | … |
| 53007 | 12393987-6039341 | -22.9 ± 0.5 | 6041 ± 171 | 4.56 ± 0.30 | 0.16 ± 0.20 | <84 | 3 | N | … | … | … | … | n | NG |
| 56267 | 12400352-6034182 | -35.4 ± 1.4 | … | … | … | … | … | N | … | … | … | … | n | … |
| 56064 | 12393989-6045334 | -4.4 ± 0.1 | 4631 ± 67 | 3.08 ± 0.07 | 0.31 ± 0.26 | … | … | N | … | … | … | … | n | … |
| 56268 | 12400374-6037050 | -36.7 ± 1.2 | … | … | … | … | … | Y | … | … | … | … | n | … |
| 56065 | 12393993-6037292 | -40.5 ± 0.3 | … | … | … | … | … | Y | … | … | … | … | n | … |
| 53151 | 12400377-6041319 | -42.4 ± 0.8 | 5540 ± 73 | 4.51 ± 0.36 | -0.08 ± 0.19 | … | … | Y | … | … | … | … | n | … |
| 56066 | 12393996-6037096 | -40.8 ± 0.1 | 4969 ± 123 | 3.16 ± 0.05 | 0.29 ± 0.26 | … | … | Y | … | … | … | … | n | … |
| 53152 | 12400385-6040480 | -36.6 ± 0.4 | 5868 ± 67 | 4.20 ± 0.43 | 0.06 ± 0.15 | … | … | Y | … | … | … | … | n | … |
| 53008 | 12394009-6034319 | -24.9 ± 0.5 | 5402 ± 150 | 4.18 ± 0.70 | -0.11 ± 0.16 | … | … | N | … | … | … | … | n | … |
| 56269 | 12400389-6035475 | -47.5 ± 0.2 | … | … | … | … | … | N | … | … | … | … | n | … |
| 56067 | 12394011-6046277 | -15.3 ± 0.1 | 5139 ± 112 | 3.54 ± 0.05 | -0.14 ± 0.17 | … | … | N | … | … | … | … | n | … |
| 53153 | 12400425-6040299 | -25.5 ± 0.3 | 5408 ± 132 | 4.34 ± 0.46 | 0.09 ± 0.13 | … | … | N | … | … | … | … | n | … |
| 56068 | 12394017-6041568 | -1.1 ± 0.2 | … | … | … | … | … | N | … | … | … | … | n | … |
| 3397 | 12400449-6036566 | -40.1 ± 0.6 | 4412 ± 132 | 2.14 ± 0.25 | 0.00 ± 0.11 | 159 ± 3 | … | Y | N | Y | Y | … | Y | … |
| 56069 | 12394018-6044112 | -41.8 ± 0.6 | … | … | … | … | … | Y | … | … | … | … | n | … |
| 53154 | 12400467-6034129 | -28.0 ± 0.3 | 5620 ± 145 | 3.82 ± 0.29 | 0.27 ± 0.17 | 70 ± 50 | 1 | N | Y | N | N | … | n | NG |
| 53009 | 12394019-6044187 | -37.2 ± 0.5 | 5824 ± 15 | 3.96 ± 0.51 | -0.05 ± 0.15 | 66 ± 65 | 1 | Y | Y | Y | Y | … | Y | … |
| 56270 | 12400470-6040429 | -35.8 ± 1.9 | … | … | … | … | … | N | … | … | … | … | n | … |
| 56070 | 12394032-6040560 | -38.8 ± 0.9 | … | … | … | … | … | Y | … | … | … | … | n | … |
| 53179 | 12400899-6040071 | -14.6 ± 0.5 | 5652 ± 150 | 4.34 ± 0.45 | 0.04 ± 0.15 | … | … | N | … | … | … | … | n | … |
| 56071 | 12394036-6041356 | -75.8 ± 0.4 | … | … | … | … | … | N | … | … | … | … | n | … |
| 53180 | 12400902-6035101 | -23.5 ± 0.2 | 5880 ± 149 | 4.37 ± 0.48 | -0.15 ± 0.18 | … | … | N | … | … | … | … | n | … |
| 56072 | 12394039-6039597 | -35.8 ± 0.6 | … | … | … | … | … | N | … | … | … | … | n | … |
| 53181 | 12400948-6035262 | -46.0 ± 0.4 | 6728 ± 118 | 4.08 ± 0.23 | 0.19 ± 0.10 | 97 ± 50 | … | N | Y | Y | Y | Y | Y | … |

| ID | CNAME | RV (km s$^{-1}$) | T$_{eff}$ (K) | logg (dex) | [Fe/H] (dex) | EW(Li)$^a$ (mÅ) | EW(Li) error flag$^b$ | Membership RV | Li | logg | [Fe/H] | Gaia study Cantat-Gaudin$^c$ | Final$^d$ | NMs with Li$^e$ |
|---|---|---|---|---|---|---|---|---|---|---|---|---|---|---|
| 3375 | 12394049-6041006 | -42.0 ± 0.4 | 4926 ± 120 | 2.82 ± 0.24 | 0.10 ± 0.11 | <9 | 3 | Y | Y | Y | Y | … | Y | … |
| 53182 | 12400962-6036092 | -14.3 ± 0.5 | 5617 ± 50 | 4.12 ± 0.93 | -0.21 ± 0.20 | … | … | N | … | … | … | … | n | … |
| 56073 | 12394054-6035278 | -40.1 ± 1.2 | … | … | … | … | … | Y | … | … | … | … | n | … |
| 56294 | 12400963-6046292 | -40.1 ± 0.1 | 5617 ± 121 | 4.54 ± 0.03 | 0.55 ± 0.20 | … | … | Y | … | … | … | … | n | … |
| 53010 | 12394057-6036027 | -40.6 ± 0.4 | 5095 ± 460 | 3.83 ± 0.45 | -0.28 ± 0.28 | … | … | Y | … | … | … | … | n | … |
| 53183 | 12400992-6040087 | -11.0 ± 0.5 | 5897 ± 92 | 4.43 ± 0.18 | 0.16 ± 0.15 | <79 | 3 | N | … | … | … | … | n | NG |
| 53011 | 12394058-6036434 | 10.7 ± 0.3 | 5857 ± 23 | 4.39 ± 0.20 | 0.19 ± 0.19 | <29 | 3 | N | … | … | … | … | n | NG |
| 56295 | 12400994-6043462 | -33.0 ± 0.3 | … | … | … | … | … | N | … | … | … | … | n | … |
| 56074 | 12394062-6041247 | -41.4 ± 0.6 | … | … | … | … | … | Y | … | … | … | … | n | … |
| 56296 | 12401007-6040187 | -12.0 ± 0.4 | … | … | … | … | … | N | … | … | … | … | n | … |
| 53012 | 12394068-6040516 | 12.3 ± 0.6 | 6285 ± 137 | 4.48 ± 0.58 | -0.20 ± 0.16 | … | … | N | … | … | … | … | n | … |
| 56297 | 12401014-6032403 | -39.5 ± 0.3 | … | … | … | … | … | Y | … | … | … | … | n | … |
| 56075 | 12394077-6033324 | 2.2 ± 0.1 | 4835 ± 75 | 3.26 ± 0.09 | 0.05 ± 0.17 | … | … | N | … | … | … | … | n | … |
| 53184 | 12401016-6033343 | -53.9 ± 0.7 | 5510 ± 172 | 4.45 ± 0.22 | -0.39 ± 0.31 | <54 | 3 | N | … | … | … | … | n | NG |
| 53029 | 12394422-6035497 | -40.5 ± 0.4 | 5307 ± 381 | 3.90 ± 1.07 | 0.16 ± 0.15 | … | … | Y | … | … | … | … | n | … |
| 53185 | 12401020-6038241 | -42.8 ± 1.3 | 6204 ± 408 | 4.65 ± 0.87 | -0.54 ± 0.61 | <104 | 3 | Y | Y | Y | N | Y | Y | … |
| 56104 | 12394427-6046251 | -32.8 ± 0.2 | 6049 ± 132 | 2.96 ± 0.05 | … | … | … | N | … | … | … | … | n | … |
| 56298 | 12401025-6031220 | -35.9 ± 2.6 | … | … | … | … | … | N | … | … | … | … | n | … |
| 53186 | 12401025-6037173 | 20.3 ± 0.4 | 5688 ± 135 | 4.22 ± 0.19 | 0.01 ± 0.20 | … | … | N | … | … | … | … | n | … |
| 56105 | 12394433-6042541 | -16.2 ± 0.4 | … | … | … | … | … | N | … | … | … | … | n | … |
| 53187 | 12401051-6033369 | -17.0 ± 1.7 | 4712 ± 491 | 4.25 ± 0.53 | -0.46 ± 0.34 | … | … | N | … | … | … | … | n | … |
| 53030 | 12394436-6039028 | -31.4 ± 0.4 | 5691 ± 277 | 4.70 ± 0.40 | 0.29 ± 0.14 | <67 | 3 | N | … | … | … | … | n | NG |
| 53188 | 12401059-6046189 | -25.2 ± 0.9 | 5436 ± 117 | 3.87 ± 1.18 | -0.25 ± 0.36 | … | … | N | … | … | … | … | n | … |
| 56106 | 12394441-6028390 | -26.6 ± 0.5 | … | … | … | … | … | N | … | … | … | … | n | … |
| 56299 | 12401068-6041219 | -29.7 ± 0.8 | … | … | … | … | … | N | … | … | … | … | n | … |
| 56107 | 12394470-6036484 | -41.3 ± 0.8 | … | … | … | … | … | Y | … | … | … | … | n | … |
| 53189 | 12401083-6036309 | -38.4 ± 0.3 | 6751 ± 87 | 4.30 ± 0.17 | 0.22 ± 0.07 | 45 ± 44 | … | Y | Y | Y | Y | N | Y | … |
| 53031 | 12394473-6036592 | -50.2 ± 0.5 | 5868 ± 369 | 4.83 ± 0.53 | 0.04 ± 0.25 | <72 | 3 | N | … | … | … | … | n | NG |
| 53190 | 12401092-6042343 | -30.1 ± 0.4 | 5391 ± 240 | 4.90 ± 0.53 | -0.05 ± 0.17 | <81 | 3 | N | … | … | … | … | n | NG |
| 3380 | 12394475-6038339 | -37.9 ± 0.6 | 4882 ± 118 | 2.77 ± 0.23 | 0.12 ± 0.10 | <8 | 3 | Y | Y | Y | Y | … | Y | … |
| 53191 | 12401098-6039364 | -18.8 ± 1.0 | 5830 ± 28 | 4.15 ± 0.63 | -0.43 ± 0.14 | … | … | N | … | … | … | … | n | … |
| 53032 | 12394484-6041059 | 43.8 ± 0.3 | 5319 ± 226 | 4.46 ± 0.20 | -0.18 ± 0.26 | … | … | N | … | … | … | … | n | … |
| 56300 | 12401099-6040509 | -35.5 ± 1.0 | … | … | … | … | … | N | … | … | … | … | n | … |
| 53033 | 12394486-6035299 | -22.3 ± 0.3 | 5120 ± 62 | 4.40 ± 0.60 | -0.23 ± 0.16 | … | … | N | … | … | … | … | n | … |
| 53192 | 12401102-6039214 | -8.5 ± 1.5 | 6495 ± 398 | 4.43 ± 0.54 | 0.07 ± 0.14 | … | … | N | … | … | … | … | n | … |
| 56108 | 12394488-6037445 | -25.1 ± 0.1 | … | … | … | … | … | N | … | … | … | … | n | … |
| 56301 | 12401122-6042437 | -82.4 ± 0.1 | … | … | … | … | … | N | … | … | … | … | n | … |
| 56109 | 12394505-6039303 | -39.0 ± 0.1 | … | … | … | … | … | Y | … | … | … | … | n | … |
| 53193 | 12401123-6033408 | 62.1 ± 0.2 | 4968 ± 74 | 2.97 ± 0.18 | -0.31 ± 0.18 | … | … | N | … | … | … | … | n | G |
| 56110 | 12394510-6040046 | -41.9 ± 0.3 | … | … | … | … | … | Y | … | … | … | … | n | … |
| 56302 | 12401141-6030427 | -38.8 ± 0.1 | 5181 ± 120 | 3.20 ± 0.05 | 0.41 ± 0.25 | … | … | Y | … | … | … | … | n | … |
| 56111 | 12394513-6035454 | -41.8 ± 0.4 | … | … | … | … | … | Y | … | … | … | … | n | … |
| 53208 | 12401406-6034047 | -55.8 ± 1.1 | 5871 ± 467 | 4.57 ± 0.46 | -0.19 ± 0.68 | <133 | 3 | N | … | … | … | … | n | NG |
| 3381 | 12394514-6038258 | -40.7 ± 0.6 | 4928 ± 119 | 2.85 ± 0.23 | 0.12 ± 0.10 | <7 | 3 | Y | Y | Y | Y | … | Y | … |
| 53209 | 12401409-6040509 | -41.5 ± 0.4 | 5693 ± 76 | 4.49 ± 0.63 | -0.14 ± 0.15 | … | … | Y | … | … | … | … | n | … |
| 56112 | 12394518-6041123 | -34.2 ± 1.1 | … | … | … | … | … | N | … | … | … | … | n | … |
| 53210 | 12401439-6041562 | -36.6 ± 0.5 | 5811 ± 55 | 4.55 ± 0.36 | -0.10 ± 0.15 | 83 ± 68 | 1 | Y | Y | Y | Y | … | Y | … |
| 53034 | 12394526-6041409 | -37.8 ± 0.6 | 6346 ± 330 | 4.30 ± 0.31 | 0.04 ± 0.32 | … | … | Y | … | … | … | … | n | … |
| 56313 | 12401449-6031471 | 19.4 ± 0.1 | … | … | … | … | … | N | … | … | … | … | n | … |
| 56113 | 12394530-6033068 | -40.6 ± 0.4 | … | … | … | … | … | Y | … | … | … | … | n | … |
| 56314 | 12401451-6038002 | -38.6 ± 0.2 | … | … | … | … | … | Y | … | … | … | … | n | … |
| 56114 | 12394531-6042292 | -26.1 ± 0.3 | … | … | … | … | … | N | … | … | … | … | n | … |
| 53211 | 12401456-6039375 | -52.7 ± 0.3 | 5549 ± 117 | 3.94 ± 0.59 | 0.01 ± 0.15 | … | … | N | … | … | … | … | n | … |
| 56115 | 12394550-6028146 | -50.1 ± 0.1 | … | … | … | … | … | N | … | … | … | … | n | … |
| 56315 | 12401460-6033018 | -35.4 ± 0.2 | … | … | … | … | … | N | … | … | … | … | n | … |
| 56116 | 12394552-6039521 | -38.4 ± 0.3 | … | … | … | … | … | Y | … | … | … | … | n | … |
| 53212 | 12401462-6039511 | 0.6 ± 0.3 | 6147 ± 81 | 4.36 ± 0.37 | 0.03 ± 0.13 | 66 ± 23 | 1 | N | … | … | … | … | n | NG |
| 56316 | 12401467-6038006 | 11.8 ± 0.9 | … | … | … | … | … | N | … | … | … | … | n | … |
| 56317 | 12401499-6040111 | -42.0 ± 0.7 | … | … | … | … | … | Y | … | … | … | … | n | … |

**Table C.17.** continued.

| ID | CNAME | RV (km s$^{-1}$) | $T_{\rm eff}$ (K) | $\log g$ (dex) | [Fe/H] (dex) | $EW(\rm Li)^a$ (mÅ) | $EW(\rm Li)$ error flag$^b$ | Membership RV | Li | $\log g$ | [Fe/H] | Gaia study Cantat-Gaudin$^c$ | Final$^d$ | NMs with Li$^e$ |
|---|---|---|---|---|---|---|---|---|---|---|---|---|---|---|
| 56318 | 12401501-6039376 | -14.3 ± 0.1 | 5311 ± 146 | 3.52 ± 0.05 | 0.05 ± 0.23 | … | … | N | … | … | … | … | n | … |
| 53213 | 12401525-6035142 | -18.3 ± 0.3 | 5371 ± 106 | 4.50 ± 0.20 | 0.11 ± 0.13 | … | … | N | … | … | … | … | n | … |
| 56320 | 12401550-6042507 | -50.2 ± 0.1 | … | … | … | … | … | N | … | … | … | … | n | … |
| 56321 | 12401557-6040413 | -13.5 ± 0.1 | … | … | … | … | … | N | … | … | … | … | n | … |
| 56322 | 12401576-6042061 | -3.9 ± 0.1 | 5411 ± 122 | 4.21 ± 0.10 | -0.10 ± 0.17 | … | … | N | … | … | … | … | n | … |
| 56323 | 12401607-6035535 | -28.2 ± 1.3 | … | … | … | … | … | N | … | … | … | … | n | … |
| 56324 | 12401611-6035127 | 12.9 ± 0.1 | 4421 ± 111 | 2.87 ± 0.05 | 0.23 ± 0.22 | … | … | N | … | … | … | … | n | … |
| 53214 | 12401630-6038551 | -45.2 ± 0.4 | 5736 ± 420 | 4.39 ± 0.28 | -0.37 ± 0.16 | <35 | 3 | N | … | … | … | … | n | NG |
| 56325 | 12401647-6043527 | -28.6 ± 0.2 | … | … | … | … | … | N | … | … | … | … | n | … |
| 53215 | 12401650-6035291 | -65.0 ± 0.6 | 5351 ± 279 | 4.63 ± 0.36 | -0.24 ± 0.27 | … | … | N | … | … | … | … | n | … |
| 56326 | 12401662-6046173 | -32.1 ± 1.4 | … | … | … | … | … | N | … | … | … | … | n | … |
| 56327 | 12401664-6040079 | -41.5 ± 0.2 | … | … | … | … | … | Y | … | … | … | … | n | … |
| 53216 | 12401667-6031050 | -14.5 ± 0.7 | 5497 ± 72 | 3.99 ± 0.69 | -0.52 ± 0.23 | <141 | 3 | N | … | … | … | … | n | NG |
| 56357 | 12402309-6042580 | -24.8 ± 0.2 | 5685 ± 127 | 3.01 ± 0.05 | 0.22 ± 0.26 | … | … | N | … | … | … | … | n | … |
| 53232 | 12402317-6036122 | -18.7 ± 0.5 | 6292 ± 60 | 3.62 ± 0.12 | 0.26 ± 0.04 | 90 ± 38 | … | N | … | … | … | … | n | NG |
| 56358 | 12402321-6041423 | -38.2 ± 2.8 | … | … | … | … | … | Y | … | … | … | … | n | … |
| 56359 | 12402335-6030222 | -20.4 ± 0.2 | … | … | … | … | … | N | … | … | … | … | n | … |
| 56360 | 12402343-6036550 | 12.0 ± 0.2 | … | … | … | … | … | N | … | … | … | … | n | … |
| 56361 | 12402358-6043587 | -20.4 ± 0.1 | 5361 ± 152 | 4.23 ± 0.19 | 0.39 ± 0.25 | … | … | N | … | … | … | … | n | … |
| 56362 | 12402385-6037581 | -10.1 ± 0.1 | … | … | … | … | … | N | … | … | … | … | n | … |
| 56363 | 12402395-6031121 | -24.8 ± 0.4 | … | … | … | … | … | N | … | … | … | … | n | … |
| 56364 | 12402425-6042334 | -12.7 ± 0.3 | … | … | … | … | … | N | … | … | … | … | n | … |
| 53233 | 12402442-6043250 | -42.8 ± 0.7 | 5878 ± 376 | 4.38 ± 0.25 | -0.10 ± 0.32 | … | … | Y | … | … | … | … | n | … |
| 56365 | 12402465-6033376 | -39.1 ± 1.0 | … | … | … | … | … | Y | … | … | … | … | n | … |
| 53234 | 12402477-6040178 | -3.6 ± 2.1 | 5795 ± 22 | 3.66 ± 0.44 | -0.94 ± 1.01 | … | … | N | … | … | … | … | n | … |
| 3400 | 12402478-6043103 | -39.1 ± 0.6 | 4590 ± 123 | 2.28 ± 0.22 | 0.03 ± 0.11 | <8 | 3 | Y | Y | Y | Y | … | Y | … |
| 56366 | 12402518-6045594 | -23.3 ± 0.2 | … | … | … | … | … | N | … | … | … | … | n | … |
| 56367 | 12402521-6036436 | 1.2 ± 0.1 | … | … | … | … | … | N | … | … | … | … | n | … |
| 53235 | 12402627-6035356 | -31.6 ± 0.4 | 5782 ± 335 | 4.23 ± 0.25 | -0.01 ± 0.36 | … | … | N | … | … | … | … | n | … |
| 56368 | 12402639-6036007 | -42.8 ± 0.4 | … | … | … | … | … | Y | … | … | … | … | n | … |
| 56369 | 12402682-6033375 | -41.9 ± 0.7 | … | … | … | … | … | Y | … | … | … | … | n | … |
| 56371 | 12402692-6034388 | -26.2 ± 0.1 | … | … | … | … | … | N | … | … | … | … | n | … |
| 53236 | 12402731-6037066 | -64.9 ± 0.7 | 5580 ± 445 | 4.67 ± 0.13 | -0.29 ± 0.33 | … | … | N | … | … | … | … | n | … |
| 53070 | 12394977-6035221 | -36.3 ± 1.8 | … | … | … | … | … | Y | … | … | … | … | n | … |
| 53082 | 12395155-6038032 | 4.0 ± 2.2 | … | … | … | … | … | N | … | … | … | … | n | … |
| 52811 | 12390893-6028011 | 16.2 ± 3.1 | … | … | … | … | … | N | … | … | … | … | n | … |
| 52812 | 12390893-6034318 | -10.1 ± 0.2 | 6424 ± 223 | 4.43 ± 0.29 | 0.30 ± 0.25 | … | … | N | … | … | … | … | n | … |
| 55815 | 12390921-6035264 | -35.4 ± 0.6 | … | … | … | … | … | N | … | … | … | … | n | … |
| 52814 | 12390957-6038567 | -19.4 ± 0.4 | 5514 ± 284 | 4.36 ± 0.48 | 0.20 ± 0.23 | … | … | N | … | … | … | … | n | … |
| 55816 | 12390961-6041520 | -39.0 ± 0.4 | … | … | … | … | … | Y | … | … | … | … | n | … |
| 55817 | 12390985-6035119 | -37.9 ± 0.7 | … | … | … | … | … | Y | … | … | … | … | n | … |
| 52815 | 12390985-6036397 | -10.1 ± 0.5 | 6130 ± 369 | 3.94 ± 0.32 | 0.43 ± 0.30 | 142 ± 64 | 1 | N | … | … | … | … | n | NG |
| 52816 | 12390993-6036333 | -45.0 ± 1.1 | 6749 ± 560 | 4.14 ± 0.32 | 0.21 ± 0.33 | … | … | N | … | … | … | … | n | … |
| 55818 | 12391000-6039306 | -40.3 ± 0.2 | … | … | … | … | … | Y | … | … | … | … | n | … |
| 3365 | 12391002-6038402 | -40.2 ± 0.6 | 4471 ± 118 | 2.14 ± 0.23 | 0.01 ± 0.10 | 8 ± 1 | … | Y | Y | Y | Y | … | Y | … |
| 52817 | 12391004-6042071 | 1.1 ± 0.4 | 5693 ± 322 | 4.07 ± 0.32 | 0.11 ± 0.17 | … | … | N | … | … | … | … | n | … |
| 55819 | 12391004-6042391 | -39.4 ± 1.7 | … | … | … | … | … | Y | … | … | … | … | n | … |
| 55820 | 12391016-6034218 | -41.0 ± 0.5 | … | … | … | … | … | Y | … | … | … | … | n | … |
| 55821 | 12391016-6038290 | -38.7 ± 0.1 | 4891 ± 107 | 3.82 ± 0.01 | 0.27 ± 0.20 | … | … | Y | … | … | … | … | n | … |
| 55822 | 12391019-6040586 | -30.8 ± 1.0 | … | … | … | … | … | N | … | … | … | … | n | … |
| 55823 | 12391037-6030333 | -49.6 ± 0.1 | 6083 ± 144 | 4.02 ± 0.01 | … | … | … | N | … | … | … | … | n | … |
| 55824 | 12391049-6038234 | -30.1 ± 0.1 | … | … | … | … | … | N | … | … | … | … | n | … |
| 55825 | 12391061-6039367 | -50.0 ± 0.3 | … | … | … | … | … | N | … | … | … | … | n | … |
| 55826 | 12391063-6042577 | -38.8 ± 0.2 | 5825 ± 113 | 3.40 ± 0.07 | -0.01 ± 0.22 | … | … | Y | … | … | … | … | n | … |
| 52818 | 12391082-6037229 | -36.9 ± 0.5 | 5991 ± 302 | 4.62 ± 0.56 | 0.09 ± 0.34 | <126 | 3 | Y | N | Y | Y | … | n | NG |
| 55853 | 12391477-6029059 | -37.5 ± 1.6 | … | … | … | … | … | Y | … | … | … | … | n | … |
| 52839 | 12391485-6037349 | 28.3 ± 0.3 | 5718 ± 20 | 4.50 ± 0.16 | 0.06 ± 0.22 | <52 | 3 | N | … | … | … | … | n | NG |
| 52840 | 12391504-6040346 | -15.5 ± 0.4 | 5862 ± 341 | 4.27 ± 0.04 | 0.23 ± 0.17 | … | … | N | … | … | … | … | n | … |
| 55854 | 12391522-6042013 | -40.1 ± 0.7 | … | … | … | … | … | Y | … | … | … | … | n | … |

Table C.17. continued.| ID | CNAME | RV (km s$^{-1}$) | $T_{\text{eff}}$ (K) | $logg$ (dex) | [Fe/H] (dex) | EW(Li)$^a$ (mÅ) | EW(Li) error flag$^b$ | Membership RV | Li | $logg$ | [Fe/H] | Gaia study Cantat-Gaudin$^c$ | Final$^d$ | NMs with Li$^e$ |
|---|---|---|---|---|---|---|---|---|---|---|---|---|---|---|
| 55855 | 12391525-6042220 | -38.4 ± 0.1 | 5413 ± 118 | 4.48 ± 0.11 | 0.33 ± 0.22 | … | … | Y | … | … | … | … | n | … |
| 55856 | 12391534-6039275 | -62.1 ± 0.2 | … | … | … | … | … | N | … | … | … | … | n | … |
| 52841 | 12391551-6035352 | -20.9 ± 0.8 | 5520 ± 274 | 3.88 ± 0.68 | 0.24 ± 0.16 | … | … | N | … | … | … | … | n | … |
| 52842 | 12391557-6035469 | 18.5 ± 0.4 | 5843 ± 175 | 4.22 ± 0.19 | -0.18 ± 0.19 | <38 | 3 | N | … | … | … | … | n | NG |
| 52843 | 12391571-6036013 | -7.5 ± 0.5 | 6505 ± 547 | 4.22 ± 0.17 | 0.18 ± 0.25 | <53 | 3 | N | … | … | … | … | n | NG |
| 55857 | 12391573-6045019 | -39.1 ± 1.7 | … | … | … | … | … | Y | … | … | … | … | n | … |
| 55858 | 12391576-6034406 | -40.1 ± 0.1 | … | … | … | … | … | Y | … | … | … | … | n | … |
| 55859 | 12391576-6036419 | -37.1 ± 0.5 | … | … | … | … | … | Y | … | … | … | … | n | … |
| 3368 | 12391577-6034406 | -38.6 ± 0.6 | 4851 ± 118 | 2.88 ± 0.23 | 0.02 ± 0.10 | 63 ± 4 | 1 | Y | Y | Y | Y | Y | Y | … |
| 52844 | 12391582-6033056 | 2.9 ± 0.7 | 5994 ± 224 | 4.08 ± 0.50 | 0.42 ± 0.13 | … | … | N | … | … | … | … | n | … |
| 55860 | 12391596-6031006 | -44.2 ± 0.1 | … | … | … | … | … | N | … | … | … | … | n | … |
| 52845 | 12391605-6038210 | -37.9 ± 0.3 | 6811 ± 83 | 4.27 ± 0.16 | 0.34 ± 0.07 | 62 ± 19 | … | Y | Y | Y | Y | … | Y | … |
| 52846 | 12391609-6038505 | -35.0 ± 0.5 | 6812 ± 107 | 3.96 ± 0.20 | 0.33 ± 0.09 | 87 ± 69 | … | N | Y | Y | Y | Y | Y | … |
| 52847 | 12391612-6039055 | -38.7 ± 0.4 | 5687 ± 355 | 4.84 ± 0.68 | -0.23 ± 0.47 | <81 | 3 | Y | Y | Y? | Y | … | Y | … |
| 55861 | 12391621-6033518 | -39.9 ± 1.1 | … | … | … | … | … | Y | … | … | … | … | n | … |
| 55862 | 12391624-6037563 | -48.2 ± 0.3 | … | … | … | … | … | N | … | … | … | … | n | … |
| 52848 | 12391636-6038535 | 13.9 ± 2.2 | 6457 ± 245 | 4.39 ± 0.52 | 0.03 ± 0.20 | … | … | N | … | … | … | … | n | … |
| 55863 | 12391648-6029091 | -42.4 ± 0.1 | 5545 ± 136 | 2.67 ± 0.13 | -0.49 ± 0.21 | … | … | Y | … | … | … | … | n | … |
| 55864 | 12391652-6036571 | -55.0 ± 0.1 | 5361 ± 99 | 4.19 ± 0.05 | 0.06 ± 0.15 | … | … | N | … | … | … | … | n | … |
| 55865 | 12391655-6028579 | -12.5 ± 0.2 | … | … | … | … | … | N | … | … | … | … | n | … |
| 52860 | 12391834-6027503 | -29.3 ± 0.9 | 5774 ± 608 | 4.82 ± 0.25 | -0.28 ± 0.36 | … | … | N | … | … | … | … | n | … |
| 52861 | 12391845-6037459 | -53.8 ± 0.7 | 5050 ± 132 | 4.39 ± 0.60 | -0.50 ± 0.65 | <79 | 3 | N | … | … | … | … | n | NG |
| 55879 | 12391859-6036303 | -43.0 ± 2.2 | … | … | … | … | … | Y | … | … | … | … | n | … |
| 52862 | 12391859-6036403 | -35.0 ± 0.4 | 5889 ± 719 | 4.45 ± 0.56 | -0.03 ± 0.51 | … | … | N | … | … | … | … | n | … |
| 55880 | 12391865-6036501 | -39.2 ± 1.1 | … | … | … | … | … | Y | … | … | … | … | n | … |
| 55881 | 12391876-6034544 | -43.2 ± 0.1 | 5595 ± 148 | 4.38 ± 0.03 | 0.27 ± 0.21 | … | … | N | … | … | … | … | n | … |
| 52863 | 12391877-6042242 | -24.8 ± 0.4 | 5699 ± 258 | 4.22 ± 0.08 | 0.23 ± 0.13 | … | … | N | … | … | … | … | n | … |
| 55882 | 12391880-6029489 | -39.7 ± 0.1 | 5539 ± 126 | 4.54 ± 0.05 | 0.58 ± 0.21 | … | … | Y | … | … | … | … | n | … |
| 55883 | 12391881-6044318 | -26.6 ± 1.4 | … | … | … | … | … | N | … | … | … | … | n | … |
| 55884 | 12391892-6033104 | -43.9 ± 0.1 | … | … | … | … | … | N | … | … | … | … | n | … |
| 52864 | 12391893-6035203 | -40.1 ± 0.6 | 6159 ± 374 | 4.58 ± 0.33 | 0.26 ± 0.26 | <66 | 3 | Y | Y | Y | Y | … | Y | … |
| 52865 | 12391899-6038328 | -53.6 ± 0.3 | 5779 ± 225 | 4.04 ± 0.31 | 0.28 ± 0.17 | … | … | N | … | … | … | … | n | … |
| 55885 | 12391901-6031583 | -43.2 ± 0.3 | … | … | … | … | … | N | … | … | … | … | n | … |
| 56078 | 12394104-6046232 | -12.4 ± 0.2 | … | … | … | … | … | N | … | … | … | … | n | … |
| 55886 | 12391903-6034333 | -38.0 ± 0.4 | … | … | … | … | … | Y | … | … | … | … | n | … |
| 56079 | 12394118-6038582 | -37.4 ± 0.8 | … | … | … | … | … | Y | … | … | … | … | n | … |
| 56080 | 12394121-6035351 | -40.4 ± 0.6 | … | … | … | … | … | Y | … | … | … | … | n | … |
| 3376 | 12394121-6040040 | -37.8 ± 0.6 | 4989 ± 117 | 2.95 ± 0.25 | 0.13 ± 0.09 | <7 | 3 | Y | Y | Y | Y | … | Y | … |
| 52866 | 12391905-6032435 | -16.8 ± 0.3 | 6171 ± 279 | 4.30 ± 0.36 | 0.09 ± 0.21 | 68 ± 24 | 1 | N | … | … | … | … | n | NG |
| 53014 | 12394124-6038014 | -36.6 ± 0.4 | 6380 ± 122 | 4.12 ± 0.12 | 0.03 ± 0.21 | … | … | Y | … | … | … | … | n | … |
| 52867 | 12391905-6035310 | -39.9 ± 0.4 | 6545 ± 345 | 4.20 ± 0.40 | 0.17 ± 0.21 | 67 ± 20 | 1 | Y | Y | Y | Y | N | Y | … |
| 56081 | 12394147-6044292 | -42.3 ± 0.1 | 5341 ± 122 | 4.36 ± 0.05 | 0.03 ± 0.24 | … | … | Y | … | … | … | … | n | … |
| 55888 | 12391908-6044506 | -40.9 ± 0.1 | 5271 ± 107 | 4.18 ± 0.04 | 0.24 ± 0.20 | … | … | Y | … | … | … | … | n | … |
| 53015 | 12394165-6039446 | -19.3 ± 0.3 | 6100 ± 33 | 4.22 ± 0.31 | -0.27 ± 0.28 | 47 ± 22 | … | N | … | … | … | … | n | NG |
| 52868 | 12391928-6041104 | -27.0 ± 1.1 | 6016 ± 95 | 4.40 ± 0.78 | -0.31 ± 0.38 | … | … | N | … | … | … | … | n | … |
| 55594 | 12381825-6042585 | -22.6 ± 0.2 | … | … | … | … | … | N | … | … | … | … | n | … |
| 53016 | 12394166-6041491 | -7.2 ± 1.1 | 5365 ± 31 | 3.07 ± 0.84 | -0.86 ± 0.21 | … | … | N | … | … | … | … | n | … |
| 55889 | 12391929-6044132 | -40.9 ± 0.1 | 5245 ± 80 | 3.98 ± 0.01 | -0.02 ± 0.17 | … | … | Y | … | … | … | … | n | … |
| 53017 | 12394168-6039176 | -38.8 ± 0.4 | 6154 ± 396 | 4.49 ± 0.15 | 0.00 ± 0.34 | <90 | 3 | Y | Y | Y | Y | … | Y | … |
| 55595 | 12381864-6034227 | -20.2 ± 0.4 | 5725 ± 125 | 3.32 ± 0.05 | 0.08 ± 0.24 | … | … | N | … | … | … | … | n | … |
| 55890 | 12391939-6030182 | -24.2 ± 0.2 | … | … | … | … | … | N | … | … | … | … | n | … |
| 56082 | 12394169-6041280 | -36.2 ± 0.1 | … | … | … | … | … | Y | … | … | … | … | n | … |
| 55596 | 12381884-6036322 | -39.1 ± 0.1 | 5215 ± 146 | 4.18 ± 0.04 | 0.10 ± 0.23 | … | … | Y | … | … | … | … | n | … |
| 55597 | 12382027-6044322 | -33.5 ± 2.7 | … | … | … | … | … | N | … | … | … | … | n | … |
| 56083 | 12394171-6044191 | -15.2 ± 0.1 | 5749 ± 134 | 3.48 ± 0.09 | -0.10 ± 0.21 | … | … | N | … | … | … | … | n | … |
| 52869 | 12391942-6035210 | -18.4 ± 0.3 | 5472 ± 102 | 4.07 ± 0.32 | 0.42 ± 0.34 | <32 | 3 | N | … | … | … | … | n | NG |
| 55598 | 12382081-6032185 | -7.9 ± 0.2 | 5913 ± 113 | 4.13 ± 0.05 | -0.04 ± 0.20 | … | … | Y | … | … | … | … | n | … |
| 56084 | 12394173-6034393 | -40.3 ± 2.2 | … | … | … | … | … | Y | … | … | … | … | n | … |
| 52870 | 12391942-6036514 | -39.2 ± 0.3 | 6530 ± 233 | 4.00 ± 0.03 | 0.14 ± 0.19 | 81 ± 31 | 1 | Y | Y | Y | Y | Y | Y | … |

Article number, page 239 of 264Gutiérrez Albarrán et al.: Calibrating the lithium–age relation I.

**Table C.17.** continued.

| ID | CNAME | RV (km s$^{-1}$) | $T_{\text{eff}}$ (K) | logg (dex) | [Fe/H] (dex) | EW(Li)$^a$ (mÅ) | EW(Li) error flag$^b$ | Membership RV | Li | logg | [Fe/H] | Gaia study Cantat-Gaudin$^c$ | Final$^d$ | NMs with Li$^e$ |
|---|---|---|---|---|---|---|---|---|---|---|---|---|---|---|
| 55599 | 12382203-6038018 | -41.8 ± 0.1 | 6089 ± 228 | 4.91 ± 0.05 | … | … | … | Y | … | … | … | … | n | … |
| 56085 | 12394173-6035267 | -39.5 ± 0.3 | … | … | … | … | … | Y | … | … | … | … | n | … |
| 55891 | 12391952-6044438 | -39.7 ± 0.7 | … | … | … | … | … | Y | … | … | … | … | n | … |
| 55600 | 12382220-6039565 | -40.2 ± 0.1 | 5241 ± 47 | 3.66 ± 0.02 | 0.50 ± 0.22 | … | … | Y | … | … | … | … | n | … |
| 56086 | 12394185-6034032 | -36.7 ± 1.5 | … | … | … | … | … | Y | … | … | … | … | n | … |
| 52889 | 12392343-6036074 | -17.9 ± 0.5 | 5773 ± 648 | 4.44 ± 0.41 | -0.02 ± 0.39 | … | … | N | … | … | … | … | n | … |
| 55601 | 12382240-6036556 | -38.5 ± 1.6 | … | … | … | … | … | Y | … | … | … | … | n | … |
| 53018 | 12394200-6034459 | -34.2 ± 0.4 | 5960 ± 190 | 4.10 ± 0.35 | -0.40 ± 0.17 | … | … | N | … | … | … | … | n | … |
| 52890 | 12392343-6037528 | -69.3 ± 0.9 | 6126 ± 227 | 4.53 ± 0.82 | -0.62 ± 0.50 | 22 ± 18 | 1 | N | … | … | … | Y | n | NG |
| 55602 | 12382279-6035046 | -40.2 ± 0.1 | 5389 ± 103 | 4.15 ± 0.05 | 0.46 ± 0.25 | … | … | Y | … | … | … | … | n | … |
| 56087 | 12394205-6036594 | -40.2 ± 0.1 | … | … | … | … | … | Y | … | … | … | … | n | … |
| 52891 | 12392344-6042543 | -29.2 ± 0.3 | 5141 ± 102 | 4.39 ± 0.68 | 0.21 ± 0.17 | … | … | N | … | … | … | … | n | … |
| 55603 | 12382281-6042360 | -3.5 ± 0.4 | … | … | … | … | … | N | … | … | … | … | n | … |
| 56088 | 12394221-6038098 | -6.0 ± 0.5 | … | … | … | … | … | N | … | … | … | … | n | … |
| 55604 | 12382287-6040575 | -44.2 ± 1.0 | … | … | … | … | … | N | … | … | … | … | n | … |
| 55922 | 12392346-6037279 | -38.2 ± 0.9 | … | … | … | … | … | Y | … | … | … | … | n | … |
| 56089 | 12394248-6035597 | -37.0 ± 0.2 | … | … | … | … | … | Y | … | … | … | … | n | … |
| 55605 | 12382333-6042241 | -38.8 ± 1.6 | … | … | … | … | … | Y | … | … | … | … | n | … |
| 52892 | 12392346-6039543 | -39.4 ± 0.4 | 5496 ± 367 | 4.34 ± 0.16 | 0.17 ± 0.31 | … | … | Y | … | … | … | … | n | … |
| 53019 | 12394255-6035033 | -14.0 ± 0.4 | 5287 ± 168 | 4.37 ± 0.46 | -0.11 ± 0.16 | … | … | N | … | … | … | … | n | … |
| 55606 | 12382355-6034040 | -18.7 ± 0.2 | 6229 ± 175 | 4.04 ± 0.06 | … | … | … | N | … | … | … | … | n | … |
| 52893 | 12392350-6040582 | -40.1 ± 0.3 | 5454 ± 62 | 3.94 ± 0.13 | -0.69 ± 0.45 | <65 | 3 | Y | Y | N | N | … | n | NG |
| 53020 | 12394266-6037161 | -57.2 ± 0.3 | 5379 ± 138 | 4.22 ± 0.20 | -0.35 ± 0.14 | … | … | N | … | … | … | … | n | … |
| 52894 | 12392354-6037014 | -16.0 ± 0.4 | 5147 ± 144 | 4.74 ± 0.15 | -0.09 ± 0.15 | … | … | N | … | … | … | … | n | … |
| 56090 | 12394294-6036126 | -37.9 ± 1.3 | … | … | … | … | … | Y | … | … | … | … | n | … |
| 55608 | 12382399-6035087 | -38.7 ± 0.6 | … | … | … | … | … | Y | … | … | … | … | n | … |
| 55923 | 12392358-6034032 | -9.3 ± 0.1 | … | … | … | … | … | N | … | … | … | … | n | … |
| 56120 | 12394560-6042006 | -16.7 ± 1.6 | … | … | … | … | … | N | … | … | … | … | n | … |
| 55609 | 12382460-6036379 | -10.6 ± 0.5 | … | … | … | … | … | N | … | … | … | … | n | … |
| 52895 | 12392366-6041524 | -22.1 ± 1.1 | 6383 ± 214 | 4.58 ± 0.54 | 0.17 ± 0.18 | <42 | 3 | N | … | … | … | … | n | NG |
| 53035 | 12394571-6039001 | -8.5 ± 1.4 | 5373 ± 147 | 3.85 ± 0.91 | -0.33 ± 0.31 | … | … | N | … | … | … | … | n | … |
| 55610 | 12382489-6040382 | -48.2 ± 0.1 | 4509 ± 111 | 2.47 ± 0.06 | 0.08 ± 0.23 | … | … | N | … | … | … | … | n | … |
| 56121 | 12394571-6042230 | -23.0 ± 0.1 | … | … | … | … | … | N | … | … | … | … | n | … |
| 52896 | 12392371-6034274 | -38.9 ± 0.4 | 6187 ± 326 | 4.72 ± 0.52 | 0.29 ± 0.25 | <120 | 3 | Y | Y | Y | Y | … | Y | … |
| 55611 | 12382549-6028310 | -21.4 ± 0.1 | … | … | … | … | … | N | … | … | … | … | n | … |
| 56122 | 12394585-6041274 | -10.1 ± 0.1 | … | … | … | … | … | N | … | … | … | … | n | … |
| 52897 | 12392386-6032508 | -40.3 ± 0.4 | 6230 ± 344 | 4.12 ± 0.04 | -0.04 ± 0.21 | 44 ± 24 | 1 | Y | Y | Y | Y | Y | Y | … |
| 3382 | 12394596-6038389 | -39.0 ± 0.6 | 4889 ± 122 | 2.75 ± 0.24 | 0.10 ± 0.11 | <6 | 3 | Y | Y | Y | Y | … | Y | … |
| 55612 | 12382576-6040413 | -36.0 ± 0.1 | 5765 ± 124 | 3.22 ± 0.05 | 0.11 ± 0.24 | … | … | N | … | … | … | … | n | … |
| 55924 | 12392392-6033176 | -38.6 ± 0.7 | … | … | … | … | … | Y | … | … | … | … | n | … |
| 56123 | 12394599-6031273 | -55.2 ± 0.2 | … | … | … | … | … | N | … | … | … | … | n | … |
| 55613 | 12382626-6038053 | -14.2 ± 0.3 | … | … | … | … | … | N | … | … | … | … | n | … |
| 52898 | 12392402-6041379 | -40.2 ± 0.5 | 5989 ± 267 | 4.36 ± 0.12 | 0.10 ± 0.28 | … | … | Y | … | … | … | … | n | … |
| 53036 | 12394601-6038476 | -40.4 ± 0.5 | 5652 ± 87 | 3.94 ± 0.23 | -0.07 ± 0.20 | … | … | Y | … | … | … | … | n | … |
| 55614 | 12382632-6044384 | -30.1 ± 0.2 | … | … | … | … | … | N | … | … | … | … | n | … |
| 55925 | 12392416-6045170 | -15.5 ± 0.2 | … | … | … | … | … | N | … | … | … | … | n | … |
| 53037 | 12394601-6042400 | 3.4 ± 0.8 | 6056 ± 211 | 4.24 ± 0.10 | -0.23 ± 0.21 | … | … | N | … | … | … | … | n | … |
| 55615 | 12382679-6030470 | -14.4 ± 2.4 | … | … | … | … | … | N | … | … | … | … | n | … |
| 52899 | 12392420-6038138 | -37.1 ± 0.7 | 6268 ± 338 | 4.09 ± 0.36 | -0.06 ± 0.19 | … | … | Y | … | … | … | … | n | … |
| 56124 | 12394608-6044088 | -15.0 ± 0.3 | … | … | … | … | … | N | … | … | … | … | n | … |
| 55616 | 12382711-6044534 | -10.1 ± 0.1 | … | … | … | … | … | N | … | … | … | … | n | … |
| 52900 | 12392423-6034256 | -45.1 ± 0.2 | 6781 ± 45 | 4.20 ± 0.09 | 0.29 ± 0.04 | 56 ± 41 | … | N | … | … | … | N | n | NG |
| 53038 | 12394609-6042251 | -32.4 ± 0.3 | 5613 ± 79 | 4.11 ± 0.42 | 0.21 ± 0.14 | … | … | N | … | … | … | … | n | … |
| 55617 | 12382718-6034384 | -41.0 ± 0.5 | … | … | … | … | … | Y | … | … | … | … | n | … |
| 55926 | 12392432-6044502 | -19.7 ± 0.1 | … | … | … | … | … | N | … | … | … | … | n | … |
| 56125 | 12394613-6031511 | -41.8 ± 1.1 | … | … | … | … | … | Y | … | … | … | … | n | … |
| 3359 | 12383597-6045242 | -40.2 ± 0.6 | 5003 ± 117 | 2.97 ± 0.24 | 0.14 ± 0.10 | <7 | 3 | Y | Y | Y | Y | … | Y | … |
| 56126 | 12394613-6036154 | -34.9 ± 0.2 | 6167 ± 150 | 4.09 ± 0.01 | … | … | … | N | … | … | … | … | n | … |
| 52901 | 12392434-6037476 | -37.3 ± 0.5 | 6271 ± 206 | 4.18 ± 0.38 | -0.07 ± 0.14 | <42 | 3 | Y | Y | Y | Y | Y | Y | … |

| ID | CNAME | RV (km s$^{-1}$) | $T_{\text{eff}}$ (K) | logg (dex) | [Fe/H] (dex) | EW(Li)$^a$ (mÅ) | EW(Li) error flag$^b$ | Membership RV | Li | logg | [Fe/H] | Gaia study Cantat-Gaudin$^c$ | Final$^d$ | NMs with Li$^e$ |
|---|---|---|---|---|---|---|---|---|---|---|---|---|---|---|
| 55640 | 12383607-6037118 | -45.2 ± 0.1 | 4723 ± 126 | 4.03 ± 0.04 | 0.23 ± 0.23 | … | … | N | … | … | … | … | n | … |
| 53039 | 12394615-6036058 | -41.1 ± 0.5 | 5887 ± 106 | 4.67 ± 0.47 | 0.09 ± 0.18 | … | … | Y | … | … | … | … | n | … |
| 52902 | 12392442-6039264 | 79.3 ± 0.7 | 5341 ± 163 | 4.02 ± 0.24 | -0.60 ± 0.27 | … | … | N | … | … | … | … | n | … |
| 55641 | 12383616-6041304 | -12.2 ± 0.2 | 5587 ± 103 | 2.59 ± 0.08 | -0.30 ± 0.17 | … | … | N | … | … | … | … | n | … |
| 53040 | 12394615-6036154 | -35.7 ± 0.2 | 6053 ± 243 | 4.20 ± 0.28 | -0.22 ± 0.16 | 52 ± 18 | 1 | N | … | … | … | … | n | NG |
| 55927 | 12392450-6035361 | -34.4 ± 0.1 | 4265 ± 105 | 1.88 ± 0.07 | -0.09 ± 0.24 | … | … | N | … | … | … | … | n | … |
| 55642 | 12383656-6038541 | -29.8 ± 0.2 | 6223 ± 135 | 3.77 ± 0.01 | … | … | … | N | … | … | … | … | n | … |
| 56127 | 12394616-6045177 | -38.3 ± 2.0 | … | … | … | … | … | Y | … | … | … | … | n | … |
| 55928 | 12392451-6040057 | 15.4 ± 0.1 | … | … | … | … | … | N | … | … | … | … | n | … |
| 3360 | 12383657-6045300 | -41.6 ± 0.6 | 4689 ± 128 | 2.52 ± 0.25 | -0.01 ± 0.12 | <16 | 3 | Y | Y | Y | Y | … | Y | … |
| 56128 | 12394621-6039263 | -41.5 ± 0.1 | … | … | … | … | … | Y | … | … | … | … | n | … |
| 55944 | 12392576-6038091 | -38.0 ± 2.1 | … | … | … | … | … | Y | … | … | … | … | n | … |
| 56129 | 12394622-6046309 | -42.8 ± 0.3 | … | … | … | … | … | Y | … | … | … | … | n | … |
| 55643 | 12383663-6034072 | 11.3 ± 0.1 | … | … | … | … | … | N | … | … | … | … | n | … |
| 55945 | 12392582-6030155 | -24.9 ± 0.4 | … | … | … | … | … | N | … | … | … | … | n | … |
| 56130 | 12394624-6037320 | -20.2 ± 0.1 | 5123 ± 98 | 4.52 ± 0.02 | 0.02 ± 0.20 | … | … | N | … | … | … | … | n | … |
| 55644 | 12383692-6032097 | -13.0 ± 0.2 | … | … | … | … | … | N | … | … | … | … | n | … |
| 55946 | 12392583-6043419 | -10.1 ± 0.1 | … | … | … | … | … | N | … | … | … | … | n | … |
| 53041 | 12394630-6038271 | -27.2 ± 0.4 | 5649 ± 303 | 4.04 ± 0.51 | -0.10 ± 0.15 | … | … | N | … | … | … | … | n | … |
| 55645 | 12383723-6040148 | -10.1 ± 0.1 | … | … | … | … | … | N | … | … | … | … | n | … |
| 3369 | 12392584-6038279 | -41.8 ± 0.6 | 5035 ± 119 | 3.14 ± 0.25 | 0.12 ± 0.10 | <11 | 3 | Y | Y | Y | Y | … | Y | … |
| 56131 | 12394633-6035189 | -30.7 ± 0.2 | … | … | … | … | … | N | … | … | … | … | n | … |
| 55646 | 12383727-6029570 | 19.7 ± 0.2 | 5633 ± 95 | 3.29 ± 0.03 | -0.16 ± 0.20 | … | … | N | … | … | … | … | n | … |
| 52912 | 12392588-6037528 | -45.3 ± 0.3 | 5645 ± 29 | 4.27 ± 0.40 | -0.11 ± 0.15 | … | … | N | … | … | … | … | n | … |
| 56132 | 12394644-6035336 | -40.2 ± 0.3 | … | … | … | … | … | Y | … | … | … | … | n | … |
| 55647 | 12383749-6031556 | -40.0 ± 0.6 | … | … | … | … | … | Y | … | … | … | … | n | … |
| 52913 | 12392596-6039462 | -23.6 ± 0.5 | 5948 ± 206 | 4.32 ± 0.86 | 0.15 ± 0.15 | … | … | N | … | … | … | … | n | … |
| 53042 | 12394647-6035378 | -40.2 ± 0.3 | 5848 ± 19 | 4.17 ± 0.49 | 0.00 ± 0.17 | 51 ± 36 | 1 | Y | Y | Y | Y | Y | Y | … |
| 55648 | 12383751-6042153 | 11.6 ± 0.1 | 6165 ± 121 | 4.02 ± 0.05 | … | … | … | N | … | … | … | … | n | … |
| 55947 | 12392601-6040541 | -25.1 ± 0.1 | … | … | … | … | … | N | … | … | … | … | n | … |
| 53043 | 12394669-6034282 | 3.8 ± 0.8 | 5315 ± 644 | 4.50 ± 1.07 | 0.32 ± 0.69 | … | … | N | … | … | … | … | n | … |
| 55649 | 12383753-6042016 | -13.9 ± 0.3 | … | … | … | … | … | N | … | … | … | … | n | … |
| 53044 | 12394669-6040041 | -4.0 ± 0.6 | 6085 ± 578 | 3.88 ± 0.25 | -0.44 ± 0.14 | 52 ± 51 | 1 | N | … | … | … | … | n | NG |
| 55948 | 12392605-6036418 | -10.1 ± 0.1 | … | … | … | … | … | N | … | … | … | … | n | … |
| 52752 | 12383778-6035494 | -3.5 ± 0.3 | 6388 ± 35 | 4.39 ± 0.26 | 0.26 ± 0.18 | … | … | N | … | … | … | … | n | … |
| 53054 | 12394810-6038131 | -40.6 ± 0.5 | 6770 ± 62 | 3.93 ± 0.12 | 0.47 ± 0.05 | … | … | Y | … | … | … | … | n | … |
| 52914 | 12392609-6028126 | -36.4 ± 3.1 | 5208 ± 168 | 5.10 ± 0.44 | -0.93 ± 0.22 | … | … | Y | … | … | … | … | n | … |
| 55650 | 12383786-6039015 | -16.7 ± 0.2 | 6191 ± 147 | 3.80 ± 0.04 | … | … | … | N | … | … | … | … | n | … |
| 56142 | 12394815-6031103 | -4.8 ± 0.1 | 5521 ± 124 | 3.80 ± 0.08 | -0.08 ± 0.19 | … | … | N | … | … | … | … | n | … |
| 55949 | 12392629-6041204 | -46.0 ± 0.1 | 6143 ± 150 | 4.91 ± 0.05 | … | … | … | N | … | … | … | … | n | … |
| 55651 | 12383790-6041139 | 9.6 ± 0.6 | … | … | … | … | … | N | … | … | … | … | n | … |
| 56143 | 12394825-6035189 | -41.0 ± 1.1 | … | … | … | … | … | Y | … | … | … | … | n | … |
| 52915 | 12392632-6039149 | -15.9 ± 0.3 | 5704 ± 285 | 4.56 ± 0.34 | 0.13 ± 0.17 | … | … | N | … | … | … | … | n | … |
| 55652 | 12383850-6029426 | -15.5 ± 0.3 | … | … | … | … | … | N | … | … | … | … | n | … |
| 52916 | 12392632-6046340 | 9.8 ± 0.4 | 4941 ± 90 | 4.39 ± 0.79 | -0.13 ± 0.20 | … | … | N | … | … | … | … | n | … |
| 55653 | 12383870-6030151 | -65.5 ± 0.4 | … | … | … | … | … | N | … | … | … | … | n | … |
| 53056 | 12394834-6036364 | -41.0 ± 0.4 | 6333 ± 352 | 4.51 ± 0.24 | -0.40 ± 0.24 | 37 ± 34 | … | Y | Y | Y | N | … | Y | … |
| 3370 | 12392636-6040217 | -40.9 ± 0.6 | 4953 ± 121 | 2.87 ± 0.23 | 0.08 ± 0.10 | 18 ± 1 | … | Y | Y | Y | Y | … | Y | … |
| 56144 | 12394838-6046517 | -10.6 ± 0.1 | 4639 ± 107 | 3.61 ± 0.05 | 0.05 ± 0.22 | … | … | N | … | … | … | … | n | … |
| 55654 | 12383913-6038125 | -25.9 ± 0.1 | … | … | … | … | … | N | … | … | … | … | n | … |
| 52917 | 12392645-6038072 | -41.4 ± 0.3 | 6619 ± 92 | 4.06 ± 0.18 | 0.29 ± 0.07 | 47 ± 29 | … | Y | Y | Y | Y | Y | Y | … |
| 56145 | 12394844-6042424 | -8.0 ± 0.2 | … | … | … | … | … | N | … | … | … | … | n | … |
| 55555 | 12383942-6035563 | -16.5 ± 0.1 | 5241 ± 118 | 4.08 ± 0.05 | 0.26 ± 0.22 | … | … | N | … | … | … | … | n | … |
| 55950 | 12392647-6037118 | -38.0 ± 0.7 | … | … | … | … | … | Y | … | … | … | … | n | … |
| 53057 | 12394846-6041120 | -33.5 ± 0.3 | 6601 ± 38 | 3.99 ± 0.10 | -0.23 ± 0.32 | … | … | N | … | … | … | … | n | … |
| 55656 | 12383979-6030211 | 9.7 ± 0.2 | 5603 ± 96 | 2.79 ± 0.05 | -0.16 ± 0.16 | … | … | N | … | … | … | … | n | … |
| 56146 | 12394851-6046298 | -50.4 ± 0.3 | … | … | … | … | … | N | … | … | … | … | n | … |
| 52918 | 12392647-6043019 | -41.5 ± 0.4 | 6436 ± 347 | 4.40 ± 0.38 | 0.23 ± 0.31 | … | … | Y | … | … | … | … | n | … |
| 55657 | 12384003-6031325 | -29.5 ± 3.0 | … | … | … | … | … | N | … | … | … | … | n | … |

**Table C.17.** continued.

| ID | CNAME | RV (km s$^{-1}$) | $T_{\rm eff}$ (K) | $logg$ (dex) | [Fe/H] (dex) | EW(Li)$^a$ (mÅ) | EW(Li) error flag$^b$ | Membership RV | Li | $logg$ | [Fe/H] | Gaia study Cantat-Gaudin$^c$ | Final$^d$ | NMs with Li$^e$ |
|---|---|---|---|---|---|---|---|---|---|---|---|---|---|---|
| 53058 | 12394852-6037478 | -42.4 ± 0.3 | 6287 ± 329 | 4.57 ± 0.57 | -0.01 ± 0.37 | 53 ± 28 | 1 | Y | Y | Y | Y | Y | Y | … |
| 52919 | 12392648-6037118 | -40.3 ± 0.3 | 6661 ± 437 | 3.85 ± 0.06 | 0.13 ± 0.21 | 75 ± 31 | 1 | Y | Y | Y | Y | Y | Y | … |
| 53059 | 12394854-6038050 | -40.7 ± 0.5 | 6052 ± 178 | 4.11 ± 0.35 | -0.03 ± 0.14 | 66 ± 60 | 1 | Y | Y | Y | Y | … | Y | … |
| 55658 | 12384012-6045225 | -35.3 ± 0.1 | … | … | … | … | … | N | … | … | … | … | n | … |
| 52920 | 12392651-6041120 | -4.1 ± 8.8 | … | … | … | … | … | N | … | … | … | … | n | … |
| 53060 | 12394863-6033383 | -24.7 ± 0.3 | 5765 ± 341 | 4.70 ± 0.57 | -0.04 ± 0.34 | 38 ± 36 | 1 | N | … | … | … | … | n | NG |
| 55659 | 12384051-6040487 | -40.8 ± 0.1 | 4885 ± 119 | 3.06 ± 0.05 | 0.28 ± 0.25 | … | … | Y | … | … | … | … | n | … |
| 56147 | 12394875-6031419 | -37.5 ± 0.2 | 6467 ± 150 | 4.24 ± 0.01 | … | … | … | Y | … | … | … | … | n | … |
| 52921 | 12392653-6039573 | -20.1 ± 0.2 | 6299 ± 78 | 4.03 ± 0.09 | 0.32 ± 0.18 | … | … | N | … | … | … | … | n | … |
| 55660 | 12384053-6046160 | -40.6 ± 3.0 | … | … | … | … | … | Y | … | … | … | … | n | … |
| 53061 | 12394885-6039251 | -40.8 ± 0.4 | 6640 ± 58 | 3.78 ± 0.11 | 0.24 ± 0.05 | 75 ± 27 | … | Y | Y | Y | Y | Y | Y | … |
| 52922 | 12392657-6035049 | -41.3 ± 0.4 | 6433 ± 51 | … | 0.02 ± 0.12 | 81 ± 45 | … | Y | Y | … | Y | Y | Y | … |
| 55700 | 12384765-6041267 | 2.6 ± 0.1 | 6063 ± 117 | 4.65 ± 0.02 | … | … | … | N | … | … | … | … | n | … |
| 53062 | 12394886-6040557 | -40.5 ± 0.4 | 6067 ± 205 | 4.31 ± 0.38 | 0.02 ± 0.19 | <50 | 3 | Y | Y | Y | Y | … | Y | … |
| 55951 | 12392659-6044181 | -34.6 ± 0.1 | 5633 ± 175 | 4.75 ± 0.08 | 0.34 ± 0.26 | … | … | N | … | … | … | … | n | … |
| 55701 | 12384777-6041043 | -37.6 ± 0.6 | … | … | … | … | … | Y | … | … | … | … | n | … |
| 56329 | 12401686-6036217 | -43.2 ± 1.3 | … | … | … | … | … | N | … | … | … | … | n | … |
| 56148 | 12394891-6036114 | -39.3 ± 1.0 | … | … | … | … | … | Y | … | … | … | … | n | … |
| 55952 | 12392671-6035423 | -39.9 ± 1.3 | … | … | … | … | … | Y | … | … | … | … | n | … |
| 3386 | 12394897-6033282 | -38.8 ± 0.6 | 4947 ± 114 | 2.81 ± 0.22 | 0.10 ± 0.09 | <9 | 3 | Y | Y | Y | Y | … | Y | … |
| 52759 | 12384791-6039273 | -17.5 ± 0.9 | 5059 ± 907 | 4.51 ± 0.31 | 0.03 ± 0.66 | … | … | N | … | … | … | … | n | … |
| 56330 | 12401692-6040054 | -38.7 ± 1.0 | … | … | … | … | … | Y | … | … | … | … | n | … |
| 55953 | 12392674-6039122 | -16.9 ± 0.2 | … | … | … | … | … | N | … | … | … | … | n | … |
| 53063 | 12394904-6035156 | 22.2 ± 0.3 | 5749 ± 55 | 4.57 ± 0.24 | 0.10 ± 0.20 | … | … | N | … | … | … | … | n | … |
| 55702 | 12384792-6036433 | -32.2 ± 0.2 | 5747 ± 114 | 3.01 ± 0.05 | 0.12 ± 0.23 | … | … | N | … | … | … | … | n | … |
| 56331 | 12401720-6042158 | -35.2 ± 0.1 | … | … | … | … | … | N | … | … | … | … | n | … |
| 3371 | 12392698-6036053 | -39.3 ± 0.6 | 4834 ± 120 | 2.76 ± 0.25 | 0.11 ± 0.10 | 19 ± 1 | … | Y | Y | Y | Y | … | Y | … |
| 53064 | 12394908-6035118 | -23.0 ± 0.3 | 7023 ± 604 | 4.17 ± 0.17 | 0.07 ± 0.36 | 52 ± 27 | 1 | N | … | … | … | … | n | NG |
| 55703 | 12384793-6041016 | -30.9 ± 0.1 | 5937 ± 136 | 3.80 ± 0.01 | 0.11 ± 0.21 | … | … | N | … | … | … | … | n | … |
| 56332 | 12401724-6037336 | -44.5 ± 0.5 | … | … | … | … | … | N | … | … | … | … | n | … |
| 52941 | 12393037-6037595 | -17.3 ± 0.6 | 5374 ± 336 | 3.89 ± 0.37 | -0.44 ± 0.44 | … | … | N | … | … | … | … | n | … |
| 53065 | 12394909-6040513 | -33.3 ± 0.5 | 6514 ± 163 | 4.82 ± 0.75 | -0.12 ± 0.24 | 31 ± 19 | … | N | Y | Y | Y | Y | Y | … |
| 52760 | 12384820-6042154 | -40.7 ± 0.7 | 5777 ± 474 | 4.46 ± 0.19 | -0.66 ± 0.21 | … | … | Y | … | … | … | … | n | … |
| 56333 | 12401729-6038033 | -37.1 ± 0.6 | … | … | … | … | … | Y | … | … | … | … | n | … |
| 52942 | 12393040-6042257 | 37.4 ± 0.3 | 5676 ± 122 | 4.73 ± 0.43 | 0.26 ± 0.16 | … | … | N | … | … | … | … | n | … |
| 56149 | 12394916-6033123 | -39.6 ± 0.5 | … | … | … | … | … | Y | … | … | … | … | n | … |
| 55704 | 12384823-6041087 | -35.6 ± 1.9 | … | … | … | … | … | N | … | … | … | … | n | … |
| 56334 | 12401739-6032143 | -42.7 ± 0.5 | … | … | … | … | … | Y | … | … | … | … | n | … |
| 55984 | 12393059-6040401 | -35.1 ± 0.1 | … | … | … | … | … | N | … | … | … | … | n | … |
| 56150 | 12394926-6040152 | -37.2 ± 1.9 | … | … | … | … | … | Y | … | … | … | … | n | … |
| 52761 | 12384833-6037371 | 55.7 ± 0.2 | 4915 ± 92 | 2.85 ± 0.24 | -0.48 ± 0.17 | … | … | N | … | … | … | … | n | G |
| 53219 | 12401747-6044074 | -40.5 ± 1.1 | 6552 ± 129 | 3.41 ± 0.47 | -0.08 ± 0.21 | … | … | Y | … | … | … | … | n | … |
| 55985 | 12393060-6032010 | -1.5 ± 0.2 | 5951 ± 145 | 3.44 ± 0.05 | 0.27 ± 0.23 | … | … | N | … | … | … | … | n | … |
| 53066 | 12394934-6036066 | -29.3 ± 0.2 | 4893 ± 93 | 3.28 ± 0.54 | 0.32 ± 0.30 | … | … | N | … | … | … | … | n | G |
| 55705 | 12384880-6030006 | -38.0 ± 0.6 | … | … | … | … | … | Y | … | … | … | … | n | … |
| 53220 | 12401759-6034564 | -41.4 ± 0.6 | 5272 ± 280 | 4.44 ± 0.31 | -0.39 ± 0.43 | … | … | Y | … | … | … | … | n | … |
| 52943 | 12393070-6041027 | -48.2 ± 0.5 | 6777 ± 61 | 3.87 ± 0.11 | 0.31 ± 0.05 | … | … | N | … | … | … | … | n | … |
| 53067 | 12394945-6033567 | -40.5 ± 0.5 | 5097 ± 86 | 4.40 ± 0.86 | -0.15 ± 0.14 | … | … | Y | … | … | … | … | n | … |
| 55706 | 12384882-6032017 | -9.8 ± 0.1 | 6303 ± 148 | 4.85 ± 0.01 | … | … | … | N | … | … | … | … | n | … |
| 56335 | 12401766-6042591 | -34.2 ± 0.1 | … | … | … | … | … | N | … | … | … | … | n | … |
| 52944 | 12393073-6030032 | -15.3 ± 0.7 | 6038 ± 167 | 4.13 ± 0.22 | 0.04 ± 0.28 | <68 | 3 | N | … | … | … | … | n | NG |
| 56180 | 12395207-6038146 | -39.9 ± 0.2 | … | … | … | … | … | Y | … | … | … | … | n | … |
| 55707 | 12384897-6037530 | -16.0 ± 0.2 | … | … | … | … | … | N | … | … | … | … | n | … |
| 53221 | 12401772-6038451 | -39.6 ± 0.6 | 5843 ± 335 | 4.45 ± 0.27 | -0.13 ± 0.26 | … | … | Y | … | … | … | … | n | … |
| 52945 | 12393076-6034595 | -9.4 ± 0.3 | 5885 ± 334 | 4.51 ± 0.32 | -0.17 ± 0.27 | 50 ± 27 | 1 | N | … | … | … | … | n | NG |
| 56181 | 12395214-6039274 | -38.5 ± 0.4 | … | … | … | … | … | Y | … | … | … | … | n | … |
| 55708 | 12384934-6033251 | -11.0 ± 0.1 | 5595 ± 128 | 3.69 ± 0.06 | 0.03 ± 0.21 | … | … | N | … | … | … | … | n | … |
| 56336 | 12401807-6041156 | -28.4 ± 0.2 | … | … | … | … | … | N | … | … | … | … | n | … |
| 55986 | 12393077-6039137 | -40.3 ± 0.2 | … | … | … | … | … | Y | … | … | … | … | n | … |

**Table C.17.** continued.

| ID | CNAME | RV (km s$^{-1}$) | $T_{\rm eff}$ (K) | $logg$ (dex) | [Fe/H] (dex) | EW(Li)[a] (mÅ) | EW(Li) error flag[b] | Membership RV | Li | $logg$ | [Fe/H] | Gaia study Cantat-Gaudin[c] | Final[d] | NMs with Li[e] |
|---|---|---|---|---|---|---|---|---|---|---|---|---|---|---|
| 53089 | 12395215-6035301 | -39.2 ± 0.5 | 5903 ± 130 | 4.24 ± 0.39 | 0.08 ± 0.16 | … | … | Y | … | … | … | … | n | … |
| 53222 | 12401834-6035587 | -38.7 ± 0.4 | 5824 ± 31 | 4.34 ± 0.67 | -0.12 ± 0.18 | <115 | 3 | Y | N | Y | Y | N | n | NG |
| 55709 | 12384997-6044027 | -28.6 ± 0.2 | 6343 ± 149 | 3.73 ± 0.01 | … | … | … | N | … | … | … | … | n | … |
| 55987 | 12393088-6039340 | -25.1 ± 0.3 | … | … | … | … | … | N | … | … | … | … | n | … |
| 55988 | 12393097-6036280 | -26.2 ± 0.2 | 6221 ± 150 | 3.77 ± 0.03 | … | … | … | N | … | … | … | … | n | … |
| 53090 | 12395217-6037138 | -20.4 ± 0.3 | 6411 ± 190 | 4.28 ± 0.33 | 0.00 ± 0.13 | … | … | N | … | … | … | … | n | … |
| 53223 | 12401842-6036534 | 67.2 ± 0.8 | 5638 ± 134 | 4.63 ± 0.39 | -0.32 ± 0.18 | … | … | N | … | … | … | … | n | … |
| 52762 | 12385042-6034342 | -3.3 ± 0.4 | 5649 ± 441 | 4.24 ± 0.37 | 0.08 ± 0.28 | <83 | 3 | N | … | … | … | … | n | NG |
| 52946 | 12393097-6042167 | -39.0 ± 0.4 | 6232 ± 253 | 4.32 ± 0.41 | -0.01 ± 0.16 | 58 ± 31 | 1 | Y | Y | Y | Y | N | Y | … |
| 56182 | 12395224-6038361 | -28.2 ± 0.6 | … | … | … | … | … | N | … | … | … | … | n | … |
| 53224 | 12401846-6037271 | -48.5 ± 0.6 | 5614 ± 171 | 4.21 ± 0.28 | -0.31 ± 0.16 | <89 | 3 | N | … | … | … | … | n | NG |
| 55710 | 12385069-6035523 | -38.9 ± 0.4 | … | … | … | … | … | Y | … | … | … | … | n | … |
| 55989 | 12393100-6045419 | -10.1 ± 0.2 | … | … | … | … | … | N | … | … | … | … | n | … |
| 56183 | 12395234-6041012 | -18.0 ± 0.2 | … | … | … | … | … | N | … | … | … | … | n | … |
| 56337 | 12401855-6035258 | 71.4 ± 0.1 | 5687 ± 90 | 4.03 ± 0.06 | -0.14 ± 0.15 | … | … | N | … | … | … | … | n | … |
| 55711 | 12385078-6036270 | -24.0 ± 0.2 | … | … | … | … | … | N | … | … | … | … | n | … |
| 55990 | 12393108-6043535 | -40.3 ± 1.0 | … | … | … | … | … | Y | … | … | … | … | n | … |
| 56184 | 12395240-6031511 | -39.2 ± 0.8 | … | … | … | … | … | Y | … | … | … | … | n | … |
| 53225 | 12401867-6033397 | -56.1 ± 1.1 | 6404 ± 97 | 4.38 ± 1.13 | 0.26 ± 0.42 | … | … | N | … | … | … | … | n | … |
| 55712 | 12385092-6039414 | -41.1 ± 0.6 | … | … | … | … | … | Y | … | … | … | … | n | … |
| 53091 | 12395242-6036179 | 18.9 ± 0.3 | 5622 ± 202 | 4.66 ± 0.37 | 0.35 ± 0.28 | … | … | N | … | … | … | … | n | … |
| 56338 | 12401870-6040159 | -24.1 ± 0.4 | … | … | … | … | … | N | … | … | … | … | n | … |
| 56185 | 12395245-6032025 | -38.8 ± 2.7 | … | … | … | … | … | Y | … | … | … | … | n | … |
| 55713 | 12385106-6035283 | -30.1 ± 0.2 | 6257 ± 107 | 4.39 ± 0.05 | … | … | … | N | … | … | … | … | n | … |
| 3372 | 12393131-6039423 | -35.3 ± 0.6 | 4988 ± 114 | 3.04 ± 0.22 | 0.13 ± 0.10 | <8 | 3 | N | … | … | … | … | n | … |
| 56339 | 12401880-6046226 | -38.0 ± 0.7 | … | … | … | … | … | Y | … | … | … | … | n | … |
| 56186 | 12395260-6044548 | -2.8 ± 0.2 | … | … | … | … | … | N | … | … | … | … | n | … |
| 55714 | 12385137-6030125 | 0.8 ± 0.2 | … | … | … | … | … | N | … | … | … | … | n | … |
| 55991 | 12393137-6048046 | -24.7 ± 0.1 | 5475 ± 102 | 3.68 ± 0.05 | 0.18 ± 0.15 | … | … | N | … | … | … | … | n | … |
| 56187 | 12395268-6041059 | -4.1 ± 0.7 | … | … | … | … | … | N | … | … | … | … | n | … |
| 53237 | 12402832-6037301 | -40.6 ± 0.3 | 6627 ± 65 | 4.07 ± 0.13 | 0.04 ± 0.05 | 49 ± 23 | … | Y | Y | Y | Y | Y | Y | … |
| 55715 | 12385138-6038002 | -39.2 ± 0.9 | … | … | … | … | … | Y | … | … | … | … | n | … |
| 55992 | 12393148-6041436 | -63.4 ± 0.3 | … | … | … | … | … | N | … | … | … | … | n | … |
| 56188 | 12395269-6031100 | -44.2 ± 1.6 | … | … | … | … | … | N | … | … | … | … | n | … |
| 53238 | 12402835-6046414 | -41.5 ± 0.8 | 5963 ± 221 | 4.66 ± 0.18 | 0.02 ± 0.18 | … | … | Y | … | … | … | … | n | … |
| 55716 | 12385147-6037456 | -33.9 ± 0.5 | … | … | … | … | … | N | … | … | … | … | n | … |
| 55993 | 12393154-6040585 | -42.9 ± 0.1 | … | … | … | … | … | Y | … | … | … | … | n | … |
| 56189 | 12395278-6037016 | -39.8 ± 0.6 | … | … | … | … | … | Y | … | … | … | … | n | … |
| 56375 | 12402899-6044393 | -37.1 ± 3.1 | … | … | … | … | … | Y | … | … | … | … | n | … |
| 55717 | 12385156-6033440 | -55.3 ± 0.1 | 5061 ± 72 | 3.51 ± 0.05 | -0.06 ± 0.19 | … | … | N | … | … | … | … | n | … |
| 53092 | 12395278-6038235 | -17.9 ± 0.3 | 5996 ± 338 | 4.56 ± 0.30 | 0.14 ± 0.23 | … | … | N | … | … | … | … | n | … |
| 52948 | 12393182-6041111 | -45.6 ± 0.5 | 5490 ± 74 | 4.17 ± 0.29 | -0.13 ± 0.19 | … | … | N | … | … | … | … | n | … |
| 53239 | 12402907-6041274 | -13.6 ± 0.6 | 5982 ± 175 | 4.16 ± 0.11 | 0.14 ± 0.16 | <92 | 3 | N | … | … | … | … | n | NG |
| 55718 | 12385157-6029428 | -10.0 ± 0.3 | 5631 ± 137 | 2.71 ± 0.05 | -0.01 ± 0.32 | … | … | N | … | … | … | … | n | … |
| 53093 | 12395279-6037015 | -39.6 ± 0.6 | 6692 ± 468 | 4.00 ± 0.08 | 0.16 ± 0.22 | 77 ± 38 | 1 | Y | Y | Y | Y | Y | Y | … |
| 52949 | 12393182-6041554 | 2.2 ± 0.3 | 5502 ± 137 | 4.36 ± 0.54 | 0.05 ± 0.12 | … | … | N | … | … | … | … | n | … |
| 56190 | 12395284-6035194 | -39.6 ± 0.4 | … | … | … | … | … | Y | … | … | … | … | n | … |
| 56376 | 12402943-6036269 | -41.6 ± 0.1 | 5289 ± 121 | 3.52 ± 0.05 | 0.39 ± 0.25 | … | … | Y | … | … | … | … | n | … |
| 55719 | 12385174-6038057 | -7.8 ± 0.1 | 5905 ± 108 | 3.62 ± 0.01 | 0.38 ± 0.14 | … | … | N | … | … | … | … | n | … |
| 55994 | 12393188-6033104 | -40.3 ± 0.1 | … | … | … | … | … | Y | … | … | … | … | n | … |
| 53094 | 12395284-6037593 | -60.1 ± 0.6 | 6831 ± 536 | 4.22 ± 0.19 | 0.21 ± 0.25 | … | … | N | … | … | … | … | n | … |
| 56377 | 12402949-6038519 | -40.6 ± 0.1 | 5175 ± 948 | 4.13 ± 0.40 | 0.04 ± 0.51 | … | … | Y | … | … | … | … | n | … |
| 52768 | 12385456-6035185 | 5.1 ± 0.6 | 5782 ± 404 | 4.63 ± 0.52 | 0.32 ± 0.38 | <104 | 3 | N | … | … | … | … | n | NG |
| 52950 | 12393188-6034508 | -49.0 ± 0.6 | 6424 ± 799 | 4.25 ± 0.14 | -0.12 ± 0.26 | <42 | 3 | N | … | … | … | … | n | NG |
| 56191 | 12395295-6042134 | 3.0 ± 0.3 | … | … | … | … | … | N | … | … | … | … | n | … |
| 56378 | 12402963-6038024 | -39.9 ± 0.2 | … | … | … | … | … | Y | … | … | … | … | n | … |
| 55739 | 12385485-6041524 | 86.9 ± 0.1 | … | … | … | … | … | N | … | … | … | … | n | … |
| 55995 | 12393192-6043181 | -25.7 ± 0.1 | 6131 ± 98 | 4.42 ± 0.06 | … | … | … | N | … | … | … | … | n | … |
| 53095 | 12395311-6042026 | -49.2 ± 0.5 | 6098 ± 138 | 4.08 ± 0.78 | -0.57 ± 0.34 | … | … | N | … | … | … | … | n | … |

**Table C.17.** continued.

| ID | CNAME | RV (km s$^{-1}$) | $T_{\mathrm{eff}}$ (K) | $\log g$ (dex) | [Fe/H] (dex) | EW(Li)$^a$ (mÅ) | EW(Li) error flag$^b$ | Membership RV | Li | $\log g$ | [Fe/H] | Gaia study Cantat-Gaudin$^c$ | Final$^d$ | NMs with Li$^e$ |
|---|---|---|---|---|---|---|---|---|---|---|---|---|---|---|
| 56379 | 12402989-6037263 | -39.6 ± 1.9 | … | … | … | … | … | Y | … | … | … | … | n | … |
| 55740 | 12385498-6046171 | -32.5 ± 0.1 | 5541 ± 150 | 3.86 ± 0.07 | 0.12 ± 0.28 | … | … | N | … | … | … | … | n | … |
| 55996 | 12393210-6042123 | -40.6 ± 0.1 | … | … | … | … | … | Y | … | … | … | … | n | … |
| 56008 | 12393387-6035411 | -5.1 ± 0.1 | … | … | … | … | … | N | … | … | … | … | n | … |
| 53096 | 12395312-6037424 | -40.7 ± 0.4 | 5469 ± 250 | 4.25 ± 0.60 | 0.03 ± 0.13 | … | … | Y | … | … | … | … | n | … |
| 53240 | 12403031-6038212 | 39.5 ± 0.9 | 5592 ± 451 | 4.28 ± 0.44 | -0.61 ± 0.56 | … | … | N | … | … | … | … | n | … |
| 53241 | 12403033-6041560 | 17.5 ± 0.8 | 5643 ± 136 | 4.29 ± 0.12 | -0.08 ± 0.33 | … | … | N | … | … | … | … | n | … |
| 53097 | 12395315-6037078 | -43.8 ± 0.4 | 5671 ± 119 | 4.05 ± 0.35 | -0.10 ± 0.15 | 140 ± 106 | 1 | N | … | … | … | … | … | … |
| 55741 | 12385514-6034324 | -29.2 ± 0.1 | … | … | … | … | … | N | … | … | … | … | n | … |
| 56009 | 12393387-6042140 | -40.6 ± 0.6 | … | … | … | … | … | Y | … | … | … | … | n | … |
| 56380 | 12403061-6036171 | -42.0 ± 0.5 | … | … | … | … | … | Y | … | … | … | … | n | … |
| 53098 | 12395324-6034337 | -2.9 ± 0.3 | 6151 ± 270 | 4.45 ± 0.14 | 0.35 ± 0.21 | 84 ± 21 | 1 | N | … | … | … | … | n | NG |
| 52769 | 12385538-6039028 | -53.6 ± 0.4 | 6702 ± 108 | 4.02 ± 0.21 | 0.27 ± 0.09 | 63 ± 30 | … | N | … | … | … | Y | n | NG |
| 52964 | 12393388-6035411 | -32.3 ± 0.5 | 6354 ± 193 | 3.02 ± 1.14 | -0.43 ± 0.50 | 81 ± 44 | 1 | N | Y | N | N | … | n | .. |
| 56381 | 12403098-6038087 | -4.0 ± 0.2 | … | … | … | … | … | N | … | … | … | … | n | … |
| 56192 | 12395336-6039474 | -40.9 ± 0.7 | … | … | … | … | … | Y | … | … | … | … | n | … |
| 52770 | 12385550-6035009 | -38.7 ± 0.4 | 6115 ± 533 | 4.19 ± 0.06 | 0.03 ± 0.50 | 62 ± 41 | 1 | Y | Y | Y | Y | … | Y | … |
| 52771 | 12385575-6037001 | -38.0 ± 0.3 | 6384 ± 48 | 3.59 ± 0.10 | 0.16 ± 0.04 | 91 ± 29 | … | Y | Y | Y | Y | Y | Y | … |
| 56010 | 12393388-6037303 | -41.3 ± 0.9 | … | … | … | … | … | Y | … | … | … | … | n | … |
| 56382 | 12403118-6044134 | -48.3 ± 0.2 | … | … | … | … | … | N | … | … | … | … | n | … |
| 53099 | 12395339-6040135 | -14.5 ± 0.6 | 5499 ± 204 | 4.21 ± 0.32 | -0.14 ± 0.17 | … | … | N | … | … | … | … | n | … |
| 52772 | 12385579-6037549 | 11.0 ± 0.4 | 5674 ± 316 | 4.29 ± 0.26 | 0.21 ± 0.16 | <101 | 3 | N | … | … | … | … | n | NG |
| 56011 | 12393392-6038504 | -35.1 ± 0.2 | … | … | … | … | … | N | … | … | … | … | n | … |
| 56204 | 12395483-6045373 | -39.8 ± 0.5 | … | … | … | … | … | Y | … | … | … | … | n | … |
| 56383 | 12403132-6036025 | -37.4 ± 0.4 | … | … | … | … | … | Y | … | … | … | … | n | … |
| 52773 | 12385607-6035045 | -12.6 ± 0.3 | 6262 ± 109 | 4.37 ± 0.40 | 0.31 ± 0.16 | … | … | N | … | … | … | … | n | … |
| 56012 | 12393405-6040457 | -42.7 ± 2.2 | … | … | … | … | … | Y | … | … | … | … | n | … |
| 56205 | 12395488-6036424 | -38.7 ± 0.2 | … | … | … | … | … | Y | … | … | … | … | n | … |
| 53242 | 12403148-6044480 | -41.7 ± 0.7 | 6187 ± 268 | 4.71 ± 0.42 | 0.23 ± 0.99 | … | … | Y | … | … | … | … | n | … |
| 52774 | 12385659-6044005 | -38.8 ± 0.3 | 6541 ± 236 | 4.40 ± 0.42 | -0.10 ± 0.14 | … | … | Y | … | … | … | … | n | … |
| 56013 | 12393408-6034297 | -39.1 ± 0.7 | … | … | … | … | … | Y | … | … | … | … | n | … |
| 56206 | 12395489-6030331 | -48.5 ± 0.1 | 4889 ± 150 | 4.64 ± 0.05 | 0.27 ± 0.24 | … | … | N | … | … | … | … | n | … |
| 53243 | 12403186-6044064 | -5.8 ± 0.3 | 4984 ± 146 | 4.46 ± 0.67 | 0.13 ± 0.23 | <122 | 3 | N | … | … | … | … | n | NG |
| 55742 | 12385669-6040591 | -17.7 ± 0.1 | 6235 ± 134 | 3.51 ± 0.05 | … | … | … | N | … | … | … | … | n | … |
| 56014 | 12393418-6034446 | -23.0 ± 0.1 | 6541 ± 134 | 3.85 ± 0.06 | … | … | … | N | … | … | … | … | n | … |
| 53112 | 12395491-6036424 | -44.4 ± 0.3 | 6707 ± 410 | 4.13 ± 0.15 | 0.07 ± 0.19 | 42 ± 16 | 1 | N | Y | Y | Y | Y | Y | … |
| 56384 | 12403192-6046285 | 5.9 ± 0.1 | … | … | … | … | … | N | … | … | … | … | n | … |
| 55743 | 12385672-6041538 | -46.8 ± 0.4 | 6275 ± 184 | 4.20 ± 0.05 | … | … | … | N | … | … | … | … | n | … |
| 53113 | 12395498-6038005 | -51.9 ± 0.3 | 5822 ± 98 | 4.11 ± 0.18 | 0.14 ± 0.16 | … | … | N | … | … | … | … | n | … |
| 56015 | 12393422-6036164 | -39.1 ± 0.2 | … | … | … | … | … | Y | … | … | … | … | n | … |
| 53244 | 12403214-6039095 | 1.5 ± 0.4 | 4603 ± 269 | 3.45 ± 0.68 | -0.09 ± 0.15 | … | … | N | … | … | … | … | n | G |
| 53114 | 12395506-6034375 | -40.5 ± 0.4 | 5812 ± 699 | 4.31 ± 0.52 | -0.10 ± 0.59 | … | … | Y | … | … | … | … | n | … |
| 55744 | 12385672-6044582 | -24.8 ± 0.3 | … | … | … | … | … | N | … | … | … | … | n | … |
| 56016 | 12393423-6038220 | 9.1 ± 6.9 | … | … | … | … | … | N | … | … | … | … | n | … |
| 53245 | 12403231-6031362 | 56.7 ± 0.9 | 5116 ± 719 | 4.01 ± 0.46 | 0.38 ± 0.71 | … | … | N | … | … | … | … | n | … |
| 53115 | 12395519-6041417 | -14.9 ± 0.7 | 6270 ± 217 | 4.42 ± 0.41 | 0.03 ± 0.38 | … | … | N | … | … | … | … | n | … |
| 55745 | 12385680-6033491 | -69.8 ± 0.1 | … | … | … | … | … | N | … | … | … | … | n | … |
| 52965 | 12393424-6036164 | -38.8 ± 0.3 | 6716 ± 353 | 4.15 ± 0.12 | 0.13 ± 0.19 | 68 ± 27 | 1 | Y | Y | Y | Y | Y | Y | … |
| 56385 | 12403233-6038411 | -21.7 ± 0.4 | … | … | … | … | … | N | … | … | … | … | n | … |
| 56207 | 12395526-6035597 | -5.0 ± 0.3 | … | … | … | … | … | N | … | … | … | … | n | … |
| 55746 | 12385680-6038350 | 1.9 ± 0.2 | … | … | … | … | … | N | … | … | … | … | n | … |
| 56017 | 12393434-6038169 | -40.6 ± 1.5 | … | … | … | … | … | Y | … | … | … | … | n | … |
| 56208 | 12395536-6037568 | -39.8 ± 0.3 | … | … | … | … | … | Y | … | … | … | … | n | … |
| 56386 | 12403240-6039510 | -12.4 ± 0.3 | … | … | … | … | … | N | … | … | … | … | n | … |
| 52775 | 12385682-6038313 | -39.1 ± 1.2 | 7163 ± 173 | 4.05 ± 0.21 | -0.29 ± 0.18 | <47 | 3 | Y | Y | Y | Y | Y | Y | … |
| 3388 | 12395554-6037268 | -39.8 ± 0.6 | 4900 ± 112 | 2.79 ± 0.23 | 0.13 ± 0.11 | <5 | 3 | Y | Y | Y | Y | … | Y | … |
| 56387 | 12403258-6030302 | -9.1 ± 0.2 | … | … | … | … | … | N | … | … | … | … | n | … |
| 52776 | 12385690-6036141 | -49.2 ± 0.7 | 5513 ± 95 | 4.07 ± 0.29 | -0.32 ± 0.12 | <129 | 3 | N | … | … | … | … | n | NG |
| 56019 | 12393449-6042316 | -5.1 ± 0.1 | … | … | … | … | … | N | … | … | … | … | n | … |

| ID | CNAME | RV (km s$^{-1}$) | $T_{\text{eff}}$ (K) | $\log g$ (dex) | [Fe/H] (dex) | EW(Li)$^a$ (mÅ) | EW(Li) error flag$^b$ | Membership | | | | Gaia study Cantat-Gaudin$^c$ | Final$^d$ | NMs with Li$^e$ |
| | | | | | | | | RV | Li | $\log g$ | [Fe/H] | | | |
|---|---|---|---|---|---|---|---|---|---|---|---|---|---|---|
| 56209 | 12395557-6033005 | 8.4 ± 0.1 | 6165 ± 119 | 4.91 ± 0.05 | … | … | … | N | … | … | … | … | n | … |
| 56388 | 12403273-6039242 | -10.1 ± 0.1 | … | … | … | … | … | N | … | … | … | … | n | … |
| 55747 | 12385690-6042076 | -39.7 ± 0.8 | … | … | … | … | … | Y | … | … | … | … | n | … |
| 56210 | 12395560-6031251 | 37.4 ± 0.2 | … | … | … | … | … | N | … | … | … | … | n | … |
| 56020 | 12393450-6041071 | -39.8 ± 0.9 | … | … | … | … | … | Y | … | … | … | … | n | … |
| 56389 | 12403308-6039170 | -29.7 ± 0.4 | … | … | … | … | … | N | … | … | … | … | n | … |
| 3389 | 12395566-6035233 | -38.2 ± 0.6 | 5992 ± 125 | 3.79 ± 0.25 | 0.20 ± 0.10 | 87 ± 2 | … | Y | Y | Y | Y | Y | Y | … |
| 52777 | 12385704-6040563 | -27.6 ± 0.3 | 5172 ± 49 | 4.34 ± 0.56 | 0.13 ± 0.19 | … | … | N | … | … | … | … | n | … |
| 56021 | 12393457-6036356 | -37.4 ± 2.8 | … | … | … | … | … | Y | … | … | … | … | n | … |
| 53252 | 12403783-6041351 | 44.4 ± 0.3 | 5859 ± 202 | 4.40 ± 0.20 | 0.09 ± 0.18 | … | … | N | … | … | … | … | n | … |
| 56211 | 12395572-6040194 | -10.3 ± 0.2 | 5667 ± 102 | 3.06 ± 0.07 | -0.55 ± 0.22 | … | … | N | … | … | … | … | n | … |
| 55748 | 12385707-6041265 | -37.5 ± 1.1 | … | … | … | … | … | Y | … | … | … | … | n | … |
| 56022 | 12393460-6040086 | 5.6 ± 0.1 | … | … | … | … | … | N | … | … | … | … | n | … |
| 52966 | 12393465-6039183 | -36.8 ± 0.4 | 5926 ± 213 | 4.05 ± 0.22 | -0.14 ± 0.23 | 78 ± 35 | 1 | Y | Y | Y | Y | … | Y | … |
| 56212 | 12395579-6035416 | -32.4 ± 0.3 | … | … | … | … | … | N | … | … | … | … | n | … |
| 53253 | 12403816-6034071 | 33.8 ± 0.7 | 5367 ± 406 | 4.24 ± 0.62 | -0.49 ± 0.73 | … | … | N | … | … | … | … | n | … |
| 55749 | 12385718-6035009 | -27.4 ± 0.4 | … | … | … | … | … | N | … | … | … | … | n | … |
| 56213 | 12395583-6032323 | 68.5 ± 0.1 | … | … | … | … | … | N | … | … | … | … | n | … |
| 52967 | 12393472-6034488 | 9.6 ± 0.5 | 6629 ± 1038 | 4.64 ± 0.41 | 0.30 ± 0.34 | <83 | 3 | N | … | … | … | … | n | NG |
| 56408 | 12403816-6037586 | -45.9 ± 0.5 | … | … | … | … | … | N | … | … | … | … | n | … |
| 53116 | 12395583-6036370 | -16.4 ± 0.6 | 5968 ± 301 | 4.12 ± 0.14 | 0.05 ± 0.19 | … | … | N | … | … | … | … | n | … |
| 56023 | 12393479-6038197 | -28.3 ± 0.1 | … | … | … | … | … | N | … | … | … | … | n | … |
| 56409 | 12403876-6039426 | -105.9 ± 1.2 | … | … | … | … | … | N | … | … | … | … | n | … |
| 56214 | 12395584-6045320 | -49.4 ± 0.2 | … | … | … | … | … | N | … | … | … | … | n | … |
| 55751 | 12385726-6044263 | -40.8 ± 0.3 | … | … | … | … | … | Y | … | … | … | … | n | … |
| 56024 | 12393480-6041303 | -38.2 ± 1.4 | … | … | … | … | … | Y | … | … | … | … | n | … |
| 56215 | 12395591-6036191 | -30.0 ± 1.3 | … | … | … | … | … | N | … | … | … | … | n | … |
| 53254 | 12403882-6042213 | -7.5 ± 0.5 | 5518 ± 67 | 4.17 ± 0.53 | 0.46 ± 0.40 | … | … | N | … | … | … | … | n | … |
| 55752 | 12385766-6040115 | -35.1 ± 0.4 | … | … | … | … | … | N | … | … | … | … | n | … |
| 52968 | 12393481-6038198 | -5.0 ± 0.3 | 6450 ± 234 | 3.99 ± 0.09 | -0.15 ± 0.12 | 79 ± 14 | 1 | N | … | … | … | … | n | NG |
| 56025 | 12393492-6036459 | -75.9 ± 0.1 | … | … | … | … | … | N | … | … | … | … | n | … |
| 53117 | 12395592-6040285 | -48.1 ± 0.7 | 6720 ± 77 | 4.02 ± 0.15 | 0.18 ± 0.06 | … | … | N | … | … | … | … | n | … |
| 56410 | 12403915-6033097 | 14.3 ± 0.3 | 5899 ± 129 | 3.23 ± 0.05 | 0.06 ± 0.27 | … | … | N | … | … | … | … | n | … |
| 56411 | 12403957-6035231 | -37.1 ± 1.3 | … | … | … | … | … | Y | … | … | … | … | n | … |
| 52793 | 12390325-6041167 | 1.8 ± 0.4 | 5359 ± 200 | 4.66 ± 0.45 | 0.00 ± 0.14 | … | … | N | … | … | … | … | n | … |
| 56216 | 12395601-6042241 | -45.0 ± 0.5 | … | … | … | … | … | N | … | … | … | … | n | … |
| 52969 | 12393494-6032262 | -37.2 ± 0.6 | 6212 ± 570 | 4.45 ± 0.57 | 0.25 ± 0.76 | … | … | Y | … | … | … | … | n | … |
| 53255 | 12403979-6046315 | -57.8 ± 1.4 | 6721 ± 705 | 4.33 ± 0.41 | -0.35 ± 0.07 | … | … | N | … | … | … | … | n | … |
| 53118 | 12395613-6037287 | -40.6 ± 0.4 | 6734 ± 62 | 3.89 ± 0.12 | 0.14 ± 0.05 | 58 ± 31 | … | Y | Y | Y | Y | Y | Y | … |
| 52794 | 12390326-6039322 | -12.8 ± 0.7 | 5891 ± 214 | 4.33 ± 0.28 | 0.13 ± 0.17 | … | … | N | … | … | … | … | n | … |
| 52995 | 12393801-6041119 | -30.0 ± 0.5 | 5912 ± 223 | 4.09 ± 0.05 | -0.11 ± 0.21 | … | … | N | … | … | … | … | n | … |
| 56217 | 12395628-6033438 | -12.7 ± 0.1 | … | … | … | … | … | N | … | … | … | … | n | … |
| 53256 | 12403990-6040452 | -46.6 ± 1.0 | 5120 ± 890 | 3.59 ± 1.44 | -0.62 ± 0.25 | … | … | N | … | … | … | … | n | … |
| 55785 | 12390329-6034470 | -33.1 ± 0.1 | … | … | … | … | … | N | … | … | … | … | n | … |
| 56047 | 12393803-6037440 | -41.9 ± 2.0 | … | … | … | … | … | Y | … | … | … | … | n | … |
| 56218 | 12395632-6035538 | -76.5 ± 1.2 | … | … | … | … | … | N | … | … | … | … | n | … |
| 56412 | 12404026-6032099 | -10.1 ± 0.1 | … | … | … | … | … | N | … | … | … | … | n | … |
| 52795 | 12390351-6043001 | -13.5 ± 0.2 | 5846 ± 73 | 3.98 ± 0.16 | 0.01 ± 0.15 | … | … | N | … | … | … | … | n | … |
| 52996 | 12393811-6040389 | -36.1 ± 0.8 | 6452 ± 106 | 3.83 ± 0.22 | 0.04 ± 0.08 | … | … | Y | … | … | … | … | n | … |
| 53135 | 12400054-6039389 | -7.5 ± 0.7 | 5250 ± 288 | 3.74 ± 0.49 | -0.30 ± 0.20 | <106 | 3 | N | … | … | … | … | n | NG |
| 56413 | 12404047-6030538 | -40.4 ± 0.2 | … | … | … | … | … | Y | … | … | … | … | n | … |
| 55786 | 12390359-6045366 | -46.6 ± 0.2 | … | … | … | … | … | N | … | … | … | … | n | … |
| 52997 | 12393819-6043009 | -39.6 ± 0.7 | 5178 ± 62 | 4.37 ± 0.83 | -0.24 ± 0.19 | … | … | Y | … | … | … | … | n | … |
| 56248 | 12400060-6029262 | -60.4 ± 0.3 | 5649 ± 129 | 2.62 ± 0.05 | 0.29 ± 0.31 | … | … | N | … | … | … | … | n | … |
| 56414 | 12404065-6042425 | -28.2 ± 1.8 | … | … | … | … | … | N | … | … | … | … | n | … |
| 56048 | 12393842-6034479 | -37.7 ± 0.6 | … | … | … | … | … | Y | … | … | … | … | n | … |
| 55787 | 12390366-6036008 | -14.9 ± 0.1 | 5835 ± 143 | 4.89 ± 0.05 | 0.25 ± 0.22 | … | … | N | … | … | … | … | n | … |
| 53136 | 12400068-6036501 | -40.5 ± 0.5 | 5238 ± 277 | 4.48 ± 0.73 | -0.21 ± 0.15 | … | … | Y | … | … | … | … | n | … |
| 56415 | 12404137-6035522 | -13.9 ± 0.3 | … | … | … | … | … | N | … | … | … | … | n | … |

**Table C.17.** continued.

| ID | CNAME | RV (km s$^{-1}$) | T$_{\text{eff}}$ (K) | logg (dex) | [Fe/H] (dex) | EW(Li)$^a$ (mÅ) | EW(Li) error flag$^b$ | Membership RV | Li | logg | [Fe/H] | Gaia study Cantat-Gaudin$^c$ | Final$^d$ | NMs with Li$^e$ |
|---|---|---|---|---|---|---|---|---|---|---|---|---|---|---|
| 52998 | 12393847-6036012 | -49.9 ± 0.4 | 5869 ± 130 | 4.14 ± 0.34 | 0.02 ± 0.16 | … | … | N | … | … | … | … | n | … |
| 56249 | 12400075-6033516 | -20.1 ± 0.1 | … | … | … | … | … | N | … | … | … | … | n | … |
| 55788 | 12390368-6038400 | -45.3 ± 0.1 | … | … | … | … | … | N | … | … | … | … | n | … |
| 52999 | 12393849-6039127 | -40.0 ± 0.4 | 6019 ± 122 | 4.22 ± 0.35 | 0.10 ± 0.14 | 80 ± 31 | 1 | Y | Y | Y | Y | … | Y | … |
| 56250 | 12400100-6035556 | -39.8 ± 0.1 | 4731 ± 110 | 3.22 ± 0.06 | 0.30 ± 0.23 | … | … | Y | … | … | … | … | n | … |
| 52796 | 12390368-6041065 | -31.1 ± 0.6 | 5730 ± 133 | 4.06 ± 0.44 | -0.05 ± 0.18 | … | … | N | … | … | … | … | n | … |
| 56417 | 12404321-6045477 | -20.9 ± 0.1 | … | … | … | … | … | N | … | … | … | … | n | … |
| 53257 | 12404368-6043366 | 13.5 ± 0.5 | 5662 ± 264 | 4.41 ± 0.10 | 0.47 ± 0.18 | … | … | N | … | … | … | … | n | … |
| 53137 | 12400101-6038232 | -46.7 ± 0.9 | 6289 ± 321 | 4.41 ± 0.34 | 0.25 ± 0.25 | 48 ± 36 | 1 | N | … | … | … | … | n | NG |
| 53000 | 12393855-6027443 | -21.0 ± 1.1 | 4926 ± 225 | 2.81 ± 0.39 | -0.55 ± 0.17 | … | … | N | … | … | … | … | n | G |
| 3393 | 12400109-6031395 | -40.1 ± 0.6 | 4888 ± 116 | 2.85 ± 0.23 | 0.13 ± 0.10 | 24 ± 1 | … | Y | Y | Y | Y | … | Y | … |
| 52797 | 12390369-6036008 | -14.8 ± 0.2 | 5770 ± 139 | 4.15 ± 0.14 | 0.37 ± 0.18 | … | … | N | … | … | … | … | n | … |
| 53258 | 12404378-6036487 | 59.8 ± 0.3 | 5586 ± 178 | 4.37 ± 0.34 | 0.41 ± 0.23 | <28 | 3 | N | … | … | … | … | n | NG |
| 56049 | 12393873-6029279 | -30.6 ± 0.1 | 5599 ± 238 | 4.91 ± 0.05 | 0.01 ± 0.19 | … | … | N | … | … | … | … | n | … |
| 56251 | 12400115-6035219 | -40.1 ± 0.1 | 4947 ± 94 | 3.37 ± 0.04 | 0.17 ± 0.20 | … | … | Y | … | … | … | … | n | … |
| 55789 | 12390372-6044074 | -11.4 ± 0.7 | … | … | … | … | … | N | … | … | … | … | n | … |
| 53259 | 12404413-6043168 | -31.9 ± 0.6 | 6181 ± 545 | 4.38 ± 0.25 | 0.20 ± 0.36 | … | … | N | … | … | … | … | n | … |
| 56050 | 12393879-6045226 | -68.2 ± 0.4 | … | … | … | … | … | N | … | … | … | … | n | … |
| 52798 | 12390375-6033110 | 0.9 ± 0.2 | 5921 ± 4 | 3.51 ± 0.10 | -0.66 ± 0.56 | 47 ± 13 | … | N | … | … | … | … | n | NG |
| 56252 | 12400115-6038527 | -40.7 ± 1.1 | … | … | … | … | … | Y | … | … | … | … | n | … |
| 56418 | 12404470-6037323 | 13.0 ± 0.1 | … | … | … | … | … | N | … | … | … | … | n | … |
| 56051 | 12393882-6042090 | -44.8 ± 0.1 | 5567 ± 105 | 4.44 ± 0.05 | 0.28 ± 0.19 | … | … | N | … | … | … | … | n | … |
| 3394 | 12400116-6035218 | -39.3 ± 0.6 | 4887 ± 122 | 2.77 ± 0.22 | 0.10 ± 0.10 | <7 | 3 | Y | Y | Y | Y | … | Y | … |
| 52799 | 12390394-6038558 | -38.5 ± 0.7 | 6690 ± 85 | 3.73 ± 0.17 | 0.01 ± 0.07 | <22 | 3 | Y | Y | Y | Y | Y | Y | … |
| 53260 | 12404608-6039269 | -35.7 ± 0.7 | 5044 ± 227 | 3.00 ± 0.42 | -0.44 ± 0.21 | … | … | N | … | … | … | … | n | G |
| 56253 | 12400118-6032492 | 60.7 ± 0.1 | … | … | … | … | … | N | … | … | … | … | n | … |
| 56052 | 12393885-6045154 | -37.1 ± 0.2 | 6311 ± 128 | 3.52 ± 0.02 | … | … | … | Y | … | … | … | … | n | … |
| 3362 | 12390409-6034001 | -39.0 ± 0.6 | 4406 ± 114 | 2.05 ± 0.24 | 0.00 ± 0.10 | <4 | 3 | Y | Y | Y | Y | … | Y | … |
| 56419 | 12404651-6032542 | -10.1 ± 0.1 | … | … | … | … | … | N | … | … | … | … | n | … |
| 56254 | 12400121-6034326 | -39.6 ± 0.8 | … | … | … | … | … | Y | … | … | … | … | n | … |
| 56053 | 12393886-6034142 | -39.6 ± 0.9 | … | … | … | … | … | Y | … | … | … | … | n | … |
| 55790 | 12390429-6039566 | -9.3 ± 0.2 | … | … | … | … | … | N | … | … | … | … | n | … |
| 53138 | 12400127-6036301 | -23.3 ± 0.3 | 6727 ± 44 | 3.82 ± 0.08 | 0.40 ± 0.04 | … | … | N | … | … | … | … | n | … |
| 56054 | 12393886-6043298 | -38.3 ± 1.3 | … | … | … | … | … | Y | … | … | … | … | n | … |
| 55791 | 12390435-6028185 | -134.0 ± 0.7 | … | … | … | … | … | N | … | … | … | … | n | … |
| 56421 | 12404805-6041259 | 45.8 ± 0.1 | 5165 ± 109 | 4.05 ± 0.05 | -0.06 ± 0.23 | … | … | N | … | … | … | … | n | … |
| 56255 | 12400128-6041004 | -37.2 ± 0.6 | … | … | … | … | … | Y | … | … | … | … | n | … |
| 53001 | 12393887-6034141 | -46.7 ± 1.3 | 6318 ± 498 | 4.77 ± 0.84 | -0.45 ± 0.77 | … | 3 | N | … | … | … | Y | n | … |
| 55792 | 12390443-6033154 | -12.2 ± 0.2 | … | … | … | … | … | N | … | … | … | … | n | … |
| 56422 | 12404809-6036471 | -14.5 ± 2.2 | … | … | … | … | … | N | … | … | … | … | n | … |
| 53139 | 12400129-6036096 | -17.0 ± 0.4 | 5338 ± 82 | 4.10 ± 0.28 | -0.12 ± 0.12 | … | … | N | … | … | … | … | n | … |
| 56055 | 12393899-6038032 | 2.8 ± 0.1 | 5577 ± 96 | 2.51 ± 0.05 | -0.17 ± 0.14 | … | … | N | … | … | … | … | n | … |
| 53140 | 12400147-6036148 | -10.1 ± 0.2 | 4763 ± 146 | 3.24 ± 0.49 | 0.27 ± 0.28 | … | … | N | … | … | … | … | n | … |
| 55793 | 12390476-6041475 | -40.3 ± 0.1 | 5489 ± 131 | 4.19 ± 0.01 | 0.19 ± 0.18 | … | … | Y | … | … | … | … | n | … |
| 53002 | 12393903-6038514 | -39.1 ± 0.4 | 6058 ± 152 | 4.16 ± 0.44 | -0.12 ± 0.14 | 92 ± 63 | 1 | Y | Y | Y | Y | N | Y | … |
| 53141 | 12400173-6036269 | -7.6 ± 0.5 | 5384 ± 191 | 4.53 ± 0.10 | -0.34 ± 0.16 | … | … | N | … | … | … | … | n | … |
| 55794 | 12390477-6045048 | -39.3 ± 1.7 | … | … | … | … | … | Y | … | … | … | … | n | … |
| 56056 | 12393907-6037309 | 11.0 ± 0.2 | … | … | … | … | … | N | … | … | … | … | n | … |
| 56256 | 12400173-6036552 | 4.1 ± 0.1 | … | … | … | … | … | N | … | … | … | … | n | … |
| 3363 | 12390478-6041475 | -40.2 ± 0.6 | 5003 ± 116 | 2.95 ± 0.22 | 0.18 ± 0.09 | <8 | 3 | Y | Y | Y | Y | … | Y | … |
| 56057 | 12393911-6035283 | -27.9 ± 0.1 | 6491 ± 149 | 4.00 ± 0.02 | … | … | … | N | … | … | … | … | n | … |
| 53142 | 12400174-6033133 | -58.2 ± 1.0 | 6012 ± 307 | 4.37 ± 0.19 | -0.55 ± 0.32 | <92 | 3 | N | … | … | … | … | n | NG |
| 55795 | 12390479-6032043 | -2.1 ± 0.2 | … | … | … | … | … | N | … | … | … | … | n | … |
| 53003 | 12393919-6038118 | -2.2 ± 0.4 | 6600 ± 344 | 4.07 ± 0.20 | -0.02 ± 0.13 | 80 ± 28 | 1 | N | … | … | … | N | n | NG |
| 56257 | 12400179-6040396 | -40.3 ± 0.6 | … | … | … | … | … | Y | … | … | … | … | n | … |
| 55796 | 12390482-6033528 | 48.1 ± 0.2 | 5851 ± 140 | 3.17 ± 0.07 | 0.32 ± 0.20 | … | … | N | … | … | … | … | n | … |
| 53143 | 12400181-6039481 | -40.1 ± 0.6 | 6050 ± 246 | 4.57 ± 0.51 | 0.10 ± 0.24 | 75 ± 52 | 1 | Y | Y | Y | Y | … | Y | … |
| 53004 | 12393922-6035550 | 0.7 ± 0.3 | 6109 ± 173 | 4.45 ± 0.27 | -0.25 ± 0.21 | 48 ± 25 | … | N | … | … | … | … | n | NG |
| 55797 | 12390517-6036213 | -33.9 ± 1.0 | … | … | … | … | … | N | … | … | … | … | n | … |

| ID | CNAME | RV (km s$^{-1}$) | $T_{\rm eff}$ (K) | $logg$ (dex) | [Fe/H] (dex) | $EW$(Li)$^a$ (mÅ) | $EW$(Li) error flag$^b$ | Membership RV | Li | $logg$ | [Fe/H] | Gaia study Cantat-Gaudin$^c$ | Final$^d$ | NMs with Li$^e$ |
|---|---|---|---|---|---|---|---|---|---|---|---|---|---|---|
| 53144 | 12400187-6037190 | -38.8 ± 0.4 | 6238 ± 112 | 4.25 ± 0.36 | 0.19 ± 0.23 | … | … | Y | … | … | … | … | n | … |
| 53005 | 12393927-6036278 | -40.2 ± 0.3 | 6049 ± 186 | 4.34 ± 0.28 | 0.02 ± 0.16 | 115 ± 46 | 1 | Y | Y | Y | Y | … | Y | … |
| 55798 | 12390528-6041138 | -10.1 ± 0.1 | … | … | … | … | … | N | … | … | … | … | n | … |
| 56258 | 12400218-6030257 | -8.7 ± 0.1 | 5355 ± 125 | 4.16 ± 0.05 | 0.53 ± 0.24 | … | … | N | … | … | … | … | n | … |
| 56058 | 12393930-6035079 | -35.5 ± 0.6 | … | … | … | … | … | N | … | … | … | … | n | … |
| 55813 | 12390867-6044467 | -31.7 ± 0.1 | 6281 ± 102 | 4.68 ± 0.05 | … | … | … | N | … | … | … | … | n | … |
| 56259 | 12400220-6040421 | -39.3 ± 0.3 | … | … | … | … | … | Y | … | … | … | … | n | … |
| 56059 | 12393940-6035044 | -38.9 ± 0.1 | … | … | … | … | … | Y | … | … | … | … | n | … |
| 52810 | 12390869-6036163 | 23.8 ± 0.5 | 5682 ± 457 | 4.38 ± 0.41 | -0.14 ± 0.52 | <120 | 3 | N | … | … | … | … | n | NG |
| 56271 | 12400486-6044164 | -40.6 ± 0.1 | 4483 ± 82 | 2.38 ± 0.05 | -0.09 ± 0.18 | … | … | Y | … | … | … | … | n | … |
| 56076 | 12394079-6036281 | -11.7 ± 0.1 | 4859 ± 81 | 3.41 ± 0.02 | 0.18 ± 0.17 | … | … | N | … | … | … | … | n | … |
| 55814 | 12390890-6034318 | -10.8 ± 0.1 | 6757 ± 136 | 4.63 ± 0.02 | … | … | … | N | … | … | … | … | n | … |
| 56272 | 12400496-6039241 | -38.1 ± 0.5 | … | … | … | … | … | Y | … | … | … | … | n | … |
| 56077 | 12394088-6043283 | -45.7 ± 0.1 | 5765 ± 130 | 2.77 ± 0.02 | 0.16 ± 0.22 | … | … | N | … | … | … | … | n | … |
| 56273 | 12400513-6038488 | -40.5 ± 0.4 | … | … | … | … | … | Y | … | … | … | … | n | … |
| 53013 | 12394103-6038054 | 2.0 ± 0.4 | 5823 ± 397 | 4.64 ± 0.32 | 0.19 ± 0.18 | … | … | N | … | … | … | … | n | … |
| 56274 | 12400515-6031251 | -81.4 ± 0.2 | … | … | … | … | … | N | … | … | … | … | n | … |
| 53155 | 12400515-6038487 | -41.4 ± 0.3 | 6628 ± 355 | 3.83 ± 0.03 | 0.10 ± 0.19 | 79 ± 23 | 1 | Y | Y | Y | Y | Y | Y | … |
| 53156 | 12400523-6034198 | 2.3 ± 0.3 | 6470 ± 143 | 4.23 ± 0.10 | -0.08 ± 0.17 | … | … | N | … | … | … | … | n | … |
| 56275 | 12400532-6032547 | -24.9 ± 0.1 | … | … | … | … | … | N | … | … | … | … | n | … |
| 56276 | 12400535-6041420 | -40.5 ± 0.3 | … | … | … | … | … | Y | … | … | … | … | n | … |
| 53157 | 12400537-6039359 | -17.5 ± 1.1 | 5394 ± 117 | 3.02 ± 0.26 | -0.58 ± 0.39 | … | … | N | … | … | … | … | n | … |
| 56277 | 12400545-6038114 | -35.0 ± 0.3 | … | … | … | … | … | N | … | … | … | … | n | … |
| 53158 | 12400561-6037144 | -23.9 ± 0.3 | 5411 ± 202 | 4.60 ± 0.56 | -0.07 ± 0.18 | 103 ± 41 | 1 | N | … | … | … | … | n | NG |
| 53159 | 12400567-6039048 | 10.1 ± 0.4 | 5298 ± 124 | 4.31 ± 0.36 | 0.24 ± 0.14 | <84 | 3 | N | … | … | … | … | n | NG |
| 53160 | 12400571-6042373 | -15.1 ± 0.2 | 6310 ± 103 | 4.43 ± 0.48 | -0.08 ± 0.32 | 49 ± 21 | … | N | … | … | … | … | n | NG |
| 56278 | 12400578-6038364 | -34.7 ± 0.5 | … | … | … | … | … | N | … | … | … | … | n | … |
| 56279 | 12400579-6035292 | -39.5 ± 0.2 | … | … | … | … | … | Y | … | … | … | … | n | … |
| 53161 | 12400590-6033264 | -44.8 ± 0.3 | 5775 ± 97 | 4.14 ± 0.37 | 0.28 ± 0.17 | … | … | N | … | … | … | … | n | … |
| 56280 | 12400598-6040042 | -44.8 ± 1.6 | … | … | … | … | … | N | … | … | … | … | n | … |
| 56281 | 12400625-6040277 | 10.3 ± 0.2 | 5455 ± 111 | 2.51 ± 0.01 | -0.39 ± 0.18 | … | … | N | … | … | … | … | n | … |
| 53162 | 12400650-6033443 | -41.5 ± 0.5 | 5616 ± 263 | 3.94 ± 0.04 | -0.07 ± 0.17 | … | … | Y | … | … | … | … | n | … |
| 56282 | 12400654-6041347 | -40.7 ± 0.2 | … | … | … | … | … | Y | … | … | … | … | n | … |
| 56283 | 12400678-6041414 | -44.1 ± 1.3 | … | … | … | … | … | N | … | … | … | … | n | … |
| 53163 | 12400693-6037388 | -32.7 ± 0.5 | 5916 ± 224 | 4.24 ± 0.05 | -0.08 ± 0.21 | … | … | N | … | … | … | … | n | … |
| 56284 | 12400705-6045432 | -42.3 ± 0.1 | 5653 ± 124 | 4.86 ± 0.05 | 0.49 ± 0.22 | … | … | Y | … | … | … | … | n | … |
| 53164 | 12400706-6037452 | 19.8 ± 0.4 | 6490 ± 299 | 4.47 ± 0.28 | 0.22 ± 0.23 | … | … | N | … | … | … | … | n | … |
| 53194 | 12401164-6038206 | -24.6 ± 0.4 | 5606 ± 254 | 4.34 ± 0.13 | 0.10 ± 0.13 | … | … | N | … | … | … | … | n | … |
| 56303 | 12401168-6034443 | 28.4 ± 0.1 | 5421 ± 114 | 4.40 ± 0.05 | -0.21 ± 0.16 | … | … | N | … | … | … | … | n | … |
| 56304 | 12401175-6040535 | -10.1 ± 0.2 | … | … | … | … | … | N | … | … | … | … | n | … |
| 56305 | 12401187-6046588 | -13.5 ± 1.5 | … | … | … | … | … | N | … | … | … | … | n | … |
| 56306 | 12401188-6042503 | 25.9 ± 0.1 | 5761 ± 149 | 4.75 ± 0.06 | 0.08 ± 0.19 | … | … | N | … | … | … | … | n | … |
| 56307 | 12401189-6035548 | -41.2 ± 0.1 | 5775 ± 89 | 4.84 ± 0.04 | 0.58 ± 0.19 | … | … | Y | … | … | … | … | n | … |
| 53195 | 12401194-6034051 | -38.7 ± 0.5 | 5693 ± 60 | 4.29 ± 0.07 | -0.05 ± 0.19 | … | … | Y | … | … | … | … | n | … |
| 53196 | 12401196-6038033 | -11.2 ± 0.5 | 6464 ± 52 | 3.65 ± 0.10 | 0.28 ± 0.04 | 72 ± 37 | … | N | … | … | … | … | n | NG |
| 53197 | 12401220-6036508 | -102.5 ± 0.6 | 5700 ± 137 | 3.75 ± 0.47 | -0.60 ± 0.31 | … | … | N | … | … | … | … | n | … |
| 53198 | 12401220-6037063 | -1.1 ± 0.5 | 5089 ± 250 | 4.10 ± 0.35 | -0.38 ± 0.45 | … | … | N | … | … | … | … | n | … |
| 56308 | 12401224-6033260 | -11.1 ± 0.2 | … | … | … | … | … | N | … | … | … | … | n | … |
| 56309 | 12401261-6033011 | -51.7 ± 0.1 | 4777 ± 101 | 3.09 ± 0.08 | 0.13 ± 0.21 | … | … | N | … | … | … | … | n | … |
| 53200 | 12401268-6035293 | -12.0 ± 0.3 | 5812 ± 215 | 4.35 ± 0.18 | -0.05 ± 0.24 | … | … | N | … | … | … | … | n | … |
| 53201 | 12401301-6032009 | -26.6 ± 0.3 | 5834 ± 378 | 4.31 ± 0.20 | 0.30 ± 0.19 | … | … | N | … | … | … | … | n | … |
| 56310 | 12401310-6045270 | -40.1 ± 0.6 | … | … | … | … | … | Y | … | … | … | … | n | … |
| 53202 | 12401318-6037507 | -40.3 ± 0.4 | 5833 ± 198 | 4.44 ± 0.31 | -0.21 ± 0.27 | <48 | 3 | Y | Y | Y | Y | … | Y | … |
| 53203 | 12401318-6039190 | -18.6 ± 0.3 | 5757 ± 261 | 4.39 ± 0.20 | 0.27 ± 0.14 | … | … | N | … | … | … | … | n | … |
| 53204 | 12401328-6038137 | -26.8 ± 0.6 | 5105 ± 926 | 4.10 ± 0.84 | -0.12 ± 0.19 | <83 | 3 | N | … | … | … | … | n | NG |
| 53205 | 12401337-6032440 | -39.4 ± 0.2 | 6352 ± 164 | 3.97 ± 0.28 | 0.21 ± 0.19 | 127 ± 45 | 1 | Y | Y | Y | Y | Y | Y | … |
| 56311 | 12401345-6046068 | -39.9 ± 0.2 | 5693 ± 125 | 3.30 ± 0.07 | 0.20 ± 0.24 | … | … | Y | … | … | … | … | n | … |
| 53206 | 12401349-6036559 | -29.2 ± 0.3 | 6214 ± 411 | 4.71 ± 0.45 | 0.01 ± 0.50 | 72 ± 52 | 1 | N | … | … | … | N | n | NG |
| 53207 | 12401378-6039529 | -20.1 ± 0.3 | 5083 ± 88 | 4.62 ± 0.58 | -0.05 ± 0.13 | … | … | N | … | … | … | … | n | … |

**Table C.17.** continued.

| ID | CNAME | RV (km s$^{-1}$) | $T_\mathrm{eff}$ (K) | logg (dex) | [Fe/H] (dex) | EW(Li)$^a$ (mÅ) | EW(Li) error flag$^b$ | RV | Membership Li | logg | [Fe/H] | Gaia study Cantat-Gaudin$^c$ | Final$^d$ | NMs with Li$^e$ |
|---|---|---|---|---|---|---|---|---|---|---|---|---|---|---|
| 56312 | 12401403-6037190 | -39.4 ± 0.1 | 5519 ± 134 | 4.25 ± 0.02 | 0.21 ± 0.21 | … | … | Y | … | … | … | … | n | … |
| 53217 | 12401670-6037140 | -18.8 ± 0.4 | 5306 ± 216 | 4.38 ± 0.47 | -0.44 ± 0.27 | … | … | N | … | … | … | … | n | … |
| 56328 | 12401670-6038445 | -22.5 ± 0.7 | … | … | … | … | … | N | … | … | … | … | n | … |
| 53218 | 12401685-6034580 | -41.2 ± 0.8 | 6332 ± 358 | 4.18 ± 0.03 | -0.10 ± 0.16 | … | … | Y | … | … | … | … | n | … |

**Notes.** $^{(a)}$ The values of EW(Li) for this cluster are corrected (subtracted adjacent Fe (6707.43 Å) line). $^{(b)}$ Flags for the errors of the corrected EW(Li) values, as follows: 1=EW(Li) corrected by blends contribution using models; and 3=Upper limit (no error for EW(Li) is given). $^{(c)}$ Cantat-Gaudin et al. (2018). $^{(d)}$ The letters "Y" and "N" indicate if the star is a cluster member or not. $^{(e)}$ 'Li-rich G', 'G' and 'NG' indicate "Li-rich giant", "giant" and "non-giant" Li field outliers, respectively.

**Table C.18.** Berkeley 44

| ID | CNAME | RV (km s$^{-1}$) | $T_{\text{eff}}$ (K) | logg (dex) | [Fe/H] (dex) | EW(Li)[a] (mÅ) | EW(Li) error flag[b] | Membership RV | Li | logg | [Fe/H] | Gaia study Cantat-Gaudin[c] | Final[d] | NMs with Li[e] |
|---|---|---|---|---|---|---|---|---|---|---|---|---|---|---|
| 3608 | 19170284+1928353 | 13.7 ± 0.7 | 5652 ± 98 | 4.21 ± 0.13 | -0.53 ± 0.19 | <73 | 3 | N | N | N | N | ... | n | NG |
| 3626 | 19170961+1929259 | -12.2 ± 0.4 | 6718 ± 385 | 4.39 ± 0.37 | 0.17 ± 0.20 | 69 ± 27 | ... | N | Y | Y | Y | Y | Y | ... |
| 3631 | 19171111+1933569 | -7.7 ± 0.3 | 5992 ± 84 | 4.39 ± 0.30 | -0.22 ± 0.20 | 62 ± 32 | ... | Y | Y | Y | N | ... | Y | ... |
| 3634 | 19171258+1931562 | -9.2 ± 0.4 | 6543 ± 304 | 4.38 ± 0.49 | 0.21 ± 0.20 | 86 ± 29 | ... | Y | Y | Y | Y | Y | Y | ... |
| 3635 | 19171260+1936298 | -8.1 ± 0.6 | 6381 ± 316 | 4.40 ± 0.35 | 0.25 ± 0.21 | 70 ± 42 | ... | Y | Y | Y | Y | Y | Y | ... |
| 3637 | 19171317+1932246 | -5.7 ± 0.5 | 6560 ± 300 | 4.49 ± 0.52 | 0.11 ± 0.20 | 71 ± 36 | ... | N | Y | Y | Y | Y | Y | ... |
| 3638 | 19171326+1934055 | -8.3 ± 0.4 | 6466 ± 132 | 4.05 ± 0.23 | 0.10 ± 0.20 | 76 ± 36 | ... | Y | Y | Y | Y | Y | Y | ... |
| 3641 | 19171454+1937098 | -6.2 ± 1.2 | 6620 ± 500 | 3.84 ± 0.23 | 0.19 ± 0.19 | <48 | 3 | N | Y | Y | Y | Y | Y | ... |
| 3643 | 19171523+1933146 | -8.9 ± 0.5 | 6272 ± 90 | 3.97 ± 0.18 | 0.07 ± 0.25 | 59 ± 30 | ... | Y | Y | Y | Y | Y | Y | ... |
| 3647 | 19171634+1933251 | -8.0 ± 1.0 | 6426 ± 326 | 4.42 ± 0.55 | 0.08 ± 0.25 | 26 ± 20 | ... | Y | Y | Y | Y | Y | Y | ... |
| 3671 | 19172749+1935144 | -7.2 ± 0.7 | 6467 ± 304 | 4.00 ± 0.15 | 0.15 ± 0.24 | 32 ± 30 | ... | Y | Y | Y | Y | Y | Y | ... |
| 3677 | 19173159+1934134 | -8.5 ± 0.7 | 6503 ± 215 | 3.57 ± 0.28 | 0.15 ± 0.17 | 92 ± 44 | 1 | Y | Y | Y | Y | Y | Y | ... |
| 3606 | 19170216+1930159 | -6.0 ± 0.5 | 6437 ± 316 | 4.24 ± 0.19 | 0.17 ± 0.20 | 59 ± 35 | 3 | N | Y | Y | Y | Y | Y | ... |
| 3609 | 19170344+1931232 | 15.3 ± 0.8 | 6268 ± 168 | 4.22 ± 0.32 | 0.06 ± 0.13 | ... | ... | N | ... | ... | ... | ... | n | ... |
| 3610 | 19170500+1929506 | -5.4 ± 0.7 | 6708 ± 521 | 4.44 ± 0.40 | 0.26 ± 0.30 | ... | ... | N | ... | ... | ... | Y | n | ... |
| 3611 | 19170520+1931147 | -8.7 ± 0.4 | 6177 ± 162 | 4.19 ± 0.16 | 0.15 ± 0.16 | ... | ... | Y | ... | ... | ... | ... | n | ... |
| 3612 | 19170549+1934008 | 62.0 ± 0.3 | 5725 ± 146 | 4.39 ± 0.19 | -0.02 ± 0.15 | ... | ... | N | ... | ... | ... | ... | n | ... |
| 3613 | 19170614+1932172 | -8.8 ± 0.6 | 6493 ± 271 | 4.24 ± 0.28 | 0.13 ± 0.21 | ... | 3 | Y | ... | ... | ... | Y | n | ... |
| 3614 | 19170626+1934590 | 6.5 ± 0.5 | 5722 ± 264 | 4.31 ± 0.24 | -0.06 ± 0.30 | ... | ... | N | ... | ... | ... | ... | n | ... |
| 3615 | 19170631+1928158 | -14.4 ± 0.3 | 5292 ± 130 | 4.08 ± 0.51 | 0.09 ± 0.13 | <102 | 3 | N | N | N | N | ... | n | NG |
| 3616 | 19170634+1933322 | -7.6 ± 0.3 | 6338 ± 166 | 4.34 ± 0.43 | 0.04 ± 0.20 | ... | 1 | Y | ... | ... | ... | ... | n | ... |
| 3617 | 19170643+1928379 | 39.4 ± 2.5 | 6465 ± 235 | 3.61 ± 0.51 | -0.36 ± 0.23 | ... | ... | N | ... | ... | ... | ... | n | ... |
| 3618 | 19170646+1933119 | -7.4 ± 0.7 | 5927 ± 124 | 4.26 ± 0.19 | -0.10 ± 0.15 | ... | ... | Y | ... | ... | ... | ... | n | ... |
| 3619 | 19170666+1932396 | -7.0 ± 0.6 | 6601 ± 341 | 4.13 ± 0.30 | 0.17 ± 0.20 | 115 ± 54 | ... | N | Y | Y | Y | Y | Y | ... |
| 3620 | 19170685+1929101 | 3.7 ± 0.6 | 5855 ± 284 | 3.97 ± 0.26 | 0.04 ± 0.17 | <98 | 3 | N | N | N | N | ... | n | NG |
| 466 | 19170732+1930555 | -7.7 ± 0.6 | 4998 ± 233 | 3.13 ± 0.55 | 0.11 ± 0.14 | 24 ± 1 | ... | Y | Y | Y | Y | ... | Y | ... |
| 3621 | 19170738+1936570 | 25.9 ± 0.5 | 6222 ± 309 | 4.55 ± 0.19 | -0.06 ± 0.25 | ... | 1 | N | ... | ... | ... | ... | n | ... |
| 3622 | 19170833+1934095 | -9.4 ± 0.5 | 6199 ± 157 | 4.86 ± 0.38 | 0.00 ± 0.35 | ... | ... | Y | ... | ... | ... | ... | n | ... |
| 3623 | 19170871+1931060 | -2.7 ± 0.4 | 6332 ± 182 | 4.38 ± 0.24 | 0.25 ± 0.15 | ... | ... | N | ... | ... | ... | ... | n | ... |
| 3624 | 19170890+1932231 | -10.1 ± 0.6 | 6311 ± 183 | 4.71 ± 0.49 | 0.20 ± 0.16 | 98 ± 29 | 1 | N | N | N | N | ... | n | NG |
| 3625 | 19170897+1936018 | -49.7 ± 0.3 | 6000 ± 250 | 4.34 ± 0.47 | -0.22 ± 0.23 | ... | 1 | N | ... | ... | ... | ... | n | ... |
| 467 | 19170911+1933256 | -9.1 ± 0.6 | 4943 ± 182 | 3.06 ± 0.45 | 0.20 ± 0.13 | 42 ± 1 | ... | Y | Y | Y | Y | ... | Y | ... |
| 3627 | 19170972+1931594 | -8.7 ± 0.8 | 6507 ± 590 | 4.35 ± 0.32 | 0.12 ± 0.22 | ... | 1 | Y | ... | ... | ... | Y | n | ... |
| 3628 | 19171000+1933207 | -8.6 ± 0.9 | 6372 ± 116 | 4.51 ± 0.50 | 0.27 ± 0.31 | ... | 1 | Y | ... | ... | ... | Y | n | ... |
| 3629 | 19171024+1935454 | 30.6 ± 0.3 | 5551 ± 149 | 4.46 ± 0.31 | -0.13 ± 0.12 | ... | 3 | N | ... | ... | ... | ... | n | ... |
| 468 | 19171032+1932507 | -9.4 ± 0.6 | 5106 ± 134 | 3.55 ± 0.27 | 0.35 ± 0.14 | 41 ± 1 | ... | Y | Y | Y | Y | ... | Y | ... |
| 3630 | 19171039+1935200 | 69.1 ± 1.1 | 5728 ± 165 | 3.96 ± 0.02 | -0.55 ± 0.45 | ... | 1 | N | ... | ... | ... | ... | n | ... |
| 3632 | 19171144+1933285 | -10.5 ± 0.8 | 6579 ± 416 | 4.18 ± 0.33 | 0.18 ± 0.20 | ... | 3 | N | ... | ... | ... | N | n | ... |
| 3633 | 19171160+1932583 | -9.4 ± 0.2 | 4806 ± 134 | 2.88 ± 0.33 | 0.29 ± 0.25 | ... | 3 | Y | ... | ... | ... | Y | n | G |
| 3636 | 19171314+1937108 | 2.3 ± 0.3 | 5892 ± 137 | 4.52 ± 0.33 | -0.18 ± 0.16 | ... | 1 | N | ... | ... | ... | ... | n | ... |
| 3639 | 19171359+1935050 | -14.9 ± 0.5 | 5822 ± 97 | 4.04 ± 0.12 | -0.29 ± 0.13 | <88 | 3 | N | N | N | N | ... | n | NG |
| 469 | 19171388+1933333 | -9.1 ± 0.6 | 4996 ± 181 | 2.79 ± 0.46 | 0.22 ± 0.14 | <17 | 3 | Y | Y | Y | Y | ... | Y | ... |
| 3640 | 19171416+1932520 | -8.1 ± 0.5 | 6453 ± 287 | 4.31 ± 0.29 | 0.26 ± 0.21 | 117 ± 45 | ... | Y | Y | Y | Y | Y | Y | ... |
| 3642 | 19171497+1931283 | 26.9 ± 0.3 | 4847 ± 378 | 3.69 ± 1.10 | -0.51 ± 0.27 | 87 ± 20 | 1 | N | N | N | N | ... | n | NG |
| 3644 | 19171538+1930591 | -9.3 ± 0.8 | 6484 ± 204 | 4.40 ± 0.44 | 0.13 ± 0.18 | ... | ... | Y | ... | ... | ... | Y | n | ... |
| 3645 | 19171555+1928451 | 14.7 ± 0.3 | 5447 ± 74 | 4.32 ± 0.09 | 0.08 ± 0.14 | ... | 1 | N | ... | ... | ... | ... | n | ... |
| 3646 | 19171588+1932063 | -9.1 ± 0.2 | 4749 ± 151 | 2.73 ± 0.29 | 0.31 ± 0.26 | ... | 1 | Y | ... | ... | ... | Y | n | G |
| 3648 | 19171680+1933005 | -7.5 ± 0.7 | 6595 ± 456 | 4.19 ± 0.30 | 0.22 ± 0.23 | ... | 3 | Y | ... | ... | ... | Y | n | ... |
| 3670 | 19172723+1932188 | -6.4 ± 0.6 | 6377 ± 176 | 4.22 ± 0.42 | 0.05 ± 0.29 | 89 ± 43 | ... | N | N | N | N | ... | n | NG |
| 3672 | 19172865+1933377 | -41.3 ± 0.4 | 5370 ± 110 | 3.83 ± 0.58 | -0.34 ± 0.16 | 83 ± 67 | 1 | N | N | N | N | ... | n | NG |
| 3673 | 19173018+1931575 | -3.4 ± 1.0 | 6274 ± 167 | 4.37 ± 0.41 | 0.00 ± 0.27 | ... | ... | N | ... | ... | ... | Y | n | ... |
| 3674 | 19173047+1930284 | 37.3 ± 0.9 | 5932 ± 120 | 4.26 ± 0.23 | -0.75 ± 0.51 | 121 ± 67 | ... | N | N | N | N | ... | n | NG |
| 3675 | 19173062+1929190 | 9.6 ± 0.2 | 4915 ± 163 | 3.17 ± 0.54 | 0.31 ± 0.28 | ... | 1 | N | ... | ... | ... | ... | n | G |
| 3676 | 19173109+1934371 | -21.1 ± 0.3 | 5273 ± 100 | 4.66 ± 0.41 | -0.04 ± 0.17 | ... | 3 | N | ... | ... | ... | ... | n | ... |
| 3678 | 19173275+1930057 | -62.9 ± 0.3 | 5194 ± 131 | 4.07 ± 0.57 | -0.23 ± 0.20 | 83 ± 29 | 1 | N | N | N | N | ... | n | NG |
| 3679 | 19173328+1933570 | 26.0 ± 0.4 | 6299 ± 60 | 4.66 ± 0.62 | -0.25 ± 0.17 | 86 ± 29 | ... | N | N | N | N | ... | n | NG |
| 3680 | 19173346+1931449 | -17.0 ± 0.3 | 5829 ± 160 | 4.51 ± 0.36 | -0.05 ± 0.12 | ... | 1 | N | ... | ... | ... | ... | n | ... |
| 3681 | 19173457+1933295 | 37.5 ± 0.6 | 5901 ± 270 | 4.53 ± 0.32 | -0.22 ± 0.28 | 101 ± 91 | 1 | N | N | N | N | ... | n | NG |
| 3596 | 19165746+1932089 | 18.8 ± 0.4 | 6207 ± 53 | 4.42 ± 0.25 | -0.18 ± 0.14 | 82 ± 64 | ... | N | N | N | N | ... | n | NG |

**Table C.18.** continued.

| ID | CNAME | RV (km s$^{-1}$) | $T_{\rm eff}$ (K) | logg (dex) | [Fe/H] (dex) | EW(Li)$^a$ (mÅ) | EW(Li) error flag$^b$ | \multicolumn{4}{c}{Membership} | Gaia study Cantat-Gaudin$^c$ | Final$^d$ | NMs with Li$^e$ |
| | | | | | | | | RV | Li | logg | [Fe/H] | | | |
|---|---|---|---|---|---|---|---|---|---|---|---|---|---|---|
| 3597 | 19165749+1934018 | 15.5 ± 0.4 | 6074 ± 207 | 4.16 ± 0.15 | -0.36 ± 0.25 | … | 1 | N | … | … | … | … | n | … |
| 3598 | 19165857+1931376 | -38.5 ± 0.3 | 5781 ± 132 | 4.29 ± 0.36 | -0.54 ± 0.22 | 61 ± 31 | … | N | N | N | N | … | n | NG |
| 3599 | 19165874+1934193 | -8.2 ± 0.2 | 4726 ± 143 | 2.72 ± 0.26 | 0.31 ± 0.25 | … | … | Y | … | … | … | Y | n | G |
| 3600 | 19165900+1934512 | -19.1 ± 0.4 | 5427 ± 219 | 3.74 ± 0.52 | -0.07 ± 0.14 | … | … | N | … | … | … | … | n | … |
| 3601 | 19165979+1929245 | -49.6 ± 0.3 | 5649 ± 201 | 4.10 ± 0.28 | 0.37 ± 0.19 | … | 1 | N | … | … | … | … | n | … |
| 3602 | 19170026+1935222 | 48.9 ± 0.4 | 6230 ± 232 | 4.28 ± 0.39 | -0.28 ± 0.15 | 57 ± 42 | 1 | N | N | N | N | … | n | NG |
| 3603 | 19170173+1930379 | 35.8 ± 0.4 | 6391 ± 280 | 4.04 ± 0.18 | 0.01 ± 0.33 | 92 ± 36 | … | N | N | N | N | … | n | NG |
| 3604 | 19170200+1933193 | -8.8 ± 0.7 | 6595 ± 539 | 4.19 ± 0.26 | 0.23 ± 0.30 | … | 3 | Y | … | … | … | Y | n | … |
| 3605 | 19170202+1936036 | 8.2 ± 0.4 | 5797 ± 168 | 4.36 ± 0.51 | -0.18 ± 0.17 | 77 ± 35 | 1 | N | N | N | N | … | n | NG |
| 3607 | 19170220+1929022 | -3.3 ± 0.5 | 5574 ± 137 | 4.44 ± 0.27 | -0.21 ± 0.17 | <140 | 3 | N | N | N | N | … | n | NG |
| 3649 | 19171714+1931191 | -25.4 ± 0.6 | 6660 ± 444 | 4.21 ± 0.36 | -0.20 ± 0.25 | … | … | N | … | … | … | … | n | … |
| 3650 | 19171774+1933344 | -8.0 ± 1.6 | 6144 ± 615 | 4.62 ± 0.41 | 0.09 ± 0.35 | <140 | 3 | Y | Y | Y | Y | Y | Y | … |
| 3651 | 19171827+1929005 | 0.5 ± 0.5 | 5918 ± 586 | 4.24 ± 0.49 | -0.12 ± 0.15 | <98 | 3 | N | N | N | N | … | n | NG |
| 3652 | 19171832+1932559 | -9.9 ± 0.6 | 6081 ± 137 | 4.37 ± 0.17 | -0.05 ± 0.21 | <135 | 3 | Y | N | N | Y | … | n | NG |
| 3653 | 19171876+1934571 | -20.3 ± 0.3 | 5868 ± 194 | 4.41 ± 0.22 | 0.32 ± 0.16 | … | … | N | … | … | … | … | n | … |
| 3654 | 19171915+1931507 | 19.4 ± 2.8 | 5605 ± 98 | 4.08 ± 0.22 | -1.02 ± 0.85 | <82 | 3 | N | N | N | N | Y | n | NG |
| 3655 | 19171972+1933243 | -6.2 ± 0.8 | 6662 ± 541 | 3.92 ± 0.15 | 0.16 ± 0.20 | … | … | N | … | … | … | Y | n | … |
| 3656 | 19172070+1928391 | 4.1 ± 1.2 | 6713 ± 532 | 4.04 ± 0.21 | 0.02 ± 0.24 | … | … | N | … | … | … | … | n | … |
| 3657 | 19172074+1935097 | 52.8 ± 1.0 | 6125 ± 219 | 4.39 ± 0.41 | 0.00 ± 0.54 | … | … | N | … | … | … | … | n | … |
| 3658 | 19172103+1931174 | -10.0 ± 0.4 | 6040 ± 212 | 4.43 ± 0.27 | 0.12 ± 0.31 | 94 ± 75 | 1 | Y | Y | Y | Y | … | Y | … |
| 470 | 19172126+1933307 | -8.7 ± 0.6 | 5455 ± 190 | 3.79 ± 0.30 | 0.46 ± 0.16 | <12 | 3 | Y | Y | N | N | … | n | NG |
| 3659 | 19172129+1933138 | 1.2 ± 0.4 | 5942 ± 260 | 4.24 ± 0.25 | 0.39 ± 0.18 | … | … | N | … | … | … | … | n | … |
| 3660 | 19172169+1930311 | 62.4 ± 0.3 | 5771 ± 62 | 4.32 ± 0.10 | 0.21 ± 0.20 | 109 ± 63 | 1 | N | N | N | N | … | n | NG |
| 3661 | 19172180+1937021 | -39.1 ± 0.3 | 5531 ± 63 | 3.96 ± 0.20 | -0.04 ± 0.14 | 60 ± 34 | 1 | N | N | N | N | … | n | NG |
| 3662 | 19172189+1931363 | 7.5 ± 0.3 | 5475 ± 56 | 4.56 ± 0.34 | 0.23 ± 0.17 | <100 | 3 | N | N | N | N | … | n | NG |
| 471 | 19172208+1933254 | -8.7 ± 0.6 | 4880 ± 222 | 2.77 ± 0.56 | 0.15 ± 0.13 | 42 ± 4 | … | Y | Y | Y | Y | … | Y | … |
| 472 | 19172255+1932012 | -8.5 ± 0.6 | 5232 ± 133 | 3.56 ± 0.24 | 0.29 ± 0.10 | 36 ± 1 | … | Y | Y | Y | Y | … | Y | … |
| 3663 | 19172345+1932033 | -7.9 ± 0.6 | 6442 ± 322 | 4.07 ± 0.21 | 0.23 ± 0.21 | … | … | Y | … | … | … | Y | n | … |
| 3664 | 19172408+1935141 | -0.3 ± 0.5 | 5607 ± 376 | 3.80 ± 0.36 | 0.05 ± 0.37 | <99 | 3 | N | N | N | N | … | n | NG |
| 3665 | 19172438+1935468 | -2.2 ± 0.4 | 5883 ± 37 | 4.53 ± 0.24 | 0.10 ± 0.13 | … | … | N | … | … | … | N | n | … |
| 3666 | 19172439+1930201 | -8.9 ± 0.9 | 6350 ± 264 | 4.50 ± 0.29 | 0.22 ± 0.33 | … | … | Y | … | … | … | … | n | … |
| 3667 | 19172557+1931290 | -33.6 ± 0.3 | 5366 ± 112 | 4.61 ± 0.30 | 0.23 ± 0.15 | … | … | N | … | … | … | … | n | … |
| 3668 | 19172671+1933482 | -16.2 ± 0.3 | 5787 ± 97 | 4.48 ± 0.30 | -0.01 ± 0.18 | 71 ± 33 | 1 | N | N | N | N | … | n | NG |
| 3669 | 19172720+1930185 | 0.2 ± 1.2 | 6132 ± 195 | 4.99 ± 0.96 | 0.14 ± 0.29 | 40 ± 28 | 1 | N | N | N | N | … | n | NG |

**Notes.** [a] The values of EW(Li) for this cluster are corrected (subtracted adjacent Fe (6707.43 Å) line). [b] Flags for the errors of the corrected EW(Li) values, as follows: 1=EW(Li) corrected by blends contribution using models; and 3=Upper limit (no error for EW(Li) is given). [c] Cantat-Gaudin et al. (2018). [d] The letters "Y" and "N" indicate if the star is a cluster member or not. [e] 'Li-rich G', 'G' and 'NG' indicate "Li-rich giant", "giant" and "non-giant" Li field outliers, respectively.

**Table C.19.** M67

| ID | CNAME | RV (km s$^{-1}$) | $T_{\text{eff}}$ (K) | $logg$ (dex) | [Fe/H] (dex) | $EW$(Li)$^a$ (mÅ) | $EW$(Li) error flag$^b$ | Membership RV | Li | $logg$ | [Fe/H] | Final$^c$ | NMs with Li$^d$ |
|---|---|---|---|---|---|---|---|---|---|---|---|---|---|
| 74 | 08505182+1156559 | 35.4 ± 0.6 | 6115 ± 113 | 3.84 ± 0.23 | -0.06 ± 0.10 | <4 | 3 | Y | Y | Y | Y | Y | … |
| 75 | 08505600+1153519 | 35.9 ± 0.6 | 6116 ± 110 | 4.17 ± 0.22 | -0.02 ± 0.10 | 40 ± 1 | … | Y | Y | Y | Y | Y | … |
| 76 | 08505891+1148192 | 35.5 ± 0.6 | 6058 ± 113 | 4.18 ± 0.22 | -0.04 ± 0.09 | 43 ± 1 | … | Y | Y | Y | Y | Y?$^e$ | … |
| 77 | 08510017+1154321 | 34.6 ± 0.6 | 5472 ± 121 | 3.82 ± 0.23 | 0.12 ± 0.10 | <4 | 3 | Y | Y | Y | N | Y | … |
| 78 | 08510080+1148527 | 34.7 ± 0.6 | 5689 ± 114 | 4.33 ± 0.23 | -0.02 ± 0.10 | 10 ± 1 | … | Y | Y | Y | Y | Y | … |
| 79 | 08510325+1145473 | 35.5 ± 0.6 | 5924 ± 117 | 3.74 ± 0.24 | 0.01 ± 0.09 | <3 | 3 | Y | Y | Y | Y | Y | … |
| 80 | 08510524+1149340 | 35.8 ± 0.6 | 6035 ± 120 | 4.14 ± 0.23 | -0.05 ± 0.10 | 44 ± 1 | … | Y | Y | Y | Y | Y | … |
| 81 | 08510838+1147121 | 33.9 ± 0.6 | 4911 ± 112 | 3.38 ± 0.23 | 0.05 ± 0.09 | <3 | 3 | Y | Y | Y | Y | Y | … |
| 82 | 08510969+1159096 | 37.0 ± 0.6 | 6067 ± 117 | 4.16 ± 0.23 | -0.05 ± 0.10 | 77 ± 2 | … | N | … | … | … | n | NG |
| 83 | 08511267+1150345 | 34.5 ± 0.6 | 6057 ± 117 | 4.09 ± 0.23 | 0.01 ± 0.10 | 48 ± 1 | … | Y | Y | Y | Y | Y | … |
| 84 | 08511799+1145541 | 52.4 ± 0.6 | 6281 ± 134 | 3.98 ± 0.28 | -0.16 ± 0.10 | … | … | N | … | … | … | n | … |
| 85 | 08511854+1149214 | 35.8 ± 0.6 | 5939 ± 117 | 3.83 ± 0.23 | -0.03 ± 0.10 | <3 | 3 | Y | Y | Y | Y | Y | … |
| 88 | 08512012+1146417 | 34.5 ± 0.6 | 6114 ± 117 | 3.86 ± 0.23 | -0.07 ± 0.09 | 16 ± 1 | … | Y | Y | Y | Y | Y | … |
| 91 | 08513045+1148582 | 42.4 ± 0.6 | 4843 ± 116 | 3.22 ± 0.24 | -0.01 ± 0.10 | 4 ± 1 | … | N | Y | Y | Y | Y | … |
| 92 | 08513322+1148513 | 34.1 ± 0.6 | 6063 ± 117 | 3.86 ± 0.24 | -0.03 ± 0.10 | <3 | 3 | Y | Y | Y | Y | Y | … |
| 93 | 08513577+1153347 | 34.5 ± 0.6 | 4868 ± 120 | 3.26 ± 0.23 | 0.00 ± 0.09 | <3 | 3 | Y | Y | Y | Y | Y | … |
| 94 | 08513740+1150052 | 32.9 ± 0.6 | 5967 ± 122 | 3.79 ± 0.22 | -0.01 ± 0.10 | 25 ± 1 | … | Y | Y | Y | Y | Y | … |
| 95 | 08514081+1149055 | 34.5 ± 0.6 | 6076 ± 114 | 4.23 ± 0.23 | -0.01 ± 0.10 | 53 ± 2 | … | Y | Y | Y | Y | Y | … |
| 96 | 08514122+1154290 | 34.0 ± 0.6 | 6111 ± 116 | 3.87 ± 0.23 | -0.03 ± 0.10 | 11 ± 1 | … | Y | Y | Y | Y | Y | … |
| 97 | 08514507+1147459 | 33.7 ± 0.6 | 4788 ± 119 | 2.98 ± 0.23 | 0.03 ± 0.10 | <2 | 3 | Y | Y | Y | Y | Y | … |
| 98 | 08514995+1149311 | 33.7 ± 0.6 | 5945 ± 113 | 3.71 ± 0.23 | -0.06 ± 0.10 | <3 | 3 | Y | Y | Y | Y | Y | … |

**Notes.** $^{(a)}$ The values of $EW$(Li) for this cluster are corrected (subtracted adjacent Fe (6707.43 Å) line). $^{(b)}$ Flags for the errors of the corrected $EW$(Li) values, as follows: 1=$EW$(Li) corrected by blends contribution using models; and 3=Upper limit (no error for $EW$(Li) is given). $^{(c)}$ The letters "Y" and "N" indicate if the star is a cluster member or not. $^{(d)}$ 'Li-rich G', 'G' and 'NG' indicate "Li-rich giant", "giant" and "non-giant" Li field outliers, respectively. $^{(e)}$ For more details about the membership of the stars listed as possible candidates, see the individual notes of Appendix A for M67.

**Table C.20.** NGC 2243

| ID | CNAME | RV (km s$^{-1}$) | $T_{\text{eff}}$ (K) | logg (dex) | [Fe/H] (dex) | EW(Li)$^a$ (mÅ) | EW(Li) error flag$^b$ | Membership RV | Li | logg | [Fe/H] | Gaia study Cantat-Gaudin$^c$ | Final$^d$ | NMs with Li$^e$ |
|---|---|---|---|---|---|---|---|---|---|---|---|---|---|---|
| 222 | 06292300-3117299 | 59.85 ± 0.57 | 5010 ± 114 | 2.55 ± 0.23 | -0.43 ± 0.10 | <5 | 3 | Y | Y | Y | Y | Y | Y | … |
| 223 | 06292939-3115459 | 59.65 ± 0.57 | 4988 ± 118 | 2.53 ± 0.22 | -0.38 ± 0.10 | <5 | 3 | Y | Y | Y | Y | … | Y | … |
| 224 | 06294108-3119059 | 80.81 ± 0.57 | … | … | … | … | … | N | … | … | … | Y | n | … |
| 225 | 06294149-3114360 | 60.15 ± 0.57 | 4741 ± 115 | 2.43 ± 0.23 | -0.38 ± 0.10 | <15 | 3 | Y | Y | Y | Y | … | Y | … |
| 226 | 06294582-3115381 | 60.79 ± 0.57 | 4953 ± 117 | 2.48 ± 0.22 | -0.38 ± 0.10 | 5 ± 1 | 1 | Y | Y | Y | Y | … | Y | … |
| 227 | 06294621-3116016 | 60.35 ± 0.57 | 5210 ± 115 | 2.89 ± 0.23 | -0.28 ± 0.11 | <9 | 3 | Y | Y | Y | N | … | Y | … |
| 228 | 06295252-3121339 | 2.77 ± 0.57 | 6155 ± 111 | 4.22 ± 0.24 | -0.19 ± 0.10 | 42 ± 2 | | N | … | … | … | … | n | NG |
| 2674 | 06290477-3121263 | 60.02 ± 0.57 | 5011 ± 120 | 2.53 ± 0.23 | -0.50 ± 0.10 | <7 | 3 | Y | Y | Y | Y | Y | Y | … |
| 2675 | 06290541-3117025 | 60.42 ± 0.57 | 4931 ± 111 | 2.51 ± 0.22 | -0.36 ± 0.09 | <9 | 3 | Y | Y | Y | Y | … | Y | … |
| 2676 | 06290934-3110325 | 61.30 ± 0.57 | 4870 ± 114 | 2.69 ± 0.23 | -0.38 ± 0.10 | 16 ± 1 | | Y | Y | Y | Y | … | Y | … |
| 2677 | 06291101-3120394 | 60.75 ± 0.57 | 4842 ± 114 | 2.40 ± 0.23 | -0.38 ± 0.09 | <4 | 3 | Y | Y | Y | Y | Y | Y | … |
| 2678 | 06292122-3118312 | 45.05 ± 0.40 | 6314 ± 235 | 3.90 ± 0.54 | -0.40 ± 0.16 | … | … | N | … | … | … | Y | n | … |
| 2679 | 06292225-3117092 | 52.06 ± 0.57 | 5318 ± 117 | 3.84 ± 0.23 | 0.06 ± 0.10 | 28 ± 1 | | N | … | … | … | … | n | NG |
| 2680 | 06292311-3126112 | 74.29 ± 0.57 | 4915 ± 119 | 3.32 ± 0.22 | -0.28 ± 0.09 | <11 | 3 | N | … | … | … | … | n | G |
| 2681 | 06292562-3115261 | 58.56 ± 0.10 | 6018 ± 144 | 3.57 ± 0.26 | -0.40 ± 0.12 | <18 | 3 | Y | Y | Y | Y | Y | Y | … |
| 2682 | 06292841-3117174 | 59.19 ± 0.57 | 4200 ± 112 | 1.38 ± 0.23 | -0.48 ± 0.09 | 17 ± 1 | 1 | Y | Y | Y | Y | … | Y | … |
| 2683 | 06293003-3122493 | 73.51 ± 0.57 | 4926 ± 145 | 4.59 ± 0.24 | -0.23 ± 0.10 | <28 | 3 | N | … | … | … | … | n | NG |
| 2684 | 06293009-3116587 | 60.28 ± 0.57 | 4682 ± 115 | 2.13 ± 0.23 | -0.41 ± 0.10 | <4 | 3 | Y | Y | Y | Y | … | Y | … |
| 2685 | 06293240-3117294 | 59.06 ± 0.57 | 4970 ± 119 | 2.51 ± 0.22 | -0.37 ± 0.09 | <5 | 3 | Y | Y | Y | Y | Y | Y | … |
| 2686 | 06293385-3107310 | 93.55 ± 0.57 | 4845 ± 115 | 2.52 ± 0.23 | -0.40 ± 0.09 | <7 | 3 | N | … | … | … | … | n | G |
| 2687 | 06293518-3117239 | 61.21 ± 0.57 | 4934 ± 115 | 2.79 ± 0.23 | -0.36 ± 0.10 | <6 | 3 | Y | Y | Y | Y | … | Y | … |
| 2688 | 06293525-3115470 | 60.17 ± 0.57 | 5029 ± 109 | 3.11 ± 0.23 | -0.34 ± 0.09 | <10 | 3 | Y | Y | Y | N | … | Y | … |
| 2689 | 06293646-3117402 | -16.05 ± 0.57 | 5181 ± 121 | 4.47 ± 0.23 | -0.09 ± 0.10 | 7 ± 1 | | N | … | … | … | … | n | … |
| 2690 | 06293684-3114423 | 61.43 ± 0.57 | 4917 ± 110 | 2.80 ± 0.23 | -0.38 ± 0.09 | 14 ± 1 | | N | … | … | … | … | n | … |
| 2691 | 06294022-3114504 | 60.09 ± 0.57 | 5586 ± 117 | 3.62 ± 0.23 | -0.28 ± 0.10 | <17 | 3 | Y | Y | Y | N | Y | Y | … |
| 2692 | 06294998-3116419 | 71.77 ± 0.57 | 5203 ± 121 | 3.46 ± 0.24 | -0.36 ± 0.10 | 12 ± 2 | … | N | … | … | … | Y | n | … |
| 2693 | 06295017-3118531 | 61.81 ± 0.57 | 5131 ± 116 | 4.49 ± 0.23 | -0.16 ± 0.10 | <4 | 3 | N | … | … | … | … | n | … |
| 2694 | 06295099-3114428 | 59.45 ± 0.57 | 4950 ± 120 | 2.91 ± 0.25 | -0.36 ± 0.09 | 32 ± 2 | … | Y | Y | Y | Y | … | Y | … |
| 39639 | 06285287-3115375 | 39.41 ± 0.27 | 6348 ± 181 | 4.39 ± 0.41 | -0.21 ± 0.13 | … | … | N | … | … | … | … | n | … |
| 39640 | 06285343-3117326 | 61.79 ± 0.80 | 5463 ± 217 | 4.05 ± 0.65 | -0.84 ± 0.23 | … | … | N | … | … | … | … | n | … |
| 39641 | 06285390-3115213 | 60.28 ± 0.86 | 5746 ± 345 | 4.79 ± 0.35 | -0.68 ± 0.29 | <107 | 3 | Y | N | Y | Y | … | n | NG |
| 39642 | 06285405-3117250 | 59.83 ± 0.49 | 6353 ± 369 | 4.68 ± 0.38 | -0.63 ± 0.40 | … | … | Y | … | … | … | Y | n | … |
| 39643 | 06285471-3119505 | 57.23 ± 0.10 | 5922 ± … | 4.25 ± … | 0.06 ± … | … | … | N | … | … | … | … | n | … |
| 39644 | 06285487-3119368 | 57.57 ± 1.04 | 6594 ± 542 | 3.30 ± 0.51 | -0.49 ± 0.08 | … | … | N | … | … | … | Y | n | … |
| 39645 | 06285570-3114594 | 60.61 ± 1.05 | 5591 ± 336 | 3.60 ± 0.33 | -1.14 ± 1.39 | … | … | Y | … | … | … | … | n | … |
| 39646 | 06285594-3114347 | 60.40 ± 0.95 | 5826 ± 590 | 3.96 ± 0.12 | -0.41 ± 0.63 | … | … | Y | … | … | … | … | n | … |
| 39647 | 06285712-3123559 | 135.22 ± 1.22 | 5601 ± 343 | 3.58 ± 0.70 | -0.44 ± 0.15 | … | … | N | … | … | … | … | n | … |
| 39648 | 06285874-3116564 | -5.65 ± 0.10 | 5796 ± … | 4.33 ± … | -0.17 ± … | … | … | N | … | … | … | … | n | … |
| 39649 | 06285934-3111507 | 58.04 ± 1.42 | 6125 ± 355 | 3.86 ± 0.46 | -0.56 ± 0.70 | … | … | Y | … | … | … | N | n | … |
| 39650 | 06285971-3125142 | -7.30 ± 0.10 | 6088 ± … | 4.18 ± 0.17 | -0.36 ± 0.13 | <62 | 3 | N | … | … | … | … | n | NG |
| 39651 | 06285975-3122340 | 59.14 ± 0.10 | 5241 ± … | 3.70 ± … | -0.38 ± … | … | … | Y | … | … | … | Y | n | … |
| 39652 | 06290031-3121040 | 59.57 ± 0.83 | 5824 ± 217 | 4.24 ± 0.56 | -0.60 ± 0.25 | … | … | Y | … | … | … | … | n | … |
| 39653 | 06290103-3118539 | 61.46 ± 1.19 | 5276 ± 107 | 4.38 ± 0.56 | -0.56 ± 0.42 | … | … | N | … | … | … | … | n | … |
| 39654 | 06290132-3110238 | 13.48 ± 0.10 | 5906 ± … | 4.34 ± … | 0.12 ± … | … | … | N | … | … | … | … | n | … |
| 39655 | 06290158-3124130 | 83.11 ± 0.96 | 6092 ± 209 | 4.37 ± 0.12 | -0.82 ± 0.42 | … | … | N | … | … | … | … | n | … |
| 39656 | 06290193-3114123 | 60.42 ± 0.16 | 5715 ± … | 4.13 ± … | -1.20 ± … | … | … | Y | … | … | … | Y | n | … |
| 39657 | 06290222-3111106 | 18.67 ± 0.11 | 6181 ± … | 4.23 ± … | -0.62 ± … | … | … | N | … | … | … | … | n | … |
| 39658 | 06290235-3116531 | 59.91 ± 0.49 | 5867 ± 123 | 4.08 ± 0.34 | -0.86 ± 0.25 | … | … | Y | … | … | … | Y | n | … |
| 39659 | 06290236-3120494 | 58.99 ± 0.70 | 5371 ± 369 | 4.25 ± 0.16 | -0.88 ± 0.60 | … | … | Y | … | … | … | Y | n | … |
| 39660 | 06290375-3117475 | 49.12 ± 0.92 | 5917 ± 245 | 3.92 ± 0.59 | -0.45 ± 0.13 | … | … | N | … | … | … | … | n | … |
| 39662 | 06290433-3118524 | 89.65 ± 0.90 | 6247 ± 229 | 4.57 ± 0.54 | -0.48 ± 0.17 | … | … | N | … | … | … | … | n | … |
| 39663 | 06290444-3113178 | 60.16 ± 0.76 | 5881 ± 115 | 4.14 ± 0.10 | -0.59 ± 0.21 | … | … | Y | … | … | … | Y | n | … |
| 39664 | 06290459-3123184 | 59.46 ± 0.45 | 6098 ± 251 | 4.96 ± 0.83 | -0.83 ± 0.34 | 64 ± 46 | 1 | Y | Y | N | Y | Y | Y | … |
| 39665 | 06290464-3109237 | 70.40 ± 2.32 | 5301 ± 162 | … | -1.74 ± 0.29 | … | … | N | … | … | … | … | n | … |
| 39666 | 06290483-3122404 | 20.92 ± 0.41 | 6061 ± 222 | 3.63 ± 0.36 | -0.42 ± 0.16 | 72 ± 45 | 1 | N | … | … | … | … | n | NG |
| 39667 | 06290493-3123537 | 58.94 ± 0.34 | 5144 ± 162 | 3.39 ± 0.78 | -0.37 ± 0.15 | … | … | Y | … | … | … | Y | n | Li-rich G |
| 39668 | 06290502-3121410 | 64.32 ± 1.59 | 6265 ± 995 | 4.48 ± 0.35 | -0.62 ± 0.41 | … | … | N | … | … | … | … | n | … |
| 39669 | 06290518-3120349 | 61.49 ± 0.59 | 6077 ± 75 | 3.86 ± 0.60 | -0.74 ± 0.43 | … | … | N | … | … | … | Y | n | … |
| 39670 | 06290540-3116540 | 113.52 ± 1.28 | 6170 ± 728 | 4.32 ± 0.59 | -0.53 ± 0.56 | … | … | N | … | … | … | … | n | … |

**Table C.20.** continued.

| ID | CNAME | RV (km s$^{-1}$) | $T_{\text{eff}}$ (K) | logg (dex) | [Fe/H] (dex) | EW(Li)$^a$ (mÅ) | EW(Li) error flag$^b$ | RV | Membership Li | logg | [Fe/H] | Gaia study Cantat-Gaudin$^c$ | Final$^d$ | NMs with Li$^e$ |
|---|---|---|---|---|---|---|---|---|---|---|---|---|---|---|
| 39671 | 06290603-3111023 | 83.10 ± 1.95 | 6231 ± 291 | 3.42 ± 0.94 | -0.65 ± 0.54 | … | … | N | … | … | … | … | n | … |
| 39672 | 06290624-3122275 | 58.51 ± 1.45 | 5849 ± 394 | 3.94 ± 1.26 | -0.56 ± 0.15 | … | … | Y | … | … | … | … | n | … |
| 39673 | 06290642-3114454 | 59.56 ± 1.33 | 6269 ± 949 | 4.00 ± 0.55 | -0.83 ± 0.21 | … | … | Y | … | … | … | … | n | … |
| 39674 | 06290678-3126061 | 58.20 ± 1.15 | 5847 ± 756 | 4.00 ± 0.59 | -0.56 ± 0.92 | … | … | Y | … | … | … | … | n | … |
| 39675 | 06290682-3119103 | 103.42 ± 0.73 | 5884 ± 42 | 3.95 ± 0.21 | -0.31 ± 0.16 | <80 | 3 | N | … | … | … | … | n | NG |
| 39676 | 06290691-3116105 | 58.83 ± 1.51 | 5858 ± 279 | 4.56 ± 0.10 | -0.27 ± 0.43 | … | … | Y | … | … | … | … | n | … |
| 39677 | 06290710-3120331 | 60.14 ± 0.16 | 6259 ± … | 4.55 ± … | -1.21 ± … | <56 | 3 | Y | Y | Y | N | Y | Y | … |
| 39678 | 06290715-3109243 | 58.94 ± 0.88 | 6099 ± 287 | 4.34 ± 0.26 | -0.53 ± 0.28 | … | … | Y | … | … | … | Y | n | … |
| 39679 | 06290720-3117351 | 59.22 ± 0.71 | 5717 ± 16 | 3.75 ± 0.69 | -0.61 ± 0.36 | <66 | 3 | Y | N | Y | Y | … | n | NG |
| 39680 | 06290737-3112013 | 59.63 ± 0.58 | 5606 ± 355 | 3.76 ± 0.54 | -0.66 ± 0.53 | 71 ± 67 | 1 | Y | N | Y | Y | Y | n | NG |
| 39681 | 06290742-3112052 | 60.64 ± 1.34 | 5196 ± 292 | 4.35 ± 0.49 | -0.55 ± 0.15 | … | … | Y | … | … | … | … | n | … |
| 39682 | 06290762-3114456 | 57.65 ± 1.32 | 5651 ± 938 | 3.91 ± 0.78 | -0.88 ± 0.80 | … | … | N | … | … | … | … | n | … |
| 39683 | 06290770-3113277 | 58.84 ± 0.15 | 6333 ± … | 4.12 ± … | -0.62 ± … | … | … | Y | … | … | … | Y | n | … |
| 39684 | 06290802-3110422 | 59.98 ± 1.03 | 6245 ± 416 | 4.40 ± 0.62 | -0.44 ± 0.70 | <88 | 3 | Y | N | Y | Y | Y | n | NG |
| 39685 | 06290849-3117120 | 59.38 ± 1.59 | 5465 ± 109 | 4.64 ± 0.36 | -0.56 ± 0.14 | … | … | Y | … | … | … | … | n | … |
| 39686 | 06290869-3118293 | 92.90 ± 0.77 | 6321 ± 432 | 4.09 ± 0.38 | -0.62 ± 0.30 | … | … | N | … | … | … | … | n | … |
| 39687 | 06290917-3113363 | 47.21 ± 0.10 | 5367 ± … | 4.50 ± … | -0.08 ± … | … | … | N | … | … | … | … | n | … |
| 39688 | 06290926-3126203 | 59.66 ± 1.04 | 5319 ± 166 | 4.29 ± 0.45 | -0.83 ± 0.97 | … | … | Y | … | … | … | … | n | … |
| 39689 | 06290941-3117365 | 70.11 ± 0.84 | 5850 ± 179 | 4.13 ± 0.63 | -0.39 ± 0.18 | … | … | N | … | … | … | … | n | … |
| 39690 | 06290969-3117478 | 59.64 ± 1.30 | 5838 ± 516 | 4.91 ± 0.27 | -0.60 ± 0.28 | … | … | Y | … | … | … | … | n | … |
| 39691 | 06290976-3108505 | 60.93 ± 0.16 | 6480 ± … | 3.85 ± … | -0.59 ± … | … | … | Y | … | … | … | Y | n | … |
| 39692 | 06291032-3118359 | 57.91 ± 1.07 | 5981 ± 380 | 3.72 ± 0.53 | -0.84 ± 0.65 | <142 | 3 | N | … | … | … | … | n | NG |
| 39693 | 06291037-3112493 | 60.45 ± 0.98 | 5663 ± 326 | 4.19 ± 0.42 | -0.98 ± 0.38 | … | … | Y | … | … | … | … | n | … |
| 39694 | 06291058-3118342 | 59.22 ± 1.12 | 5737 ± 901 | 4.97 ± 0.49 | -0.73 ± 1.37 | … | … | Y | … | … | … | … | n | … |
| 39695 | 06291083-3115565 | 63.75 ± 0.60 | 5822 ± 183 | 3.97 ± 0.25 | -0.32 ± 0.29 | … | … | N | … | … | … | … | n | … |
| 39696 | 06291113-3121461 | 61.80 ± 0.77 | 5806 ± 165 | 4.22 ± 0.32 | -0.62 ± 0.41 | … | … | N | … | … | … | Y | n | … |
| 39697 | 06291125-3119328 | 60.53 ± 1.07 | 5795 ± 353 | 4.81 ± 0.20 | -1.05 ± 0.28 | <82 | 3 | Y | N | N | N | Y | n | NG |
| 39698 | 06291166-3120246 | 74.02 ± 1.87 | 6053 ± 212 | 4.25 ± 0.10 | -0.77 ± 0.56 | … | … | N | … | … | … | … | n | … |
| 39699 | 06291191-3119484 | 86.68 ± 1.19 | 5529 ± 177 | 4.10 ± 0.24 | -0.23 ± 0.28 | … | … | N | … | … | … | … | n | … |
| 39700 | 06291192-3116568 | 58.35 ± 0.99 | 5920 ± 265 | 3.36 ± 0.20 | -0.52 ± 0.28 | … | … | Y | … | … | … | … | n | … |
| 39701 | 06291215-3117198 | 60.22 ± 0.93 | 5424 ± 107 | 4.32 ± 0.58 | -0.70 ± 0.30 | <91 | 3 | Y | N | N | Y | … | n | NG |
| 39702 | 06291222-3114487 | 59.52 ± 0.81 | 6435 ± 719 | 3.77 ± 0.39 | -0.80 ± 0.06 | … | … | Y | … | … | … | … | n | … |
| 39703 | 06291224-3115358 | 59.71 ± 0.83 | 5747 ± 159 | 4.03 ± 0.75 | -0.59 ± 0.26 | 92 ± 77 | 1 | Y | N | N | Y | … | n | NG |
| 39704 | 06291227-3116081 | 88.20 ± 0.75 | 5924 ± 694 | 4.32 ± 0.61 | -0.80 ± 0.87 | … | … | N | … | … | … | … | n | … |
| 39705 | 06291261-3117208 | 59.90 ± 0.13 | 6516 ± … | 4.64 ± … | -0.90 ± … | … | … | Y | … | … | … | Y | n | … |
| 39706 | 06291281-3117566 | 58.36 ± 0.77 | 6182 ± 453 | 4.34 ± 0.85 | -0.61 ± 0.34 | … | … | Y | … | … | … | … | n | … |
| 39707 | 06291286-3120428 | 60.11 ± 0.15 | 6487 ± … | 4.17 ± … | -0.46 ± … | <54 | 3 | Y | Y | Y | Y | Y | Y | … |
| 39708 | 06291289-3118590 | 60.04 ± 1.06 | 5768 ± 368 | 4.35 ± 1.06 | -0.53 ± 0.27 | … | … | Y | … | … | … | … | n | … |
| 39709 | 06291305-3109173 | 59.54 ± 0.11 | 5999 ± … | 3.76 ± … | -0.59 ± … | … | … | Y | … | … | … | Y | n | … |
| 39710 | 06291306-3108543 | 68.20 ± 0.70 | 6182 ± 21 | 4.03 ± 1.11 | -0.58 ± 0.30 | … | … | N | … | … | … | … | n | … |
| 39711 | 06291316-3117168 | 98.65 ± 0.90 | 4836 ± 284 | 4.47 ± 0.46 | -0.59 ± 0.26 | … | … | N | … | … | … | … | n | … |
| 39712 | 06291319-3120349 | 59.31 ± 0.98 | 5308 ± 106 | 4.00 ± 0.30 | -1.08 ± 0.64 | … | … | Y | … | … | … | … | n | … |
| 39713 | 06291358-3116426 | 57.32 ± 0.80 | 6228 ± 170 | 3.49 ± 0.27 | -0.56 ± 0.37 | … | … | N | … | … | … | N | n | … |
| 39714 | 06291371-3121010 | 59.25 ± 0.10 | 6054 ± … | 3.92 ± … | -0.69 ± … | 107 ± 34 | … | Y | N | Y | Y | Y | n | NG |
| 39715 | 06291382-3122382 | 60.64 ± 1.00 | 6317 ± 952 | 4.24 ± 0.31 | -0.75 ± 0.51 | … | … | Y | … | … | … | … | n | … |
| 39716 | 06291385-3115157 | 62.39 ± 0.24 | 5125 ± 66 | 4.44 ± 0.29 | 0.04 ± 0.14 | … | … | N | … | … | … | … | n | … |
| 39717 | 06291418-3121224 | 32.77 ± 0.43 | 6283 ± 338 | 4.96 ± 0.80 | -0.45 ± 0.22 | … | … | N | … | … | … | … | n | … |
| 39718 | 06291421-3121302 | 59.19 ± 1.13 | 5986 ± 686 | 3.98 ± 0.26 | -0.82 ± 1.40 | … | … | Y | … | … | … | … | n | … |
| 39719 | 06291454-3125598 | 60.61 ± 1.12 | 5708 ± 119 | 3.67 ± 0.46 | -0.44 ± 0.19 | … | … | Y | … | … | … | … | n | … |
| 39720 | 06291459-3110249 | 64.41 ± 1.94 | 5928 ± 298 | 4.38 ± 1.28 | -0.68 ± 0.26 | … | … | N | … | … | … | … | n | … |
| 39721 | 06291475-3118080 | 59.77 ± 1.07 | 5903 ± 402 | 4.60 ± 0.74 | -1.07 ± 0.25 | … | … | Y | … | … | … | Y | n | … |
| 39722 | 06291485-3112161 | 59.99 ± 0.63 | 6057 ± 86 | 4.01 ± 0.48 | -0.62 ± 0.20 | … | … | Y | … | … | … | Y | n | … |
| 39723 | 06291487-3123004 | 55.20 ± 0.77 | 5697 ± 215 | 4.31 ± 0.31 | -0.74 ± 0.20 | … | … | N | … | … | … | … | n | … |
| 39724 | 06291541-3109243 | 73.40 ± 1.05 | 6252 ± 447 | 4.44 ± 0.39 | -0.56 ± 0.59 | … | … | N | … | … | … | … | n | … |
| 39725 | 06291559-3119252 | 60.01 ± 1.17 | 5975 ± 455 | 3.66 ± 0.48 | -0.55 ± 0.48 | … | … | Y | … | … | … | … | n | … |
| 39726 | 06291560-3118064 | 70.72 ± 9.75 | … | … | … | … | … | N | … | … | … | … | n | … |
| 39727 | 06291569-3120403 | 59.58 ± 0.92 | 5964 ± 493 | 4.39 ± 0.27 | -0.69 ± 0.49 | … | … | Y | … | … | … | … | n | … |
| 39728 | 06291589-3117220 | 65.59 ± 1.86 | 5732 ± 1078 | 4.31 ± 0.42 | -0.77 ± 1.12 | … | … | N | … | … | … | … | n | … |
| 39729 | 06291623-3115259 | 59.91 ± 0.97 | 5928 ± 177 | 4.16 ± 0.22 | -0.55 ± 0.14 | … | … | Y | … | … | … | … | n | … |

**Table C.20.** continued.

| ID | CNAME | RV (km s$^{-1}$) | $T_{\text{eff}}$ (K) | logg (dex) | [Fe/H] (dex) | EW(Li)$^a$ (mÅ) | EW(Li) error flag$^b$ | Membership RV | Membership Li | Membership logg | Membership [Fe/H] | Gaia study Cantat-Gaudin$^c$ | Final$^d$ | NMs with Li$^e$ |
|---|---|---|---|---|---|---|---|---|---|---|---|---|---|---|
| 39730 | 06291628-3116465 | 58.71 ± 0.10 | 5116 ± … | 3.53 ± … | -0.38 ± … | … | … | Y | … | … | … | Y | n | … |
| 39731 | 06291633-3114184 | 68.78 ± 3.03 | 5742 ± 149 | 2.95 ± 0.96 | -1.00 ± 0.16 | … | … | N | … | … | … | Y | n | … |
| 39732 | 06291634-3109469 | 59.05 ± 0.61 | 6087 ± 30 | 3.83 ± 0.30 | -0.67 ± 0.29 | <64 | 3 | Y | Y | Y | Y | Y | Y | … |
| 39733 | 06291635-3124509 | 59.84 ± 0.87 | 6112 ± 187 | 3.99 ± 0.49 | -0.40 ± 0.22 | … | … | Y | … | … | … | … | n | … |
| 39734 | 06291637-3112439 | 60.03 ± 0.99 | 5903 ± 369 | 3.69 ± 0.80 | -0.66 ± 0.22 | … | … | Y | … | … | … | … | n | … |
| 39735 | 06291637-3114105 | 59.91 ± 1.22 | 5148 ± 282 | 3.89 ± 1.08 | -0.57 ± 0.16 | … | … | Y | … | … | … | … | n | … |
| 39736 | 06291646-3118502 | 61.00 ± 1.56 | 5018 ± 99 | 4.16 ± 0.44 | -0.71 ± 0.49 | … | … | Y | … | … | … | … | n | … |
| 39737 | 06291657-3116370 | 60.77 ± 0.83 | 5685 ± 280 | 3.97 ± 0.26 | -0.66 ± 0.18 | <95 | 3 | Y | N | N | Y | Y | n | NG |
| 39738 | 06291670-3123333 | 27.69 ± 0.29 | 5941 ± 175 | 4.11 ± 0.14 | 0.19 ± 0.14 | … | … | N | … | … | … | … | n | … |
| 39739 | 06291677-3118367 | 57.93 ± 0.63 | 6253 ± 370 | 3.79 ± 0.46 | -0.69 ± 0.26 | <80 | 3 | N | … | … | … | Y | n | NG |
| 39740 | 06291690-3113190 | 59.97 ± 1.24 | 5748 ± 575 | 4.18 ± 0.19 | -0.71 ± 0.75 | … | … | Y | … | … | … | … | n | … |
| 39741 | 06291697-3116549 | 59.86 ± 0.13 | 6316 ± … | 3.92 ± … | -0.68 ± … | … | … | Y | … | … | … | Y | n | … |
| 39742 | 06291698-3119231 | 59.66 ± 0.70 | 5813 ± 170 | 4.97 ± 0.19 | -1.03 ± 0.76 | 102 ± 63 | 1 | Y | N | N | N | Y | n | NG |
| 39743 | 06291708-3117123 | 159.60 ± 1.27 | 5458 ± 319 | 3.51 ± 1.23 | -1.52 ± 0.29 | … | … | N | … | … | … | … | n | … |
| 39744 | 06291723-3123199 | 59.99 ± 0.13 | 6605 ± … | 4.28 ± … | -0.47 ± … | … | … | Y | … | … | … | Y | n | … |
| 39745 | 06291741-3116223 | 64.12 ± 0.54 | 5996 ± 163 | 4.54 ± 0.54 | -0.59 ± 0.19 | … | … | N | … | … | … | Y | n | … |
| 39746 | 06291748-3115507 | 59.74 ± 0.26 | 6136 ± 210 | 3.73 ± 0.28 | -0.56 ± 0.18 | … | … | Y | … | … | … | Y | n | … |
| 39747 | 06291778-3120097 | 58.33 ± 0.15 | 6382 ± … | 4.41 ± … | -0.72 ± … | … | … | Y | … | … | … | Y | n | … |
| 39748 | 06291795-3118491 | 58.78 ± 0.69 | 5916 ± 93 | 3.74 ± 0.37 | -0.72 ± 0.37 | 84 ± 63 | 1 | Y | N | Y | Y | Y | n | NG |
| 39749 | 06291805-3115374 | 178.25 ± 0.12 | 5236 ± … | 4.46 ± … | -0.59 ± … | … | … | N | … | … | … | … | n | … |
| 39750 | 06291835-3120339 | 58.02 ± 0.43 | 6305 ± 90 | 3.95 ± 0.45 | -0.48 ± 0.30 | … | … | Y | … | … | … | … | n | … |
| 39751 | 06291837-3123407 | 61.46 ± 0.55 | 6416 ± 364 | 4.32 ± 0.30 | -0.44 ± 0.13 | … | … | N | … | … | … | Y | n | … |
| 39752 | 06291841-3119086 | 59.07 ± 0.13 | 6433 ± … | 3.83 ± … | -0.46 ± … | … | … | Y | … | … | … | Y | n | … |
| 39753 | 06291853-3111192 | 78.29 ± 0.28 | 6274 ± 175 | 4.01 ± 0.08 | -0.32 ± 0.17 | … | … | N | … | … | … | … | n | … |
| 39754 | 06291855-3113305 | 71.22 ± 0.11 | 6130 ± … | 3.55 ± … | -0.76 ± … | … | … | N | … | … | … | Y | n | … |
| 39755 | 06291867-3114580 | 529.79 ± 0.21 | 6399 ± 375 | 4.56 ± 0.34 | -0.64 ± 0.20 | … | … | N | … | … | … | Y | n | … |
| 39756 | 06291887-3116125 | 59.45 ± 0.16 | 6496 ± … | 4.56 ± … | -0.67 ± … | … | … | Y | … | … | … | Y | n | … |
| 39757 | 06291903-3122135 | 59.45 ± 0.14 | 6300 ± … | 4.41 ± … | -0.60 ± … | 68 ± 60 | … | Y | Y | Y | Y | Y | Y | … |
| 39758 | 06291924-3114106 | 59.43 ± 0.15 | 6404 ± … | 4.46 ± … | -0.74 ± … | … | … | Y | … | … | … | Y | n | … |
| 39759 | 06291929-3125331 | 58.74 ± 0.15 | 6289 ± … | 3.61 ± … | -0.73 ± … | … | … | Y | … | … | … | Y | n | … |
| 39760 | 06291932-3112417 | 60.37 ± 0.16 | 6324 ± … | 4.59 ± … | -0.65 ± … | … | … | Y | … | … | … | Y | n | … |
| 39761 | 06291949-3121072 | 59.60 ± 0.30 | 6381 ± 353 | 3.78 ± 0.33 | -0.62 ± 0.22 | … | … | Y | … | … | … | Y | n | … |
| 39762 | 06291961-3124046 | 38.53 ± 0.11 | 6124 ± … | 4.05 ± … | -0.62 ± … | … | … | N | … | … | … | … | n | … |
| 39763 | 06291976-3115235 | 43.50 ± 0.76 | 5794 ± 302 | 4.87 ± 0.61 | -0.47 ± 0.18 | … | … | N | … | … | … | … | n | … |
| 39764 | 06291982-3119377 | 59.56 ± 0.67 | 6132 ± 350 | 4.23 ± 0.34 | -0.45 ± 0.26 | … | … | Y | … | … | … | Y | n | … |
| 39765 | 06292006-3116113 | 57.94 ± 0.14 | 6485 ± … | 4.35 ± … | -0.53 ± … | … | … | N | … | … | … | Y | n | … |
| 39766 | 06292007-3122084 | 13.55 ± 0.10 | 6040 ± … | 4.26 ± … | -0.26 ± … | … | … | N | … | … | … | … | n | … |
| 39767 | 06292011-3120208 | 61.08 ± 1.08 | 5797 ± 57 | 3.24 ± 0.17 | -0.60 ± 0.29 | … | … | N | … | … | … | … | n | … |
| 39768 | 06292012-3123041 | 52.32 ± 0.44 | 6142 ± 212 | 4.40 ± 0.31 | -0.34 ± 0.17 | 96 ± 48 | 1 | N | … | … | … | … | n | NG |
| 39769 | 06292021-3119511 | 58.14 ± 1.17 | 5525 ± 200 | 4.41 ± 0.22 | -0.62 ± 0.30 | … | … | Y | … | … | … | … | n | … |
| 39770 | 06292053-3115259 | 57.93 ± 0.96 | 5404 ± 298 | 3.94 ± 0.61 | -0.67 ± 0.32 | … | … | N | … | … | … | … | n | … |
| 39771 | 06292061-3115489 | 59.28 ± 0.34 | 6361 ± 572 | 4.66 ± 0.34 | -0.54 ± 0.31 | … | … | Y | … | … | … | Y | n | … |
| 39772 | 06292068-3125104 | 154.10 ± 0.25 | 5376 ± 169 | 4.35 ± 0.24 | -0.62 ± 0.29 | … | … | N | … | … | … | … | n | … |
| 39773 | 06292091-3122407 | 59.87 ± 0.99 | 5614 ± 89 | 4.24 ± 0.14 | -0.54 ± 0.16 | … | … | Y | … | … | … | … | n | … |
| 39774 | 06292092-3117508 | 59.34 ± 0.12 | 6461 ± … | 3.77 ± … | -0.54 ± … | … | … | Y | … | … | … | Y | n | … |
| 39775 | 06292101-3112342 | 62.11 ± 1.09 | 5988 ± 500 | 4.72 ± 0.63 | -0.76 ± 0.66 | … | … | N | … | … | … | … | n | … |
| 39776 | 06292105-3115298 | 57.97 ± 0.34 | 6429 ± 203 | 4.29 ± 0.20 | -0.61 ± 0.19 | … | … | N | … | … | … | Y | n | … |
| 39777 | 06292114-3112055 | 58.09 ± 0.79 | 6245 ± 509 | 4.78 ± 0.47 | -0.56 ± 0.27 | <110 | 3 | Y | N | N | Y | Y | n | NG |
| 39778 | 06292118-3119301 | 60.95 ± 0.10 | 5123 ± … | 4.51 ± … | 0.38 ± … | … | … | Y | … | … | … | … | n | … |
| 39779 | 06292133-3118094 | 59.82 ± 0.11 | 6298 ± … | 3.84 ± 0.22 | -0.59 ± 0.23 | 67 ± 42 | … | Y | Y | Y | Y | Y | Y | … |
| 39780 | 06292174-3113119 | 58.68 ± 1.89 | 6136 ± 1559 | 5.18 ± 0.25 | -0.81 ± 0.13 | … | … | Y | … | … | … | … | n | … |
| 39781 | 06292178-3118037 | 60.26 ± 0.50 | 5953 ± 248 | 4.20 ± 0.36 | -0.65 ± 0.24 | … | … | Y | … | … | … | … | n | … |
| 39782 | 06292181-3115163 | 497.86 ± 0.42 | 6426 ± 444 | 4.31 ± 0.12 | -0.55 ± 0.17 | … | … | N | … | … | … | Y | n | … |
| 39783 | 06292191-3115322 | 61.35 ± 1.04 | 5973 ± 251 | 4.21 ± 0.15 | -0.45 ± 0.24 | … | … | N | … | … | … | … | n | … |
| 39784 | 06292192-3119193 | 60.18 ± 0.40 | 6288 ± 235 | 4.39 ± 0.35 | -0.55 ± 0.12 | … | … | Y | … | … | … | Y | n | … |
| 39785 | 06292199-3122007 | 57.66 ± 0.98 | 5789 ± 188 | 4.13 ± 0.22 | -0.75 ± 0.20 | … | … | N | … | … | … | Y | n | … |
| 39786 | 06292203-3109261 | 29.65 ± 0.25 | 5232 ± 32 | 4.31 ± 0.25 | -0.24 ± 0.15 | … | … | N | … | … | … | … | n | … |
| 39787 | 06292207-3121303 | 57.57 ± 0.52 | 6200 ± 227 | 3.88 ± 0.21 | -0.59 ± 0.19 | … | … | N | … | … | … | Y | n | … |
| 39788 | 06292213-3116520 | 58.72 ± 0.11 | 5874 ± … | 3.56 ± … | -0.61 ± … | … | … | Y | … | … | … | … | n | … |

| ID | CNAME | RV (km s$^{-1}$) | $T_{\text{eff}}$ (K) | logg (dex) | [Fe/H] (dex) | EW(Li)$^a$ (mÅ) | EW(Li) error flag$^b$ | RV | Membership Li | logg | [Fe/H] | Gaia study Cantat-Gaudin$^c$ | Final$^d$ | NMs with Li$^e$ |
|---|---|---|---|---|---|---|---|---|---|---|---|---|---|---|
| 39789 | 06292219-3114218 | 55.88 ± 0.63 | 6125 ± 228 | 4.10 ± 0.18 | -0.64 ± 0.23 | ... | ... | N | ... | ... | ... | ... | n | ... |
| 39790 | 06292226-3110270 | 59.80 ± 1.25 | 6113 ± 1044 | 4.13 ± 0.33 | -0.45 ± 0.09 | ... | ... | Y | ... | ... | ... | ... | n | ... |
| 39791 | 06292236-3124298 | 30.88 ± 0.26 | 5297 ± 46 | 3.87 ± 0.22 | 0.12 ± 0.17 | 39 ± 38 | 1 | N | ... | ... | ... | ... | n | NG |
| 39792 | 06292240-3118364 | 59.91 ± 0.73 | 5945 ± 191 | 4.44 ± 0.43 | -0.63 ± 0.27 | ... | ... | Y | ... | ... | ... | Y | n | ... |
| 39793 | 06292250-3123093 | 58.19 ± 0.90 | 5958 ± 423 | 3.92 ± 0.04 | -0.63 ± 0.38 | ... | ... | Y | ... | ... | ... | ... | n | ... |
| 39794 | 06292255-3109159 | 16.81 ± 0.69 | 5798 ± 492 | 4.35 ± 0.25 | -0.45 ± 0.15 | ... | ... | N | ... | ... | ... | ... | n | ... |
| 39795 | 06292259-3116231 | 61.02 ± 0.92 | 5443 ± 225 | 5.13 ± 0.49 | -0.44 ± 0.25 | ... | ... | N | ... | ... | ... | ... | n | ... |
| 39796 | 06292272-3113279 | 60.60 ± 0.39 | 6269 ± 398 | 4.29 ± 0.18 | -0.66 ± 0.16 | ... | ... | Y | ... | ... | ... | Y | n | ... |
| 39797 | 06292284-3120160 | 60.17 ± 1.05 | 5626 ± 103 | 4.94 ± 0.31 | -0.68 ± 0.51 | ... | ... | Y | ... | ... | ... | ... | n | ... |
| 39798 | 06292304-3112445 | 52.66 ± 1.65 | 5871 ± 698 | 4.04 ± 0.36 | -0.86 ± 1.06 | ... | ... | N | ... | ... | ... | ... | n | ... |
| 39799 | 06292306-3122164 | 59.96 ± 0.36 | 6292 ± 404 | 4.19 ± 0.06 | -0.63 ± 0.22 | ... | ... | Y | ... | ... | ... | Y | n | ... |
| 39800 | 06292327-3117062 | 59.62 ± 1.77 | 6126 ± 653 | 4.28 ± 0.05 | -0.98 ± 1.31 | ... | ... | Y | ... | ... | ... | Y | n | ... |
| 39801 | 06292336-3110504 | 59.83 ± 0.16 | 6372 ± ... | 3.71 ± ... | -0.66 ± ... | ... | ... | Y | ... | ... | ... | Y | n | ... |
| 39802 | 06292337-3113157 | 497.59 ± 0.31 | 6344 ± 234 | 4.36 ± 0.23 | -0.65 ± 0.17 | ... | ... | N | ... | ... | ... | Y | n | ... |
| 39803 | 06292339-3114230 | 59.13 ± 0.84 | 5344 ± 187 | 3.99 ± 0.88 | -0.60 ± 0.22 | ... | ... | Y | ... | ... | ... | ... | n | ... |
| 39804 | 06292346-3115114 | 499.38 ± 0.37 | 6185 ± 317 | 4.74 ± 0.82 | -0.96 ± 0.20 | ... | ... | N | ... | ... | ... | ... | n | ... |
| 39805 | 06292351-3118320 | 59.97 ± 0.39 | 6197 ± 290 | 4.29 ± 0.21 | -0.75 ± 0.29 | ... | ... | Y | ... | ... | ... | Y | n | ... |
| 39806 | 06292363-3117081 | 58.49 ± 0.98 | 4907 ± 288 | 4.41 ± 0.82 | -0.80 ± 0.22 | ... | ... | Y | ... | ... | ... | ... | n | ... |
| 39807 | 06292377-3124164 | 59.13 ± 0.18 | 6320 ± ... | 4.38 ± ... | -0.72 ± ... | ... | ... | Y | ... | ... | ... | Y | n | ... |
| 39808 | 06292392-3111240 | 62.55 ± 1.07 | 5426 ± 28 | 3.48 ± 0.27 | -0.64 ± 0.21 | ... | ... | N | ... | ... | ... | ... | n | ... |
| 39809 | 06292398-3121282 | 61.11 ± 0.44 | 6074 ± 303 | 4.32 ± 0.25 | -0.54 ± 0.20 | ... | ... | N | ... | ... | ... | Y | n | ... |
| 39810 | 06292401-3118305 | 26.93 ± 0.48 | 6395 ± 387 | 4.33 ± 0.28 | -0.64 ± 0.31 | ... | ... | N | ... | ... | ... | Y | n | ... |
| 39811 | 06292402-3118050 | 59.05 ± 1.10 | 6345 ± 1054 | 4.08 ± 0.59 | -0.83 ± 0.63 | ... | ... | Y | ... | ... | ... | ... | n | ... |
| 39812 | 06292408-3120385 | 59.62 ± 0.67 | 5770 ± 134 | 4.22 ± 0.61 | -0.79 ± 0.24 | ... | ... | Y | ... | ... | ... | ... | n | ... |
| 39813 | 06292409-3114428 | 60.91 ± 1.25 | 5960 ± 418 | 4.64 ± 0.11 | -0.50 ± 0.22 | ... | ... | Y | ... | ... | ... | ... | n | ... |
| 39814 | 06292412-3122509 | 33.35 ± 0.10 | 5795 ± ... | 4.36 ± ... | -0.51 ± ... | ... | ... | N | ... | ... | ... | ... | n | ... |
| 39815 | 06292419-3122421 | 59.03 ± 1.15 | 5031 ± 219 | 3.64 ± 1.49 | -0.73 ± 0.20 | ... | ... | Y | ... | ... | ... | ... | n | ... |
| 39816 | 06292424-3121063 | 61.77 ± 1.42 | 5172 ± 303 | 3.41 ± 0.83 | -0.31 ± 0.29 | ... | ... | N | ... | ... | ... | ... | n | Li-rich G |
| 39817 | 06292430-3123351 | 59.08 ± 1.08 | 5225 ± 114 | 4.26 ± 0.73 | -1.01 ± 0.25 | ... | ... | Y | ... | ... | ... | ... | n | ... |
| 39818 | 06292435-3116596 | 58.86 ± 0.49 | 6173 ± 372 | 4.64 ± 0.43 | -0.54 ± 0.18 | ... | ... | Y | ... | ... | ... | ... | n | ... |
| 39819 | 06292436-3119330 | 61.99 ± 0.56 | 5675 ± 123 | 3.95 ± 0.16 | -0.48 ± 0.27 | <52 | 3 | N | ... | ... | ... | Y | n | NG |
| 39820 | 06292458-3109210 | 25.44 ± 0.25 | 5699 ± 55 | 4.27 ± 0.29 | -0.38 ± 0.15 | ... | ... | N | ... | ... | ... | ... | n | ... |
| 39821 | 06292461-3117055 | 62.06 ± 0.50 | 6531 ± 339 | 4.09 ± 0.23 | -0.49 ± 0.14 | ... | ... | N | ... | ... | ... | Y | n | ... |
| 39822 | 06292465-3116292 | 58.32 ± 0.69 | 5506 ± 190 | 4.22 ± 0.38 | -0.52 ± 0.26 | ... | ... | Y | ... | ... | ... | Y | n | ... |
| 39823 | 06292470-3115303 | 60.42 ± 0.31 | 6398 ± 188 | 3.99 ± 0.08 | -0.46 ± 0.23 | ... | ... | Y | ... | ... | ... | Y | n | ... |
| 39824 | 06292482-3117556 | 59.10 ± 0.44 | 6493 ± 235 | 4.13 ± 0.25 | -0.53 ± 0.17 | ... | ... | Y | ... | ... | ... | Y | n | ... |
| 39825 | 06292487-3118011 | 59.75 ± 0.43 | 6510 ± 490 | 4.32 ± 0.13 | -0.64 ± 0.19 | ... | ... | Y | ... | ... | ... | Y | n | ... |
| 39826 | 06292489-3125364 | 54.34 ± 6.49 | ... | ... | ... | ... | ... | N | ... | ... | ... | ... | n | ... |
| 39827 | 06292514-3121429 | 58.64 ± 1.44 | 5446 ± 106 | 4.05 ± 0.72 | -0.69 ± 0.62 | ... | ... | Y | ... | ... | ... | ... | n | ... |
| 39828 | 06292527-3113565 | 61.60 ± 0.56 | 5656 ± 78 | 4.66 ± 0.59 | -0.77 ± 0.23 | <69 | 3 | N | ... | ... | ... | ... | n | NG |
| 39829 | 06292528-3117084 | 61.31 ± 1.04 | 5100 ± 345 | 4.79 ± 0.67 | -0.79 ± 0.31 | ... | ... | N | ... | ... | ... | ... | n | ... |
| 39830 | 06292538-3117157 | 61.57 ± 1.14 | 5467 ± 119 | 4.51 ± 0.12 | -0.66 ± 0.30 | ... | ... | N | ... | ... | ... | ... | n | ... |
| 39831 | 06292540-3113218 | 46.98 ± 0.13 | 6180 ± ... | 4.38 ± ... | -0.24 ± ... | 71 ± 57 | ... | N | ... | ... | ... | ... | n | NG |
| 39832 | 06292554-3121380 | 59.01 ± 0.81 | 5732 ± 11 | 4.30 ± 0.19 | -0.78 ± 0.48 | ... | ... | Y | ... | ... | ... | ... | n | ... |
| 39833 | 06292558-3113428 | 10.34 ± 4.11 | ... | ... | ... | ... | ... | N | ... | ... | ... | ... | n | ... |
| 39834 | 06292559-3116070 | 61.15 ± 0.88 | 6228 ± 580 | 4.04 ± 0.02 | -0.65 ± 0.28 | ... | ... | N | ... | ... | ... | Y | n | ... |
| 39835 | 06292569-3116238 | 61.85 ± 0.39 | 6181 ± 465 | 3.72 ± 0.37 | -0.77 ± 0.19 | <73 | 3 | N | ... | ... | ... | Y | n | NG |
| 39836 | 06292580-3117375 | 67.08 ± 1.82 | 5171 ± 57 | 3.66 ± 0.72 | -0.70 ± 0.49 | ... | ... | N | ... | ... | ... | ... | n | ... |
| 39837 | 06292583-3107574 | 61.19 ± 0.45 | 6504 ± 506 | 4.35 ± 0.22 | -0.71 ± 0.25 | ... | ... | N | ... | ... | ... | Y | n | ... |
| 39838 | 06292595-3119295 | 57.68 ± 1.07 | 5400 ± 105 | 2.84 ± 0.49 | -0.88 ± 0.24 | ... | ... | N | ... | ... | ... | ... | n | ... |
| 39839 | 06292599-3115451 | 61.09 ± 0.75 | 6375 ± 473 | 4.54 ± 0.48 | -0.54 ± 0.16 | ... | ... | N | ... | ... | ... | Y | n | ... |
| 39840 | 06292602-3115214 | 61.36 ± 0.75 | 5481 ± 167 | 4.55 ± 0.52 | -0.66 ± 0.26 | <83 | 3 | N | ... | ... | ... | ... | n | NG |
| 39841 | 06292602-3118247 | 57.58 ± 0.53 | 6321 ± 76 | 4.91 ± 0.17 | -1.21 ± 0.95 | ... | ... | N | ... | ... | ... | ... | n | ... |
| 39842 | 06292609-3119014 | 59.42 ± 1.26 | 5861 ± 225 | 4.13 ± 0.64 | -0.29 ± 0.16 | ... | ... | Y | ... | ... | ... | ... | n | ... |
| 39843 | 06292610-3114477 | 61.73 ± 0.40 | 6304 ± 45 | 4.38 ± 0.65 | -0.78 ± 0.28 | ... | ... | N | ... | ... | ... | Y | n | ... |
| 39844 | 06292622-3117119 | 59.57 ± 0.40 | 6317 ± 237 | 4.23 ± 0.11 | -0.44 ± 0.26 | ... | ... | Y | ... | ... | ... | Y | n | ... |
| 39845 | 06292623-3124476 | 103.49 ± 1.30 | 6045 ± 724 | 4.85 ± 0.77 | -0.57 ± 0.26 | ... | ... | N | ... | ... | ... | ... | n | ... |
| 39846 | 06292626-3112134 | 60.39 ± 0.49 | 6383 ± 500 | 4.26 ± 0.37 | -0.59 ± 0.19 | ... | ... | Y | ... | ... | ... | Y | n | ... |
| 39847 | 06292631-3117307 | 60.09 ± 0.85 | 6090 ± 346 | 4.15 ± 0.23 | -0.65 ± 0.34 | ... | ... | Y | ... | ... | ... | ... | n | ... |

**Table C.20.** continued.

| ID | CNAME | RV (km s$^{-1}$) | $T_{\rm eff}$ (K) | logg (dex) | [Fe/H] (dex) | EW(Li)$^a$ (mÅ) | EW(Li) error flag$^b$ | RV | Membership Li | logg | [Fe/H] | Gaia study Cantat-Gaudin$^c$ | Final$^d$ | NMs with Li$^e$ |
|---|---|---|---|---|---|---|---|---|---|---|---|---|---|---|
| 39848 | 06292642-3109317 | 61.36 ± 0.55 | 6261 ± 157 | 4.57 ± 0.58 | -0.44 ± 0.18 | 87 ± 44 | 1 | N | … | … | … | Y | n | NG |
| 39849 | 06292642-3114380 | 64.33 ± 1.69 | 6004 ± 776 | 4.85 ± 0.23 | -0.60 ± 0.47 | … | … | N | … | … | … | Y | n | … |
| 39850 | 06292642-3125436 | 55.04 ± 1.11 | 5533 ± 158 | 3.66 ± 0.28 | -0.37 ± 0.21 | … | … | N | … | … | … | … | n | … |
| 39851 | 06292643-3115445 | 53.92 ± 1.62 | 6099 ± 188 | 4.38 ± 0.66 | -1.35 ± 0.36 | … | … | N | … | … | … | Y | n | … |
| 39852 | 06292644-3117485 | 59.84 ± 0.67 | 6394 ± 253 | 4.65 ± 0.17 | -0.79 ± 0.58 | … | … | Y | … | … | … | Y | n | … |
| 39853 | 06292651-3122208 | 57.81 ± 0.55 | 6144 ± 98 | 4.07 ± 0.25 | -0.46 ± 0.28 | 66 ± 51 | 1 | N | … | … | … | Y | n | NG |
| 39854 | 06292683-3116180 | 59.49 ± 1.16 | 5120 ± 170 | 4.34 ± 0.56 | -0.63 ± 0.30 | … | … | Y | … | … | … | … | n | … |
| 39855 | 06292686-3112265 | 60.23 ± 0.63 | 6561 ± 464 | 4.05 ± 0.11 | -0.63 ± 0.29 | 62 ± 48 | 1 | Y | Y | Y | Y | Y | Y | … |
| 39856 | 06292689-3122001 | 92.59 ± 1.02 | 6201 ± 614 | 4.35 ± 0.63 | -0.55 ± 0.42 | … | … | N | … | … | … | … | n | … |
| 39857 | 06292710-3118285 | 102.68 ± 0.28 | 6244 ± 171 | 3.98 ± 0.58 | -0.57 ± 0.20 | … | … | N | … | … | … | Y | n | … |
| 39858 | 06292722-3109419 | 61.03 ± 1.03 | 6122 ± 544 | 3.88 ± 0.04 | -0.54 ± 0.46 | … | … | N | … | … | … | … | n | … |
| 39859 | 06292725-3116018 | 534.77 ± 0.21 | 5955 ± 71 | 4.84 ± 0.19 | -0.71 ± 0.20 | <78 | 3 | N | … | … | … | Y | n | NG |
| 39860 | 06292733-3114196 | -3.06 ± 0.12 | 5932 ± … | 4.49 ± … | -0.31 ± … | … | … | N | … | … | … | … | n | … |
| 39861 | 06292742-3117534 | 59.25 ± 0.11 | 6395 ± … | 3.67 ± … | -0.77 ± … | … | … | Y | … | … | … | Y | n | … |
| 39862 | 06292750-3116049 | 59.08 ± 1.05 | 6033 ± 444 | 4.11 ± 0.24 | -0.47 ± 0.56 | … | … | Y | … | … | … | … | n | … |
| 39863 | 06292757-3115034 | 59.42 ± 0.17 | 6427 ± … | 4.75 ± … | -0.62 ± … | … | … | Y | … | … | … | Y | n | … |
| 39864 | 06292774-3115104 | 59.83 ± 1.25 | 5761 ± 528 | 4.12 ± 0.06 | -0.61 ± 0.42 | … | … | Y | … | … | … | … | n | … |
| 39865 | 06292784-3114404 | 59.85 ± 0.48 | 5613 ± 86 | 4.41 ± 0.27 | -0.33 ± 0.17 | <111 | 3 | Y | N | N | N | Y | n | NG |
| 39866 | 06292787-3126276 | 57.59 ± 1.37 | 6088 ± 227 | 3.28 ± 0.29 | -0.79 ± 0.25 | <109 | 3 | N | … | … | … | N | n | … |
| 39867 | 06292788-3116258 | 59.89 ± 0.79 | 5262 ± 235 | 3.95 ± 0.39 | -0.62 ± 0.32 | … | … | Y | … | … | … | … | n | … |
| 39868 | 06292790-3119120 | 57.99 ± 0.64 | 6185 ± 358 | 5.06 ± 0.78 | -0.63 ± 0.52 | … | … | N | … | … | … | Y | n | … |
| 39869 | 06292806-3115454 | 60.66 ± 0.76 | 5757 ± 128 | 5.03 ± 0.62 | -0.72 ± 0.34 | … | … | Y | … | … | … | … | n | … |
| 39870 | 06292843-3113102 | 61.46 ± 0.43 | 5919 ± 82 | 4.49 ± 0.35 | -0.60 ± 0.28 | 73 ± 54 | 1 | N | … | … | … | Y | n | NG |
| 39871 | 06292843-3116472 | 59.11 ± 0.33 | 6309 ± 146 | 4.08 ± 0.17 | -0.65 ± 0.38 | … | … | Y | … | … | … | Y | n | … |
| 39872 | 06292846-3120350 | 58.80 ± 0.15 | 6327 ± … | 4.41 ± … | -0.64 ± … | … | … | Y | … | … | … | Y | n | … |
| 39873 | 06292857-3114367 | 57.17 ± 0.59 | 6178 ± 439 | 4.32 ± 0.39 | -0.57 ± 0.27 | … | … | N | … | … | … | Y | n | … |
| 39874 | 06292859-3112192 | 58.69 ± 0.11 | 5522 ± … | 3.57 ± … | -0.36 ± … | … | … | Y | … | … | … | Y | n | … |
| 39875 | 06292865-3116068 | 59.25 ± 0.15 | 6418 ± … | 4.02 ± … | -0.61 ± … | … | … | Y | … | … | … | Y | n | … |
| 39876 | 06292873-3115511 | 62.44 ± 0.69 | 6008 ± 187 | 4.73 ± 0.56 | -0.56 ± 0.24 | … | … | N | … | … | … | Y | n | … |
| 39877 | 06292880-3116339 | 59.29 ± 0.34 | 6095 ± 348 | 3.81 ± 0.60 | -0.77 ± 0.42 | … | … | Y | … | … | … | Y | n | … |
| 39878 | 06292900-3112472 | 72.50 ± 0.74 | 5563 ± 272 | 4.05 ± 0.24 | -1.36 ± 2.19 | … | … | N | … | … | … | … | n | … |
| 39879 | 06292931-3117438 | 59.37 ± 0.25 | 5385 ± 69 | 3.63 ± 0.55 | -0.41 ± 0.16 | … | … | Y | … | … | … | Y | n | … |
| 39880 | 06292933-3111563 | 10.56 ± 0.44 | 5969 ± 173 | 4.26 ± 0.18 | -0.51 ± 0.16 | … | … | N | … | … | … | … | n | … |
| 39881 | 06292933-3121354 | 60.05 ± 0.10 | 5116 ± … | 3.42 ± … | -0.36 ± … | … | … | Y | … | … | … | Y | n | G |
| 39882 | 06292944-3120358 | 56.63 ± 1.24 | 5219 ± 50 | 3.60 ± 0.69 | -0.56 ± 0.26 | … | … | N | … | … | … | … | n | … |
| 39883 | 06292944-3121577 | 59.11 ± 0.10 | 5940 ± … | 3.61 ± … | -0.53 ± … | … | … | Y | … | … | … | Y | n | … |
| 39884 | 06292946-3115026 | 59.81 ± 0.44 | 6466 ± 351 | 4.73 ± 0.14 | -0.85 ± 0.23 | … | … | Y | … | … | … | Y | n | … |
| 39885 | 06292950-3118547 | 58.19 ± 1.48 | 5917 ± 205 | 4.14 ± 0.90 | -0.39 ± 0.23 | … | … | Y | … | … | … | … | n | … |
| 39886 | 06292961-3115264 | 57.55 ± 0.27 | 6280 ± 391 | 4.24 ± 0.08 | -0.71 ± 0.28 | … | … | N | … | … | … | Y | n | … |
| 39887 | 06292961-3115529 | 59.63 ± 0.55 | 6359 ± 275 | 4.67 ± 0.33 | -0.71 ± 0.53 | <87 | 3 | Y | N | N | Y | Y | n | NG |
| 39888 | 06292977-3109439 | 60.69 ± 0.13 | 6336 ± … | 3.86 ± … | -0.66 ± … | … | … | Y | … | … | … | Y | n | … |
| 39889 | 06292978-3116232 | 59.83 ± 0.24 | 5039 ± 241 | 3.50 ± 0.64 | -0.38 ± 0.19 | … | … | Y | … | … | … | Y | n | … |
| 39890 | 06292979-3115339 | 60.45 ± 0.47 | 6269 ± 441 | 4.40 ± 0.24 | -0.72 ± 0.19 | … | … | Y | … | … | … | Y | n | … |
| 39891 | 06292989-3120327 | 59.15 ± 0.15 | 6339 ± … | 4.95 ± … | -0.83 ± … | … | … | Y | … | … | … | Y | n | … |
| 39892 | 06292993-3119032 | 60.06 ± 0.74 | 5399 ± 573 | 3.24 ± 0.16 | -0.84 ± 0.86 | … | … | Y | … | … | … | … | n | … |
| 39893 | 06292997-3116336 | 58.15 ± 0.60 | 6422 ± 85 | 4.06 ± 0.65 | -0.64 ± 0.20 | … | … | Y | … | … | … | Y | n | … |
| 39894 | 06292998-3127384 | 117.28 ± 0.32 | 5600 ± 144 | 4.44 ± 0.20 | -0.53 ± 0.23 | … | … | N | … | … | … | … | n | … |
| 39895 | 06293015-3111401 | 59.03 ± 1.23 | 6694 ± 1777 | 4.05 ± 0.33 | -0.60 ± 0.08 | … | … | Y | … | … | … | … | n | … |
| 39896 | 06293016-3115084 | 59.04 ± 0.81 | 5149 ± 446 | 3.95 ± 0.50 | -0.69 ± 0.36 | <102 | 3 | Y | N | N | Y | … | n | NG |
| 39897 | 06293018-3116297 | 61.30 ± 0.67 | 6215 ± 219 | 4.47 ± 0.15 | -0.59 ± 0.15 | … | … | N | … | … | … | Y | n | … |
| 39898 | 06293037-3117409 | 60.08 ± 0.50 | 6144 ± 288 | 3.85 ± 0.23 | -0.58 ± 0.34 | 93 ± 88 | 1 | Y | N | Y | Y | Y | n | NG |
| 39899 | 06293041-3112365 | 60.08 ± 0.11 | 6285 ± … | 3.68 ± … | -0.75 ± … | … | … | Y | … | … | … | Y | n | … |
| 39900 | 06293043-3118331 | 59.85 ± 1.09 | 5716 ± 66 | 3.91 ± 0.20 | -0.52 ± 0.12 | <94 | 3 | Y | N | N | Y | … | n | NG |
| 39901 | 06293044-3116377 | 59.44 ± 0.97 | 6029 ± 359 | 4.13 ± 0.34 | -0.46 ± 0.37 | 126 ± 110 | 1 | Y | N | N | Y | Y | n | NG |
| 39902 | 06293044-3120148 | 56.62 ± 0.42 | 6436 ± 395 | 4.79 ± 0.71 | -0.41 ± 0.19 | … | … | N | … | … | … | Y | n | … |
| 39903 | 06293049-3109112 | 59.73 ± 0.65 | 6097 ± 295 | 4.07 ± 0.05 | -0.86 ± 0.73 | 86 ± 80 | 1 | Y | N | N | N | Y | n | NG |
| 39904 | 06293049-3123545 | 33.99 ± 0.10 | 6063 ± … | 4.33 ± … | -0.48 ± … | 68 ± 20 | … | N | … | … | … | … | n | NG |
| 39905 | 06293051-3116029 | 55.81 ± 0.56 | 6095 ± 192 | 4.46 ± 0.55 | -0.58 ± 0.27 | <112 | 3 | N | … | … | … | … | n | NG |
| 39906 | 06293054-3117457 | 60.00 ± 0.51 | 6328 ± 132 | 4.54 ± 0.18 | -0.72 ± 0.22 | <55 | 3 | Y | Y | Y | Y | Y | Y | … |

| ID | CNAME | RV (km s$^{-1}$) | $T_{\text{eff}}$ (K) | logg (dex) | [Fe/H] (dex) | EW(Li)$^a$ (mÅ) | EW(Li) error flag$^b$ | Membership RV | Li | logg | [Fe/H] | Gaia study Cantat-Gaudin$^c$ | Final$^d$ | NMs with Li$^e$ |
|---|---|---|---|---|---|---|---|---|---|---|---|---|---|---|
| 39907 | 06293058-3118383 | 47.95 ± 0.77 | 5534 ± 214 | 4.26 ± 0.54 | -0.82 ± 0.37 | … | … | N | … | … | … | Y | n | … |
| 39908 | 06293080-3116325 | 58.07 ± 0.97 | 6211 ± 384 | 4.47 ± 0.26 | -0.59 ± 0.23 | … | … | Y | … | … | … | Y | n | … |
| 39909 | 06293088-3126134 | 73.78 ± 0.26 | 5656 ± 82 | 3.99 ± 0.21 | -0.53 ± 0.22 | … | … | N | … | … | … | … | n | … |
| 39910 | 06293089-3118366 | 62.11 ± 0.17 | 6453 ± … | 4.43 ± … | -0.46 ± … | … | … | N | … | … | … | Y | n | … |
| 39911 | 06293094-3115433 | 59.06 ± 0.32 | 6447 ± 448 | 3.70 ± 0.38 | -0.74 ± 0.40 | … | … | Y | … | … | … | Y | n | … |
| 39912 | 06293112-3116434 | 60.99 ± 0.42 | 6099 ± 285 | 4.17 ± 0.04 | -0.58 ± 0.14 | … | … | Y | … | … | … | Y | n | … |
| 39913 | 06293125-3115429 | 59.10 ± 0.41 | 6331 ± 269 | 4.41 ± 0.39 | -0.72 ± 0.21 | … | … | Y | … | … | … | Y | n | … |
| 39914 | 06293135-3120311 | 61.96 ± 0.46 | 5496 ± 81 | 4.28 ± 0.31 | -0.59 ± 0.24 | 63 ± 54 | 1 | N | … | … | … | N | n | NG |
| 39915 | 06293145-3121105 | 58.84 ± 0.53 | 6239 ± 202 | 3.99 ± 0.15 | -0.59 ± 0.16 | <52 | 3 | Y | Y | Y | Y | Y | Y | … |
| 39916 | 06293154-3120207 | 58.62 ± 1.03 | 5420 ± 217 | 3.73 ± 0.07 | -0.50 ± 0.16 | … | … | Y | … | … | … | … | n | … |
| 39917 | 06293155-3116065 | 59.91 ± 0.47 | 6265 ± 390 | 4.38 ± 0.62 | -0.77 ± 0.20 | … | … | Y | … | … | … | … | n | … |
| 39918 | 06293155-3117462 | 59.62 ± 0.40 | 6590 ± 535 | 4.48 ± 0.57 | -0.44 ± 0.18 | … | … | Y | … | … | … | Y | n | … |
| 39919 | 06293163-3116535 | 47.17 ± 0.50 | 6325 ± 278 | 4.81 ± 0.25 | -0.91 ± 0.46 | … | … | N | … | … | … | Y | n | … |
| 39920 | 06293164-3117263 | 60.82 ± 0.88 | 6216 ± 453 | 4.43 ± 0.69 | -0.41 ± 0.19 | … | … | Y | … | … | … | Y | n | … |
| 39921 | 06293165-3125143 | 59.81 ± 1.20 | 5847 ± 498 | 3.78 ± 0.39 | -0.72 ± 0.37 | … | … | Y | … | … | … | … | n | … |
| 39922 | 06293166-3115337 | 59.48 ± 0.25 | 5329 ± 178 | 3.70 ± 0.59 | -0.41 ± 0.17 | … | … | Y | … | … | … | Y | n | … |
| 39923 | 06293170-3117563 | 59.33 ± 0.45 | 6455 ± 213 | 4.40 ± 0.31 | -0.62 ± 0.25 | … | … | Y | … | … | … | Y | n | … |
| 39924 | 06293174-3116392 | 60.01 ± 0.18 | 6235 ± … | 3.36 ± … | -0.89 ± … | … | … | Y | … | … | … | Y | n | … |
| 39925 | 06293174-3117072 | 55.08 ± 0.25 | 5781 ± 116 | 4.46 ± 0.12 | -0.09 ± 0.12 | … | … | N | … | … | … | … | n | … |
| 39926 | 06293188-3118043 | 59.24 ± 0.28 | 5963 ± 166 | 4.09 ± 0.37 | -0.87 ± 0.31 | … | … | Y | … | … | … | Y | n | … |
| 39927 | 06293194-3119535 | 59.02 ± 1.10 | 5872 ± 112 | 4.27 ± 0.31 | -0.51 ± 0.15 | … | … | Y | … | … | … | … | n | … |
| 39928 | 06293204-3120036 | 59.80 ± 0.46 | 6117 ± 276 | 4.09 ± 0.25 | -0.43 ± 0.21 | 71 ± 44 | 1 | Y | Y | N | Y | Y | Y | … |
| 39929 | 06293206-3113257 | 57.44 ± 1.05 | 5788 ± 141 | 3.86 ± 0.15 | -0.48 ± 0.18 | … | … | N | … | … | … | … | n | … |
| 39930 | 06293207-3117253 | 58.99 ± 1.59 | 6559 ± 701 | 4.24 ± 0.36 | -0.63 ± 0.16 | … | … | Y | … | … | … | Y | n | … |
| 39931 | 06293208-3110172 | 132.60 ± 0.10 | 5643 ± … | 4.38 ± … | -0.22 ± … | … | … | N | … | … | … | … | n | … |
| 39932 | 06293208-3117100 | 60.07 ± 0.58 | 5775 ± 68 | 4.04 ± 0.03 | -0.58 ± 0.20 | … | … | Y | … | … | … | Y | n | … |
| 39933 | 06293212-3114272 | 59.98 ± 0.45 | 5722 ± 91 | 4.06 ± 0.34 | -0.57 ± 0.22 | … | … | Y | … | … | … | … | n | … |
| 39934 | 06293239-3118580 | 55.53 ± 0.97 | 5752 ± 574 | 4.42 ± 0.10 | -0.69 ± 0.24 | … | … | N | … | … | … | … | n | … |
| 39935 | 06293241-3116442 | 59.29 ± 1.04 | 4986 ± 91 | 3.36 ± 0.20 | -0.98 ± 0.20 | … | … | Y | … | … | … | … | n | G |
| 39936 | 06293245-3119391 | 58.85 ± 1.05 | 5463 ± 174 | 4.43 ± 0.31 | -0.59 ± 0.32 | … | … | Y | … | … | … | … | n | … |
| 39937 | 06293258-3118087 | 58.49 ± 0.16 | 6392 ± … | 3.49 ± … | -0.56 ± … | … | … | Y | … | … | … | Y | n | … |
| 39938 | 06293263-3112447 | 93.92 ± 0.58 | 5835 ± 64 | 4.22 ± 0.06 | -0.20 ± 0.24 | … | … | N | … | … | … | … | n | … |
| 39939 | 06293264-3116237 | 124.41 ± 2.58 | 5286 ± 172 | 2.65 ± 0.21 | -0.40 ± 0.15 | … | … | N | … | … | … | Y | n | … |
| 39940 | 06293267-3121261 | 58.50 ± 1.06 | 5602 ± 37 | 4.25 ± 1.34 | -0.62 ± 0.21 | … | … | Y | … | … | … | … | n | … |
| 39941 | 06293272-3115396 | 59.15 ± 0.15 | 6201 ± … | 4.68 ± … | -0.82 ± … | … | … | Y | … | … | … | Y | n | … |
| 39942 | 06293276-3112483 | 534.79 ± 0.20 | 6432 ± 145 | 3.53 ± 0.10 | -0.77 ± 0.24 | … | … | N | … | … | … | Y | n | … |
| 39943 | 06293282-3121153 | 59.06 ± 0.46 | 6078 ± 221 | 4.03 ± 0.10 | -0.48 ± 0.21 | 103 ± 53 | 1 | Y | N | N | Y | Y | n | NG |
| 39944 | 06293285-3118310 | 59.33 ± 0.11 | 5530 ± … | 3.59 ± … | -0.46 ± … | … | … | Y | … | … | … | Y | n | … |
| 39945 | 06293294-3118198 | 55.91 ± 1.71 | 4774 ± 156 | 1.76 ± 0.63 | -1.17 ± 0.16 | … | … | N | … | … | … | … | n | … |
| 39946 | 06293298-3114235 | 60.90 ± 1.32 | 5088 ± 200 | 3.46 ± 0.96 | -0.73 ± 0.23 | … | … | Y | … | … | … | … | n | G |
| 39947 | 06293304-3116035 | 109.58 ± 1.19 | 5850 ± 204 | 3.74 ± 0.56 | -0.70 ± 0.26 | … | … | N | … | … | … | … | n | … |
| 39948 | 06293305-3110392 | 59.81 ± 1.11 | 5908 ± 200 | 3.46 ± 0.57 | -0.59 ± 0.23 | … | … | Y | … | … | … | … | n | … |
| 39949 | 06293316-3118591 | 59.11 ± 0.34 | 6354 ± 294 | 4.80 ± 0.50 | -0.62 ± 0.27 | … | … | Y | … | … | … | … | n | … |
| 39950 | 06293321-3109302 | 60.57 ± 0.51 | 6364 ± 144 | 4.45 ± 0.15 | -0.51 ± 0.19 | … | … | Y | … | … | … | Y | n | … |
| 39951 | 06293335-3114058 | 59.25 ± 1.38 | 5376 ± 37 | 4.30 ± 0.21 | -0.71 ± 0.30 | … | … | Y | … | … | … | … | n | … |
| 39952 | 06293343-3116238 | 61.03 ± 0.12 | 6250 ± … | 3.72 ± … | -0.48 ± … | … | … | N | … | … | … | … | n | … |
| 39953 | 06293343-3119172 | 57.79 ± 1.52 | 5775 ± 579 | 4.51 ± 0.82 | -0.70 ± 1.02 | … | … | N | … | … | … | … | n | … |
| 39954 | 06293347-3120023 | 59.85 ± 0.31 | 6469 ± 354 | 4.33 ± 0.51 | -0.52 ± 0.14 | … | … | Y | … | … | … | Y | n | … |
| 39955 | 06293349-3115552 | 57.34 ± 1.06 | 5400 ± 201 | 4.15 ± 0.80 | -0.88 ± 0.24 | … | … | N | … | … | … | … | n | … |
| 39956 | 06293352-3117009 | 56.92 ± 0.67 | 5650 ± 92 | 4.67 ± 0.55 | -0.71 ± 0.31 | … | … | N | … | … | … | Y | n | … |
| 39957 | 06293357-3117583 | 69.67 ± 0.80 | 5863 ± 376 | … | -2.61 ± 0.05 | … | … | N | … | … | … | … | n | … |
| 39958 | 06293359-3116152 | 62.22 ± 0.73 | 5547 ± 278 | 3.57 ± 0.66 | -0.65 ± 0.28 | … | … | N | … | … | … | Y | n | … |
| 39959 | 06293366-3113363 | 59.69 ± 0.63 | 6098 ± 223 | 3.87 ± 0.37 | -0.60 ± 0.24 | … | … | Y | … | … | … | Y | n | … |
| 39960 | 06293368-3122596 | 24.95 ± 0.10 | 5307 ± … | 4.51 ± … | 0.17 ± … | … | … | N | … | … | … | … | n | … |
| 39961 | 06293380-3118158 | 60.99 ± 0.63 | 5974 ± 113 | 4.26 ± 0.26 | -0.62 ± 0.28 | … | … | Y | … | … | … | Y | n | … |
| 39962 | 06293381-3116031 | 59.82 ± 0.29 | 6405 ± 422 | 3.97 ± 0.23 | -0.58 ± 0.20 | … | … | Y | … | … | … | Y | n | … |
| 39963 | 06293388-3116410 | 57.63 ± 0.60 | 6049 ± 168 | 4.29 ± 0.17 | -0.58 ± 0.25 | … | … | N | … | … | … | Y | n | … |
| 39964 | 06293394-3120406 | 59.90 ± 1.09 | 5889 ± 485 | 4.34 ± 0.52 | -0.44 ± 0.41 | … | … | Y | … | … | … | … | n | … |
| 39965 | 06293398-3115493 | 59.75 ± 0.20 | 6242 ± … | 4.40 ± … | -0.64 ± … | … | … | Y | … | … | … | … | n | … |

**Table C.20.** continued.

| ID | CNAME | RV (km s$^{-1}$) | $T_{\text{eff}}$ (K) | logg (dex) | [Fe/H] (dex) | EW(Li)$^a$ (mÅ) | EW(Li) error flag$^b$ | Membership RV | Li | logg | [Fe/H] | Gaia study Cantat-Gaudin$^c$ | Final$^d$ | NMs with Li$^e$ |
|---|---|---|---|---|---|---|---|---|---|---|---|---|---|---|
| 39966 | 06293401-3114168 | 60.12 ± 1.67 | 5071 ± 212 | 4.16 ± 0.63 | -0.62 ± 0.29 | … | … | Y | … | … | … | … | n | … |
| 39967 | 06293413-3120542 | 60.04 ± 0.40 | 5842 ± 62 | 4.60 ± 0.33 | -0.34 ± 0.16 | 74 ± 59 | 1 | Y | N | Y | N | Y | n | NG |
| 39968 | 06293415-3117423 | 58.30 ± 0.22 | 6431 ± 514 | 3.77 ± 0.39 | -0.71 ± 0.24 | … | … | Y | … | … | … | Y | n | … |
| 39969 | 06293420-3114518 | 61.19 ± 0.91 | 5791 ± 32 | 4.10 ± 0.06 | -0.49 ± 0.29 | … | … | N | … | … | … | … | n | … |
| 39970 | 06293425-3117276 | 60.37 ± 0.34 | 6081 ± 321 | 4.10 ± 0.11 | -0.54 ± 0.27 | 59 ± 32 | 1 | Y | Y | Y | Y | Y | Y | … |
| 39971 | 06293426-3120324 | 57.58 ± 0.88 | 5641 ± 100 | 3.68 ± 0.45 | -0.41 ± 0.30 | … | … | N | … | … | … | … | n | … |
| 39972 | 06293433-3113290 | 59.67 ± 0.15 | 6444 ± … | 4.15 ± … | -0.75 ± … | … | … | Y | … | … | … | … | n | … |
| 39973 | 06293436-3121473 | 72.49 ± 0.91 | 5919 ± 188 | 3.94 ± 0.47 | -0.59 ± 0.32 | … | … | N | … | … | … | Y | n | … |
| 39974 | 06293456-3115507 | 59.14 ± 0.56 | 5779 ± 68 | 4.29 ± 0.37 | -0.52 ± 0.22 | … | … | Y | … | … | … | … | n | … |
| 39975 | 06293456-3118372 | 61.45 ± 1.03 | 5265 ± 332 | 4.59 ± 0.14 | -0.61 ± 0.28 | … | … | N | … | … | … | … | n | … |
| 39976 | 06293457-3114048 | 59.93 ± 0.69 | 6008 ± 251 | 4.10 ± 0.52 | -0.69 ± 0.26 | … | … | Y | … | … | … | Y | n | … |
| 39977 | 06293462-3112533 | 91.94 ± 1.62 | 5759 ± 533 | 4.27 ± 0.08 | -0.32 ± 0.82 | … | … | N | … | … | … | … | n | … |
| 39978 | 06293474-3120203 | 59.46 ± 0.78 | 5736 ± 202 | 4.38 ± 0.24 | -0.70 ± 0.35 | <68 | 3 | Y | N | Y | Y | Y | n | NG |
| 39979 | 06293475-3113291 | 534.73 ± 0.21 | 6059 ± 250 | 4.11 ± 0.05 | -0.81 ± 0.56 | … | … | N | … | … | … | Y | n | … |
| 39980 | 06293478-3117392 | 34.71 ± 0.27 | 5737 ± 30 | 4.15 ± 0.06 | -0.63 ± 0.22 | … | … | N | … | … | … | … | n | … |
| 39981 | 06293481-3118335 | 64.66 ± 0.46 | 6089 ± 222 | 4.22 ± 0.09 | -0.61 ± 0.25 | … | … | N | … | … | … | … | n | … |
| 39982 | 06293488-3117306 | 58.09 ± 0.33 | 6340 ± 397 | 4.02 ± 0.26 | -0.56 ± 0.13 | … | … | Y | … | … | … | Y | n | … |
| 39983 | 06293499-3118141 | 75.04 ± 0.73 | 5524 ± 61 | 4.33 ± 0.37 | -0.34 ± 0.21 | <128 | 3 | N | … | … | … | … | n | NG |
| 39984 | 06293502-3119198 | 58.63 ± 0.20 | 5082 ± 224 | 3.54 ± 0.53 | -0.27 ± 0.21 | … | … | Y | … | … | … | Y | n | … |
| 39985 | 06293510-3120190 | 60.18 ± 0.11 | 6096 ± … | 3.79 ± … | -0.77 ± … | … | … | Y | … | … | … | Y | n | … |
| 39986 | 06293533-3113408 | 60.11 ± 0.90 | 5625 ± 80 | 4.40 ± 0.27 | -0.79 ± 0.22 | … | … | Y | … | … | … | … | n | … |
| 39987 | 06293538-3111092 | 58.59 ± 1.24 | 6620 ± 641 | 4.28 ± 0.53 | -0.54 ± 0.09 | … | … | Y | … | … | … | Y | n | … |
| 39988 | 06293552-3120515 | 58.46 ± 0.43 | 6692 ± 169 | 4.70 ± 0.41 | -0.58 ± 0.20 | … | … | Y | … | … | … | Y | n | … |
| 39989 | 06293561-3117560 | 501.52 ± 0.50 | 6411 ± 41 | 3.98 ± 0.75 | -0.51 ± 0.17 | … | … | N | … | … | … | Y | n | … |
| 39990 | 06293564-3121176 | 59.01 ± 0.94 | 5495 ± 115 | 4.66 ± 0.37 | -0.52 ± 0.17 | <69 | 3 | Y | N | Y | Y | N | n | NG |
| 39991 | 06293565-3108124 | 61.37 ± 0.52 | 6167 ± 122 | 3.98 ± 0.91 | -0.74 ± 0.29 | … | … | N | … | … | … | … | n | … |
| 39992 | 06293583-3117489 | -4.40 ± 8.16 | 4910 ± 854 | … | -0.18 ± 0.17 | … | … | N | … | … | … | … | n | … |
| 39993 | 06293583-3117532 | 60.11 ± 0.84 | 6394 ± 437 | 3.79 ± 0.39 | -0.40 ± 0.26 | … | … | Y | … | … | … | Y | n | … |
| 39994 | 06293586-3116195 | 59.76 ± 0.36 | 6363 ± 388 | 4.20 ± 0.17 | -0.46 ± 0.14 | … | … | Y | … | … | … | Y | n | … |
| 39995 | 06293587-3112204 | 46.04 ± 0.10 | 5504 ± … | 4.36 ± … | 0.02 ± … | … | … | N | … | … | … | … | n | … |
| 39996 | 06293589-3116494 | 60.16 ± 0.74 | 5217 ± 262 | 4.26 ± 0.92 | -0.87 ± 0.39 | <81 | 3 | Y | N | N | N | Y | n | NG |
| 39997 | 06293598-3117477 | 63.56 ± 0.13 | 6549 ± … | 4.36 ± … | -0.76 ± … | … | … | N | … | … | … | Y | n | … |
| 39998 | 06293600-3116342 | 59.68 ± 0.11 | 6350 ± … | 4.18 ± … | -0.92 ± … | … | … | Y | … | … | … | Y | n | … |
| 39999 | 06293603-3116392 | 59.47 ± 0.32 | 6374 ± 378 | 4.37 ± 0.32 | -0.60 ± 0.27 | … | … | Y | … | … | … | Y | n | … |
| 40000 | 06293609-3113470 | 39.34 ± 0.69 | 5358 ± 218 | 4.21 ± 0.30 | -0.20 ± 0.29 | … | … | N | … | … | … | … | n | … |
| 40001 | 06293614-3118450 | 61.96 ± 1.04 | 5448 ± 42 | 3.75 ± 0.67 | -0.56 ± 0.34 | … | … | N | … | … | … | … | n | … |
| 40002 | 06293623-3122540 | 96.03 ± 0.62 | 5594 ± 74 | 3.99 ± 0.55 | -0.31 ± 0.13 | … | … | N | … | … | … | … | n | … |
| 40003 | 06293627-3115559 | 59.91 ± 0.31 | 6337 ± 204 | 4.41 ± 0.35 | -0.49 ± 0.19 | … | … | Y | … | … | … | Y | n | … |
| 40004 | 06293629-3113402 | 59.86 ± 0.81 | 5102 ± 392 | 3.31 ± 1.59 | -0.69 ± 0.43 | … | … | Y | … | … | … | … | n | Li-rich G |
| 40005 | 06293629-3117126 | 59.76 ± 0.11 | 6227 ± … | 4.62 ± … | -0.77 ± … | … | … | Y | … | … | … | Y | n | … |
| 40006 | 06293630-3116579 | 59.89 ± 0.34 | 6180 ± 219 | 4.57 ± 0.53 | -0.62 ± 0.27 | … | … | Y | … | … | … | Y | n | … |
| 40007 | 06293633-3119024 | 61.24 ± 0.56 | 5576 ± 103 | 4.77 ± 0.34 | -0.39 ± 0.20 | <92 | 3 | N | … | … | … | Y | n | NG |
| 40008 | 06293640-3118334 | 58.67 ± 0.13 | 6430 ± … | 4.12 ± … | -0.66 ± … | … | … | Y | … | … | … | Y | n | … |
| 40009 | 06293641-3119453 | 59.21 ± 0.15 | 6547 ± … | 4.13 ± … | -0.59 ± … | … | … | Y | … | … | … | Y | n | … |
| 40010 | 06293642-3120032 | 59.91 ± 0.12 | 6392 ± … | 4.29 ± … | -0.53 ± … | … | … | Y | … | … | … | Y | n | … |
| 40011 | 06293652-3117157 | 59.83 ± 0.79 | 6128 ± 396 | 3.91 ± 0.19 | -0.28 ± 0.42 | … | … | Y | … | … | … | … | n | … |
| 40012 | 06293659-3117302 | 59.45 ± 0.11 | 6349 ± … | 3.56 ± … | -1.01 ± … | … | … | Y | … | … | … | Y | n | … |
| 40013 | 06293662-3117519 | 59.90 ± 0.34 | 6511 ± 248 | 4.53 ± 0.27 | -0.60 ± 0.15 | … | … | Y | … | … | … | Y | n | … |
| 40014 | 06293664-3115415 | 27.93 ± 0.37 | 6084 ± 190 | 4.23 ± 0.13 | -0.62 ± 0.56 | 67 ± 47 | 1 | N | … | … | … | … | n | NG |
| 40015 | 06293667-3119238 | 60.54 ± 1.27 | 6282 ± 842 | 4.79 ± 0.77 | -0.58 ± 0.19 | … | … | Y | … | … | … | … | n | … |
| 40016 | 06293676-3116466 | 60.13 ± 0.81 | 4999 ± 664 | 3.83 ± 0.45 | -0.71 ± 0.41 | … | … | Y | … | … | … | … | n | … |
| 40017 | 06293678-3117064 | 54.51 ± 0.36 | 6381 ± 264 | 4.64 ± 0.37 | -0.52 ± 0.23 | … | … | N | … | … | … | … | n | … |
| 40018 | 06293682-3115515 | 57.22 ± 0.37 | 6238 ± 28 | 4.38 ± 0.89 | -0.71 ± 0.26 | … | … | N | … | … | … | Y | n | … |
| 40019 | 06293685-3121341 | 93.32 ± 0.24 | 5163 ± 40 | 4.36 ± 0.58 | 0.17 ± 0.17 | … | … | N | … | … | … | … | n | … |
| 40020 | 06293696-3115413 | 53.10 ± 0.32 | 6390 ± 34 | 4.39 ± 0.84 | -0.69 ± 0.21 | … | … | N | … | … | … | Y | n | … |
| 40021 | 06293698-3110026 | 59.57 ± 0.89 | 5765 ± 55 | 3.65 ± 0.68 | -0.63 ± 0.24 | … | … | Y | … | … | … | Y | n | … |
| 40022 | 06293700-3115233 | 59.20 ± 0.90 | 6120 ± 31 | 4.89 ± 0.20 | -0.41 ± 0.19 | … | … | Y | … | … | … | … | n | … |
| 40023 | 06293709-3116127 | 59.52 ± 0.11 | 6387 ± … | 4.38 ± … | -0.71 ± … | … | … | Y | … | … | … | Y | n | … |
| 40024 | 06293709-3116258 | 59.49 ± 0.42 | 6353 ± 219 | 5.03 ± 0.62 | -0.68 ± 0.18 | … | … | Y | … | … | … | … | n | … |

| ID | CNAME | RV (km s$^{-1}$) | $T_{\text{eff}}$ (K) | logg (dex) | [Fe/H] (dex) | EW(Li)$^a$ (mÅ) | EW(Li) error flag$^b$ | RV | Membership Li | logg | [Fe/H] | Gaia study Cantat-Gaudin$^c$ | Final$^d$ | NMs with Li$^e$ |
|---|---|---|---|---|---|---|---|---|---|---|---|---|---|---|
| 40025 | 06293710-3116332 | 57.61 ± 0.54 | 5768 ± 73 | 4.65 ± 0.38 | -0.47 ± 0.24 | … | … | N | … | … | … | Y | n | … |
| 40026 | 06293711-3118415 | 58.71 ± 0.15 | 6368 ± … | 3.96 ± … | -0.72 ± … | … | … | Y | … | … | … | Y | n | … |
| 40027 | 06293713-3117413 | 57.55 ± 0.30 | 6323 ± 342 | 4.34 ± 0.21 | -0.79 ± 0.19 | … | … | N | … | … | … | Y | n | … |
| 40028 | 06293714-3118200 | 60.00 ± 0.74 | 6071 ± 282 | 5.00 ± 0.54 | -0.54 ± 0.22 | … | … | Y | … | … | … | … | n | … |
| 40029 | 06293730-3111487 | 59.33 ± 1.02 | 4870 ± 327 | 3.83 ± 1.14 | -0.61 ± 0.33 | … | … | Y | … | … | … | … | n | … |
| 40030 | 06293739-3122078 | 26.05 ± 0.39 | 5947 ± 205 | 4.43 ± 0.28 | -0.51 ± 0.21 | … | … | N | … | … | … | … | n | … |
| 40031 | 06293743-3108124 | 58.97 ± 0.89 | 5593 ± 115 | 4.58 ± 0.23 | -1.14 ± 0.47 | <116 | 3 | Y | N | Y | N | Y | n | NG |
| 40032 | 06293746-3113353 | 57.25 ± 0.80 | 6098 ± 272 | 4.40 ± 0.54 | -0.38 ± 0.28 | … | … | N | … | … | … | … | n | … |
| 40033 | 06293746-3118006 | 62.86 ± 0.66 | 6140 ± 387 | 4.34 ± 0.38 | -0.57 ± 0.31 | … | … | N | … | … | … | Y | n | … |
| 40034 | 06293750-3116229 | 63.30 ± 0.27 | 6097 ± 178 | 3.87 ± 0.13 | -0.63 ± 0.24 | … | … | N | … | … | … | … | n | … |
| 40035 | 06293754-3117075 | 59.84 ± 0.48 | 5986 ± 21 | 4.67 ± 0.54 | -0.62 ± 0.21 | 77 ± 63 | 1 | Y | N | Y | Y | Y | n | NG |
| 40036 | 06293771-3117486 | 59.06 ± 0.16 | 6266 ± … | 3.98 ± … | -0.92 ± … | … | … | Y | … | … | … | Y | n | … |
| 40037 | 06293778-3117226 | 60.07 ± 0.38 | 6197 ± 236 | 4.72 ± 0.73 | -0.76 ± 0.22 | … | … | Y | … | … | … | Y | n | … |
| 40038 | 06293786-3114410 | 15.29 ± 2.25 | 5000 ± 73 | 4.15 ± 1.32 | -1.30 ± 0.24 | … | … | N | … | … | … | … | n | … |
| 40039 | 06293790-3116279 | 59.69 ± 0.67 | 6123 ± 361 | 4.33 ± 0.14 | -0.65 ± 0.35 | <74 | 3 | Y | N | Y | Y | … | n | NG |
| 40040 | 06293793-3116491 | 58.05 ± 0.81 | 5631 ± 186 | 4.40 ± 0.24 | -0.44 ± 0.19 | … | … | Y | … | … | … | … | n | … |
| 40041 | 06293796-3116368 | 59.70 ± 0.28 | 6456 ± 503 | 4.08 ± 0.10 | -0.63 ± 0.21 | … | … | Y | … | … | … | Y | n | … |
| 40042 | 06293802-3118489 | 52.66 ± 0.61 | 6123 ± 295 | 4.43 ± 0.45 | -0.71 ± 0.33 | … | … | N | … | … | … | Y | n | … |
| 40043 | 06293807-3119083 | 60.27 ± 0.69 | 5266 ± 510 | 4.05 ± 0.56 | -0.39 ± 0.46 | … | … | Y | … | … | … | … | n | … |
| 40044 | 06293808-3118265 | 59.83 ± 0.54 | 6375 ± 264 | 3.73 ± 0.36 | -0.88 ± 0.34 | … | … | Y | … | … | … | Y | n | … |
| 40045 | 06293812-3117234 | 59.57 ± 0.35 | 6316 ± 370 | 3.73 ± 0.50 | -0.69 ± 0.20 | … | … | Y | … | … | … | Y | n | … |
| 40046 | 06293816-3119020 | 59.77 ± 0.46 | 6135 ± 157 | 4.70 ± 0.53 | -0.86 ± 0.57 | … | … | Y | … | … | … | Y | n | … |
| 40047 | 06293824-3117528 | 61.18 ± 0.72 | 6490 ± 132 | 4.46 ± 0.43 | -0.76 ± 0.23 | … | … | N | … | … | … | Y | n | … |
| 40048 | 06293827-3122547 | 57.59 ± 1.07 | 5806 ± 20 | 4.50 ± 0.33 | -0.44 ± 0.26 | … | … | N | … | … | … | … | n | … |
| 40049 | 06293832-3112569 | 58.92 ± 0.98 | 6158 ± 669 | 4.79 ± 0.14 | -0.84 ± 0.26 | … | … | Y | … | … | … | … | n | … |
| 40050 | 06293833-3116374 | 70.78 ± 1.00 | 6368 ± 480 | 3.87 ± 0.74 | -0.68 ± 0.29 | … | … | N | … | … | … | Y | n | … |
| 40051 | 06293834-3114114 | 61.14 ± 0.99 | 5507 ± 155 | 3.93 ± 0.21 | -0.70 ± 0.43 | … | … | N | … | … | … | … | n | … |
| 40052 | 06293839-3116473 | 60.04 ± 0.37 | 6099 ± 238 | 4.39 ± 0.25 | -0.61 ± 0.20 | … | … | Y | … | … | … | Y | n | … |
| 40053 | 06293869-3114348 | 59.13 ± 0.30 | 6446 ± 190 | 3.72 ± 0.51 | -0.59 ± 0.25 | … | … | Y | … | … | … | Y | n | … |
| 40054 | 06293895-3117049 | 58.28 ± 0.95 | 6038 ± 157 | 4.17 ± 0.65 | -0.40 ± 0.37 | 107 ± 104 | 1 | Y | N | N | Y | Y | n | NG |
| 40055 | 06293897-3115061 | 61.82 ± 0.15 | 6213 ± … | 4.39 ± … | -0.66 ± … | … | … | N | … | … | … | … | n | … |
| 40056 | 06293901-3115545 | 70.96 ± 0.74 | 5837 ± 372 | 4.46 ± 1.02 | -0.69 ± 0.26 | … | … | N | … | … | … | … | n | … |
| 40057 | 06293902-3121395 | 58.08 ± 0.81 | 5812 ± 107 | 3.75 ± 0.77 | -0.62 ± 0.18 | … | … | Y | … | … | … | Y | n | … |
| 40058 | 06293907-3117588 | 59.33 ± 0.11 | 6224 ± … | 3.56 ± … | -0.97 ± … | … | … | Y | … | … | … | Y | n | … |
| 40059 | 06293913-3117288 | 58.83 ± 0.36 | 6189 ± 279 | 4.64 ± 0.33 | -0.67 ± 0.25 | 68 ± 43 | … | Y | Y | Y | Y | Y | Y | … |
| 40060 | 06293915-3116229 | 57.04 ± 0.35 | 6230 ± 201 | 4.51 ± 0.41 | -0.63 ± 0.27 | … | … | N | … | … | … | Y | n | … |
| 40061 | 06293916-3112270 | 66.61 ± 1.26 | 5559 ± 109 | 4.30 ± 0.12 | -0.47 ± 0.18 | … | … | N | … | … | … | … | n | … |
| 40062 | 06293925-3124056 | 58.20 ± 0.81 | 5817 ± 259 | 3.94 ± 0.44 | -0.77 ± 0.45 | … | … | Y | … | … | … | … | n | … |
| 40063 | 06293930-3116257 | 59.20 ± 0.42 | 6241 ± 73 | 4.24 ± 0.36 | -0.46 ± 0.29 | 49 ± 38 | 1 | Y | Y | Y | Y | … | Y | … |
| 40064 | 06293950-3117146 | 502.19 ± 0.39 | 6523 ± 83 | 4.37 ± 0.13 | -0.57 ± 0.22 | … | … | N | … | … | … | Y | n | … |
| 40065 | 06293952-3118056 | 58.15 ± 0.42 | 6814 ± 904 | 4.85 ± 0.72 | -0.63 ± 0.22 | … | … | Y | … | … | … | Y | n | … |
| 40066 | 06293968-3114599 | 61.07 ± 0.89 | 5127 ± 199 | 4.39 ± 0.19 | -0.69 ± 0.33 | … | … | N | … | … | … | … | n | … |
| 40067 | 06293971-3115444 | 60.59 ± 0.41 | 6418 ± 312 | 4.33 ± 0.47 | -0.55 ± 0.19 | … | … | Y | … | … | … | Y | n | … |
| 40068 | 06293973-3122528 | 58.79 ± 1.05 | 5867 ± 72 | 4.16 ± 0.52 | -0.56 ± 0.16 | … | … | Y | … | … | … | … | n | … |
| 40069 | 06293977-3112259 | 60.34 ± 0.65 | 5711 ± 229 | 3.94 ± 0.45 | -0.69 ± 0.40 | … | … | Y | … | … | … | Y | n | … |
| 40070 | 06293983-3116457 | 59.91 ± 0.50 | 6145 ± 349 | 4.36 ± 0.30 | -0.37 ± 0.21 | 86 ± 67 | 1 | Y | N | Y | Y | Y | n | NG |
| 40071 | 06293985-3120369 | 103.86 ± 0.45 | 6126 ± 464 | 4.45 ± 0.24 | -0.69 ± 0.34 | … | … | N | … | … | … | … | n | … |
| 40072 | 06293986-3118064 | 59.92 ± 0.43 | 6150 ± 328 | 4.00 ± 0.18 | -0.63 ± 0.19 | … | … | Y | … | … | … | Y | n | … |
| 40073 | 06293991-3115557 | 59.13 ± 0.59 | 5244 ± 441 | 3.75 ± 0.50 | -0.58 ± 0.25 | <67 | 3 | Y | N | Y | Y | … | n | NG |
| 40074 | 06294010-3119296 | 59.86 ± 1.08 | 5540 ± 24 | 4.35 ± 0.18 | -0.41 ± 0.15 | … | … | Y | … | … | … | … | n | … |
| 40075 | 06294017-3117395 | 58.06 ± 0.21 | 6295 ± 208 | 4.17 ± 0.18 | -0.68 ± 0.23 | … | … | Y | … | … | … | Y | n | … |
| 40076 | 06294018-3121310 | 58.37 ± 0.16 | 6358 ± … | 4.32 ± … | -0.62 ± … | … | … | Y | … | … | … | Y | n | … |
| 40077 | 06294027-3120547 | 60.47 ± 0.32 | 6444 ± 262 | 4.54 ± 0.33 | -0.73 ± 0.17 | … | … | Y | … | … | … | Y | n | … |
| 40078 | 06294029-3117033 | 59.87 ± 0.34 | 6414 ± 309 | 4.38 ± 0.19 | -0.67 ± 0.18 | … | … | Y | … | … | … | Y | n | … |
| 40079 | 06294034-3126041 | 50.67 ± 0.60 | 6062 ± 76 | 4.49 ± 0.47 | -0.42 ± 0.25 | 66 ± 59 | 1 | N | … | … | … | … | n | NG |
| 40080 | 06294055-3122075 | 60.16 ± 0.15 | 6275 ± … | 4.53 ± … | -0.78 ± … | … | … | Y | … | … | … | Y | n | … |
| 40081 | 06294056-3114331 | 57.35 ± 0.40 | 5843 ± 91 | 4.67 ± 0.45 | -0.64 ± 0.28 | … | … | N | … | … | … | N | n | … |
| 40082 | 06294056-3115352 | 58.77 ± 0.50 | 5901 ± 56 | 4.00 ± 0.40 | -0.46 ± 0.31 | … | … | Y | … | … | … | … | n | … |
| 40083 | 06294060-3113101 | 472.20 ± 1.43 | 5424 ± 382 | 2.14 ± 1.04 | -1.42 ± 0.40 | … | … | N | … | … | … | … | n | … |

**Table C.20.** continued.

| ID | CNAME | RV (km s$^{-1}$) | $T_{\text{eff}}$ (K) | logg (dex) | [Fe/H] (dex) | EW(Li)$^a$ (mÅ) | EW(Li) error flag$^b$ | RV | Membership Li | logg | [Fe/H] | Gaia study Cantat-Gaudin$^c$ | Final$^d$ | NMs with Li$^e$ |
|---|---|---|---|---|---|---|---|---|---|---|---|---|---|---|
| 40084 | 06294062-3115157 | 76.98 ± 0.38 | 6171 ± 247 | 4.04 ± 0.45 | -0.67 ± 0.17 | … | … | N | … | … | … | … | n | … |
| 40085 | 06294094-3118333 | 59.81 ± 0.37 | 6190 ± 327 | 4.36 ± 0.23 | -0.63 ± 0.27 | … | … | Y | … | … | … | Y | n | … |
| 40086 | 06294102-3114126 | 61.11 ± 0.69 | 5764 ± 97 | 4.41 ± 0.34 | -0.47 ± 0.13 | … | … | N | … | … | … | Y | n | … |
| 40087 | 06294118-3122435 | 58.35 ± 0.73 | 5917 ± 242 | 4.59 ± 0.17 | -0.72 ± 0.31 | … | … | Y | … | … | … | Y | n | … |
| 40088 | 06294122-3114505 | 60.51 ± 0.32 | 5642 ± 578 | 4.03 ± 0.72 | -1.08 ± 0.20 | … | … | Y | … | … | … | Y | n | … |
| 40089 | 06294129-3117234 | 58.69 ± 1.17 | 5432 ± 186 | 4.08 ± 0.01 | -0.97 ± 0.65 | … | … | Y | … | … | … | … | n | … |
| 40090 | 06294143-3114442 | 60.21 ± 1.47 | 5224 ± 114 | 4.54 ± 0.39 | -0.79 ± 0.33 | … | … | Y | … | … | … | … | n | … |
| 40091 | 06294144-3119298 | 67.45 ± 0.14 | 6320 ± … | 4.31 ± … | -0.74 ± … | … | … | N | … | … | … | Y | n | … |
| 40092 | 06294146-3118062 | 60.94 ± 0.66 | 4877 ± 305 | 3.63 ± 0.39 | -0.55 ± 0.32 | … | … | Y | … | … | … | … | n | … |
| 40093 | 06294151-3119072 | 59.63 ± 0.38 | 6363 ± 303 | 5.04 ± 0.71 | -0.75 ± 0.19 | … | … | Y | … | … | … | Y | n | … |
| 40094 | 06294161-3110388 | 60.64 ± 1.21 | 5561 ± 81 | 4.28 ± 0.39 | -0.58 ± 0.31 | … | … | Y | … | … | … | … | n | … |
| 40095 | 06294169-3117528 | 59.07 ± 0.36 | 6585 ± 717 | 4.48 ± 0.29 | -1.00 ± 1.00 | … | … | Y | … | … | … | … | n | … |
| 40096 | 06294178-3111321 | 4.24 ± 0.28 | 6037 ± 239 | 4.46 ± 0.29 | -0.03 ± 0.21 | … | … | N | … | … | … | … | n | … |
| 40097 | 06294179-3111370 | 51.27 ± 0.35 | 6269 ± … | 3.91 ± … | -0.62 ± … | … | … | N | … | … | … | Y | n | … |
| 40098 | 06294181-3117562 | 60.12 ± 0.80 | 5542 ± 53 | 4.88 ± 0.22 | -0.95 ± 0.36 | … | … | Y | … | … | … | … | n | … |
| 40099 | 06294188-3115475 | 60.31 ± 0.24 | 5371 ± 146 | 3.69 ± 0.44 | -0.40 ± 0.19 | … | … | Y | … | … | … | … | n | … |
| 40100 | 06294191-3118309 | 58.87 ± 1.62 | 5639 ± 193 | 4.20 ± 0.46 | -0.63 ± 0.21 | … | … | Y | … | … | … | … | n | … |
| 40101 | 06294193-3114447 | 53.49 ± 1.21 | 5599 ± 177 | 3.78 ± 0.85 | -0.53 ± 0.49 | … | … | N | … | … | … | … | n | … |
| 40102 | 06294200-3118513 | -21.50 ± 0.86 | 5477 ± 62 | 3.61 ± 0.52 | -0.57 ± 0.18 | … | … | N | … | … | … | … | n | … |
| 40103 | 06294204-3116389 | 63.21 ± 0.64 | 6165 ± 284 | 4.52 ± 0.26 | -0.75 ± 0.35 | … | … | N | … | … | … | Y | n | … |
| 40104 | 06294223-3110302 | 72.33 ± 0.73 | 6142 ± 127 | 4.18 ± 0.92 | -0.59 ± 0.21 | … | … | N | … | … | … | … | n | … |
| 40105 | 06294231-3116504 | 59.39 ± 0.35 | 6356 ± 387 | 4.25 ± 0.09 | -0.73 ± 0.30 | … | … | Y | … | … | … | Y | n | … |
| 40106 | 06294244-3121457 | 60.17 ± 0.84 | 5965 ± 369 | 3.82 ± 0.28 | -0.37 ± 0.22 | <97 | 3 | Y | N | N | Y | … | n | NG |
| 40107 | 06294245-3118340 | 59.47 ± 0.35 | 6389 ± 306 | 4.19 ± 0.32 | -0.51 ± 0.15 | 60 ± 55 | … | Y | Y | Y | Y | Y | Y | … |
| 40108 | 06294253-3116001 | 60.43 ± 0.86 | 6240 ± 160 | 4.08 ± 0.39 | -0.39 ± 0.21 | <79 | 3 | Y | Y | N | Y | Y | Y | … |
| 40109 | 06294255-3120527 | 34.95 ± 0.29 | 6130 ± 241 | 4.09 ± 0.09 | -0.80 ± 0.22 | … | … | N | … | … | … | Y | n | … |
| 40110 | 06294260-3116473 | 61.19 ± 0.87 | 5725 ± 278 | 4.45 ± 0.47 | -0.77 ± 0.39 | 39 ± 39 | 1 | Y | Y | Y | Y | Y | Y | … |
| 40111 | 06294267-3115096 | 54.98 ± 0.26 | 5756 ± 41 | 3.62 ± 0.31 | -0.56 ± 0.16 | … | … | N | … | … | … | Y | n | … |
| 40112 | 06294267-3116548 | 61.14 ± 1.38 | 5538 ± 334 | 3.86 ± 0.31 | -0.62 ± 0.23 | … | … | N | … | … | … | … | n | … |
| 40113 | 06294269-3116325 | 59.74 ± 0.37 | 6595 ± 211 | 3.58 ± 0.32 | -0.78 ± 0.34 | … | … | Y | … | … | … | Y | n | … |
| 40114 | 06294272-3116272 | 61.25 ± 1.09 | 5774 ± 295 | 4.53 ± 0.15 | -0.41 ± 0.65 | … | … | N | … | … | … | … | n | … |
| 40115 | 06294298-3115222 | 59.88 ± 0.40 | 5920 ± 132 | 4.25 ± 0.11 | -0.58 ± 0.23 | 96 ± 47 | 1 | Y | N | N | Y | Y | n | NG |
| 40116 | 06294299-3125187 | 27.09 ± 0.10 | 5562 ± … | 4.48 ± … | 0.15 ± … | … | … | N | … | … | … | … | n | … |
| 40117 | 06294328-3116438 | 59.68 ± 0.55 | 6464 ± 541 | 4.22 ± 0.06 | -0.55 ± 0.22 | … | … | Y | … | … | … | Y | n | … |
| 40118 | 06294334-3119149 | 60.00 ± 1.07 | 5823 ± 596 | 4.26 ± 0.20 | -0.53 ± 0.88 | … | … | Y | … | … | … | … | n | … |
| 40119 | 06294339-3111529 | 59.26 ± 0.12 | 6376 ± … | 4.20 ± … | -0.56 ± … | … | … | Y | … | … | … | Y | n | … |
| 40120 | 06294348-3121501 | 61.15 ± 0.36 | 6134 ± 414 | 4.36 ± 0.21 | -0.19 ± 0.27 | … | … | N | … | … | … | … | n | … |
| 40121 | 06294348-3122289 | 58.85 ± 0.70 | 5728 ± 93 | 4.10 ± 0.31 | -0.47 ± 0.13 | <97 | 3 | Y | N | N | Y | … | n | NG |
| 40122 | 06294350-3112147 | 50.42 ± 1.24 | 5792 ± 26 | 5.08 ± 0.52 | -0.15 ± 0.16 | … | … | N | … | … | … | … | n | … |
| 40123 | 06294381-3110186 | 56.73 ± 0.13 | 6361 ± … | 4.08 ± … | -0.67 ± … | … | … | N | … | … | … | … | n | … |
| 40124 | 06294396-3122271 | 59.85 ± 0.13 | 6331 ± … | 4.29 ± … | -0.66 ± … | … | … | Y | … | … | … | Y | n | … |
| 40125 | 06294403-3108576 | 59.95 ± 0.55 | 6215 ± 291 | 4.07 ± 0.23 | -0.65 ± 0.18 | 72 ± 60 | 1 | Y | N | N | Y | Y | n | NG |
| 40126 | 06294403-3109309 | 59.90 ± 0.62 | 5986 ± 208 | 3.94 ± 0.62 | -0.70 ± 0.35 | … | … | Y | … | … | … | Y | n | … |
| 40127 | 06294409-3112099 | 57.13 ± 1.28 | 5540 ± 240 | 3.55 ± 0.76 | -0.54 ± 0.32 | … | … | N | … | … | … | … | n | … |
| 40128 | 06294409-3116276 | 67.13 ± 1.55 | 5590 ± 385 | 4.01 ± 0.22 | -0.91 ± 0.26 | … | … | N | … | … | … | … | n | … |
| 40129 | 06294417-3115254 | 61.02 ± 0.50 | 5799 ± 47 | 4.70 ± 0.55 | -0.53 ± 0.24 | 73 ± 60 | 1 | N | … | … | … | Y | n | NG |
| 40130 | 06294418-3111012 | 60.05 ± 0.65 | 5511 ± 256 | 3.94 ± 0.23 | -0.65 ± 0.26 | … | … | Y | … | … | … | Y | n | … |
| 40131 | 06294419-3117461 | 57.69 ± 1.20 | 6147 ± 645 | 4.07 ± 0.17 | -0.69 ± 0.88 | … | … | N | … | … | … | … | n | … |
| 40132 | 06294427-3109142 | 76.83 ± 0.41 | 5895 ± 204 | 3.79 ± 0.26 | -0.64 ± 0.38 | … | … | N | … | … | … | … | n | … |
| 40133 | 06294428-3111271 | 59.73 ± 0.80 | 5245 ± 193 | 3.91 ± 0.61 | -0.61 ± 0.18 | … | … | Y | … | … | … | … | n | … |
| 40134 | 06294435-3125185 | 82.61 ± 0.25 | 4979 ± 142 | 3.25 ± 0.17 | -0.13 ± 0.19 | … | … | N | … | … | … | N | n | G |
| 40135 | 06294436-3110077 | 82.08 ± 0.56 | 5622 ± 91 | 4.04 ± 0.25 | -0.30 ± 0.31 | <66 | 3 | N | … | … | … | … | n | NG |
| 40136 | 06294452-3112345 | 62.16 ± 1.99 | 5153 ± 115 | 4.63 ± 0.36 | -0.68 ± 0.16 | … | … | N | … | … | … | … | n | … |
| 40137 | 06294458-3108070 | 87.71 ± 0.11 | 5799 ± … | 4.29 ± … | -0.37 ± … | … | … | N | … | … | … | … | n | … |
| 40138 | 06294464-3114501 | 59.76 ± 1.00 | 5764 ± 390 | 3.74 ± 0.67 | -0.58 ± 0.28 | … | … | Y | … | … | … | … | n | … |
| 40139 | 06294464-3118288 | 62.54 ± 0.87 | 6055 ± 356 | 4.80 ± 0.60 | -0.54 ± 0.36 | 162 ± 117 | 1 | N | … | … | … | Y | n | NG |
| 40140 | 06294478-3110041 | 60.98 ± 0.40 | 5760 ± 79 | 3.87 ± 0.23 | -0.41 ± 0.16 | 39 ± 37 | 1 | Y | Y | Y | Y | Y | Y | … |
| 40141 | 06294509-3113167 | 59.74 ± 0.47 | 5799 ± 26 | 4.31 ± 0.42 | -0.36 ± 0.15 | … | … | Y | … | … | … | Y | n | … |
| 40142 | 06294511-3110190 | 66.38 ± 0.11 | 5748 ± … | 4.19 ± … | -0.59 ± … | 72 ± 25 | … | N | … | … | … | … | n | NG |

**Table C.20.** continued.

| ID | CNAME | $RV$ (km s$^{-1}$) | $T_{\text{eff}}$ (K) | $logg$ (dex) | [Fe/H] (dex) | $EW$(Li)$^a$ (mÅ) | $EW$(Li) error flag$^b$ | RV | Membership Li | $logg$ | [Fe/H] | Gaia study Cantat-Gaudin$^c$ | Final$^d$ | NMs with Li$^e$ |
|---|---|---|---|---|---|---|---|---|---|---|---|---|---|---|
| 40143 | 06294515-3114098 | 59.25 ± 0.34 | 6349 ± 480 | 4.60 ± 0.34 | -0.61 ± 0.23 | … | … | Y | … | … | … | Y | n | … |
| 40144 | 06294518-3115267 | 86.53 ± 0.80 | 4991 ± 132 | 4.21 ± 0.85 | -0.36 ± 0.34 | … | … | N | … | … | … | … | n | … |
| 40145 | 06294523-3110010 | 59.77 ± 0.86 | 5404 ± 353 | 3.62 ± 0.72 | -0.72 ± 0.48 | 131 ± 96 | 1 | Y | N | Y | Y | … | n | NG |
| 40146 | 06294525-3115105 | 58.99 ± 0.42 | 5745 ± 87 | 4.80 ± 0.57 | -0.65 ± 0.34 | 82 ± 50 | 1 | Y | N | Y | Y | Y | n | NG |
| 40147 | 06294532-3117189 | 30.22 ± 1.19 | 5189 ± 377 | 3.67 ± 0.97 | -0.71 ± 0.32 | … | … | N | … | … | … | … | n | … |
| 40148 | 06294547-3111301 | 62.85 ± 1.38 | 5641 ± 257 | 4.41 ± 0.45 | -0.48 ± 0.63 | … | … | N | … | … | … | … | n | … |
| 40149 | 06294559-3111549 | 59.90 ± 0.66 | 5786 ± 267 | 4.14 ± 0.53 | -0.52 ± 0.25 | <73 | 3 | Y | N | N | Y | … | n | NG |
| 40150 | 06294572-3113423 | 59.59 ± 0.72 | 6049 ± 237 | 4.42 ± 0.44 | -0.59 ± 0.23 | … | … | Y | … | … | … | … | n | … |
| 40151 | 06294584-3118132 | 61.69 ± 1.26 | 5866 ± 378 | 4.44 ± 0.51 | -0.60 ± 0.45 | … | … | N | … | … | … | … | n | … |
| 40152 | 06294586-3113468 | 59.87 ± 0.55 | 6307 ± 276 | 4.88 ± 0.35 | -0.59 ± 0.38 | … | … | Y | … | … | … | Y | n | … |
| 40153 | 06294592-3114150 | 59.64 ± 0.98 | 5378 ± 75 | 3.58 ± 0.72 | -0.74 ± 0.44 | … | … | Y | … | … | … | … | n | … |
| 40154 | 06294607-3116255 | 59.05 ± 0.42 | 6526 ± 252 | 4.15 ± 0.40 | -0.66 ± 0.21 | … | … | Y | … | … | … | Y | n | … |
| 40155 | 06294610-3115138 | 58.54 ± 1.13 | 5792 ± 490 | 3.97 ± 0.53 | -0.36 ± 0.67 | … | … | Y | … | … | … | … | n | … |
| 40156 | 06294620-3121269 | 58.92 ± 1.07 | 5668 ± 257 | 4.80 ± 0.47 | -0.66 ± 0.27 | … | … | Y | … | … | … | … | n | … |
| 40157 | 06294642-3108494 | 61.53 ± 1.37 | 6245 ± 792 | 4.62 ± 0.18 | -0.80 ± 1.05 | … | … | N | … | … | … | … | n | … |
| 40158 | 06294648-3119491 | 59.76 ± 0.37 | 6554 ± 495 | 4.15 ± 0.19 | -0.60 ± 0.17 | … | … | Y | … | … | … | Y | n | … |
| 40159 | 06294661-3118537 | 53.03 ± 0.30 | 6357 ± 401 | 4.26 ± 0.07 | -0.57 ± 0.19 | … | … | N | … | … | … | … | n | … |
| 40160 | 06294667-3118184 | 534.95 ± 0.23 | 6224 ± 418 | 4.36 ± 0.15 | -0.57 ± 0.24 | 59 ± 48 | … | N | … | … | … | Y | n | NG |
| 40161 | 06294700-3116219 | 58.24 ± 0.30 | 6166 ± 102 | 3.94 ± 0.05 | -0.64 ± 0.52 | … | … | Y | … | … | … | Y | n | … |
| 40162 | 06294711-3123586 | 60.34 ± 0.92 | 5867 ± 448 | 4.26 ± 0.49 | -0.70 ± 0.20 | … | … | Y | … | … | … | … | n | … |
| 40163 | 06294718-3118095 | 59.33 ± 0.14 | 6636 ± … | 4.58 ± … | -0.69 ± … | … | … | Y | … | … | … | Y | n | … |
| 40164 | 06294759-3118297 | 56.83 ± 0.99 | 5937 ± 41 | 3.22 ± 0.58 | -0.63 ± 0.18 | … | … | N | … | … | … | Y | n | … |
| 40165 | 06294760-3111267 | 60.10 ± 0.10 | 5927 ± … | 3.98 ± … | -0.41 ± … | … | … | Y | … | … | … | … | n | … |
| 40166 | 06294760-3117344 | 51.18 ± 0.69 | 5792 ± 74 | 3.72 ± 0.48 | -0.03 ± 0.19 | <104 | 3 | N | … | … | … | … | n | NG |
| 40167 | 06294764-3114109 | 59.33 ± 1.58 | 5725 ± 19 | 4.40 ± 0.15 | -0.76 ± 0.27 | … | … | Y | … | … | … | … | n | … |
| 40168 | 06294773-3114452 | 59.36 ± 0.28 | 6280 ± 375 | 3.85 ± 0.28 | -0.55 ± 0.13 | <146 | 3 | Y | N | Y | Y | Y | n | NG |
| 40169 | 06294776-3117193 | 60.54 ± 0.29 | 5673 ± 105 | 3.78 ± 0.70 | -0.65 ± 0.33 | … | … | Y | … | … | … | … | n | … |
| 40170 | 06294783-3119016 | 59.31 ± 0.90 | 5400 ± 315 | 3.08 ± 0.17 | -0.71 ± 0.40 | … | … | Y | … | … | … | … | n | … |
| 40171 | 06294792-3112015 | 54.02 ± 0.37 | 6456 ± 318 | 4.43 ± 0.29 | -0.47 ± 0.24 | … | … | N | … | … | … | Y | n | … |
| 40172 | 06294794-3120174 | 58.54 ± 0.33 | 6346 ± 437 | 3.72 ± 0.37 | -0.79 ± 0.36 | … | … | Y | … | … | … | Y | n | … |
| 40173 | 06294800-3117576 | 59.18 ± 0.79 | 6003 ± 196 | 4.38 ± 0.23 | -0.59 ± 0.16 | … | … | Y | … | … | … | … | n | … |
| 40174 | 06294806-3119414 | 60.45 ± 0.41 | 6448 ± 252 | 4.86 ± 0.51 | -0.59 ± 0.23 | … | … | Y | … | … | … | Y | n | … |
| 40175 | 06294809-3119276 | 63.63 ± 0.26 | 6000 ± 120 | 4.05 ± 0.17 | -0.61 ± 0.22 | … | … | N | … | … | … | … | n | … |
| 40176 | 06294813-3120141 | 59.84 ± 0.31 | 6458 ± 341 | 4.28 ± 0.17 | -0.58 ± 0.13 | … | … | Y | … | … | … | Y | n | … |
| 40177 | 06294825-3112564 | 47.80 ± 0.78 | 5489 ± 228 | 4.33 ± 0.05 | -0.64 ± 0.25 | … | … | N | … | … | … | … | n | … |
| 40178 | 06294831-3111284 | 59.41 ± 0.25 | 5323 ± 68 | 4.44 ± 0.31 | -0.43 ± 0.33 | … | … | Y | … | … | … | … | n | … |
| 40179 | 06294837-3122376 | 58.78 ± 0.24 | 5075 ± 255 | 3.45 ± 0.58 | -0.36 ± 0.19 | … | … | Y | … | … | … | Y | n | G |
| 40180 | 06294888-3115571 | 58.27 ± 0.68 | 6061 ± 363 | 5.21 ± 0.77 | -0.55 ± 0.32 | <99 | 3 | Y | N | N | Y | Y | n | NG |
| 40181 | 06294898-3120014 | 59.95 ± 0.77 | 5676 ± 141 | 3.68 ± 0.54 | -0.22 ± 0.28 | … | … | Y | … | … | … | … | n | … |
| 40182 | 06294899-3117478 | 59.86 ± 0.74 | 5883 ± 401 | 3.56 ± 0.47 | -0.45 ± 0.42 | … | … | Y | … | … | … | … | n | … |
| 40183 | 06294931-3112223 | 60.46 ± 0.16 | 6302 ± … | 3.92 ± … | -0.58 ± … | … | … | Y | … | … | … | Y | n | … |
| 40184 | 06294944-3117140 | 75.90 ± 0.54 | 5365 ± 325 | 4.31 ± 0.86 | -0.40 ± 0.13 | … | … | N | … | … | … | … | n | … |
| 40185 | 06294971-3114345 | 59.76 ± 0.86 | 6060 ± 390 | 4.72 ± 0.18 | -1.27 ± 0.20 | … | … | Y | … | … | … | Y | n | … |
| 40186 | 06294983-3113196 | 59.80 ± 0.41 | 6295 ± 245 | 4.62 ± 0.53 | -0.76 ± 0.24 | … | … | Y | … | … | … | Y | n | … |
| 40187 | 06294986-3123051 | 59.70 ± 1.66 | … | … | … | … | … | Y | … | … | … | N | n | … |
| 40188 | 06294996-3119496 | 59.05 ± 0.56 | 6202 ± 303 | 4.63 ± 0.52 | -0.48 ± 0.22 | … | … | Y | … | … | … | Y | n | … |
| 40189 | 06294999-3117496 | 59.77 ± 0.56 | 5558 ± 271 | 3.36 ± 0.50 | -0.63 ± 0.32 | … | … | Y | … | … | … | … | n | … |
| 40190 | 06295000-3114520 | 60.01 ± 0.99 | 5224 ± 82 | 4.23 ± 0.25 | -0.80 ± 0.30 | … | … | Y | … | … | … | … | n | … |
| 40191 | 06295001-3122236 | 60.34 ± 0.45 | 6395 ± 65 | 4.22 ± 0.15 | -0.61 ± 0.19 | … | … | Y | … | … | … | Y | n | … |
| 40192 | 06295007-3120256 | 57.92 ± 1.04 | 5472 ± 27 | 4.26 ± 1.34 | -0.55 ± 0.40 | … | … | N | … | … | … | … | n | … |
| 40193 | 06295025-3117568 | 57.60 ± 1.12 | 5906 ± 341 | 4.44 ± 0.13 | -0.63 ± 0.59 | … | … | N | … | … | … | … | n | … |
| 40194 | 06295025-3121199 | 59.04 ± 0.57 | 6010 ± 311 | 4.57 ± 0.47 | -0.46 ± 0.25 | 100 ± 85 | 1 | Y | N | Y | Y | … | n | NG |
| 40195 | 06295077-3109225 | 44.09 ± 0.58 | 5758 ± 53 | 3.97 ± 0.26 | -0.30 ± 0.22 | … | … | N | … | … | … | … | n | … |
| 40196 | 06295105-3116234 | 61.92 ± 0.45 | 6162 ± 229 | 4.69 ± 0.45 | -0.65 ± 0.29 | … | … | N | … | … | … | Y | n | … |
| 40197 | 06295138-3118432 | 60.84 ± 0.86 | 5724 ± 143 | 4.00 ± 0.04 | -0.50 ± 0.15 | … | … | Y | … | … | … | … | n | … |
| 40198 | 06295144-3125172 | 64.34 ± 0.74 | 6058 ± 239 | 3.80 ± 0.31 | -0.36 ± 0.24 | <159 | 3 | N | … | … | … | N | n | NG |
| 40199 | 06295150-3120463 | 59.54 ± 0.53 | 6105 ± 121 | 4.21 ± 0.13 | -0.40 ± 0.18 | … | … | Y | … | … | … | … | n | … |
| 40200 | 06295151-3111488 | 71.44 ± 0.26 | 5719 ± 99 | 4.50 ± 0.21 | -0.16 ± 0.13 | … | … | N | … | … | … | … | n | … |
| 40201 | 06295158-3110104 | 58.31 ± 1.10 | 5298 ± 112 | 3.63 ± 1.38 | -0.78 ± 0.25 | … | … | Y | … | … | … | … | n | … |

**Table C.20.** continued.

| ID | CNAME | RV (km s$^{-1}$) | $T_{\text{eff}}$ (K) | logg (dex) | [Fe/H] (dex) | EW(Li)$^a$ (mÅ) | EW(Li) error flag$^b$ | Membership RV | Li | logg | [Fe/H] | Gaia study Cantat-Gaudin$^c$ | Final$^d$ | NMs with Li$^e$ |
|---|---|---|---|---|---|---|---|---|---|---|---|---|---|---|
| 40202 | 06295158-3118142 | 61.25 ± 0.60 | 6410 ± 383 | 3.93 ± 0.72 | -0.77 ± 0.26 | … | … | N | … | … | … | Y | n | … |
| 40203 | 06295200-3114195 | 58.93 ± 0.13 | 6366 ± … | 4.49 ± … | -0.81 ± … | … | … | Y | … | … | … | Y | n | … |
| 40204 | 06295211-3114433 | 46.36 ± 0.80 | 5854 ± 209 | 4.21 ± 0.55 | -0.54 ± 0.32 | <104 | 3 | N | … | … | … | … | n | NG |
| 40205 | 06295215-3116527 | 4.37 ± 0.24 | 4877 ± 200 | 4.60 ± 0.25 | -0.50 ± 0.23 | … | … | N | … | … | … | … | n | … |
| 40206 | 06295289-3119220 | 59.49 ± 0.11 | 6126 ± … | 4.19 ± … | -0.62 ± … | 119 ± 42 | … | Y | N | Y | Y | Y | n | NG |
| 40207 | 06295290-3117471 | 57.25 ± 0.33 | 6384 ± 89 | 4.34 ± 0.17 | -0.66 ± 0.21 | … | … | N | … | … | … | … | n | … |
| 40208 | 06295310-3119338 | 58.06 ± 0.21 | 6089 ± 226 | 4.05 ± 0.13 | -0.59 ± 0.13 | 101 ± 19 | … | Y | N | N | Y | Y | n | NG |
| 40209 | 06295328-3117472 | 60.01 ± 0.48 | 6498 ± 367 | 4.57 ± 0.27 | -0.52 ± 0.17 | … | … | Y | … | … | … | Y | n | … |
| 40210 | 06295329-3121393 | 188.33 ± 0.10 | 5048 ± … | 3.52 ± … | -0.28 ± … | … | … | N | … | … | … | … | n | … |
| 40211 | 06295335-3117218 | 61.51 ± 1.07 | 5798 ± 559 | 3.92 ± 0.35 | -0.80 ± 0.82 | … | … | N | … | … | … | … | n | … |
| 40212 | 06295339-3115199 | 62.51 ± 0.90 | 6379 ± 282 | 3.91 ± 0.30 | -0.37 ± 0.21 | … | … | N | … | … | … | Y | n | … |
| 40213 | 06295362-3118557 | 59.73 ± 0.68 | 5683 ± 60 | 4.36 ± 0.34 | -0.64 ± 0.23 | <99 | 3 | Y | N | N | Y | Y | n | NG |
| 40214 | 06295363-3121374 | 57.42 ± 1.48 | 6026 ± 321 | 3.88 ± 0.65 | -0.80 ± 0.25 | … | … | N | … | … | … | Y | n | … |
| 40215 | 06295370-3127225 | 44.39 ± 0.32 | 6088 ± 336 | 4.59 ± 0.36 | 0.05 ± 0.28 | 69 ± 42 | 1 | N | … | … | … | … | n | NG |
| 40216 | 06295384-3115532 | 59.70 ± 0.50 | 6538 ± 198 | 4.08 ± 0.26 | -0.50 ± 0.21 | <55 | 3 | Y | Y | Y | Y | Y | Y | … |
| 40217 | 06295399-3120163 | 42.51 ± 0.28 | 6035 ± 170 | 4.34 ± 0.18 | -0.07 ± 0.13 | … | … | N | … | … | … | … | n | … |
| 40218 | 06295402-3115016 | 501.33 ± 0.40 | 6350 ± 172 | 4.95 ± 0.57 | -0.78 ± 0.17 | … | … | N | … | … | … | … | n | … |
| 40219 | 06295438-3114273 | 59.47 ± 2.86 | 5704 ± 164 | 4.61 ± 0.61 | -0.73 ± 0.20 | … | … | Y | … | … | … | … | n | … |
| 40220 | 06295442-3119045 | 59.57 ± 0.11 | 5368 ± … | 3.64 ± … | -0.49 ± … | … | … | Y | … | … | … | Y | n | … |
| 40221 | 06295455-3124484 | 150.62 ± 1.49 | 6300 ± 866 | 3.58 ± 1.56 | -1.10 ± 0.29 | … | … | N | … | … | … | … | n | … |
| 40222 | 06295459-3119321 | 59.37 ± 0.99 | 5106 ± 517 | 3.48 ± 0.36 | -1.02 ± 0.66 | … | … | Y | … | … | … | … | n | Li-rich G |
| 40223 | 06295465-3113489 | 81.83 ± 2.73 | 5713 ± 349 | 3.52 ± 1.32 | -1.34 ± 0.26 | … | … | N | … | … | … | … | n | … |
| 40224 | 06295469-3120266 | 59.01 ± 0.64 | 5808 ± 55 | 3.92 ± 0.40 | -0.61 ± 0.36 | … | … | Y | … | … | … | Y | n | … |
| 40225 | 06295488-3117129 | 59.33 ± 0.81 | 5264 ± 232 | 4.79 ± 0.48 | -0.79 ± 0.47 | … | … | Y | … | … | … | … | n | … |
| 40226 | 06295522-3121153 | 50.69 ± 0.25 | 6052 ± 194 | 4.39 ± 0.14 | 0.09 ± 0.13 | … | … | N | … | … | … | … | n | … |
| 40227 | 06295536-3126300 | 53.25 ± 0.11 | 5769 ± … | 3.76 ± … | -0.16 ± … | … | … | N | … | … | … | … | n | … |
| 40228 | 06295604-3118564 | 59.02 ± 0.85 | 6234 ± 481 | 4.67 ± 0.40 | -0.66 ± 0.30 | … | … | Y | … | … | … | Y | n | … |
| 40229 | 06295613-3113435 | 125.09 ± 0.54 | 5789 ± 48 | 4.33 ± 0.36 | -0.44 ± 0.13 | … | … | N | … | … | … | … | n | … |
| 40230 | 06295613-3118030 | 59.48 ± 1.35 | 5799 ± 379 | 3.93 ± 0.61 | -0.59 ± 0.48 | … | … | Y | … | … | … | … | n | … |
| 40231 | 06295620-3121142 | 60.65 ± 0.62 | 6170 ± 118 | 4.53 ± 0.41 | -0.33 ± 0.24 | 82 ± 72 | 1 | Y | N | Y | N | Y | n | NG |
| 40232 | 06295639-3108264 | 52.94 ± 0.11 | 5735 ± … | 3.48 ± … | -0.40 ± … | … | … | N | … | … | … | … | n | … |
| 40233 | 06295652-3115489 | 59.64 ± 0.80 | 5170 ± 364 | 3.85 ± 0.53 | -0.51 ± 0.29 | <70 | 3 | Y | N | N | Y | … | n | NG |
| 40234 | 06295654-3112341 | 60.28 ± 0.12 | 6493 ± … | 4.87 ± … | -0.58 ± … | … | … | Y | … | … | … | Y | n | … |
| 40235 | 06295661-3114230 | 59.86 ± 1.57 | 5301 ± 78 | 3.69 ± 0.53 | -0.64 ± 0.28 | … | … | Y | … | … | … | … | n | … |
| 40236 | 06295669-3121581 | 48.19 ± 0.70 | 5558 ± 179 | 3.57 ± 0.32 | -0.72 ± 0.32 | <129 | 3 | N | … | … | … | Y | n | NG |
| 40237 | 06295700-3114076 | 74.99 ± 0.69 | 5918 ± 95 | 4.53 ± 0.35 | -0.50 ± 0.20 | … | … | N | … | … | … | … | n | … |
| 40238 | 06295722-3111447 | 53.77 ± 0.83 | 5969 ± 239 | 4.49 ± 0.47 | -0.70 ± 0.35 | … | … | N | … | … | … | … | n | … |
| 40239 | 06295734-3118040 | 61.39 ± 1.00 | 5112 ± 355 | 4.63 ± 0.12 | -0.68 ± 0.48 | … | … | N | … | … | … | … | n | … |
| 40240 | 06295734-3124306 | 41.75 ± 0.10 | 5523 ± … | 4.40 ± … | -0.29 ± … | … | … | N | … | … | … | … | n | … |
| 40241 | 06295735-3116031 | 61.21 ± 0.56 | 6542 ± 453 | 4.34 ± 0.35 | -0.70 ± 0.32 | … | … | N | … | … | … | Y | n | … |
| 40242 | 06295743-3123095 | 58.06 ± 1.61 | 5591 ± 387 | 4.78 ± 0.29 | -0.74 ± 0.35 | … | … | Y | … | … | … | … | n | … |
| 40243 | 06295746-3111377 | 64.74 ± 1.38 | 5210 ± 638 | 4.99 ± 0.32 | -1.14 ± 1.03 | … | … | N | … | … | … | … | n | … |
| 40244 | 06295763-3111525 | 45.41 ± 0.98 | 4878 ± 859 | 3.04 ± 1.47 | -0.93 ± 0.28 | … | … | N | … | … | … | … | n | Li-rich G |
| 40245 | 06295768-3111125 | 57.93 ± 1.08 | 5824 ± 364 | 4.36 ± 0.44 | -0.69 ± 0.33 | … | … | N | … | … | … | … | n | … |
| 40246 | 06295768-3121395 | 59.05 ± 0.35 | 6556 ± 226 | 4.20 ± 0.06 | -0.50 ± 0.14 | … | … | Y | … | … | … | Y | n | … |
| 40247 | 06295788-3114386 | 60.06 ± 0.40 | 6434 ± 324 | 4.74 ± 0.22 | -0.62 ± 0.23 | … | … | Y | … | … | … | Y | n | … |
| 40248 | 06295807-3114354 | 20.97 ± 0.26 | 5916 ± 240 | 4.03 ± 0.14 | -0.28 ± 0.32 | … | … | N | … | … | … | … | n | … |
| 40249 | 06295818-3120414 | 57.73 ± 1.15 | 5338 ± 382 | 3.55 ± 0.89 | -0.56 ± 0.36 | … | … | N | … | … | … | … | n | … |
| 40250 | 06295827-3120314 | 41.85 ± 0.81 | 5652 ± 232 | 3.69 ± 0.69 | -0.37 ± 0.21 | … | … | N | … | … | … | … | n | … |
| 40251 | 06295830-3124458 | 98.35 ± 1.79 | … | … | … | … | … | N | … | … | … | … | n | … |
| 40252 | 06295839-3125236 | 58.62 ± 1.32 | 5712 ± 401 | 3.18 ± 1.28 | -0.73 ± 0.48 | … | … | Y | … | … | … | Y | n | … |
| 40253 | 06295865-3116081 | 318.57 ± 0.26 | 6434 ± 384 | 4.30 ± 0.23 | -0.41 ± 0.19 | <69 | 3 | N | … | … | … | Y | n | NG |
| 40254 | 06295868-3120051 | 59.17 ± 0.54 | 6318 ± 339 | 4.27 ± 0.07 | -0.59 ± 0.26 | … | … | Y | … | … | … | … | n | … |
| 40255 | 06295886-3110129 | 59.98 ± 0.14 | 6180 ± … | 3.67 ± … | -0.76 ± … | … | … | Y | … | … | … | Y | n | … |
| 40256 | 06295899-3119073 | 116.26 ± 0.74 | 5792 ± 152 | 4.44 ± 0.41 | -0.66 ± 0.37 | … | … | N | … | … | … | … | n | … |
| 40257 | 06295899-3125095 | 57.38 ± 2.22 | 5111 ± 184 | 3.08 ± 0.72 | -0.73 ± 0.18 | … | … | N | … | … | … | … | n | Li-rich G |
| 40258 | 06295926-3120012 | 55.65 ± 0.53 | 6139 ± 352 | 4.94 ± 0.50 | 0.16 ± 0.32 | … | … | N | … | … | … | … | n | … |
| 40259 | 06295933-3109220 | 63.65 ± 0.26 | 5318 ± 66 | 4.55 ± 0.32 | 0.01 ± 0.12 | … | … | N | … | … | … | … | n | … |
| 40260 | 06300010-3122222 | 56.98 ± 1.32 | 5863 ± 190 | 4.58 ± 0.18 | -0.60 ± 0.37 | … | … | N | … | … | … | N | n | … |

**Table C.20.** continued.

| ID | CNAME | RV (km s$^{-1}$) | $T_{\text{eff}}$ (K) | logg (dex) | [Fe/H] (dex) | EW(Li)$^a$ (mÅ) | EW(Li) error flag$^b$ | Membership RV | Li | logg | [Fe/H] | Gaia study Cantat-Gaudin$^c$ | Final$^d$ | NMs with Li$^e$ |
|---|---|---|---|---|---|---|---|---|---|---|---|---|---|---|
| 40261 | 06300032-3115551 | 72.26 ± 3.25 | 5062 ± 275 | 5.70 ± 0.56 | -1.21 ± 0.29 | … | … | N | … | … | … | … | n | … |
| 40262 | 06300040-3113197 | 59.97 ± 0.49 | 6369 ± 25 | 3.97 ± 0.06 | -0.71 ± 0.36 | … | … | Y | … | … | … | … | n | … |
| 40263 | 06300128-3118278 | 59.86 ± 0.15 | 6405 ± … | 4.43 ± … | -0.68 ± … | … | … | Y | … | … | … | … | n | … |
| 40264 | 06300129-3114535 | 61.57 ± 1.90 | 5189 ± 355 | 3.09 ± 0.86 | -0.26 ± 0.24 | … | … | N | … | … | … | … | n | Li-rich G |
| 40265 | 06300149-3117256 | 100.24 ± 0.40 | 6027 ± 171 | 3.84 ± 0.50 | -0.64 ± 0.34 | … | … | N | … | … | … | … | n | … |
| 40266 | 06300156-3122145 | 60.05 ± 0.70 | 6243 ± 346 | 4.28 ± 0.65 | -0.65 ± 0.30 | 51 ± 37 | 1 | Y | Y | Y | Y | Y | Y | … |
| 40267 | 06300194-3112256 | 99.95 ± 0.11 | 5878 ± … | 3.81 ± … | -0.42 ± … | … | … | N | … | … | … | … | n | … |
| 40268 | 06300234-3114249 | 76.37 ± 1.00 | 5957 ± 131 | 4.14 ± 0.30 | -1.02 ± 0.20 | … | … | N | … | … | … | … | n | … |
| 40269 | 06300257-3115594 | 47.16 ± 0.33 | 5985 ± 125 | 4.09 ± 0.17 | -0.36 ± 0.18 | 66 ± 26 | 1 | N | … | … | … | … | n | NG |
| 40270 | 06300356-3122506 | 59.71 ± 0.60 | 6442 ± 109 | 3.82 ± 0.25 | -0.48 ± 0.16 | … | … | Y | … | … | … | Y | n | … |
| 40271 | 06300387-3114117 | 59.67 ± 0.52 | 6194 ± 348 | 3.71 ± 0.52 | -0.68 ± 0.33 | 71 ± 49 | 1 | Y | Y | Y | Y | Y | Y | … |
| 40272 | 06300394-3119560 | 63.37 ± 0.59 | 6473 ± 211 | 4.35 ± 0.17 | -0.63 ± 0.20 | <78 | 3 | N | … | … | … | Y | n | NG |
| 40273 | 06300395-3116321 | 60.75 ± 1.43 | 5961 ± 372 | 3.47 ± 0.70 | -0.66 ± 0.19 | … | … | Y | … | … | … | … | n | … |
| 40274 | 06300401-3118197 | -2.05 ± 0.27 | 5379 ± 43 | 4.48 ± 0.40 | -0.13 ± 0.13 | … | … | N | … | … | … | … | n | … |
| 40275 | 06300416-3117553 | 59.78 ± 0.33 | 6413 ± 334 | 4.10 ± 0.08 | -0.57 ± 0.17 | … | … | Y | … | … | … | Y | n | … |
| 40276 | 06300458-3113454 | 80.38 ± 0.62 | 6105 ± 247 | 4.04 ± 1.00 | -0.57 ± 0.36 | … | … | N | … | … | … | … | n | … |
| 40277 | 06300465-3120584 | 59.19 ± 0.69 | 6012 ± 250 | 4.40 ± 0.29 | -0.42 ± 0.24 | … | … | Y | … | … | … | … | n | … |
| 40278 | 06300474-3109526 | 136.90 ± 1.77 | 5596 ± 84 | 2.63 ± 0.56 | -0.75 ± 0.11 | … | … | N | … | … | … | … | n | … |
| 40279 | 06300490-3116038 | 139.82 ± 1.08 | 4973 ± 453 | 4.26 ± 1.22 | -0.46 ± 0.39 | … | … | N | … | … | … | … | n | … |
| 40280 | 06300496-3122166 | 59.60 ± 1.00 | 6513 ± 749 | 4.23 ± 0.35 | -0.71 ± 0.07 | … | … | Y | … | … | … | Y | n | … |
| 40281 | 06300511-3118158 | 60.41 ± 0.52 | 6437 ± 140 | 4.36 ± 0.33 | -0.61 ± 0.20 | … | … | Y | … | … | … | Y | n | … |
| 40282 | 06300518-3113594 | 59.09 ± 0.38 | 6303 ± 130 | 4.41 ± 0.19 | -0.65 ± 0.20 | … | … | Y | … | … | … | … | n | … |
| 40283 | 06300534-3108170 | 17.45 ± 0.11 | 5720 ± … | 4.18 ± … | -0.51 ± … | … | … | N | … | … | … | … | n | … |
| 40284 | 06300741-3115425 | 95.84 ± 1.65 | 5844 ± 437 | 4.02 ± 0.27 | -0.57 ± 0.39 | <111 | 3 | N | … | … | … | … | n | NG |
| 40285 | 06300762-3119409 | 91.95 ± 1.78 | 6164 ± 177 | … | -1.81 ± 0.13 | … | … | N | … | … | … | … | n | … |
| 40286 | 06300775-3117395 | 59.57 ± 0.54 | 6453 ± 378 | 4.21 ± 0.22 | -0.52 ± 0.13 | … | … | Y | … | … | … | Y | n | … |
| 40287 | 06300786-3116076 | 48.24 ± 0.27 | 6056 ± 157 | 4.16 ± 0.09 | -0.58 ± 0.20 | 70 ± 27 | 1 | N | … | … | … | … | n | NG |
| 40288 | 06300789-3120212 | 145.55 ± 2.88 | 6639 ± 813 | 3.91 ± 0.70 | -0.46 ± 0.15 | … | … | N | … | … | … | … | n | … |
| 40289 | 06300827-3107135 | 135.71 ± 0.12 | … | … | … | … | … | N | … | … | … | … | n | … |
| 40290 | 06300834-3124531 | 42.75 ± 8.14 | … | … | … | … | … | N | … | … | … | … | n | … |
| 40291 | 06300931-3123526 | 75.28 ± 0.10 | 5634 ± … | 3.73 ± … | 0.01 ± … | … | … | N | … | … | … | … | n | … |
| 40292 | 06300933-3118470 | 86.18 ± 0.24 | 5423 ± 58 | 3.87 ± 0.08 | 0.14 ± 0.15 | … | … | N | … | … | … | … | n | … |
| 40293 | 06301059-3110152 | -1.57 ± 0.27 | 6005 ± 167 | 3.99 ± 0.14 | -0.13 ± 0.19 | 70 ± 28 | 1 | N | … | … | … | … | n | NG |
| 40294 | 06301081-3114484 | 14.39 ± 0.50 | 5649 ± 82 | 4.18 ± 0.63 | -0.35 ± 0.14 | … | … | N | … | … | … | … | n | … |
| 40295 | 06301144-3120541 | 57.03 ± 0.25 | 5029 ± 106 | 2.88 ± 0.33 | -0.19 ± 0.22 | … | … | N | … | … | … | … | n | … |
| 40296 | 06301145-3121392 | 59.95 ± 0.65 | 5997 ± 213 | 4.23 ± 0.10 | -0.45 ± 0.18 | <75 | 3 | Y | N | Y | Y | … | n | NG |
| 40297 | 06301152-3116556 | 7.50 ± 0.42 | 6132 ± 258 | 4.48 ± 0.34 | -0.56 ± 0.18 | … | … | N | … | … | … | … | n | … |
| 40298 | 06301274-3121243 | 93.40 ± 0.48 | 5342 ± 95 | 3.97 ± 0.12 | -0.07 ± 0.27 | … | … | N | … | … | … | … | n | … |
| 40299 | 06301447-3116514 | 33.46 ± 0.72 | 5737 ± 195 | 3.60 ± 0.54 | -0.08 ± 0.37 | 107 ± 102 | 1 | N | … | … | … | … | n | NG |
| 40300 | 06301454-3112159 | 62.14 ± 1.60 | 5370 ± 168 | 3.90 ± 0.64 | -0.93 ± 0.53 | … | … | N | … | … | … | … | n | … |
| 40301 | 06301486-3121190 | 72.69 ± 1.12 | 6248 ± 624 | 5.12 ± 0.24 | -0.92 ± 0.46 | … | … | N | … | … | … | … | n | … |
| 40302 | 06301507-3119573 | 60.04 ± 0.99 | 6383 ± 51 | 4.05 ± 0.63 | -0.44 ± 0.15 | <114 | 3 | Y | N | Y | Y | Y | n | NG |
| 40303 | 06301601-3121058 | 60.89 ± 2.29 | 6746 ± 534 | 4.90 ± 0.64 | -0.34 ± 0.11 | … | … | Y | … | … | … | Y | n | … |
| 40304 | 06301712-3118499 | 50.13 ± 0.61 | 5432 ± 116 | 4.90 ± 0.54 | -0.09 ± 0.15 | … | … | N | … | … | … | … | n | … |
| 40305 | 06301718-3116308 | 19.19 ± 0.28 | 5589 ± 316 | 4.47 ± 0.34 | 0.19 ± 0.12 | … | … | N | … | … | … | … | n | … |
| 40306 | 06301788-3114465 | 59.67 ± 1.26 | 6286 ± 986 | 4.63 ± 0.40 | -0.78 ± 0.81 | … | … | Y | … | … | … | … | n | … |
| 40307 | 06301835-3115073 | 59.07 ± 1.07 | 5750 ± 117 | 3.33 ± 0.40 | -1.06 ± 0.48 | … | … | Y | … | … | … | … | n | … |
| 40308 | 06301948-3121152 | 71.02 ± 0.97 | 6126 ± 290 | 3.22 ± 0.56 | 0.08 ± 0.44 | … | … | N | … | … | … | … | n | … |
| 40309 | 06301963-3115458 | 60.41 ± 0.96 | 6226 ± 274 | 3.17 ± 0.28 | -0.53 ± 0.20 | … | … | Y | … | … | … | Y | n | … |
| 40310 | 06301988-3114595 | 25.32 ± 1.08 | 6227 ± 382 | 3.94 ± 0.46 | -0.26 ± 0.32 | … | … | N | … | … | … | … | n | … |
| 40311 | 06302138-3118357 | 67.94 ± 1.40 | 6309 ± 941 | 2.98 ± 0.84 | -0.74 ± 0.88 | … | … | N | … | … | … | … | n | … |
| 40312 | 06302244-3115290 | 135.58 ± 1.52 | 5364 ± 462 | 4.55 ± 0.43 | -1.37 ± 2.03 | … | … | N | … | … | … | … | n | … |
| 40313 | 06302365-3117468 | 36.14 ± 1.22 | 5824 ± 242 | 3.27 ± 0.80 | -0.44 ± 0.20 | … | … | N | … | … | … | … | n | … |
| 40314 | 06302518-3116561 | 75.13 ± 1.69 | 5860 ± 321 | 3.06 ± 0.43 | -0.47 ± 0.16 | … | … | N | … | … | … | … | n | … |
| 40315 | 06302690-3115007 | 80.57 ± 2.06 | 5044 ± 329 | 1.45 ± 1.12 | -1.35 ± 0.31 | … | … | N | … | … | … | … | n | … |
| 40316 | 06292507-3117388 | 59.37 ± 0.16 | … | … | … | … | … | Y | … | … | … | Y | n | … |
| 40317 | 06292945-3116320 | 59.81 ± 0.15 | … | … | … | … | … | Y | … | … | … | Y | n | … |
| 40318 | 06293793-3116405 | 63.05 ± 0.19 | … | … | … | … | … | N | … | … | … | … | n | … |
| 40319 | 06293914-3112421 | 60.92 ± 0.15 | … | … | … | … | … | Y | … | … | … | Y | n | … |

**Table C.20.** continued.

| ID | CNAME | RV (km s$^{-1}$) | $T_{\text{eff}}$ (K) | logg (dex) | [Fe/H] (dex) | EW(Li)$^a$ (mÅ) | EW(Li) error flag$^b$ | Membership RV | Membership Li | Membership logg | Membership [Fe/H] | Gaia study Cantat-Gaudin$^c$ | Final$^d$ | NMs with Li$^e$ |
|---|---|---|---|---|---|---|---|---|---|---|---|---|---|---|
| 40320 | 06294026-3116221 | 60.45 ± 0.16 | … | … | … | … | … | Y | … | … | … | Y | n | … |
| 40321 | 06294770-3117520 | 59.27 ± 0.14 | … | … | … | … | … | Y | … | … | … | Y | n | … |
| 54398 | 06291452-3114042 | 526.19 ± 0.21 | … | … | … | … | … | N | … | … | … | … | n | … |
| 54399 | 06292307-3116161 | 75.17 ± 0.20 | … | … | … | … | … | N | … | … | … | … | n | … |
| 54400 | 06292430-3118536 | 500.23 ± 0.42 | … | … | … | … | … | N | … | … | … | Y | n | … |
| 54401 | 06292458-3120281 | 41.30 ± 0.22 | … | … | … | … | … | N | … | … | … | … | n | … |
| 54402 | 06292559-3115261 | 58.54 ± 0.21 | … | … | … | … | … | Y | … | … | … | … | n | … |
| 54403 | 06292642-3117482 | 534.73 ± 0.21 | … | … | … | … | … | N | … | … | … | Y | n | … |
| 54404 | 06292930-3121353 | 59.58 ± 0.20 | … | … | … | … | … | Y | … | … | … | Y | n | … |
| 54405 | 06292978-3115339 | 60.59 ± 0.34 | … | … | … | … | … | Y | … | … | … | … | n | … |
| 54406 | 06292998-3117009 | 58.86 ± 0.25 | … | … | … | … | … | Y | … | … | … | … | n | … |
| 54407 | 06293001-3121436 | 77.17 ± 0.20 | … | … | … | … | … | N | … | … | … | … | n | … |
| 54408 | 06293033-3121118 | -20.62 ± 0.25 | … | … | … | … | … | N | … | … | … | … | n | … |
| 54409 | 06293153-3116064 | 58.63 ± 0.27 | … | … | … | … | … | Y | … | … | … | … | n | … |
| 54410 | 06293236-3114287 | 51.86 ± 0.27 | … | … | … | … | … | N | … | … | … | … | n | … |
| 54411 | 06293255-3118264 | 318.56 ± 0.24 | … | … | … | … | … | N | … | … | … | Y | n | … |
| 54412 | 06293267-3115140 | 529.86 ± 0.20 | … | … | … | … | … | N | … | … | … | Y | n | … |
| 54413 | 06293414-3117236 | 58.28 ± 0.33 | … | … | … | … | … | Y | … | … | … | Y | n | … |
| 54414 | 06293463-3118231 | 500.96 ± 0.74 | … | … | … | … | … | N | … | … | … | Y | n | … |
| 54415 | 06293494-3115501 | 529.88 ± 0.20 | … | … | … | … | … | N | … | … | … | Y | n | … |
| 54416 | 06293521-3115159 | 319.83 ± 0.21 | … | … | … | … | … | N | … | … | … | Y | n | … |
| 54417 | 06293628-3116379 | 60.37 ± 0.28 | … | … | … | … | … | Y | … | … | … | Y | n | … |
| 54418 | 06293639-3119453 | 53.41 ± 0.40 | … | … | … | … | … | N | … | … | … | … | n | … |
| 54419 | 06293640-3114333 | 557.46 ± 0.88 | … | … | … | … | … | N | … | … | … | Y | n | … |
| 54420 | 06293863-3116472 | 58.54 ± 0.20 | … | … | … | … | … | Y | … | … | … | Y | n | … |
| 54421 | 06293870-3117026 | 61.37 ± 1.57 | … | … | … | … | … | N | … | … | … | Y | n | … |
| 54422 | 06293963-3115550 | 90.00 ± 0.36 | … | … | … | … | … | N | … | … | … | Y | n | … |
| 54423 | 06294134-3112318 | 12.45 ± 0.21 | … | … | … | … | … | N | … | … | … | … | n | … |
| 54424 | 06294335-3117314 | 58.74 ± 0.29 | … | … | … | … | … | Y | … | … | … | Y | n | … |
| 54425 | 06294394-3121077 | 8.41 ± 0.20 | … | … | … | … | … | N | … | … | … | … | n | … |
| 54426 | 06294491-3118191 | 555.05 ± 0.24 | … | … | … | … | … | N | … | … | … | Y | n | … |
| 54427 | 06294786-3119182 | 46.57 ± 0.47 | … | … | … | … | … | N | … | … | … | … | n | … |
| 54428 | 06294809-3114302 | 67.22 ± 0.36 | … | … | … | … | … | N | … | … | … | Y | n | … |
| 54429 | 06294938-3117510 | 59.72 ± 0.44 | … | … | … | … | … | Y | … | … | … | Y | n | … |
| 54430 | 06295470-3113177 | 34.72 ± 0.20 | … | … | … | … | … | N | … | … | … | … | n | … |
| 54431 | 06295582-3113517 | 57.85 ± 0.35 | … | … | … | … | … | N | … | … | … | Y | n | … |
| 54432 | 06295663-3121499 | -530.21 ± 0.72 | … | … | … | … | … | N | … | … | … | Y | n | … |
| 54433 | 06295702-3115162 | 527.23 ± 0.34 | … | … | … | … | … | N | … | … | … | … | n | … |
| 54434 | 06295743-3113002 | 14.82 ± 0.20 | … | … | … | … | … | N | … | … | … | … | n | … |
| 54435 | 06295771-3121405 | 542.19 ± 4.14 | … | … | … | … | … | N | … | … | … | … | n | … |

**Notes.** $^{(a)}$ The values of EW(Li) for this cluster are corrected (subtracted adjacent Fe (6707.43 Å) line). $^{(b)}$ Flags for the errors of the corrected EW(Li) values, as follows: 1=EW(Li) corrected by blends contribution using models; and 3=Upper limit (no error for EW(Li) is given). $^{(c)}$ Cantat-Gaudin et al. (2018). $^{(d)}$ The letters "Y" and "N" indicate if the star is a cluster member or not. $^{(e)}$ 'Li-rich G', 'G' and 'NG' indicate "Li-rich giant", "giant" and "non-giant" Li field outliers, respectively.